# Belle II Technical Design Report

Belle II

Edited by: Z. Doležal and S. Uno

KEK Report 2010-1

Belle II Secretariat
Belle group, Physics Division I,
KEK(High Energy Accelerator Research Organization)
Institute of Particle and Nuclear Studies
1-1 Oho, Tsukuba-shi, Ibaraki-ken, 305-0801, Japan
Phone : +81-298-64-5342
Fax : +81-298-64-5340

Edited by: Z. Doležal and S. Uno

T. Abe,[55] I. Adachi,[13] K. Adamczyk,[37] S. Ahn,[23] H. Aihara,[55] K. Akai,[13] M. Aloi,[30] L. Andricek,[29] K. Aoki,[13] Y. Arai,[13] A. Arefiev,[20] K. Arinstein,[3,40] Y. Arita,[31] D. M. Asner,[42] V. Aulchenko,[3,40] T. Aushev,[20] T. Aziz,[51] A. M. Bakich,[50] V. Balagura,[20] Y. Ban,[44] E. Barberio,[30] T. Barvich,[22] K. Belous,[18] T. Bergauer,[17] V. Bhardwaj,[43] B. Bhuyan,[14] S. Blyth,[35] A. Bondar,[3,40] G. Bonvicini,[59] A. Bozek,[37] M. Bračko,[28,21] J. Brodzicka,[37] O. Brovchenko,[22] T. E. Browder,[11] G. Cao,[16] M.-C. Chang,[6] P. Chang,[36] Y. Chao,[36] V. Chekelian,[29] A. Chen,[34] K.-F. Chen,[36] P. Chen,[36] B. G. Cheon,[10] C.-C. Chiang,[36] R. Chistov,[20] K. Cho,[23] S.-K. Choi,[9] K. Chung,[49] A. Comerma,[1] M. Cooney,[11] D. E. Cowley,[42] T. Critchlow,[42] J. Dalseno,[29,52] M. Danilov,[20] A. Dieguez,[1] A. Dierlamm,[22] M. Dillon,[30] J. Dingfelder,[2] R. Dolenec,[21] Z. Doležal,[4] Z. Drásal,[4] A. Drutskoy,[5] W. Dungel,[17] D. Dutta,[14] S. Eidelman,[3,40] A. Enomoto,[13] D. Epifanov,[3,40] S. Esen,[5] J. E. Fast,[42] M. Feindt,[22] M. Fernandez Garcia,[46] T. Fifield,[30] P. Fischer,[12] J. Flanagan,[13] S. Fourletov,[2] J. Fourletova,[2] L. Freixas,[1] A. Frey,[8] M. Friedl,[17] R. Frühwirth,[17] H. Fujii,[13] M. Fujikawa,[33] Y. Fukuma,[13] Y. Funakoshi,[13] K. Furukawa,[13] J. Fuster,[57] N. Gabyshev,[3,40] A. Gaspar de Valenzuela Cueto,[45] A. Garmash,[3,40] L. Garrido,[1] Ch. Geisler,[8] I. Gfall,[17] Y. M. Goh,[10] B. Golob,[26,21] I. Gorton,[42] R. Grzymkowski,[37] H. Guo,[48] H. Ha,[24] J. Haba,[13] K. Hara,[31] T. Hara,[13] T. Haruyama,[13] K. Hayasaka,[31] K. Hayashi,[13] H. Hayashii,[33] M. Heck,[22] S. Heindl,[22] C. Heller,[29] T. Hemperek,[2] T. Higuchi,[13] Y. Horii,[54] W.-S. Hou,[36] Y. B. Hsiung,[36] C.-H. Huang,[6] S. Hwang,[23] H. J. Hyun,[25] Y. Igarashi,[13] C. Iglesias,[47] Y. Iida,[13] T. Iijima,[31] M. Imamura,[31] K. Inami,[31] C. Irmler,[17] M. Ishizuka,[33] K. Itagaki,[54] R. Itoh,[13] M. Iwabuchi,[60] G. Iwai,[13] M. Iwai,[13] M. Iwasaki,[13] M. Iwasaki,[55] Y. Iwasaki,[13] T. Iwashita,[33] S. Iwata,[56] H. Jang,[23] X. Ji,[16] T. Jinno,[31] M. Jones,[11] T. Julius,[30] T. Kageyama,[13] D. H. Kah,[25] H. Kakuno,[55] T. Kamitani,[13] K. Kanazawa,[13] P. Kapusta,[37] S. U. Kataoka,[32] N. Katayama,[13] M. Kawai,[13] Y. Kawai,[13] T. Kawasaki,[38] J. Kennedy,[11] H. Kichimi,[13] M. Kikuchi,[13] C. Kiesling,[29] B. K. Kim,[23] G. N. Kim,[25] H. J. Kim,[25] H. O. Kim,[25] J.-B. Kim,[24] J. H. Kim,[23] M. J. Kim,[25] S. K. Kim,[49] K. T. Kim,[24] T. Y. Kim,[10] K. Kinoshita,[5] K. Kishi,[33] B. Kisielewski,[37] K. Kleese van Dam,[42] J. Knopf,[12] B. R. Ko,[24] M. Koch,[2] P. Kodyš,[4] C. Koffmane,[29] Y. Koga,[31] T. Kohriki,[13] S. Koike,[13] H. Koiso,[13] Y. Kondo,[13] S. Korpar,[28,21] R. T. Kouzes,[42] Ch. Kreidl,[12] M. Kreps,[22] P. Križan,[26,21] P. Krokovny,[3,40] H. Krüger,[2] A. Kruth,[2] W. Kuhn,[7] T. Kuhr,[22] R. Kumar,[43] T. Kumita,[56] S. Kupper,[21] A. Kuzmin,[3,40] P. Kvasnička,[4] Y.-J. Kwon,[60] C. Lacasta,[57] J. S. Lange,[7] I.-S. Lee,[10] M. J. Lee,[49] M. W. Lee,[25] S.-H. Lee,[24] M. Lemarenko,[2] J. Li,[11] W. D. Li,[16] Y. Li,[58] J. Libby,[15] A. Limosani,[30] C. Liu,[48] H. Liu,[16] Y. Liu,[36] Z. Liu,[16] D. Liventsev,[20] A. Lopez Virto,[46] Y. Makida,[13] Z. P. Mao,[16] C. Mariñas,[57] M. Masuzawa,[13] D. Matvienko,[3,40] W. Mitaroff,[17] K. Miyabayashi,[33] H. Miyata,[38] Y. Miyazaki,[31] T. Miyoshi,[13] R. Mizuk,[20] G. B. Mohanty,[51] D. Mohapatra,[58] A. Moll,[29,52] T. Mori,[31] A. Morita,[13] Y. Morita,[13] H.-G. Moser,[29] D. Moya Martin,[46] T. Müller,[22] D. Münchow,[7] J. Murakami,[33] S. S. Myung,[49] T. Nagamine,[54] I. Nakamura,[13] T. T. Nakamura,[13] E. Nakano,[41] H. Nakano,[54] M. Nakao,[13] H. Nakazawa,[34] S.-H. Nam,[23] Z. Natkaniec,[37] E. Nedelkovska,[29] K. Negishi,[54] S. Neubauer,[22] C. Ng,[55] J. Ninkovic,[29] S. Nishida,[13] K. Nishimura,[11] E. Novikov,[20] T. Nozaki,[13] S. Ogawa,[53] K. Ohmi,[13] Y. Ohnishi,[13] T. Ohshima,[31] N. Ohuchi,[13] K. Oide,[13] S. L. Olsen,[49,11] M. Ono,[13] Y. Ono,[54] Y. Onuki,[54] W. Ostrowicz,[37] H. Ozaki,[13] P. Pakhlov,[20] G. Pakhlova,[20] H. Palka,[37] H. Park,[25] H. K. Park,[25] L. S. Peak,[50] T. Peng,[48] I. Peric,[12] M. Pernicka,[17] R. Pestotnik,[21] M. Petrič,[21] L. E. Piilonen,[58] A. Poluektov,[3,40] M. Prim,[22] K. Prothmann,[29,52] K. Regimbal,[42] B. Reisert,[29] R. H. Richter,[29] J. Riera-Babures,[45] A. Ritter,[22] A. Ritter,[29] M. Ritter,[29] M. Röhrken,[22] J. Rorie,[11] M. Rosen,[11] M. Rozanska,[37] L. Ruckman,[11] S. Rummel,[27] V. Rusinov,[20] R. M. Russell,[42] S. Ryu,[49] H. Sahoo,[11] K. Sakai,[38] Y. Sakai,[13] L. Santelj,[21] T. Sasaki,[13] N. Sato,[13] Y. Sato,[54] J. Scheirich,[4] J. Schieck,[27] C. Schwanda,[17] A. J. Schwartz,[5] B. Schwenker,[8] A. Seljak,[21] K. Senyo,[31] O.-S. Seon,[60] M. E. Sevior,[30] M. Shapkin,[18] V. Shebalin,[3,40] C. P. Shen,[11] H. Shibuya,[53] S. Shiizuka,[31] J.-G. Shiu,[36] B. Shwartz,[3,40] F. Simon,[29,52] H. J. Simonis,[22] J. B. Singh,[43] R. Sinha,[19] M. Sitarz,[37] P. Smerkol,[21] A. Sokolov,[18] E. Solovieva,[20] S. Stanič,[39] M. Starič,[21] J. Stypula,[37] Y. Suetsugu,[13] S. Sugihara,[55] T. Sugimura,[13] K. Sumisawa,[13] T. Sumiyoshi,[56] K. Suzuki,[31] S. Y. Suzuki,[13] H. Takagaki,[56] F. Takasaki,[13] H. Takeichi,[31] Y. Takubo,[54] M. Tanaka,[13] S. Tanaka,[13] N. Taniguchi,[13] E. Tarkovsky,[20] G. Tatishvili,[42] M. Tawada,[13] G. N. Taylor,[30] Y. Teramoto,[41] I. Tikhomirov,[20] K. Trabelsi,[13] T. Tsuboyama,[13] K. Tsunada,[31] Y.-C. Tu,[6] T. Uchida,[13] S. Uehara,[13] K. Ueno,[36] T. Uglov,[20] Y. Unno,[10] S. Uno,[13] P. Urquijo,[30] Y. Ushiroda,[13] Y. Usov,[3,40] S. Vahsen,[11] M. Valentan,[17] P. Vanhoefer,[29] G. Varner,[11] K. E. Varvell,[50] P. Vazquez,[47] I. Vila,[46] E. Vilella,[1] A. Vinokurova,[3,40] J. Visniakov,[47] M. Vos,[57] C. H. Wang,[35] J. Wang,[44] M.-Z. Wang,[36] P. Wang,[16] A. Wassatch,[29] M. Watanabe,[38] Y. Watase,[13] T. Weiler,[22] N. Wermes,[2] R. E. Wescott,[42] E. White,[5] J. Wicht,[13] L. Widhalm,[17] K. M. Williams,[58] E. Won,[24] H. Xu,[16] B. D. Yabsley,[50] H. Yamamoto,[54] H. Yamaoka,[13] Y. Yamaoka,[13] M. Yamauchi,[13] Y. Yin,[48]

H. Yoon,[23] J. Yu,[23] C. Z. Yuan,[16] Y. Yusa,[58] D. Zander,[22] M. Zdybal,[37] Z. P. Zhang,[48] J. Zhao,[48] L. Zhao,[48] Z. Zhao,[48] V. Zhilich,[3,40] P. Zhou,[59] V. Zhulanov,[3,40] T. Zivko,[21] A. Zupanc,[22] and O. Zyukova[3,40]

[1]*Universidad de Barcelona, Barcelona (not a member of the Belle II collaboration)*
[2]*Physikalisches Institut der Universität Bonn, Bonn*
[3]*Budker Institute of Nuclear Physics, Novosibirsk*
[4]*Faculty of Mathematics and Physics, Charles University, Prague*
[5]*University of Cincinnati, Cincinnati, Ohio 45221*
[6]*Department of Physics, Fu Jen Catholic University, Taipei*
[7]*Justus-Liebig-Universität Gießen, Gießen*
[8]*Georg-August-Universität Göttingen, Göttingen*
[9]*Gyeongsang National University, Chinju*
[10]*Hanyang University, Seoul*
[11]*University of Hawaii, Honolulu, Hawaii 96822*
[12]*Institut für Technische Informatik der Universität Heidelberg*
[13]*High Energy Accelerator Research Organization (KEK), Tsukuba*
[14]*Indian Institute of Technology Guwahati, Guwahati*
[15]*Indian Institute of Technology Madras, Chennai*
[16]*Institute of High Energy Physics, Chinese Academy of Sciences, Beijing*
[17]*Institute of High Energy Physics, Vienna*
[18]*Institute of High Energy Physics, Protvino*
[19]*Institute of Mathematical Sciences, Chennai*
[20]*Institute for Theoretical and Experimental Physics, Moscow*
[21]*J. Stefan Institute, Ljubljana*
[22]*Institut für Experimentelle Kernphysik, Karlsruher Institut für Technologie, Karlsruhe*
[23]*Korea Institute of Science and Technology Information, Daejeon*
[24]*Korea University, Seoul*
[25]*Kyungpook National University, Taegu*
[26]*Faculty of Mathematics and Physics, University of Ljubljana, Ljubljana*
[27]*Ludwig Maximilians University and Excellence Cluster Universe, Munich*
[28]*University of Maribor, Maribor*
[29]*Max-Planck-Institut für Physik, Munich*
[30]*University of Melbourne, School of Physics, Victoria 3010*
[31]*Nagoya University, Nagoya*
[32]*Nara University of Education, Nara*
[33]*Nara Women's University, Nara*
[34]*National Central University, Chung-li*
[35]*National United University, Miao Li*
[36]*Department of Physics, National Taiwan University, Taipei*
[37]*H. Niewodniczanski Institute of Nuclear Physics, Krakow*
[38]*Niigata University, Niigata*
[39]*University of Nova Gorica, Nova Gorica*
[40]*Novosibirsk State University, Novosibirsk*
[41]*Osaka City University, Osaka*
[42]*Pacific Northwest National Laboratory, Richland, Washington 99352*
[43]*Panjab University, Chandigarh*
[44]*Peking University, Beijing*
[45]*Universitat Ramon Llull, Barcelona (not a member of the Belle II collaboration)*
[46]*Instituto de Fisica de Cantabria, Santander (not a member of the Belle II collaboration)*
[47]*Universidad de Santiago de Compostela, Santiago de Compostela (not a member of the Belle II collaboration)*
[48]*University of Science and Technology of China, Hefei*
[49]*Seoul National University, Seoul*
[50]*School of Physics, University of Sydney, NSW 2006*
[51]*Tata Institute of Fundamental Research, Mumbai*
[52]*Excellence Cluster Universe, Technische Universität Munich, Garching*
[53]*Toho University, Funabashi*
[54]*Tohoku University, Sendai*
[55]*Department of Physics, University of Tokyo, Tokyo*
[56]*Tokyo Metropolitan University, Tokyo*
[57]*IFIC, CSIC-UVEG, Valencia (not a member of the Belle II collaboration)*
[58]*IPNAS, Virginia Polytechnic Institute and State University, Blacksburg, Virginia 24061*
[59]*Wayne State University, Detroit, Michigan 48202*
[60]*Yonsei University, Seoul*

# Contents

# Chapter 1

# Motivation and Overview

## 1.1 Introduction

By November 2009, the Belle detector [1], operating at the asymmetric electron positron collider KEKB [2], had accumulated a data sample with an integrated luminosity of 1000 $\mathrm{fb}^{-1}$. This achievement completes all of the technical milestones put forward in 1999 for Belle and KEKB. In a decade of operation, measurements arising from KEKB and Belle have offered important insights into the flavor structure of elementary particles, especially the violation of $CP$ symmetry in the quark sector. Numerous results of the Belle collaboration, and its companion collaboration BaBar [3], operating at the PEP-II $B$-meson factory at SLAC, have confirmed with good precision the theoretical predictions of the Standard Model. These experimental tests culminated in the 2008 Nobel Prize for physics awarded to M. Kobayashi and T. Maskawa for their theory of $CP$ violation [4].

With the much larger data sample that will become available at SuperKEKB, a Super B factory level upgrade of the KEKB collider, a new panorama of measurements in heavy flavor physics will be possible. These studies will provide an important and unique source of information on the details of new physics processes that are expected to be uncovered at hadron colliders in the coming years. An upgrade of the Belle experiment, Belle II, was first presented in a detailed Letter of Intent in March 2004 [5], followed by a supplemental report in October 2008 [6]. After receiving strong encouragement from the KEK management and its review committees, the Belle II collaboration was formed in December 2008 and a formal management structure (described in Chapter 15) was implemented in July 2009.

This document gives a full account of the detector design details and illustrates the progress and changes since the 2008 report. The document is structured as follows. In the remainder of this overview, we will give a short description of the Belle II physics motivation and reach, and introduce the highlights of the Belle II detector. Chapter 2 discusses the SuperKEKB accelerator design. The remaining chapters describe the detector components in detail. Emphasis is placed on the detector construction, and on the changes in detector component performance, background levels and physics performance as compared to the Letter of Intent and its 2008 supplement. Finally, there are chapters devoted to the time schedule, and to the management structure of the experiment.

## 1.2 Physics Motivation

### 1.2.1 Moving beyond the Standard Model

The Standard Model (SM) of interactions among elementary particles is—at the current level of experimental precision and at the energies reached so far in man-made particle accelerators — one of the best verified physics theories. Despite its tremendous success in describing the basic forces of nature — recognized by several Nobel prizes — many fundamental questions remain unanswered within the SM.

It is unclear why there should be only three generations of elementary fermions. Considering only its known content, there is a prejudice that the SM has too many free parameters. The masses and mixing parameters of the SM bosons and fermions are apriori unknown and must therefore be determined experimentally. The origin of mass is resolved within the SM by spontaneous breaking of the electroweak symmetry, resulting in a yet-unobserved neutral Higgs boson. In order to preserve the unitarity of the theory its mass should lie at or below the TeV energy scale. This, however, requires an unnatural cancellation (fine tuning) among the parameters of the SM. Several solutions for this hierarchy problem are proposed, the best known being the existence of new supersymmetric particles with masses on the TeV scale, or the existence of unobserved extra spatial dimensions. In these and other New Physics (NP) scenarios, new generations, new particles and new processes arise quite naturally.

Studies of symmetries have often illuminated our understanding of nature. For example, there is an observed hierarchy in the fermion masses and some NP models have attempted to exploit new symmetries to explain this hierarchy. At the cosmological scale, there is a serious unresolved problem with the matter-antimatter asymmetry in the universe. While the violation of $CP$ symmetry is a necessary condition for the evolution of a matter-dominated universe, as was first noted by A. Sakharov in 1967 [7], the observed $CP$ violation within the quark sector that originates from the complex phase of the Cabibbo-Kobayashi-Maskawa (CKM) matrix is many orders of magnitude too small to explain the dominance of matter in the universe. Hence, there must exist undiscovered sources of the $CP$ asymmetry responsible for the matter-dominated Universe. As a final example, the elements of the CKM matrix, describing the fractions of charged $W^{\pm}$ boson decays to quarks of specific flavors, exhibit a roughly diagonal hierarchy, even though the SM does not require this. This may indicate the presence of a new mechanism, such as a flavor symmetry, that exists unbroken at a higher energy scale.

Considering the open questions that in the SM remain unanswered, it is fair to conclude that the present theory is an extremely successful *but phenomenological* description of subatomic processes at the energy scales up to $\mathcal{O}(0.1\ \text{TeV})$.

The future experiments in high energy physics are designed to address the above questions through searches of NP using complementary approaches. One approach is at the energy frontier with the main representatives being the ATLAS and the CMS experiments at the Large Hadron Collider (LHC) at CERN. The second is the rare/precision frontier, exemplified by the LHCb experiment at the LHC, the BESIII experiment at the BEPCII charm factory, the planned SuperB factory in Italy, and the Belle II experiment at SuperKEKB.

At the energy frontier, the LHC experiments will be able to discover new particles produced in proton-proton collisions at a center-of-mass energy of up to 14 TeV. Since the constituent gluons or quarks interact in these collisions, only a fraction of the center-of-mass energy is available to produce such new particles, and the mass reach is limited to $\mathcal{O}(1\ \text{TeV}/c^2)$. Sensitivity to the direct production of a specific new particle depends on the cross section and on the size of the data sample (i.e., on the luminosity of the accelerator and performance of the detector).

At the rare/precision frontier, observable signatures of new particles or processes can be obtained through measurements of flavor physics reactions at lower energies and evidence of a deviation from the SM prediction. (Here, "rare" and "precision" refer to processes that are strongly suppressed or allowed, respectively, in the SM.) An observed discrepancy can be interpreted in terms of NP models. This is the approach of Belle II.
Apart from being a complementary approach to the direct high energy searches, the precision frontier has unprecedented sensitivity to the effects of NP. The sensitivity depends on the strength of the flavor violating couplings of the NP. The mass reach for new particle/process effects can be as high as $\mathcal{O}(100\ \mathrm{TeV}/c^2)$ if the couplings are enhanced compared to the SM. In the most pessimistic Minimal Flavor Violation case, where the NP flavor violating processes (such as neutral meson oscillations) are a consequence of the same Yukawa couplings as in the SM, SuperKEKB and Belle II would still be able to observe the effects of so far unknown particles up to $\mathcal{O}(1\ \mathrm{TeV}/c^2)$ [8]. Again, sensitivity to the contribution of a new particle or process to a particular flavor physics reaction depends on the NP model and on the size of the data sample. The reach of various colliders in searching for NP is illustrated in Fig. 1.1.

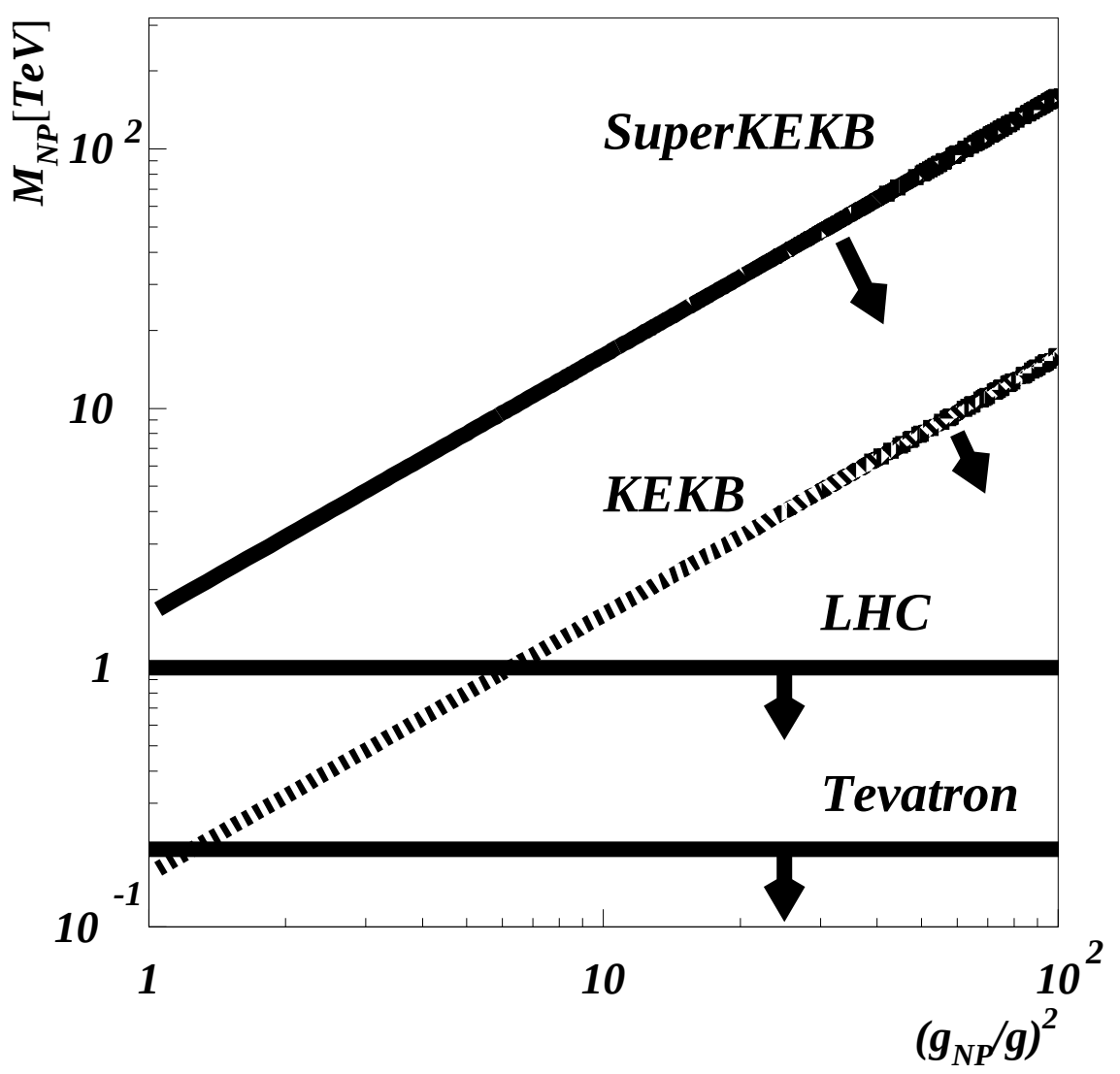

*Figure 1.1: Illustrative region of sensitivity to NP as a function of the flavor violating couplings (relative to the SM) in the indirect searches at KEKB and SuperKEKB, and direct searches at LHC and Tevatron.*

The value of the high-energy and rare/precision frontiers is associated with the complementary direct *vs.* indirect nature of the contribution of new particles or processes to the ensemble of available measurements and the distinct predictions from NP models in these two regimes. The processes in which unknown particles are expected to be observed are different in most of the cases between the energy and precision frontier experiments.
Belle II and SuperKEKB, described in this report, will exploit our strengths at the rare/precision frontier by moving beyond a simple observation of a NP effect to its detailed characterization through overconstraining measurements in several related flavor physics reactions. This is also the reason for the existence of several experiments in the precision frontier with, to a large extent, non-overlapping and thus complementary programs. In Sec. 1.2.3.2, we briefly address the complementarity between the LHCb and Belle II.

### 1.2.2 NP-sensitive physics at Belle

To illustrate the expected sensitivity of Belle II and SuperKEKB to New Physics phenomena, we examine some of the major scientific breakthroughs at the Belle experiment and then identify where Belle II can extend our reach.

- 2002: Belle observed mixing-induced time-dependent $CP$ violation (TCPV) in the system of neutral $B$ mesons and provided a measurement of one of the angles of the unitarity triangle (UT) $\sin 2\phi_1$ [9] through a careful analysis of $B^0 \to J/\psi K^0$ and related decays. Since then, the determination of this parameter has become a precision measurement (5.5% relative accuracy, which is further reduced by averaging measurements performed by different experiments in different decay modes).
- 2003: Belle presented evidence of TCPV in $B^0 \to \pi^+\pi^-$ decays and a measurement of the UT angle $\phi_2$ [10]. Nowadays, $\phi_2$ is also measured precisely (19% relative accuracy in the $B^0 \to \rho^+\rho^-$ decay mode, which is further reduced by averaging measurements performed by different experiments in different decay modes).
- In the same year, Belle measured TCPV in the penguin-dominated modes $B^0 \to \phi K_S$, $K^+K^-K_S$ and $\eta' K_S$ and observed that the parameter $\sin 2\phi_1$ deviated somewhat from the value measured in tree-dominated $B^0 \to J/\psi K^0$ decays [11]. The discrepancy has become less significant with more data, but is still statistically limited. The greater sensitivity of Belle II will permit us to discern the presence of NP effects in these penguin modes.
- In 2003 Belle discovered a new resonance named $X(3872)$ [12], produced in $B$ decays in a similar manner as charmonia, but exhibiting properties that made its classification as a conventional meson quite difficult. Many properties and decay modes of the $X(3872)$ state have since been measured, and it has been joined by a large number of other interesting new particles with similar puzzling properties. Studies of these states at Belle II, as well as possible discoveries of new ones, will help to extend our understanding of quantum chromodynamics.
- 2004: Data revealed for the first time direct $CP$ violation (DCPV) in $B$ decays to $\pi^+\pi^-$ and $K^+\pi^-$ [13, 14].
- In the same year, Belle collaborators performed the first measurement of the last of the unitarity triangle angles, $\phi_3$, using a new method: a time-independent measurement of the Dalitz distribution in $B \to D^{(*)}K$ decays [15]. This approach has proven to be the most sensitive method to determine $\phi_3$ and will be exploited by Belle II.
- 2005: The first measurements of TCPV in $B^0 \to K^{*0}\gamma$ were reported [16], where the constituent $b$ quark decays into an $s$ quark and a photon. The $CP$ asymmetry in this decay is strongly suppressed by fermion helicity conservation in the SM, but can be dramatically enhanced in some NP models. Measurements with improved precision will be an important tool in the search for new processes with Belle II.
- 2006: The purely leptonic decay $B^+ \to \tau^+\nu_\tau$, was observed for the first time [17]. The branching fraction is sensitive to the value of the least well known CKM matrix element, $V_{ub}$ and the wavefunction overlap of the B meson at the origin. In NP scenarios, it can be enhanced by the contribution of a charged Higgs boson.

- 2007: Belle and BaBar found the first evidence for the phenomenon of mixing in the system of neutral charmed mesons $D^0$ [18, 19]. The large rate can be accomodated by NP but may also be explained by difficult to calculate long-distance strong interaction effects in the SM. In contrast, CP violation in the D meson system with a magnitude comparable to the current experimental sensitivity is an unambiguous signature of NP. With the sample of charmed meson decays collected by Belle II it will be possible to clarify the situation.

- 2008: A measurement of the branching fraction and photon energy spectrum for fully inclusive $B \to X_s\gamma$ decays was performed with an energy threshold of 1.7 GeV [20]. This result is in agreement with the SM and provides strong constraints on charged Higgs bosons. The shape of the spectrum helps in the determination of the b quark mass and $|V_{ub}|$.

- In the same year, a measurement of DCPV in $B^+ \to K^+\pi^0$ proved it to be different than the same quantity in $B^0 \to K^+\pi^-$ decays [21], contrary to the naive SM expectation. In combination with other $B \to K\pi$ measurements and with the larger Belle II data set, the validity of the SM can be tested in a model-independent way.

- 2009: Rare $B \to K^*\ell^+\ell^-$ penguin decays were measured with improved precision [22]. The forward-backward asymmetry of the leptons is a sensitive probe of possible NP contributions. With current data samples, the asymmetry appears to deviate from the SM expectation, but the difference is statistically limited. The greater sensitivity of Belle II will resolve whether the discrepancy is due to NP effects or not.

### 1.2.3 NP-sensitive physics at Belle II

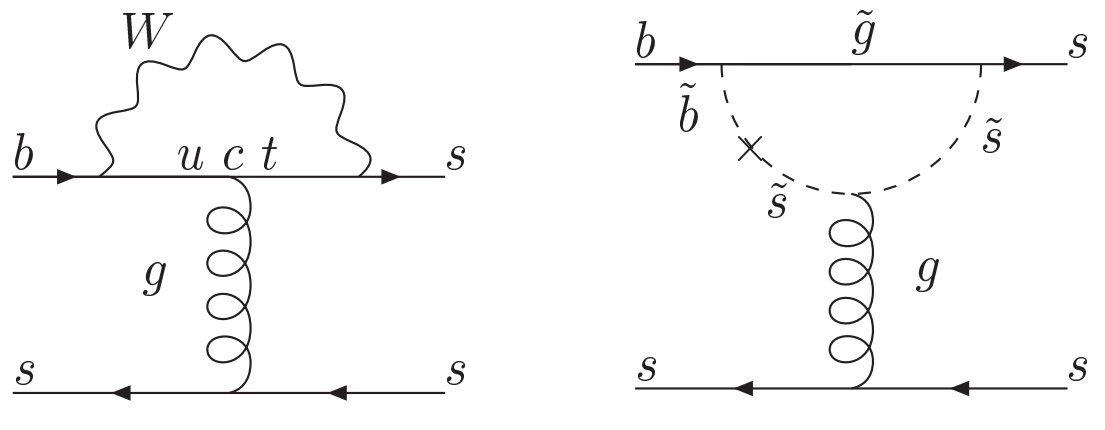

*Figure 1.2: The SM contribution (left) and the gluino–down squark contribution (right) to the $b \to s\bar{s}s$ transition.*

Here, in order to illustrate the unprecedented sensitivity to the presence of NP effects at the rare/precision frontier that the improved performance of the Belle II detector and the much larger data set will offer, we expand on few of the above examples of measurements: $b \to s\bar{s}s$, $b \to s\gamma$, $B \to \tau\nu$, and $B \to K\pi$ decays.
The examples presented are just a few examples in the broad physics program of Belle II, which brings the searches of New Physics performed at the existing $B$ factories to a completely new level of sensitivity, and significantly extends the reach of present experimental efforts in a way complementary to the energy frontier experiments. The real value of the Super $B$ factory is in its ability to perform measurements in all fields of heavy flavor physics, extending from $B^{0(\pm)}$ meson decays to $B_s^{(*)}$ meson decays, charm physics, $\tau$ lepton physics, spectroscopy, and pure electroweak measurements. A large number of planned measurements will over-constrain the parameter space of the SM as well as its extensions and will shed light on the nature of NP signatures that may be observed directly at the LHC. A detailed study of correlations among various observables, sensitivities and expected physics output can be found in Ref. [23]. Tables 1.1 and 1.2 summarize these studies.

#### 1.2.3.1 $b \to s\bar{s}s$ decays

The value of $\sin 2\phi_1$ as measured in $B^0 \to \phi K_S$ and similar $b \to s$ transitions differs slightly from the value measured in $B \to J/\psi K_S$ decays, the current world average difference being $\Delta S \equiv \sin 2\phi_1^{\phi K_S} - \sin 2\phi_1^{J/\psi K^0} = 0.22 \pm 0.17$ [25]. The former decays proceed through $b \to ss\bar{s}$ underlying quark process, possible only through the loop processes shown in Fig. 1.2 (left), and the latter through the $b \to cs\bar{c}$ tree diagram. While the CKM matrix elements included in the amplitudes of these decays are approximately real, the possibility of $B^0 - \bar{B}^0$ mixing before the decay introduces an additional factor $(V_{tb}^* V_{td})^2 \propto e^{-2i\phi_1}$. Hence, the decay time distribution of both decays is sensitive to $\sin 2\phi_1$, and the difference in the value measured in the two decays is expected to vanish within small corrections, $\Delta S = 0.03 \pm {}^{0.01}_{0.04}$ [26]. However, NP particles can contribute in the loop of $B^0 \to \phi K_S$, as illustrated in Fig. 1.2 (right), and change the expectation for $\Delta S$.
In $B \to K^+K^-K_S$ decays, $\phi K_S$ is one of several intermediate resonant contributions to the final state. In order to determine the value of $\sin 2\phi_1^{\phi K_S}$, one has to perform a decay time dependent Dalitz plot analysis, where the accuracy of $K^+K^-K_S$ vertex determination and the particle identification for the suppression of backgrounds are crucial. These are achieved in Belle II with the vertex detector (Chapters 4 and 5) and the particle identification system (Chapters 7 and 8). The expected dependence on integrated luminosity of the $\Delta S$ sensitivity from these and related decays is shown in Fig. 1.3 [23]. With $\mathcal{L} = 10$ ab$^{-1}$ of data, the experimental and theoretical uncertainties will be comparable.

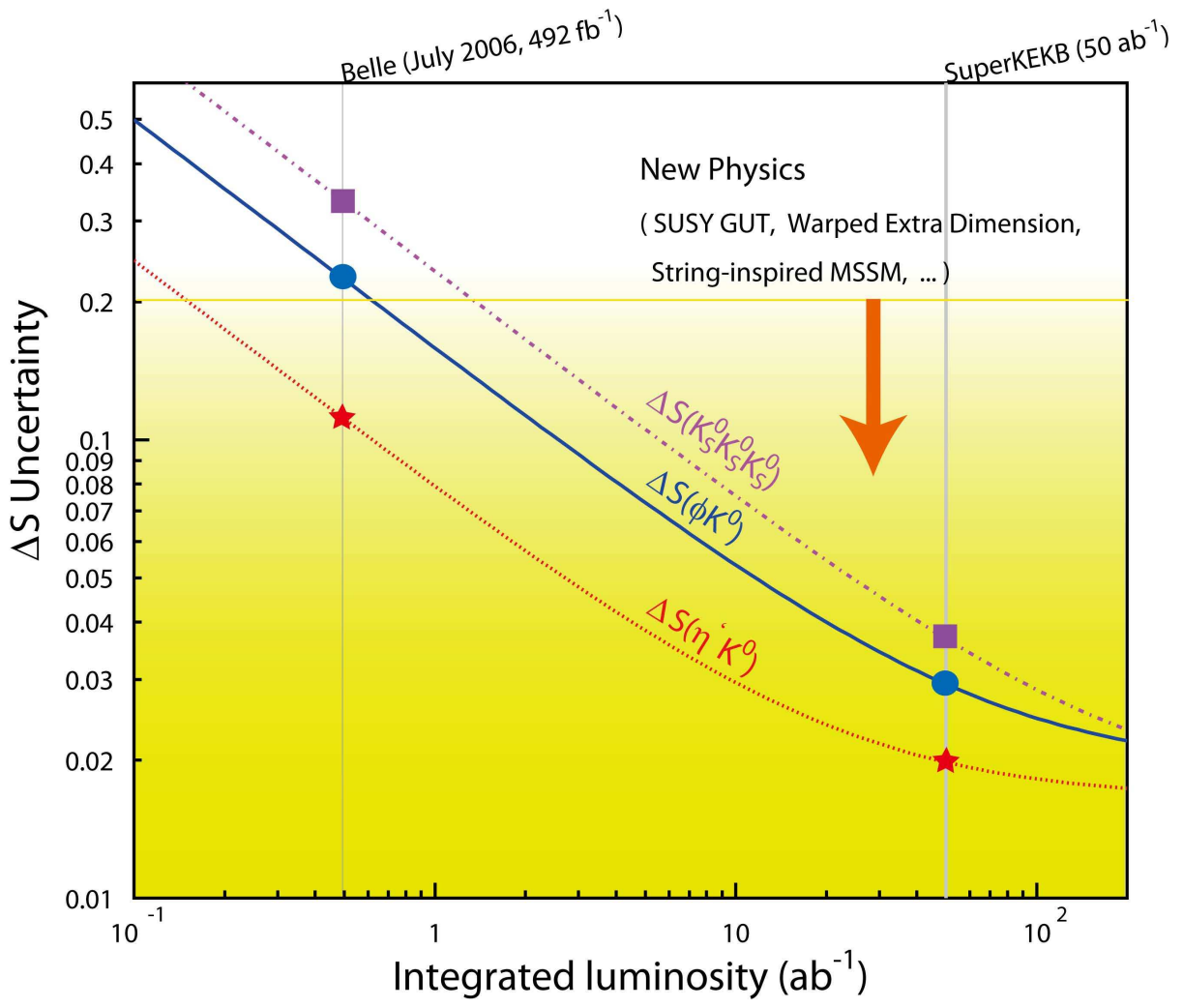

*Figure 1.3: Expected precision of $\Delta S$ measurements as a function of integrated luminosity [23].*

#### 1.2.3.2 $b \to s\gamma$ decays

Radiative decays $b \to s\gamma$ are sensitive probes of right-handed weak currents, which are absent by fiat in the SM. The helicity structure of the effective Hamiltonian that describes this loop process allows only for $b_R \to s_L\gamma_L$ and $b_L \to s_R\gamma_R$ decays, where the subscript denotes the handedness of the particle. The amplitude of the former (latter) process depends on the helicity flip and is proportional to $m_b$ ($m_s$). In mesons the $b_L \to s_R\gamma_R$ transition, for example, can proceed directly or via $\bar{B}^0 \to B^0$ mixing; the interference leads to a small time dependent $CP$ asymmetry that is proportional to $m_s/m_b$. In various NP models (e.g. Left-right symmetric models) the right-handed currents are not suppressed and can lead to a sizable $CP$ asymmetry. A prominent example of such radiative quark transitions is the decay $B^0 \to K_S\pi^0\gamma$. Within the SM, the decay time dependent $CP$ asymmetry in this decay is estimated to be $S \approx -2(m_s/m_b)\sin 2\phi_1 \approx -0.04$; some SM predictions allow for a value of $|S|$ up to 0.1 [27, 28]. On the other hand, in L-R symmetric models, the asymmetry can be as large as $S \approx 0.67\cos 2\phi_1 \approx 0.5$.
The decay-time dependence in $B^0 \to K_S\pi^0\gamma$ is measured through reconstruction of the $B$ meson decay vertex using only pions from $K_S \to \pi^+\pi^-$ decay that are constrained to the $e^+e^-$ interaction region profile. The Belle II vertex detector (Chapters 4 and 5) will improve the vertex position resolution and, more importantly, increase the reconstruction efficiency of $K_S$ decays with charged pion hits in the silicon detectors.
The expected accuracy of the $S(K_S\pi^0\gamma)$ measurements is shown in Fig. 1.4 [23]. With a data set corresponding to 50 ab$^{-1}$, the sensitivity of the measurement will reach the SM value and thus cover a range of NP predictions.

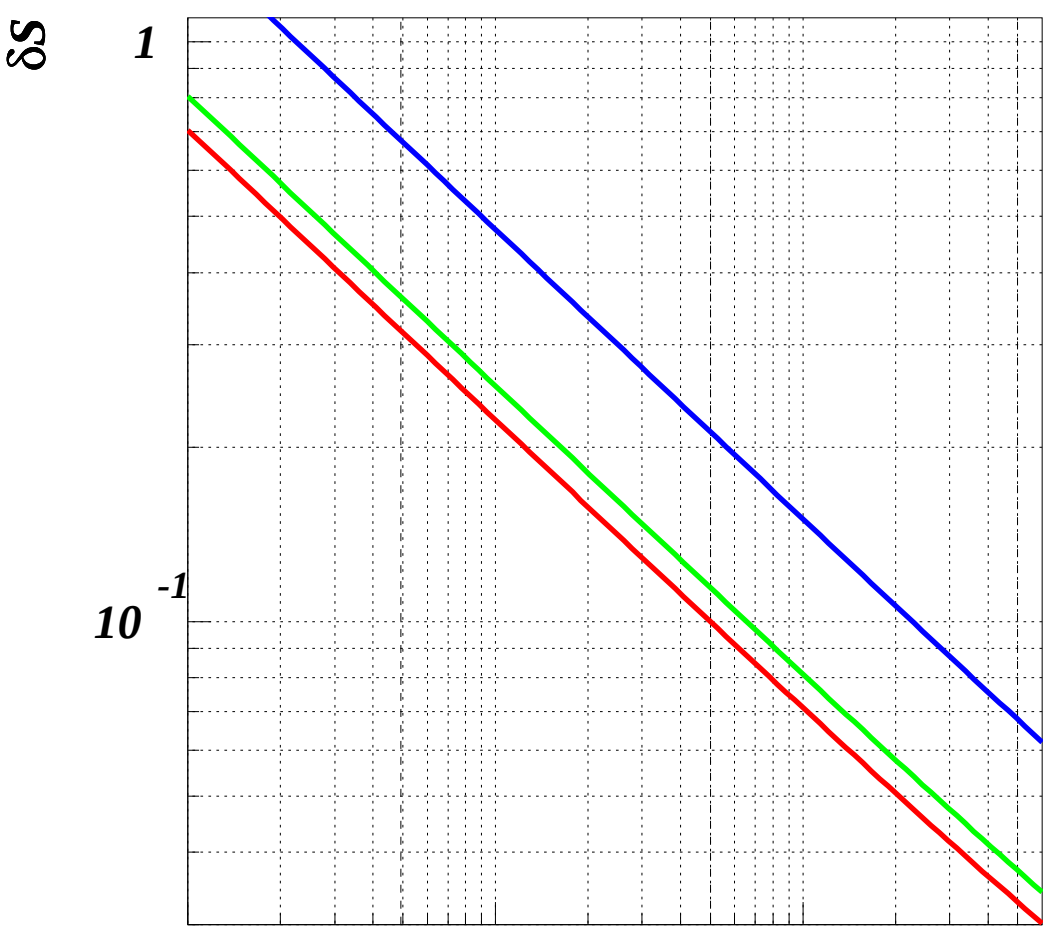

*Figure 1.4: Expected precision of the TCPV asymmetry measurements for $B \to K^{*0}(982)\gamma$ (green), other $K_S\pi^0\gamma$ (blue) and all $K_S\pi^0\gamma$ final state (red) as a function of integrated luminosity [23].*

The measurement of the $S(K_S\pi^0\gamma)$ also nicely illustrates the complementarity among the precision frontier experiments, specifically between Belle II and LHCb. The sensitivity of Belle II for the $S(K_S\pi^0\gamma)$ measurement cannot be reached by the LHC experiment. On the other hand (neglecting the possibility of dedicated longer data taking periods at the $\Upsilon(5S)$ resonance), the LHCb experiment can perform a much more precise measurement of the $B_s \to \mu^+\mu^-$ decay rate. When interpreting both measurements within the Minimal Supersymmetric Standard Model in the mass insertion approximation (MIA), the contours of the MIA parameter and $\tan\beta$ are shown in Fig. 1.5 [23]. While the measurement of the $\mathcal{B}(B_s \to \mu^+\mu^-)$ at the level of $10^{-9}$ yields a rather loose constraint on the MIA parameter, a combination with the measurement of the $S(K_S\pi^0\gamma)$ with a precision of 0.1 results in a tight constraint in both dimensions.

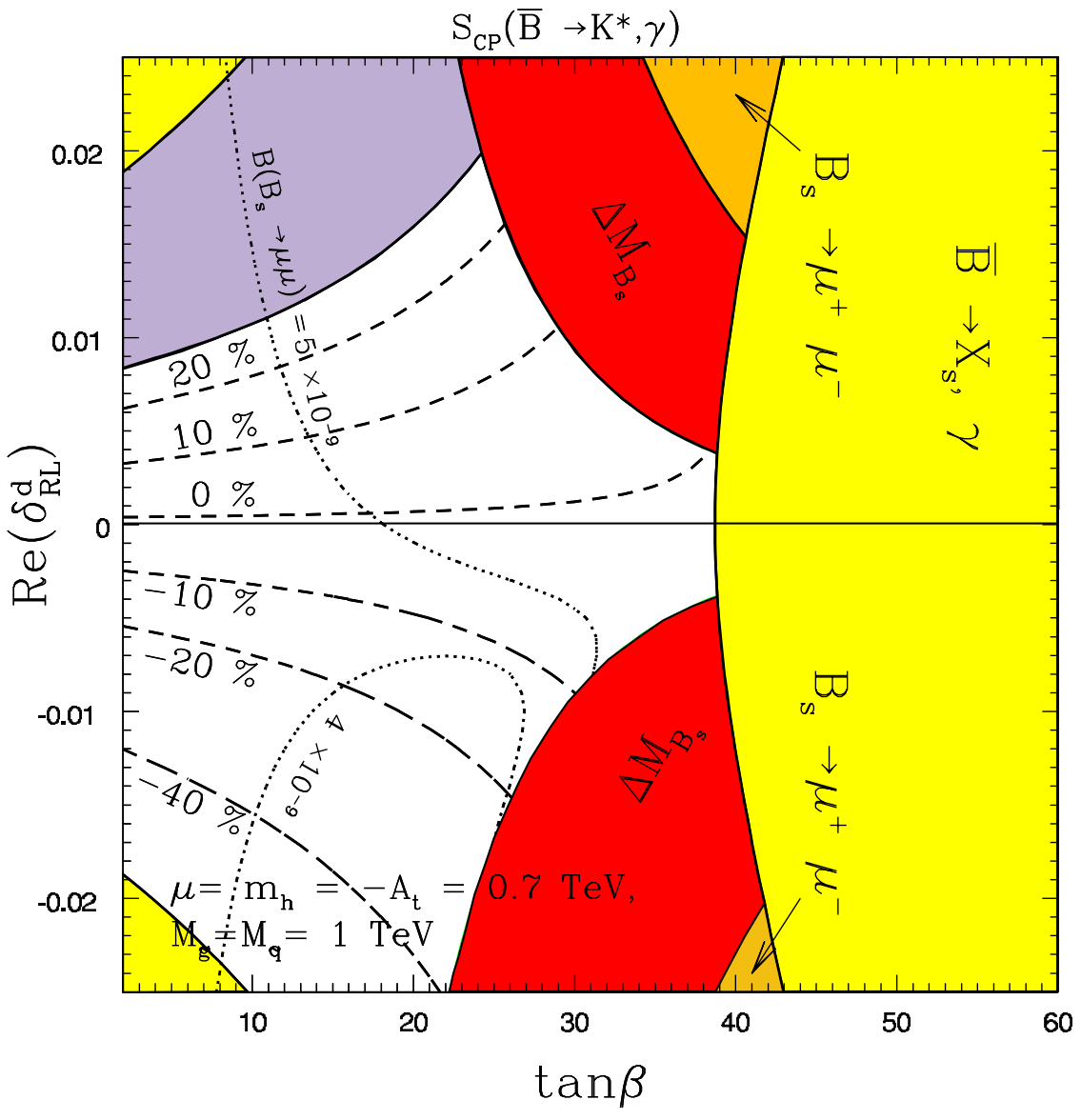

*Figure 1.5: Constraints on the MIA parameter Re($\delta^d_{RL}$) and $\tan\beta$ from the measurement of the $S(K_S\pi^0\gamma)$ and $\mathcal{B}(B_s \to \mu^+\mu^-)$ [23]. Contours of $S(K_S\pi^0\gamma)$ are shown in the intervals of 0.1.*

#### 1.2.3.3 $B \to \tau\nu$

One of the outstanding problems in particle physics is the question of the origin of masses and the related Higgs boson(s). The SM incorporates a single neutral Higgs boson. The Higgs sector of various extensions of the SM is richer, with charged Higgs bosons possible as well. In the Type II Two Higgs Doublet Models (2HDM-II), the charged Higgs boson $H^\pm$ behaves like the charged weak bosons $W^\pm$ apart from its couplings to fermions, which are proportional to their masses. The contribution of $H^\pm$ can thus be expected in all charged weak current processes, especially those involving heavy fermions. A typical example is the purely leptonic decay of charged $B$ mesons, $B^+ \to \tau^+\nu_\tau$, where, in the 2HDM-II models, the contribution of $H^+$ is expected to be largest due to the masses of the $\tau$ lepton and the $b$ quark.

The effect of a possible charged Higgs boson on the partial leptonic decay width of $B$ mesons is given by

$$\Gamma(B^+ \to \tau^+\nu_\tau) = \Gamma^{\mathrm{SM}}(B^+ \to \tau^+\nu_\tau)[1 - (m_B^2/m_H^2)\tan^2\beta]^2 \quad , \tag{1.1}$$

where $\Gamma^{\mathrm{SM}}(B^+ \to \tau^+\nu_\tau)$ denotes the SM partial decay width, and $\tan\beta$ denotes the ratio of the vacuum expectation values of the two Higgs fields and is a free parameter of the models. The leptonic decay width can thus be suppressed or — if the $H^\pm$ contribution is dominant — enhanced compared to the SM value.

Experimentally, of the leptonic branching fraction measurement [17] consists of (partial) reconstruction of the accompanying $B$ meson in the event, called the tagging $B$ meson ($B_{\mathrm{tag}}$). $B_{\mathrm{tag}}$ can be fully reconstructed in a number of hadronic decays (hadronic tagging) or partially reconstructed in semileptonic decays (semileptonic tagging), where the hadronic system (and the charged lepton) of the final state is detected while the neutrino escapes the detection. The hadronic tagging method has better purity in the $B_{\mathrm{tag}}$ sample, but suffers from a lower efficiency compared to semileptonic tagging. The remaining particles in the event are assigned to the signal $B$ meson ($B_{\mathrm{sig}}$); if they are consistent with a possible $\tau$ decay, the undetected part of the event consists of one or more neutrinos from (semi)leptonic decays. The signature of such event is thus a little or no residual energy detected in the electromagnetic calorimeter, after removing the contributions from the particles used in the reconstruction of $B_{\mathrm{tag}}$ and the $\tau$ from $B_{\mathrm{sig}} \to \tau\nu_\tau$. The resulting distribution of the residual calorimeter energy for the measurement based on the semileptonic tagging [29] is shown in Fig. 1.6 (left). The peaking component at low energy is the signal of $B \to \tau\nu_\tau$. The leptonic branching fraction is found to be $\mathcal{B}(B \to \tau\nu_\tau) = (1.65 \pm {}^{0.38}_{0.37} \pm {}^{0.35}_{0.37}) \times 10^{-4}$.

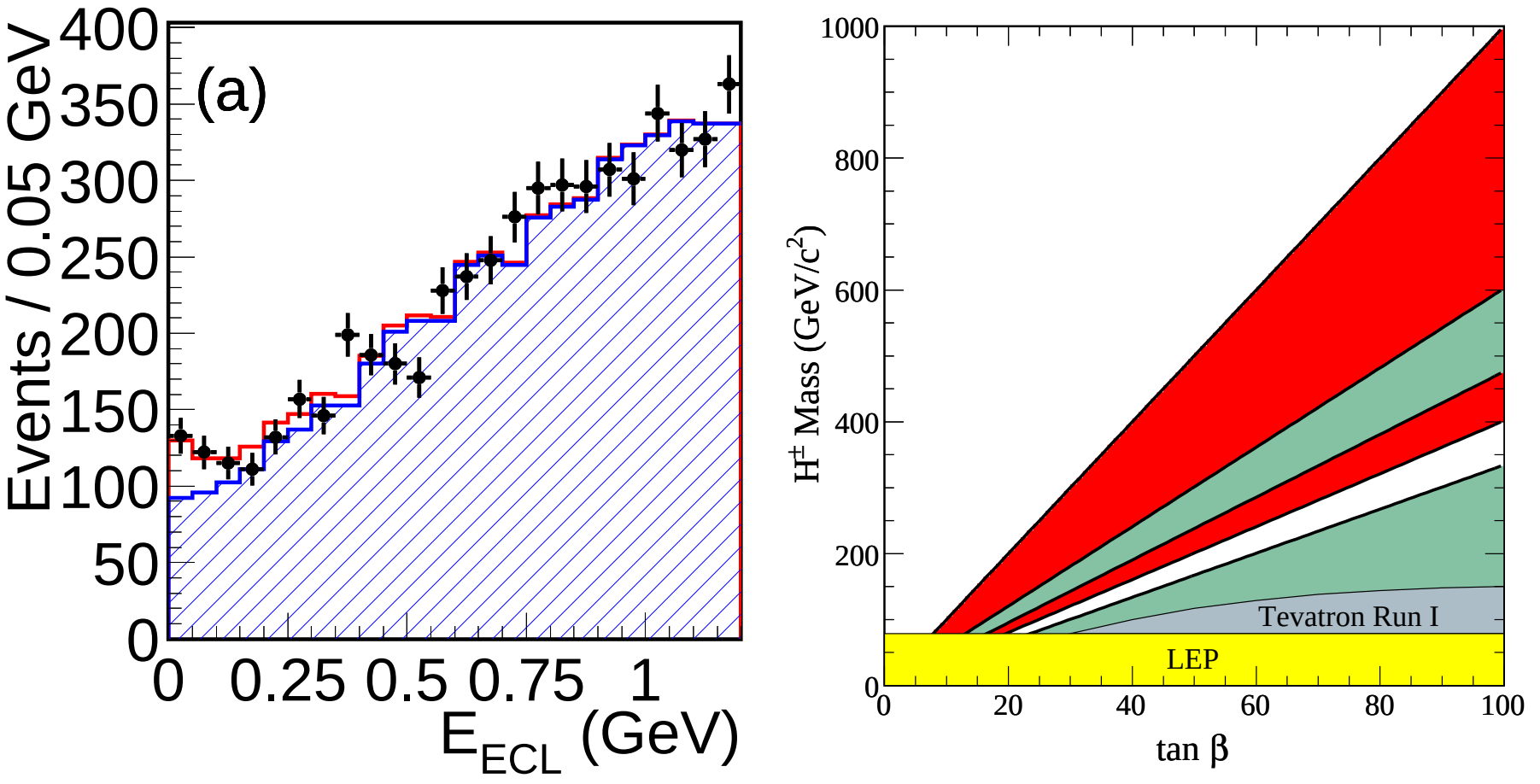

*Figure 1.6: Left: Distribution of residual energy in the calorimeter for semileptonically tagged $B \to \tau\nu_\tau$ candidate events [29]. The dashed histogram represents the background, and the solid red histogram the result of the fit that includes a signal component peaking at null value. Right: 5 $\sigma$ discovery sensitivity region for the charged Higgs boson from $B \to \tau\nu_\tau$ branching fraction measurement at an integrated luminosity of 5 ab$^{-1}$ (red). Other colored regions denote the presently excludedcurrent exclusion regions.*

Excellent performance of the electromagnetic calorimeter is crucial for the described measurement. The Belle II calorimeter (Chapter 9) maintains this performance even with the more severe backgrounds expected at SuperKEKB.
The statistical and systematic uncertainty of the result are almost equal in magnitude. However, the main sources of the systematic error—statistics of the control samples for the shape of the residual energy distribution and efficiency of the tagging—will decrease with the increased integrated luminosity of the available data sample. Assuming the theoretical uncertainties on the CKM element $|V_{ub}|$ and $B$ meson decay constant $f_B$ will be reduced to 5% in a few years, we obtain the five standard deviations discovery sensitivity region at $\mathcal{L} = 5$ ab$^{-1}$ in the $(\tan\beta, m_H)$ plane shown in Fig. 1.6 (right).

#### 1.2.3.4 $B \to K\pi$

Charmless 2-body $B$ meson decays are another example of rare SM processes in which the possible contribution of NP could be large enough to be observed in the future. The decays $B \to K\pi$ proceed through a tree diagram depicted in Fig. 1.7 (left) but are suppressed by the small CKM matrix element $|V_{ub}|$. Thus, the contribution of the loop penguin diagram is of similar magnitude. The interference of the two leads to a direct $CP$ asymmetry of $A^f_{CP} = [\Gamma(\bar{B} \to \bar{f}) - \Gamma(B \to f)]/[\Gamma(\bar{B} \to \bar{f}) + \Gamma(B \to f)]$. As suggested in diagrams of Fig. 1.7 (left), the main processes underlying the $B \to K\pi$ decays are the same and equal for neutral and charged $B$ mesons. Neglecting additional diagrams contributing to $B^+$ decays only (and expected to be much smaller than the shown contributions), the asymmetries $A^{K^+\pi^0}_{CP}$ in $B^\pm \to K^\pm\pi^0$ decays and $A^{K^+\pi^-}_{CP}$ in $B^0(\bar{B}^0) \to K^\pm\pi^\mp$ decays are expected to be the same. However, a precise measurement by Belle [21] showed a significant difference between the two, $\Delta A = A^{K^+\pi^0}_{CP} - A^{K^+\pi^-}_{CP} = 0.164 \pm 0.035 \pm 0.013$. The asymmetry in the number of reconstructed signal decays can be observed visually in Fig. 1.7 (right). The difference could be due to the neglected diagrams contributing to charged $B$ meson decays, for which the theoretical uncertainty is still rather large, or to some unknown NP effect that violates isospin. In Ref. [30], the author proposes a test of sum rule for NP free of theoretical uncertainties. The sum rule reads

$$A^{K^+\pi^-}_{CP} + A^{K^0\pi^+}_{CP} \frac{\mathcal{B}(B^+ \to K^0\pi^+)\tau_{B^0}}{\mathcal{B}(B^0 \to K^+\pi^-)\tau_{B^+}} = A^{K^+\pi^0}_{CP} \frac{2\,\mathcal{B}(B^+ \to K^+\pi^0)\tau_{B^0}}{\mathcal{B}(B^0 \to K^+\pi^-)\tau_{B^+}} + A^{K^0\pi^0}_{CP} \frac{2\,\mathcal{B}(B^0 \to K^0\pi^0)}{\mathcal{B}(B^0 \to K^+\pi^-)}\ , \tag{1.2}$$

where $\mathcal{B}(B \to f)$ denotes the corresponding branching fraction and $\tau_{B^0(B^+)}$ lifetimes of neutral and charged $B$ mesons.

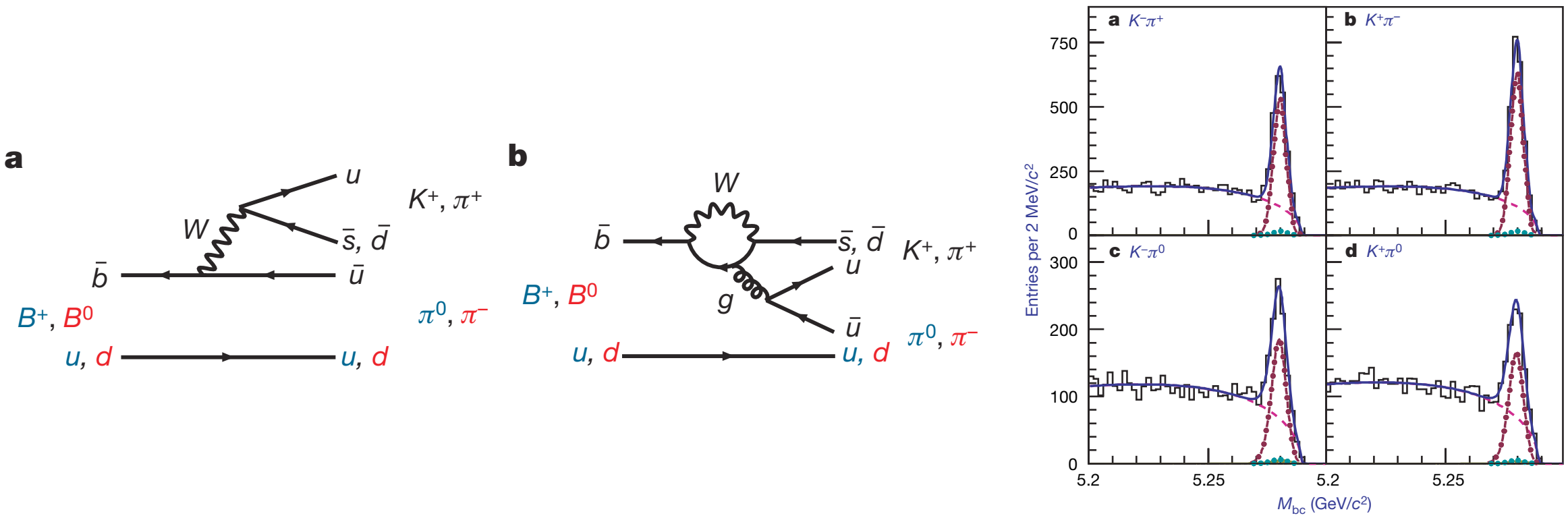

*Figure 1.7: Left: Tree and penguin diagrams contributing to $B \to K\pi$ decays [21]. Right: The reconstructed signal of $\bar{B}^0 \to K^-\pi^+$ (top left), $B^0 \to K^+\pi^-$ (top right), $B^- \to K^-\pi^0$ (bottom left) and $B^+ \to K^+\pi^0$ (bottom right) [21].*

By measuring all the observables in the above equation, one can test the prediction of the SM. Using the current world average values for the corresponding quantities [25], the isospin sum rule can be presented as a diagonal band in the plane of $A^{K^0\pi^0}_{CP}$ vs. $A^{K^0\pi^+}_{CP}$ (see Fig. 1.8 (left)). The slope of this dependence is determined by the precisely known branching fractions and lifetimes, and the uncertainty of the offset is mainly due to $A^{K^+\pi^0}_{CP}$ and to a much lesser extent to $A^{K^+\pi^-}_{CP}$.

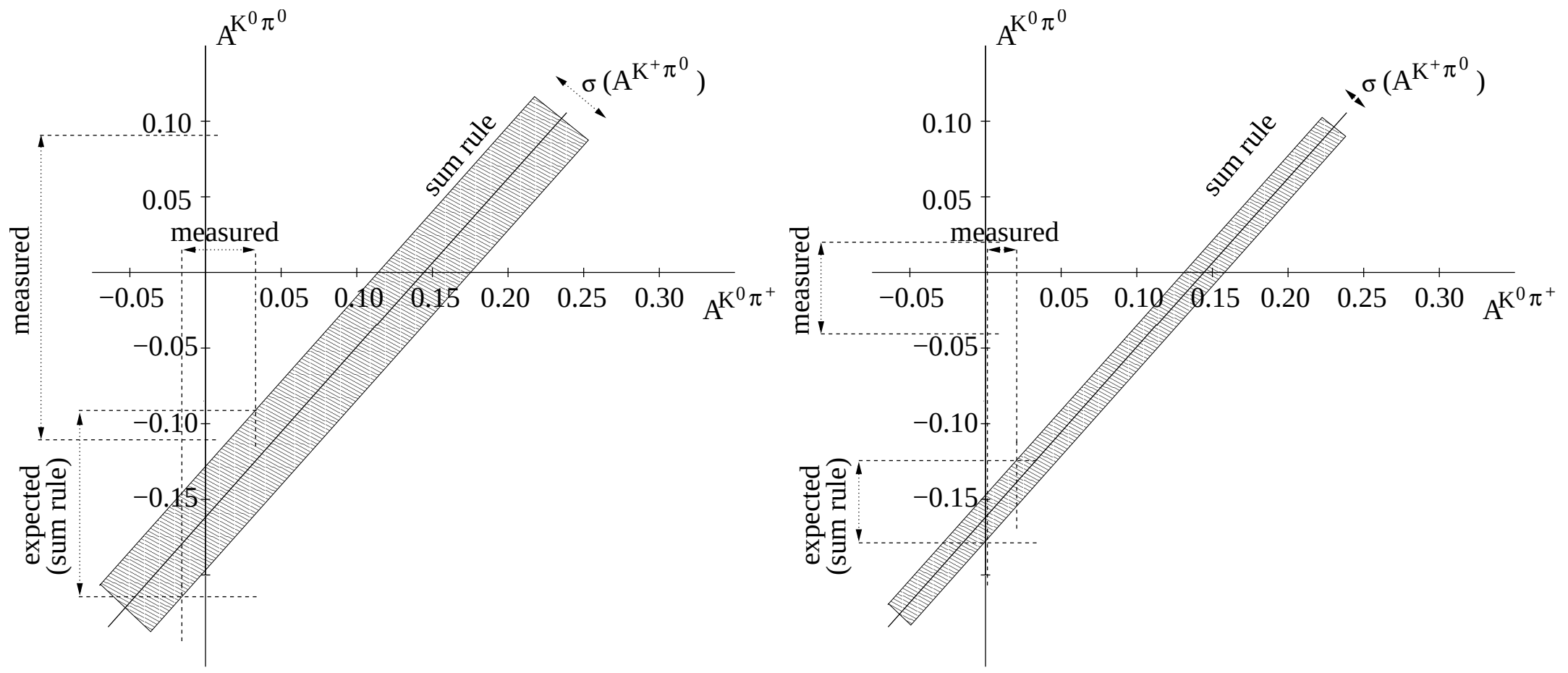

*Figure 1.8: Left: Sum rule from Eq. 1.2 [30] shown as $A_{CP}^{K^0\pi^0}$ dependence on $A_{CP}^{K^0\pi^+}$. The experimental values of various observables are taken from Ref. [25] and the hatched region represents the expectation from the sum rule. Right: The same sum rule with the current central values of observables and the accuracies expected with $\mathcal{L} = 50\,ab^{-1}$.*

Several measurements must be accomplished in order to perform the above test. The most demanding is the measurement of the all-neutral final state $K^0\pi^0$. In the decay-time dependent measurement, the vertex reconstruction is based on charged pions from the neutral kaon decays (as in $B^0 \to K_S\pi^0\gamma$ decays) and depends crucially on a vertex detector with a large radial acceptance. Reconstruction of the neutral pion ($\pi^0 \to \gamma\gamma$) requires very good electromagnetic calorimetry. For the final states with charged kaons and pions, an excellent separation between the two particle species must be provided by the particle identification system.
The main systematic uncertainty contributions (tag side interference) in the recent measurement of $B^0 \to K^0\pi^0$ [31] are expected to be reduced as the data sample increases. The expected sensitivity of the sum rule test with the integrated luminosity of 50 ab$^{-1}$ is illustrated in Fig. 1.8 (left). We have conservatively scaled only statistical uncertainties on $CP$ asymmetries of $K^0\pi^+$ and $K^+\pi^0$ final states. If one assumes the current central values of observables and the expected accuracy a discrepancy in the expectation of Eq. 1.2 can be clearly established.

### 1.2.4 Expected sensitivities for various observables

The summary of the expected sensitivities for individual observables at selected integrated luminosities is shown in Tables 1.1 and 1.2. A more detailed overview of physics program to be fulfilled at the SuperKEKB can be found in Ref. [23].

| Observable | Belle 2006 | Belle II/SuperKEKB | | LHCb† | |
|---|---|---|---|---|---|
| | ($\sim$0.5 ab$^{-1}$) | (5 ab$^{-1}$) | (50 ab$^{-1}$) | (2 fb$^{-1}$) | (10 fb$^{-1}$) |
| Hadronic $b \to s$ transitions | | | | | |
| $\Delta\mathcal{S}_{\phi K^0}$ | 0.22 | 0.073 | 0.029 | | 0.14 |
| $\Delta\mathcal{S}_{\eta' K^0}$ | 0.11 | 0.038 | 0.020 | | |
| $\Delta\mathcal{S}_{K_S^0 K_S^0 K_S^0}$ | 0.33 | 0.105 | 0.037 | - | - |
| $\Delta\mathcal{A}_{\pi^0 K_S^0}$ | 0.15 | 0.072 | 0.042 | - | - |
| $\mathcal{A}_{\phi\phi K^+}$ | 0.17 | 0.05 | 0.014 | | |
| $\phi_1^{eff}(\phi K_S)$ Dalitz | | 3.3° | 1.5° | | |
| Radiative/electroweak $b \to s$ transitions | | | | | |
| $\mathcal{S}_{K_S^0 \pi^0 \gamma}$ | 0.32 | 0.10 | 0.03 | - | - |
| $\mathcal{B}(B \to X_s\gamma)$ | 13% | 7% | 6% | - | - |
| $A_{CP}(B \to X_s\gamma)$ | 0.058 | 0.01 | 0.005 | - | - |
| $C_9$ from $A_{FB}(B \to K^*\ell^+\ell^-)$ | - | 11% | 4% | | |
| $C_{10}$ from $A_{FB}(B \to K^*\ell^+\ell^-)$ | - | 13% | 4% | | |
| $C_7/C_9$ from $A_{FB}(B \to K^*\ell^+\ell^-)$ | - | | 5% | | 7% |
| $R_K$ | | 0.07 | 0.02 | | 0.043 |
| $\mathcal{B}(B^+ \to K^+\nu\nu)$ | †† $< 3\ \mathcal{B}_{\rm SM}$ | | 30% | - | - |
| $\mathcal{B}(B^0 \to K^{*0}\nu\bar{\nu})$ | †† $< 40\ \mathcal{B}_{\rm SM}$ | | 35% | - | - |
| Radiative/electroweak $b \to d$ transitions | | | | | |
| $\mathcal{S}_{\rho\gamma}$ | - | 0.3 | 0.15 | | |
| $\mathcal{B}(B \to X_d\gamma)$ | - | 24% (syst.) | | - | - |
| Leptonic/semileptonic $B$ decays | | | | | |
| $\mathcal{B}(B^+ \to \tau^+\nu)$ | $3.5\sigma$ | 10% | 3% | - | - |
| $\mathcal{B}(B^+ \to \mu^+\nu)$ | †† $< 2.4\mathcal{B}_{\rm SM}$ | 4.3 ab$^{-1}$ for $5\sigma$ discovery | | - | - |
| $\mathcal{B}(B^+ \to D\tau\nu)$ | - | 8% | 3% | - | - |
| $\mathcal{B}(B^0 \to D\tau\nu)$ | - | 30% | 10% | - | - |
| LFV in $\tau$ decays (U.L. at 90% C.L.) | | | | | |
| $\mathcal{B}(\tau \to \mu\gamma)$ [$10^{-9}$] | 45 | 10 | 5 | - | - |
| $\mathcal{B}(\tau \to \mu\eta)$ [$10^{-9}$] | 65 | 5 | 2 | - | - |
| $\mathcal{B}(\tau \to \mu\mu\mu)$ [$10^{-9}$] | 21 | 3 | 1 | - | - |
| Unitarity triangle parameters | | | | | |
| $\sin 2\phi_1$ | 0.026 | 0.016 | 0.012 | $\sim$0.02 | $\sim$0.01 |
| $\phi_2$ ($\pi\pi$) | 11° | 10° | 3° | - | - |
| $\phi_2$ ($\rho\pi$) | $68° < \phi_2 < 95°$ | 3° | 1.5° | 10° | 4.5° |
| $\phi_2$ ($\rho\rho$) | $62° < \phi_2 < 107°$ | 3° | 1.5° | - | - |
| $\phi_2$ (combined) | | 2° | $\lesssim 1°$ | 10° | 4.5° |
| $\phi_3$ ($D^{(*)}K^{(*)}$) (Dalitz mod. ind.) | 20° | 7° | 2° | 8° | |
| $\phi_3$ ($DK^{(*)}$) (ADS+GLW) | - | 16° | 5° | 5-15° | |
| $\phi_3$ ($D^{(*)}\pi$) | - | 18° | 6° | | |
| $\phi_3$ (combined) | | 6° | 1.5° | 4.2° | 2.4° |
| $\|V_{ub}\|$ (inclusive) | 6% | 5% | 3% | - | - |
| $\|V_{ub}\|$ (exclusive) | 15% | 12% (LQCD) | 5% (LQCD) | - | - |
| $\bar{\rho}$ | 20.0% | | 3.4% | | |
| $\bar{\eta}$ | 15.7% | | 1.7% | | |

*Table 1.1: Summary of sensitivity studies (I) [23]. Branching fraction limits in the table are at the 90% confidence level. † Values for LHCb are statistical only and are taken from Ref. [24] unless otherwise stated. †† $\mathcal{B}_{SM}$ represents the expected branching fraction in the SM; $\mathcal{B}_{SM} = 5 \times 10^{-6}$ for $\mathcal{B}(B^+ \to K^+\nu\nu)$, $7 \times 10^{-6}$ for $B^0 \to K^{*0}\nu\bar{\nu}$ and $7.07 \times 10^{-7}$ for $\mathcal{B}(B^+ \to \mu\nu)$ are used in this table.*

| Observable | Belle | Belle II/SuperKEKB | | LHCb† | |
|---|---|---|---|---|---|
| | | | | ($2\ \text{fb}^{-1}$) | ($10\ \text{fb}^{-1}$) |
| $B_s$ physics | ($25\ \text{fb}^{-1}$) | ($5\ \text{ab}^{-1}$) | | | |
| $\mathcal{B}(B_s \to \gamma\gamma)$ | $< 8.7 \times 10^{-6}$ | $0.25 \times 10^{-6}$ | | - | - |
| $\Delta\Gamma_s^{CP}/\Gamma_s$ ($Br(B_s \to D_s^{(*)} D_s^{(*)})$) | 3% | 1% (model dependency) | | - | - |
| $\Delta\Gamma_s/\Gamma_s$ ($B_s \to f_{CP}$ t-dependent) | - | 1.2% | | - | - |
| $\phi_s$ (with $B_s \to J/\psi\phi$ etc.) | - | - | - | 0.02 | 0.01 |
| $\mathcal{B}(B_s \to \mu^+\mu^-)$ | - | | | $6\ \text{fb}^{-1}$ for $5\sigma$ discovery | |
| $\phi_3$ ($B_s \to KK$) | - | | | 7-10° | |
| $\phi_3$ ($B_s \to D_s K$) | - | | | 13° | |
| $\Upsilon$ decays | ($3\ \text{fb}^{-1}$) | ($500\ \text{fb}^{-1}$) | | | |
| $\mathcal{B}(\Upsilon(1S) \to$ invisible) | $< 2.5 \times 10^{-3}$ | $< 2 \times 10^{-4}$ | | | |
| | ($\sim 0.5\ \text{ab}^{-1}$)‡ | ($5\ \text{ab}^{-1}$) | ($50\ \text{ab}^{-1}$) | | |
| Charm physics | | | | | |
| $D$ mixing parameters | | | | | |
| $x$ | 0.25% | 0.12% | 0.09% | | 0.25%†† |
| $y$ | 0.16% | 0.10% | 0.05% | | 0.05%†† |
| $\delta_{K\pi}$ | 10° | 6° | 4° | | |
| $\lvert q/p \rvert$ | 0.16 | 0.1 | 0.05 | | |
| $\phi$ | 0.13 rad | 0.08 rad | 0.05 rad | | |
| $A_D$ | 2.4% | 1% | 0.3% | | |
| New particles$^\aleph$ | | | | | |
| $\gamma\gamma \to Z(3930) \to D\bar{D}^*$ | | $> 3\sigma$ | | | |
| $B \to KX(3872)(\to D^0\bar{D}^{*0})$ | | 400 events | | | |
| $B \to KX(3872)(\to J/\psi\pi^+\pi^-)$ | | 1250 events | | | |
| $B \to KZ^+(4430)(\to \psi'\pi^+)$ | | 1000 events | | | |
| $e^+e^- \to \gamma_{\rm ISR}Y(4260)(\to J/\psi\pi^+\pi^-)$ | | 3000 events | | | |
| Electroweak parameters | | ($\sim 10\ \text{ab}^{-1}$) | | | |
| $\sin^2\Theta_W$ | - | $3 \times 10^{-4}$ | | | |

*Table 1.2: Summary of sensitivity studies (II) [23]. †Values for LHCb are taken from Ref. [24] unless otherwise stated. ‡For $D$ mixing parameters the world average of results from [25] is quoted. ††LHCb sensitivities on $x$ and $y$ are estimated from sensitivities on $x'^2$, $y'$ and $y_{CP}$ [32] assuming $x = y = 0.80\%$ and $\delta_{K\pi} = 25° \pm 6°$. $^\aleph$Due to a large number of various possible measurements we only list expected signal yields for a few interesting processes.*

## 1.3 The Belle II overview

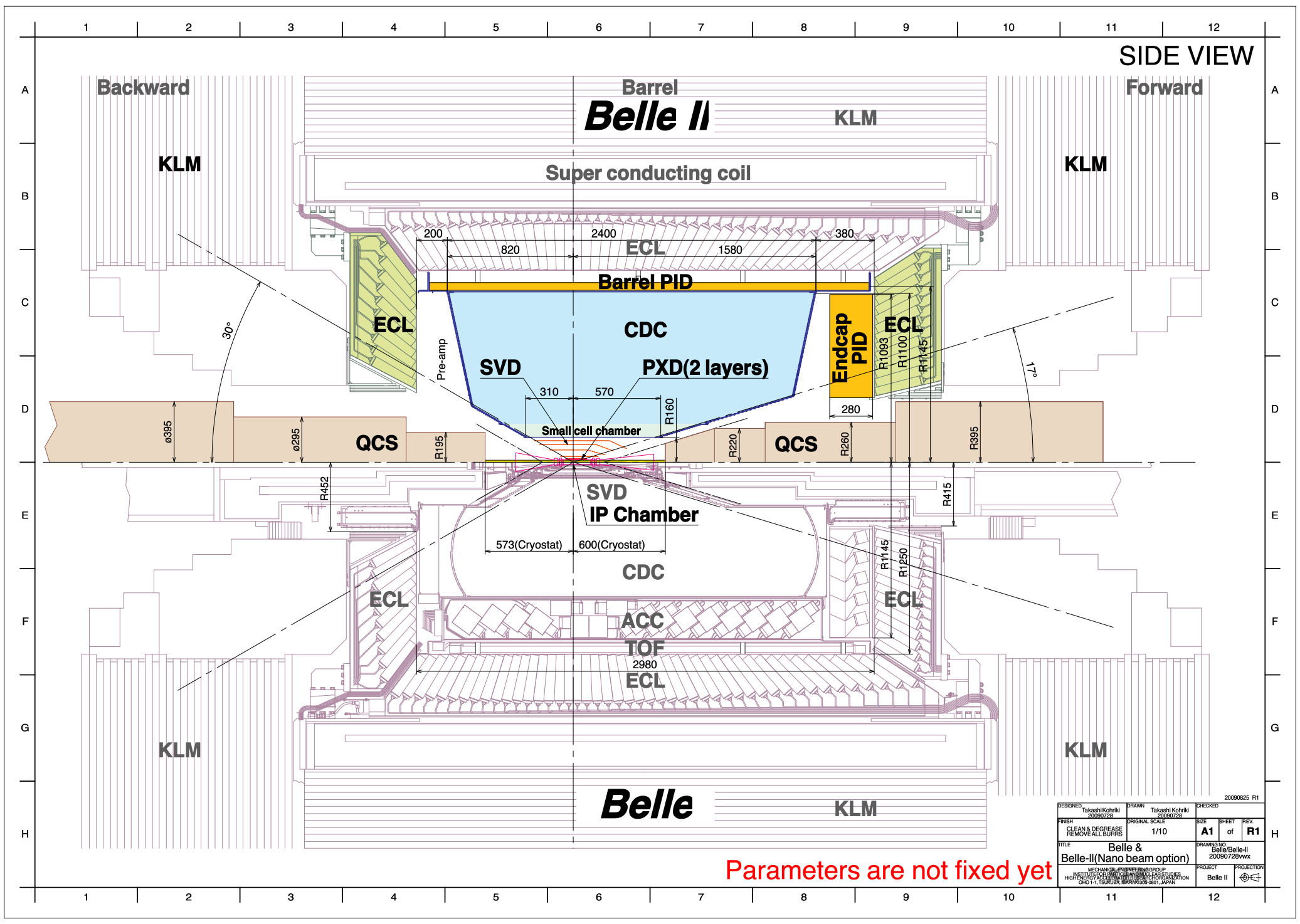

*Figure 1.9: Upgraded Belle II spectrometer (top half) as compared to the present Belle detector (bottom half).*

The design of the Belle II detector follows to a large extent the scheme discussed in the Letter of Intent [5] and its 2008 supplement, Design Study Report [6], with one notable exception: a pixel detector now appears in the innermost part of the vertex detector. Other modifications are due to the change in the accelerator design from the high current version to the "nano-beam" collider, and are associated with the larger crossing angle, the need to have the final quadrupoles as close as possible to the interaction point, and the smaller beam energy asymmetry (7 GeV/$c$ on 4 GeV/$c$ instead of 8 GeV/$c$ on 3.5 GeV/$c$).
For the Belle II detector, our main concern is to maintain the current performance of Belle in an environment with considerably higher background levels. As discussed in detail in the 2008 Design Report [6], we evaluate the possible degradation of the performance in a high-background environment by extrapolating from the present operating conditions of KEKB and Belle by accounting for the scaling of each component of background with the higher currents, smaller beam sizes and modified interaction region. From these studies, we assume a conservative factor of twenty increase in the background hit rate. The physics event rate will be about 50 times higher.
The following changes to Belle will maintain a comparable or better performance in Belle II:

- just outside the beam pipe, the silicon strip detector is replaced by a two-layer silicon pixel detector based on the DEPFET technology;
- the silicon strip detector extends from just outside the pixel detector to a larger radius

than in Belle;

- the readout of the silicon strip detector is based on the APV25 chip with a much shorter shaping time than the present read-out based on the VA1TA chip;
- the central tracking device—a large volume drift chamber—has smaller drift cells than in Belle, starts just outside the expanded silicon strip detector, and extends to a larger radius;
- the completely new particle identification devices—higher performance yet more compact—in the barrel and endcap regions are of the Cherenkov imaging type, with very fast read-out electronics;
- the electronics of the electromagnetic calorimeter is of the wave-form-sampling type; a replacement of the endcap scintillator crystal (CsI(Tl)) with a faster and radiation tolerant version (e.g. pure CsI) is considered as an upgrade option;
- the barrel part of the muon and $K_L$ detector is still equipped with RPCs, but with some inner layers perhaps operating in proportional mode, while the endcap part is replaced with scintillators instrumented with silicon photomultipliers;
- the new data acquisition system meets the requirements of a considerably higher event rates.

In several aspects, Belle II will offer considerably better performance than Belle:

- the vertex resolution is improved by the excellent spatial resolution of the two innermost pixel detector layers;
- the efficiency for reconstructing $K_S$ decays to two charged pions with hits in the silicon strip detector is improved because the silicon strip detector occupies a larger volume;
- the new particle identification devices in the barrel and endcap regions extend the very good pion/kaon separation to the kinematic limits of the experiment;
- the new electronics of the electromagnetic calorimeter considerably reduce the noise pile up, which is of particular importance for missing-energy studies.

The expected performance of the upgraded spectrometer is summarized in Table 1.3. In Chapters 4–12, we discuss the details for each individual component.

*Table 1.3: Expected performance of components of the Belle II spectrometer.*

| Component | Type | Configuration | Readout | Performance |
|---|---|---|---|---|
| Beam pipe | Beryllium double-wall | Cylindrical, inner radius 10 mm, 10 $\mu$m Au, 0.6 mm Be, 1 mm coolant (paraffin), 0.4 mm Be | | impact parameter resolution $\sigma_{z_0} \sim 20$ $\mu$m (PXD and SVD) |
| PXD | Silicon pixel (DEPFET) | Sensor size: 15×100 (120) mm$^2$ pixel size: 50×50 (75) $\mu$m$^2$ 2 layers: 8 (12) sensors | 10 M | |
| SVD | Double sided Silicon strip | Sensors: rectangular and trapezoidal Strip pitch: 50(p)/160(n) - 75(p)/240(n) $\mu$m 4 layers: 16/30/56/85 sensors | 245 k | |
| CDC | Small cell drift chamber | 56 layers, 32 axial, 24 stereo r = 16 - 112 cm - 83 $\leq z \leq$ 159 cm | 14 k | $\sigma_{r\phi} = 100$ $\mu$m, $\sigma_z = 2$ mm $\sigma_{p_t}/p_t = \sqrt{(0.2\% p_t)^2 + (0.3\%/\beta)^2}$ $\sigma_{p_t}/p_t = \sqrt{(0.1\% p_t)^2 + (0.3\%/\beta)^2}$ (with SVD) $\sigma_{dE/dx} = 5\%$ |
| TOP | RICH with quartz radiator | 16 segments in $\phi$ at $r \sim 120$ cm 275 cm long, 2 cm thick quartz bars with 4x4 channel MCP PMTs | 8 k | $N_{p.e.} \sim 20$, $\sigma_t = 40$ ps K/$\pi$ separation : efficiency $> 99\%$ at $< 0.5\%$ pion fake prob. for $B \to \rho\gamma$ decays |
| ARICH | RICH with aerogel radiator | 4 cm thick focusing radiator and HAPD photodetectors for the forward end-cap | 78 k | $N_{p.e.} \sim 13$ K/$\pi$ separation at 4 GeV/$c$: efficiency 96% at 1% pion fake prob. |
| ECL | CsI(Tl) (Towered structure) | Barrel: $r = 125$ - 162 cm End-cap: $z =$ -102 cm and +196 cm | 6624 1152 (F) 960 (B) | $\frac{\sigma E}{E} = \frac{0.2\%}{E} \oplus \frac{1.6\%}{\sqrt[4]{E}} \oplus 1.2\%$ $\sigma_{pos} = 0.5$ cm/$\sqrt{E}$ (E in GeV) |
| KLM | barrel: RPCs | 14 layers (5 cm Fe + 4 cm gap) 2 RPCs in each gap | $\theta$: 16 k, $\phi$: 16 k | $\Delta\phi = \Delta\theta = 20$ mradian for $K_L$ $\sim 1$ % hadron fake for muons |
| | end-caps: scintillator strips | 14 layers of $(7-10) \times 40$ mm$^2$ strips read out with WLS and G-APDs | 17 k | $\Delta\phi = \Delta\theta = 10$ mradian for $K_L$ $\sigma_p/p = 18\%$ for 1 GeV/$c$ $K_L$ |

# Chapter 2

# SuperKEKB

## 2.1 Machine Parameters

### 2.1.1 Nano-Beam Scheme

The KEKB B-Factory will be upgraded to SuperKEKB using the same tunnel as KEKB [1]. The upgrade is based on the "Nano-Beam" scheme, which was first proposed for the Super B factory in Italy [2]. The basic idea of this scheme is to squeeze the vertical beta function at the IP ($\beta_y^*$) by minimizing the longitudinal size of the overlap region of the two beams at the IP, which generally limits the effective minimum value of $\beta_y^*$ through the "hourglass effect." Figure 2.1 shows a schematic view of the beam collision, which is a plane figure, in the Nano-Beam scheme. The size of the overlap region $d$, which is considered to be the effective bunch length for the Nano-Beam scheme, is much smaller than the bunch length ($\sigma_z$). The length $d$ is determined by the horizontal half crossing angle ($\phi$) and the horizontal beam size at the IP ($\sigma_x^*$) via the following equation:

$$d \cong \frac{\sigma_x^*}{\phi}. \tag{2.1}$$

The hourglass condition in the Nano-Beam scheme is expressed as

$$\beta_y^* > d, \tag{2.2}$$

instead of that for a usual head-on collision of

$$\beta_y^* > \sigma_z. \tag{2.3}$$

To shorten the length $d$, a relatively large horizontal crossing angle and extremely small horizontal emittances and horizontal beta functions at the IP for both beams are required.
The luminosity of a collider is expressed by the following formula, assuming flat beams and equal horizontal and vertical beam sizes for two beams at the IP:

$$L = \frac{\gamma_\pm}{2er_e}\left(\frac{I_\pm \xi_{y\pm}}{\beta_{y\pm}^*}\right)\left(\frac{R_L}{R_{\xi_y}}\right), \tag{2.4}$$

where $\gamma$, $e$ and $r_e$ are the Lorentz factor, the elementary electric charge and the electron classical radius, respectively. The suffix $\pm$ specifies the positron (+) or the electron (−). The parameters $R_L$ and $R_{\xi_y}$ represent reduction factors for the luminosity and the vertical beam-beam parameter, which arise from the crossing angle and the hourglass effect. The ratio of these parameters is usually not far from unity. Therefore, the luminosity is mainly determined by the three fundamental parameters; *i.e.* the total beam current ($I$), the vertical beam-beam parameter ($\xi_y$) and the vertical beta function at the IP ($\beta_y^*$). The choice of these three parameters, the beam energy and the luminosity is shown in Table 2.1 together with those of present KEKB.
For the vertical beam-beam parameter $\xi_y$, we assume the same value of 0.09 as has been achieved at KEKB. The vertical beta functions at the IP for SuperKEKB are smaller by almost by a factor of 20 than those of the present KEKB owing to the adoption of the Nano-Beam scheme. Assuming these parameters, we need to double the total beam currents compared with those of the present KEKB to achieve the luminosity goal of SuperKEKB, $8 \times 10^{35}\text{cm}^{-2}\text{s}^{-1}$. The machine parameters of SuperKEKB including the three fundamental parameters are shown in Table 2.2. In the following, it is shown how these parameters are determined.

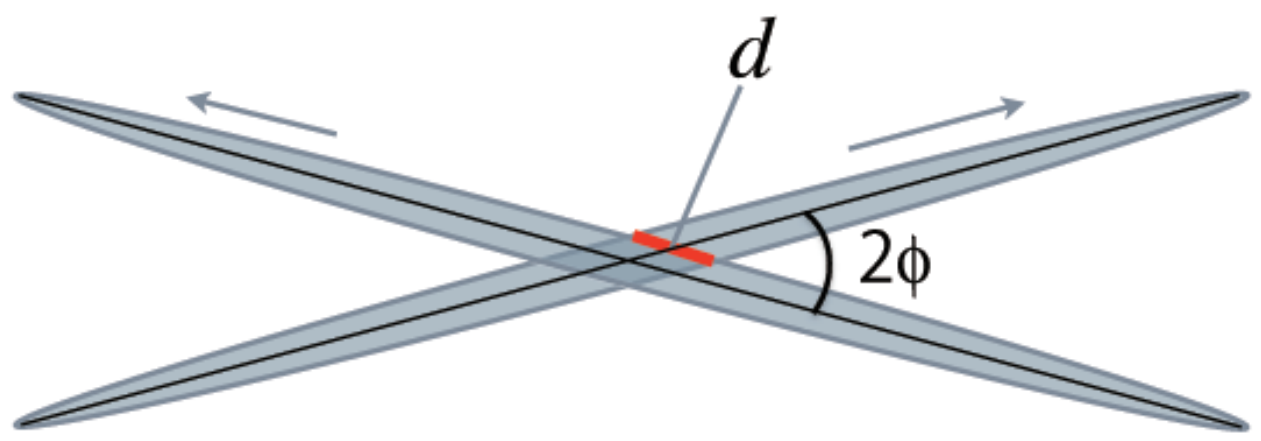

*Figure 2.1: Schematic view of beam collision in the Nano-Beam scheme.*

| | KEKB Achieved | SuperKEKB |
|---|---|---|
| Energy (GeV) (LER/HER) | 3.5/8.0 | 4.0/7.0 |
| $\xi_y$ | 0.129/0.090 | 0.090/0.088 |
| $\beta_y^*$ (mm) | 5.9/5.9 | 0.27/0.41 |
| $I$ (A) | 1.64/1.19 | 3.60/2.62 |
| Luminosity ($10^{34}\text{cm}^{-2}\text{s}^{-1}$) | 2.11 | 80 |

*Table 2.1: Fundamental parameters of SuperKEKB and present KEKB.*

### 2.1.2 Machine parameters of SuperKEKB

#### 2.1.2.1 Emittance, crossing angle, beta functions at the IP

To realize the Nano-Beam scheme, the effective bunch length $d(= \sigma_x^*/\phi)$ should be small. Of the two parameters of $\sigma_x^*$ and $\phi$, having a smaller $\sigma_x^*$ is more important than having a larger $\phi$, since it becomes difficult to obtain the design beam-beam parameter if we decrease $d$ only by enlarging $\phi$. In the Nano-Beam scheme, each particle in a bunch interacts with only a small portion of the other colliding bunch. To obtain the design value of $\xi_y$, extremely small horizontal and vertical beam sizes are needed.
In the optics design of SuperKEKB, we have made efforts to decrease the horizontal emittance while preserving as much as possible of the present lattice . The design values of the horizontal emittance shown in Table 2.2, which are smaller by a factor 5 or 10 than those of the present KEKB, include some enlargement due to intra-beam scattering. We are continuing the process of optimizing machine parameters, such as the horizontal emittance, etc., as shown in Table 2.3. The optics of the IR are also undergoing further optimization.
The horizontal beta functions at the IP are also very small compared with those of the present KEKB, the typical value of which is 1.2 m. Even with the very small horizontal emittances and the horizontal beta functions at the IP, a rather low $x - y$ coupling of approximately 0.25-0.4 % is needed to obtain $\xi_y$ of 0.09.
The half crossing angle $\phi$ is 41.5 mrad, which is about 4 times larger than that of the present KEKB. This choice of $\phi$ also contributes to decreasing the effective bunch length $d$. However, the design value of $\phi$ is determined mainly by considerations related to the optics of the interaction region (IR), magnet design, and the detector background. With a large crossing angle, the final focus quadrupole magnets can be independent for the two beams, which has the merit of reducing the detector background due to synchrotron radiation. Another merit of a larger crossing angle is that the final focus quadrupole magnets can be placed closer to the IP, which contributes to widening the dynamic aperture.
Dynamic aperture is one of the most serious issues for SuperKEKB in the Nano-Beam scheme. A narrow dynamic aperture shortens the beam lifetime from the Touschek effect and the lost particles cannot be replenished by the injector if the lifetime is too short. With the parameters in Table 2.2, the effective bunch length $d$ is $\sim$ 0.25 mm and $\sim$ 0.19 mm for the LER and the HER, respectively. From the viewpoint of the hourglass condition, values of $\beta_y^*$ even smaller than those in Table 2.2 are possible . However, the achievable values of $\beta_y^*$ in SuperKEKB are restricted more strictly by dynamic aperture than by the hourglass effect.

| | | LER (e+) | HER (e-) | units |
|---|---|---|---|---|
| Beam Energy | $E$ | 4 | 7 | GeV |
| Half Crossing Angle | $\phi$ | 41.5 | | mrad |
| Horizontal Emittance | $\varepsilon_x$ | 3.2(2.7) | 2.4(2.3) | nm |
| Emittance ratio | $\varepsilon_y/\varepsilon_x$ | 0.40 | 0.35 | % |
| Beta Function at the IP | $\beta_x^*/\beta_y^*$ | 32 / 0.27 | 25 / 0.41 | mm |
| Horizontal Beam Size | $\sigma_x^*$ | 10.2(10.1) | 7.75(7.58) | $\mu$m |
| Vertical Beam Size | $\sigma_y^*$ | 59 | 59 | nm |
| Betatron tune | $\nu_x/\nu_y$ | 45.530/45.570 | 58.529/52.570 | |
| Momentum Compaction | $\alpha_c$ | $2.74 \times 10^{-4}$ | $1.88 \times 10^{-4}$ | |
| Energy Spread | $\sigma_\varepsilon$ | $8.14(7.96) \times 10^{-4}$ | $6.49(6.34) \times 10^{-4}$ | |
| Beam Current | $I$ | 3.60 | 2.62 | A |
| Number of Bunches/ring | $n_b$ | 2503 | | |
| Energy Loss/turn | $U_0$ | 2.15 | 2.50 | MeV |
| Total Cavity Voltage | $V_c$ | 8.4 | 6.7 | MV |
| Synchrotron Tune | $\nu_s$ | -0.0213 | -0.0117 | |
| Bunch Length | $\sigma_z$ | 6.0(4.9) | 5.0(4.9) | mm |
| Beam-Beam Parameter | $\xi_y$ | 0.0900 | 0.0875 | |
| Luminosity | $L$ | $8 \times 10^{35}$ | | cm$^{-2}$s$^{-1}$ |

*Table 2.2: Machine Parameters of SuperKEKB. Values in parentheses denote parameters at zero beam currents.*

| | | LER (e+) | HER (e-) | units |
|---|---|---|---|---|
| Beam Energy | $E$ | 4 | 7 | GeV |
| Half Crossing Angle | $\phi$ | 41.5 | | mrad |
| Horizontal Emittance | $\varepsilon_x$ | 3.2(2.3) | 5.1(5.0) | nm |
| Emittance ratio | $\varepsilon_y/\varepsilon_x$ | 0.27 | 0.25 | % |
| Beta Function at the IP | $\beta_x^*/\beta_y^*$ | 32 / 0.27 | 25 / 0.31 | mm |
| Horizontal Beam Size | $\sigma_x^*$ | 10.2(10.1) | 11.2(11.1) | $\mu$m |
| Vertical Beam Size | $\sigma_y^*$ | 48.3 | 61.8 | nm |
| Betatron tune | $\nu_x/\nu_y$ | 45.530/45.570 | 45.530/43.570 | |
| Momentum Compaction | $\alpha_c$ | $3.28 \times 10^{-4}$ | $4.36 \times 10^{-4}$ | |
| Energy Spread | $\sigma_\varepsilon$ | $8.23(7.96) \times 10^{-4}$ | $5.85(5.77) \times 10^{-4}$ | |
| Beam Current | $I$ | 3.60 | 2.60 | A |
| Number of Bunches/ring | $n_b$ | 2500 | | |
| Energy Loss/turn | $U_0$ | 2.13 | 2.07 | MeV |
| Total Cavity Voltage | $V_c$ | 9.64 | 12.0 | MV |
| Synchrotron Tune | $\nu_s$ | -0.0250 | -0.0245 | |
| Bunch Length | $\sigma_z$ | 6.0(5.0) | 5.0(4.9) | mm |
| Beam-Beam Parameter | $\xi_y$ | 0.0886 | 0.0830 | |
| Luminosity | $L$ | $8 \times 10^{35}$ | | cm$^{-2}$s$^{-1}$ |

*Table 2.3: Example of further optimization of machine parameters. Values in parentheses denote parameters at zero beam currents.*

#### 2.1.2.2 Beam energy

In SuperKEKB, the beam energies have been changed from the present values of 3.5 and 8.0 GeV to 4.0 and 7.0 GeV. In the Nano-Beam scheme, emittance growth due to intra-beam scattering and the short beam lifetime due to the Touschek effect are very serious problems, particularly in the LER. The increase in the beam energy of the LER from 3.5 to 4.0 GeV helps mitigate these problems. In addition, the decrease in the beam energy of the HER from 8.0 to 7.0 GeV is beneficial in obtaining a lower emittance. The impact of this change of the beam energy asymmetry on the physics sensitivity is discussed elsewhere in this report.

#### 2.1.2.3 Beam-beam parameter

As a design value for $\xi_y$, we assumed the value of 0.09, which was actually achieved at KEKB. This value was achieved by using the crab cavities, which effectively enable a head-on collision at the IP with beams crossing at an angle. On the other hand, SuperKEKB will adopt a large crossing angle. Therefore, we need a careful study of the effect of the crossing angle on the achievable value of $\xi_y$.
In the case of a large crossing angle, there exists another kind of hourglass effect. A particle with a finite horizontal offset at the IP collides with (the center of) the other beam at the place where the vertical beta function $\beta_y$ is larger than its minimum value $\beta_y^*$. The difference of $\beta_y$ from its minimum value depends on the horizontal offset. This shift of the collision point from the vertical waist position depending on the horizontal offset creates another kind of hourglass effect. This secondary hourglass effect could possibly degrade the beam-beam performance and lower the achievable value of $\xi_y$. It is known that this effect can be avoided by using the so-called “crab waist” scheme [2].
The crab waist scheme shifts the vertical waist position using sextupole magnets so that the vertical waistline of one beam is aligned along the trajectory of the other beam at the IP. Beam-beam simulations have been done based on a strong-weak model to investigate the beam-beam performance in the Nano-Beam scheme.
The simulations showed that the effectiveness of the crab waist scheme depends on the machine parameters, and that the luminosity improvement with the crab waist scheme is only about 10% with the SuperKEKB parameters in Table 2.2, as shown in Fig. 2.2. The simulation also showed that the design value of 0.09 for $\xi_y$ is achievable with the design parameters without the crab waist scheme. A tune survey was also done in the beam-beam simulations to find the best working point. The fractional parts of the betatron tunes shown in Table 2.2 were determined by the simulation to maximize $\xi_y$. The optimum horizontal tune is not as close to the half integer as is the case in KEKB.

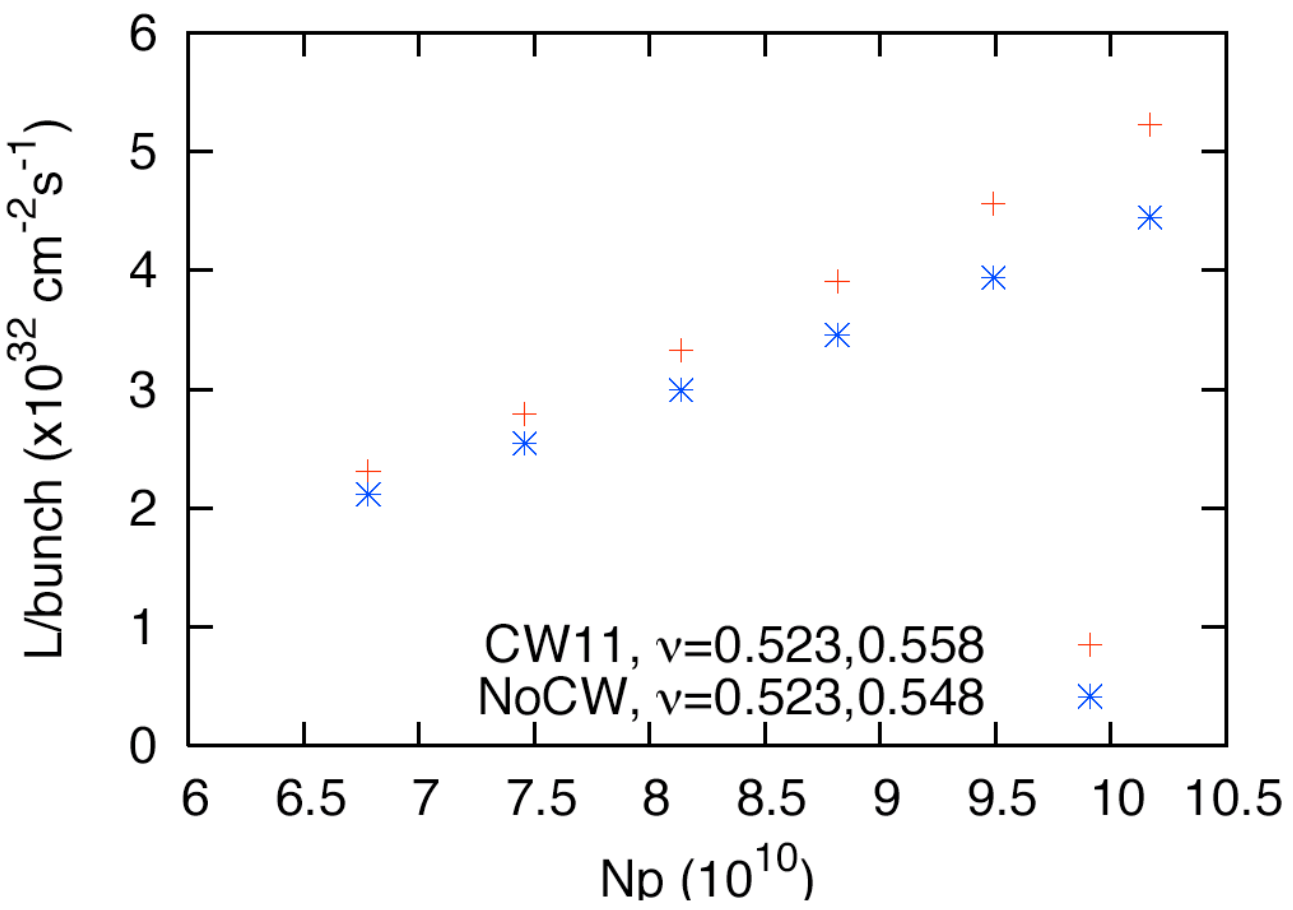

*Figure 2.2: The strong-weak beam-beam simulation with and without crab waist.*

In the present design of SuperKEKB, we do not employ the crab waist scheme. However, we are still considering the crab waist scheme as a backup option, because the simulation has predicted that use of the crab waist scheme gives larger tolerances for machine errors and larger choices of tunes. To give a more realistic estimation of the beam-beam performance, a strong-strong simulation is underway.

#### 2.1.2.4 Beam current and beam current related parameters

To achieve the target luminosity of $8 \times 10^{35}\mathrm{cm}^{-2}\mathrm{s}^{-1}$ with the design values of $\beta_y^*$ and $\xi_y$ in Table 2.2, beam currents of 3.60 and 2.62 A are needed for the LER and HER, respectively. These currents are about twice as high as those in KEKB. We have been conducting R&D on hardware components, such as the vacuum system and the RF system, assuming the even higher beam currents of 9.4 A and 4.1 A in the LER and HER, respectively. These currents had been assumed before we adopted the Nano-Beam scheme. The designs resulting from this long-term R&D program can also be used at the relatively low currents that are now required.
The number of bunches per ring ($n_b$) is 2503, which implies that every other RF buckets is filled with beam. If we decrease $n_b$ while maintaining the total beam currents, which implies higher bunch currents, then we can obtain the same luminosity with a larger x-y coupling. This will somewhat reduce the difficulty of the optics design or relax the requirement for the $x-y$ coupling correction in actual beam operation. However, higher bunch currents bring other problems such as difficulty in handling higher HOM power, single bunch instabilities such as the micro-wave instability or emittance growth due to intra-beam scattering. With the design value of $n_b$ in Table 2.2, it has been determined that these problems are manageable.
As for the bunch length $\sigma_z$, a shorter bunch is preferable, since with a longer bunch length we need to decrease the x-y coupling or increase the bunch currents to compensate for lower particle densities in the beam overlap region at the IP. However, several bunch lengthening effects, such as potential-well distortion, micro-wave instability and intra-beam scattering, prevent us from achieving a short bunch length. The calculations of $\sigma_z$ considering these effects with a wakefield that includes the effect of the coherent synchrotron radiation (CSR) showed that the design values of $\sigma_z$ in the Table 2.2 are attainable. The RF voltages in Table 2.2 were adjusted so that the design value for $\sigma_z$ is obtained at the design bunch currents with the effects mentioned above.
We have also calculated the effect of the micro-wave instability at the design bunch currents to asses the impact on the energy spread $\sigma_\varepsilon$. Although some enlargement of $\sigma_\varepsilon$ is expected, it is still tolerable.

## 2.2 Lattice Design

### 2.2.1 Beam-optical parameters

As shown in Table 2.2 of Section 2.1.2, low emittances and small beta functions at the IP are necessary to achieve the design luminosity of $8\times10^{35}$ cm$^{-2}$s$^{-1}$. Two rings of 4 GeV and 7 GeV have been designed to achieve the low emittance. A new Final Focus (FF) section with local chromaticity correction sections has also been designed to strongly squeeze the colliding beams in both the horizontal and vertical planes at the IP.
The LER has 4 arcs and 4 straight sections. Each arc is approximately 540 m long and consists of 6 normal cells and one modified cell. The FF section is placed in the Tsukuba straight section, containing the IP. A vertical chicane to transport one beam over the other ring is placed in the straight section diametrically opposite the IP, the Fuji straight section. This straight section is also utilized to adjust betatron tunes. RF cavities are also placed in the Fuji straight section. The wiggler magnets, which adjust the emittance and damping time, are placed in the other two straight sections (Oho and Nikko), and RF cavities are also installed on the upstream side of each wiggler section.
The geometry of the HER is the same as that of the LER. The straight section on the opposite side of the IP is utilized to adjust betatron tunes. RF cavities are placed in the other two straight sections and there are no wiggler magnets in those sections.
The quadrupole magnets from KEKB are reused as much as possible. However, the main dipole magnets are replaced by new ones in order to achieve low emittance. To reduce the emittance one order of magnitude below that of KEKB, the arc cells of both rings have been redesigned. The emittance is analytically given by

$$\varepsilon_x = \frac{C_q \gamma^2}{2\pi\rho_0^2} \oint_{Bend} H ds, \tag{2.5}$$

where

$$H = \gamma_x \eta_x^2 + 2\alpha_x \eta_x \eta_x' + \beta_x \eta_x'^2. \tag{2.6}$$

In the case of the LER, the length of the main dipole magnet is changed from 0.89 m to 4 m to make the bending radius longer. The wiggler sections have also been redesigned along with the arc cells. The period of the wiggler magnets is reduced to half of that of KEKB by adding new wiggler magnets.
In the case of the HER, the number of cells for each arc is increased from 7 to 9 to keep the horizontal dispersion small and to reduce the emittance. The main dipole magnets, which are 5.9 m long, are replaced with ones 3.8 m long. Other straight sections outside the IR are almost identical to those of KEKB.
The bunch length at zero beam current (nominal) is expressed as

$$\sigma_z = \frac{C\alpha_p}{2\pi\nu_s}\sigma_\delta \propto \sqrt{\frac{\alpha_p E^3}{V_c}}, \tag{2.7}$$

where

$$\sigma_\delta = \gamma\sqrt{\frac{C_q}{2\rho}} \tag{2.8}$$

$$\nu_s \propto \sqrt{\frac{\alpha_p V_c}{E}}. \tag{2.9}$$

Since the design of arc cells is dedicated to achieve extremely low emittance, the momentum compaction factor, $\alpha_p$, cannot easily be used to adjust the bunch length. Therefore, the total RF voltage, $V_c$, is adjusted to achieve the design bunch length. The nominal bunch length is determined by the luminosity requirement, with bunch lengthening due to intra-beam scattering, the wake field effects from the ring impedance, and the effect of CSR at the design beam current. Skew sextupole magnets have also been considered to correct chromatic $x-y$ coupling at the IP, due to their demonstrated effectiveness in improving the luminosity at KEKB [3, 4]. This technique is expected to give a dynamic aperture larger than that achievable without skew sextupoles.

### 2.2.2 Arc cell

The arc lattice consists of $2.5\pi$ unit cells. Each arc cell includes two non-interleaved sextupole pairs for the chromaticity correction. Two sextupole magnets in a pair are connected with a pseudo $-I$ transform,

$$\begin{pmatrix} -1 & 0 & 0 & 0 \\ m_{21} & -1 & 0 & 0 \\ 0 & 0 & -1 & 0 \\ 0 & 0 & m_{43} & -1 \end{pmatrix}. \tag{2.10}$$

In principle, nonlinearities in the sextupole magnets are cancelled in each pair. The arc unit cells in the LER and HER are shown in Figs. 2.3 and 2.4, respectively. There are 7 families of quadrupole magnets in a cell in the LER and 6 families in the HER with 6 constraints: 4 for the pseudo $-I$ conditions of two sextupole pairs, and 2 to adjust $\varepsilon_x$ and $\alpha_p$ by adjusting the horizontal dispersion at the dipole magnets. We reduced the number of quadrupole families in the arc cell for the HER in order to reduce the number of new magnets to be installed as much as possible.

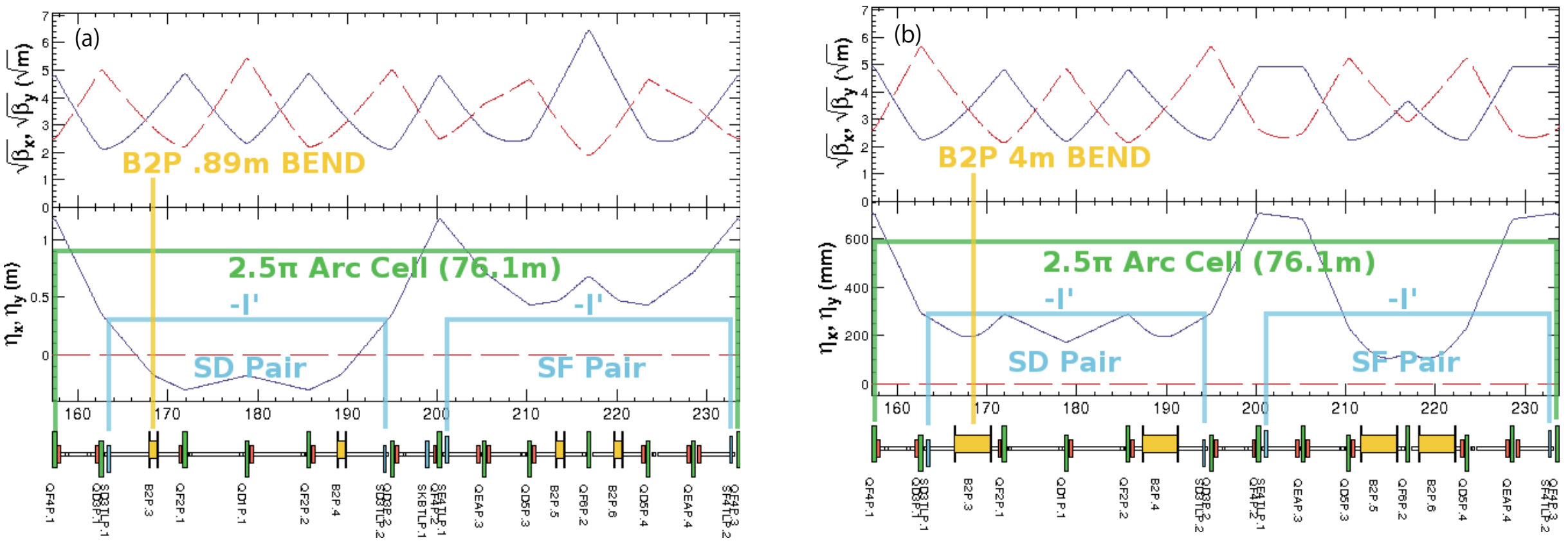

*Figure 2.3: Arc unit cell in the LER. (a) KEKB-LER and (b) SuperKEKB-LER.*

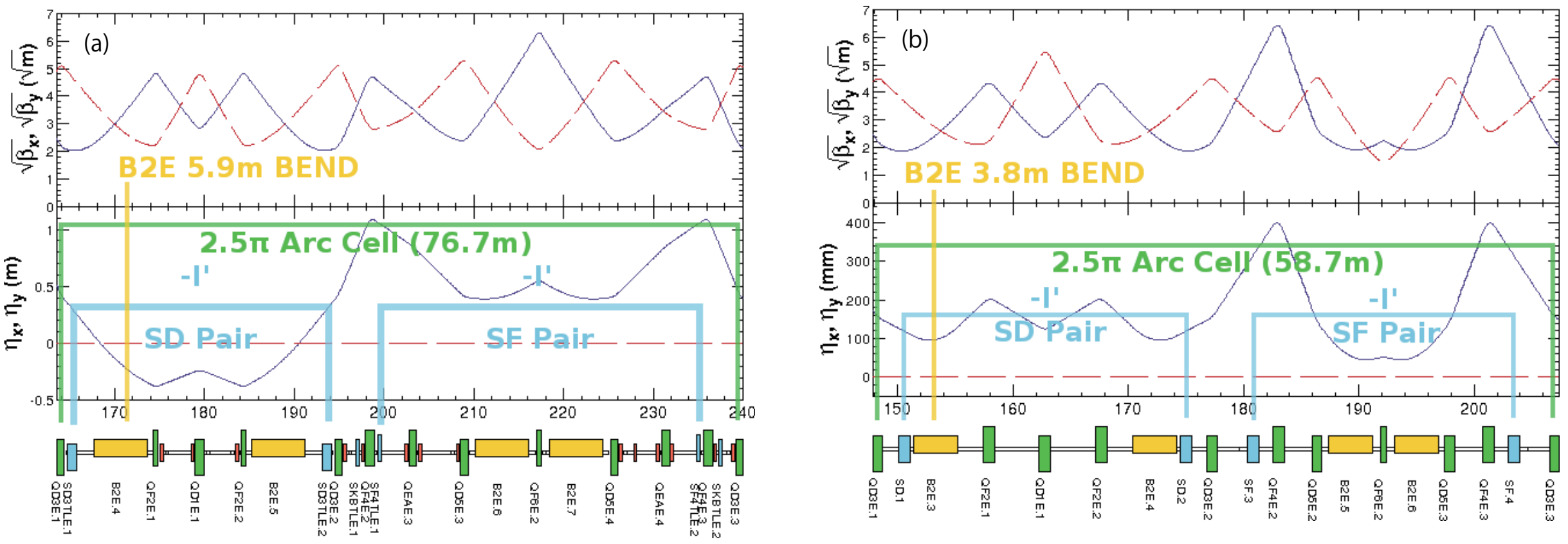

*Figure 2.4: Arc unit cell in the HER. (a) KEKB-HER and (b) SuperKEKB-HER.*

### 2.2.3 Interaction region

The FF section is designed to achieve very small beta functions at the IP, small nonlinearities with non-interleaved sextupole pairs, and corrections of the large chromaticity generated in the IR. The nominal values for the beta functions at the IP are 32 mm in the horizontal plane and 270 $\mu$m in the vertical plane for the LER, 25 mm in the horizontal plane and 410 $\mu$m in the vertical plane for the HER, respectively.

To squeeze the beta functions, doublets of vertical focusing quadrupole magnets (QC1s) and horizontal focusing quadrupole magnets (QC2s) are utilized. The distance between the center of the first quadrupole and the IP is 0.92 m for QC1LP and 0.91 m for QC1RP in the LER and 1.46 m for QC1LE and 1.38 m for QC1RE in the HER. The beams in the LER and HER collide at the IP with a horizontal crossing angle of 83 mrad.

In the FF section, bending magnets are placed on both sides of the IP to make dispersion in the region of the local chromaticity corrections (LCC). One family of sextupole pairs in the vertical plane (Y-LCC) or two families of sextupole pairs both in the horizontal and vertical planes (X-LCC and Y-LCC) are used. A chicane type of LCC is adopted in the LER to increase geometrical flexibility. In the case of the HER, a chicane type is adopted for the X-LCC and an arc type is adopted for the Y-LCC to suppress the emittance generation at the LCC region. The betatron phase advance between the center of QC1 and the closest sextupole magnet to the IP in a Y-LCC pair is $\pi$ in the vertical plane and the phase advance between QC2 and the closest sextupole magnet to the IP in X-LCC is $2\pi$ in the horizontal plane. When the LCC scheme is Y-LCC only, the closest sextupole pairs (SF2TL and SF2TR) to the IP in the arc section are used in anway similar to that of X-LCC. The IR lattice is shown in Figs. 2.5, 2.6, and 2.7.
The solenoid field of Belle II is fully compensated with compensation solenoids on each side of the IP:

$$\int_{IP} B_z(s)ds = 0. \tag{2.11}$$

The remaining $x-y$ coupling components, and the horizontal and vertical dispersions are corrected to be zero at the IP, and are localized in the IR on each side of the IP. In order to make this possible, skew quadrupole magnets, as well as horizontal and vertical dipole magnets are used to adjust these optical parameters and to satisfy geometrical constraints.
The beam trajectory becomes a three-dimensional twist. Therefore, the $x-y$ axis rotates around the beam axis along the beam trajectory. The rotation angle can be calculated as

$$\theta = \frac{1}{2B\rho}\int_{IP} B_z(s)ds. \tag{2.12}$$

Final focusing quadrupole magnets (QC1 and QC2) within the solenoid field also rotate around the beam axis and the optical parameters can be optimized to eliminate $x-y$ coupling at the IP. The angle between the solenoid axis and each beam axis of the LER and HER is 41.5 mrad, which is optimized by the vertical emittance due to synchrotron radiation coming from the fringe field of the solenoid magnet. The vertical emittance is estimated approximately as

$$\varepsilon_y \propto \left(\frac{p}{\rho}\right)^2 \int H(s)ds \propto B_x^4(s), \tag{2.13}$$

where

$$B_x(s) \simeq -\frac{x}{2}B_z'(s) \simeq -\frac{s\phi}{2}B_z'(s). \tag{2.14}$$

To reduce the vertical emittance due to the solenoid fringe field, the half crossing angle, $\phi$ and/or the rate of change of $B_z$ should be decreased.

### 2.2.4 Dynamic aperture

The dynamic aperture is defined by requiring stability in 1000 turns with synchrotron oscillation and without radiation damping, which gives almost the same result in one transverse damping time with radiation damping. The dynamic aperture is estimated numerically with six-dimensional tracking simulations using *SAD*[5], which is an integrated code for optics design and particle tracking. In the case of the Nano-Beam scheme, the dynamic aperture will be reduced because of the nonlinear fringe and kinematic terms in the IR, and also strong sextupole magnets. The aperture for the initial action $J_{y0}$ is written as [6]

$$J_{y0} = \frac{\beta_y^{*2}}{(1-\frac{2}{3}Kl^{*2})l^*}A(\mu_y), \tag{2.15}$$

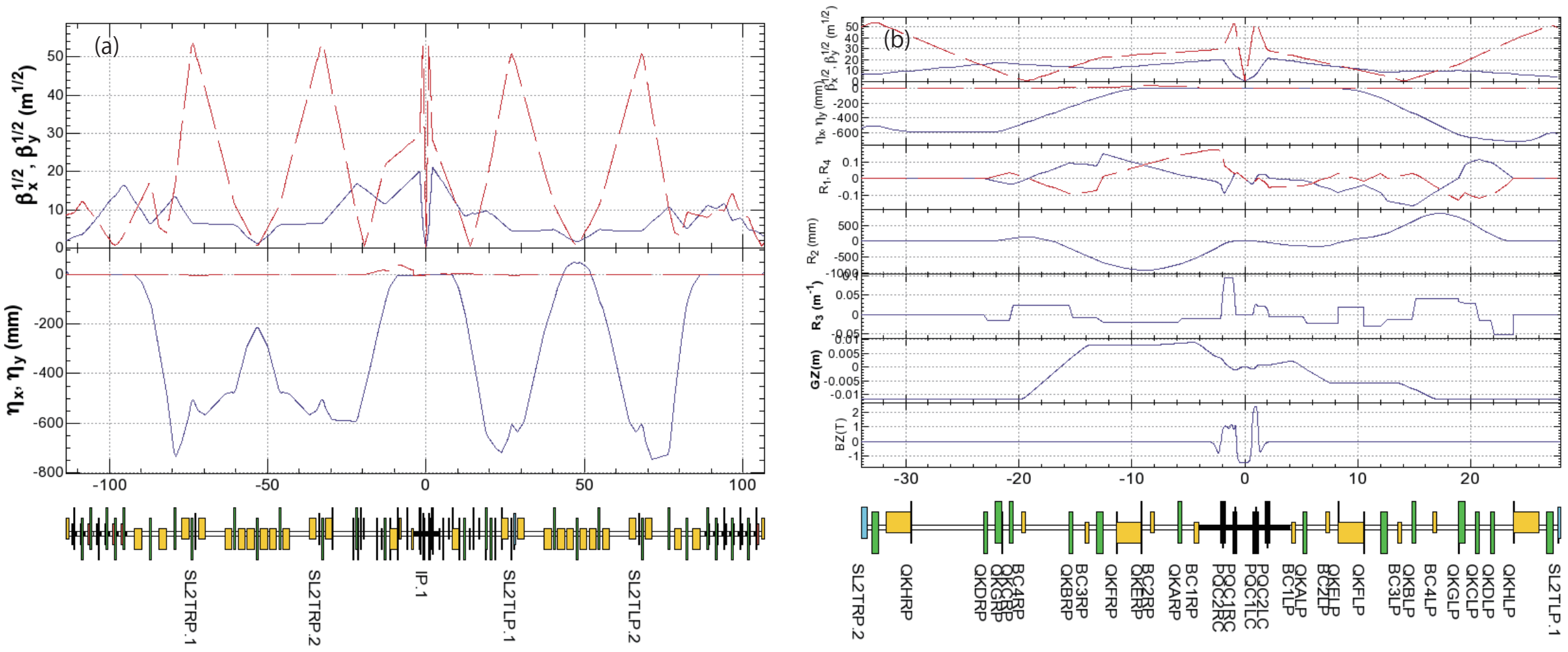

Figure 2.5: *One family of sextupoles in LCC. (a) LER lattice design in the IR and (b) the region for adjusting of $x-y$ coupling and dispersions.*

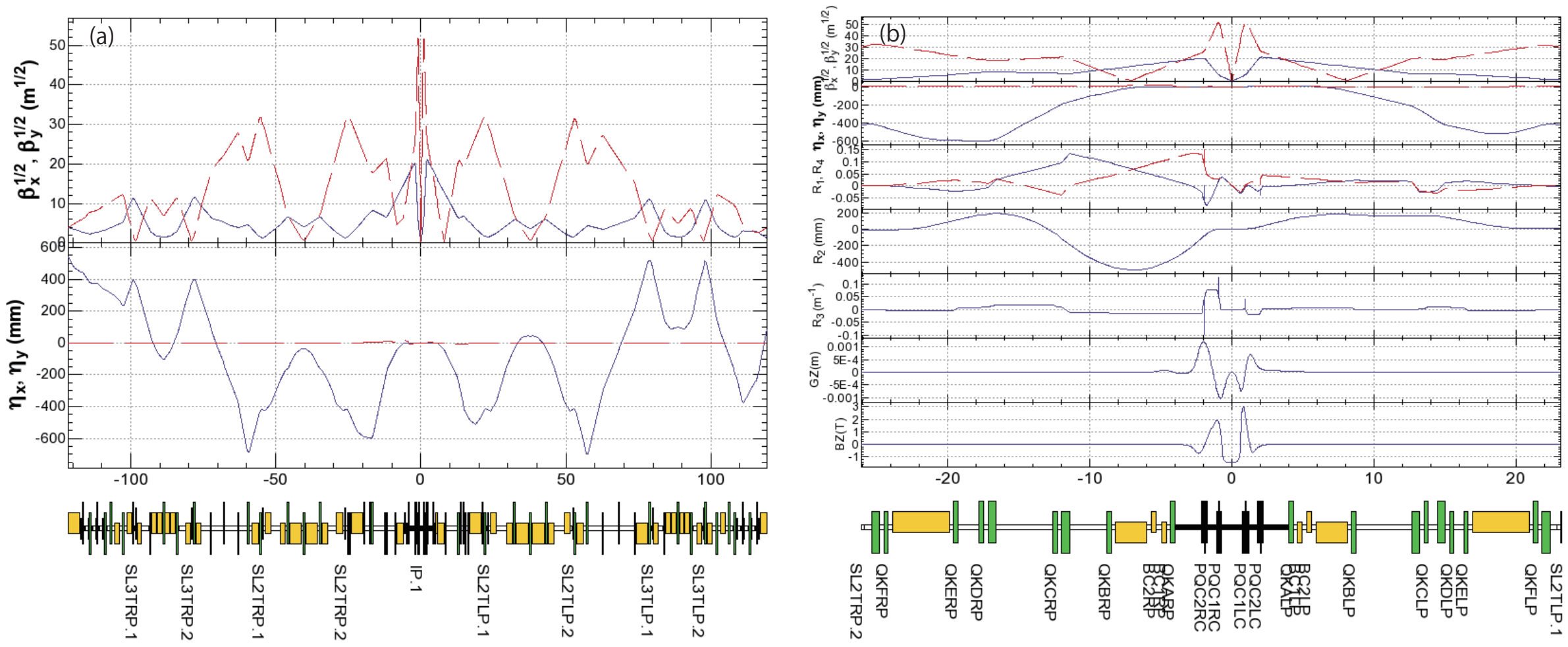

Figure 2.6: *Two families of sextupoles in LCC. (a) LER lattice design in the IR and (b) the region for adjusting of $x-y$ coupling and dispersions.*

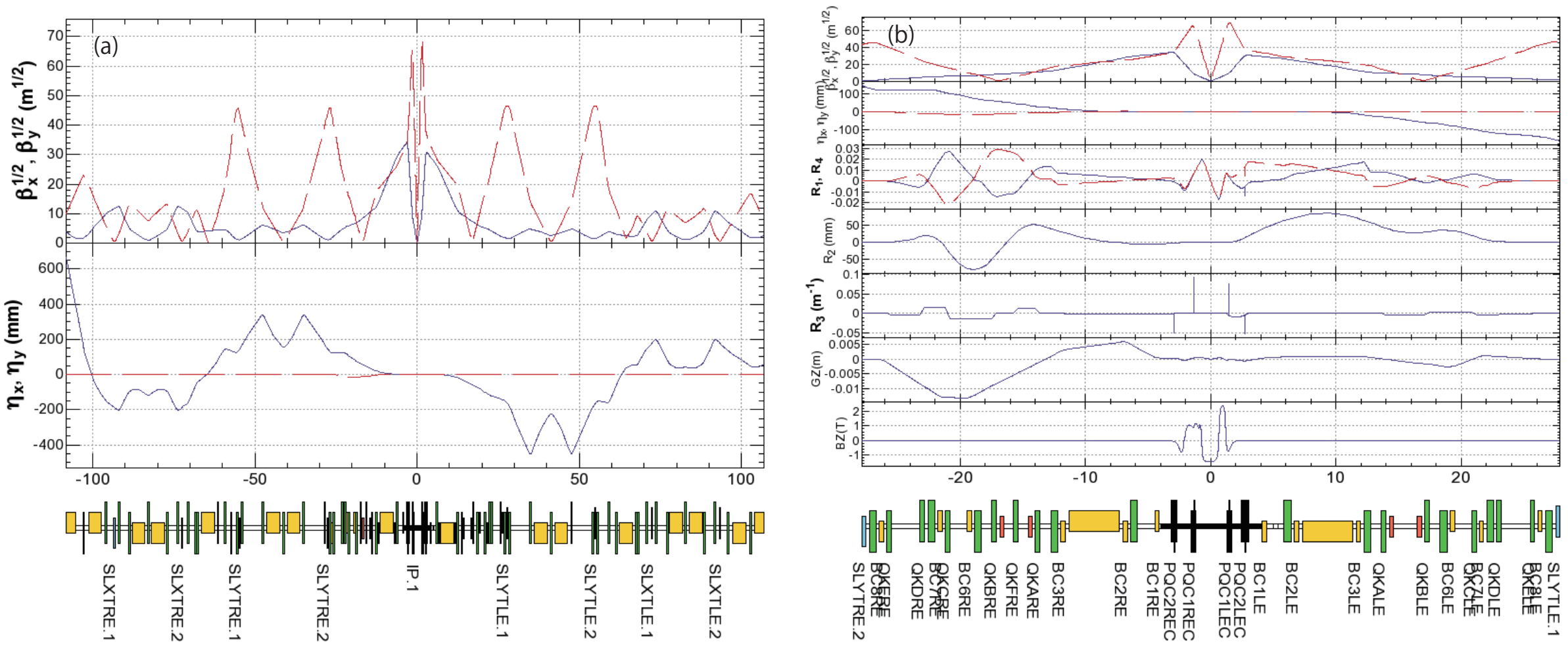

*Figure 2.7: Two families of sextupoles in LCC. (a) HER lattice design in the IR and (b) the region for adjusting of $x - y$ coupling and dispersions.*

where $K$ is a magnetic-field strength and the sign of $K$ is negative for the final quadrupole magnet. This equation implies that the dynamic aperture decreases as the beta function at the IP is squeezed and as the distance ($l^*$) between the IP and the final quadrupole is increased. The dynamic aperture affects both Touschek lifetime and injection efficiency of the beams. The target Touschek lifetime is 600 sec for the LER and HER. The beam lifetime must be larger than 181 sec for the LER and 105 sec for the HER at least, which is calculated from the injection rate assuming 100% injection efficiency. The requirements for dynamic aperture for betatron injection is shown in Table 2.4. The lattice optimization for the dynamic aperture is still in progress for both LER and HER.

*Table 2.4: Required transverse apertures ($A_x$ and $A_y$) and momentum aperture ($\Delta p/p_0$) for betatron injection.*

| | | **LER** | **HER** | Unit |
|---|---|---|---|---|
| Beam Energy | $E$ | 4 | 7 | GeV |
| Horizontal | $A_x$ | $5.11\times10^{-7}$ | $3.77\times10^{-7}$ | m |
| | injection error $2J_x$ | $4.19\times10^{-7}$ | $3.32\times10^{-7}$ | m |
| Vertical | $A_y$ | $1.10\times10^{-8}$ | $1.46\times10^{-8}$ | m |
| | injection error $2J_y$ | 0 | 0 | m |
| Longitudinal | $\Delta p/p_0$ | 0.21 | 0.24 | % |

## 2.3 Magnets

### 2.3.1 Magnet replacements

A major change in the magnet system will be required. The main dipole magnets in the LER and HER will be replaced. The LER dipoles become longer (from an effective length of 0.89 m to ~4 m) and the HER dipoles become shorter (from 5.91 m to 3.8 m). More dipole, quadrupole and sextupole magnets are needed in the HER since the number of cells in the arc sections is by about 30% larger than in the KEKB lattice, as shown in Fig. 2.8. More wiggler magnets, with shorter pole lengths, are needed in the Nikko and Oho straight sections. The layout of the wiggler section magnets will change as shown in Fig. 2.9. Vertical steering magnets need to be replaced by new magnets with wider gaps to account for the antechambers. The magnetic properties of the steel used for the magnets should be the same as that used for the KEKB magnets, as some of the magnets will be reused.
There may be a need for new, stronger sextupole magnets. One way to increase the field is to replace the power supplies and run the magnets at higher currents. The excitation curve of a KEKB sextupole magnet is shown in Fig. 2.10, where the integrated sextupole field is plotted against the current. Since a clear saturation is seen, replacement by new sextupole magnets is preferable to achieve much higher fields.

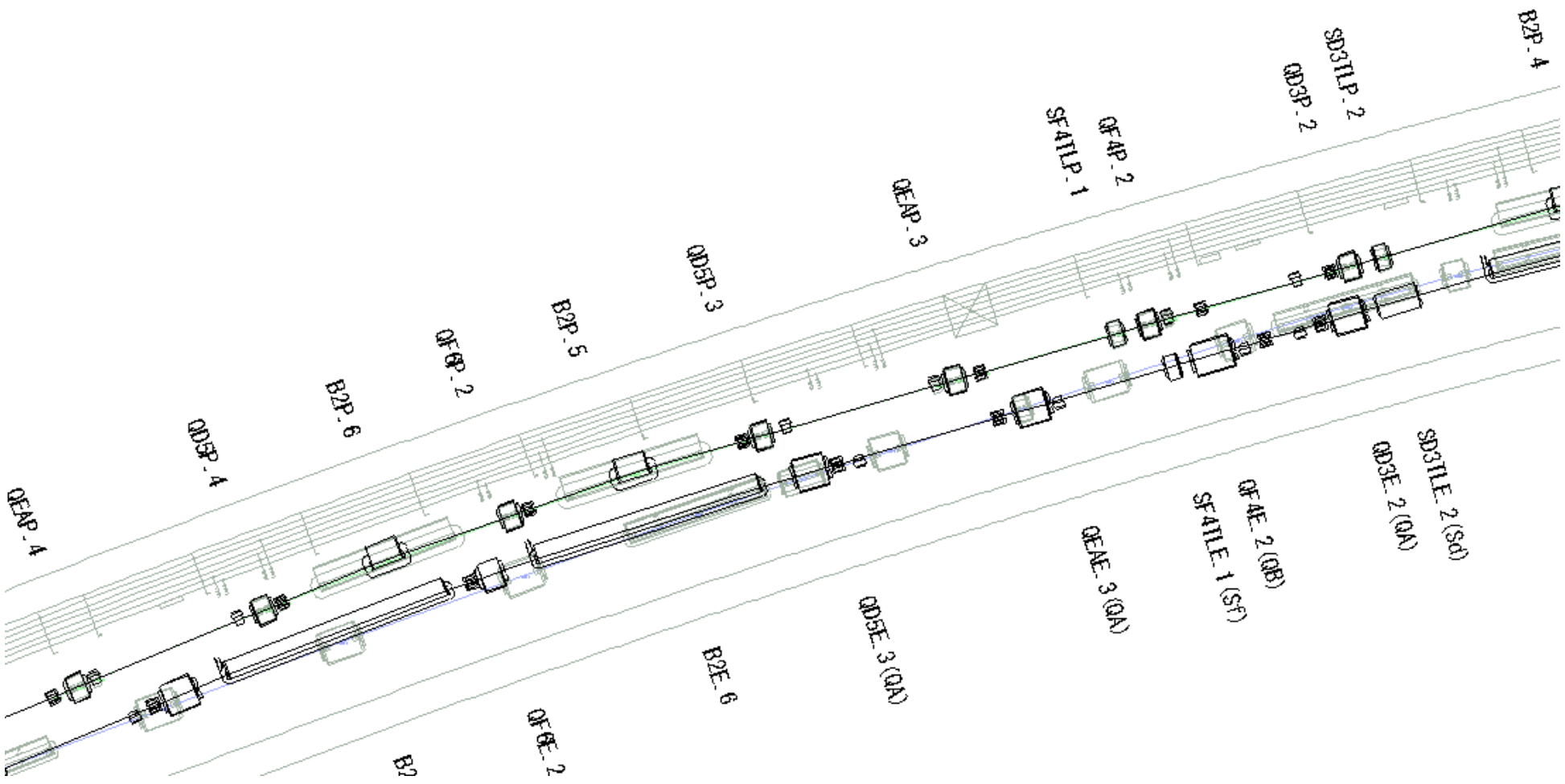

*Figure 2.8: Magnet layout in the north arc section. The present KEKB magnet layout and the new magnet layout are represented with black and grey lines, respectively. The LER dipole magnets (B2Ps) become longer while the HER dipole magnets (B2Es) become shorter. The HER arc lattice changes result in more cells and more magnets being needed.*

### 2.3.2 Field measurements

Precise magnetic field measurements need to be performed for all new magnets. Some of the recycled magnets also need to be measured, as they will be operated at higher currents than in KEKB. A new flip-flop coil system needs to be made for measuring the new dipole magnets. The probes of the two harmonic coil systems need refurbishment, most likely to be replaced by new sensors. The total number of magnets to be measured is estimated to be close to one thousand. The tolerances on the multipole errors are expected to be the same as those of the KEKB magnets [7].

### 2.3.3 Magnet installation and alignment

First, almost all of the magnets will need to be removed from the beam line; whether they will be moved to outside of the tunnel depends of the construction scenario. Most of the HER quadrupole magnets will be reused, at new positions in the tunnel. Most of the base-plates, which support the magnets in the arc sections, will also be removed. A survey of monuments for the magnet alignment will follow the floor refurbishment.
Though the LER arc lattice does not change, the LER magnet positions are shifted toward the outside of the tunnel to compensate for the circumference change, which comes from the new orbit in the interaction region. Realignment of all the magnets will be necessary in both the LER and the HER. The alignment network is expected to be improved with more monuments being added to the tunnel walls. Alignment of the magnets and mechanical stability of magnets will be very important to approach single-beam emittance coupling values of 0.3-0.4%. The movers of the sextupole magnets might be replaced by more rigid supports to reduce vibration.

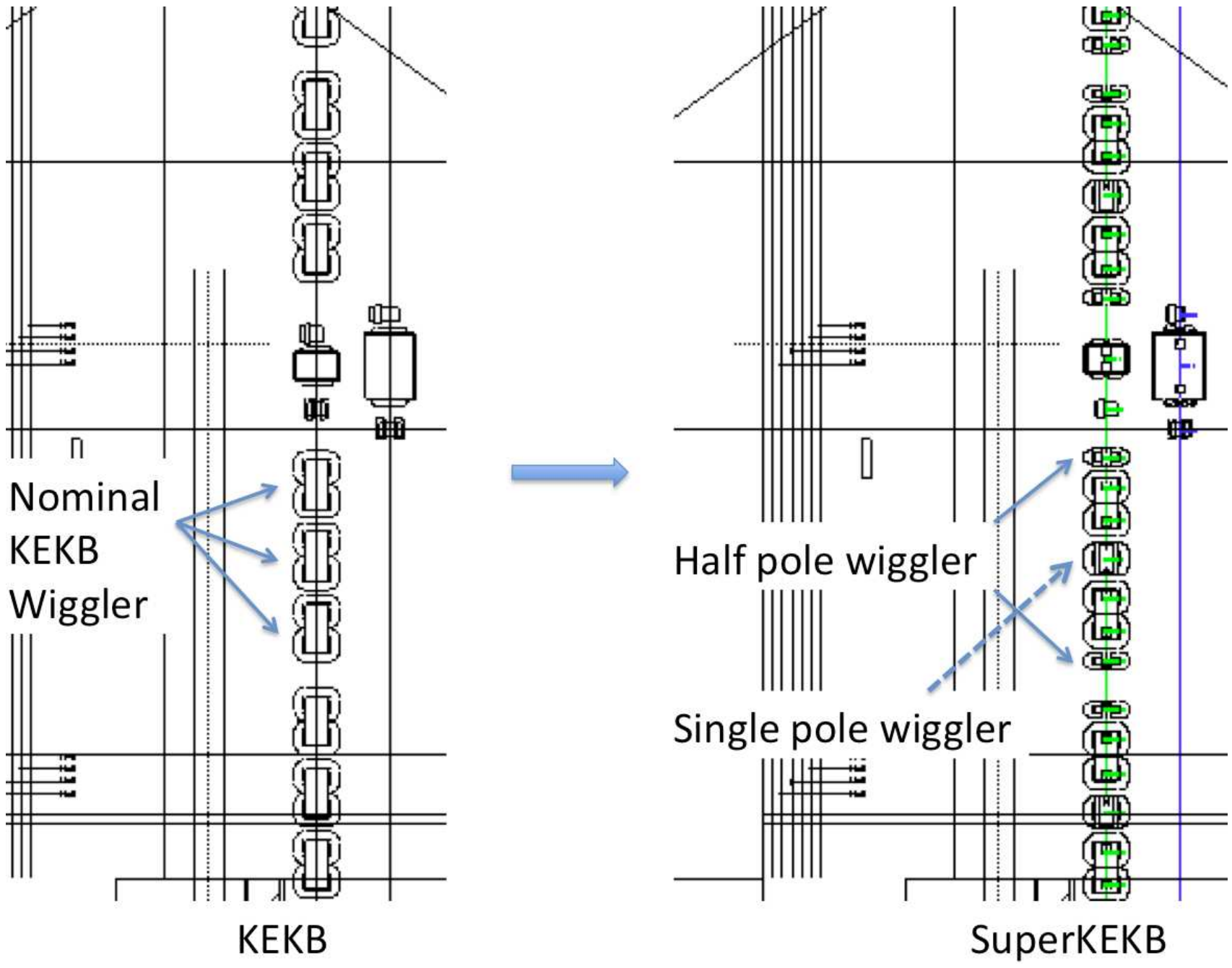

*Figure 2.9: Wiggler section layouts for KEKB (left) and SuperKEKB (right).*

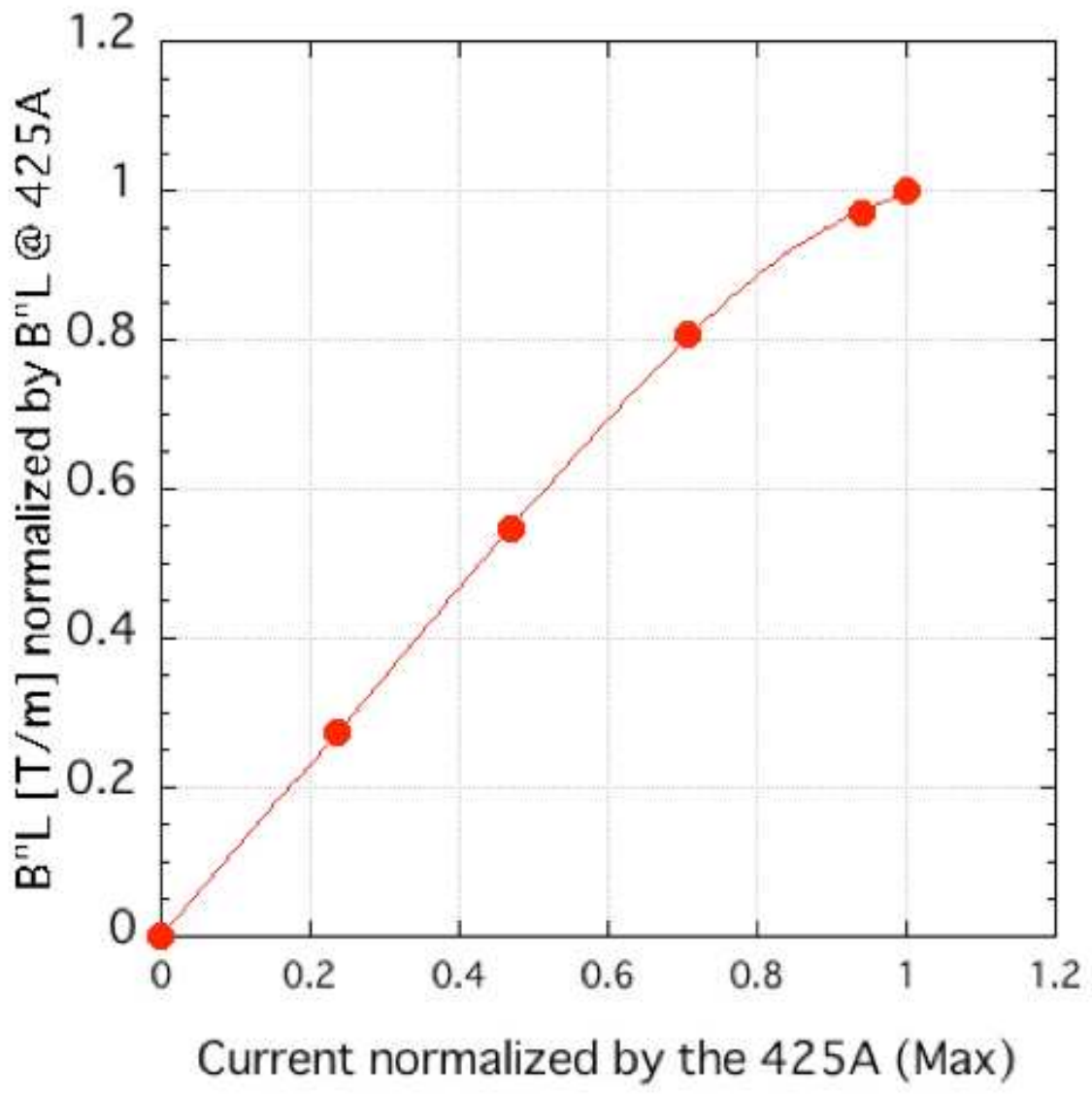

*Figure 2.10: The excitation curve of the KEKB sextupole magnet.*

### 2.3.4 Utilities and facilities

A larger capacity water cooling system is required to operate the increased number of magnets. The number of water pump systems will be doubled from the present four to eight, as is indicated in Fig. 2.11. The plumbing work outside and inside the tunnel will be arranged and carried out by the Facilities Department.

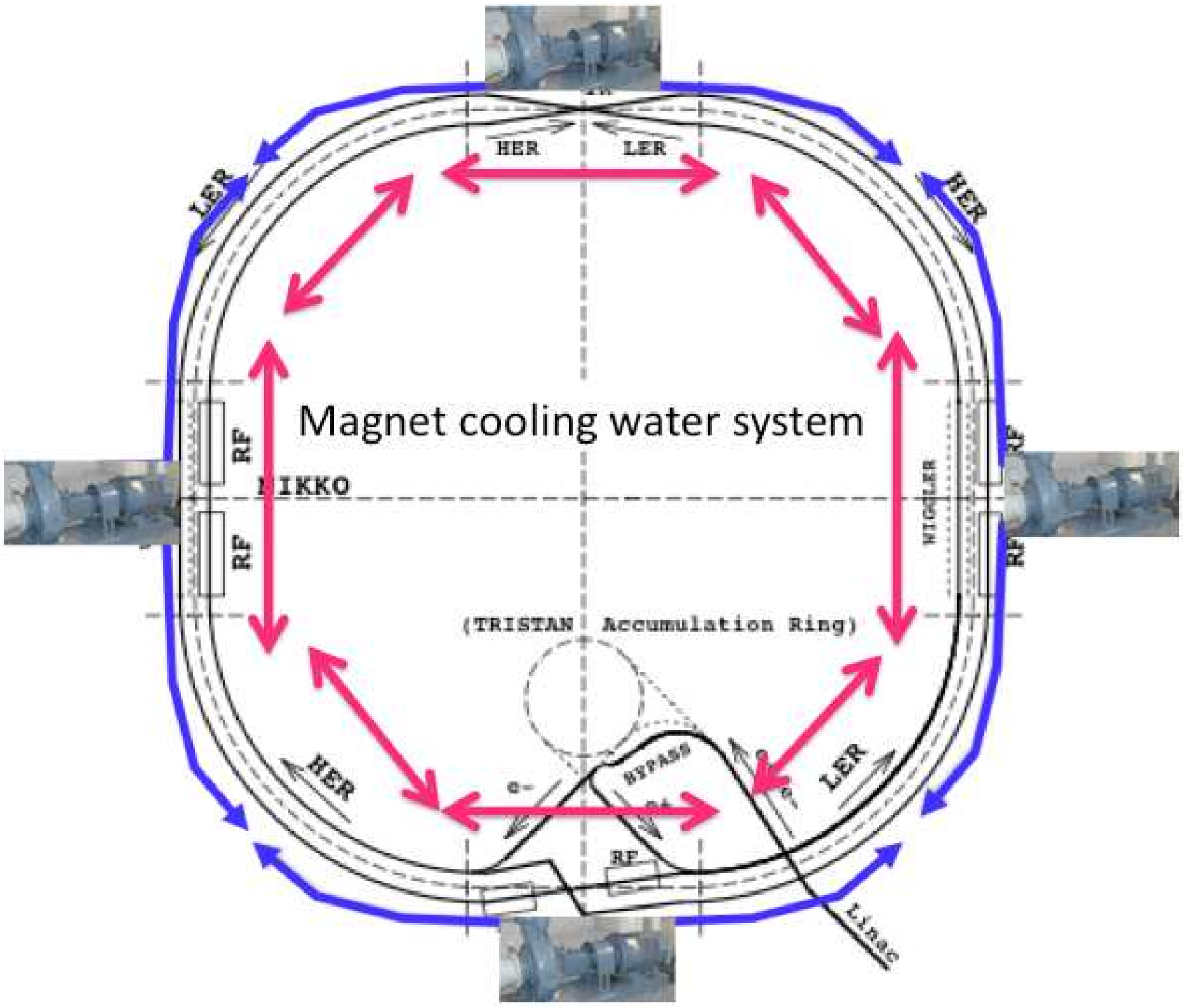

*Figure 2.11: Cooling water system.*

### 2.3.5 Power supplies

More power supplies are needed, as the number of magnet families increases. A request for more space in the power supply buildings has been made. There will be more cables from the buildings to the tunnel, which may require adding more thru holes from the power supply building to the tunnel. New power supplies are needed to run at higher currents and/or with different specifications, which include the power supplies for the new IR superconducting magnets. The design work of the new power supplies can start when the magnet design is fixed, after the machine lattice is finalized. Some very old power supplies need to be replaced: most likely the power supplies for the LER and HER dipole magnets will be replaced. Cabling in the tunnel needs to be (mostly) removed and reorganized.
Since the amount of work and effort of redoing the magnet and power supply systems for SuperKEKB is massive, the magnet workload may become the critical path item for the project's schedule.

## 2.4 RF

### 2.4.1 RF Overview

The present RF system for KEKB consists of 20 ARES cavities for the LER and 12 ARES and 8 superconducting (SC) cavities for the HER, powered by 25 klystrons in total [8]. Beam currents of up to 2.0 A in the LER and 1.45 A in the HER have been stored.
Table 2.5 gives the RF-related machine parameters and the design RF parameters for SuperKEKB. In each ring the design beam current is twice that of the present KEKB, and the required beam power in the LER is three times larger. On the other hand, the total RF voltage is low, particularly in the HER, about half that of the KEKB HER. The low RF voltage with higher beam current and larger beam power makes beam-loading effects much more serious in SuperKEKB. If the present RF system is used with the current configuration, several problems arise. First, the input coupling of the ARES needs to be increased to an unacceptably high level to obtain the optimum coupling at the design beam current. Second, the input couplers of the SC cavities also need to be replaced with more strongly coupling ones to reduce the loaded-Q values by a factor of three or four. This work can cause possible air dust contamination on the SC cavity surface. The tips of the couplers also may have heating problems. Third, the detuning frequency of the SC cavities will become comparable with the revolution frequency, which makes the growth time of the -1 mode longitudinal coupled-bunch instability associated with the accelerating mode on the order of 100 microseconds.
To solve these problems, two measures will be taken. One is that the RF system configuration with the ARES cavities will be changed from the present scheme where one klystron drives two ARES cavities to a different scheme where one klystron drives one ARES. For this, we add klystrons, but remove some of the ARES cavities. The high power RF and LLRF system layout will be changed accordingly. Each RF station will be operated at around 800 kW, well below the saturation level of the klystrons. At this power the klystrons, power supplies and high-power RF system can be operated stably. In this way the power delivered by each ARES cavity to the beam can be increased from 220 kW up to 600 kW, while the total RF voltage is kept low.
Another measure is to introduce a new operation scheme for the SC stations, the so-called "Reverse Phase Operation (RPO)." Figure 2.12 shows a phasor diagram of the cavity voltage for RPO scheme. The phase of a part of SC cavities are set on the time-rising side (reverse phase), while others on the time-descending side (normal phase). With this scheme a low total RF voltage is obtained, while each cavity can be operated at a high voltage. Beam power is shared by all cavities, including the reverse-phase ones. Merits of the RPO scheme are: (1) no need to change the input coupling, (2) the detuning frequency is kept sufficiently small, and (3) the impedance of the reverse-phase cavities cancels that of normal phase ones. The -1 mode instability growth rate becomes acceptably small. Harmful gap transient effects also cancel out. By changing to the 1:1 scheme for the ARES stations and introducing the RPO for the SC stations, the required beam parameters can be satisfied, as shown in Table 2.5. The system configuration change is summarized in Table 2.6.
In the HER, another scheme with 14 ARES cavities alone without SC cavities would also work.

*Table 2.5: RF parameters for SuperKEKB.*

| | **LER** | **HER** | | **unit** |
|---|---|---|---|---|
| Beam energy | 4.0 | 7.0 | | GeV |
| Beam current | 3.60 | 2.62 | | A |
| Number of bunches | 2503 | 2503 | | |
| Energy loss/turn | 2.15 | 2.50 | | MV |
| Momentum compaction | 2.74 | 1.88 | | $\times 10^{-4}$ |
| Radiation loss | 7.74 | 6.55 | | MW |
| Total loss factor | 25 | 40 | | V/pC |
| Parasitic loss | 1.30 | 1.10 | | MW |
| Total beam power | 9.04 | 7.65 | | MW |
| Total RF voltage | 8.4 | 6.7 | | MV |
| Cavity type | ARES | ARES | + SCC | |
| Number of cavities | 18 | 8 | 8 | |
| | | | (3 for RPO†) | |
| RF voltage/cavity | 0.467 | 0.48 | 1.3 | MV |
| Wall loss/cavity | 131 | 139 | - | kW |
| Beam power/cavity | 502 | 556 | 400 | kW |
| Input coupling | 4.84 | 5.01 | - | |
| Loaded $Q$ value | 1.9 | 1.8 | $\sim 5$ | $\times 10^4$ |
| Klystron output | 677 | 744 | 428 | kW |
| Detuning frequency | 28.1 | 18.4 | 47.4 | kHz |

†Reverse Phase Operation (RPO)

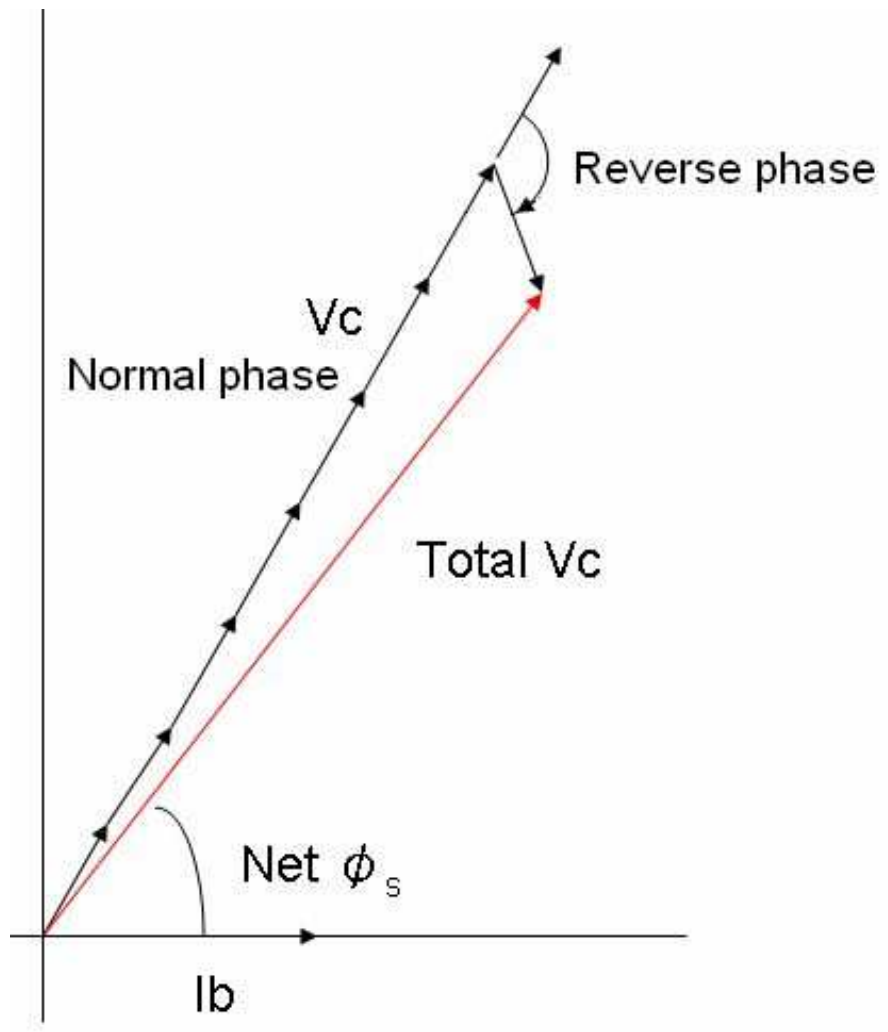

*Figure 2.12: Phasor diagram of the reverse phase operation.*

However, we choose the SC and ARES hybrid system as the baseline scheme for the following reasons: (1) the existing SC cavities and cryogenic system can be used without large changes. (2) We can continue progress on KEK's world-leading technology and maintain experience with SC cavities for high current applications; KEKB-type cavities have been successfully operating in BEPC-II at IHEP, and are also being considered for adoption at other laboratories. (3) The RPO scheme has been tested at KEKB, and no serious problem with this scheme has been found so far. Thus we consider the 14 ARES scheme as a back-up option. Details of the beam test results for RPO will be given in Section 2.4.3.

*Table 2.6: Reinforcement of RF stations by changing to 1:1 configuration. (Numbers listed are number of klystrons/number of cavities.)*

| **Building** | **KEKB** | | **SuperKEKB** | | **Change** |
|---|---|---|---|---|---|
| | **LER** | **HER** | **LER** | **HER** | |
| **(ARES cavity stations)** | | | | | |
| D4 | - | 3/6 | - | 6/6 | add 3 klystrons |
| D5 | - | 4/6 | 6/6 | - | add 2 klystrons, convert from HER to LER |
| D7 | 5/10 | - | 6/6 | - | add 1 klystron, remove 4 ARES |
| D8 | 5/10 | - | 6/6 | - | add 1 klystron, remove 4 ARES |
| D11 | - | - | - | 2/2 | add 2 klystrons, install 2 ARES |
| **(SC cavity stations)** | | | | | |
| D10 | - | 4/4 | - | 4/4 | no change |
| D11 | - | 4/4 | - | 4/4 | no change |

#### 2.4.1.1 Bunch gap transient

In KEKB, owing to the high stored energy of the ARES and SC cavities, transient phase modulation and longitudinal position shift along the bunch train due to an abort gap is small, about 3 to 5 degrees. No luminosity degradation along a train related to this effect has been observed. Calculation and measurements show good agreement [9]. In SuperKEKB, since the beam current is twice as high, the gap length has to be reduced to less than half, from 500ns to 200ns, to keep the transient phase modulation as small as in KEKB. For this, the rise time of the abort kicker needs to be shortened.

#### 2.4.1.2 Coupled-bunch instabilities

Table 2.7 shows a summary of growth time of coupled-bunch instabilities related with the RF cavities. The fastest growth time of longitudinal instabilities caused by HOM in the ARES cavities is 15 ms, which is faster than the radiation damping time. A longitudinal bunch-by-bunch feedback system is needed for SuperKEKB, in addition to the transverse feedback. The -1 mode instability will be safely cured by the -1 mode damper implemented in the RF system.

#### 2.4.1.3 RF system for Damping Ring

According to the design parameters of the Damping Ring, the required RF voltage and beam power is 0.261 MV and less than 10 kW, respectively. This can be provided by one RF station with a klystron output power of less than 150 kW. The HPRF and LLRF components can be compatible with those of the LER and HER. However, because of tight space constraints in the tunnel and no need for the energy-storage cavity, a new accelerating cavity based on the ARES design, but without a storage cavity, will be used.

Table 2.7: Instability due to RF cavities and cure.

| Direction | Ring | Cause | Frequency (MHz) | Growth time (ms) | Cure |
|---|---|---|---|---|---|
| Longitudinal | LER | ARES-HOM | 1850 | 15 | BbyB FB |
| | | ARES-zero/pi | 504 | 29 | BbyB FB |
| | | -1 mode | 508.79 | 5 | -1 mode damper |
| Longitudinal | HER | ARES-HOM | 1850 | 75 | slower than $\tau_{rad,L}$ † |
| | | SCC-HOM | 1018 | 58 | slower than $\tau_{rad,L}$ |
| | | -1 mode | 508.79 | 11 | -1 mode damper |
| Transverse | LER | ARES-HOM | 633 | 9 | BbyB FB |
| Transverse | HER | ARES-HOM | 633 | 48 | slower than $\tau_{rad,T}$ †† |
| | | SCC-HOM | 688 | 14 | BbyB FB |

† $\tau_{rad,L}$ and †† $\tau_{rad,T}$ are longitudinal and transverse radiation damping times, respectively.

### 2.4.2 ARES cavity

The ARES cavity is a three-cavity system operated in the $\pi/2$ mode, in which an accelerating cavity is resonantly coupled with an energy storage cavity via a coupling cavity between [10]. The name "ARES" stands for "Accelerator Resonantly coupled with Energy Storage," which represents its RF schematic. Fig. 2.13 shows a schematic 3D view of the ARES cavity developed based on a conceptual demonstrator named "ARES96" [11], together with an equivalent kinematic model consisting of three coupled pendulums.

The energy storage cavity, corresponding to the left-side pendulum with the very large mass, is operated in the TE013 mode and functions as a kind of electromagnetic flywheel to stabilize the $\pi/2$ accelerating mode against heavy beam loading on the accelerating cavity, which corresponds to the right-side pendulum. The accelerating cavity itself is a HOM-damped structure, carefully designed to be smoothly embedded into the whole ARES scheme. Four rectangular waveguides are directly brazed to the upper and lower sides of the accelerating cavity. Each HOM waveguide is terminated with two bullet-shaped SiC ceramic absorbers directly cooled by water flowing inside. The beam pipes at both ends of the accelerating cavity are Grooved Beam Pipes (GBPs) [12] with two grooves at the upper and lower sides of each circular pipe. In each groove, eight SiC tiles are arranged in a line and are brazed to a water-cooled copper plate. Details of the HOM absorbers for the ARES cavity system are found in [13]. The coupling cavity, corresponding to the central pendulum, is the keystone of the ARES cavity system, and is equipped with an antenna-type coupler [14] to damp the parasitic 0 and $\pi$ modes. RF power is fed through an input coupler [15] installed on the storage cavity, being transferred from a rectangular waveguide, via a doorknob transformer, through a coaxial line with an RF window, and finally to a coupling loop at the end.

At present, there are 32 ARES cavities being stably operated in KEKB, 20 for the LER and 12 for the HER, respectively. With our experience accumulated through KEKB machine operations so far, we are reasonably confident that the present ARES cavity system can be operated, even without upgrades to the ARES cavity itself, for SuperKEKB using the Nano-Beam scheme. In SuperKEKB, 26 out of the 32 ARES cavities will be reused, 18 for the LER and 8 for the HER, respectively, where every ARES cavity is fed with RF power up to 800 kW at maximum from one klystron capable to deliver 1 MW CW.
However, the increased HOM power for SuperKEKB is our primary concern. Table 2.8 summarizes the HOM power projections for the ARES cavity when being operated in the SuperKEKB LER, together with the results of calorimetric measurements in the KEKB LER. The total HOM power per cavity is estimated to be 17 kW for the design total beam current of 3.6 A consisting of 2503 bunches with a bunch length of 6 mm. As for the HOM waveguide, the estimated power is 3.3 kW per waveguide, which is 66 % of the handling capability of 5 kW determined based on the results of high power tests with an L-band klystron. As for the GBP, the estimated power is 0.93 kW per groove, which is close to the power handling capability of 1 kW with a margin of about 8 % left.

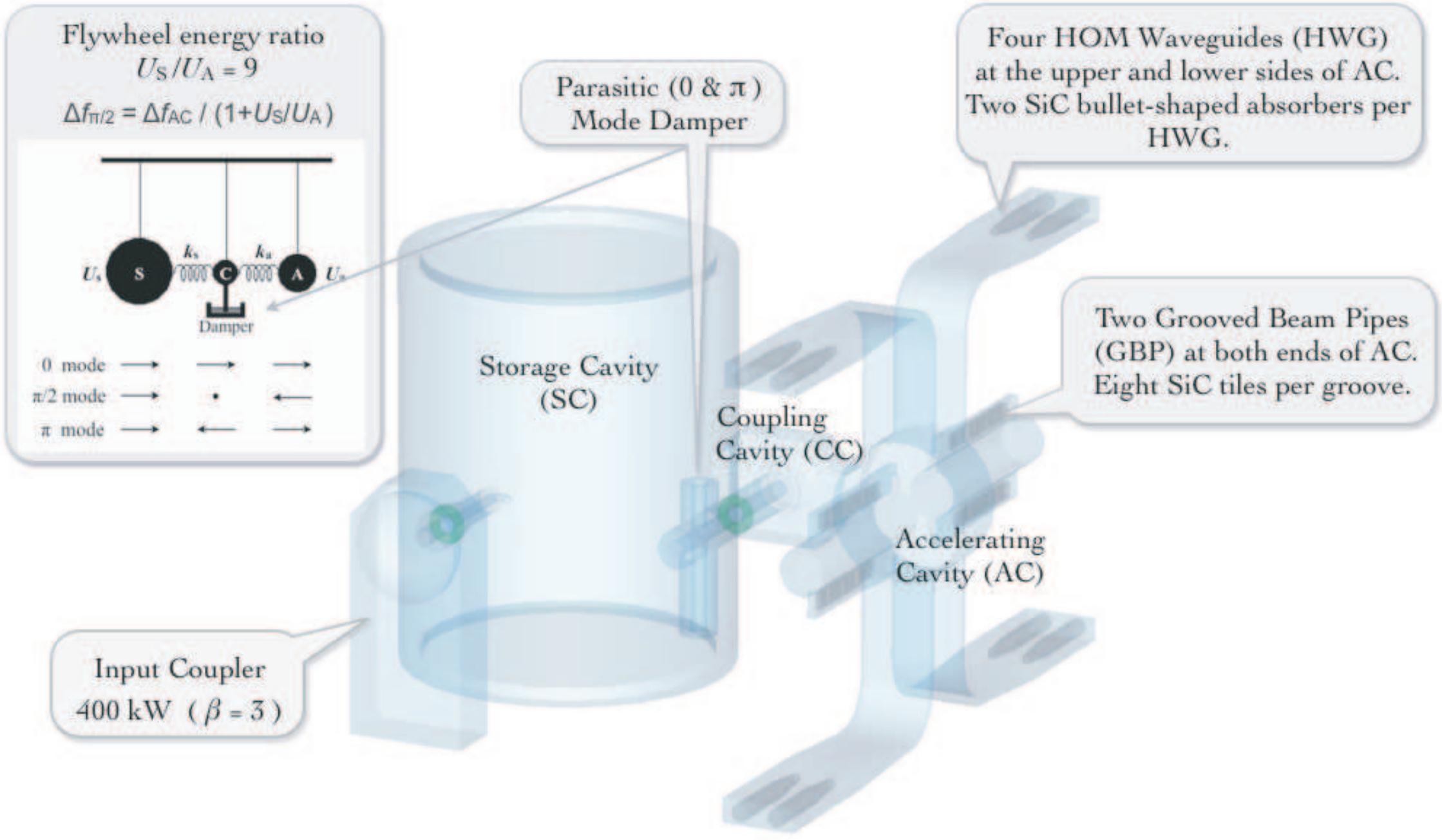

*Figure 2.13: A schematic 3D view of the ARES cavity.*

*Table 2.8: HOM power projections for SuperKEKB LER.*

| | **KEKB LER Sep. 21,2004** | **SuperKEKB LER** | **Verified††** |
|---|---|---|---|
| Beam current [A] | 1.6 | 3.6 | - |
| Number of bunches | 1293 | 2503 | - |
| Bunch length [mm] | 7 | 6 | - |
| Loss factor [V/pC] | 0.40 (0.39†) | 0.44 | - |
| $P_{HOM}$ /ARES [kW] | 5.4† | 17 | - |
| $P_{HOM}$ /HWG [kW] | 1.05† | 3.3 | 5.0 |
| $P_{HOM}$ /Groove [kW] | 0.3† | 0.93 | 1.0 |

†Calorimetric measurement.
††RF power handling capability verified at 1.3 GHz.

As for the input coupler, the power handling capability over 800 kW has already been demonstrated at a high-power test stand [16]. However, multipactoring discharge in the coaxial section has been observed in about 10% of the input couplers in use in KEKB. To eliminate this concern for the high power operation of the couplers for SuperKEKB, fine circumferential grooving has been applied to the inside of the outer conductor, and successfully tested on the high-power test stand, and used in actual KEKB operation [17]. In addition, the input coupling factor needs to be increased up to 6 for SuperKEKB, about two times as large as in the current type, and has been demonstrated over 6 up to 8 in low power tests. Currently, a prototype coupler is under fabrication, and will be tested on the high-power test stand in April 2010. It will be installed on an operational ARES cavity for May.

### 2.4.3 Superconducting cavity

Eight superconducting accelerating cavities in KEKB-HER have been operated stably for more than ten years. These cavities will be reused in SuperKEKB. Two major modifications are needed to satisfy the requirements for SuperKEKB. One is to improve the HOM dampers to absorb higher HOM power. The other is introducing Reverse Phase Operation (RPO), described in Sec. 2.4.1.

#### 2.4.3.1 HOM damper

The design bunch length, 5 mm in the HER, generates 46 kW of HOM power in one cavity at 2.6 A with 2500 bunches. The HOM power has to be absorbed in the ferrite HOM dampers attached to the two beam pipes on either side of the cavity. The most serious issue is the outgassing from the ferrite material. The outgassing condenses on the cavity wall and can trigger a breakdown. To suppress outgassing, the surface temperature of the ferrite material must be decreased. Two improvements are being developed for this. One is a thinner ferrite damper with a thickness of 3 mm (the present damper is 4 mm thick). A prototype of the 3 mm thick damper was fabricated and high-power tested. The surface temperature was reduced by 25%. Another is a double cooling channel structure for more efficient cooling of the ferrite surface. A prototype damper with a double cooling channel structure is being fabricated, and will be tested. This is expected to further reduce the surface temperature by 17%. With these improvements, the HOM damper is expected to absorb the required HOM power without generating serious outgassing.

#### 2.4.3.2 Beam test for the Reverse Phase Operation

Feasibility studies for the RPO scheme were conducted in KEKB several times in 2009. The parameters related to the RF voltage and the synchronous phase, such as the synchrotron tune, the bunch length and the beam loading in the reversed cavity, were measured. The results showed that the total RF voltage obtained was as expected by calculation. Transient behavior of the cavity voltage, tuner phase, the input power and reflected power were measured when one of the cavities tripped. The transient behaviors are well understood by taking the beam-loading effects into account, and the beam was normally aborted by the abort system. The increase of the cavity voltage and the reflected power before the beam abort were acceptable.

After the dedicated machine studies, the RPO scheme was tried in beam collision operation during the physics run for five days. A beam current of up to 1.2 A was stored, and an RF power of more than 300 kW was delivered to the beam in each cavity. The RPO scheme operated stably during this period. Thus the feasibility of the RPO scheme was demonstrated, and this scheme will be used in SuperKEKB.

## 2.5 Beam instrumentation

### 2.5.1 Beam position monitor system

About 510 beam position monitors (BPMs) will be installed in the HER and 460 in the LER. The BPM system will have four functions based on discussions so far: 1) precise slow orbit measurement with a repetition rate of several Hz, 2) fast orbit measurement for orbit stabilization feedback with a repetition rate of several kHz, 3) turn by turn orbit measurement of a pilot bunch that is located at the end of a bunch train for the optics measurement during collision and 4) orbit measurement near the interaction point to maintain stable collision.

We need to take into account the following factors for the system design: 1) a large beam current, about twice that of KEKB, which could cause poor electrical contact of a button electrode, large signal power to a detector and movement of the BPM due to thermal deformation of the beam chamber, 2) use of antechambers whose lowest cut-off frequency of about 900 MHz is below the detection frequency of the present KEKB system, and 3) maximal use of the existing BPM system to minimize costs. Based on the above considerations the following efforts are in progress.

A button type pickup electrode with a small diameter of 6 mm was developed. It has a pin type inner conductor to ensure tight electrical connection. The electrode is attached to the beam chamber with a flange for ease of removal in case of problems with the electrode, and for ease of TiN coating of the chamber. It has already been tested at KEKB. No problem has been found so far.

A narrow-band super-heterodyne detector with a detection frequency of 509 MHz is under development based on a detector used at KEKB. Switch modules can be used in front of the detector to cover four electrodes with one detector.

A medium-band detector with a repetition rate of 5 kHz is also being developed for orbit stabilization feedback and optics measurement during collision. It has two signal paths: detection of 509 MHz frequency component via direct sampling followed by a digital filter, and log-ratio detection with a fast gate for optics measurement. Cost is a factor limiting full use of the medium-band detector. It will be evaluated in detail after a prototype of the medium-band detector is tested. A combination of narrowband detectors together with medium-band detectors could be a good compromise of cost and performance. For example, the medium-band detectors can be used for BPMs near sextupoles where optics will be sensitive to orbit deviations.

A special detector for the orbit measurement for the collision feedback is also being developed. A displacement sensor that measures the displacement between a quadrupole or a sextupole and a nearby BPM caused by movement of a chamber will be used in SuperKEKB. About 230 sensors are already installed in KEKB. A special reference target and sensor arms made of ceramic with a very low thermal expansion coefficient of less than $3 \times 10^{-6}\,\mathrm{K}^{-1}$ is under development for use at positions where the optics are sensitive to the orbit position.

### 2.5.2 Bunch-by-bunch feedback system

The following considerations are taken into account to design the bunch-by-bunch feedback system that damps coherent betatron and synchrotron oscillations: 1) a longitudinal feedback system, which is not used in routine operation at KEKB, is required in the LER of SuperKEKB because the growth time of longitudinal coupled-bunch instabilities caused by higher order modes of the ARES cavities is estimated to be 15 ms, which is shorter than the longitudinal radiation damping time of 19 ms, 2) noise in the transverse feedback system should be minimized to reduce the blowup of the beam size during collision as suggested by beam-beam simulation, and

3) vacuum components such as kickers, power cables, feedthroughs and BPM electrodes should withstand large beam currents.
Two sets of short strip line kickers will be installed in each ring for the transverse feedback system. The maximum kicker voltage is 8.9 kV in both rings. The expected damping time is 0.27 ms for the LER and 0.45 ms for the HER for a maximum oscillation amplitude of 0.3 mm. The expected damping time is comparable to the estimated instability growth time of electron cloud instability (ECI) in the LER and fast ion instability (FII) in the HER. The study of ECI and FII has just started. More study is needed to determine target values of the damping time. The results of further study might have an impact on the design of the feedback system. A simulation and a recent beam study at KEKB showed that the noise signal in the transverse feedback system affected the luminosity. Considering the very small beam size at the IP, the feedback noise should be reduced as much as possible. Development of low-noise front-end electronics with comb filters has started.
Two DAFNE type kickers will be installed in the LER for the longitudinal feedback system. The total kicker voltage is 3.2 kV. The expected damping time is 10 ms for an energy deviation of $4\text{x}10^{-4}$.
The iGp digital signal processing system will be used in both the longitudinal and transverse feedback systems. The system is developed under US-Japan collaboration between KEK and SLAC. FIR filters with maximum 8 taps are available. It has a down-sampling function for the longitudinal feedback system. Transient analysis of the bunch oscillation is possible for the study of beam instabilities. The system was tested successfully at KEKB, the KEK-PF and other laboratories.

### 2.5.3 Synchrotron light monitor

#### 2.5.3.1 Visible light monitor

The vertical beam size at the monitor source point is about 20 $\mu$m in both LER and HER. Measurement of the size is possible with interferometers, though it is at the limit of the resolution. An extraction mirror is installed in an antechamber slot to reduce the impedance. The limitation on the resolution is due to interferometer slit separation, which is limited by the slot height of the antechambers; this appears difficult to increase due to the presence of upstream quads.
Heat load at high beam currents causes distortion of the extraction mirrors. Input power to the mirrors is about 50% higher than that in KEKB if the bending radius of the light source bend in the LER is doubled. Investigation of new mirror structures, e.g. thin Be on thin water-cooled Cu plate, is underway in order to reduce the distortion further.
Several options for the visible light monitors, e.g., a standard interferometer using $\sigma$ polarization, an interferometer using $\pi$ polarization and a vertical polarization monitor as used at PSI, are considered for applicability to different types of measurements. The longitudinal bunch profiles will be measured by streak-cameras.

#### 2.5.3.2 X-ray beam size monitor

An X-ray beam size monitor will be introduced to provide high-resolution bunch-by-bunch measurement capability. The system has no extraction mirror; thus, low beam current dependence of the measurement is expected. We are developing a system based on coded aperture imaging, which was developed by X-ray astronomers, using a mask to modulate the incoming light. The large open-aperture of the mask of 50% gives high flux throughput for bunch-by-bunch measurements. The X-ray sources will be the last arc-bends located immediately upstream of the straight section (LER: Fuji, HER: Oho). Research and development of the monitors is in progress under collaboration with Cornell University, University of Hawaii and SLAC. Beam size measurements down to 10-15 $\mu$m have been demonstrated so far at CesrTA at Cornell.

#### 2.5.3.3 Beamstrahlung Monitor

We have a plan to install a beamstrahlung monitor developed by G. Bonvicini originally for use at CESR. The monitor uses the relative strengths of horizontal and vertical polarization components of wide-angle beamstrahlung to diagnose the quality of collisions, allowing a direct evaluation of parameters such as beam separation and relative beam sizes at the collision point. Estimation of signal and background level shows that a much stronger signal than that seen at CESR is expected.

#### 2.5.3.4 Direct detection X-ray monitor

This monitor is under consideration for supplementary size measurement. The basic principle of the monitor is to use X-rays penetrating the chamber walls downstream of bends directly to measure beam emittance. The system is very simple, with no optics elements required. It requires a small thin region (~1 mm thick) on the outer wall of the antechamber to extract the X-rays. Candidate locations for the monitor are found in the local chromaticity correction sections around the interaction point with huge beta functions and no overlap light from upstream bends.

### 2.5.4 Gated measurement system for tune, orbit and longitudinal phase of a bunch

The measurement of tune, orbit and longitudinal phase of a bunch proved to be a very useful diagnostics tool in KEKB. It is used for tune stabilization feedback, the measurement of tune shift along a train due to electron clouds, the estimation of beam-beam kick and so on. The system will also be used in SuperKEKB. A fast gate module has already been developed to select individual bunch signals separated by one RF bucket of 2 ns. A test to excite betatron oscillation of a particular bunch was successfully carried out with a phase-locked loop. As a feasibility test of the optics measurement during collision, the turn-by-turn orbit of a bunch located 6 ns after a bunch train was measured with a gated circuit. The experiment showed that the system had the sensitivity needed to measure $x - y$ coupling parameters near the IP.

## 2.6 Vacuum System

### 2.6.1 Issues for vacuum system

The design of the vacuum system for SuperKEKB is based on our operational experiences at KEKB, but also introduces some novel concepts at the same time, in order to meet the challenging beam parameters to achieve an unprecedented luminosity.

The first major issue for the vacuum system comes from the high beam currents. The synchrotron radiation (SR) power and the photon density are consequently high, and the resultant heat and gas loads are also large. Another issue comes from the high bunch current and the short bunch length. Higher order modes (HOM) are likely to be excited and the loss factors of various vacuum components need to be minimized to keep the HOM power losses as low as possible. Beam impedance is also a key issue for maintaining small beam emittances. The electron cloud effect in the positron ring (LER) and the fast ion effect in the electron ring (HER) are more severe than before. Described here are the basic vacuum system designs mainly for arc sections of the rings. A design for the IR will be presented elsewhere.

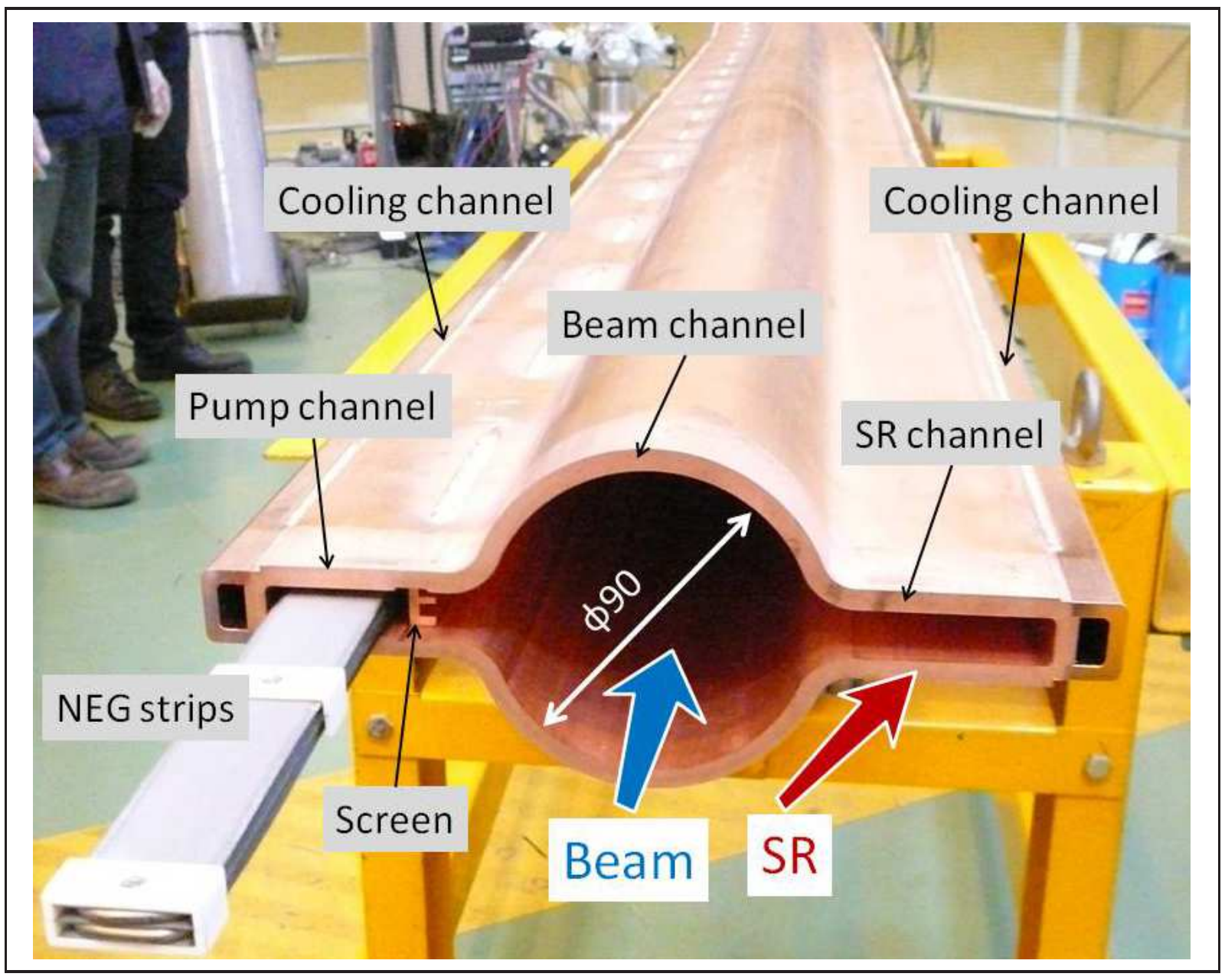

*Figure 2.14: Beam chamber with antechambers for the LER.*

### 2.6.2 Beam chamber

Beam chambers for the arc sections have an antechamber structure [18]. A beam chamber consists of a beam channel and two antechambers at both sides. A schematic view of an antechamber for the LER is shown in Fig. 2.14 . The beam goes through a beam channel, and the SR passes through an antechamber located at the outside of the ring (SR channel) and hits the side wall of the antechamber. The beam chamber should be nearly circular in order to minimize the incoherent tune shift due to the image charge.

Pumps are provided in an antechamber (pump channel) at the inside of the ring. The pump channel is connected through a screen with many small holes. Two cooling channels are provided on the outside of the antechambers. A beam position monitor (BPM) section will be fabricated in a block, as in the present KEKB, with the same cross section as that of the beam chamber. When a single-pipe chamber is used in some sections, in a straight section for example, tapers are required at the transitions from an antechamber structure to a single-pipe one. The cross section, as shown in Fig. 2.14, fits within the present magnets except for vertical correctors.

One advantage of the antechamber scheme is that the power density of the SR can be reduced. Since the incident point of SR on the side wall is far from the emitting point, the incident SR power density is diluted. The antechamber scheme also has the advantage of a small beam impedance. Since the pumping channels are located in the antechamber, the pumping holes have little effect on the beam. Photon masks are also placed in the SR channel. For the LER, the antechamber structure is very important in reducing the effects of the photoelectrons, as described later.

In the case of the HER, if an antechamber structure with a horizontal half-width of 90 mm is used, the maximum SR power density at a beam current of 2.62 A is approximately 11 kW/m (19 W/mm$^2$), which is almost the same level to the case of the present KEKB HER (1.4 A).

The wiggler section in the LER also requires the antechamber structure to reduce the SR power density down to 20 kW/m (20 W/mm$^2$) by using a beam chamber with a half-width of 110 mm. Copper (Oxygen free copper, OFC) is therefore a suitable material in these cases. No lead shielding is required for copper. Copper beam chambers with antechambers can be formed by a cold-drawn method.
On the other hand, for the LER arc section, the SR power density is approximately 2.6 kW/m (2.5 W/mm$^2$) at a beam current of 3.6 A for a beam chamber with a half width of 110 mm. In this case, an aluminum-alloy beam chamber can be used. Manufacturing and welding are easier than for copper. Aluminum-alloy will be also used for special and complex chambers in straight sections. Such beam chambers with antechambers can be formed by an extrusion method.

### 2.6.3 Electron cloud issues in the LER

Here the countermeasures against the electron cloud issue in the LER are discussed. The single bunch instability due to the electron cloud is a serious problem for SuperKEKB, and more thorough countermeasures than ever before are required [19]. From simulation, the average electron density should be less than approximately $1 \times 10^{11}$ electrons/m$^3$. The antechamber scheme is effective in reducing the photoelectrons in a beam duct, which are seeds of the electron cloud.
In the high bunch-current regime, however, the secondary electrons due to electron impact play a major role in forming the electron cloud. Copper is said to have a relatively low secondary electron yield (SEY), but realizing the target electron density will be hard with the use of only copper beam chambers. A solenoid field is very effective at preventing multipactoring, and has been used successfully at KEKB [20]. The electron density around the beam is decreased by several orders of magnitude when the solenoid field is applied to a copper beam chamber. The solenoid, however, is usable only in a drift region (field free region).
Another way to suppress the secondary electron emission is to coat the inner surface with some materials having a low SEY, such as TiN, NEG or graphite. These coatings are usable for both drift regions and inside magnets. A TiN coating has been used in the LER beam chamber of PEP-II [21] and the coating technique is well developed. The measured electron densities for beam chambers with TiN coatings decreased to one half or one third of that for a bare copper beam chamber [22]. It should be noted that a bare aluminum alloy has a much higher SEY than copper and any coatings are indispensable if aluminum-alloy beam chambers are used.
Recently, it was found that a grooved surface and a clearing electrode (Fig. 2.15) are very useful to suppress the electron cloud, especially in the dipole field of a bending magnet or a wiggler magnet [23, 24]. The former reduces SEY geometrically, and the latter absorbs the electrons around the beam orbit with a static electric field. They can reduce the electron density by one or two orders of magnitude compared to that for a bare copper beam chamber. The current plan will use clearing electrodes in wiggler magnets, where the beam chamber is straight. The application of the grooved surface in a bending magnet is under consideration. Impedance issues have to be considered carefully for both cases.

### 2.6.4 Pumps

The target vacuum pressure in the ring is on the order of $10^{-7}$ Pa on average, which guarantees a beam lifetime determined by the vacuum pressure longer than the luminosity lifetime, and sufficiently long to be compatible with the injection scheme. From the viewpoint of ion instabilities in the electron ring (HER), an acceptable pressure is also on the order of $10^{-7}$ Pa on average, but is less than $10^{-6}$ Pa locally. A pressure of less than $1\times10^{-7}$ Pa will be required around the IR, especially just upstream of Belle II.
To achieve the target pressure, a linear pumping speed of approximately 0.1 $m^3s^{-1}m^{-1}$ is required, assuming a photo-desorption coefficient of $1\times10^{-6}$ molecules photon$^{-1}$. To evacuate the long, narrow beam chambers effectively, a distributed pumping scheme will be adopted [25]. The main pump is a strip-type non-evaporable getter (NEG), ST707 (SAES GETTERS Co. Ltd.). To evacuate non-active gases and to assist evacuation at relatively high-pressure regimes, sputter ion pumps are provided as auxiliary pumps. The rough-pumping system consists of a turbo-molecular pump and a dry pump. The system is completely oil-free. Ports for sputter ion pumps or rough pumps are located at the bottom of the pump channel.

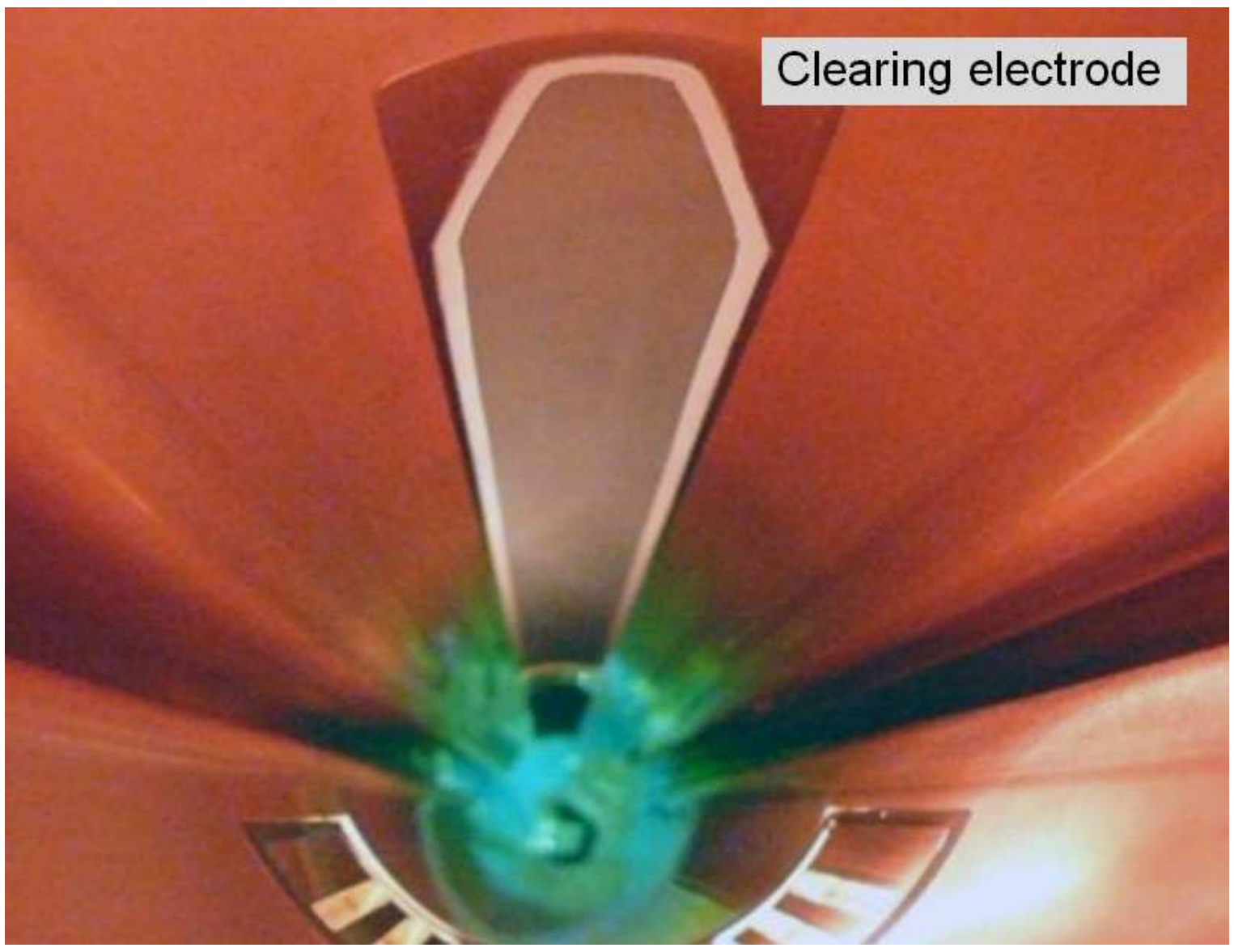

*Figure 2.15: Clearing electrode in a beam chamber.*

### 2.6.5 Other components

Bellows chambers are installed between adjacent beam chambers to ease beam chamber installation and to absorb any thermal deformation. Since the beam current is large, heating of bellows due to HOM, especially TE-mode-like HOM, will become a serious problem. The present design is equipped with a comb-type RF-shield structure that has a higher thermal strength and a lower impedance than that used before [26]. The RF shield consists of thin, interlocking comb-teeth. This shield is applicable to various cross sections of the beam chamber. The comb-type RF-shield will also be used for gate valves (Fig. 2.16).
An MO-type flange will be used as a connection flange [27]. The flange has a square edge to maintain a smooth current flow across a copper gasket as well as vacuum tightness for an ultra-high vacuum. Adaptation to a complicated cross section is possible. A copper-alloy and an aluminum-alloy flange are now available, which make the manufacturing process of beam chambers simple and also relaxes the heating problems of flanges due to wall losses.
To protect Belle II from damage due to spent particles and to reduce the background noise, a movable mask (collimator) system will be installed in each ring. The first issue for the movable mask is the HOM generated at the mask head. A chamber-type movable mask that has been used in KEKB, which has a trapped-mode-free structure, will be employed for SuperKEKB in the commissioning stage, at least [28]. HOM dampers will have to be included with the movable mask system. Another problem is the development of grooves on the mask head. An effective way to avoid the production of grooves is to use a light material as the mask head, such as graphite, with a thickness of about one radiation length. A new, improved movable mask system will be required to handle high current beams in the future.

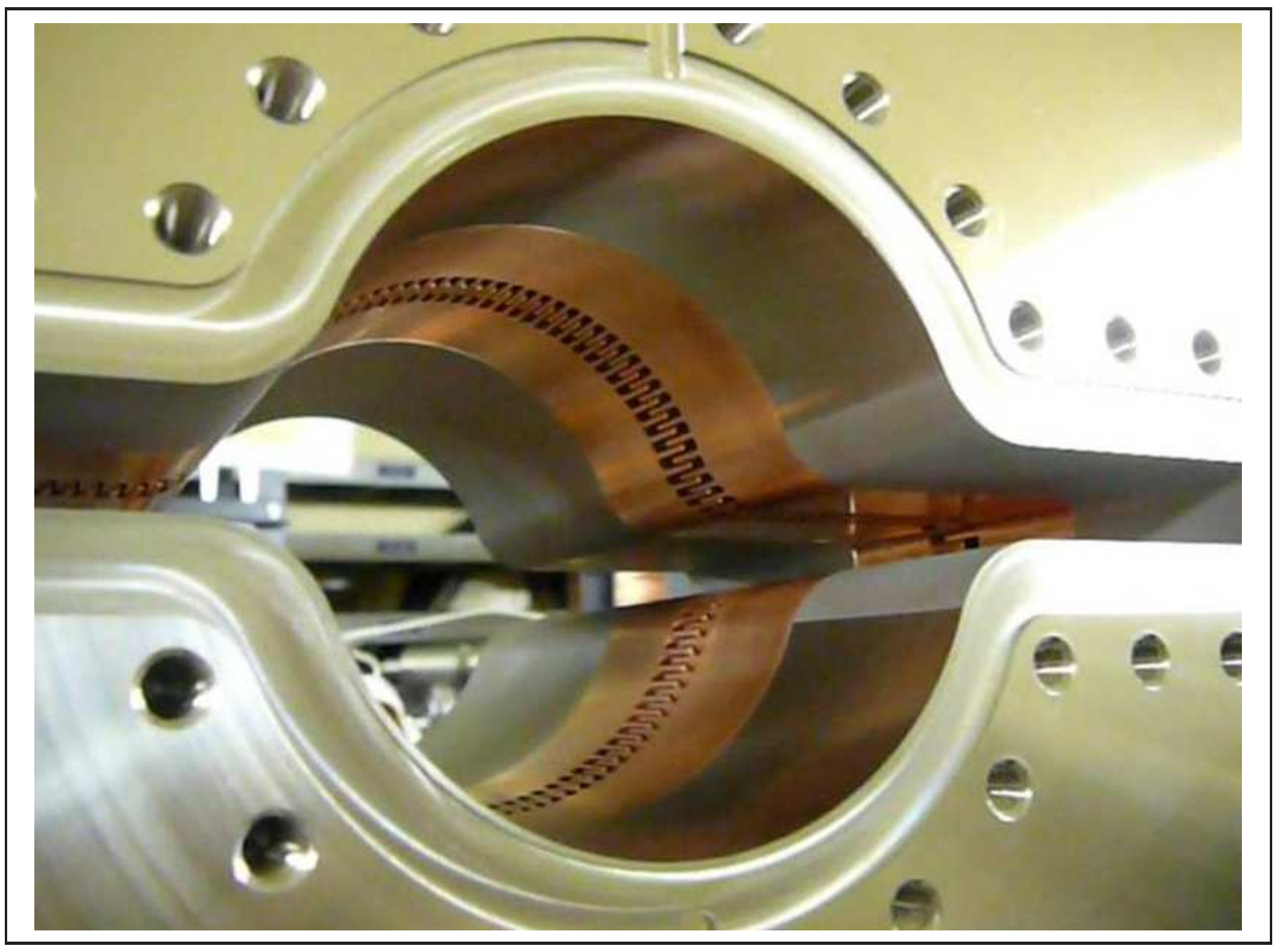

*Figure 2.16: Comb-type RF shield for gate valves.*

The design of the control and monitoring system follows that of the present system. Vacuum gauges are cold cathode gauges (CCG) and are positioned about every 10 m. The compressed air system will also be reused to actuate the gate valves. The flow rates of the cooling water and the temperatures of the vacuum components must be monitored. Highly reliable alarm and interlock systems are required.

The cooling power should be improved since the total SR power increases by a factor of two. The power loss in both rings is estimated to be approximately 14 MW in total. On the other hand, the present cooling capacity is about 8 MW in total for the vacuum system. The length of one cooling loop in the tunnel or the layout of the loop should also be optimized.

## 2.7 Injector Linac

### 2.7.1 Requirements and upgrade strategy

The present KEKB injector linac supplies 8.0 GeV electrons to the HER and 3.5 GeV positrons to the LER. The electron beam for the HER (1 nC, 2 bunches) is generated in the pre-injector at the beginning of the linac. Electrons for positron production (10 nC, 2 bunches) are also generated in the same pre-injector and accelerated up to 4 GeV to irradiate a tungsten target located in the middle of the linac to generate positrons. The positrons (1 nC, 2 bunches) are extracted by a capture section and accelerated up to 3.5 GeV in the latter half of the linac. The HER and LER injections are switched pulse-by-pulse to keep the stored currents constant.

After the transition of the SuperKEKB design strategy from the high current option to the nano-beam (low-emittance) option, the upgrade scheme of the injector linac has been revised to meet the new requirements of the storage rings. Table 2.9 shows a comparison of KEKB and SuperKEKB machine parameters concerning the injector upgrade. The most significant changes are the beam emittance and the intensity. More than one order of magnitude lower emittances are required for both electrons and positrons. Due to the doubled stored currents and the beam lifetimes as short as 10 minutes in SuperKEKB, injection beam intensities should be increased.

| | KEKB | | SuperKEKB | |
|---|---|---|---|---|
| | $e^+$ | $e^-$ | $e^+$ | $e^-$ |
| Beam energy (GeV) | 3.5 | 8.0 | 4.0 | 7.0 |
| Stored current (mA) | 1600 | 1200 | 3600 | 2620 |
| Beam lifetime (min) | 150 | 200 | 10 | 10 |
| Bunch charge (nC) | 10.0/1.0 | 1.0 | 10.0/4.0 | 5.0 |
| Number of bunches | 2 | 2 | 2 | 2 |
| Beam emittance $\gamma\epsilon_{(1\sigma)}$ ($\mu$m) | 2100 | 300 | 10 | 20 |
| Energy spread $\sigma_E/E$ (%) | 0.125 | 0.05 | 0.07 | 0.08 |
| Bunch length $\sigma_z$ (mm) | 2.6 | 1.3 | 0.5 | 1.3 |

Table 2.9: KEKB/SuperKEKB machine parameters

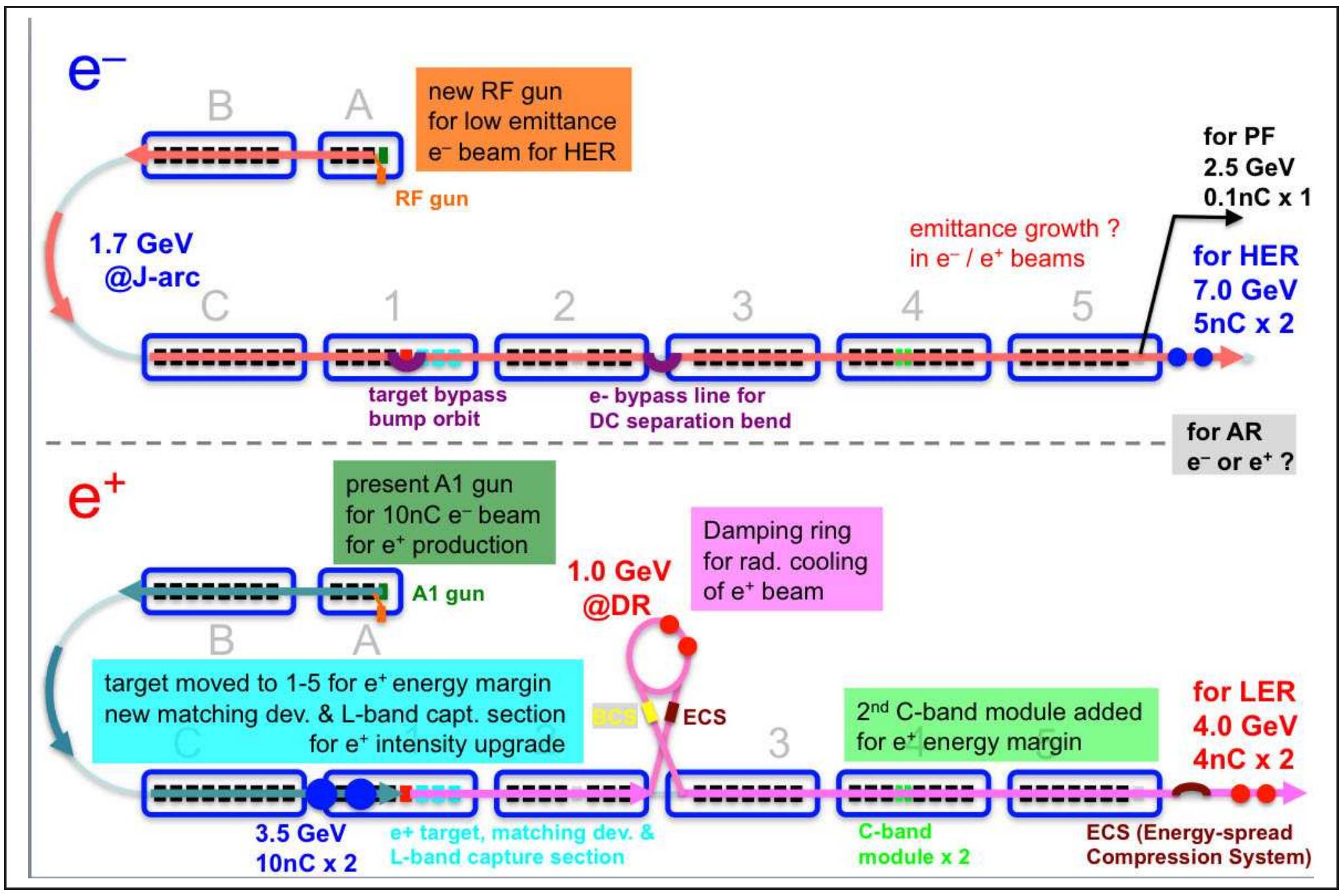

Figure 2.17: SuperKEKB injector

To achieve this smaller emittance and higher intensity of the electrons, we adopt a photocathode RF gun in a new pre-injector. A low-emittance and short-pulse beam will be generated by irradiating a small-size cold cathode with short-pulse laser photons. This beam is accelerated to 7.0 GeV in the linac and injected into the HER. For primary electrons that are used to generate positrons, the present pre-injector will be used. This is because the required electron intensity for positron production is much higher than the injection beam and it is difficult to achieve with a photocathode RF gun. Thus, these two pre-injectors will be switched pulse-by-pulse for electron and for positron injections.

To increase positron beam intensity, we adopt a new matching device and a L-band capture section to increase positron acceptance. By replacing the existing matching device (air-core pulse coil) with a flux concentrator or a superconducting solenoid, a wider positron energy acceptance will be achieved. Replacing the present S-band (2856 MHz) capture section with an L-band (1298 MHz) will give us a larger beam aperture in the accelerating structure and longer RF wave length for larger phase space acceptance. To obtain low emittance positrons, we will build a damping ring to reduce the emittance with radiation damping. The damping ring will be constructed in a location where 1 GeV positrons can be injected from the linac. Positrons are re-injected from the damping ring to the linac two-beam pulse period after the injection to make the emittance sufficiently low. Details of the damping ring are described elsewhere (Section 2.8).

### 2.7.2 Positron Source Upgrade

Since the positrons for the LER are produced as secondary particles from the converter target, the emittance is much larger than the electron beam and the intensity is limited by the acceptance of the capture section. We will enlarge the acceptance to increase the intensity for SuperKEKB. The emittance of the positrons from the capture section will be larger than that required for injection in the nano-beam scheme. To compensate for this, a damping ring (DR) is introduced. The beam energy at the DR is designed to be 1.0 GeV. The DR will be located on one side of the linac, between Sectors 2 and 3, as shown in Fig. 2.17, where there is suitable space to construct beam transport lines from and to the DR. To increase the energy margin for DR injection, considering the lower acceleration field in the L-band modules, the positron station including the target and the capture section will be moved approximately 50 meters upstream to have a longer positron acceleration region. Low emittance positrons after the damping in the DR are re-injected into the linac and accelerated up to 4.0 GeV in Sectors 3, 4 and 5. To increase the energy margin for LER injection, a second C-band accelerator module will be added in the Sector 4 in 2010.

In the present capture section, a short air-core pulse solenoid (2 T) is used as a matching device to focus positrons and an S-band accelerating field captures them in a long DC solenoid (0.4 T). With a short and sharp field distribution of the pulse solenoid, this matching device has a monochromatic energy acceptance. We will replace the present solenoid with an adiabatic matching device that has a wider energy acceptance, with a higher peak field and a more gradual field distribution. One of the candidates for this device is a flux-concentrator type of pulse solenoid, which can yield a very high field by an intense induced current inside a copper conductor. We are developing a prototype in collaboration with BINP in Russia, which has already achieved 10 T as a peak field strength. The solenoid has a certain amount of non-axial field component that kicks positrons and creates beam loss. The observed field distribution of the prototype is rather sharp and the energy acceptance was not sufficiently wide. Positron yield including these effects will be evaluated by a simulation study and by a beam study with the prototype, which will be installed in the KEKB linac in 2010. The other candidate is a super-conducting solenoid. Though it has the merit that an ideal adiabatic field distribution can be created by an appropriate coil design, quenching of the magnet by heating with particles from the target can be a problem in operation. Preliminary irradiation test of a sample super-conducting solenoid with KEKB linac beams has shown that the quenching is not sensitive to the very high heat density of the pulsed beams. Thus we can design a super-conducting solenoid for SuperKEKB with a quench limit determined by the average heating from the beams and by the field strength in the coil wire. A design study of a prototype solenoid is underway and beam irradiation studies will be performed with the prototype. Final decision on the matching device will be done after the evaluation of the studies and the tests of the flux-concentrator and the super-conducting solenoid.
To enlarge the positron acceptance, we will introduce L-band accelerating structures in the capture section. By replacing the present S-band accelerating structures with the L-band structures, we will enlarge the aperture diameter from 20 mm to 30 mm, which corresponds to 2 times larger transverse phase-space acceptance. Though the acceleration field gradient is lower in the L-band (10 MV/m) compared to the S-band (14 MV/m), a longer capture section can compensate for this. Since the wave length of the L-band RF is twice as long as that of the S-band, the longitudinal acceptance will increase as well. The positron intensity is expected to increase by more than a factor of four by introducing the adiabatic matching device and the L-band capture section. The L-band system is also introduced in the three subsequent accelerator modules for the quadrupole focusing system to have comparable transverse acceptance to the capture section. Development work on components for the L-band modules is underway. A klystron of 40 MW output power and the first prototype L-band accelerating structure are under fabrication and high-power tests will be performed in a test stand. Accelerating structures in a solenoidal field and in intense radiation from a converter target are sensitive to RF breakdowns. After studies to investigate the breakdown properties of the structures, the final design of the L-band accelerating structures will be decided.

### 2.7.3 Electron Source Upgrade

In the present KEKB linac pre-injector, a triode-type thermionic electron gun with a dispenser cathode of 200 mm$^2$ size generates an electron beam. Electrons are extracted from the cathode by a electric field of 200 kV accelerating voltage. A minimum pulse length (typically 1 ns) of the beam from the cathode is limited by a grid pulser signal and the beam intensity is controlled by a grid pulse and a DC bias voltage. The electron beam pulse is compressed by RF buncher cavities (two sub-harmonic bunchers of 114 MHz, 571 MHz, a pre-buncher and a buncher of 2856 MHz) to yield a single bunched beam of 10 ps length (FWHM) with no satellite bunches, to fulfill the longitudinal acceptance requirement of the HER and LER. To double the injection current to the rings, a second beam pulse is generated 96 ns after the first pulse by a grid pulse signal from a second pulser. This second beam pulse is also compressed by the RF bunchers into a single bunch. These two bunches, at a 96 ns interval, are accelerated in the same RF pulse. Bunch intensities are switched for each operation mode, with 1 nC for the HER injection, 10 nC for positron generation and 0.1 nC for synchrotron facilities (PF and AR). The typical normalized emittance of electrons from the pre-injector is 100 $\mu$m or less.
For HER injection in SuperKEKB, the electron emittance is required to be less than 20 $\mu$m. In addition, the bunch intensity is required to be 5 nC, which is five times the present value. To fulfill the emittance and the intensity requirements, we will develop a new pre-injector with a photocathode RF gun. For generating a low-emittance beam, the cathode size should be small as 7 mm$^2$ to make a low intrinsic emittance, and the acceleration field gradient should be high (more than 100 MV/m) to suppress space charge effects in the low velocity region. An RF field in a cavity is used to achieve this high gradient instead of a static field. To make a compact new pre-injector that can generate single bunch beams without RF buncher cavities, a photoemission cathode with a short laser pulse of 10 ps order is used. Eliminating bunchers is advantageous in avoiding emittance growth as well.
To achieve the specified beam intensity with this small cathode size, a sufficiently high intensity laser should be used considering a quantum efficiency of the cathode material. Taking a photocathode in KEK-ATF as an example [29], a $Cs_2Te$ cathode of 1% quantum efficiency and a 266 nm laser of 3 $\mu$J/bunch intensity are used to generate a 1.6 nC/bunch beam. Considering the short lifetime of the $Cs_2Te$ cathode, a more tolerant material like copper is preferred for long-term operation in SuperKEKB. Yet, since the quantum efficiency of copper is very low ($3 \times 10^{-5}$), the laser intensity should be very high. For the SLAC-LSLC injector [30], a 255 nm laser of 500 $\mu$J/bunch is used to irradiate a copper cathode to generate a beam with 1 nC/bunch intensity. To achieve 5 nC beam in this scheme, 2500 $\mu$J/bunch laser is required and it is beyond the power level of lasers available now. Our idea to overcome this difficulty is improve the quantum efficiency by two orders of magnitude by adopting 205 nm wavelength instead of 260 nm, as suggested by the test data of wavelength dependence measured at CERN [31]. A higher field gradient in the RF cavity helps to enhance the quantum efficiency. In this wavelength with an assumed 170 MV/m field gradient, the required laser power is expected to be less than 100 $\mu$J/bunch.

According to an estimation of contributions to the beam emittance by thermal effects, rf field effects and space charge effects, it is suggested that the space charge effect is dominant for typical designs of RF gun cavities. This effect can be suppressed with a higher field gradient. Thus we adopt a SPring8-type half-cell cavity [32] which can yield the highest field gradient among candidate cavity designs. This type of cavity, operated at an S-band frequency, is expected to achieve a 170 MV/m field gradient.

Even though the new pre-injector can yield a low-emittance beam, wake field effects during acceleration can degrade the emittance. Care should be taken with the longitudinal charge distribution in the beam bunch and of the design of the beam focusing system to suppress the emittance growth.

Detailed design study and prototype development of the photocathode RF gun will be performed in the coming few years.

## 2.8 Damping Ring

The injection aperture of the LER is 0.5 $\mu$m. Assuming a septum width of 4 mm and a beta function at the injection point of 120 m, the emittance of the injected beam must be less than 4 nm. The emittance of the positron beam, which is generated with a conventional target, is as large as 2.1 $\mu$m at an energy of 1 GeV. Thus a damping ring is necessary to reduce the emittance by a factor of 130.

The intensity of the injected beam, on the other hand, has to be larger than 4 nC per Linac-pulse in order to maintain the current of the LER, whose lifetime is expected to be as short as 600 sec. For a high-intensity positron beam, the Damping Ring (DR) therefore serves as a collector ring that accepts the beam with a large energy spread and a large transverse emittance.

The DR is placed at the end of Sector 2 of the Linac, at an energy of 1 GeV. Parameters of the incoming beam to the DR are shown in Table 2.10. One of the nice features is an energy compression system (ECS) incorporated in the injection line to the DR. The ECS reduces the energy spread of the beam from $\pm 5\%$ to $\pm 1.5$ %, which is the acceptance of the DR.

Parameters of the DR are shown in Table 2.11. We adopt a FODO cell with a reverse bend as the arc cell. This structure allows us to get low momentum compaction while preserving the short damping time. The injected bunches stay for two Linac pulses (40 ms) before extraction. We assume 8 nC per bunch as the design intensity, which gives us a great operability in the case of shorter life time in the LER. The bunch length of the extracted beam is compressed from 5 mm to 0.5 mm by the BCS system incorporated in the RTL line.

The dynamic aperture is vitally important to accommodate the large emittance and large energy spread of the beam. Figure 2.18 shows the results of a tracking simulation.

Table 2.10: Parameters of the injected beam.

| | | before ECS | after ECS |
|---|---|---|---|
| Energy | (GeV) | 1.0 | |
| Repetition frequency | (Hz) | 50 | |
| Emittance | ($\mu$m) | 2.1 | |
| Energy spread* | (%) | ± 5 | ± 1.5 |
| Bunch length* | (mm) | ± 8.7 | ± 28.2 |
| Number of bunches per pulse | | 2 | |
| Bunch spacing | (ns) | 98 | |
| Bunch charge | (nC) | 8 | |

* Full width

Table 2.11: Parameters of the Damping Ring

| | | | | |
|---|---|---|---|---|
| Energy | | 1.0 | | GeV |
| Circumference | | 135 | | m |
| Number of bunch trains | | 2 | | |
| Number of bunches/train | | 2 | | |
| Maximum stored curent* | | 70.8 | | mA |
| Energy loss per turn | | 0.0714 | | MV |
| Horizontal damping time | | 12.66 | | ms |
| Injected-beam emittance | | 2.1 | | $\mu$m |
| Stored current | 0 | 35.4 | 70.8 | mA |
| Equilibrium emittance | 11.7 | 13.0 | 14.0 | nm |
| Coupling | | 10 | | % |
| Emittance at ejection(hor.) | 15.3 | 16.6 | 17.6 | nm |
| Emittance at ejection(ver.) | 4.8 | 5.0 | 5.1 | nm |
| Energy band-width of injected beam | | ±1.5 | | % |
| Energy spread | 0.0526 | 0.0535 | 0.0542 | % |
| Bunch length | 5.2 | 5.3 | 5.4 | mm |
| Momentum compaction factor | | 0.00191 | | |
| Cavity voltage for 1.5 % bucket-height | | 0.257 | | MV |
| RF frequency | | 509 | | MHz |
| Inner diameter of chamber | | 32 | | mm |
| Bore diameter of magnets | | 44 | | mm |

* 8 nC.bunch

In Fig. 2.18(a), relative-strength errors of quadrupoles ($1 \times 10^{-3}$) and sextupoles ($2 \times 10^{-3}$), transverse misalignments of quadrupoles/sextupoles (0.15 mm), rotation error (0.3 mrad), and BPM offset of 0.15 mm are assumed. In Fig. 2.18(b), the systematic high-order allowed multipole error of quadrupoles and sextupoles are added, where the strength of the multipole field error of 0.06% and 0.23% at the radius of 22 mm were assumed for the first allowed multipole and higher multipoles, respectively. The dynamic aperture is sensitive to the systematic high-order multipoles. They can be corrected by end-shimming after field measurement.
The electron cloud density near the beam is estimated to be $0.7 \times 10^{14}$ m$^{-2}$, which is close to the threshold of electron-cloud instability of $1.57 \times 10^{14}$ m$^{-2}$, assuming a secondary electron emission coefficient $\delta$ of 1.2. Methods for mitigation of the effect is under investigation.

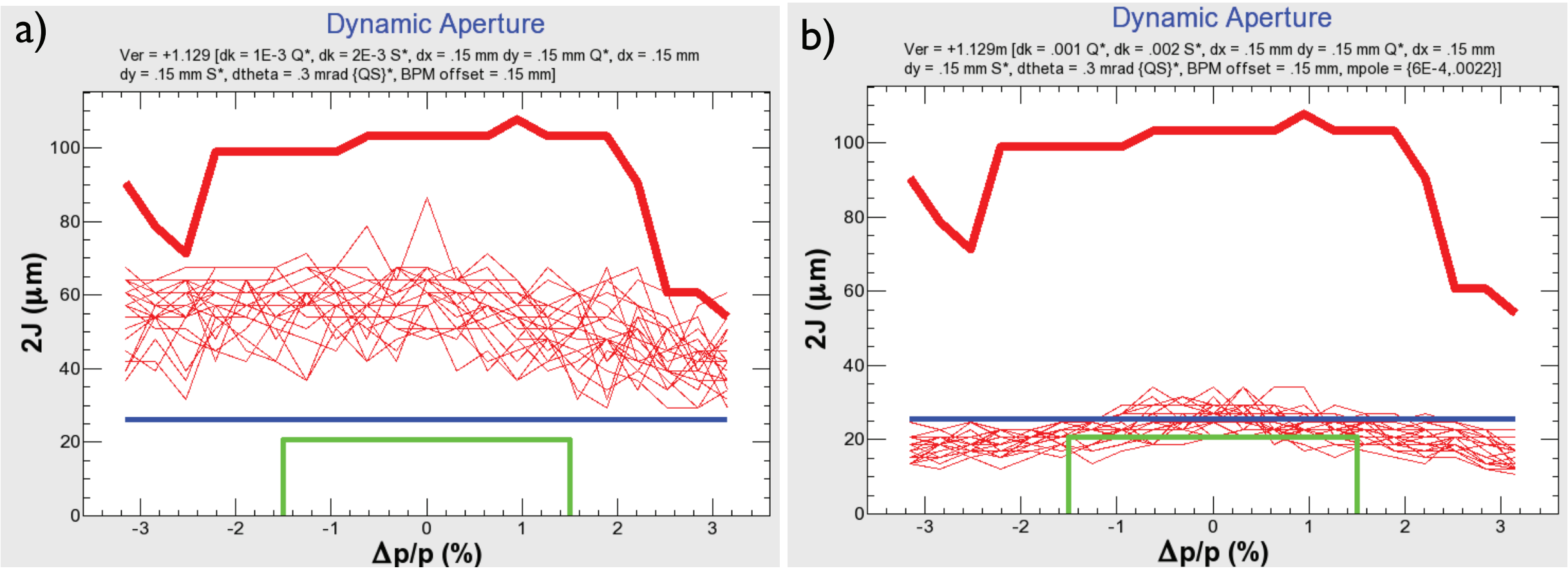

*Figure 2.18: Tracking results of dynamic aperture. The horizontal axis is the initial momentum deviation and the vertical axis is the maximum action of the survived particle after 4000 turns. The green line corresponds to the injected beam. The thick red line is the result for the case of no machine error while the thin red lines are for the case of machine errors. (a) Machine errors include relative strength error of quads ($1 \times 10^{-3}$) and sexts ($2 \times 10^{-3}$), transverse misalignments of quads/sexts (0.15 mm), rotation error (0.3 mrad), and BPM offset of 0.15 mm. (b) Systematic higher-order allowed multipole error of quads and sexts are added to the errors of (a).*

## 2.9 Control

The design of the control system for SuperKEKB is basically an extension of the present KEKB control system. Its software is based on EPICS (Experimental Physics and Industrial Control System). EPICS is an open source software toolkit used to construct distributed control systems mainly for accelerators and other large physics experiments. In the EPICS system, the control system hardware consists of the OPI (Operator Interface) and IOC (Input Output Controller). The OPI is a workstation which can run various EPICS tools. The IOC is a frontend computer which has interfaces to the various components of the accelerator. Many IOCs and OPIs are connected each other through a LAN (local area network).
In the present control system, VME board computers with VxWorks realtime operating systems are mainly used as the IOCs. There are roughly 90 VME IOCs for the LER, HER and beam transport lines. Because these VME computers were installed more than 10 years ago and have become outdated and are expected to be hard to maintain for the next decade, we have the following plans.

1. Upgrade the VME CPU board
2. Introduce new type IOCs other than VME-based one

Most of the current VME CPU board are Force PowerCore-6750. We are planning to replace them with Emerson MVME5500 (or MVME4100). The replacements have already started partially. We are continuing the replacements. At the same time it is necessary to upgrade VxWorks and EPICS for these IOCs. VxWorks will be upgraded from 5.3 to 5.5 (or 6.7 or higher, which we are evaluating). EPICS will be upgraded from R3.13.1 to R3.14.
It is also required to replace the outdated field buses, especially the CAMAC system, which has been used for more than 20 years and has become hard to maintain. ARCNET has been used in the magnet power supply control system. The PSICM (Power Supply Interface Controller Module) was designed with the ARCNET interface in the present system. The number of magnet power supplies is expected to increase in SuperKEKB. Thus, we have started developing ePSICM (Ethernet-based Power Supply Interface Controller Module) for the additional magnet power supplies. The ePSICM is the newer version of the PSICM using an Ethernet interface instead of ARCNET.
We are developing various types of EPICS IOC-embedded equipment. In our new configuration, EPICS software is installed more on the frontend side. It makes the controlled equipment itself being IOC. In our design plan the following new type of IOCs are being developed.

1. F3RP61, a CPU module running Linux for YOKOGAWA FA-M3 PLC;
2. EPICS embedded oscilloscope, which is a Tektronix DPO7104 with Windows embedded;
3. The prototype of the ePSICM with EPICS embedded;
4. Embedded EPICS on micro-TCA cards for new LLRF system.

The F3RP61 is the key product of the new control system. Generally speaking, most of the control system does not require high speed nor complicated data processing, but it needs to handle various kinds of equipment. For such purpose the F3RP61 is suitable. We have installed EPICS on it to be a light weight IOC. Compared to the VME IOC it is cheaper, smaller and less power consuming. It has no fans and is sufficiently reliable. It reduces not only hardware cost but also the cost of software development. User space programming on Linux and simple synchronous I/O access make software development much easier. Although graceful handling of high-density signal cable is required, we are considering the F3RP61 as the first candidate of the VME/CAMAC alternatives.
Fast control is also expected to be required in some cases. We are preparing for such requirements with a new event-based control system. The OPIs and network are going to be upgraded. We are introducing Linux Blade Servers as OPI. The replacement of the network core switch and edge switches are planned and partially in progress. In the new network system we are introducing redundant configuration, VLAN, and potentially 10Gbps connections for future purposes.

## 2.10 Schedule

A tentative schedule for the SuperKEKB construction is shown in Fig. 2.19. The Injector upgrade and Damping Ring construction are shown in the upper part of the figure, and the Main Ring construction is shown in the lower part. The construction will start after the final run of the present KEKB operation, which runs through June, 2010. The commissioning of SuperKEKB will begin in April 2014.

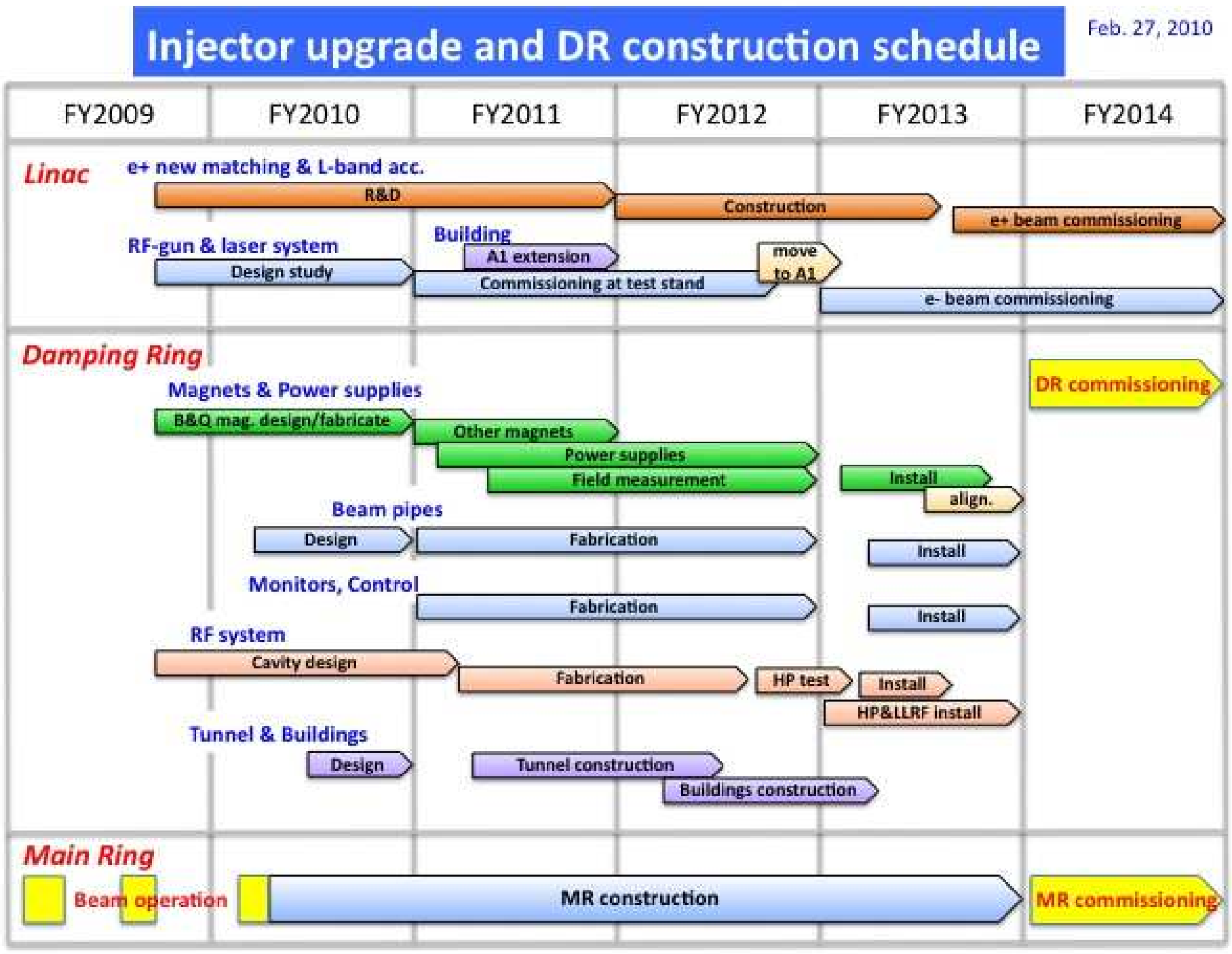

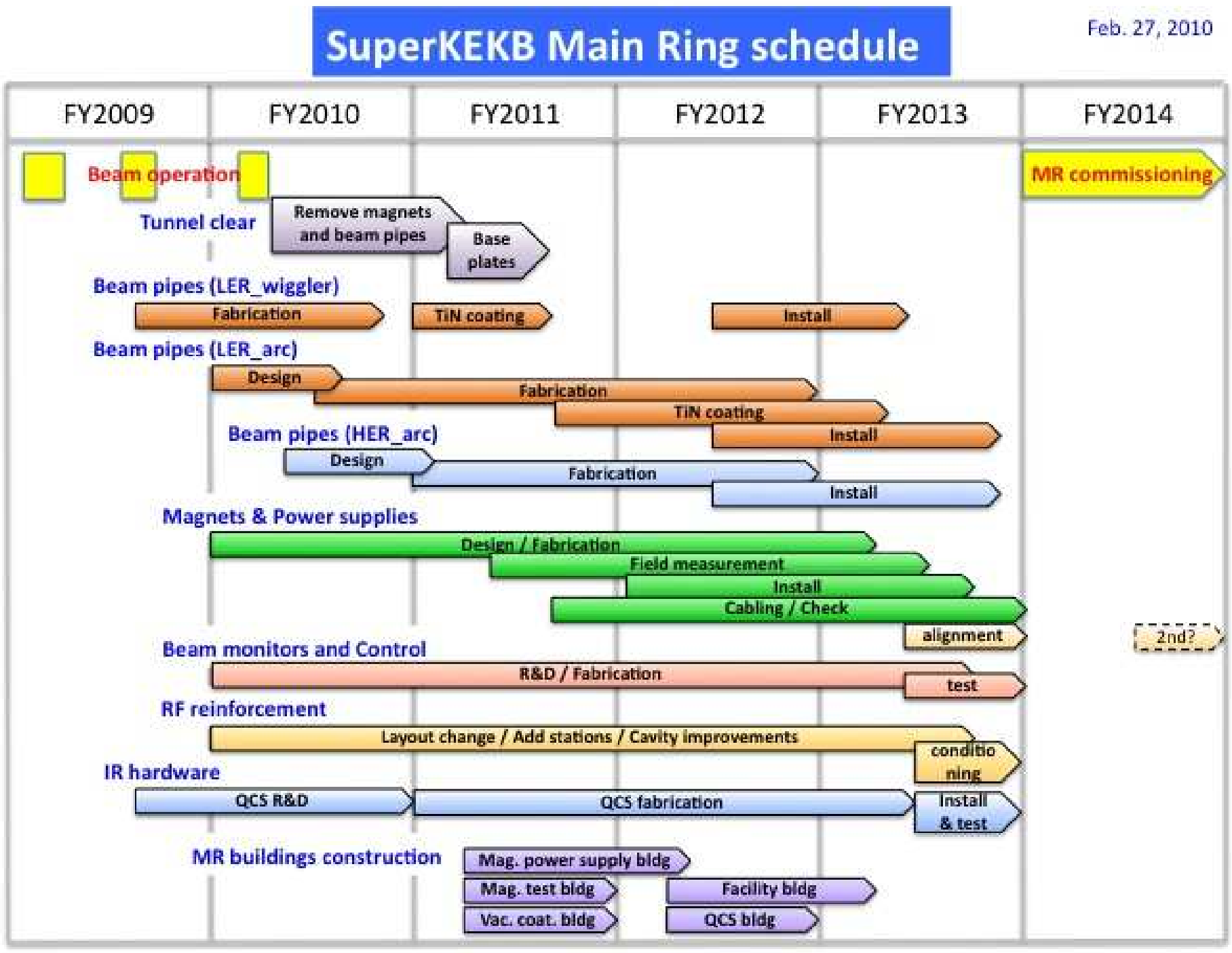

Figure 2.19: Schedules for injector upgrade and damping ring (upper) and Main Ring (lower).

# Chapter 3

# IR design

## 3.1 Interaction Region

The 7 GeV electrons stored in the high-energy ring (HER) and the 4 GeV positrons in the low-energy ring (LER) collide at one interaction point (IP) with a non-zero crossing angle of 83 mrad. The large crossing angle helps to separate the two beams quickly and allows the final quadrupole magnets to be closer to the IP in order to achieve small $\beta$-functions at the IP. Some pertinent parameters are listed in Table 3.1.

*Table 3.1: Beam parameters of SuperKEKB.*

| | LER | HER | |
|---|---|---|---|
| Energy | 4.0 | 7.0 | GeV |
| $\beta_x^*$ / $\beta_y^*$ | 32 / 0.27 | 25 / 0.41 | mm |
| $\epsilon_x$ / $\epsilon_y$ | 3.2 / 12.8 | 2.4 / 8.4 | nm / pm |
| Coupling | 0.4 | 0.35 | % |
| Horizontal beam size at IP | 10.2 | 7.75 | $\mu$m |
| Vertical beam size at IP | 0.059 | 0.059 | $\mu$m |
| Beam-beam parameter | 0.09 | 0.0875 | |
| Bunch length | 6 | 5 | mm |
| Beam current | 3.6 | 2.6 | A |
| Betatron tune ($\nu_x/\nu_y$) | 45.53 / 45.57 | 58.53 / 52.57 | |
| Synchrotron tune | -0.023 | -0.012 | |
| Number of bunches | 2503 | | |
| Crossing angle | 83 | | mrad |
| Luminosity | $8 \times 10^{35}$ | | $\mathrm{cm}^{-2}\mathrm{s}^{-1}$ |

A schematic layout of the beam line near ithe IP is shown in Fig. 3.1. This figure shows five superconducting quadrupole magnets (QC1LP, QC1LE, QC1RP, QC1RE and QC2RP) and three permanent quadrupole magnets (QC2LP, QC2LE and QC2RE). The final vertical and horizontal focusing fields will be provided by the QC1 and QC2 magnets in each ring, respectively. These superconducting magnets overlap with the solenoids (S-L and S-R) to compensate for the 1.5 T field created by the Belle solenoid. The field axis of the compensation solenoids is aligned to the detector solenoid. The superconducting quadrupole and solenoid magnets on each side of the IP are contained in a common cryostat, respectively. These magnets have a warm bore vacuum chamber connected to the cryostats via bellows. QC2LP, QC2LE and QC2RE, which are located outside of cryostats, are designed as Halbach-type permanent magnets. The cryostats will be small and the assembly of the vacuum chamber is simplified by using permanent magnets. The parameters for these magnets are summarized in Table 3.1.
The inclination of the HER and LER orbits relative to the solenoid axis, which is 41.5 mrad, respectively, is chosen to minimize the vertical emittance degradation for the HER in the fringe field of the anti-solenoid. Because it is not easy to rotate the HER orbit without degradation of its performance, rotation of the detector is under consideration at present.

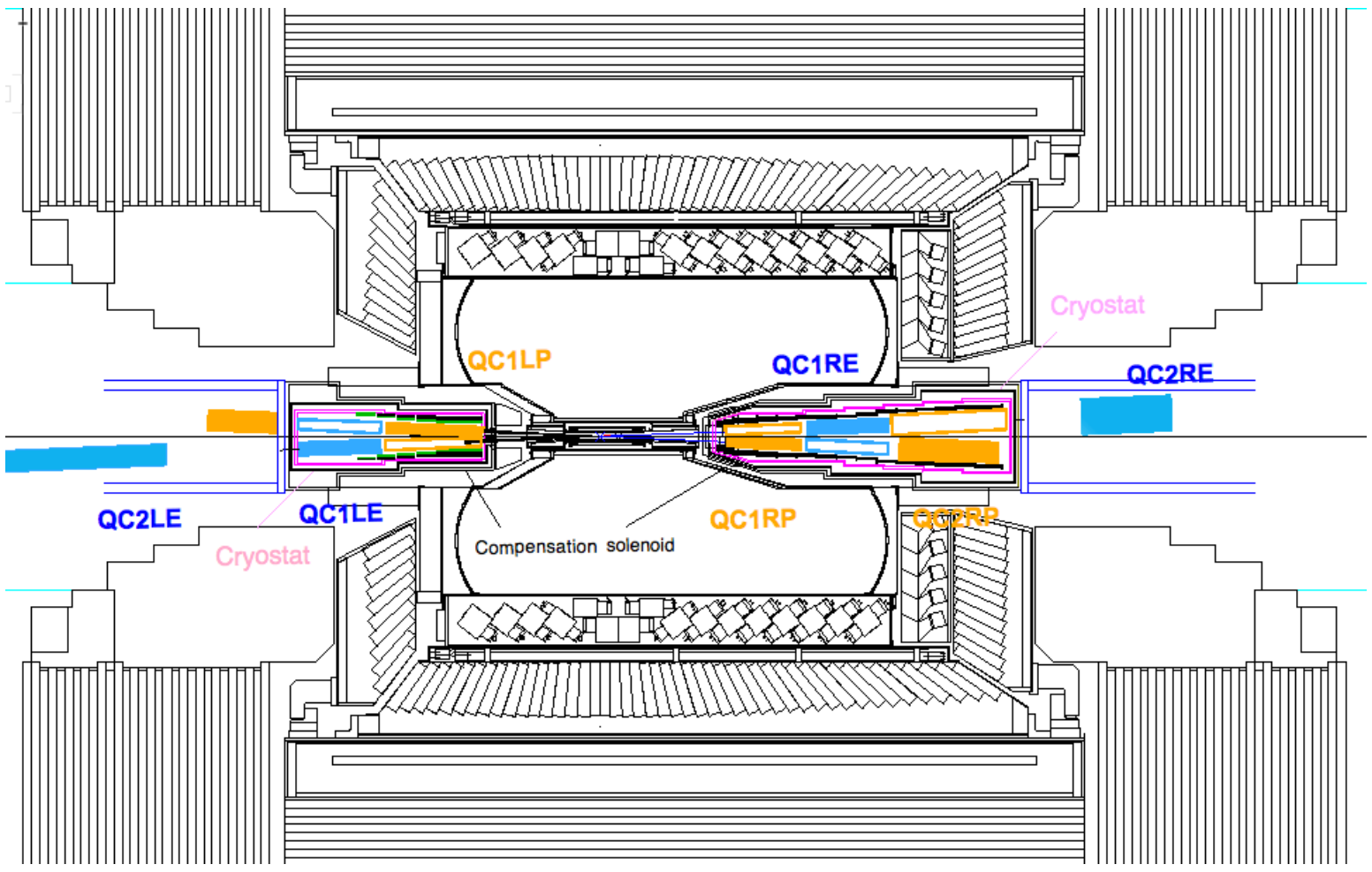

*Figure 3.1: Schematic layout of the magnets and beam line near the interaction point.*

*Table 3.2: Locations of the magnet center from IP, effective length, field gradients, magnet type of final focus magnets and inner diameters of vacuum chamber are shown.*

| Magnet | $z$ (m) | $L_{\rm eff}$ (m) | Field gradient (T/m) | Focus | Type | I.D. of vac. ch. (mm) |
|---|---|---|---|---|---|---|
| QC2LE | -2.90 | 0.6 | 23.6 | Horiz $e^-$ | Permanent | 70 |
| QC2LP | -1.98 | 0.35 | 31.3 | Horiz $e^+$ | Permanent | 60 |
| QC1LE | -1.46 | 0.36 | 72.3 | Vert $e^-$ | S.C. | 30 |
| QC1LP | -0.92 | 0.39 | 58.7 | Vert $e^+$ | S.C. | 20 |
| QC1RP | 0.91 | 0.28 | 80.1 | Vert $e^+$ | S.C. | 20 |
| QC1RE | 1.38 | 0.36 | 72.8 | Vert $e^-$ | S.C. | 30 |
| QC2RP | 1.94 | 0.35 | 31.2 | Horiz $e^+$ | S.C. | 60 |
| QC2RE | 2.93 | 0.4 | 32.3 | Horiz $e^-$ | Permanent | 70 |

The beam stay-clear for the QC1 and QC2 magnets has been determined by considering the requirements for injection. The acceptance required for beam injection is assumed as $2J_x = 5 \times 10^{-7}$ (m) for the horizontal direction and $2J_y = 2 \times 10^{-8}$ (m) for the vertical direction. Dynamic effects are ignored because the nominal horizontal betatron tune is far from the half-integer resonance. The inner diameter of the vacuum chamber is 20 mm for QC1RP and QC1LP and 30 mm for QC1RE and QC1LE.

### 3.1.1 IR magnets

The final focusing system for the two beams consists of five superconducting quadrupoles, three permanent quadrupoles, two superconducting compensation solenoids, and 32 superconducting corrector coils. In the following sections, the magnet designs and specifications are described.

**QC1RP and QC1LP**

QC1RP and QC1LP have the most stringent space constraints since these magnets are located closest to the IP with a narrow gap of the two beams. They provide the final focusing for the LER ($e^+$) beam. The cross section of QC1RP is shown in Fig. 3.2. The magnets QC1RP and QC1LP have the same cross section. The design parameters of these quadrupoles are listed in Table 3.3. The magnet inner radii are 22 mm, and the magnetic lengths of QC1RP and QC1LP are 0.2967 m and 0.3695 m, respectively. QC1RP and QC1LP generate integral field gradients of 22.43 and 22.91 T at the operating currents of 1510.7 A and 1232.3 A, respectively. Combining their magnetic fields together with the solenoid and the compensation solenoids, the maximum fields in the magnets are calculated to be 2.70 T and 3.98 T for QC1RP and QC1LP. These operating points correspond to 66% and 80% of the critical conditions for superconductivity, respectively.

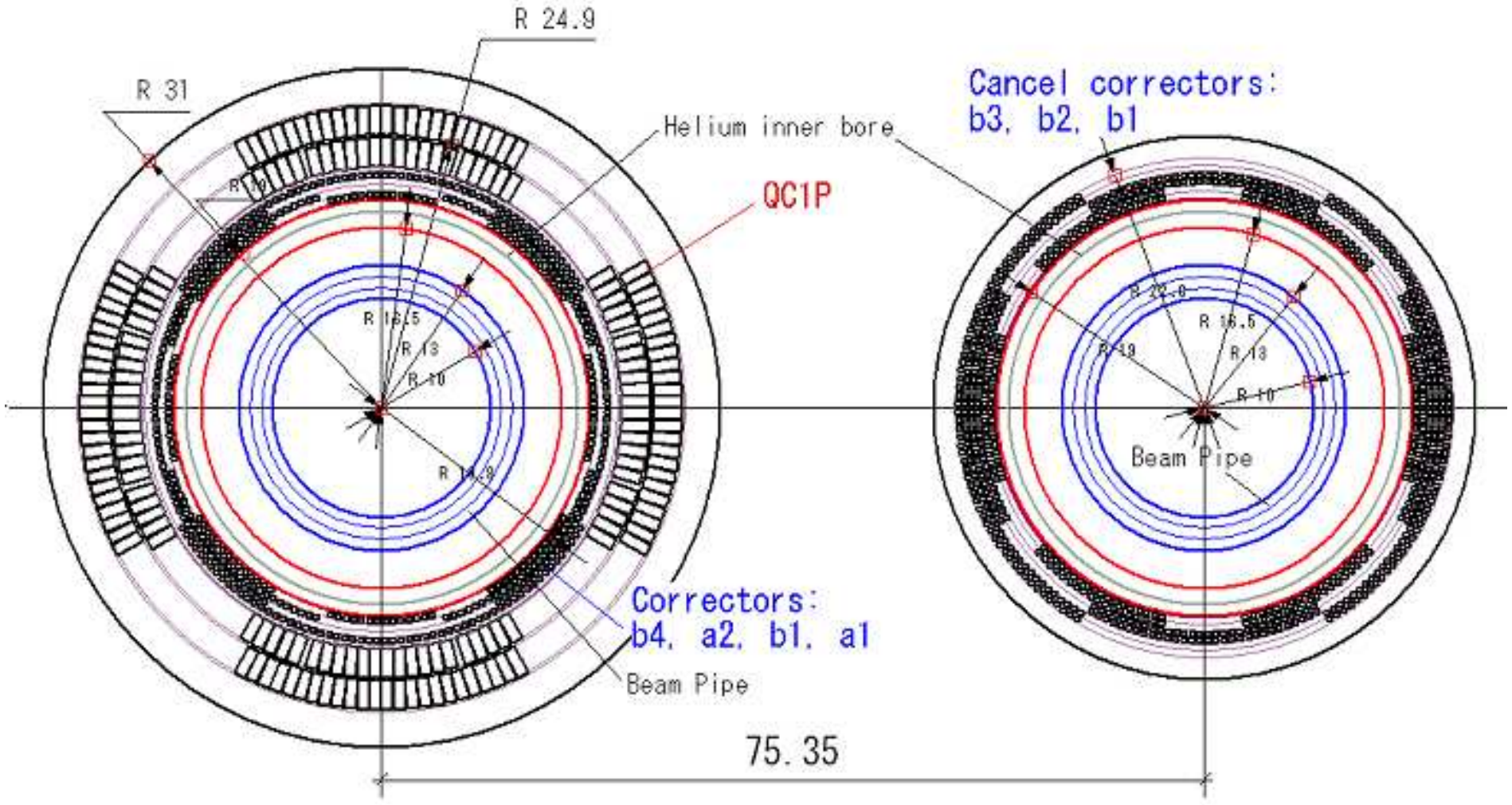

*Figure 3.2: Cross section of QC1RP and the cancel corrector coils of the QC1RP leakage field.*

Within the magnet bore, the beam pipe is located and should be kept at room temperature in the vacuum of the cryostat. Since the distance between the beam pipe and the helium vessel at 4.5 K is only 3.5 mm, the cryostats must be fabricated accurately and have good thermal performance.

Inside the magnet bores, four types of superconducting corrector coils are designed. The skew and normal dipoles ($a_1$ and $b_1$ in Fig. 3.2) align the magnetic centers of QC1R/LP. The skew quadrupole is for the alignment of the mid-plane of the quadrupole. The octupole ($b_4$) is required to improve the dynamic aperture.

Since QC1RP and QC1LP are iron-free magnets, their magnetic fluxes pass also through the neighboring HER beam line. These leakage fields have a harmful influence on the beam dynamics of the electron beam, and consist mainly of dipole, quadrupole and sextupole components ($b_1$, $b_2$ and $b_3$). To cancel out these fields, the correction coils $b_1$, $b_2$ and $b_3$ for the electron beam line have been designed, as shown in Fig. 3.2.

Table 3.3: Design parameters of QC1RP and QC1LP

| | QC1RP | QC1LP |
|---|---|---|
| Coil inner radius, mm | 22.00 | |
| Coil outer radius, mm | 27.55 | |
| Spec. of integral field, T | 22.43 | 22.91 |
| Field gradient, T/m | 75.61 | 62.00 |
| Effective magnetic length, m | 0.2967 | 0.3695 |
| Magnet current, A | 1510.7 | 1232.3 |
| Magnetic field by Belle and comp. sol., T | 1.2 | 3.0 |
| Max. field in the coil without solenoid field, T | 2.24 | 1.84 |
| Max. field in the coil with solenoid field, T | 2.70 | 3.98 |
| Operating point with respect to $B_c$ at 4.4 K | 66% | 80% |

**QC1RE, QC1LE and QC2RP**

The design concepts for QC1RE and QC1LE shown in Fig. 3.3 are the same as for QC1RP and QC1LP. They are designed to operate under nearly the same solenoid fields of 1.2 T and 1.0 T, respectively, and have identical magnet configurations. Their parameters are listed in Table 3.4. The inner radii and the effective magnetic lengths are 28.25 mm and 0.3596 m, respectively. To generate the field gradients of 72.91 T/m and 72.38 T/m in QC1RE and QC1LE, the operating currents are designed to be 1814.6 A and 1801.4 A. These operating points correspond to 76% and 75% of the critical conditions, respectively. The field correction scheme for QC1RE/QC1LE is the same as for QC1R/LP; it consists of 7 superconducting multi-pole coils. The design concept for QC2RP is same as QC1 magnets, and thus it is assembled in a common cryostat with QC1RP and QC1RE.

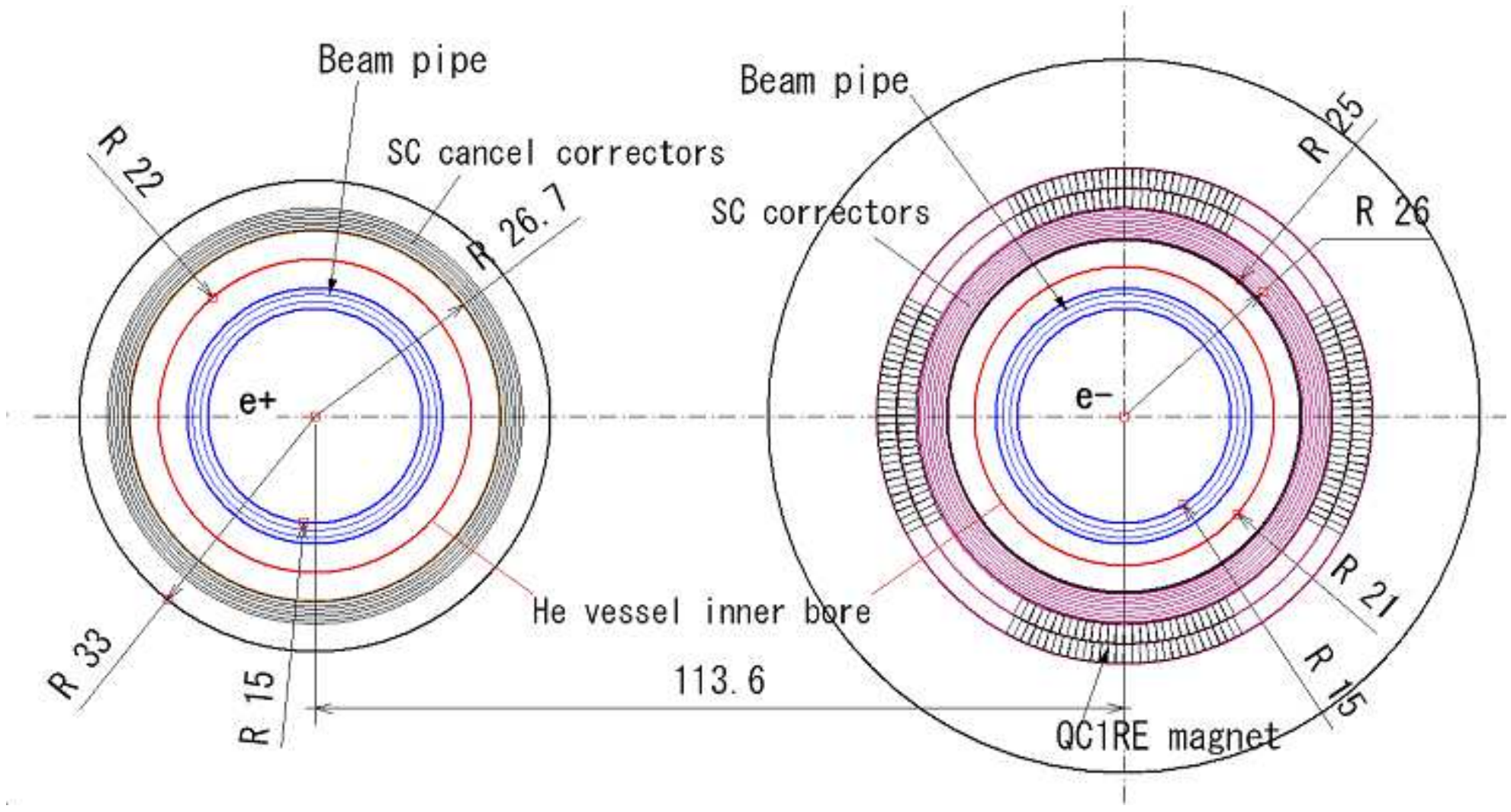

Figure 3.3: Cross section of QC1RE and the cancel corrector coils of the QC1RE leakage field.

**QC2RE, QC2LP and QC2LE**

Table 3.4: *Design parameters of QC1RE and QC1LE.*

| | QC1RE | QC1LE |
|---|---|---|
| Coil inner radius, mm | 28.25 | |
| Coil outer radius, mm | 33.80 | |
| Spec. of integral field, T | 26.22 | 26.03 |
| Field gradient, T/m | 72.91 | 72.38 |
| Effective magnetic length, m | 0.3596 | 0.3596 |
| Magnet current, A | 1814.6 | 1801.4 |
| Magnetic field by Belle and comp. sol., T | 1.2 | 1.0 |
| Max. field in the coil without solenoid field, T | 2.89 | 2.87 |
| Max. field in the coil with solenoid field, T | 3.28 | 3.15 |
| Operating point with respect to $B_c$ at 4.4 K | 76% | 75% |

QC2RE, QC2LP and QC2LE are designed as permanent magnets, and are located behind the cryostats. Their parameters are listed in Table 3.5. The field strength required for the permanent material is at most 0.9 T, and Samarium-Cobalt was selected due to its reduced sensitivity to radiation. The remnant field of Samarium-Cobalt ranges from 1.04 to 1.12 T. Study of the permanent magnet design is on-going.

Table 3.5: *Design parameters of QC2RE, QC2LP and QC2LE*

| | Integral field gradient (T) | Magnet length (m) | Required field gradient (T/m) | Beam pipe outer radius (m) | Field at coil inner radius (T) |
|---|---|---|---|---|---|
| QC2RE | 12.91 | 0.6 | 21.52 | 0.04 | 0.86 |
| QC2LP | 10.96 | 0.45 | 24.36 | 0.035 | 0.85 |
| QC2LE | 14.13 | 1.0 | 14.13 | 0.04 | 0.57 |

**Compensation solenoids**

From recent studies of beam dynamics, it is found that the fringe fields of the compensation solenoids, S-R and S-L, increase the vertical beam emittance. To reduce this effect, the compensation solenoids are designed to be segmented into small coil pieces. The solenoids house superconducting quadrupoles and correctors in their magnet bore, and their axes are coincident with the solenoid axis, as shown in Fig. 3.4. The coil pieces have gradually decreasing number of turns along the axis from the IP to have a slow gradient of the solenoid fringe field along the solenoid axis, shown in Fig. 3.5. According to these field profiles, the compensation coils are subject to repulsive electro-magnetic forces of 31.2 kN and 26.2 kN against the solenoid field. These forces are at the same level as that of the KEKB S-L magnet, and they will be taken care of in the cryostat and the support structure designs. The peak fields with the solenoid, ESR and ESL, are -2 T and -3 T, respectively. These peak fields overlap with the fields of QC1R/LP, and the operating points of S-R/S-L are calculated with these field profiles.

### 3.1.2 Vacuum chamber and IR assembly

The vacuum system for the region within 4m of the IP has to fulfill some prescribed conditions. The inner diameter of the beam duct in the region of the QC1 magnet is 2 cm. The inner surface of the beam ducts must be plated with 50 $\mu$m of gold. The beam ducts next to the IP chamber are inside a cryostat and have a small (insulation-vacuum) gap to the liquid helium vessel. The essential components to be incorporated in the beam ducts are:

1. Beam position monitors to control the beam orbit. These must have a clearly defined position reference to provide reliable information;

2. Bellows units that tolerate an independent motion between the various units of the vacuum chamber and absorb engineering errors in the fabrication and also deformations during assembly and operation;

3. Vacuum pumps to realize the local pressure without affecting the overall average pressure of the ring;

4. A structure to provide cooling of synchrotron radiation and heating due to the wall currents etc. so that the temperatures of the ducts are acceptable for the duct itself as well as outside components.

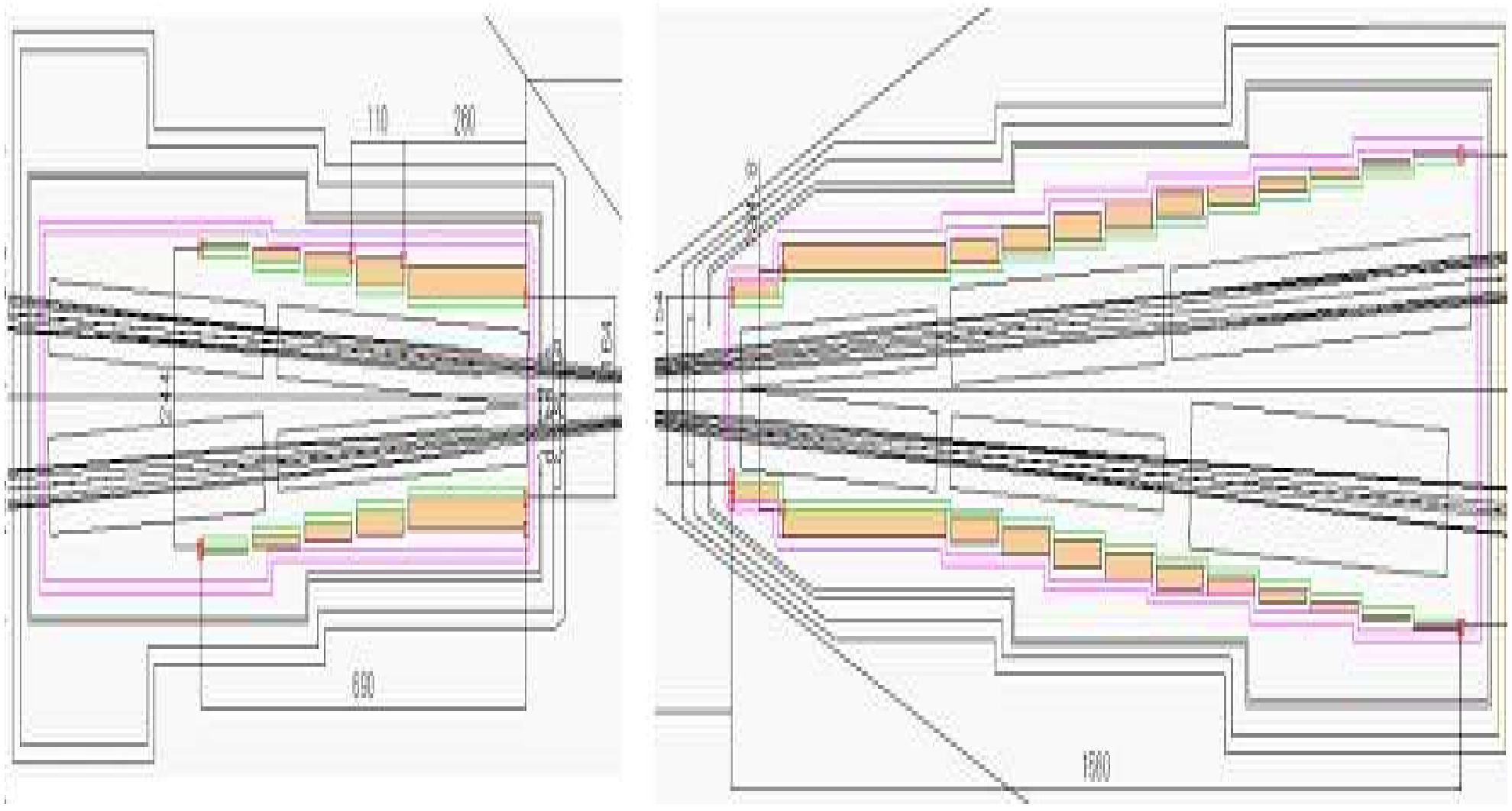

*Figure 3.4: Compensation solenoids in the cryostats.*

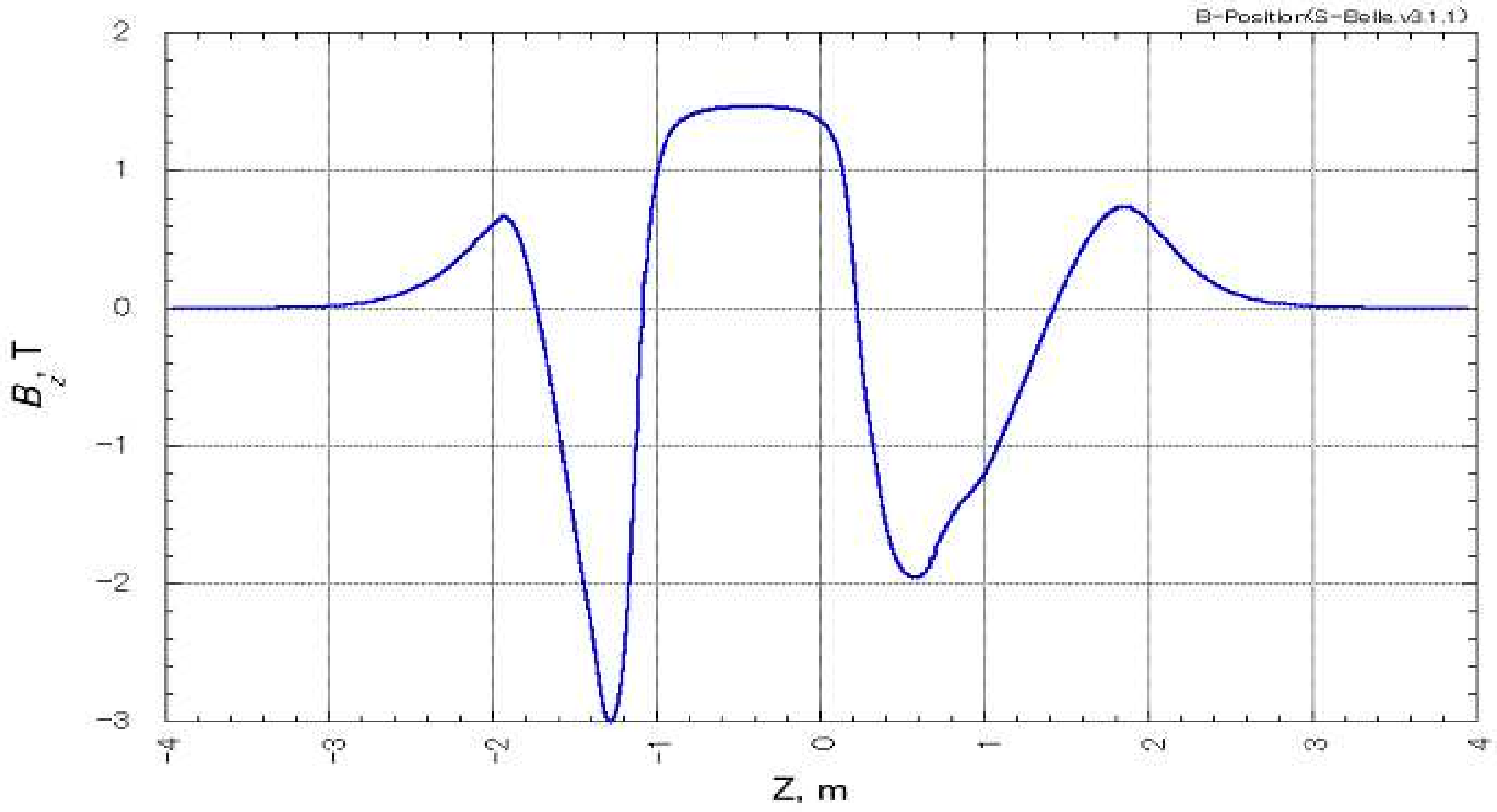

*Figure 3.5: Solenoid field profile along the solenoid axis.*

Figure 3.6 shows a rough sketch of the vacuum system for the right hand side. The center of the ring is towards the bottom of the page. The dimension of the cryostat is determined by cryogenic considerations. At the end of the IP chamber, the ducts for the positron beam and the electron beam are completely separated. The inner diameter of the beam duct is designed so as not to trap the electromagnetic field created by the circulating beams, especially at the Y-shaped structure of the IP chamber. The presently adopted diameter for the central part of the IP chamber is 2 cm. The inner diameter of the beam duct increases from 2 cm after the QC1RP magnet region to 3 cm in the QC1RE region, and to 6 cm in the QC2LP region. The inner surface of the duct will have small masks to prevent scattered photons from reaching the IP. The material of the beam duct in the cryostat is OFC clad with stainless steel outside. Details of the IP chamber are discussed in Sec. 3.3.

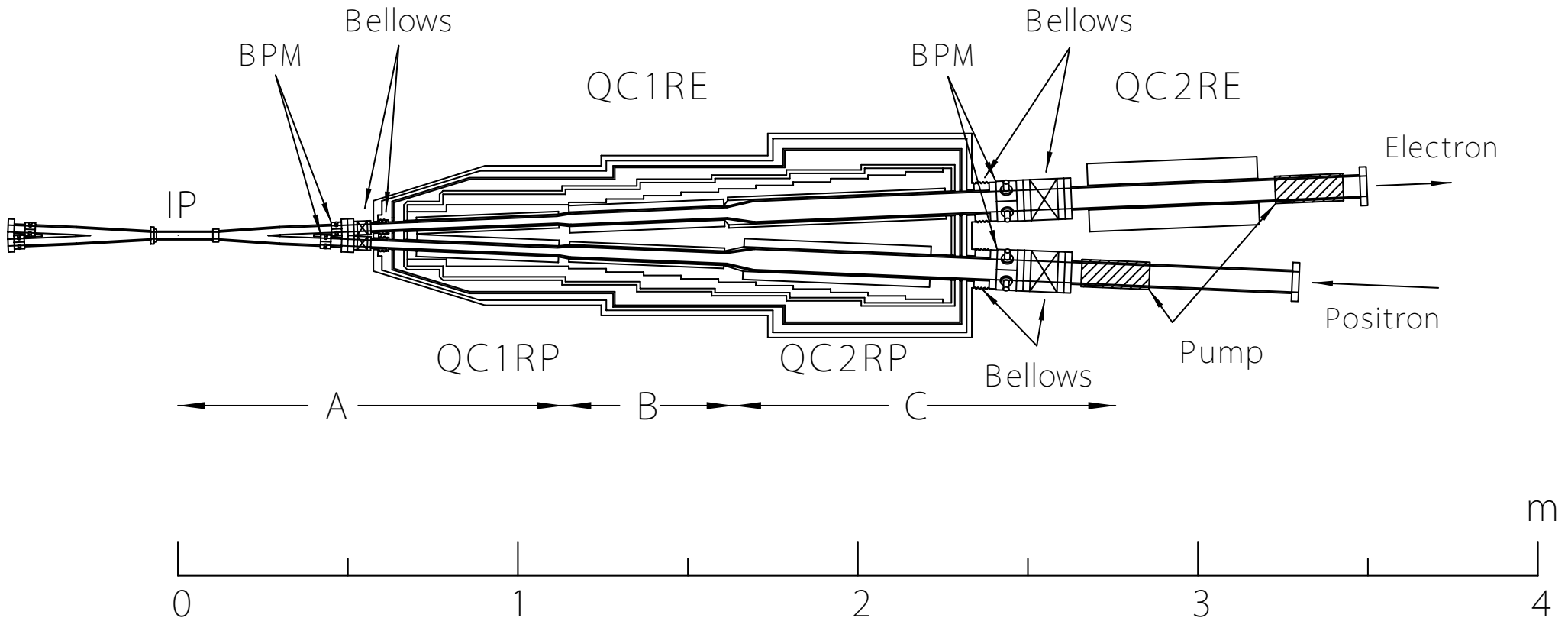

*Figure 3.6: Sketch of the vacuum system for the right hand part of the IR (top view).*

Possible locations of the beam position monitor (BPM) are outside the ends of each cryostat. In the front position, the BPM will be placed on the IP chamber. A trial model pick-up electrode for the small diameter duct is shown in Fig. 3.7. In the rear position, the BPM will be placed on the beam ducts just outside of the cryostat. Bellows for the beam ducts are inserted between the IP chamber and the cryostat ducts, and between the cryostat ducts and the outer beam ducts. Bellows are also used to connect the cryostat ducts to the body of the cryostat. There is no space for vacuum pumps until the end of the cryostat. The pumping speed of the pump will be typically 0.05 $m^3s^{-1}$, which is limited by geometrical constraints. The distance of the pumps from the IP and their pumping speed are not so different from the present KEKB situation. However, the very small conductance of new beam pipes produces a large pressure difference between the IP and the pump. The heat load due to synchrotron radiation from the last bend and the ohmic loss of the wall current is estimated along the beam duct for the incoming positron in Table 3.6. The sections A, B, and C given in the table are shown in Fig. 3.6. There is an additional heat load due to the electromagnetic wave produced by the non-uniform structure of the duct. The amount of this heat is 100 W and distributed along the ducts. For the section A, the heat load is about 0.5 kW. If we allow a temperature rise of 2 K for the cooling water with a velocity of less than 2 $ms^{-1}$, a cross section of the cooling channel of more than 30 $mm^2$ is necessary. Figure 3.8 shows a cross section through the beam pipe with the water channels indicated. However, the temperature rise of the duct must be checked for realistic water flow conditions.

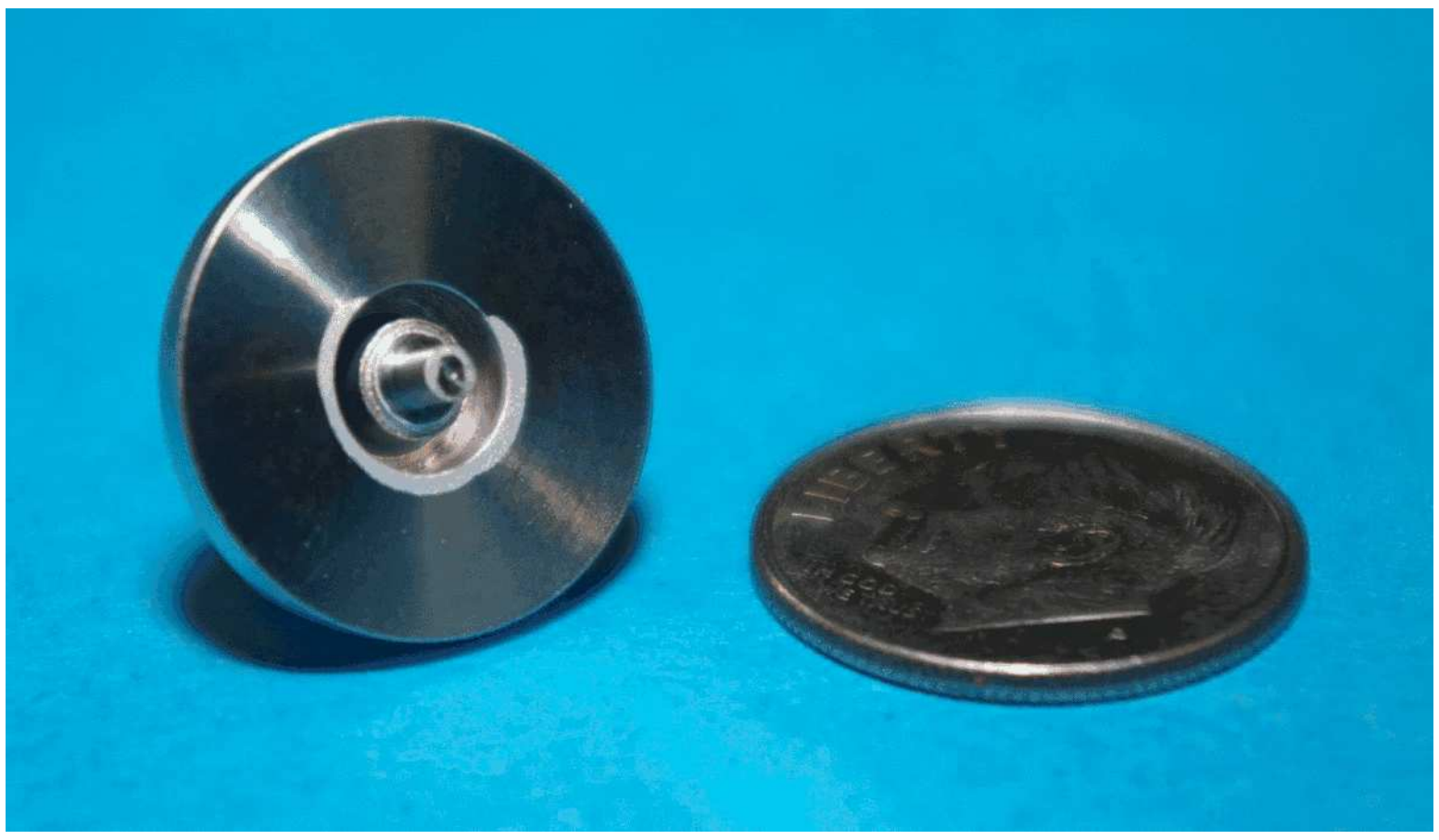

*Figure 3.7: Test model of a pick-up electrode for the beam position monitor at the IP chamber.*

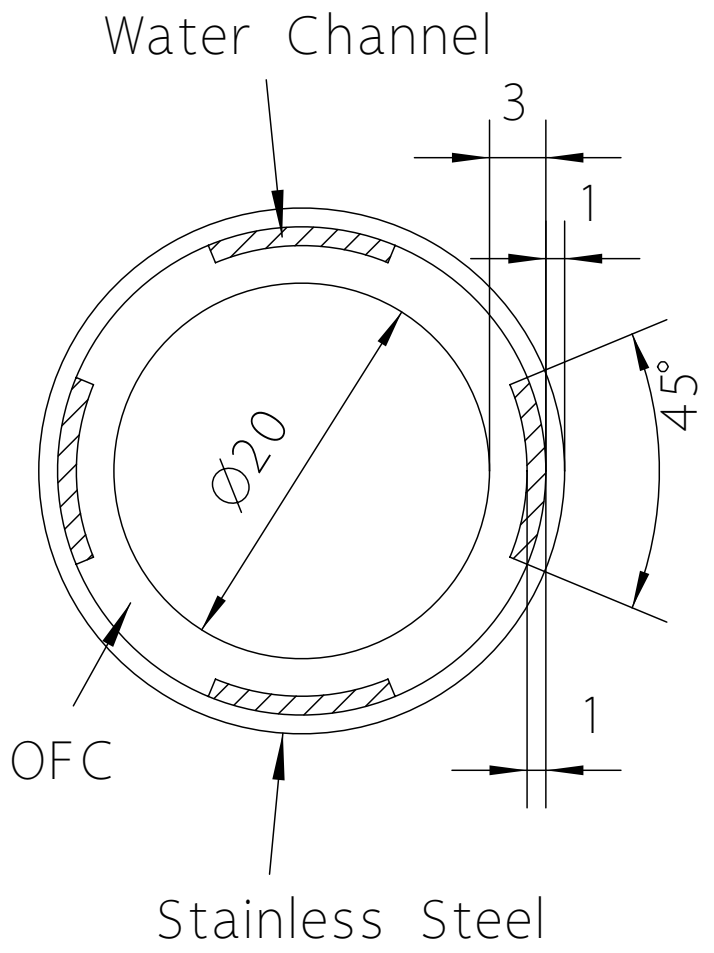

*Figure 3.8: Cross section of the cryostat beam duct close to the IP (see text).*

*Table 3.6: Heat loads and ohmic losses along the beam duct.*

| Section | A | B | C |
|---|---|---|---|
| Length (m) | 1.14 | 0.5 | 1.12 |
| ID (cm) | 2 | 3 | 6 |
| SR Power (W) | 240 | 340 | 395 |
| Ohmic loss (Au) (W) | 250 | 72 | 81 |

The pressure around the IP is estimated for sections A, B, and C of the beam duct. The main gas load is due to photon-desorption by synchrotron radiation. The pressure around the IP is determined mainly by this part of the beam duct. An approximate estimate of the pressure can be made by considering a model pipe where both ends, one at IP and the other at the pump port, are closed. The pressure at the pump is determined by the total gas load in the pipe and the pumping speed. The pressure difference between the IP and the pump is determined by the gas load and the conductance of each segment and is independent of the pressure at the pump. Assuming a sufficiently scrubbed value of the photo-desorption coefficient of $10^{-5}$, the pressure at the IP and at the pump is estimated as $6 \times 10^{-5}$ and $6 \times 10^{-6}$ Pa, respectively, for the design current. Though the pressure is acceptable from the view point of the detector background according to recent studies at KEKB, the pressure is high enough to affect the average pressure in the ring. A sketch of the left hand side vacuum system is shown in Fig. 3.9. The design is similar to the right hand side (see Fig. 3.6). However, in this case, the space between the cryostat and the QC2LP magnet is very tight. Further iteration is necessary to design this part.

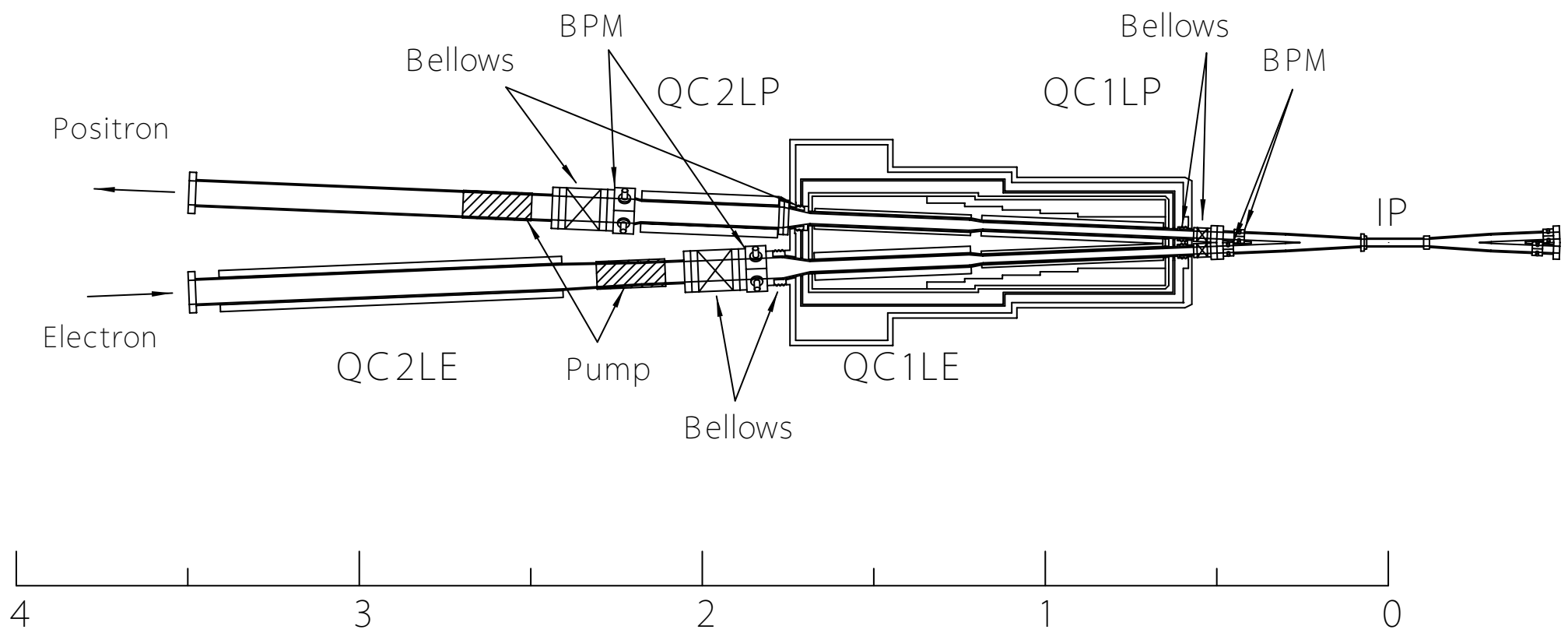

*Figure 3.9: Sketch of the vacuum system for the left hand side of the IR (top view).*

The connection between the IP chamber and the cryostat duct is a difficult issue to be solved. One idea is to fabricate the cryostat so that it provides channels in the body for tightening rods to access the screws on the connecting flanges (Fig. 3.10). For this purpose, the liquid helium vessel has narrow open tubes for the rods and the body of the cryostat has holes that can be closed. This idea necessarily prefers a symmetrical design of the cryostat with respect to the beam orbits. In this case, the bellows between the IP chamber and the cryostat chamber are located on the cryostat chamber. Another idea for the connection is described in Sec. 3.3. The type of connection has some impact on the design of the IP chamber.

## 3.2 Beam-induced background

In SuperKEKB, the expected luminosity will be 40 times higher than at KEKB, due to the higher beam currents and smaller interaction point (IP) beam sizes. These features cause higher beam-induced background in the Belle II detector. To assure stable detector operations at such high luminosity, an interaction-region (IR) design based on beam background estimations inferred from the present level is important.

In this section, we discuss the expected beam-induced background at SuperKEKB. Note that all expectations strongly depend on the beam optics. We expect to update the numerical results below after fixing the beam optics and the interaction region design.

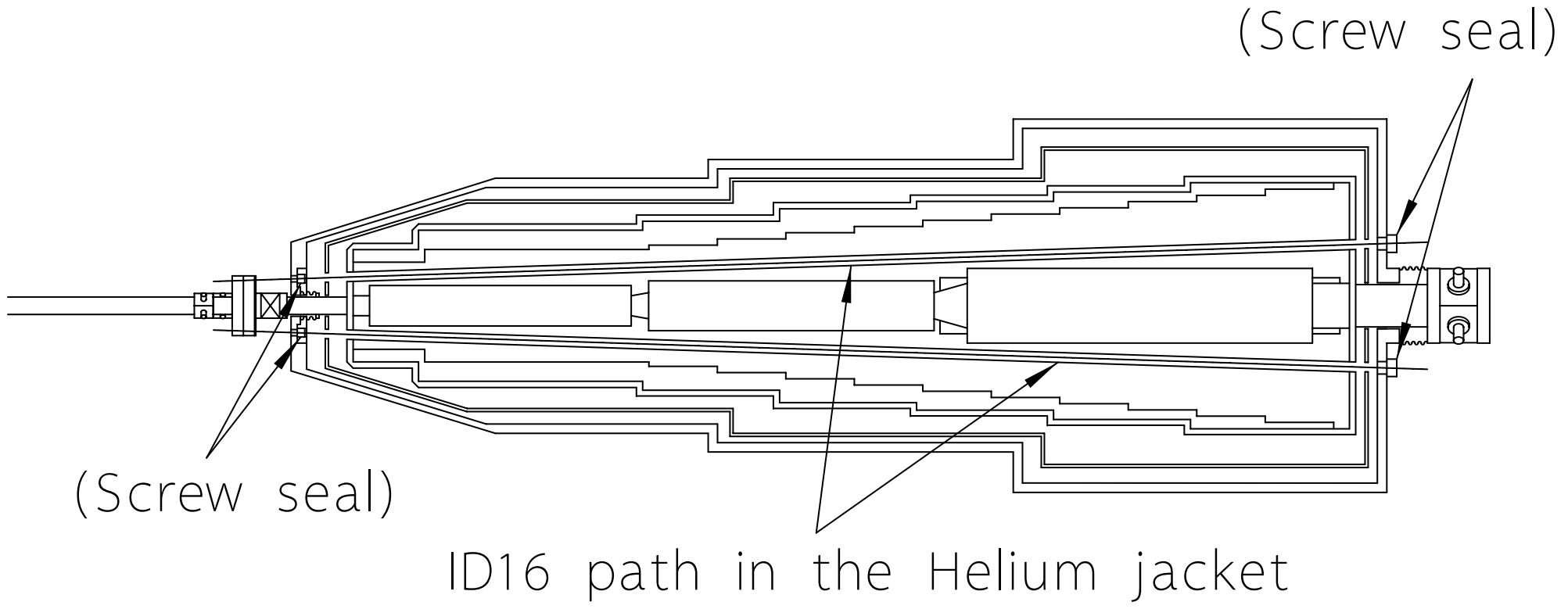

*Figure 3.10: Suggested modification of the cryostat to tighten screws for the flanges in front of the cryostat (side view).*

### 3.2.1 Evaluation of possible beam-induced backgrounds

In this section, we describe the possible background sources in SuperKEKB: synchrotron radiation (SR) from the high energy ring (HER) upstream direction, backscattering of SR from HER downstream, scattering of the beam on residual gas,[1] Touschek scattering, radiative Bhabha scattering, and electron-positron pair production via the two photon process $e^+e^- \to e^+e^-e^+e^-$.

**• SR from upstream (SR Upstream)**

It is important to evaluate the upstream SR background to design the IP chamber to protect the inner detectors (PXD and SVD). Here we design the IP chamber to avoid direct SR hits from the HER in the detector. Since the SR background level depends on the HER current, beam-optics (magnet positions, magnetic field strength, and beam orbits), and the geometry of the IR components, we perform GEANT4-based beam-line simulations. We find no direct SR hits from HER are expected with the current beam optics and IP chamber design.[2]

**• Backscattering of SR from downstream (SR Backscattering)**

The final focusing magnet for the HER downstream side is called QCS-R, where R stands for the right hand side in the schematic drawing. The magnet QCS-R provides the final focusing of the LER beam. In the current KEKB case, it also works as a bending magnet to separate the outgoing HER beam from the LER. Strong SR is then emitted because of this bending. On the other hand, in the SuperKEKB case, two separate quadrupole magnets are located on both the R and L sides, and beam orbits of both incoming and outgoing beams are on center in the magnets. We therefore expect much lower background than at KEKB. Based on the optics calculations, we expect 1/800 of the current KEKB backscattering SR at SuperKEKB. To confirm this, careful simulation studies will also be necessary after fixing the beam optics.

[1] In what follows, we will use the expression 'beam-gas scattering' or just 'beam-gas' for this kind of scattering.

[2] Recently, we find there are direct SR hits from HER to the IP chamber. We have developed an IP chamber design to take care of these SR hits.

- **Beam-Gas scattering**

Beam-gas scattering (bremsstrahlung and Coulomb scattering) changes the momenta of beam particles, which then hit the walls of vacuum chambers and magnets. Shower particles are produced and are one of the major sources of the beam-induced background. The size of this background depends on the beam current, the vacuum pressure in the ring, and the strength of the magnets.
In SuperKEKB, the beam currents will be ∼2 times higher than in KEKB. Except for the IR, the vacuum and the overall magnet-field strength will be the same as the current levels. We therefore expect the same order of magnitude (a few times higher) beam-gas background dose. However, the vacuum level at the IR (±2 m from IP) will be 100 to 1000 times higher than at present, and the magnetic fields of the final focus magnets are much higher; therefore, it is important to evaluate these effects. In addition, in the SuperKEKB IR, due to the limited space around the IP, the thickness of the IP chamber mask that shields the shower particle backgrounds arising from beam-gas and Touschek will be only 1-2$X_0$, so a careful beam-gas background evaluation is necessary.

- **Touschek scattering**

Touschek scattering is intra-bunch scattering. It changes the momenta of beam particles so that they hit the vacuum chamber and magnet walls. Shower particles are then produced. This background is proportional to the beam bunch current, the number of bunches, and the inverse of the beam size. In SuperKEKB, the beam size will be ∼1/20 of that at present, therefore the Touschek background is expected to be the major background source.
The contribution from the HER can be ignored, because the rate of Touschek scattering is proportional to $E^{-3}$, where $E$ is the beam energy, and also to the bunch current density, which is less than in the LER because the HER current is smaller than the LER current.
Based on the expected beam lifetime, we expect the Touschek background from the LER to be 20-30 times that of KEKB. A detailed Touschek background estimation based on the beam-line simulation is important.

- **Radiative Bhabha scattering**

The rate of the radiative Bhabha events is proportional to the luminosity. Photons from the radiative Bhabha scattering propagate along the beam axis direction and interact with the iron of the magnets. In these interactions, neutrons are copiously produced via the giant photo-nuclear resonance mechanism [1]. These neutrons are the main background source for the outermost detector, the $K_L$ and muon detector (KLM) in the instrumented return yoke of the spectrometer. Detailed simulation studies are important to design the IR.
In addition, in a radiative Bhabha event, both electron and positron energies decrease. If we employ the shared QCS magnets for incoming and outgoing beams as in KEKB, the scattered particles are over-bent by the QCS magnets. The particles then hit the wall of magnets where electromagnetic showers are generated. In the SuperKEKB case, we use two separate quadrupole magnets and both orbits for incoming and outgoing beams are centered in the Q-magnets. We therefore expect the radiative Bhabha background due to over-bent electrons and positrons will be small. Based on a beam optics calculation, we estimate 1/40 of the current KEKB background in SuperKEKB.

- **Electron-positron pair production via two photon process**

In SuperKEKB, we locate the pixel detector close to the IP. Therefore, we need to evaluate the very low momentum electron-positron pair backgrounds produced via the two-photon process $e^+e^- \to e^+e^-e^+e^-$.
Based on MC simulations, we estimate that there are 900 to possibly 14000 $e^+e^-$ pairs in each event in the first layer of the PXD, 1.3 cm from IP. The rate drops roughly as $1/r^2$. These tracks will result in an occupancy of between 0.1 and 1.5% for the first layer of the PXD. Here we assume the number of pixel in the first layer to be 3.2M and an integration time of 20 $\mu$s. The cluster size is 3 pixels per track. These track numbers translate to a track rate of 700 kHz/cm$^2$ to 10 MHz/cm$^2$. For the first PXD layer, we expect a hit rate of 45 M to 670 M tracks per second.
The estimation strongly depends on how the MC simulations are done. In the last part of the 2010 KEKB/Belle run we carried out a series of tests to measure this background.
To understand the particle shower background features, we performed background studies during the 2009 fall and 2010 spring Belle runs. To evaluate the beam-gas scattering, we carried out vacuum bump studies, in which we changed the vacuum level of each section of the ring. To study the Touschek effect, we vary the overall beam size. In these studies, we use either single LER or HER beams.

- **Vacuum Bump study**

To study beam-gas background, we vary the vacuum level of one of the sections in the HER or LER ring. We compare the SVD PIN diode current, CDC current, or TOF hit rate and vacuum level for each section. All of the above sub-detector background levels depend on the vacuum level of the last arc or the straight sections just upstream. However, there is no strong correlation between the detector background level and vacuum levels in other sections. This study shows that the vacuum level between the last arc and the upstream straight section of the IR determines the beam-gas background level.
We also vary the vacuum level of the IP in this background study. Figure 3.11 shows the results for (a) CDC current, (b) SVD PIN diode, (c) HER beam life time, and (d) vacuum level of the IP. This study indicates that a bad IP vacuum level affects the beam lifetime, but does not degrade the detector background level.

- **Touschek study**

To study the Touschek effect, we vary the HER or LER beam size. Figure 3.12 shows scatter plots of (a) 1/beamsize vs 1/lifetime, and (b) 1/lifetime vs CDC current. Since CDC current is proportional to the lifetime, we estimate the expected Touschek background level based on the expected beam lifetime, as shown in Sec. 3.2.1.

## 3.3 IP-chamber design

Figure 3.13 shows a rough sketch of the current IP chamber design for SuperKEKB. As shown in the figure, there are crotch structures on both the left and right sides, since there are two separate QCS magnets. The radius of the IP chamber is 10 mm, to avoid a resonant cavity structure, since the radius of the QCS beam-pipe is 10 mm. The central part is made with double-walled beryllium, and the crotch parts is tantalum, to shield against particle shower

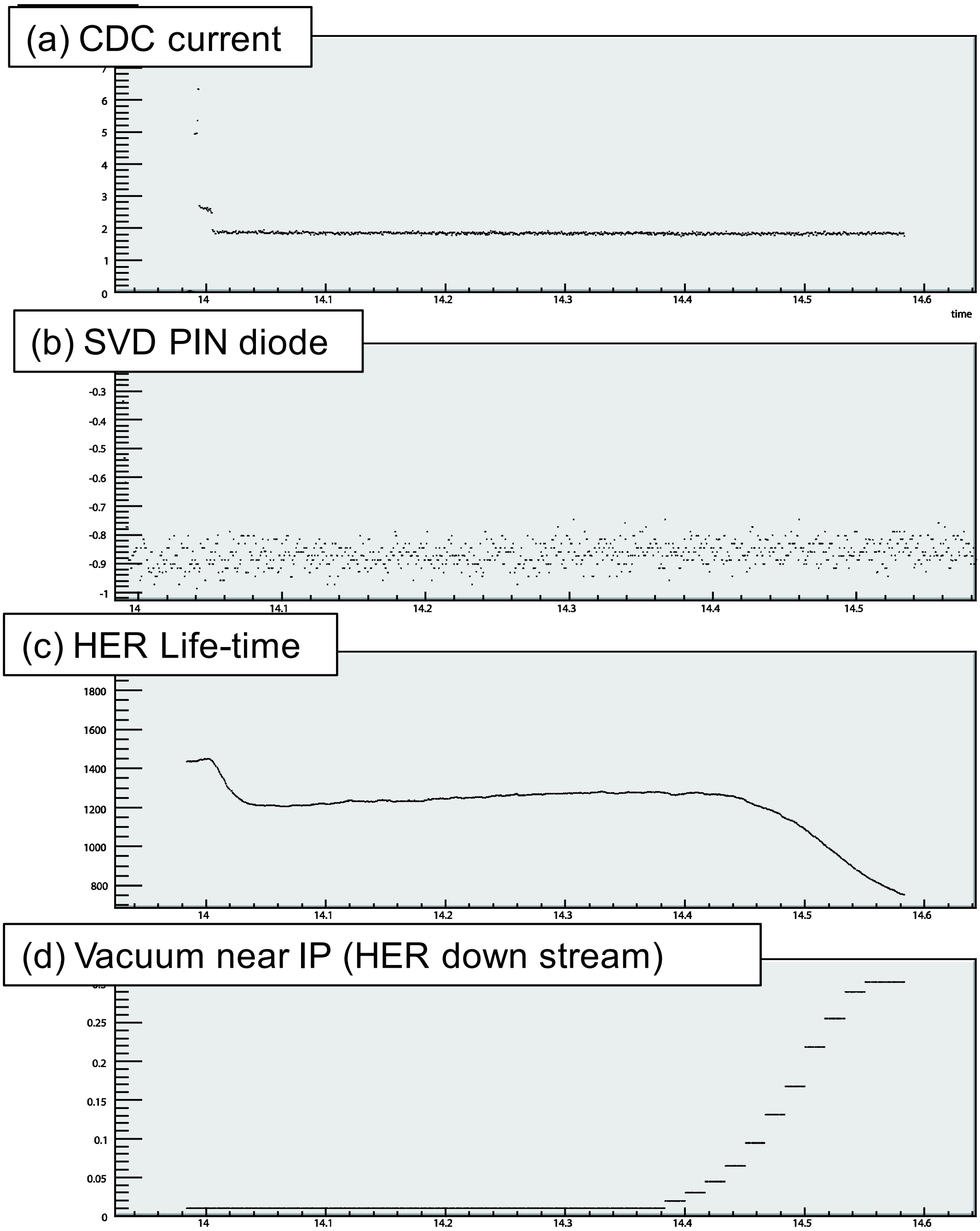

*Figure 3.11: Beam backgrounds for various vacuum conditions at the IP.*

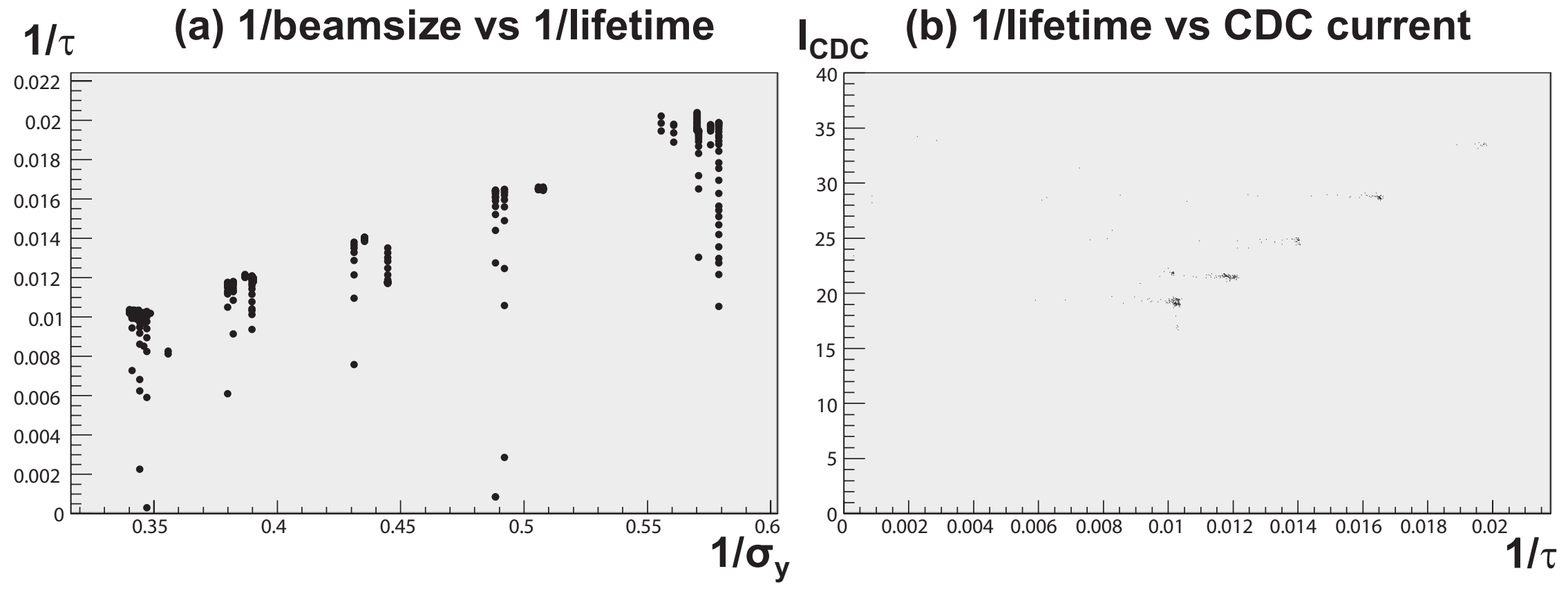

*Figure 3.12: Results of the LER Touschek background study. (a) 1/beamsize vs 1/lifetime, and (b) 1/lifetime vs CDC current.*

*Table 3.7: Basic parameters for the central Be part of the IP chamber.*

| | | |
|---|---|---|
| Gold plate | Thickness | 10 $\mu$ m |
| Inner Be pipe | Inner radius | 10.0 mm |
| | Thickness | 0.6 mm |
| Gap for coolant | Thickness | 1.0 mm |
| Outer Be pipe | Outer radius | 12.0 mm |
| | Thickness | 0.4 mm |

backgrounds. Table 3.7 summarizes the current basic parameters for the IP chamber Be-part design.

Cooling of synchrotron radiation (SR) and the heat due to wall current etc. is also necessary. The central beryllium part and crotch tantalum parts are cooled with paraffin and water, respectively. In the SuperKEKB IP chamber, beam-position monitors are mounted for fast and precise beam feedback.

In the Belle IP chamber, there are SR masks to prevent direct SR hits the central IP chamber part. In SuperKEKB, the polarity of the last bend is designed so that the SR fan from the bend does not directly hit the central part of the IP chamber, as shown in Fig. 3.14. Therefore we do not have a mask structure, because the SR is blocked by the crotch.

As described in Sec. 3.1.2, the connection between the IP chamber and the QCS cryostat is one of the key issues in the IR design. In KEKB, the IP chamber and the QCS beam pipes without QCS cryostat ducts were directly connected first; the QCS beam pipes and QCS cryostat ducts were then connected. In SuperKEKB, due to the limited space in the IR, the QCS cryostat ducts and the QCS beam pipes are integrated. Therefore we must connect the IP chamber and QCS cryostat ducts directly. Some possibilities for this connection are discussed in Sec. 3.1.2. Another idea to connect the IP chamber and QCS cryostat ducts is shown in Fig. 3.15. Here, we use a quick disconnect system to connect the IP chamber and cryostat ducts via universal joints. It is important to test these IR assembly methods for the IR design.

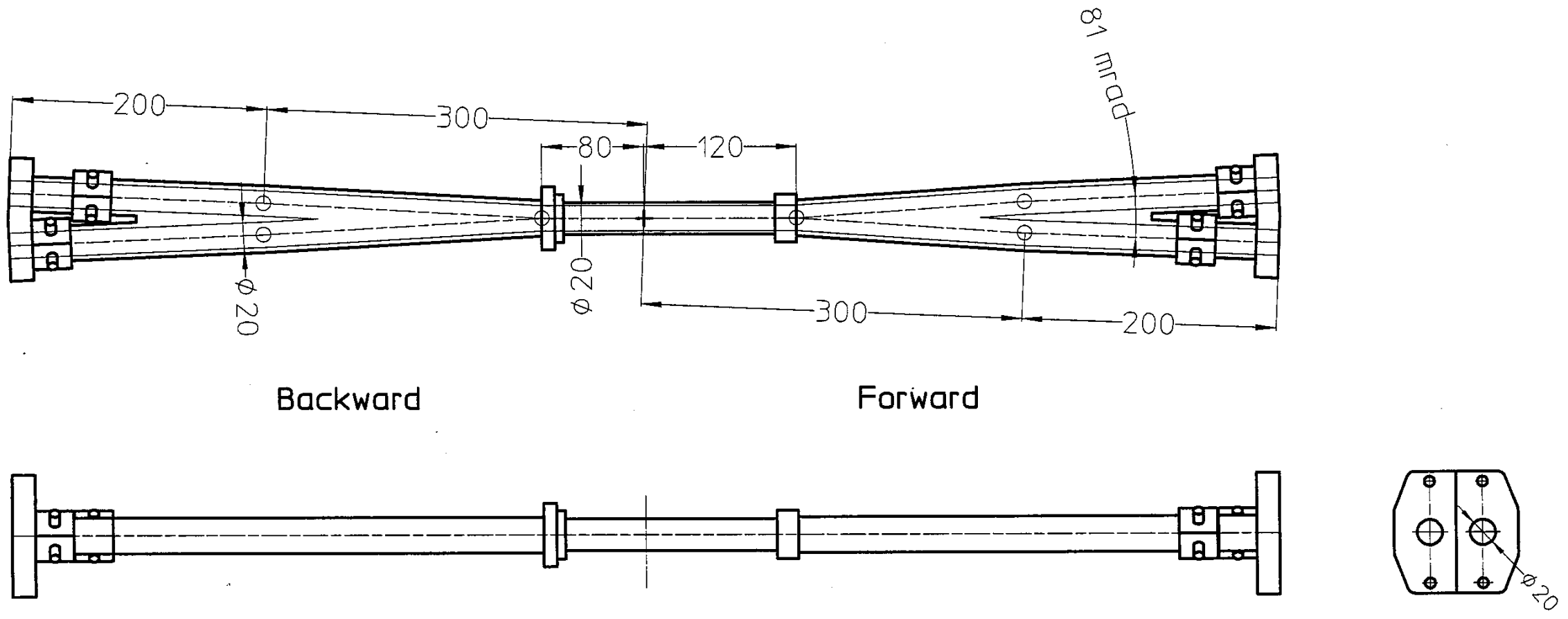

*Figure 3.13: Picture of IP chamber.*

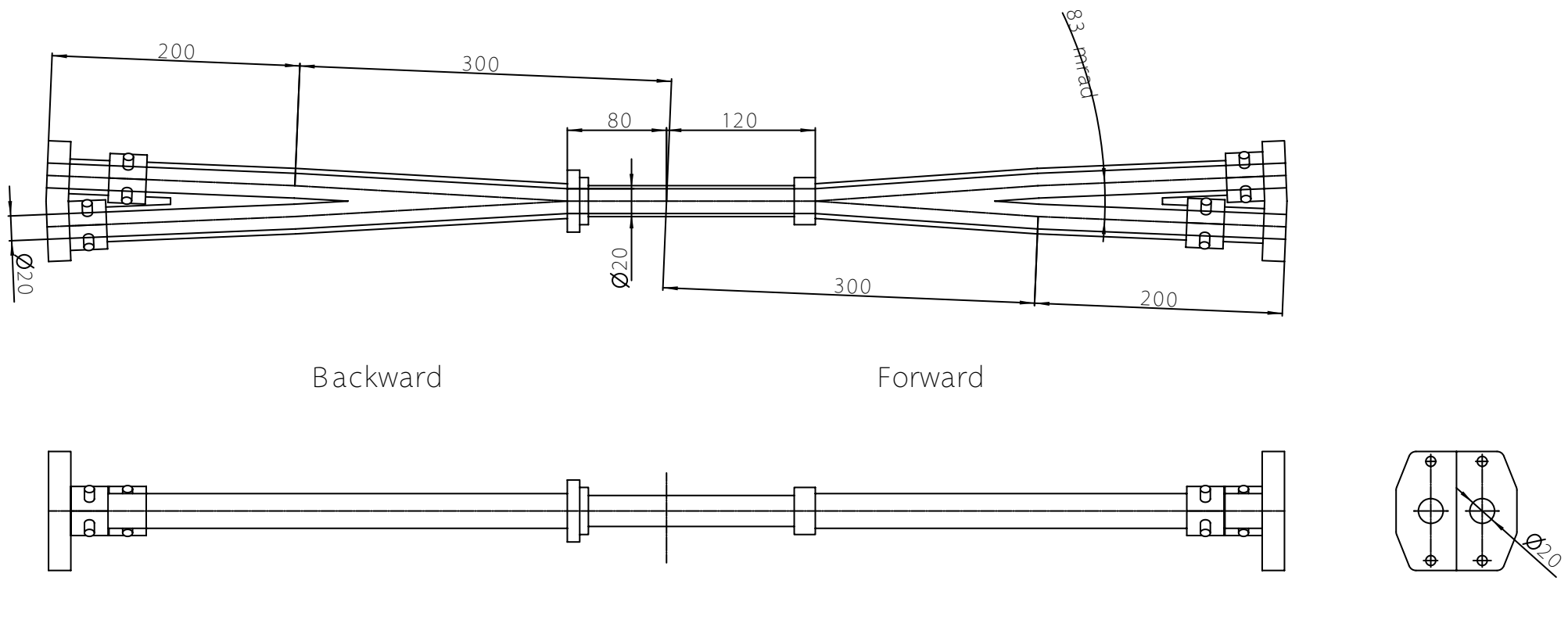

*Figure 3.14: SR fan and IP chamber structure.*

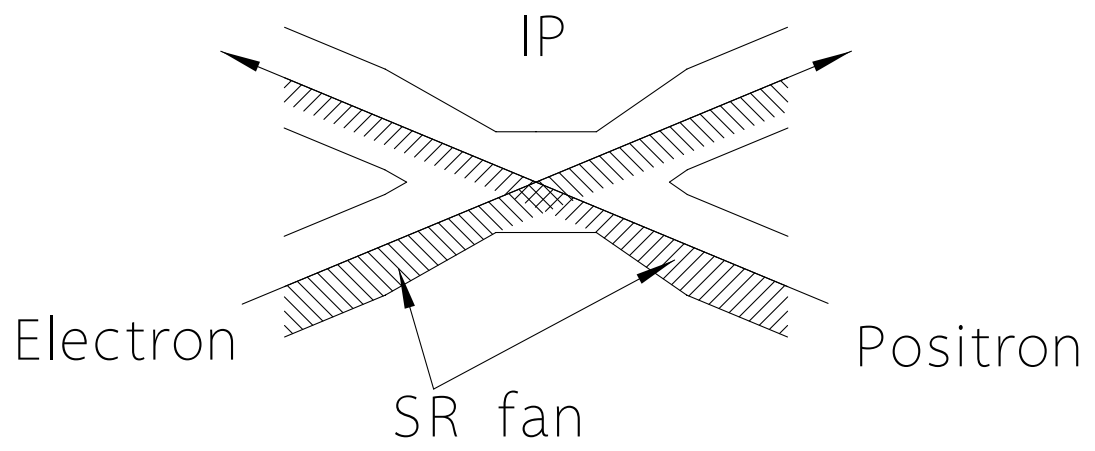

*Figure 3.15: SR fan and IP chamber structure.*

As the central part is made of thin beryllium tubes, special care is necessary. The synchrotron radiation from the HER and LER will not hit this part directly. We minimize the HOM (higher-order-mode) heat generation with a smooth transition from the crotch to the beryllium part. The largest heat source in this section is caused by the image current. Based on the calculation in Ref. [2], the generated heat values by LER and HER beam currents are 217 and 149 W/m, respectively. As the length of the beryllium part is 22 cm, the total heat will be 81 W.
Since beryllium is reactive, we would like to avoid cooling by water. Instead, a normal paraffin liquid such as $C_{10}H_{22}$ is the candidate coolant. As the heat capacity of $C_{10}H_{22}$ is around 2 J/K/g, we need to flow about 300 cm$^3$/min of paraffin to keep the temperature rise below 10°C. To ensure a safety margin, we need to design the cooling system so that a paraffin flow of 1000 cm$^3$/min is allowed. In addition, the pressure drop of the liquid should be less than 0.1 MPa to avoid breakdown of the cylinder. According to experience with the IP chamber in KEKB, this flow rate can be realized with a 1 mm gap between the inner and outer cylinders. The liquid flow speed amounts to 10 m/s and the flow should be turbulent to allow efficient heat exchange between the wall and the liquid. The paraffin liquid should be dehydrated and the water contamination should be monitored in order to avoid corrosion.
Temperature monitor sensors should be distributed over the chamber and the interlock system should work during accelerator operations. Depending on the HOM mode, there could be large amounts of heat dissipation in certain limited locations. Such temperature rises should be detected by the temperature monitor to prevent fatal accidents.

## 3.4 Vibration measurements

Because the beam size of SuperKEKB is expected to be very small, beam oscillation amplitudes must be kept much smaller than are required at present. A large vibration amplitude results in luminosity degradation. In order to evaluate the present vibrations around the IP, measurements were carried out on the KEKB tunnel floor, on some of the magnets in the interaction region (IR), on magnet supports, on movers and on the Belle detector.
Servo accelerator sensors, MG-102, made by Tokkyokiki Corporation, were used for the vibration measurements. An output of 1 V corresponds to 1 gal with our setup. The frequency range of the sensors is from 0.1 to 400 Hz.
Several kinds of measurements were carried out. They are:

1. Vibration measurement at the Belle detector;
2. Vibration measurement around the IR region in the KEKB tunnel;
3. The effect of the air conditioner mounted on the Belle electronics hut;
4. Coherence between the two sides of the KEKB tunnel at the IP.

### Vibration measurement on the Belle detector

Measurement locations include the floor just under the Belle detector, the Belle support table, the middle and the top positions of the Belle end-yoke and the top of the barrel-yoke. Three axes of vibration, North-South (NS), East-West(EW) and Up-Down (UD), were measured simultaneously using three sensors.
Power spectral densities (PSD) at the floor and the top of the end-yoke are shown in Fig. 3.4. Some typical resonant peaks were measured on the floor. There are resonances at 0.3–0.4 Hz due to the effect of micro-seismic motions in the horizontal direction and at around 3 Hz due to the resonance of gravel soil in the vertical direction.

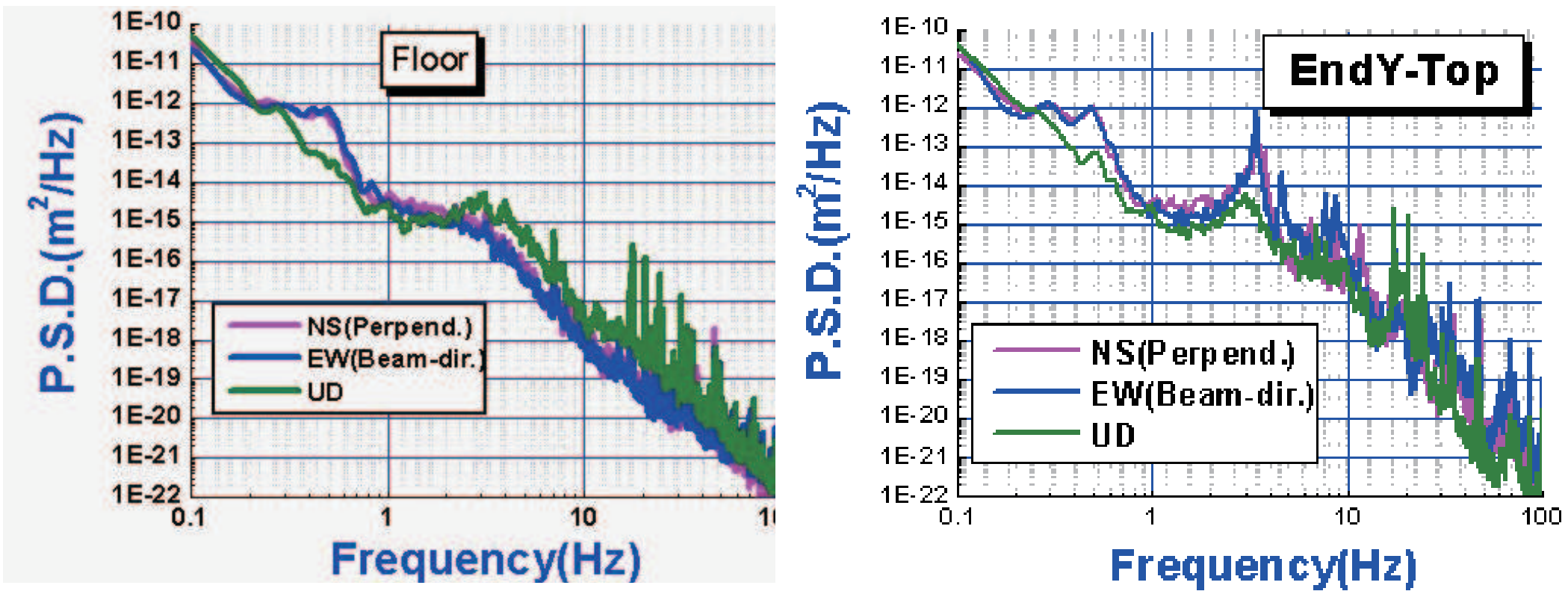

*Figure 3.16: Measurement results at the Belle detector: (a) Vibration on the floor, and (b) Vibration on the top of the end-yoke.*

All measurement data show a unique resonant peak at 4 Hz, mainly in the horizontal direction. This resonance becomes larger when going from the bottom of the end yoke toward the top. The vibration on the top of the barrel yoke is almost the same level as at the top of the end yoke. This indicates that the Belle detector is oscillating horizontally. Though the integrated amplitude above 1 Hz is around 60 nm on the floor, the integrated amplitude in the horizontal direction grows to about 250nm to 350nm at the top of the end yoke. The amplitude in the vertical direction does not show a large difference. The larger vibration on the top can probably be explained by the support configuration of the end yoke and the Belle detector.

### Vibration measurements in the KEKB tunnel

Vibrations of the KEKB IR magnets were also measured. Measurement positions are the QC magnets placed on the support table called the "QCS boat," the QCS boat itself, the QCS movable table and the KEKB tunnel floor. The QCS boat is placed on the QCS movable table. The QCS movable table is mounted on rails that are fixed to the floor.

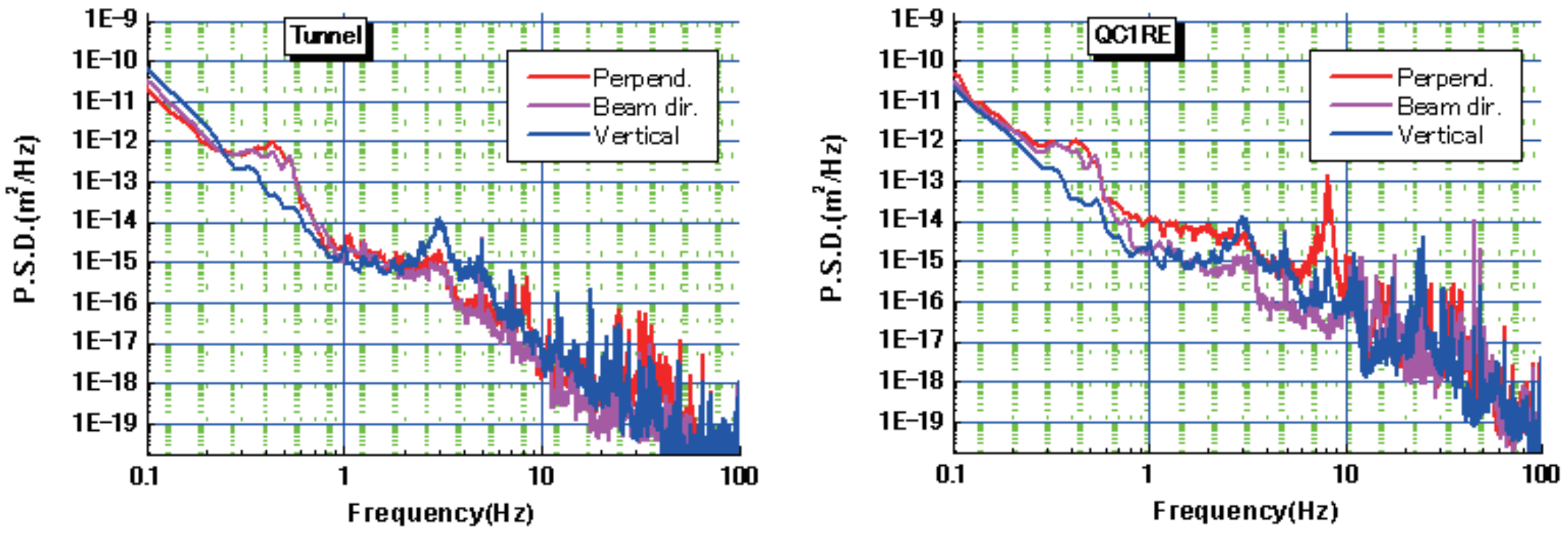

*Figure 3.17: Measurement results around the KEKB. (a) Vibration on the floor, and (b) Vibration on the QC1RE.*

A unique resonant peak at 8 Hz was measured with other two typical resonances at 0.3 Hz and 3 Hz, as shown in Fig. 3.4. The vibration amplitude is larger than that on the floor. This 8 Hz resonance was confirmed by a hammering test where all components were vibrating horizontally with the same phase.
The integrated amplitude in the horizontal direction was measured to be around 60 nm on the floor, growing to about 240 nm on the QC1 magnet. The amplitudes of the vibrations are similar on the B4 floor and the KEKB tunnel floor.

### Other measurements

Coherence between the two sides of the Belle detector was measured (Fig. 3.4). Sensors were placed on the left and right sides of the KEKB tunnel and data were taken simultaneously on both sides. The distance between the two locations was about 8 m. The result showed no coherence between the left and right sides, except for resonant frequencies of around 0.3 Hz and 3 Hz.
The effect of the air conditioner (AC), which is mounted on the Belle electronics hut, was also investigated. Vibrations were measured at both sides of tunnel with the air conditioner turned on and off. The effect of the AC turned out to be small.

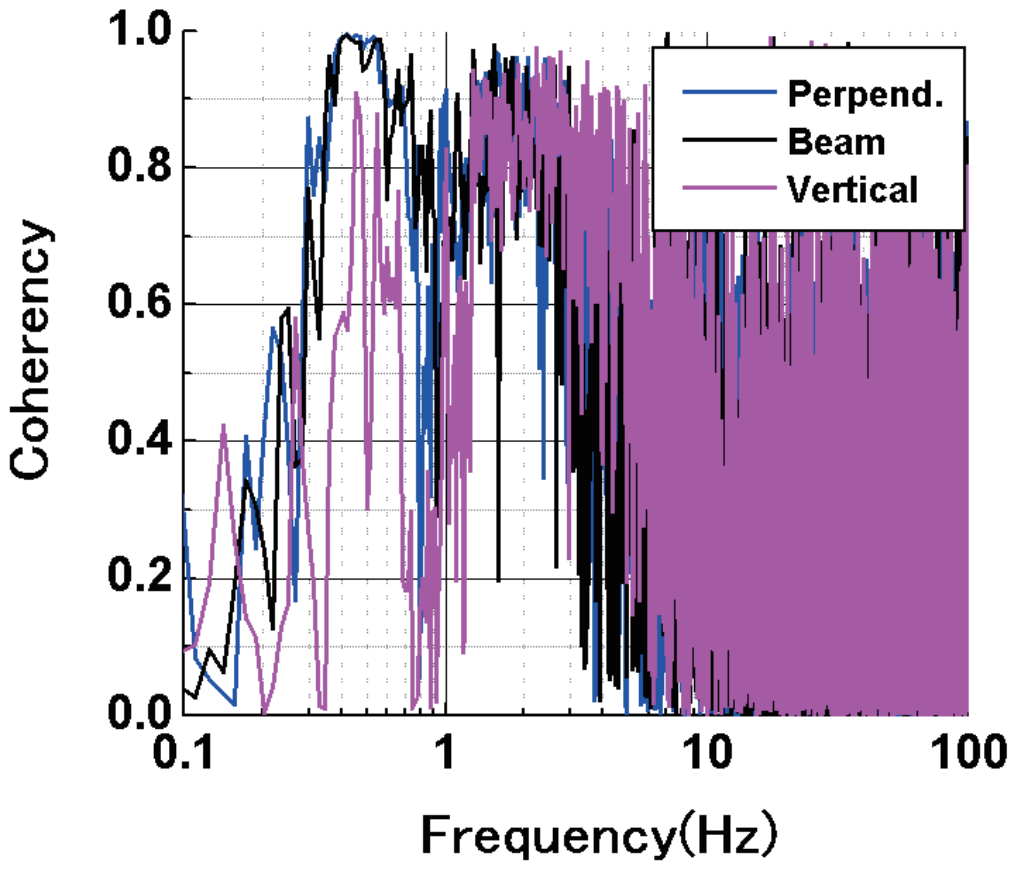

*Figure 3.18: Vibration coherency between left and right sides in Tsukuba Hall.*

### Summary

Vibrations of the Belle detector, the tunnel floor, and the magnets and their supports are summarized in Table 3.4. The floor vibration is magnified by the Belle detector and becomes as large as ∼300 nm at the top of the Belle detector. There is no coherence between the left and right sides of the IP. The effect of the AC being turned on was measured to be small. The influence of the helium flow in the cryogenic system is scheduled to be measured.
A collision orbital feedback system, similar to the "iBump" feedback [3] is required to maintain good collision conditions. In order to monitor the collision, vertical and horizontal beam-beam kicks are monitored using BPM data. The new "iBump" system will probably be configured with iron core magnets, and with faster BPM readouts. The target rate of the new system is 50 Hz, which should be sufficient to compensate for the vibration measured at the KEKB tunnel floor and magnets.

*Table 3.8: Integrated vibration amplitudes*

| Integrated amplitude (nm) | | | | | | |
|---|---|---|---|---|---|---|
| | > 1Hz | | | > 10Hz | | |
| | Perpend | Beam | Vertical | Perpend | Beam | Vertical |
| Barrel Top | 196 | 301 | 93 | 18 | 12 | 9 |
| End-yoke Top | 248 | 354 | 80 | 25 | 17 | 20 |
| Endyoke Middle | 204 | 254 | 121 | 14 | 27 | 19 |
| Belle stand | 105 | 69 | 71 | 13 | 11 | 13 |
| B4 floor | 50 | 46 | 67 | 4 | 3 | 9 |
| KEKB floor | 55 | 45 | 68 | 10 | 5 | 9 |
| Magnet table | 90 | 50 | 76 | 12 | 16 | 19 |
| QCS boat | 250 | 60 | 118 | 15 | 21 | 30 |
| QC1RE | 241 | 77 | 112 | 52 | 50 | 46 |

# Chapter 4

# Pixel Detector (PXD)

At the high luminosities envisaged for SuperKEKB, the detectors close to the beam pipe are faced with extremely high hit rates, caused by beam-related background—for example, the Touschek effect—and by low-momentum-transfer QED processes. Beam-related backgrounds are discussed in detail in Chapter 3, and the QED backgrounds (mostly photon-photon reactions) are discussed below. Such backgrounds need to be considered carefully when designing the first few layers of the vertex detector.

In the nano-beam option chosen for the SuperKEKB machine (Chapter 2), the beampipe radius in the interaction region will only be about 10 mm. This is good news for the physics related to vertex reconstruction, but is a challenge for the vertex detector itself because the background increases roughly with the inverse square of the radius. The innermost layers of a high precision vertex detector can no longer be realized by strip detectors due to the large occupancy, defined as the fraction of channels hit in each triggered event: the large strip occupancy at SuperKEKB luminosities makes the reconstruction of the $B$-decay vertices impossible. The solution is to use pixel sensors rather than strips for the innermost layers, which have a much larger number of channels and therefore a much smaller occupancy. Strip detectors are safe at radii beyond 40 mm at SuperKEKB luminosities.

This scheme—pixels followed by strips—has been successfully applied for the detectors at the LHC. However, at the lower energies of the SuperKEKB machine, the LHC silicon detectors are too thick: the relatively large amount of material necessary to make these pixel sensors work (thick sensors for large signals, equipped with amplifier electronics and digital logic for the signal extraction located above the sensors, requiring active cooling) would cause too much multiple scattering for a precise reconstruction of $B$-decay vertices at our energies.

For Belle II, we therefore propose a different pixel detector concept, the PXD, based on the DEPFET (DEPleted Field Effect Transistor) technology which allows for very thin (50 micron) sensors. In this concept, the readout electronics, which needs active cooling, are located outside the acceptance region and will therefore not contribute to the multiple-scattering material budget. The sensors themselves will consume very little power so that air cooling is sufficient. Radiation hardness clearly is an issue, but our irradiation tests show that the DEPFET can be engineered in an inherently radiation-hard technology.

After a general description of the layout of the PXD in Sec. 4.1, we describe the individual components of this detector. These are the DEPFET sensors (Sec. 4.2), the readout ASICs mounted on the ends of the sensors (Sec. 4.3), and the construction and assembly of an entire module including the readout electronics (Sec. 4.4). We then address the studies on radiation hardness, pointing towards a refined technology step with thin oxides (Sec. 4.5). Quality control of the sensors, electronics and modules are described in Sec. 4.6, and the foreseen scheme for the data acquisition is sketched in Sec. 4.6. The mechanical support of the sensors and the cooling of the ASICs is engineered in an integral design, described in Sec. 4.8, followed by a discussion on the necessary interfaces to the rest of the detector and the services to the PXD (Sec. 4.9). The DEPFET sensors, initially planned for experiments at a future linear collider, have been qualified intensively in high-energy test beams (Sec. 4.10), and the expected performance in the environment of SuperKEKB is shown in Sec. 4.11. Finally, in Sec. 4.12, we describe the work packages attributed to the various collaborating institutions and sketch the schedule for realizing the PXD up to the commissioning phase at the accelerator. We conclude this chapter with a summary.

## 4.1 Layout

The DEPFET pixel (Sec. 4.2) consists of a fully depleted silicon substrate and is equipped with a p-channel MOSFET structure with an internal gate where the electrons liberated by traversing charged particles are collected. The internal gate modulates the current through the MOSFET at readout time. The DEPFET pixel sensor is a monolithic structure, with current-digitizing electronics at the ends of the sensor, outside of the acceptance region.

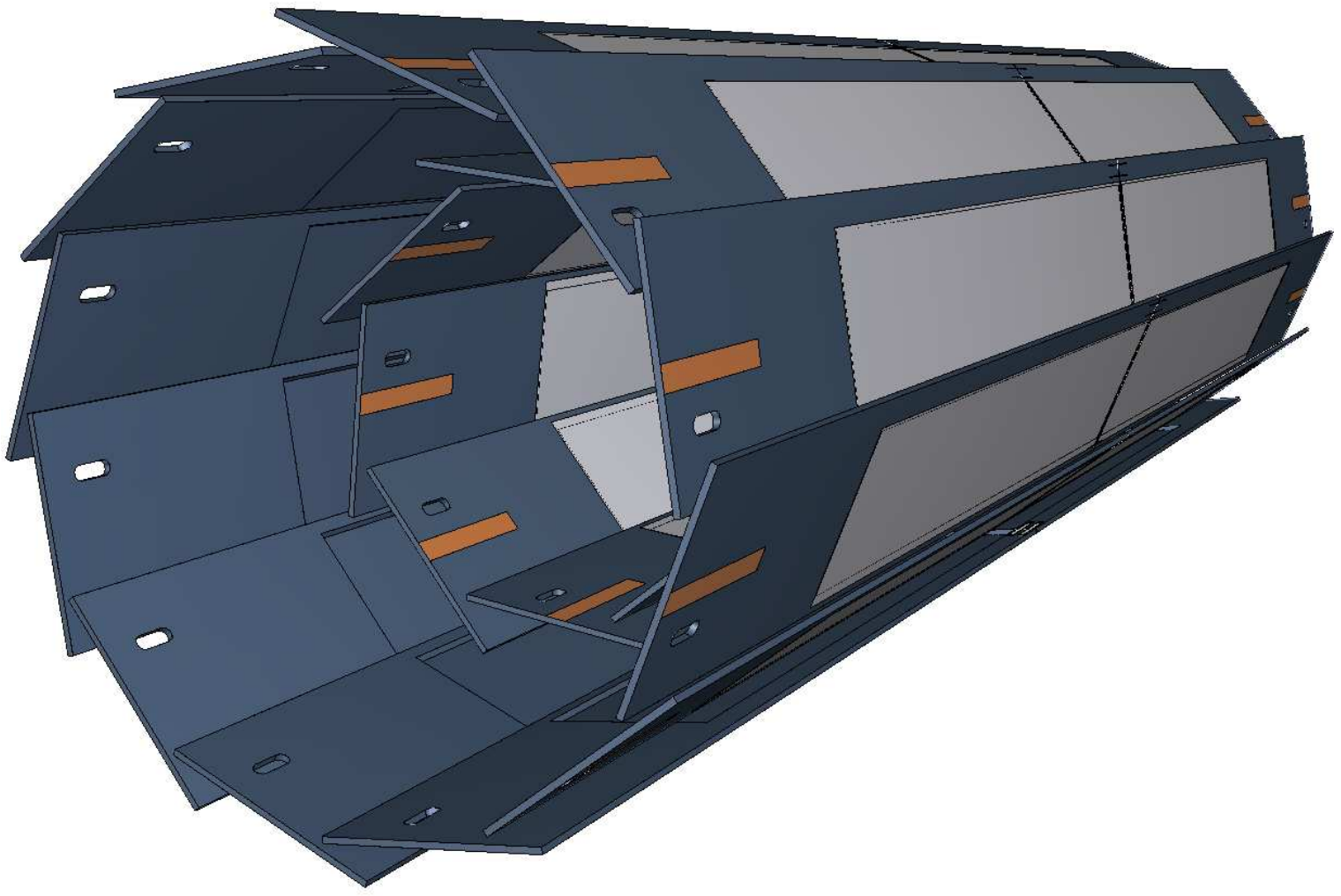

*Figure 4.1: Schematic view of the geometrical arrangement of the sensors for the PXD. The light grey surfaces are the sensitive DEPFET pixels, which are thinned to 50 microns and cover the entire acceptance of the tracker system. The full length of the outer modules is 174 mm.*

Due to its internal amplification, the DEPFET sensor can be made very thin (down to 50 microns), minimizing the multiple scattering (see Sec. 4.4). Since the DEPFET pixels are "on" only during the readout, it is a low power device and no active cooling is necessary for the pixel sensor itself. The readout electronics consists of three types of ASICs (Sec. 4.3): The "Drain Current Digitizers" (DCD), which digitize the MOSFET currents from a row of pixels; the "Digital Handling Processor" (DHP), which does the zero-suppression of the "empty" pixels; and the "SWITCHERs," which switch on a pixel row to send the currents to the DCD. While the SWITCHERs are located along the side of the DEPFET sensor on a 2 mm wide unthinned rim, the DCD and the DHP are located at the ends of the sensors outside of the acceptance region (Fig. 4.10). Active cooling (Sec. 4.8) is needed for the DCDs and the DHPs.

The PXD consists of two layers of sensors, with radii at 14 mm and 22 mm. The inner radius leaves sufficient space for possible variations of the final beampipe layout (Ch. 3). A schematic drawing of the sensor arrangement is shown in Fig. 4.1. The inner layer consists of 8 planar sensors ("ladder"), each with a width of 15 mm, and a sensitive length of 90 mm. The outer layer consists of 12 modules with a width of 15 mm and a sensitive length of 123 mm. The sensitive lengths in each of the layers are determined by the required angular acceptance of the tracker (Ch. 6), i.e., a polar angle range of 17 degrees (forward) to 150 degrees (backward). The sensors are mounted on an integrated support and cooling structure, held by screws. The support on the backward side can slide on the beampipe to compensate for thermal expansion of the beampipe and the beampipe supports.

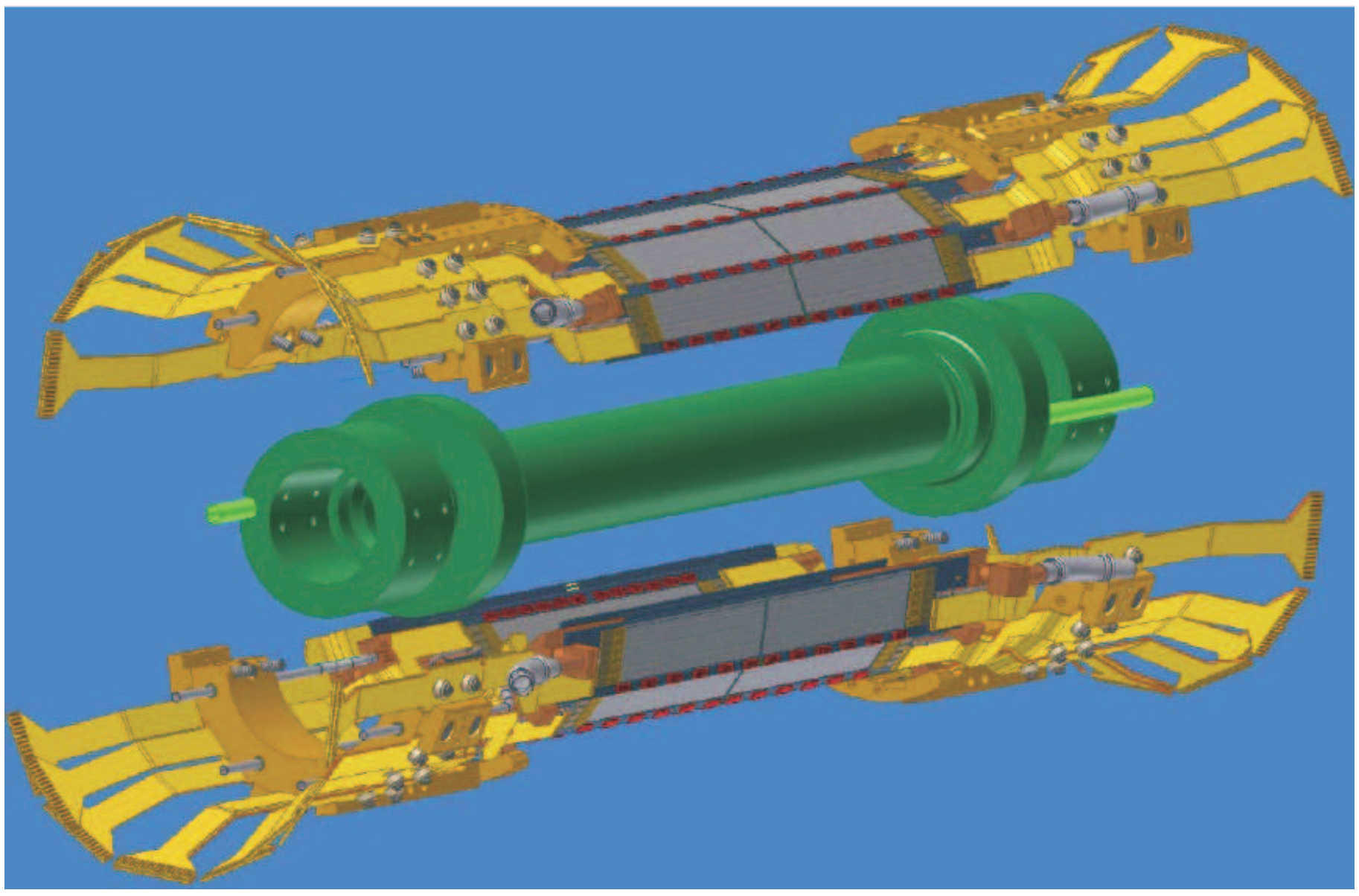

*Figure 4.2: Layout of the complete support structure with mounted sensor ladders and services to the outside. The two half-shells are supported on the beampipe (shown in green). They are displaced from their mounting positions in the picture for clarity.*

At the end of the active part of the ladder ("end-of-stave"), there is 25 mm of room on each side for the DCD and the DHP ASICs. A total of four DCDs and four DHPs are bump-soldered on each end-of-stave. From there, a multi-layer Kapton cable of 6 mm width guides the signals to the "Data Handling Hybrid" (DHH) (Sec. 4.7), where the data stream is fed via optical fiber to the data acquisition system. The PXD is read out in rolling shutter mode, with a speed of 100 ns per pixel row. Four rows are read out in parallel, so that the total readout time of 1600 pixel rows (800 on each side) is about 20 microseconds for an entire frame. Several frames are stored in a ring buffer within the DHP, waiting for a external trigger to initiate the frame readout.
An exploded view of the entire pixel detector, up to the end of the Kapton cable, is shown in Fig. 4.2. There is an 16 mm radial gap between the PXD and the innermost layer of the silicon strip detector (Ch. 5).

## 4.2 DEPFET Sensors

The DEPFET (**DEP**leted **F**ield **E**ffect **T**ransistor) is a semiconductor detector concept that combines detection and amplification within one device. It was invented in 1987 by Josef Kemmer and Gerhard Lutz of the MPI for Physics [1]. After a period of principal investigations, the move of the MPI-Semiconductor Laboratory into new facilities provided the technological infrastructure to fabricate large scale detectors. In recent years, the sensor development was driven by an intensive R&D and prototyping for x-ray imagers [2] and the ILC vertex detector [3].

### 4.2.1 Operation Principles

A cross section through the device is shown in Fig. 4.3. A p-channel MOSFET or JFET (junction field effect transistor) is integrated onto a silicon detector substrate, which becomes fully depleted by a sufficiently high negative voltage to a $p^+$ contact on the back side. A potential minimum is formed by sideward depletion [4], which is shifted directly underneath the transistor channel at a depth of about $1\,\mu$m by an additional phosphorus implantation underneath the external gate. Incident particles generate electron-hole pairs within the fully depleted bulk. While the holes drift to the back contact, electrons are accumulated in the potential minimum, called the "Internal Gate." When the transistor is switched on, the electrons modulate the channel current. The readout is non-destructive and can be repeated many times.

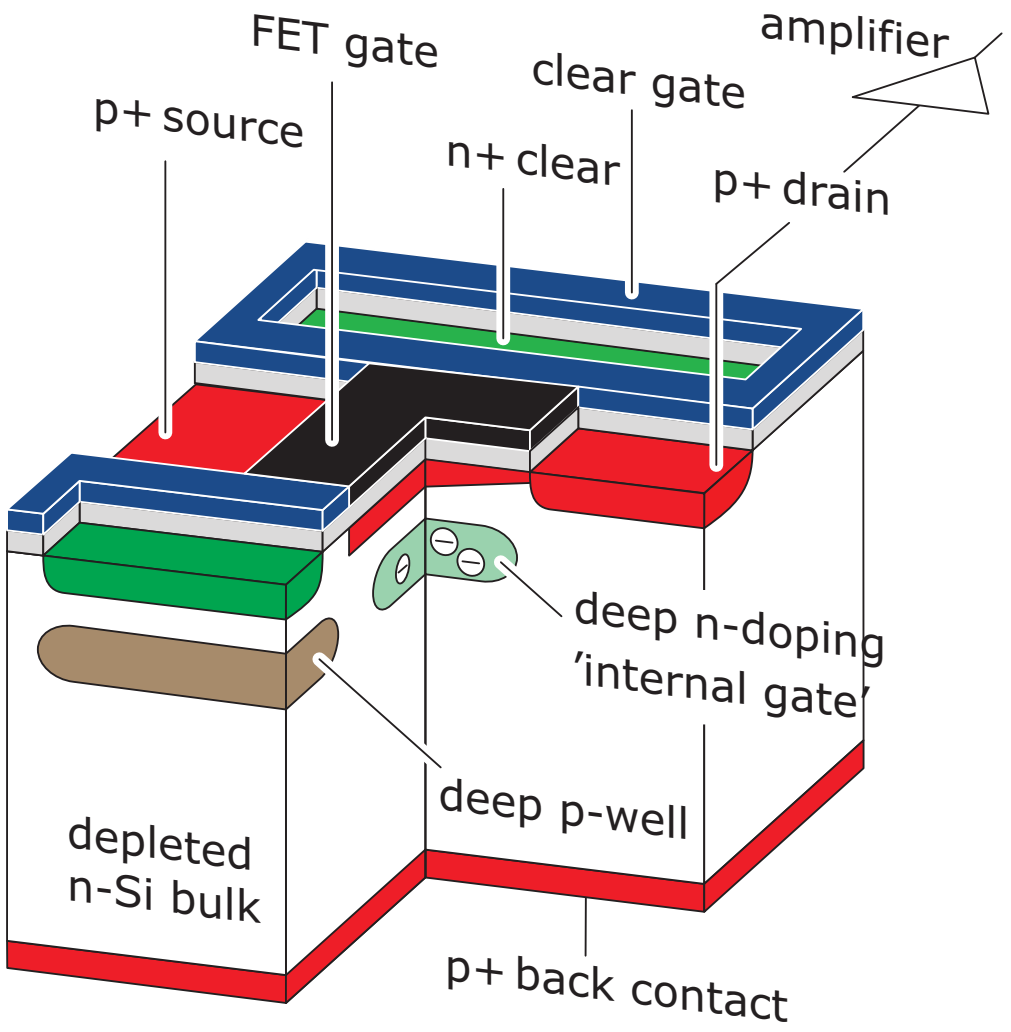

*Figure 4.3: Operating principle of a DEPFET*

The removal of the signal charge and thermally generated electrons from the internal gate is called "Clear." A neighboring $n^+$ contact is pulsed at a positive voltage providing a punch-through into the internal gate. Any reset noise is avoided if the entire charge is removed. An advantage of the DEPFET device is the amplification of the signal charge just above the position of its generation, thus avoiding any lateral charge transfer where losses could occur. The most important feature of the DEPFET is the very small capacitance of the internal gate, resulting in a very low noise performance even at room temperature.

DEPFET structures can be operated individually as integrated on-chip amplifiers—for instance in the readout nodes of a silicon drift detector—or collectively as an active pixel detector.

Figure 4.4 shows the scheme of a DEPFET pixel matrix that is operated with a row-by-row access via the external gates, also called "rolling shutter mode." Here, the external gates switch the pixel on and off while the signal amplification is achieved via the internal gate. There is no current flow in the non-selected DEPFET rows so that the array consumes only very little power. The readout amplifiers at the end of the columns can be connected either to the sources (source follower) or the drains (drain readout) of the DEPFETs. For collider applications, we use the faster drain (current) readout. A measurement cycle consists of a collection phase (Collect), the readout (Read) and the reset of the internal gate (Clear). A pedestal subtraction is realized either by a consecutive Read–Clear–Read sequence (double sampling) or by a faster Read–Clear procedure (single sampling), where cached pedestal values are subtracted in the DHP readout chip (Sec. 4.3). Only during this readout mode is the DEPFET (partially) insensitive. Hence the dead time is only 100 ns within the 20 $\mu$s readout cycle (0.5%).

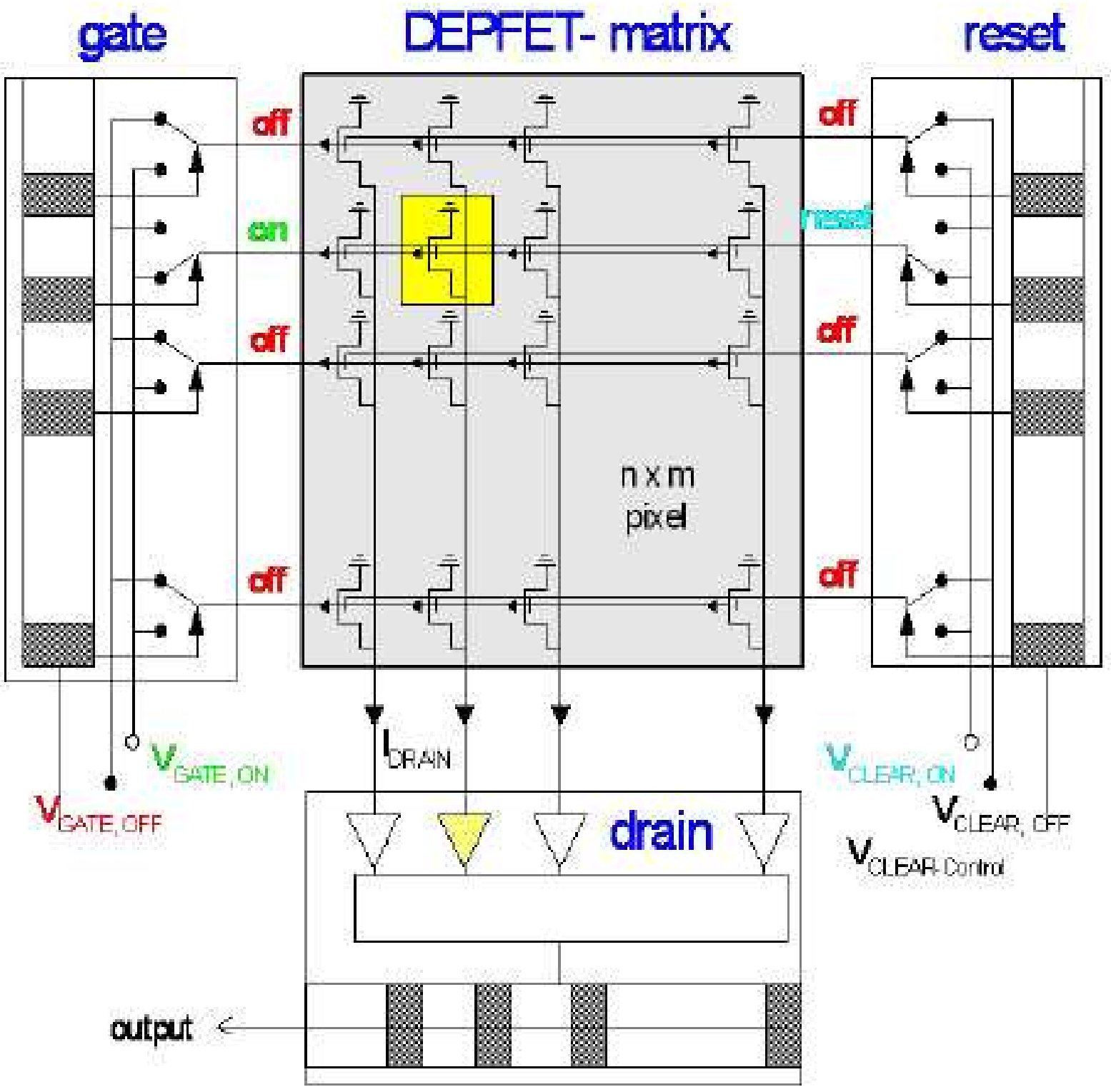

*Figure 4.4: DEPFET matrix operated in rolling shutter mode*

The figure of merit is the internal amplification of the DEPFET. It is defined as the current response to an electron collected in the Internal Gate.

$$g_q = \Delta I_d / e^- \tag{4.1}$$

Note that $g_q$ is directly proportional to the signal-to-noise ($S/N$) of the system, because the noise of the readout ASICs dominates over the noise of the DEPFET. The relationship between $g_q$ and the technology and design parameter can be derived [5] from the simplified equation 4.2

$$g_q = g_m / C_{int} = \sqrt{2 \frac{I_d \mu}{L^3 W C_{ox}}} \tag{4.2}$$

where $g_m$ is the conductance of the external gate, $C_{int}$ is the capacitance of the internal gate, $C_{ox}$ is the sheet capacitance of the gate oxide, $I_d$ the drain current, $\mu$ the carrier mobility, and $L$ and $W$ are the length and width, respectively, of the gate. The gate length $L$ has the biggest impact on $g_q$ (see Fig. 4.5, left side). Prototypes with the standard channel geometry have a $g_q$ of about $400\,\mathrm{pA}/e^-$, but DEPFETs with a scaled channel geometry show even better results. For instance, the 128x128 test matrix H09 with $20 \times 20\,\mu\mathrm{m}^2$ pixel size investigated during the last test beam at CERN in 2009 (Sec. 4.10) has a $g_q$ of $560\,\mathrm{pA}/e^-$. This was achieved by a modest decrease of the DEPFET gate length $L$ from $5\,\mu$m to $4\,\mu$m.

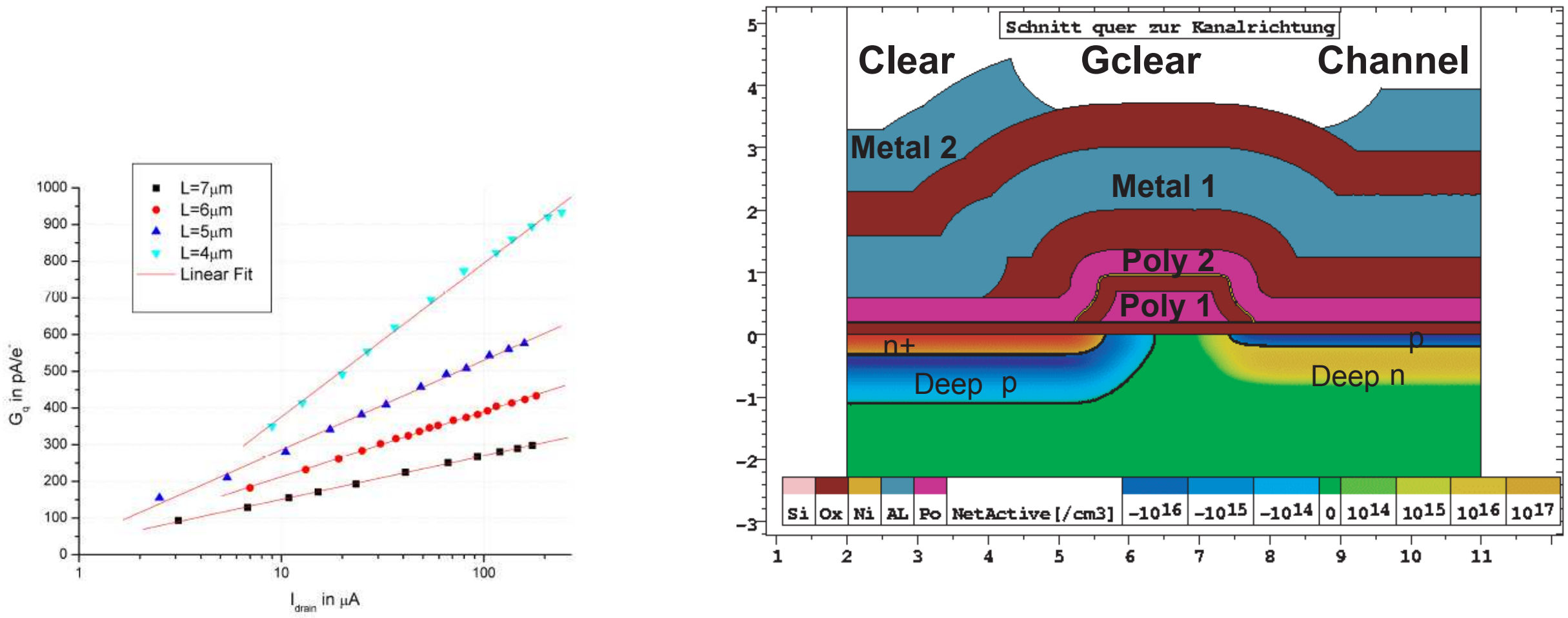

*Figure 4.5: (Left) Internal amplification $g_q$ vs drain current and channel length $L$ illustrating the scaling potential of the DEPFET [5]. (Right) Cross section through the Clear - Clear Gate - Internal Gate simulated with Synopsis TCAD-DIOS [6] ($x$ and $y$ scales in $\mu$m).*

### 4.2.2 Technology Extensions for Belle II Vertexing

For DEPFET applications in ILC and astrophysics experiments, a new double-sided DEPFET technology was developed. This CCD-like technology was specially tailored to the functional DEPFET requirements. It contains two polysilicon and two metal layers and is the world's most complex technology on high ohmic wafers. The cross section through the clear and channel region of the DEPFET (Fig.4.5, right side) gives an impression of the DEPFET technology.

Two major changes must be incorporated to meet the Belle II requirements. The separately developed thinning-technology module (Sec. 4.4) is implemented into the currently processed DEPFET prototype run, called PXD6. PXD6 is the first technology run that produces DEPFETs on $50\,\mu$m thick substrates. In parallel, the smaller PXD-TO project aims to optimize a radiation-hard gate insulator that is able to cope with the harsh backgrounds near the Belle II beamline. The starting point for this development is a gate dielectric that shows an acceptable flatband voltage shift of 3 V measured on MOS capacitors after 10 MRad irradiation (Sec. 4.5). However, PXD-TO contains not only MOS capacitors but also transistors with "DEPFET-equivalent" doping profiles and will test a larger parameter space in order to find an even more radiation-hard dielectric. The resulting technology module, "Thin Oxides," will be implemented in the final Belle II production. The reduction in $g_q$ expected from the increased gate capacitance will be compensated by a slight reduction of the gate length $L$ according to equation 4.2.

Two wafers of the PXD6 batch are produced with plasma etched polysilicon to get smoother and smaller transistor gates. If this development succeeds, we can increase the $S/N$ by taking advantage of the scaling potential of the device (Fig. 4.5). Finally, a third metal layer, made of copper, must be implemented to provide the under-bump metal necessary to bump-bond the ASICs and to reduce the voltage drops on the all-silicon module (Sec. 4.4). The DEPFETs for the Belle II vertex detector will be processed on wafer bonded SOI material with a specially developed technology including 18 photo-lithographic mask steps, 9 ion implantations, two polysilicon layers, and three metal layers.

### 4.2.3 Prototype Designs for Belle II

Due to the similar requirements, the Belle II DEPFETs are developed based on the ILC devices [7]. However, the size of the ILC pixel is in the range of only $25 \times 25\,\mu\text{m}^2$ while the pixel sizes envisaged for the inner and outer layers of the Belle II vertex detector are $50 \times 50\,\mu\text{m}^2$ and $50 \times 75\,\mu\text{m}^2$, respectively. The most challenging requirement is the short frame readout time of $20\,\mu$s, corresponding to 12.5 ns per row for 1600 pixel rows in total. Such a short readout time would, however, be incompatible with the intrinsic settling times on the long readout lines and control lines of the matrix and would introduce additional high-bandwidth noise. In order to relax the speed requirement, we split the matrix to have readout electronics on both ends and we implement in addition a high degree of parallelization: Four rows are read out simultaneously, which multiplies the number of drain lines and readout channels by the same factor (Fig. 4.6, left side). The combined factor of eight leads to a row processing time of 100 ns, which is manageable.

The increase of pixel size over the ILC design implies an increased charge collection time, i.e., the drift time that the charge needs to reach the internal gate. An efficient drift structure is implemented to increase the electric field, especially at the pixel edges. Device simulations show that drift times can be kept below 60 ns with the favored design, which should be short enough to suppress ballistic deficits. The readout itself is more critical. Using the fast single sampling readout mode, we must sample and reset the signal within the 100 ns window. To provide enough time for signal settling and sampling, the reset (Clear) time must be less than 20 ns. As in the ILC designs, a narrow channel width is needed to reduce the time for the electrons to drift from the internal gate into the clear. Therefore, the small linear basic transistor structure [7] is adopted from the proven ILC designs, while the clear and drift regions are expanded to fill the enlarged pixel area.

The baseline design comprises a DEPFET structure with individual source and drain regions embedded in drift and clear regions (see Fig. 4.6, right side). This approach has fast charge collection times and is suitable for both pixel sizes to be used in Belle II. A blow-up of a successfully tested ILC design [8] was chosen as a fall-back solution. Here, each contact region (drain, clear, source) is shared by two neighboring pixels, which leads to a very compact design. The disadvantage is a longer collection time for the charge generated at the edge of the pixel.

Although all devices were designed by using technology and device simulations tools [6, 8, 9], there are features that are unpredictable. In particular, production yield and radiation hardness may be influenced drastically by minor design changes. Therefore, we will test several designs, concepts and options. We have numerous smaller wire-bonded test matrices on the PXD6 wafer in order to be able to select the best option based on laser and radiation tests.

### 4.2.4 Power consumption of a DEPFET matrix

Due to the row-wise addressing, the power consumption depends only on the number of columns:

$$P = 2 \cdot I_d \cdot V_{ds} \cdot n_{col} = 2(100\,\mu\text{A})(5\,\text{V})(800) = 0.8\,\text{W} \tag{4.3}$$

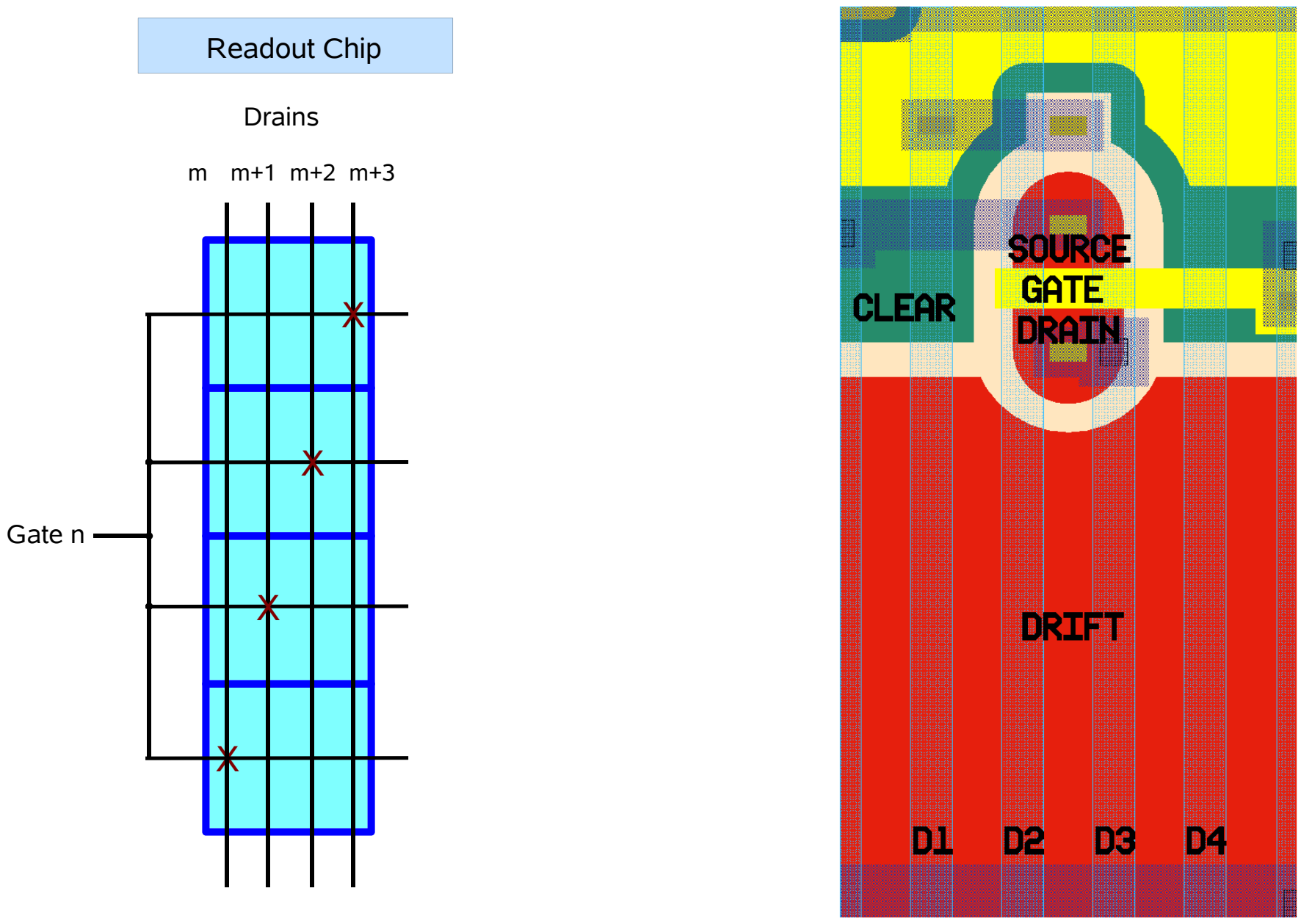

*Figure 4.6: Left: Principle of 4-fold parallel readout and resulting matrix organization for the DEPFET baseline design. Right: a single pixel.*

| | Collection | Read | Clear |
|---|---|---|---|
| $V^{*}_{Gate}$ | $3V$ | $-2V$ | $-2V$ |
| $V^{*}_{Clear}$ | $2V$ | $2V$ | $15V$ |
| $V^{**}_{Drain}$ | $-5V$ | | |
| $V_{ClearGate}$ | $-2V$ | | |
| $V_{Bulk}$ | $10V$ | | |
| $V^{***}_{Back}$ | $-10V ... -50V$ | | |

* control lines (transferred via Switcher)
** provided by DCD input stage
*** depending on detector thickness and contact mode

*Table 4.1: Operation voltages of a DEPFET matrix with respect to source ($0V$)*

The factor of two results from splitting the matrix into two readout directions (Sec. 4.1). The power necessary to ramp up and down the gate and clear lines is mainly consumed in the SWITCHERs (Sec. 4.3). Compared with these contributions, we can neglect the power consumption caused by the leakage current: this is at most a few mW even after irradiation.

### 4.2.5 Matrix Operation Voltages

Table 4.1 summarizes the operating voltages of the matrices. Note that the gate and clear gate voltages may have to be shifted to more negative values to compensate for radiation-induced oxide damage during operation.

## 4.3 Electronics (ASICs)

As illustrated in Fig. 4.4, the DEPFET pixel modules are read out in a rolling shutter mode: a matrix segment consisting of four multiplexed rows (Fig. 4.6) is selected by pulling the gate line to a negative potential using the SWITCHER chip (see below). The selected DEPFET pixels send currents to the vertically connected drain lines. These currents are processed at the bottom of the matrix by the Drain Current Digitizer (DCD) chips. The DCD performs an immediate digitization of the current with 8 bit resolution and sends the data serially through many low-swing single ended lines to a third chip, the Data Handling Processor (DHP), which buffers and analyzes the digital data stream and performs a zero suppression. The remaining data are then sent to the off-module data handling hybrid DHH (Fig. 4.22).
Three different ASIC types are, therefore, used for the matrix readout:

- The SWITCHER chips select and clear the DEPFET four-row segments by generating voltage swings of 10–20 V. In order to reduce the inactive module border, both functionalities are merged in a single chip, as shown in Fig. 4.7. The channel selection is done internally by a simple shift register. Precise timing is provided by additional STROBE input signals.
- The DCD chips process the currents in the columns and digitize the analog values.
- The DHP chips perform data processing, compression, buffering and fast serialization and send the data off module to the DHH.

All chips are mounted directly on the detector module using bump-bonding and flip-chip techniques, leading to all-silicon modules.

### 4.3.1 Readout Cycle

The readout cycle starts with the selection of a DEPFET four-row segment by a SWITCHER chip. The corresponding SWITCHER channel is enabled by setting a single enable bit in its shift register and by confirming the selection using the fast STROBEGATE signal. All SWITCHER control signals are generated by a master DHP. The low voltage control signals are translated internally to the required higher gate-on voltages that steer large MOS switches. These devices pull the gates of a complete DEPFET row (with a capacitance of as much as 50 pF) to the GATEON potential. The DEPFET currents flow then through the column lines into the DCDs. The current receivers in the DCDs are based on transimpedance amplifiers. They hold the column-line potentials at a constant level (virtual ground) to make the readout speed more insensitive to the large column-line capacitances (50 pF). Since the potentials of the lines are constant, the signal currents flow directly into the DCDs rather than charging the line capacitances. The current receivers can also amplify the current signals and act as low-pass filters to reduce the noise. The amplified and shaped signals are then processed by current-mode cyclic ADCs. Two are placed in each DCD channel to achieve the necessary sampling frequency of at least 10 MHz. The shaping time and the gain of the current receivers are programmable: a larger shaping time leads to a better S/N and a lower readout speed, while a larger gain leads to an increased S/N and a decreased input-signal dynamic range.
The currents in the columns consist of a standing current of 50–100 $\mu$A, required to operate the DEPFET devices, and a small signal component that depends on the charge gain $g_q$ of the devices. Due to device mismatch in the matrix, the standing currents are not identical for all devices, leading to *pedestal fluctuations*. Digitizing signals *plus* the large pedestals would require a very high resolution ADC with correspondingly high area and power consumption. The pedestals must therefore be subtracted prior to digitization. Two methods are implemented:

- *Double sampling:* A clear pulse sent to the DEPFETs during current readout removes charges in the internal gate. Subtracting the current before and after this pulse leaves only the signal component, as the pedestal is constant. This Read–Clear–Read sequence is implemented by providing a current storage cell with a sufficient dynamic range that holds $-(\text{pedestal} + \text{signal})$ in the first phase. The pedestal signal after the clearing is subtracted by simply adding the drain current and the stored value.
  Because the DCD must take two samples during each readout cycle, the shaping time of the current receiver must be below 35 ns for an 80 ns readout cycle. Using such a readout mode simplifies the DHP design since no digital pedestal correction is needed (see below). However, the short sampling and shaping times and the fact that two samples are taken in each readout cycle lead to an increased readout noise that may reduce the theoretical advantange, in terms of $1/f$ noise cancellation, of the double correlated sampling.

- *Single sampling:* A constant current is subtracted from the drain current and the signal is directly digitized. Clearing can be done after this. There is no need to wait for signal settling after the clear, so that this Read–Clear procedure is faster. The ADC would still have to accommodate the full pedestal current dispersion. To reduce the required range, a 2 bit DAC in each channel is available to subtract varying currents. If we assume a 50 $\mu A$ DEPFET bias current and a conservative 20% pedestal variation among the pixels, the required dynamic range would be 10 $\mu$A. With the highest possible gain for the DCD, we have an ADC range of 4 $\mu$A and a full range (with DACs) of 16 $\mu$A. Note, however, that the gain of the transimpedance amplifiers in the current version of the DCD can be programmed to lower values, yielding an even higher range of the ADCs, which can be as high as 64 $\mu$A. The control words for these DACs can be different for each pixel. They are stored in the DHP and are sent serially from the DHP to the DCD.
  The shaping time of the current receiver can be longer in this mode. Assuming a readout cycle of 80 ns and a clear time of 20 ns, the shaping time can be 60 ns. Under such conditions, the noise introduced by the current receiver (its input transistor) will be lower than the noise introduced by the ADC (typically, rms(ADC) is 40 nA).
  The use of two-bit pedestal correction in the DCD will not be enough to reduce the remaining part of the pedestal dispersion to a level that allows an effective signal recognition and zero suppression. For this reason, an additional pedestal current compensation will be performed in the DHP on the digital codes received from the DCD. The DHP will also calculate the average of the signals received from every DEPFET row-segment and subtract this average part from the signals itself and in this way compensate for any average-signal (common mode) fluctuation. The low-frequency signal fluctuations can be eliminated by updating the pedestal values frequently.

After the readout cycle the DEPFET is switched off till the next cycle, 20 $\mu$s later. Because of this frame readout mode, all signals arriving within the 20 $\mu$s between two readout cycles are stored (integrated). However, this might cause some increased dead time due to the SuperKEKB injection mode. Injection will be every 20 ms (50 Hz). Since the injected bunches are noisy, the trigger will be vetoed during the passage of these bunches for a couple of $\mu$s until the noise has calmed down about 3 ms after injection. However, due to the long integration time, it is not possible to veto the DEPFET during the passage of the noisy bunches. In the worst case, the DEPFET data will be flushed with noise hits during the 3 ms, corresponding to a dead time of 15%. A possible solution would be to decrease the injection rate while injecting more bunches each time.

### 4.3.2 SWITCHER

The SWITCHER steering chips are mounted on the $\sim 2$ mm inactive rim of the module. These chips provide fast ($\sim 10$ ns into 50 pF) voltage pulses of up to 20 V amplitude to activate gate rows and to clear the internal DEPFET gates. The SWITCHER is implemented in the AMS 0.35 $\mu$m HV technology using special design techniques to achieve radiation hardness (enclosed NMOS gates, guard rings). One SWITCHER will be able to control 32 different DEPFET matrix segments (row groups). Each channel provides a clear- and a gate-driver. Channels are selected via a simple shift register with serial input to (output from) the previous (next) chip. Figure 4.7 shows a preliminary layout of the chip with an approximate size of $2 \times 3.6\,\mathrm{mm}^2$.
All required SWITCHER control signals and power must run on the 2 mm-wide module rim. The number of I/O signals has therefore been minimized and an unusual bump pad geometry has been chosen to simplify routing on the module. To simplify system design, the SWITCHERs have a floating digital ground and use fast, low power level shifters. A JTAG interface with boundary scan logic allows for interconnectivity tests on the module.

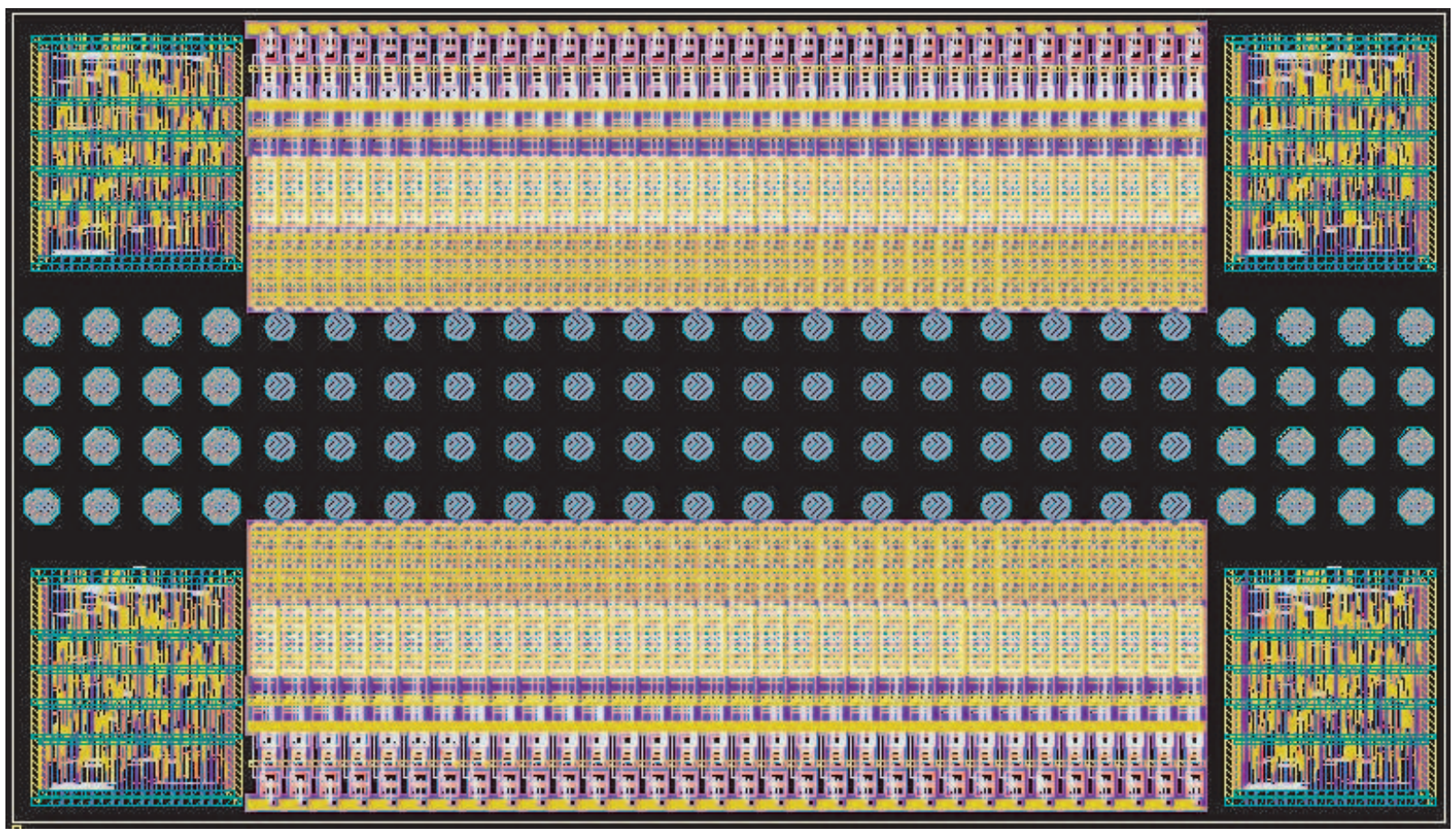

*Figure 4.7: Preliminary layout of the SWITCHER chip. The $4\times 16$ bump pads for the gate/clear outputs are in the center; additional $4 \times 4$ pads for power and control are at top and bottom.*

The SWITCHER optimized for the Belle II requirements ("SWITCHERB") has been submitted in February 2010. The existing prototype ("SWITCHER 4") has been successfully tested. Radiation hardness of at least 37 MRad has been demonstrated.

### 4.3.3 DCD

The currents generated by a selected DEPFET four-row segment are routed out via column lines and read out by the DCD chips placed at the end of each half-module matrix. Each DCD has 256 analog inputs. The DCD is implemented in the UMC 0.18 $\mu$m CMOS technology using special radiation hard design techniques. The chip occupies an area of $3.2 \times 5\,\mathrm{mm}^2$. Chip size and input pitch have been adapted to the Belle II requirements ("DCDB"). DCDB uses bump bonding with bumps on the UMC technology provided by EuroPractice. Figure 4.8 shows the DCDB layout and the organization of the pads.

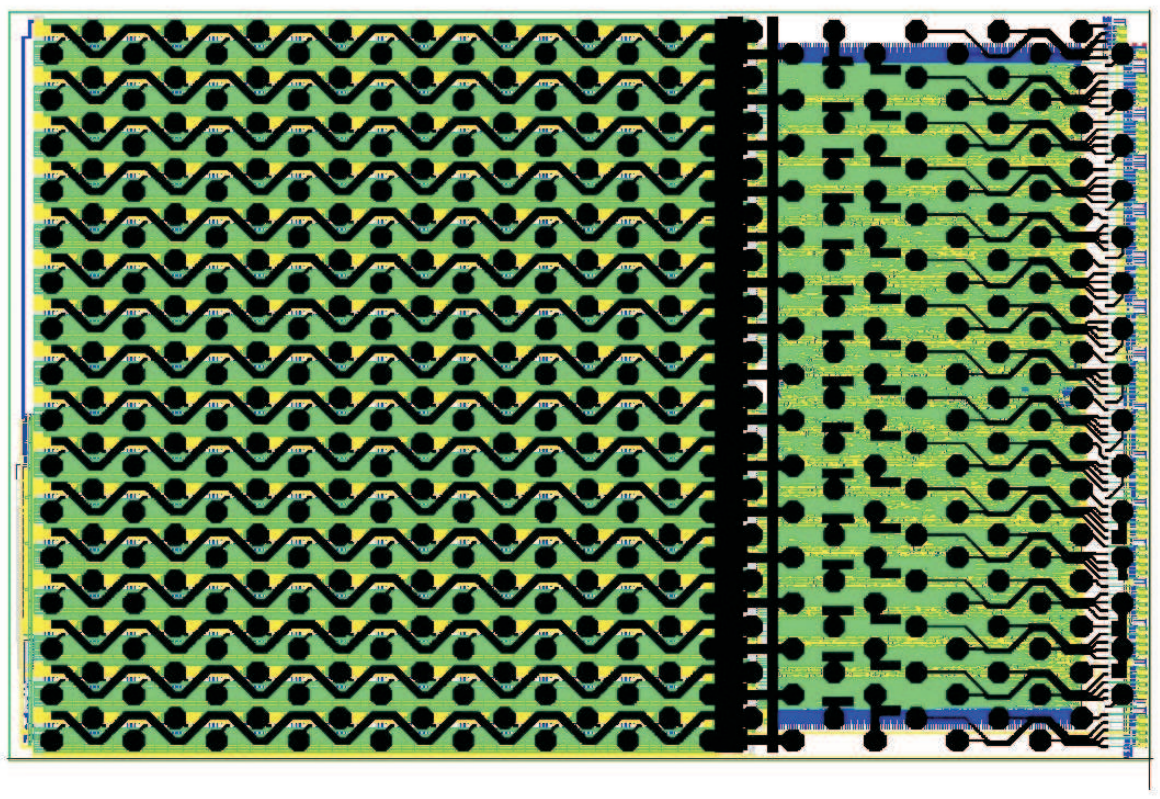

*Figure 4.8: Layout of the DCDB chip. The upper part contains the 256 inputs; the lower part contains the digital logic and digital input / output pads.*

| | Noise | | | |
|---|---|---|---|---|
| Readout Mode | Receiver | Sampling cell | ADC | Total |
| Double corr. sampling | $\sqrt{2}$·60 nA | 40 nA | 40 nA | 102 nA |
| Single sampl. w/o ped. corr. | 25 nA | 0 | 40 nA | 47 nA |
| Single sampl. with ped. corr. | 25 nA | 0 | 10 nA | 30 nA |

*Table 4.2: Expected noise contributions by readout mode, for a readout cycle time of 80 ns.*

Each channel consists of an input stage, two ADCs and digital logic. The analog input stage keeps the column line potential constant (necessary for fast readout), compensates for DEPFET pedestal current variations, amplifies the signal, and provides shaping for noise reduction. For these purposes, the input stage offers programmable gain and bandwidth, a two-bit DEPFET pedestal current compensation (using digital data from the DHP) and sampling of the pedestal current. The analog signal is digitized using current-mode cyclic ADCs, two of them placed in each channel. A large synthesized digital block decodes and derandomizes the ADC raw data, which are then transmitted in a well sorted sequence to the DHP chips using fast parallel 8-bit digital outputs. The digital timing and data protocols are tuned for rapid communication with the DHP. Several operating modes using single sampling or double correlated sampling are possible.

Due to the expected quite significant power dissipation of up to 600 mA per chip, voltage drops on the internal power buses are a significant concern. This is addressed by several methods: the use of separate power buses for sensitive parts, high power supply rejection, massive decoupling, and the use of the extra redistribution metalization of the bumping technology for additional power routing.

The DCD chip optimized for Belle II (DCDB) has already been submitted, received and is being characterized. A very similar prototype chip ("DCD2") has been successfully tested and used to readout a prototype DEPFET matrix at nearly full speed with a noise value smaller than 100 nA (see Fig. 4.9). Radiation tolerance of at least 7 MRad has been proven. Table 4.2 summarizes the noise figures that are expected in the various readout modes for the latest version of DCD.

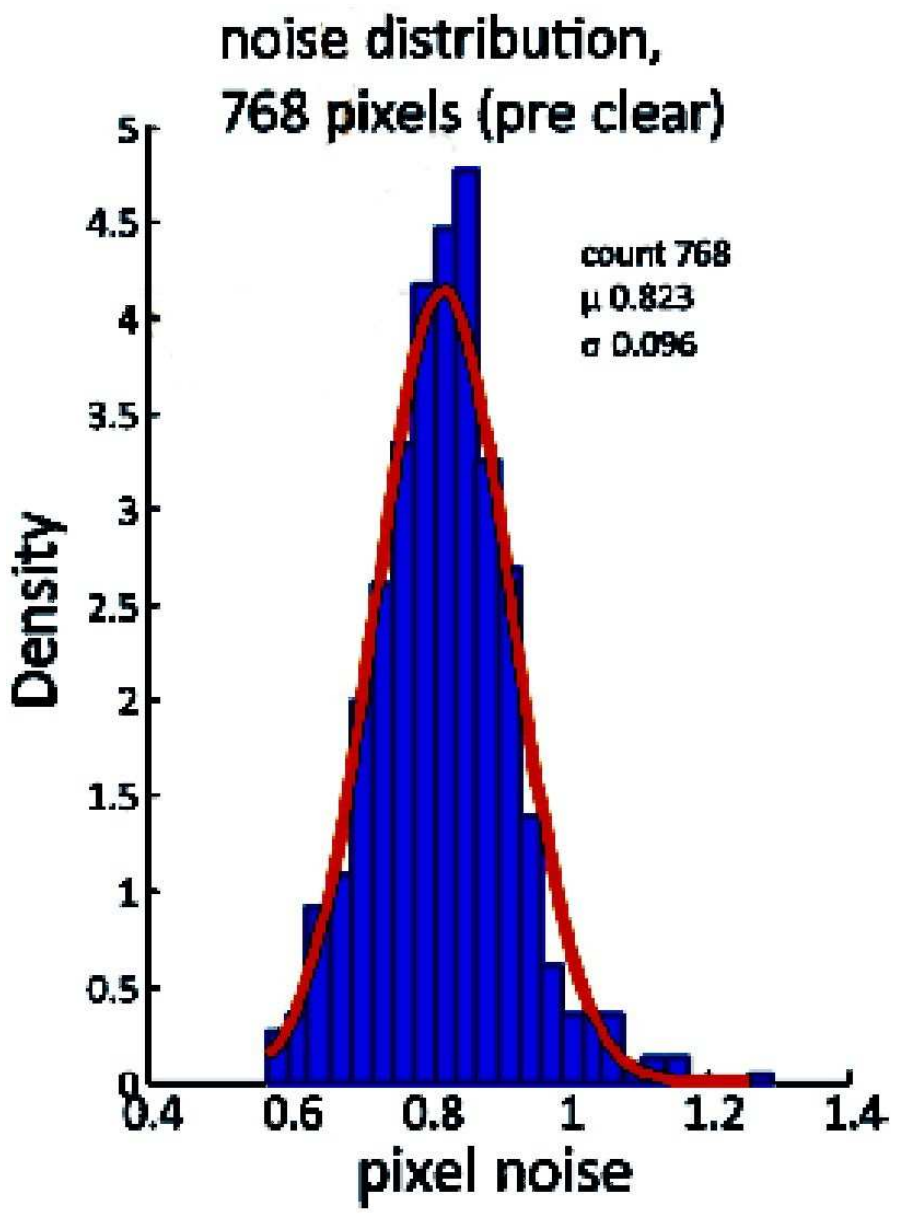

*Figure 4.9: Noise when reading out a DEPFET matrix with DCD2 at 11 MHz (in ADC counts, one ADC count corresponding to approximately 100 nA).*

### 4.3.4 DHP

The data handling processor (DHP) reduces the data rates produced by the DCDs. This is achieved by zero-suppressing the DCD data and by reading out only triggered data. To allow for efficient zero-suppression, the raw data is corrected for common mode noise and pedestal fluctuations. The latter is an offset in the signal data and is not removed by the DCD when running in single sampling mode (Sec. 4.3.3). These individual offset values will be continuously updated within the DHP to compensate for low frequency fluctuations and drifts.
The common mode refers to an offset that is found in all signals sampled at the same time. This offset is time dependent and therefore different for each sampling interval. To remove this offset, the mean signal value of all simultaneously sampled channels is calculated inside the DHP and subtracted from the original input signals. The zero-suppressed data, together with a time stamp, are stored in hit buffers and selected for readout upon receipt of a trigger. The triggered hit data are transmitted to the data handling hybrid (DHH) with high speed links running at 1.25 Gbit/s per chip, resulting in an overall maximum data rate of 5 Gbit/s per half-module. In addition to the data processing task, the DHP will also serve as a module controller by generating all fast timing signals for synchronization of DCD and SWITCHER and by providing a slow control JTAG interface. The DHP chip will be implemented in the IBM 90 nm CMOS technology to accommodate the memory needed for pedestal and frame storage.

### 4.3.5 Bumping and Flipping

To make the module as compact as possible, all chips are bumped and flip chip mounted directly onto the silicon substrate. This procedure requires two major steps: the *bumping*, i.e., the deposition of (in our case) solder spheres onto the chip pads and the *flipping*, i.e., placing the chips upside down onto the substrate and melting the solder by a temperature cycle. For both steps, the chip/substrate surface must be suitable for soldering: they must have an under-bump metalization (“UBM”).
For the substrate, this is assured by the post-processing technology used in the MPI *Halbleiterlabor*, HLL, as explained in Sec. 4.4. The DCDB and DHP chips can be ordered with bumps (and UBM) being placed by the vendors or subcontracted companies. Presently, the pitch for these bumps is limited to 200 $\mu$m, which is a serious boundary condition for the designs. The DCDB chip was ordered with such bumps. For the SWITCHER, there seems to be no possibility to get bumped chips from the *Multi-Project Wafers* (MPW). We will therefore use a backup technology that was developed in Heidelberg: In the first step, gold studs are placed on the SWITCHER pads, acting as a UBM for the solder. In the second step, individual solder balls are placed by a solder bumping machine (available in Heidelberg). This technique can accommodate the SWITCHER's pitch of 150 $\mu$m. The technology has been successfully tried out on dummy chips, with solder balls being placed by the vendor of the machine.
The flipping can be done on a manual machine with very high precision available in Heidelberg, or with an automated machine that will be available at HLL. Other involved institutes have (or will have) flipping capabilities as well.

## 4.4 Module Description and Interconnections

The DEPFET pixel array is made on thin (about 75 $\mu$m) detector grade silicon supported by a directly bonded silicon frame of about 400 $\mu$m thickness. The dimensions of the ladders and the overall layout of the PXD are described in detail in Sec. 4.8. The readout electronics, the lines for power, data, and slow control are placed on both short sides of the ladder, outside the sensitive volume; the steering chips for the row-wise read out are attached on the thick frame along the long side of the ladder. The circuit paths for these steering chips are integrated on the support frame of the sensor module. Figure 4.10 shows the sketch of a module for the inner layer of the PXD. In the following, we describe the thinning technology and the interconnection details between sensor and read-out chip, as well as the off-ladder interconnection at the ends of the ladders.

### 4.4.1 Thinning Technology

Back-thinning of microelectronic chips is widely used in semiconductor industry. However, these technologies are not easily applicable for fully depleted sensors with an electrically active back side. DEPFETs are FETs on fully depleted bulk with a deep-$n$ implant under the channel. They have a structured implant, contacts, and metalization at the back side. Conventional thinning, i.e., grinding or chemical mechanical polishing (CMP) of the back side, is usually done after the top side processing is finished. The processing steps at the back side (implantation and laser-assisted implant activation or diffusion) would then have to be done with a thin and fragile wafer, a procedure that is extremely difficult and cost-intensive.
To avoid these critical steps, we have chosen another way to produce ultra thin sensors (see [10]). Figure 4.11 illustrates our approach to fabricating such thin devices with a minimum of processing steps after thinning. The feasibility of this approach has been shown with the production of 50 $\mu$m thin PiN diodes. The volume-generated reverse current of under 100 pA/cm$^2$ achieved with these thin devices is extremely encouraging. The mechanical stability of the diced chips is sufficient for safe handling and mounting. There is no distortion visible on the 50 $\mu$m thin large area silicon membrane with single-sided aluminum metalization.
The same technology of deep anisotropic etching is used to reduce the material in the supporting frame along the long sides of the ladder. In this way, the contribution of the silicon frames can be reduced by roughly 30%. Figure 4.12 shows the result of the technology simulation with a close-up of the perforated frame. The assessment of the overall material budget of the first layer is seen in Table 4.3, including all major contributions (Si, SWITCHER, bumps, Cu interconnection, and frame) and with selected thicknesses of the sensitive layer. The overall material budget in the baseline configuration (Option 1 in Table 4.3) is 0.19% of a radiation length. The material contributions are calculated under the following assumptions:

- Sensitive area of the first ladder: $1.25 \times 9\,\text{cm}^2$
- Support frame: $0.1 + 0.2\,\text{cm}$
- Switcher-Sensor interconnect: Au stud bumps, $\phi = 48\,\mu\text{m}$
- Cu Layer: $t = 3\,\mu\text{m}$, 50% coverage in sensitive region
- Switcher dimensions: $0.15 \times 0.36\,\text{cm}^2$
- Number of Switchers: 8
- Material reduction by frame etching: 1/3

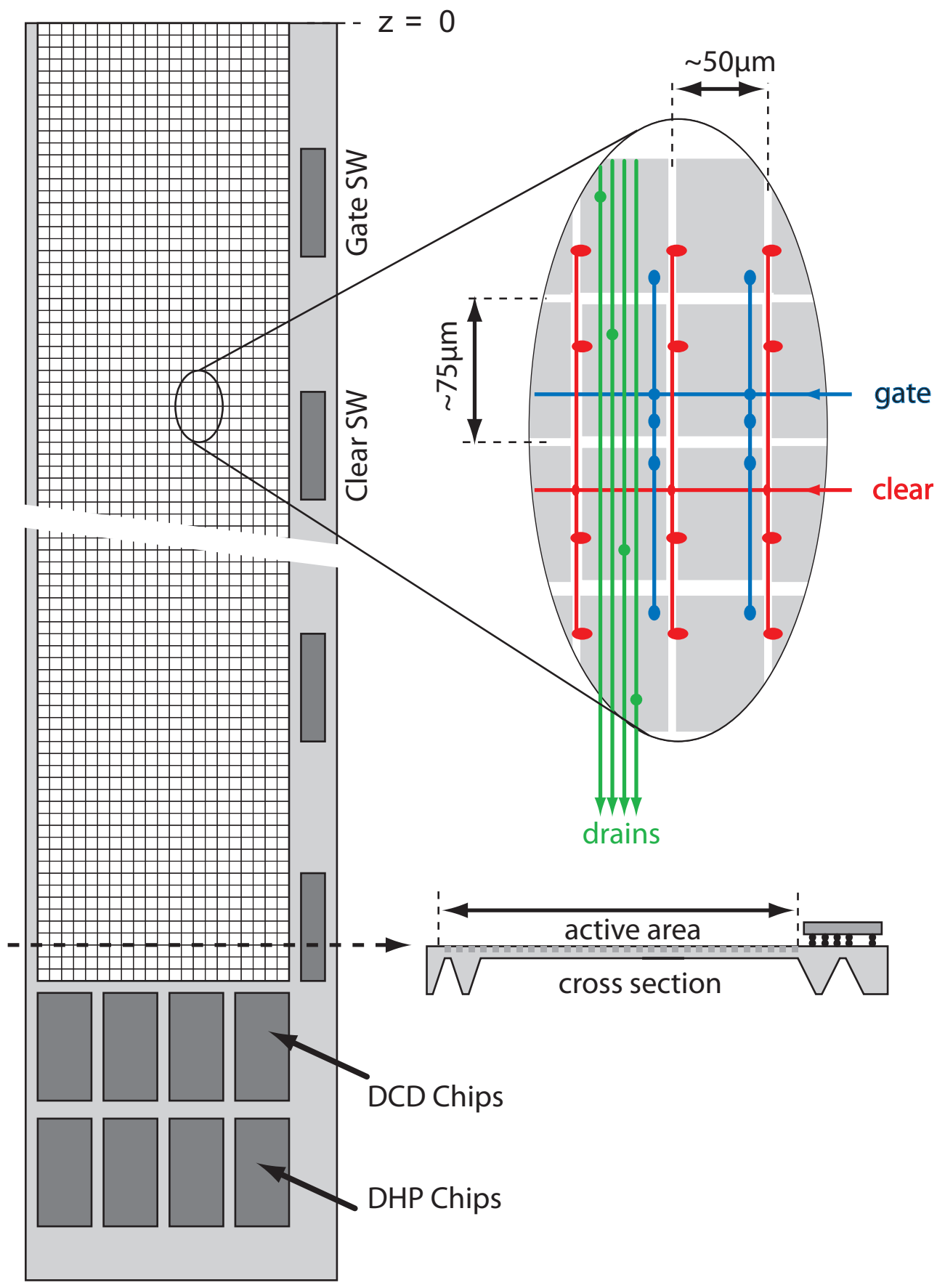

*Figure 4.10: Sketch of a first layer module with thinned sensitive area supported by a silicon frame, illustrating the DEPFET ladder concept.*

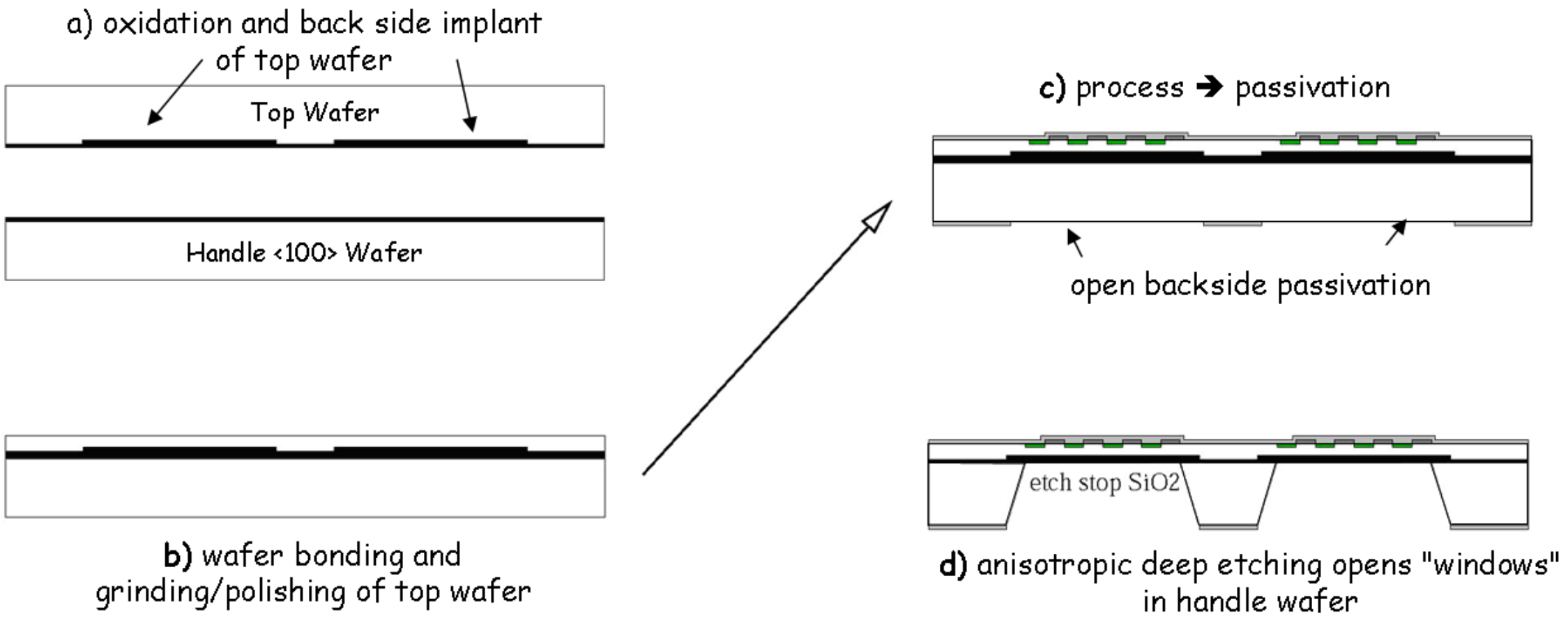

*Figure 4.11: The process sequence for production of thin silicon sensors with electrically active back side implant. (a) It starts with the oxidation of the top and handle wafer and the back side implantation for the sensor devices. (b) After direct wafer bonding, the top wafer is thinned and polished to the desired thickness. (c) The processing of the devices on the top side of the wafer stack is done on conventional equipment; the openings in the back side passivation define the areas where the bulk of the handle wafer will be removed. (d) The bulk of the handle wafer is removed by deep anisotropic wet etching and the etch process stops at the silicon oxide interface between the two wafers.*

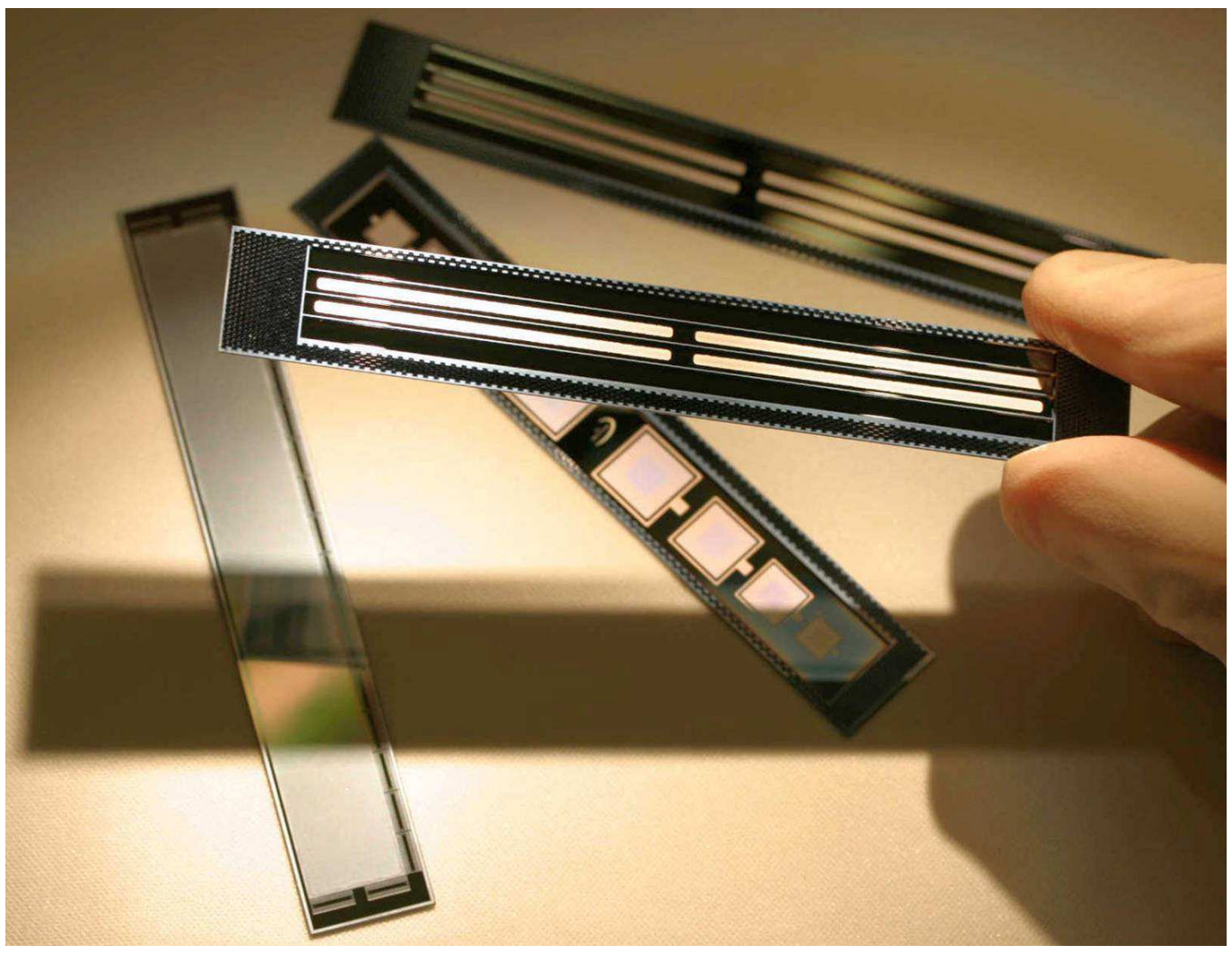

*Figure 4.12: Full size electrically active samples made using the process of Fig. 4.11.*

The largest contribution to the overall material budget is the support frame, as seen in the breakdown of the various components in Fig. 4.13.

| Option | Sensitive Thickness ($\mu$m) | Switcher Thickness ($\mu$m) | Frame Thickness ($\mu$m) | Total Thickness (% of $X_0$) |
|---|---|---|---|---|
| 1 | 50 | 500 | 450 | 0.16 |
| 2 | 50 | 500 | 400 | 0.15 |
| 3 | 50 | 200 | 450 | 0.15 |
| 4 | 50 | 200 | 400 | 0.14 |
| 5 | 75 | 500 | 450 | 0.19 |
| 6 | 75 | 500 | 400 | 0.18 |
| 7 | 75 | 200 | 450 | 0.18 |
| 8 | 75 | 200 | 400 | 0.17 |

*Table 4.3: Material assessment of the DEPFET Ladder (L1), averaged over sensitive area.*

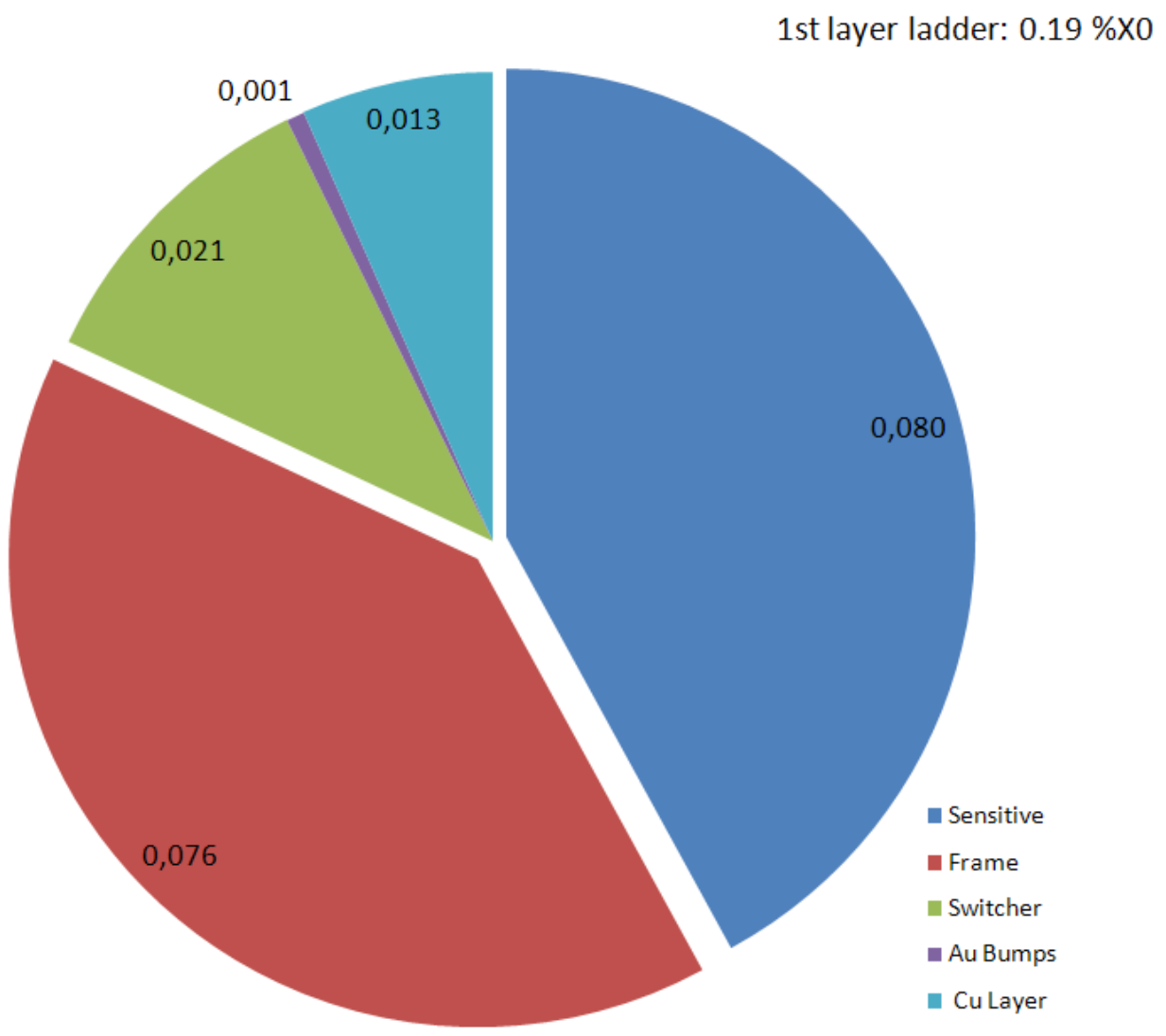

*Figure 4.13: Breakdown of the material contribution of various ladder components.*

### 4.4.2 Low-Mass Joint Between Half-Ladders

The ladders for the second layer of the PXD are too long to fit on a 150 mm wafer. These ladders are split into two halves that are joined using a low-mass joint that maintains mechanical stability. As shown in Fig.4.14, the support frame is laser-cut at one end of the ladder, leaving a narrow band ($\approx 250\,\mu$m) of inactive silicon as a support. In addition, there are V-grooves etched in the support frame perpendicular to the laser-cut line. This etching is done simultaneously with the back-thinning of the handle wafer. The two half-ladders are then aligned to each other and glued. The mechanical stability is provided by inserts (ceramic or silicon) into the V-grooves, again attached with the appropriate adhesive (Araldite 2011).

This technology is indispensable for the second layer. Although an entire first-layer ladder could fit on a 150 mm wafer, yield considerations compel us to use this low-mass joint there as well. Simulations show that the small insensitive region ($\approx 500\,\mu$m) and the little extra material does not deteriorate the overall performance of the PXD (see Sec. 4.11).

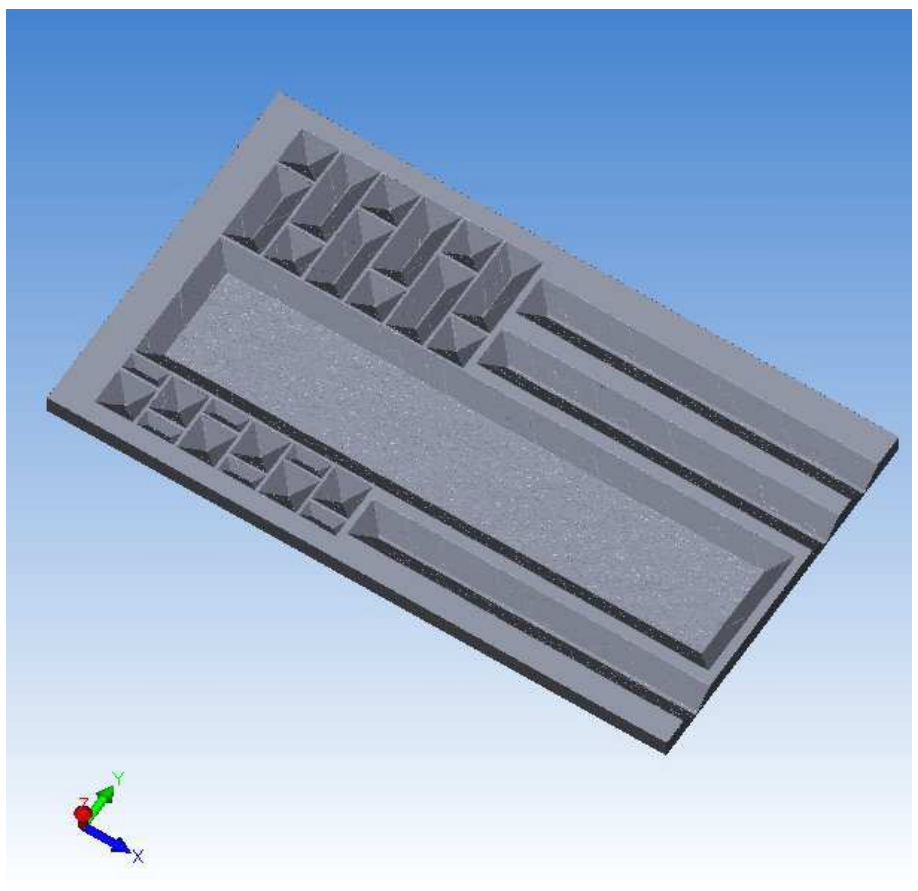

*Figure 4.14: Perforation of the silicon frame of the ladder for further material reduction (technology simulation result, not to scale).*

### 4.4.3 On- and Off-Ladder Interconnection

As described in section 4.3, there are three types of ASICs placed directly on the DEPFET ladder. The interconnection between these ASICs and the DEPFET ladder is done by conventional flip-chip bonding of the solder-bumped ASICs to the landing pads on the ladder. The bumping of DCD and DHP is done by the manufacturer using standard wafer-level bumping technologies. The bumping of the SWITCHER chips has to be done on the chip level using a solder-jet technology provided by the company Pac-Tech.
On the other side of the DEPFET ladder, an appropriate landing pad with solder-wettable metallurgy has to be prepared. This metallization is done during the processing of the DEPFET wafers by the electroplating of copper and tin. This process provides a third low-impedance wiring layer on the DEPFET ladder.
The data transmission, slow control of the ASICs, and power supply connections are realized by a three-layer impedance-controlled Kapton cable glued directly to the end of the ladder. The electrical connections are done by conventional wire bonds from the pads on the Kapton cable to three banks of wire-bond pads on the inactive part of the sensor ladder. A PCB serves as a patch panel to connect to the more remotely located DHH (4.9). It needs some circuitry for impedance matching of the HF signal lines to twisted pair cables upstream and overvoltage protection for the power lines.

## 4.5 Radiation Hardness

To date, we have no reliable simulations of the expected irradiation level. We do not know the exact composition of the background nor the spectrum of the particles and photons. Extrapolations from Belle are difficult, since important machine parameters will change. An educated guess would suggest 1–2 MRad/y. Pending more information, we want to assure that PXD components survive at least 10 Mrad.
Rough estimates can be derived from the expected occupancy. Assuming an occupancy of 0.5% from charged particles in pixels of $50 \times 50\,\mu\text{m}^2$ and $20\,\mu\text{s}$ integration time, the absorbed dose is 2.6 Mrad in one year ($10^7$ s), assuming a particle flux of about $10\,\text{MHz/cm}^2$. Assuming that most of the particles are low energy electrons just above the cutoff given by the magnetic field (6 MeV), they have a damage factor of 0.08 compared to 1 MeV neutrons. Hence, the expected NIEL damage is in the order of $10^{13}$ n/cm$^2$ (1 MeV neutron equivalent).

### 4.5.1 DEPFET Irradiations

The DEPFET is a MOSFET device and suffers mainly from threshold voltage shifts due to an increase of oxide charges and interface states in the gate oxide. In addition, some increase of noise due to interface traps and a reduction of the amplification is expected. NIEL damage increases the leakage current and contributes to the shot noise. In case significant shot noise is generated, it could be reduced by lowering the sensor temperature. Since there is no charge transfer over long distances during the operation of DEPFET matrices, trap generation by NIEL is of minor importance for this kind of device.

#### 4.5.1.1 Ionizing Radiation Damage

All MOS technologies are inherently susceptible to ionizing radiation. The main total ionizing dose effects are:

- A shift of the threshold voltage to more negative values caused by radiation induced charge build up in the oxide;
- Build up of states at the interface between Si and $SiO_2$, resulting in an increased sub-threshold slope and possibly a higher $1/f$ noise of the transistors;
- Reduction of the transconductance ($g_m$) due to a lower mobility of the charge carriers in the channel after irradiation.

The degradation of MOS transistors with a certain thickness of the gate dielectrics (about 200 nm in the case of the present DEPFET prototypes) for a given total ionizing dose in the oxide depends on the technology and the biasing conditions during irradiation.
During several irradiation campaigns at an x-ray source in Karlsruhe, DEPFET devices and MOS test structures were irradiated up to 10 MRad. During normal operation the DEPFET is in "charge collection mode," i.e. fully depleted bulk with empty internal gate and switched off by applying a positive gate voltage with respect to the source. The transistors of a row are only switched on during the short readout period. The time ratio between "off" and "on" state $t_{on}/t_{off}$ in the vertex detector (assuming a 1600 pixel array read out at both sides) is in the order of 0.1%. Thus, the irradiation of the test devices was done with the transistors in "off" state with an empty internal gate to test for the radiation tolerance in this most frequent operation mode.
The dose rate of the X-ray source ranged from 73 kRad($SiO_2$)/h up to 625 kRad($SiO_2$)/h. The dosimetry was provided via the generated photo current in a diode in depletion mode and using the spectrum of the x-ray source. Also, this dosimetry was cross checked with RadFET monitoring devices. The input characteristic of the devices were measured immediately (approximately 1 min) after each irradiation period and the threshold voltage was extracted by a quadratic extrapolation of the $I_{Drain}(V_{Gate})$ curve to $I_{Drain} = 0$.
Figure 4.15 shows the threshold voltage shifts directly after irradiation. The irradiation was stopped after 10 Mrad($SiO_2$) and the devices were annealed at room temperature. The threshold voltage shift after 300 h annealing at room temperature is reduced from 16.5 V to 12.5 V.

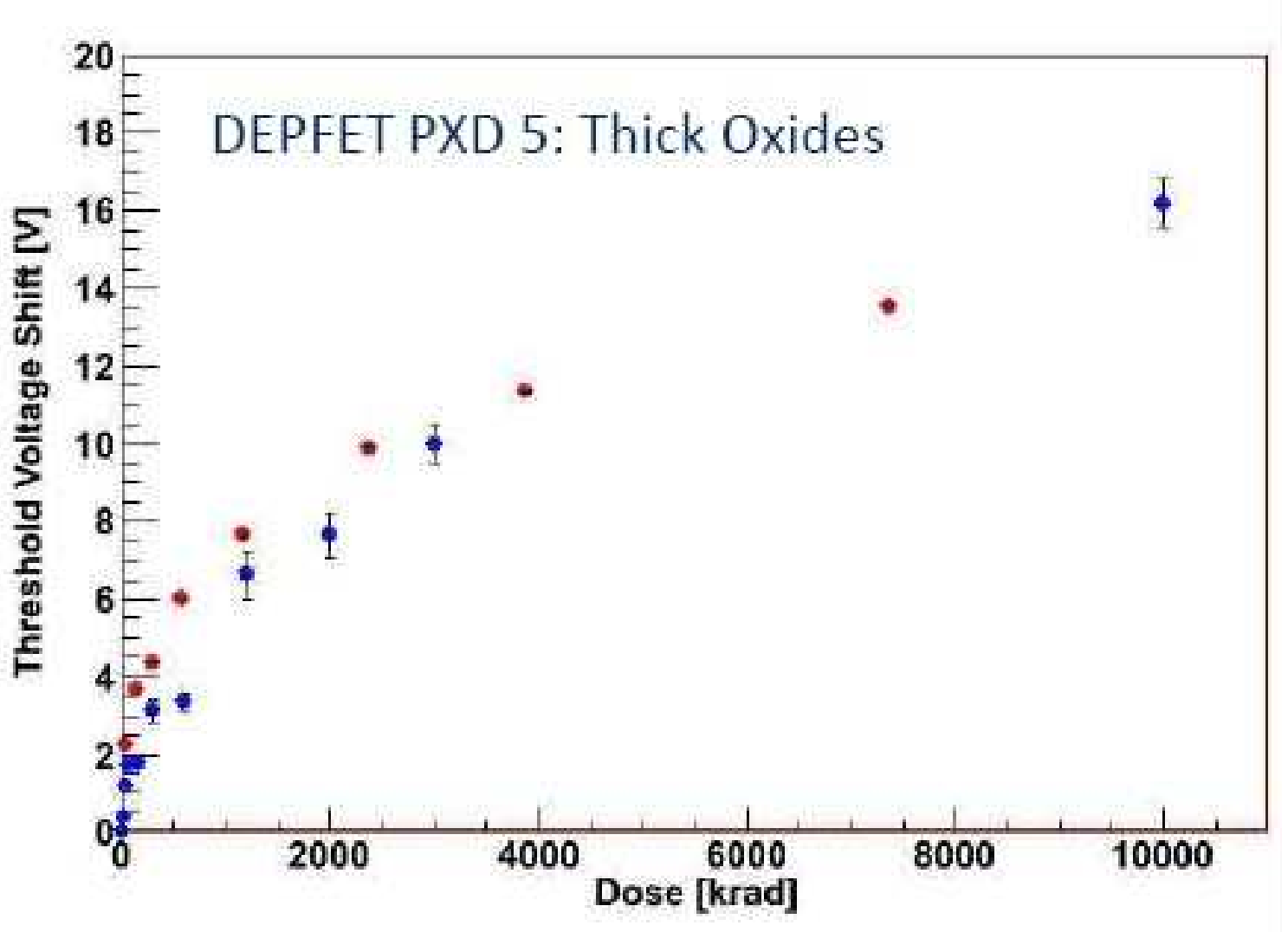

*Figure 4.15: Threshold shift of DEPFETs irradiated up to 10 Mrad, before annealing. The red dots indicate a device with the gate voltage adjusted to optimal transistor performance while the blue dots show a device with 0 V gate voltage all the time.*

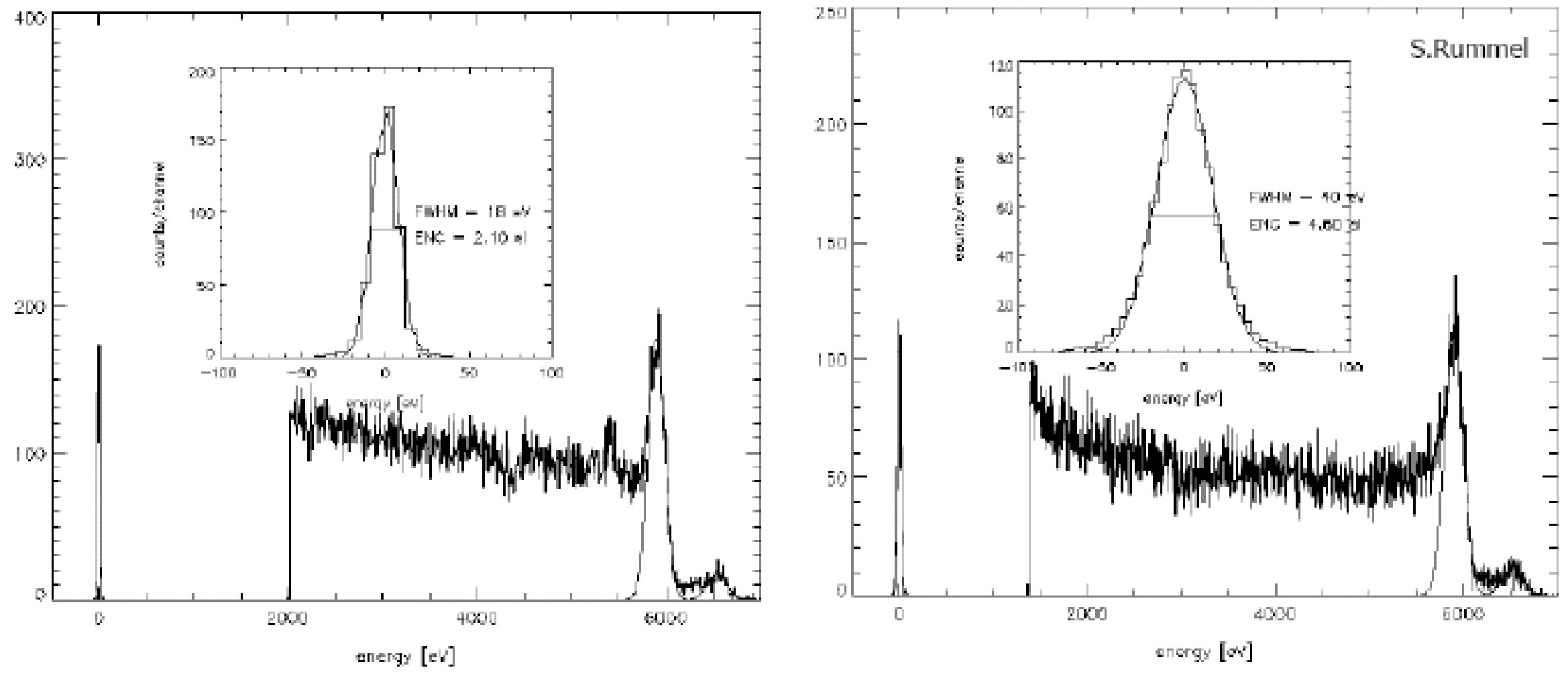

*Figure 4.16: $^{55}$Fe spectrum taken with a linear MOS-type DEPFET. Left figure: before irradiation. Right figure: after irradiation (4 Mrad).*

The most important issue is the spectroscopic performance of the DEPFET after irradiation. Fig.4.16 shows an $^{55}$Fe spectrum taken with a DEPFET before and after irradiation. The equivalent noise charge of the pedestal peak changed from 2.1 $e^-$ to 4.6 $e^-$ (rms), attributed to the increase of the low frequency noise which does not depend on the shaping time. Even after this high total ionizing dose of 4.1 MRad(Si), the DEPFET shows only a moderate noise increase of about 6 electrons equivalent noise charge. Hence it can be concluded that the DEPFETs are radiation tolerant to doses up to 10 Mrad. However the large threshold voltage shifts demand large corrections of the operation parameters. These corrections may be outside the range of the control electronics. In case of inhomogeneous irradiations across a module it might be impossible to find optimal parameters for the complete sensor. Hence it is highly desirable to reduce these large shifts.

#### 4.5.1.2 Irradiation of MOS test devices with thin oxides

A well known remedy for effects of oxide damage in MOS devices is the reduction of the oxide thickness and the addition of a nitride layer which partly compensates the damage in the oxide. However, the thickness of the oxide is not a free parameter but it also affects properties like the internal amplification $g_q$ which is expected to vary depending on the oxide thickness $d$ like $\sqrt{d}$. In order to evaluate this possibility test diodes with thinner oxides (100 nm) and Nitride layers have been produced and irradiated up to 10 Mrad. The results are shown in Fig. 4.17.

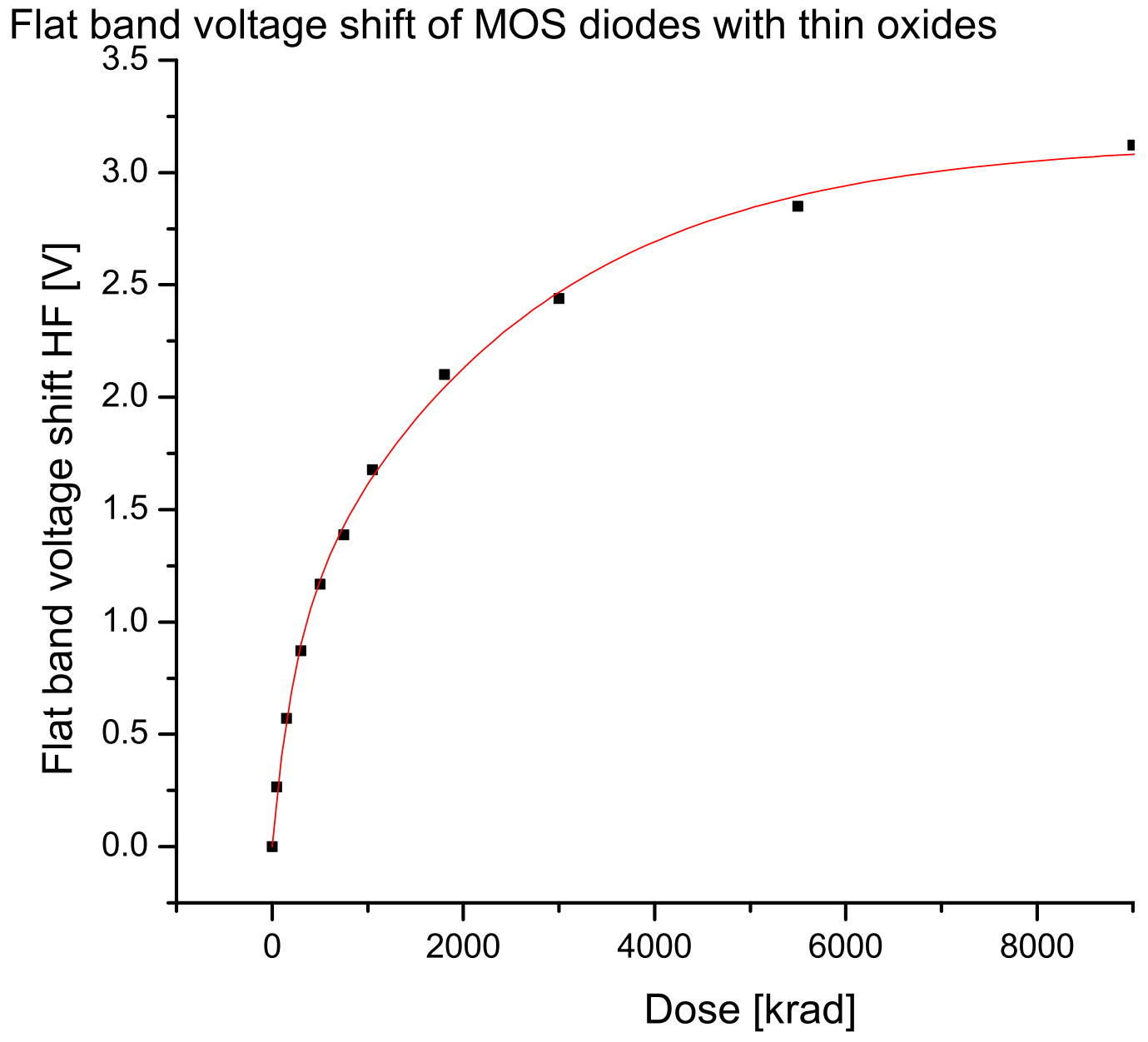

*Figure 4.17: Flatband voltage shift of diodes with thin oxide.*

Indeed the shifts of the flat band voltage, which correspond to the threshold voltage shifts in MOSFETs, are dramatically reduced. Only 3 V were observed which is reduced to 2.2 V after 500 h of annealing compared to 12 V with thick oxides. Such shifts can be compensated with the existing SWITCHER electronics and do not pose a serious challenge for the operation of the DEPFET sensors. Further tests are planned optimizing the oxide-nitride combination.

### 4.5.2 Proton and Neutron Irradiation

In addition to the already presented irradiations with photons, DEPFET devices have also been irradiated at the LBNL Cyclotron with protons and neutrons. The protons had an energy of 30 MeV and the irradiation was carried out with all DEPFET terminals grounded. The achieved proton fluence was $1.2 \times 10^{12}$ p/cm$^2$ and the total ionizing dose at this fluence was calculated to be 283 krad(Si). For the neutron irradiation the fluence was $1.6 \times 10^{11}$ n/cm$^2$ with the energy of the neutrons ranging from 1 to 20 MeV. This is summarized in Table 4.4.
The IV characteristic of DEPFETs after these irradiations is shown in Fig. 4.18. This confirms the shift of the threshold voltage observed after gamma irradiation. The increased subthreshold slope after ionizing radiation is due to the build-up of interface traps. Also the reduced transconductance of the proton irradiated sample suggests this interpretation. The expected higher level of the low frequency noise is verified by the spectral analysis of the noise. The results confirm the radiation damage models of MOS devices, which predict that neutron irradiation does not degrade the subthreshold slope and has no influence on threshold voltage.

| irradiation | dose | flux | $\Delta V_{th}$ | $g_m$ | $I_{leak}$ |
|---|---|---|---|---|---|
| x-ray | 10 Mrad | 0 | 16V | -25% | |
| neutron | 0 | $2.4 \times 10^{11}$ n/cm$^2$ | 0 | - | 1.4 pA |
| proton | 283 krad | $3 \times 10^{12}$ n/cm$^2$ | 5V | -15% | 26 pA |

*Table 4.4: Results of DEPFET irradiations*

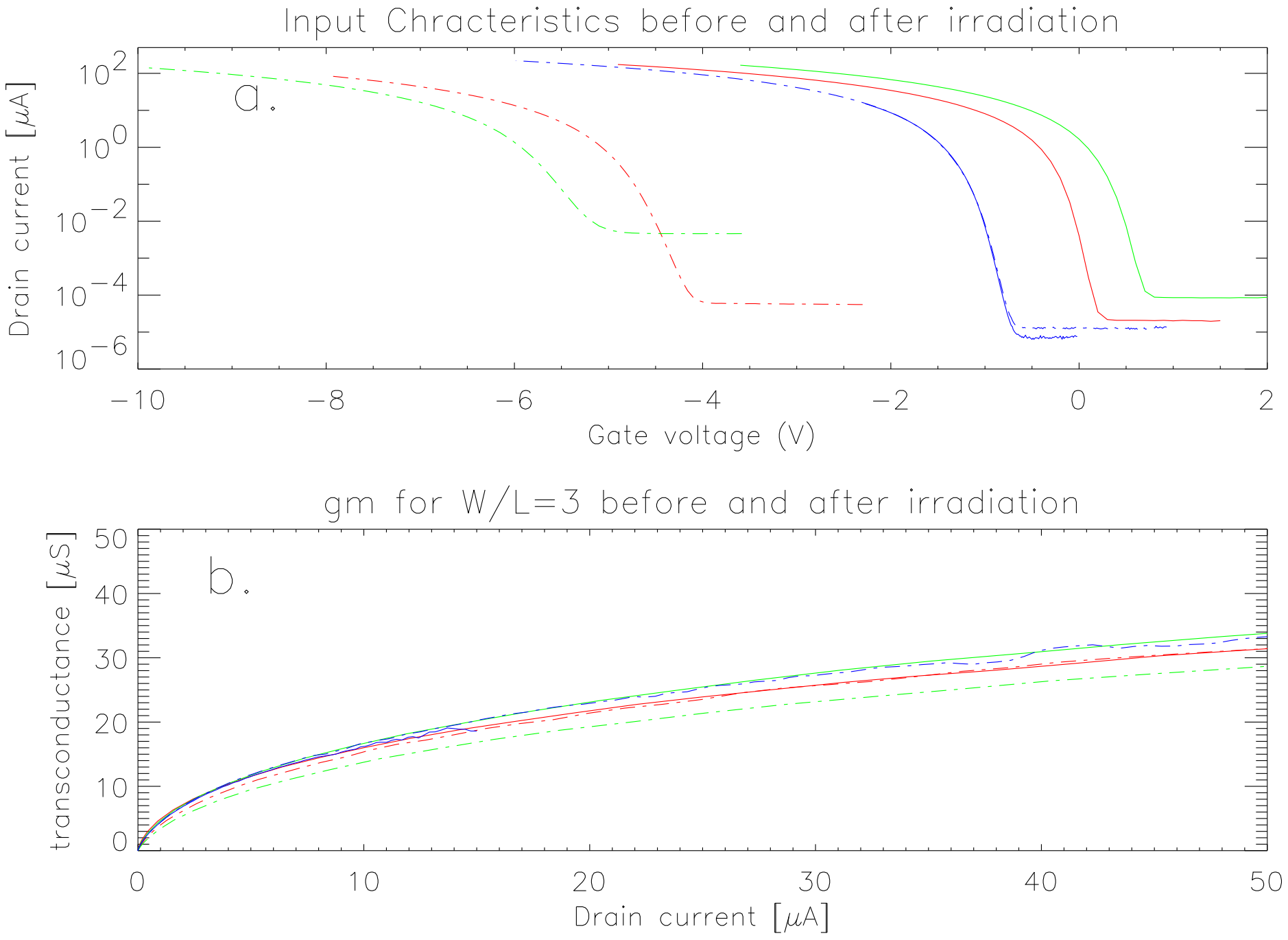

*Figure 4.18: (a) Drain current-gain voltage characteristics of three DEPFETs before (solid lines) and after (dashed lines) irradiation with 912 krad $^{60}$Co (red), $1.2 \times 10^{12}$ $cm^{-2}$ 30 MeV protons (green), and $1.6 \times 10^{11}$ $cm^{-2}$ 1–20 MeV neutrons (blue). (b) Transconductance normalized to $W/L = 3$ of the transistors before and after irradiation, using the same color code.*

However, the bulk damage due to non-ionizing energy loss (NIEL) of neutrons and protons increases the bulk generated leakage current flowing into the internal gate and hence the shot noise of the device.
At SuperKEKB, the shot noise contribution will be proportional to the readout time of a complete matrix. The cell accumulates leakage current between two consecutive readouts. This will lead to an offset that can be corrected. However, the fluctuations of this contribution lead to noise. For a readout cycle of $20\,\mu$s, the noise contribution would be 78 electrons for a NIEL damage corresponding to a flux of $10^{13}\,\text{n/cm}^2$.

| Component | Process | max | Dose | Result | comment |
|---|---|---|---|---|---|
| DCD | UMC 180 nm | 10 Mrad | 7.5 Mrad | ok | DSM CMOS |
| SWITCHER4 | AMS 0.35 HV | 10 Mrad | 36 Mrad | ok | |
| DHP | IBM 90 nm | 10 Mrad | | | DSM CMOS |
| Optolink | | ? Mrad | | | |
| DHH ASIC | | ? Mrad | | | |
| Regulator | | ? Mrad | | | |

*Table 4.5: Irradiations of PXD electronics components*

#### 4.5.2.1 Summary and Future R&D

We conclude from these results that the DEPFET, using a technology with thin oxides, is remarkably radiation tolerant beyond the doses expected at SuperKEKB. The threshold voltage shift of only $-3\,$V after 10 Mrad can be handled by a simple readjustment of the gate voltage needed to switch on the DEPFET for readout. The shot noise due to leakage currents can be controlled by the sensor temperature. Assuming a bulk leakage current of $4.9 \times 10^{-11}$ A at room temperature, corresponding to $10^{13}\,\text{n/cm}^2$, a surface current of $1.2 \times 10^{-11}$ A, the total noise, including DEPFET channel noise and DCD noise, would reach 200 electrons (or a S/N of 20:1) at a temperature of 27°C.[1] As a next step, we plan to test various combinations of thin oxides and nitride layers to minimize the threshold voltage shift. In addition, we will perform irradiations with low energy electrons, which are probably the dominant background at SuperKEKB.

### 4.5.3 Electronics

Similar to the DEPFET, the CMOS ASICs suffer from threshold voltage shifts. Fortunately, due to the extremely thin oxides used in the deep submicron (DSM) CMOS processes used for most PXD ASICs, these shifts are very small and the ASICs turn out to be inherently radiation tolerant [11, 12, 13]. The expected total ionizing dose at SuperKEKB does not challenge the current deep sub-micron CMOS technologies as long as the designers use radiation hard layouts for the transistors. An exception is the Belle II SWITCHER, which uses a $0.35\,\mu$m HV-CMOS technology with rather thick oxides of the HV-transistors. The design therefore uses devices that are based on thin oxides and uses enclosed structures and guard rings throughout. A prototype of the Belle II SWITCHER, SWITCHER4, has been exposed to 36 Mrad with no observed degradation.
An overview of the ASICs used for the PXD and the status of radiation tests is given in Table 4.5.

Another item is the tolerance against single event upset (SEU). Some protection is implemented in the DHH:

[1] The temperature dependence of the leakage current is

$$I(T) = I_0 \left(\frac{T}{T_0}\right)^2 \exp\left(\frac{E_g}{2kT_0} - \frac{E_g}{2kT}\right) \tag{4.4}$$

where $I_0$ is the current at reference temperature $T_0$ (in Kelvins), $T$ the operation temperature (in Kelvins), $E_g$ the band gap energy (1.1eV), and $k$ the Boltzmann constant.

- Configuration registers use triple redundancy with majority voting;
- Data buffers have error detection and correction that is able to correct for one bit error in a 32 bit word and detect two bit errors in a word;
- The PLL is designed like the ATLAS FE-I4 chip [14], which is able to work in a much more challenging environment than that of SuperKEKB.

The Switcher and DCD will use triple redundancy with majority voting for all configuration bits. The SEU rate can be measured by reading back all three bits via JTAG.

### 4.5.4 Other Components

In principle, other components like glues or thermal grease may be subject to radiation damage. Since members of our collaboration have experience with the construction of silicon detectors for the LHC, the use of LHC grade materials is foreseen, thus ensuring sufficient radiation hardness.

## 4.6 Quality Control and Assurance

In the following section, general aspects of quality control via testing and pre-commissioning are discussed, and the transition from wafer-level testing of matrices to pre-testing of ASICs to full functionality testing is outlined. Finally, the ladder qualification and planned quality assurance procedures are depicted.

### 4.6.1 Production quality characterization

#### 4.6.1.1 Sensors

Production quality tests are done via static measurements of test structures that are homogeneously distributed over the wafer surface. Testing of the DEPFET matrices starts on the wafer level via static measurements where the basic transistor characteristics are being measured. Technology related defects of the matrices can be discovered this way and the structures preselected according to the static measurement results then undergo testing in a dynamic operation. Static measurements will be done on the wafer level. A semiautomatic probe station is used for static $I/V$ and $C/V$ measurements. The probe station is integrated into the semiconductor parameterization system and the system allows the selection of structures of interest on the wafer and measurement of those in an automatic mode. Special devices designed for the characterization of the production quality assurance as well as small size matrices ($128 \times 16$ pixels) can be measured in this manner. Static measurement of full size module matrices can be performed using a specially designed prober card.

#### 4.6.1.2 ASICs

The final ASICs will be available by March 2011 and they will be subject to quality test using a probe station. If problems are observed, there will be enough time for resubmission and retesting without impact on the overall schedule.
Dedicated needle cards will be used to contact the I/O and power pads. The chips will be designed in a way to allow testing of most important features using a minimal number of I/O pads. For instance, in the case of the DCDB chip, only the pads at the outer pad-row will be used: the JTAG pads, clock and synchronization input, eight output pads and a test signal pad that allows emulating DEPFET input currents and testing of all analog channels without contacted DEPFET matrix. A special needle card for the pads with bumps will be used. In the case of the SWITCHERB chip, the clear and gate outputs of all 32 channels will be multiplexed to a monitor pad that can be contacted by a needle card.

#### 4.6.1.3 Belle II PXD ladders

Once the pixel layout is fixed and the final production of the Belle II PDX matrices is done, the matrices to be installed in the detector have to be selected. The static measurements will be performed in the same way as planned for the prototype production. Since the steering and the readout chips are to be bump-bonded onto the rim of the matrices, the characterization of the matrices as well as the chips has to be done independently so that pre-characterized chips will be bump-bonded onto the selected matrices.

### 4.6.2 Dynamic tests

#### 4.6.2.1 Sensors

Structures preselected according to the static measurements undergo testing in a dynamic operation. The test system for the DEPFET characterization has followed the evolution of the DEPFET matrices from ILC-type to Belle II-type matrices.

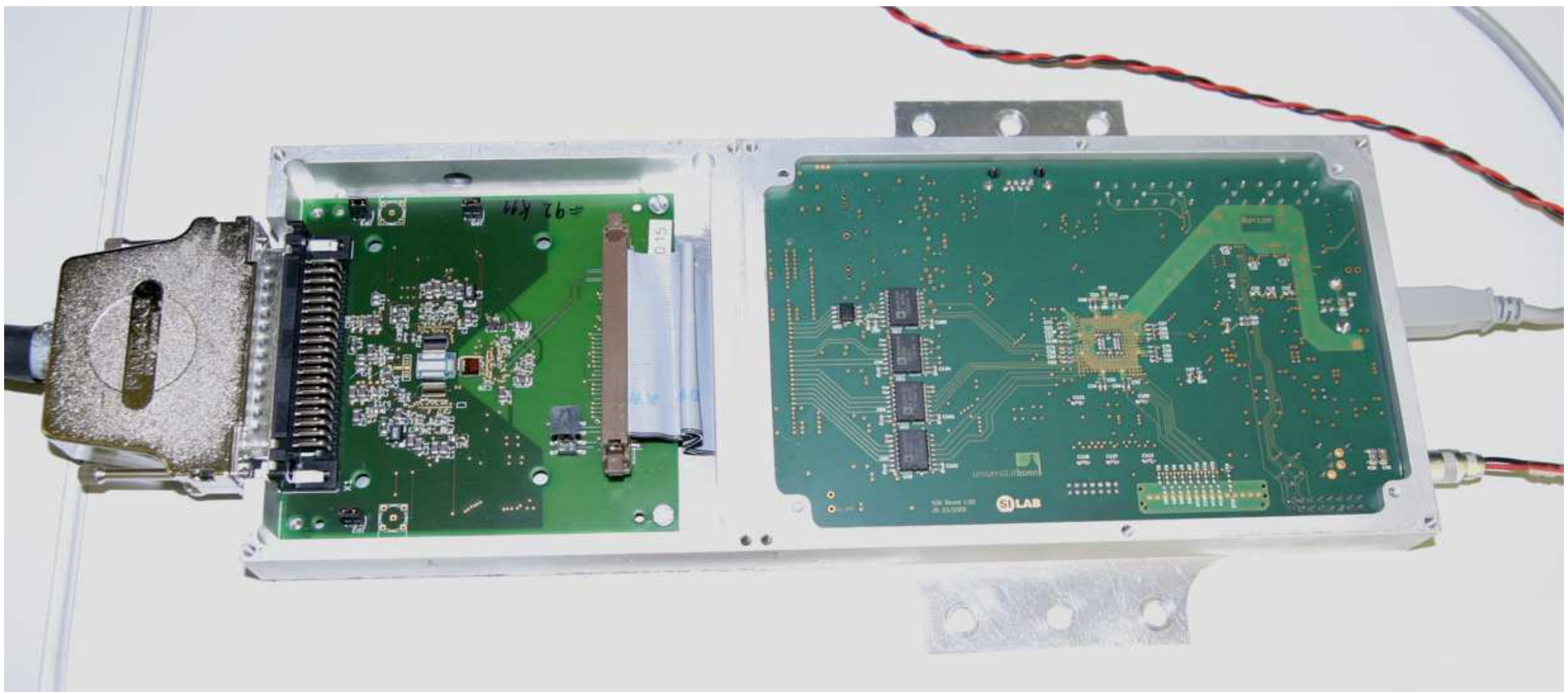

*Figure 4.19: View of the S3B system for testing the ILC matrices.*

For the characterization of matrices for ILC applications, the S3B system was developed and most of the measurements shown in this report have been obtained using this system (Fig. 4.19). This system was made for the testing of the $64 \times 128$-pixel matrices with one CURO II readout chip [15] and controlled by two SWITCHER III chips [15], one for the DEPFET gate control and the second for the clear pulses. The matrix and the ASICs are housed on a PCB board or hybrid, where the output current of the CURO is converted into a voltage signal by an external low noise and fast transimpedance amplifier. The hybrid is connected to a second board where the voltage signal from the hybrid is digitized by a 14-bit ADC and stored in SRAM cells for a subsequent readout. This board contains a SPARTAN 3 FPGA that is in charge of the slow control and readout of the chips, the timing, and the data transfer to the PC via USB. Special multi-channel power supplies have been built to conveniently provide all voltages. LEMO connectors are used to trigger the system and to interface to the data acquisition during the beam tests. Everything is housed in a compact metal frame with standardized supports to simplify alignment in the test beam.
An intermediate setup is being developed for testing the Belle II prototype matrices of the PXD6 production. This test system is schematically shown in Fig. 4.20. For the operation of standard $128 \times 16$-pixel matrices, one SWITCHER-B chip (SWB) for gate and clear control and one DCD-B and a DCDRO chip[2] for the readout is used. (A description of the chips is given in Sec. 4.3.) The matrix is bonded on a separate ceramic carrier accommodated with a pin connector. An adapter board is designed to accommodate conversion from bump-bonded chips (SWB, DCDB and DCDRO) to wire-bonded matrices. The hybrid is connected to a readout board via a flat band cable. This board contains the drivers for the SWB, the memory, and a VIRTEX4 FPGA to generate all timing and control signals. The data transfer to a PC is done via a single USB2.0 link. A multi-pin connector is used to feed all required supply voltages.

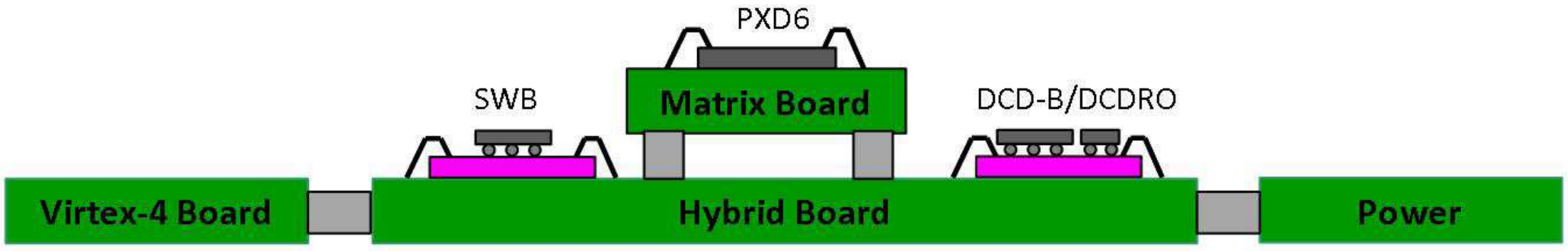

*Figure 4.20: Schematics of the planned test system for the PXD6 matrix qualification.*

This system allows simple and fast exchange of small prototype matrices due to its modular design. For testing of the half modules, adapter boards are not required since the design is intended for bump-bonded steering and readout chips. The same FPGA board or DHP chip (Sec. 4.3) can be used for the readout of the signals from the modules.
Final testing of the ladders for the Belle II PXD detector will be performed using the final versions of the SWB, DCDB, DHP, and DHH chips. The full DEPFET matrix characterization is performed combining the source and the laser scan results.

**4.6.2.1.1 Source tests** A number of tests are necessary to characterize the DEPFET module in terms of noise, gain, etc. For an absolute measurement of the noise and $g_q$, spectra using a number of radioactive sources are measured (Fig. 4.21, left). The measurements are usually performed using $^{55}$Fe, $^{109}$Cd, and $^{133}$Ba irradiation. A typical value measured from such measurements for a thick matrix is $g_q = 0.280\,\mathrm{nA}/e^-$. This value is smaller than that usually obtained from measurements on single DEPFETs and small arrays with $^{55}$Fe irradiation. This effect is understood and is due to the incomplete clear process in the used test system. The system noise was measured to be 15 ADC units for the S3B system, which corresponds to 90 nA. As explained in Sec. 4.3, the performance of the new readout chip and steering chip should lead to an improved S/N ratio.

**4.6.2.1.2 Laser tests** Laser tests are done to study inhomogeneities of the matrix response and can be performed nowadays at several collaboration institutes using laser systems with different wavelengths: 682 nm, 810 nm, and 1055 nm. The laser is triggered either by an internally generated signal in the S3B board or by an external pulse generator that triggers both the laser and the acquisition system. The laser spot is focused on the matrix backplane with spot sizes which vary, depending on the setup, from 2.5 $\mu$m to 5 $\mu$m. The DEPFET module is mounted on a motor-controlled $xy$-stage with steps of the order of 1 $\mu$m, allowing a full scan of the matrix surface so that the charge collection mechanism can be studied. Figure 4.21 (right) shows an example of the variations of the pixel signal across the matrix. Variations of about $\pm 5\%$ between the mean cluster signals can be observed. The reasons for these variations in pixel gain are not fully understood and are under investigation. Even so, they can be corrected in the analysis and therefore pose no harm to the final performance of the detector.

[2] As the DCDB has non-standard low swing single ended signal outputs, it cannot be connected directly to an FPGA. To be able to test the chip without the foreseen DHP, the DCDRO support chip has been designed to convert the signals to the LVDS standard.

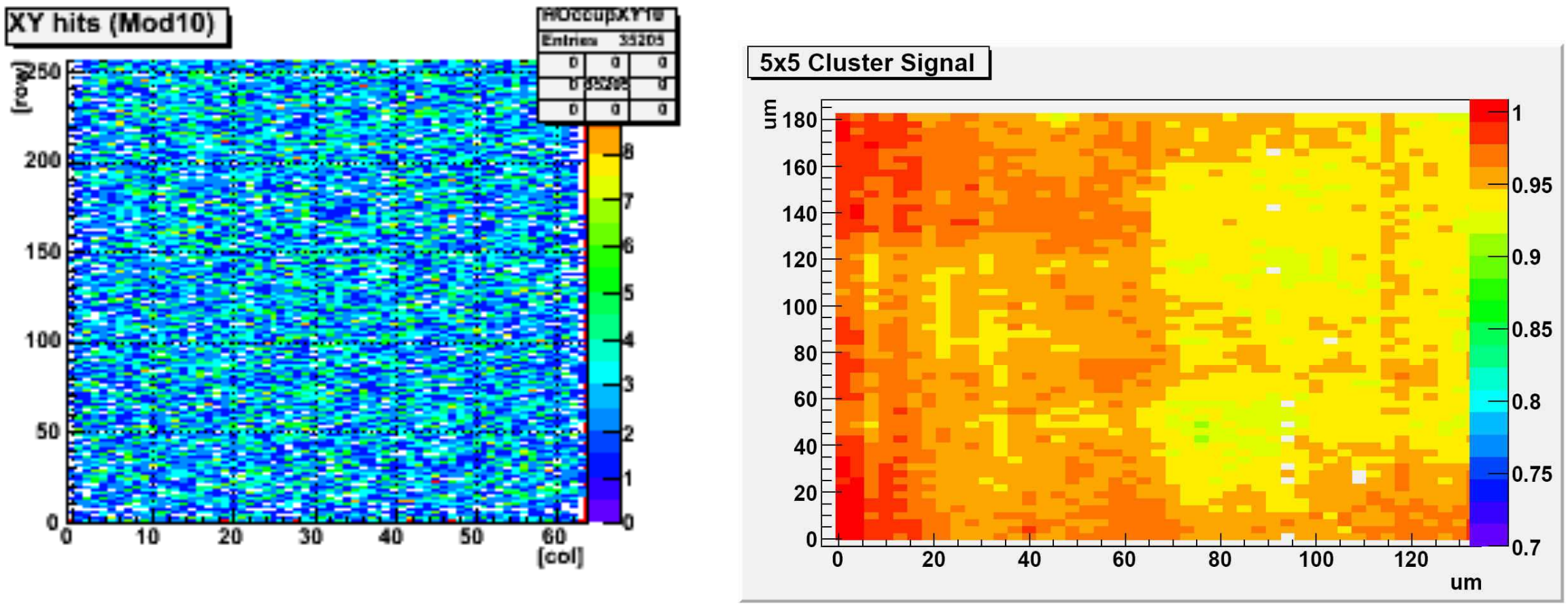

*Figure 4.21: Left: Hit map for an ILC-type matrix during the source testing. The CURO chip is located at the bottom. Right: Result of a laser scan over an ILC-type matrix, using the common-mode correction method. The $5 \times 5$ cluster signal is plotted versus the laser position $(x, y)$. Steps are $3\,\mu m$ in both directions. The CURO readout chip is located on the right side (readout direction from left to right).*

#### 4.6.2.2 Belle II PXD ladders

To perform a functionality test of the ladder, a sensor probing with the probe card is planned. All contacts are provided via needle contacts placed onto the bump-bond pads, and the steering of the matrix is performed via an external instrument. In this way, a full functionality test can be done. The results of these scans can be used for the selection of the matrices for the ladder construction. Bump-bonding of the chips is done sequentially with a functionality test performed after every bonded chip. Once all the chips are in place and the flex cable connected, the final characterization of the ladder performance occurs, using both source and laser tests.

### 4.6.3 Belle II PXD commissioning

Once all the pre-tests are performed and the selection of ladders to be used in the detector is made, the ladders are mounted onto the support structure and the final commissioning is performed using source. The source positioning tube through which the source is moved in the defined way is placed around the PXD so that the performance of all ladders in their mount positions can be studied. All components of the PXD system (cooling system, power supplies and DAQ) are commissioned at this stage.

Once the detector has passed all tests, it will be disassembled and the individual ladders will be shipped to KEK. Final commissioning of the reassembled detector will happened there. Pretesting of the individual ladders with the source irradiation is planned to check the functionality of the ladders at KEK before final assembly. Additional tests are planned once the ladders are mounted onto the supporting structure for the integration in the Belle II. Finally, the full system test will be repeated once the detector is in place. After the test of the full Belle II PXD detector, its commissioning together with the rest of the Belle II detector will take place.

### 4.6.4 Production Yield

From past experience with devices of similar size and complexity (for high energy physics and astrophysical applications) we conservatively expect a yield for the DEPFET production of 50%. In similar semiconductor devices, the HLL MPI routinely obtains a better than 80% yield. The expected yield of the flip chipping is high due to the relaxed pitch of the ASICs. The yield for the full ladder assembly is also assumed to be 50% (again conservative: for the ATLAS pixel detector a yield of 90% was reached in the assembly procedure). In total, we count on a yield of working ladders of better than 25%. With this yield, we will produce 40 complete and functioning ladders (20 are needed for the PXD detector).

## 4.7 Data Acquisition and Slow Control

### 4.7.1 Overview

The architecture of the PXD readout system is shown in Fig. 4.22. The patch panel is an interface adapter card between the Kapton flex cable ($\simeq$ 20 cm length) and the differential cable ($\simeq$ 10 m length, probably LVDS). The DHH is described in Sec. 4.7.2 and the COMPUTE NODE in Sec. 4.7.4.

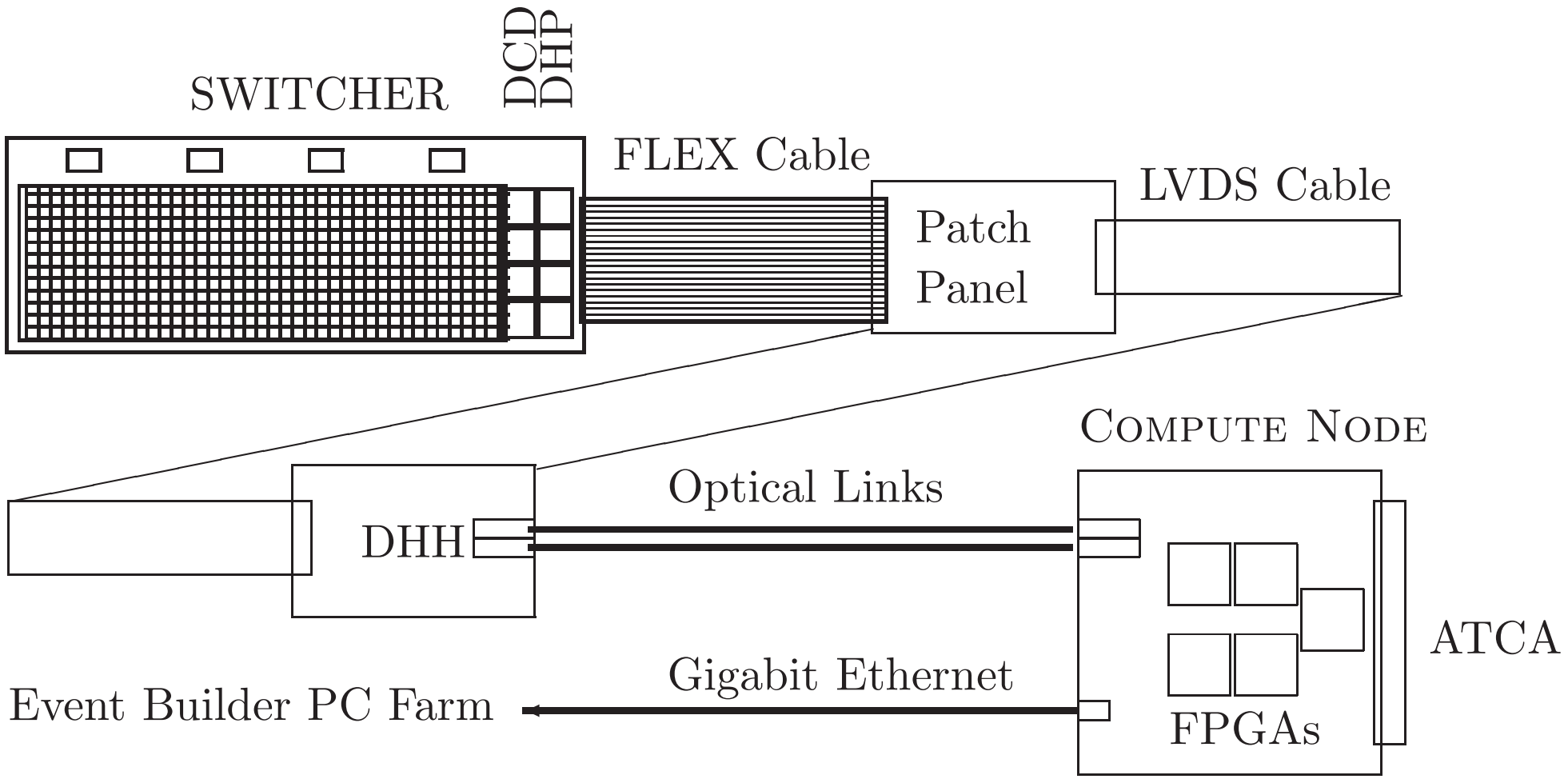

*Figure 4.22: Architecture of the PXD readout system.*

### 4.7.2 Data Transmission - Patch Panel and Data Handling Hybrid (DHH)

The Data Transmission System is the interface of the DEPFET modules to the outside world, i.e., power supplies and DAQ system. Each half ladder will be connected via the Kapton flex to a small Patch Panel (PP) that, in turn, connects to the Data Handling Hybrid (DHH). The PP will be located very close to the modules to keep the Kapton flex cable as short as possible (20 cm); this minimizes the degradation of the fast output signals from the DHP. The PP is the cable adapter for the fast signals from the DHP (4 signals with 1.25 Gbit/s). In addition, the PP performs power filtering, impedance matching and possibly over-voltage protection. Signals and power will then be routed via shielded twisted pair copper cables to the DHH located in the docks (or perhaps outside the detector). The DHH is connected via optical links to the Compute Nodes, to the PXD power supplies, and to the Belle II GDL (Global Decision Logic) trigger and master clock. Its main components are an FPGA and Finisar optical transceivers. The tasks of the DHH are the following:

- generate the DHP system clock from the Belle II clock distribution system,
- receive the four fast signals coming from the DHP in an AURORA Gbit link receiver,
- multiplex the four DHP outputs per half-ladder to one optical link (5 Gbit/s),
- provide a JTAG master for the slow control of the DHP,
- send the slow control data to the Compute Nodes via the optical link,
- decode slow control commands from the external Belle II slow control panel, and
- regulate the voltage.

### 4.7.3 Data Rate Estimate

For the following estimate, we assume a trigger rate of 30 kHz, corresponding to the highest luminosity of $\mathcal{L}$=0.8×$10^{35}$ $cm^{-2}$ $s^{-1}$. For an estimate of the required PXD readout data rate, we assume 40 PXD half ladders with 250 × 800 (or 0.2 × $10^6$) pixels each and 8 × $10^6$ pixels in total. The average cluster size is expected to be 2 pixels. Given a readout time of 20 $\mu$s, the average occupancy is $\simeq$ 1%, corresponding to $\simeq 8 \times 10^4$ fired pixels per frame. This occupancy still does not take into account any background. For a frame rate of 50 kHz and a trigger rate of 30 kHz, the expected reduction factor on the DHH can be calculated using Poisson statistics to a factor of 2.2. This amounts to a fired-pixel rate of $\simeq 1.8 \times 10^9$ Hz. Multiplying by a data size of 4 bytes per pixel for the encoded position and ADC charge information, we estimate a total data rate of $\simeq$ 58 GBit/s for the complete PXD. With 40 optical links, this implies $\simeq$ 1.44 GBit/s (or $\simeq$ 180 MByte/s) per optical link. However, there will be a significant contribution to the occupancy by radiative QED events of the type $e^+e^- \to e^+e^-\gamma$, beam gas events, synchrotron radiation from the upstream and downstream dipoles, and the Touschek effect. For the background, we take into account a factor of three safety margin, i.e., the system will be prepared to cope with a maximum occupancy of 3%. If the background increases the occupancy to values above 3%, PXD subevents will have to be truncated. Precise track vertex reconstruction in events with such high occupancy degrades in any case because of too high combinatorics and subsequently an increased fraction of wrong PXD hit-to-track assignments.

### 4.7.4 Readout System

The hardware platform for the PXD readout is the Compute Node (CN) and is shown in Fig. 4.23 (left side). The 14-layer printed circuit board has been developed by IHEP Beijing and the II. Physics Department of Giessen University. Each CN has five XILINX Virtex-4 FX-60 FPGAs, chosen for their high computing performance and their links for high bandwidth data transfer (RocketIO). Fig. 4.24 shows the schematic block diagram of the Compute Node. On each board, all FPGAs are interconnected by point-to-point links (see below for details) for event building. The programming of the FPGAs in VHDL is being done using XILINX ISE (Integrated Software Environment) Vers. 10.1 framework. Each Virtex-4 FX60 FPGA has two 300 MHz PowerPCs implemented as core; these are used for slow control purposes but not for algorithms. In the current design, the PowerPCs run Linux 2.6.27. In addition, each FPGA has 2 GB of DDR2 memory attached. The CN is designed as a board of the ATCA (Advanced Telecommunications Computing Architecture) standard. The ATCA shelf is shown in Fig. 4.23 (right side). In an ATCA shelf with a full mesh backplane, point-to-point pairwise connections between CNs are wired. This avoids any bus arbitration. In addition to the high computing performance, the CN also provide high bandwidth interconnections. *(a)* All five FPGAs are connected pairwise (on the board) by one 32-bit general purpose bus (GPIO) and one full duplex RocketIO link. *(b)* Four of the five FPGAs have two RocketIO links routed to the front panel using Multi-Gigabit Transceivers (MGT) for optical links. *(c)* One of the five FPGAs serves as a router and has 16 RocketIO links through the full mesh backplane to all the other compute nodes in the same ATCA shelf. *(d)* All five FPGAs have a Gigabit Ethernet Link routed to the front panel. With the current design, the input bandwidth in one ATCA crate is $\leq 35$ GBit/s (14 CN, eight optical links each, operating at $\leq 2.5$ Gbit/s). The slow control tasks (e.g., power negotiation) between the CN and the ATCA shelf are based upon IPMI (Intelligent Platform Management Interface), implemented by an ATMEL ATmega2560 micro-controller on a CN add-on card that was developed at Giessen.

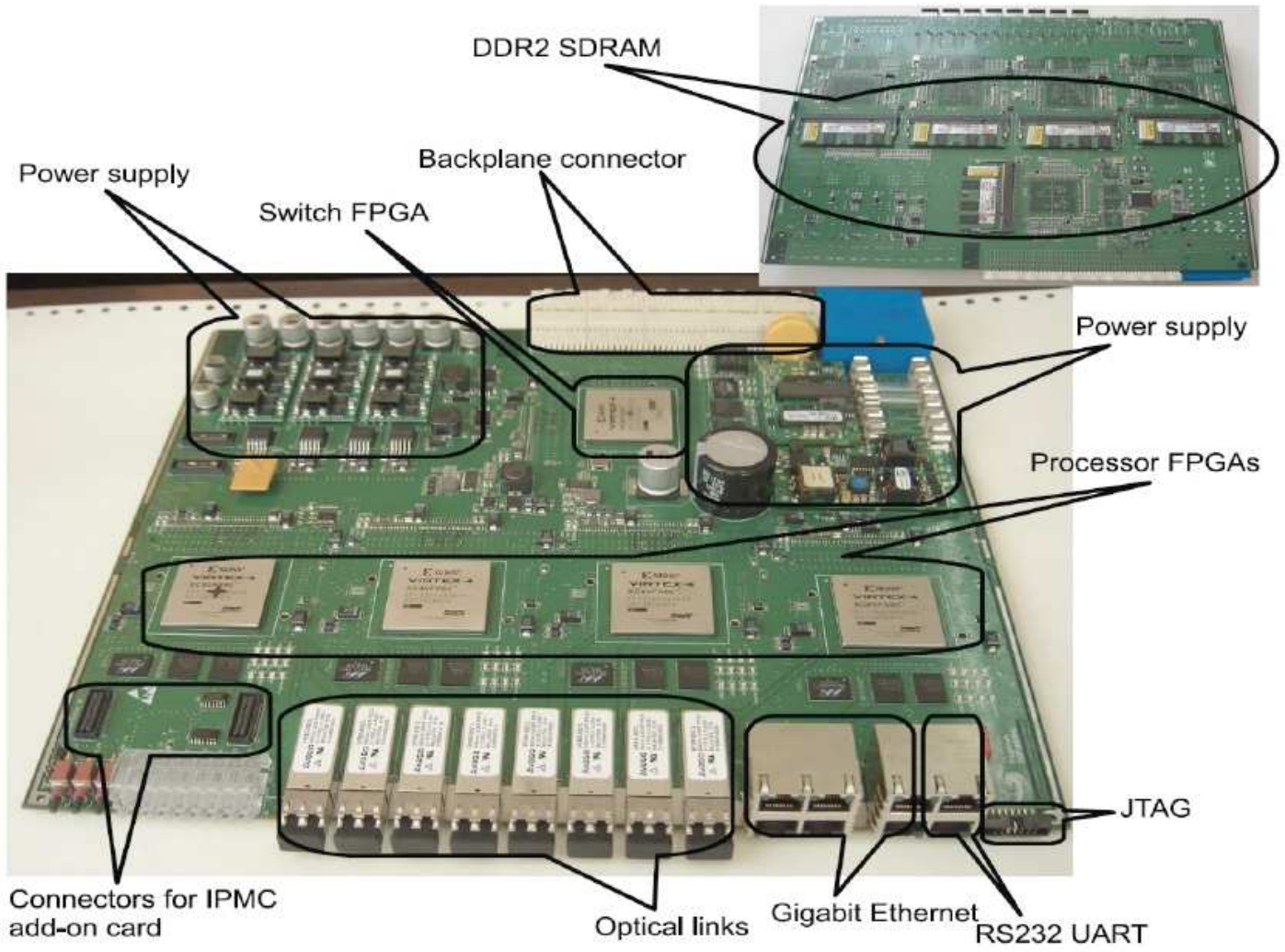

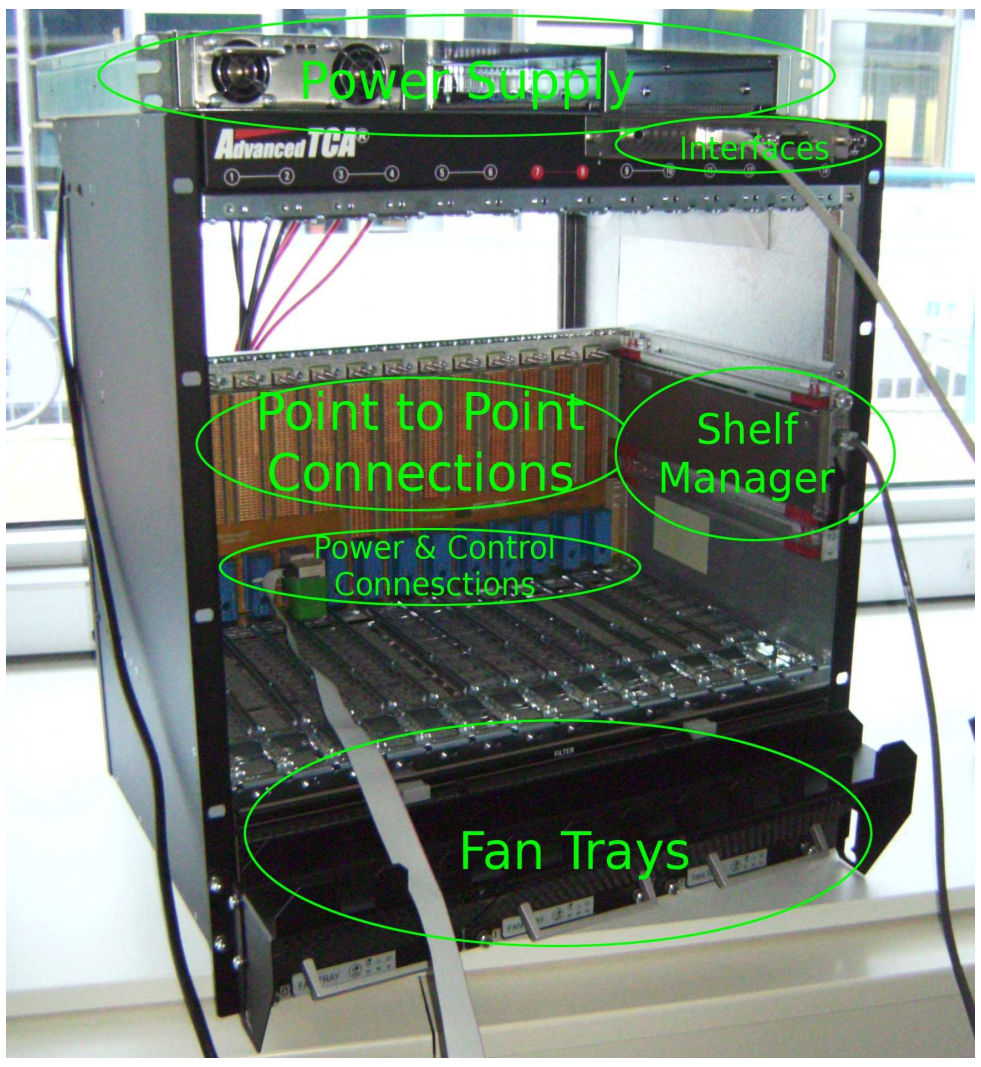

*Figure 4.23: Left: Photo of a Compute Node (CN) prototype. Right: Photo of an ATCA Shelf.*

The performance of the compute nodes was measured in several tests:

- Concerning the data input of the compute nodes, the RocketIO transmission over the optical links were operated stably in long term tests at a sustained data rate of $\simeq 1.6$ GBit/s, limited by the usage of the SFP transceivers. For the PXD readout, the compute nodes will be upgraded to SFP+ transceivers, allowing to operate at the full RocketIO transmission rate of up to 6.5 Gbit/s.

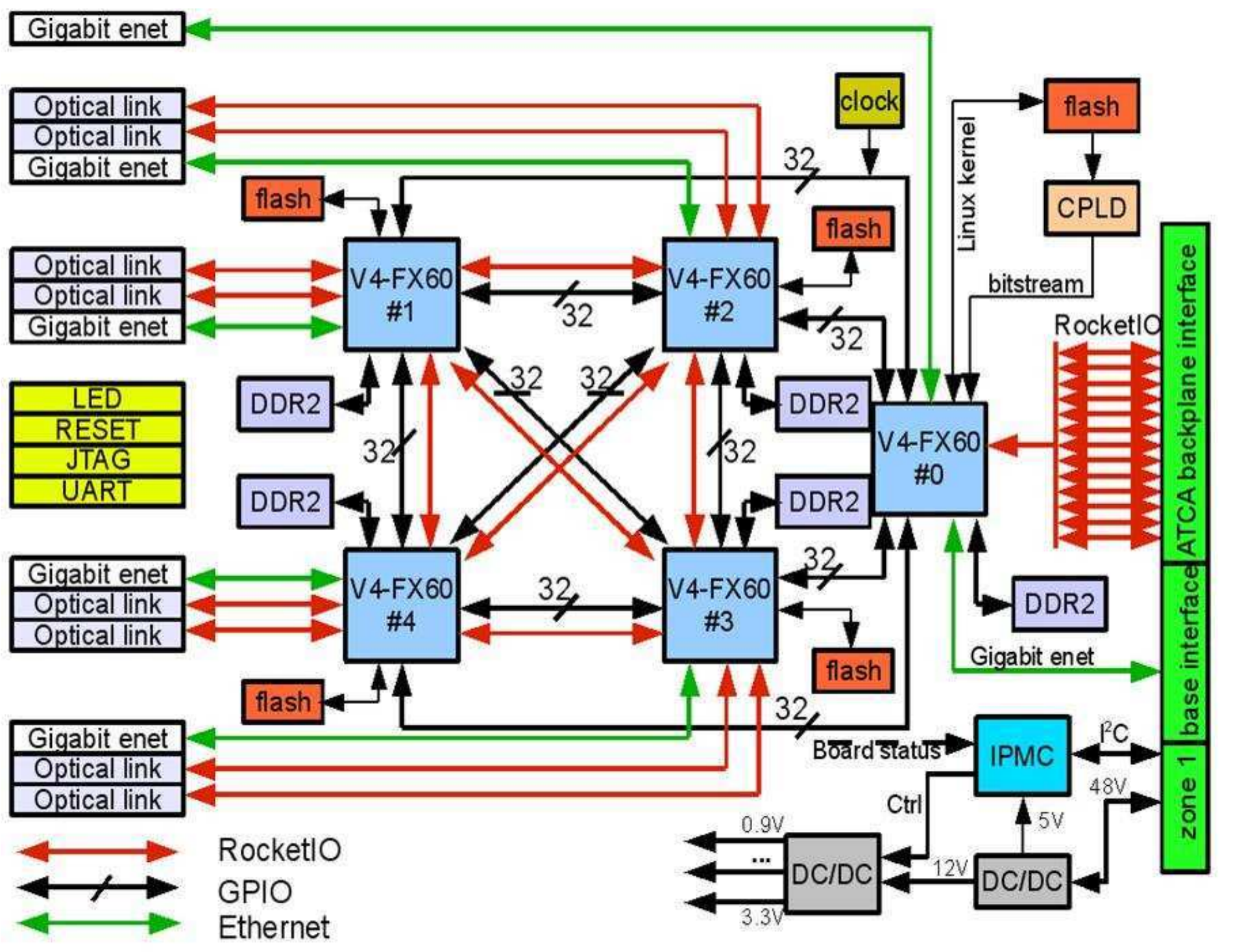

*Figure 4.24: Schematic block diagram of the* COMPUTE NODE.

- Concerning the data output of the compute nodes, the gigabit ethernet was tested by using the PowerPC405, which is embedded on the Virtex-4 FPGA. Data rates of $\simeq$ 212 Gbit/s ($\simeq$ 26.5 MByte/s) from PowerPC to PC and $\simeq$ 241 Gbit/s ($\simeq$ 30.1 MByte/s) from PC to PowerPC were achieved. For the final readout system, the FPGA, not the PowerPC, will be used for the gigabit ethernet transmission. This requires implementation of the UDP transmission protocols in VHDL. Tests by XILINX indicate that data rates of $\geq$ 100 MByte/s are reachable.

- As a complete system test, a readout chain for one readout channel in the sequence data transmitter → optical link → FPGA → write to RAM → read from RAM → gigabit ethernet → PC was tested. For this purpose, pseudo-random data were generated on the transmitter side and checked by a logic analyzer program on the PC. No bit error was observed during a long-term test of $\simeq$ 150 hour duration.

### 4.7.5 Data Reduction

The high data rate of $\simeq$ 2.4 Gbit/s ($\simeq$ 300 MByte/s) per optical link must be reduced by about a factor of $\simeq$ 10 to match the gigabit ethernet bandwidth from the compute nodes to the event builder farm and to avoid saturation of the event builder. There are at least two options for the data reduction: First, a track finder and track fitter using the PXD and the SVD raw data will be programmed in VHDL on the Virtex-4 FPGAs. Second, the compute nodes will buffer PXD events and wait for fitted tracks, provided by the HLT (High Level Trigger). In both options, the fitted tracks are extrapolated to the PXD modules, and hits that are not matched to any track (generated by QED, beam-gas background, or synchrotron radiation) are discarded.
In the reduction of the raw PXD data stream by a factor of $\simeq 10$, data of physics interest should be kept and only background-induced data should be discarded. Since the PXD has only two layers, the distinction between physics and background hits relies on the extrapolation of tracks reconstructed in the relatively quiet SVD through the PXD. Using the data from the SVD alone, a fast track finding procedure is performed. These tracks are then propagated back through the PXD ladders. Around each intersection point of a track with a PXD ladder, a region of interest ("ROI") in the plane of the ladder is defined. If a fired PXD pixel lies inside an ROI, it is kept; otherwise, it is discarded.

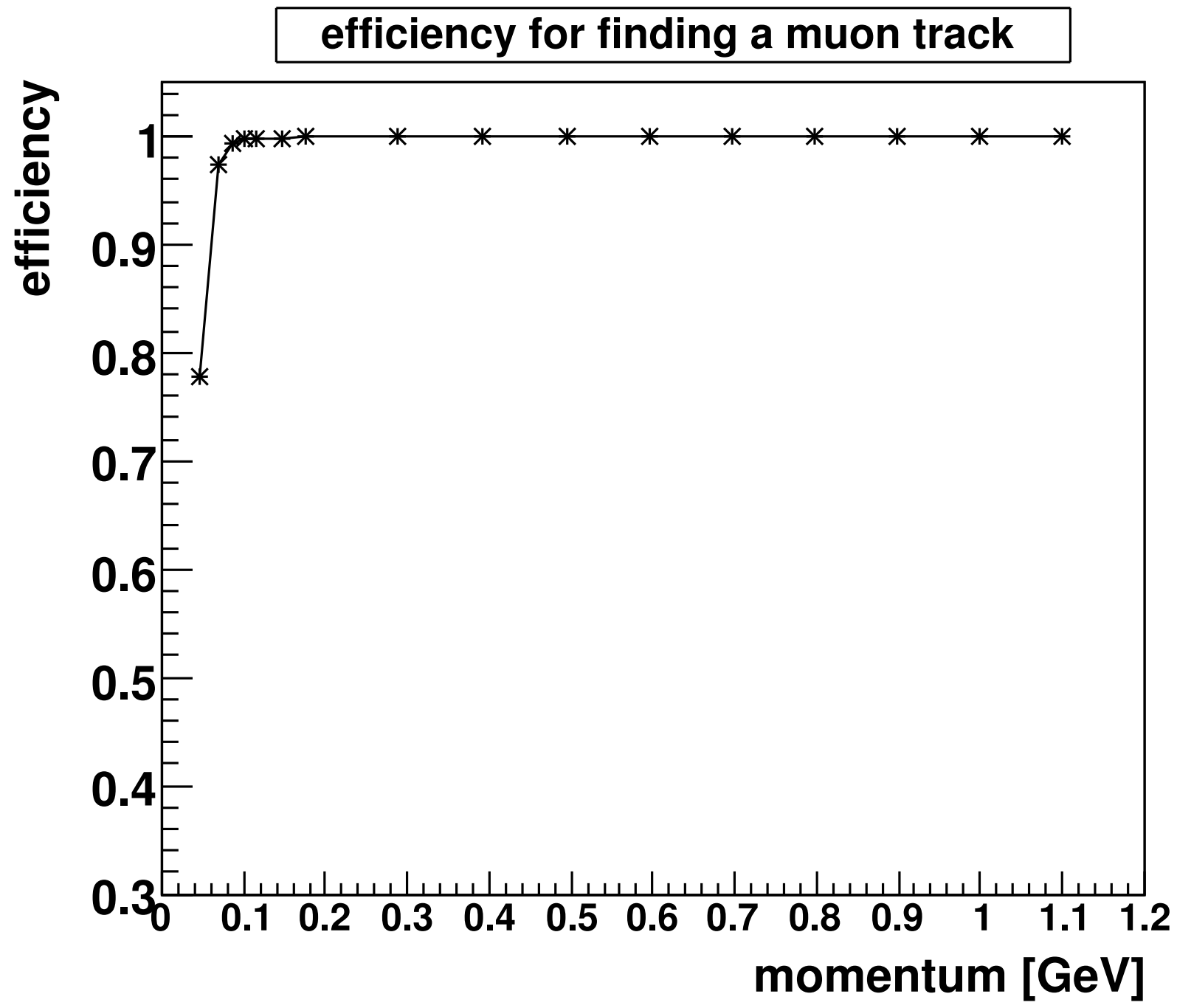

*Figure 4.25: Track-finding efficiency of the Hough transform algorithm in $r$–$z$ space, based only on the SVD information. From the tracks found, the ROIs for the PXD are constructed.*

Since the SVD-only track finding has to be done early in the DAQ chain, it has to be very fast; therefore, advanced track finding algorithms cannot be applied. In particular, a full three-dimensional track finding procedure is too costly. Instead, the SVD hits are projected into $r$–$z$ space, where $z$ is the axial coordinate of the hit and $r$ is its distance from the $z$-axis. Two types of tracks are handled by this data reduction method: straight tracks (for particles above a tunable momentum cutoff) and helical tracks (for particles below the cutoff). Straight (helical) tracks appear then as straight lines (sine curves) in $r$–$z$ space. Tests show that a reasonable cutoff parameters is 1 GeV.
The Hough transform technique finds the tracks in $r$–$z$ space. It determines the parameters of the tracks by transforming each hit from $r$–$z$ space to a two dimensional parameter space. Depending on the track type, each SVD hit is represented by a line or a sine curve in the parameter space.

SVD hits originating from a common track are then given by the intersection of their lines/sine curves in the parameter space. A robust method of determining these intersections is given by a space subdivision scheme. It recursively subdivides the parameter space in smaller areas, yielding as final output areas containing the intersections and therefore the final track parameters. From these parameters, the ROIs on the PXD ladders can be calculated by transforming the Hough-transform parameters back to the track representation in $r$–$z$ space. The resulting ROIs appear in a PXD layer as complete rings in $r$–$\phi$ with a finite width along $z$. A preliminary estimate of the track-finding efficiency of the Hough algorithm is shown in Fig. 4.25. One can see that excellent efficiency is obtained down to very low momenta. High efficiency is essential, since all PXD hits are removed that are not located in a ROI defined by the tracks found by the Hough algorithm.

The data reduction factor can be improved by introducing an additional subdivision of the $r$–$\phi$ plane into wedge-shaped sectors. Such a subdivision is very well suited to the parallelization of the ROI-finding computations. Two types of sectors are considered: sectors with straight edges (for energetic particles) and with curved edges (for soft particles). An arbitrary combination of sector types is possible to optimize the track finding efficiency. Since each sector only covers a small part of the PXD, ring-like ROIs are avoided, resulting in a better data reduction factor. First tests show that a reduction factor of 15-20 is achievable with this refinement.

### 4.7.6 Backup Solution of the PXD Readout System

In the baseline option of the PXD readout, the CN receives data from the SVD and performs ROI-finding in the FPGA equipped on it. This approach is well suited to separate PXD readout from other sub-systems as well as possible, which consequently results in a good scalability of the system. However, this option requires (1) a signal branching system of SVD to the main stream and to the PXD, and (2) partial event building system of the branched signal to match to the number of CN inputs (40). The option also requires special skill to develop and maintain the FPGA codes for the ROI-finding.

As an fallback solution, we investigate a backup option for the ROI-finding by the FPGA with a high level trigger farm (HLT) consisting of $\mathcal{O}(10)$ units of PC farms. In this backup option, the HLT runs the software for the ROI-finding. Figure 4.26 shows the comparison of the baseline and the backup option.

In case of the backup option, we need to consider a latency for the HLT decision. According to our experience in Belle, the online track reconstruction on one Intel Xeon $2 \times 3.4$ GHz PC takes typically less then 1 second and up to 10 seconds for harsh events. Accounting for CPU power evolution, we expect $\sim 5$ seconds for the HLT latency at most. In addition, the few PXD events that can not be processed within 5 seconds will be accepted by default, and their full data will be stored.

We have two sub-options for the backup option. In the following subsections, we describe each of them. A final decision for the choice of the readout system will be made in spring 2011 based upon prototype results, in order to provide sufficient time for any required hardware development.

#### 4.7.6.1 ATCA-based readout system with HLT

As well as the main option, this sub-option employs the ATCA boards to read out the PXD. The data from the PXD is read by 40 ATCA boards, which are housed in an ATCA crate. Each ATCA board is connected to each DHH and receives data from the DHH over the RocketIO link. The ATCA board has an Ethernet (or GbE) port to communicate with the HLT about the HLT decision. For this sub-option, the ATCA readout system needs to hold PXD data for up to 5-second HLT latency, which corresponds to 3–5 GB at most per node. There exist two solutions for this: the first is to redesign the CN to equip the large memory, and the second is to purchase and install a commercially available memory unit of ATCA. At this stage, the first solution seems more feasible.

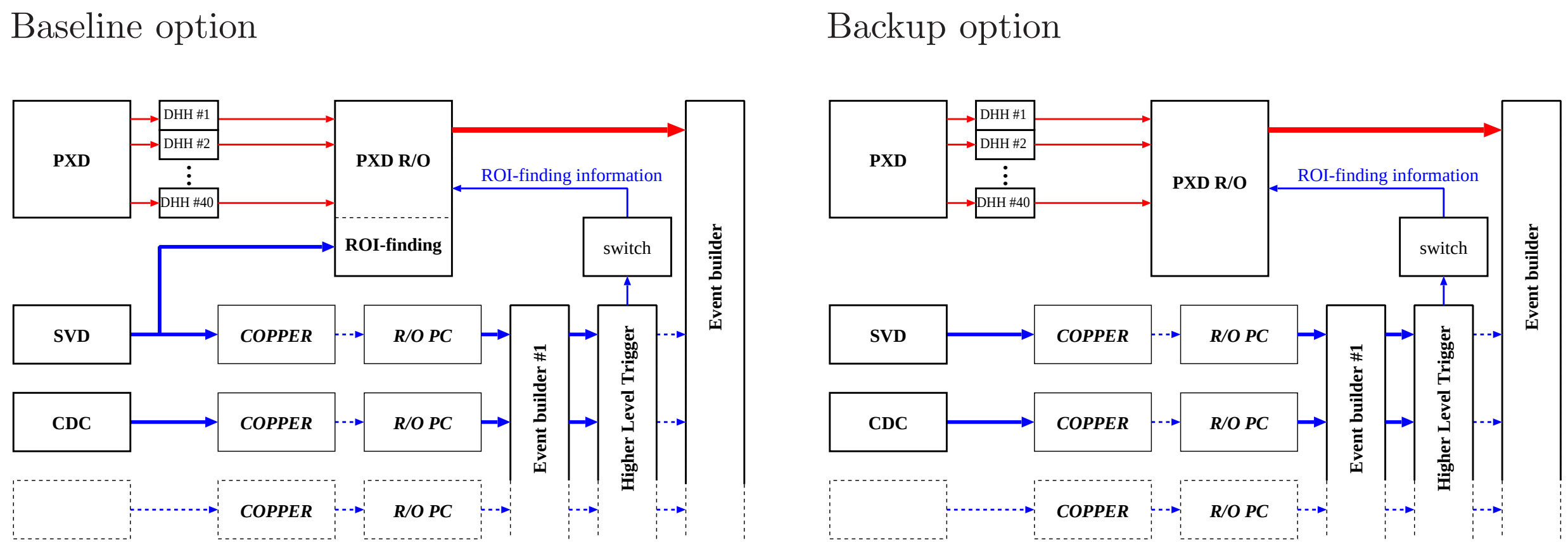

Figure 4.26: Block diagrams of the baseline and backup options of the ROI-finding scheme.

#### 4.7.6.2 PC-based readout system with HLT

In this sub-option, the DHH outputs the data to a PC over the RocketIO link instead of the ATCA board. An internal block diagram of the PXD-readout PC is shown in Figure 4.27. To accommodate the 600 MB/s rate of the RocketIO input, the internal link of the PC between the RocketIO interface and the processor should be larger than this throughput. The PCI-express (PCIe) allows 5 Gbps data transfer per link. Two or more PCIe links can be occupied by a single PCIe card to multiply the data transfer speed. There are commercially available PC mainboards with PCIe slots connected to 8-fold PCIe links, which enable 40 Gbps per slot in total. As for the HLT latency, the PC can house physical memory of $\mathcal{O}(10)$ GB at a reasonable price. The ROI-finding information is given to the PC from the HLT over the GbE. Data size reduction is performed by the processor in the PC. The reduction can be realized in software, which is an advantage of this sub-option from the viewpoint of the development cost and maintainability. The PXD data are integrated by 40 PXD-readout PCs in total. All output from the PXD-readout PCs are transferred to the second level event builder.

A critical path of this sub-option is the development of the RocketIO card on the PCIe. Tokyo Electron Device Ltd. provides an evaluation board of Virtex5 on the PCIe with optical link connectors [16]. The board provides 3 Gbps RocketIO link with firmware and interfacing firmware to the PCIe. The design of our own RocketIO PCIe card with much higher link speed, using this board as a reference, is an option.

One of the differences between the ATCA-based and the PC-based system is the different handling of the interrupts that are generated at the end of data transfers of the small incoming data packages ($\leq$100 kBytes). DMA of large data blocks is not an option because of non-implemented large scale buffer on the DHH. The handling of these interrupts in the operating system of the PCs could lead to a quite high load on the processors. In the ATCA-based system, there is no operating system, so this problem is nonexistent. A prototype system of the PC-based approach

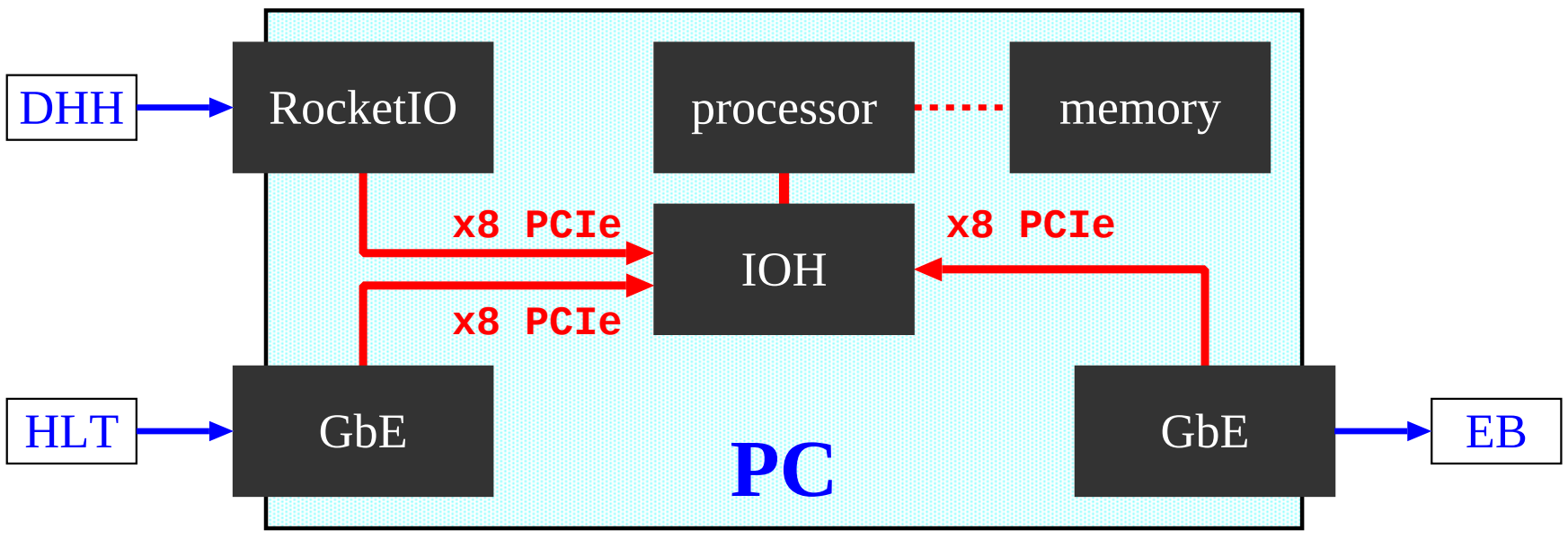

*Figure 4.27: The internal block diagram of the PXD-readout PC.*

will provide measurements of the CPU load.

#### 4.7.6.3 Communication with the HLT

Figure 4.28 shows data stream and communication diagram between the PXD-readout system and the HLT.
Event records stored in the ATCA/PC memory are sorted by event number since they directly come from the DHH. On the other hand, the HLT outputs are not sorted, since the time to make ROI-decision by the HLT varies event-by-event. Because the event building system, which receives data from the ATCAs/PCs, does not expect event records disordered, the outputs from the ATCA/PC should be sorted by event number. To sort the HLT decision by event number, we install a 'sorter PC' at the final output of the HLT.
In the following, we describe successful operation and faulty operations.

- Success
  If the HLT decision is made within the predefined HLT latency ($\sim 5$ seconds), the HLT-decision packet (an example is shown in Figure 4.29) is broadcast to all of ATCAs/PCs.

- Fault (event recored expiration in the sorter PC)
  After the sorter PC broadcasts the $n$-th HLT decision successfully, the sorter PC waits for the $(n + 1)$-th HLT decision. If the $(n + 1)$-th decision does not arrive within the HLT latency, the sorter PC generates a special packet of the HLT decision to assume all data as the ROI. This action will take place irrespective of the arrival of $(n + 2)$-th event. If the $(n + 1)$-th HLT decision comes later, that decision is discarded.

- Fault (event record expiration in the ATCA/PC)
  After the ATCA/PC sends out the $n$-th event record to the event building system successfully, the ATCA/PC waits for the $(n + 1)$-th HLT decision from the sorter PC. If the $(n + 1)$-th decision does not arrive within the HLT latency, the ATCA/PC will send the event record to the event building system without any data size reduction. If the $(n+1)$-th HLT decision comes later, that decision is discarded.

Figure 4.29 shows an example packet of the HLT decision to be broadcast to the ATCAs/PCs from the sorter PC.

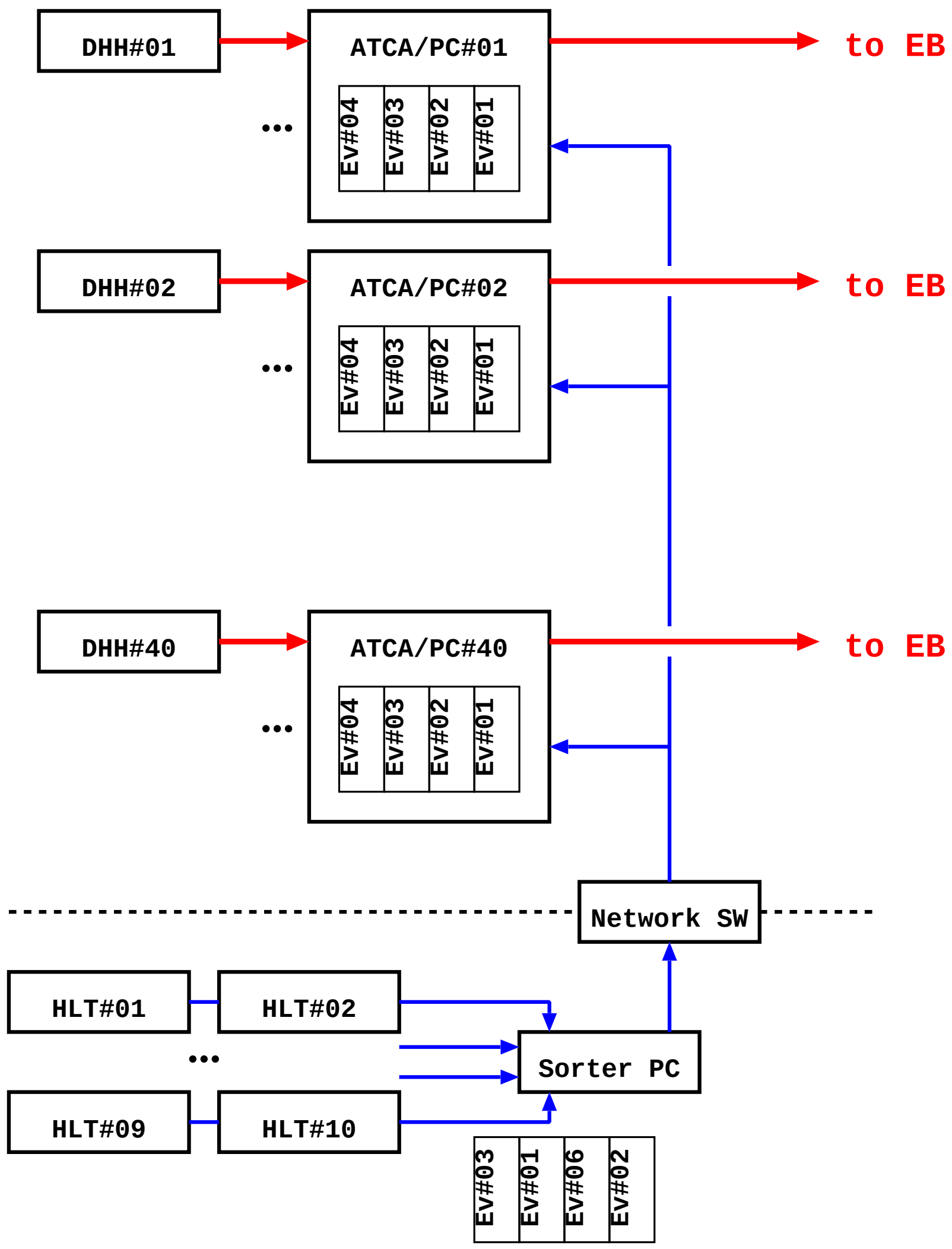

*Figure 4.28: The data stream and communication diagram between the PXD-readout system and the HLT.*

## 4.8 Mechanical Support and Cooling

The mechanical structure of the PXD strongly depends on the geometry of the beamline. Since the design of the interaction region and the beampipe for the nano-beam option is still evolving, the PXD mechanical design has to accommodate (among other things) geometric changes in the length of the straight part of the beampipe as well as the location and diameter of flanges that serve as anchor points for the PXD. Therefore, the numbers given here are subject to change.
A very important parameter still under discussion is the angle of the beamline with respect to the axis of the Belle II solenoid. In case this angle is not zero, the PXD loses its cylindrical symmetry because it has to cover the specified acceptance region. As a consequence, the mechanical construction becomes more complicated and, concurrently, the active area of the detector modules needs to be longer. On the other hand, the most important parameter for the PXD mechanical design—the radius of the beamline at the interaction point—has settled to a stable value ($r_{inner} = 10\,\text{mm}$). This means that a close-to-final mechanical design can be developed while keeping in mind that minor changes will surely be made as the IR design is finalized. The dimensions of the mechanical support structure for the case that the beampipe is parallel to the Belle II solenoid is shown in Fig. 4.30.

| |
|---|
| Grand header / event ID |
| DHH#01 header / word count |
| Extract region [000] xxx ... yyy |
| Extract region [001] xxx ... yyy |
| Extract region [002] xxx ... yyy |
| DHH#02 header / word count |
| Extract region [000] xxx ... yyy |
| Extract region [001] xxx ... yyy |
| ⋮ |
| DHH#40 header / word cound |
| Extract region [000] xxx ... yyy |
| Extract region [001] xxx ... yyy |
| Extract region [002] xxx ... yyy |
| Grand trailer / checksum |

*Figure 4.29: The data stream and communication diagram between the PXD-readout system and the HLT.*

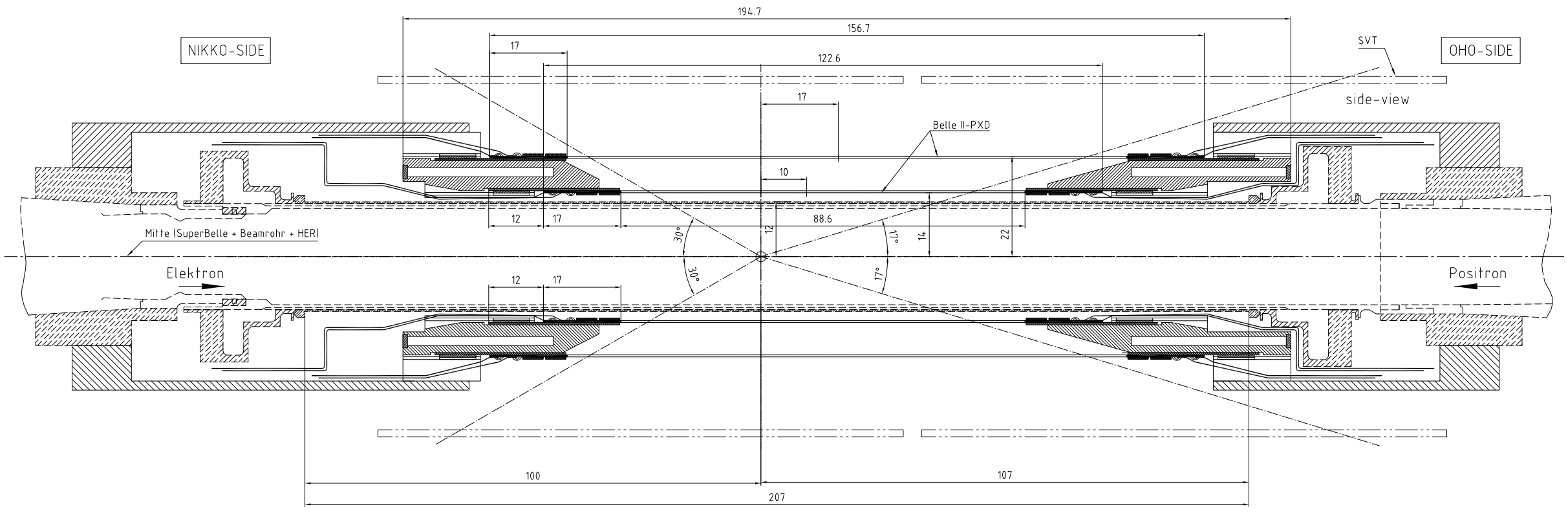

*Figure 4.30: Overview of the detector layout.*

## 4.8.1 General Concept

The PXD is a barrel-only system consisting of two cylindrical active detector layers coaxial with the beam line (Fig. 4.1). These layers consist of eight or twelve detector modules which overlap in $\phi$ in such a way that the insensitive balcony area in one module is covered by the sensitive pixel area of the neighboring module. The overall coverage is nearly 100% within the acceptance angle. *In situ* geometric alignment is done continuously using high-$p_t$ tracks that pass through the overlap region.

The inner layer is placed as close to the beamline as possible at $r = 13$ mm. The outer layer is at a radius of $r = 22$ mm. The envisaged spatial resolution is $\approx 10\,\mu$m. The support structure must be robust enough to prevent short-term spatial movement of the mounted PXD, but does not need an absolute mounting precision of better than $\approx 0.1$ mm because continuous position calibration and monitoring will be achieved with energetic charged tracks.
An online alignment system for each module is considered to be unnecessary even though a monitoring of the position of the support structure with respect to the neighboring silicon strip detector system is highly desirable. A system based on fiber Bragg gratings is under study.

### 4.8.2 Detector Modules

The detector modules consist mainly of the DEPFET silicon sensor (thickness of 400 $\mu$m) with integrated electrical lines and pitch adaptors. All necessary chips to operate the DEPFET (switcher chips, digitizers and data processors) are bump-bonded to the wafer surface. The sensor is thinned to 75 $\mu$m for the full sensitive area (Sec. 4.4). Only the end sections of the sensor that are outside the acceptance region exhibit the full thickness for reasons of rigidity and bondability. Each module consists of two half-modules that are connected with a joint piece of silicon material glued to the bottom side of the module.

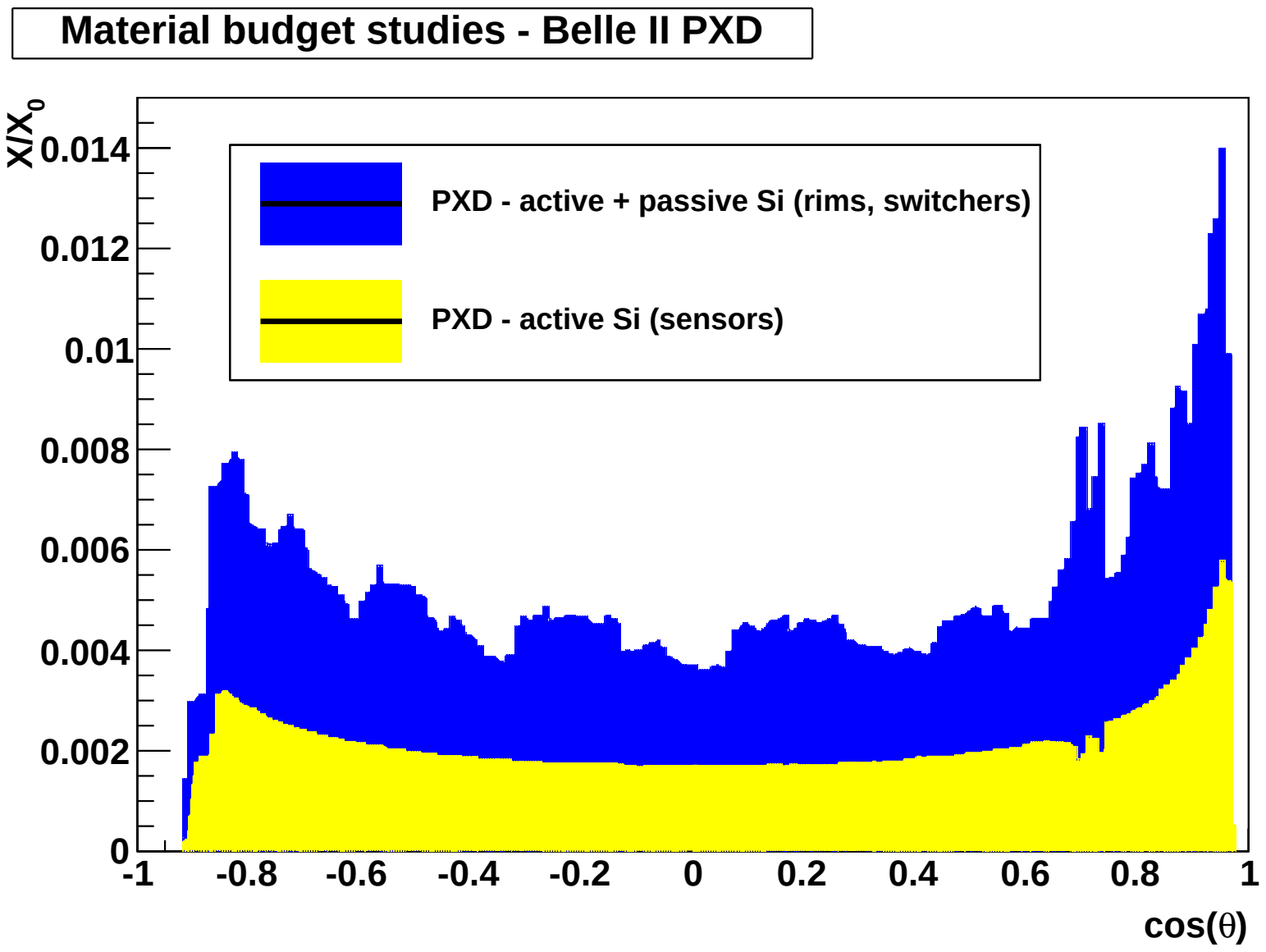

*Figure 4.31: Distribution of expected material as a function of* $\cos(\theta)$.

| layer | radius | wafer size (mm) | | sensitive region (mm) | | no. of pixels | | pixel pitch ($\mu$m) | |
|---|---|---|---|---|---|---|---|---|---|
| | | length | width | length | width | in $z$ | in $\phi$ | in $z$ | in $\phi$ |
| 1 | 13 | 131 | 15 | 90 | 12.5 | 1600 | 250 | 56 | 50 |
| 2 | 22 | 174 | 15 | 123.45 | 12.5 | 1600 | 250 | 77 | 50 |

*Table 4.6: PXD module parameters*

In the baseline mechanical design, we attach the silicon wafers directly to the support structure, using either screws or clamps. The necessary holes in the wafer are produced by laser cutting. Since this is a rather challenging approach, a fall-back solution is under study, wherein extension bars of CVD-diamond ($0.4 \times 7 \times 30\,\mathrm{mm}^3$) are glued at both ends to the backside of the modules. These bars are clamped to the support structure with springs.
Table 4.6 shows the main parameters of the modules. Figure 4.31 shows the amount of material as a function of $\cos\theta$.
Electrical connections are realized with triple-layer Kapton cables, glued and wire-bonded at both ends of the module. These cables emerge from the detector volume along the beamline (Fig. 4.32).

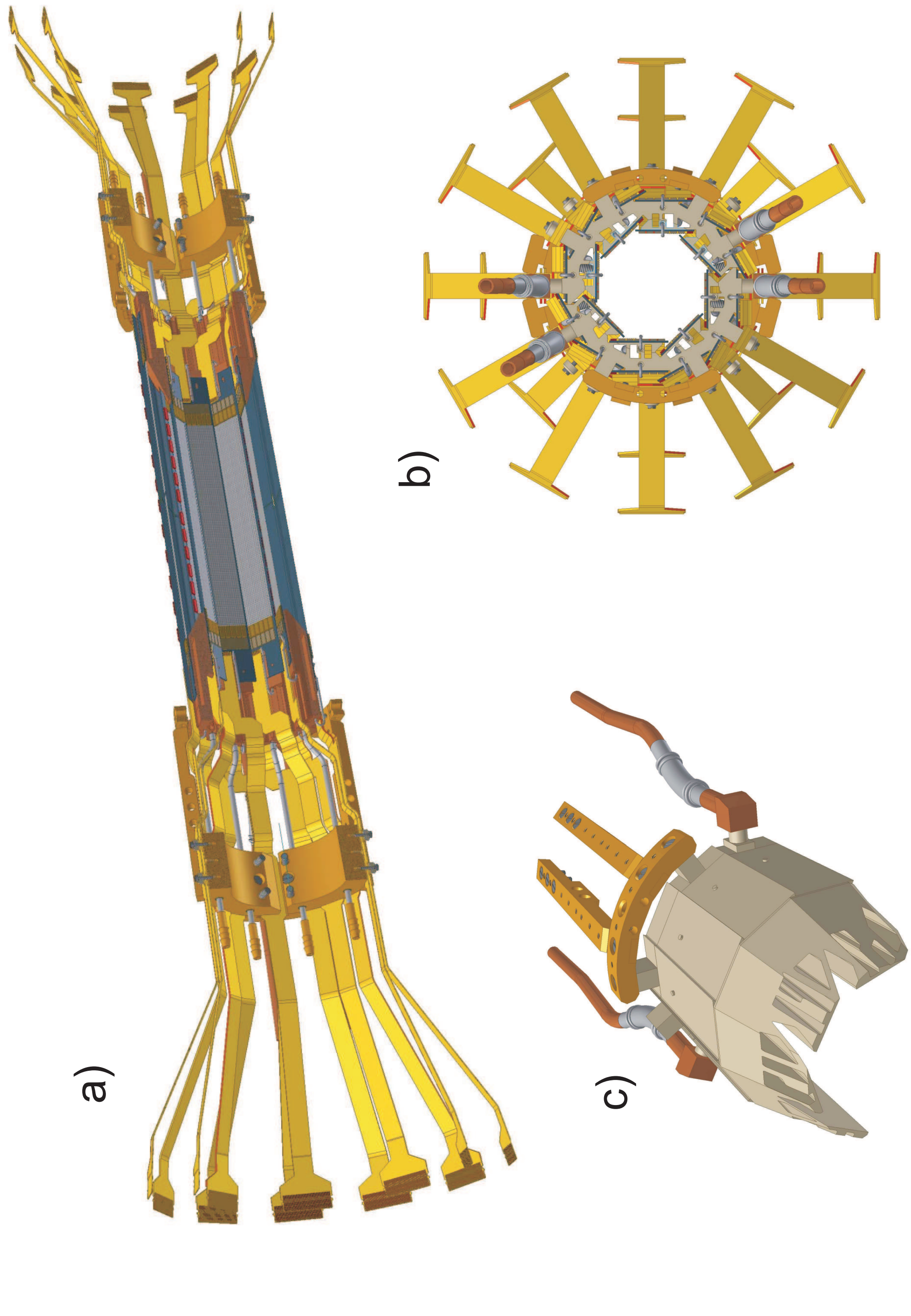

*Figure 4.32: (a) 3D view into the PXD detector with the integrated support and cooling structure, including the Kapton cables to the outside services. (b) View along the beampipe. (c) Detail of the support structure holding the sensors.*

### 4.8.3 Support Structure

The support structure (Fig. 4.32) is a formidable engineering challenge. Since it must also serve as the heat sink for the detector modules, it has an integrated cooling channel that can carry evaporative $CO_2$ as the coolant, delivering cooling capacity as close as possible to the power dissipating chips. The many holes in the material result in suboptimal heat transport, but they are necessary to enable the flow of cold air between the two sensor layers (Fig. 4.32 b). The ratio between hole sizes and heat-conducting material is still being optimized. Copper is used rather than aluminum because of its better thermal conduction and expansion coefficient.
The support structure itself is mounted to the structural parts of the beam pipe far from the interaction point. The two end rings are not identical due to the asymmetry of the Belle II acceptance. The support structure is designed as two half shells that are clamped together during the installation of the detector to allow the mounting on the assembled beam pipe. The two mechanically independent half shells are completely assembled prior to installation.
The detector modules are fixed to grooves in the support structure by springs. Precision holes in the wafers and pins on the support structure on one side provide the required mechanical accuracy. On the other detector side, long holes in the wafers that can slide on pins on the support structure enable the system to yield to thermal movements. The pressure applied by the fixation springs is adjusted to provide the best possible thermal coupling of the modules to the support structure while allowing thermal expansion and contraction without mechanical deformations. Due to the small coefficient of thermal expansion of $2.0 \times 10^{-6}\,\mathrm{K}^{-1}$ for silicon, the expected contraction will be $\approx 50\,\mu$m when the temperature is lowered from room temperature to the operating temperature of slightly below 0°C. On the other hand, when the electrical power of the DEPFET system is turned on, the thermal expansion will be of $\approx 100\,\mu$m.
In parallel, we are evaluating an alternative solution wherein both sides of the silicon wafer are fixed with screws and the whole support structure yields to thermal expansion.

### 4.8.4 The cooling concept

The cooling of the PXD system must balance the conflicting demands of robust heat dissipation with a minimal material budget (driven by the desire for a small multiple Coulomb scattering contribution to the charged-track vertex resolution).

- The thermal power of the PXD system amounts to a total of 360 W. The twenty detector modules contribute with 18 W each: There are four DCD chips and four DHP chips on each side of the module; each chip dissipates about 1 W. The DEPFET itself produces about 1 W distributed homogeneously over the whole area. The switcher chips on the balcony are turned off for most of the time and contribute only moderately to the heat production (1 W on average) (Fig. 4.33).
- The beampipe as the central part within the cylindrical pixel volume has its own cooling system and will probably be kept at a temperature slightly above the dew point.

- There is limited available space in the innermost region of Belle II for services, tubing, etc.
- One boundary condition stems from the detector properties: the temperature of the silicon sensor should not exceed 30° C while the chip's temperature should be kept below 60° C to prevent electromigration.
- For simplicity and to economize on space, a common cold dry volume for both the PXD and SVD is under study. In this design, the forced convection, which is needed to cool down the center of the ladders, is provided for both detector systems, and temperatures below the external dew point can be operated inside this volume.

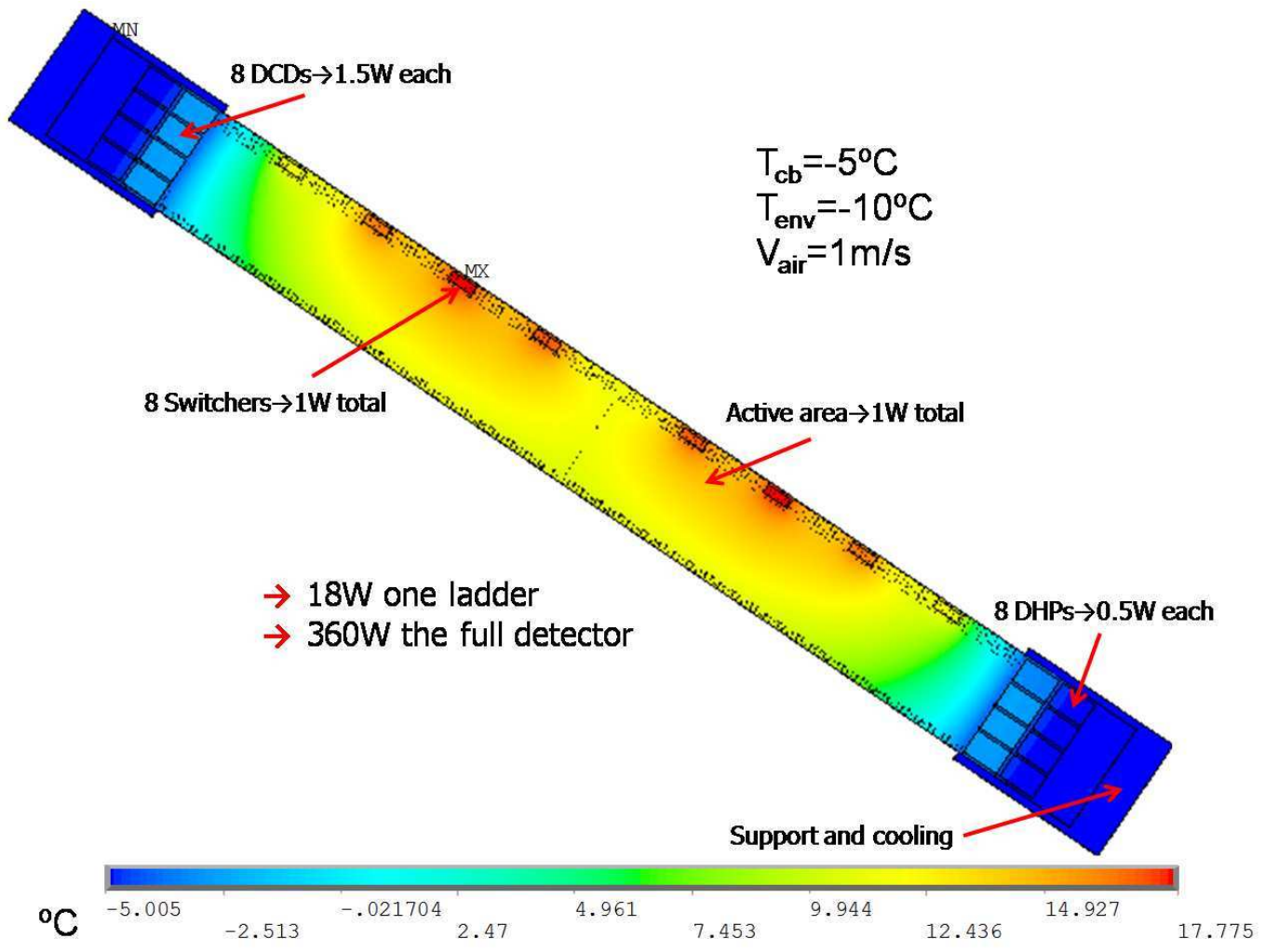

*Figure 4.33: Parameters for the simulation studies. The color code shows the temperature distribution along the ladder as a result from a finite-element simulation with a set of reasonable boundary conditions.*

In our present design, the thermal power of the DEPFET sensor is dissipated by heat conduction through the silicon material itself, supported by a moderate cold air flow ($\approx -5°$ C at a speed of $\approx 1$ m/s) through the full tracker volume. The thermal power produced by the readout chips is absorbed from the support structure, which acts as a heat sink.

### 4.8.5 Cooling simulations and tests

The main feature of the design is the absence of any cooling-related material within the acceptance region. To verify its validity, detailed thermal simulation studies have been performed. The full complexity of the pixel system, however, exceeds the available computing power, so single modules have been simulated. The expected power consumption of the chips is shown in Fig. 4.33.
The thermal models are validated with experimental studies in which the heat generation sources are simulated with resistors and heat foils glued on silicon. These measurements give confidence in the reliability of the simulations. Figure 4.34 shows the good agreement of simulation and experiment.

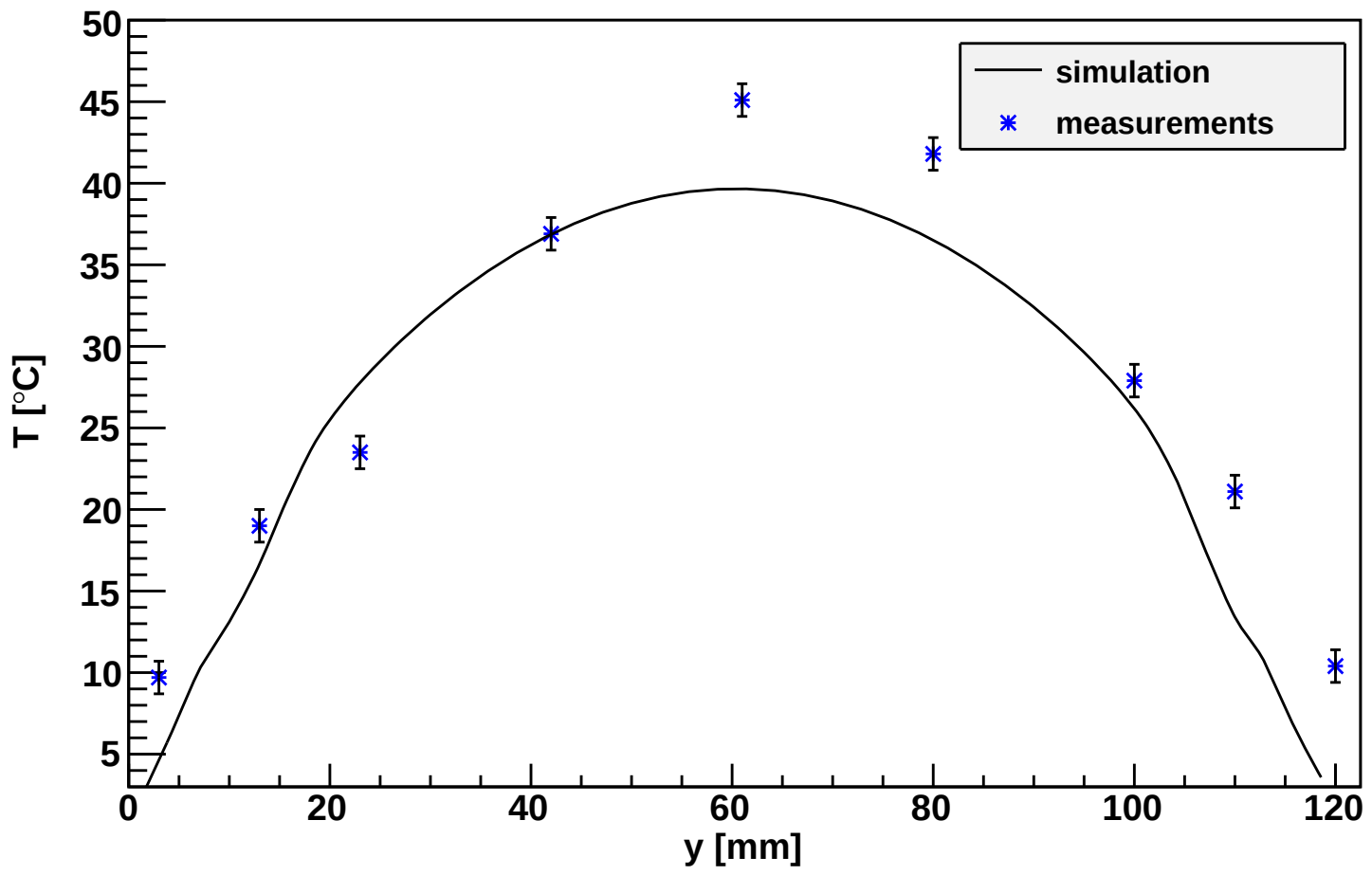

*Figure 4.34: Comparison of simulated and measured data. The diagram shows the temperature profile along the middle axis of a DEPFET module with* 2 mm *CVD extension on both sides. The ambient temperature is* $20^\circ$ C*, the support structure is kept at* $-10^\circ$ C*.*

## 4.9 Interface to Belle II

The PXD needs the following services from the outside:

- Supply voltages for the DEPFET sensors, the readout and control ASICs and auxiliary electronics in the DHH;
- Digital data I/O for configuration and readout, clock signals, trigger signals and readout protocol (electrical);
- Optical fibers for the readout;
- Pipes for the liquid coolant;
- Pipes for the air coolant.

Since the space in the IR region is very limited, only very thin cables are allowed close to the detector. It is therefore foreseen to bring the thick power and optical IO fibers only up to a distance of 2m from the PXD to the data handling hybrid (Sec. 4.7.2). From there, thin power cables and thin twisted pair cables for the digital I/O go to a patch panel. From there, flat Kapton cables are used for the last 20–50 cm. If possible, the patch panel should house only passive components for impedance matching and overvoltage protection. These services will be described in more detail in the following sections.

### 4.9.1 Kapton Links

Each module is serviced with a custom made Kapton cable connecting the ladder to a patch panel. The Kapton cable is fixed on the module (glued and wire bonded). The cable has to be pre-shaped to follow the complicated geometry through the support structure. To fit through the gaps in the support structure and on the available space on the module, its width is limited to 7 mm. The large number of electrical lines needed requires a three-layer tape. The total length is limited by the production process (lithography) and the damping of the HF signals to approximately 50 cm.

| Nr | short name | name | V | $I_{max}$ | comment |
|---|---|---|---|---|---|
| 0 | common | common digital power | | | |
| 0.1 | DVDD 1.8 | DVDD DCD, JTAG SW, DVDD IO, DHP | 1.8V | 1A | |
| 0.1.1 | sDVDD 1.8 | sense DVDD 1.8 | | | |
| 0.2 | DGND | comon digital ground | 0 | 1.4A | |
| 0.2.1 | sDGND | sense DGND | | | |
| 1 | DCD | DCD power | | | |
| 1.1 | AVDD DCD | DCD analog supply | 1.8V | 2.3A | - |
| 1.1.1 | sAVDD DCD | sense line AVDD DCD | | | |
| 1.2 | AGND DCD | DCD analog ground | 0 | -2.3A | - |
| 1.2.1 | sAGND DCD | sense line AGND DCD | | | |
| 1.3 | REFIN DCD | referance of memory cell | 1.1V | 200 mA | - |
| 1.4 | AmpLow | analog ground for amplifier | 0.35V | -1.52A | |
| 2 | SW | switcher power | | | |
| 2.1 | DVDD SW | switcher digital supply | 3.3V | 4mA | ? |
| 2.1.1 | sDVDD SW | sense of DVDD SW | | | - |
| 2.2 | CLEAR on SW | clear on | 17V | 27mA | - |
| 2.2.1 | sCLEAR on | sense line of CLEAR on | | | |
| 2.3 | CLEAR off SW | clear off | 8V | -27mA | - |
| 2.3.1 | sCLEAR off | sense line of clear off | | | |
| 2.4 | sub | substrate | lowest | | local? |
| 2.5 | GATE on SW | gate on | 4V | 27mA | - |
| 2.5.1 | sGATE on | sense line of gate on | | | |
| 2.6 | GATE off SW | gate off | 13V | 27mA | - |
| 2.6.1 | sGATE off | sense line of gate off | | | |
| 3 | DHP | DHP power | | | |
| 3.1 | DVDD DHPCORE | DHP digital supply | 1V | 500mA? | - |
| 3.1.1 | sDVDD DHP | sense of DVDD DHPCORE | | | |
| 3.2 | DVDD DHPGP | supply for gigabit link | 1.2V | | |
| 4 | DEPFET | DEPFET sensor | | | |
| 4.1 | VSOURCE | source | 7V | 100mA | |
| 4.1.1 | sVSOURCE | sense line of source | | | |
| 4.2 | VCCG | common clear gate | 7V | | |
| 4.3 | VBULK | bulk | 17V | | |
| 4.3 | VBP | back plane | -20V | | |
| 4.4 | VGUARD | guard ring | | | |

*Table 4.7: Power lines per module*

| Nr | short name | name | | | | comment |
|---|---|---|---|---|---|---|
| 1 | GCK | system clock | 42.3MHz(?) | | | LVDS |
| 2 | FCK | frame clock | 99.2kHz | | | LVDS |
| 3 | TRG | trigger | 10-30Hz | | | LVDS |
| 4 | TMS | JTAG mode select | | | | LVDS |
| 5 | TCK | JTAG clock | | | | LVDS |
| 6 | TDI | JTAG data in | | | | LVDS |
| 7 | TDO | JTAG data out | | | | LVDS |
| 8 | RST | reset | | | | LVDS |
| 9 | DO(4) | data out (4) | | | | CML |

*Table 4.8: I/O lines per module (DHH/patch panel/module)*

### 4.9.2 Power Cables

The static voltages for the DEPFET biasing and control can be supplied by very thin cables (0.8 $mm^2$, 150 $\Omega$/km). The same is true for the sense wires. Analog and digital power of the ASIC has higher currents, up to 2.3 A. Here, wires of 10 $mm^2$ (5 $\Omega$/km) and 3.1 $mm^2$ (20 $\Omega$/km) will be used. The voltage drop of up to 350 mV requires sensing and regulation in the power supplies. The cable servicing one module has a diameter of 5 mm (a cross sectional area of 20 $mm^2$), including insulation and shielding.

### 4.9.3 Data links and fibers

Data, trigger and clock signals are transferred from the module to the patch panel in the Kapton tapes described above. From there to the DHH, twisted pair cables will be used. Special attention is needed to limit damping and signal distortion in these links, either by impedance matching in the patch panels or active repeaters. The lines needed are listed in Table 4.8. Data transfer between the DHH and the ATCA rack and the distribution of the clock and trigger signal to the DHH will be done optically.

### 4.9.4 Cooling tubes

The PXD needs dry-air cooling to cope with the power dissipated by the sensors and switcher chips without adding extra material in the acceptance region and active cooling to remove the heat generated by the electronics (DCD and DHP) on the module ends.

#### 4.9.4.1 Air cooling

Heat removal from the sensors and SWITCHER chips will require air cooling. The power dissipated in the active region is about 20 W. According to simulations, the air velocity should be 1 m/s with a temperature of about $-5°$C . These calculations do not include the effect of heat radiation from one layer to the other; therefore, the required air temperature may be lower. The PXD support structure will have 1 mm holes on top of each module support to allow the air to flow into the active region. This will require 20 of those pipes on each side that should provide the the required volumetric flow, which will be about 6 $m^3$/h. Air may need to be pumped out on one of the sides. It is not yet clear how to bring the air into the system. Although the flow rate is modest, there are risks of vibrations, given the thickness of the sensors, and detailed studies are being carried out.

A −5°C dry-air flow will presumably require a cold, dry volume that might include the SVD.

#### 4.9.4.2 Active cooling at the module end

The PXD power dissipated on each end of the detector is 180 W. Heat removal will happen on the support structure, which is divided into two halves, each having its own built-in cooling channel. That means that each of the two cooling circuits has to remove 90 W.
We have chosen a two-phase evaporative cooling system based on $CO_2$. According to the thermal simulations, the temperature at the module cooling contact should be 5°C. Simulations and measurements have shown that the temperature difference between the coolant and the module end can be as high as 20°C, requiring a coolant temperature of less than −15°C. For both cooling circuits, corresponding to the two shells of the support structure, the required mass flow would be 0.65 g/s for $CO_2$ with a pressure of 23 bar. In this case the required inner diameter of the pipe would be 0.9 mm, although this needs to be fine-tuned to control the pressure drop. The return lines will need to have a larger diameter to reduce the return line pressure drop. Each half of the support structure will, thus, be connected to a large diameter (outer diameter $\geq 6$ mm) exhaust pipe and a relatively flexible capillary (outer diameter $\sim 2$ mm).
One can design a warm pipe cooling system by introducing a counter-current heat exchanger between the incoming condensed liquid and the outgoing evaporated fluid. The location of the heat exchanger is not yet determined but it would need to be as close as possible to the PXD support structure to minimize the length of the capillaries.

### 4.9.5 Grounding

A PXD ladder is made from two monolithic silicon modules glued together. An electrical contact between the modules cannot be excluded. In this case, the silicon bulk would be electrically connected while the electronics and DEPFET transistors are serviced independently from two sides. Hence there is the danger of a large ground loop. In order to minimize this risk, the main ground of the PXD system must be close to the modules with all power lines and I/O and clock lines floating. The support structures at both ends need a solid ground connection. The grounding scheme is not yet fully engineered; this will to be done together with the SVD and CDC, where some interference could be expected.

## 4.10 Test beam results

Beam tests of DEPFET prototypes form an important complement to the laboratory test procedures described in Sec. 4.6. Their main aim is to characterize the response of DEPFET prototypes to minimum-ionizing particles, thus validating the results obtained with source or laser tests, and to establish the spatial resolution of the devices.

### 4.10.1 Testbeams of DEPFET prototypes

Since 2005, DEPFET prototypes have been submitted to multiple beam tests in a 6 GeV electron beam at DESY, 24 GeV proton beam at the CERN PS and 120 GeV pion beam at the CERN SPS. Sophisticated beam test infrastructure has been developed. For a detailed description of these tests, the reader is referred to several published test beam reports [17, 18, 19]. The write-up of beam tests performed in 2008 and 2009 is in progress [20].

#### 4.10.1.1 Testbeam setup

Several DEPFET devices, or modules, were prepared for each beam test and installed on remote-controlled $xy$-stages in the beam area. The outermost modules constitute the beam telescope required to characterize the spatial resolution of the innermost devices (DUTs). All modules are powered using custom DEPFET power supplies developed by Bonn University. A pair of scintillators read out by photomultiplier tubes is used to generate a trigger for the readout system.
A photo of the setup for the 2009 beam test at the H6 line of the CERN SPS is shown in Fig. 4.35.

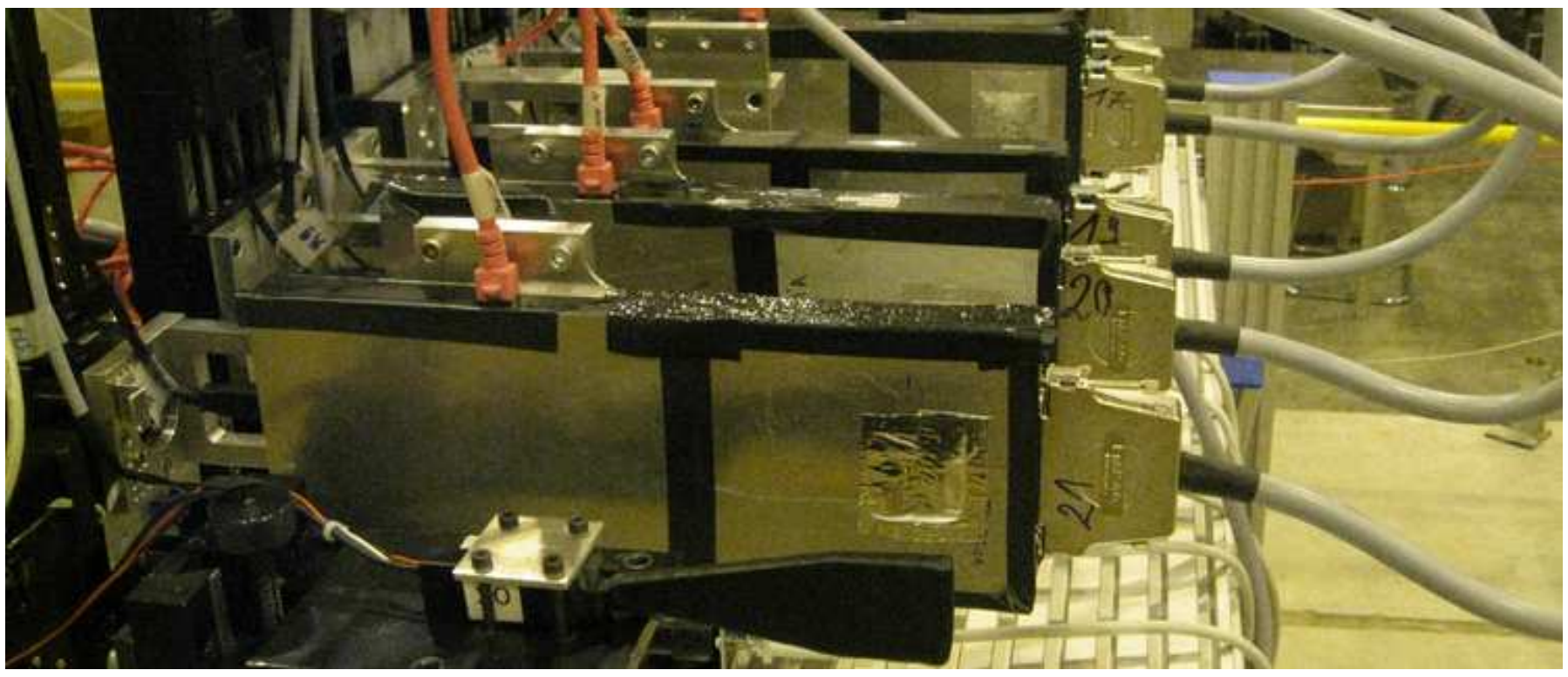

*Figure 4.35: A photo of the DEPFET beam telescope in the H6 area at the CERN SPS.*

#### 4.10.1.2 DEPFET prototypes under test

The DEPFET active pixel sensors submitted to beam tests are read out using the S3B system, described in Sec. 4.6 (Fig. 4.19). The PXD5 sensors tested so far are 450 $\mu$m thick DEPFET sensors with $64 \times 256$ or $64 \times 128$ pixels. The issue of thinning will be addressed in the PXD6 production currently in progress. Typical pixel sizes in these designs for the ILC vertex detector range from $20 \times 20\,\mu\text{m}^2$ to $24 \times 32\,\mu\text{m}^2$, whereas the pixels on Belle II ladders are $50 \times 50\,\mu\text{m}^2$ and $50 \times 75\,\mu\text{m}^2$. Also, the read-out system differs in several important aspects from the Belle II PXD ladders. The CURO front end ASIC [21, 22] is being phased out in favor of the DCD chip described in Sec. 4.3 and a design iteration of the switcher is foreseen.
Prior to the beam period, all modules were characterized using source and laser tests. An analysis of the former provides a determination of the internal gain of the first amplification stage in the sensor (the quantum gain $g_q$, described in Sec. 4.2). The latter gives a first idea of the uniformity of the charge collection over the sensor active area (Sec. 4.6).

### 4.10.2 Results on PXD5 prototypes

DEPFET beam tests have led to a wealth of results. A discussion of all these is beyond the scope of this document and the reader is referred to the references cited previously for a more exhaustive discussion. Here, the most important findings are given. The focus is on two basic aspects of the DEPFET performance: signal collection and in-pixel amplification in the DEPFET sensors and the spatial resolution of the DUTs.

#### 4.10.2.1 Characterization of the signal collection

The most straightforward measurement is that of the signal charge distribution. Pedestal-subtracted and common-mode corrected analog data are presented to a clustering algorithm with a seed threshold of $7\sigma$ and a neighbor threshold of $2.6\sigma$. The resulting cluster signal (*Landau*) distribution is shown in Fig. 4.36 (left).

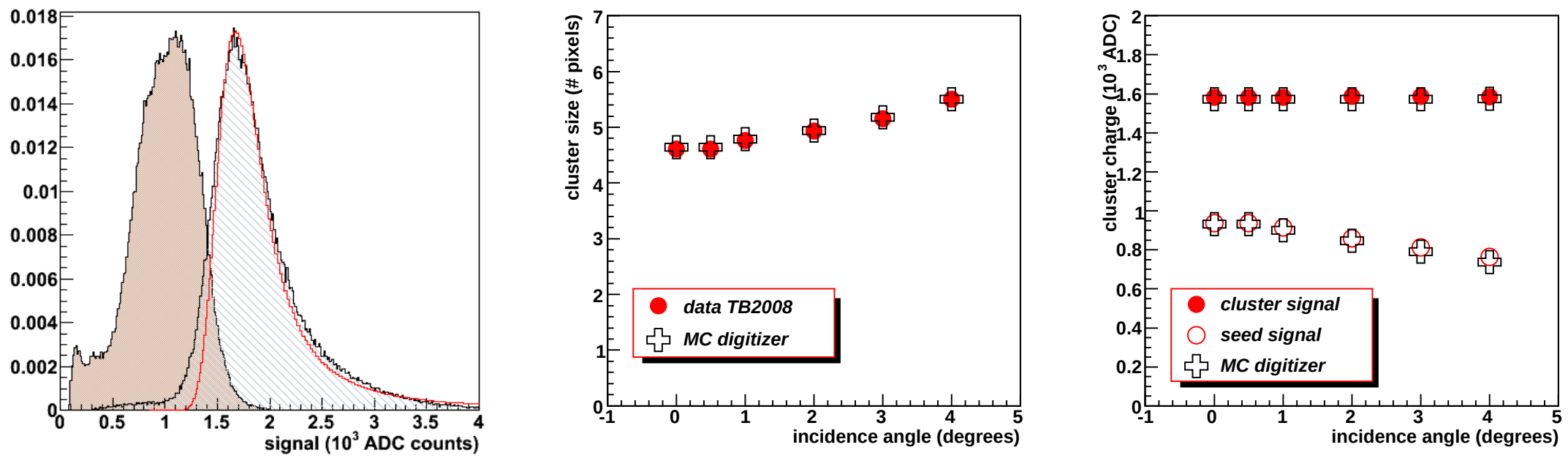

*Figure 4.36: Left: the seed (filled) and cluster signal distribution for 120 GeV pions perpendicularly incident on the small-pixel DUT. Center: cluster size vs. incidence angle. Right: seed (open circles) and cluster signal (filled circles) vs. incidence angle. The cross markers correspond to the result of a MC model that is discussed in Sec. 4.10.3.*

The distribution is found to be in excellent agreement with the most commonly-used models [23, 24, 25, 26] in the range up to several MIPs. This agreement corroborates the excellent performance of the DEPFET system: noise and gain variations have a very small effect on the width of the distribution. The saturation of the DEPFET internal gate for a signal of approximately 70,000 electrons affects the high signal tail (not visible on this linear scale).
The distribution of the signal on the seed pixel is also shown in Fig. 4.36. On average, two thirds of the signal is contained in the seed pixel. For perpendicularly incident MIPs, the average cluster size (the number of pixels with a signal exceeding $2.6\sigma$) is approximately 4.5. For these small-pixel devices, charge sharing between pixels increases strongly with incidence angle, leading to an increase in cluster size and a decrease in the fraction of the signal in the seed pixel.
The most probable signal ranges from 1600 to over 3000 ADC counts (ADU) in different sensor designs. The noise level in these devices is approximately 13 to 14 ADU. Thus, these devices have a very comfortable signal-to-noise ratio of well over 100. The most probable signal is used to determine in-pixel gain $g_q$.[3] The results for the most important DEPFET designs are summarized in Table 4.9. These results are in good agreement with the expectations from the source tests.
Two design variants tested in 2009 prove to yield a very significant increase in the in-pixel amplification. In a DEPFET device with a modified clear-gate structure known as Capacitively Coupled Clear Gate (CCCG), a 40% increase of the gain is observed when compared to the same sensor design with the usual Common Clear Gate design. The measurements on a PXD5 sensor with a short gate length ($L_{\rm gate} = 4\,\mu$m) confirm, on a large-scale device, the strong dependence of the gain observed in single-pixel measurements in Fig. 4.5. This module has shown excellent performance throughout the test beam, while operated with an 80% higher $g_q$ than the standard sensors with a gate length of 5 $\mu$m. This increase in gain translates to a proportional enhancement of the signal-to-noise ratio, crucial to the operation of very thin sensors.

[3] The in-pixel gain $g_q$, in nA/$e^-$, equals $\mu \times (7.7\,\text{nA/ADU})/(E_{\rm MP}/(3.6\,e^-/\text{keV})$, where $\mu$ is the most probable value in ADU, 7.7 nA/ADU corresponds to the measured gain of the CURO/S3B system, and $E_{\rm MP}$ is the most probable energy deposition in 450 $\mu$m of silicon by 120 GeV pions: $E_{\rm MP} = 131$ keV [23, 24, 25, 26] corresponding to 36.4 k$e^-$

Table 4.9: The in-pixel gain of different DEPFET prototypes.

| sensor type, matrix & pixel size | | | results | |
|---|---|---|---|---|
| name | row × column | ($\mu$m$^2$) | $g_q$ (pA/$e^-$) | noise ($e^-$) |
| telescope TB2008 | $64 \times 128$ | $32 \times 24$ | $320 \pm 20$ | $320 \pm 20$ |
| DUT 2008 | $64 \times 128$ | $24 \times 24$ | 360 | 284 |
| telescope TB2009 | $64 \times 256$ | $32 \times 24$ | $394 \pm 14$ | $290 \pm 2$ |
| DUT CCCG | $64 \times 256$ | $32 \times 24$ | 507 | 225 |
| DUT short gate | $64 \times 256$ | $20 \times 20$ | 655 | 174 |

The uniformity of the response over the DEPFET sensor has been measured. A spread of 5% is observed when comparing the response of the over 8000 pixels of a sensor. The response is, moreover, found to depend very slightly on the position inside the pixel. The RMS in this case is approximately 3%. None of these effects is expected to affect the resolution significantly.

#### 4.10.2.2 Spatial resolution of small-pixel DEPFETs

The spatial resolution of the DUT is evaluated by comparison with the prediction of the beam telescope. The position of a high-momentum pion is predicted using the information of a multi-module DEPFET telescope with sub-micron precision. The resolution of the DUT is extracted from the residual distribution using the method described in Ref. [18] that takes into account the finite telescope precision and multiple scattering. The spatial resolution of the 2008 DUT with $24 \times 24\,\mu\text{m}^2$ pixels is thus found to be:

$$\sigma_x = (1.3 \pm 0.2)\,\mu\text{m}, \quad \sigma_y = (1.2 \pm 0.1)\,\mu\text{m} \tag{4.5}$$

This excellent resolution establishes the DEPFET as a firm candidate for high-precision vertex detectors.

### 4.10.3 Comparison with a detailed DEPFET model

These tests of small-pixel DEPFET prototypes have allowed a detailed understanding of the characteristics of MIP signals in these devices. The test beam results are a crucial input to validate the model of the DEPFET response. This model, described in more detail in Sec. 4.11, is used as the *digitizer* stage of Monte Carlo studies to predict the performance of Belle II vertex detector ladders and in the Belle II physics studies.

A detailed comparison has been performed of the model predictions with test beam measurements under different incidence angles. Excellent agreement is found for the cluster size and fraction of the signal in the seed pixels, as shown in Fig. 4.36. The agreement in the $\eta$ distribution[4] of Fig. 4.37 (left) is an even better indication that charge sharing between neighboring pixels is described correctly.

[4] $\eta$ is defined as $\pm S_N/S_N + S_S$, where $S_S$ and $S_N$ are the seed and highest-neighbor signal, respectively, and the sign is positive (negative) if the highest neighbor is located to the right (left) of the seed pixel.

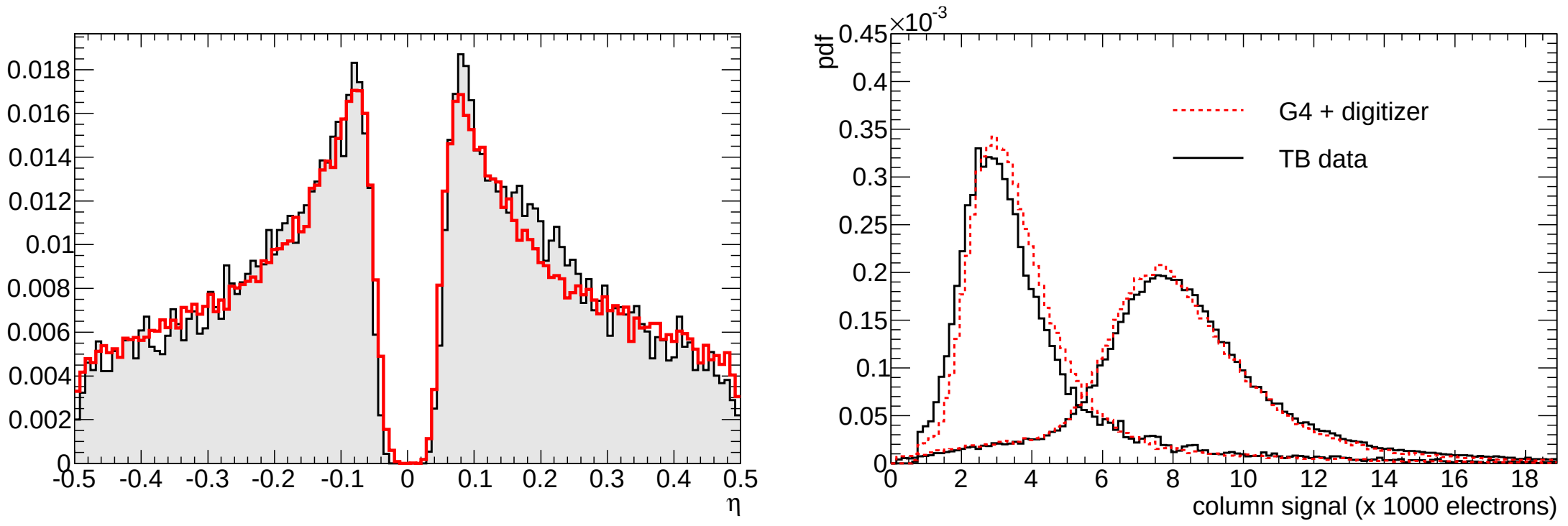

*Figure 4.37: Comparison of measured beam test data (black) to the prediction of the digitizer (red). Left: the $\eta$ distribution. Right: the column signal distribution from TB data (black continuous line) and MC (red dashed line) for incidence angles of $12°$ and $36°$ with respect to the perpendicular. (See text in Ssec. 4.10.3 for the explanation).*

To measure the signal distribution for a thin sensor, data was taken with particles incident at a shallow angle. Since the particle trajectory projects onto a large number of pixel columns, the signal collected in each column corresponds to the energy deposited along a small fraction of the trajectory. The signal distributions in the rightmost plot of Fig. 4.37 for sensor rotations of $12°$ and $36°$ with respect to perpendicular incidence correspond to a sensor thickness of $112\,\mu$m and $40\,\mu$m, respectively. The excellent agreement observed in both cases demonstrates that the Landau fluctuations in thin sensors are taken into account correctly by the digitizer model.

#### 4.10.3.1 Discussion

Beam tests of DEPFET prototypes have been vital to prove the principle of DEPFET in-pixel amplification. Small-pixel prototypes have demonstrated excellent signal-to-noise ratio, achieving an in-pixel gain of over $600\,\mathrm{pA}/e^-$. Moreover, the DEPFET response is found to be uniform within 5% over the sensor and within the elementary pixel cell. The spatial resolution of a single small-pixel DEPFET sensor is found to approach the micron level (Eq. 4.5).
Test beam results are a crucial input to the estimate of the DEPFET performance in Belle II. Due to the differences in sensor design, the extrapolation must be performed by a careful Monte Carlo simulation. The measured response of the DEPFET to high-energy particles from an accelerator has been used to validate a detailed model of the DEPFET response (the DEPFET digitizer described in Sec. 4.11). Thus, the MC studies of the DEPFET vertex detector performance rest on a solid empirical basis.

## 4.11 Expected Performance of the PXD

Here, we present the expected performance of the PXD in the harsh background environment of SuperKEKB. The situation is particularly severe for the PXD since it is always live and thus integrates all impinging particles during the readout interval of $20\,\mu$s.

### 4.11.1 QED Background

According to preliminary Monte Carlo calculations (see below), the expected background arising from QED processes will be very large and may even surpass in rate the background from the machine (beam-gas reactions, Touschek effect, and synchrotron radiation). Apart from the Touschek effect, the machine background is not expected to be significantly larger than at KEKB. The main reason is that the SuperKEKB beam currents are only double those of KEKB, whereas all processes proportional to the luminosity, such as QED reactions, will increase relative to KEKB by a factor of 40 at design luminosity.

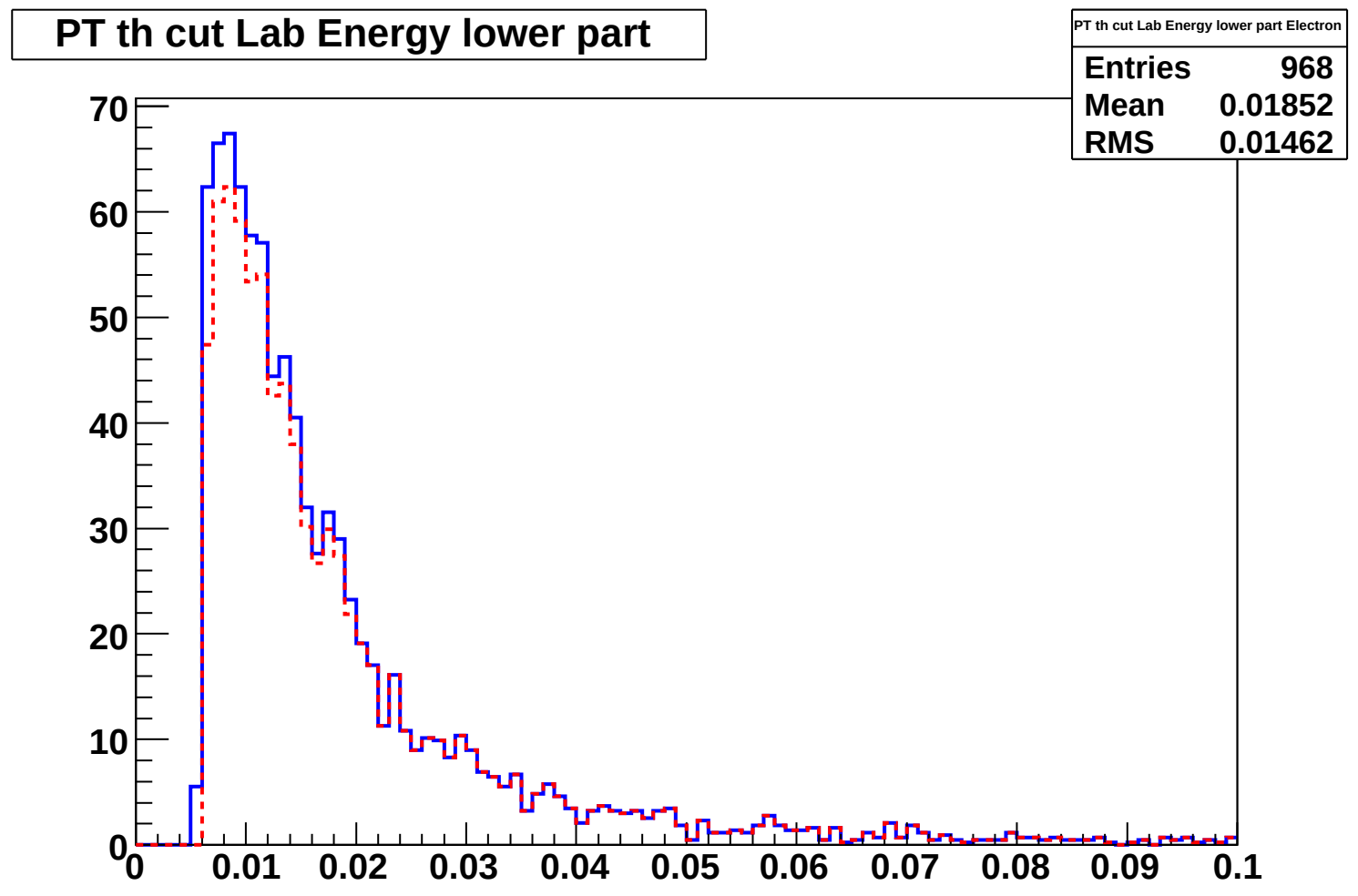

*Figure 4.38: Energy distribution of secondary electrons from $\gamma\gamma$ interactions (in GeV). The solid blue line is from the KoralW generator, while the dashed red line is from the BDK generator (see text).*

Several Monte Carlo generators (BDK [27], KoralW [28], and Racoon [29]) were used to estimate the background. Since the QED processes (mostly $\gamma\gamma$ reactions) making up the background for the PXD are very low in energy, electrons and positrons above a transverse momentum of only 6 MeV will reach the first layer of the PXD. So far, there are no direct measurements of such low-momentum particles in an $e^+e^-$ collider. As a consequence, the predictions from the various generators may not all give a consistent estimate for the rate of particles into the PXD. To have a conservative estimate, the generator with the largest particle flux has been chosen. In Fig. 4.38, the low energy part of the corresponding spectrum from electrons and positrons, limited to the Belle II acceptance (polar angles of $17^\circ$ to $150^\circ$) is shown. In the figure, the spectrum has been cut off at a transverse momentum of 5 MeV, since charged particles with lower energy will not reach the PXD due to the 1.5 T solenoidal field. From this spectrum and the cross section for the reaction $\gamma\gamma \to e^+e^-$ around a center-of-mass energy of about 10.5 GeV, the occupancy for the first layer of the PXD is below 0.5%.

### 4.11.2 General Strategy

The evaluation of the performance of the PXD is carried out by simulating single track events as well as more realistic physics benchmark processes. In the former, muons are generated at the unsmeared $e^+e^-$ interaction point with selected values of track momenta, $p$, and polar angles, $\theta$, uniformly distributed in azimuth, $\phi$. For an evaluation of physics events with a more realistic track occupancy, we choose the reaction $\Upsilon(4S) \to B^0\bar{B}^0$ where one of the $B$s decays via a $CP$ eigenstate and the other through a state tagging the $b-$ (or $\bar{b}-$)quark of the entangled $B$ mesons. For the $CP$ side, we choose the "golden" mode $B^0 \to J/\psi\, K_S$ with $J/\psi \to \mu^+\mu^-$ and $K_S \to \pi^+\pi^-$, while for the tag side we simulate the decay chain $B_{\mathrm{tag}} \to D^{*\pm}\pi^{\mp}$ with $D^{*+} \to D^0\pi^+_{\mathrm{slow}} \to K^-\pi^+\pi^+_{\mathrm{slow}}$. Thus, the decay muons of the $J/\psi$ provide high momentum tracks with a transverse momentum, $p_t$, up to 2.5 GeV/$c$, while the decay particles of the $D^{*\pm}$ provide tracks at medium ($K$, $\pi$) and low ($\pi_{\mathrm{slow}}$) momenta of only a few tens of MeV.
The performance of the PXD is then characterized by the track impact parameter resolution in the $rz$ projection, $\Delta z$, as well as in the $r\phi$ plane, i.e., the resolution of the distance of closest approach, $\Delta d_0$. In addition to the track impact parameters, the full physics events are also used to evaluate the PXD performance in terms of $z$-vertex resolution of the $CP$- and tag-side decays. Furthermore, the full physics events allow us to evaluate the quality of the $J/\psi$, $D^{*\pm}$ and $D^0$ candidate reconstruction in terms of the width of the signal peak and the background level under the peak.

### 4.11.3 Event Generation, Simulation and Reconstruction

For our optimization studies, we take advantage of the ILC framework, which has been developed to guide the detector research and design for the future linear collider [30]. Thanks to its modular design, the ILC framework proved itself to be flexible enough so that it could be adopted quickly to the special needs for the Belle II PXD optimization study.
The event simulation is implemented in the GEANT4 [26] based Mokka package. For the single track events, we use the built-in particle gun; for the physics events, we use our privately developed interface to read events in HEPEvt format. In this way, the EvtGen [31] generator, which is well established in Belle, can serve as a source of events. After EvtGen, the events are boosted according to the Belle II beam configuration, and constitute the input to our simulation step.
Within Mokka, the generated particles are traced through the tracking volume. The full detector simulation takes into account the interactions with the detector material, e.g., the production of secondary particles and the energy deposition in the active volumes of the detector.
Once the simulation step is finished, the events are handed over to the Marlin program for reconstruction, which is handled by a sequence of so-called Marlin processors. The digitization step converts the energy deposits into hit pixels (PXD), strips (SVD) and wires (CDC). The implementation of the PXD digitizer follows the procedure adopted in the simulation/reconstruction software of the CMS experiment at the LHC.
In the simulation, a charged particle traversing the active volume of the PXD sensors has its trajectory divided into segments (so-called ionization points) whose charge is smeared according to the Landau distribution. Ionization points are then drifted to the electrodes, subjected to the Lorentz shift in the presence of the Belle II magnetic field and smeared by a Gaussian distribution to account for diffusion. For each electrode, one gets a digit, with either a vanishing signal or non-zero signal, depending on how many ionization points contribute to the signal. Finally, the digits are converted into real hits.
For the clustering of hit pixels, either a center of gravity or an analog head-tail algorithm (for cluster multiplicities larger than two hit pixels) is used. Before the actual start of the clustering algorithm, the pixel sensors are populated with possible additional hits. These originate from signal hits from the other tracks in the event, from additional tracks due to QED background events, or from electronic noise.

After the digitization, hits from PXD, SVD and CDC are fed into another Marlin processor that performs the track pattern recognition and the track fitting. For the study presented here, we used the standard tracking processor as implemented in the ILC framework that is based on the tracking software package of the Delphi Collaboration.
Once tracks are found in real physics events, they are combined to decay candidates in another Marlin processor. In the current study, candidates of $J/\psi \rightarrow \mu\mu$ and $D^0 \rightarrow K\pi$ decays are formed from all possible track pair combinations with a total charge of zero. No particle ID information has been used so far. The lists of track combinations are then fed into the next processor that fits the decay candidate vertex using the RAVE library [32] (developed for CMS). The whole chain of the event generation, detector simulation, and reconstruction up to the reconstructed decay candidates has been verified against the the equivalent chain using the Belle experiment's software framework (BASF). Track impact parameter resolutions were determined in single track events and full physics events, and are found to agree well within the LCIO for Belle II and the BASF frameworks. For the full physics events, the track impact parameter and the vertex resolution for $J/\psi$ agree well.

### 4.11.4 Simulation Scenarios

For the baseline beam configuration, we assume that the nano-beam option 2 will be realized: 4 GeV positrons collide with 7 GeV electrons at an crossing angle of 83 mrad. The tilt angle of the lower energy positron beam with respect to the direction of the axis of the solenoid has a non-vanishing component only in the horizontal plane; its value is 15.55 mrad.
Since the Delphi tracking code, as implemented in the ILC framework, only allows for rotationally symmetric geometries with a magnetic-field axis that coincides with the $z$ axis, we are forced to assume that the beam pipe axis and the central axis of the PXD coincide with this $z$ axis. This is not necessarily the case for the final design of the interaction region; however, for the detector optimization, for which we compare the relative performance of various detector scenarios in exactly the same environment, possible tilts of the symmetry axis of the PXD and the beam pipe with respect to the magnetic-field axis can be safely neglected.
Guided by preparatory studies, we decided upon the basic parameters of a baseline PXD that would have a good chance to meet the physics requirements of Belle II. This design consists of two layers of DEPFET pixel sensors, with the inner (outer) layer containing 8 (12) ladders at a radius of 14 mm (22 mm). Both layers have $250 \times 1600$ pixels, corresponding to a pixel size of $50 \times 50\,\mu\text{m}^2$ on the inner layer and $50 \times 75\,\mu\text{m}^2$ on the outer layer. In the outer layer only, the ladder is built from two DEPFET pixel sensors, glued end to end (Sec. 4.8). The sensor thickness was a parameter in the study.
The baseline scenario is evaluated against a number of scenarios in which only one of the parameters characterizing the baseline detector is varied at a time. For example, we studied the reduction of the inner radius to 13 mm, the use of 50 or 75 $\mu$m-thick sensors, the reduction in the number of pixel rows from 1600 to 800 (to permit the full readout of the PXD within 10 $\mu$s), and the splitting of the inner-layer ladders into two end-to-end sensors as in the outer layer (to simplify the final production, but at the cost of $\sim 500\,\mu$m-wide dead regions). Finally, we also consider an "optimal" detector scenario, which might still be feasible, although difficult in light of the mechanical considerations. In this scenario, we reduce the radius of the inner layer to 13 mm and assume 2000 pixel rows for both the inner and outer layer, while keeping the sensitive length in each layer constant.

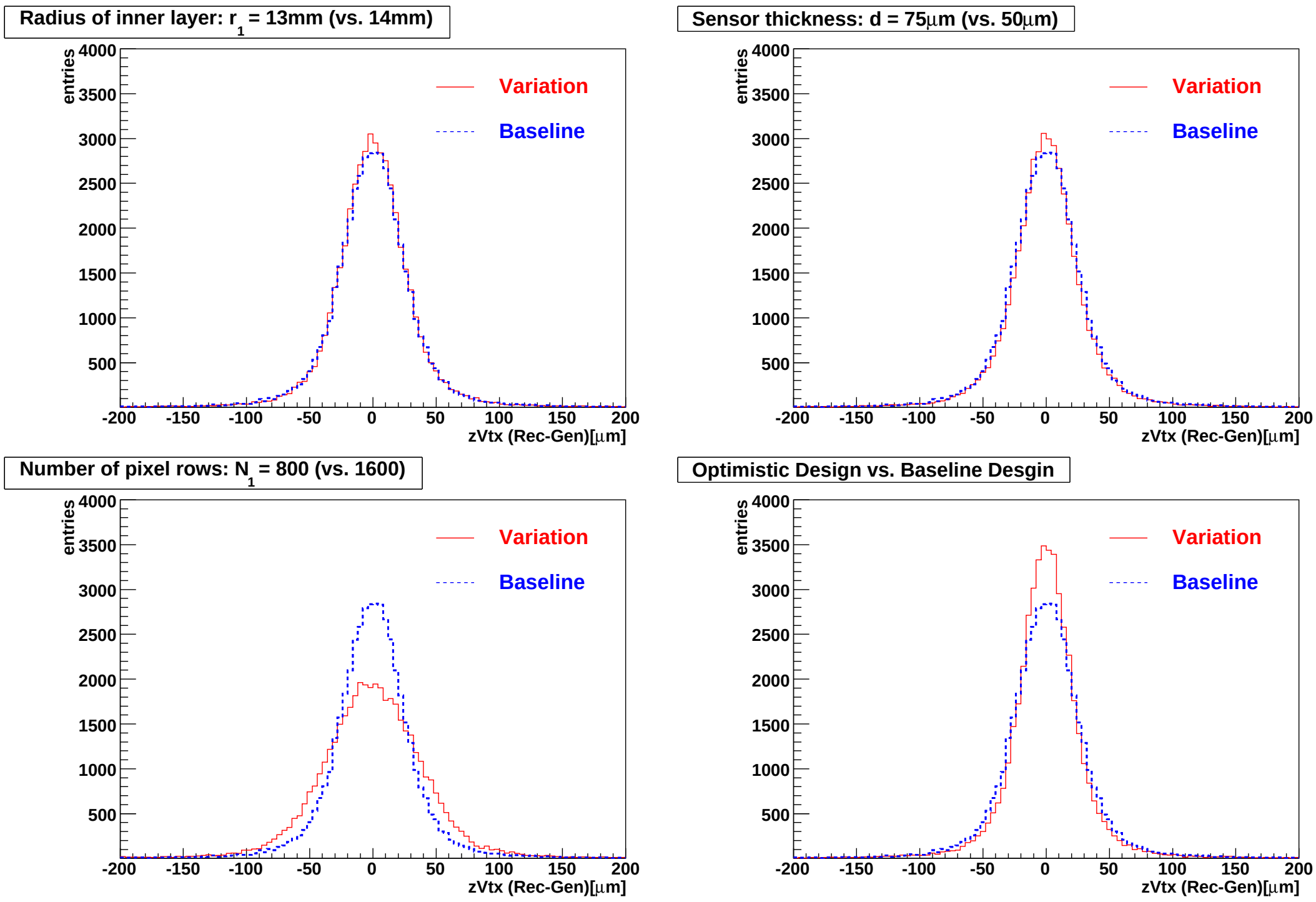

*Figure 4.39: The $J/\psi$ z-vertex resolution of the baseline detector (blue) compared to selected alternate scenarios (red).*

### 4.11.5 Results of the Simulation

A comparison of the $z$-vertex resolutions of $J/\psi$ in the different scenarios is shown in Fig. 4.39. A reduction of the number of pixel rows clearly worsens the resolution. A slight improvement is found when the inner radius is reduced to 13 mm.

Surprisingly, increasing the thickness of the sensor from 50 to 75 $\mu$m does not noticeably degrade the resolution, as one might naively expect. The additional multiple scattering is outweighed by the increasing charge sharing between neighboring pixels, which reduces the fraction of single-pixel clusters whose hit-position resolution is worse than for multiple-pixel clusters. (For multiple-pixel clusters, more advanced position reconstruction algorithms such as center-of gravity or analog head-tail can be applied.) Hence 75 $\mu$m-thickness was chosen for the baseline design.

Splitting the inner-layer ladders into two end-to-end sensors has no visible effect (therefore not shown in Fig. 4.39). As expected, the best performance is seen for the so-called "optimal" detector layout.

For muons from $J/\psi$ decays, tracks with large transverse momenta dominate the sample. As an alternative, we also studied the tag-side $B$ decay vertex that is reconstructed from the vertex and four-momentum of the $D^0$ alone by extrapolating the $D^0$ from its decay vertex back towards the beam line. The same trends as for the $J/\psi$ vertex resolutions are observed, though less prominently than in the former case due to the lower momenta of the kaons and pions. The PXD adds only slightly more material to a particle that has already traversed the thicker beam pipe. (The thickness of the Belle beampipe is about $0.7\% X_0$, while that of a single PXD layer is only $0.19\% X_0$.) The conclusion obtained from the decay vertex resolutions are supported by the track impact parameter resolution. Figure 4.40 shows the track $z_0$ impact parameter resolutions of muons from $J/\psi$ decays. Results from single-track events are consistent with those shown in this figure.

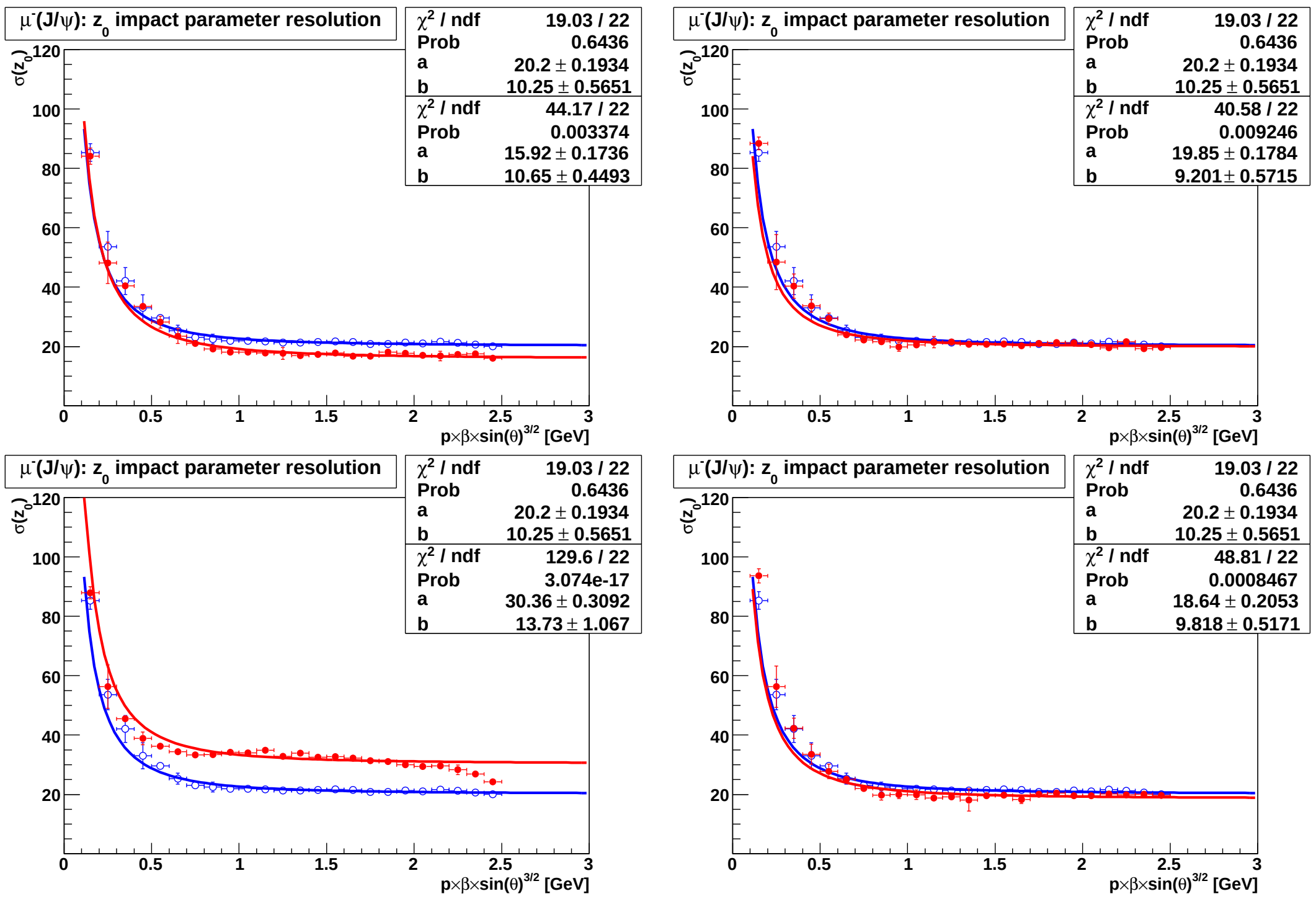

*Figure 4.40: Comparison of the track impact parameter resolution, $\Delta z_0$. The blue open circles and line represent the baseline detector, while the red points and line indicate the varied detector scenarios.*

To verify that the Belle II detector with the DEPFET PXD as its innermost element meets the physics requirements of precision reconstruction of vertices and track impact parameters, we compared the Belle II baseline detector to a simulation of the tracking volume of the Belle detector, both within the ILC framework. The muon-track $z_0$ impact parameter resolution is shown in Fig. 4.41, demonstrating the superior performance of the Belle II tracking systems. At momenta below 800 MeV, an improvement of almost a factor of two is expected for Belle II.

### 4.11.6 Conclusion

A simulation chain for the evaluation of various detector options has been established. We studied both clean physics signal events and events where we mixed in the expected QED background and electronic noise. The results of the studies with and without background lead to the same conclusions. Findings of the comparison of baseline and alternate PXD scenarios provide a clear guideline to further improve the performance of the PXD. Already, the baseline scenario fulfills the physics requirements, as is demonstrated by comparing the performance of the Belle II and Belle tracking systems.

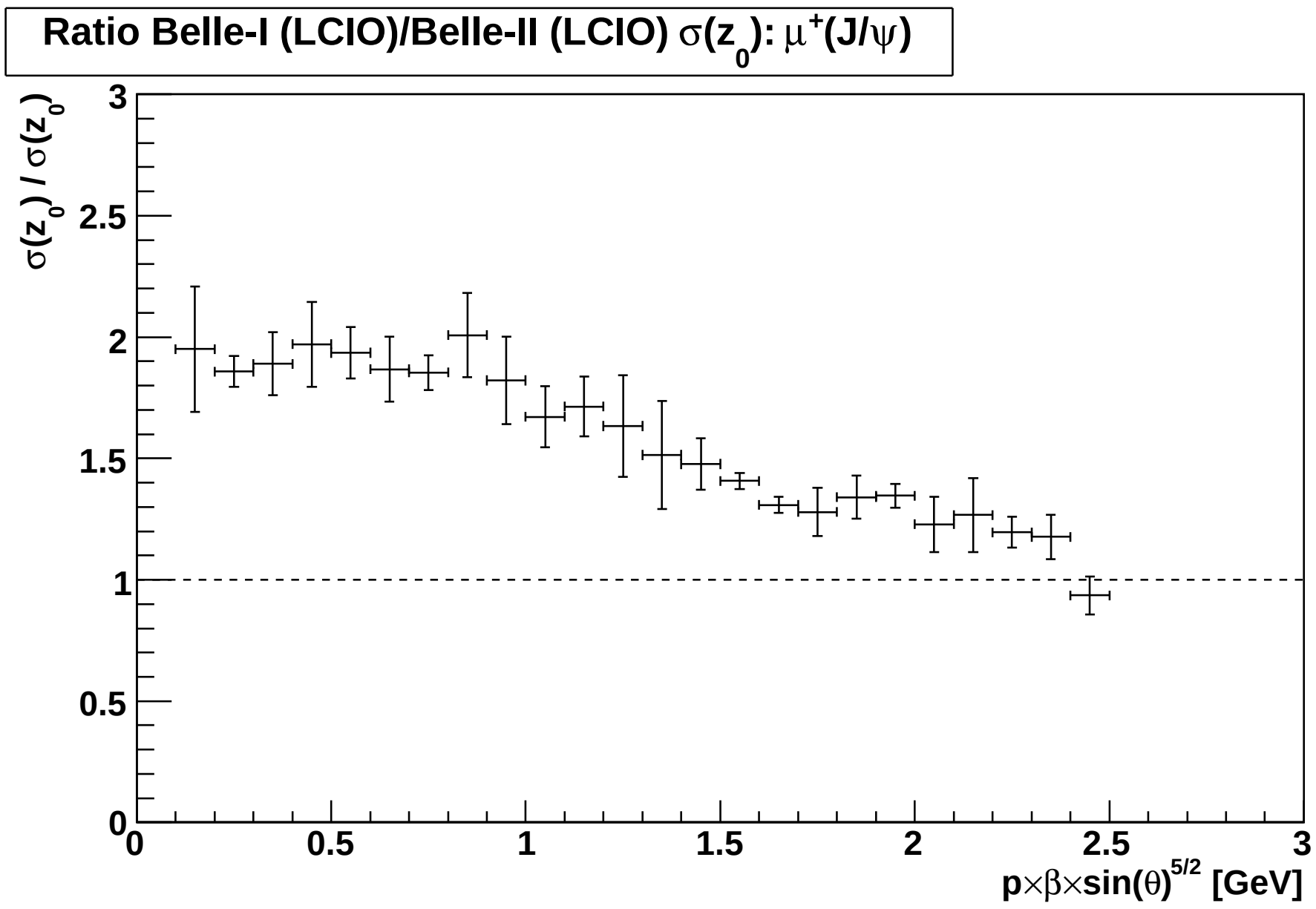

*Figure 4.41: Belle over Belle II ratio of the $z_0$ track impact parameter resolutions as function of the pseudo-momentum, $p \times \beta \times \sin^{5/2}\theta$. At low momenta, the impact parameter resolution of Belle II is about twice as good as that of Belle.*

## 4.12 Project Organization

The institutes participating in the construction of the Belle II PXD are

1. Charles University, Prague, Czech Republic* (PRA)
2. Physikalisches Institut der Universität Bonn, Germany* (BON)
3. Institut für Kernphysik, Karlsruhe, Germany* (KAR)
4. II Physikalisches Institut der Universität Göttingen, Germany (GOE)
5. Max-Planck-Institut für Physik, Munich, Germany* (MPI)
6. Institut für Technische Informatik der Universität Heidelberg, Germany* (HEI)
7. Ludwig-Maximilians-Universität Munich, Germany (LMU)
8. Technische Universität, Munich, Germany (TUM)
9. II Physikalisches Institut der Universität Gießen, Germany* (GIE)
10. IFIC Valencia, Spain (IFV)
11. University of Barcelona, Spain (UBA)
12. University Ramon Llull, Barcelona, Spain (URL)
13. University of Santiago de Compostela, Spain (USC)

| Id | work package | lead institute | participants |
|---|---|---|---|
| 1.0 | DEPFET Sensors | | |
| 1.1 | parameter definition | MPI | PRA |
| 1.2 | sensor development | MPI | |
| 1.3 | ASIC Development | | |
| 1.3.1 | Switcher | HEI | |
| 1.3.2 | DCD | HEI | |
| 1.3.3 | DHP | BON | UBA, MPI |
| 1.3.4 | Interconnections | BON | MPI, USC, URL |
| 1.4 | Module design | | |
| 1.4.1 | sensor Ladder | MPI | HEI, BON, IFV, CNM, IFB |
| 1.4.2 | Kapton Flex | BON | LMU, URL |
| 1.4.3 | DHH | TUM | BON, URL |
| 1.5 | Mechanical design | MPI | KAR, VIE, IFV, KEK |
| 1.6 | Thermal Issues | KAR | MPI, VIE, KRA, IFV, IFB |
| 1.7 | System | | |
| 1.7.1 | Data Acquisition | GIE | TUM, MPI, KRA, GOE, URL, KEK |
| 1.7.2 | Power supplies | LMU | KRA, KEK, USC |
| 1.7.3 | Cooling plant | KEK | KAR, IFV |
| 2.0 | Test facilities | | |
| 2.1.1 | Test beam setup | IFV | KAR, BON, VIE, IFV, IFC |
| 2.1.2 | Beam tests analysis | PRA | URL, CNM, IFB, USC, MPI, HEI |
| 2.1.3 | Lab tests | MPI | (all) |
| 2.2 | Thermal tests | KAR | MPI, VIE, IFV, IFC |
| 2.3 | Mechanical mockup | | MPI |
| 2.4 | Irradiation tests | | |
| 3.0 | Integration | | |
| 4.0 | Operation Issues | | |

*Table 4.10: Work packages and assignments to the collaborating institutions*

14. Instituto de Fisica de Cantabria, Santander, Spain (IFC)

15. Centro Nacional de Microelectronica, Barcelona, Spain (CNM)

16. Henryk Niewodniczanski Institute of Nuclear Physics, Cracow, Poland* (KRA)

The institutions marked with * are already members of Belle II; the other will apply once national funding is secured. In addition the project is supported by other Belle II collaboration members like the HEPHY, Vienna, Austria (VIE) and KEK, Japan (KEK).

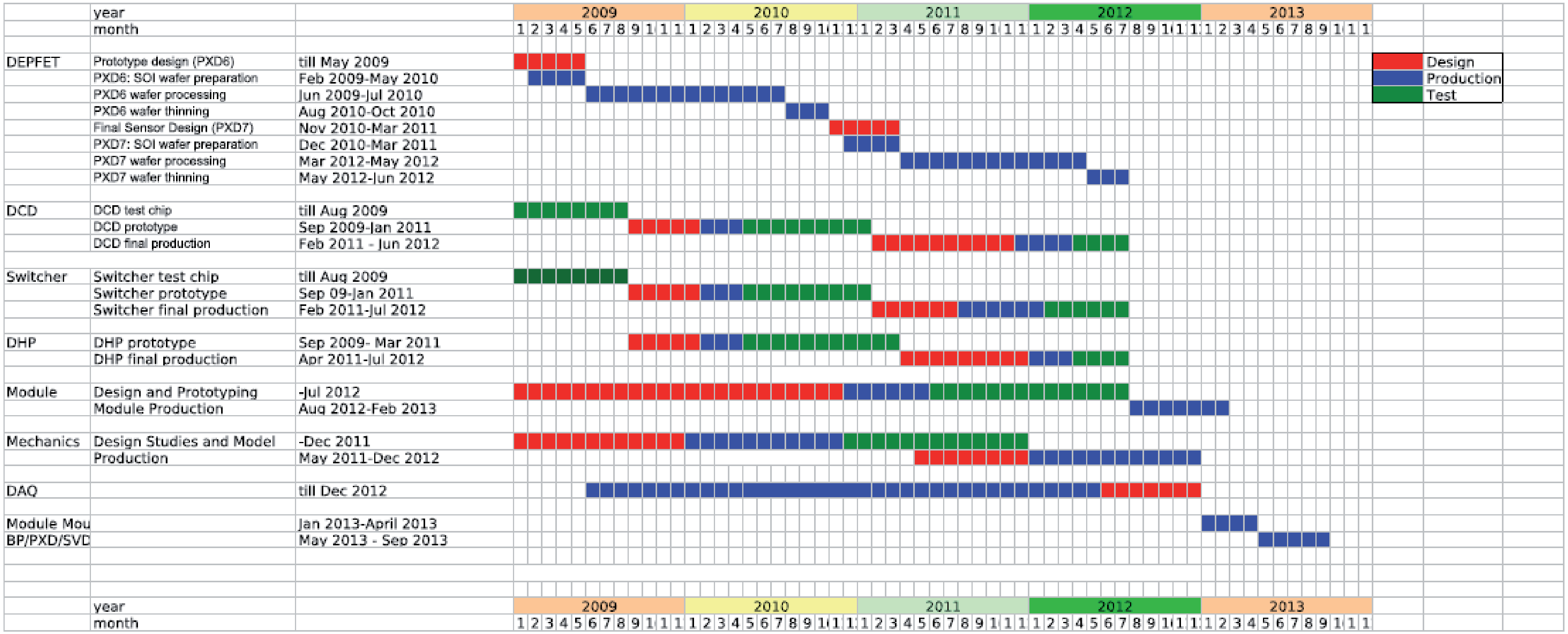

*Figure 4.42: Schedule of the PXD construction.*

The project is led by a project leader assisted by a technical coordinator. Representatives of all institutions form an institutional board, with its chair person.
The project is organized in work packages which are managed by lead institutions. Table 4.10 lists the work packages and the participating institutions. The schedule of the PXD production and installation is shown in Fig. 4.42. The schedule is mainly driven by the production of the DEPFET sensors. This happens in two stages: a prototype production (PXD6) in order to test certain design and technology variants, and the final-design production. A production cycle of a DEPFET takes approximately 16 months, due to the very complex processing (more than 80 process steps). Driven by the installation date (summer 2013), the DEPFET sensors need to be ready for assembly mid 2012. In order to achieve this, sensor production needs to start April 2011. By then, all technical developments explained in Sec. 4.2 (thin oxides, shorter gate length) need to be finished and all geometrical parameters need to be finalized. The schedule does not leave room for a 'final' prototype run (the PXD6 prototypes do not have final geometry and thickness) but has enough contingency to cope with downtime of the production line. The development and production of the control and readout electronics is done in parallel. Since ASIC productions have a much faster turnaround, typically three months, these activities are not on a critical path.

### 4.12.1 Upgrade Option with SOI Pixel Technology

The baseline design with the DEPFET technology is well studied and we are confident that the DEPFET will work from the startup of the Belle II experiment. However, the readout rate of the DEPFET is limited to 20 $\mu$s per frame, even with four-fold parallel readout. Depending on the background conditions, this may lead to a high occupancy and make it difficult to find the vertex positions when the SuperKEKB luminosity is at its design value.
There is an advanced pixel detector development project, based at KEK, using the Silicon-On-Insulator (SOI) technology [33, 34]. With this technology, complex CMOS circuits can be implemented in each pixel, so that intelligent data handling such as sparse data scan, pipelining, clustering etc. become possible. In addition, in the SOI pixel detector, all the peripheral circuits can be integrated within the chip, so much finer segmentation becomes possible without increasing materials. This will lead to a higher data acquisition bandwidth and a faster frame rate.
The basic structure of the SOI pixel detector is shown in Fig. 4.43. It uses bonded SOI wafers: the bottom wafer is high-resistivity Si and works as radiation sensor, and CMOS circuits are implemented in the top wafer. The process is being developed based on the 0.2 $\mu$m CMOS Fully-Depleted SOI process of OKI Semiconductor Co. Ltd. A major issue in the use of the SOI material as a radiation sensor is the back gate effect. The potential applied to the sensor part will affect the transistor characteristics since both are in close vicinity. This issue was recently solved by introducing a Buried p-Well (BPW) process. Figure 4.44 shows the transistor drain current and gate voltage characteristics variation due to the back side voltage. By adding the BPW layer under the transistors, the back gate effect is completely suppressed.
The SOI pixel detector development is conducted through Multi Project Wafer (MPW) submissions, which include many designs in a mask. Although a detailed readout scheme is not yet determined, we designed a test chip (called SBPIX1) for the Belle II experiment. The block diagram of the chip and an image taken with laser light are shown in Fig. 4.45. We are pursuing a possible upgrade with the SOI detector about 5 years after Belle II startup when the luminosity will have reached its maximum.

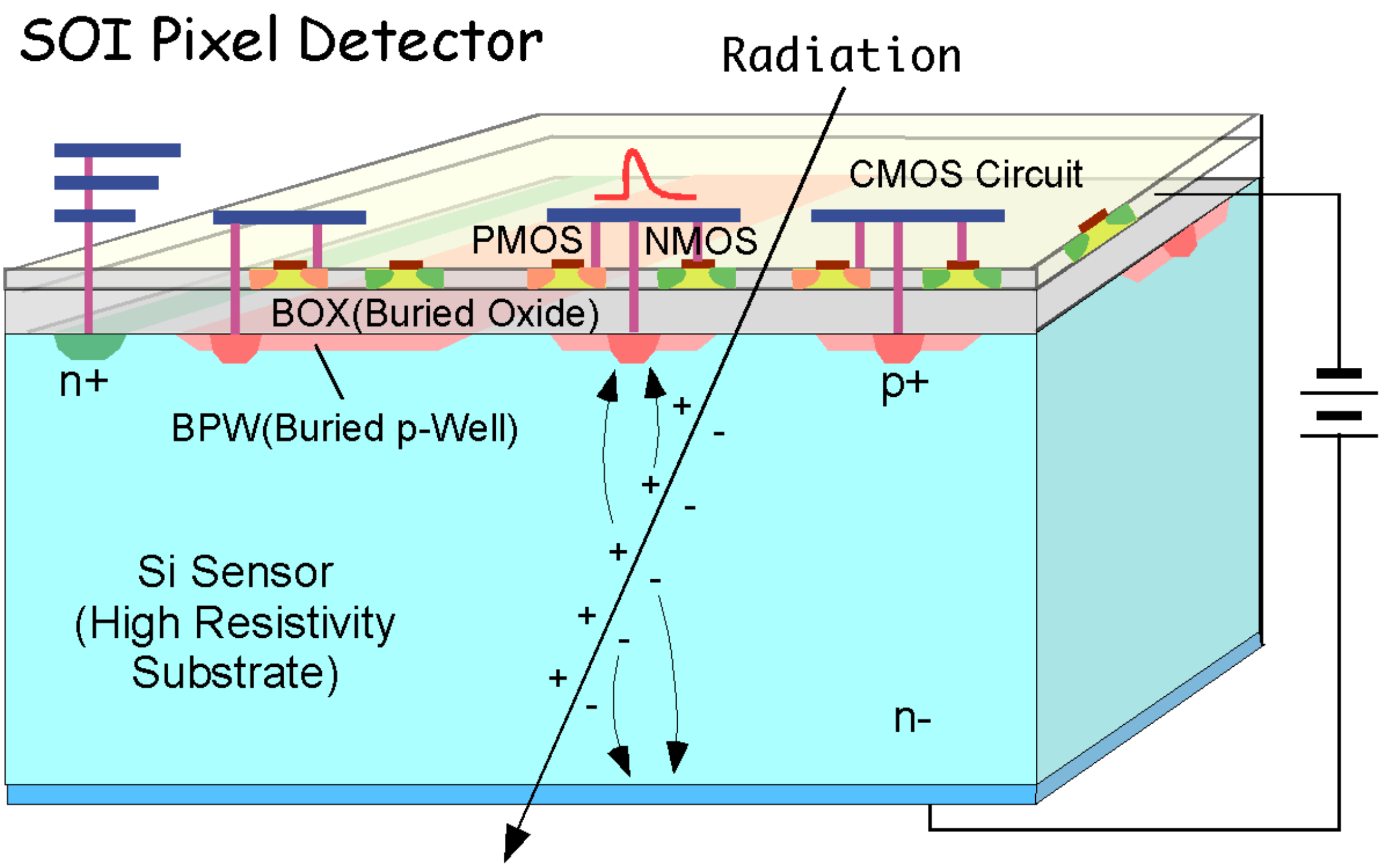

*Figure 4.43: The basic structure of the SOI pixel detector.*

## 4.13 Summary

The Pixel Detector of Belle II will be optimized for precision vertex reconstruction of B-meson decays. The DEPFET sensors used in the pixel detector are monolithic all-silicon sensors without need of additional support and cooling material in the active region of the detector. The sensors are thinned to 75 $\mu$m resulting in a radiation length of about 0.19% $X_0$. This is possible due the low noise offered by the DEPFET technology allowing a high signal to noise ratio even with thin a sensor, and the low power density of 0.1 W/cm$^2$ of the matrix operated in a rolling shutter readout mode. The location close to the beam, 14 mm, results in an extremely high background rate and considerable radiation damage. The first is addressed by a fine granularity, with a pixel size of 50 $\mu$m $\times$ 50 $\mu$m and a high readout rate of 50 kHz, which should limit the background occupancy to $1 - 2\%$. The DEPFET and the associated ASIC chips will be engineered to withstand up to 10 MRad without significant deterioration of the performance, which should allow operation in Belle II for a minimum of five years. The high background occupancy results in an extremely high data volume. The information of the outer tracking devices is used to filter the PXD data in special processors, reducing the data volume to a manageable level without loss of efficiency.
In summary, the Belle II PXD will be one of the most advanced pixel detectors ever developed for a particle physics experiment.

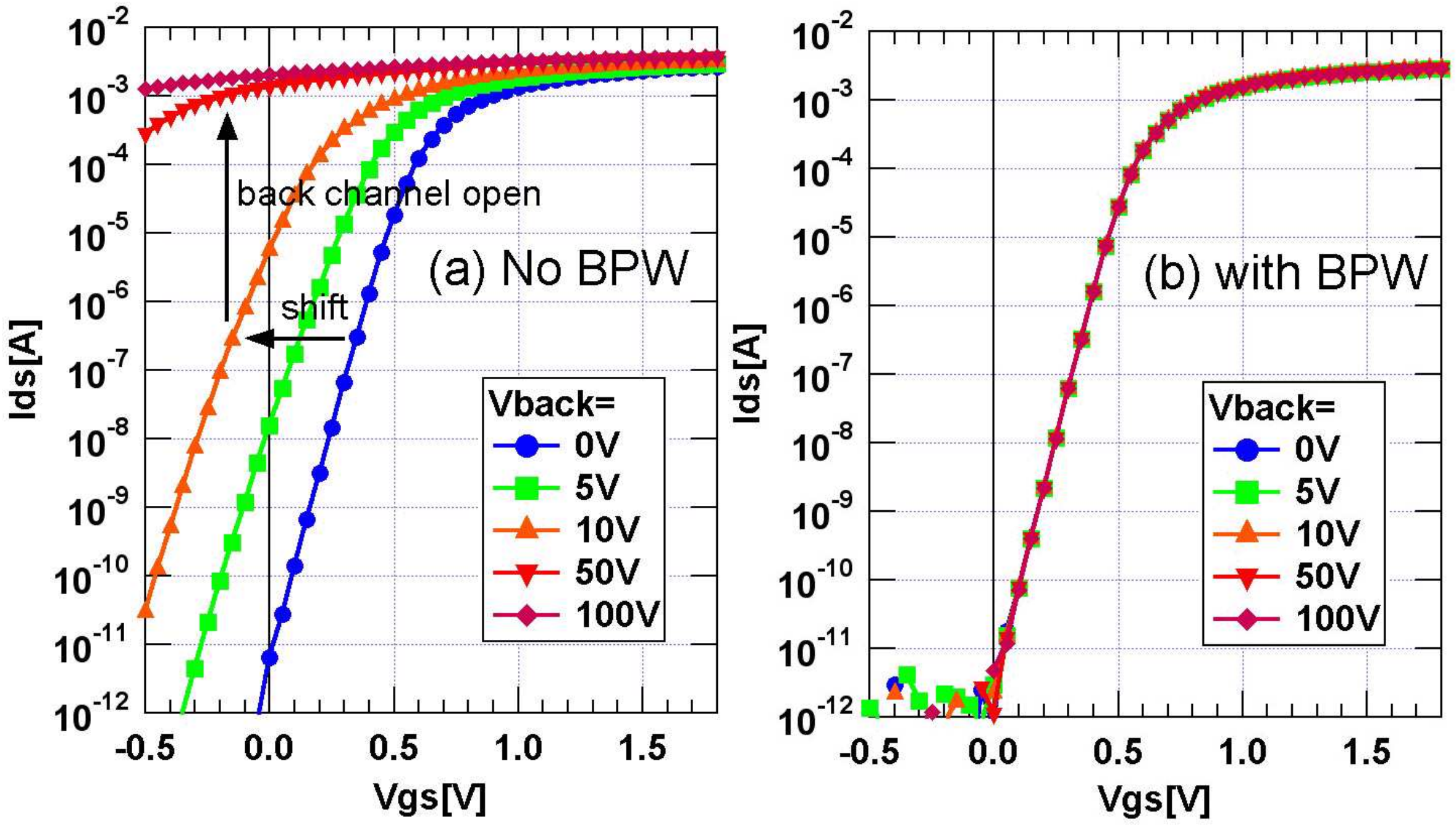

*Figure 4.44: The transistor drain current and gate voltage characteristics variation due to the back side voltage.*

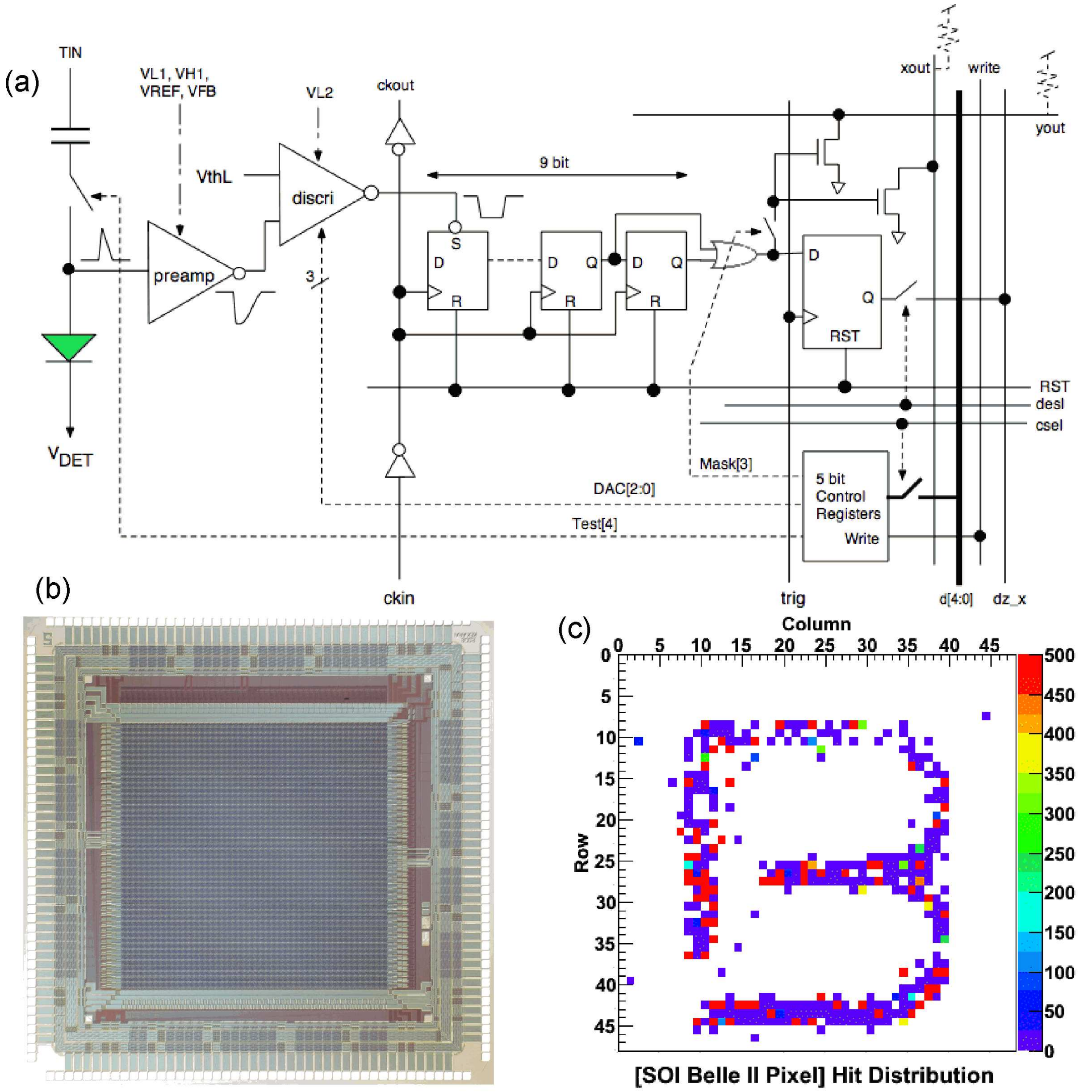

*Figure 4.45: The block diagram of the chip and an image taken with laser light.*

# Chapter 5

# Silicon Vertex Detector (SVD)

## 5.1 Introduction

The main purpose of the Belle II SVD,[1] together with the PXD (Ch. 4) and CDC (Ch. 6) is to measure the two $B$ decay vertices for the measurement of mixing-induced $CP$ asymmetry. In addition, the SVD measures vertex information in other decay channels involving $D$-meson and $\tau$-lepton decays.
The design of the Belle II SVD inherits the good characteristics of the Belle vertex detector: low mass, high precision, immunity to background hits, radiation tolerance and long-term stability. It is designed with silicon strip sensors to avoid the huge channel count of pixels without compromising the vertex-detection capability of Belle II.

### 5.1.1 Vertexing in the Belle II environment

As explained in Ch. 2, SuperKEKB uses a low-emittance collision scheme with electron and positron beams colliding at 7 GeV and 4 GeV, respectively. The Lorentz boost factor of the center-of-mass system is $\beta\gamma = 0.28$, about two-thirds of that in Belle; this results in less separation between the $B$ vertices in an event. However, the beam pipe at the interaction region has a radius of 10 mm, two-thirds of that in KEKB, so Belle II's vertex measurement performance with the SVD and PXD is as good as or better than in Belle.
Due to the higher beam current and luminosity, the detectors are required to operate at up to 30 times larger beam background and up to 30 kHz trigger rate.

### 5.1.2 Requirements and Constraints

The main characteristics of the Belle II SVD are the following:

- It covers the full Belle II angular acceptance of $17° < \theta < 150°$. The inner radius is 38 mm and outer radius is 140 mm; these are determined by the radii of the PXD and CDC.
- It improves the quality of the reconstruction of charged tracks compared to Belle's SVD2.
- It provides data to extrapolate the tracks reconstructed in the CDC to the PXD with high efficiency; this is the strategy we employ to determine a vertex.

[1] This nomenclature is chosen to avoid confusion with the historic SVD1 and SVD2 as well as the SVD2.5 and SVD3 projects of the Belle experiment.

- From the Belle experience, the hit occupancy[2] should be less than 10% in order to assure the association of correct hits in the vertex detector with tracks reconstructed in the drift chamber. Therefore, a proper combination of sensor geometry and readout scheme is necessary. In addition, information from the Belle II SVD will be very useful in eliminating background hits in the PXD.
- In combination with the PXD, it is able to reconstruct low-$p_t$ tracks, down to a few tens of MeV/c, that do not leave enough (or any) hits in the CDC. This is particularly important for the efficient reconstruction of the $D^*$ daughters that tag the flavor of the parent $B$ meson.
- It is able to reconstruct $K_S$ mesons that decay outside of the PXD volume. This is important for channels such as $B \to K^*\gamma$ or $B \to K_S K_S K_S$ where the only charged tracks are the $K_S$ daughter pions.
- It operates efficiently and with very low dead-time in the high beam-background and trigger-rate environment of SuperKEKB and Belle II. The background rate may be 30 times higher that in KEKB, and the expected maximum average trigger rate is 30 kHz.
- It is mechanically stable, has low mass, and operates reliably.
- Its heat load is dissipated efficiently by its cooling system.
- Its associated software environment robustly maintains peak performance through ongoing and frequent calibrations, alignments, and monitoring.

### 5.1.3 Belle SVD2 Limitations

In the SuperKEKB environment, the estimated occupancy of the SVD2 is shown—with large uncertainty—in Fig. 5.1 as a function of the sensor radius. Following the SVD2 design, a sensor size of $2.5 \times 7.5\,\mathrm{cm}^2$ and VA1TA [1] readout is assumed here, where the VA1TA is operated with 800 ns shaping time. It is obvious that the hit occupancy is too high by at least a factor of ten. For this reason, the Belle II SVD's innermost sensor layer is placed at a radius of 3.8 cm. The background level at radii as large as $\sim 10\,\mathrm{cm}$ is estimated to be too high for a conventional drift chamber, so the outermost sensor layer is at a radius of 14 cm.

### 5.1.4 Approach

To fulfill the above requirements, our design consists of four layers of double-sided silicon strip detectors (DSSDs) fabricated from six-inch wafers. It would be prohibitive in cost and channel count to instrument this volume with pixels.

To suppress the background hits, a readout chip with a fast shaping time of $\mathcal{O}(50\,\mathrm{ns})$ is indispensable. In 2003, an assessment of readout ASICs was done [2], and the APV25 [3], which was originally developed for the CMS silicon tracker, was chosen for the future vertex detector readout. The APV25 chip (Fig. 5.2) consists of 128 identical channels of low-noise preamplifiers followed by a shaper stage (50 ns nominal peaking time, tunable). Each channel has a 192-cell deep analog pipeline (ring buffer) with an index FIFO that can label up to 32 cells pending for subsequent readout, an analog pulse-shape processor (APSP) that will not be utilized here, and an analog multiplexer for the output.

[2]The occupancy is defined as the average percentage of strips with a signal above threshold at any trigger.

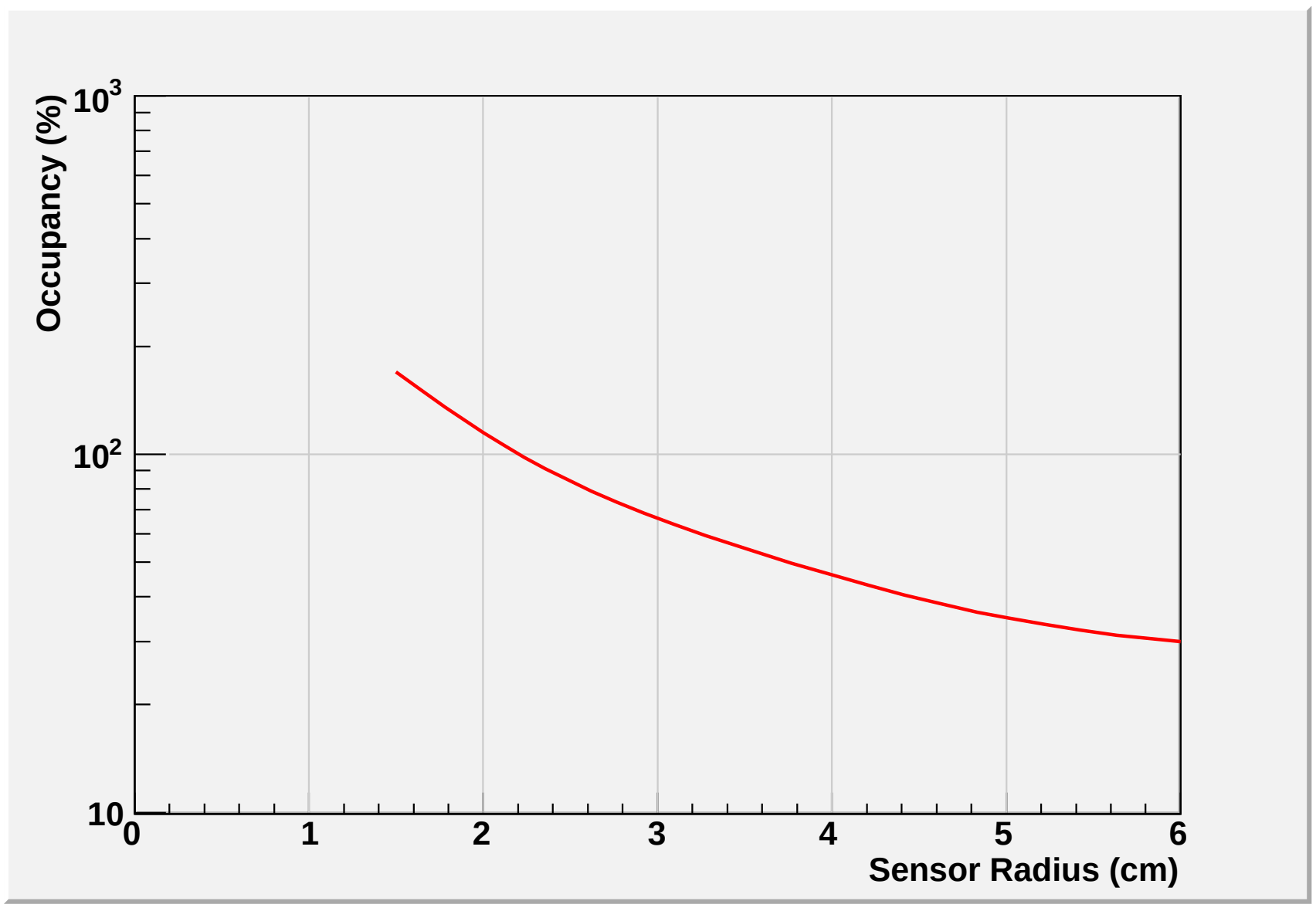

*Figure 5.1: Hit occupancy of the SVD2 expected with SuperKEKB.*

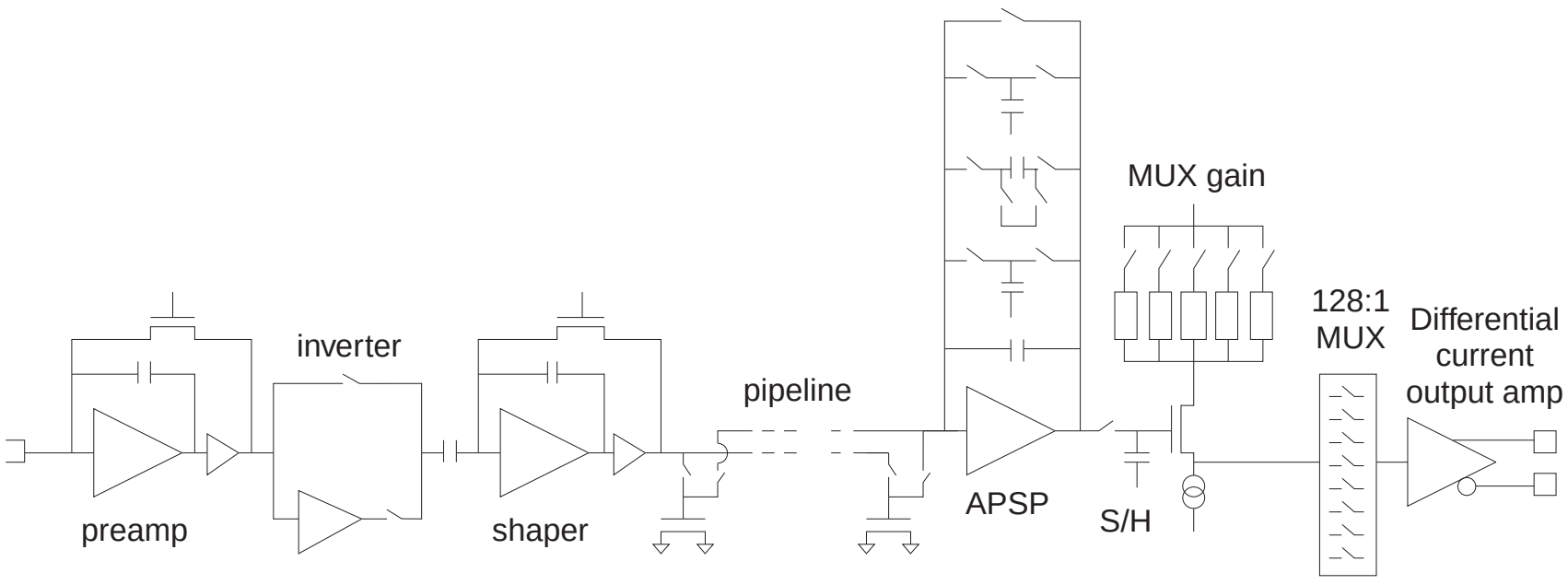

*Figure 5.2: Building blocks of one of the 128 channels of the APV25 front-end readout chip.*

In addition to the short shaping time, the operation of the pipeline and the analog readout scheme were found to be fully compatible with Belle trigger and data acquisition scheme. The APV25 chip is radiation hard up to an integrated dose in excess of 30 MRad. Intensive evaluation has been done and the results demonstrate that, even now, the APV25 is a perfectly suitable chip for the Belle II SVD.

Progress has been made regarding the DSSD sensor fabrication technology. In the LOI (2003) design, the DSSDs were assumed to be produced from four-inch wafers. Now, the six-inch wafer is standard and will be used here. This reduces the overall amount of structural elements and thus the material budget.

Due to the fast shaping and noise considerations, the input capacitance for the APV25 chips must be minimized. To achieve this, APV25 hybrids will be mounted on the DSSDs. After the optimization of hybrid and ladder structure, the material budget per DSSD ladder is 0.57% of radiation length, where the DSSD sensor itself contributes 0.32%.

### 5.1.5 Detector Layout

The Belle II SVD is composed of four layers, with sensor radii listed in Table 5.1. The geometry is mainly driven by the achievable sensor size in six-inch wafer technology (Sec. 5.2) and how to fit integer multiples thereof within the acceptance region.

| **Layer** | **Radius** (mm) | **Ladders** | **Sensors /ladder** | **Sensors** | **RO chips /sensor** | **RO chips** |
|---|---|---|---|---|---|---|
| 6 | 140 | 17 | 5 | 85 | 10 | 850 |
| 5 | 115 | 14 | 4 | 56 | 10 | 560 |
| 4 | 80 | 10 | 3 | 30 | 10 | 300 |
| 3 | 38 | 8 | 2 | 16 | 12 | 192 |
| Sum | | 49 | | 187 | | 1902 |

*Table 5.1: Geometrical properties of the Belle II SVD. As the sensors are flat, the radii given here are the minimum distances from the beam axis. "RO chips" stands for the number of APV25 readout chips. For coherence with the PXD layer numbering scheme, the innermost strip layer here is numbered 3.*

The polar angular acceptance ranges from 17° to 150°, which is asymmetric to account for the forward boost of the center-of-mass frame. The radial coverage is almost double that of the Belle SVD2 and hence would require a significantly increased number of wafers in a traditional cylindrical geometry. To avoid this, slanted sensors are used in the forward region, resulting in a lantern shape (Fig. 5.3).

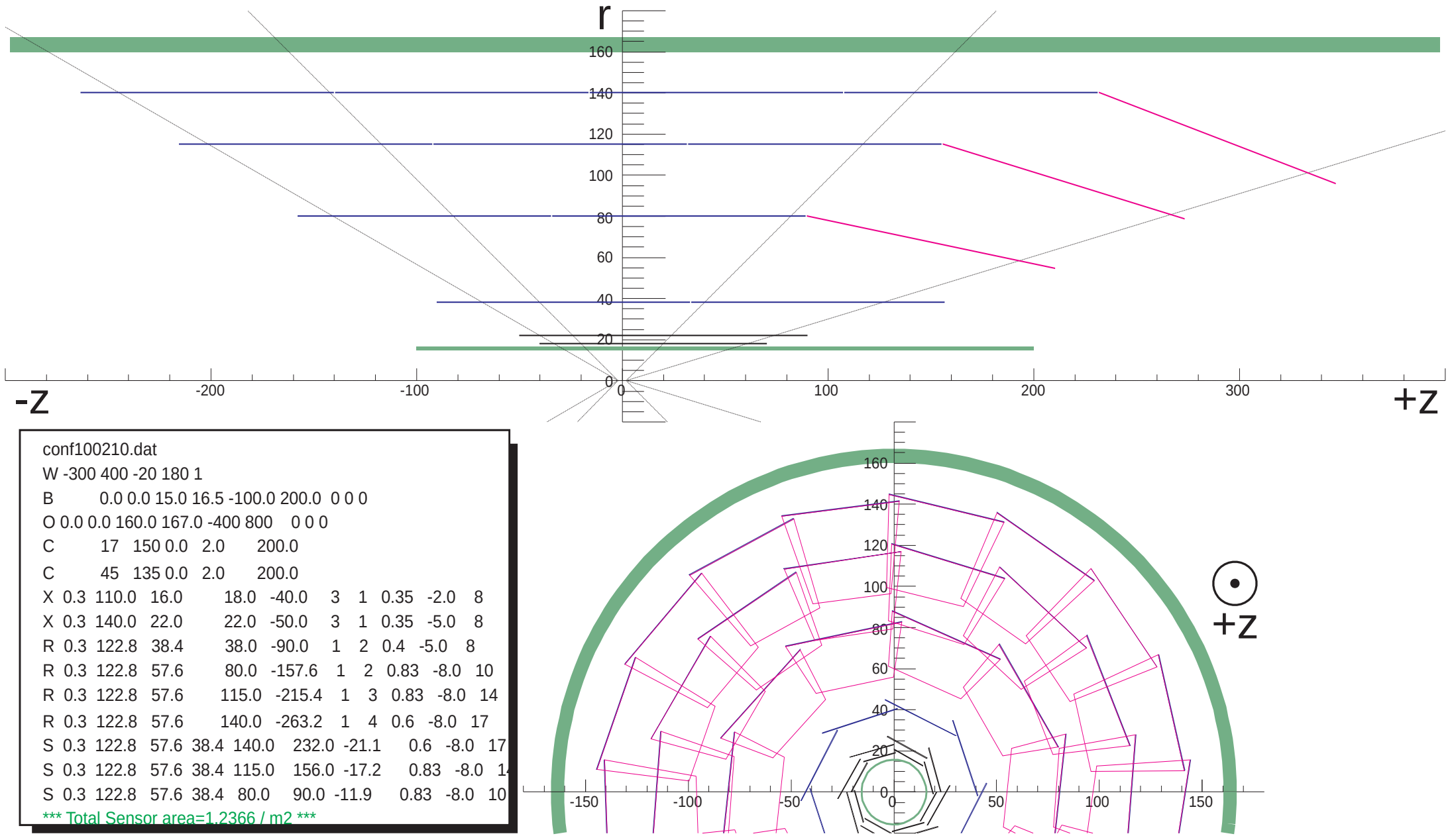

*Figure 5.3: Configuration of the four strip layers, with slanted sensors in the forward region, and the two PXD layers. All dimensions are in mm.*

The overlap between adjacent sensors accounts for about 8 to 10% of the sensor area (depending on the layer). Due to the relatively low energy of the machine, most particle tracks originating from collisions will be deflected by multiple scattering and thus cannot be used for high precision alignment, which is only available from highly energetic cosmic muons. As their rate is limited, we prefer to have a rather comfortable overlap to facilitate alignment, even though this slightly increases the overall material budget.

The slanted forward region requires sensors of trapezoidal shape, and only one design is foreseen to be used in all layers. The slant angle varies in such a way that the non-parallel strips of those sensors all intersect with the beam pipe. In reality, this condition cannot exactly be met due to the windmill structure (see below).

While this design is nearly final, there are a few minor issues to be optimized. In the outermost layer, 17 ladders would be sufficient for full circular coverage and sufficient overlap, but an even number might be beneficial from the mechanics point of view; a similar consideration applies to the innermost layer (Sec. 5.3). Furthermore, there is still a degree of freedom in how to rotate the four layers with respect to each other. In Fig. 5.3, all layers are arranged in order to have a common overlap at the 12 o'clock position. One may argue that this is less than optimal for the material budget; on the other hand, the overlap occurs only in very narrow azimuthal angular range and, in fact, can help a great deal in the alignment with cosmic muons (hence the vertical position).

All rectangular silicon sensors are double-sided with the long strips on the $p$-side, parallel to and facing the beam axis ($z$). The short $n$-side strips along $r$–$\phi$ are located on the sensor face towards the outside. The slanted sensors are similar, except that the long strips are obviously not parallel to $z$ anymore, but (almost) intersect at the beam axis. Figure 5.4 shows the dimensions of the three different sensor types together with their pitches on the $n$-side (top) and $p$-side (bottom), respectively, and the number of readout chips needed for each sensor. All DSSDs will be connected to APV25 front-end chips (Sec. 5.3), reading out 128 strips each.

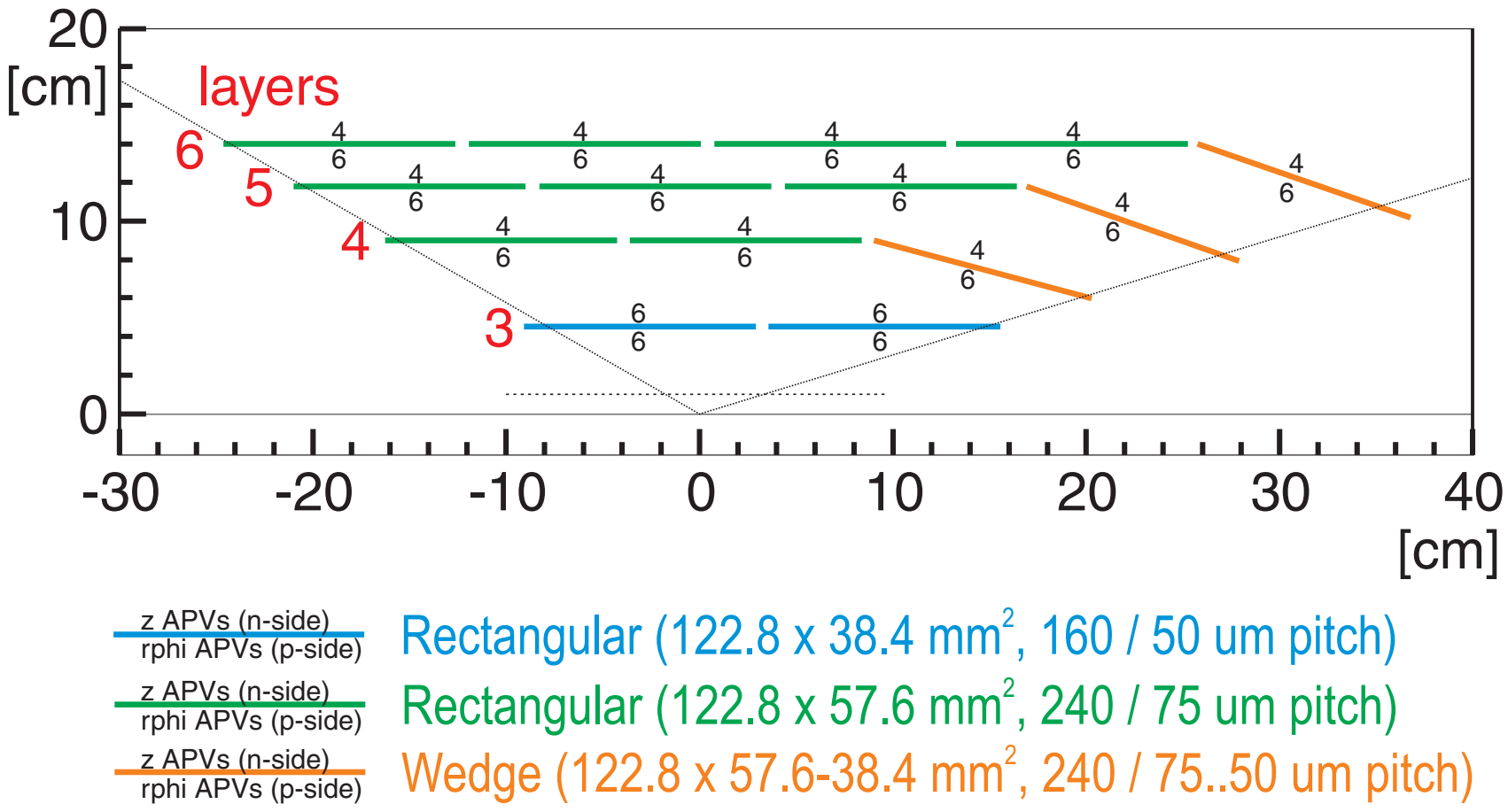

*Figure 5.4: Schematic configuration showing the three different sensor geometries and the number of APV25 readout chips for each detector.*

The solenoid's central magnetic field vector $\vec{B}$ coincides with the $z$ axis. This leads to a Lorentz force on electrons and holes in the sensor, as shown in the left half of Fig. 5.5. Consequently, the sensors should be tilted in such a way that the spread is minimized. According to Ramo's theorem [4], the motion of both carrier types (electrons and holes) induces the same currents in both electrodes ($p$- and $n$-sides). Even though only holes spread out over the segmented $p$-side strips, electrons moving towards the $n$-side induce image currents in the $p$-side strips. Due to their higher mobility, the Hall angle and thus deflection of electrons is about three times larger than that of holes. Consequently, the tilt of the sensor plane should reduce the electron spread, at the cost of an increased hole spread (right half of Fig. 5.5). This conclusion has been confirmed by a numerical simulation based on Ramo's theorem.

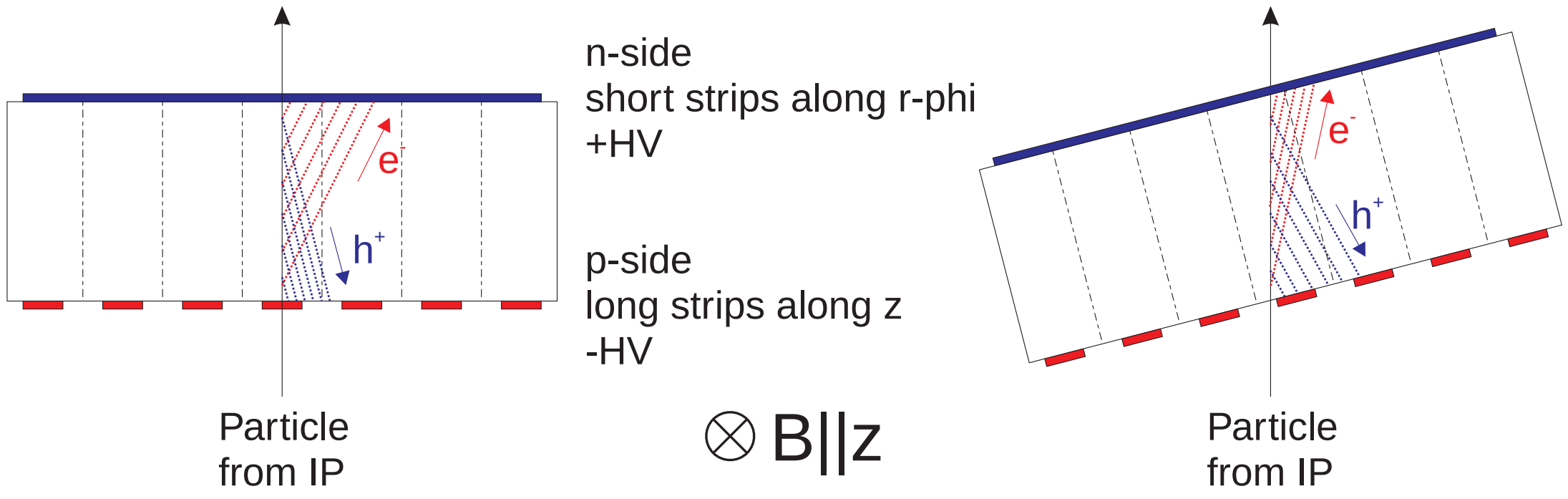

*Figure 5.5: Sensor geometry and magnetic field. Left: In the presence of a magnetic field, electrons and holes are deflected by the Lorentz force. Right: The overall charge spread can be minimized by a tilt that reduces the electron deflection.*

The Origami chip-on-sensor concept (Sec. 5.3.1.1), which will be applied to all sensors except those located at the edge of the acceptance, uses pitch adapters bent around the sensor edge. These bent fanout pieces can only be applied at the outer sides in a windmill structure, as shown schematically in Fig. 5.6. Together with the readout direction, this unambiguously determines the layout of each Origami hybrid (Sec. 5.4.2).

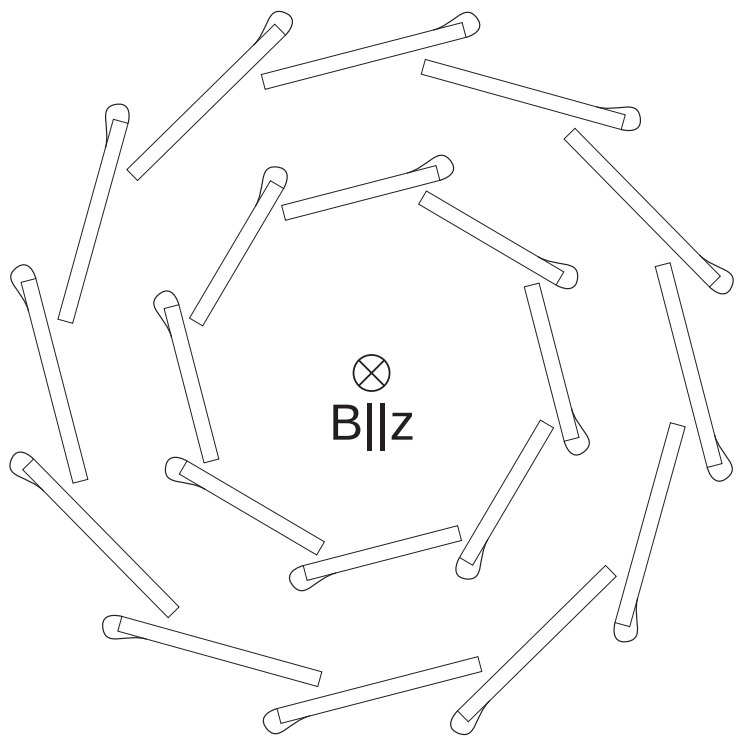

*Figure 5.6: Schematic cross-section of a windmill structure with Origami modules. The flex pitch adapters, which are bent around the sensor edges, can only be placed on the outer edge of each sensor.*

Further details on the mechanical layout and the cooling pipes attached to the Origami modules are provided in Sec. 5.3.

### 5.1.6 Readout Chain

The electronic readout chain of the Belle II SVD is based on previous developments for DSSD readout with the APV25 chip: the APVDAQ system (2005) and its successor, the SVD3 prototype readout system (2007). The latter, described in Sec. 5.7.2, was intended for a partial upgrade of the SVD2 that was never implemented. Our design builds on it, with a few changes to accommodate the significantly larger number of channels and radiation issues.
Figure 5.7 symbolically shows the electronics chain. APV25 chips read out the silicon sensors and their analog output is transmitted to junction boxes placed about 2 m away, at the outer walls of the CDC, approximately where the DOCK boxes were located for the SVD2 [5]. The main function of those junction boxes is to avoid having 10 m long cables directly attached to the Belle II SVD assembly during installation and to detach the short cables that move with the SVD from the long and permanently installed cables that lead to the outside.

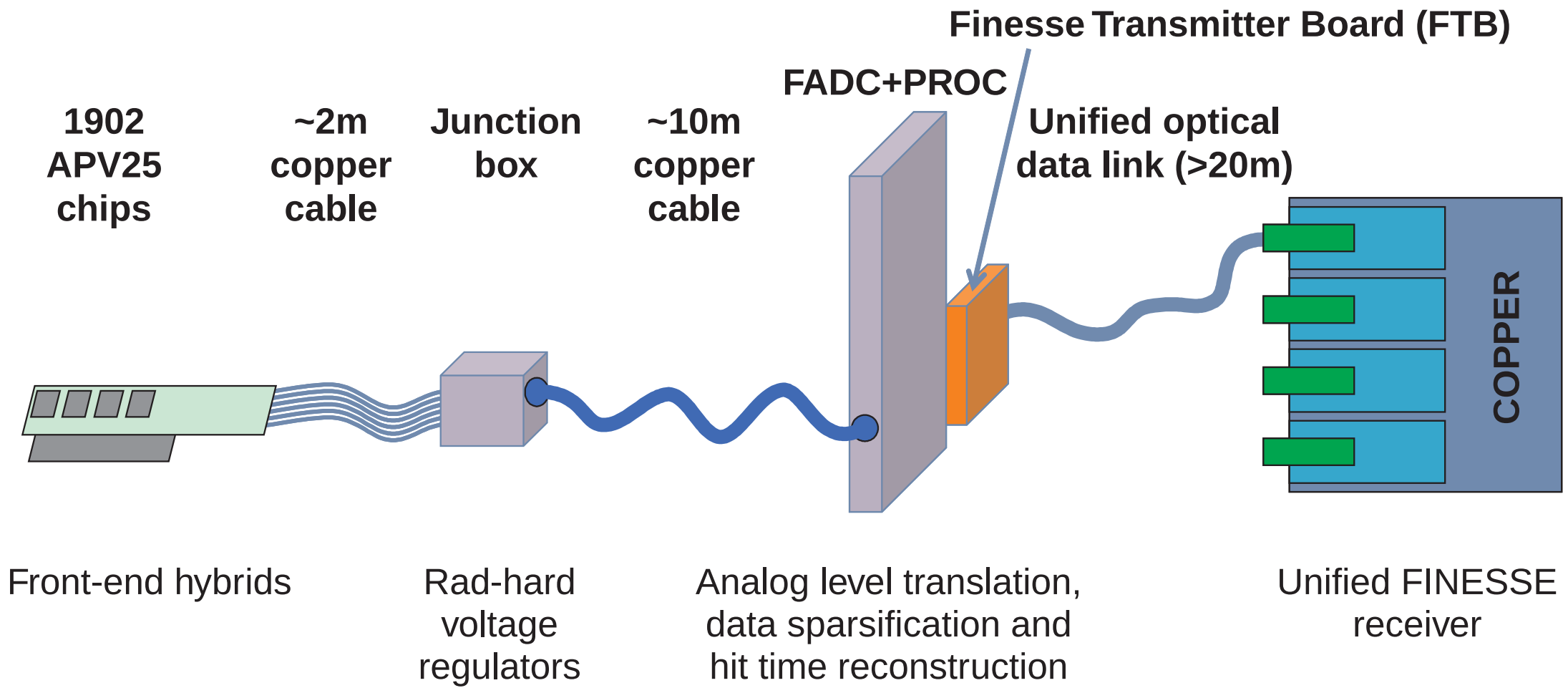

*Figure 5.7: Schematic view of the Belle II SVD readout chain.*

In early 2009, several dosimeters were installed around the forward DOCK boxes and an integrated dose rate of about 5 kRad/ab$^{-1}$ was measured. Extrapolation to SuperKEKB and Belle II is not trivial but, naively, we expect 250 kRad if we simply scale by luminosity to 50 ab$^{-1}$. This is a deadly dose for most commercial off-the-shelf components. Fortunately, the APV25 chip can easily drive long cables: this was positively tested with 12 m and there was no indication whatsoever that this was close to the limit. Consequently, we will push the repeater part farther down the chain and integrate it into the FADC+PROC (Flash Analog-to-Digital Converter and Processing) modules.
Apart from the cable patch panel, the junction box will also contain voltage regulators for the front-end hybrids to protect them from transient overvoltages; these can appear on long cables with remote sensing in case of a sudden load drop that would occur, for example, when the APV25 chips are reset. We intend to use RHFL4913 regulators by ST, which were developed for the ATLAS experiment and were successfully tested up to 100 MRad (Sec. 5.4.3).
The FADC+PROC units will be implemented as 9U VME boards and contain the analog level translation part, flash ADCs and one or more FPGAs that handle the APV25 data processing, including data reordering, pedestal subtraction, a two-pass common mode correction, zero suppression (sparsification) and hit time finding. Apart from the last item, these data processing functions are already implemented in the SVD3 readout system and were verified in the lab as well as in beam tests. Further details are given in Sec. 5.4.4.
Finally, the processed data will be propagated to Finesse Transmitter Boards (FTB), which send the information to the COPPER system using unified optical data links. In parallel, a replica optical output will be implemented to make the data available for online data reduction of the PXD system. The destination of that second link may either be the FPGA-based Giessen Box or a computer system based on CPUs or GPUs (Ch. 4). More details on the data links are discussed in Sec. 5.4.5.
Each FADC+PROC unit handles the input of 24 APV25 chips (which may either be six hybrids with four chips each or vice versa) and thus we need 80 boards in total, which will completely fill up four crates (leaving one slot each for the crate controller). Due to an imbalance in readout channels on forward and backward sides, however, this is more likely to be $3 + 2 = 5$ crates in total. Clearly, the number of FADC+PROC boards also defines the number of FTBs and furthermore the number of optical links and COPPER boards. Due to the data rate limitations, a single COPPER board might not digest more than one FADC+PROC output, in which case 80 COPPER boards will be needed (Ch. 13).

## 5.2 Sensors

The design of the silicon sensors is driven by the fact that multiple scattering is the most important issue for the Belle II experiment. Thus, the largest available sensors have to be used and the design has to be double-sided to achieve a low material budget. Nowadays, typical wafer processing facilities are able to produce six-inch wafers, resulting in a maximum sensor size of approximately 12 cm length and a width of 6 cm. The sensors will be made from $n$-type bulk with high resistivity and will have a thickness of about 300 $\mu$m, which is a standard material available from industry.
Hamamatsu Photonics (HPK), the Japanese company that provided the SVD2 DSSDs on four-inch wafers, had decided some time ago to abandon the production of doubled-sided sensors. Therefore, a market survey was started to investigate alternative vendors. Discussions were started with the following companies:

- Canberra (Belgium/US)
- irst-Trento (Italy)
- Micron Semiconductor (UK)
- SINTEF (Norway)

After short negotiations, irst-Trento and Canberra dropped out, because they were not able to produce sensors with the requested specifications. More detailed discussions started with Micron Semiconductor and SINTEF with the aim to proceed to a prototype batch and the full production later on.
Meanwhile, HPK announced that it would restart the production of DSSDs. A test production was launched with HPK for the Belle II SVD barrel sensors, while Micron Semiconductor will produce some slanted (trapezoidal) sensors for the forward part.
It is commonly agreed that the sensors for both regions need to be AC-coupled with integrated bias resistors. The designs of barrel and forward parts will be described in the next two sections.

### 5.2.1 Barrel Sensors

Two different rectangular sensors for the barrel part are produced by HPK on 320 $\mu$m thick six-inch wafers. As can be seen in Fig. 5.4, the small sensors used in layer 3 differ from the large ones used in layers 4 to 6 in both pitch and number of channels. Figure 5.8 shows the outer dimensions of the sensors (dark blue) and the dimensions of the active area (light blue, not to scale, approximated values). Note that the small rectangular sensors have not yet been ordered, so the values of active and total areas are extrapolated, awaiting confirmation by the vendor.
The large sensor will be used with conventional readout hybrids as well as with special hybrids for chip-on-sensor readout (Sec. 5.3.1.1). To this end, an additional row of readout pads (red) is foreseen on the $p$-side.
Table 5.2 states the main sensor parameters. The electrical parameters, given in Table 5.3, are valid for both the small and the large sensors.

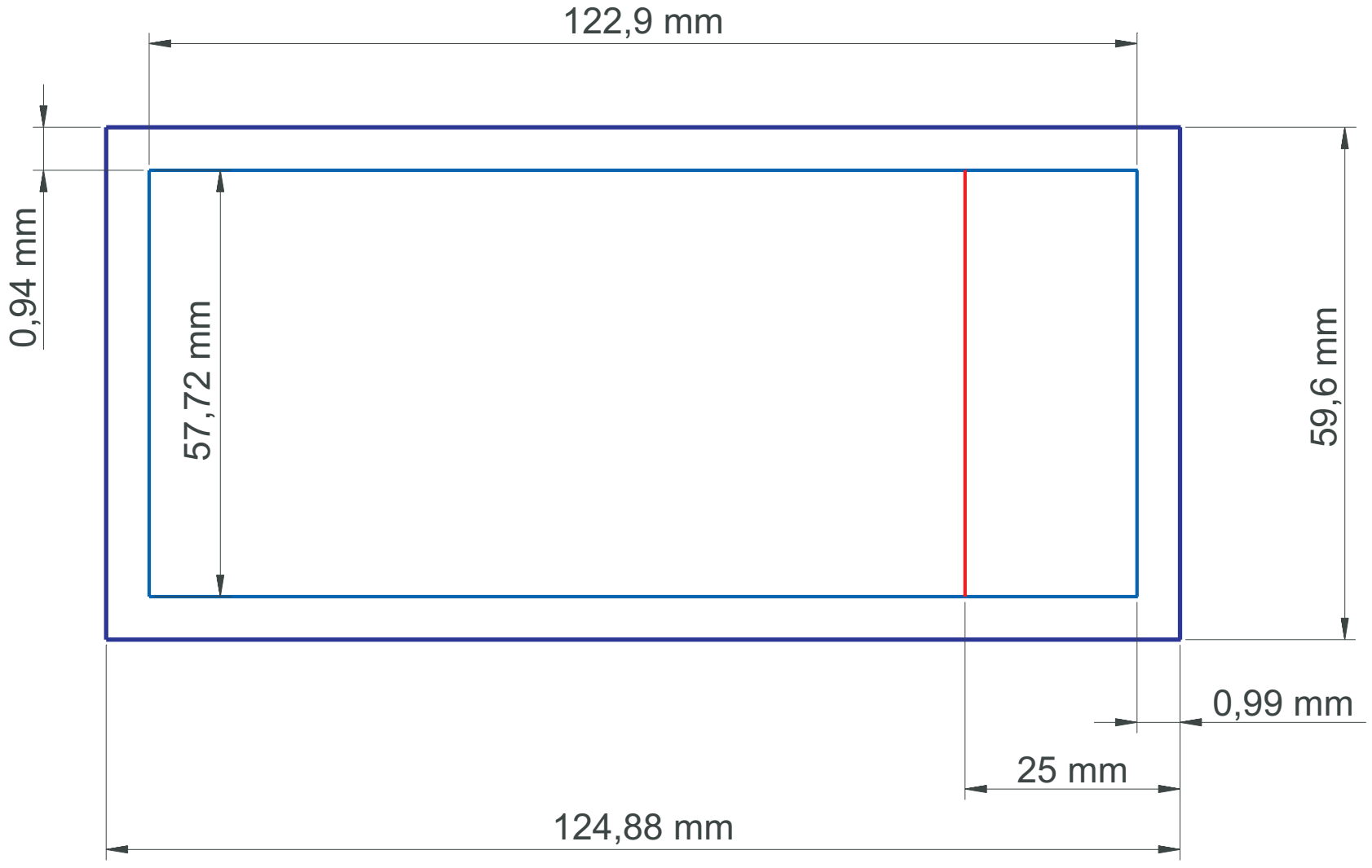

*Figure 5.8: Geometric dimensions (not to scale) of the rectangular sensor for layers 4 to 6. Dark blue: outer dimensions. Light blue: active area. Red: additional pad row.*

| **Quantity** | **Large sensor** | **Small sensor** |
|---|---|---|
| # strips $p$-side | 768 | 768 |
| # strips $n$-side | 512 | 768 |
| # intermediate strips $p$-side | 767 | 767 |
| # intermediate strips $n$-side | 511 | 767 |
| Pitch $p$-side | 75 $\mu$m | 50 $\mu$m |
| Pitch $n$-side | 240 $\mu$m | 160 $\mu$m |
| Area (total) | 7442.85 mm$^2$ | 5048.90 mm$^2$ |
| Area (active) | 7029.88 mm$^2$ (94.5%) | 4737.80 mm$^2$ (93.8%) |

*Table 5.2: Basic parameters of the rectangular sensors.*

| Quantity | Value |
|---|---|
| Base material | Si $n$-type 8 kΩcm |
| Full depletion voltage FD | $< 120$ V |
| Breakdown voltage | $\geq$ FD + 50 V |
| Polysilicon resistor | 4 MΩ (min.), 10 MΩ (typ.) |
| Coupling capacitance | $> 100$ pF |
| Breakdown voltage of AC coupling | $> 20$ V |
| Bias leak current at FD | 1 $\mu$A (typ.), 10 $\mu$A (max.) |

*Table 5.3: Electrical parameters of the rectangular sensors.*

### 5.2.2 Forward Sensors

To cover the forward acceptance region while minimizing the material, the sensors are slanted to form a conical shape. Therefore, the individual sensors are trapezoidal.
Figure 5.9 shows the outer dimensions of the sensor (dark blue) and the dimensions of the active area (light blue, not to scale, approximated values). Tables 5.4 and 5.5 give the sensor parameters.
The strips of the $p$-side (junction side) feature a variable pitch, giving them a fan shape. The strips of the $n$-side (Ohmic side) are parallel and perpendicular to the central strip of the $p$-side. Electrical strip separation is achieved by $p$-stop blocking technique [6], where a combination between the conventional method and the atoll method is used (Fig. 5.10).
The first batch of sensors is manufactured in Sussex, England, at Micron Semiconductor Limited, on 300 $\mu$m thick six-inch wafers. To comply with their design rules, the design features a 10-fold multi-guard-ring structure on both $p$- and $n$-sides.

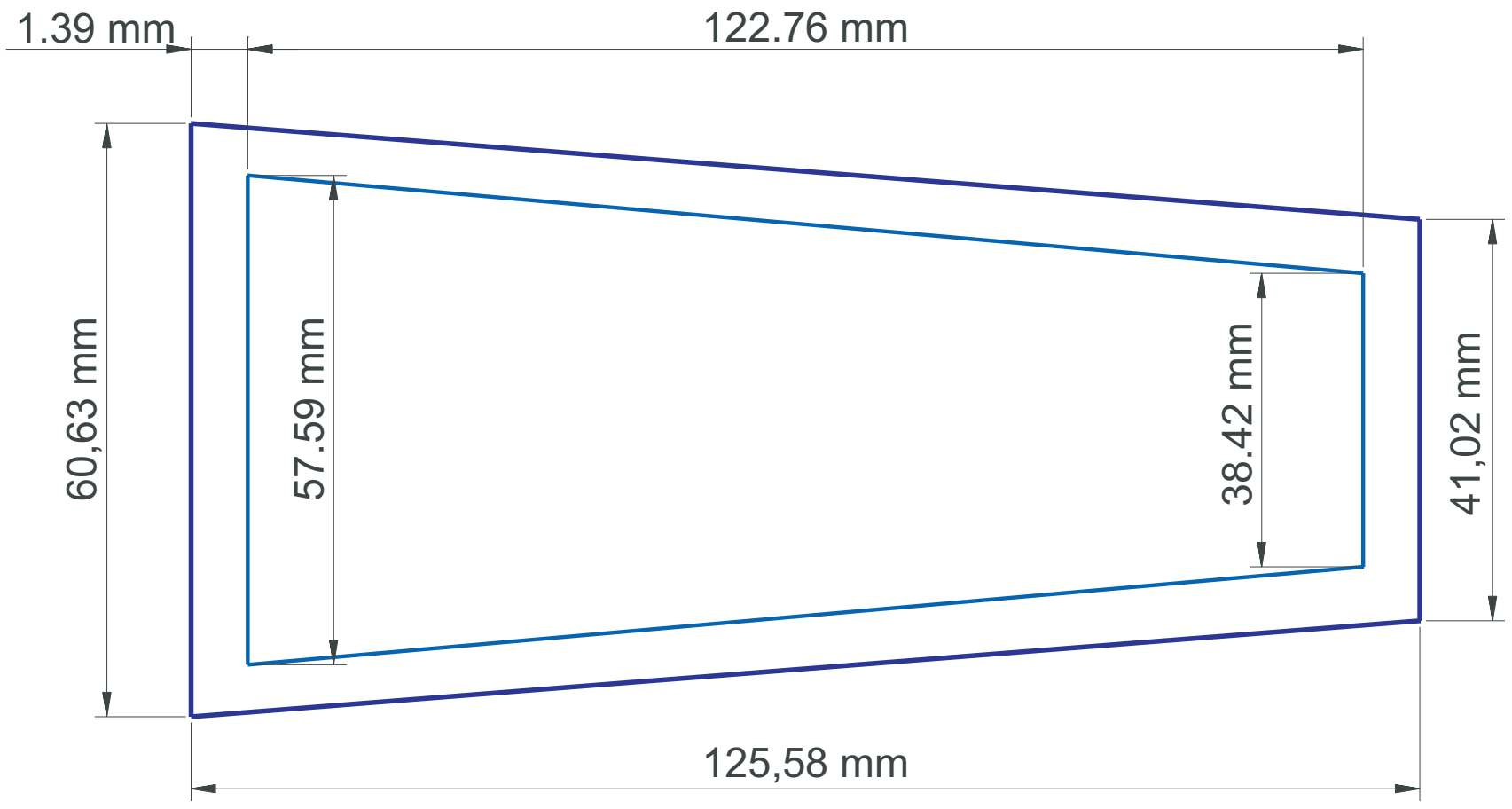

*Figure 5.9: Geometric dimensions (not to scale) of the trapezoidal sensor. Dark blue: outer dimensions. Light blue: active area.*

### 5.2.3 Radiation

Radiation damage on a silicon device by photons is dominated by surface effects rather than bulk damage. This results in the localization of the radiation defects particularly in the $Si-SiO_2$ interface. In contrast to irradiation with hadrons, no type inversion of the silicon bulk material is expected. To avoid charge-up effects on bare sensors, the DSSDs need to be continuously biased and read out during irradiation tests.

| Quantity | Value |
|---|---|
| # strips $p$-side | 768 |
| # strips $n$-side | 512 |
| # intermediate strips $p$-side | 767 |
| # intermediate strips $n$-side | 511 |
| Pitch $p$-side | $75 \ldots 50\,\mu$m |
| Pitch $n$-side | $240\mu m$ |
| Area (total) | $6382.6\,\mathrm{mm}^2$ |
| Area (active) | $5890\,\mathrm{mm}^2$ (92.3%) |
| Slant angles | Layer 6: 21.1°<br>Layer 5: 17.2°<br>Layer 4: 11.9° |

*Table 5.4: Basic parameters of the trapezoidal sensor.*

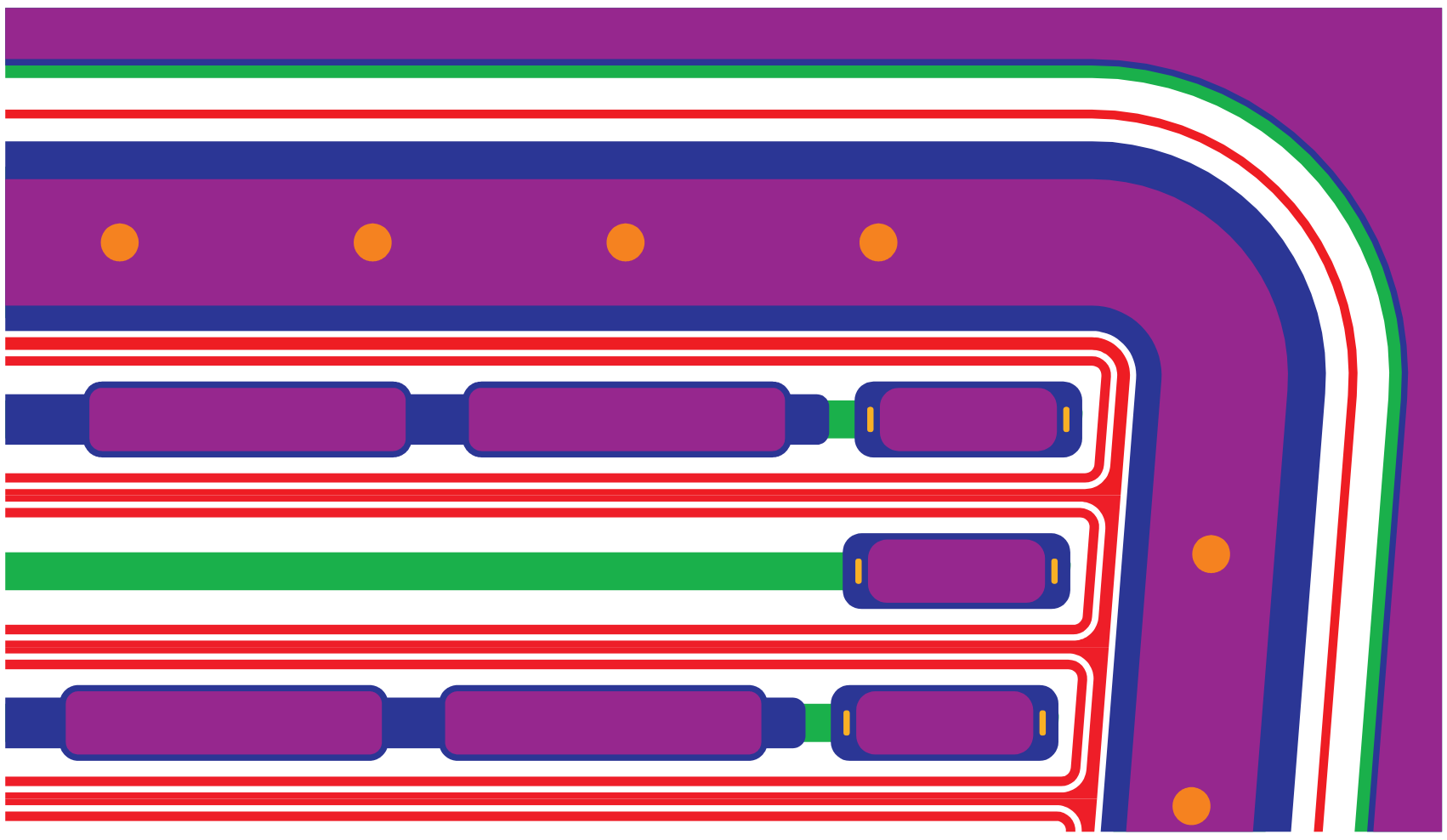

*Figure 5.10: Combined $p$-stop structure for electrical strip separation on the $n$-side.*

| Quantity | Value |
|---|---|
| Base material | Si $n$-type 8 kΩcm |
| Full depletion voltage FD | 40 V (typ.), 70 V (max.) |
| Operation voltage | FD . . . 2 × FD |
| Breakdown voltage | ≥ 2.5 × FD |
| Polysilicon resistor | 10 MΩ (min.), 15 ± 5 MΩ (max.) |
| Interstrip resistance, $p$-side | 100 MΩ (min.), 1 GΩ (typ.) |
| Interstrip resistance, $n$-side | 10 MΩ (min.), 100 MΩ (typ.) |

*Table 5.5: Electrical parameters of the trapezoidal sensor.*

In the SVD2, the radiation dose of the silicon sensors can be estimated using RadFET radiation monitors, which are positioned nearby [7]. From Fig. 5.11, one can extract a total dose of approximately 55 krad for layers 2 and 3, which corresponds to the innermost DSSD layer of the Belle II SVD. Taking some uncertainties about the position of the RadFETs in respect to the DSSDs into account, the dose is estimated to be approximately 90 krad per $ab^{-1}$ of integrated luminosity. Therefore, one can extrapolate the radiation dose of the innermost DSSD to approximately 4.5 MRad for the projected lifetime of Belle II (50 $\mathrm{ab}^{-1}$ integrated luminosity). Adding a safety margin, the DSSDs should tolerate a total dose of 10 MRad.

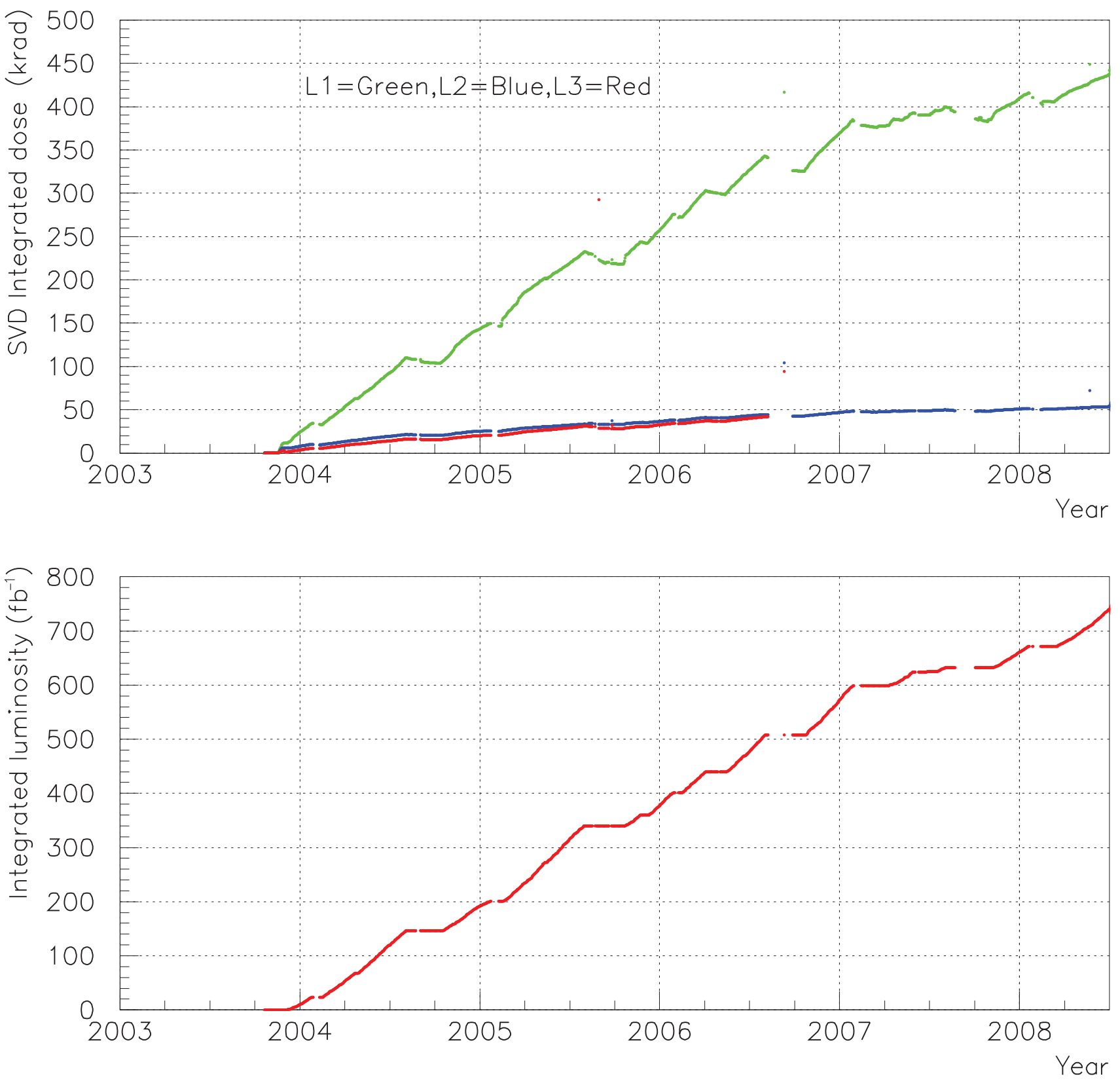

*Figure 5.11: Top: total measured dose in krad of layers 1, 2 and 3 of the SVD2. Bottom: integrated luminosity.*

Contact with a gamma irradiation facility has already be established in order to perform irradiation tests. This facility (SCK-CEN, Mol, Belgium) is equipped with a strong $^{60}$Co source, where the dose of 10 MRad can be accumulated in only four hours. This irradiation test will be performed once prototype modules equipped with sensors from the prototype batches become available.

### 5.2.4 Quality Assurance

The silicon sensors will be tested thoroughly before being mounted onto ladders. The vendors must always test 100% of the sensors including strip scans. For our quality assurance, one has to distinguish between tests on prototype sensors and full production. For the first sensors from the prototype batches, 100% will be tested. The testing effort can be reduced subsequently during full production, depending on the number of defects and reliability of the vendor test results. However, we will always test at least 10% of the sensors.

For the initial production lot, each sensor will be tested first by applying a voltage ramp while measuring its dark current and total capacitance (IV and CV curves). This results in information about breakthrough voltage and full depletion voltage. The amount of dark current at a certain voltage is a measure of the quality of processing. Moreover, strip scans will be performed, where the polysilicon resistors, the coupling capacitances, the currents through the dielectric of the coupling capacitance and the dark current of every individual strip will be tested. This has to be done on each strip of both, $p$-(junction) and $n$-(ohmic), sides of the sensors.

As a next step, long-term stability tests will be performed, during which the sensors will be biased for a couple of weeks. This results in the evolution of the dark current, which should be absolutely stable with time. In addition, similar measurements will be performed during and after thermal cycle tests to simulate aging effects.

Apart from the tests of the silicon sensors itself, measurements on test structures will be performed.[3] Destructive tests, such as on the coupling oxide, can be performed to determine its breakthrough voltage.

Due to the uncertainty in radiation background, irradiation tests of the sensors have to be performed with electron and photons. This will be done at a later stage, when the sensors are already assembled into modules so that they can be read out during the irradiation.

During mass production of the main batches, the tests done by the vendors will provide the numbers and positions of bad strips. Therefore, the collaboration can reduce its own strip measurements and sample approximately 10% of all sensors; these tests are very time consuming and not without risk of damage to the sensors—with scratches, for example. However, tests of IV and CV curves have to be carried out on every sensor to determine the dark current, breakthrough voltage and full depletion voltage.

## 5.3 Mechanics

The mechanical structure is determined by the requirements of B physics. As the Belle II collider operates at relatively low energies and requires a high precision track measurement, a very low mass but stiff mechanical structure is necessary. This task is further complicated by the use of slanted sensors in the forward region (Sec. 5.1.5). In this design, the angle between the slanted and horizontal sensors has to be transferred to the supporting structure, which leads to a torque on the structure due to the weight loads from the vertical sensors.

As mentioned in Sec. 5.1.5, there are two possible ladder counts in layer 6: 17 or 18. Eighteen ladders would allow the assembled barrel to split symmetrically in halves, thus it will likely ease the installation process. A larger overlap between the sensors ($\Delta\phi$) offers some data redundancy—as two sensors may be hit in the overlapping region—and also more data for position alignment at the cost of material budget. On the other hand, seventeen ladders are sufficient for precise track measurement and cosmic muon alignment, with lower cost and a reduced channel count and associated lower data rate. The benefits of either design are at hand and simulations will have to be conducted in order to find the proper solution.

Similar considerations hold for layer 3, where the mechanical design showed that eight ladders pose a space issue in the azimuthal view for a sensor width of 38.4 mm (active area). Again, simulations will have to determine whether the overlap is sufficient with only seven ladders.
The ladders for layers 4 and 5 will be similar in design to the ladder 6 (Fig. 5.12), but the angles and the numbers of mounted Origami hybrids are different. This allows us to use the same construction and assembly methods for those three layers. The ladders of layer 3 will have to fit in close proximity to the PXD. To have enough space for this layer a different support structure is needed, because the "ribs underneath the sensors" design could interfere with the PXD. Layers 3 and 4 are planned to share a common endring to save space or if possible a structure can be implemented into the masks to embed the third layer.

[3]For Micron, we have placed a large set of standard test structures onto the wafer. This was not possible for HPK and therefore these tests are limited to the test structures provided by them voluntarily.

### 5.3.1 Ladder Design

The ladders are designed with a restrictive material budget in mind. The averaged material budget is close to 0.57% and the supporting structure accounts only for 0.08%, averaged over the whole sensor width. The Origami concept offers few possibilities to attach a support structure, which is why the sensor on rib design was chosen. In this concept, two composite sandwich carbon fiber ribs (SCFR) are glued to the Origami sensor area in such a way that no structure interferes with the bond wires on the sensor side and shorts are avoided. This is realized by implementing a 2 mm gap between the top edge of the carbon fiber (CF) ply and the sensor. Cutting rectangle silhouettes into the Rohacell core will create the sensor distancing structure (SDS). The rib thickness is 3.13 mm including the CF layers (65 $\mu$m each), and the height is 6.5 mm without the SDS. Other important dimensions are listed in Table 5.6 (length and height being measured from one mounting point to the other). Table 5.1 give the ladder counts and radii.

| **Layer** | **Ladder length (mm)** | **Ladder height (mm)** | **Angle (°)** |
|---|---|---|---|
| 6 | 645.3 | 50 | 21.1 |
| 5 | 515.6 | 37 | 17.2 |
| 4 | 390.4 | 25 | 11.9 |
| 3 | 262 | 0 | 0 |

*Table 5.6: Current ladder dimensions. Ladder 3 does not have a slant angle nor height due to its straight design.*

At both ends, the ribs are reinforced with an additional layer of CF to have more stiffness in the area that is being bolted to the endring mount structure (EMS). Shims are glued into holes at the end of the rib to spread the clamping load of the bolts and also offer a high precision structure to position the SCFR on the EMS. These shims allow the compensation of mechanical imperfections the laminate surface naturally possesses, as they will be individually fitted with filler glue to ensure parallel positioned ribs.
Bolts mechanically couple the ribs to the EMS so that the handling of the assembly becomes easier. Attaching these SCFR to the EMS will be the first step in the assembly procedure. The two components together form an integral unit that cannot be disassembled. The EMS offer all the mechanical support the ribs need and will provide features for positioning the ladders and also a location to stow the hybrid boards. These blocks will not be actively cooled due to their small volume with little room for channels or tubes but they will offer some degree of cooling infrastructure for the APV25 chips attached to the hybrid board.
In the SVD2, a CF bridge connects the two ribs to improve the stiffness. For the Belle II SVD, no such structure is needed, thus reducing the material budget. Mechanical simulations have shown that the DSSD itself is stiff enough to distribute the loads between the ribs. Araldite glue will be used to ensure a good and strong bond between the sensor and the SDS.
Figure 5.13 shows the current development status of a ladder from layer 6. The ladders from layers 4 and 5 will have the same EMS.

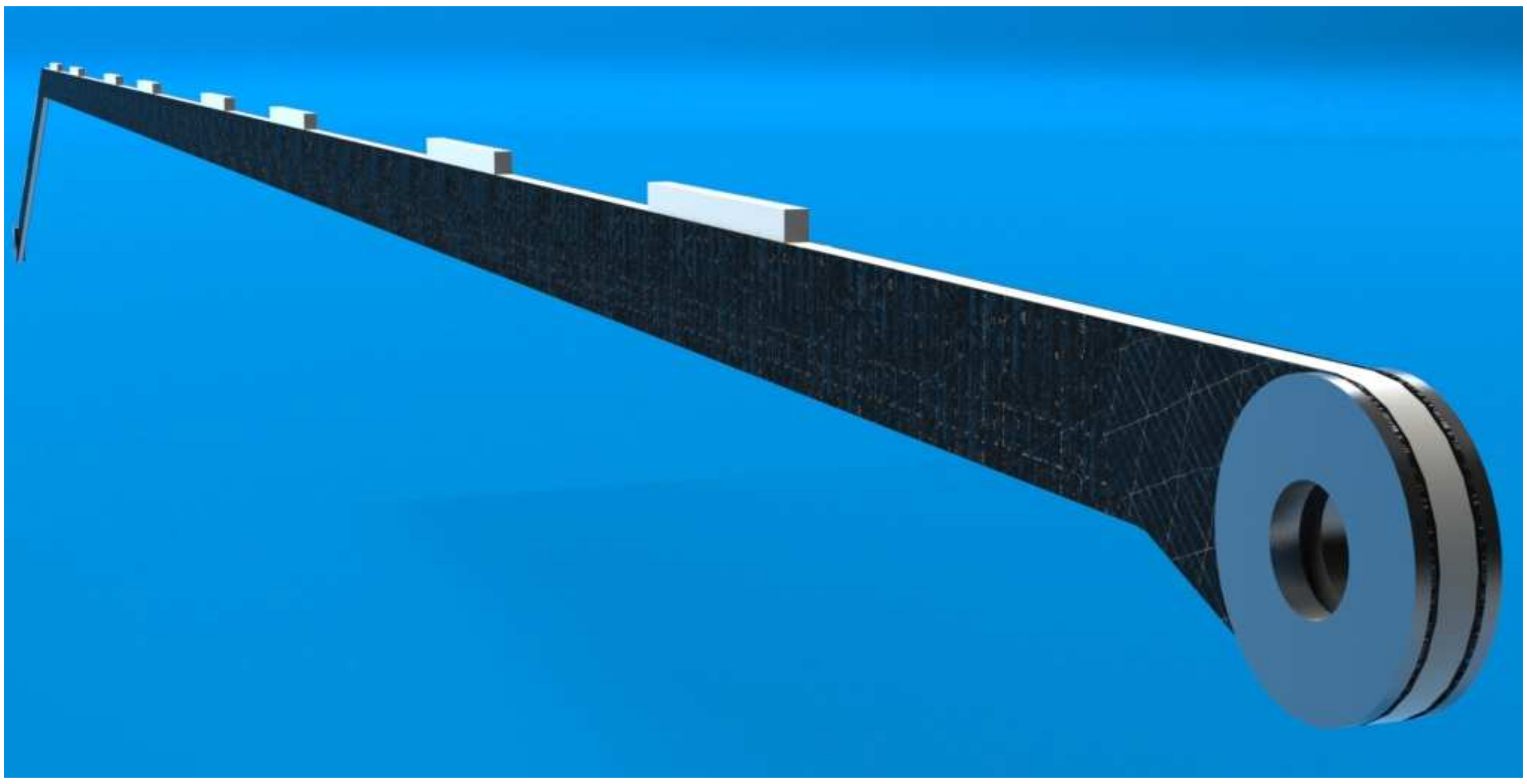

*Figure 5.12: Composite sandwich rib of layer 6.*

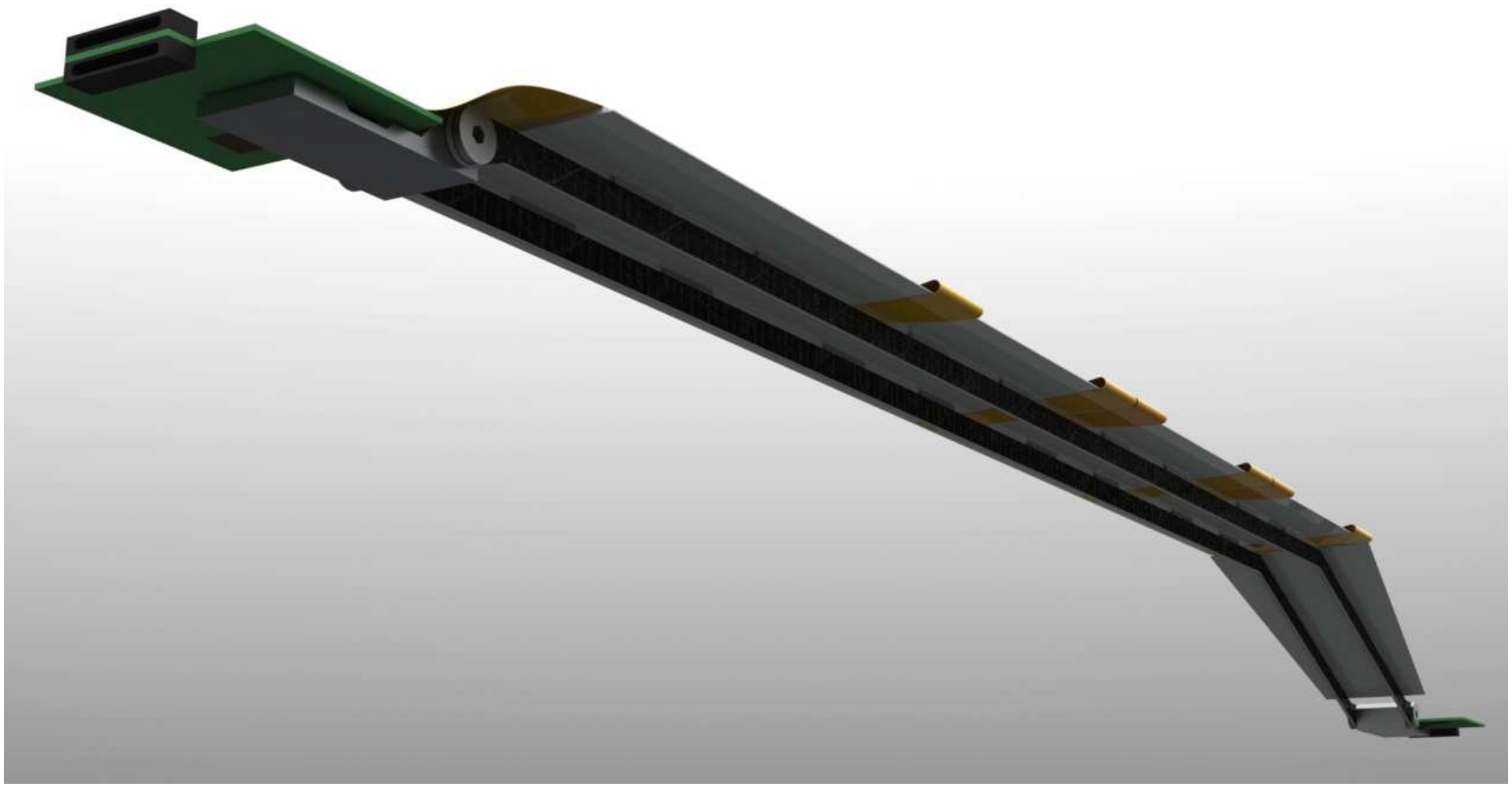

*Figure 5.13: Ladder of layer 6 with Origami sensors, hybrid boards, and end ring mounts.*

The mechanical stability was simulated and the results show a maximum gravitational sag of 108 $\mu$m when the ladder is loaded from above (horizontal position; 12 o'clock) and 48 $\mu$m when loaded from the side (vertical position; 9 o'clock). Initially, the sag was designed to be equal for each direction of load on the structure, resulting in a Rohacell core width of about 2 mm—however, such a thickness is not available in the market. As Rohacell is a very light-weight material and its contribution is negligible in the overall material budget, the standard thickness of 3 mm was chosen, resulting in a stronger structure when side-loaded thanks to the properties of sandwich composites. It would be possible to also reduce the gravitational sag in case of horizontal position, but this would require thicker CF and therefore increase the material budget in a non-negligible way.
The results for the simulations of gravitational sag are shown in Fig. 5.14 and Fig. 5.15 for horizontal and vertical positions, respectively. An ambient temperature of 25° C was assumed for these static simulations. Further mechanical and thermal behavior will be studied on full scale ladder mockups.

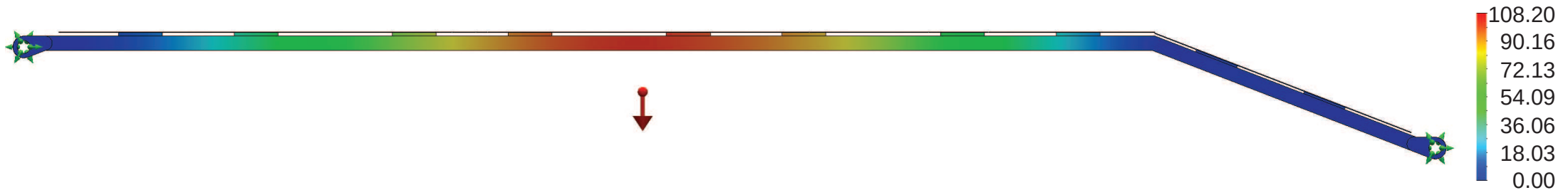

*Figure 5.14: Ladder 6 with gravitational sag (sensors only) in horizontal position.*

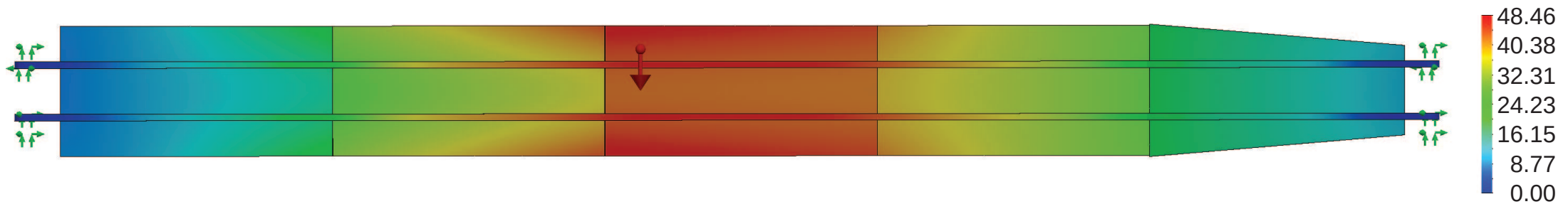

*Figure 5.15: Ladder 6 with gravitational sag (sensors only) in vertical position.*

In general, CTE (coefficient of thermal expansion) mismatch between various materials can lead to thermally induced stress in composite objects. In our case, however, there is a layer of Rohacell foam between the CF ribs and the sensors; and another layer between the sensors and the Kapton hybrids. Thus, we do not expect thermally induced mechanical stress in the ladders, as small variations in length will be absorbed by the Rohacell layers.

#### 5.3.1.1 Origami Chip-on-Sensor Concept

The APV25 front-end chip will be used to read out the strip sensors of the Belle II SVD. Unfortunately, its fast shaping is associated with higher susceptibility to noise, which mainly results from the capacitive load of the amplifier inputs. Hence, the APV25 chips have to be located as close to the sensor strips as possible. The SVD2 scheme, where up to three sensors are ganged (concatenated) and read out together from the edge of the ladder, cannot be used for the Belle II SVD. In order to achieve a reasonable signal-to-noise ratio, all sensors need to be read out individually. On the other hand, the material budget has to be kept as low as possible. To fulfill both requirements, the so-called *Origami chip-on-sensor* concept is used to read out the inner sensors, while conventional hybrids, located outside the acceptance, are foreseen at the edge sensors.
In the Origami scheme, the APV25 chips of both $r$–$\phi$ and $z$ sides are placed on a single flexible circuit, mounted onto the $n$-side of the sensor. The flex-hybrid is made of two Kapton and three copper layers with thicknesses of 25 $\mu$m and 10 $\mu$m, respectively. The (short) strips on the top side of the sensor (measuring the $z$ coordinate) are connected by a pitch adapter, which is either integrated into the hybrid itself or made as a separate two-layer piece, depending on the capability of the manufacturer. The channels of the opposite side (long strips for $r$–$\phi$ measurement) are attached by small flexible fanouts wrapped around the edge of the sensor, hence the name *Origami*. All connections between flex pieces, sensor, and APV25 chips are made by wire bonds.

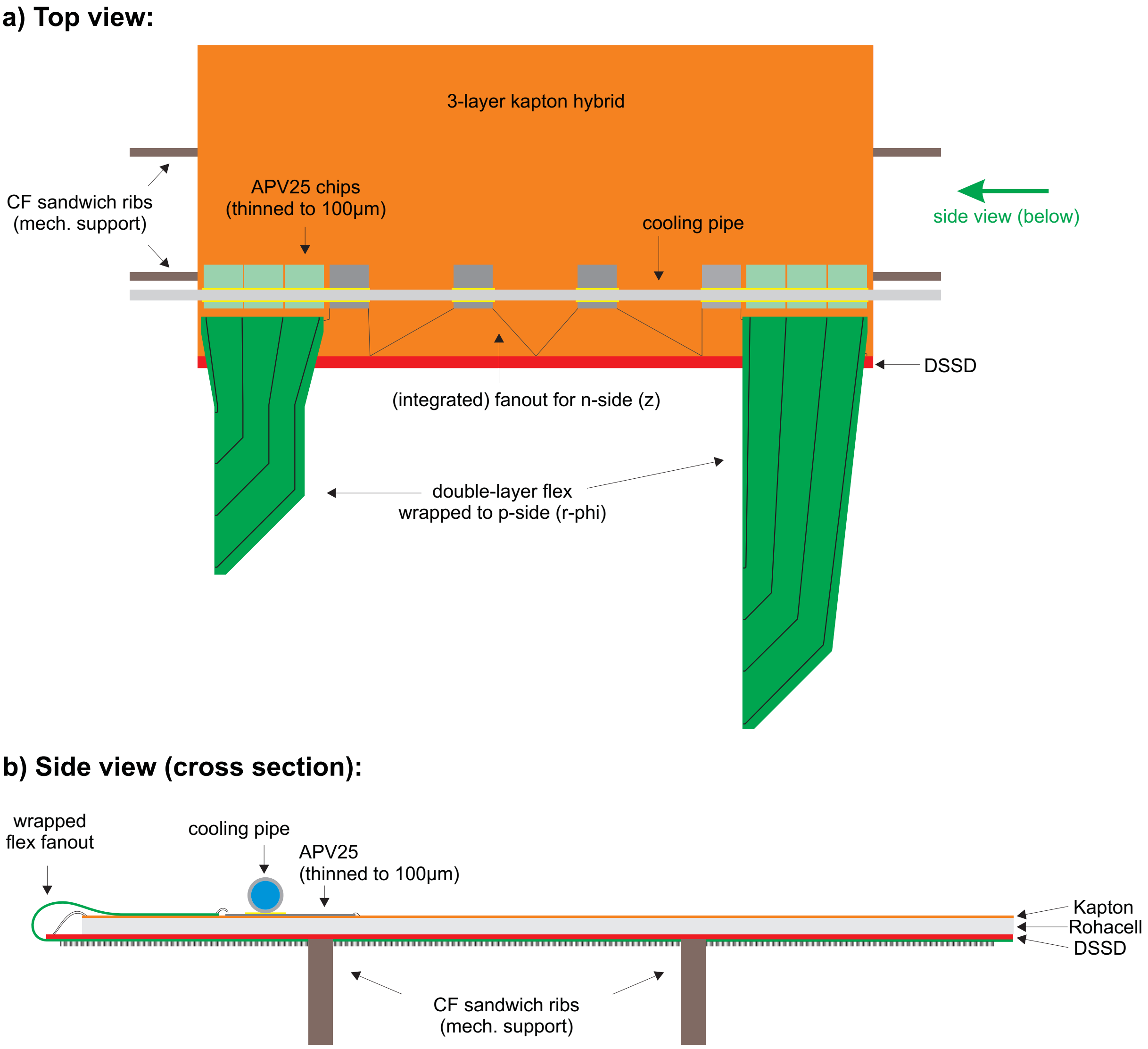

*Figure 5.16: Top and side views of the Origami chip-on-sensor concept for a 6" DSSD. Each view is to scale (but not the same scale). a) Top view: The six APV25 chips that read out the strips on the opposite side are shown in green for clarity and the flex pieces to be wrapped around the edges are unfolded. b) Side view: The wrapped flex, which connects the strips of the bottom side, is located at the left edge.*

Figure 5.16 shows drawings of top and side views of an Origami chip-on-sensor module, where only the fraction of the hybrid containing the readout chips is depicted. Depending on the location in the Belle II SVD, the flex circuit is extended either to the left or the right end of the ladder, where connectors are located outside the acceptance. Prototype flexes, designed for a four-inch wafer DSSD, are shown in Fig. 5.17.

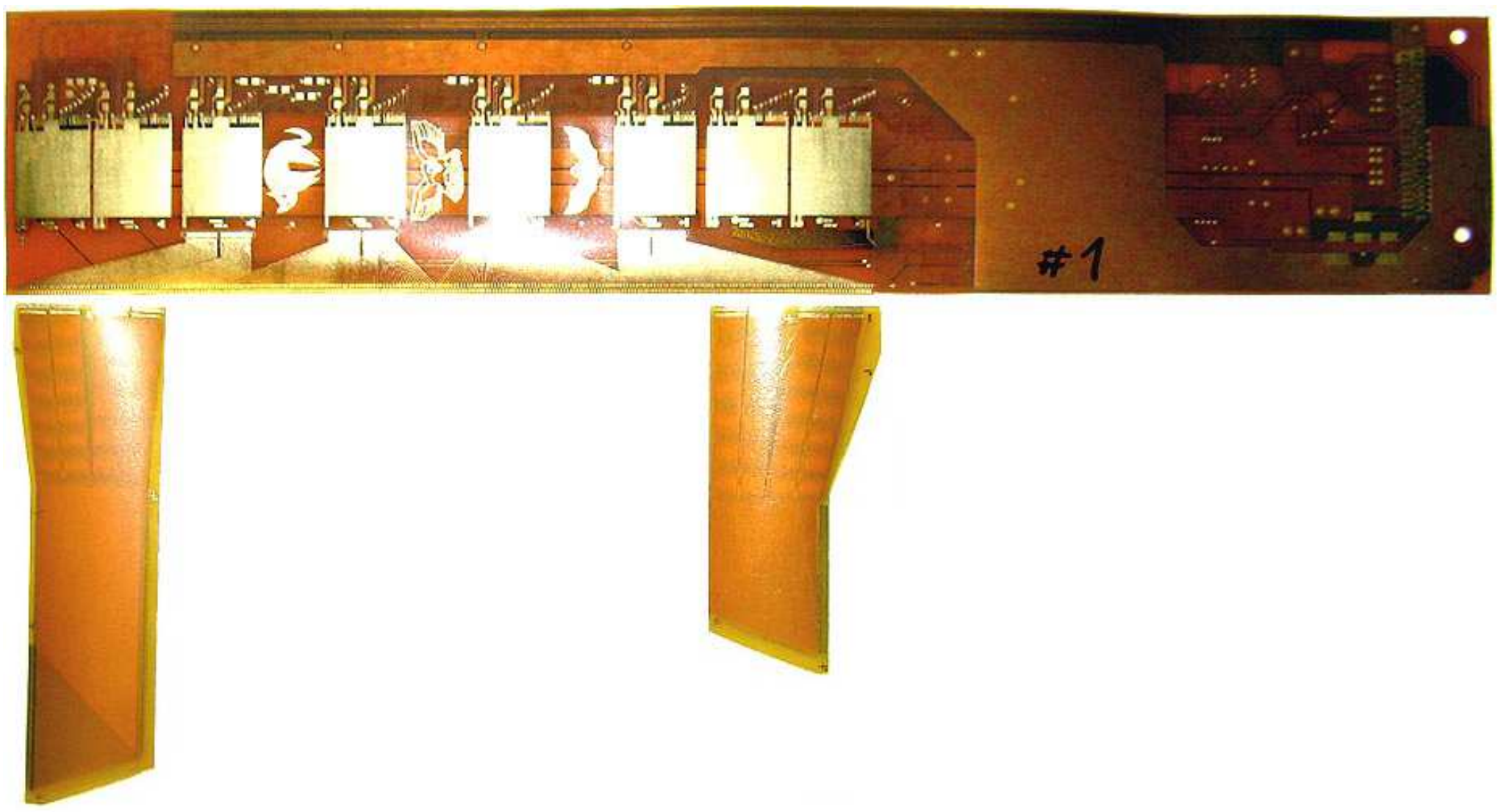

*Figure 5.17: Prototypes of an Origami hybrid and the two flex fanouts, designed for a four-inch wafer DSSD, shown approximately in full size.*

Thermal and electrical insulation between hybrid and sensor is given by a 1 mm thick sheet of low mass, but rigid foam (Rohacell). Nonetheless, the power dissipation of each chip is about 350 mW and thus sufficient cooling of the APVs is required. Since all front-end chips are arranged in a row, cooling can be done by a single thin pipe. A detailed discussion of the cooling system is given in Sec. 5.3.3.
It is clear that using such a hybrid inevitably increases the material budget in the sensitive volume, but there is no alternative solution, particularly in the outer layers, to ensure reasonable signal-to-noise ratio (SNR) with fast shaping. To achieve the lowest possible material budget, the APV25 chips will be thinned down to approximately 100 $\mu$m. The calculated average material budget is 0.57% $X_0$, where the majority is contributed by the sensor itself (Sec. 5.8.1).

#### 5.3.1.2 Materials and Properties

To have the lowest possible material budget, lightweight materials are used. For the sandwich ribs, Rohacell HF 71 is used combined with plies of high modulus CF and aircraft grade epoxy resin. The thermal properties of the materials used do not match: there is a significant coefficient of thermal expansion (CTE) mismatch between the Rohacell ($\approx 30 \times 10^{-6}\,\mathrm{K}^{-1}$) and the silicon sensor ($\approx 2.6 \times 10^{-6}\,\mathrm{K}^{-1}$). To reduce the negative effect of expansion, the contact areas are kept relatively short (20 mm). The SCFR are less susceptible to CTE mismatch because the CF layer, with a CTE of $-0.5 \times 10^{-6}\,\mathrm{K}^{-1}$, expands by $9.8 \times 10^{-6}$ m (at $\Delta T = -30^\circ$ K, distributed over 650 mm) when cooled down, while Rohacell shrinks $-5.8 \times 10^{-4}$ m (at $\Delta T = -30^\circ$ K, distributed over 650 mm). This is a small mismatch that is expected to compensate itself and pose little risk of misaligning the sensors.
A failure mode for sandwich composites is delamination. This can be caused by various factors such as excessive transverse loads [8] or by a bad contact between the layers due to contamination. First tests have shown that excessive bending quickly leads to the expected delamination, which is due to buckling of the extremely thin CF layer. A delaminated rib does not offer the required structural integrity and can lead to unwanted effects. It is rarely as easy as in Fig. 5.18 to spot a delamination. Proper care and handling of the ribs is of the utmost importance in order to avoid this failure.

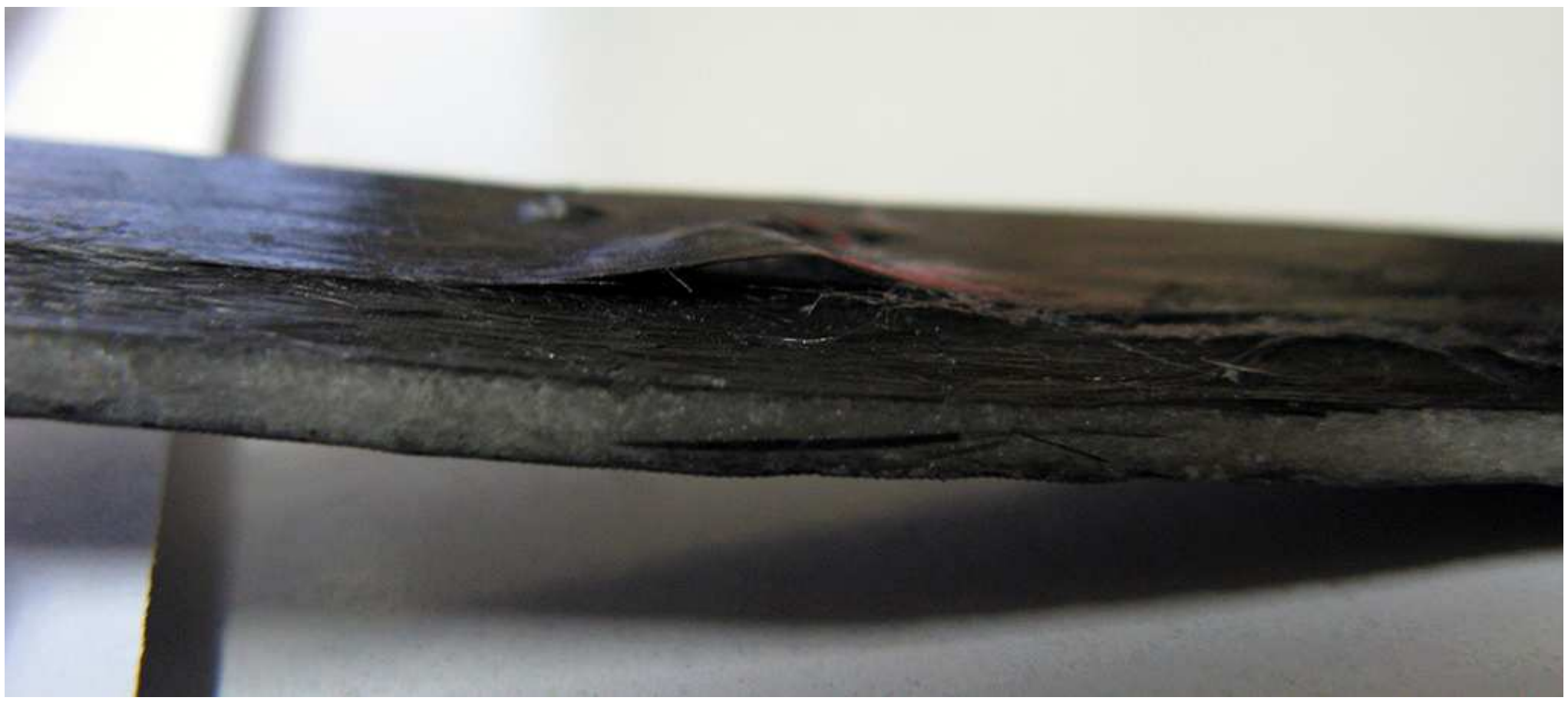

*Figure 5.18: Delamination caused by applying excessive forces on both ends.*

### 5.3.2 Endrings

The final endring design is still pending. The endrings will be joined and mounted on the beam masks as symmetrical or asymmetrical halves, depending on the number of ladders in the respective layer. The main focus of the endring design will be to free up as much space for cable routing and cooling lines as possible while maintaining mechanical stability. The current approach to achieve this is to cut out thin spokes that transfer the ladder loads onto the beam masks. The open areas between the spokes will be aligned with the underlying ladder in the azimuthal direction to ensure conflict-free cable routing. In Fig. 5.19, a possible solution is shown. The endring design on the slanted sensors side of the barrel is particularly challenging as little space is available in this region to make cutouts big enough to route cables through.

### 5.3.3 Cooling

The SVD will be cooled with either a dual phase $CO_2$ system or a glycol-water mixture as a backup option. The decision on the system will depend on the outcome of our R&D efforts using a $CO_2$ blow system and will be made together in a common effort with the PXD group. $CO_2$ systems offer low temperature cooling down to $-20^\circ$ C and beyond, which improves the signal-to-noise ratio significantly at a reasonable material budget (Sec. 5.8.1). Material in the active area is kept minimal with a tube diameter of 1.4 mm and a wall thickness of 0.05 mm. Those small tubes serve as the evaporator lines, where a relatively low mass flow and a pressure of 1.7 MPa are expected. Finite element method (FEM) simulations show that even at 8 MPa the tubes withstand the pressure (Fig. 5.20). Under normal conditions, this pressure should never be reached inside the evaporator. The pressure rating was chosen as it stands for the nominal pressure—including a safety factor—in the supply tubes where only fluid $CO_2$ can be found. The simulation results will have to be verified by pressure tests and leak tests of the assembled cooling circuit.

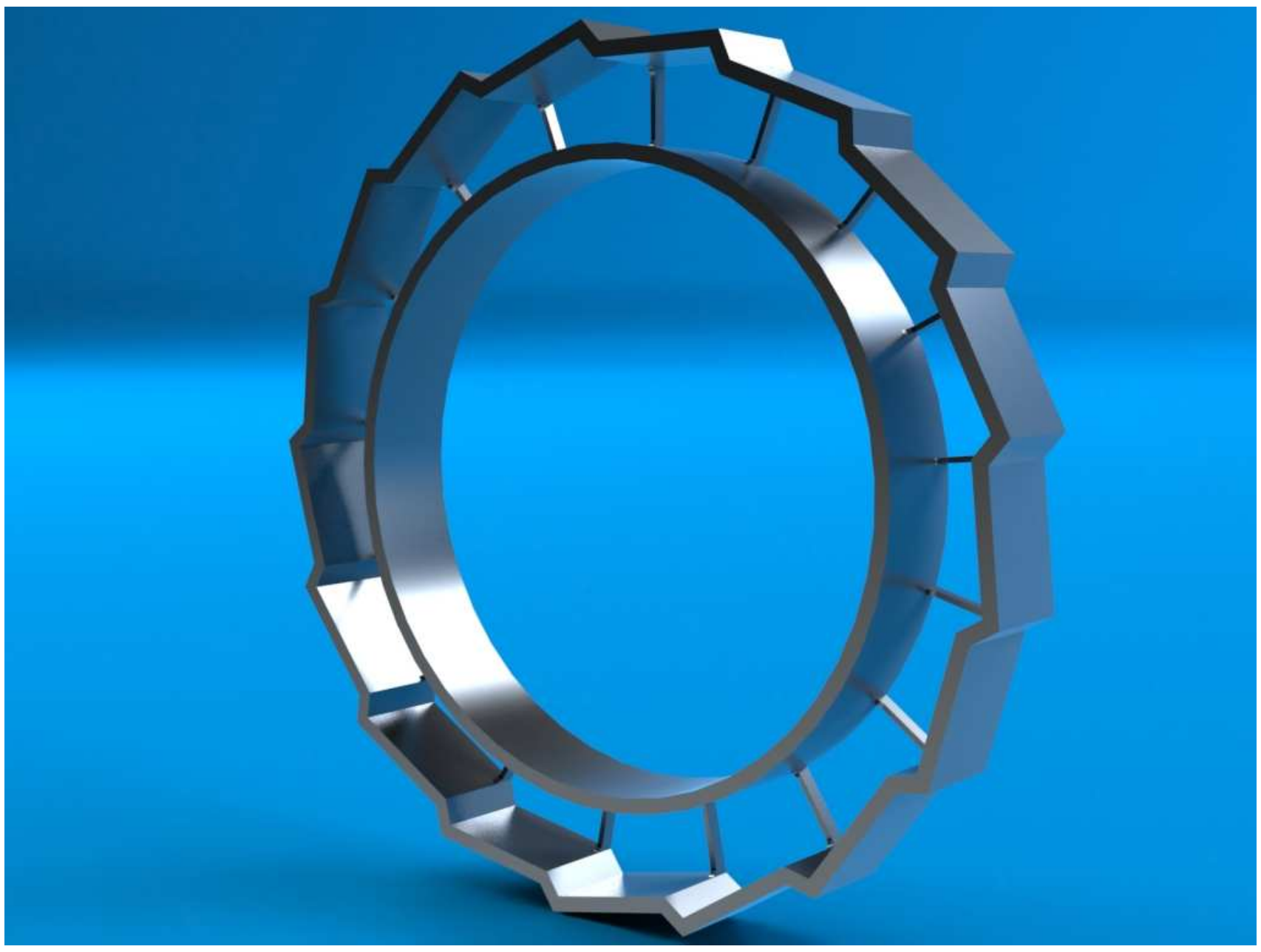

*Figure 5.19: Layer 6 forward endring design study, not split in halves and without alignment tabs for the EMS.*

The required wall thickness $t$ (in mm) is related to the operating pressure $P$ (in MPa) by:

$$t = \frac{P \cdot D_0}{2\sigma_a \eta + 0.8P}(1 + \frac{D_0}{4R}) \quad (5.1)$$

where $D_0$ is the outer diameter [mm], $\sigma_a$ is the tensile strength [N/mm$^2$], $\eta$ is the welding coefficient (typically between 0.8 and 1, depending on the weld quality) and $R$ is the minimal bending radius [mm]. Putting the proper values and estimating a weld quality of 1 (seamless tubes), the resulting thickness is $t = 0.018$ mm for an expected pressure of 1.98 MPa and a tube diameter of 1.4 mm with a minimal bending radius of 5 mm. This implies that tubes with a wall thickness of 0.05 mm will offer enough safety margin to operate the system under regular conditions.

From the mechanical point of view, it is important to reduce the number of inlets to save space. This is done by cooling two ladders with one cooling loop. As many as four ladders could be connected by one loop. Connecting two or more ladders requires a detachable cooling circuit to ease up maintenance and reduce costs if ladders are to be replaced.

#### 5.3.3.1 Cooling Materials

The Belle II SVD cooling circuit is part of a high pressure system that needs to withstand pressures up to 80 bar under regular operation. To withstand such pressure, the use of stainless steel is favored. Aluminum cooling pipes are not a good option as they would require thicker tube walls to sustain the pressure and also would be susceptible to corrosion in copper-containing environments [9]. The fact that the absence of copper cannot be guaranteed, combined with the increased thickness, are sufficient reasons not to use aluminum.

One point of concern is the proper heat conduction from the APV25 chips to the cooling tube. Thermal conductive paste tends to change its properties under radiation and therefore an ap-propriate alternative is needed. Thermal Pyrolytic Graphite (TPG) [10] and carbon foam are two current options that have to be evaluated for their performance. As TPG is difficult to handle, carbon foam is the preferred thermo-coupling material.

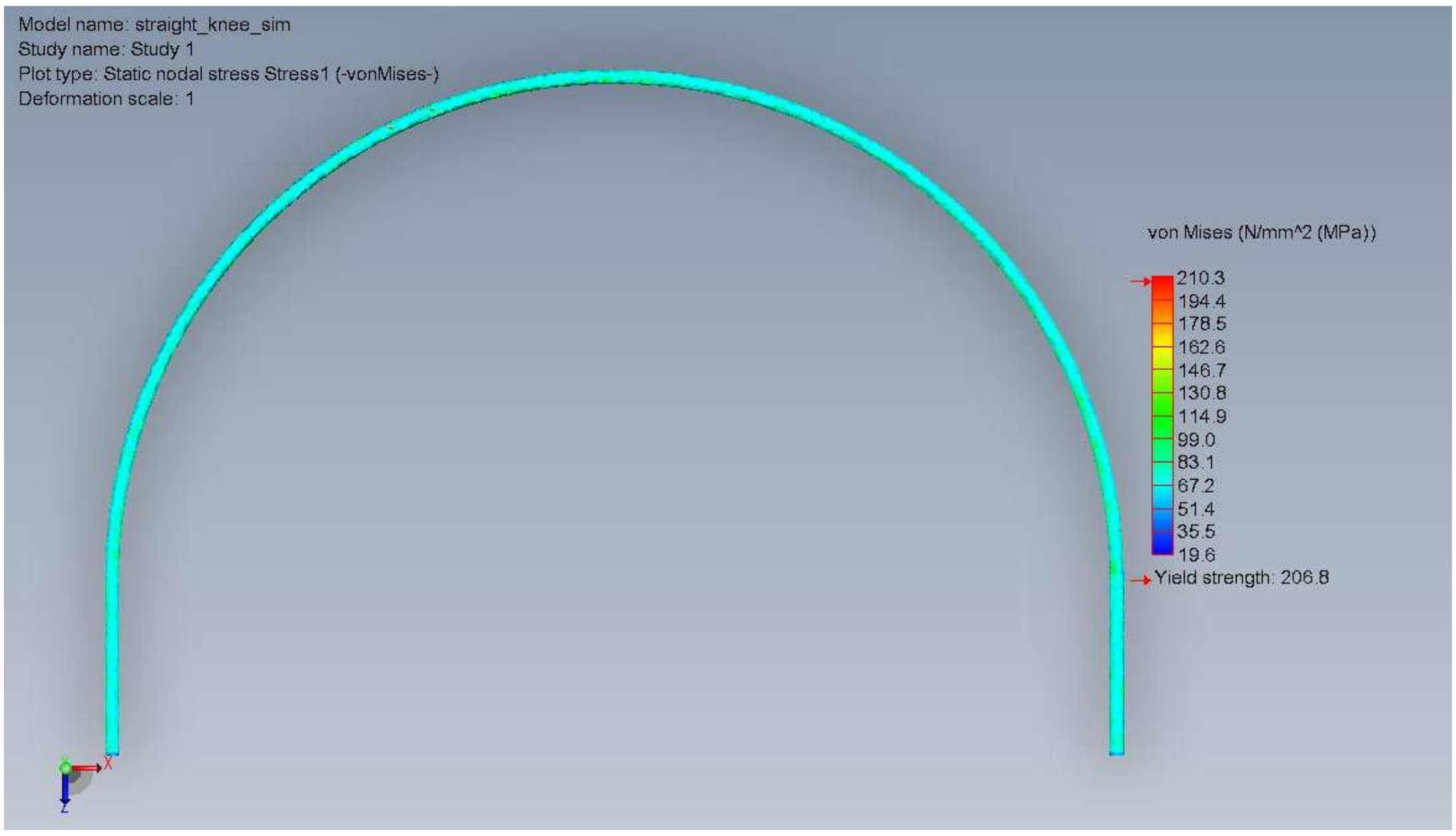

*Figure 5.20: Maximum stress from pressure on the tube end that connects two ladders in one cooling loop at* 8 MPa.

### 5.3.4 Assembly

To assemble the Belle II SVD, we will use the concept of cylindrically arranged DSSD ladders, which are mounted onto endrings. From the third to the sixth layer, one ladder consists of 2, 3, 4 or 5 sensors, respectively. In all but the innermost layer, the most forward detector is slanted. Due to lack of space, we plan to attach a common cooling pipe onto two ladders, which means that either the smallest removable unit is a double-ladder or, preferably, the cooling pipe is detachable and thus not glued onto the Origami hybrids.

#### 5.3.4.1 Procedure

The following assembly procedure is in a very early stage and basically shows the steps required to build a ladder. It strongly depends on the design of the mechanical support structure, which is itself not finalized yet. The procedure is based on the experience gained from the construction of the first Origami module prototype in summer 2009 [11]. Below, the steps required to build a ladder of the outermost—and most complex—layer are described. We plan to make a more precise definition of the procedure and build a prototype of such a ladder in late 2010 or early 2011, depending on the availability of components. The ladder of the outermost layer consists of four rectangular (barrel) and one slanted trapezoidal DSSD. The rectangular and trapezoidal edge sensors will be read out by conventional hybrids, whereas the Origami concept is foreseen with the three inner barrel sensors.
The assembly steps are listed below:

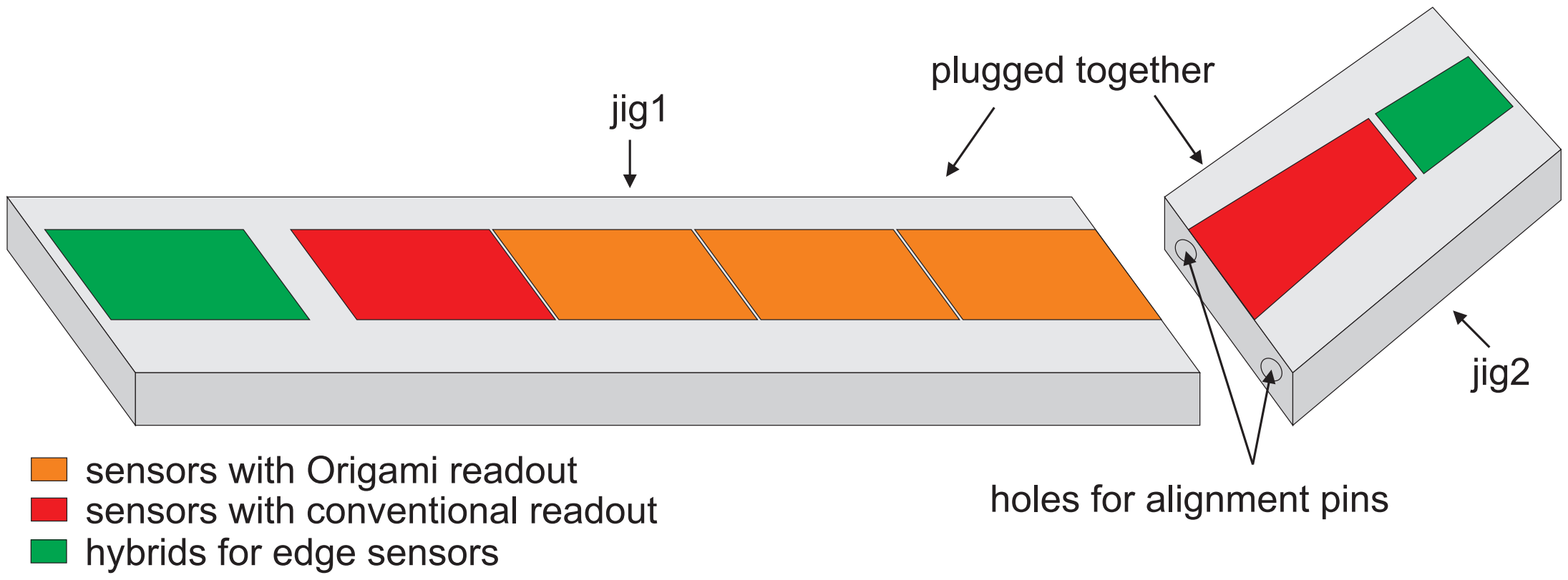

*Figure 5.21: Sketch of assembly jigs 1 and 2 before they are plugged together.*

1. The four rectangular (barrel) sensors are placed with the $p$-side (long strips) facing up onto a jig ("jig1") and aligned to each other using an optical position measurement system and an alignment tool that allows very precise movement. For each sensor, this jig provides a separate inlay, made of porous stone, to lock it in position individually after alignment. Furthermore, the pre-assembled hybrid for the edge sensor is placed on the jig and aligned to the sensor.

2. The Origami fanouts and the pitch adapter for the edge DSSD is attached using a two-component epoxy glue, followed by wire bonding between sensors and flexes.

3. Since wire bonding is only possible on horizontal surfaces, it is necessary to assemble the trapezoidal sensors with their flexes and hybrids separately, again by gluing and wire bonding. The resulting module is then placed onto a further jig ("jig2"), which can be plugged onto jig1 at the required angle between the straight and slanted sensors by using precise alignment pins, as shown in Fig. 5.21.

4. The two support ribs are placed on another jig ("jig3"), which is used to precisely adjust the distance between them. The endring mounts can then be attached to the ribs.

5. The support structure, still held by jig3, is turned around and glued onto the sensors. The correct position is assured by alignment pins between jig3 and jigs 1 and 2.

6. After curing of the glue, all three jigs together are turned around and jigs 1 and 2 are removed. The ladder is then presented on jig3 with the $n$-side (short strips) of the sensors facing up.

7. The Rohacell foam sheet is glued onto the sensors.

8. Starting with the central sensor, the Origami hybrids, which have been populated with AVP25 chips and the other electronic components in advance, are glued onto the top side.

9. This step is followed by wire bonding between the short sensor strips and the Origami hybrids.

10. Finally, the flexible fanouts are bent around the sensors edges, glued onto the hybrids and connected to the APV chips by wire bonding. A customized micro-positioner which was already used for prototyping (see Fig. 5.38), is used to bend and align the flexes.

The above procedure contains several steps in which parts are glued together. To achieve the highest possible mechanical precision, the glue must be applied by using a method that ensures exact positioning of the glue as well as a well-defined uniform thickness of the resulting adhesive film. Usually, this is done by a dispensing robot, but such machines are expensive and require the complex and exhausting tuning of several parameters such as air pressure, motion speed, and dispenser tip diameter to achieve the desired quality.

A much simpler, but equally precise, method is to apply the glue by using a stamp, which matches the form of the gluing area and has a grid of small bosses on its surface. To ensure an exact adhesive amount, a small portion of glue is uniformly distributed on the floor of a basin by using a mask with the required thickness. The stamp is then put into this glue bath and finally placed at the desired adhesive area. This concept was successfully applied at Paul Scherrer Institute (PSI, in Villigen, Switzerland) to assemble the CMS pixel detector [12]. Thanks to its simplicity and proven effectiveness, this method will be adopted for the assembly of the Belle II SVD ladders.

#### 5.3.4.2 Testing

To ensure high quality and reliability, each component of the Belle II SVD will be tested before being mounted onto the ladder. This includes the sensors (Sec. 5.2), the thinned and diced APV25 chips, the Origami hybrids, as well as all the other parts. As the ladder assembly is a very complex task with many sophisticated steps, accompanying electrical testing is foreseen from the very beginning of the process. Therefore, we need a test setup, which can be built partially from already existing prototypes of the hardware and software. It will allow the identification of errors such as shorted or open wire bonds. The results of these tests will be stored in a common ladder-assembly database that can be consulted later, when each part is attached to the SVD ladders.

Fully assembled ladders will be tested with a source and an $xy$-stage such that the functionality of every strip can be verified. Such a test will take a couple of hours to allow for scanning and to accumulate a statistically representative signal distribution for each channel. An automatic setup will be built to perform such tests in an unattended mode (e.g., overnight) and store the results in a database.

### 5.3.5 Installation and Commissioning

The detector ladders will be mounted onto the endrings in a clean environment at KEK, preferably in the Tsukuba hall building. A rendering of the fully assembled Belle II SVD is shown in Fig. 5.22. Each ladder will be tested before and after assembly, and eventually the whole detector will be powered as well as the cooling system. Four layers of silicon detectors are almost impossible to penetrate using particles from radioactive sources—at best multiple scattering will cause zigzag paths rather than tracks. Thus, the only possibility for a full SVD test will be to use cosmics by triggering on the coincidence of scintillator paddles placed on top and beneath the SVD assembly. The expected rate of cosmic events will be on the order of a couple per minute, but those will have perfectly straight tracks and thus can easily be used to check the SVD-internal alignment. With long-term measurements, the rate should be sufficient to at least verify the functionality of every channel.

Once the ex situ tests are finished and the PXD is mounted onto the beam pipe, the Belle II SVD assembly will be inserted into and then bolted to the CDC. Even though Belle II SVD and PXD are separate units, it will not be possible to insert or remove the PXD after the SVD is installed without first disassembling the SVD.

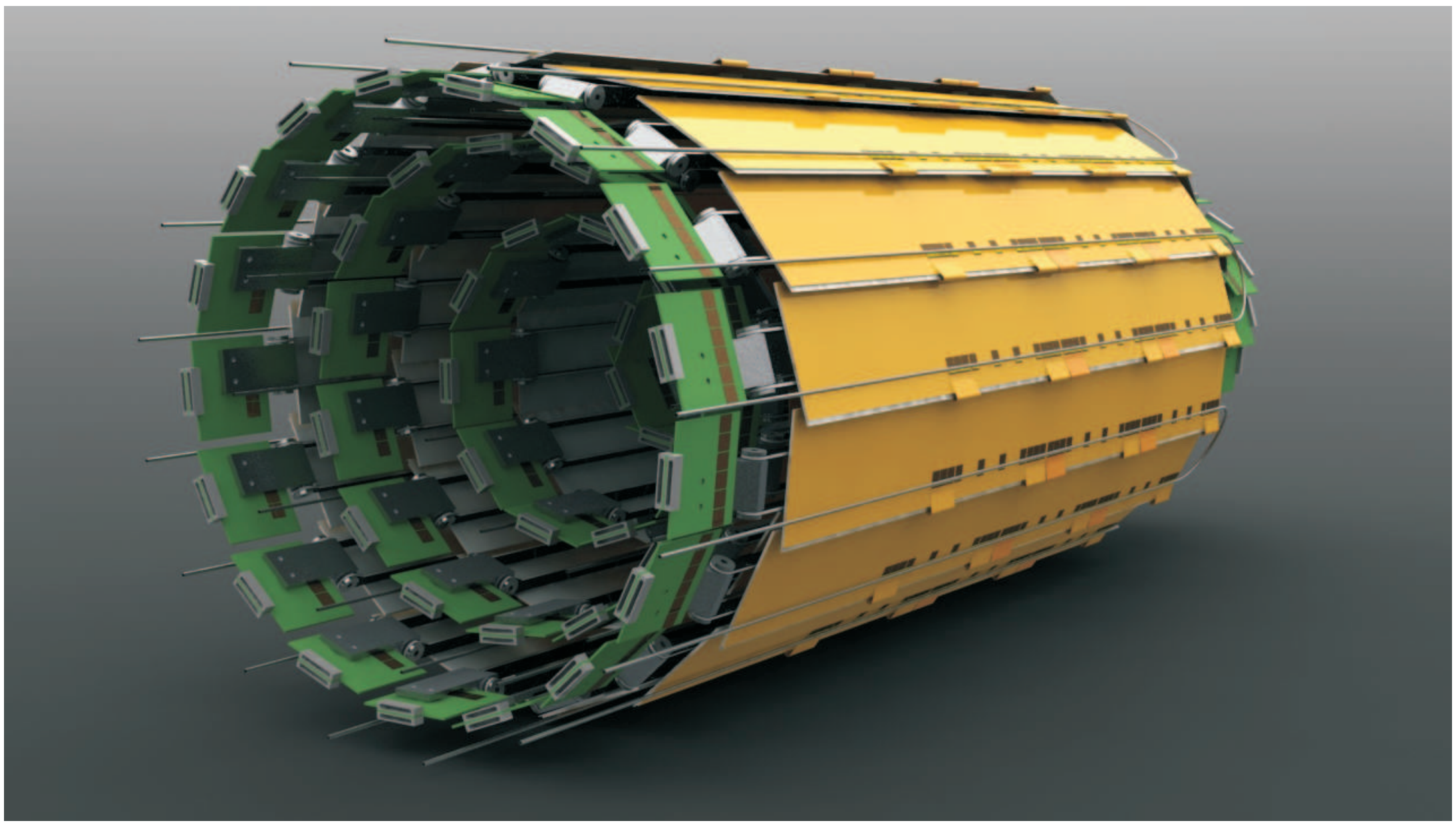

*Figure 5.22: Belle II SVD barrel showing all four layers, cooling lines and hybrids (not connected to Origami modules).*

After installation and cabling, the Belle II SVD will again be tested. Once its functionality is verified in a standalone fashion, its controls and data outputs can be integrated into the Belle II infrastructure.

## 5.4 Electronics

This section describes the components of the electronic readout chain for the Belle II SVD. Generally, its status is well advanced, as we reuse components developed for other experiments, have put several years of R&D into the development of the system, and have successfully built several prototypes.

### 5.4.1 APV25 Readout Chip

The APV25 [3] is a low-noise front-end amplifier chip in $0.25\,\mu$m CMOS technology originally developed for the CMS Experiment at CERN. Its final version, APV25S1, was released in 2000 and thoroughly tested until the installation of about 70,000 devices in the CMS Tracker.
We already purchased approximately 5,000 chips for the Belle II SVD. All of them are wafer tested. About half are already diced, and the remainder were delivered in a total of eight wafers (produced in 8" technology) with the aim of thinning them to minimize the material budget of the Origami modules. As a test, one such wafer has been thinned down to a nominal value of $100\,\mu$m and then diced. The actual thickness was measured to be $106.6 \pm 3.2\,\mu$m and the procedure had a mechanical yield of 98.4%. Up to now, about 20% of the thinned APV25 chips have been electrically tested and all of them performed very well with no difference between them and unthinned APV25 chips. Hence, we will thin all the chips to be attached to the Origami modules inside the polar-angle acceptance; the conventional hybrids at the edges, which reside outside the sensitive volume, will use standard APV25s with a thickness of $325 \pm 25\,\mu$m.

#### 5.4.1.1 Features

The APV25 has a shaping time of 50 ns (adjustable), a 192-cell deep analog pipeline, and an operating rate of 40 MHz; it is known to function over a wide range of clock frequencies. While it is not surprising that the APV25 is fully functional at 31.8 MHz, a couple of chips were also successfully tested up to 80 MHz.
The internal structure of the APV25 was already shown in Fig. 5.2. The usual preamp/shaper architecture can be seen on the left, where an optional inverter is placed between these blocks. As the supply rails are limited to 2.5 V in the 0.25 $\mu$m technology, this switch is introduced to optimize the dynamic range depending on the polarity of the signal, i.e., reading out positive ($p$-side, inverter on) or negative ($n$-side, inverter off) detector currents. Thus, a linear range of approximately $-2$ to $+7$ MIPs (referred to a standard 300 $\mu$m-thick sensor) can be achieved for both polarities.
The shaper output is written to a pipeline of 192 cells at the clock frequency. In fact, the pipeline is implemented as a ring buffer memory with an additional FIFO of 32 words to label memory addresses that are requested for output by pending triggers. Until those data are read out, the tagged cells are skipped in the write cycle. Consequently, the available pipeline depth can vary between 160 and 192 cells, depending on the amount of pending triggers. By multiplication with the clock period, this translates to the maximum trigger latency time.
After the pipeline, the APV25 has an analog pulse shape processor (APSP). This is, in fact, a switched capacitor filter that can perform a so-called *deconvolution* (see below) or, in a different configuration, simply pass on the pipeline contents. Finally, the strip data are multiplexed through three hierarchical stages and sent to the differential-current-mode output.
Various bias voltages and currents as well as general parameters of the APV25 can be configured through its $I^2C$ interface. Clock and trigger signals are received by dedicated differential inputs. The trigger line also accepts special 3-bit symbols, namely 100=trigger, 110=calibration request and 101=soft reset. Consequently, two actual triggers must be at least 3 clock cycles apart; otherwise, they would be misinterpreted as a special symbol.

#### 5.4.1.2 Operation Modes

In the CMS experiment, the APV25 is operated with a 40 MHz clock that is synchronous with the bunch crossings of the LHC. Thus, the phase can be adjusted to always sample the peak signal of the shaping curve if operating in the so-called *peak mode*.
Moreover, the APV25 can also use three consecutive samples around the peak to perform a *deconvolution* [13] with its internal APSP circuit. This operation essentially reverses the shaping function, so that the train of output samples has just a single non-zero value and thus the corresponding bunch crossing can be identified unambiguously at the cost of a moderate noise increase. Operation in this *deconvolution mode* requires a clock frequency of 40 MHz and clock-synchronous particle signals to properly adjust the phase.
Unfortunately, the deconvolution mode cannot be in the Belle II SVD, because the SuperKEKB bunch crossings occur almost continuously (precisely speaking, every second or third period of the machine RF at about 508 MHz). Moreover, the APV25 clock frequency in the SVD is not the same as in CMS: By default, it will operate at RF/16 (about 31.8 MHz) or, perhaps, RF/12 (about 42.4 MHz).
Conveniently, the APV25 allows us to read out multiples of three consecutive samples without applying the *deconvolution*. By taking six samples in this *multi-peak mode*, we can not only take care of the non-constant phase shift between bunch crossings and APV clock, but also process the data outside the APV25, thus identifying hits that belong to the triggered event and discarding off-time background (Sec. 5.4.4.4). Nonetheless, the readout system will be built in such a way that operation with between three samples and one will be possible simply by switching some configuration bits.
The choice of the APV25 clock frequency has two immediate consequences for the system. First, it defines the temporal length of the pipeline, which has a minimum effective length of 160 cells. This corresponds to either 5.04 $\mu$s (at 31.8 MHz) or 3.78 $\mu$s (at 42.4 MHz) and defines the maximum trigger latency. To implement a 3D track trigger (Ch. 12), which takes some computation time, the earlier is preferred from the viewpoint of the Trigger Group. On the other hand, a faster readout speed—and therefore a shorter overall trigger latency—makes it less likely that the APV25 FIFO will fill (and therefore introduce dead-time into the data acquisition system).
Assuming *multi-peak mode* operation with six samples read out per trigger, a simulation was done to obtain the dead-time caused by a) the minimum distance of two triggers allowed by the APV25 (six clocks) and b) the condition of a full FIFO with too many events pending for readout. The average trigger rate was a tunable parameter, with triggers following an exponential waiting-time distribution (i.e., Poisson distributed within fixed time intervals).

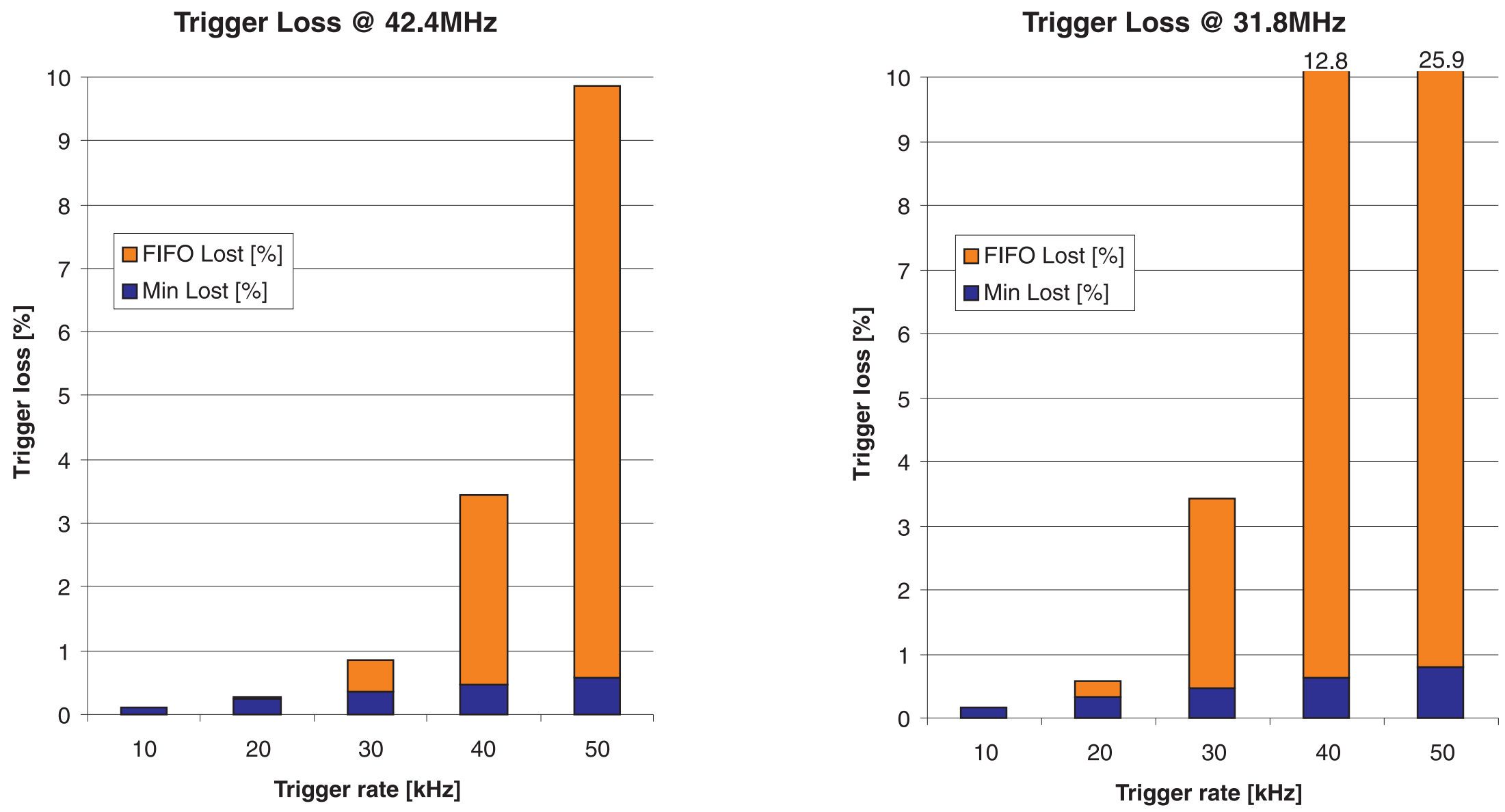

*Figure 5.23: Dead-time caused by the APV25 at two different clock speeds and various average trigger rates.*

The results of this simulation are shown in Fig. 5.23. Obviously, the dead-time caused by a) is essentially a linear function of APV25 clock and average trigger frequencies. The FIFO-full condition, however, reveals a strongly nonlinear dependence: at a trigger rate of 30 kHz, the dead time fraction is only 0.87% at a clock frequency of 42.4 MHz, but rises to 3.43% at 31.8 MHz. Clearly, the choice of APV25 clock speed is a trade-off between trigger latency and dead-time budget: from the DAQ point of view, the faster readout is preferable. Table 5.7 summarizes the behavior of the AVP25 depending on its readout clock.

| $f_{CLK}$ (MHz) | **Min. trigger distance (ns)** | **Max. trigger latency ($\mu$s)** | **Dead-time a)** | **Dead-time b)** | **Dead-time a) + b)** |
|---|---|---|---|---|---|
| 31.8 | 189 | 5.04 | 0.47% | 2.96% | 3.43% |
| 42.4 | 142 | 3.78 | 0.36% | 0.51% | 0.87% |

*Table 5.7: Summary of APV25 properties as a function of the readout clock. Dead-times are given for a Poisson-distributed average trigger rate of* 30 kHz.

### 5.4.2 Hybrids

The APV25 chip needs a limited number of external connections: naturally, several bond pads are needed for powering of the +2.5 V, +1.25 V and GND lines. These are not only needed on the rear end of the chips, but also on the front-side, which holds the input pads for 128 strips. Moreover, the rear side has differential inputs for clock and trigger, and a differential current mode output. Single-ended lines are needed for the $I^2C$ bus and the reset input. In addition, there are a few pads to define the chip's address within the $I^2C$ bus.
Numerous PCB hybrids were produced in the past years for the readout of four APV25 chips and used with prototype sensors. Fig. 5.24 shows such a test hybrid equipped with four APV25 chips. Apart from the chips and a connector (which is rather bulky in this case), there are a few resistors needed for line termination and a special power line as well as some decoupling capacitors. To the left of the connector, there are some parts related to the filtering of the bias line that supplies HV to the sensor. This (simple) hybrid concept has been well established and proven to work reliably in the lab as well as in several beam tests. So far, about 100 hybrids were produced and tested using this design.

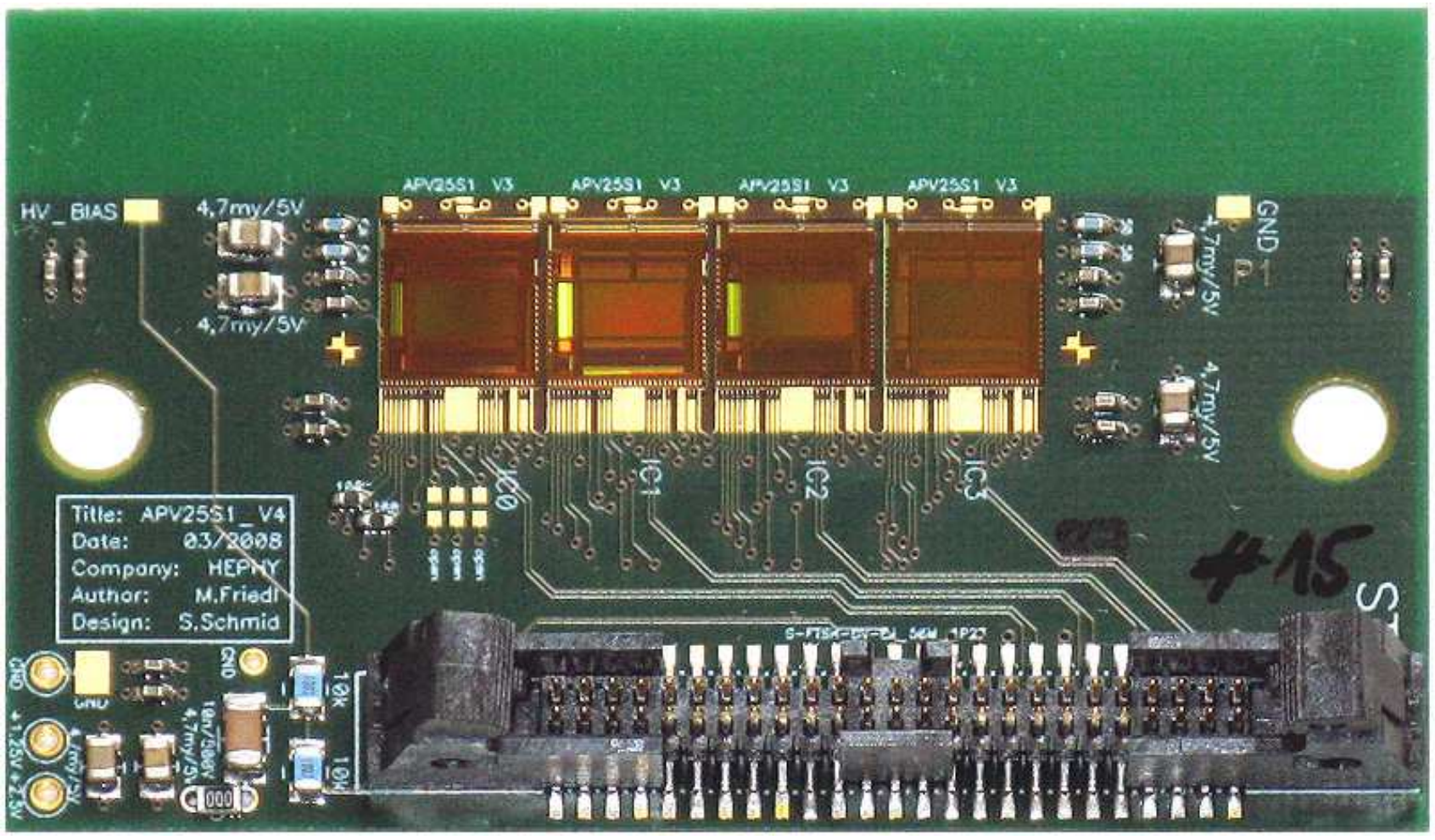

*Figure 5.24: PCB hybrid used for APV25 and sensor tests.*

In the Belle II SVD, there are two different implementations of hybrids. One is the Origami concept (Sec. 5.3.1.1), which uses the same schematics and has already been verified on a prototype. The other type, which will also be derived from the existing prototypes, is intended for the edge sensors. There, we will probably use a double-sided version of the hybrid with four or six APVs, depending on layer and *p*- or *n*-side readout. As the electrical design is flexible, the actual shape will be determined mostly by the mechanical layout. As each APV25 chip dissipates about 350 mW, this heat needs to be taken away by a thermal connection to cooling blocks at the endrings, which is another mechanical constraint.

#### 5.4.2.1 Connectors and Cables

Both for the Origami as well as for the PCB hybrids, small connectors will be used, which should have the following properties:

- Small feature size
- Mechanically locked
- Gold plated contacts
- Free of halogens
- 50 (or more) pins

So far, only one connector type has been identified that fulfills all these requirements: this is the Nanonics type STL051L2HN (part no. 2-1589483-6), already used in the SVD2.
For the cord, a halogen-free (Polyolefin) 0.05" pitch twisted-pair cable (3M type 79992-25P-270A) will be used that has already been verified in prototype systems in both lab and beam tests.
On the other end of the cable, the space constraints are not as stringent as on the detector side. Consequently, a cheaper connector is used that fulfills all the above requirements except for the small feature size. The PCB part of this connector is also shown in Fig. 5.24; it is Samtec type FTSH-125-01-L-DV-EJ-K-A.

#### 5.4.2.2 Radiation

The hybrids—at least the Origami types—are subjected to the same radiation dose as the sensors, which is expected to be $\approx 4.5$ MRad as extrapolated from the SVD2 (Sec. 5.2.3). The APV25 is made from $0.25\,\mu$m CMOS technology and was tested to be radiation tolerant up to at least 30 MRad, which is far beyond what we can expect for the Belle II SVD.
Various types of SMD resistors and capacitors were tested for LHC applications to integrated doses that far exceed the Belle II case and found to be compliant. Thus, we conclude that those parts are not critical for application in the SVD. The same is true for Kapton- and PCB-based circuits.

### 5.4.3 Junction Box

The junction box will sit in a location that was previously taken by the so-called DOCK boxes containing repeater boards. In Belle II, this location will only contain connectors, voltage regulators, and passive components.

#### 5.4.3.1 Functionality

The junction box joins the short cables attached to the Belle II SVD to the permanently installed hybrid cables, which run from here to the FADC+PROC boards located between 8 and 10 m away, outside of the Belle II detector. For the benefit of signal integrity, the disruption of impedance will be minimized, and thus the same cables will be used for the long leg of that connection. On the other hand, those cables present an ohmic impedance that is not critical for the signal path but is harmful for power connections.
Thus, the power will be brought into the junction boxes using different cables with lower resistance and limited by voltage regulators in the junction box. In the absence of such local voltage limiters, transient overvoltage can occur in case of load drop-off, e.g., when a reset signal is sent in common to a group of APV25 chips. As the $0.25\,\mu$m CMOS process specifies +2.7 V as the "absolute maximum" limit for the supply voltage, it is important to adhere to this boundary even for short-term transients. This threshold will be observed by using voltage regulators in the junction box as well as decoupling capacitors, which also damp voltage spikes.
A side benefit of using separate cables to bring in the supplies is that more leads of the hybrid type cable become available for signal transmission. Thus, the plan is to merge two hybrid cables into one single cable of the very same type for transmission to the FADC+PROC. This concept still needs to be proven by prototypes.

#### 5.4.3.2 Location, Mechanics and Cooling

The junction box is located at the outer wall of the CDC around the QCS magnets. This location is outside the acceptance and cooling services are available there.
The mechanical layout of the junction boxes will follow the present concept of the DOCK boxes, and cooling will be needed for the voltage regulators. It is expected that the total cooling power for both backward and forward parts will be less than 300 W and thus lower than the power dissipation of the present DOCK boxes.

#### 5.4.3.3 Radiation

To quantify the radiation to which the SVD2 DOCK boxes are exposed, dosimeters were placed at the forward side in early 2009. It was found that the total dose, extrapolated to Belle II, does not allow us to use commercial off-the-shelf parts there (Sec. 5.1.6). In particular, a repeater, which would use such components, cannot be placed there.
Consequently, we decided to use only connectors and passive parts there, as well as radiation-hard voltage regulators. Several such devices exists, and one candidate is the RHFL4913 type (previously called LHC4913) by ST that was developed for the ATLAS experiment. It has now become a commercial device manufactured by ST, and is specified to withstand 300 krad. Even though this is considered sufficient for our needs, tests performed by ATLAS found that it tolerates an even higher dose of 100 MRad [14].

### 5.4.4 FADC+PROC

Each FADC+PROC board (9U VME) receives the outputs of 24 AVP25 chips in either $6 \times 4$ or $4 \times 6$ configuration, depending on the hybrid location. Consequently, about 80 such modules will be needed, which are spread over several crates (Sec. 5.1.6).
In contrast to existing prototypes (Sec. 5.7.2) with a separate repeater box close to the front-end, the FADC+PROC will incorporate that functionality, which is essentially the level translation of front-end signals from the bias voltage down to ground-bound low voltages. A sketch of the FADC+PROC as a composite object of the existing repeater and FADC is given in Fig. 5.25.
As the APV25 chips have to drive their (differential) analog signal output all the way to the new FADC+PROC without any intermediate buffer, this path has to be kept as short as possible. Measurements have shown that a distance of 12 m is completely acceptable. More detailed studies were done earlier with an 8.7 m cable, resulting in no effect on the SNR within the repeat accuracy (±3%). Crosstalk was investigated with three (twisted-pair) cables stacked on top of each other over a length of 1.8 m (including a total of 20 cm of flat regions). An internal calibration signal of maximum amplitude (corresponding to about 10 MIPs) was injected in the center cable and the relative amplitude in the adjacent cables was found to be $< 0.2\%$ and thus barely visible. As it is quite unlikely that cables will be precisely stacked over a long distance in the future installation, we conclude that long hybrid cables are no problem for the signal quality. We assume that an overall distance of about 10 m is sufficient to lead the cables out to the FADC+PROC crates located on the top or at the side of the Belle II detector. These possible locations, referring to the Belle detector, are shown in Fig. 5.26. The total space needed for the FADC+PROC boards is one rack each in forward and backward sides, each containing two or three crates, respectively. This does not include the space needed for front-end power supplies, which should preferably (but not necessarily) be located nearby.

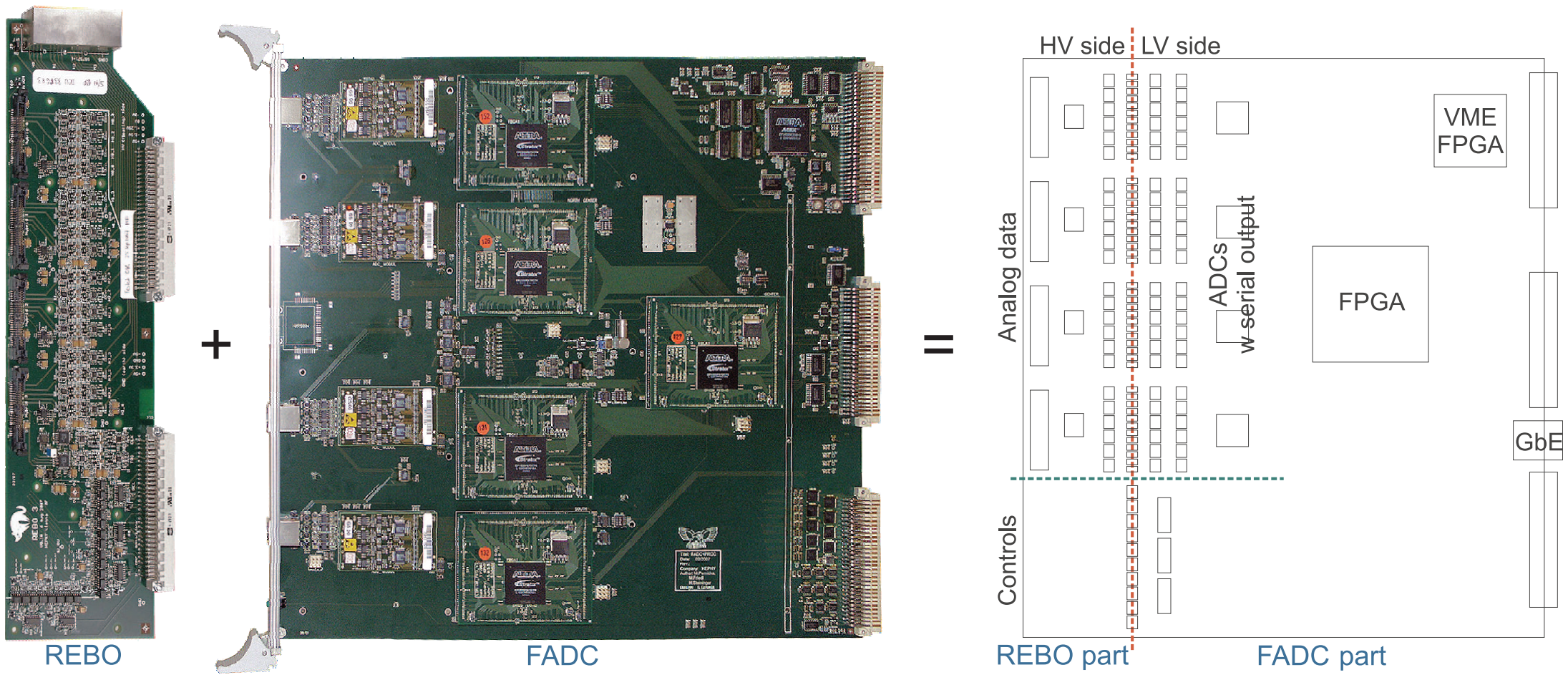

*Figure 5.25: The combination of existing repeater (REBO) and FADC constitutes the future FADC+PROC module (9U VME) with integrated voltage level translation.*

#### 5.4.4.1 Analog Level Translation

Control signals sent to the front-end as well as analog data received from there need to undergo a level shift from the bias voltages (±40 V in case of the SVD2, possibly higher in the Belle II SVD) to ground-bound signals. This is done by using capacitive AC coupling for fast signals or optocouplers for slow (quasi-static) signals.
The advantage of optocouplers is that they allow very high isolation voltages and static levels, but they are limited in speed and linearity. Thus, they are used to transfer digital slow controls, namely the $I^2C$ lines and the reset signal. Capacitive coupling, however, is used for fast, differential signals, namely, the clock and trigger (digital) and the signal output (analog). As the capacitors need to charge with the DC bias voltage, the latter shall always be ramped up and down slowly, which is also required by the silicon sensors. Nonetheless, a protection circuit on both HV and LV sides is foreseen in the translation schematics to absorb sudden voltage excursions that could otherwise damage the integrated circuits. The existing repeater prototype boards are designed for level shifts of up to ±100 V, which will also be sufficient for the Belle II SVD.

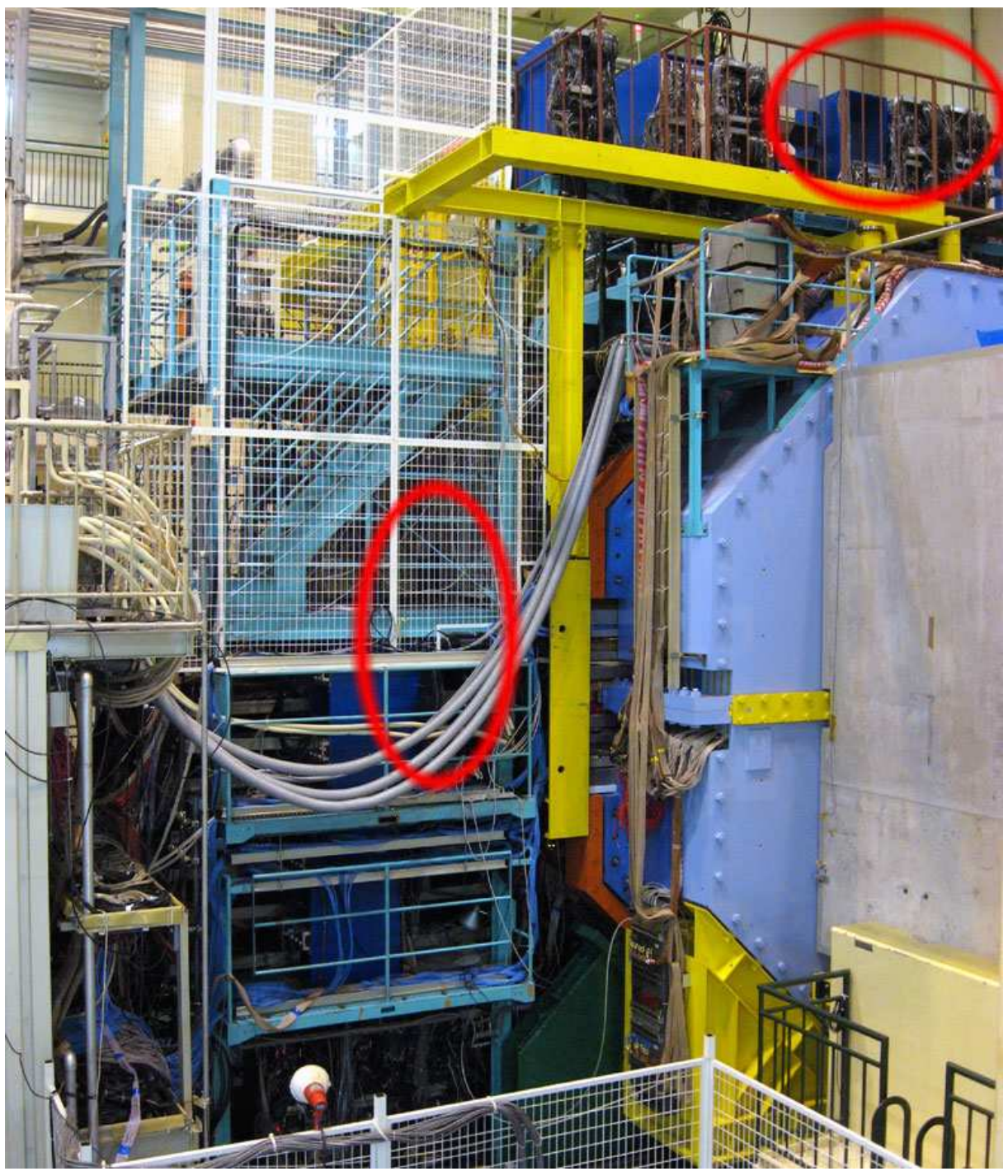

*Figure 5.26: Potential locations for the FADC racks (one each needed in forward and backward sides).*

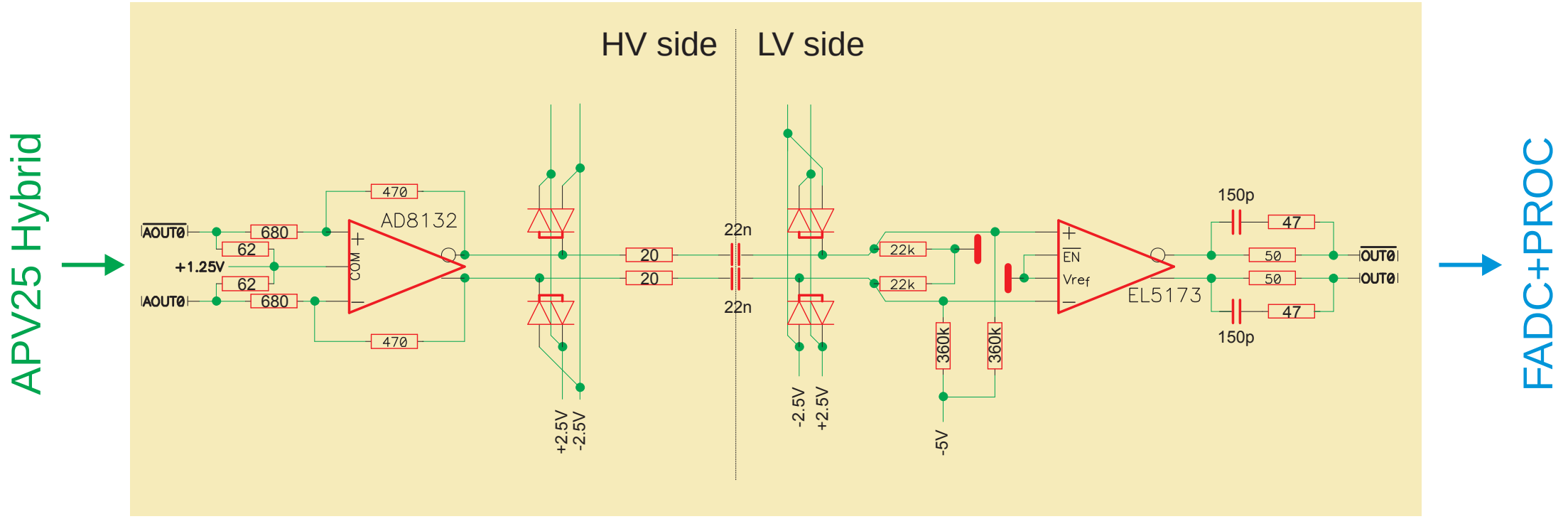

*Figure 5.27: Schematics of the analog voltage level translation (capacitive AC coupling) for the APV25 signal output.*

Figure 5.27 shows the circuit diagram for one APV25 channel as it is implemented in the repeater board. In the very center, there are 22 nF capacitors that block the DC bias level. They are surrounded by clamping diodes and damping resistors that protect the amplifiers. On the HV side, the AD8132 in fact is an attenuator rather than an amplifier and thus may be omitted to save space and power (to be tested). On the LV side, however, the EL5173 is essential, as it has a high input impedance and thus primarily acts as a buffer.
The capacitive coupling is realized as a high-pass filter circuit with a time constant of approximately 450 $\mu$s. Figure 5.28 shows the unaltered output of the high-pass filter as recorded by the FADC+PROC. In the idle state, the APV25 outputs are mostly at the digital baseline with a single tick mark (logic high) every 35 clocks. Thus, the DC level after the high-pass filter will settle slightly above that baseline. When data frames are sent, the channel pedestal values are significantly higher than the logic baseline and thus lead to an exponential discharge with the time constant of the high-pass filter. However, as the duration of six frames is very short (only 26 $\mu$s at a clock frequency of 31.8 MHz) compared to the RC time constant, this decay is barely visible. After the last frame, the baseline is slightly lower than before the data set and will recover with the same time constant.

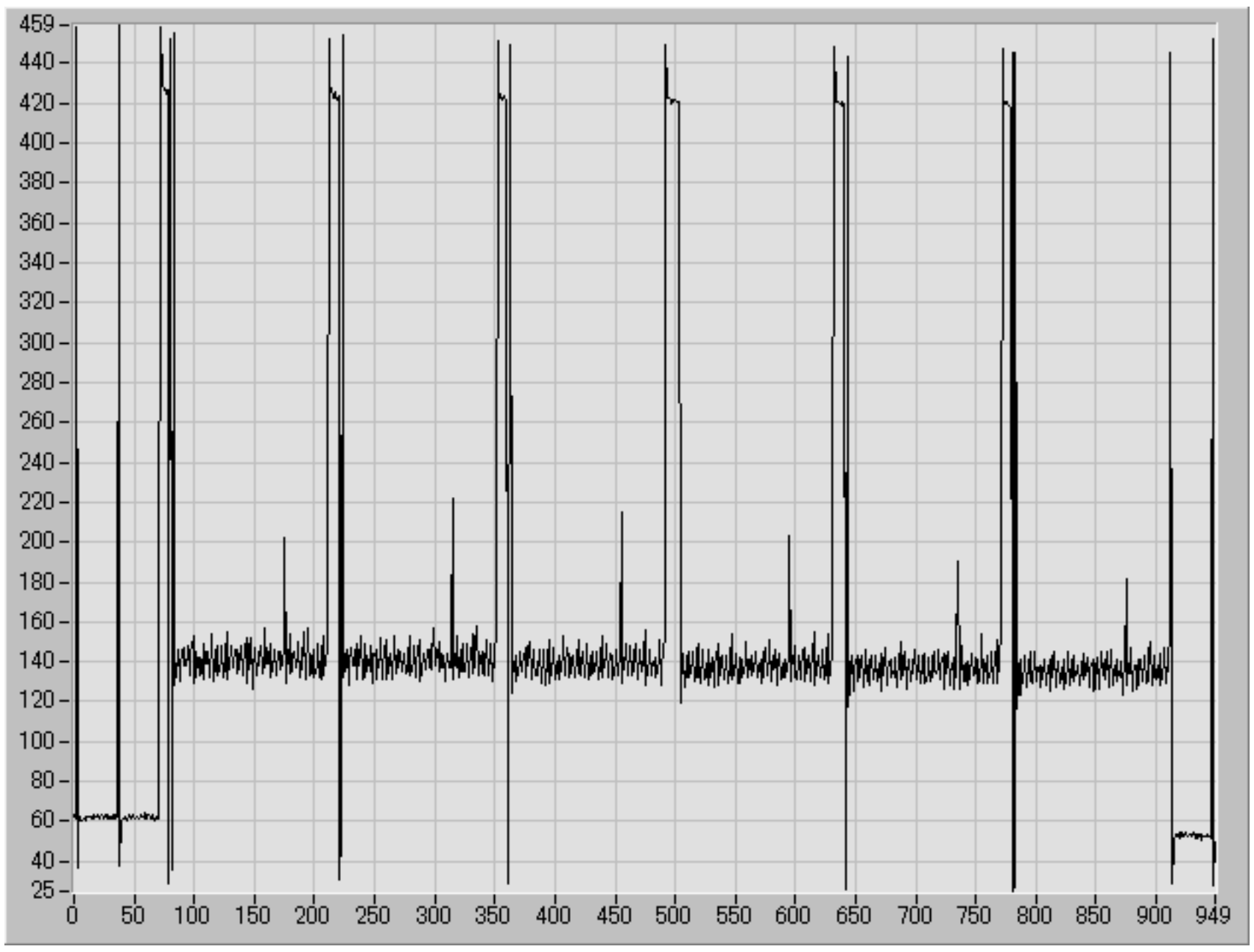

*Figure 5.28: Scope snapshot of the output of six consecutive APV25 samples embedded in data frames (abscissa: time in clock periods; ordinate: amplitude in ADC counts). One input channel has a particle hit and clearly reveals the shaped signal waveform.*

As the time constant is much larger than the duration of six frames, we can approximate the exponential decay by a straight line in that regime and thus the common-mode correction, which is performed after digitization in the FADC+PROC, compensates the individual offsets for each sample as long as the slope is (approximately) linear. In principle, the time constant of the high-pass filter could be further increased by using larger capacitors, but this also implies that more energy would be stored in those devices and thus jeopardize the functionality of the protection circuits. On the other hand, increasing the resistor value is also limited by the input impedance of the EL5173 amplifier that is placed in parallel. The present device values were found to be a good trade-off between the different requirements and were demonstrated to work very well in numerous tests both in the lab as well as in beam.

#### 5.4.4.2 Digitization

In the prototype system, dual flash ADCs (Analog Devices AD9218, rated for 65 MS/s) with parallel 10-bit output are used. As the number of channels per board will increase with the Belle II SVD FADC+PROC, it is intended to use devices with higher density and serialized output. One candidate for such an ADC is the AD9212, which accommodates eight independent channels in a package that is only slightly larger ($9 \times 9\,\mathrm{mm}^2$ compared to $8 \times 7\,\mathrm{mm}^2$ for the AD9218). In addition to the considerably higher device density, the space needed for lines on the PCB is also significantly reduced. On the other hand, as signal transmission happens at much higher speed, signal integrity considerations become much more important.

#### 5.4.4.3 Processing and Sparsification

Once digitized, the APV25 data streams are sent to one or more FPGAs, which will process them as shown in Fig. 5.29. In the beginning, there is a frame detection circuit such that the logic knows when a valid APV25 data block arrives. This is necessary as the APV25 data output is locked to its internal APSP clock, which is 1/35 of the external clock (which is also the frequency of tick marks appearing at the APV25 output in idle condition). Thus, the output data not only arrive with some delay after sending a trigger, but also with a jitter of 35 clock cycles. Once the header of a frame is detected, its contents and the subsequent channel data can be extracted. Due to the construction of the APV25 output multiplexer, the natural strip order becomes scrambled and needs to be disentangled. This is done by dual-port memories, which have a different address mapping for write and read operations so that the order is restored. There are two identical such memories for each input channel on the FADC+PROC which are mutually operated such that subsequent data frames can be processed independently.

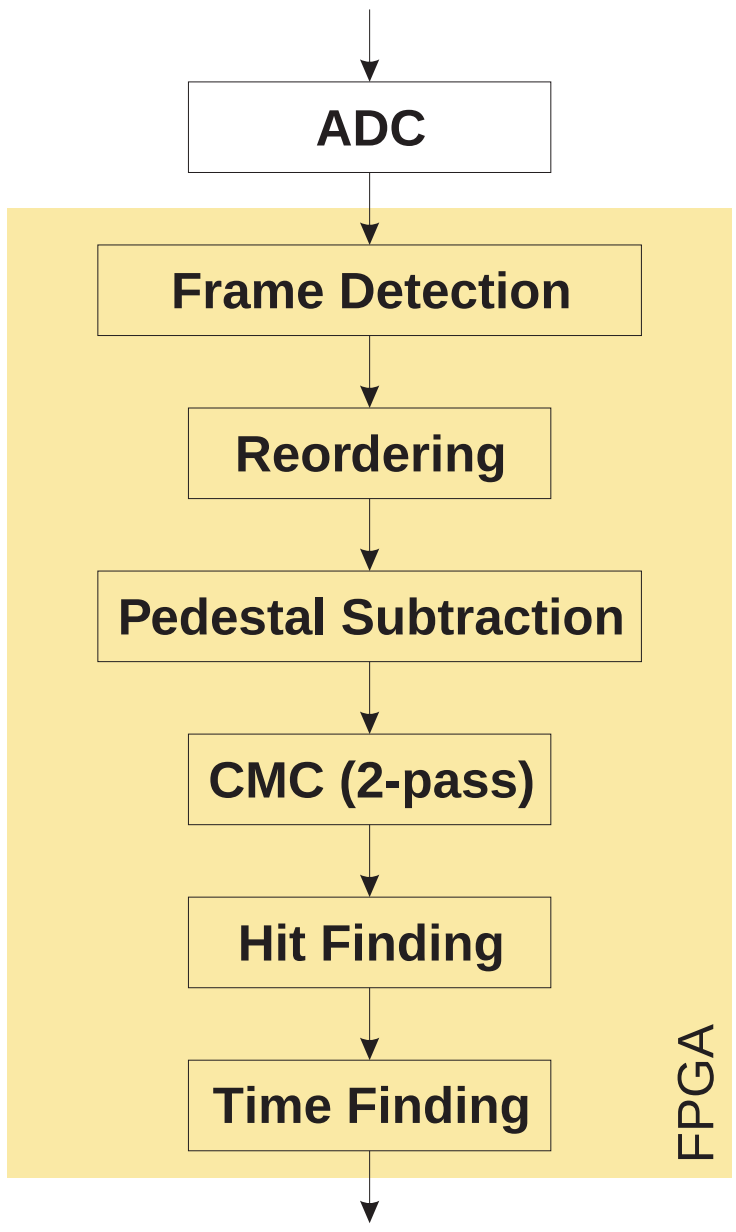

*Figure 5.29: Signal processing chain in the FADC+PROC.*

After reordering, several integer calculations happen. First, the pedestals of each strip, which are stored in memory, are subtracted from each strip's data value. Then, a common-mode correction is performed in two passes. Essentially, all strip signals below a certain stored threshold are summed and divided by their count. The resulting value is subtracted from each strip signal in the next step, where the same procedure is applied once again. Finally, each resulting strip value is compared to an individually predefined threshold and labeled as a hit if applicable in order to discard empty strips later on (zero suppression).

#### 5.4.4.4 Hit-Time Finding

By replacing the VA1TA readout chips of the SVD2 (800 ns shaping time) by the AVP25 for the Belle II SVD (50 ns shaping time), the sensitive time window and thus the occupancy is already considerably reduced. Depending on the actual shaper output waveforms and noise thresholds, the gain might not necessarily be exactly be 16, the ratio of shaping times. In fact, the time over threshold has to be considered in both cases, and was found to be 2000 ns vs. 160 ns, respectively, yielding an improvement factor of 12.5.
Depending on the estimated background in the Belle II SVD, this improvement alone should satisfy the requirement of having less than 10% occupancy even in the innermost layer. From Fig. 5.1, we can extract an occupancy projection of approximately 50% for slow VA1TA shaping at that radius using DSSDs from four-inch wafers. By scaling this figure to account for the six-inch wafer size, the number of readout strips, and the faster shaping time in the Belle II SVD, we estimate an occupancy of about 6.7%. This number includes a large uncertainty, as the actual Belle II background conditions are unknown.
To cope with higher occupancy than expected, we reduce the sensitive time window by employing the "multi-peak mode" (Sec. 5.4.1.2) of the APV25. Multiple samples along the shaped waveform allow for the off-detector reconstruction of the actual peaking time and, in conjunction with precise trigger timing, enable us to discard off-time background hits [15]. Measurements with prototype modules and readout electronics show that the RMS timing precision is of the order of a few nanoseconds (Sec. 5.7.3); thus, the sensitive time window can be set to about 20 ns, provided that the SVD system receives a precise trigger time with an accuracy of a few nanoseconds. This would allow an overall occupancy reduction of up to a factor of 100 compared to the SVD2, as shown in Fig. 5.30, depending on the precision of the trigger timing and the signal-to-noise ratio.
The hit time finding procedure has been verified in several beam tests using an offline numerical waveform fit with two parameters, peak timing and amplitude. Figure 5.31 shows an example for this waveform fit by two different methods. The exponential fit follows the theoretical output of the CR-RC shaper, which is $y \propto \frac{t}{\tau}\exp(-\frac{t}{\tau})$, and does not exactly represent the real waveform. The shape determined from internal calibration of the APV25 obviously fits better, because it has been obtained by measuring the shaper output when applying a well defined stimulus to each input. Amplitude and timing of that test signal can be varied to sample the full waveform with sufficient precision.
Using a numeric fit function gives high precision, but cannot be applied online because it would require a large amount of computing power and/or time. To accomplish hit-time finding in real-time, so that off-time background hits can be discarded immediately, the method has to be fast and reliable. FPGA logic and memory can be used to obtain the required information within a short and constant period of time.
The FPGA implementation of hit-time finding will first extract the three highest samples (triplet) of each measured set of six consecutive samples (sixlet). It has been shown that such a triplet around the peak contains the essential timing information and thus delivers almost the same precision with numeric fitting as the full sixlet. A predefined lookup table is used to find the peak timing and quality information (similar to the $\chi^2$ of numerical fitting) of the triplet data set. Different implementations of such a lookup table were evaluated by software and found to give similar results close to the output of a numeric fit applied to all samples. We assume that clustering before hit finding will improve the results, especially with tilted hits where the cluster signal scales with the number of strips while the noise only increases with the square root, leading to a higher overall cluster signal-to-noise compared to individual strips. (This needs to be confirmed by actual measurements.) In any case, the FADC+PROC output will contain individual strip signals, as those will later be needed to calculate the exact hit position by means of center-of-gravity or other methods.
The contents of the lookup table for hit-time finding can, in principle, be generated by software. However, it is important to include noise fluctuations, since the measured sixlets will be contaminated by random excursions. The best method to obtain the data is to use the internal calibration of the APV25, where particle signals are emulated by artificial stimuli, as mentioned above. With high statistics of triplets recorded in this mode, it is possible to construct a lookup table with intrinsic noise variations as these appear during calibration in the same way as they happen with particle signals. If needed, extra noise can be added by software to further broaden the distributions and thus ensure that all possible triplets are covered by the lookup table.

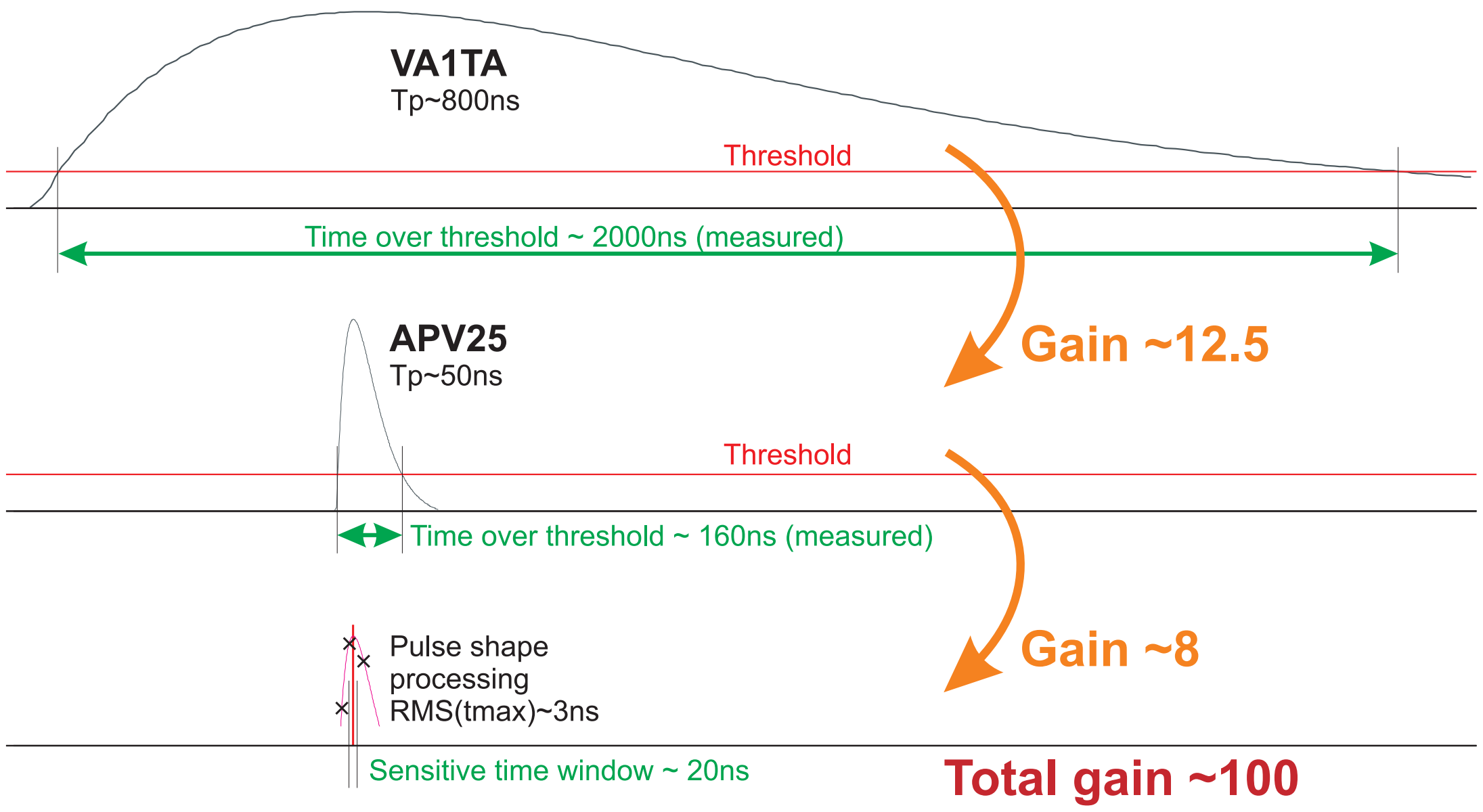

*Figure 5.30: Occupancy reduction potential with the APV25 and hit time finding.*

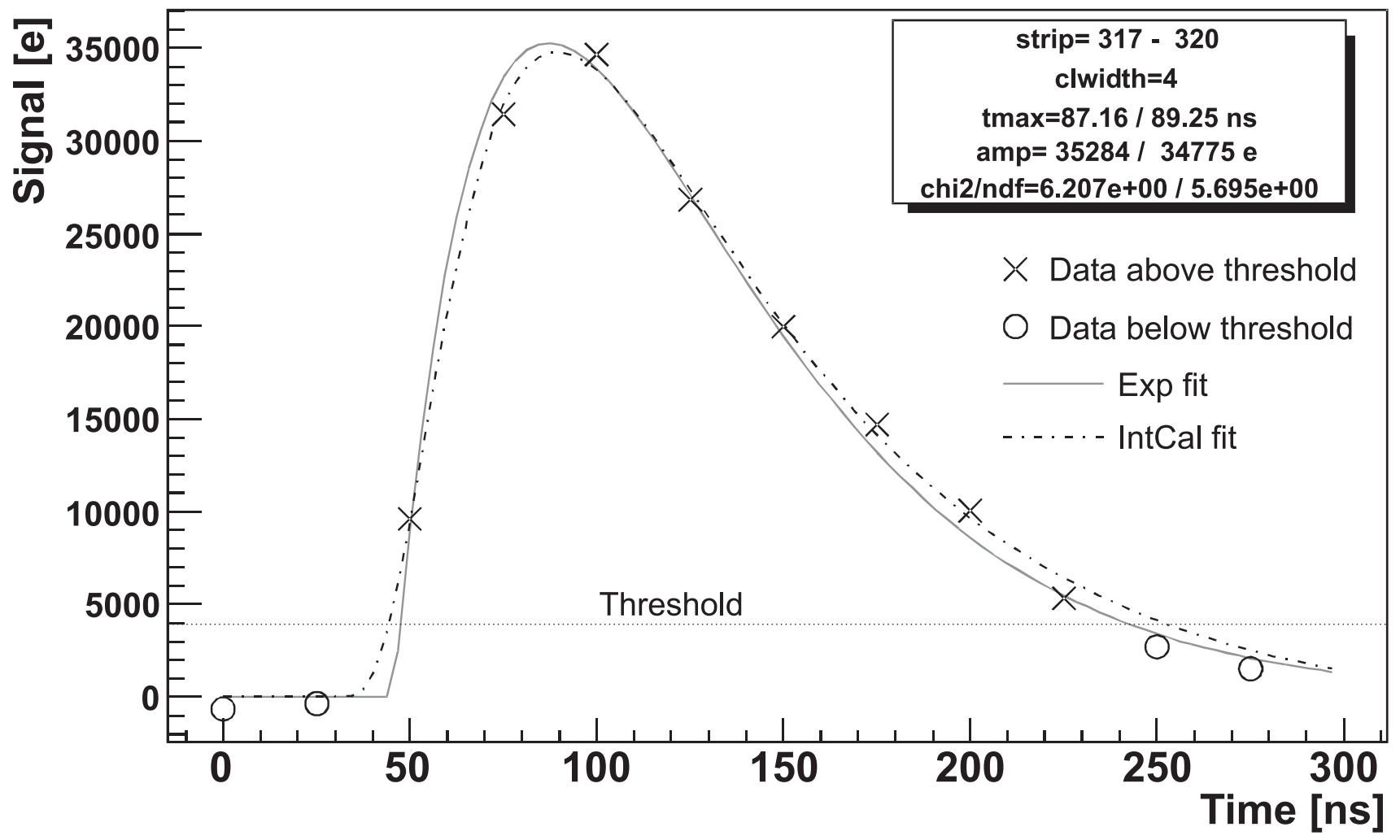

*Figure 5.31: Hit time finding by a numerical fit using two different functions. Apparently, the IntCal fit delivers a better result.*

#### 5.4.4.5 Data Rates

The output data rate depends on various conditions, but the system must be designed to handle the worst case. We assume an instantaneous luminosity of $8 \times 10^{35}\,\mathrm{cm}^{-2}\mathrm{s}^{-1}$ and an average trigger rate of 30 kHz for the following considerations.
Referring to Fig. 5.1, we extract an occupancy of 6.7% in the innermost layer. (See Sec. 5.4.4.4 for additional information.) For outer layers, we extrapolate this number using an $1/r^2$ law and adjust to the larger sensor size and number of APV25 chips for readout. Following these calculations, Table 5.8 shows the number of struck strips for each layer.

| Layer | Radius (mm) | Occupancy | RO chips | RO strips | Hit strips |
|---|---|---|---|---|---|
| 6 | 140 | 0.9% | 850 | 108800 | 967 |
| 5 | 115 | 1.3% | 560 | 71680 | 944 |
| 4 | 80 | 2.7% | 300 | 38400 | 1045 |
| 3 | 38 | 6.7% | 192 | 24576 | 1647 |
| Average / Sum | | 1.9% | 1902 | 243456 | 4602 |

*Table 5.8: Expected worst case occupancy for the Belle II SVD. “RO chips” stands for the APV25 and “RO strips” is the previous number multiplied by 128 strips per APV25.*

We consider four levels of data processing related to hit time finding:

1. no hit time finding, 6 samples per hit
2. no hit time finding, 1 sample per hit
3. online reduction by hit time finding, 6 samples per hit
4. online reduction by hit time finding, 1 sample per hit

In the first and third cases, sixlets of all found hits are transferred to the output. Otherwise, only the peak sample of each hit is propagated, which should normally be sufficient for further analysis. The latter two options include the possibility that off-time hits are immediately discarded in the FADC+PROC and thus are not propagated. Consequently, the last case provides the smallest data set.
Even though the data format is not yet defined, we can assume that 12 bytes are needed to transmit a full sixlet, while 4 bytes are sufficient for a single sample. The data are zero-suppressed in both cases, such that the position information is included in these data sizes. The size of the overhead (such as header and trailer) can be neglected for the given conditions.
Table 5.9 summarizes the resulting data rates for each of the four cases mentioned above. When online data reduction is performed by hit-time finding, we assume that the average occupancy of 1.9% is reduced by a factor of five. This includes a safety margin compared to the gain of (up to) eight, described in Sec. 5.4.4.4. For these calculations, we recall that the Belle II SVD has 80 FADC+PROC boards in total, each with its own data link to COPPER/FINESSE (Sec. 5.4.5).

| Case | Average occupancy | Data size/ channel (B) | Event size (B) | Total data rate (B/s) | Data rate/ link (B/s) |
|---|---|---|---|---|---|
| 1 | 1.9% | 12 | 53.9k | 1.54G | 19.7M |
| 2 | 1.9% | 4 | 18.0k | 527M | 6.6M |
| 3 | 0.4% | 12 | 10.8k | 316M | 3.9M |
| 4 | 0.4% | 4 | 3.6k | 105M | 1.3M |

*Table 5.9: Expected worst case data rates for Belle II SVD, depending on the level of online data processing.*

Assuming that the full sixlet information is not needed in the further data analysis, we can specify case 2 as the baseline design with the improvement potential of online hit time finding, leading to a reduced data rate as stated for case 4.

### 5.4.5 COPPER/FINESSE Data Transfer

The main view of the data stream propagation from the FADC+PROC to the central DAQ is shown in Fig. 5.32.

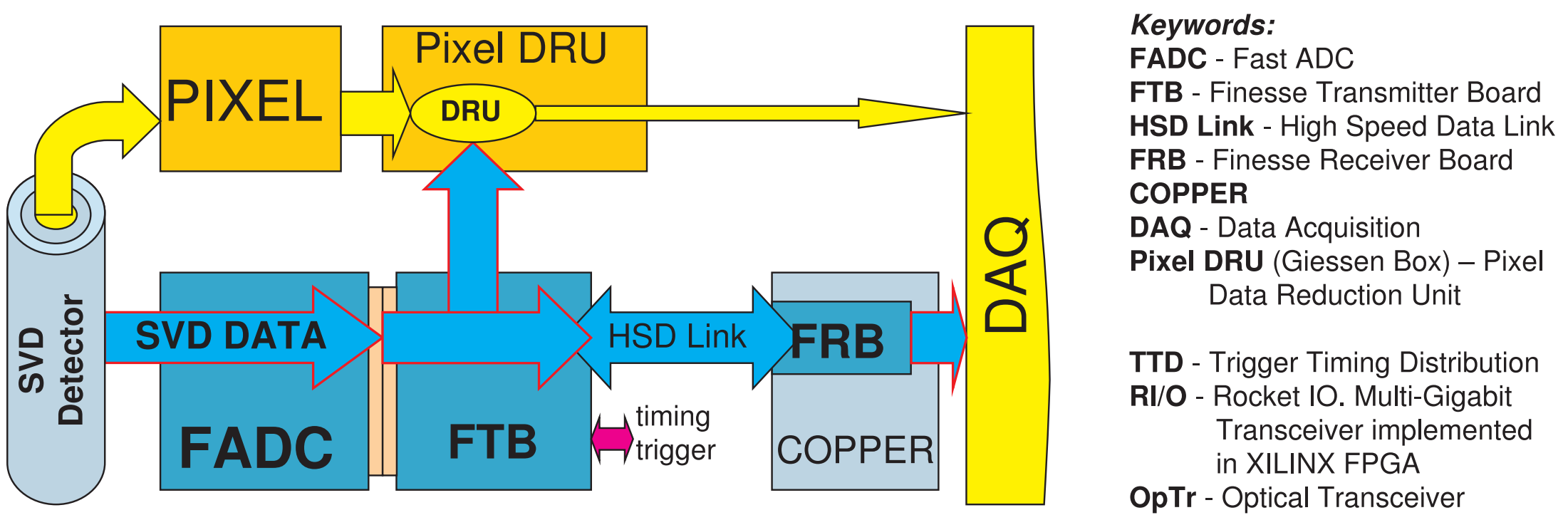

*Figure 5.32: SVD main data stream.*

The dedicated FTB (Finesse Transmitter Board) module is the main device for data streaming.

Its functionality is shown in Fig. 5.35. A general description of the two data streams—the main stream to the DAQ and the additional one to the Pixel DRU (Data Reduction Unit, a.k.a. "Giessen box")—is provided below. A description of the other signals from/to the FTB can be found in Sec. 5.4.6.

#### 5.4.5.1 Link to DAQ

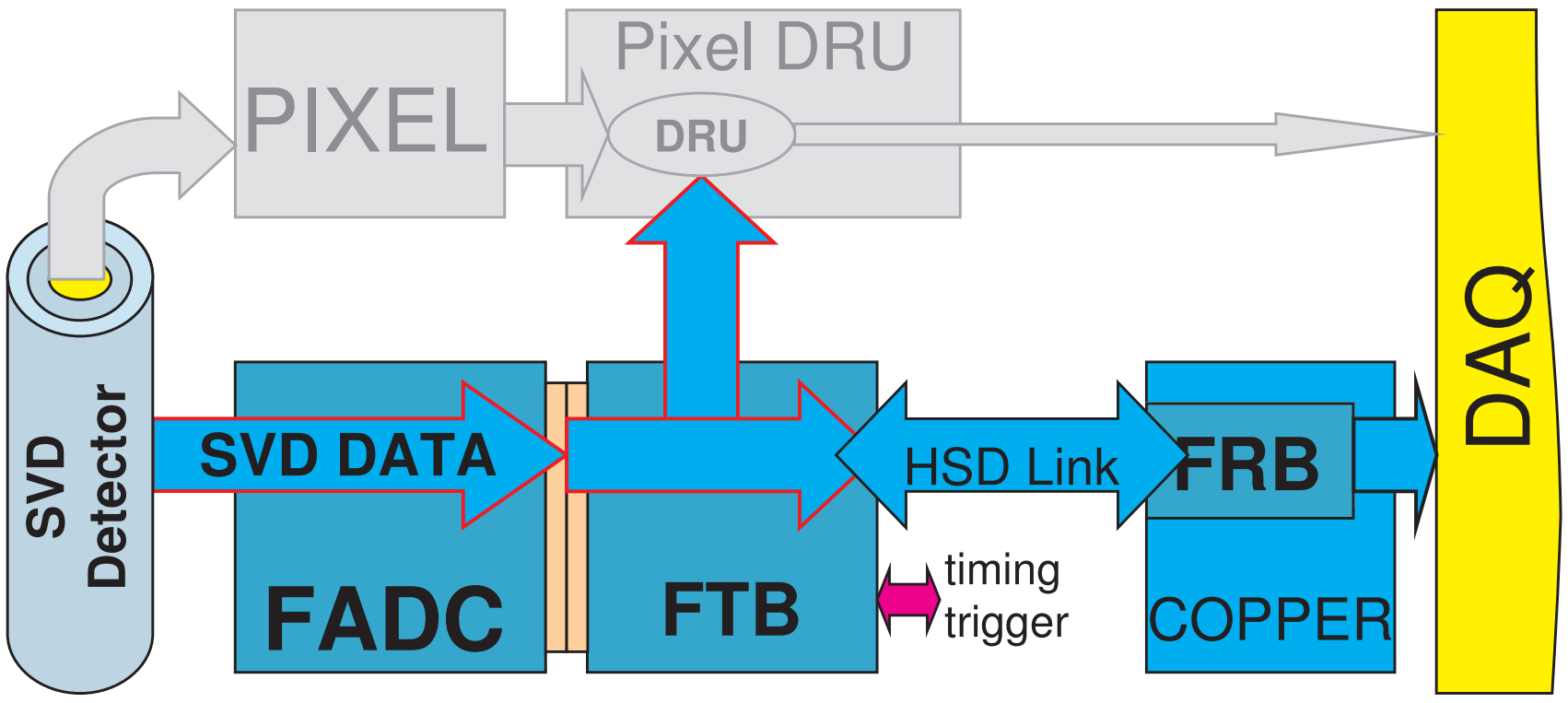

*Figure 5.33: Link to DAQ.*

The main data stream contains the Belle II SVD detector readout data sent from the FADC+PROC to the DAQ (Fig. 5.33). Data are prepared in the FADC+PROC as a parallel, four-byte-wide bus, clocked at 31.8 MHz (or 42.4 MHz) and sent to the FPGA on the FTB. This Xilinx device performs CRC input data checking and adds additional information such as CRC for the output data. Then, the prepared data are written to a FIFO buffer at 42.4 MHz. A Rocket I/O module (implemented inside the FPGA) reads out data, serializes and sends them via a fast optical link to the FINESSE Receiver Board (FRB) in the COPPER module. Another Rocket I/O FPGA module on the FRB converts back the data to a parallel four-byte bus, which is then written to a FIFO on the COPPER board. A processor on COPPER reads the data and transfers it to the central DAQ via Ethernet link.
The High Speed Data (HSD) link connects the FPGA Rocket I/O on the FTB to the Rocket I/O on the FRB in the COPPER board. It is a universal part common for the majority of Belle II subdetectors. Eighty sets of FTB boards, one for each FADC+PROC module, are used to transfer the information of the whole Belle II SVD. At the other end of the HSD, one COPPER board can hold up to four FRBs, so that at least 20 modules are needed. However, the bandwidth of COPPER is limited by its PCI bus, so each HSD link might terminate into its own COPPER board; thus, up to 80 COPPER boards might be needed.

#### 5.4.5.2 Link to Pixel Data Reduction Unit

A large amount of data is read out by the pixel detector (PXD), because it has a long integration time. The PXD group plans to use the SVD data (which, in contrast to the PXD, have a very fine time granularity but less position sensitivity) to discard the majority of off-time hits in the PXD. A dedicated Data Reduction Unit (DRU) is intended to fulfill this purpose, and the SVD data are sent not only to the DAQ, but also to the DRU in parallel (Fig. 5.34). Based on tracks found in the SVD data, this device will identify regions of interest in the pixel detector and propagate only the relevant data to the DAQ, discarding other hits.

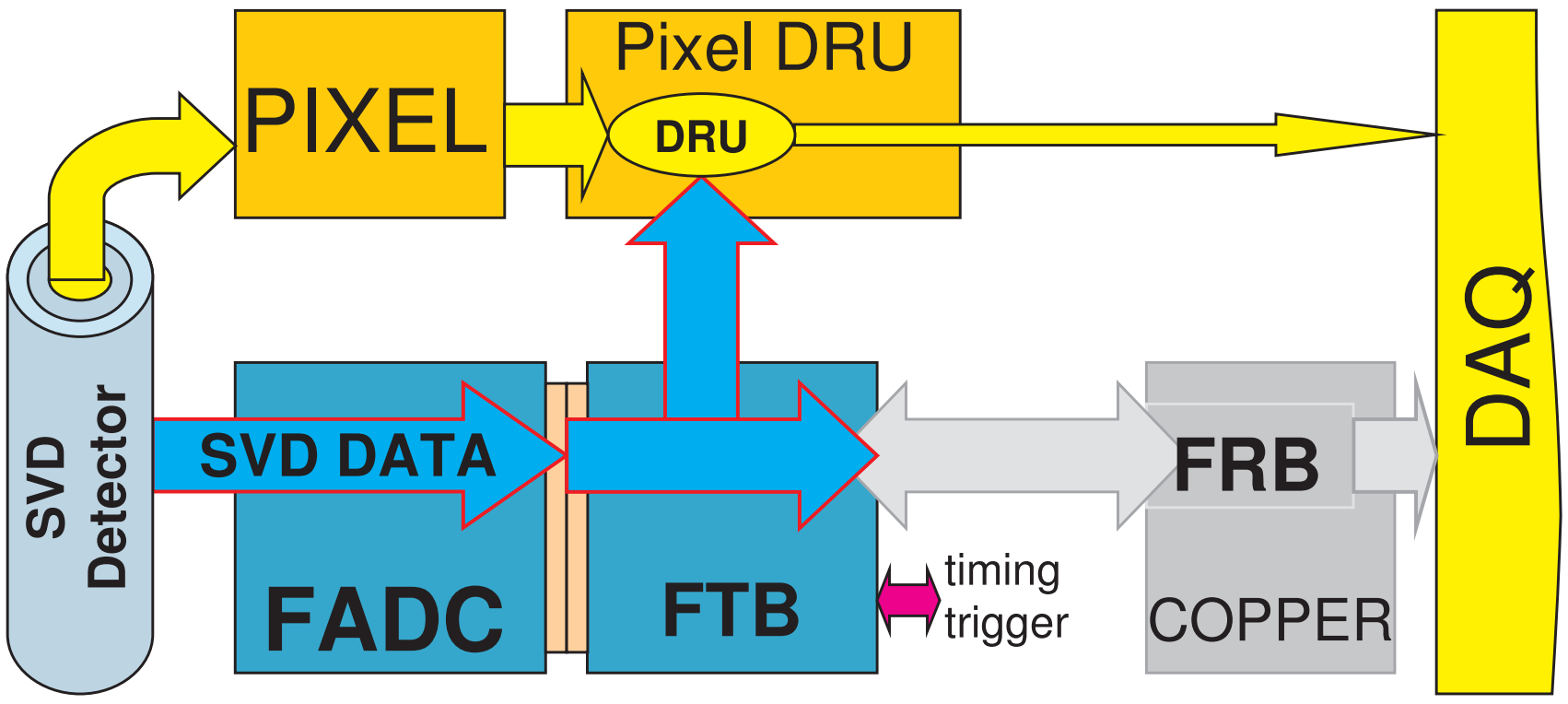

*Figure 5.34: Link to pixel data reduction unit.*

The data stream from the FTB to the pixel DRU is unidirectional without any possibility for (partial) re-sending in case of bad transmission. The only possibility of data quality checking is the CRC code appended to each data block. For transmission, a fast optical link (also Rocket I/O in FPGA) will be used.

### 5.4.6 Clock, Trigger and Control

There are basically two options for the distribution scheme of clock, trigger and controls. Either these signals are propagated to each FTB and sent to all FADC+PROC boards from there, or they are passed on to the central FADC Controller module and distributed from there. The latter option appears preferable in the sense that not only the global clock, trigger, and controls need to be propagated to each FADC+PROC board, but so do SVD-internal control signals (such as $I^2C$ slow controls). Moreover, only that scheme allows consistent timing throughout the system and also allows a central trigger throttling logic to be included in the controller. On the other hand, a low jitter clock will be needed by the FTB boards to be used for serialization of the data stream, which calls for an immediate clock distribution to each FTB. Thus, the final system will be a mixture of both options, which also provides some freedom in choosing one or the other path.
The global control signals are RCLK (127 MHz), Trigger, TAG[7:0] (trigger number) and GRST (global reset). In addition, a ready signal and the status of FADC+PROC and FTB modules are sent to the Trigger Timing Distribution (TTD) system. Differential LVDS levels will be used for transmission.
Inside the FTB, signals from the TTD are divided into two streams. One stream is connected directly to the differential inputs of the FPGA. The other goes through buffers to each FADC module (shown on Fig. 5.35) or to the FADC Controller. The same FTB board will be used for connection to both FADC and Controller modules, but the DAQ part firmware of the latter will be disabled and no optical transceivers need to be equipped there.

#### 5.4.6.1 FADC Controller

A total of 80 FADC+PROC boards are used for the Belle II SVD, spread over five crates. In order to control all those distributed boards, a unique controller module is used that connects to a custom backplane in each crate, thus reaching every FADC+PROC module. The task of the controller is to distribute configuration signals as well as clock and trigger signals. The latter is not the same as received from the TTD system, but needs to be refined before sending it to the APV25 chips at the front-end. Moreover, local (SVD-internal) triggers may be sent, for example, to perform pedestal and noise evaluation or internal calibration. The APV25 clock can either be 31.8 MHz or 42.4 MHz, but the latter is needed in either case by the FTB boards. Both clocks are distributed by the controller to all FADC+PROC modules, and the APV25 clock is selected there according to a configuration register.
The controller unit is also the natural location for throttling the trigger to prevent a buffer overflow of the APV25 when there are too many triggers pending for readout. Such a solution was also adopted in CMS, where a special module called the “APV Emulator” has an FPGA implementation of the APV25 internal state machine and thus provides real-time information about the current APV25 condition. As all APV25 chips operate synchronously, it is sufficient to have one single emulator that can throttle further triggers in a busy condition.
Moreover, such an “APV Emulator” can be used to predict the header of each APV25 data frame, depending on the history of reset (synchronization), clock and trigger signals. Every header contains an encoded 8-bit pipeline address that denotes the APV-internal storage cell number from which the data were taken. As this is the only time-dependent piece of the data frame, it should be verified in order to check that a certain data frame belongs to a particular trigger. Otherwise, it would be almost impossible to detect missing or excessive frames in the continuous (dead-time-free) DAQ scheme and those would obviously spoil all subsequent data.
The controller also sends configuration information to each FADC+PROC board and, through those, to the front-end chips. It also distributes fast controls such as enable signals for digitization or requests for copying data blocks to spy memories. Finally, it measures the precise trigger timing (probably obtained by a direct link from the global decision logic, GDL) and propagates that information to each FADC+PROC to (optionally) perform online data reduction.
The design of the FADC Controller module is based on the “NECO” (New Controller) and “SVD3_Buffer” boards as well as the backplane of the existing prototype readout system (Sec. 5.7.2), which essentially cover the same tasks.

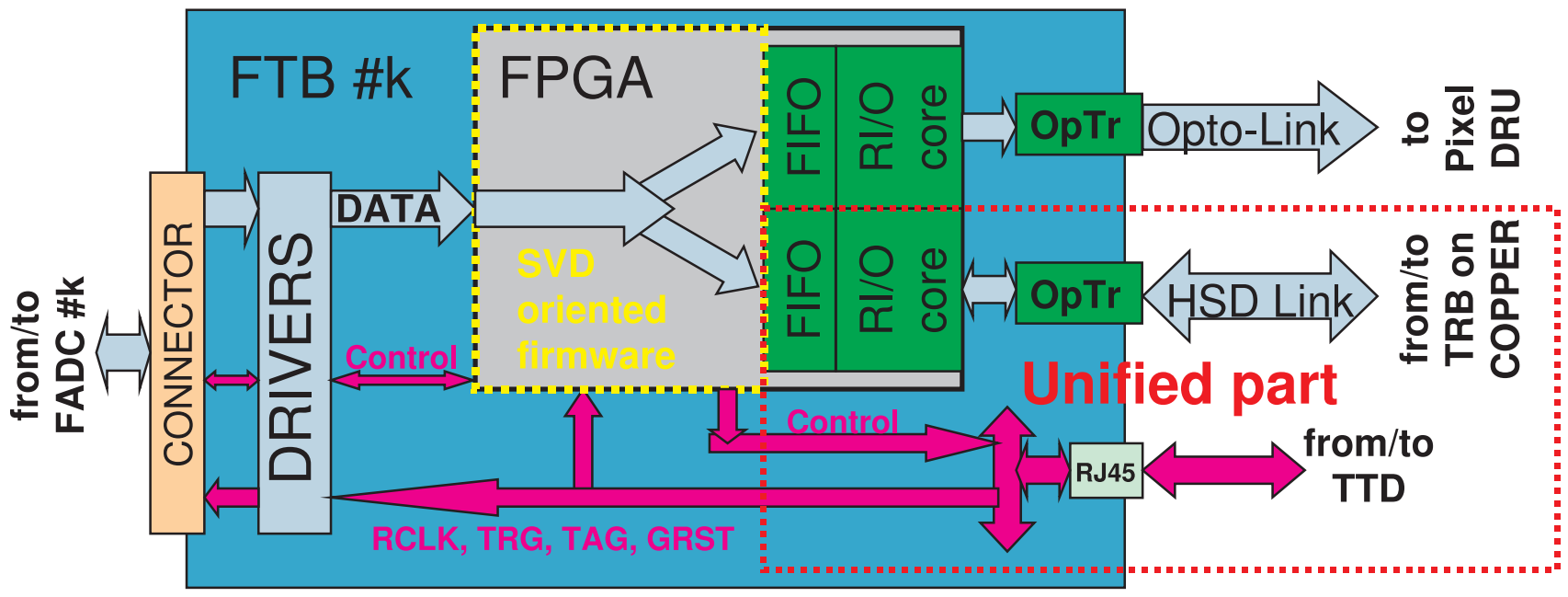

*Figure 5.35: Generic FTB module with full DAQ and TTD paths. Depending on whether the FTB is used with an FADC or the FADC Controller, only one of those paths will be used.*

## 5.5 Environmental Monitoring

The environment monitor for the SVD consists of temperature, humidity and instantaneous radiation sensors. The monitors for low- and high-voltage supplies are built into the power supply units. The leakage current monitor for each DSSD sensor resides inside the readout electronics.

### 5.5.1 Temperature and Humidity

The APV25 chips in the Belle II SVD will consume about 700 W in total at full operation. The stable operation of the cooling system is essential to maintain high performance of the detector system.
As the readout chips are mounted on the DSSD sensors with Kapton flex circuits, a temperature sensor should be included on each hybrid and read out by the front-end readout system.
To prevent failure of the cooling operation, a redundant interlock on hardware level will be prepared and installed at a very early stage of the SVD commissioning.
Moreover, it is important to ensure a dry atmosphere, as the ambient temperature in the SVD volume will be lower than the dew point in the experimental hall (up to 13° C in the past). This will be done by several humidity sensors placed in suitable locations to be determined once the mechanical design of the SVD is frozen. As humidity sensors are bulkier than the temperature probes, it is not useful to mount them on each hybrid, but rather place a few of them in the endring regions.

### 5.5.2 Instantaneous Radiation

The instantaneous radiation monitor protects the SVD and PXD from dangerous radiation damage and diagnoses the background sources under normal operation. Silicon PIN diodes with $4 \times 6 \times 0.3\,\mathrm{mm}^3$ volume, which were used in the Belle IR region, are capable of measuring dose levels above 0.1 mRad/s. The characteristics of PIN diodes are, however, sensitive to the total dose. To perform a reliable measurement of dose level over several years of operation, we use diamond sensors, which are very robust against radiation effects.
The primary purpose of the instantaneous dose measurement is accomplished by generating a beam-abort request to the accelerator control when the dose rate exceeds a threshold level in a very short time. As the revolution period of SuperKEKB is $10\,\mu$s, an extremely fast response is not necessary.

### 5.5.3 Integrated Dose

In the harsh radiation environment of SuperKEKB, the integrated dose monitor is necessary to predict the lifetime of the detector system. In the SVD2, RADFETs [16, 17] are used for this purpose. It is known, however, that the effect of integrated dose depends on the radiation dose rate, the particle type (electron, photon, proton, or neutron), and energy. This is true even for the sensors of the radiation monitor. Therefore, it is best if the integrated radiation effect can be monitored by the sensor itself. In the Belle II SVD, the leakage current of each sensor is monitored and used to calculate the total radiation dose.

### 5.5.4 Detector Leakage Current

The detector leakage current and the full depletion voltage are important properties in the lifetime of a DSSD sensor. As each DSSD sensor is read out by a hybrid, the individual detector currents can be monitored. The full depletion voltage can be estimated by the bias voltage dependence of the leakage current and the noise level of the APV25 output.
The individual detector current of each DSSD is permanently measured by a dedicated circuit on each FADC+PROC module. As the high voltages are split into positive and negative supplies, two readings are available for each sensor on different FADC+PROC boards. Occasionally, high voltage sweeps are performed to obtain the depletion voltage by measuring the noise level of the APV25 chips. Alternatively, the full depletion can also be obtained by the amplitude of particle (or cosmic) signals as a function of the bias voltage.

## 5.6 Power Supplies

The low voltage power supply (LVPS) levels for the readout chips are put on top of the detector bias voltages, so that the AC coupling in the DSSD do not have to withstand the full bias voltage. In exchange for the complex power supply design, the probability of pin-holes is kept low and, if pin-holes appear, the risk of malfunctioning of the APV25 chip is reduced significantly.
Achieving this goal requires several independent LVPS units with relatively small current. Table 5.10 shows the properties of the power supply crate that was used for the SVD2, which consists of 11 independent LVPS units and two bias voltage channels. Each unit has individual voltage/current setting, interlock control and monitors. As we have 10 such crates, we have 110 independent LVPS available.

| **Voltage (V)** | **Current (A)** | **Units /crate** | **Total current (A)** |
|---|---|---|---|
| 10 | 10 | 2 | 20 |
| 10 | 5 | 9 | 45 |
| 100 | 0.001 | 2 | 0.002 |

*Table 5.10: Properties of the SVD2 power supply system.*

As mentioned already in Sec. 5.4.4, the Belle II SVD power supplies should ideally be located next to the FADC+PROC crates (see Fig. 5.26 for potential locations). This eases the three-fold connectivity between power supplies, FADC+PROC boards and front-end and reduces the possibility of ground loops.

### 5.6.1 Low Voltage Power Supply (LVPS)

The APV25 chip is operated with 2.5 V and 1.25 V supplies. According to [18], the typical supply currents for 2.5 V and 1.25 V are 0.116 A and 0.052 A, respectively. The total supply currents are summarized in Table 5.11. From this table, we find that a LVPS system similar to that used in the SVD2 can satisfy the requirements.

| **Layer** | **# of ladders** | ***p*-side Chips/ladder** | ***p*-side total** | ***n*-side Chips/ladder** | ***n*-side total** |
|---|---|---|---|---|---|
| 3 | 8 | 12 | 96 | 12 | 96 |
| 4 | 10 | 18 | 180 | 12 | 120 |
| 5 | 14 | 24 | 336 | 16 | 224 |
| 6 | 17 | 30 | 510 | 20 | 340 |
| Total | | | 1122 | | 780 |
| Total current @ 2.5 V | | | 131 A | | 91 A |
| Total current @ 1.25 V | | | 59 A | | 41 A |

*Table 5.11: Number of APV25 chips and necessary supply currents.*

The digital clock transmitters and analog signal receivers on the FADC+PROC are also supplied by floating low voltages, which can be supplied from the same LVPS system or be provided by DC-DC converters placed on the FADC+PROC boards in order to avoid ground loops and reduce the potential power loss in long supply cables.

### 5.6.2 High Voltage Power Supply (HVPS)

The main component of the radiation background in the Belle II SVD from SuperKEKB arises from electromagnetic showers. As a result, the DSSD sensors suffer from surface damage: the detector leakage current increases but saturates at less than $1\,\mu\text{A/cm}^2$, while the full depletion voltage does not change significantly. Thus, the HVPS system for the SVD can be designed to deliver power to the sensors corresponding to at most 10 mA at $\pm 100$ V. However, to cope with variation of damage and characteristics of DSSDs, the HVPS system is divided into several independently controllable units. Such products are available on the market.
In order to control the SVD performance, a reliable HVPS and monitoring of bias voltages and leakage currents are necessary.

## 5.7 Prototyping

Several years ago, an intermediate upgrade of the SVD2 was planned for Belle, to cope with the expected boost in luminosity and SVD2 strip occupancy associated with the installation of the crab-crossing scheme in KEKB. Consequently, a replacement of the innermost layer by modules with APV25 readout was envisaged and initially called SVD2.5. Later, this plan was extended to the second layer and named SVD3. In the end, the luminosity increase with crab cavities was less than expected and, thus, the intermediate vertex-detector upgrade was suspended.
Nonetheless, significant R&D effort was put into the planned upgrade and this is now proving useful for the design of the Belle II SVD. In fact, many building blocks devised for SVD2.5 and SVD3 will be reused for Belle II. These developments include several double-sided sensors and readout modules of both conventional as well as chip-on-sensor type. Some of those modules are presented in Sec. 5.7.1. Moreover, two readout systems were constructed: a compact one for lab and beam tests and a larger, modular one that was intended for the SVD3. The Belle II SVD readout system resembles the latter.

### 5.7.1 Sensors and Modules

In the course of our R&D effort during the past six years, we built several sensor modules with different configuration and evaluated them at various beam tests. Below, the design of the three most interesting modules, the JP_Module, the Flex_Module, and the recently finished Origami module prototype, are described. The performance of those modules is discussed in Sec. 5.7.3. For all three modules, APV25 readout and the same type of DSSDs were used. The sensor is made from a four-inch wafer, has a size of $79.6 \times 28.4\,\text{mm}^2$, and was originally designed for the intermediate SVD2.5 upgrade. The readout pitch of $p$-side (long) strips is $50\,\mu$m, while that of $n$-side (short) strips is 152 $\mu$m. On each sides, a sensor has 512 readout strips plus 511 intermediate strips.
The so-called *JP_Module* (Fig. 5.36) is based on the design used for SVD2, where up to three sensors are ganged (concatenated) and read out from the edges of the ladders. The "JP" in the name stands for Japanese design. This module consists of two sensors, whose strips are concatenated by wire bonds on the $p$-side and single layer Kapton flexes on the $n$-side. Aiming to compare the signal-to-noise ratio between a single and two ganged sensors, we only ganged 384 of the 512 strips on each side, which means that one APV25 chip is attached only to a single sensor, while the other three have both detectors attached to their inputs.

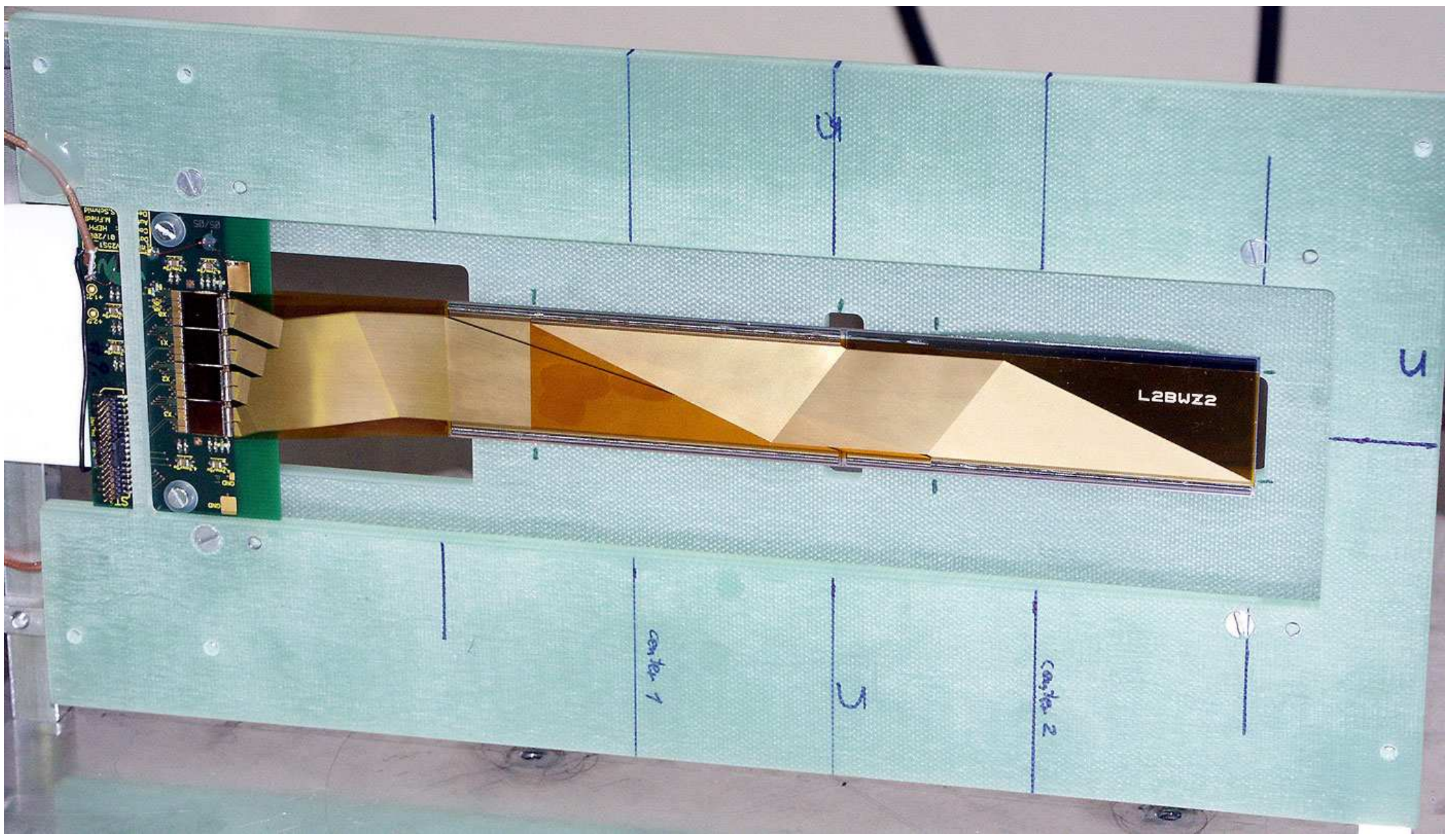

*Figure 5.36: JP_Module made of two ganged double-sided four-inch sensors.*

We introduced the chip-on-sensor concept with the *Flex_Module* (Fig. 5.37), where the readout chips are placed onto a thin double-layer flexible circuit placed on top of the sensor. As described in Sec. 5.3.1.1, this method is required to achieve a reasonable signal-to-noise ratio when fast front-end amplifiers are used for readout. In the case of the Flex_Module, the chip-on-sensor concept is only implemented on the $n$-side, while the $p$-side is read out by a conventional hybrid attached to the edge of the sensor. Moreover, we used a relatively thick carbon fiber tube for cooling, which has more material budget than desirable. Nonetheless, the main goal of this module was to demonstrate the feasibility of the chip-on-sensor concept with APV25 readout.
Encouraged by the excellent performance of the Flex_Module (Sec. 5.7.3) and keeping the material budget in mind, we wanted to extend the chip-on-sensor idea to both sides of a DSSD without doubling everything. This resulted in the Origami chip-on-sensor concept (Sec. 5.3.1.1), where both sides can be read out by a single three-layer hybrid placed on the sensor face with the short strips. The connection to the long strips of bottom side is done by Kapton fanouts that are wrapped around the edge of the sensor. By building the *Origami module* in 2009, we have shown that the concept [19] is feasible, even though its assembly is a complex task. Thanks to the use of micro-positioners equipped with a custom vacuum nozzle as shown in Fig. 5.38, wrapping the flexes around the sensors edge turned out to be surprisingly easy. A detailed description of the assembly procedure can be found in Ref. [11]. The final four-inch wafer Origami module prototype is depicted in Fig. 5.39. A simple, 5 mm high rib made of G10 was used as the support structure. In contrast to the Flex_Module, a thin aluminum pipe with a diameter of 2 mm and a wall thickness of 0.2 mm was used for cooling, which is much closer to the final design.
In spring 2010, a first batch of six-inch wafer DSSDs, designed for the outer barrel layers (see Sec. 5.2.1) of the Belle II SVD, were delivered by HPK. In parallel, a final-size prototype of the central Origami hybrid of the outermost layer (i.e., the longest of all variants), was produced by the Japanese Taiyo Industrial Co., along with pitch adapters to be bent around the edge of the sensor. In Fig. 5.40, these parts are arranged in the way they will be put together. The assembly of such a single-sensor module is underway.

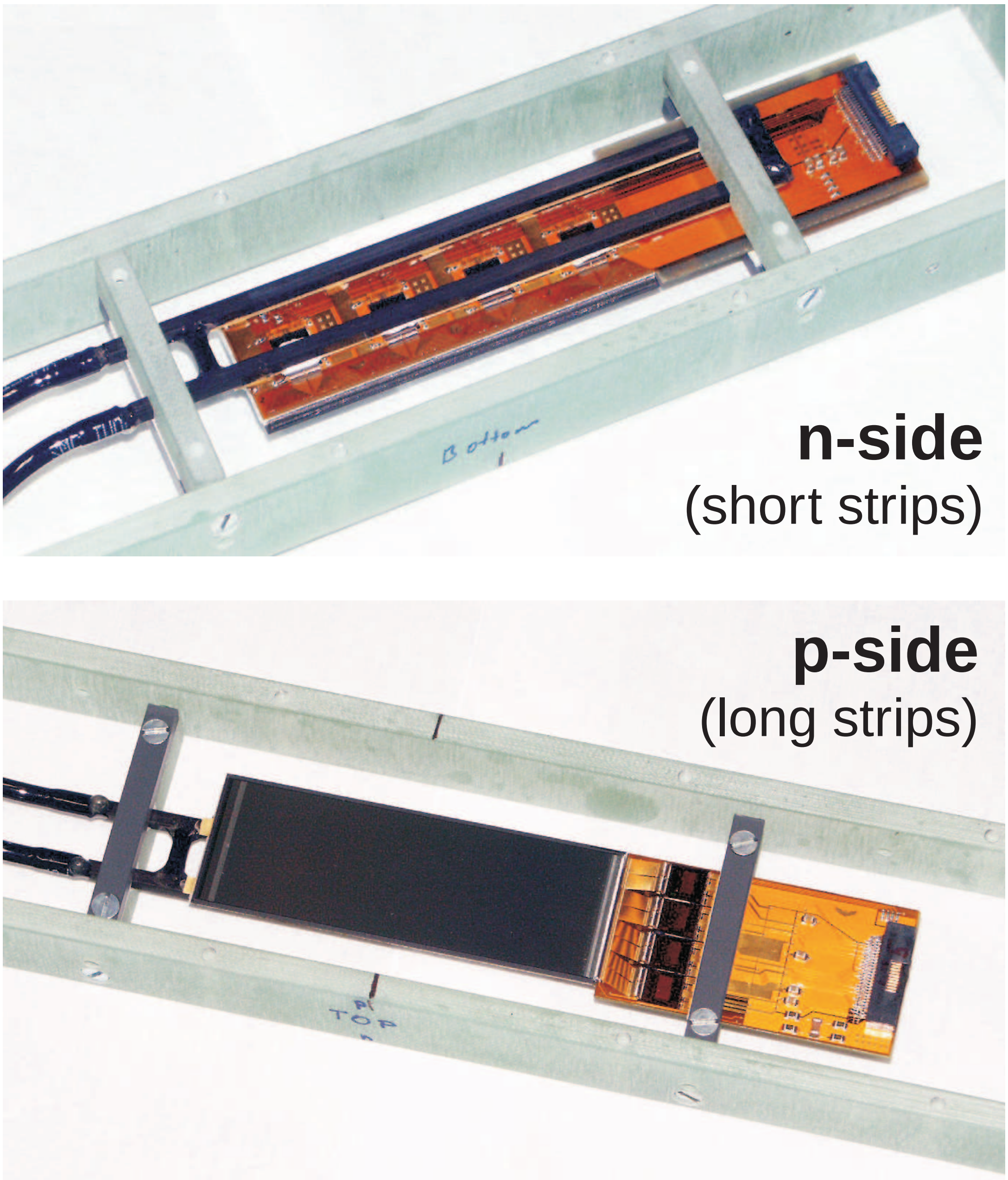

*Figure 5.37: Top and bottom views of the Flex_Module prototype. The n-side (top) is built according to the chip-on-sensor concept, while the p-side (bottom) is conventionally read out from the edge.*

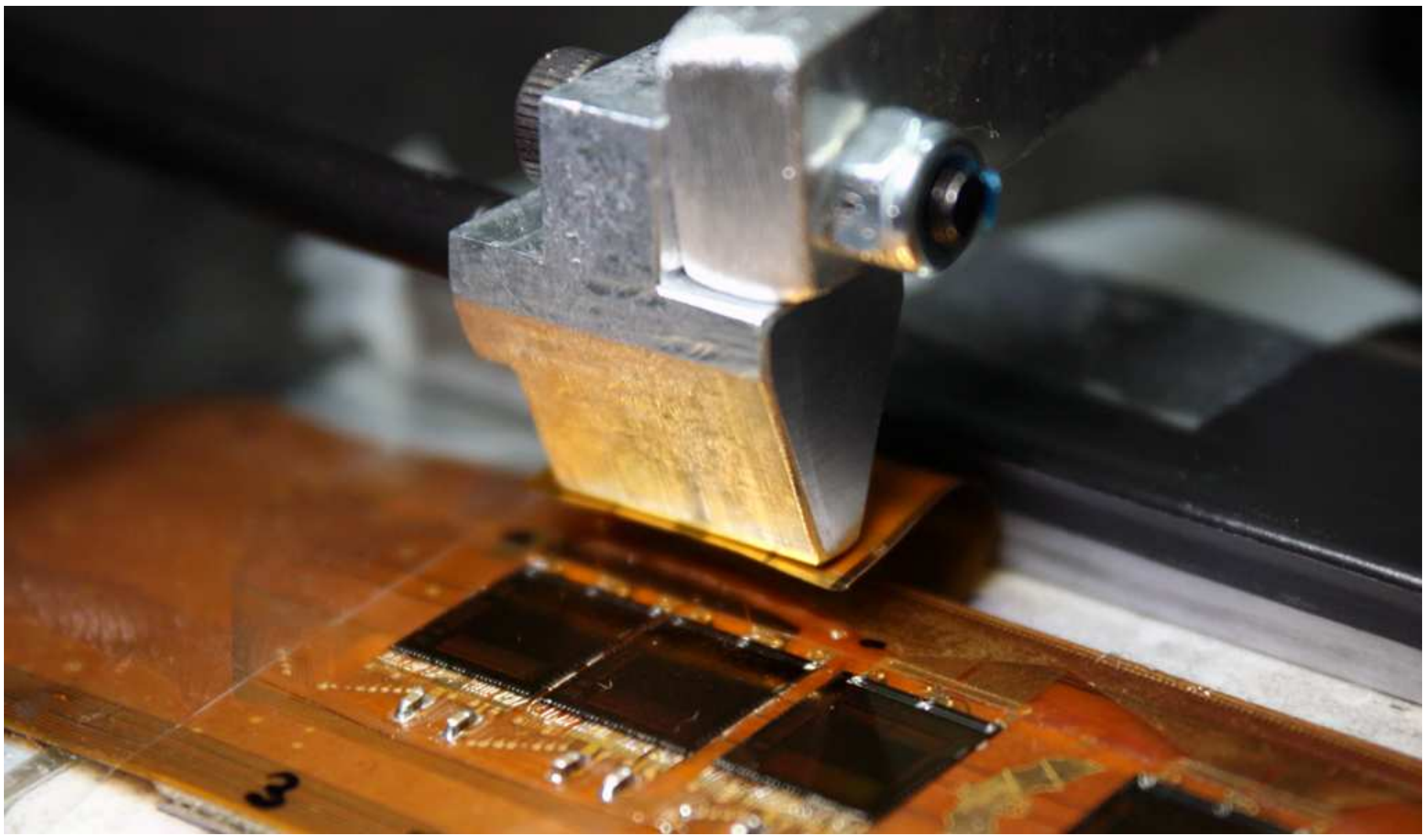

*Figure 5.38: Origami module: Bending and positioning of the flex fanouts.*

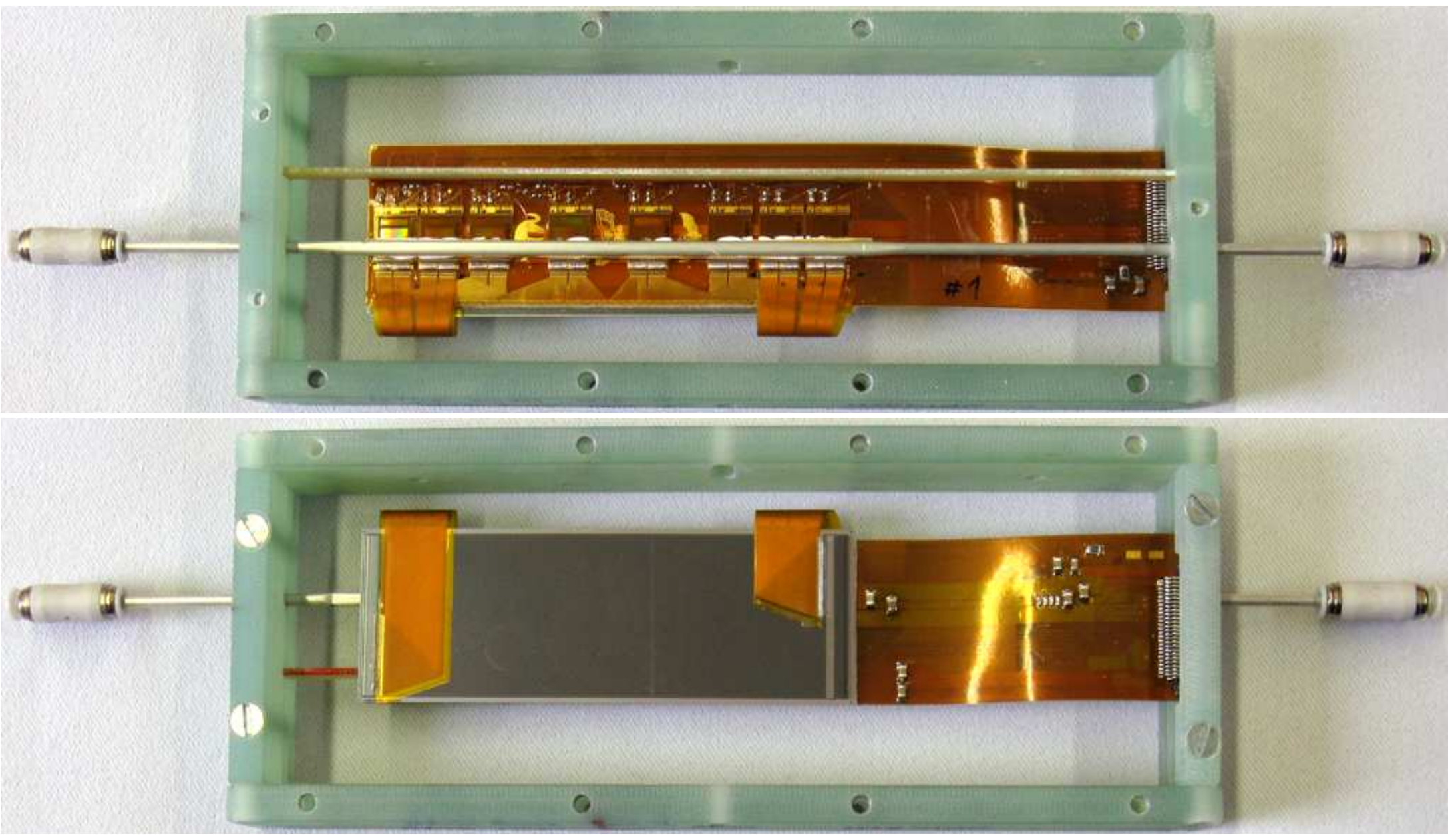

*Figure 5.39: Top and bottom views of the prototype Origami module (using a four-inch wafer DSSD), built into a frame for beam tests.*

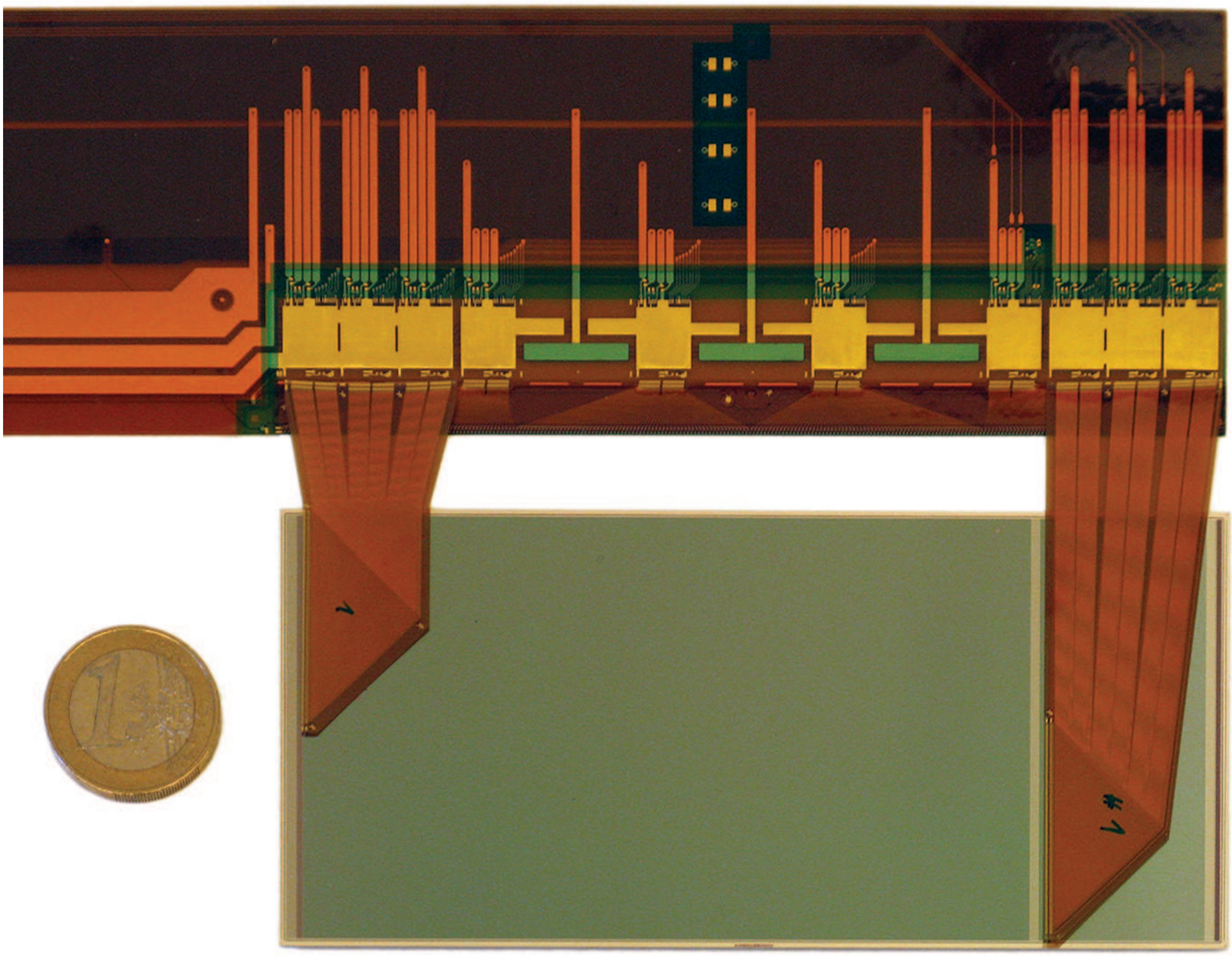

*Figure 5.40: Six-inch wafer DSSD ($\approx 125\times60\,\mathrm{mm}^2$), Origami hybrid ($\approx 445\times57\,\mathrm{mm}^2$; only about a third is shown here) and pitch adapters arranged in the Origami style, but not folded. The objects are shown approximately in full size. Except for the 1 Euro coin ($\approx 23\,\mathrm{mm}$ diameter), all pieces were manufactured in Japan.*

### 5.7.2 Electronics

#### 5.7.2.1 APVDAQ

In 2005, we developed a compact test system for readout of single- or double-sided silicon detectors. It consists of one or more 6U VME modules, each connected to one hybrid with up to four APV25 chips, and one repeater board for each hybrid between front-end and back-end, which performs the voltage level translation as described in Sec. 5.4.4. Further information about the APVDAQ system can be found in Refs. [20] and [21].

#### 5.7.2.2 SVD3 Prototype Readout System

The SVD3 system was designed to read out 384 APV25 chips in the inner two layers of the detector. Each FADC module has inputs for 16 chips, and thus 24 boards were needed in total, spanning over two 9U VME crates.
Figure 5.41 shows a block diagram of the system, which is very similar to what will be implemented for the Belle II SVD. A master controller ("NECO" for new controller) is located in one of the crates and connected to a "SVD3_Buffer" board located in each of the crates. The latter board drives the control signals onto a custom-made backplane bus that runs along the P3 connectors in the VME crate. Each FADC module had a PCI-Link plug-on board transmitting data to the unified COPPER/FINESSE DAQ system. Clock, trigger, and controls were transmitted to the controller through a dedicated link. Precise trigger timing information (for online data reduction in conjunction with hit time finding) was intended to be provided directly from the global decision logic (GDL). On the front-end side, repeater boxes ("DOCK") containing "MAMBO" and "REBO" boards are responsible for buffering and voltage level translation.

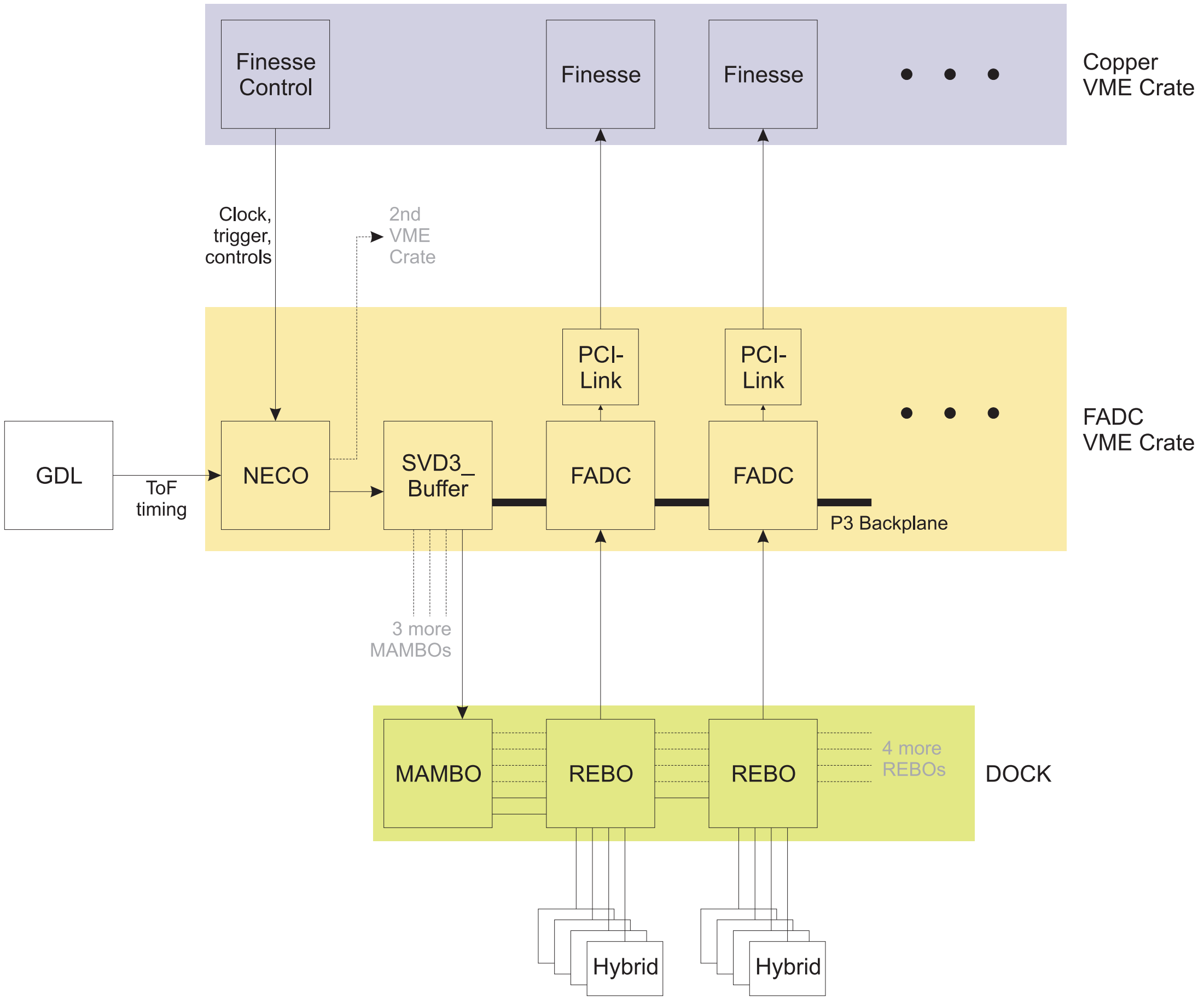

*Figure 5.41: Readout concept for the SVD3 system.*

A "REBO" (repeater board) and an FADC board are shown in Fig. 5.25. Further information and photos of the SVD3 system can be found in Ref. [22].
The Belle II SVD readout system closely resembles the SVD3 design, but differs in some details like the granularity of the FADC+PROC boards (24 instead of 16 inputs) and the integration of the voltage level translation. Nonetheless, many of the building blocks of the SVD3 prototype system will also be used in Belle II.

### 5.7.3 Beam Tests

Aiming to evaluate prototypes of the readout electronics, the sensor modules, and the hit-time finding procedure (Sec. 5.4.4), we performed several beam tests at PSI, CERN, and KEK. Since 2007, the prototype readout system was used in four beam tests and continuously operated for approximately 1,000 hours. During this period of operation it has shown excellent performance and reliability.
The setup of the most recent test (August 2009) in an SPS beamline at CERN is depicted in Figure 5.42. The modules described above, in particular the Origami prototype, were exposed to a 120 GeV proton/pion beam and tested with the full readout chain.

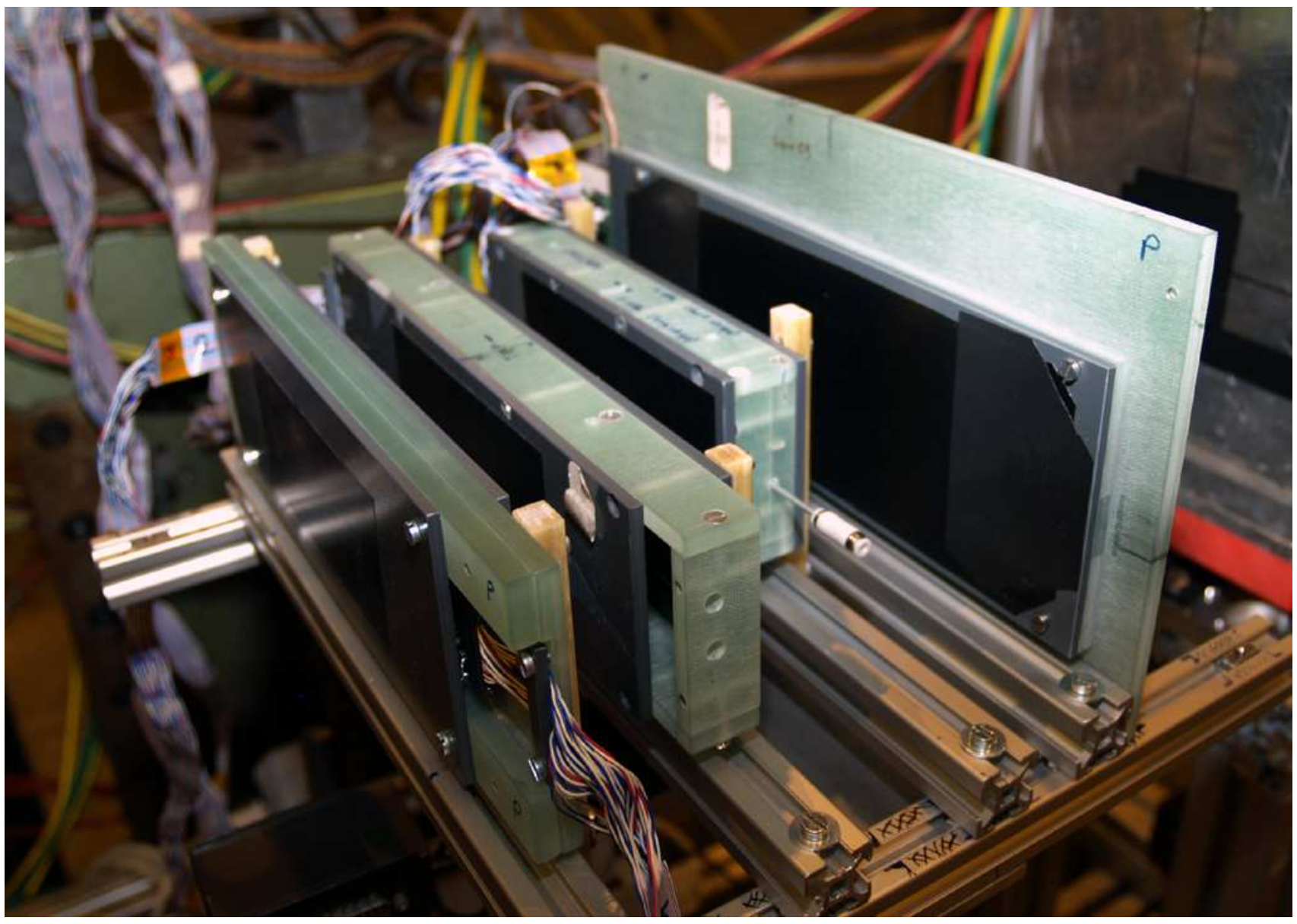

*Figure 5.42: Setup of the beam test at CERN in August 2009, from left to right: Micron module (equipped with a sensor by Micron Semiconductor), Flex_Module, Origami module, and JP_Module.*

Figure 5.43 shows the signal distribution measured on the $p$-side of the Origami module. It can be well fitted by a Landau distribution convoluted with a Gaussian component to account for electronic noise and intrinsic detector fluctuations, shown in red.
The performance of the sensor modules is shown in Table 5.12, where the measured most probable (MP) cluster SNR is compared for both $p$- and $n$-sides, respectively. It is obvious that the achieved SNR of the JP_Module for two ganged sensors is unacceptably low (even though these are only four-inch sensors), while the values of the single sensor readout are satisfactory. Also, the other modules, which read out a single sensor, demonstrated good results.
Compared to the JP_Module, both Flex_Module and Origami prototype revealed similar SNR results for the $p$-sides. On the $n$-side, however, the latter two perform much better thanks to the short connection through the integrated pitch adapters.
These measurements clearly show that it is not possible to operate the APV25 chip with more than one sensor (like on the JP_Module/ganged) due to SNR degradation, especially when they are made from six-inch wafers. Consequently, individual readout of all sensors using the chip-on-sensor concept, as it was demonstrated with Flex_Module and the Origami prototype, is mandatory for Belle II SVD.
Another interesting outcome of the beam tests is the efficiency and precision of the hit-time finding procedure, which is used in the Belle II SVD to discard off-time background hits. A description of this method is given in Sec. 5.4.4.4. For the beam tests, the hit-time reconstruction was done offline by a numeric fit using the pulse shape previously determined by internal calibration of the APV25 chips. As time reference, the distance between the incoming trigger signal and the following clock edge was measured by a time to digital converter (TDC), which is integrated in the controller board (NECO). In Fig. 5.44, the results of several beam tests and different prototype modules are plotted. Clearly, there is a double-logarithmic correlation between the RMS time resolution and the cluster SNR. A time resolution of between 2 and 4 ns can be expected for a cluster SNR range of 18 to 12.

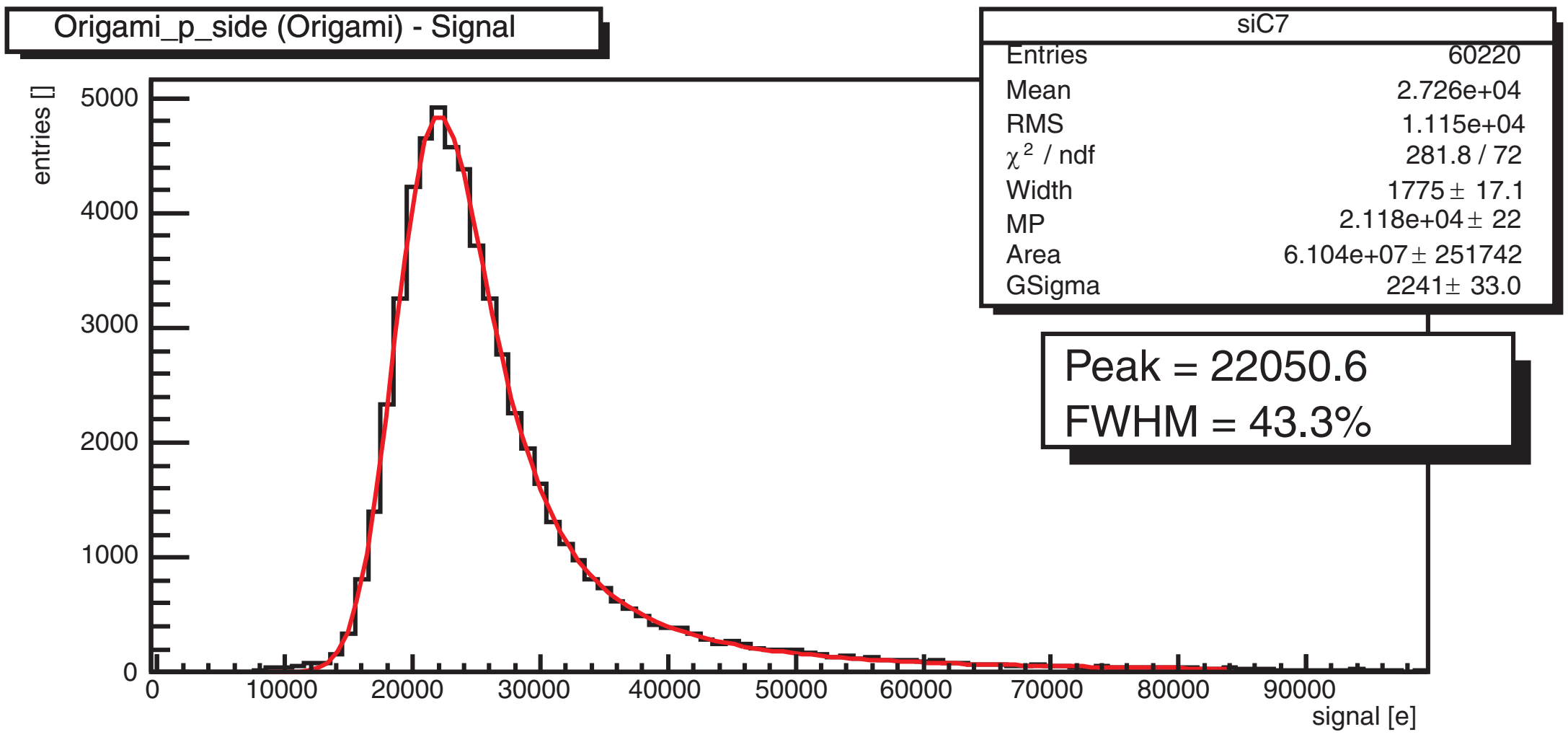

*Figure 5.43: Signal distribution of the Origami module (p-side) and a Landau*Gauss fit (red line).*

| **Module** | **JP ganged** | | **JP single** | | **Flex** | | **Origami** | |
|---|---|---|---|---|---|---|---|---|
| **Sensor side** | **p** | **n** | **p** | **n** | **p** | **n** | **p** | **n** |
| Cluster SNR: | 9.4 | 10.1 | 13.1 | 13.9 | 13.8 | 18.4 | 12.8 | 18.5 |

*Table 5.12: Cluster SNR of the Belle II SVD prototype modules, measured at beam tests at CERN in 2008 and 2009.*

### 5.7.4 Lab Tests

Before and after the beam tests, all prototypes of the electronics as well as all sensor modules were extensively tested in the laboratory. In one of these tests, the effect of cooling on the SNR of the Origami module was measured. To achieve this, the cooling pipe of the Origami module was connected to a chiller, which was used to cool the liquid.
The temperature of the coolant was observed by three thermal sensors attached to the cooling pipe of the module at three different spots. Figure 5.45 shows these locations. One is mounted on the left side at the entry of the coolant, one is placed in the center of the module, and the last one at the exhaust to the right end of the hybrid. The air temperature inside the frame was measured by an additional thermal sensor. A mixture of water and ethylene-glycol was used as coolant. During this test, the module was slightly flushed with dehumidified air through a hole of the frame to avoid condensation. With this (simple) setup, a moderate liquid flow of about $1\,\mathrm{m}\ell/\mathrm{s}$ could be achieved.
In Table 5.13, the results for the uncooled case and with coolant temperatures of 8.7° C and −8.5° C, respectively, are compared. Even though sufficient SNR can be achieved by liquid cooling at a moderate temperature, it is obvious that an improvement of about 20% is possible with sub-zero temperatures. This result is a good motivation to use a $CO_2$ cooling system as described in Sec. 5.3.3, even though it is more complex than single-phase liquid cooling. A dual-phase system would allow us to achieve even lower temperatures, e.g., −20° C, by using very thin tubes, and thus maintaining a small material budget.

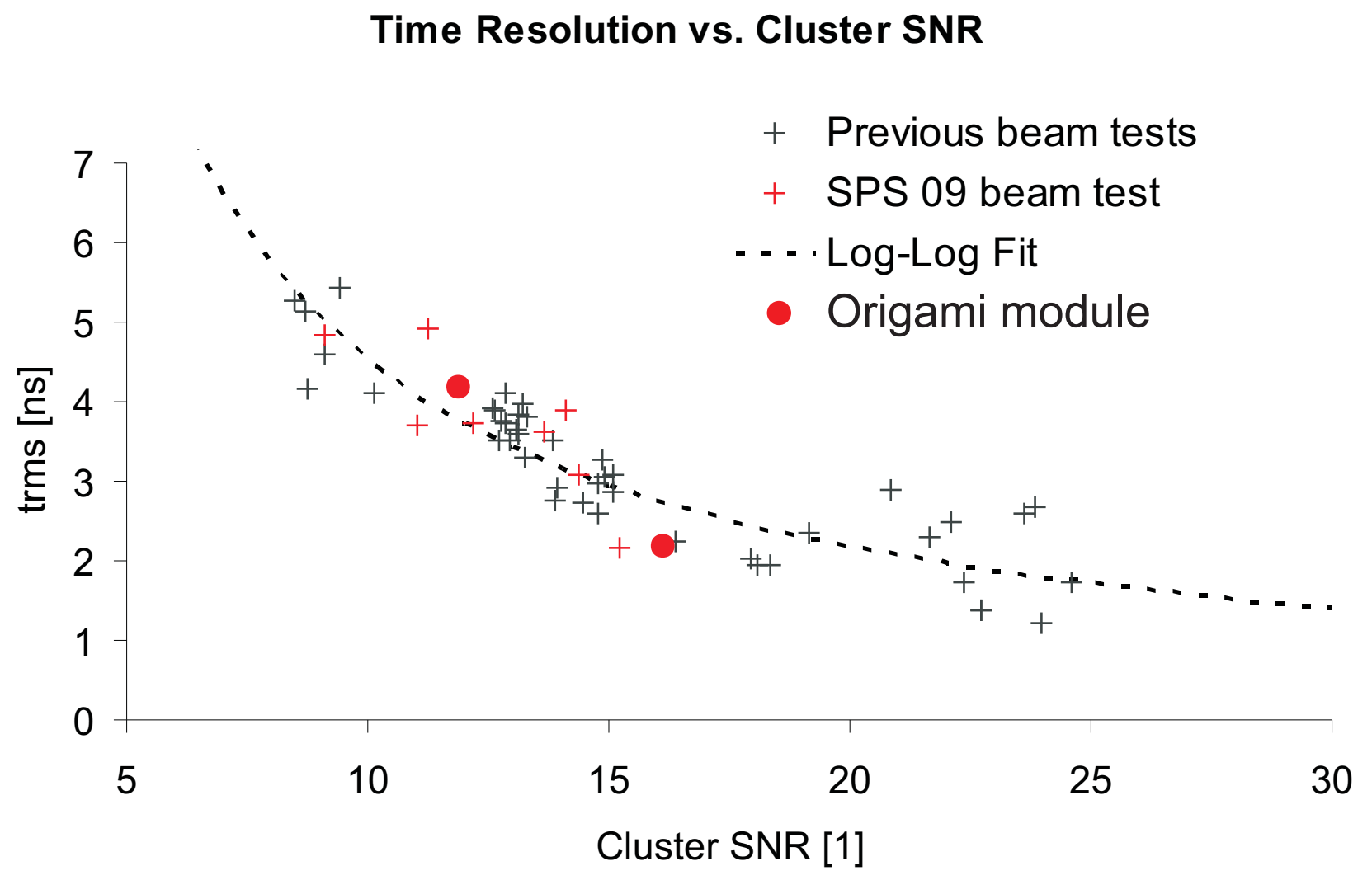

*Figure 5.44: Measured time resolution vs. cluster signal-to-noise ratio of various Belle II SVD prototype modules, measured in several beam tests.*

| **APV25 Surface Temperature** | **Cluster SNR** | | **Improvement over the uncooled condition** | |
|---|---|---|---|---|
| (°C) | $p$-side | $n$-side | $p$-side | $n$-side |
| (uncooled) 67.7: | 11.0 | 14.6 | | |
| 8.7: | 12.2 | 17.2 | 11% | 18% |
| -8.5: | 13.3 | 17.7 | 20% | 21% |

*Table 5.13: Cluster SNR of the Origami module with and without cooling, respectively, measured with a* $^{90}$Sr *source at two different temperatures of the coolant.*

## 5.8 Expected Performance

The design goals of the Belle II SVD and the properties of its components are described in detail in the earlier sections of this chapter. The overall system feasibility was demonstrated by the successful operation of the SVD3 readout system in conjunction with several prototype modules both in the lab and in beam tests (Sec. 5.7.3). Performance details of key aspects of the future SVD will be described in the following sections.

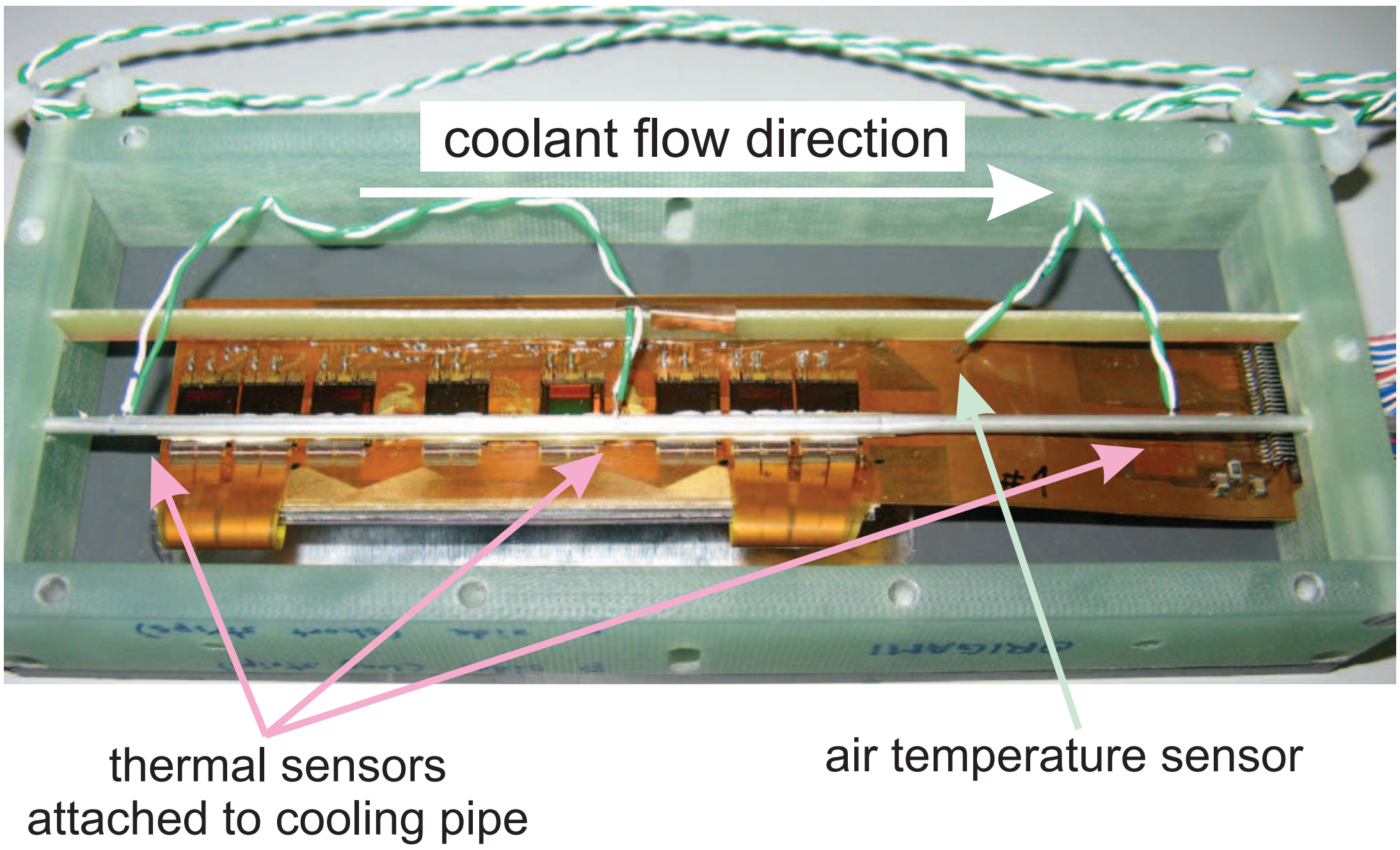

*Figure 5.45: Origami module with thermal sensors attached to the cooling pipe at three spots. The fourth probe was used to measure the air temperature inside the module frame.*

### 5.8.1 Material Budget

To reduce multiple scattering especially for low energy particles, the detector mass needs to be as low as possible. There is a high component density on the Origami modules, so it is not trivial to minimize the material budget. Steps to reduce it are to thin out the APVs as well as the sensors as much as possible. Another factor is the mechanical structure and the cooling circuit. Both have a large impact and need to be carefully dimensioned in order to keep the material budget as low as possible. The sandwich carbon-fiber ribs (SCFR) are a good approach to reduce the budget as well as the use of stainless steel tubes and $CO_2$ as the primary choice of cooling fluid.
The total average material budget is calculated to be 0.568% $X_0$ over the width of the Origami module, as listed in Table 5.14. In this calculation, the sensor is the largest component with 0.32% $X_0$, followed by the copper signal lines (0.068% $X_0$), the mechanical structure (0.022% $X_0$) and cooling.
The averaged material budget was calculated using the formula:

$$X_{0_{rel}} = \sum_m \frac{t_m \cdot w_m}{X_{0m} \cdot w_{total}}$$

| **Layer** | **Material** | $X_0$ **(mm)** | **Thickness (mm)** | **Averaged** $X/X_0$ |
|---|---|---|---|---|
| Sensor | Silicon | 93.7 | 0.3 | 0.320% |
| Isolation | Rohacell HF 71 (Degussa) | 5450.0 | 1 | 0.018% |
| Hybrid | Polyimide (3 layers of 50um each) | 300.0 | 0.15 | 0.022% |
| | Copper (3 layers of 5um each) | 14.0 | 0.015 | 0.068% |
| | Nickel (top: 1.3um) | 14.3 | 0.0013 | 0.006% |
| | Flash Gold (top: 0.4um) | 3.4 | 0.0004 | 0.007% |
| Flexes | Polyimide (1 layer of 25um) | 300.0 | 0.025 | 0.005% |
| | Copper (1 layer of 5um) | 14.0 | 0.005 | 0.010% |
| | Nickel (top: 1.3um) | 14.3 | 0.0013 | 0.002% |
| | Flash Gold (top: 0.4um) | 3.4 | 0.0004 | 0.003% |
| 10 × APV25 | Silicon | 93.7 | 0.1 | 0.008% |
| SMDs | SMD | 50.0 | 0.4 | 0.002% |
| Sil-Pad | Sil-Pad 800 (Bergquist) | 200.0 | 0.127 | 0.003% |
| Pipe | Aluminum (D=2.0mm, wall=0.2mm) | 89.0 | 0.56 | 0.021% |
| Sandwich | Carbon Fibre (0.065mm) | 280.0 | 6.5 | 0.010% |
| Filling | Rohacell HF 71 (Degussa) | 5450.0 | 6.5 | 0.012% |
| Glue | Araldite 2011 | 335.0 | 0.2 | 0.030% |
| Cooling | $H_2O$ | 360.5 | 1.26 | 0.021% |
| | | | **Total** | **0.568%** |

*Table 5.14: Averaged material budget of a single Origami cross section using an aluminum tube with water cooling. The overall figure becomes 0.548% in case of* $CO_2$ *cooling.*

In this equation, $t_m$ (in mm) stands for the thickness of material $m$m, $X_{0m}$ (in mm) denotes its radiation length, $w_m$ represents its component width, and $w_{total}$ is the overall width (59.6 mm). Table 5.14 uses water as coolant with an aluminum tube. Stainless steel, as used for the $CO_2$ option, has a 5 times smaller $X_0$ (17.55 mm for ANSI 304) compared to aluminum (89 mm); on the other hand, the steel tubes have a 4 times thinner wall thickness and a 40% smaller diameter. Taking this into account, the overall material budget for both pipes is about the same, as shown in Tab. 5.15.

| **Material** | **Dimensions** | **Avg.** $X/X_0$ |
|---|---|---|
| Aluminum | $2 \times 0.2\,\mathrm{mm}^2$ | 0.021% |
| Stainless Steel | $1.5 \times 0.05\,\mathrm{mm}^2$ | 0.018% |

*Table 5.15: Material budget comparison for the cooling pipes: aluminum versus stainless steel.*

The situation for $H_2O$ and $CO_2$ is slightly different, as $CO_2$ is in a dual-phase state, and thus the radiation length depends on the location in the tube: at the entrance of a cooling pipe, the medium is (mostly) liquid and gradually evaporates toward the end of the line, where the composition is a mixture of liquid and gas. Assuming a (constant) $CO_2$ temperature of $-20°$ C and an entrance vapor saturation of 8% (corresponding to a density of 400 kg/m$^3$), we obtain a radiation length of 905 mm, which is almost three times higher than that of water. At the exit, we can assume a vapor fraction of 75% (or 75 kg/m$^3$), leading to a radiation length of 4827 mm. In average, the $CO_2$ radiation length is 2464 mm and thus about a factor of seven better than water. In addition, the $CO_2$ pipe diameter is a bit smaller, so the contribution of $CO_2$ becomes negligible compared to water, as shown in Tab. 5.16.

| **Material** | **Cross-Section** | **Avg.** $X/X_0$ |
|---|---|---|
| Water | $2.01\,mm^2$ | 0.021% |
| $CO_2$ | $1.54\,mm^2$ | 0.003% |

*Table 5.16: Material budget comparison for the coolant: water versus* $CO_2$*.*

Another issue not considered so far is that pure water cannot be used at sub-zero temperatures, but needs to be mixed with an anti-freeze agent such as glycol, which has a slightly lower radiation length than pure water and thus would add a bit to the overall material budget. The same is true for another commonly used cooling agent, $C_3F_8$, and that would also require thicker tubes to achieve higher flow rates in order to achieve the same cooling effect as $CO_2$.
To summarize the cooling medium discussion, we conclude that $CO_2$ offers the best performance to material budget ratio, even though the overall material contribution of the cooling is not very significant. Future R&D will show whether mixed-phase cooling is worth the effort.
A low level material budget distribution has been prepared for reference, assuming that a particle traverses the sensors perpendicularly. The result, shown in Fig. 5.46, is calculated taking the dimensional parameters of the main components into account. The distribution of the material budget appears quite inhomogeneous at perpendicular view, which in reality will be far less severe. Due to the curved particle trajectories and the small carbon-fiber layer width, it is very unlikely for a particle to completely traverse the thin $3\%\,X_0$ peaks. This leads to the conclusion that further homogenization of the cross section—such as distributing the support structure over the sensor width instead of having SCFR—will not be necessary to obtain good physics results.

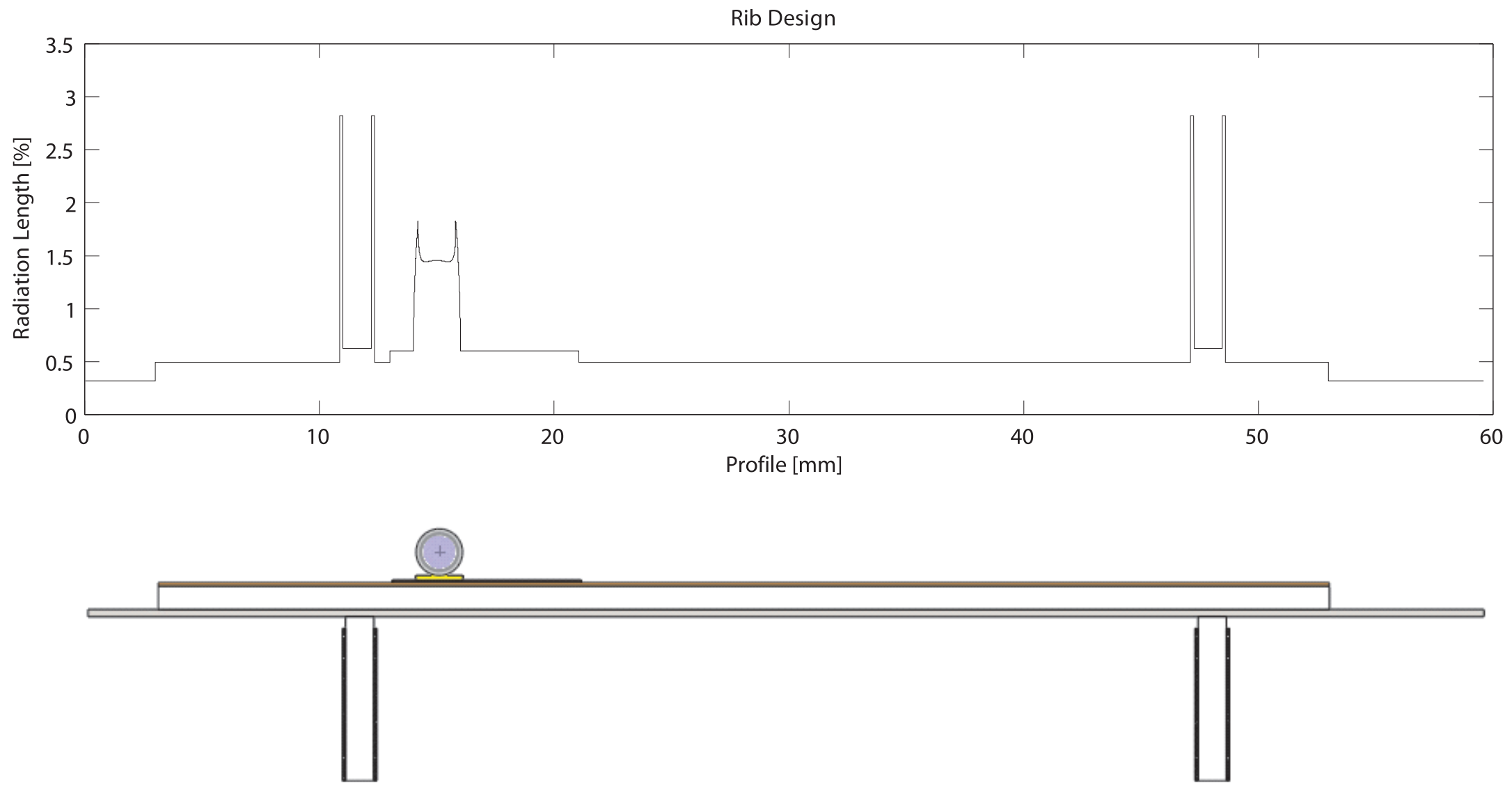

*Figure 5.46: Detailed material budget distribution along the cross-section of the module.*

The integration of the Belle II SVD mechanics in the simulations for the material budget $X/X_0$ vs. pseudorapidity $\eta = -\ln[\tan(\theta/2)]$ is currently under way. Preliminary results in Fig. 5.47 show the full integration of the beam pipe and the PXD as well as the active area (silicon sensors only) of the SVD. "Blind" material such as the structure and readout electronics have yet to be added to the simulation, thus it is likely the SVD contribution will nearly double.

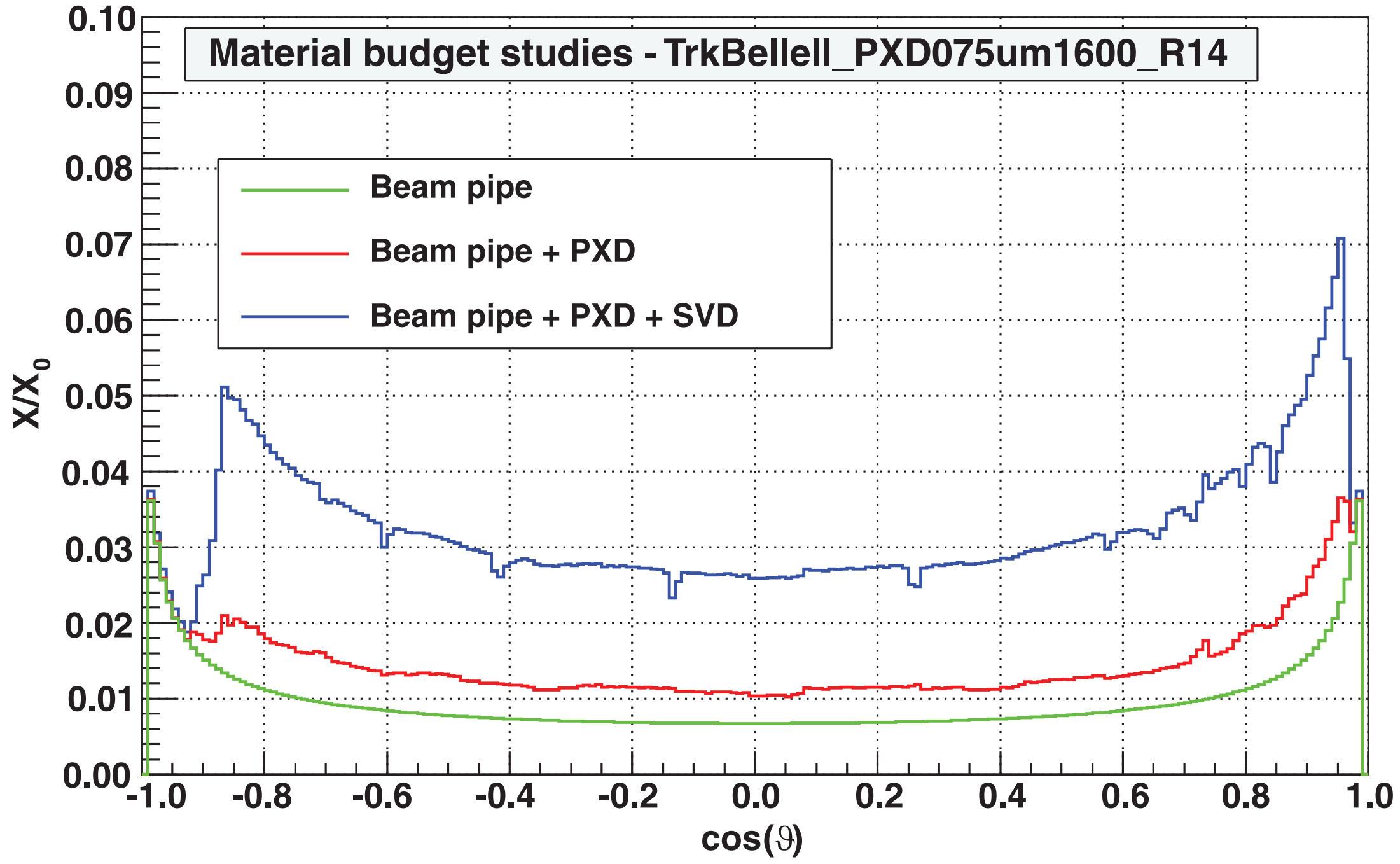

*Figure 5.47: Material budget (SVD: sensors only) vs.* $\cos(\theta) = \tanh(\eta)$. *The forward region extends to* $\cos(17°) = 0.96$, *while the backward acceptance is limited to* $\cos(150°) = -0.87$.

### 5.8.2 Intrinsic Resolution

The intrinsic resolution as well as occupancy of the future vertex detector—PXD and SVD—has a natural impact on tracking and vertexing efficiency, and thus on the overall physics performance. In particular, high resolution in $z$ is very important. On the other hand, too precise a geometrical resolution, which is correlated with the number of read-out channels, may automatically lead to big cluster sizes and thus to high occupancy. This would lead to more SVD ghost-hits (mismatched 2D spatial information combined from $r$–$\phi$ and $z$ channels in double-sided strip detectors).

To determine the expected Belle II SVD resolution, cluster sizes, impact parameter resolution, and vertex resolution, several full simulation studies have been performed. The full simulation chain consists not only of detailed implementation of the PXD and SVD geometries in the Geant4 toolkit (as well as the SVD2 geometry for comparison), but also of the simulation of all in-detector physics processes (i.e., signal processing and digitization) that typically occur in such detector devices. As the design is still ongoing, not all the parameters are known; thus, typical values have been used instead. For more details, see the simulation section in Ch. 14.

The expected intrinsic SVD resolutions in both $r$–$\phi$ and $z$ directions are shown in Fig. 5.48 and the cluster sizes in Fig. 5.49 as a function of the particle incident angle in the forward region. All results are symmetric around 90 degrees (which denotes the radial direction), except for the slanted sensors in the forward region which have no such counterpart in the backward side. The increase of spatial resolution in $z$ towards 90 degrees is due to the fact that there is a probability of single-strip clusters with such perpendicular hits and thus the exact hit position cannot be determined due to the absence of charge sharing among neighboring strips. The detailed studies of impact parameter resolution as well as vertex resolution for the Belle tracker (SVD2 and CDC) and the Belle II tracker (PXD, SVD, and CDC) are presented in Sec. 4.11.

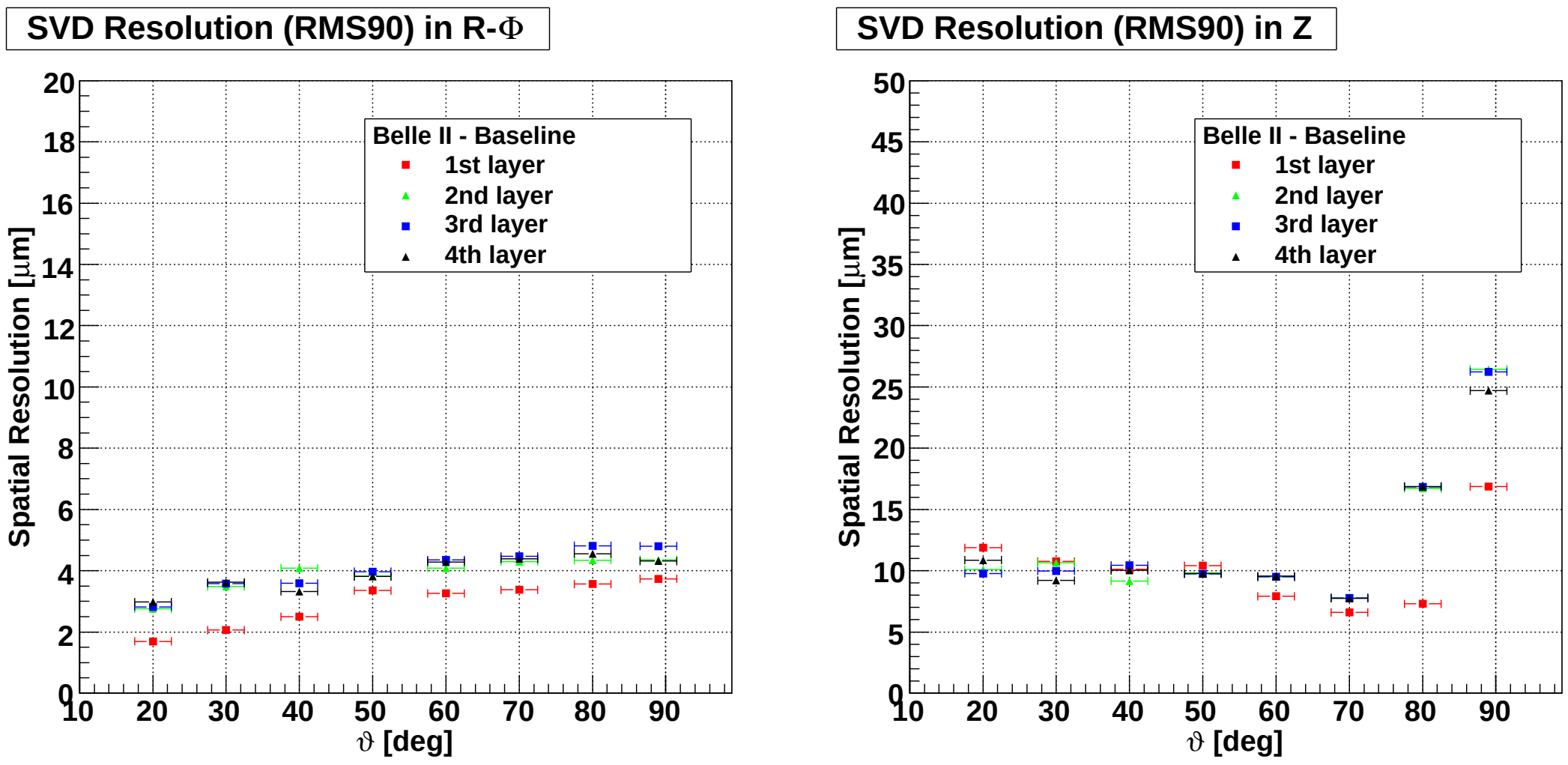

*Figure 5.48: Expected intrinsic resolution in $r$–$\phi$ and $z$ direction for 0.5 GeV single muons.*

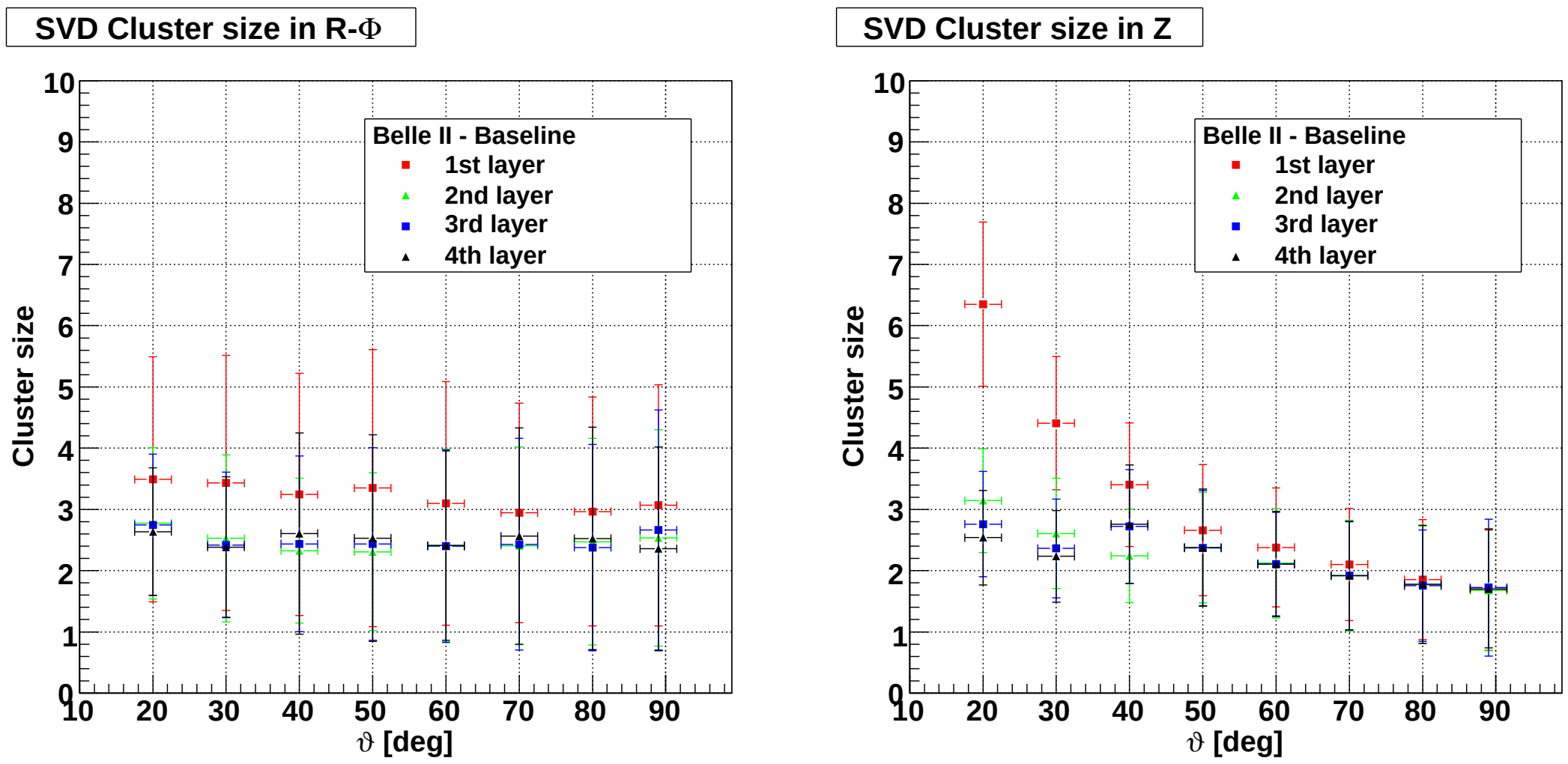

*Figure 5.49: Expected cluster sizes in $r$–$\phi$ and $z$ directions for 0.5 GeV single muons.*

## 5.9 Software

### 5.9.1 Online

The Belle II SVD online software duties consist of online control systems, data acquisition and monitoring. Online alignment and track finding is managed by the tracking group since it is an issue common for all three tracking detectors.

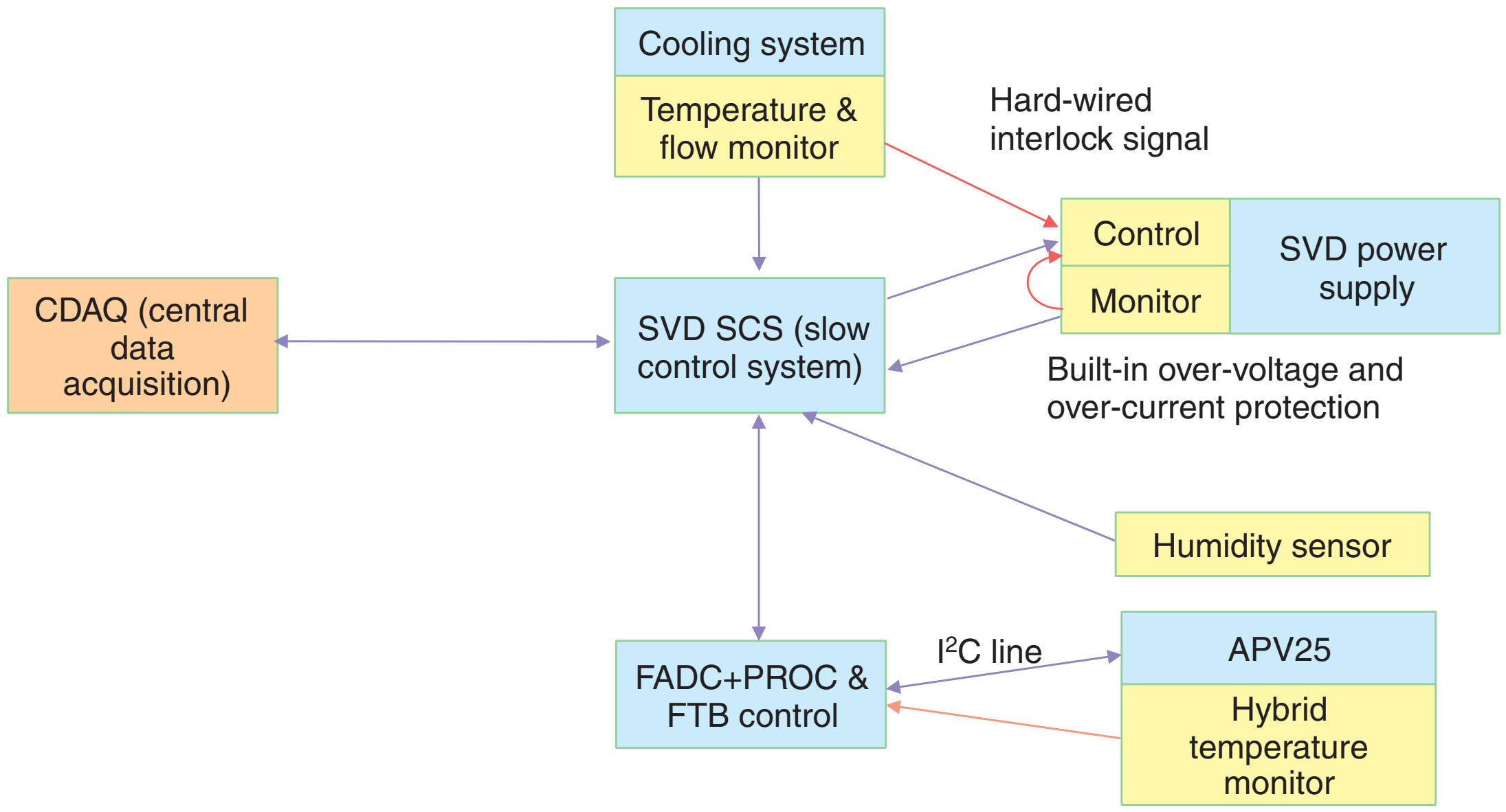

*Figure 5.50: Block diagram of the slow control system of the Belle II SVD.*

#### 5.9.1.1 Control

The online control system of the Belle II SVD will be based on the SVD2 online control scheme. It includes the DAQ control (start, stop and pause of the local SVD DAQ), SVD low and high voltage control and a restart procedure of the SVD subsystem. These functions have to be integrated with the central data acquisition system (CDAQ) of Belle II.

The function of the SVD slow control system (SCS; Fig. 5.50) is to run the SVD according to the operation mode of the Belle II CDAQ. SCS has software links to CDAQ, FADC+PROC, FTB, COPPER, FINESSE, SVD power supplies, the monitor system of the cooling system as well as front-end temperature and humidity sensors. A typical data acquisition is initiated as follows. CDAQ sends the "run start" signal to each detector subsystem. SCS first turns on the low voltage power supply (LVPS) for the APV25 chips and the initialization of APV25 is done using the $I^2C$ link via the FADC+PROC boards. Once the APV25 chips are active, SCS ramps up the the bias voltage for the DSSD sensors. In parallel, the initialization of DAQ system (FADC+PROC, FTB and Finesse boards) will be done. The SVD then becomes ready to accept the trigger signal from the global trigger decision logic (GDL). Once the data acquisition is started, SCS samples monitor data from FADC+PROC and FTB in order to confirm the healthy operation of the Belle II SVD. In addition to the CDAQ, the SVD stand-alone data acquisition for debugging and calibration of SVD is done under the control of SCS.

Another function of the SCS is to protect the Belle II SVD from hardware problems. In particular, a failure of the cooling system can potentially destroy the SVD system in a short time. Therefore, the SVD power supplies should be shut down immediately if such a problem occurs. For this purpose, the SCS continuously monitors the temperature and flow rate of the coolant. In addition, the humidity in the detector region is measured in order to protect the SVD from dew or moisture following a failure of the dry air system. The temperature of the hybrid will be monitored in order to detect blocking of coolant flow in the front-end circuit. If any anomalous behavior is detected, the SCS will shutdown the SVD power supplies and inform CDAQ that the data acquisition has to be stopped due to a SVD problem.

A hard-wired interlock will also provided in order to protect the Belle II SVD even if the SCS software is not working properly. The relay output of the cooling system failure should be directly connected to the shutdown input of the SVD power supplies. These supplies must also have internal over-voltage and over-current protection to prevent damage to the front-end in case of anomalous function.

#### 5.9.1.2 Data Acquisition

The data acquisition scheme is presented in Sec. 5.4.4. After processing and sparsification, the data will be sent via optical links to the common Belle II data acquisition system. In addition, we should be able to provide the SVD information for PXD online data reduction. Several possibilities are under investigation: dedicated FPGA based boards (a.k.a. the Giessen box), CPU-based boards in which Belle II SVD-only hits will be used to find a track and then be extrapolated to the PXD detector, or eventually at the online farm level where dedicated GPU boards can find and extrapolate tracks based on Belle II SVD and CDC information. The tracking group is studying the track finding algorithm schemes.

#### 5.9.1.3 Monitoring

Belle II SVD monitoring tasks include the observation of radiation levels, temperatures, humidity values and detector leakage currents. These issues are introduced in Sec. 5.5. In addition, we monitor average occupancy as well as noise and pedestal levels for every readout channel. The necessary procedures will be implemented to measure and monitor those properties. The occupancy level can be obtained directly from the data during a normal run. Pedestals and noise levels will be obtained for each channel during startup of a run (as this is necessary to program the threshold memories in the FADC+PROC boards), but they could also be extracted regularly in a special procedure when we will be able to read data without zero suppression. This should be done automatically as often as possible without interfering with normal data taking and would allow to monitor changes in running conditions and the eventual appearance of problematic channels to be masked from further analysis. The task of online software will be to store necessary information in the Belle II database.

### 5.9.2 Offline

The Belle II SVD offline software tasks include internal alignment of the detector, local pattern recognition and track reconstruction, and the implementation of the SVD into the Belle II detector simulation software.

#### 5.9.2.1 Alignment

SVD alignment corrects the detector geometry for small misplacements of the sensitive elements in the detector. The actual sensor positions can be determined in the lab by a mechanical survey but the ultimate precision of the device can only be achieved with track-based alignment.

An alignment procedure for the Belle II SVD is now under development. First studies have used the Millipede/Millipede II algorithm [23] and cosmic muons. Possible worries are the required frequency of the realignment procedure which depends on the mechanical stability of the mounting. Another issue are the so-called weak modes, i.e., parameters that are only weakly constrained by the alignment procedure but that might bias the physics output. Ultimately, it would be desirable to have several alignment methods for debugging and cross-validation.

#### 5.9.2.2 Local reconstruction

This task involves decoding of the raw data, clustering, and calculation of the hits in each Belle II SVD layer. These hits are associated into track candidates for global track finding.
For the latter, an "inwards/outwards" strategy is currently under study: track candidates in the CDC are extrapolated inwards to the SVD and hits which can be associated to a track candidate are removed. The remaining hits are searched for low momentum track candidates.

#### 5.9.2.3 Simulation

The simulation task includes the implementation of the Belle II SVD geometry into the Geant4-based detector simulation software. The task of the SVD-specific software is also to simulate the in-detector physics processes and the detector electronics, taking into account the magnetic field (i.e., charge collection, signal processing, and digitization). This simulation includes the drift of the electron-hole pairs in the silicon, the collected charge at each strip, and random noise.

## 5.10 Organization

### 5.10.1 Institutes and Responsibilities

The design of the Belle II SVD has been established by discussion among the following institutions (in alphabetic order):

- HEPHY (Vienna, Austria)
- Jozef Stefan Institute (Ljubljana, Slovenia)
- Karlsruhe Institute of Technology (Germany)
- KEK (Tsukuba, Japan)
- Kyungpook National University (Daegu, South Korea)
- Max-Planck-Institut für Physik (Munich, Germany)
- University of Melbourne (Australia)
- Henryk Niewodniczanski Institute of Nuclear Physics (Krakow, Poland)
- Niigata University (Japan)
- University of Nova Gorica (Slovenia)
- University of Sydney (Australia)
- Tata Institute of Fundamental Research (Mumbai, India)
- Tohoku University (Sendai, Japan)
- University of Tokyo (Japan)

As shown in Sec. 5.10.2, the critical path of this project is the ladder assembly, which requires skillful handling of fragile sensors with high precision. The ladder assembly will be shared by several institutes. The final ladder mounting to the support system and system test will be done at KEK and thus we need to concentrate the human resources at the last stage of the construction in KEK.

### 5.10.2 Schedule

The schedule of the Belle II SVD construction is shown in Fig. 5.51. We assume that the production of DSSD sensors starts from the second half of 2010. The critical path in the present schedule is whether enough parts can be prepared for the ladder assembly in 2011. Another critical path is the preparation of the support system, also scheduled in 2011. Once ladder mounting is finished, the next important step is the system test, where the whole Belle II SVD system (ladder, data acquisition and cooling system) will be integrated and the overall performance is examined before the system is installed to the Belle II structure.

Depending on the readiness of the pixel detector, an integration test will be also necessary. Once the SVD system is installed, further tests with other subdetectors and the trigger and DAQ systems will be done before the beam operation starts.

The Belle II SVD will be installed within the CDC in November 2013. The information from the SVD will be helpful for the accelerator commissioning, scheduled for April 2014.

### 5.10.3 Summary

The Belle II Silicon Vertex Detector will consist of four layers of double-sided silicon detectors made from six-inch wafers. The SVD layers are located at radii of 38, 80, 115 and 140 mm from the beam pipe, between the PXD and the CDC.

In order to cope with the 40-fold increase in luminosity, the Belle II SVD will use the APV25 readout chip with a shaping time of 50 ns, which will, together with online hit time reconstruction at a precision of 2 . . . 4 ns, give an occupancy of a few percent at most. Eighty "FADC+PROC" 9U VME modules, distributed over five crates, will be connected to the front-end, providing both controller and readout functionality.

The front-end modules will be cooled either by $CO_2$ or by a conventional single-phase liquid system. Much effort is put into the design of the mechanical structure as well as the routing of cables and pipes in a very compact space.

The "Origami" chip-on-sensor concept has been proposed for the construction of modules to ensure low-mass readout, achieving an averaged material budget of only 0.57% $X_0$ (or 0.55% $X_0$ in case of $CO_2$ cooling), while maintaining a good cluster signal-to-noise ratio in the range of 12 . . . 18 despite of using large sensors. As each sensor will be read out individually, the total number of readout chips will be 1902, more than double the number in the Belle SVD2.

Fully functional prototypes of detector and electronics components have been built and tested, including a four-inch wafer Origami module that performed well in the lab and in a beam test.

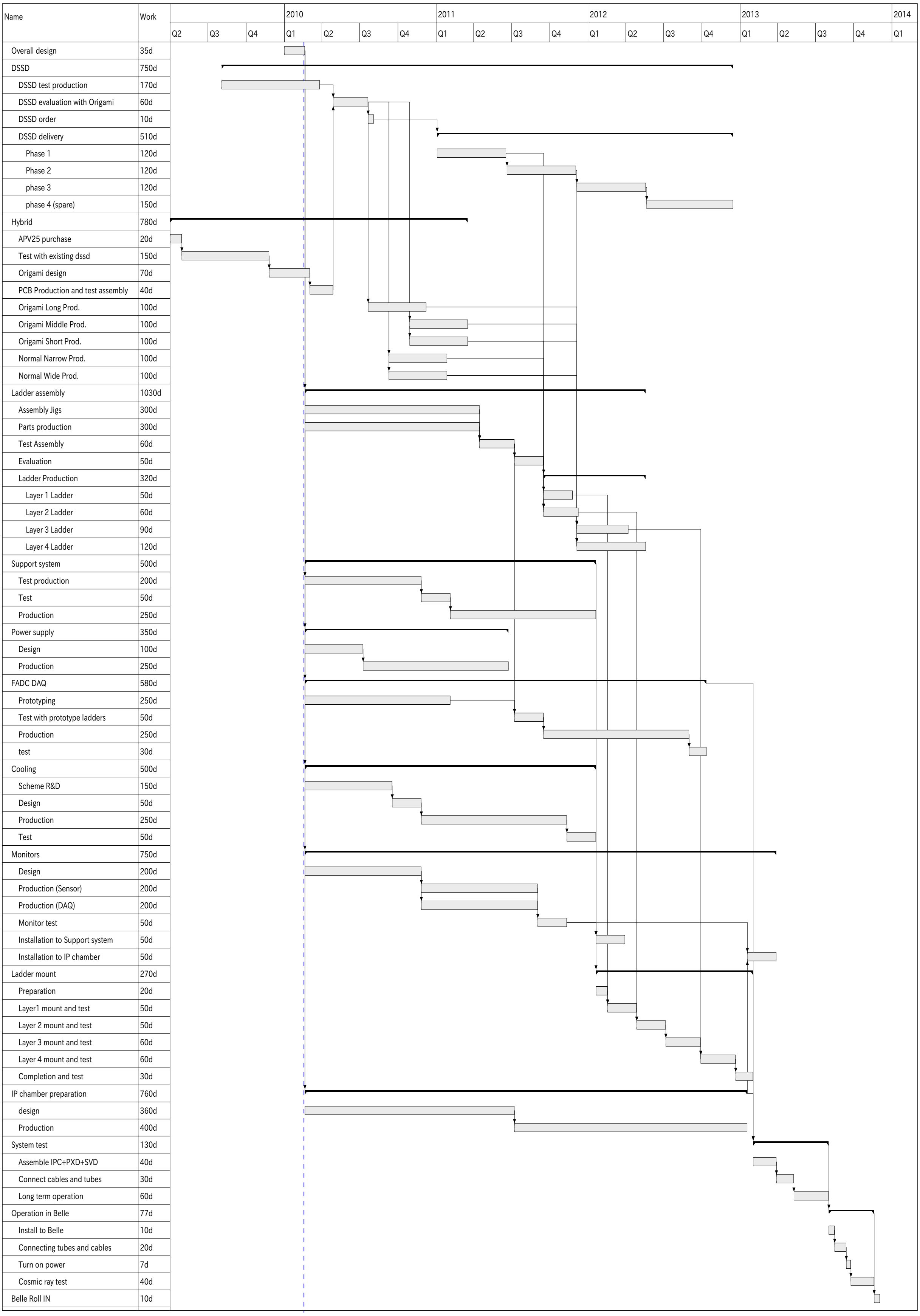

*Figure 5.51: The schedule of the Belle II SVD construction.*

# Chapter 6

# Central Drift Chamber (CDC)

## 6.1 Overview

In the Belle II detector, the central drift chamber (CDC) plays three important roles. First, it reconstructs charged tracks and measures their momenta precisely. Second, it provides particle identification information using measurements of energy loss within its gas volume. Low-momentum tracks, which do not reach the particle identification device, can be identified using the CDC alone. Finally, it provides efficient and reliable trigger signals for charged particles.
As described in the LOI [1], the Belle CDC has worked well for more than ten years without any serious problems. Therefore, the Belle II CDC follows the global structure of its predecessor for the material of the major parts, the superlayer wire configuration, the cell structure, the wire material, and the gas mixture. The main parameters are listed in Table 6.1, together with those of the Belle CDC for comparison. The main differences between the two designs are described in the following paragraphs.

*Table 6.1: Main parameters of the Belle CDC and the CDC upgrade for Belle II.*

| | Belle | Belle II |
|---|---|---|
| Radius of inner cylinder (mm) | 77 | 160 |
| Radius of outer cylinder (mm) | 880 | 1130 |
| Radius of innermost sense wire (mm) | 88 | 168 |
| Radius of outermost sense wire (mm) | 863 | 1111.4 |
| Number of layers | 50 | 56 |
| Number of sense wires | 8,400 | 14,336 |
| Gas | He–$C_2H_6$ | He–$C_2H_6$ |
| Diameter of sense wire ($\mu$m) | 30 | 30 |

First, the new readout electronics system must handle higher trigger rates with less deadtime. The front-end electronics are located near the backward endplate and send digital signals to the electronics hut through optical fibers. The front end uses a new ASIC chip that incorporates an amplifier, shaper, and discriminator. The drift time is measured with a TDC that is implemented in an FPGA. A slow FADC (around 30 MHz) measures the signal charge, and is controlled by the same FPGA. The ASIC chips, the FADC, and the FPGA are mounted on a single board, which is described in Sec. 6.4.
Second, to avoid the high-radiation and high-background region near the interaction point and to provide more space for the SuperSVD, the CDC inner radius is 160 mm. Since the Belle II barrel particle identification device is more compact than in Belle, the CDC outer radius is 1130 mm (tentative value). A new wire configuration fills the modified volume.
Third, the CDC generates three-dimensional trigger information. A $z$ trigger for charged particles is essential to reduce background without sacrificing physics events. In the original Belle CDC, there was a cathode chamber with three strip layers in the innermost region [2]. This chamber reduced the charged trigger rate by a factor of three. When the SVD2 was installed in Belle, the cathode chamber was removed to make space. In principle, the SVD2 would have been able to provide the $z$-trigger information. Fortunately, the charged-track trigger rate was low enough that this was not needed for background suppression. However, we expect that the availability of a $z$ trigger will be important for Belle II, particularly during the early stages when the beam backgrounds are expected to be high. The $z$ trigger will be based on a 3D tracking method implemented in an FPGA using axial and stereo wires. (Charge division between the two ends of a sense wire is a possible alternative to provide the $z$ information.) This approach is robust against high beam background and does not require additional material. The 2D and 3D charged-track triggers are described in Ch. 12.

## 6.2 Structure

### 6.2.1 Wire configuration

We retain the square cell and the superlayer wire configuration of the Belle CDC. There are six layers in each superlayer to make track segment finding easier; this is particularly valuable for the stereo superlayers. (There are only 3 or 4 layers in a Belle stereo superlayer.) The innermost superlayer has two additional layers that contain active guard wires. Even though the performance of these two layers is compromised by the high occupancy from beam backgrounds and the wall effect, the remaining six layers ensure that the innermost superlayer performs as well as the others. The innermost and outermost superlayers contain axial ("A") layers, to match the shape of the inner and outer cylinders. The intervening superlayers alternate between stereo ("U" or "V") and axial layers. In total, there are 9 superlayers (AUAVAUAVA) and 56 layers. The radial cell size is 10 mm for the innermost superlayer and $\sim 18.2$ mm for the other superlayers.
From the trigger group's simulation studies of the 3D trigger, they find a better $z$ resolution using an alternate wire configuration (AUVAUVAUVA) in which the stereo superlayers are adjacent. If this result is confirmed and is essential for the $z$-trigger performance, we will re-consider the overall wire configuration. In addition, the barrel PID group will soon decide the final PID configuration, which will allow the outer radius of the CDC to be fixed. It is likely that the final wire configuration will be slightly modified as a result.
The number of cells in each layer is chosen according to the following considerations. We require multiples of 32 to match the number of electronics channels and trigger segments. For the innermost superlayer, a smaller azimuthal cell size is required to reduce the occupancy in the face of the large beam background. The lower limit is determined by the size of the feed-throughs. The optimal configuration has 160 cells are selected with the minimum azimuthal cell size of only 7 mm. To realize such a small cell size, the innermost superlayer is implemented separately as a so-called small-cell chamber that is then attached to the rest of the CDC. The overall wire configuration is shown in Table 6.2 and Fig. 6.1.
The stereo angles are listed in Table 6.2. A larger stereo angle provides better $z$ resolution, but a large variation in the radial cell size along the $z$ direction occurs in the boundary region between axial and stereo superlayers. To obtain a 60 mrad stereo angle, a special technique is adopted without adding insensitive regions: we string field wires in the transitions with half of the stereo angle and we adjust the radial positions at both endplates around the transitions. The same method is used in the Belle CDC [3]. The sense wire is only $\sim$ 1 mm closer to the field wire in this case, so that a large gain variation is avoided.

*Table 6.2: Configuration of the CDC sense wires.*

| superlayer type and No. | No. of layers | Signal cells per layer | radius (mm) | Stereo angle (mrad) |
|---|---|---|---|---|
| Axial 1 | 8 | 160 | 168.0 – 238.0 | 0. |
| Stereo U 2 | 6 | 160 | 257.0 – 348.0 | 45.4 – 45.8 |
| Axial 3 | 6 | 192 | 365.2 – 455.7 | 0. |
| Stereo V 4 | 6 | 224 | 476.9 – 566.9 | -55.3 – -64.3 |
| Axial 5 | 6 | 256 | 584.1 – 674.1 | 0. |
| Stereo U 6 | 6 | 288 | 695.3 – 785.3 | 63.1 – 70.0 |
| Axial 7 | 6 | 320 | 802.5 – 892.5 | 0. |
| Stereo V 8 | 6 | 352 | 913.7 – 1003.7 | -68.5 – -74.0 |
| Axial 9 | 6 | 384 | 1020.9 – 1111.4 | 0. |

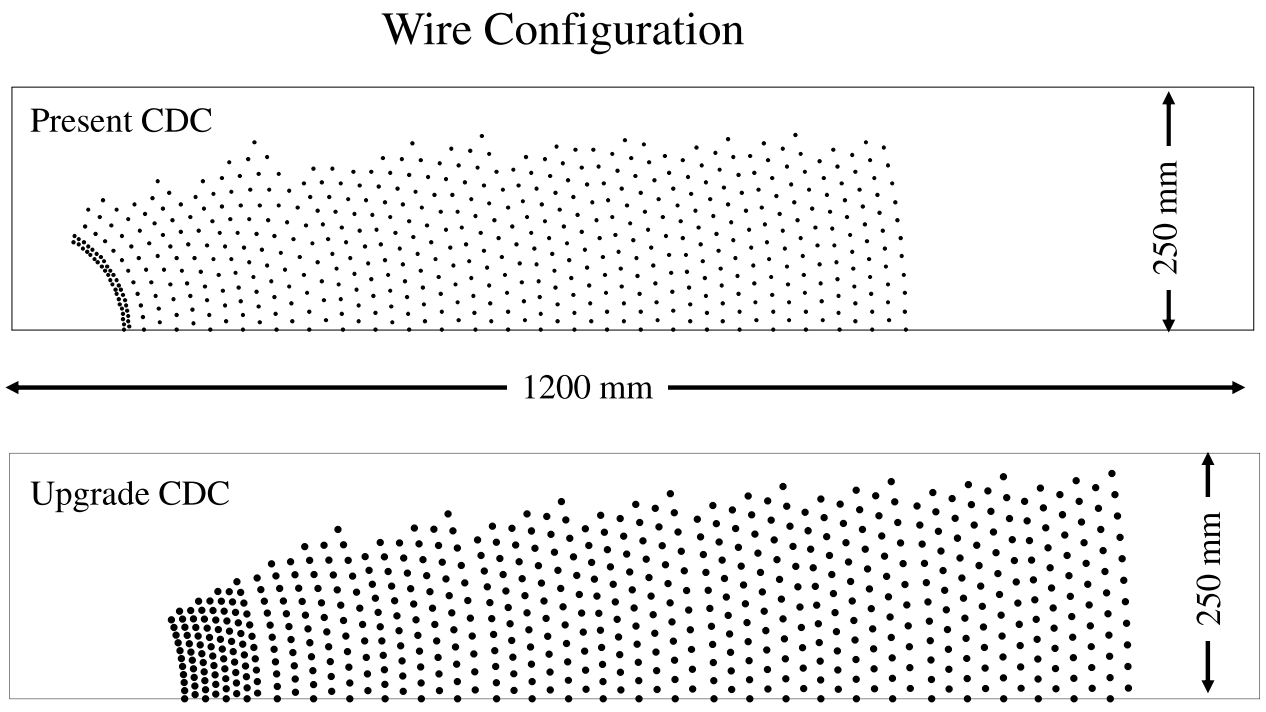

*Figure 6.1: Wire configuration of the Belle and Belle II drift chambers.*

The sense and field wire properties and counts are shown in Table 6.3. The properties are inherited from the Belle CDC, where there were no serious problems during more than ten years of operation. The counts are about a factor of 1.7 greater than in the Belle CDC. The 30 $\mu$m-diameter sense wires will operate at a slightly higher operating voltage so that the stronger electric field in the drift region reduces the maximum drift time. The aluminum field wires are unplated to avoid unnecessary material and to lower the cost.

*Table 6.3: Wire parameters in the CDC.*

| | Sense | Field |
|---|---|---|
| Material | Tungsten | Aluminum |
| Plating | Gold | No |
| Diameter ($\mu$m) | 30 | 126 |
| Tension (g) | 50 | 80 |
| Number of wires | 14,336 | 42,240 |

### 6.2.2 Mechanical design

There are three main structural components of the CDC (Fig. 6.2): a thin carbon-fiber reinforced plastic (CFRP) inner cylinder, two aluminum endplates, and a CFRP outer cylinder similar to the one in the Belle CDC.

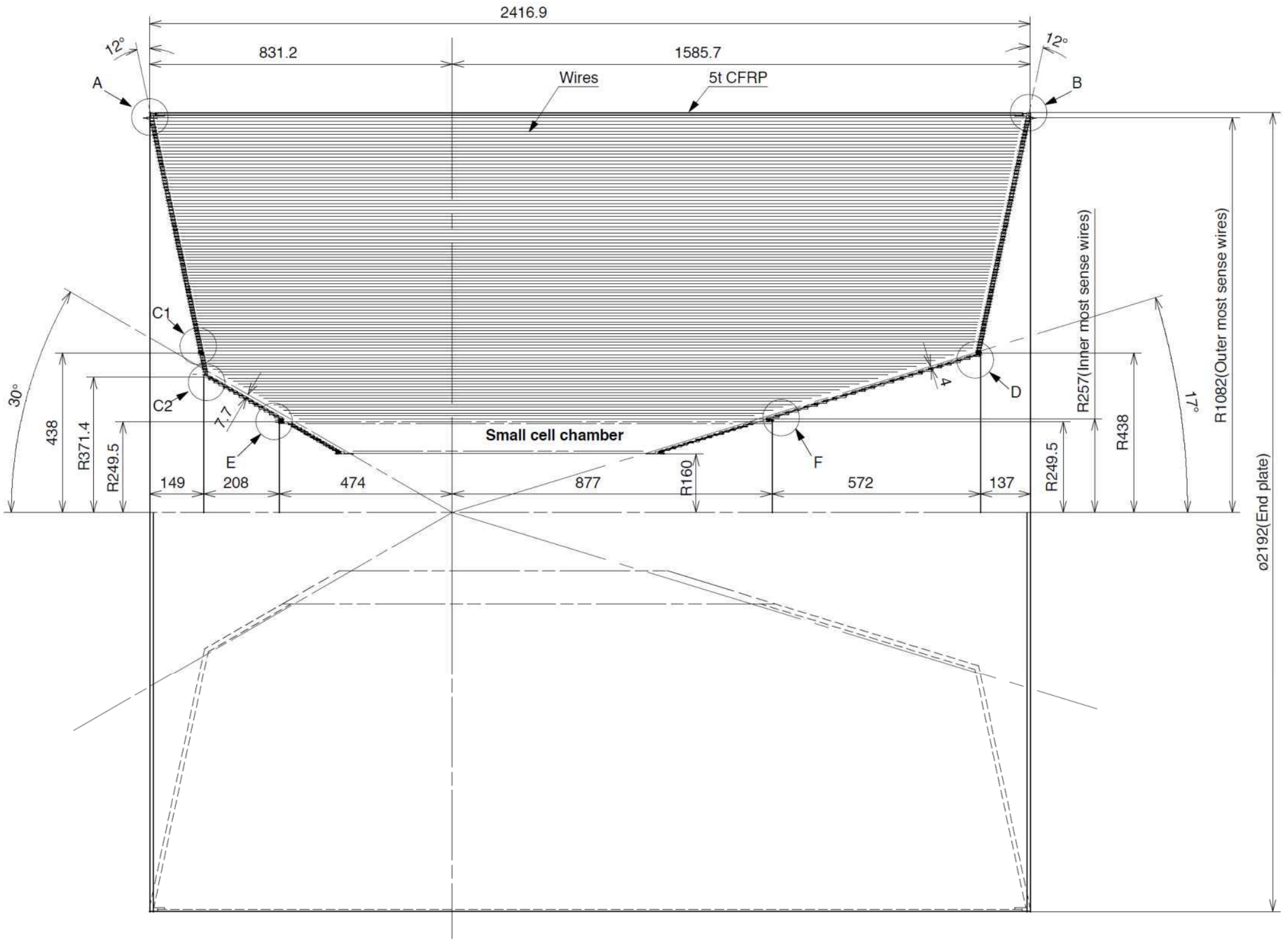

*Figure 6.2: Main structure of the CDC.*

The outer cylinder, with a thickness of 5 mm, supports most of the wire tension of about 4 tonnes. The inner cylinder should be thin (0.5 mm thickness) to minimize material, but it should also support the wire tension for the small-cell chamber (i.e., the innermost superlayer), which has its wires strung independently before installation into the main chamber. Tapered aluminum endplates are used for the outer region to reduce the deformation caused by wire tension, while conical aluminum endplates are used for the inner region to match the detector's polar angular acceptance of 17°–150°. (The endplates of the small-cell chamber are conical as well.) The tapered shape in the outer region reduces the deformation by a factor of two compared to the

Belle CDC.
As shown in Fig. 6.3, a step structure is machined in all three endplate sections to allow easier and more precise drilling of the many holes for the wire feedthroughs. This avoids the difficulty encountered in the Belle CDC, where it was quite hard and time-consuming to drill the holes through the long depth in its conical endplate. The step structure here gives a drilling depth of 10 mm for every endplate section; this is shown in Figs. 6.3 and 6.4.
The endplates for the main and conical parts are machined and drilled separately. Just before the wire stringing, the two parts bolted together, as shown in Figs. 6.3 and 6.5. The forward- and backward-connected endplates are attached on the outer CFRP cylinder (Fig. 6.3), at which time these endplates are precisely aligned.

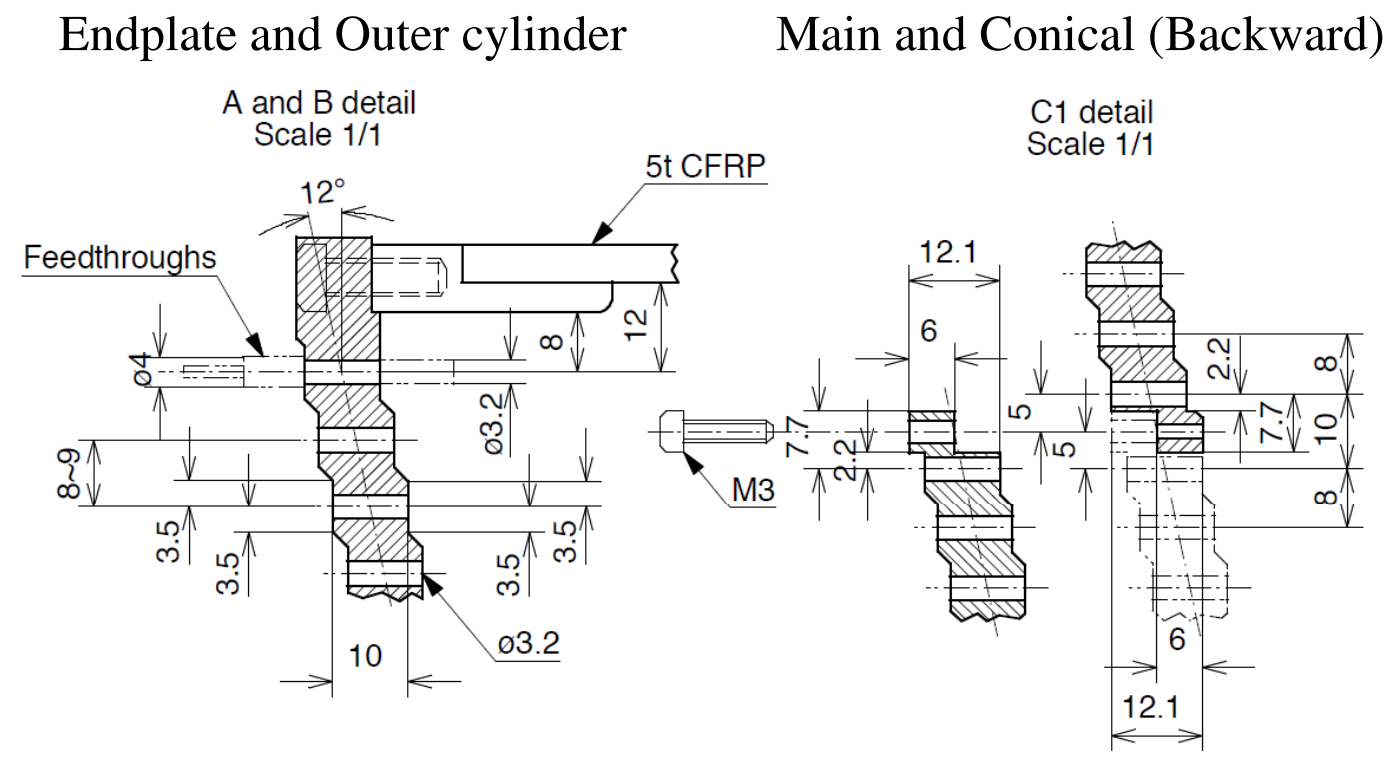

*Figure 6.3: Close-up view in the connection region between the outer cylinder and the main endplate.*

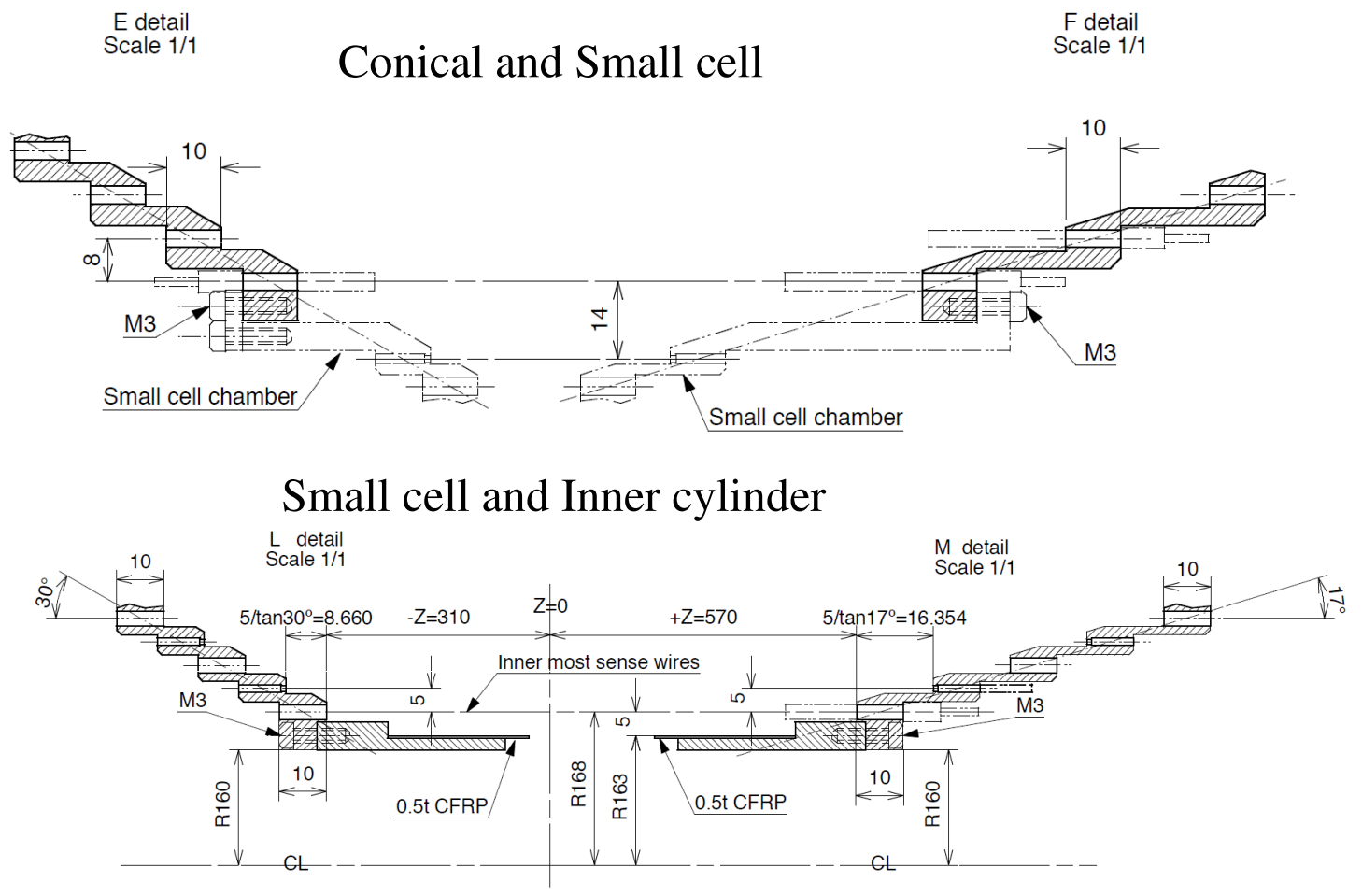

*Figure 6.4: Close-up view around the small cell chamber.*

The tension for the sense wires is the same (50 g) as in the Belle CDC. However, the tension of the field wires is reduced to 80 g from 120 g to reduce the deformation of the endplates. The gravitational sags for the sense and field wires differ by 85 $\mu$m. We estimated the maximum distortion in the $x$–$t$ relation function using a Garfield simulation, and found a difference of at most 20 $\mu$m at the cell edge. The total wire tension is 4.1 tonnes (*vs* 3.4 tonnes for the Belle CDC). Using this wire tension and the described structure, a Finite Element Method calculation shows that the maximum stress of 31 MPa at outer edge of the endplate is smaller than the allowed limit of 107 MPa. In addition, the maximum deformation is 3 mm (*vs* 4 mm for the Belle CDC) at the inner edge of the endplate. We take account of this deformation in the design of the small cell chamber. Pre-stressing of the endplates is necessary when stringing the wires for the main and conical parts. The thin CFRP inner cylinder supports 370 kg of wire tension, which is a factor of six lower than the buckling load of 2300 kg.

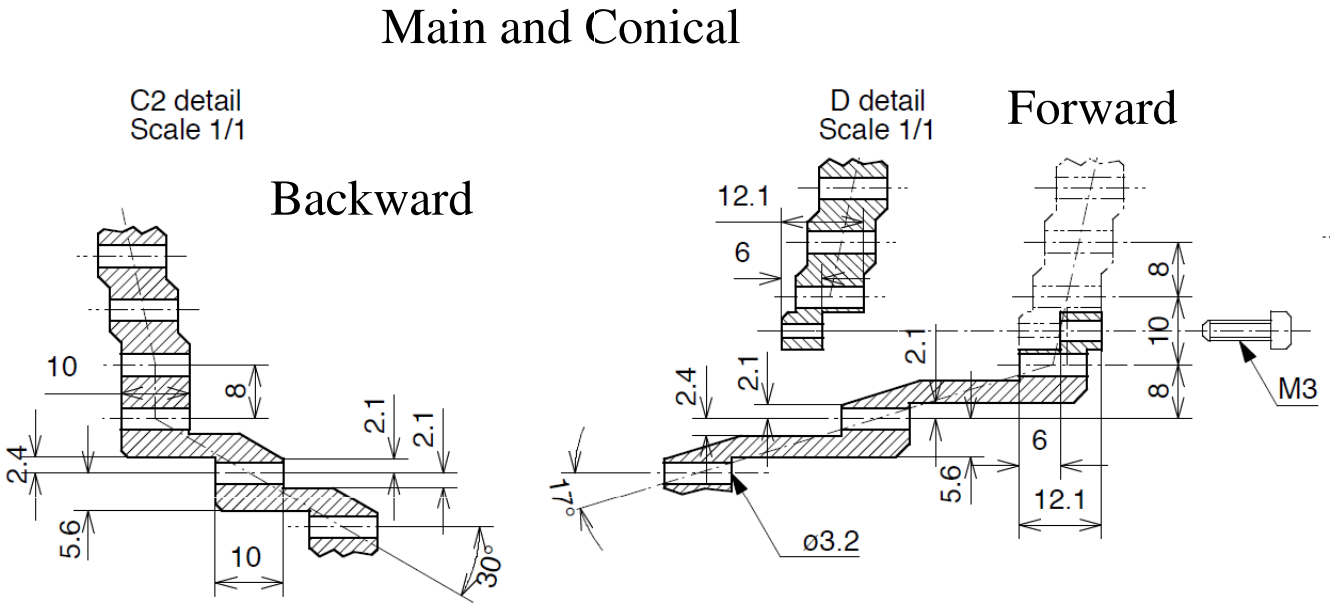

*Figure 6.5: Close-up view around the connection region between the main and conical endplates.*

The wire stringing is performed separately for the main chamber (including its conical part) and the small-cell chamber. The small-cell chamber has no outer cylinder and is small, so it is easy to string its wires horizontally from the inside on a table without any special jigs. The wire stringing for the main chamber is more challenging. The outer cylinder, already in place during the stringing operation, is constructed in one piece to be strong enough to support the large wire tension and to maintain the operating gas pressure absolutely constant. However, we need to observe directly each wire as it is being strung. We will string the wires vertically from the outside, while a person stands inside the chamber to observe each wire as it placed and to make any needed adjustments.

The feedthrough is used to fix the wires and to ensure insulation between the wire's high voltage and the endplate's ground. The shape of the feedthrough is the same as in the Belle CDC. However, the feedthrough material is changed from Delrin to Noryl due to the latter's more reliable insulation performance at high voltage. Noryl is used successfully in the Belle small-cell chamber, installed in 2003. For our small-cell chamber, we do not have enough space to use feedthroughs for the field wires: only aluminum pins are used to hold the wire tension. The pin is attached directly on the endplate, as was done in the Belle small-cell chamber.

There is space between each endplate and its thin aluminum cover to contain the feedthroughs, high voltage cables, signal cables, and readout electronics. All front-end electronics are located on the backward side to reduce the material in the forward side. The forward side is used only for the connection of high voltage cables. If we adopt the charge-division method for the $z$ trigger, a small amount of additional electronics will be located on the forward side, but just outside the detector acceptance.
The CDC is supported by the outer detector. We are considering a huge cylinder, like the Belle inner detector support (IDS), for the forward side. It connects the CDC forward endplate to the forward inner-detector support flange, which is located inside the barrel ECL's inner cylinder. For the backward side, the support of the CDC is slightly more complicated. A similar backward support cylinder is connected to the CDC backward endplate prior to installation. The outer radius of this support cylinder is smaller than inner radius of the barrel PID. This support cylinder is attached to the backward inner-detector flange though several separate jigs just after the CDC has been installed. The details will be fixed after the barrel PID option is selected.

## 6.3 Gas system

We would have preferred to use a new gas mixture with faster drift velocity to reduce the maximum drift time in each cell. Several candidates have been studied, but none has been found that is better than the Belle gas mixture of 50% helium – 50% ethane, so we will use this. Its performance has been established in the Belle CDC: low radiation length, good position resolution, good energy loss resolution, low cross section for synchrotron radiation X-rays, and little radiation damage.
The gas distribution system will be similar to that used in Belle. Separated pure-gas bottles are located in a gas stock room. Gas mixing is performed using two mass flow controllers in a room on the ground level. The mixed gas is fed into the detector through a gas system that is located on the roof of the electronics hut. The gas system consists of a circulation pump, flow controller, pressure controller, and oxygen filter (Fig. 6.6). A relatively modest flow rate ($2\,\ell$/min, which corresponds to one volume every two days) is necessary to remove oxygen from the CDC gas volume. Fresh gas is fed in at only 1/10 of the circulation flow rate to reduce cost. An oil free metal bellow pump maintains the gas circulation; the existing pump has been used for more than ten years without any problem. A small amount of hydrogen gas is mixed into the oxygen filter as a platinum catalyst.

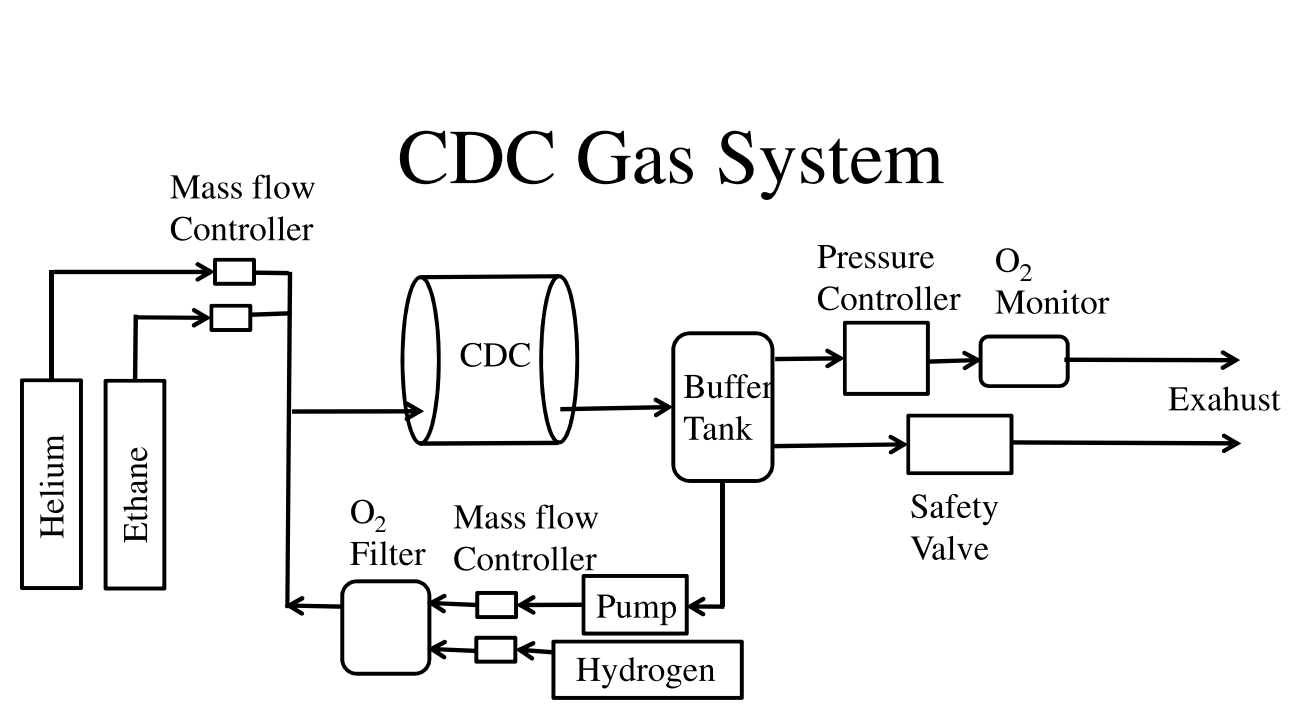

*Figure 6.6: Gas system for the CDC.*

## 6.4 Readout electronics

### 6.4.1 Overview

The CDC front-end readout system is comprised of front-end digitizer boards on the CDC endplate and signal/power distributors. Requirements for the system are listed in Table 6.4. The system transfers digitized analog signals and associated timing signals to the DAQ system in the electronics hut via the Rocket IO link. Synchronization signals (i.e., trigger, system clock, event tag, etc.) from/to the DAQ system are fanned out on signal/power distributors and transferred to each front-end-digitizer board via Rocket IO. The front-end-digitizer boards are mounted near the CDC endplate to reduce channel-to-channel cross-talk, cross-talk from other subdetector systems, and the number of cables to the electronics hut.

*Table 6.4: Requirements for the CDC front-end readout system.*

| | |
|---|---|
| Number of channels | 48 ch/board |
| Total channel | 15k |
| Board size | $15 \times 30\,\mathrm{cm}^2$ |
| Trigger latency | $5\,\mu$s max |
| Single channel hit rate | $< 1\,$MHz (average) |
| Dynamic range | 2 pC max |
| Timing resolution | 1 ns |
| Voltage resolution | 10 bits for 2 pC |
| IO interface | Rocket IO |

### 6.4.2 Front-end digitizer board

The block diagram of the digitizer board is shown in Fig. 6.7. The board consists of front-end ASICs (DC-FEAT), ADC, DAC, FPGA and Rocket IO links. Six DC-FEAT (Drift Chamber FrontEnd for Analog and Timing measurements) ASICs process the current signal from the drift chamber to optimize several requirements: signal-to-noise ratio, dynamic range, timing resolution, and power consumption. After the analog output is digitized by a 10 bit 30 MHz ADC, the digital data are fed into a ring buffer in the FPGA. In parallel, after the timing signal from the DC-FEAT is fed into a TDC with 1 ns resolution in the FPGA, the digitized timing data is stored in the ring buffer. A trigger from DAQ system asserts data transfer from the ring buffer to a local readout buffer. In the local buffer, data suppression and formatting are done to minimize deadtime during the Rocket IO data transfer to the electronics hut. In the alternate data acquisition mode, all data in the ring buffer is transferred directly to the electronics hut for detector calibration and debugging. In both cases, the DAQ system can control the buffer depth to keep data during the trigger decision and an acquisition window for ADC data via Rocket IO.

### 6.4.3 ASIC design

Figure 6.8 shows a block diagram of the CDC front-end electronics. The output of the preamplifier (operating in current mode) is split into two signals. One is for a shaper to measure $\mathrm{d}E/\mathrm{d}x$ and the other is for a comparator to measure the timing. The gain for the analog measurement is designed to maintain a dynamic range of 2 pC, while the gain for the comparator is set to optimize overdrive characteristics and reduce power consumption as shown in Fig. 6.9.

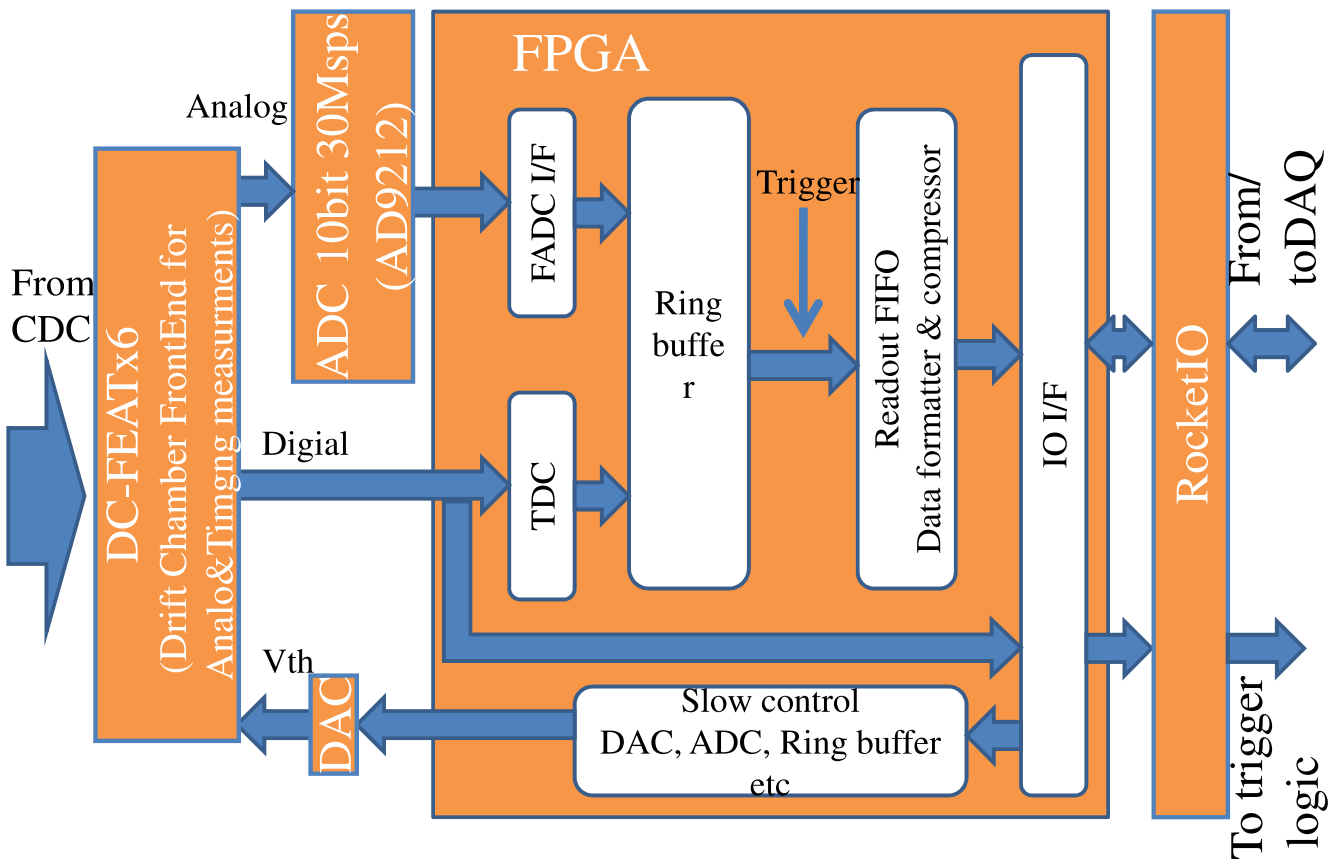

*Figure 6.7: Block diagram of the CDC readout system.*

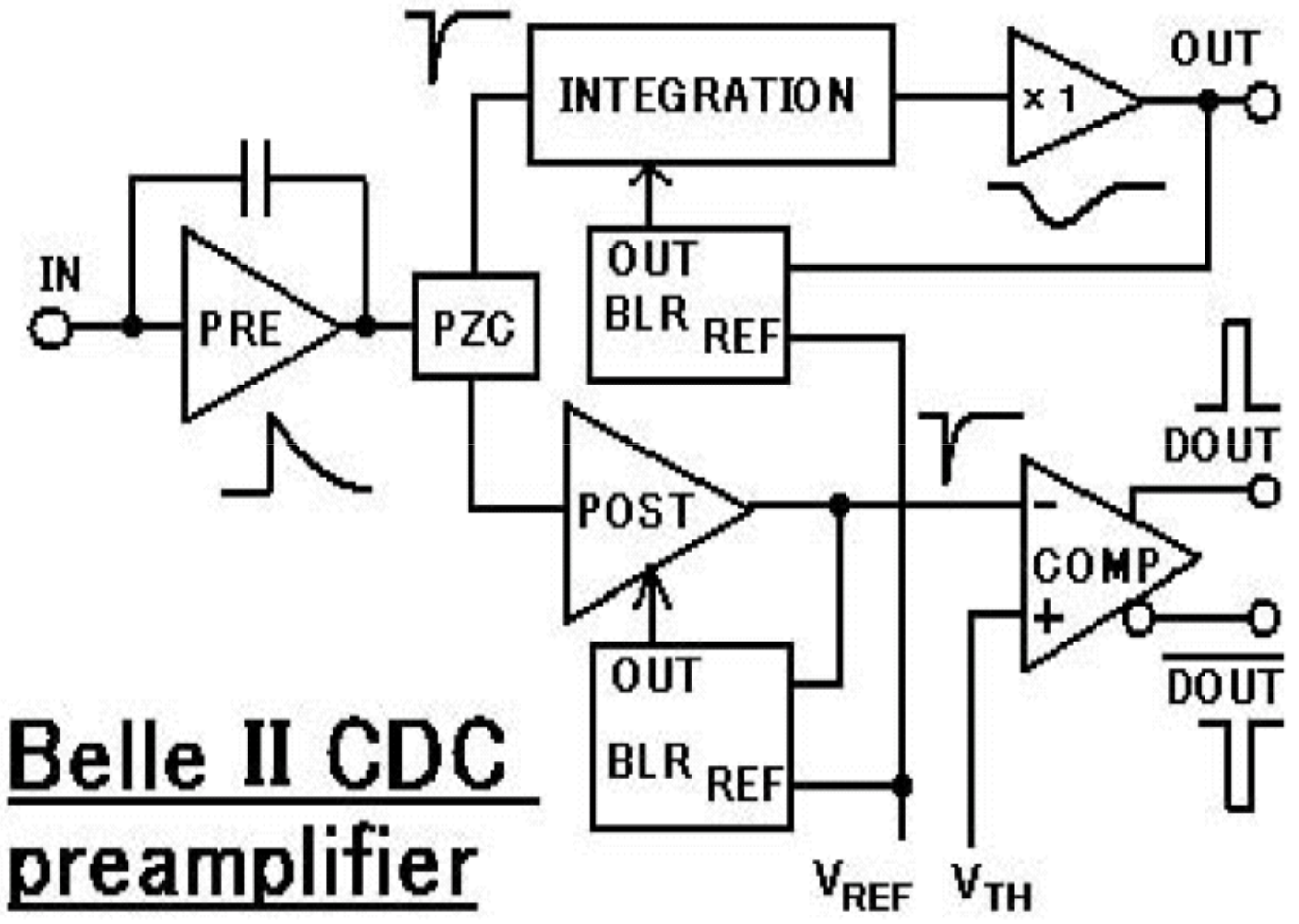

*Figure 6.8: Block diagram of the CDC front-end electronics.*

The overdrive characteristics of the comparator in the simulation are shown in Fig. 6.10. The simulation shows that the time walk is less than 500 ps, which is sufficient for our application. The specification of the DC-FEAT is summarized in Table 6.5. The DC-FEAT was developed by using the NJR 0.8 $\mu$m BiCMOS process. The transition frequency of the NPN (PNP) is 8 GHz (5.5 GHz). The layout of the chip and its package are shown in Figs. 6.11 and 6.12.

### 6.4.4 System and FPGA design

A prototype board was developed to test the functionality of the CDC front-end readout system and to confirm operation of the whole system with a CDC prototype chamber. Figure 6.13 shows the prototype board. ASB (Amplifier Shaper Buffer) chips are used instead of DC-FEATs in the prototype. Although there are some differences between the specification of the ASB and the requirements for DC-FEAT (i.e., the number of channels in a chip, absence of a discriminator, etc.), we developed a daughter board to meet our requirements. Figure 6.14 shows the analog output from the daughter board. We furthermore use SiTCP (an Ethernet processor) instead of Rocket IO for the data transfer, since there was no specification for the data transfer interface at the time of the production of our prototype board. We tested the entire functionality—including the digital data transfer for the trigger system—with the drift chamber except for the data transfer link. The data acquisition sequences are shown in Fig. 6.15. As already mentioned, two acquisition modes are implemented in the FPGA and were tested using a 100 Mbps Ethernet interface.

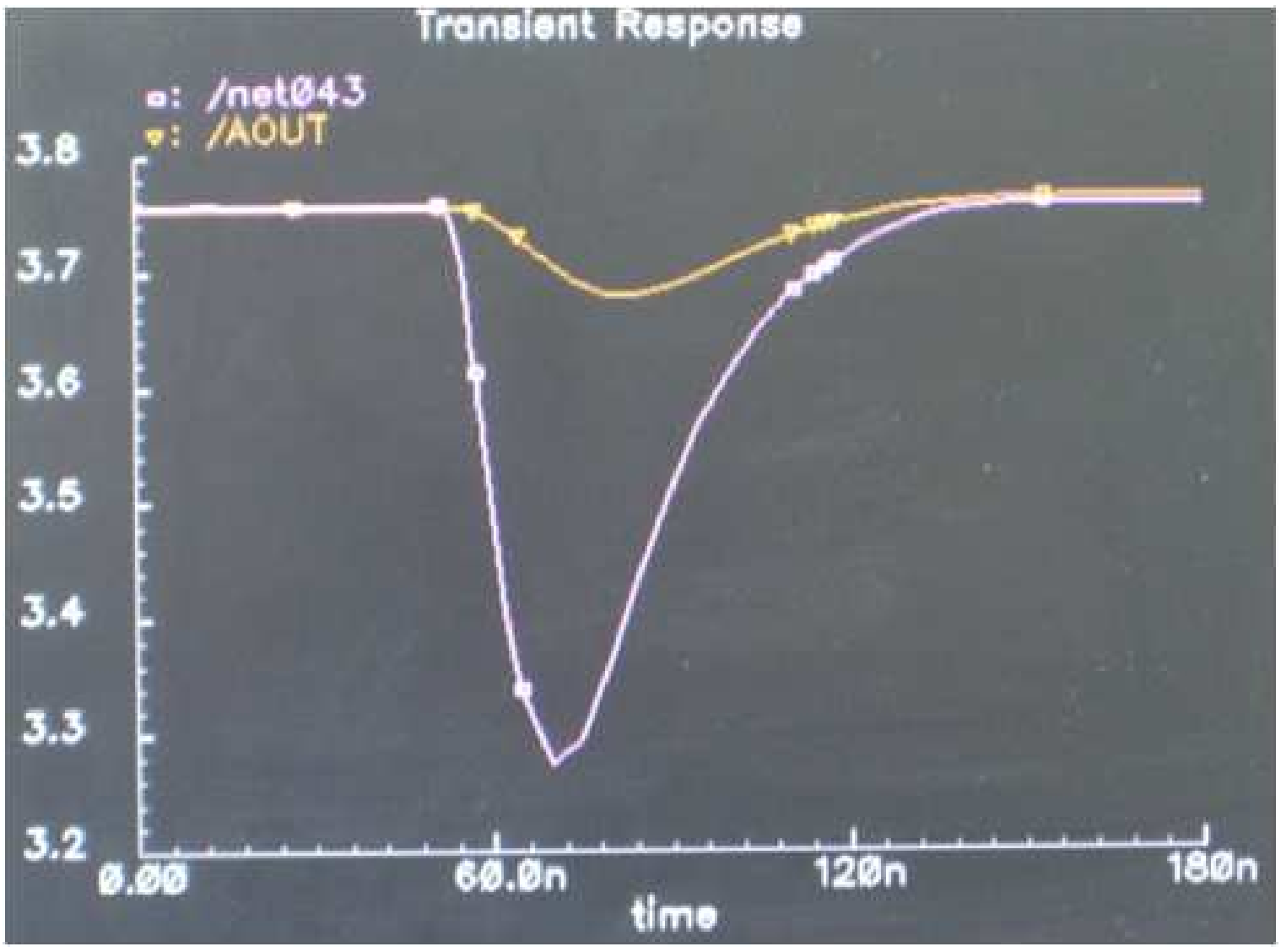

*Figure 6.9: Expected pulse shape with electronics circuit simulation.*

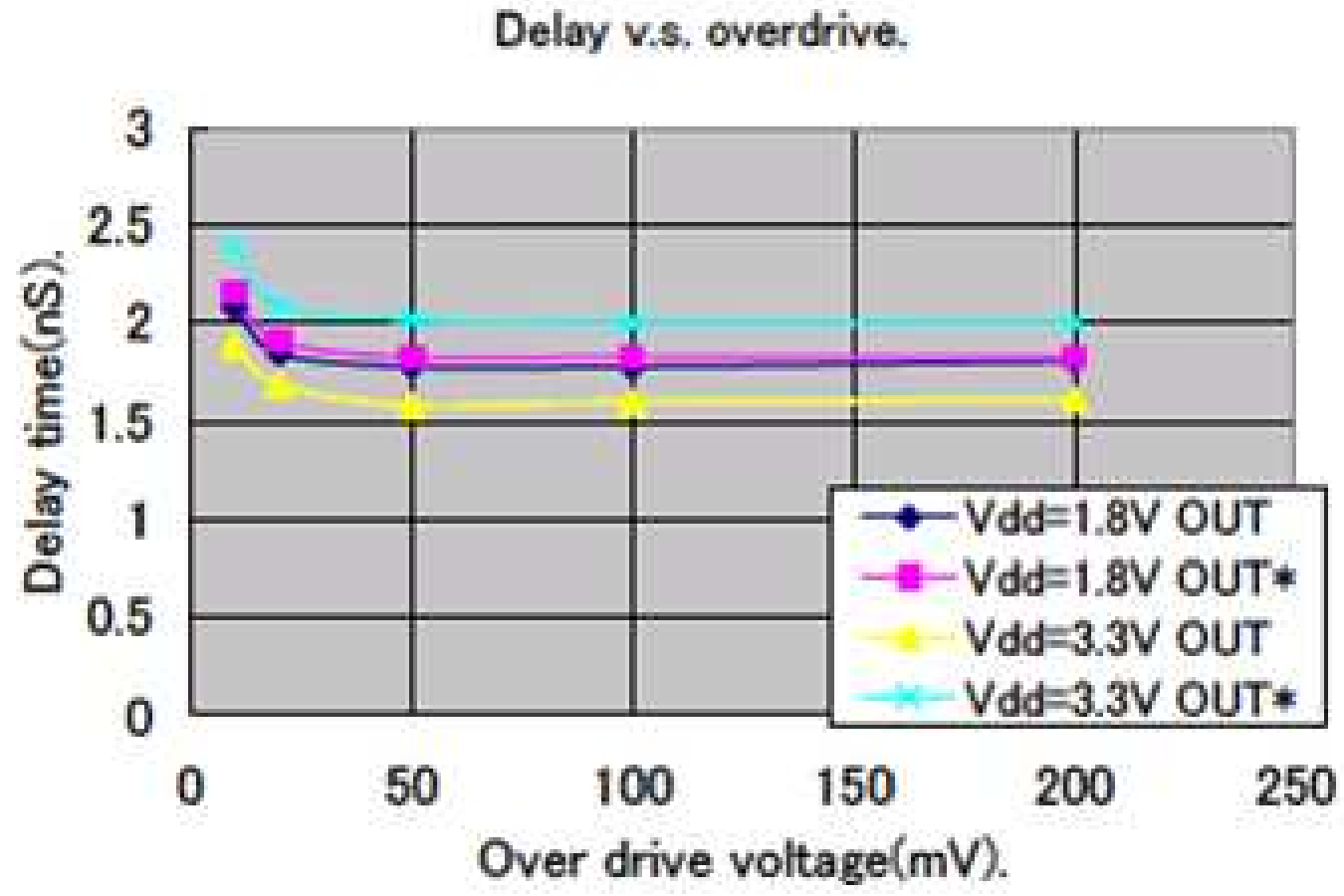

*Figure 6.10: Overdrive characteristics for the comparator.*

*Table 6.5: Specification of the CDC front-end ASIC.*

| | |
|---|---|
| Signal processing | $1/t$ cancellation |
| Gain for analog measurement | 1.1 V/pC |
| Analog output | Single ended (8 mA max) |
| Dynamic range | 2 pC |
| Gain for timing measurement | 15 V/pC |
| Overdrive voltage for $\geq$ 10 mV | $\leq$ 400 ps |
| Digital output | CMOS (single/differential) |
| No. of channels | 8 |
| Chip size | $4 \times 4\,\mathrm{mm}^2$ |
| Process | NJR 0.8 $\mu$m BiCMOS |

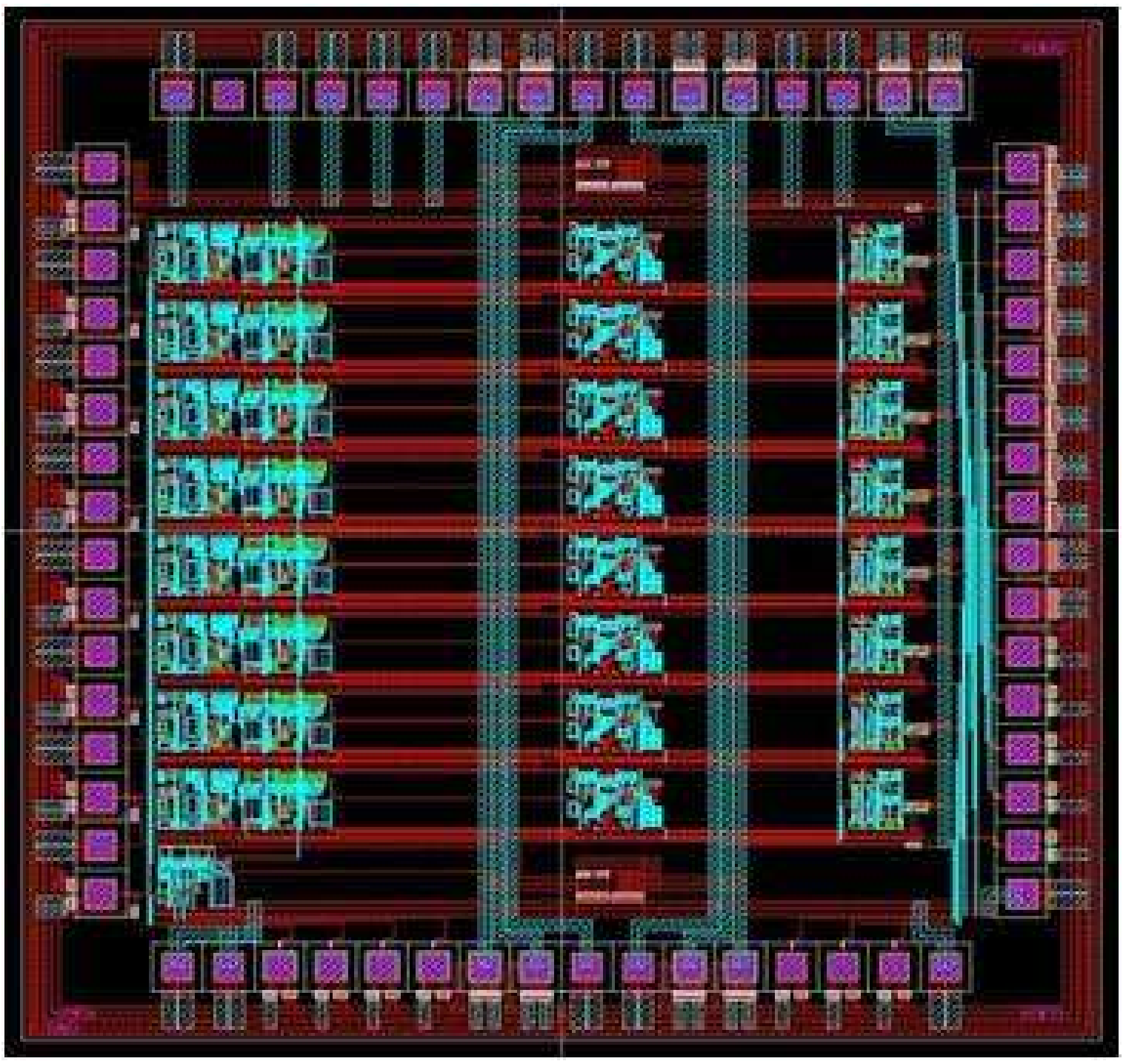

*Figure 6.11: Layout of the ASIC chip.*

### 6.4.5 Schedule

The first production prototype will be tested with the CDC prototype by the middle of 2010. At that time, we will determine where modifications of the DC-FEAT specification (i.e., gain and shaping time) are necessary. The production of the DC-FEAT will be finished by the end of FY 2010. In parallel, we will determine the number of links for trigger data transfer and the specification of DAQ interface in detail by the end of FY2010. We will be ready for mass-production at the beginning of FY2011.

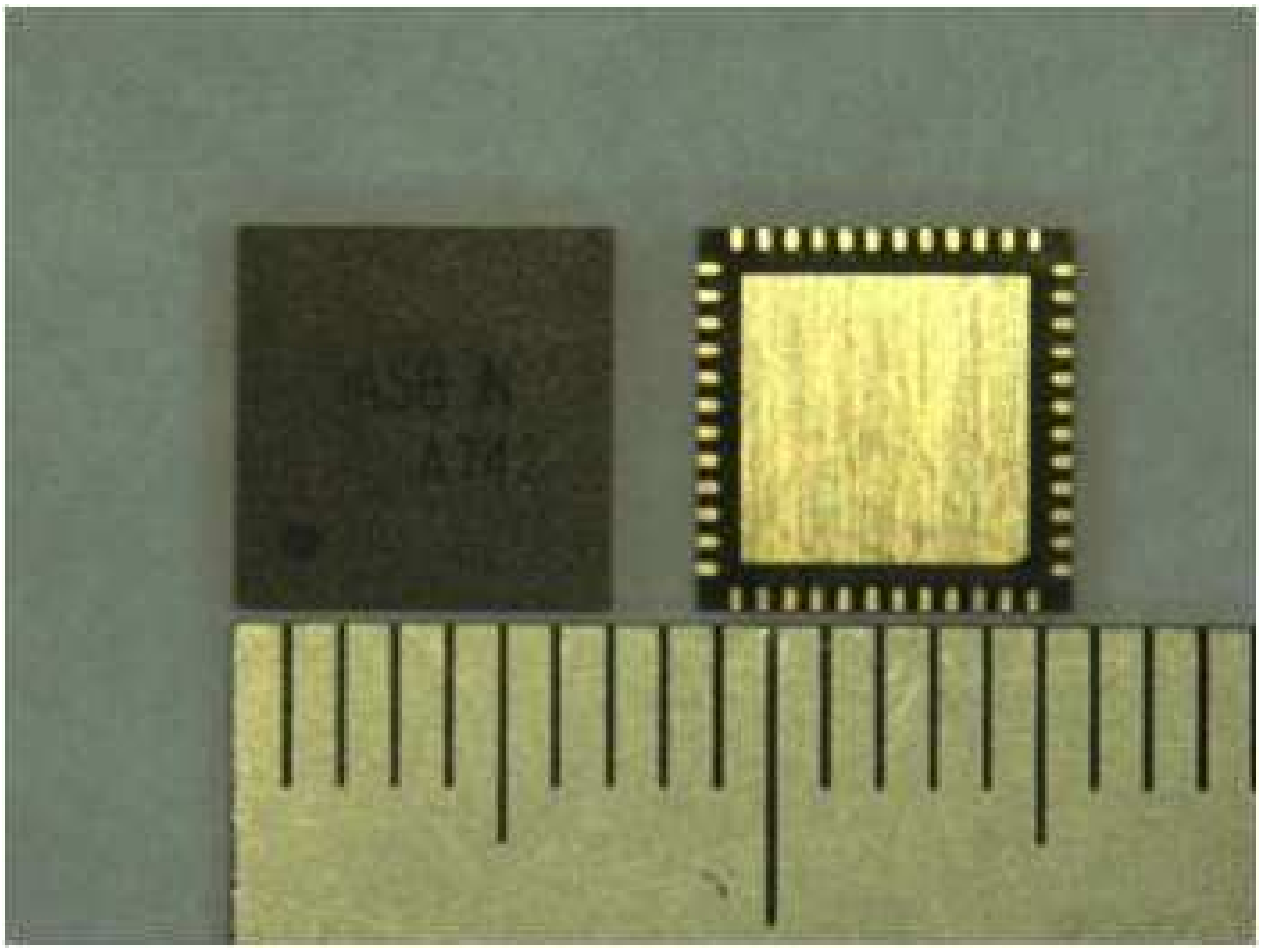

*Figure 6.12: Photo of the prototype ASIC chip.*

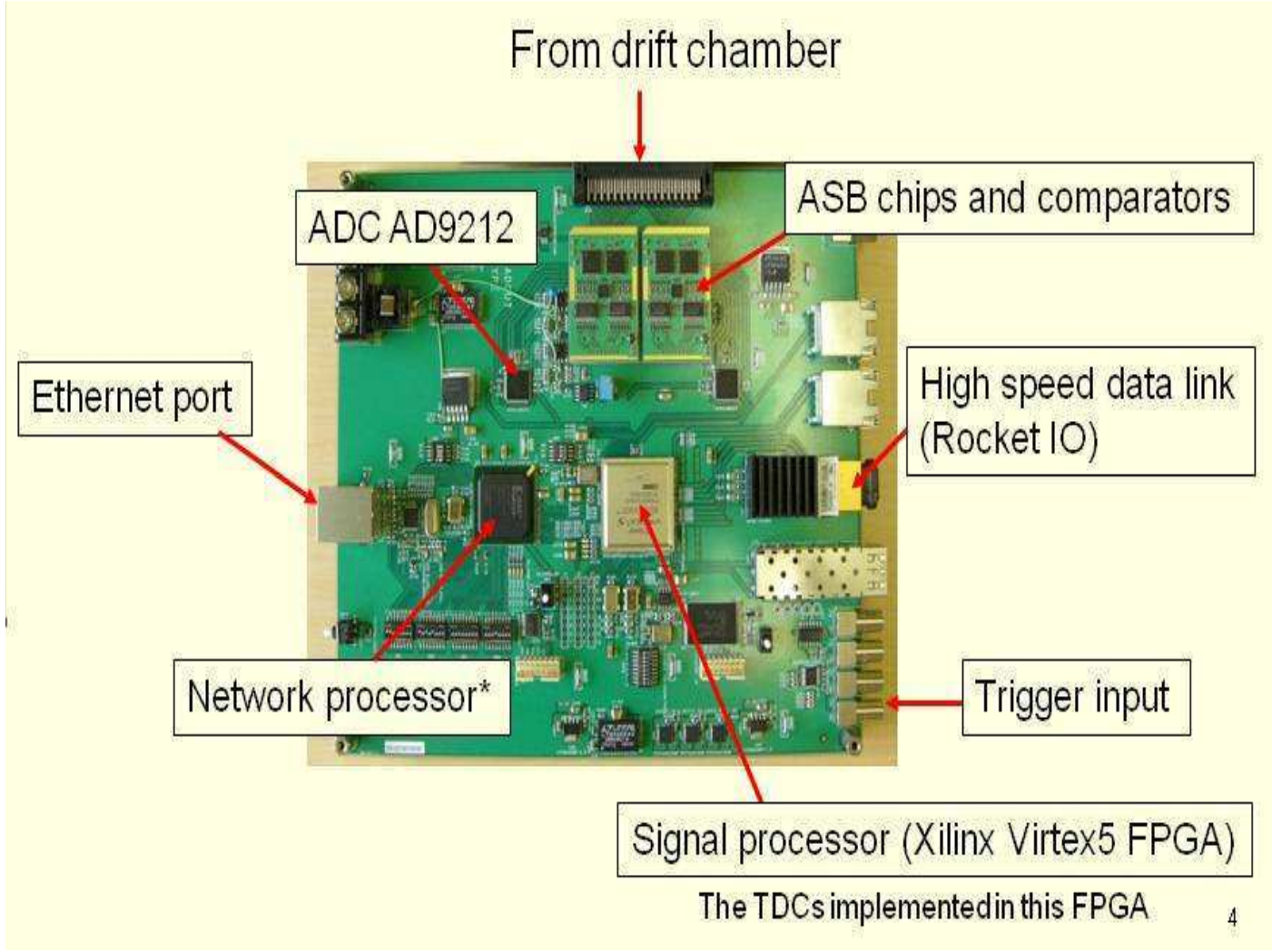

*Figure 6.13: Prototype for the CDC readout electronics board.*

## 6.5 Beam Test for a Prototype readout board

A test was carried out at the Fuji test beam line of KEK, using a 2 GeV/$c$ electron beam to study the performance of the prototype board. The readout board was connected to the test chamber, which is about 30 cm long and has five layers. The cell size is $15 \times 15$ mm and the wires are the same as those in the Belle CDC. A 50% helium – 50% ethane gas mixture is used. The sense wire high voltage of 2.3 kV was adjusted to obtain the same gain as is typical in the Belle CDC. The chamber and board were operated without any magnetic field. Figure 6.16 is a schematic view of the setup on the beam line.

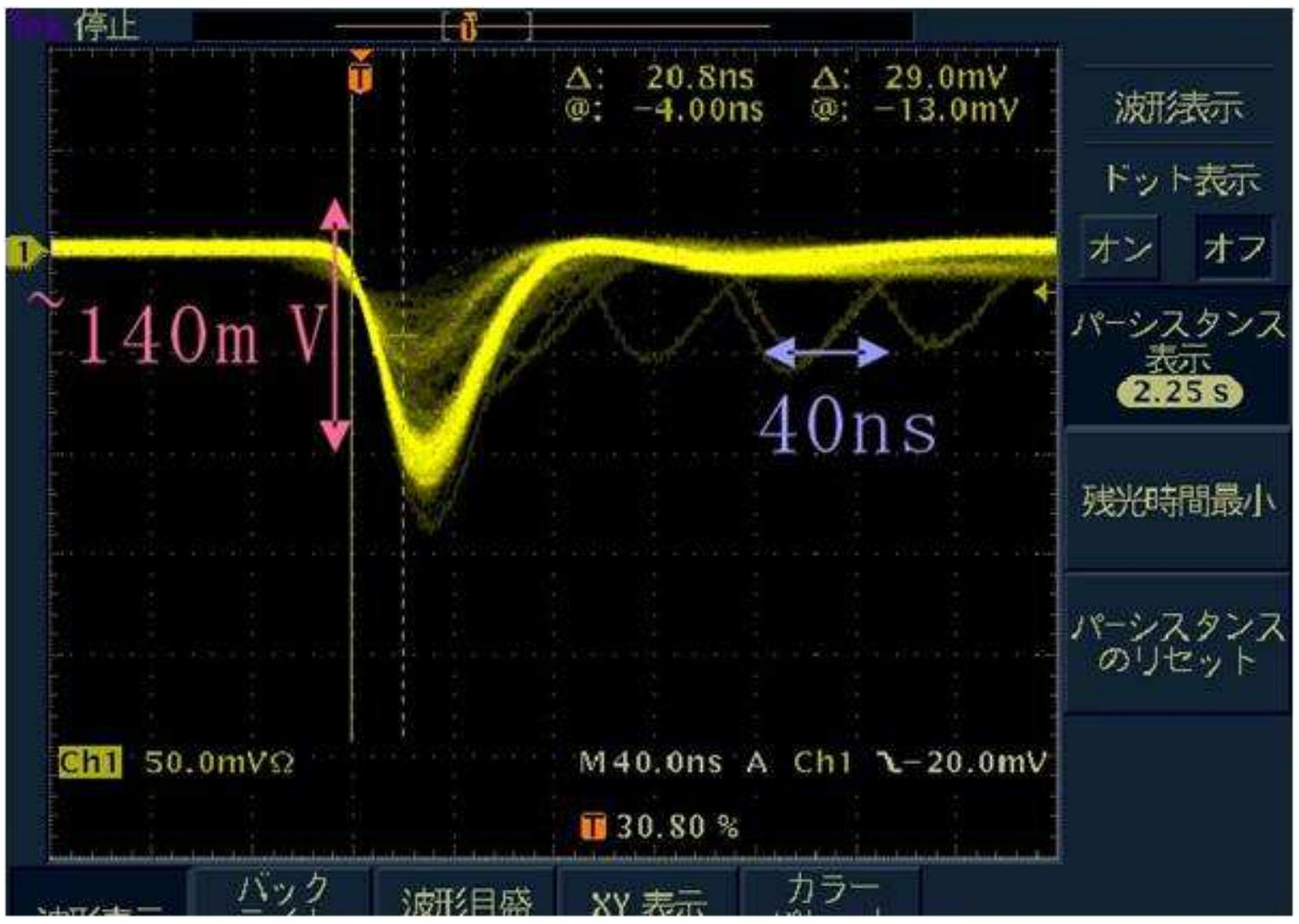

*Figure 6.14: Pulse shape for the prototype CDC ASIC chip.*

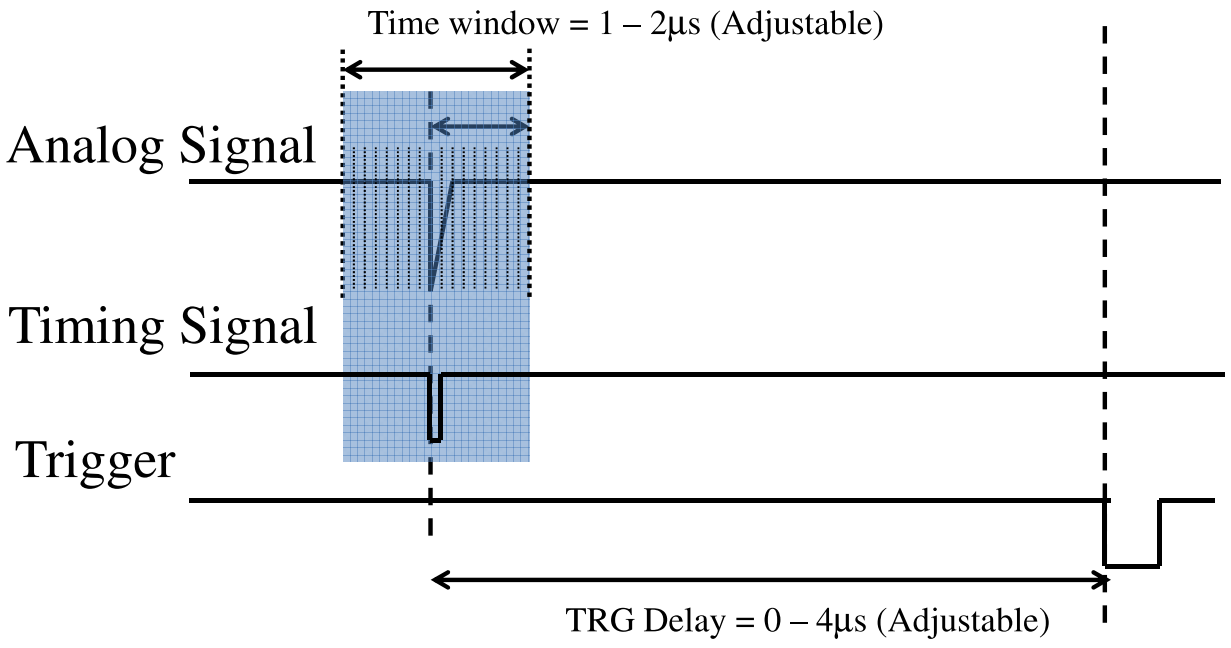

*Figure 6.15: Data acquisition time sequence.*

### 6.5.1 Specification

The board has an analog pre-amplifier and signal shaper, a 30 MHz sampling FADC, and a 1 ns-resolution TDC, which is implemented in an FPGA. Fast Ethernet is used for the data transfer.

#### 6.5.1.1 Readout mode

The board has two readout modes, raw data mode and suppressed data mode. The raw data mode was used for debugging and an initial test: all ADC and TDC data are transferred. The suppressed data mode is the normal mode of operation. Summed ADC and TDC data are transferred if a requirement is satisfied. The sparsified data size is small, typically several tens of bytes.

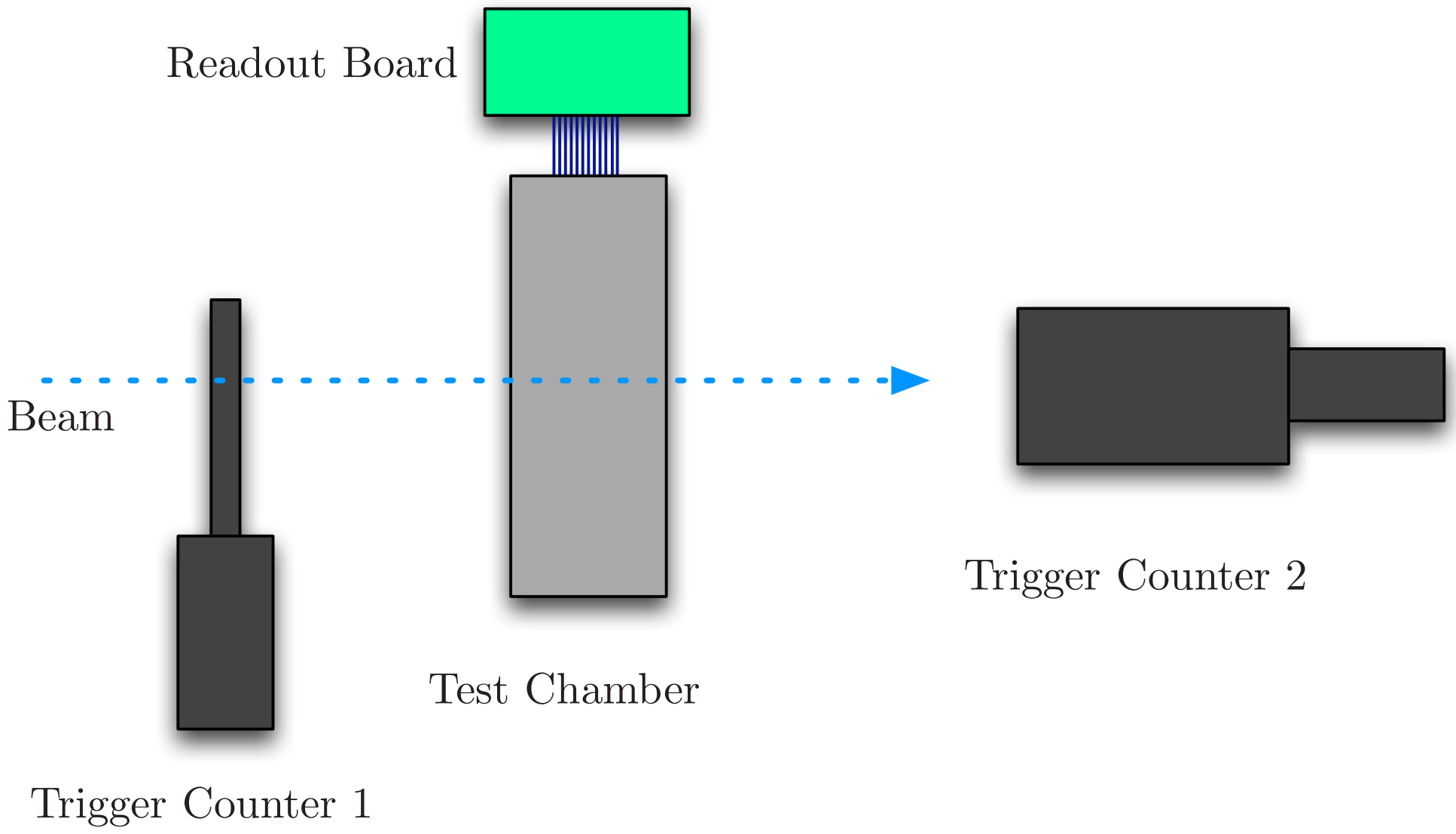

*Figure 6.16: Beam test setup at Fuji test beam line, with a 2 GeV/c electron beam. Trigger counter 1 is a plastic scintillator and a PMT; Trigger counter 2 is lead glass and a PMT.*

## 6.5.2 Performance

### 6.5.2.1 Efficiency

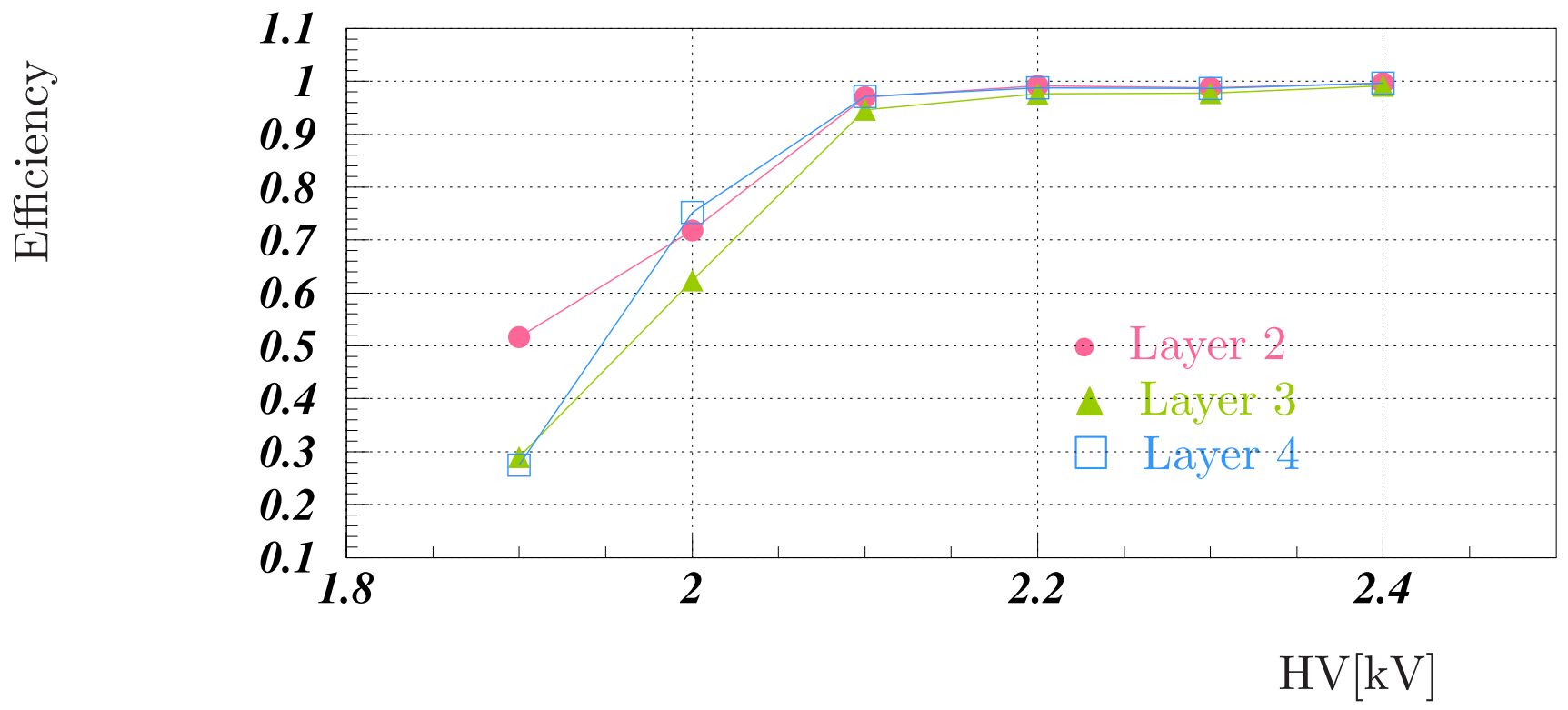

*Figure 6.17: Efficiencies for layers 2, 3 and 4 as a function of high voltage.*

Figure 6.5.2.1 shows the efficiency curves for layers 2, 3 and 4 as a function of the high voltage. Two outer-layer hits are required in the calculation of the efficiency. The efficiency is adequate in the region above 2.2 kV.

### 6.5.2.2 Drift time measurement

The $x$–$t$ (time-to-distance) relation is extracted as polynomial function or linear function, depending on the drift time region. Figure 6.5.2.2 shows this relation (left) and the residual distribution (right). The measured spatial resolution is about 100 $\mu$m.

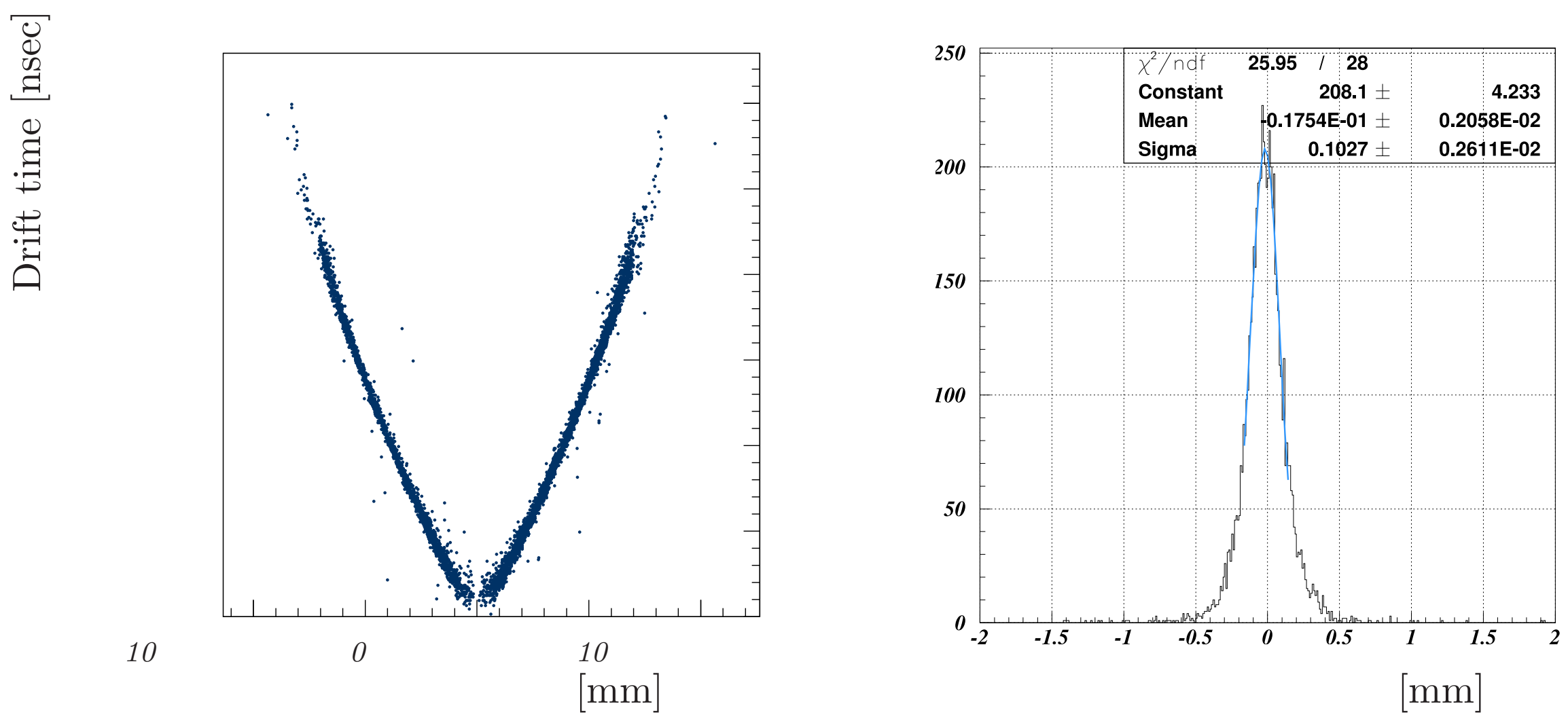

*Figure 6.18: Left: the measured $x$–$t$ relation. Right: the difference between the measured and parameterized relation.*

#### 6.5.2.3 d$E$/d$x$ measurement

The specific ionization, d$E$/d$x$ , is measured by summing a 30 MHz sampling FADC bin by bin. We sum FADC values that are larger than a fixed threshold. The threshold corresponds to the TDC threshold ($V_{\rm th}$) in this analysis. To obtain the d$E$/d$x$ resolution, we take the truncated mean to minimize the contribution of the Landau tail. In this analysis, 80% truncated means are used. The d$E$/d$x$ resolution was found to be 11.9% for nine layers at an incident angle of 90°. Figure 6.5.2.3 shows the distribution of the FADC sum at an incident angle of 90°. In the analysis, three events are associated and treated as one event. We reject the outer two layers, which have open cells.
The incident-angle dependence of d$E$/d$x$ resolution was measured. Figure 6.5.2.3 shows the d$E$/d$x$ resolution for nine layers with an incident angle of 90°, 60° or 30°. The results agree with the expectation.

#### 6.5.2.4 Event Buffer

The board has an event buffer that allows storage of at most four events in the suppressed data mode. In the raw data mode, there is no buffering, so about 30 in $10^4$ (0.3%) events were lost at an event rate of $\sim$ 20 Hz. Under the same conditions, all events were received on the suppressed data mode.

#### 6.5.2.5 Suppressed data mode

We compared the raw data mode with the suppressed data mode under the same conditions. Figure 6.5.2.5 shows the typical distribution of the FADC sum and the TDC for both modes. The value of the TDC is calculated as the difference between the time at which the trigger was received and the time when the signal exceeded threshold. The trigger is delayed by $\sim 2\,\mu$s relative to the signal.

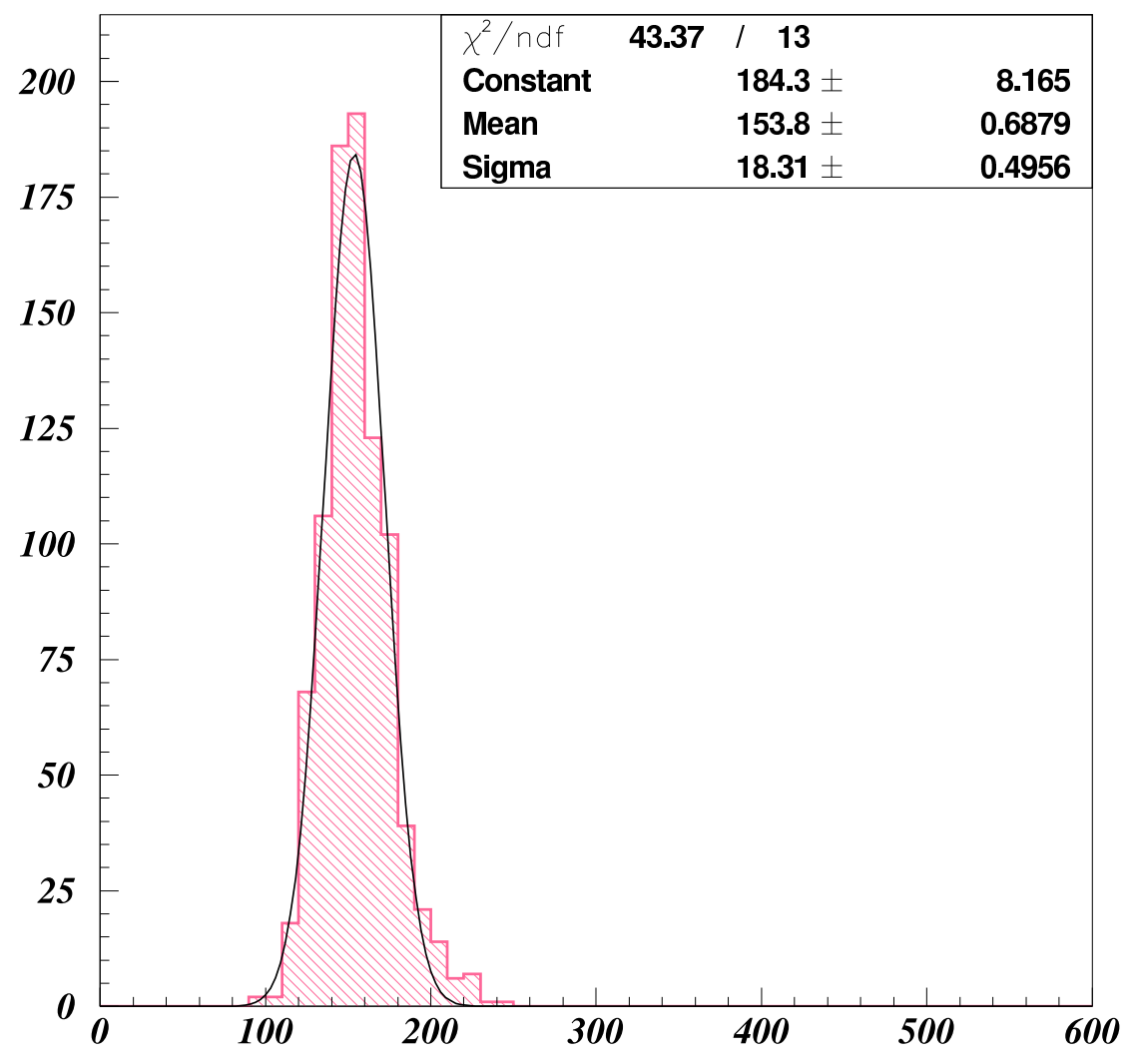

*Figure 6.19: The distribution of the FADC sum at an incident angle of* 90°.

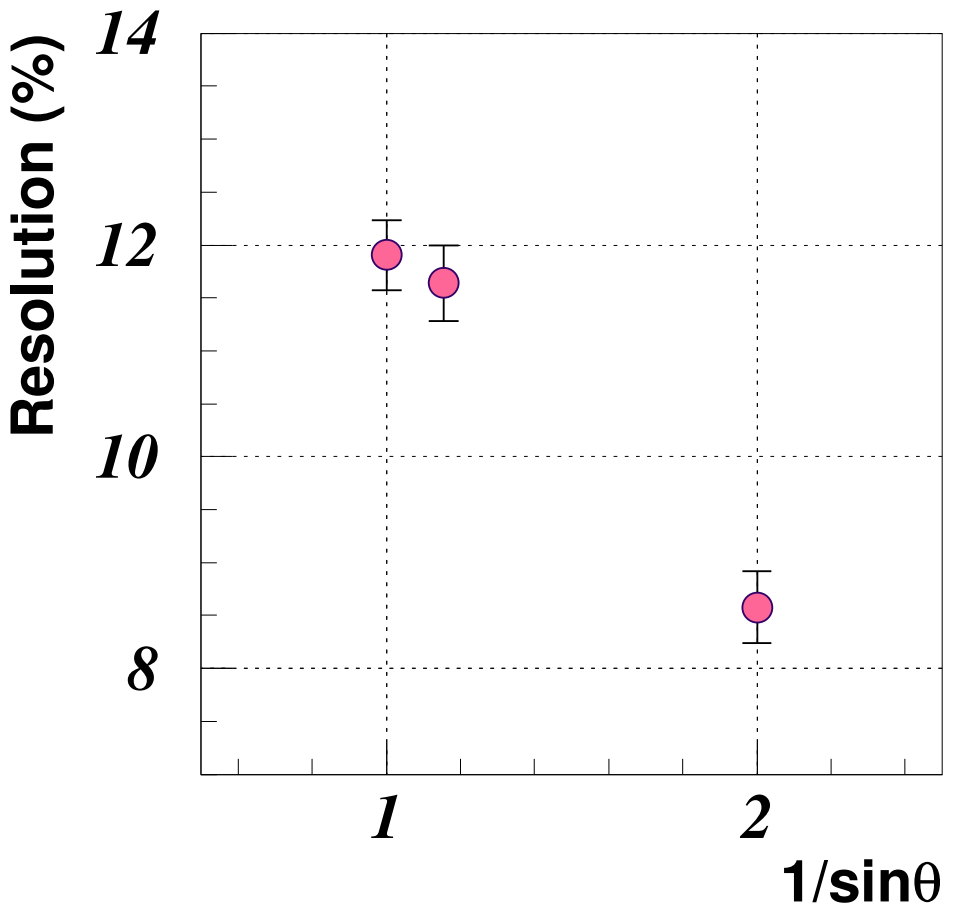

*Figure 6.20:* $\mathrm{d}E/\mathrm{d}x$ *resolution as a function of* $1/\sin\theta$ *at incident angles of 90°, 60° and 30°.*

## 6.6 Expected performance

We achieved good performance throughout the beam test using the test chamber and the prototype readout electronics board. The obtained position resolution and energy loss resolution for each cell are similar to or better than those achieved in the Belle CDC. Therefore, similar or better performance is expected without beam background.

The larger beam background in Belle II might degrade the CDC tracking performance. We assume 20 times higher beam background than in Belle and estimate the effect using a GEANT simulation. Only a 10% loss is expected for $B \to J/\psi K_S$ events. For $B \to D^* D^*$ events, a significant efficiency loss is expected without any corrective measures. However, new software and standalone tracking in the PXD and SVD will improve the overall efficiency dramatically so that even better performance can be obtained [4].

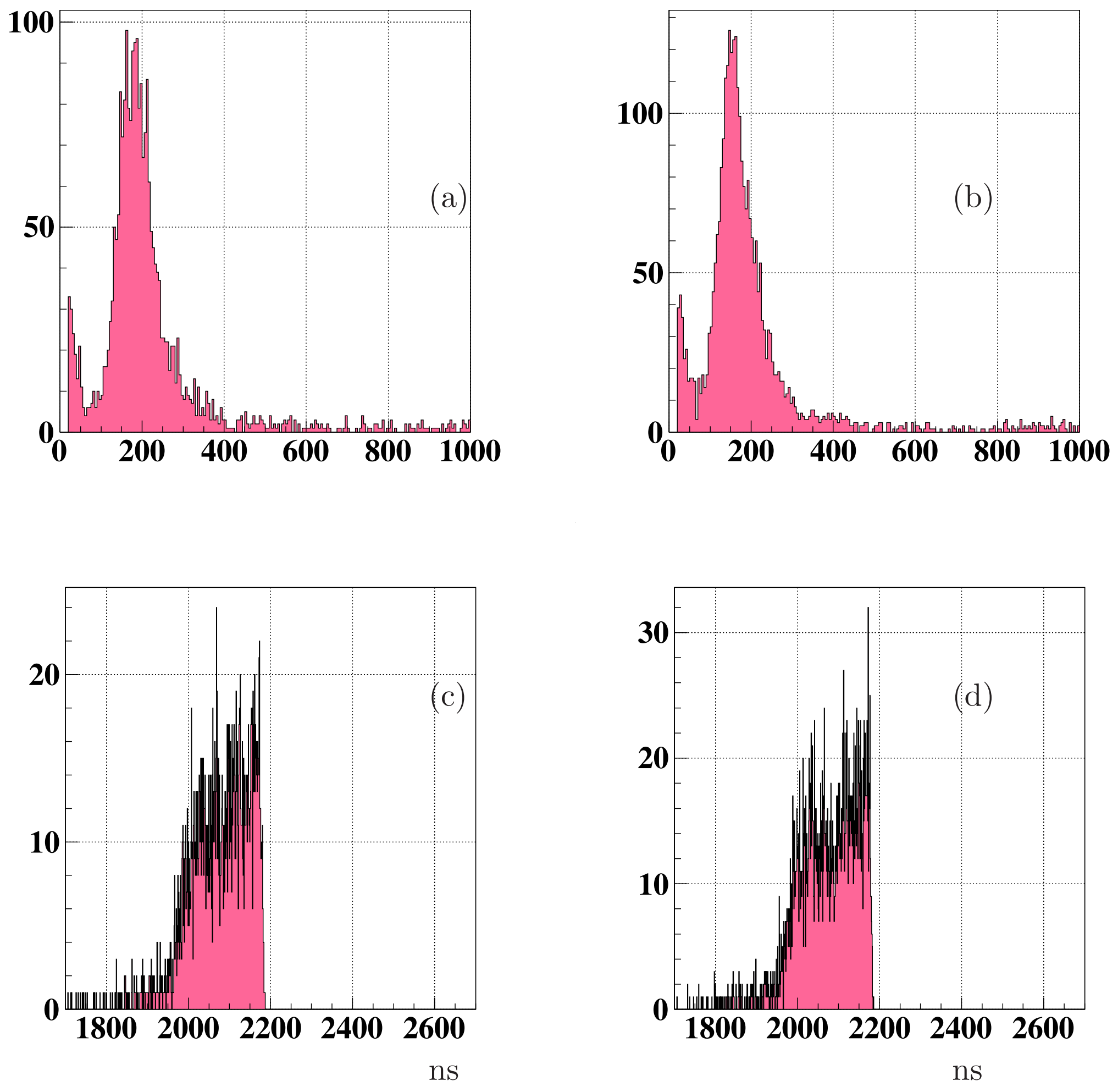

*Figure 6.21: The typical distributions of the FADC sum and the TDC. (a) FADC sum in the raw data mode, (b) FADC sum in the suppressed data mode, (c) TDC in the raw data mode, and (d) TDC in the suppressed data mode.*

# Chapter 7

# Particle Identification - Barrel

## 7.1 Introduction

To extend our physics reach, we would like to improve the $K/\pi$ separation capability of the spectrometer by upgrading the particle identification (PID) system. An upgrade of the system is also compulsory to cope with the higher background environment. Finally, to improve the calorimeter response to electromagnetic particles, we would like to to reduce the amount of PID material and make it more uniform.

In the barrel region of the spectrometer, the present time-of-flight and aerogel Cherenkov counters are replaced with a Time-Of-Propagation (TOP) counter [1], whose conceptual overview is shown in Fig. 7.1. In this counter, the time of propagation of the Cherenkov photons internally reflected inside a quartz radiator is measured (Fig. 7.2). The Cherenkov image is reconstructed from the 3-dimensional information provided by two coordinates $(x, y)$ and precise timing, which is determined by micro-channel plate (MCP) PMTs at the end surfaces of the quartz bar. The array of quartz bars surrounds the outer wall of the CDC; they are divided into 16 modules in $\phi$ in the baseline geometry.

The PID power is limited by the broadening of the time resolution due to the chromaticity of Cherenkov photons. To minimize the chromatic effect, we introduce a focusing system [2], which divides the ring image according to the wavelength of Cherenkov photons.

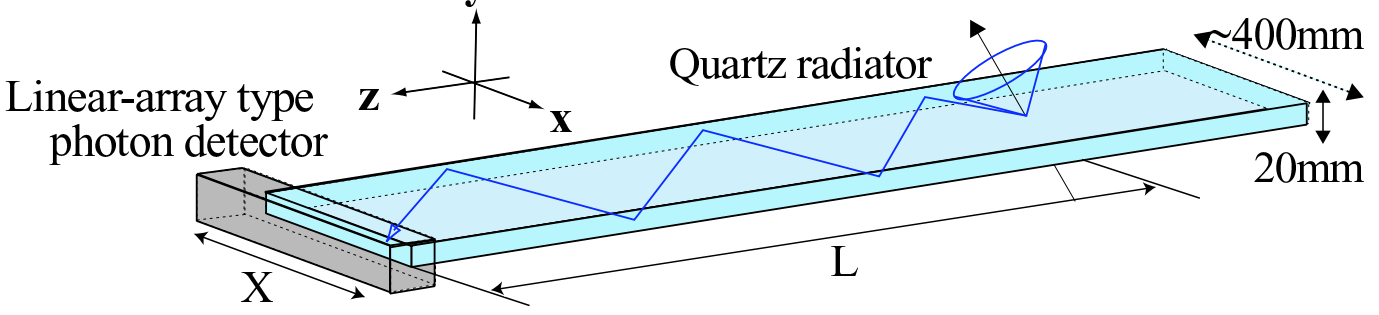

*Figure 7.1: Conceptual overview of TOP counter.*

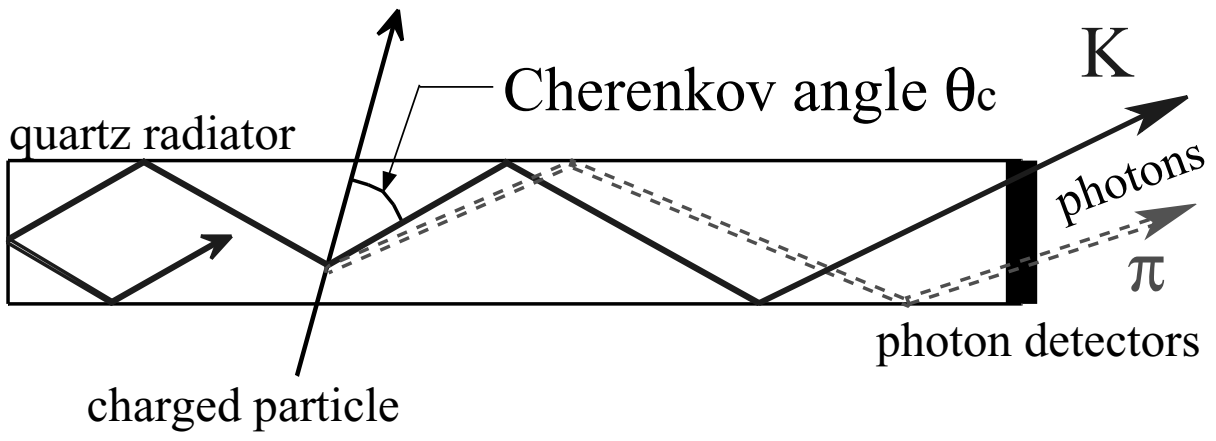

*Figure 7.2: Schematic side-view of TOP counter and internal reflecting Cherenkov photons.*

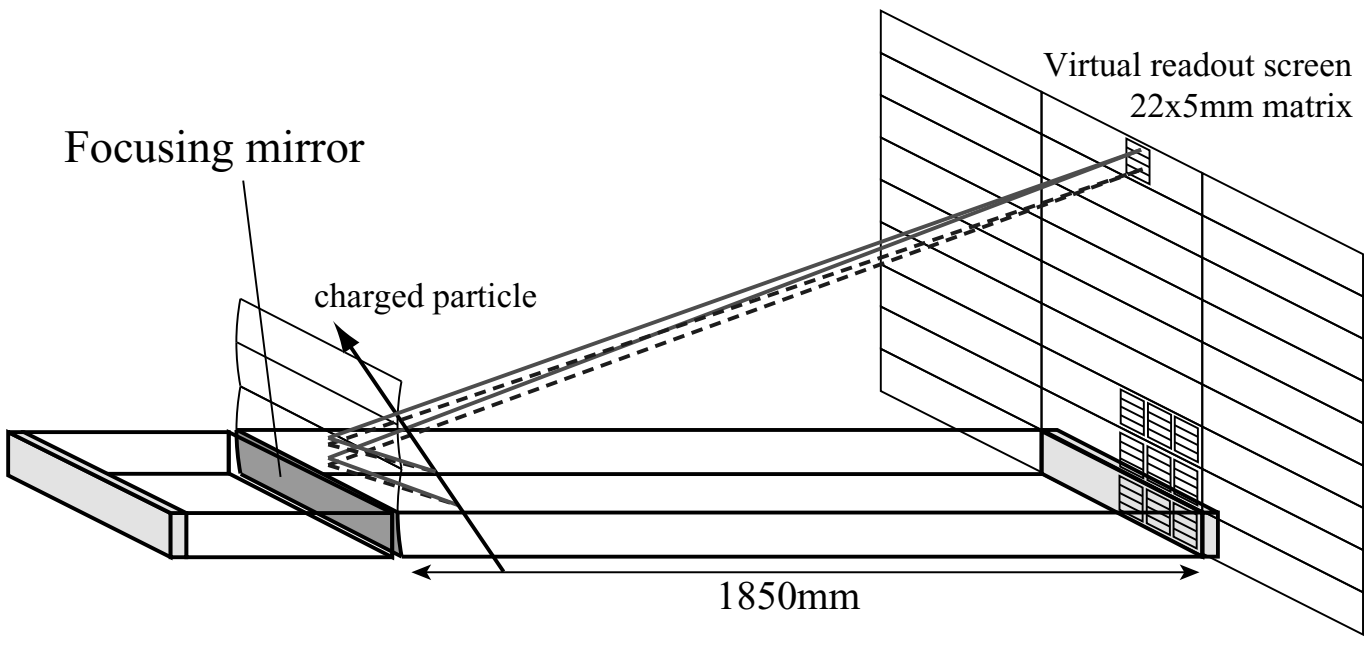

*Figure 7.3: The principle of the focusing scheme in the TOP counter. The virtual extension of the focal surface and of the photon detector plane are shown by the dashed curves and lines.*

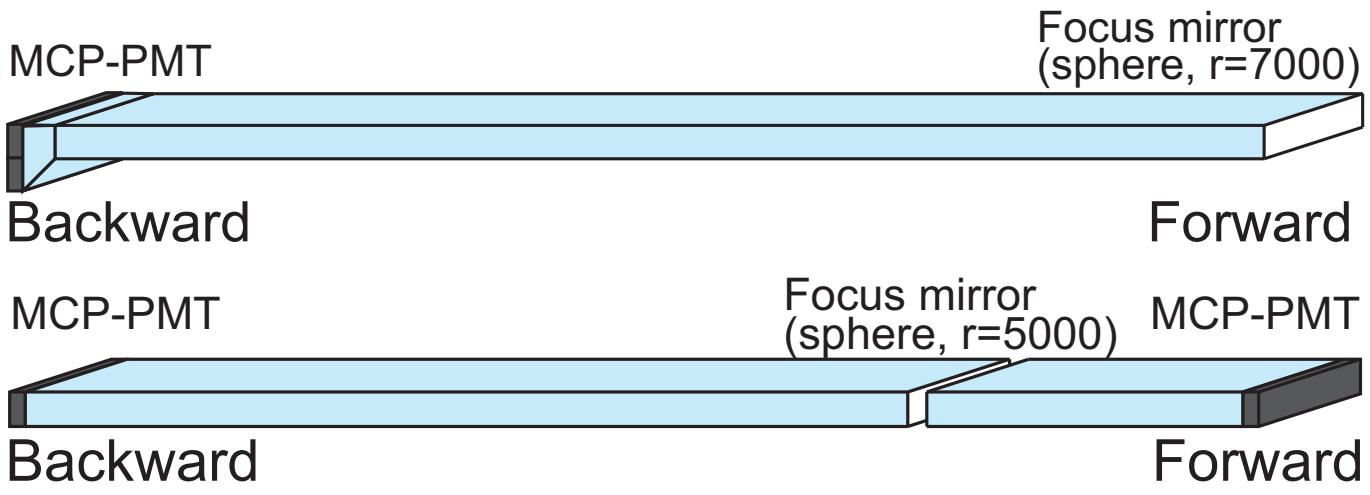

*Figure 7.4: Conceptual sketches of the 1-bar and 2-bar TOP baseline designs.*

Figure 7.3 shows the schematic set-up. A focusing mirror is introduced at the end of the quartz bar. The PMT is aligned to measure the position of the photon impact in $x$ and $y$. The cross section of the quartz bar is a rectangle. We can therefore expand the light trajectory into the mirror-image region and create a virtual readout screen. The Cherenkov photons with different $\theta_c$ will focus onto the different PMT channels; we thereby obtain $\lambda$ information from the $y$ detection position through $\theta_c$. The focusing TOP reconstructs the partial ring image from 3-dimensional information: time, $x$ and $y$.

## 7.2 Detector configuration

We studied two options for the detector configuration as shown in Fig. 7.4. For the 1-bar option, we employ an expansion prism on the bar end and expand the partial ring image onto the readout plane. It clarifies the ring image and improves the discrimination of wavelength. It also increases the number of effective photons slightly. For the 2-bar option, in order to reduce temporal degradation due to chromatic dispersion, each radiator bar is subdivided into two pieces at $z \sim 1070$ mm, where the "short" bar with $\sim 750$ mm length is used in the forward direction, and the "long" bar of $\sim 1850$ mm length is used in the backward direction. The short bar is instrumented by PMTs at the forward edge and the long bar has PMTs at the backward edge. In both options, a wavelength filter ($\lambda \gtrsim 400$ nm) is introduced in front of the PMTs to further reduce the chromaticity.

As discussed in Section 7.5, the performance estimates for the 1-bar and 2-bar options are similar for the physics cases considered, though the 2-bar option shows slightly better performance than the 1-bar option in general. The performance is affected by non-ideal measurements. The 1-bar configuration is less sensitive to the event timing degradation, while the 2-bar is less sensitive to degradation of incident angle determination in the tracking. Practically, the 1-bar case has an advantage, since it shows larger forward acceptance and requires readout only in the backward direction, so that we can avoid the crowded forward cable routing. Therefore, we chose the 1-bar configuration as a baseline, with the 2-bar as a backup option.

## 7.3 Components

### 7.3.1 Quartz radiator

The quality of the synthetic fused silica (quartz) optical components is important for the proper performance of the TOP detector. The radiator has three components: a long bar for radiating Cherenkov light and propagating this light via total internal reflection to the bar end, where the MCP-PMTs are mounted; a spherical mirror mounted on the forward end of the bar for focusing the light; and, for the 1-bar design, a prism that attaches to the backward end of the bar and allows the Cherenkov ring image to expand before the photons are recorded by the PMTs.

#### 7.3.1.1 Quartz bar

To propagate Cherenkov photons inside the radiator and preserve the Cherenkov ring image, the sides must be flat and parallel to a very tight tolerance. Our baseline specification for flatness is less than $10\lambda$ over the entire bar length ($\sim$ 1.2m). The roughness must be less than 5 Å r.m.s. to minimize photon loss. According to the G Kirchhoff equation, the reflectivity depends on the surface roughness. In the case of 100 bounces on the surface, the total reflectivity is 0.97 and 0.87 for the roughness of 5 and 15 Å, respectively. The chamfer size should be as small as possible, to reduce photon loss at the bar edges. Table 7.1 lists our current baseline bar specifications. With these specifications, the number of photons reaching the phototubes after total internal reflection is $\sim$100.
We have received two prototype bars of dimensions $400 \times 915 \times 20$ mm$^3$ fabricated by Okamoto Optics, Inc. in Japan, and we have ordered another (longer and wider) bar from Zygo Corporation in the US. The Okamoto bar is shown in Fig. 7.5. Using this bar, we have measured the timing resolution and the number of photons detected for internally reflected laser light and the performance obtained was satisfactory.

*Table 7.1: Baseline quartz bar specifications.*

| | |
|---|---|
| Size | $(440 \pm 0.15) \times (1200 \pm 0.5) \times (20 \pm 0.10)$ mm$^3$ |
| Material | Synthetic fused silica (e.g., Corning 7980, Shin-etsu Suprasil) |
| Index tolerance | $\pm0.001$ |
| Flatness | $10\lambda$ over full aperture |
| Roughness (r.m.s.) | 5 Å |
| Angle between planes | $(90 \pm 1/60)$ degree |
| Chamfer size | $< 0.20$ mm |

#### 7.3.1.2 Focusing mirror and other optical components

The focusing mirror is spherical with a focal length of $(\cos\langle\theta\rangle)^{-1} \times$ (bar length), where $\langle\theta\rangle$ is the mean angle of incidence of tracks on the mirror surface for decays of interest. For a bar of length 2.5 m, this focal length corresponds to a radius of curvature of about 7 m. At present, we have received a 40-cm width prototype mirror with a 5-m radius of curvature from Okamoto Optics and a 44-cm wide prototype mirror with a 7-m radius of curvature from Optical Solutions, Inc. (OSI) in the US. We have tested the mirror from Okamoto in the laboratory and determined the peak position of the image to be within 1 mm of expectation. We have recently received the mirror from OSI are starting to evaluate it.

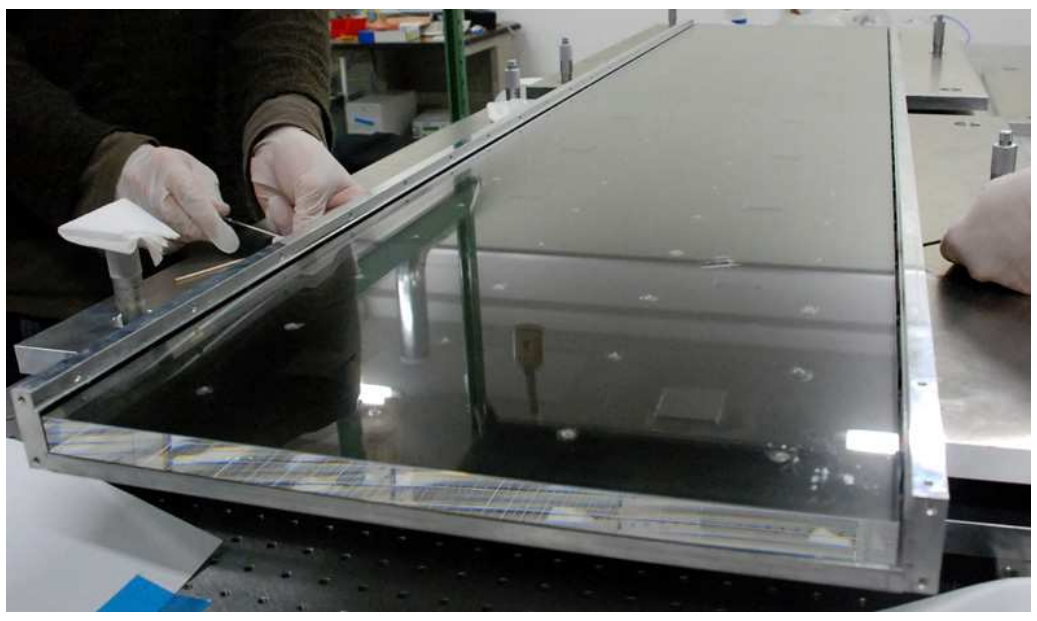

*Figure 7.5: Photograph of the Okamoto prototype quartz radiator bar.*

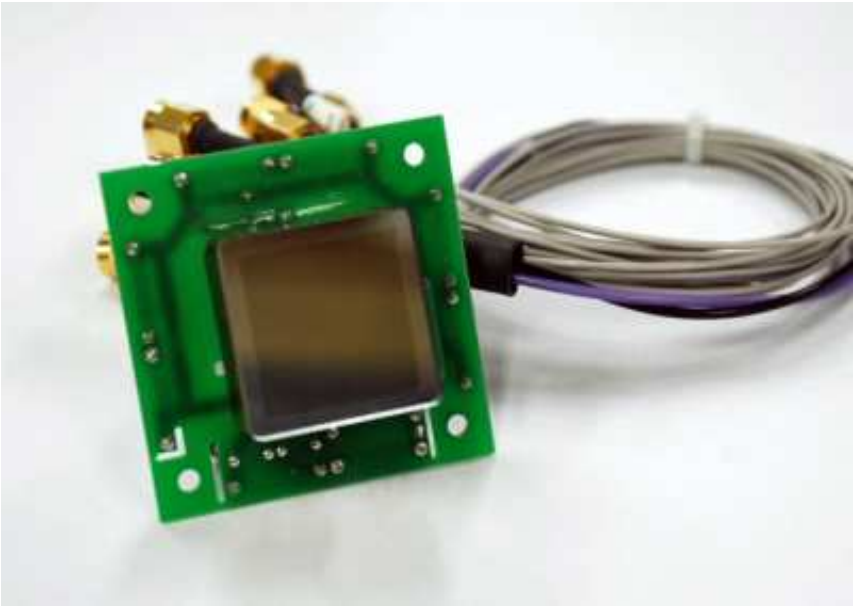

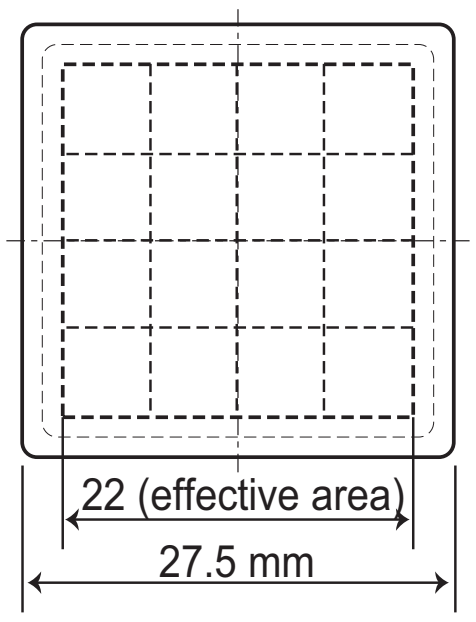

*Figure 7.6: SL-10 prototype MCP-PMT photograph (left) and anode structure diagram (right).*

To attach the MCP-PMTs to the expansion prism and that to the radiator bar, we plan to use one or more of the following methods: a UV-cured optical adhesive such as Norland NOA 61, 63; a two-component optical epoxy such as Epotek 301-2; or a high-grade silicone optical grease such as Bicron BC-630. The last option allows for the possibility of easily separating the components if desired, e.g., for replacing PMTs in-situ. We used NOA 63 for the prototype, which works well as shown in the beam test result (Sec. 7.4).

### 7.3.2 Photon detector

A TOP counter requires the precise timing measurement of single photons at the end of its quartz bar radiator. MCP-PMTs have excellent timing and gain performance. We have investigated the single-photon detection performance of several MCP-PMTs [3]. Transit time spreads (TTS) of $\sim 30$ ps at the gain of $10^6$ for single photo-electrons, even in the presence of a 1.5 T magnetic field, have been observed.

We have developed square-shape MCP-PMTs [4], denoted SL-10, to improve the packing efficiency when the MCP-PMTs are abutted on the end of the quartz bar. The basic SL-10 specifications are summarized in Table 7.2, and a photograph and diagram of the anode structure is shown in Figure 7.6. The SL-10 has a $4 \times 4$ anode array, a multi-alkali photocathode, two MCP plates with 10 $\mu$m pore size, and an aluminum layer on the second MCP to protect against ion feedback.

*Table 7.2: Hamamatsu SL-10 MCP-PMT specifications.*

| | |
|---|---|
| PMT size | 27.5 × 27.5 × 15.6 $mm^3$ |
| Effective area | 22 × 22 $mm^2$ |
| Photocathode | Multi-alkali |
| Number of MCP layers | 2 |
| MCP pore diameter | 10 $\mu$m |
| MCP aperture | ∼60% |
| MCP bias angle | 13° |
| Al protection layer | on second MCP |
| Anode pattern | 4 × 4 |
| Anode width | 5.3 mm |
| Anode gap | 0.3 mm |
| Maximum supplied HV | 3.5 kV |

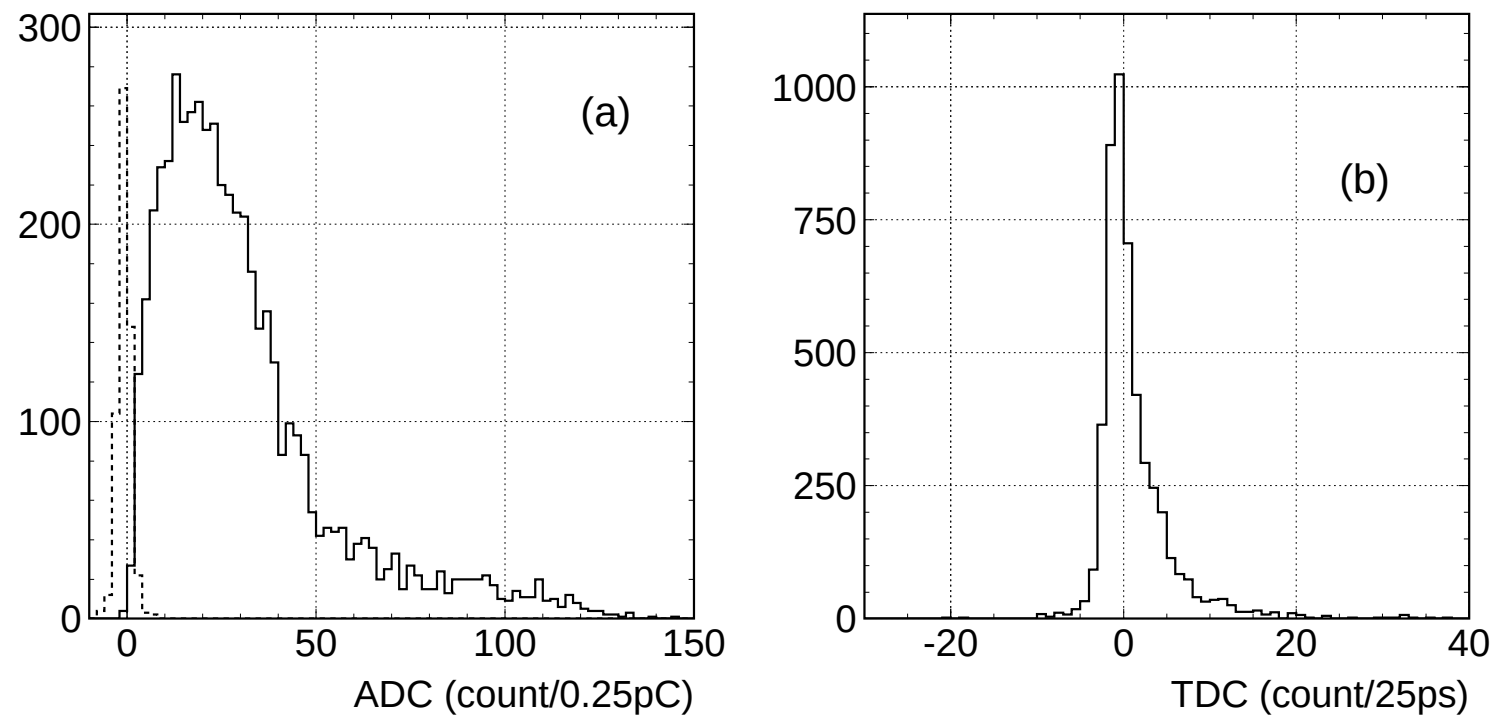

*Figure 7.7: (a) ADC and (b) TDC distributions for SL-10 single photon detection. The dashed histogram is a pedestal distribution. The TDC distribution is corrected for amplitude (ADC) discriminator threshold timing dependence.*

Figure 7.7 shows ADC and TDC distributions for the SL-10 MCP-PMT response to single photons as generated by a fast, pulsed laser (Advanced Laser Diode Systems; PiL040). From the ADC distribution, the gain is evaluated to be $1.2 \times 10^6$. After ADC amplitude correction, a single photon timing resolution of 31 ps is seen in the TDC data. The tail of the TDC distribution is due to the bounce of secondary electrons on the MCP surface, and is taken into account in the simulation programs.
Figure 7.8 shows the quantum efficiency (QE) distribution as a function of wavelength. The baseline MCP-PMT has a multi-alkali photo-cathode with the QE of 18% at 400 nm.
Long term MCP-PMT stability is also evaluated under high background rates. The main source of background is expected to be due to spent electrons from the beam. The number of the background photons is estimated using a Belle spent-electron simulation. The consistency between this generator and the data is evaluated using the trigger hit rate of the existing Belle TOF counter. At a luminosity of $10^{34}cm^{-2}s^{-1}$, the trigger hit rate is measured to be 187 kHz, while the simulator indicates a 400–500 kHz rate. For such background estimates, agreement within a factor of 2–3 is respectable, and our simulation estimate is on the conservative side. Using this generator, the Cherenkov photon rate is estimated. Optical photons are produced in the radiator due to the Cherenkov radiation generated by relativistic electrons from spent electron shower gammas. The detected photon rate is estimated to be 300 kHz/(TOP module). Scaling this estimate up by a factor of 20 to account for the increased background at Belle II, a photon hit rate of 68 kHz/cm$^2$ on the photon detector window is predicted, corresponding to an output charge from anodes of 120 mC/cm$^2$/year, assuming a gain of $10^6$.

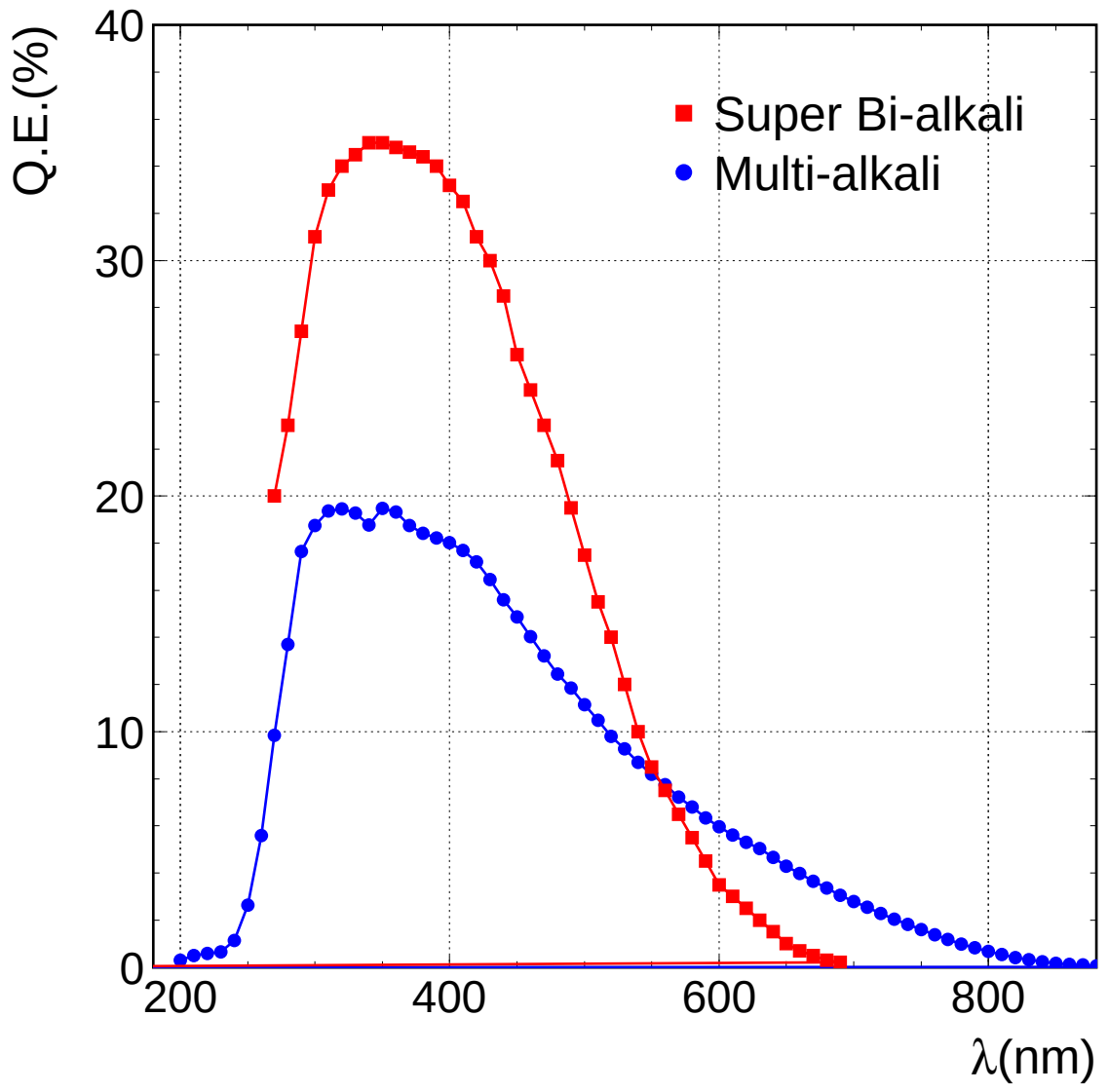

*Figure 7.8: Quantum efficiency distributions as a function of wavelength.*

Stability is determined by LED light pulse illumination. Test results for a round-shape MCP-PMT are summarized in Refs. [4] and [5], which indicate that stable gain and TTS can be obtained by changing the supplied HV. The quantum efficiency, however, is the critical item. With an aluminum protection layer, the round-shape MCP-PMT demonstrated sufficient lifetime for ten Belle II years of operation. We have also checked these parameters for the SL-10 MCP-PMT. Figure 7.9 shows the relative efficiency as a function of integrated anode charge output. Adequate lifetime (over at least three Belle II years) is also seen for the SL-10 MCP-PMT after improvement through the modification of the inner MCP structure and a cleaning method that avoids neutral gas and ion feedback from the second MCP to photocathode.

The baseline MCP-PMT described above has shown adequate performance. To provide additional operating margin, it is desired to improve the number of detected photons. Therefore, we are developing an MCP-PMT with a super bi-alkali (SBA) photo-cathode in collaboration with Hamamatsu Photonics. Figure 7.8 compares the quantum efficiency (QE) distribution as a function of wavelength for multi-alkali and SBA. The multi-alkali QE is measured using a prototype MCP-PMT, while the SBA QE curve is from a Hamamatsu data sheet. Before starting mass production, we will check the long term stability of the photo-cathode.

We also maintain the option of using a Photonis MCP-PMT. The status of a tube capable of operating in a 1.5T magnetic field is described further in Ch. 8.

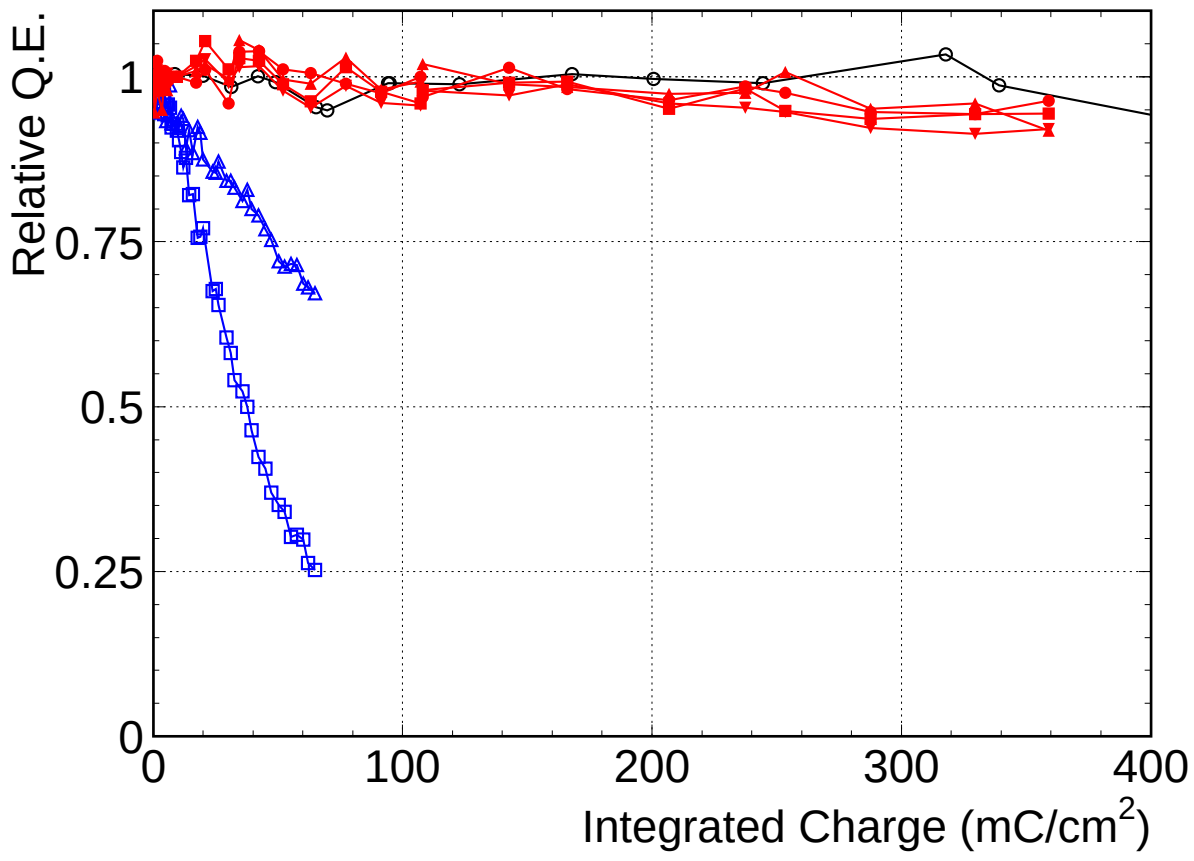

*Figure 7.9: Relative QE as a function of integrated output charge for round-shape MCP-PMT (○) and SL-10 MCP-PMT (□ and △ before improvements; ●, ▲, ▼, and ■ after improvements).*

### 7.3.3 Electronics

#### 7.3.3.1 Fast Timing with Waveform Sampling

Upgrade of the barrel PID device for the Belle II detector relies heavily on precision time recording of individual Cherenkov photon signals. This places severe demands on the transit time performance of photodetectors that need to operate in a 1.5T magnetic field, as well as precision single photo-electron signal recording. Key to this latter challenge, in a compact and cost-effective method, has been the development of low-power GHz analog bandwidth, high-performance waveform recording [6, 7, 8] Application Specific Integrated Circuits (ASICs). In contrast with previous GHz rate sampling devices such as the ATWD [9], having higher analog input bandwidth exploits the full benefit of GHz sampling. Such waveform sampling has been successfully flown twice on the ANITA long duration balloon payload [10, 11], where the excellent timing thus obtained ($\leq$ 30 ps) was able to provide sub-degree pointing resolution to an in-ice transmitter located more than 100 km away. Subsequently this "oscilloscope on a chip" technology (Fig. 7.10) has been explored for use in several Belle II subdetectors.
This development is well aligned with recent developments in high-density, high precision timing photodetectors. Direct integration of the sampling and digitization ASICs with both traditional vacuum-based (Micro-Channel Plate, Hybrid Photo-Diode) and solid state (Geiger-mode APD) photodetectors have demonstrated that lower cost, higher quantum efficiency, and improved transit-time-spread (TTS) photon detection modules are possible. Fig. 7.11 illustrates that without these improvements, the performance of large-scale Time-Of-Flight (TOF) systems for precision spectrometers has stalled at a resolution of about 100 ps.
While electronics have been available for decades that can readily measure signal times to 25 ps or less, cost and lack of direct integration with the detector sensor elements has limited system performance as illustrated. This is also true for the Time-of-Flight system of the Belle detector [13, 14]. In recent years, studies have explored mating of the photodetectors with the electronics [15]. Depending upon the timing and amplitude resolution required, a number of options for signal triggering, time encoding, and digitization have been explored [16, 17, 18]. The results are reference designs for high speed waveform sampling for iTOP precision timing, slower speed (1–2 GSa/s) sampling for drift chamber and KLM endcap scintillator readout, and direct FPGA conversion for the KLM barrel RPC readout.

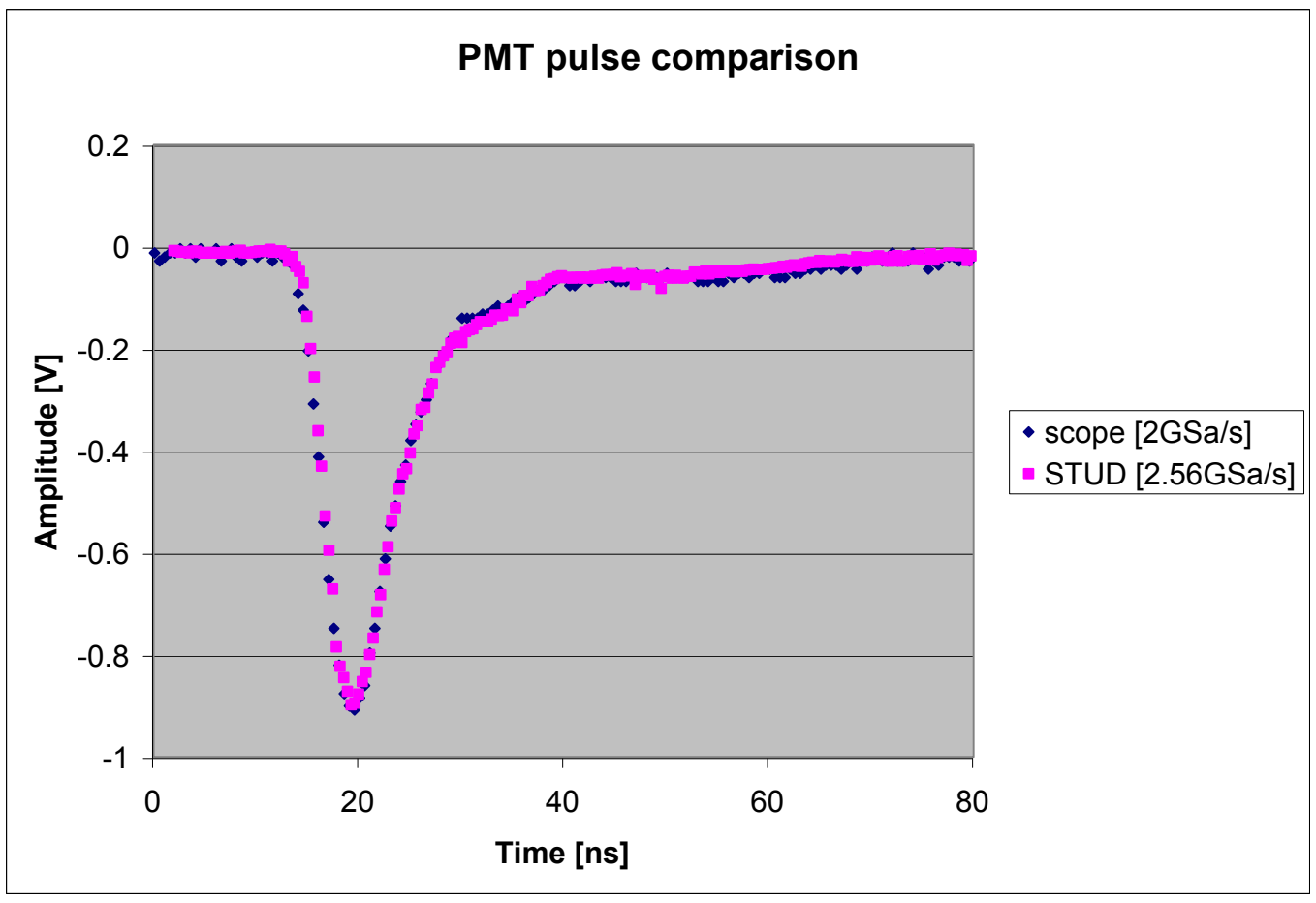

*Figure 7.10: Recording of a fast Belle fine-mesh Photo-Multiplier Tube (PMT) signal with a high-speed digital oscilloscope and a prototype sampling ASIC (SalSA Transient UHF Digitizer [STUD]).*

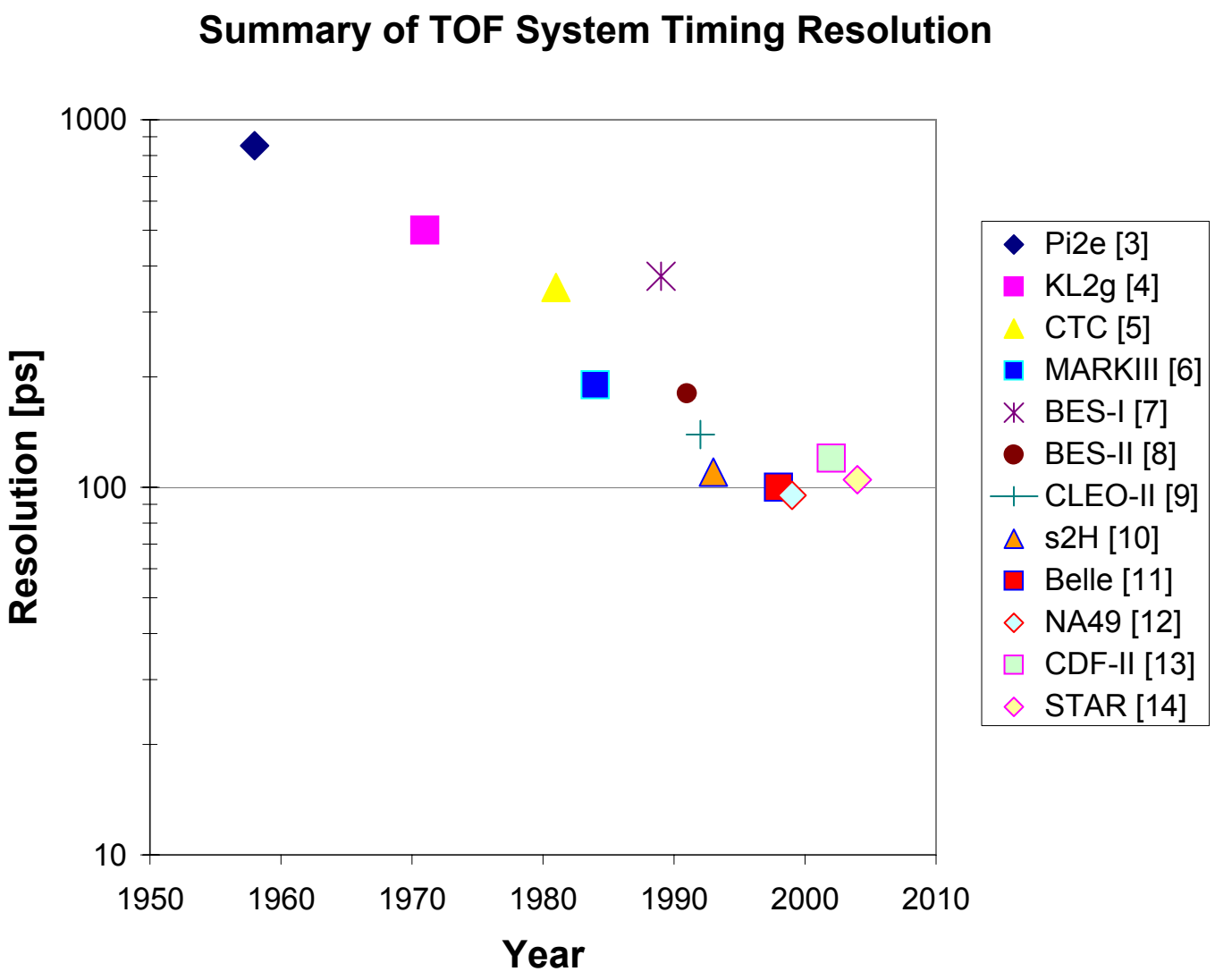

*Figure 7.11: A historical comparison of large Time-Of-Flight system timing performance, where progress has largely stagnated. The reference numbers in the legend are found in Ref. [12]. To do significantly better, improved techniques, photodetectors, and electronics are needed.*

Based on these studies, it was determined that waveform sampling—if the sample buffer depth can be extended—is the most attractive approach for readout of a Cherenkov timing device. Therefore, a deeper sampling ASIC, designated the Buffered LABRADOR (or BLAB), was fabricated and evaluated [19], a die photograph of which is seen in Fig. 7.12.

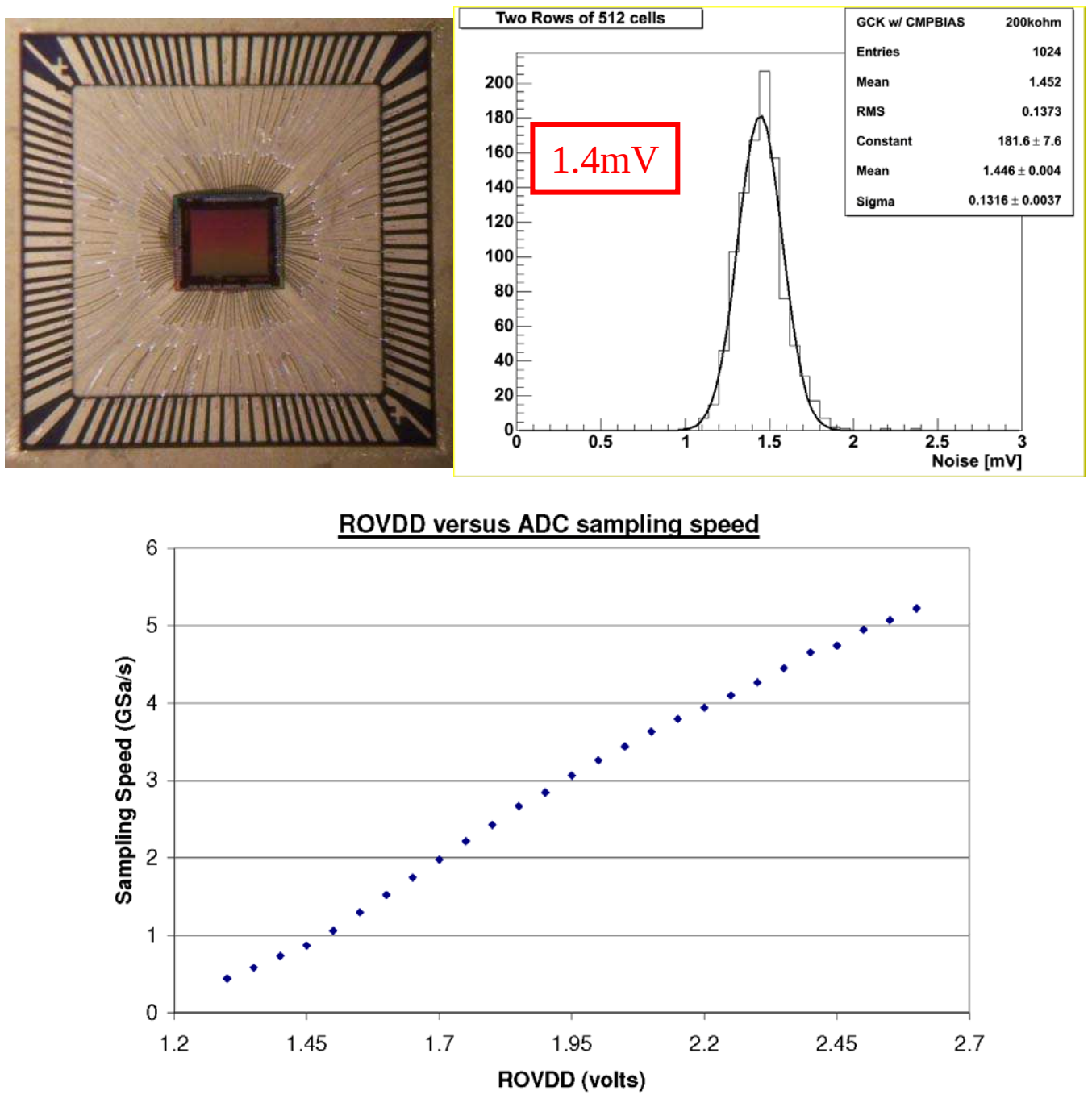

*Figure 7.12: Upper left, a die photograph of the BLAB1 ASIC (3 mm × 3 mm), a single-channel demonstrator with 65,536 storage cells fabricated in the TSMC 0.25 $\mu$m process [19]. At right and below are the measured noise and adjustable sampling speed, respectively.*

This compact ASIC demonstrated excellent performance during testing. Sampling rates of over 5 GSa/s were measured, while single samples could be recorded with over 1.4 V dynamic range and 1.4 mV of noise, or 10 real bits per sample. A 16-channel prototype was subsequently assembled and tested with the fast, focusing DIRC (fDIRC) prototype [20] at SLAC. Excellent beam test results were obtained [21], and a complete readout system has been constructed based on the next generation of this ASIC.

Further testing of this ASIC indicated that through careful calibration and attention to reference timing distribution, sub-10 ps timing can be obtained [12] between two BLAB1 ASICs recording single photon signals. This measured time difference is plotted in Fig. 7.13. If the errors were uncorrelated, the actual timing would be better per single measurement by a factor of $\sqrt{2}$.

Research continues toward the goal of obtaining something like a few ps of timing resolution, where recent studies [22] are encouraging if a sufficient signal-to-noise ratio can be realized, and systematic errors are kept under control. However, such exquisite resolution is not strictly necessary for the TOP application, as systematic errors in determination of the collision time [23] are expected to limit timing resolution at the 25 ps level, as discussed later. Maintaining such timing resolution under operating conditions is a significant challenge, as we have learned from over a decade of continual calibration of the Belle TOF. As in the latter case, there is no need to maintain tens of ps timing over the 5 $\mu$s L1 trigger latency: all hits are recorded with respect to the collision time, minimizing timing errors.

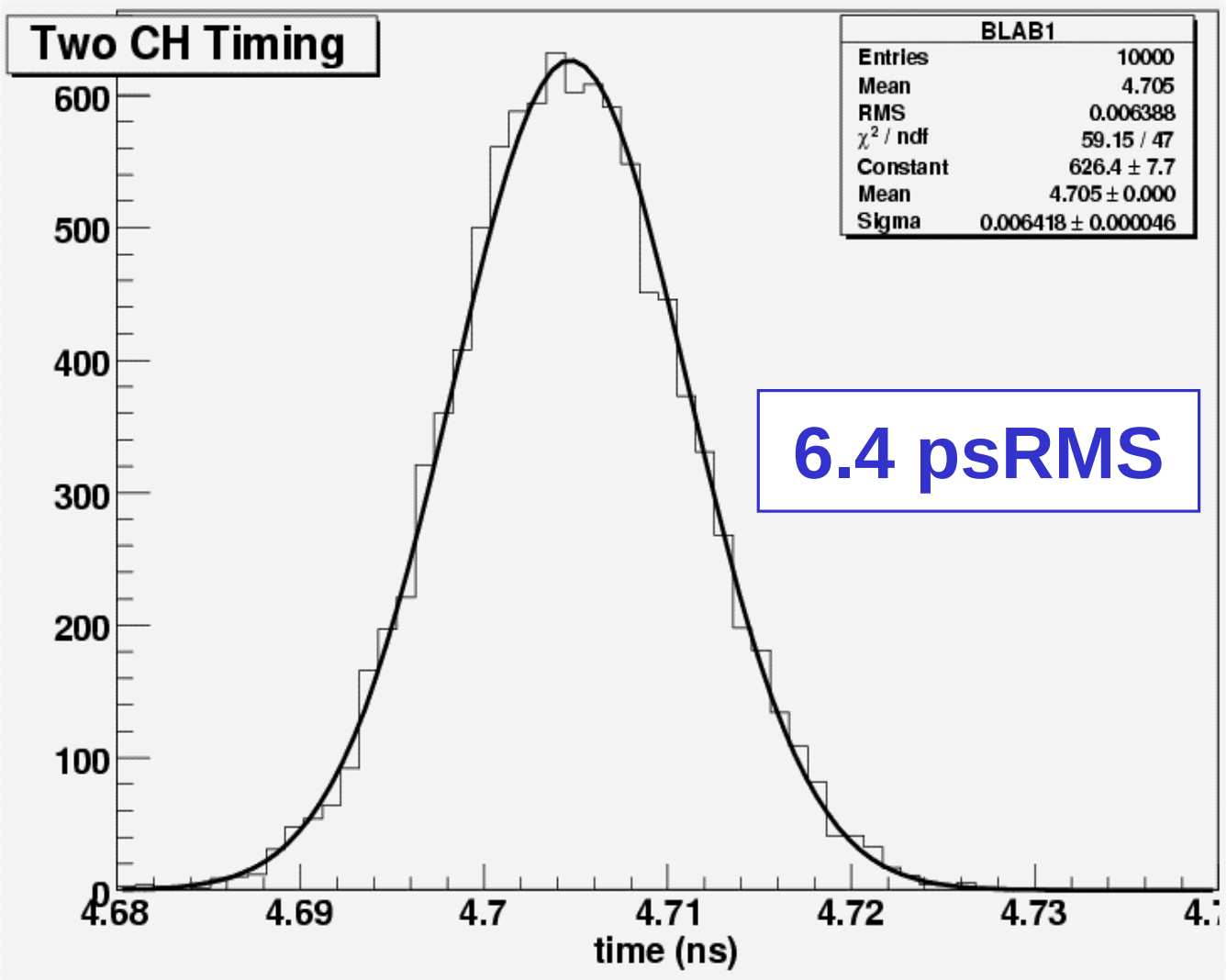

*Figure 7.13: Example timing resolution that has been obtained with low-power CMOS waveform sampling [12].*

A third generation Buffered LABRADOR ASIC (BLAB3) is the baseline for readout of either barrel PID configuration. The specifications for this device are listed in Table 7.3.

*Table 7.3: Specifications for the baseline BLAB3 readout ASIC.*

| Parameter | Value | Comment |
|---|---|---|
| Channels/BLAB3 | 8 | die size constraint |
| Sampling speed | 4 | Giga-samples/second (GSa/s) |
| Samples/channel | 32768 | allows $\geq$5 $\mu$s L1 trig latency |
| Amplifier gain | 60 | voltage (3 k$\Omega$ TIA) |
| Trigger channels | 8 | for hit matching/zero suppression |
| Effective resolution | $\approx$9 | bits (12/10 bit logging) |
| Sample convert window | 64 | samples ($\approx$16 ns) |
| Readout granularity | 1 | sample, random access |
| Readout time | $1 + n \cdot 0.02$ | $\mu$s to read $n$ samples (same window) |
| Sustained L1 rate | 30 | kHz (multi-buffer) |

#### 7.3.3.2 Integrated High-Speed Readout

To build a viable, large-scale system, it is necessary to process, sparsify and collect the data generated by the devices mentioned above. Recently, the fDIRC prototype was moved into a cosmic test stand at SLAC and outfitted with 450 channels of BLAB ASIC-based waveform digitization electronics, as shown in Fig. 7.14.

Commissioning of this readout system [24] was performed in 2009 and the results of one year of operation will be reported at a later date. A similar system, consisting of a few hundred channels of cosmic test stand of an imaging Time Of Propagation (iTOP) prototype, has also been commissioned [25]. An advantage of this compact readout system is that it is easily and cost-effectively scaled up to several hundred thousand channels.
In terms of the Belle II DAQ infrastructure, Hawaii has been collaborating with colleagues in Japan and China on the fiber-optic based upgrade of data acquisition platform, named COPPER [26]. As part of this effort, a set of custom FINESSE cards [27] have been developed at Hawaii that digitize, store and match (to event or track) hits in the data stream within the COPPER context. This new architecture is presented in Fig. 7.15, where prototypes of each of the components have been fabricated and large scale system performance tests have begun.

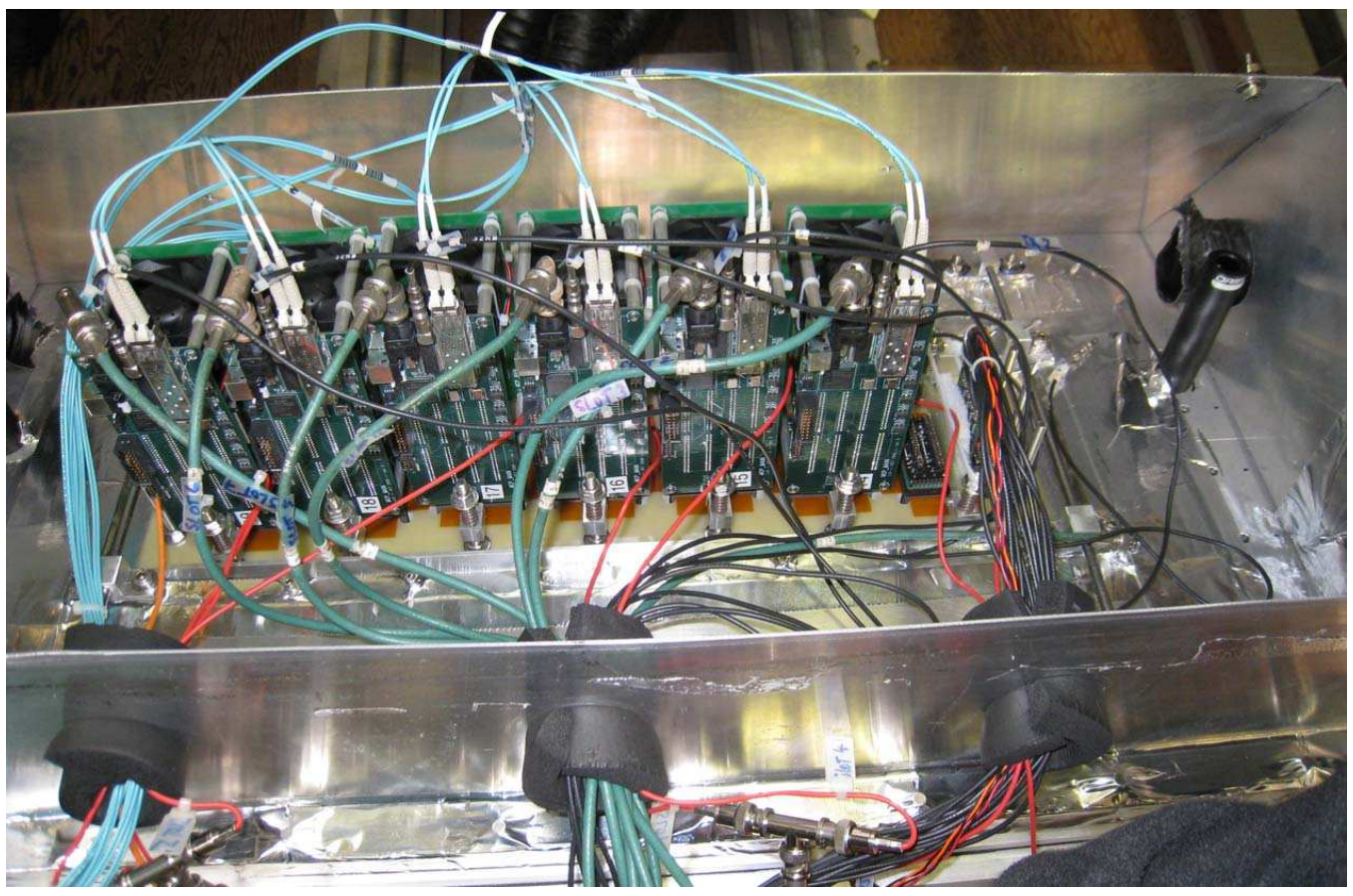

*Figure 7.14: An example of highly integrated photodetector readout, where each of the 64-channel PMTs is instrumented with its own readout module, consisting of a group of four BLAB2 ASICs, and the data is collected via fiber-optic link.*

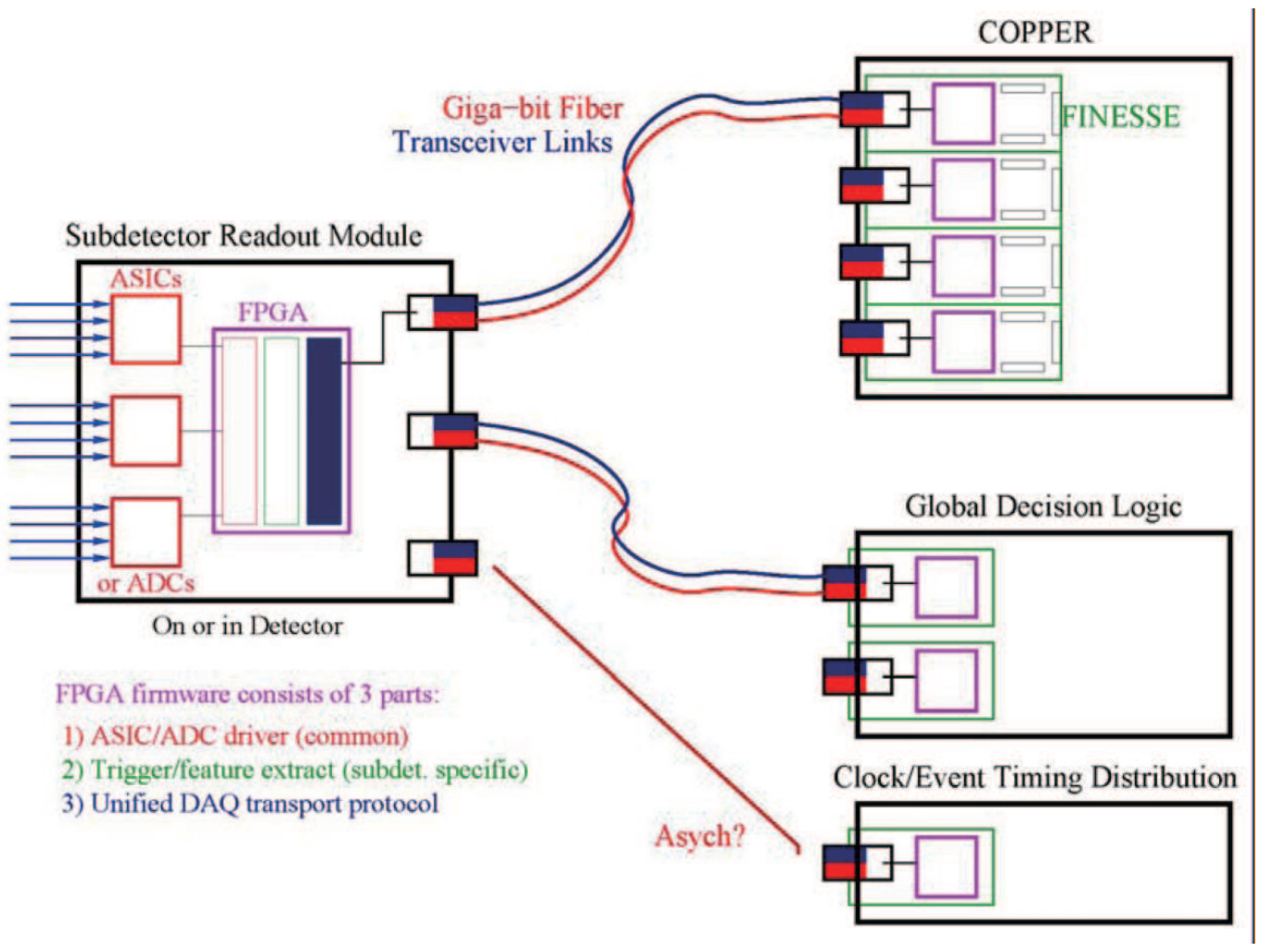

*Figure 7.15: Block diagram of the Belle II Trigger and DAQ system. Optical fiber links are used to convey trigger and front-end digitized data, significantly reducing the cable infrastructure, as well as improving signal recording fidelity.*

Both of the baseline detector configurations (1-bar and 2-bar) presented in Sec. 7.2 utilize the same number of 16-channel SL-10 PMTs (512). Therefore, apart from differences in mechanical layout, both configurations have 8k channels of readout electronics. Table 7.4 summarizes the readout system requirements. Details of the trigger processing are provided in Ch. 12.

*Table 7.4: Specifications for the barrel PID readout electronics.*

| Parameter | Value | Comment |
|---|---|---|
| Total electronics channels | 8k | either 1-bar or 2-bar |
| Number of BLAB3 ASICs | 1k | 8 channels/ASIC |
| Number of channels/SRM | 64 | 8 BLAB3 ASICs |
| Number of SRM | 128 | Subdetector Readout Modules |
| Bi-directional fiber links/SRM | 1+1 | DAQ/Trigger (see relevant Chapters) |
| Total DAQ/Trigger links | 128 | 10% bandwidth at full luminosity |
| Number FINESSE | 64 | 2 fiber links (COPPER limited) |
| Number COPPER | 16 | COPPER bus limited |
| Average size/event | 4 | kByte (2.5% occupancy) |

Qualified radiation tolerant components are prohibitively expensive, difficult to access due to export restrictions, and often many generations behind the technology forefront. Therefore, the Belle II DAQ group has undertaken a program to qualify all needed components in the environment in which they will be used. These include Gigabit fiber transceivers, programmable logic devices, and front-end ASICs. Initial tests have been carried out at particularly high radiation areas of the existing KEKB tunnel, near the Belle detector. Operation to $\mathcal{O}(100\,\text{kRad})$ dose has been successfully demonstrated for both fiber optic transceiver links, as well as a least one target programmable logic device. Testing in the KEKB environment is preferred over large radioactive source or reactor tests, as the particle species and energy are representative of those that will be seen in the Belle II detector. At this point, these initial tests indicate that, for detectors located at a radius larger than the silicon tracking layers, the radiation tolerance of the components tested thus far is adequate. However, the qualification of every part to be placed inside the detector—considering a reasonable operational safety margin—will be undertaken.

### 7.3.4 Structure

#### 7.3.4.1 Design strategy

The TOP counter will be installed between the ECL support structure inner surface and the CDC outer cover. A 3D conceptual drawing of the integrated TOP detector and CDC is shown in Fig. 7.16. For compelling financial reasons, the existing Belle barrel ECL structure will be reused in the Belle II detector. Therefore, the outer envelope on the radial position and the $z$-length of the TOP counter is fixed by the existing barrel ECL support structure and its integral support fixtures. However, with the construction of a new CDC (Ch. 6), an optimization of the TOP–CDC boundary region can be considered. Under these boundary conditions, a series of studies of how best to integrate the TOP detector array into the available space, while maximizing the expected performance, has been performed.

Upon closer scrutiny, a number of severe design constraints impact the detector integration. One of the most important issues is to have access to the PMTs after their installation. At this point, it is not evident that the MCP PMTs can survive the expected integrated charge antipated for ∼10 years of operation. It is therefore desirable to have a mechanism to access the PMTs. Openings in the support structure, referred to as "PMT access windows," can be used for maintenance and PMT replacement. Due to various geometrical constraints in the integrated Belle II detector system, provision for such access is not trivial.
Another important issue is to minimize the material budget of the entire structure. To perform well, the TOP counter needs a precise determination of the incident angle and the impact position of a candidate particle on the detector. Multiple Coulomb scattering in excess structural material degrades knowledge of these track extrapolated parameters and has a correspondingly deleterious impact on particle identification.

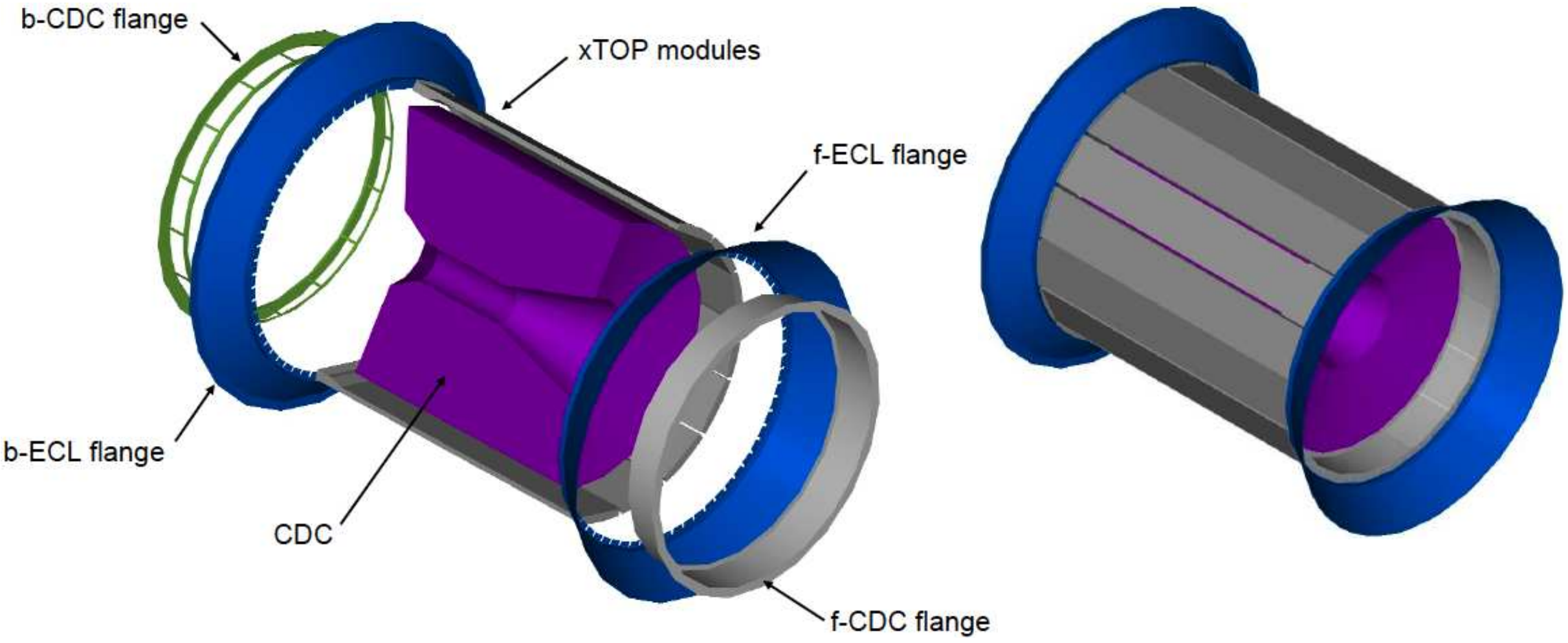

*Figure 7.16: A 3D conceptual rendering of the TOP detector integrated together with the CDC. f-ECL/b-ECL flange shows the structure of ECL conical part. f-CDC/b-CDC flange is the support of CDC from ECL flanges.*

#### 7.3.4.2 Requirements on the structure

Due to emission angle differences, a longer Cherenkov light propagation path should provide a larger time-of-propagation difference between different particle species of the same momentum. However, chromatic dispersion of the emitted spectrum of Cherenkov light increases as the propagation length increases. Studies indicate that the optimal $z$-length of a TOP module is ∼1850 mm [28]. For comparison, the $z$-distance for spanning the polar angle range of 32.2–128.7 degrees between the forward and the backward ECL endcaps, at the nominal TOP radius, is 2980 mm. Given this long $z$-distance to be covered, compared with the ideal TOP bar length, there is no obvious constraint on the detector $z$-length in terms of performance. However, a strong geometrical constraint comes from the flanges attached to the BECL container, which cannot be removed without completely disassembling the barrel ECL, an operation that will not be performed.
The inner radius of the BECL container is 1250 mm and represents the absolute maximum outer TOP radius. Existing fixtures attached to the container and a need for installation clearance will reduce the final TOP counter radius slightly. The CDC outer radius (and corresponding TOP inner radius) is flexible since it will already be much larger than the 880 mm of the current

Belle detector [29]. The mechanical width and thickness of a single TOP module is dictated by a trade-off between technical constraints on the quartz bar production, distance from the ECL, sufficient light yield, and quartz radiator thickness. As there is no strong optimum in these parameters, a quartz bar width of 400–500 mm and thickness ∼20 mm has been considered. Given the quartz bar dimensions and the available radial space, 12–18 modules can be arranged to surround the CDC volume. The number of TOP modules was chosen to be 18 in the LoI [30], but subsequently changed to 16 to better match the azimuthal symmetry of the existing BECL flange. The resulting azimuthal segmentation is 22.5 degrees per module.
Two considerations constrain the TOP geometrical envelope. First, a larger TOP outer radius makes the gap between the ECL and the TOP ($\Delta_{\rm ECL}$) smaller. A 20-mm-thick quartz radiator corresponds to 16% of a radiation length and causes $\gamma$-conversions ∼6% of the time for normal incidence energetic photons. These $e^{\pm}$ pairs will curl in the 1.5 T magnetic field; depending on their polar angle and $\Delta_{\rm ECL}$, some of these $e^{\pm}$ may miss the ECL or may be detected in distinct CsI clusters. This effect has been studied in GEANT simulation, one specific result of which is shown in Fig. 7.17 for 100 MeV photons (the typical minimum energy for "good gammas" in the barrel ECL). As clearly indicated, a larger outer radius (i.e., a smaller $\Delta_{\rm ECL}$) increases the efficiency of detection of these low energy photons, which are vital to many physics analyses.

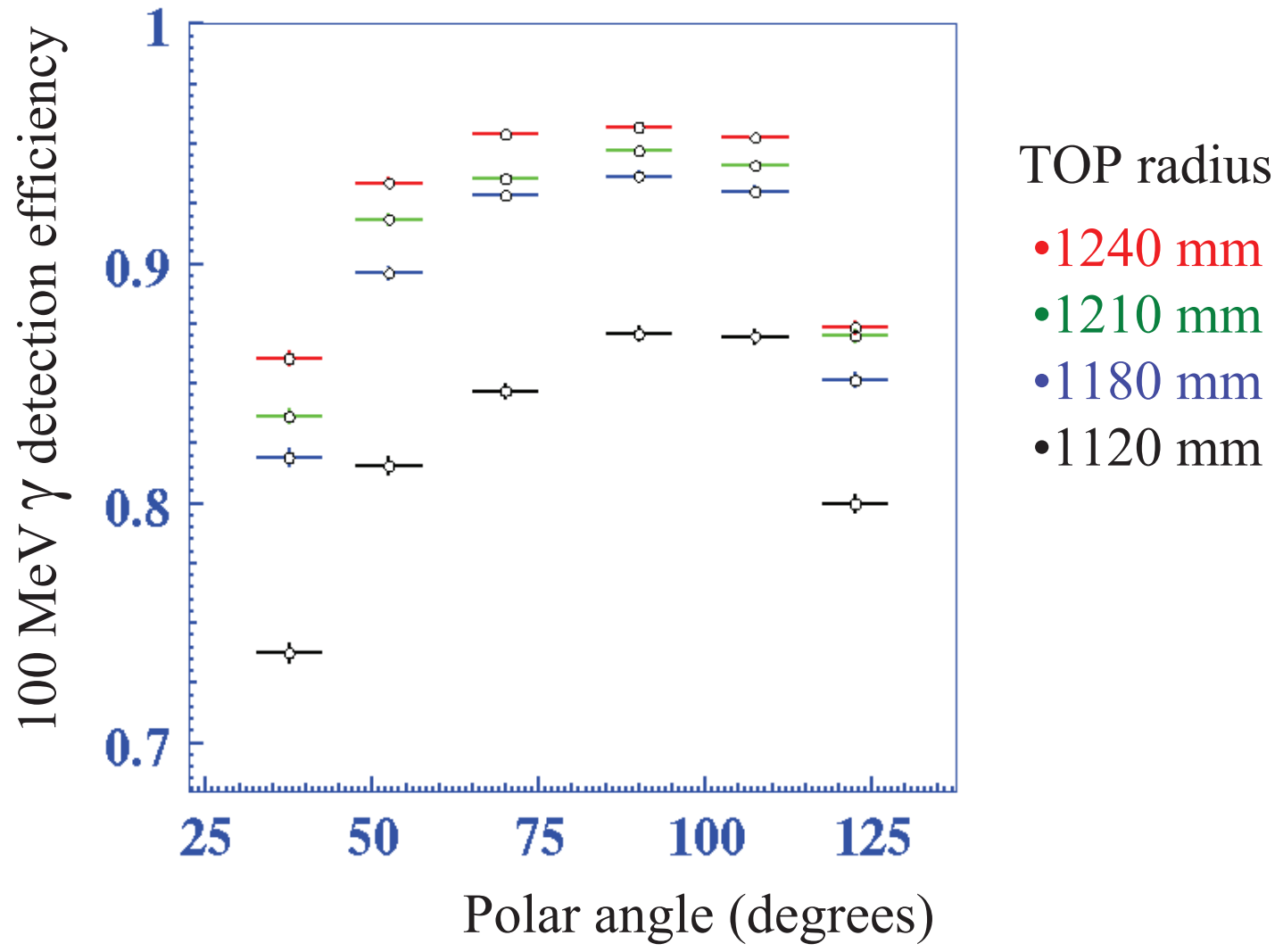

*Figure 7.17: A GEANT study of the impact of TOP outer radius for 100 MeV photons as a function of polar angle.*

Second, Belle's forward-endcap PID device (the Aerogel Cherenkov Counter) may need to be retained due to budget and resource limitations; the new endcap PID device would then be installed at a later stage. To accommodate this scenario, a generous clearance between the forward inner-detector support cylinder and the EACC ($\Delta_{\rm ACC}$) is needed to assure mechanical compatibility with the existing EACC structure.

#### 7.3.4.3 Module container

A TOP module consists of a quartz radiator for the generation and propagation of the Cherenkov light and a readout block for photon detection and signal readout. These components are housed in the quartz bar box (QBB) and the readout enclosure, respectively.
The QBB, together with the TOP structure, is required to have sufficient rigidity so that the quartz bar can be properly supported. Specifically, the quartz should be subjected to at most a tolerable deflection. Referring to a past study for the BaBar DIRC [31], the maximum deflection allowed is set to 100 $\mu$m. A quartz bar with dimensions 400–500 mm (wide) $\times$ 20 mm (thick) $\times$ 2870 mm (long) weighs 51–63 kg. To satisfy the maximum deflection requirement without the need for an excessive material budget, the QBB design uses honeycomb panels for the inner and the outer faces and solid aluminum sides. The honeycomb panel sandwiches a honeycomb core between aluminum skins.
TOP performance simulation studies (Sec. 7.5) indicate that a resolution of $\leq$ 0.5 mrad is desirable and $\leq$ 1.5 mrad is essential. The 5-mm-thick CFRP used for the CDC outer cover already contributes a multiple scattering mean deflection angle of $\sim$0.7 mrad for a normally incident $\pi^{\pm}$ at 4 GeV/$c$. Therefore, minimizing the structural support material is important. The stiffness of the QBB is dominated by the thickness of the honeycomb panel, whereas the aluminum skins are the dominant contribution to the multiple scattering. Thus, a thicker honeycomb core with thinner aluminum skins is preferable for a fixed honeycomb panel thickness. Based on deflection calculations and considerations for practical handling of the QBB, the thickness of the honeycomb panel is chosen to be 10 mm, including two 0.3-mm-thick aluminum skins, and that of the aluminum side is chosen to be 2 mm. Although this aluminum side thickness does not contribute significantly to either the stiffness of the QBB or the multiple scattering for the TOP measurement, it does affect azimuthal hermeticity. On the other hand, the stiffness of these walls is dictated by the need to support many threaded plungers, used for positioning and supporting the quartz bar in the QBB. These sides must withstand the spring load, while maintaining an air gap of 2 mm to the quartz radiator. It is necessary to verify the aluminum side thickness using simulation (FEM) or a mechanical prototype. The current design of the QBB for the 2-bar configuration is shown in Fig. 7.18, where the readout enclosures are included but not detailed. The dimensions of the QBB are summarized in Table 7.5.

*Table 7.5: Module dimensions*

| Component | Width [mm] | Thickness [mm] | Length [mm] |
|---|---|---|---|
| Quartz | 443 | 20 | $\sim$2750 |
| QBB | 451 | 44 | 2870 |
| Honeycomb panel | 451 | 10 (core: 9.4, skin: 0.3$\times$2) | 2870 |
| Al side | 2 | 44 | 2870 |

The readout enclosure houses the readout block and includes an array of readout units. The readout unit consists of a group of PMTs and their readout printed circuit boards (PCBs). The readout enclosure is mounted on the end of the QBB and provides a flush optical connection between the quartz bar end and the readout array. Figure 7.19 illustrates conceptual 3D drawings of the readout arrays; the double-row and the single-row arrays will be used for the 1-bar and the 2-bar schemes, respectively. A 3D drawing of the readout block mounted at the quartz bar end is also shown in Fig. 7.19. The readout enclosure has the PMT access window at its inner surface. Its rear end must accommodate the cabling for the PCBs and piping for the cooling. It also has to maintain the optical connection between the readout array and the quartz bar end as well as the light shield. All of these are strongly correlated with the PCB dimensions derived from the electronics design, which is in progress.

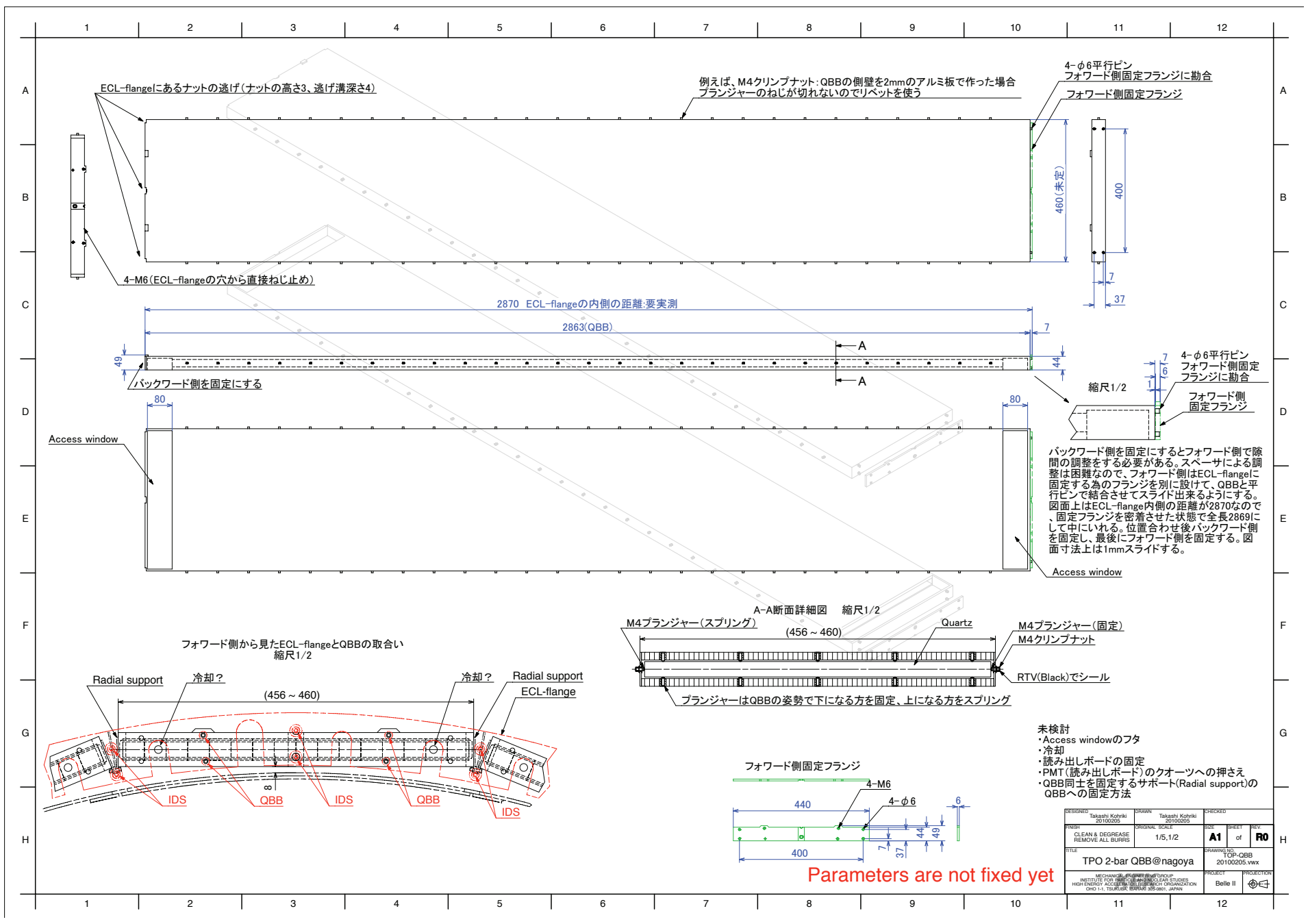

*Figure 7.18: A drawing of the QBB design.*

#### 7.3.4.4 Baseline design

The $rz$-views of the baseline structure designs are shown in Figs. 7.20 and 7.21 for the 2-bar and the 1-bar schemes, respectively. The corresponding $r\phi$-views are shown in Fig. 7.22. The 2-bar scheme has a single row of readout at both the forward and the backward ends of the QBB, each row consisting of 16 PMTs. The 1-bar scheme has an expansion volume instrumented by a double-row readout array at the backward end of the QBB, consisting of 32 PMTs. The difference between the two schemes is the application of an expansion volume in the 1-bar case; this has an impact on the mechanics around the boundary between the TOP and the CDC.
The radial geometries are summarized in Table 7.6, where the entries are the maximum radius available to the TOP counter ($R_0$), providing for a minimum ECL ($\Delta_{\mathrm{ECL}}$) and the maximum EACC ($\Delta_{\mathrm{EACC}}$) clearance, as indicated. Values reported in this table take into account the measured deformation of the BECL container and the fixtures attached to its inner surface. In the 2-bar scheme, the CDC clearance ($\Delta_{\mathrm{CDC}}$) is derived from the thickness of the CDC Support Cylinder [CDC-SC] (4 mm), an installation clearance (5 mm) and a 3-mm buffer space. In the 1-bar scheme (values shown in parentheses), the values are derived from the backward end of the QBB, where the expansion volume and the double-row readout array are equipped. The CDC outer radius ($R_{\mathrm{CDC}}$) cannot be larger than the inner radius of the CDC-SC, due to the existence of wire feedthroughs and preamplifier boards. The resulting CDC radial clearance ($\Delta_{\mathrm{CDC}}$) is 49 mm in the 1-bar scheme. Although this $\Delta_{\mathrm{CDC}}$ value seems somewhat large, even a few mrad scattering in the CDC outer cyclinder will lead to a sub-mm error in the projected TOP impact position—much smaller than typical helix-fit projection errors. The radial configuration in the forward end is almost identical for the two configurations. The only difference is the existence of the readout block, which requires having cut-outs for the PMT access windows on the forward CDC-SC. These cut-outs will be covered by 4-mm-thick aluminum covers to reinforce the stiffness of the CDC-SC, leading to the 4-mm difference in the EACC clearance ($\Delta_{\mathrm{EACC}}$) between the two schemes. This $\Delta_{\mathrm{EACC}}$ annular region is reserved for the cabling and piping of the inner sub-detectors.

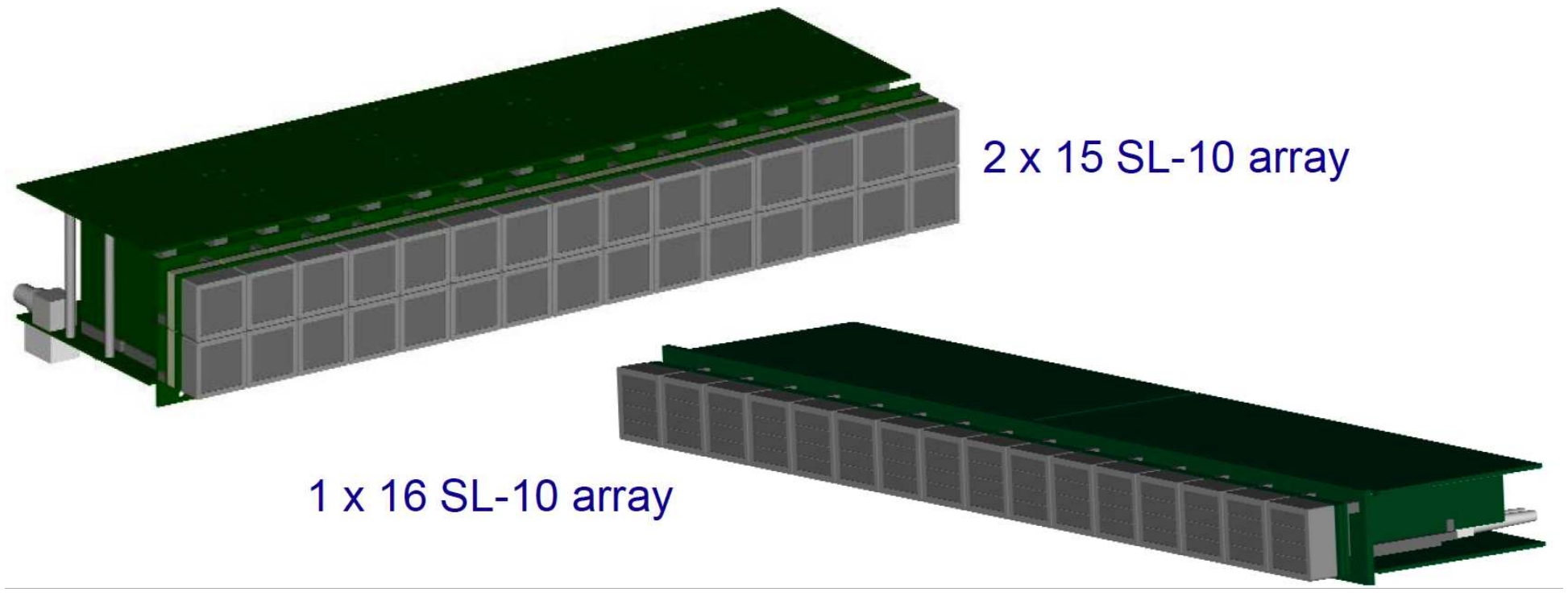

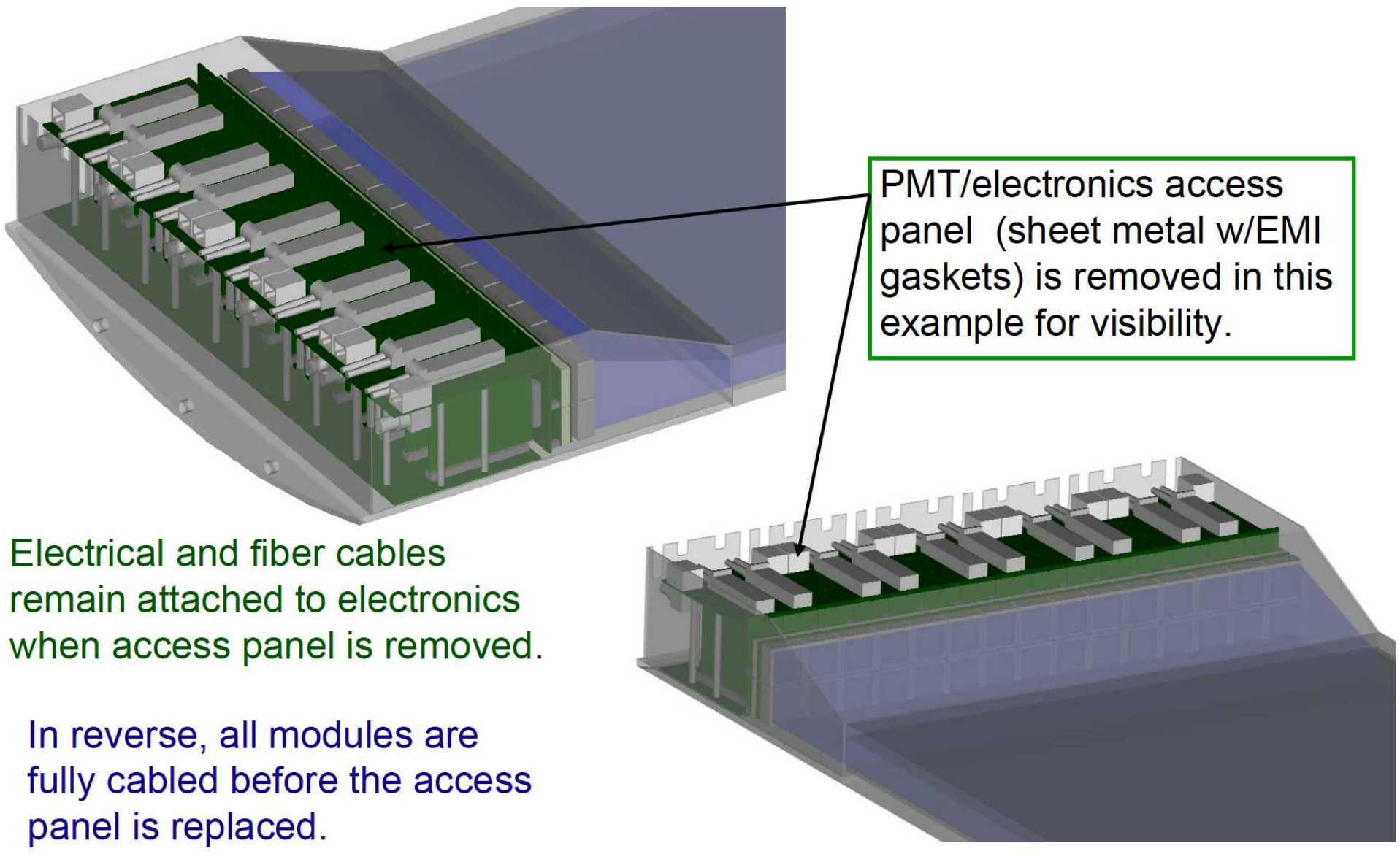

*Figure 7.19: Readout array illustrations. Top: two arrays of readout units corresponding to the 1-bar and 2-bar cases. Bottom: detailed three-dimensional rendering of the readout units inside electro-mechanical enclosures.*

*Table 7.6: Geometrical parameters of the current mechanical designs. Tabulated values refer to the 2-bar (1-bar, if different) detector configuration.*

| Parameter | Symbol | Value |
|---|---|---|
| Maximum TOP counter radius | $R_0$ | 1243 mm |
| ECL inner radius | $R_{\mathrm{ECL}}$ | 1250 mm |
| TOP-ECL clearance | $\Delta_{\mathrm{ECL}}$ | 40 mm |
| Azimuthal acceptance loss | $\eta_\phi$ | 7% |
| CDC outer radius | $R_{\mathrm{CDC}}$ | 1166 (1120) mm |
| TOP-CDC clearance | $\Delta_{\mathrm{CDC}}$ | 12 (24/58) mm |
| Fwd inner-detector-support radius | $R_{\mathrm{IDS}}$ | 1166 (1162) mm |
| EACC outer radius | $R_{\mathrm{EACC}}$ | 1145 mm |
| TOP-EACC clearance | $\Delta_{\mathrm{EACC}}$ | 17 (21) mm |

In both designs, the fraction of the insensitive azimuthal coverage ($\eta_\phi$) is not small and approximately 7% at the center radius of the QBB in the 2-bar scheme. To increase the azimuthal coverage, a staggered module arrangement was considered in the LoI [30]. Unfortunately, this staggered arrangement doubles the material budget in the overlap region, makes the structural design more complex, and increases the azimuthally-dependent worst-case spacing to both the ECL and CDC.
An approach applicable to the current module arrangement is to maximize the quartz bar width, with a starting point of 451 mm in Table 7.5. If the need for middle support can be mitigated, a quartz bar width of 460 mm would be possible, reducing $\eta_\phi$ to ∼3%. Since the width of the PMT array is rigidly quantized, some area of the quartz bar end is typically not fully covered by the PMTs. The covered fraction is estimated to be 96% in the latter bar width case. The typical number of Cherenkov photons detected is ∼20 and a 4% loss does not significantly affect detector performance, while the azimuthal track coverage can be increased by 4%. Such a configuration, though with different module dimensions, has already been evaluated in a beam test [32] and it performed as expected, as reported in the next section.

## 7.4 Prototype test

We have produced a prototype TOP counter to check the effects of chromatic dispersion and to demonstrate overall timing performance [32]. This prototype is comprised of a quartz radiator with a spherical focusing mirror (5-m radius) and 12 square-shape MCP-PMTs with a multi-alkali photo-cathode (Hamamatsu R10754X-00-L4). Two quartz bar pieces and the mirror were glued together for a total radiator size of $1850 \times 400 \times 20$ mm$^3$. To reduce chromatic dispersion effects, a wavelength high-pass filter of $\lambda > 400$ nm was inserted between the radiator and the MCP-PMTs. Operation of the MCP-PMTs demonstrated stable gain ($\sim 10^6$) and good time resolution ($< 40$ ps) for all channels. The average QE observed was 17% at 400 nm. We also developed constant-fraction-discriminator (CFD) modules as the high-speed readout for this test; these use a fast amplifier ($\mu$PC2710 MMIC, 1 GHz, $\times 20$ gain) and a comparator (180-ps

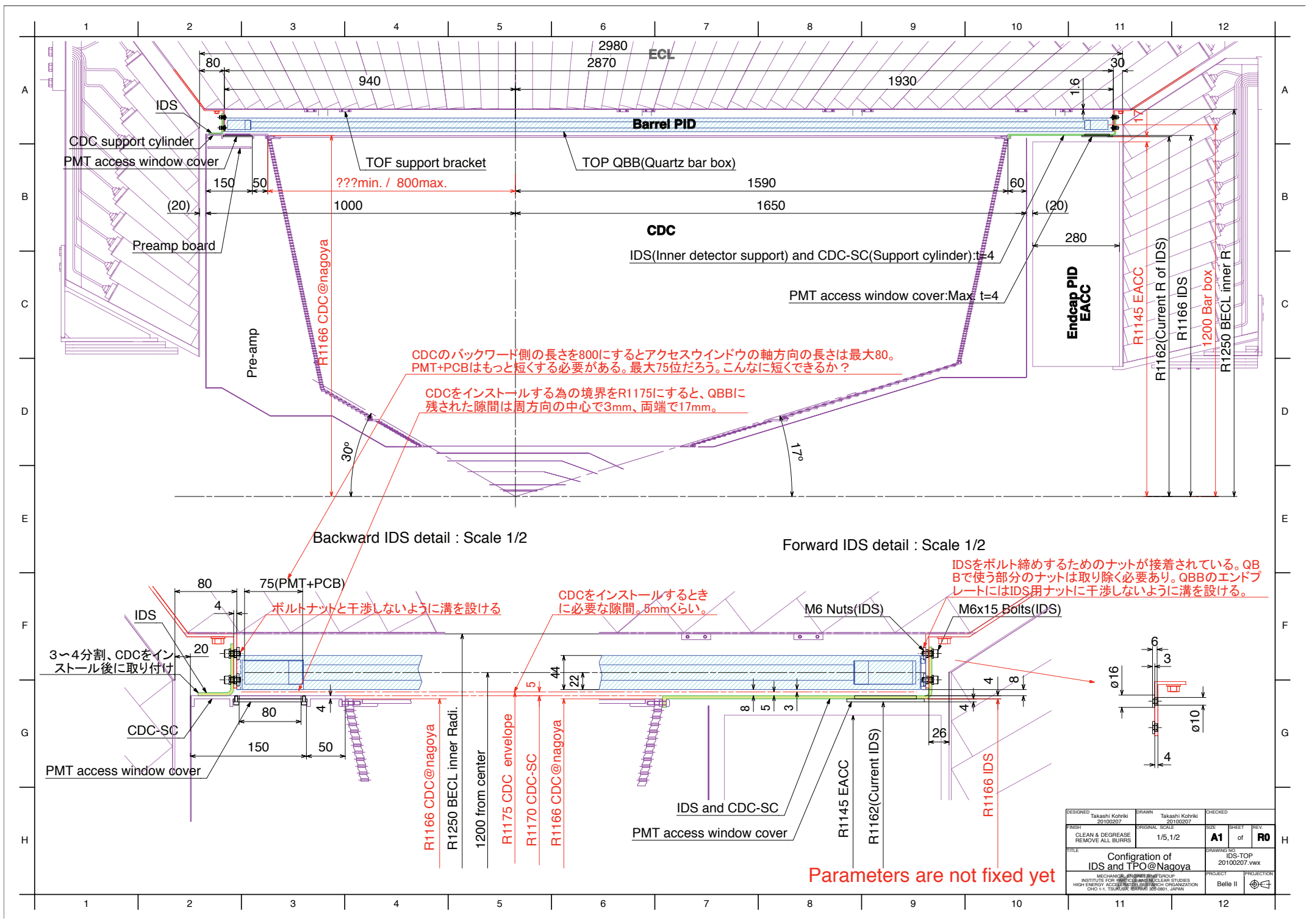

Figure 7.20: *rz*-view drawing of the current 2-bar scheme baseline design.

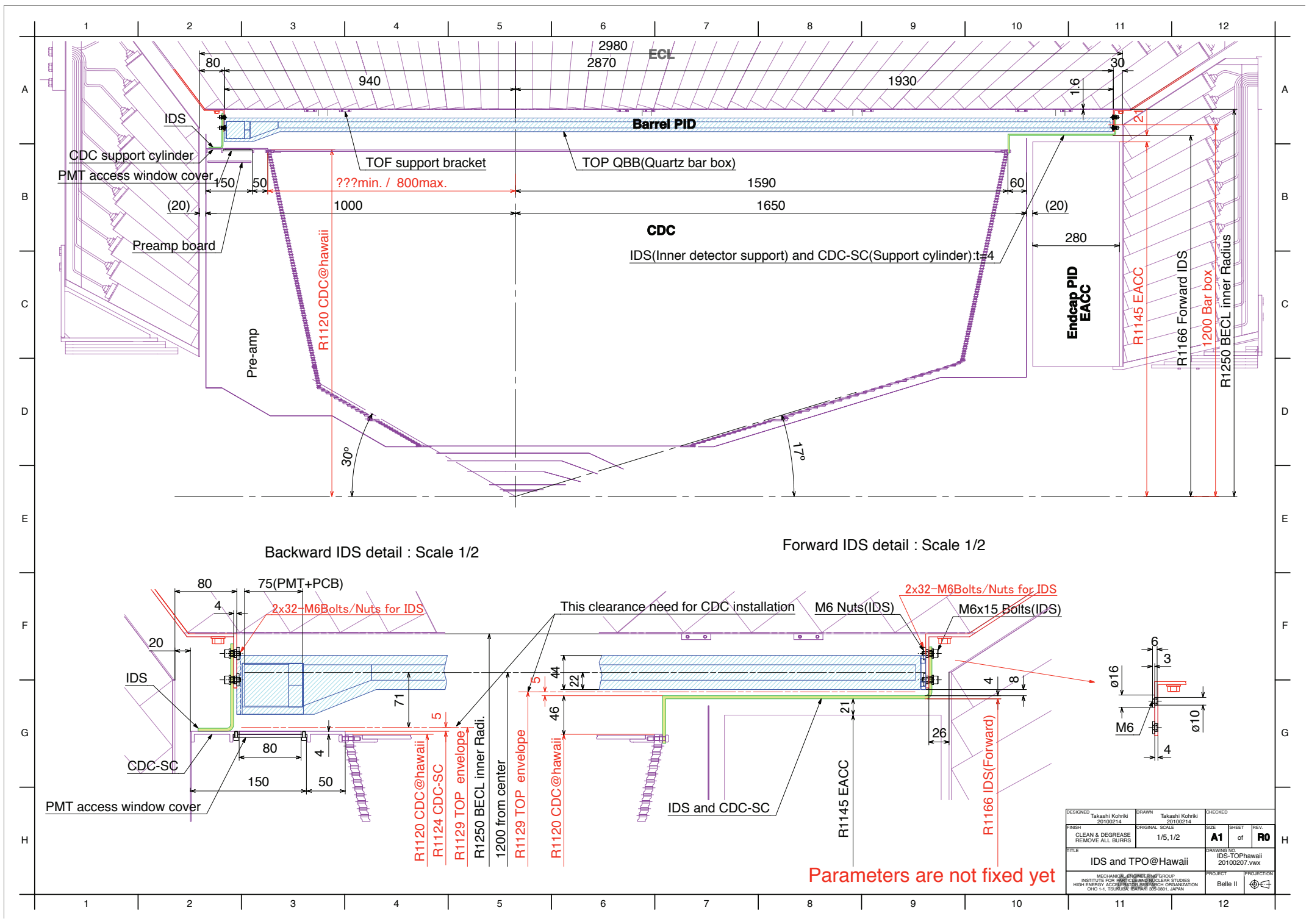

Figure 7.21: *rz*-view drawing of the current 1-bar scheme baseline design.

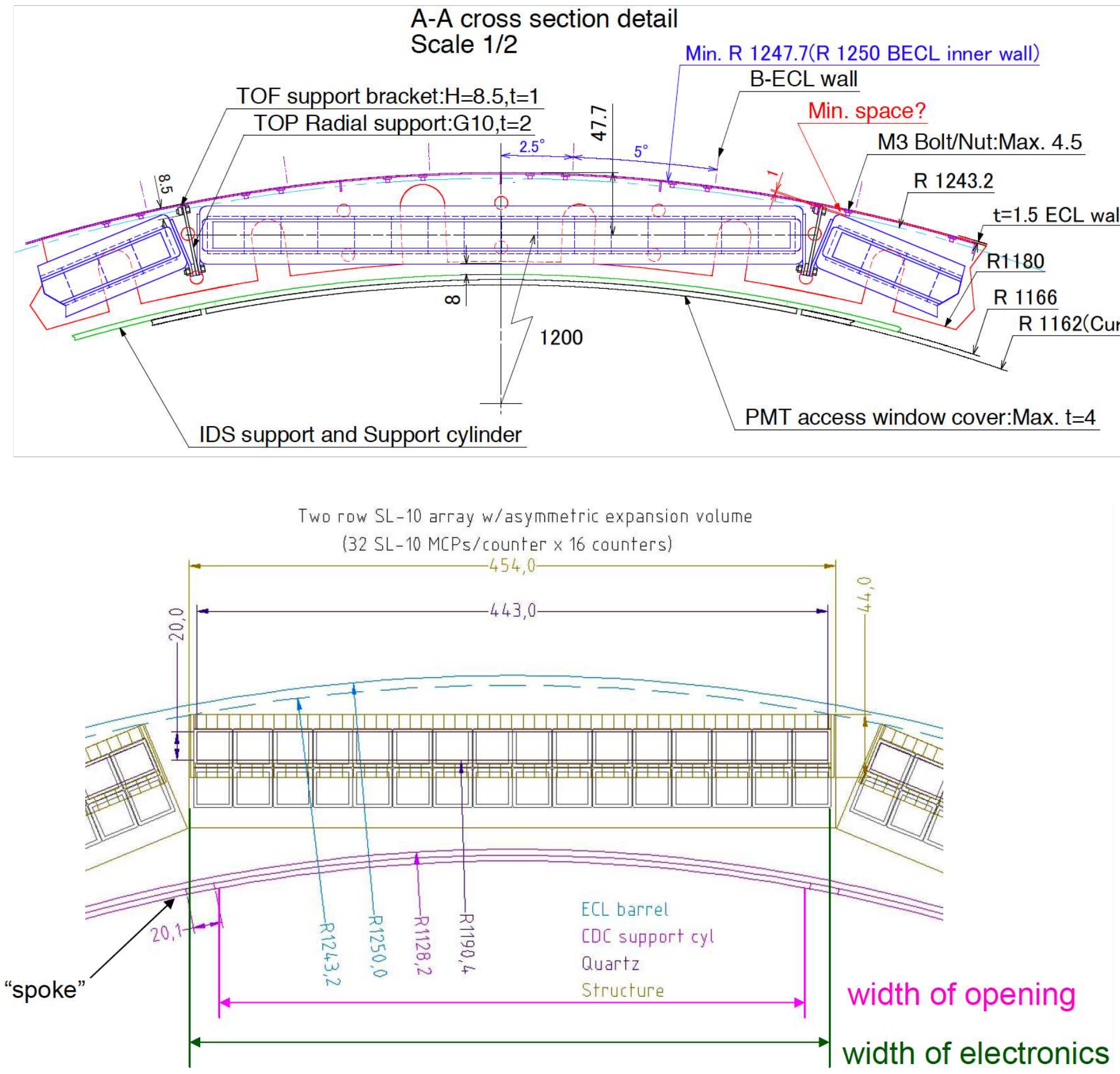

*Figure 7.22: $r\phi$-view drawings of the current baseline designs for the 2-bar (upper) and the 1-bar (lower) concepts.*

propagation delay). The CFD time resolution is $\sim 5$ ps for test pulses.

Using a prototype counter, we performed beam tests with 2 GeV/$c$ electrons at the KEK Fuji test beam line in June and December 2008. The TOP counter was located between trigger scintillation counters and tracking chambers. To determine the beam timing precisely, we put a timing counter [33] along the beam line; this consisted of a small quartz radiator (10 mm$^{\phi}$ $\times$10 mm$^{L}$) and a round MCP-PMT (Hamamatsu R3809-50-11X). The time resolution obtained for these start counters was determined to be 14.8 ps during the beam test.

Figure 7.23(a) shows a partial ring image obtained during the beam test. The beam was normally incident at the center of the radiator. The distance between this incident position and the nearest MCP-PMT was 358 mm. A clear partial ring image was obtained, as predicted by simulation. Figure 7.23(b) shows the number of detected photons for the normal-incidence case and the match to our simulation prediction. Figure 7.23(c) is the TDC distribution for an anode at the center of the readout plane. The distance between the incident position and this MCP-PMT was 875 mm. A comparison was made of the time resolutions obtained during the beam test with the resolution expected from simulation, which included PMT resolution and the effect of chromatic dispersion. The time resolution of the first peak was $(76.0 \pm 2.0)$ ps and $(77.7 \pm 2.3)$ ps for data and simulation, respectively, indicating consistent results between data and simulation. The background tail under the peak is due to Cherenkov photons generated by $\delta$-rays and electron showers from upstream. These effects are included in the simulation.

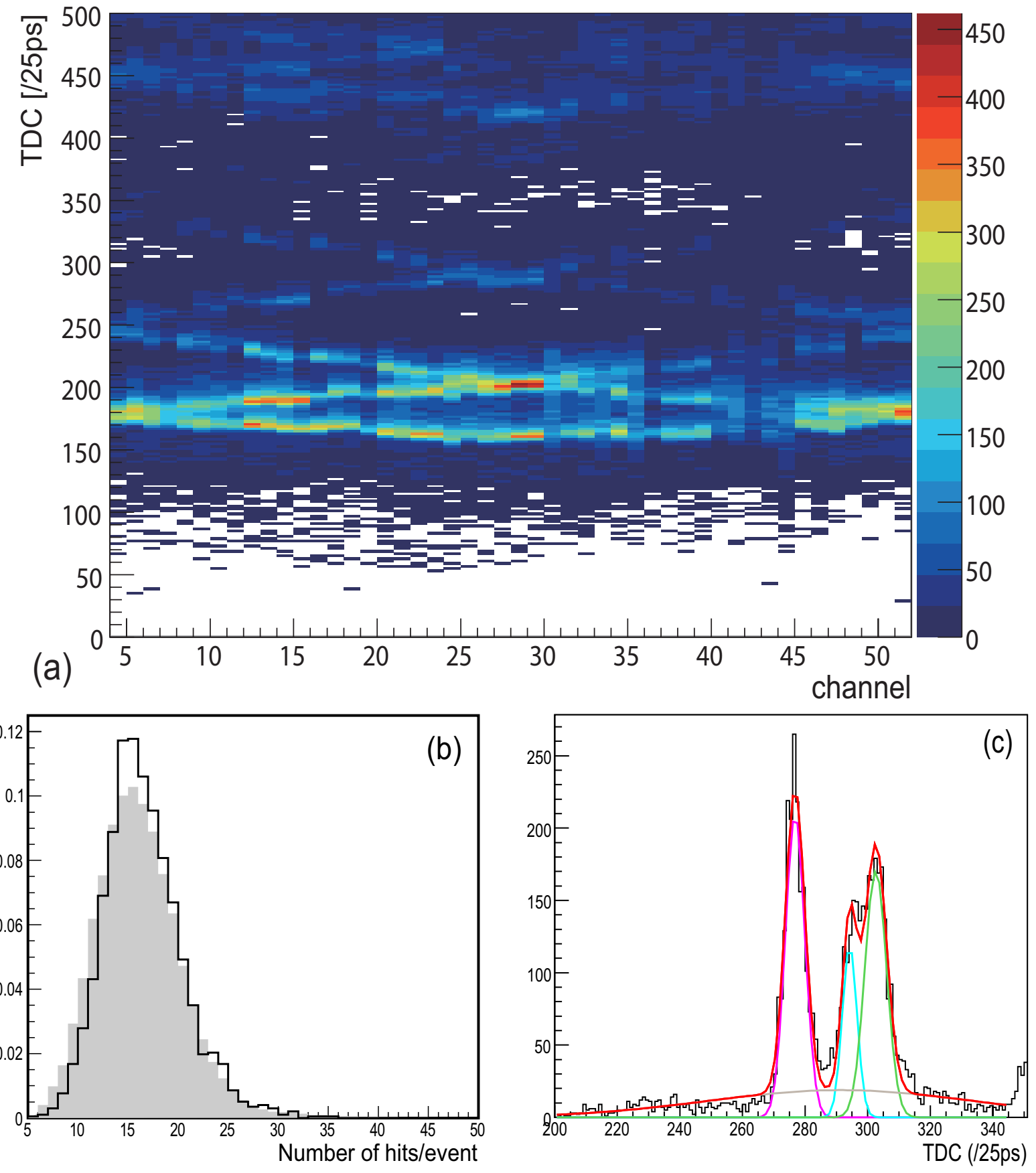

*Figure 7.23: Beam test results. (a) Partial ring image from about 350,000 events, (b) number of detected photons, and (c) TDC distribution of an anode at the center of the quartz end. The colored scale in (a) corresponds to the number of events in the bins. The solid and shaded histograms in (b) correspond to data and simulation, respectively. In (c), the histogram is the data, and the curves show the fitted Gaussian and background functions.*

## 7.5 Expected performance

### 7.5.1 Simulation methods

To study expected TOP performance in various configurations, three independent Monte Carlo simulation studies have been undertaken:

- **Nagoya** simulations utilize the Belle Geant3-based Monte Carlo, GSIM, to determine the trajectory of primary charged particles, as well as any secondary charged tracks produced by electromagnetic and hadronic interactions as the primaries traverse the quartz. The threshold energy for propagation of electrons is 0.2 MeV; this is smaller than the Cherenkov radiation threshold. Portions of charged tracks that pass through the TOP radiator bars are then passed to a stand-alone code to simulate all optical processes occurring in the counters (Cherenkov emission, photon propagation with the velocity variation on the wavelength, total internal reflection, optical attenuation, photon detection at the MCP-PMTs, etc).

- **Ljubljana** simulations utilize a stand-alone simulation that tracks a charged primary through the detector volume and handles optical processes in the counters. No attempt is made to handle secondary tracks.
- **Hawaii** simulations utilize a stand-alone Geant4-based program to fire primary tracks, produce secondary particles resulting from electromagnetic interactions in the bar, and carry out all optical processes for all charged particles. The threshold energy for propagation of electrons is about 0.5 MeV, which is determined by the minimum step length.

All programs model the multiple scattering effect inside quartz. Simulation inputs for the TTS and QE distributions are taken from MCP-PMT prototype measurement data, as described in Sec. 7.4. Consistency between simulation output and measurement is confirmed by checking the GSIM program against beam test data.

### 7.5.2 Reconstruction

Existing studies have thus far focused primarily on $K/\pi$ separation. To determine expected efficiencies and fake rates, the detected photons for each track are tested against probability distribution functions (PDFs) for each particle hypothesis ($\mathcal{P}^K(x,t)$ and $\mathcal{P}^\pi(x,t)$).
The source of the PDFs differs by simulation group. In the methods employed by Nagoya and Hawaii, the PDFs are based upon a large number of events using single tracks at a particular momentum, impact position, and angle on the quartz bar. The Ljubljana reconstruction utilizes the known properties and distributions of Cherenkov radiation to form analytically calculated PDFs [34]. Typical distributions used to construct the PDF are shown in the left of Fig. 7.24.

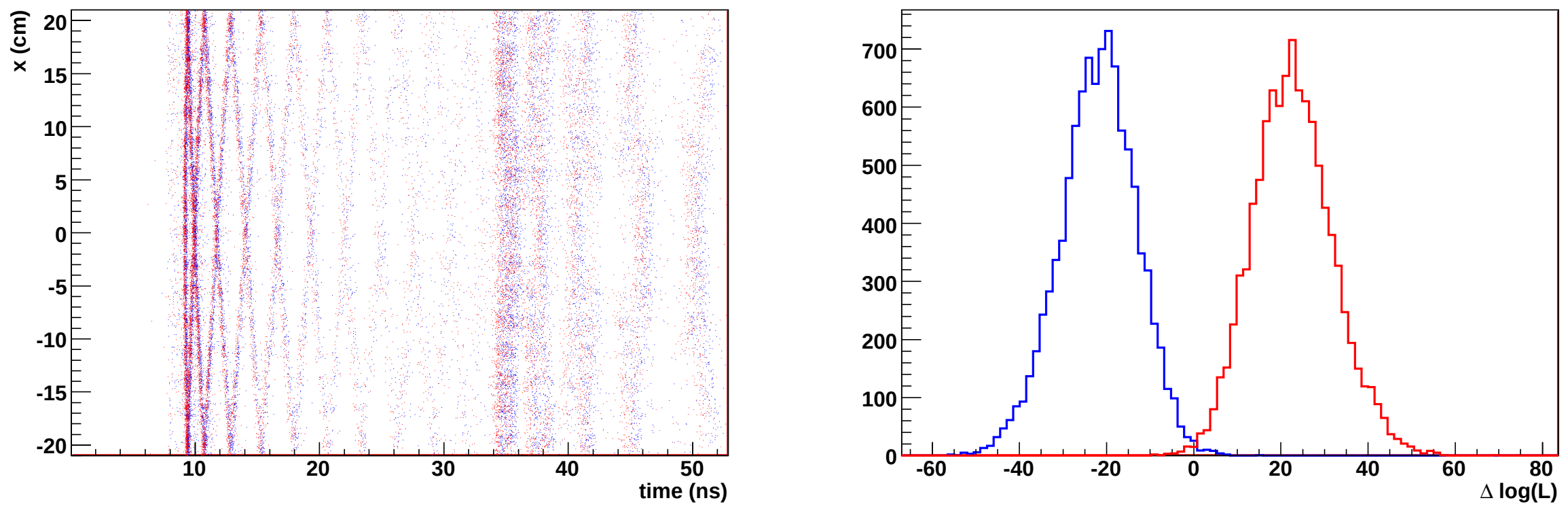

*Figure 7.24: (Left) Sample distributions for an ensemble of 500 tracks of detected photon positions, x, and times, t, for a 2-bar counter with 3 GeV/c pions (red) and kaons (blue) at normal incidence on the quartz bar. (Right) Distributions of $\Delta \log \mathcal{L}$ for pions (red) and kaons (blue), corresponding to the distributions on the left.*

From the PDFs, a likelihood is determined for a simulated primary charged particle:

$$\mathcal{L}^{K,\pi} = \prod_i \mathcal{P}_i^{K,\pi}(x,t) \tag{7.1}$$

where the index $i$ runs over each detected photon in the event. Typically the log of the likelihood is used as it is more computationally stable. The difference between the likelihood under each hypothesis is calculated.

$$\Delta \log \mathcal{L} = \log \mathcal{L}^{\pi} - \log \mathcal{L}^{K} \tag{7.2}$$

If this log-likelihood difference is positive, the particle is classified as a pion, and if negative, a kaon. Example distributions of log likelihood can be found in Fig. 7.24. These likelihood-based classifications can be compared with the known species of particle simulated to determine the fractions of correctly identified kaons and incorrectly identified pions.

Because the Nagoya and Hawaii methods utilize PDFs determined from Monte Carlo, generating a PDF for a single combination of track momentum, impact angle, impact position, and particle species takes a significant amount of time and computational resources. As such, the expected efficiencies and fake rates can be calculated only for a limited number of track parameters.

In contrast, the reconstruction procedure utilized in the studies by Ljubljana makes use of analytically calculated PDFs, thus drastically reducing the computation required to make a performance estimate and allowing much finer sampling of performance across the parameter space for potential charged tracks. The potential drawback to this method is that these PDFs may lack some of the details seen under realistic operating conditions (e.g., secondaries produced by the initial charged track).

As the method utilizing Monte Carlo based PDFs is unrealistic to apply in the final detector configuration, there is an ongoing effort to determine suitable reconstruction methods for the final system, for example by adapting the analytical reconstruction method to the Geant-based Monte Carlo data.

### 7.5.3 Simulated configurations

Performance of the 1-bar and 2-bar configurations have been studied. These studies use the detector dimensions described previously. For each configuration, both multi-alkali and super-bialkali photocathodes have been studied.

### 7.5.4 Included uncertainties

Various non-idealities in reconstruction, detector geometry, and actual detector performance (both within the TOP and other Belle II measurements) will degrade ideal TOP performance. We describe here the primary uncertainties that have been studied and their baseline values for the performance estimates.

#### 7.5.4.1 Tracking uncertainty

The performance estimates include event-by-event variations in each track's impact angle and position, relative to their nominally generated values. These variations are due to the tracking resolution and multiple scattering in the CDC outer wall and inner wall of the quartz bar support structure. These estimates reflect the resolution expected from the Belle II tracking systems and algorithms.

At the inner surface of the TOP counter, each track can be defined by four parameters: incident position $(z, x)$ and incident angle $(\theta, \phi)$. Track uncertainties can thus be defined as the uncertainties of these four respective parameters; $\sigma_z$, $\sigma_x$, $\sigma_\theta$ and $\sigma_\phi$. Estimates of these uncertainties have been obtained using the Belle Geant3 framework, with the existing Belle ACC removed.

These results are shown in Table 7.7. As a cross check, results obtained in this way are roughly consistent with an analytic calculation of multiple scattering, which includes the material of the CDC outer cylinder and the TOP honeycomb frame.
The nominal values used for the performance estimates are:

- $\sigma_z = 1.4$ mm
- $\sigma_x = 1.0$ mm
- $\sigma_\theta = 1.5$ mrad
- $\sigma_\phi = 1.5$ mrad

*Table 7.7: Tracking resolution estimates using Belle GSIM MC, with the ACC removed.*

| p [GeV/$c$] | particle | $\sigma_{\rm dz}$ [mm] | $\sigma_{\rm dx}$ [mm] | $\sigma_\theta$ [mrad] | $\sigma_\phi$ [mrad] |
|---|---|---|---|---|---|
| 2 | $\pi$ | 2.0 | 0.57 | 2.6 | 3.6 |
| 2 | $K$ | 1.8 | 0.49 | 2.6 | 3.3 |
| 3 | $\pi$ | 1.4 | 0.45 | 1.8 | 2.5 |
| 3 | $K$ | 1.3 | 0.39 | 1.8 | 2.3 |
| 4 | $\pi$ | 1.3 | 0.42 | 1.5 | 2.0 |
| 4 | $K$ | 1.2 | 0.33 | 1.4 | 2.0 |

#### 7.5.4.2 Event start time ($t_0$) jitter

As the TOP requires high performance timing resolution, it is particularly sensitive to timing uncertainties. One such uncertainty is in the event start time, $t_0$, which determines the time of the $e^+e^-$ collision. This quantity can be estimated based upon numerous studies with the Belle TOF system. Reported results of one such study [23], as well as estimates of performance improvements for Belle II, have been used to estimate the start time uncertainty, $\sigma_{t_0}$.
For the purposes of TOP reconstruction, the target value for the final $t_0$ resolution is $\sigma_{t_0} = 25$ ps, and this is the baseline value used in simulations to estimate detector performance. From experience, to achieve this level of precision will require significant and concerted calibration efforts. Within the uncertainties of the studies performed, a more pessimistic event timing outcome is also possible and thus $\sigma_{t_0} = 50$ ps has also been studied. This will be summarized in a more detailed note on TOP simulations and is an important consideration in the final detector configuration choice.

#### 7.5.4.3 PMT efficiency degradation

As evident in Fig. 7.9, MCP-PMT efficiency degrades over their lifetime due to aging effects. Furthermore, variations in the production process may cause some PMTs to exhibit lower efficiencies. To account for these effects, the performance estimates can be made using the nominal efficiencies, and repeated with efficiencies at 80% of their nominal values.

#### 7.5.4.4 Beam background rate

The detected photon rate due to beam background is estimated to be 300 kHz/(TOP module) at the luminosity of $10^{34}\,\mathrm{cm}^{-2}\mathrm{s}^{-1}$, as described in Sec. 7.3.2. Scaling by a factor of 20 and assuming a time window of $\sim$100 ns for each TOP reconstruction, the number of background photons is estimated as 4.8 [(photons detected)/(TOP module)/(event)].
Baseline estimates of TOP performance use estimates of 0–1 detected background photons per TOP module per event. However, backgrounds as high as about 7 detected photons per module per event have been studied, and show only minor ($\leq$ 1%) degradation to efficiencies and fake rates, as seen in Figures 7.25 and 7.26.

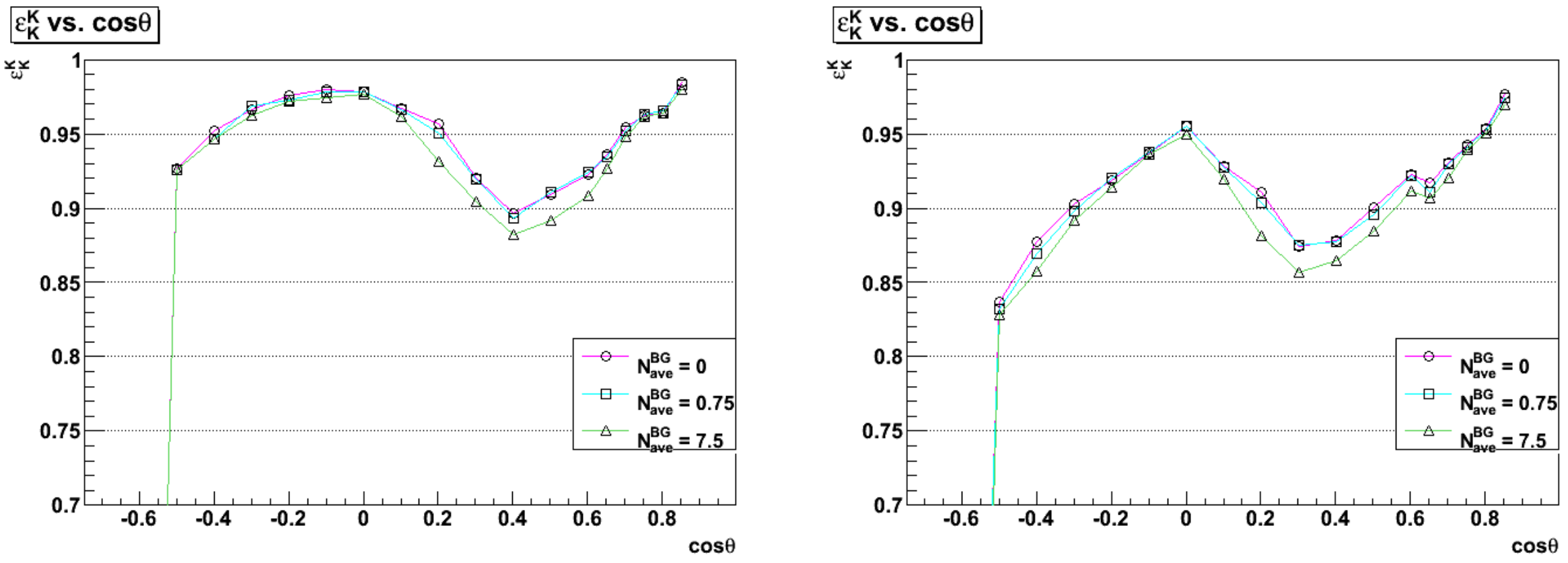

*Figure 7.25: Kaon efficiencies estimated using GSIM for the TOP detector at 3 (left) and 4 (right) GeV/c momentum with varying numbers of background photons from spent electrons.*

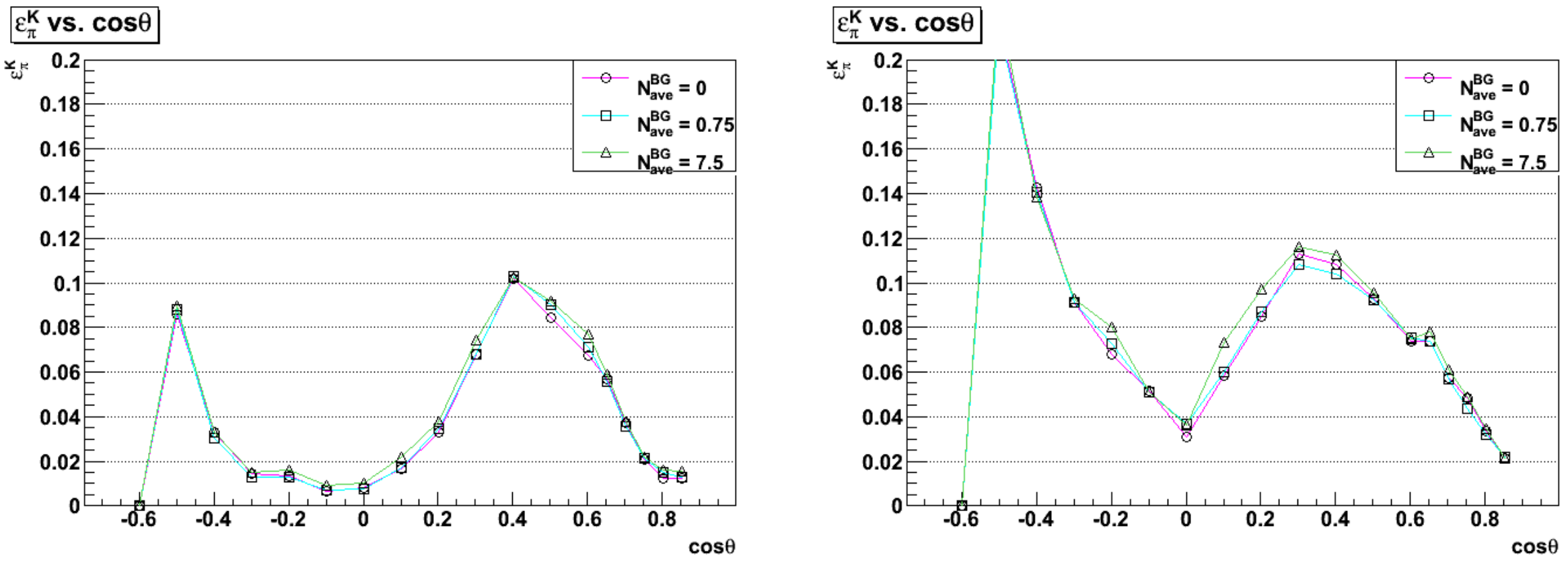

*Figure 7.26: Kaon fake rate estimated using GSIM for the TOP detector at 3 (left) and 4 (right) GeV/c momentum with varying numbers of background photons from spent electrons.*

#### 7.5.4.5 Mirror misalignment

Estimates of misalignment parameters from prototype testing in Nagoya indicate that, due to manufacturing limits, the mirror's center of curvature may be displaced from its nominal position by approximately 0.3 mm. During the process of gluing the mirror to the quartz bar, rotations of order 0.2 mrad may be introduced. These imperfections were added into the simulations, and there was little noticeable impact on the overall performance. As such, we report here only the results assuming perfect mirrors, and leave the results of simulated misalignment studies for a detailed TOP simulation document.

#### 7.5.4.6 ECL Backsplash

The contribution of albedo from the calorimeter ("backsplash") is suppressed by the 1.5T solenoidal magnetic field. Nevertheless, preliminary studies indicate that the effect could be significant and further study is needed.

### 7.5.5 Performance estimates

#### 7.5.5.1 Estimated performance by momentum, polar angle

Kaon efficencies and pion fake rates are tabulated for various track momenta and polar angles. The results of the three independent simulations and reconstruction methods are shown in Figs. 7.25–7.28. The three methods show reasonable agreement, particularly in terms of reproducing similar features as a function of track impact angle.

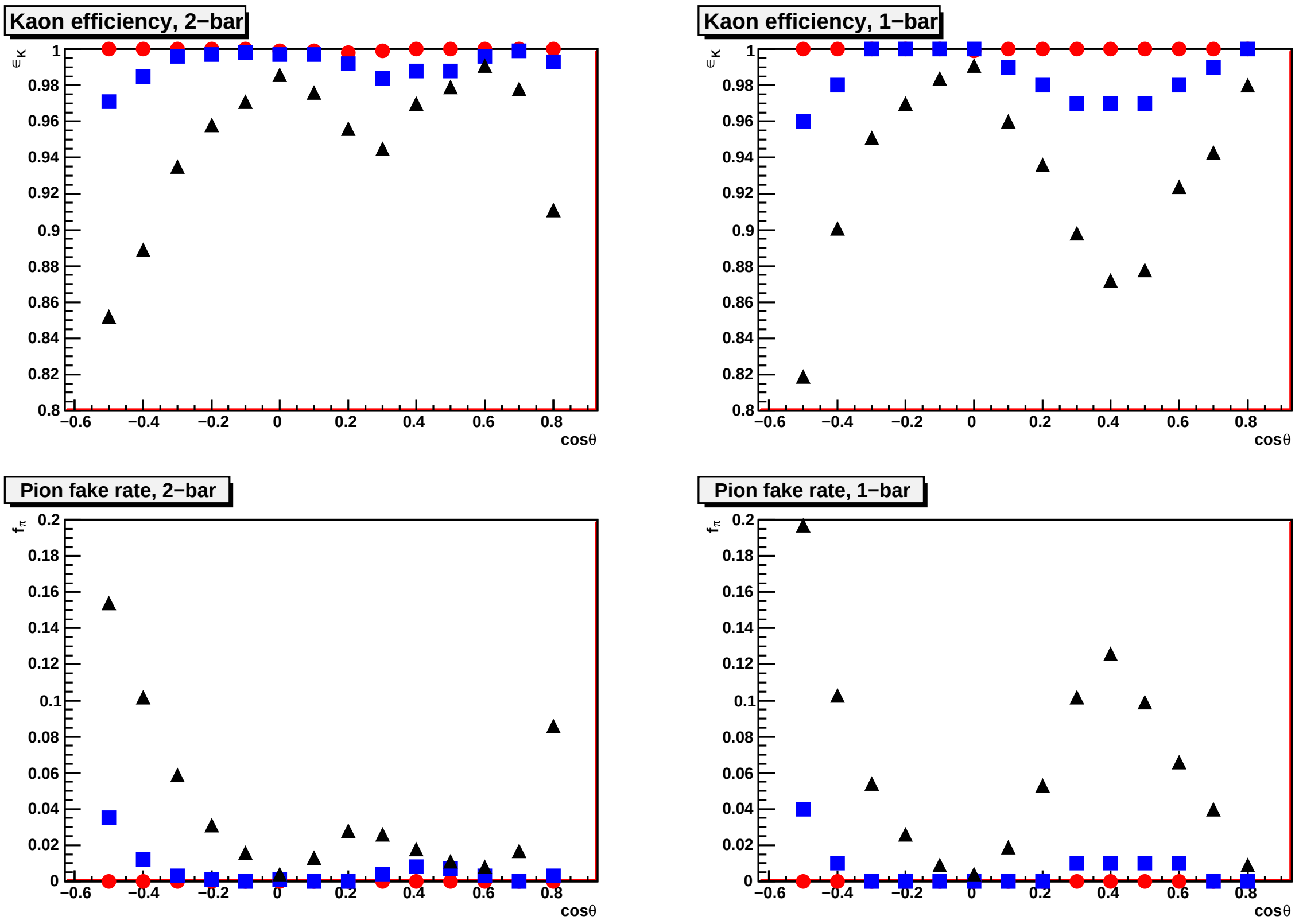

*Figure 7.27: Kaon efficiencies (top row) and pion fake rates (bottom row) for the 2-bar (left column) and 1-bar (right column) configurations, estimated from a Geant4-based simulation and reconstruction. Results are plotted for 1.5 GeV/c (●), 2.5 GeV/c (■), and 3.5 GeV/c (▲).*

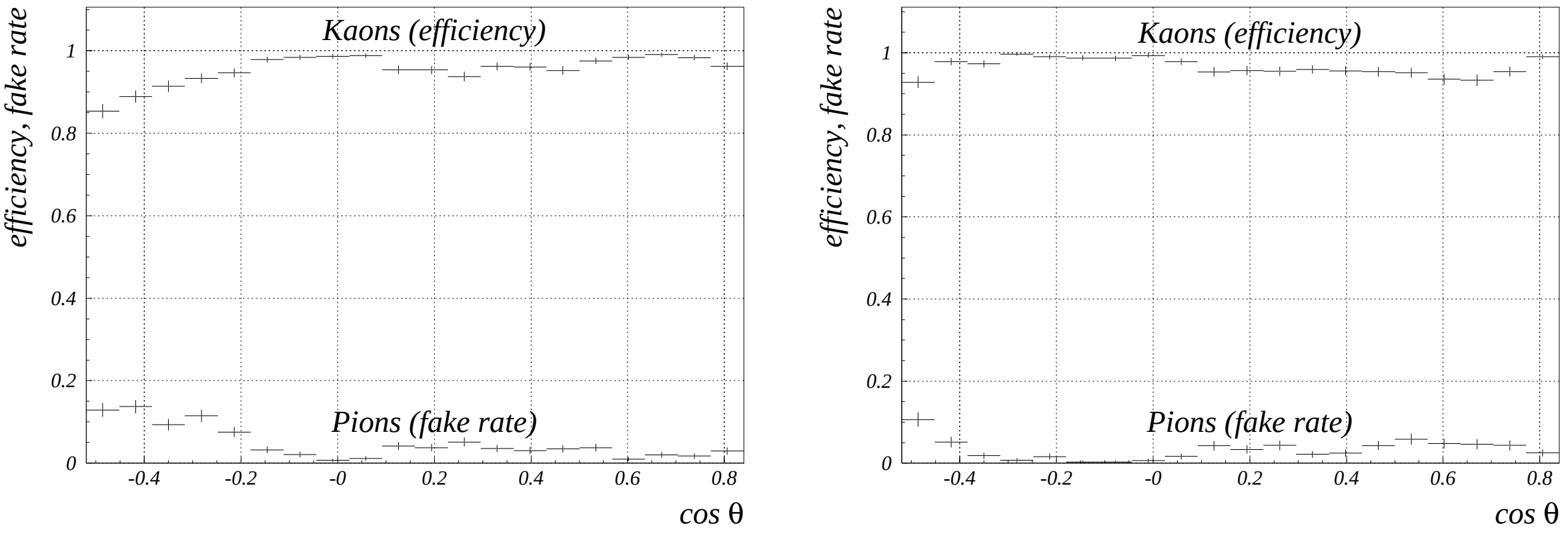

*Figure 7.28: Kaon efficiency and pion fake rates for the 2-bar (left) and 1-bar (right) configurations, based on a stand-alone simulation and reconstruction. Results are plotted for 3 GeV/c.*

#### 7.5.5.2 Estimated performance in $B \to \pi\pi$ and $B \to \rho\gamma$

Two $B$ decay modes that are expected to significantly benefit from an improved PID system are $B \to \pi\pi$ and $B \to \rho\gamma$, distinguished from the background modes of $B \to K\pi$ and $B \to K^*\gamma$, respectively. Estimates of the efficiencies and fake rates in these modes are computed in one of two ways. For the Geant-based simulations, the efficiencies and fake rates as a function of momentum and track polar angle are coarsely sampled, as in Figs. 7.26 and 7.27. The phase space of charged tracks from each decay mode is then used to perform a weighted average of efficiencies and fake rates. For the analytical reconstruction method, charged tracks from the appropriate decay mode are produced using a Monte Carlo generator, and the overall efficiencies and fake rates are determined for these tracks. In both cases, the overall efficiency and fake rate values are reported only over the acceptance of the TOP counters. A summary table of the results for each simulation group can be seen in Tables 7.8 and 7.9.

*Table 7.8: Efficiencies and fake rates obtained from simulation for $B \to \pi\pi$, under the set of assumptions described in the text. "MA" and "SBA" indicate the multi-alkali and super-bialkali photocathode, respectively. The GSIM simulations include secondaries from both electromagnetic and hadronic interactions, the Geant4 simulations include secondaries from electromagnetic interactions, and the stand-alone simulations do not include secondaries.*

| | | $\pi$ efficiency (%) | | | $K$ fake rate (%) | | |
|---|---|---|---|---|---|---|---|
| Geometry | Condition | GSIM | Geant4 | stand-alone | GSIM | Geant4 | stand-alone |
| 1-bar | MA | 92.8 | 96.1 | 92.6 | 7.2 | 4.5 | 4.0 |
| 2-bar | MA | 96.2 | 97.6 | 96.5 | 3.8 | 2.7 | 3.8 |
| 1-bar | SBA | 95.6 | 98.1 | 94.8 | 4.4 | 2.8 | 2.5 |
| 2-bar | SBA | 97.7 | 98.5 | 97.2 | 2.3 | 1.9 | 2.7 |
| 1-bar | MA 50ps | 91.3 | 94.7 | 90.6 | 8.7 | 5.8 | 6.0 |
| 1-bar | MA 2mrad | 92.2 | – | 91.4 | 7.9 | – | 4.8 |
| 1-bar | MA 80% | 91.3 | 95.0 | 90.7 | 8.7 | 5.6 | 5.3 |
| 1-bar | SBA 80% | 94.6 | 97.5 | 93.6 | 5.4 | 3.5 | 3.3 |
| 2-bar | MA 50ps | 93.2 | 93.9 | 91.6 | 6.8 | 6.9 | 8.9 |
| 2-bar | MA 2mrad | 95.7 | – | 95.9 | 4.3 | – | 4.2 |
| 2-bar | MA 80% | 95.2 | 96.9 | 95.3 | 4.8 | 3.4 | 4.8 |
| 2-bar | SBA 80% | 97.1 | 98.2 | 96.8 | 2.9 | 2.2 | 3.4 |

*Table 7.9: Efficiencies and fake rates obtained from simulation for $B \to \rho\gamma$, under the set of assumptions described in the text. "MA" and "SBA" indicate the multi-alkali and super-bialkali photocathode, respectively. The GSIM simulations include secondaries from both electromagnetic and hadronic interactions, the Geant4 simulations include secondaries from electromagnetic interactions, and the stand-alone simulations do not include secondaries.*

| | | $\pi$ efficiency (%) | | | $K$ fake rate (%) | | |
|---|---|---|---|---|---|---|---|
| Geometry | Condition | GSIM | Geant4 | stand-alone | GSIM | Geant4 | stand-alone |
| 1-bar | MA | 97.6 | 99.5 | 98.5 | 1.3 | 0.9 | 0.8 |
| 2-bar | MA | 98.4 | 99.7 | 99.4 | 1.1 | 0.5 | 0.6 |
| 1-bar | SBA | 98.6 | 99.9 | 99.1 | 0.9 | 0.4 | 0.4 |
| 2-bar | SBA | 98.9 | 99.9 | 99.6 | 1.0 | 0.2 | 0.4 |
| 1-bar | MA 50ps | 97.3 | 99.1 | 98.1 | 1.5 | 1.0 | 1.1 |
| 1-bar | MA 2mrad | 97.5 | – | 98.3 | 1.3 | – | 1.0 |
| 1-bar | MA 80% | 97.0 | 99.1 | 97.9 | 1.8 | 1.3 | 1.1 |
| 1-bar | SBA 80% | 98.3 | 99.7 | 99.0 | 1.0 | 0.6 | 0.6 |
| 2-bar | MA 50ps | 97.6 | 98.9 | 98.3 | 1.5 | 1.3 | 1.9 |
| 2-bar | MA 2mrad | 98.4 | – | 99.3 | 1.2 | – | 0.7 |
| 2-bar | MA 80% | 98.1 | 99.5 | 99.2 | 1.3 | 0.8 | 0.9 |
| 2-bar | SBA 80% | 98.7 | 99.9 | 99.6 | 1.0 | 0.3 | 0.6 |

# Chapter 8

# Particle Identification: End-cap

## 8.1 Introduction

Identification of charged particles over the full kinematic range is one of the basic requirements for Belle II. In the forward endcap, the proximity-focusing Aerogel Ring-Imaging Cherenkov detector (ARICH) has been designed to separate kaons from pions over most of their momentum spectrum (Fig. 1.9) and to provide discrimination between pions, muons and electrons below 1 GeV/$c$.
The ARICH elements (Fig. 8.1) are:

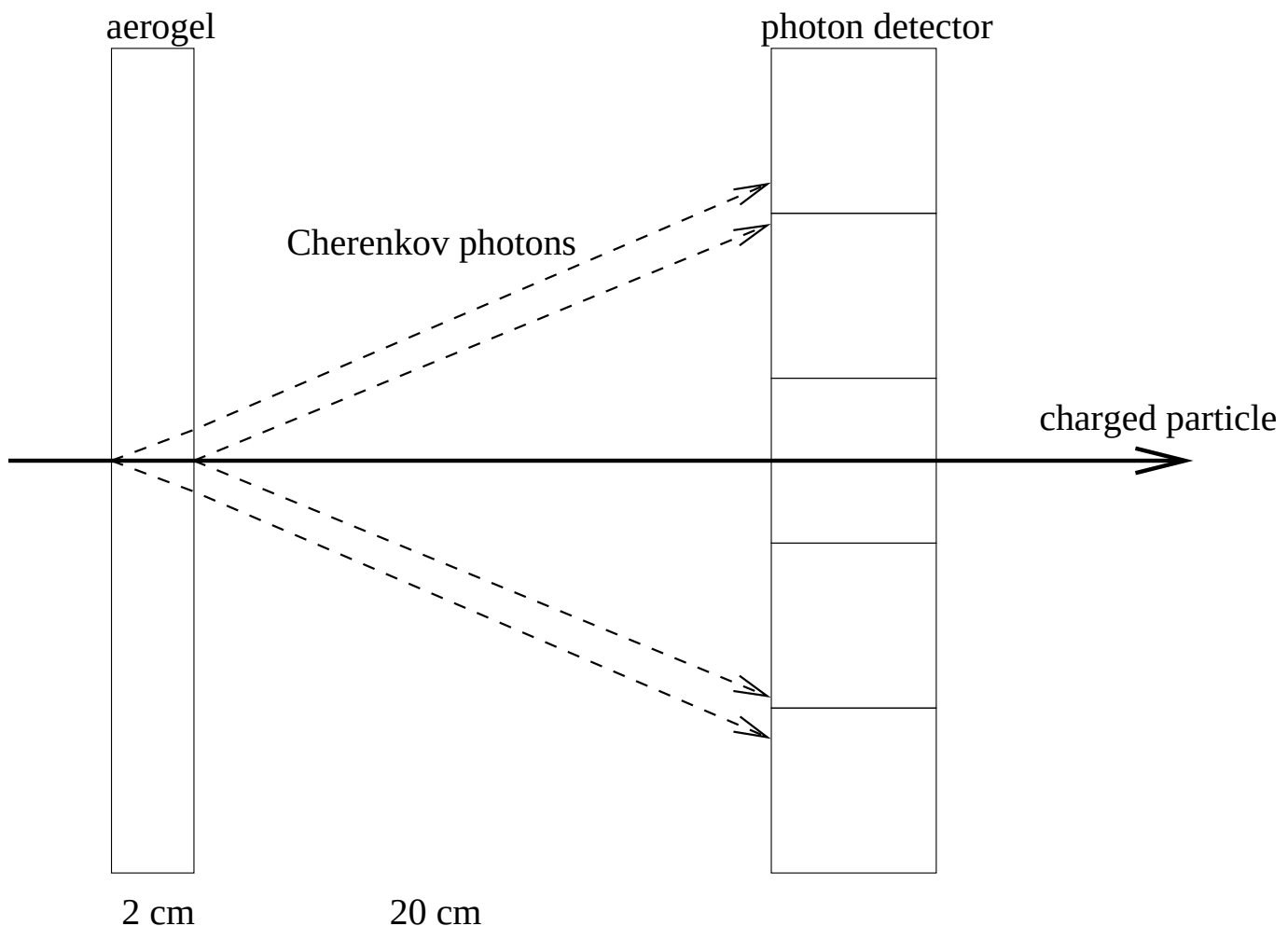

*Figure 8.1: Proximity focusing ARICH - principle*

- an aerogel *radiator* where Cherenkov photons are produced by charged particles,
- an *expansion volume* to allow Cherenkov photons to form rings on the photon detector surface,
- an array of *position sensitive photon detectors*, that is capable of detecting single photons in a high magnetic field with high efficiency and with good resolution in two dimensions, and
- a *read-out system* for the photon detector.

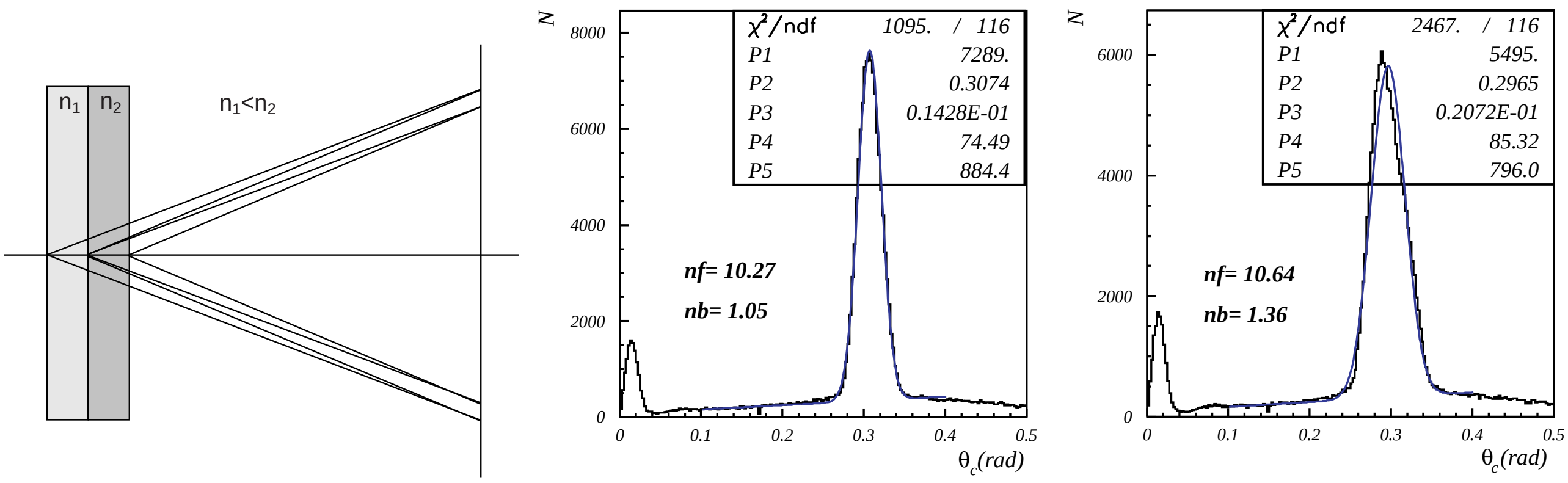

*Figure 8.2: Left: Proximity focusing RICH with an inhomogeneous aerogel radiator in the focusing configuration. Center: the distribution of Cherenkov photon hits vs. Cherenkov angle for a focusing configuration with $n_1 = 1.046$ and $n_2 = 1.056$. Right: the corresponding distribution for a 4-cm homogeneous radiator.*

The design choices are governed by the following criteria:

- To achieve the necessary performance, enough photons (about 10) have to be detected for each ring image for at least one of the particle species. This requirement fixes the length of the aerogel radiator to several centimeters.
- The required resolution in the measurement of the Cherenkov angle is achievable only for an expansion gap of about 20 cm and a radiator thickness that does not exceed a few centimeters, with a photon detector granularity of a few millimeters.

As already discussed in the LoI [1], a prototype of the counter showed excellent performance both in on-the-bench and in beam tests. However, two major issues remained: the need to increase the number of detected Cherenkov photons and the development of a detector for single photons that would reliably work in the high magnetic field of Belle II. Both problems were solved in a satisfactory manner.
The key parameter in the performance of a RICH counter is the Cherenkov angle resolution per charged particle $\sigma_{\text{track}} = \sigma_\theta/\sqrt{N}$. With a longer radiator, the number of detected photons increases, but in a proximity focusing RICH the single photon resolution degrades because of the emission point uncertainty. For Belle II, the optimal thickness is around 20 mm [1, 2, 3]. However, in the R&D phase following the LoI, we have found a solution to this limitation. The problem is solved if a proximity focusing RICH with a non-homogeneous radiator is employed [3, 4, 5, 6]. By appropriately choosing the refractive indices of consecutive aerogel radiator layers, one may achieve overlapping of the corresponding Cherenkov rings on the photon detector (Fig. 8.2) [6]. This is equivalent to focusing of the photons within the radiator, and eliminates or at least considerably reduces the spread due to emission point uncertainty. Note that such a tuning of refractive indices for individual layers is only possible with aerogel, which may be produced with any desired refractive index in the range 1.01-1.2 [7].
In Fig. 8.2, we compare the data for two 4-cm thick radiators: one with aerogel tiles of equal refractive index (n = 1.046), the other with the focusing arrangement ($n_1 = 1.046, n_2 = 1.056$). The improvement is clearly visible. The single photon resolution $\sigma_\theta = 14.3$ mrad for the dual radiator is considerably better than the corresponding value for the single refractive index radiator ($\sigma_\theta = 20.7$ mrad), while the number of detected photons is the same in both cases.

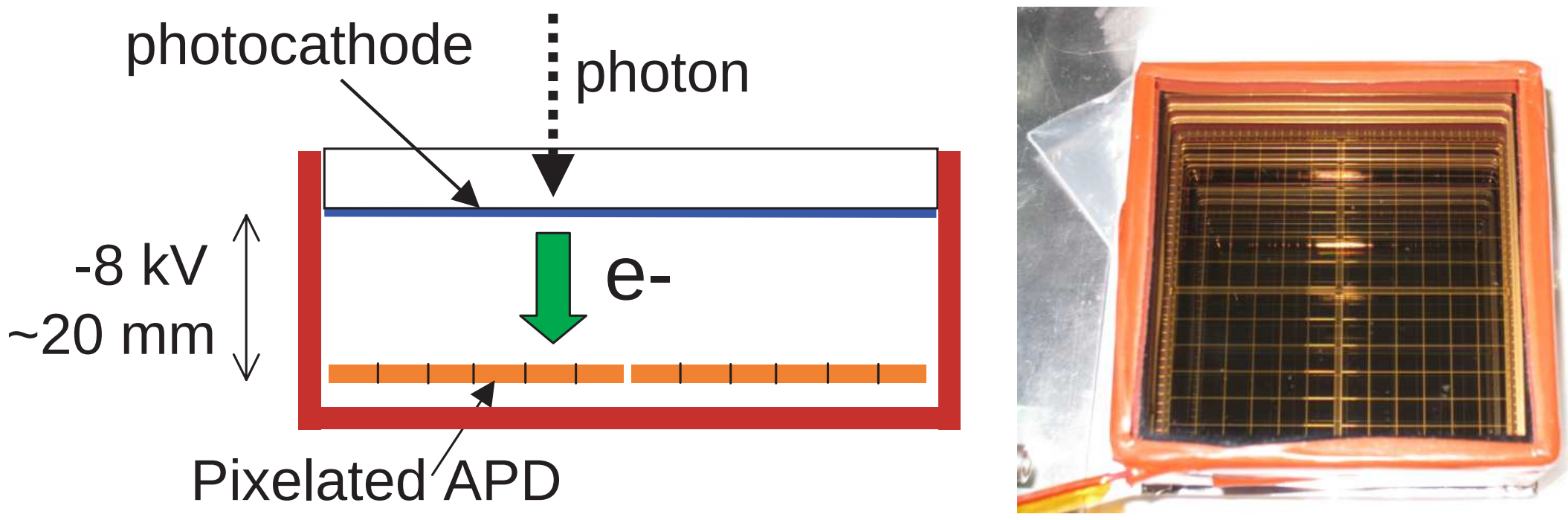

*Figure 8.3: Schematic drawing (left) and a photograph (right) of the HAPD, hybrid avalanche photon detector.*

The second open issue—development of a reliable sensor of single photons—has also been solved in a satisfactory manner, as reported in Secs. 8.2.2 and 8.5.2. As the baseline detector, a hybrid avalanche photo-detector (HAPD by Hamamatsu Photonics) of the proximity focusing type was selected, while the Photonis MCP-PMT is kept as a backup option.

## 8.2 Basic Detector Elements

### 8.2.1 Aerogel Radiator

The silica aerogel Cherenkov radiator should be highly transparent in order not to lose photons inside the medium via Rayleigh scattering or absorption. To obtain well separated patterns from kaons and pions at 4 GeV/$c$, a refractive index near the range 1.04–1.06 is required. In our design, two 20-mm thick layers of silica aerogel with refractive indices of 1.055 and 1.065 are employed as Cherenkov radiators. As discussed above, the indices are chosen so that the rings from the two aerogel layers overlap on the photon detector plane.

The optical quality of the aerogel can be characterized as $T = T_0 \exp(-d/\Lambda(\lambda))$, where $T_0$ and $T$ are light intensities before and after passing through the aerogel tile of thickness $d$, and $\Lambda$ is the transmission length at wavelength $\lambda$. The value of $\Lambda$ at $\lambda = 400$ nm is required to exceed 40 mm for both layers to guarantee high transparency.

The shape of one piece of the radiator is hexagonal in the baseline design. Thanks to the hydrophobic property of our aerogel material, we can cut them from square tiles using a water-jet device without degrading the optical characteristics. At present, the tile dimension is assumed to be $160 \times 160 \times 20$ mm$^3$. In this case, about 300 square-shaped tiles are needed for each refractive index, (600 tiles for two layers) to cover the entire radiator area.

Further R&D to increase the photoelectron yield as well as to establish technical procedures to maintain uniform quality during large-scale production is still ongoing.

### 8.2.2 Photon Detector

The square-shaped HAPD produced by Hamamatsu Photonics (HPK) is the baseline photon detector. This device consists of a vacuum tube with enclosed solid state sensor of the avalanche photo-diode (APD) type (Fig. 8.3 (right)). Figure 8.3 (left) shows a schematic view of the HAPD operation, while typical specifications are summarized in Table 8.1.

Cherenkov photons enter through the entrance window and generate photoelectrons from a bialkali photocathode. Photoelectrons are accelerated along the electric field, where a typical voltage is 7–10 kV, and are directed onto the avalanche photodiode. In the APD, an additional gain of ∼40 is obtained when a bias voltage is provided. The photon detector consists of four APD chips, each of which is pixelated into 6×6 pads for a position measurement, and each pad has a size of 4.9×4.9 $mm^2$.
Fig. 8.4 (left) shows the electron bombardment gain as a function of HV. Typically, the gain is about 1500 at −8 kV. The avalanche gain and leakage current in a typical APD are given in Fig. 8.4 (right). The leakage current is around a few $\mu$A for 36 channels, and an avalanche gain of 50 is obtained at 330 V [8].

| | |
|---|---|
| package size | 72×72 $mm^2$ |
| # of pixels | 12×12 (6×6/APD chip) |
| pixel size | 4.9×4.9 $mm^2$ |
| effective area | 67 % |
| typical QE | 25 % |
| gain | $\sim 10^5$ |
| weight | 220 g |

*Table 8.1: HAPD specifications.*

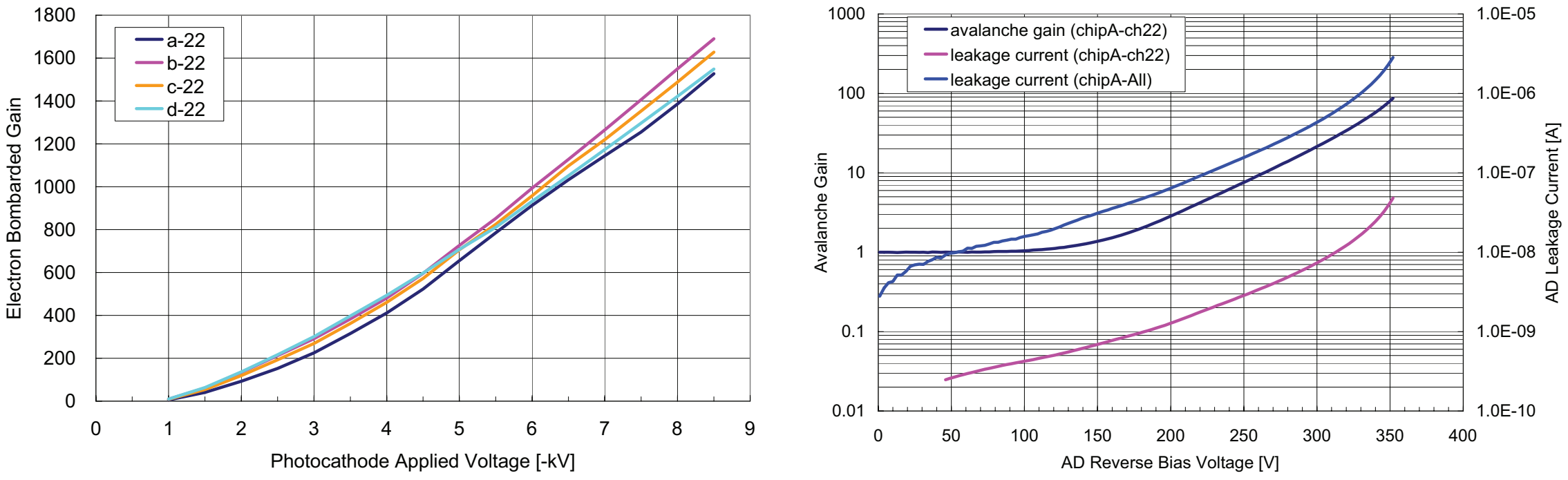

*Figure 8.4: Electron bombardment gain (left) as a function of photocathode voltage and APD gain and leakage current as a function of APD reverse bias voltage (right). Both are taken from HPK data sheets.*

### 8.2.3 Read-Out System

Because of the large number of photon detector read-out channels, ∼ 80k in total, a dedicated front-end read-out system based on an ASIC (Application Specific Integrated Circuit) chip is employed. In the ASIC chip, the input signal is processed in three successive steps: a preamplifier, a shaper and a comparator. The output signal is fed to an external FPGA. The gain and the shaping time are variable, and can be externally set. The threshold to the comparator is common to all the channels in a chip, but the offset of the signal is adjustable for each channel. A schematic of the ASIC and its layout are shown in Fig. 8.5. Table 8.2 summarizes the specifications of the present ASIC chip.
After the comparator, only the on/off information is registered in the FPGA to conform to the overall Belle II DAQ scheme. In our design, one FPGA manipulates all signals of the 144 channels from one HAPD. The clock signal inside the FPGA runs at 64 MHz and the latency and the time window are programmable.
The ASIC and the FPGA chips are encapsulated in a printed circuit board directly attached to the HAPD output pins. The data merger may be situated inside the ARICH container to make full use of the bandwidth in optical fibers as well as to reduce the number of fibers, depending on the available space.

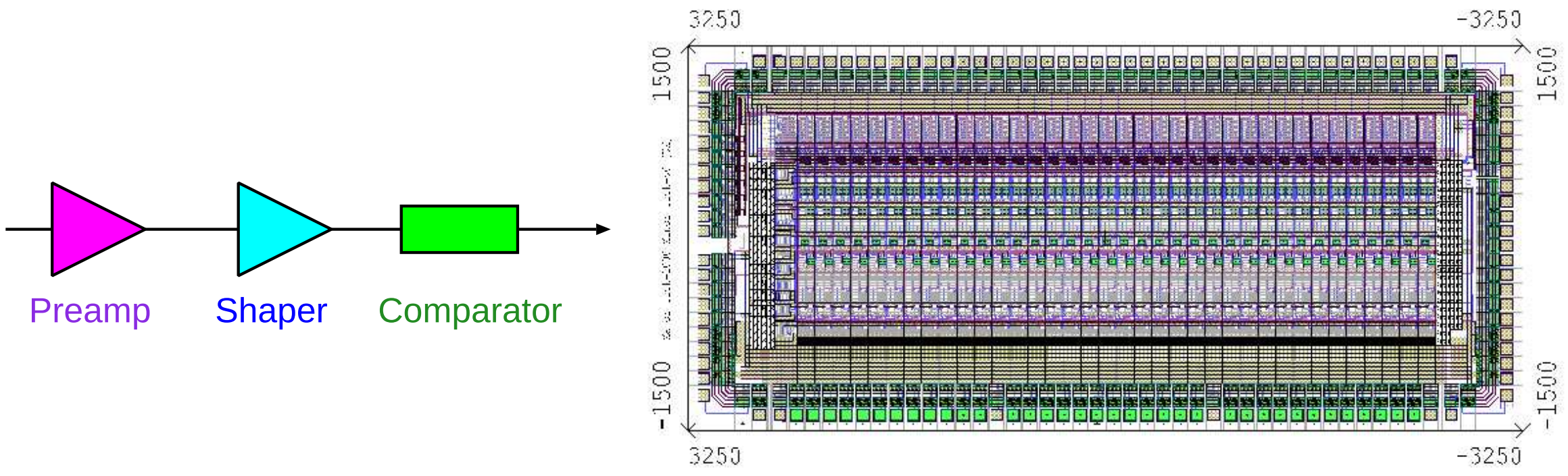

*Figure 8.5: Schematic and layout of the ASIC (SA02).*

| | |
|---|---|
| process | TSMC CMOS 0.35 $\mu$m |
| chip size | $3.0 \times 6.5\,\text{mm}^2$ |
| noise level | 1200 $e^-$ at 80 pF |
| amplification | 4 steps: 18–70 mV/fC |
| peaking time | 4 steps: 250–1000 ns |
| offset adjustment | coarse: 60 mV/ch; fine: 5 mV/ch |

*Table 8.2: Specifications of the present ASIC (SA02).*

## 8.3 Mechanical Design

A schematic view of the mechanical structure is shown in Fig. 8.6. The RICH container consists of two cylinders with tentative inner and outer radii of 410 mm and 1140 mm, respectively. The two cylinders are connected by reinforcement bars in the radial direction at the back plane. The front surface, which is mounted on both inner and outer cylinders, is a thin Al plate. The back-end plane is machined so that HAPDs with readout electronics modules are mounted on it. The photon detector support plane is further reinforced by few-cm high radially oriented septum walls of 0.5-mm thick Al. The septums define six identical azimuthal sectors, each organized as a honeycomb-like structure. The container housing the ARICH counter is supported by the inner cylinder of the E-CsI container in the same way as the present end-cap aerogel Cherenkov counter. The HAPDs are arranged in 9 concentric rings in the radial direction; in total, 540 sensors of this type are required. The breakdown of sensors per ring is listed in Table 8.3. The minimum distance between HAPDs as well as between HAPD and aluminum walls is assumed to be $\sim 2$ mm.

| Layer # | 1 | 2 | 3 | 4 | 5 | 6 | 7 | 8 | 9 |
|---|---|---|---|---|---|---|---|---|---|
| # of HAPDs | 36 | 42 | 48 | 54 | 60 | 66 | 72 | 78 | 84 |

*Table 8.3: Number of HAPDs for the nine concentric rings.*

We are considering mounting a segmented flat mirror on the inner side of the outer cylinder to reflect Cherenkov photons, which would otherwise be absorbed at the cylinder walls, to the area covered by photo-sensors. As discussed in Sec. 8.6, such a mirror would allow for a continuous transition in PID coverage from the barrel to the end-cap boundary region.
The aerogel radiators are tiled as shown in Fig. 8.6. They are assumed to be tiled on a thin aluminum plate and black paper will be inserted into the gaps between the tiles to absorb photons hitting the radiator boundary.

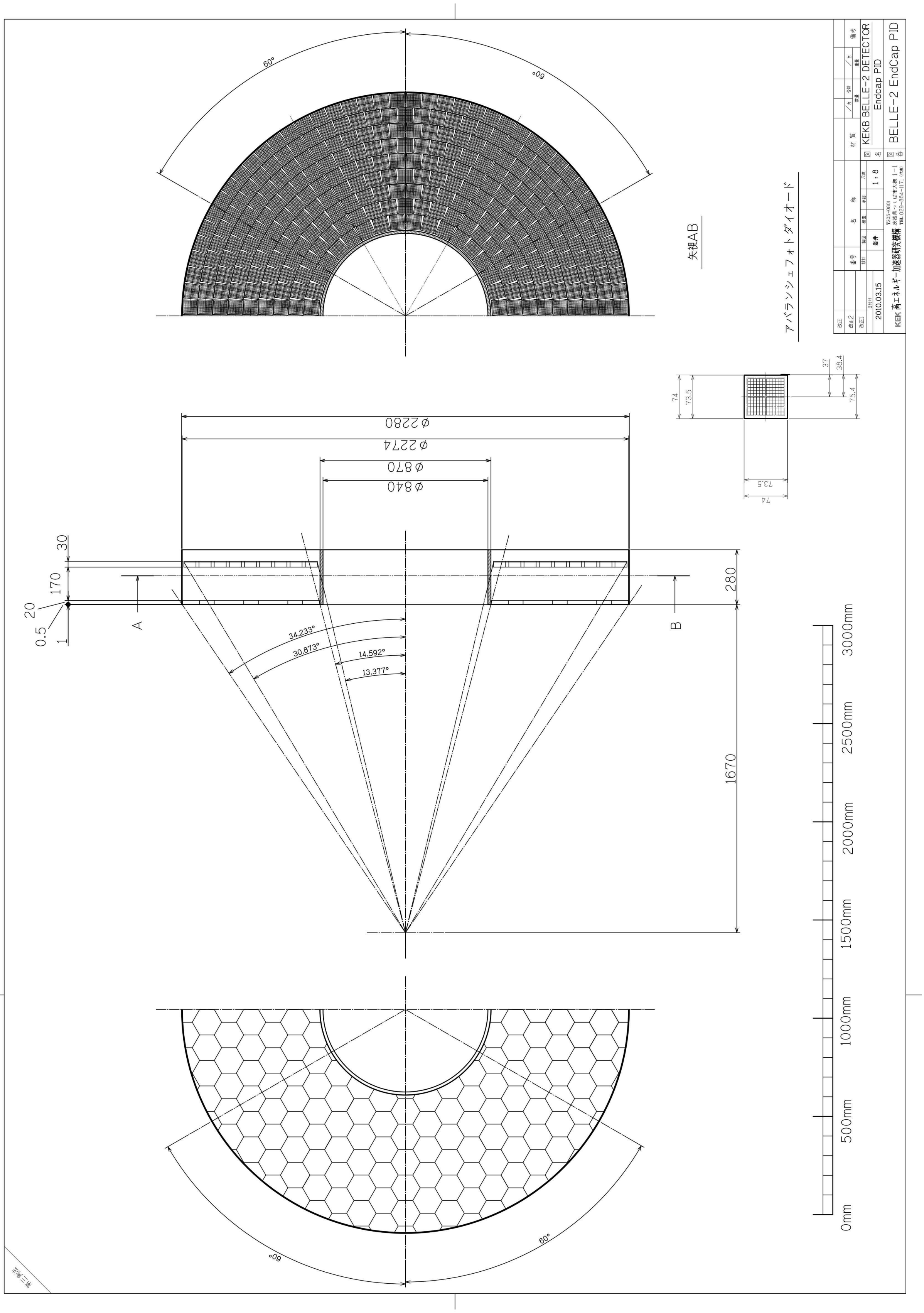

*Figure 8.6: Schematic drawing of the RICH mechanical structure.*

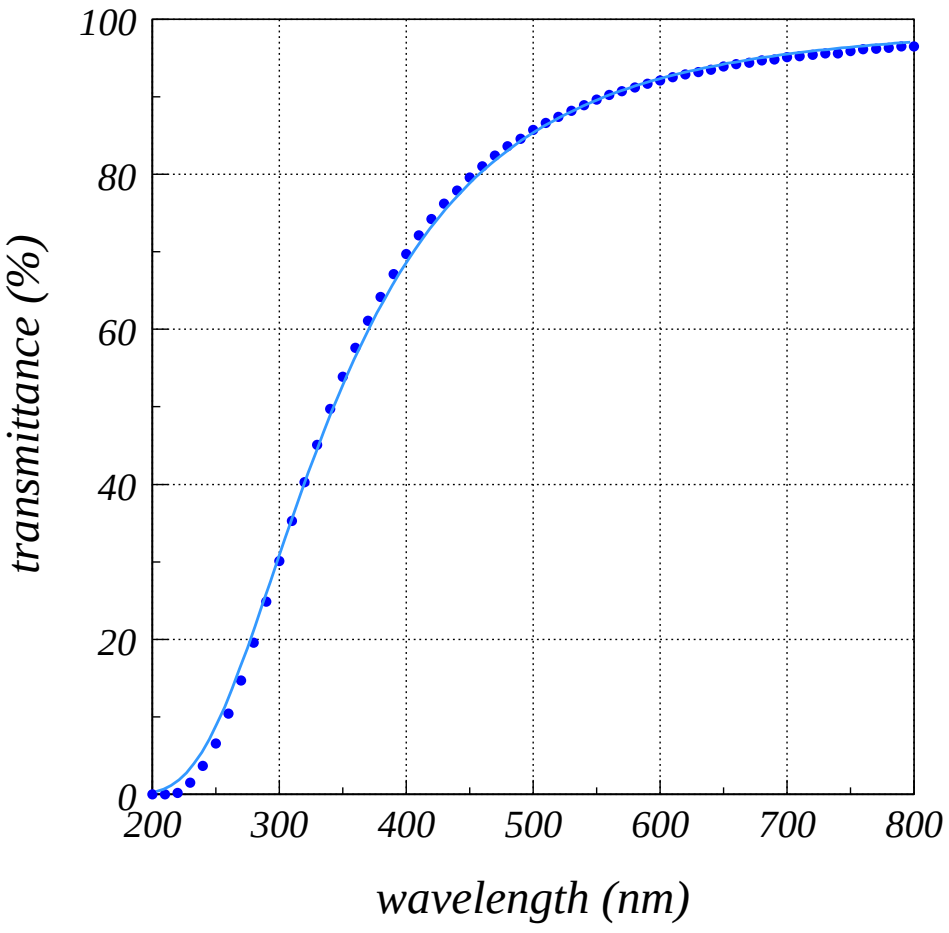

*Figure 8.7: Transmittance of 20-mm thick aerogel with $n = 1.065$ as a function of wave-length(nm). The solid curve is fit used to determine the clarity.*

## 8.4 Cabling and Other Services

One HV cable per HAPD is assumed at the moment, although the final decision might depend on the cost. For the bias voltage, each APD needs a separate cable since the avalanche gain in the HAPD is highly sensitive to the bias voltage. On the other hand, several guard voltage channels may be combined. In the readout system, three fiber cables will be used: two for readout and one for triggering.
In addition to these cables, two cooling pipes (in/out) and one nitrogen pipe per sector are required. Monitoring cables for temperature and humidity are also necessary.
The joint station has to be located at the back plane of the E-CsI container. In this station, all cables and pipes have to be pluggable to permit retraction of the end-cap apparatus.

## 8.5 R&D Results

### 8.5.1 Aerogel Radiator

Since 2008, we have worked to improve the optical quality of the aerogel radiators, particularly those with refractive index greater than 1.055 [7, 9]. Recently, a new production technique, the so called “pin-hole drying” method, was invented and successfully tested [10]. This method allows us to reach a transmission length of more than 45 mm $\lambda = 400$ nm for $n > 1.055$ samples. Such a performance was not achieved with the previous method. Figure 8.7 shows the measured transmittance of an $n = 1.065$ aerogel sample with 20 mm thickness. These data were fitted with the Hunt formula $T = A\exp(-d\,C/\lambda^4)$, where $C$ is the clarity of the aerogel with thickness $d$, $\lambda$ is the photon wavelength, and $A$ is the absorption (assumed to be independent of wavelength). We obtained $C = 0.00048\,\mu\text{m}^4/\text{mm}$ and $A = 0.9934$. The corresponding transmission length was calculated to be $\Lambda = 55.2$ mm at $\lambda = 400$ nm.
In parallel, an effort to produce large aerogel tiles was also made. Larger radiator tiles are important to reduce the loss of the Cherenkov photons at the radiator boundaries. Figure 8.8 shows a photograph of a $180 \times 260 \times 20$ mm$^3$ aerogel tile with transmission length of about 40 mm at 400 nm. Further optimization of the supercritical drying process to suppress cracks inside the samples is underway.

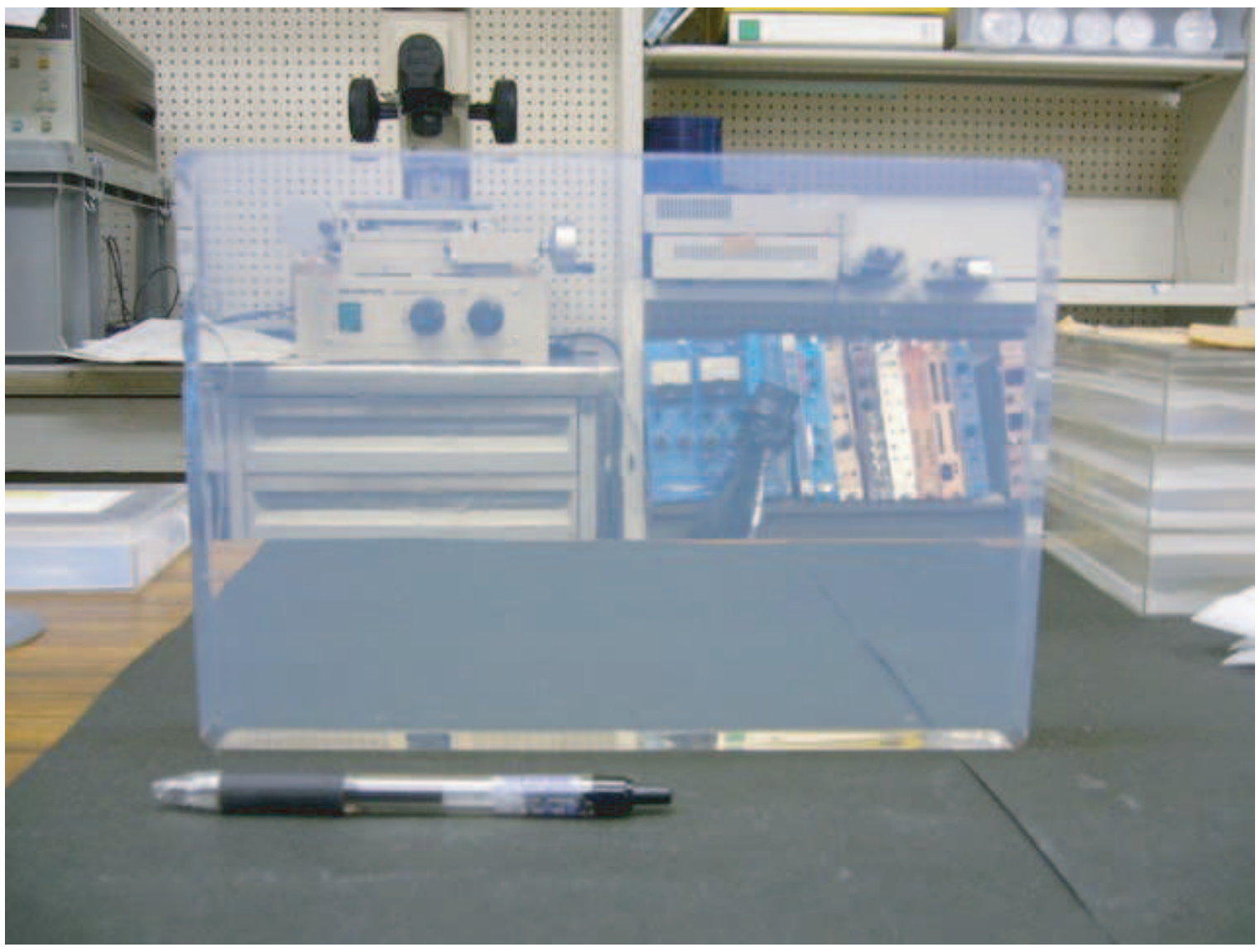

*Figure 8.8: A large-size (180× 260× 20 mm$^3$) aerogel tile.*

### 8.5.2 Photon Sensor

#### 8.5.2.1 Hybrid Avalanche Photo-Detector (HAPD)

• **Noise level and gain** As a first check, we measured the noise level of the signal channel by applying a bias voltage. The noise level was estimated as the r.m.s. value, $\sigma_{\text{pedestal}}$, obtained by fitting a gaussian function to the pedestal distribution. We tested one channel for each diode chip in the HAPD; the obtained values are plotted in Fig. 8.9. As can be seen in the figure, the noise level quickly decreases at a bias voltage of 50 V, where individual channels become electrically disconnected. Minimum noise is obtained around 300 V, and then gradually increases due to leakage current. We confirmed that the diode behavior for all four chips, with the bias voltage turned on, was consistent with expectations.
Gains and S/N for HAPDs were then checked using a blue LED light source. A typical pulse height distribution is plotted in Fig. 8.10. In this figure, a clear single photon peak can be observed, where the typical gain is $G \simeq 5{\times}10^4$. The S/N ratio was then calculated as the ratio $G/\sigma_{\text{pedestal}}$; We obtained S/N $\sim 15$ from these data.
The responses from all HAPD channels were studied using the ASIC readout electronics system. In this system, the digitized hit information is stored if the signal from individual pixels exceeds the threshold value (for a detailed description see Sec. 8.5.3).

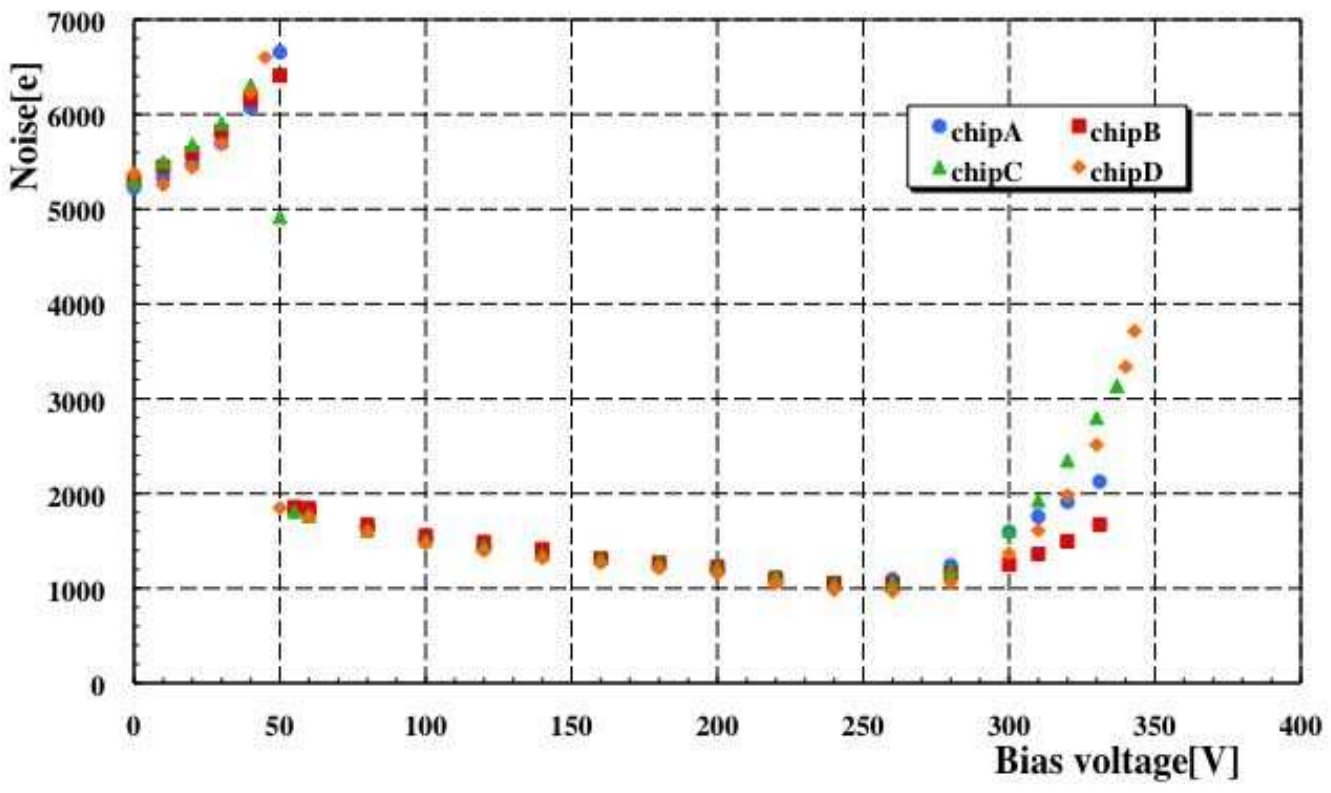

*Figure 8.9: Noise level as a function of bias voltage (V).*

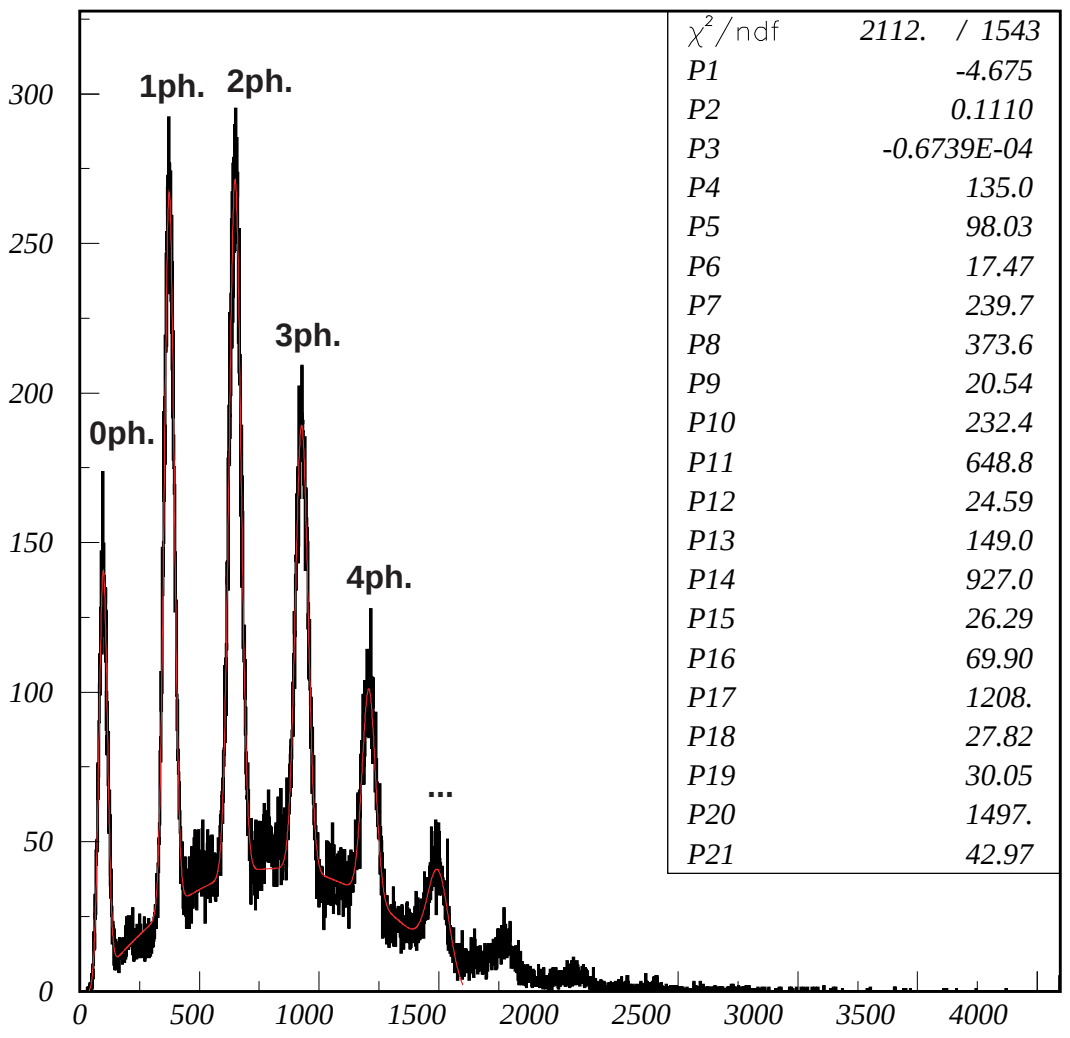

*Figure 8.10: HAPD pulse height distribution.*

The noise level from each pad was first checked by varying the threshold value applied to the ASIC chips. The standard deviation of the noise value was then calculated. The threshold level was set to $4\sigma$ of the noise distribution, which roughly corresponds to 0.5 photoelectrons. After this procedure, a light source was placed on a two dimensional displacement stage controlled by a personal computer. The detector was illuminated at different positions and the hit information from all HAPD pixels was recorded; the light intensity was adjusted so that on average $\sim$ 0.4 photoelectrons per pulse were detected [11]. Figure 8.11 shows the result of such a two-dimensional scan. In this figure, all the hit counts are accumulated, where $x$ and $y$ positions are taken from the incident light beam position. Several pads are missing because of dead channels either on the APD or the ASIC chip. A high detection efficiency was observed over entire active area. Image is distorted at the edges due to the nonuniform electric field. Distortion will disappear when HAPD is operated in the magnetic field.

• **Quantum efficiency (QE)** A high QE is always beneficial to photon detection. In recent years, Hamamatsu has provided photomultiplier tubes with very high peak QE, greater than 30% ("Super Bialkali") and 40% ("Ultra Bialkali") [12]. On the other hand, fabrication of the HAPD is different from conventional PMT production since the transfer method has to be employed. This could be one of the barriers for a direct application of the high QE technology. Last year, the producers tried to adapt the “Super Bialkali” technique to the HAPD photocathode production process and a high QE sample was successfully fabricated for the first time. This sample was tested using a Xenon lamp in the laboratory, and a QE exceeding 30% at 400 nm was confirmed, as shown in Fig. 8.12, where a comparison to a conventional bialkali photocathode is shown. From this measurement, 32% peak quantum efficiency at 360 nm was obtained.
We expect that a higher QE HAPD can be produced in a stable way as the technology improves and that 30% QE on average will be achieved in mass production.

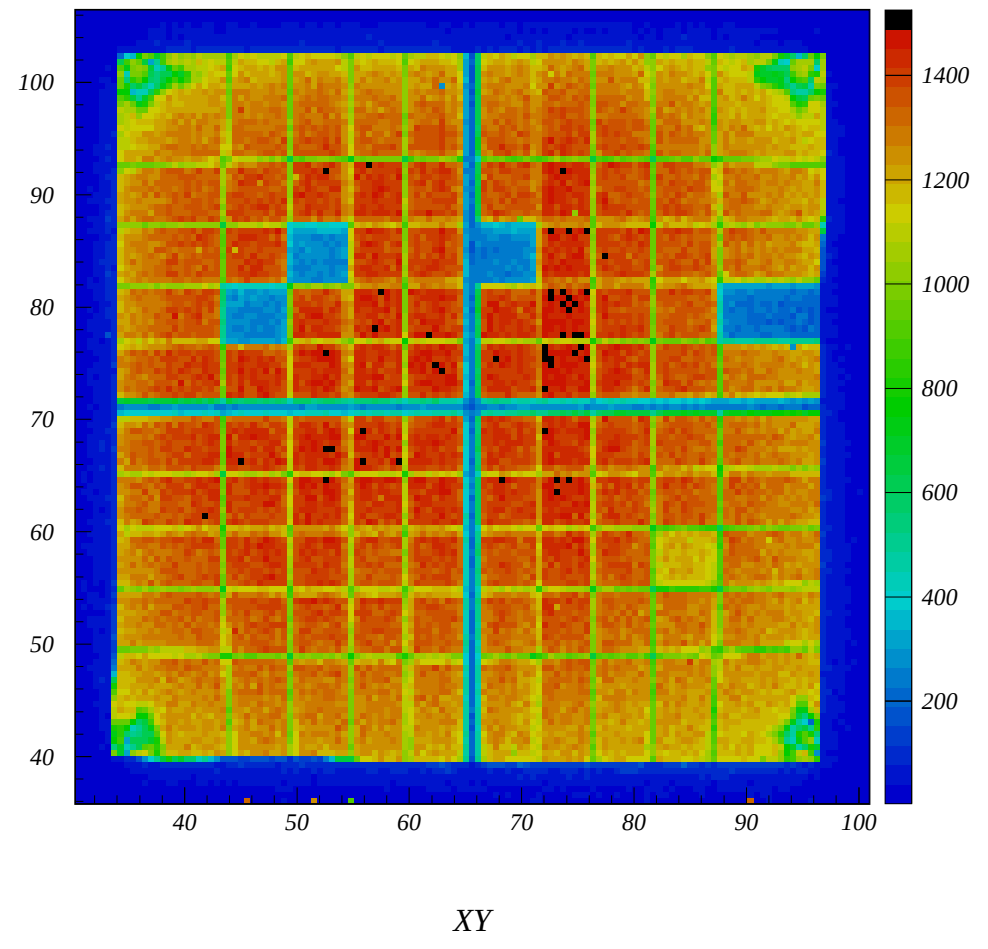

*Figure 8.11: Hit responses in 144 channels by scanning a light beam across the HAPD surface.*

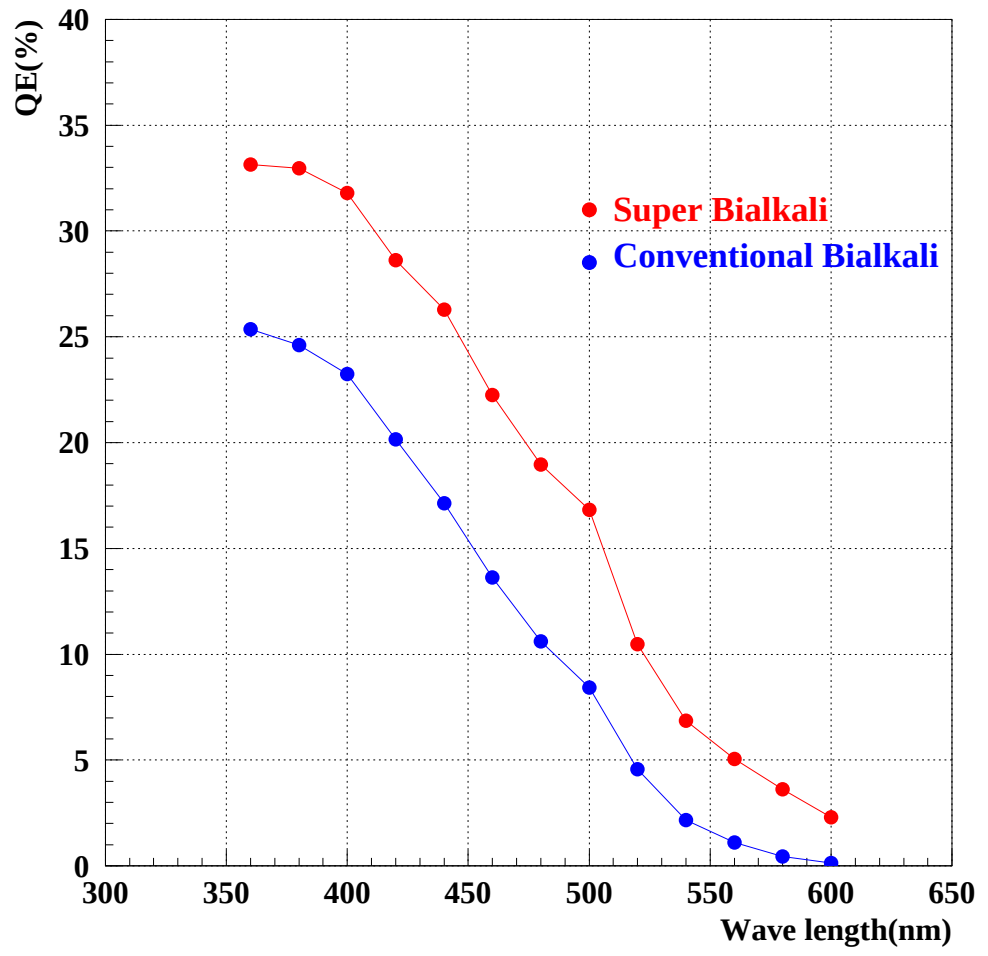

*Figure 8.12: QE of the “Super Bialkali” photocathode in the HAPD as a function of wave length measured with a Xenon lamp. Measurement of the QE values for a conventional bialkali photocathode is also shown for comparison.*

• **Magnetic field immunity** The behavior in a magnetic field was studied using the "Ushiwaka" magnet at the KEK PS (Proton Synchrotron). The scanning stage in two directions was attached to the magnet so that the HAPD performance in the presence of a magnetic field could be examined. The experimental setup including the readout system is identical to what was used in the bench tests. After adjusting the threshold value for all pads, the hit information from the HAPD was measured by varying the incident light position. The hit distributions for 12 channels of one HAPD row, with and without a magnetic field, are plotted in Fig. 8.13.

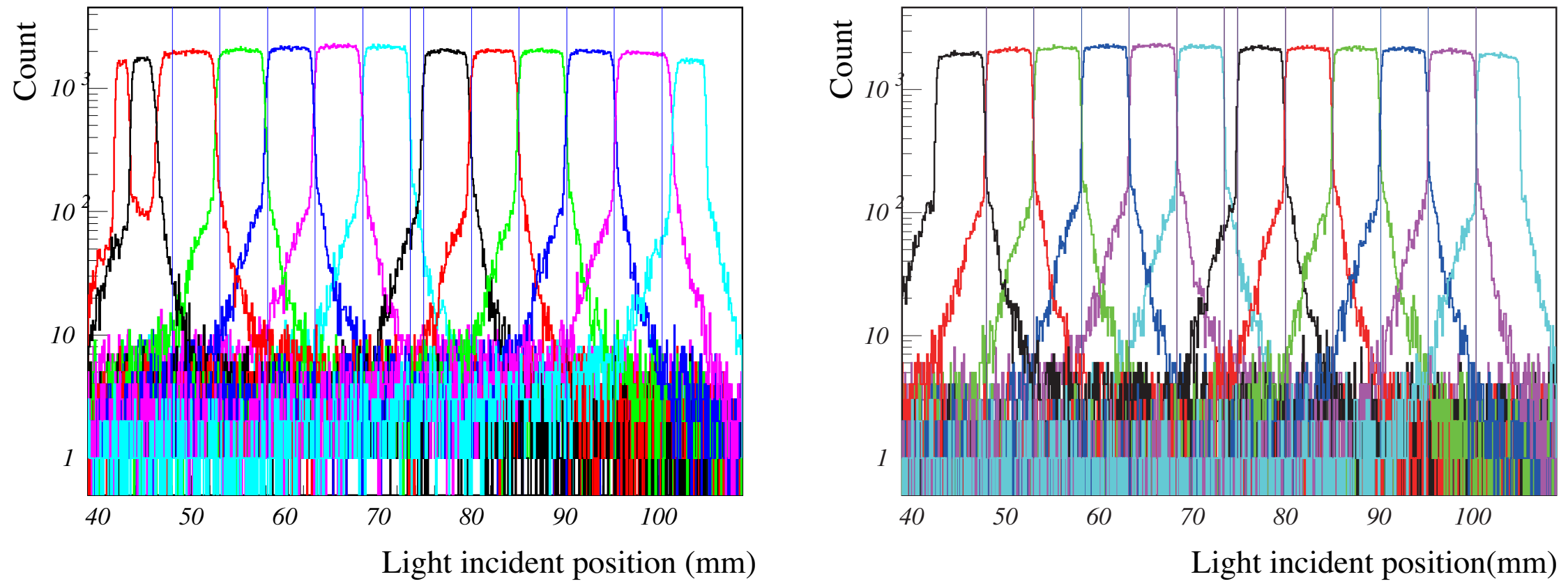

*Figure 8.13: One dimensional scan across the HAPD surface with 0 T (left) and 1.5 T (right) magnetic fields.*

Two improvements in this distribution are apparent when the magnetic field is on. One is an elimination of the strange hit behavior close to the side of the detector, which can be seen in Fig. 8.13 (left) without the magnetic field. This was eliminated in the right figure; it is caused by a distortion in electric field near the detector side wall, illustrated in Fig. 8.14. There is a 4-mm wide metal tape on the side wall, covering the connection between the entrance window and the detector housing made of ceramic, with the HV to the window supplied through this metal component. Since the electric potential of the window and of the metal tape are the same, the electric field is distorted in this area. However, once the magnetic field is turned on, the photoelectrons propagate along the magnetic field lines, on a helix with a Larmor radius of the order of 1 $\mu$m.

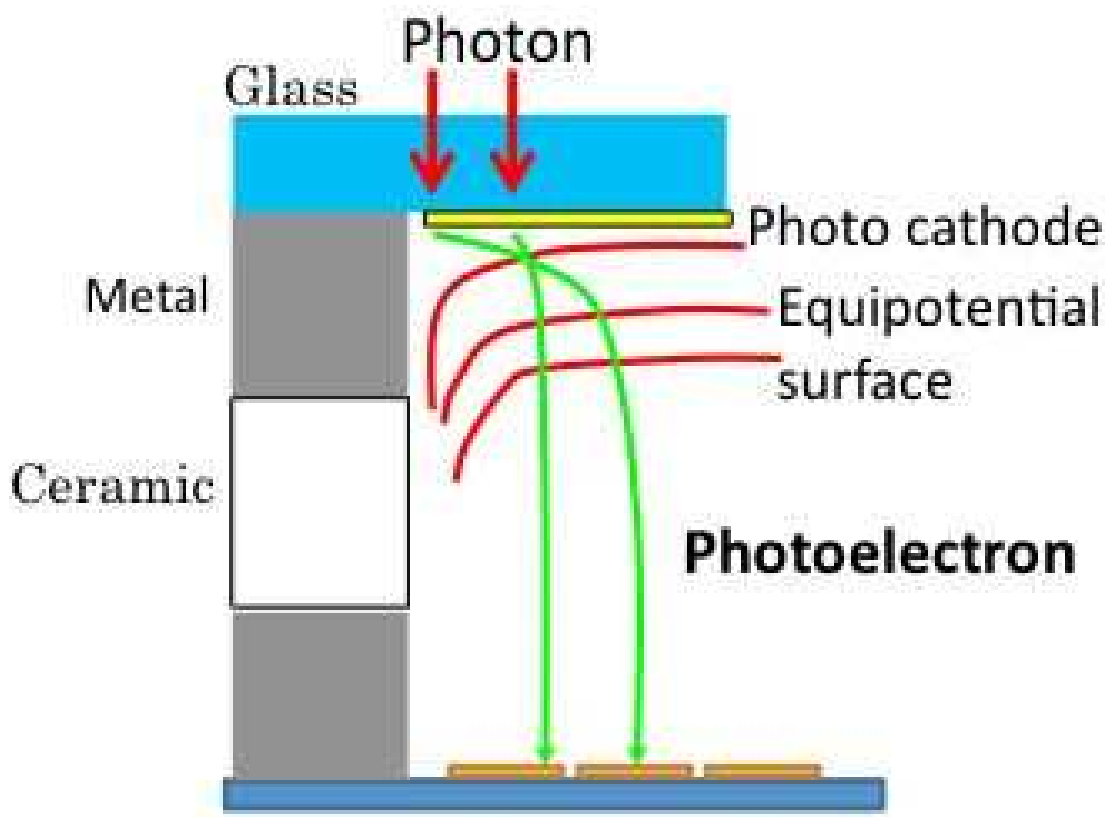

*Figure 8.14: Schematic view of the electric field in the HAPD.*

The other effect is a considerable reduction of the tails observed in the hit distribution. This effect is thought to be due to photoelectrons scattered on the APD surface, which bounce into neighboring APD pads (back-scattering).
To study this effect in detail, the hit correlation between two pads is investigated. Two hits are required, with one detected in a particular pad and the other in any other pad. We then plotted the location of the second hit relative to the first (Fig. 8.15). From this plot it is evident that backscattered photoelectrons can generate signals on other pads within 40 mm, in good agreement with expectations. Note, however, that this effect is considerably reduced in a 1.5 T magnetic field (Fig. 8.15 right). The remaining spread is thought to be due to incident-light reflections.
Based on these studies, we conclude that the single photon sensitivity in the HAPD is not degraded and that a 1.5-T magnetic field is actually beneficial for sensor performance [13].

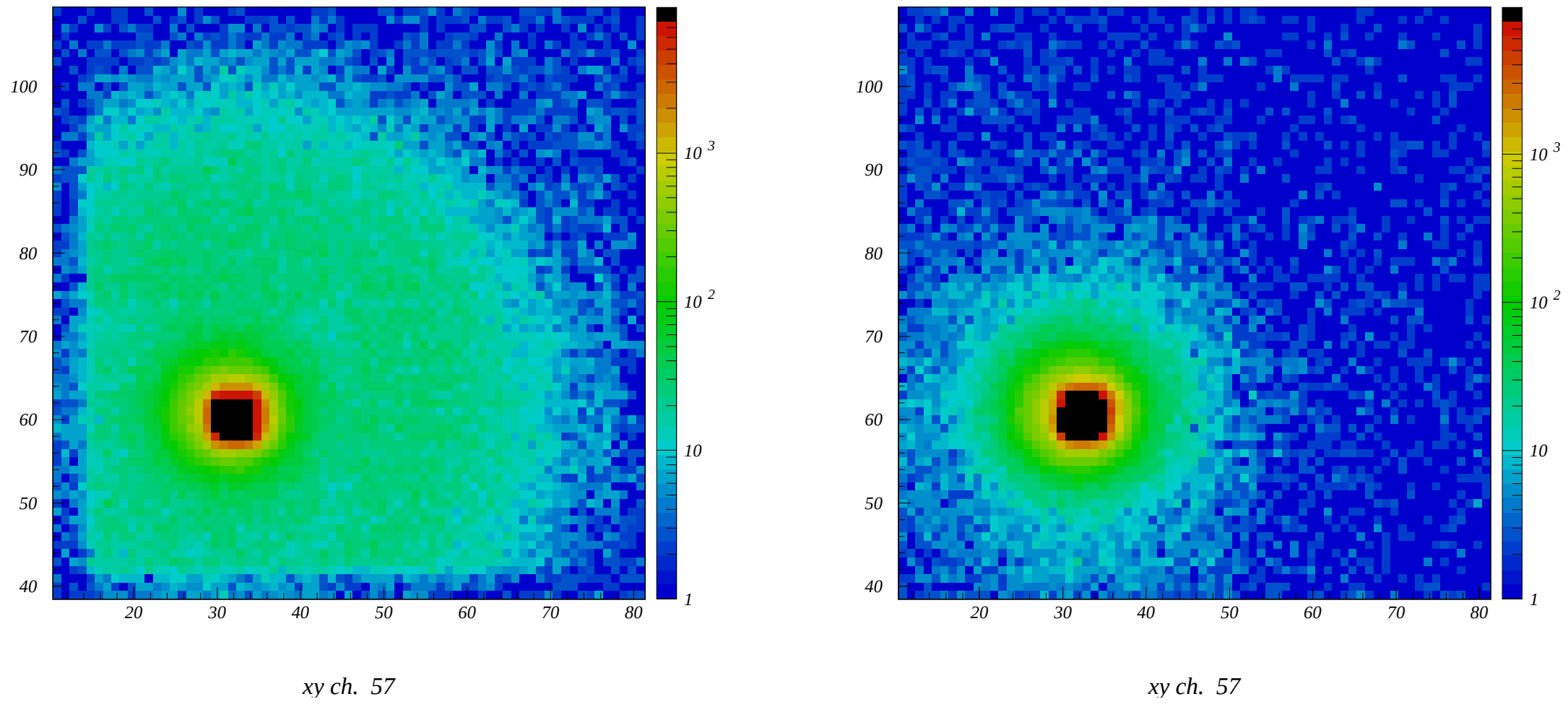

*Figure 8.15: Back-scattering electron effect observed with 0 T (left) and 1.5 T (right) magnetic fields.*

• **Aging test** Effects of aging of a HAPD have been studied in order to evaluate and possibly predict its performance under conditions of elevated irradiation by light inside the Belle II spectrometer.
The test apparatus (Fig. 8.16) consists of an HAPD and a reference photomultiplier tube (HPK R1463), both mounted on a computer-controlled, mobile XY stage. In standard operation, the entire HAPD surface is illuminated by 470-nm light from a light emitting diode (LED).
Measurements of the photon detection efficiency (PDE) and quantum efficiency (QE) have been performed at the beginning and the end of the aging process (Fig. 8.17). An additional PDE measurement was performed during aging after an irradiation corresponding to about nine years of operation in Belle II (corresponding to $3\times10^{12}$ detected photoelectrons per $cm^2$). For the PDE and QE measurements, we used tungsten and deuterium lamps, both with a monochromator (SP-2155 PI Acton) and a reference PMT. Measurements of QE and PDE show no significant degradation of performance after an accumulated irradiation corresponding to about 27 years of operation in Belle II ($\approx 10^{13}$ detected photoelectrons per $cm^2$).

• **Radiation damage** From a study of leakage current increase in the photodiodes used in the Belle ECL [14], severe neutron damage is expected in the Belle II environment. We estimate

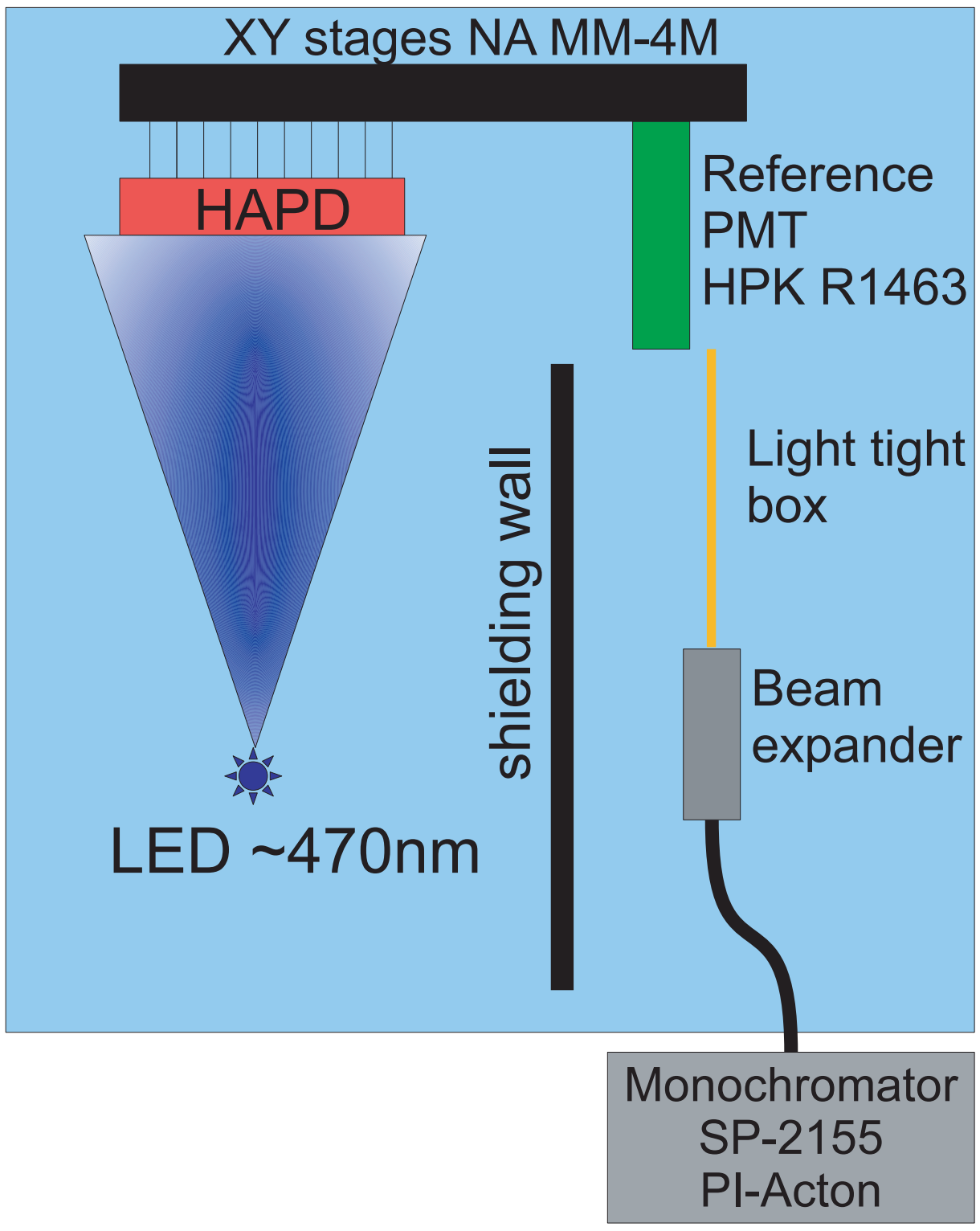

Figure 8.16: Schematic drawing of the aging test setup.

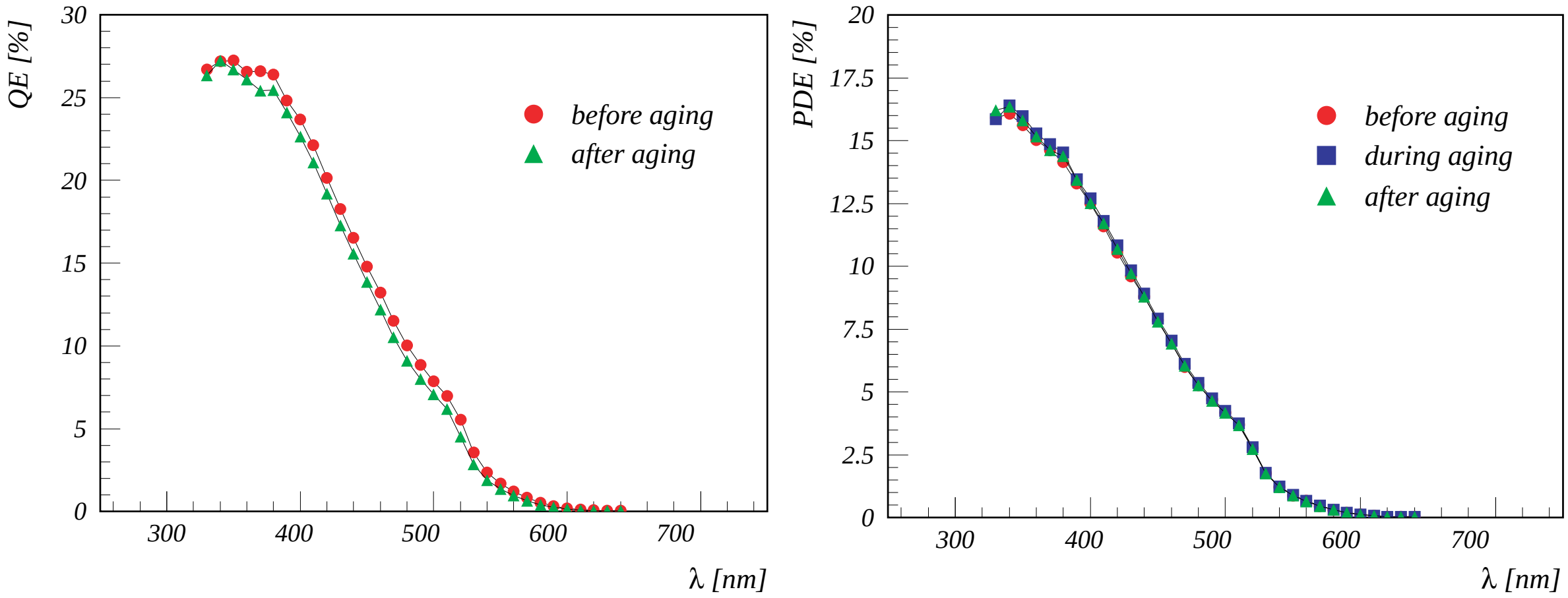

Figure 8.17: QE and PDE of the HAPD measured before, during and after the aging test.

that the neutron fluence will be $10^{11}$ neutrons/cm$^2$/year after the accelerator upgrade. To study the impact of neutrons on the photo-sensor performance, several HAPDs were irradiated with up to $5 \times 10^{11}$ neutrons/cm$^2$ in the "Yayoi" nuclear reactor facility at the University of Tokyo [15]. This dose corresponds to 5 years of operation in Belle II. Note that the average energy of the neutron flux in the Yayoi is 370 keV. To convert to a 1-MeV equivalent flux, an ELMA diode [16] was also irradiated in the same way for calibration. The fluence of $5 \times 10^{11}$ neutrons/cm$^2$ in the Yayoi facility was found to be equivalent to a fluence of $2 \times 10^{11}$ 1-MeV neutrons/cm$^2$. In this section, the neutron fluence with the Yayoi energy spectrum is employed unless explicitly stated otherwise.
The first comparison in QE before and after the neutron irradiation of $5 \times 10^{11}$ neutrons/cm$^2$

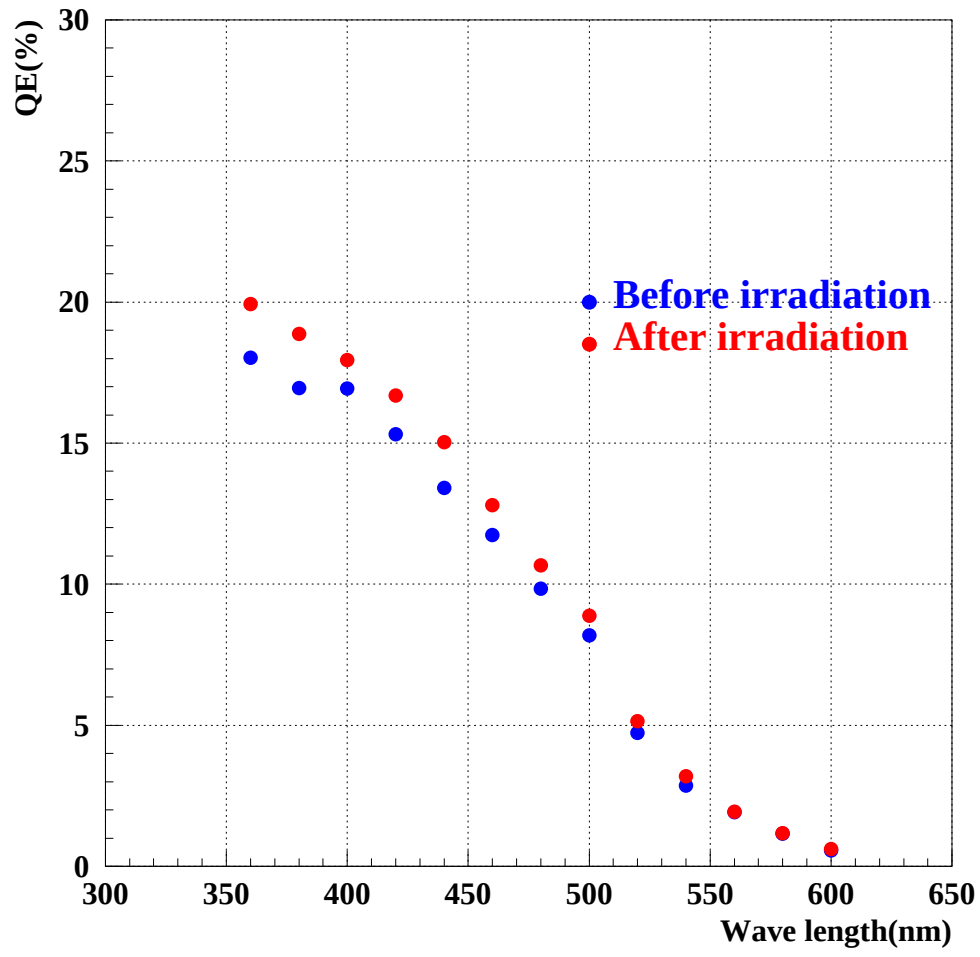

*Figure 8.18: Comparison in QE before and after exposure of $5{\times}10^{11}$ neutrons/cm$^2$.*

was made, and the results are plotted in Fig. 8.18. As can be seen in this figure, no degradation was observed. On the other hand, a significant increase in APD leakage current was measured (Fig. 8.19). As can been seen in this figure, the leakage current increases linearly as a function of neutron exposure, reaching more than 10 $\mu$A in one APD chip. Noise measurements showed a large width of the pedestal peak. In addition, the increase in leakage current caused a drop in actual bias voltage, resulting in an avalanche gain reduction. Because of these phenomena, the post-irradiation single-photon signal is barely separated from noise, as shown in Fig. 8.20 (left). The width of the noise level, $\sigma_{\text{noise}}$, is related to the readout electronics parameters as $\sigma_{\text{noise}} \propto \sqrt{I_{\text{leak}} \times T_{\text{shaping}}}$, where $I_{\text{leak}}$ and $T_{\text{shaping}}$ represent the leakage current and signal shaping time in the readout electronics system, respectively. Therefore, a shorter shaping time in signal processing allows us to compensate for the degradation due to noise.
The pulse height distribution plotted in Fig. 8.20 (left) shows a S/N ratio of $\sim 3$ after exposure to $5\times10^{11}$ neutrons/cm$^2$. However, with a considerably shorter shaping time of 20 ns in the readout system we could suppress the deterioration due to higher noise. Furthermore, by applying an increased HV of $-8.5$ kV, a S/N ratio of $\sim 7$ was achieved, and detection of a clear single photon signal was possible (Fig. 8.20 right).
In conclusion, by optimization of the shaping time parameter in the electronics readout, radiation damage up to $5 \times 10^{11}$ neutrons/cm$^2$ is manageable [13].
In the present ASIC readout, the shaping time is one of the adjustable parameters, and it can be varied in the range from 250 ns to 1 $\mu$s. A 1 $\mu$s shaping time has been used as the default set up, for instance in the beam test described later. It should be noted that excessively short shaping times may emphasize the intrinsic APD noise rather than reduce noise from radiation damage. From this viewpoint, further studies to optimize the shaping time should be done.
As an alternative approach, the possibility of suppressing the increase of leakage current after neutron irradiation is also investigated. The leakage current is proportional to the APD volume, so it can be reduced if a thinner APD layer can be produced. The first APD samples with a thinner layer structure were recently delivered and were irradiated at Yayoi in January 2010. Detailed tests are now in progress.

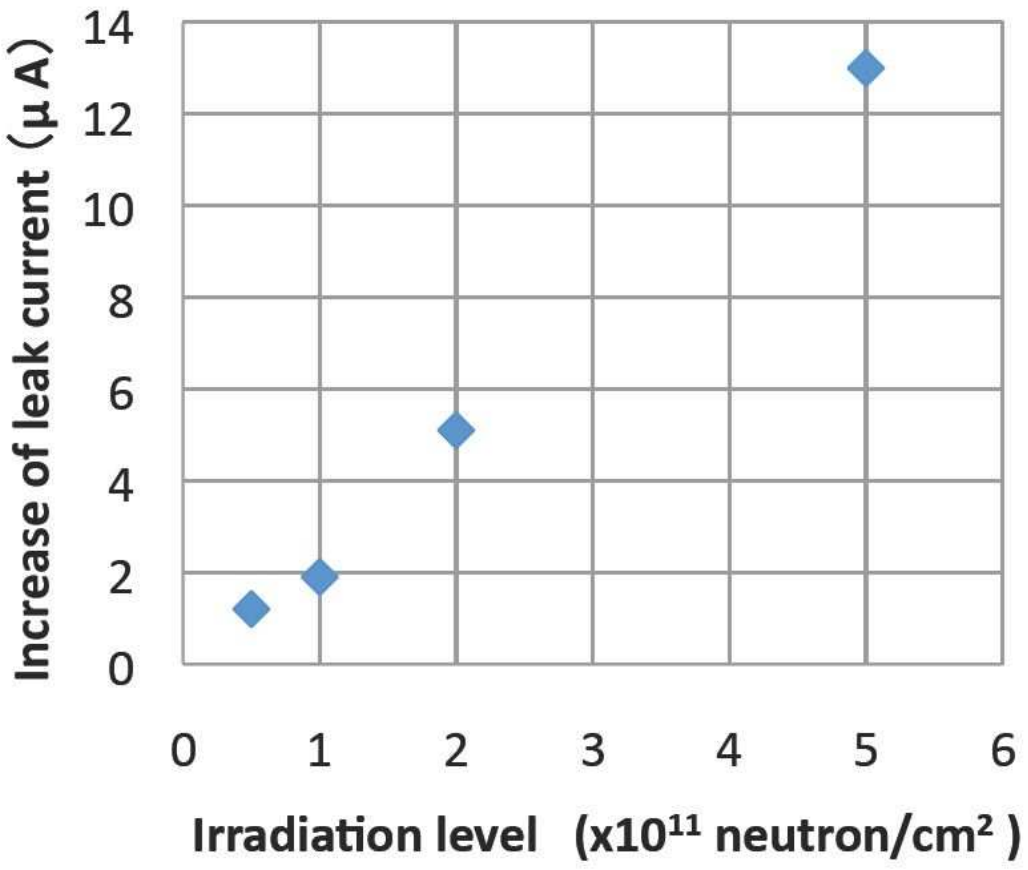

*Figure 8.19: Leakage current increase in one APD as a function of neutron dose.*

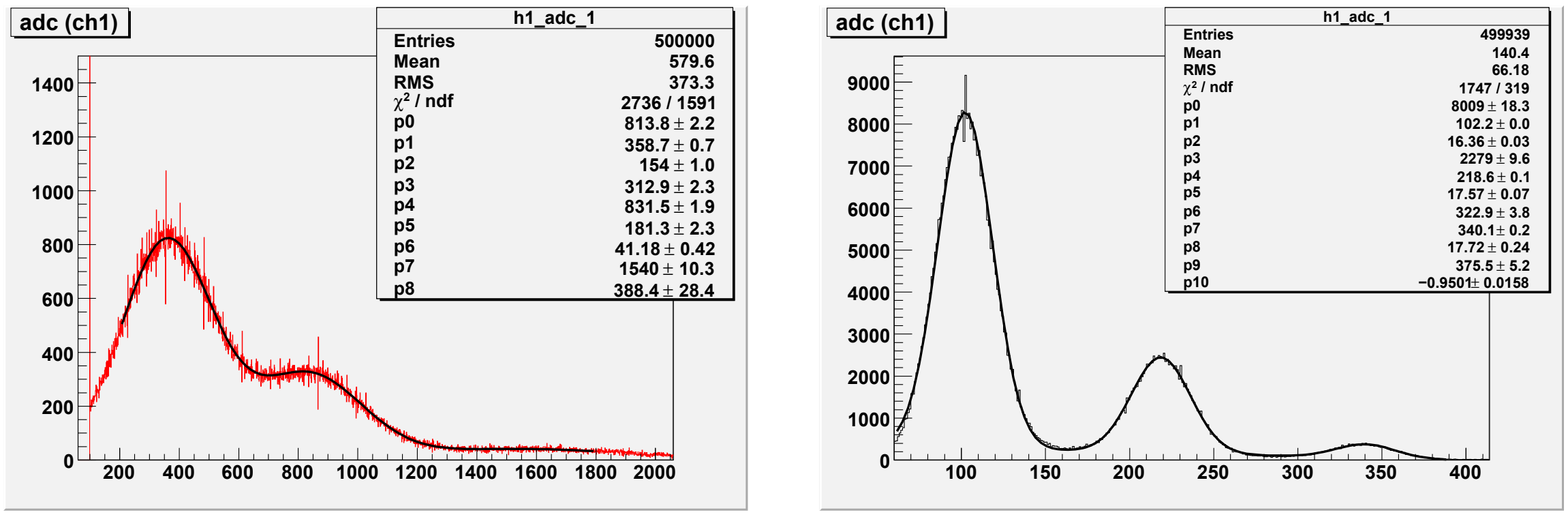

*Figure 8.20: Pulse height distributions of the HAPD after exposure to $5\times10^{11}$ neutrons/cm$^2$ using a 1 $\mu$s shaping time with HV of $-7$ kV (left) and a 20 ns shaping time with $-8.5$ kV (right).*

#### 8.5.2.2 Micro-channel plate photomultiplier tube (MCP-PMT)

As an alternative photo sensor, a micro-channel plate photomultiplier tube (MCP-PMT) has been studied. The characteristics of the sensor, Photonis/Burle 85112, are listed in Table 8.4. The tube has two micro-channel plates with 10-$\mu$m pores, and 64 (=8×8) anode pads at 6.5 mm pitch [17]. It has a bialkali photocathode on the inside of a 2-mm thick quartz window, separated by 6 mm from the microchannel plates.

| | |
|---|---|
| package size | $59 \times 59\,\mathrm{mm}^2$ |
| # of pixels | 64 ($8 \times 8$ array) |
| pixel size | $5.9 \times 5.9\,\mathrm{mm}^2$ @ 6.5 mm pitch |
| effective area | 80% |
| typical peak QE | ≈25% |
| gain | $\approx 0.6 \times 10^6$ |
| weight | 130 g |

*Table 8.4: MCP-PMT specifications.*

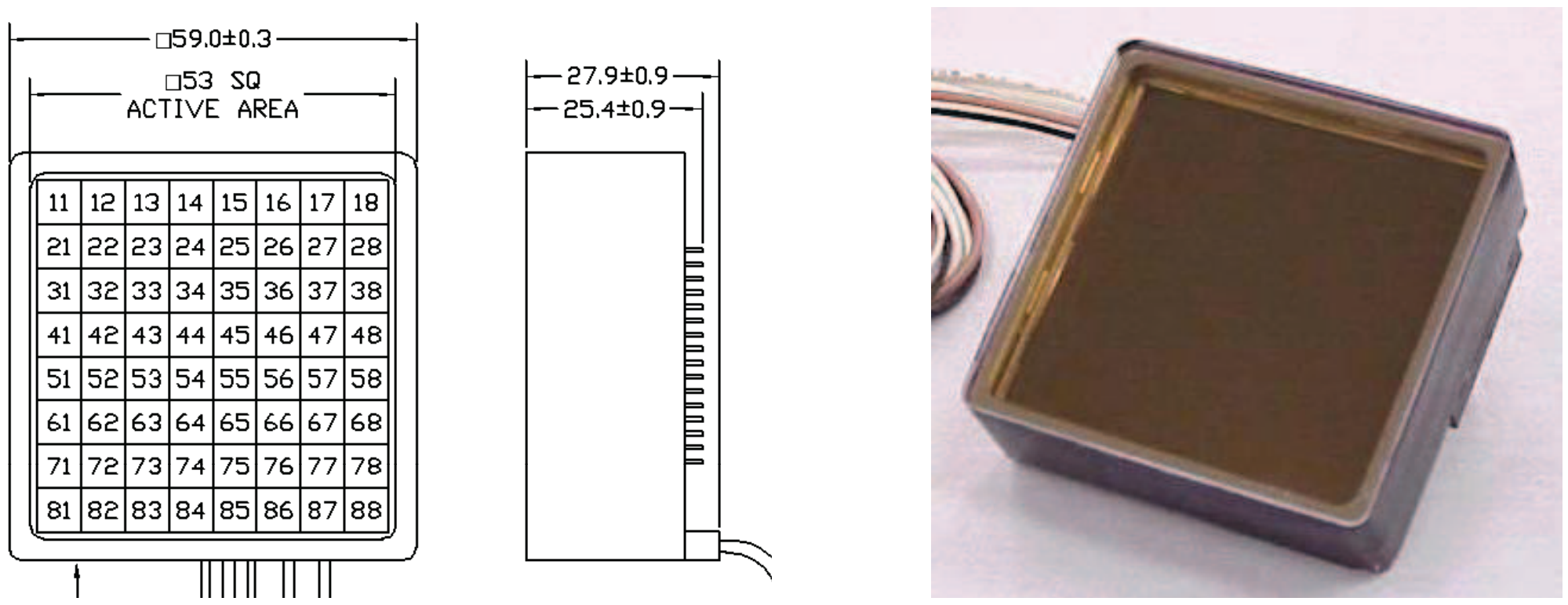

*Figure 8.21: Schematic drawing and photograph of the MCP-PMT.*

Single photon detection

Extensive tests have been performed on the bench to test the position and timing resolution for single photons [18]. A picosecond laser (a fast laser with picosecond timing resolution) was used as a light source. Light from the laser was focused onto the surface of the MCP-PMT to form a spot of $\approx 10\,\mu$m diameter. The light spot could be positioned at a desired location by a 3D moving stage to investigate the uniformity of response. A timing resolution better than 40 ps RMS can be obtained for prompt signals, as shown in Fig. 8.22. Effects of charge sharing (Fig. 8.22) and photoelectron backscattering, which can affect the position resolution for single photon detection, were also observed. Due to charge sharing, double counting appears in a ≈2 mm wide region at the pad boundaries. Photoelectron backscattering has a longer range of about 1 cm. Both effects are expected to be strongly reduced when the MCP-PMT is operated in a magnetic field.

• **Operation in magnetic field** An advantage of the MCP-PMT over conventional PMTs is its ability to operate in a high magnetic field. Operation up to 1.5 T is possible if the diameter of pores is 10 $\mu$m or less [19]. Tests of a sample with 10-$\mu$m pores were performed in the magnetic field using the same setup as for the tests of the HAPD. The variation of the gain as a function of the magnetic field up to 1.5 T is shown in Fig. 8.23.

To achieve the same gain and the same single photon detection efficiency at 1.5 T as with no magnetic field, the high voltage needs to be increased by about 200 V (Fig. 8.23). A position sensitivity scan was also performed at 1.5 T to assess the effect of magnetic field on charge sharing and photoelectron backscattering. From the result of the scan (Fig. 8.24), we conclude that the range of both effects is strongly limited and that the position resolution in a magnetic field will improve.

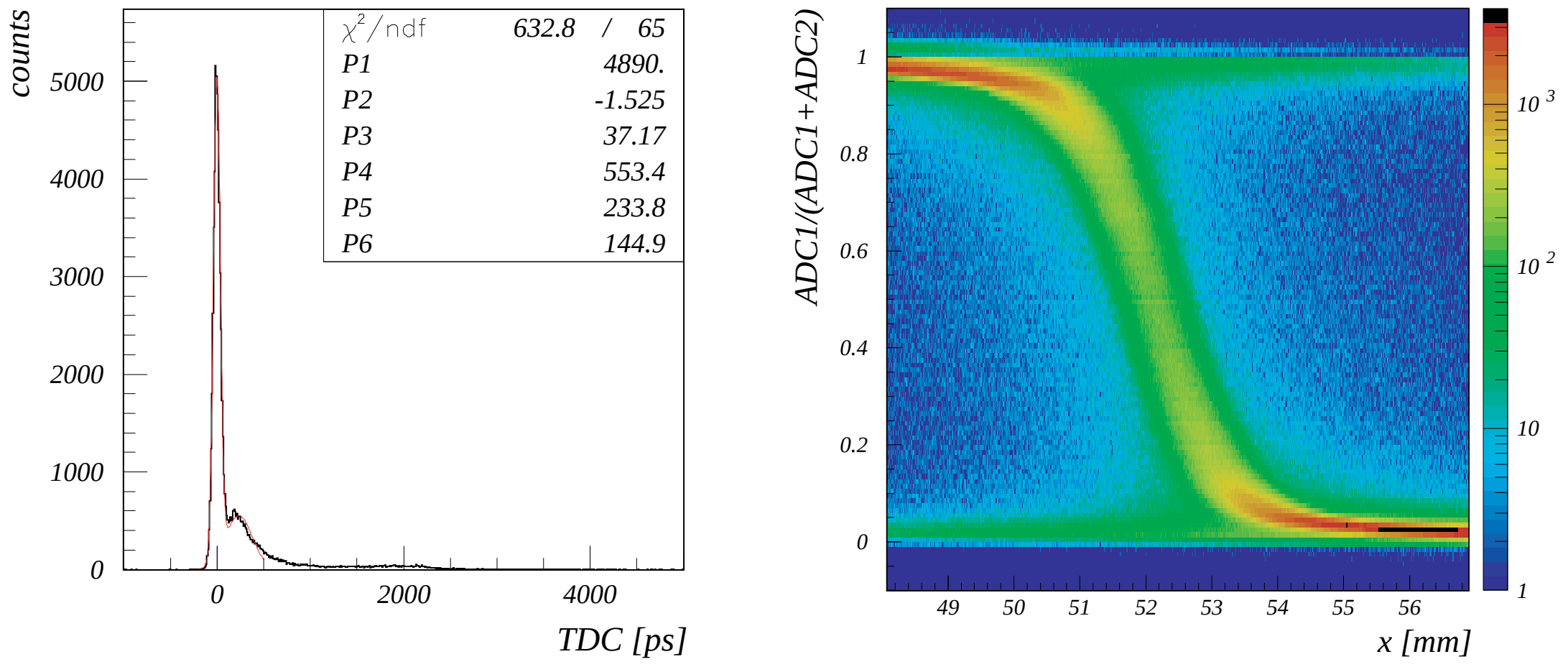

Figure 8.22: Left: timing distribution measured for single photons. Right: position dependence of the signal fraction appearing on the left of two pads, with the boundary between the pads located at $x = 52$ mm.

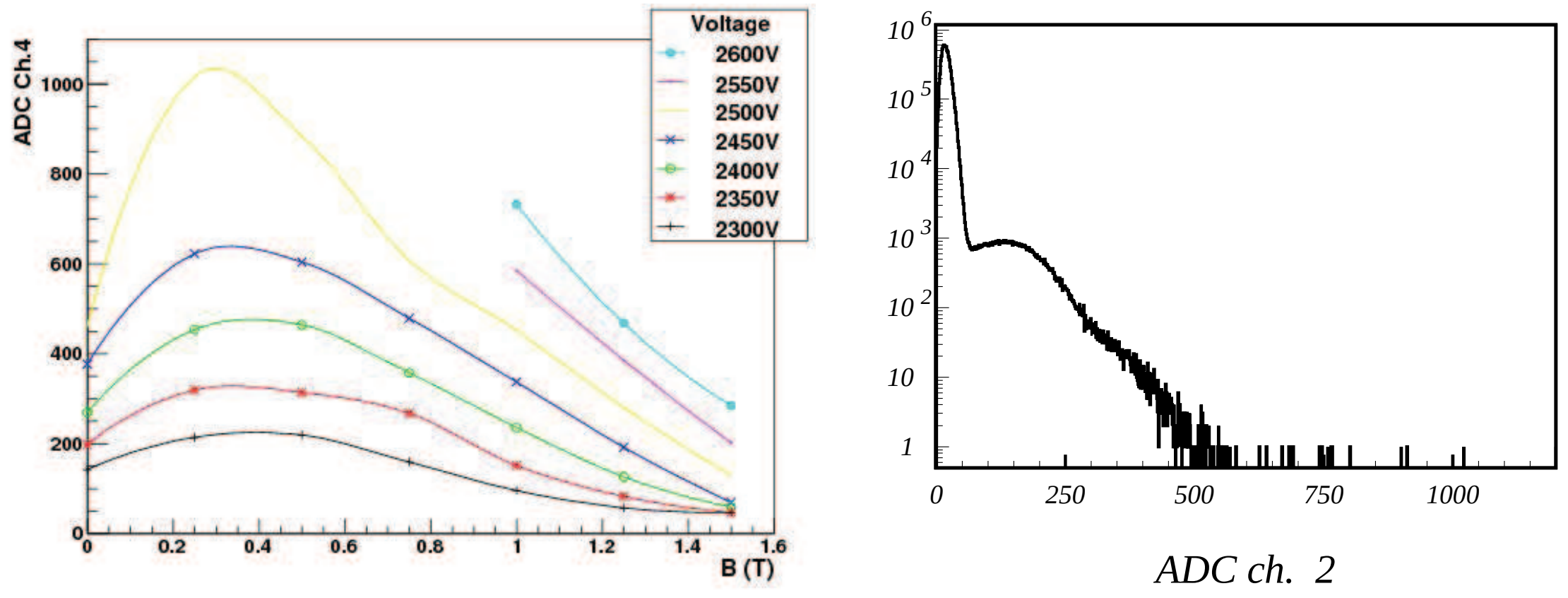

Figure 8.23: Left: variation of the single-photoelectron pulse height as a function of the magnetic field. Right: pulse height distribution for single photons at 1.5 T and HV of 2500 V.

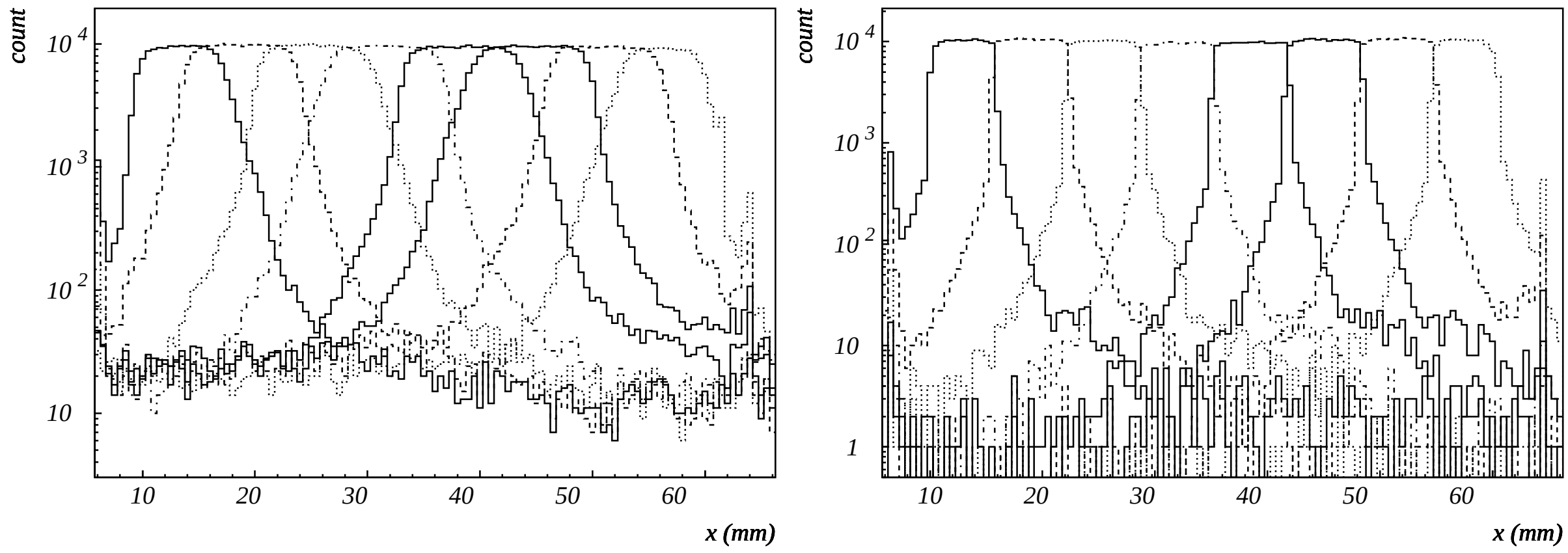

Figure 8.24: Total count rate of one row of channels measured without magnetic field (left) and with a magnetic field of 1.5 T (right).

- **Aging test** To check the stability of the photon detection efficiency during the lifetime of the experiment, an aging test was performed in a similar way as with the HAPD. The total accumulated anode charge during the aging test was ≈400 mC/cm$^2$, after which we observed a drop in PDE of the order of 10%. Since such a value of accumulated charge corresponds to ten years of operation of Belle II, the observed drop in photon detection efficiency is acceptable.

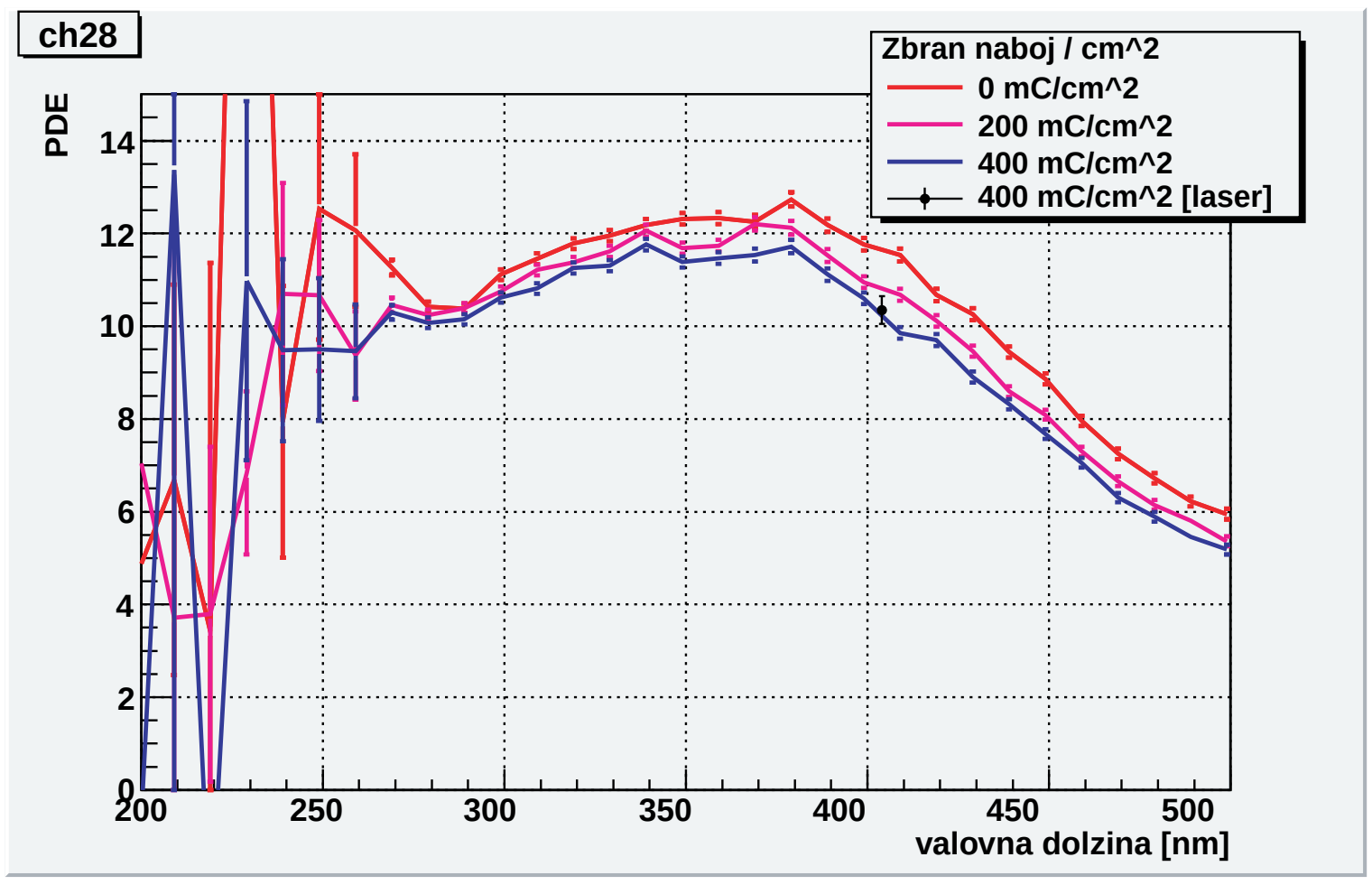

*Figure 8.25: Measured PDE before the MCP-PMT aging and after 200 mC/cm$^2$ and 400 mC/cm$^2$, respectively.*

<u>Beam test</u>
In a series of beam tests between 2004 and November 2009 [20], we measured the performance of the Photonis multichannel-plate PMTs. The experimental arrangement was similar to the one used for the HAPD, as described in section 8.5.4, except that a single MCP PMT was used in the tests.
From the distribution of hits and clusters with respect to the corresponding Cherenkov angle, we obtain the resolution ($\sigma$ of the fitted Gaussian) as well as the average number of hits (pads) or clusters per incident charged particle (Fig.8.26). The number of detected Cherenkov photons is obtained by normalising a Poissonian distribution at zero hits and, as expected, is found to be in between the number of clusters and the number of hits. For the present MCP PMT, with a geometrical acceptance of 14% (for a single MCP PMT), we detected 1.4 clusters/track, corresponding to 3.1 Cherenkov photons/track. For full coverage of the photon detector plane with MCP PMTs, the acceptance would be 80% and one extrapolates to 8 clusters/ring, or to about 17 detected Cherenkov photons per ring. The angular resolution for single Cherenkov photons as obtained from the measurement is 14.8 mrad.
As Cherenkov photons are emitted promptly with respect to the charged particle traversing the radiator, they may be used for timing purposes. The time distribution of Cherenkov photons emitted in the MCP-PMT window was measured relative to a start signal provided by another MCP-PMT with a quartz radiator [21]. The standard deviation of the distribution for single photons from the aerogel is found to be 50 ps. For about 10 detected photons, one may therefore expect $\sigma \sim 20$ ps. Since a large number of Cherenkov photons ($\approx 20$) are emitted in the MCP-PMT entry window (Fig.8.27), a TOF measurement is possible even for particles below Cherenkov threshold in aerogel. The timing capability of the apparatus is illustrated by the measured time distribution of Cherenkov light generated by beam particles in the glass window of the MCP PMT. One clearly separates pions from protons at 2 GeV/c, even for the very short flight path (65 cm) of the present measurement.

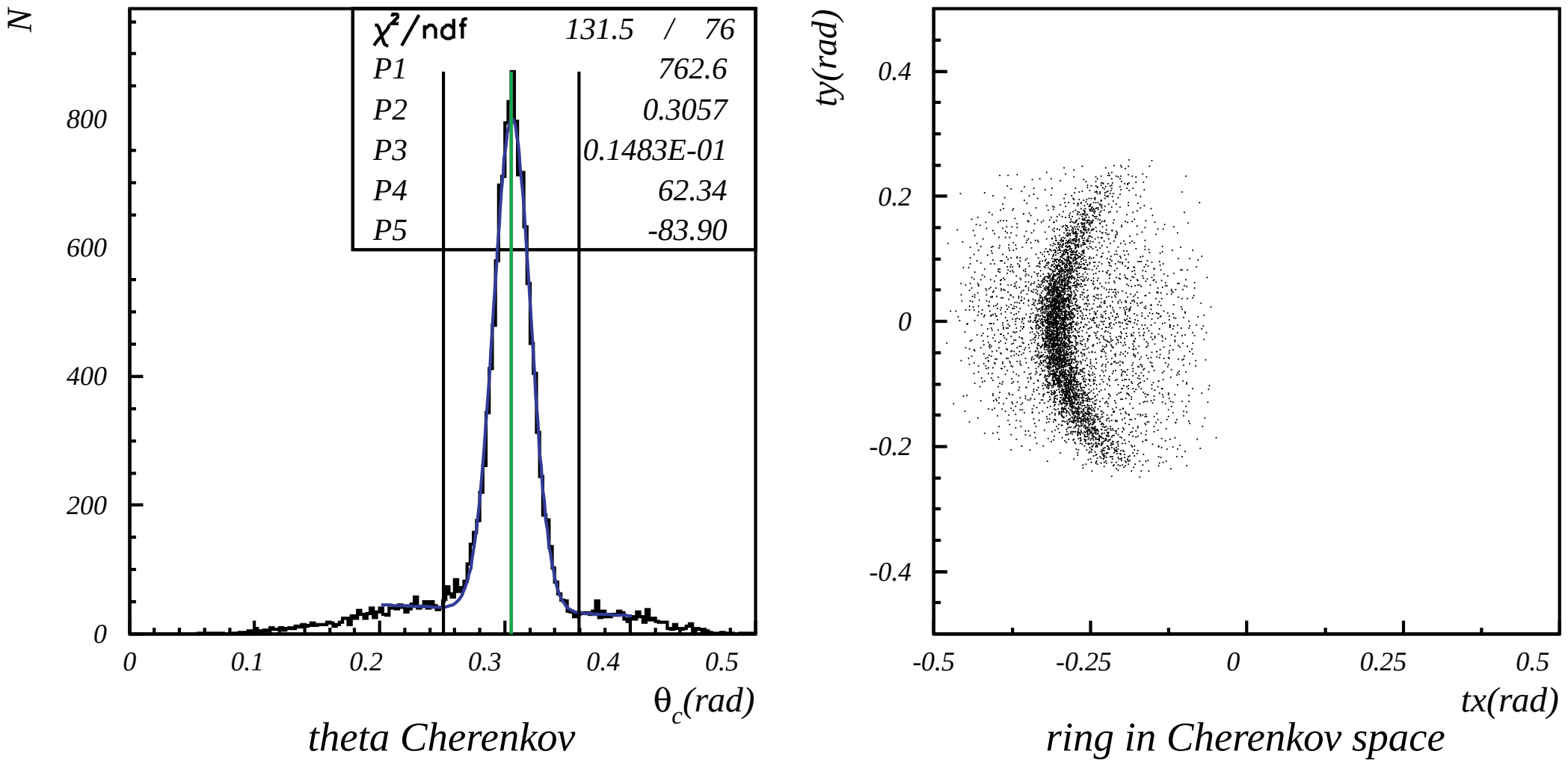

Figure 8.26: Beam test of the MCP PMT as a detector of Cherenkov photons. Left: the distribution of Cherenkov angles corresponding to center of gravity of hits in clusters as recorded for 4 GeV/c pions. Right: accumulated hits in Cherenkov space.

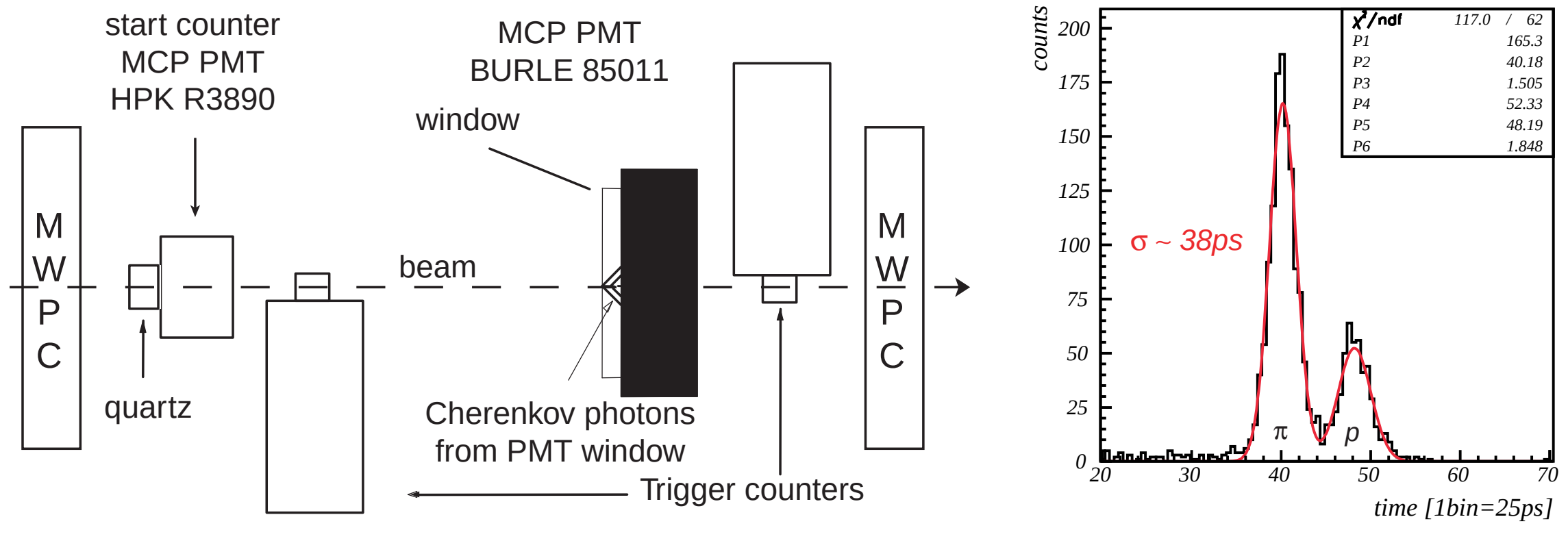

Figure 8.27: Beam test of the TOF capabilities of a RICH counter with MCP-PMTs as photo-sensor. Left: The experimental set-up; Cherenkov photons emitted in the window of the MCP-PMT are used for measuring the time of arrival. The start signal is provided by another MCP-PMT. Right: the distribution of the time of flight as measured for the Cherenkov photons from the MCP-PMT window, shown for 2 GeV/c pions and protons.

### 8.5.3 Read-out Development

We first produced several versions of ASICs for prototyping at the VLSI Design and Education Center (VDEC) of the University of Tokyo using the ROHM CMOS 0.35 $\mu$m process [22]. The ASIC consists of a charge-sensitive preamplifier, a shaper and a variable gain amplifier (VGA), followed by a comparator for the digitization of analog signals to on/off hit information, and a digital shift register for pipelined readout. Each chip has 18 channels, so we need eight chips to read out one HAPD. The threshold voltage for the comparator is common to all channels, but the offset voltage can be adjusted channel-by-channel. Therefore, we can effectively vary the threshold voltage for each channel.

The target noise level is 1200 $e^-$ at 80 pF, which corresponds to a signal-to-noise ratio of 10, assuming an HAPD gain of 12000. After four iterations of test production since 2003, the noise level of the final evaluation ASIC is measured to be around 1900$e^-$, which is higher than the target value of 1200$e^-$ but is still small enough for the detection of single photons with an HAPD. We have confirmed that noise in all HAPD channels can be suppressed by adjusting the offset values. We then performed a beam test of a prototype aerogel RICH in 2008. Here, we read out six HAPDs using 48 prototype ASICs, and succeeded in observing a clear Cherenkov ring, as mentioned in Sec. 8.5.4.

The remaining issues revolve around compatibility with the Belle II central DAQ scheme and the management of all the readout channels. To meet these requirements, we have started developing a new second-generation ASIC, which consists of a preamplifier, shaper and comparator, in which the digital readout elements are excluded from the ASIC; instead, the hit information is passed through output pins, and is subsequently manipulated by an external FPGA. The amplification factor and the shaping time are variable, and the offset voltage for each channel is adjustable as well. They can be controlled by setting external parameters. The first version ASIC of this type, called SA01, was produced at MOSIS with the TSMC CMOS 0.35 $\mu$m process. This SA01 chip handles only 12 channels since the primary purpose for this iteration is just to check basic performance.

Figure 8.28 shows the dependence of the measured noise level in the SA01 on input capacitance. The noise level at 80 pF is around 1200$e^-$, which is better than the ASICs in the previous generation.

Figure 8.29 shows the signal pulse height measured before the comparator as a function of input charge with four different gain settings. We find good linearity when the input charge is small, but saturation is observed when the input charge exceeds 60,000 electrons, even when running at the lowest gain. This is because the HAPD gain is assumed to be 12000 at the design stage, while recent HAPD samples exhibit excellent performance with a gain of more than 40000. This mismatch, related to the input signal saturation, is fixed in the SA02 by reducing the preamplifier gain to one-fourth. In the beam test in 2009, we succeeded in reading out HAPD signal with SA01. Details of the beam test are discussed in Sec. 8.5.4.

Based on feedback from the SA01 results, a second iteration—the SA02—was made with the same process at TSMC. Aimed at higher density, the SA02 chip is capable of 36 channels, and an appropriate gain is designed. The SA02 is packaged in a 160-pin QFP for the first time.

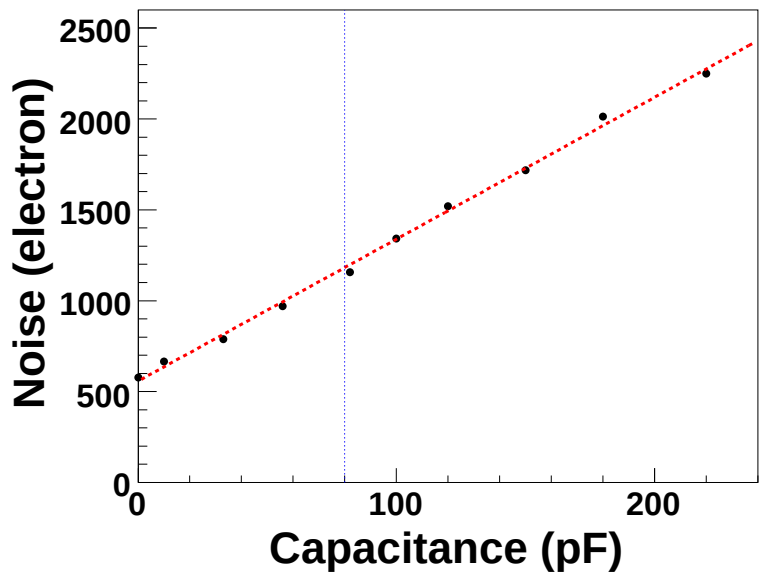

*Figure 8.28: The dependence of the SA01 noise level on detector capacitance, where the HAPD corresponds to 80 pF capacitance.*

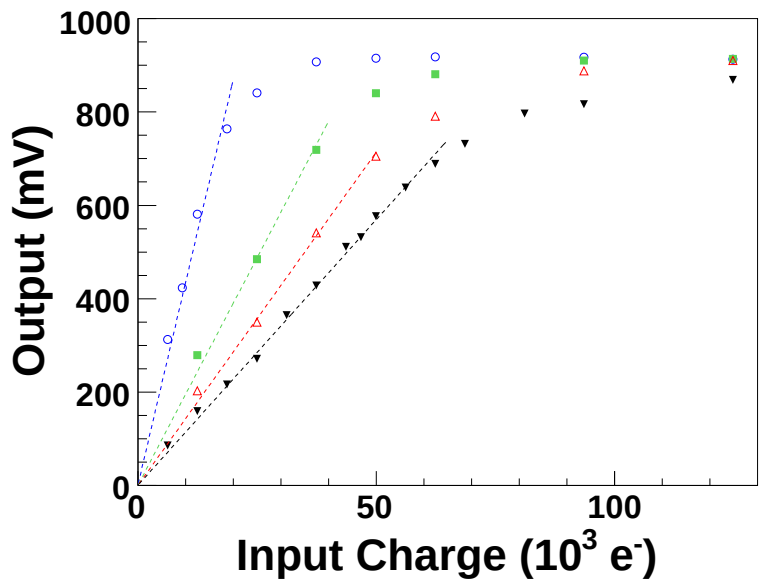

*Figure 8.29: Measured linearity of SA01. The horizontal axis is the input charge from the test pulse, while the vertical axis indicates the pulse height before the comparator. Measurements with four different gain settings are shown.*

However, the size of this package, around 30 mm × 30 mm, is too large for our application: we need a smaller package to make the front-end electronics compact. One of the candidate technologies is Low-Temperature Co-fired multi-layer Ceramic (LTCC) substrates. The size of an LTCC substrates that we produced with SA02 is 1.3 mm × 1.3 mm (Fig. 8.30). The performance of this LTCC substrate will be tested.

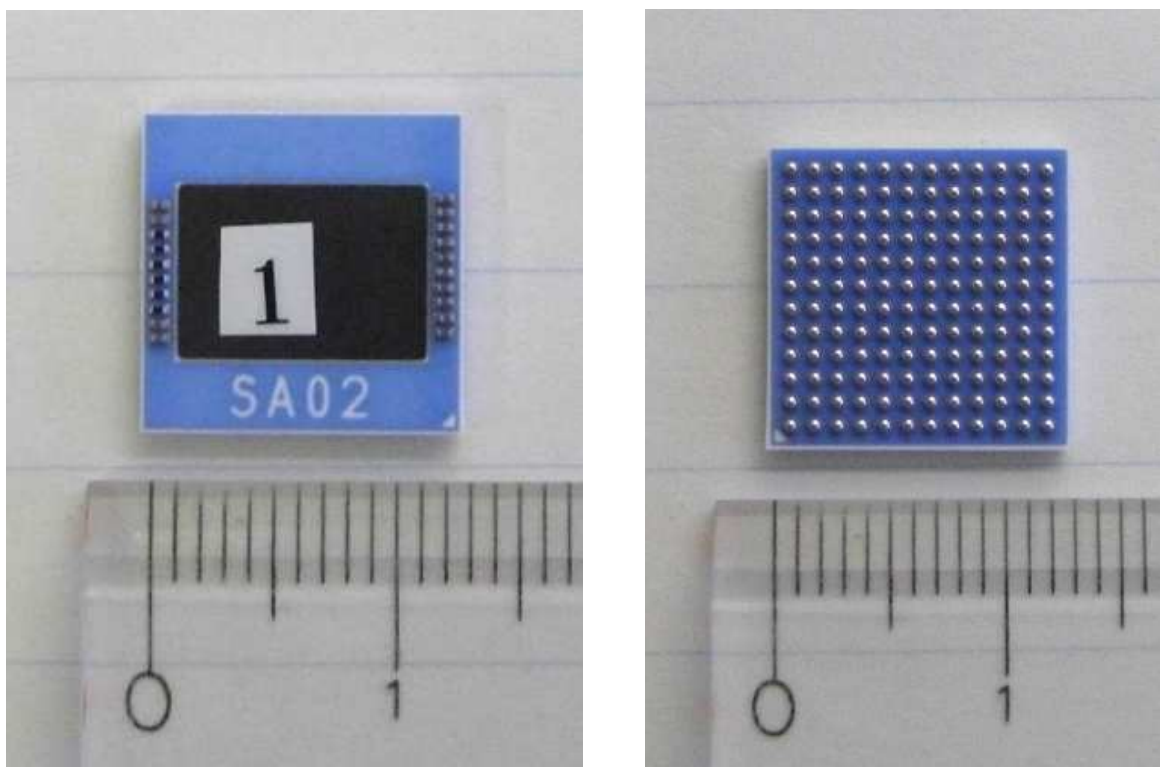

*Figure 8.30: SA02 chip with LTCC substrates (left: front view, right: rear view).*

We then need to design readout electronics that would fit in the small space behind the HAPD. ASICs and the FPGA will be placed on a readout board attached to the rear of the HAPD. We may put another patch board, connected to several readout boards, to collect the data and send it to the electronics hut via an optical fiber.

### 8.5.4 Test Beam Experiment

#### 8.5.4.1 Experimental Setup

The first beam tests of a prototype RICH counter with HAPDs, equipped with the ASIC readout system, were performed in March and June 2008. We used the Fuji test beam line at KEK, where a 2.0 GeV/$c$ electron beam was available. In these beam tests, we successfully observed a Cherenkov ring image for the first time with HAPDs and ASICs. In November 2009, we carried out another more realistic test beam experiment with an improved prototype RICH counter that incorporated the aerogel radiators produced with the "pin-hole drying" method, new HAPDs with the "Super Bialkali" photocathode, and SA01 ASICs. The setup of the prototype RICH counter is shown in Fig. 8.31.

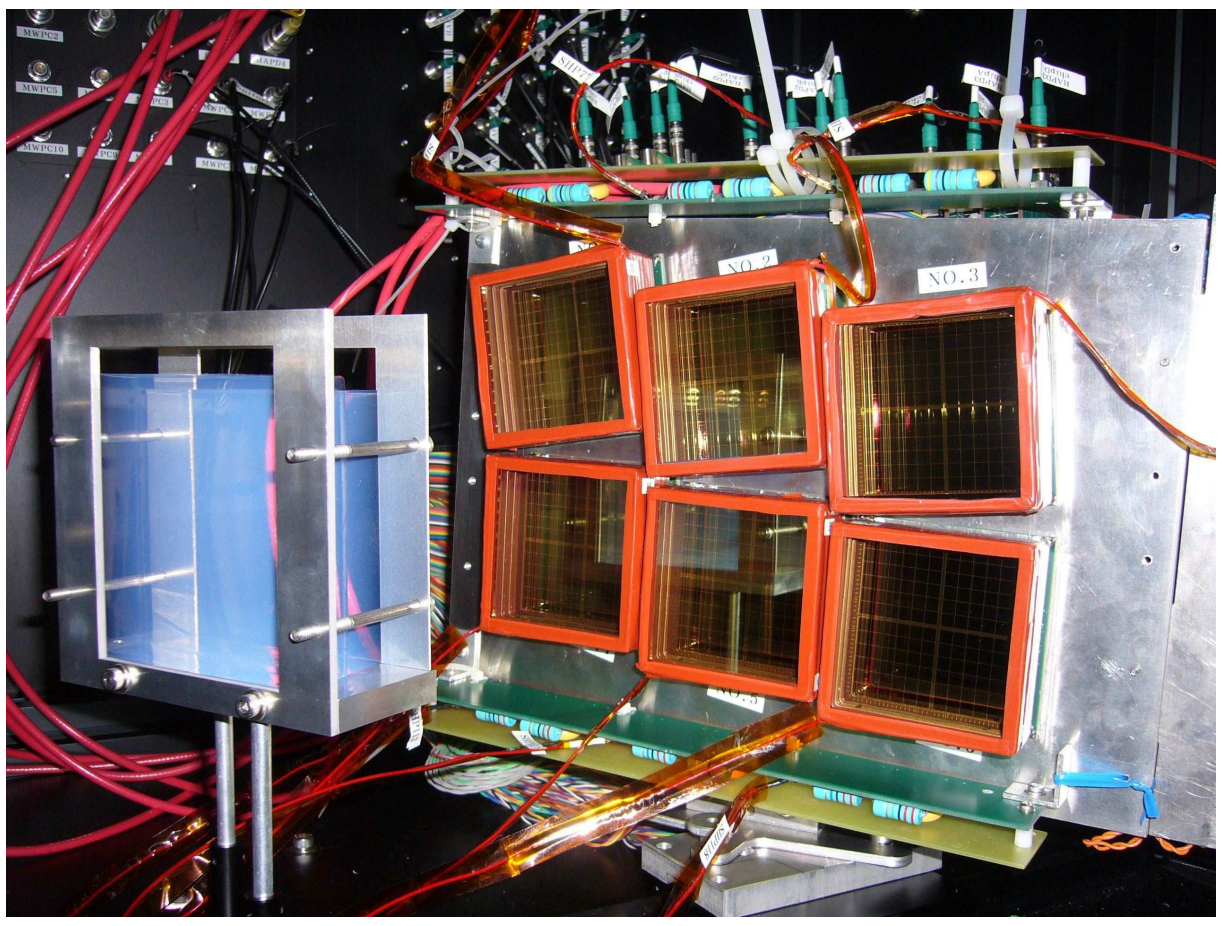

*Figure 8.31: Prototype RICH counter used in the test beam experiment in November 2009.*

Six HAPDs were arranged in a $2 \times 3$ array, as illustrated in Fig. 8.32. The HAPD geometry was taken from the innermost two layers of one sector of the final photon detector sextant. Table 8.5 summarizes the HAPD QE used in this beam test. The average QE taking into account the geometrical acceptance and the distribution of tracks on the photon detector is 24% at $\lambda = 400$ nm.
The front-end electronics system, equipped with the ASIC chips, was directly attached to the back plane of the HAPD. The upper three HAPDs are read out by 36 chips of the ASIC SA01, and the bottom three are read out by the ASIC chips of the old version used in the 2008 beam test for the purpose of comparison. The aerogel radiator was located 200 mm from the surface of the HAPD, which is the same configuration as in the final detector.
The track parameters were determined by two multi-wire proportional chambers (MWPCs) placed upstream and downstream of the prototype counter. The size of the sensitive area of the

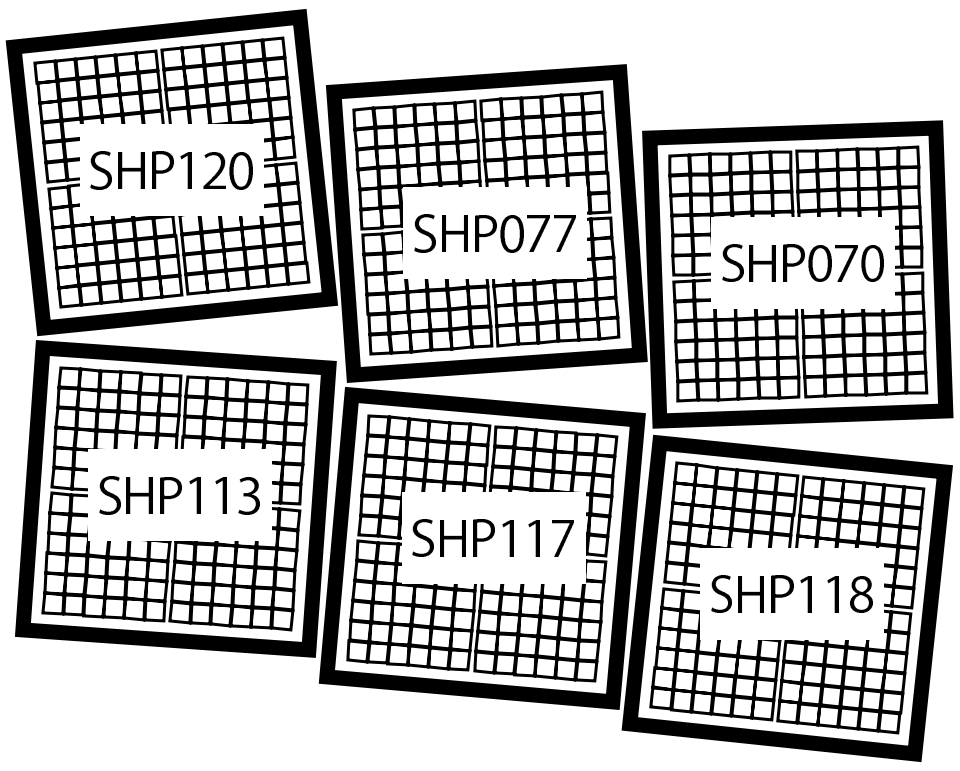

*Figure 8.32: HAPDs of the prototype RICH counter.*

| serial # | QE (%) at $\lambda = 400$ nm |
|---|---|
| SHP120 | 20.5 |
| SHP077 | 22.0 |
| SHP070 | 25.0 |
| SHP113 | 21.1 |
| SHP117 | 30.1 |
| SHP118 | 27.4 |

*Table 8.5: Summary of HAPD QE (at 400 nm) used in the beam test.*

MWPC is about $5 \times 5$ cm. Figure 8.33 shows the distribution of the measured track position on the radiator plane with the HAPD positions overlayed.

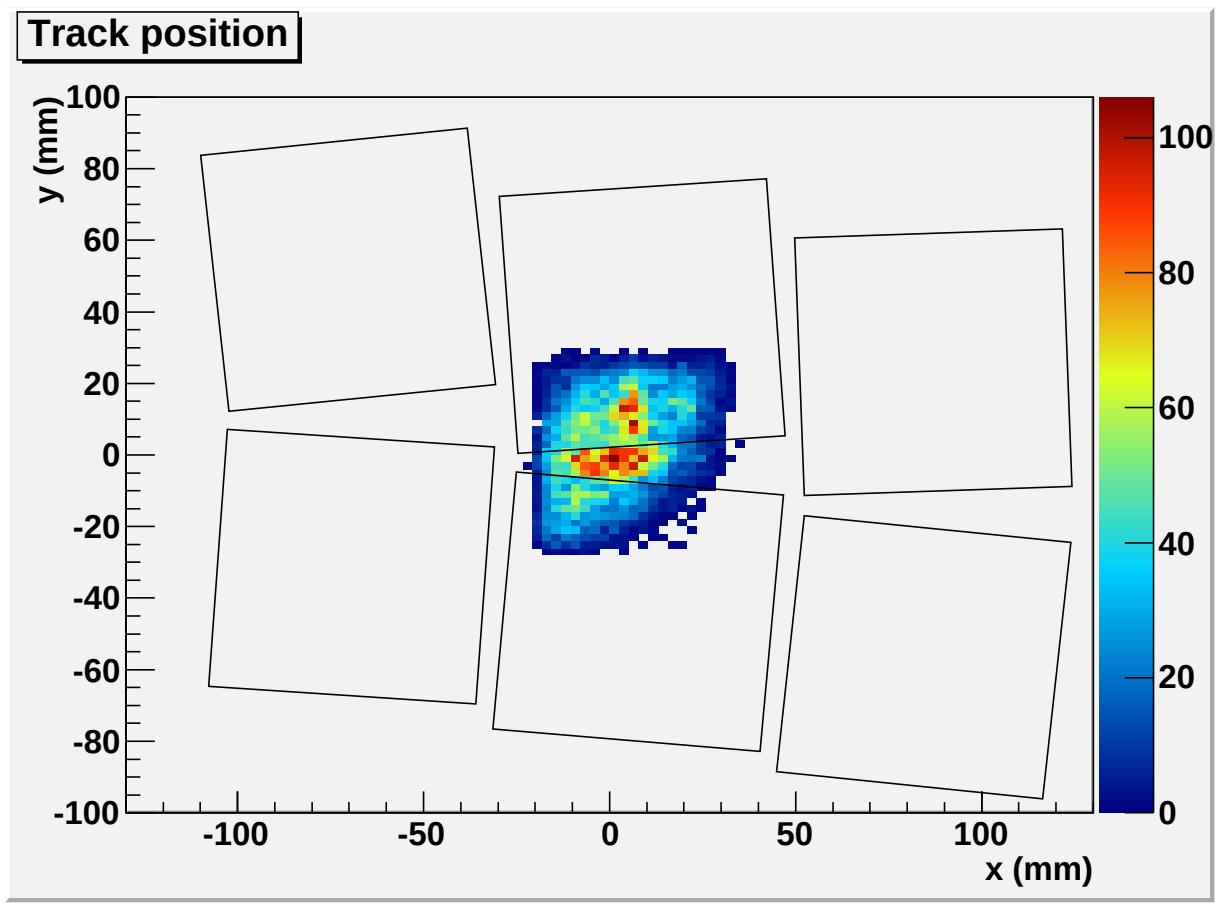

*Figure 8.33: Track position on the radiator plane. Boxes indicate the HAPD positions.*

#### 8.5.4.2 Photoelectron yield and Cherenkov angle resolution

In this beam test, two aerogel radiator layers of $n = 1.054$ and 1.065, each 20 mm thick, were used to produce a Cherenkov ring image on the HAPD window plane. Their transmission lengths at $\lambda = 400$ nm are 47.8 mm ($n = 1.054$) and 55.2 mm ($n = 1.065$). Figure 8.34 shows the accumulated hit image and the Cherenkov angle distribution for the tracks passing through the center of the prototype RICH counter ($-10 < x < 10$ mm and $-10 < y < 5$ mm) perpendicular to the radiator and to the photon detector plane. By fitting this distribution with a gaussian, the single photon resolution was measured to be 13.5 mrad. The photoelectron yield ($N_{pe}$) was calculated by integrating the fitted function over a $3\sigma$ area; a one dimensional polynomial was assumed for the background contribution. We obtained $N_{pe} = 15.3$ with this configuration, and a simple calculation indicates that a $6.7\sigma$ $\pi/K$ separation can be achieved at a momentum of 4 GeV/$c$ . The background contribution was found to be about two photoelectrons, which is sufficiently small. Only a few HAPDs used in the experiment were equipped with a "Super Bialkali" photocathode. In HAPD mass production, the QE of "Super Bialkali" HAPDs is expected to be improved further to an average peak value of 30%, with a resulting $\sim 30\%$ increase in photoelectron yield.
The expected number of photoelectrons can be calculated from the radiator properties, the

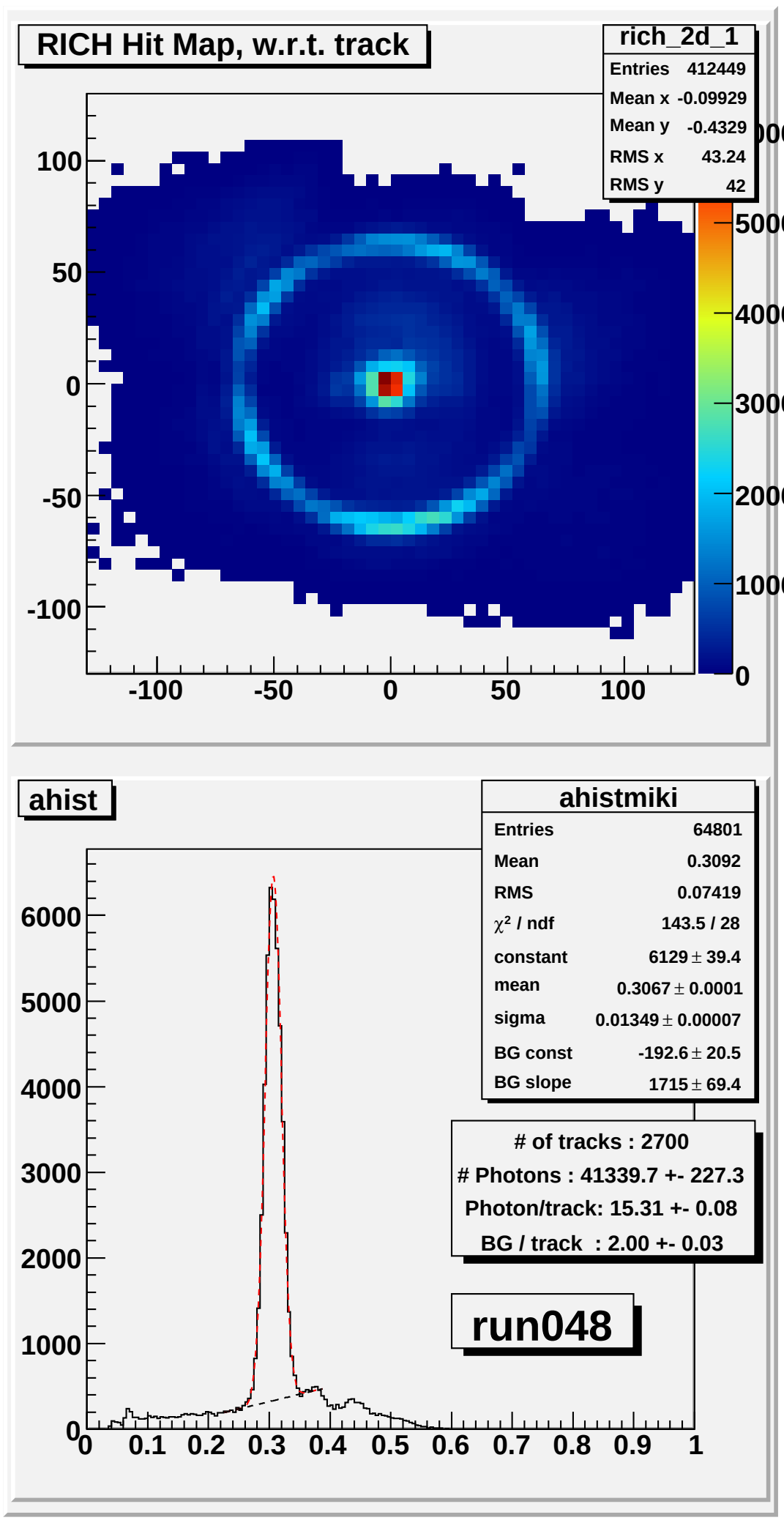

*Figure 8.34: Cherenkov image and Cherenkov angle distribution.*

HAPD QE, and the acceptance of the prototype RICH counter as

$$N_{phot} = N_1 + N_2, \tag{8.1}$$

$$N_1 = 2\pi\alpha \sin^2\theta_{c1}\,\epsilon_a \int \exp\left(-\frac{d_2}{\Lambda_2(\lambda)\cos\theta_{c1}}\right) \times\Lambda_1(\lambda)\cos\theta_{c1}\left(1-\exp\left(-\frac{d_1}{\Lambda_1(\lambda)\cos\theta_c 1}\right)\right)\epsilon_q(\lambda)\,\lambda^{-2}d\lambda, \tag{8.2}$$

$$N_2 = 2\pi\alpha \sin^2\theta_{c2}\,\epsilon_a \int \Lambda_2(\lambda)\cos\theta_{c1}\left(1-\exp\left(-\frac{d_2}{\Lambda_2(\lambda)\cos\theta_{c2}}\right)\right)\epsilon_q(\lambda)\,\lambda^{-2}d\lambda, \tag{8.3}$$

where the indices 1 and 2 indicate the upstream and the downstream aerogel, respectively, $\lambda$ is the wavelength of the Cherenkov photon, $\alpha$ is the fine structure constant, $\theta_c$ is the Cherenkov angle, $\Lambda(\lambda)$ is the transmission length of the aerogel, $\epsilon_q$ is the QE of HAPDs, and $\epsilon_a$ is the acceptance of the setup including the geometrical acceptance and other effects. The acceptance of the setup is estimated to be 40%, taking into account the APD sensitive region for the ring image (61%), the fraction of events without electron backscattering on the APD surface (80%), the counting loss of photoelectrons hitting the same channel (90%), and the fraction of active channels (96%). The expected numbers of photoelectrons as a function of the incident photon wavelength are shown in Fig. 8.35 for the case of 100% acceptance. They are $N_1 = 5$, $N_2 = 12$, so that $N_{phot} = 17$, which is within 10% of the measured result.

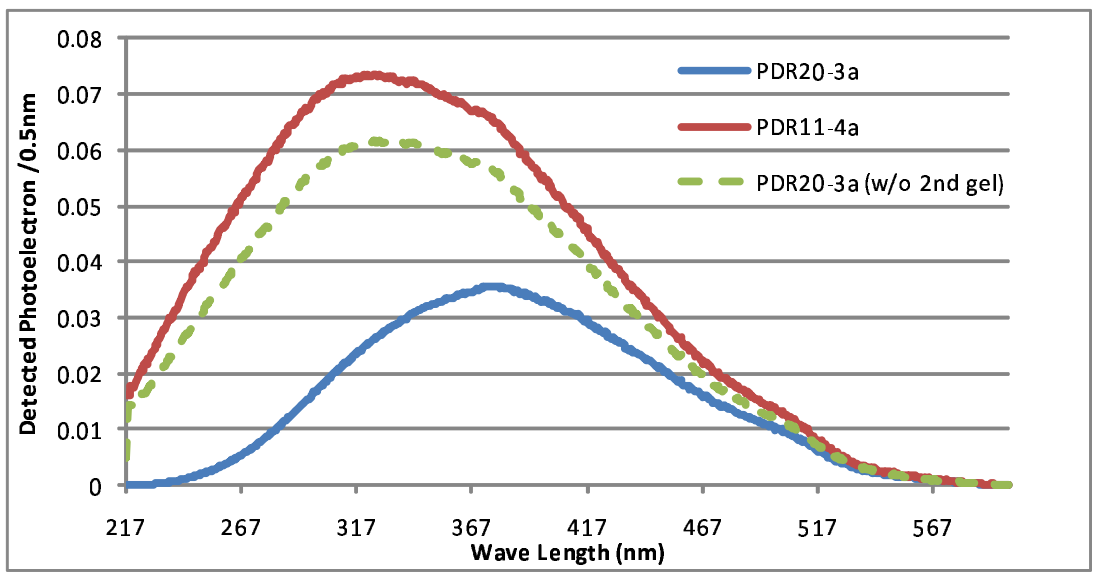

*Figure 8.35: Expected number of photoelectrons depending on the incident photon wavelength for the case of 100% acceptance. The upstream aerogel is PDR20-3a ($n$ = 1.054) and the downstream aerogel is PDR11-4a ($n$ = 1.065).*

#### 8.5.4.3 Performance for inclined tracks

In the Belle II environment, the charged tracks from the interaction region hitting the ARICH are inclined with polar angles between 15 and 30 degrees. Hence, we measured the RICH performance for tracks inclined by 15 and 30 degrees in the test beam experiment. Figures 8.36 and 8.37 show the Cherenkov ring image and Cherenkov angle distribution. In these measurements, the tracks detected in the full acceptance of the MWPCs are used. For tracks inclined at an angle of 15 degrees, we obtained $N_{pe} = 14.5$ and a single photon resolution of 14.9 mrad, which corresponds to $5.8\sigma$ $\pi/K$ separation at 4 GeV/$c$. For 30-degree inclination, the fit to the test beam result gives $N_{pe} = 13.4$ and a single photon resolution of 16.7 mrad, corresponding to $5.0\sigma$ $\pi/K$ separation. The detected number of photoelectrons is smaller than that for 15 degrees because about 15% of the ring image is outside of the acceptance of the $2 \times 3$ HAPD array. Corrected for the acceptance loss, the result for 30 degrees corresponds to $N_{pe} = 15.8$ and $5.5\sigma$ $\pi/K$ separation at 4 GeV/$c$. We have therefore confirmed that the prototype RICH counter achieves more than $5\sigma$ $\pi/K$ separation.

## 8.6 Expected Performance

### 8.6.1 Number of Photons

The expected number of Cherenkov photons was calculated by combining the measured values for quantum efficiency $\epsilon_q(\lambda)$ (Fig. 8.12) and for aerogel transmission $T(\lambda)$ (Fig. 8.7),

$$N_{phot} \approx 370\text{cm}^{-1}\, d\, \epsilon_e\, \epsilon_a\, \epsilon_t\, \sin^2\theta_c \int \epsilon_q(\lambda)\, T(\lambda)\, \lambda^{-2}\, d\lambda, \tag{8.4}$$

where $d$ is the radiator thickness (2 cm for single aerogel layer), $\epsilon_e$ is the single photoelectron detection efficiency (estimated to be 0.80 for the HAPD), and $\epsilon_a$ is the active surface fraction of the photon detector (0.67 in the case of HAPD); $\epsilon_t = 0.9$ is the fraction of the surface covered with photon detector modules in a particular tiling scheme (Fig. 8.6). The resulting expected average number of detected Cherenkov photons from both layers is 20.

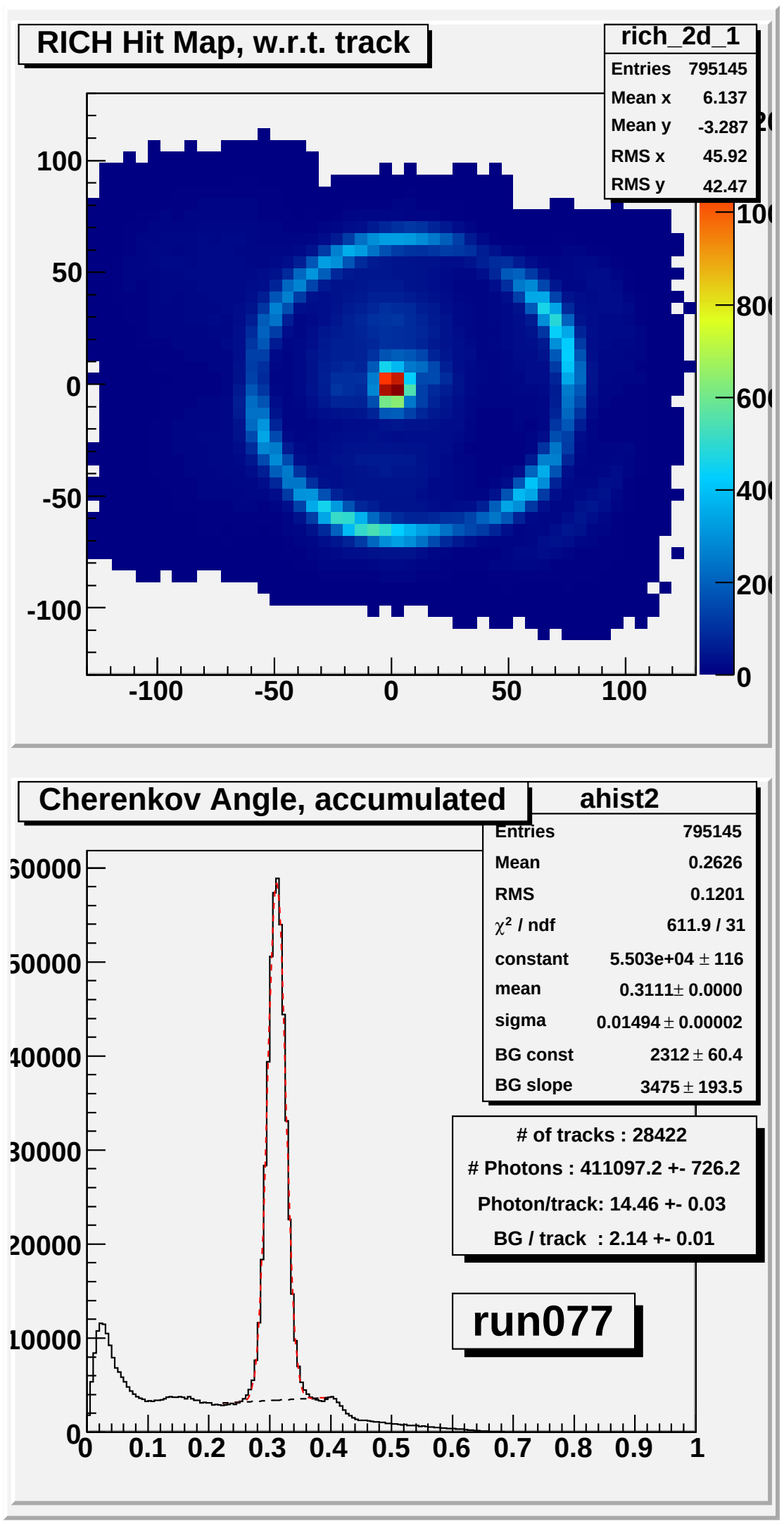

Figure 8.36: Cherenkov image and Cherenkov angle distribution for tracks inclined at 15 degrees.

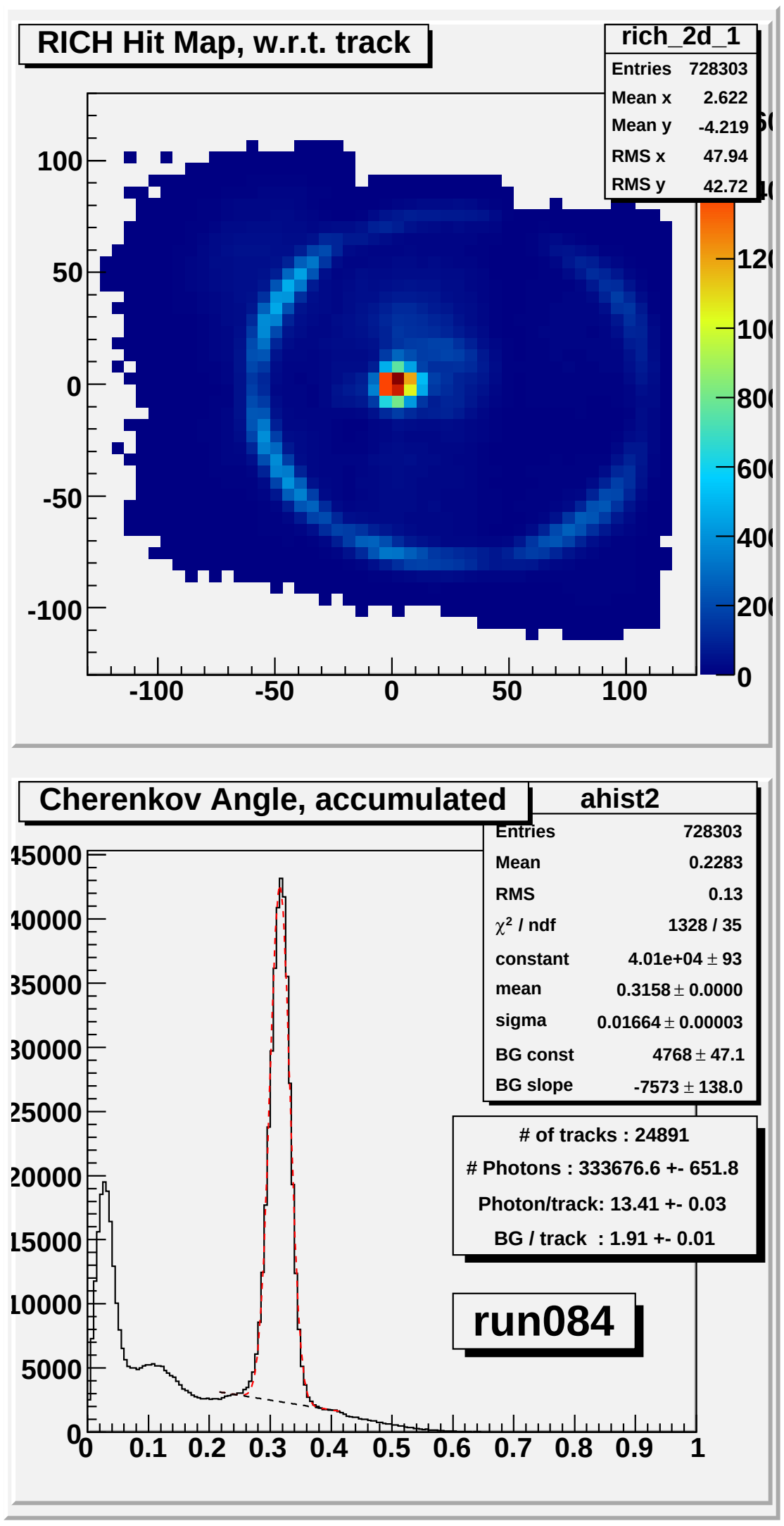

Figure 8.37: Cherenkov image and Cherenkov angle distribution for tracks inclined at 30 degrees.

### 8.6.2 Resolution of Cherenkov angle

From the photon hit coordinate as measured by the photon detector and from the particle trajectory as given by the tracking chamber, the photon direction is reconstructed to obtain the Cherenkov angle and thus the velocity $\beta c$ of the particle. Sources of error in the measurement with a single photon are as follows. The finite radiator effective thickness, $d = 2$ cm, contributes $\sigma_\theta^{emp} = d/(\ell\sqrt{12})\sin\theta_c\cos\theta_c = 9.8\,\mathrm{mrad}$. The second contribution comes from the finite coordinate resolution of the photon detector $\sigma_\theta^{pix} = a/(\ell\sqrt{12})\cos^2\theta_c = 6.6\,\mathrm{mrad}$, with a pad size of $a = 4.9$ mm and the radiator to photon detector distance of $\ell = 19$ cm. The chromatic error, i.e., the variation of the refractive index over the energy range of Cherenkov photons that are detected by the photon detector, is determined from the measured variation of refractive index with wavelength, the transmission of aerogel and the photo-detector quantum efficiency. The resulting r.m.s. spread $\sigma_n$ of the refractive index can be determined, from which $\sigma_\theta^{dis} = \sigma_n/(n\tan\theta_c) = 2\,\mathrm{mrad}$ is obtained. From the experience with the HERMES RICH [23], we conclude that the contribution due to the imperfections and inhomogeneity of the radiator are about 2 mrad. Finally, the contribution arising from the error in track parameters as determined by the tracking system and extrapolated to the radiator is expected to amount to $\sigma_\theta^{tr} = 6\,\mathrm{mrad}\sqrt{1 + (1.75\mathrm{GeV}/c/p)^2}$.

Finally, the combined single-photon error is obtained by summing the above errors in quadrature. For 4 GeV/$c$ momentum tracks this gives $\sigma_\theta = 14\,\mathrm{mrad}$. Assuming 20 detected photons for a $\beta = 1$ particle, the measurement precision is

$$\sigma\theta^{(N)} = \frac{\sigma_\theta}{\sqrt{N}} = 3.1\,\mathrm{mrad}. \tag{8.5}$$

In the absence of backgrounds, the proposed counter would thus enable better than $5\sigma$ pion-kaon separation at the kinematic limit of 4 GeV/$c$. Pions would be separated by $4\sigma$ from electrons up to about 1 GeV/$c$.

### 8.6.3 Background

Since the number of photons per ring can be rather low, backgrounds could degrade the performance as discussed above. The following sources of background expected in the RICH photon detector were considered:

**1.** Rayleigh-scattered Cherenkov photons from the same charged particle.

**2.** Cherenkov photons emitted by the same charged particle in the photon detector window.

**3.** Cherenkov photons emitted by other charged charged particles in the event.

**4.** Cherenkov photons emitted by products of a primary photon conversion in the material in front of the RICH radiator and in the radiator itself.

**5.** Beam related background hits.

**6.** Electronic noise.

The levels of backgrounds (1) and (2) were estimated from the test beam data (Fig. 8.34) and turned out to be the dominant source. The levels of backgrounds (3) and (4) were studied by computer simulation. Because of the rather small diameter of Cherenkov rings (12 cm) as compared to the typical distance between charged particle impact points on the aerogel radiator, background (3) does not contribute significantly. The level of source (5) was estimated from the background rate in the Belle EACC, and scaled by a factor of 20 as expected from the background calculations. This rough estimate yields values on the order of 100 Hz/cm$^2$. Finally, the electronic noise was estimated by assuming a conservative value of $10^{-3}$ for the probability of a noise hit. In total, we expect $\approx$0.7 smoothly distributed background hits within a $\pm 3\sigma_\theta$ annulus of the 4 GeV/$c$ pion ring.

### 8.6.4 Particle Identification Capabilities

The particle identification capabilities of the counter were evaluated by using simulated data. The backgrounds that are not included in the simulation (Rayleigh-scattered Cherenkov photons from the same charged particle, Cherenkov photons emitted by the same charged particle in the photon detector window, beam related background hits, and electronic noise) were added according to observed or expected rates. In the analysis of simulated events, the likelihood for the observed hit pattern is calculated for each hypothesis, and momentum-dependent selection criteria are chosen for a given misidentification probability [24]. The resulting identification efficiency for kaons is shown in Fig. 8.38; as expected, we observe a clear advantage of using a focusing radiator compared to a homogenous one.

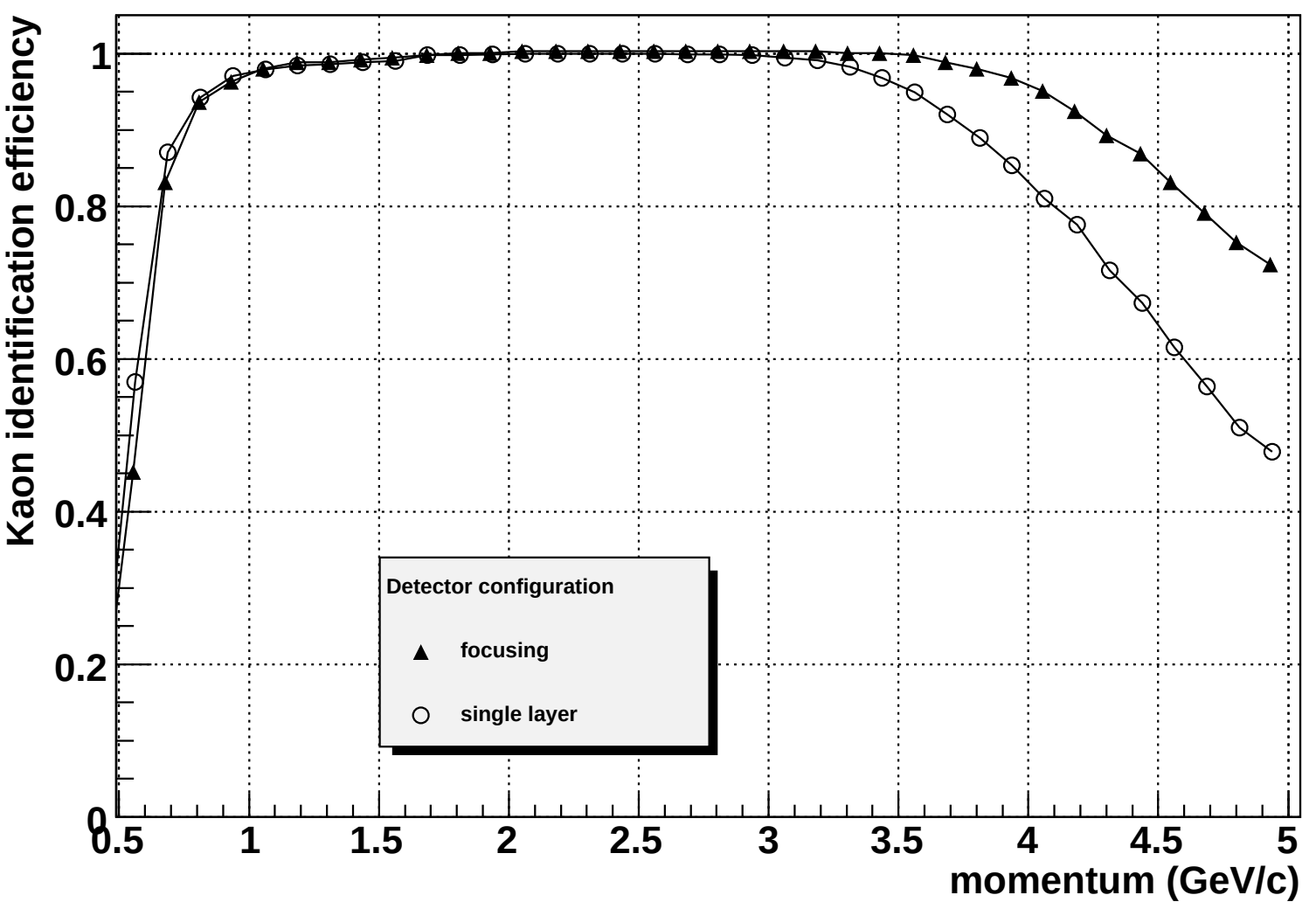

*Figure 8.38: Particle identification capability of the aerogel RICH counter: kaon efficiency for 1% pion misidentification probability. The focusing radiator configuration (▲) is compared to the homogenous radiator (○).*

We have also investigated the performance of the counter in the region close to the boundary of the barrel PID device. As can be seen from Fig. 8.39, a sizable fraction of Cherenkov photons emitted by charged particles close to this boundary do not hit the photon detector. This problem could in principle be overcome by a planar mirror as illustrated in Fig. 8.40.

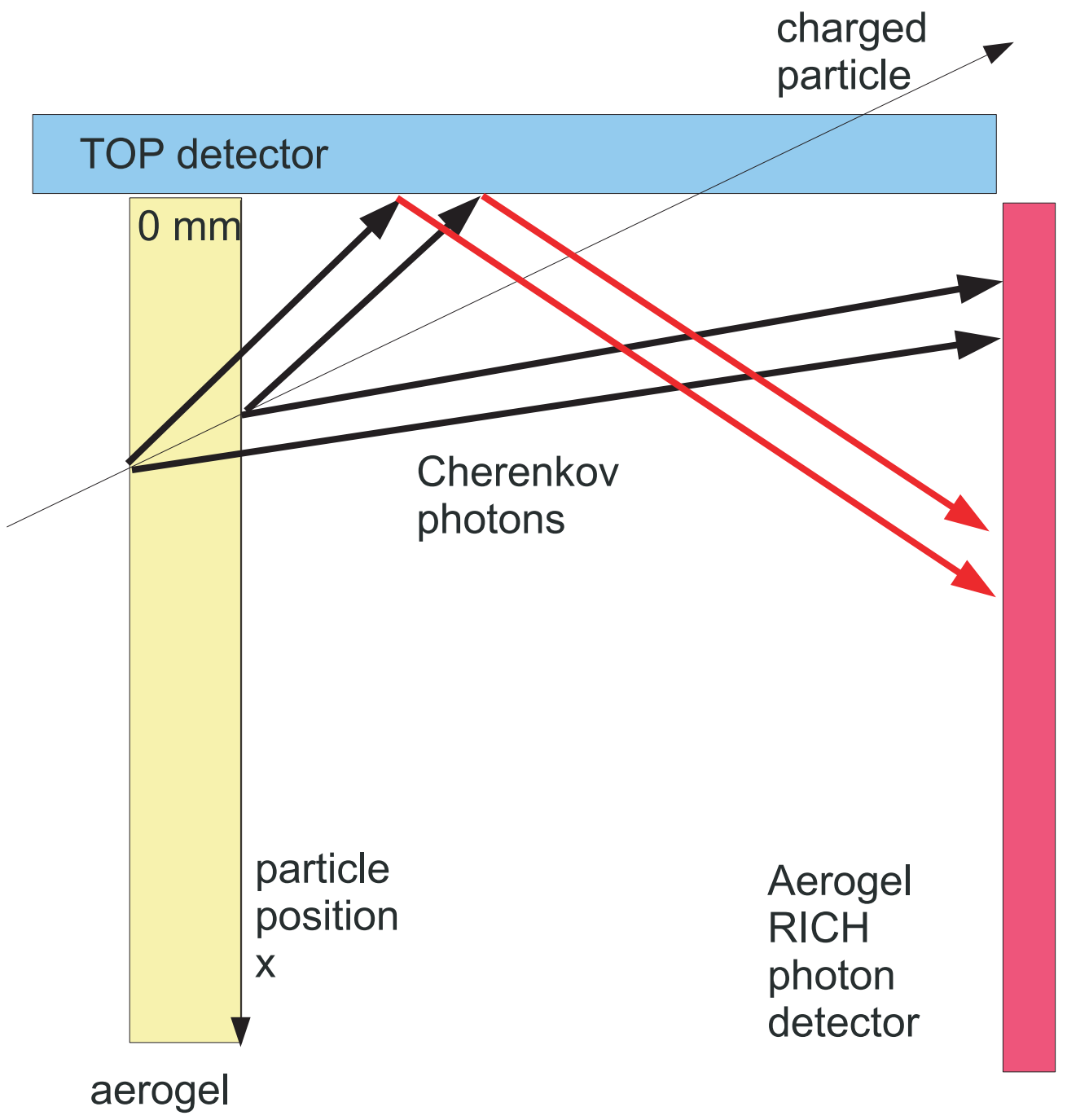

*Figure 8.39: Propagation of Cherenkov photons of a charged particle hitting the aerogel radiator in the vicinity of the boundary to the barrel part, for the case with an additional planar mirror.*

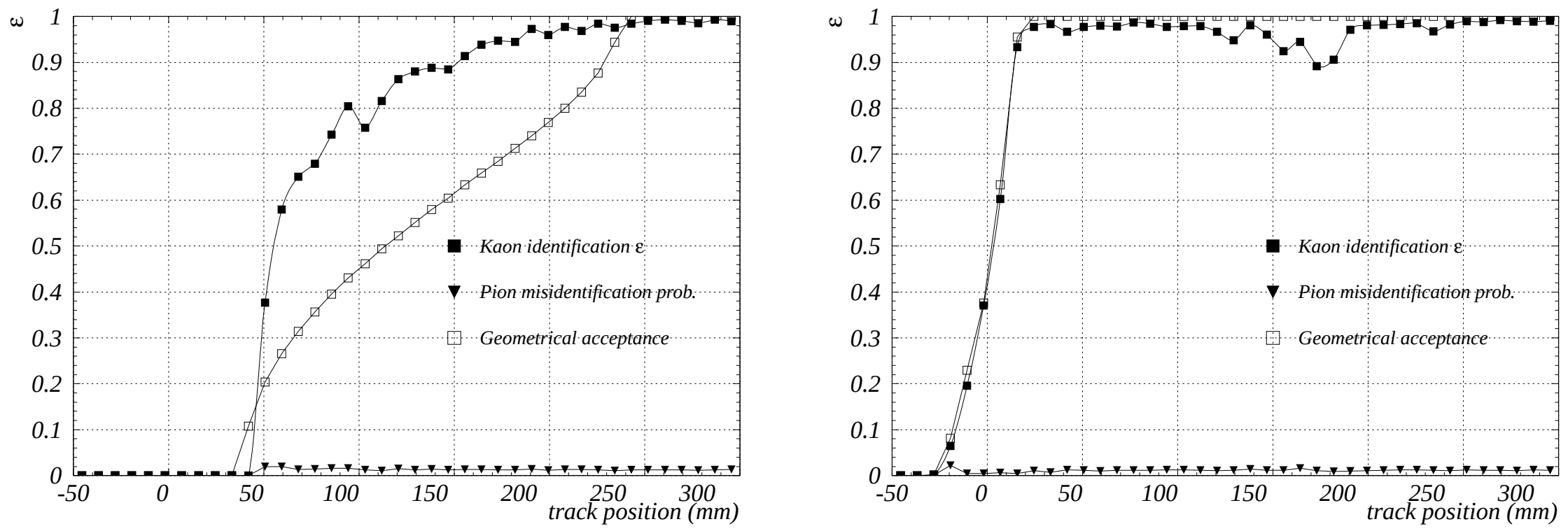

*Figure 8.40: Particle identification capability in the vicinity of the boundary to the barrel PID: kaon efficiency for 1% pion misidentification probability. Cases with an additional planar mirror (right) and without it (left) are shown as function of the distance from the boundary.*

### 8.6.5 Physics Impact Study in $B \to \rho\gamma$ Analysis

The physics impact of the ARICH is estimated by the analysis of $B \to \rho\gamma$ for two forward-endcap configurations:

(F1) the Belle PID (EACC and $dE/dx$). This corresponds to the scenario where the EACC is used in Belle II; this is the reference configuration.

(F2) the ARICH detector.

The simulation for the (F1) case is done using Belle's gsim with the Belle geometry, whereas the (F2) case is simulated with fsim [25].
In addition to the end-cap configuration, the following barrel-PID cases are evaluated:

(B1) the present Belle PID (BACC, TOF and $dE/dx$); this is the reference, corresponding to (F1) in the end-cap.

(B2) the focusing TOP detector.

The (B1) and (B2) simulations are performed using gsim and fsim, respectively.
We reconstruct $B^0 \to \rho^0\gamma$ and $B^+ \to \rho^+\gamma$ events from a $B \to \rho\gamma$ signal Monte Carlo (MC) sample. The $B \to K^*\gamma$ process is also simulated as the physics background channel, with the events subjected to the same analysis program. The selection criteria are similar to those in the recent Belle analysis [26]. For simplicity, the contribution from the $q\bar{q}$ background is assumed to be independent of the PID configuration and is fixed to 2000 events for 7.5 ab$^{-1}$. We estimate the figure of merit (FOM) from the number of signal ($B \to \rho\gamma$) events $N_S$ and background ($B \to K^*\gamma$ and $q\bar{q}$ background) events $N_B$ in the signal region. The signal region is defined by $M_{\rm bc} > 5.27$ GeV/$c^2$ and $\Delta E_{\rm min} < \Delta E < 0.08$ GeV, where $\Delta E_{\rm min}$ is chosen for each PID configuration to maximize the FOM.

| Barrel | Forward | $N_S$ | $N_B$ | FOM | $\Delta E_{\rm min}$ [ GeV] |
|---|---|---|---|---|---|
| B1 | F1 | 987 | 5242 | 12.5 | −0.25 |
| B1 | F2 | 1032 | 5026 | 13.3 | −0.25 |
| B2 | F1 | 982 | 2865 | 15.8 | −0.30 |
| B2 | F2 | 1027 | 2651 | 16.9 | −0.30 |

*Table 8.6: Number of signal ($N_S$) and background ($N_B$) events, figure of merit (FOM), and the $\Delta E$ lower limit $\Delta E_{\rm min}$ chosen for each barrel and forward PID configuration for $B^0 \to \rho^0\gamma$ analysis at 7.5 ab$^{-1}$. See the text for an explanation of the configuration label.*

| Barrel | Forward | $N_S$ | $N_B$ | FOM | $\Delta E_{\rm min}$ [ GeV] |
|---|---|---|---|---|---|
| B1 | F1 | 1152 | 3006 | 17.9 | −0.30 |
| B1 | F2 | 1194 | 2981 | 18.5 | −0.30 |
| B2 | F1 | 1141 | 2490 | 18.9 | −0.30 |
| B2 | F2 | 1183 | 2465 | 19.6 | −0.30 |

*Table 8.7: Number of signal ($N_S$) and background ($N_B$) events, figure of merit (FOM), and the $\Delta E$ lower limit $\Delta E_{\rm min}$ chosen for each barrel and forward PID configuration for $B^+ \to \rho^+\gamma$ analysis at 7.5 ab$^{-1}$. See the text for an explanation of the configuration label.*

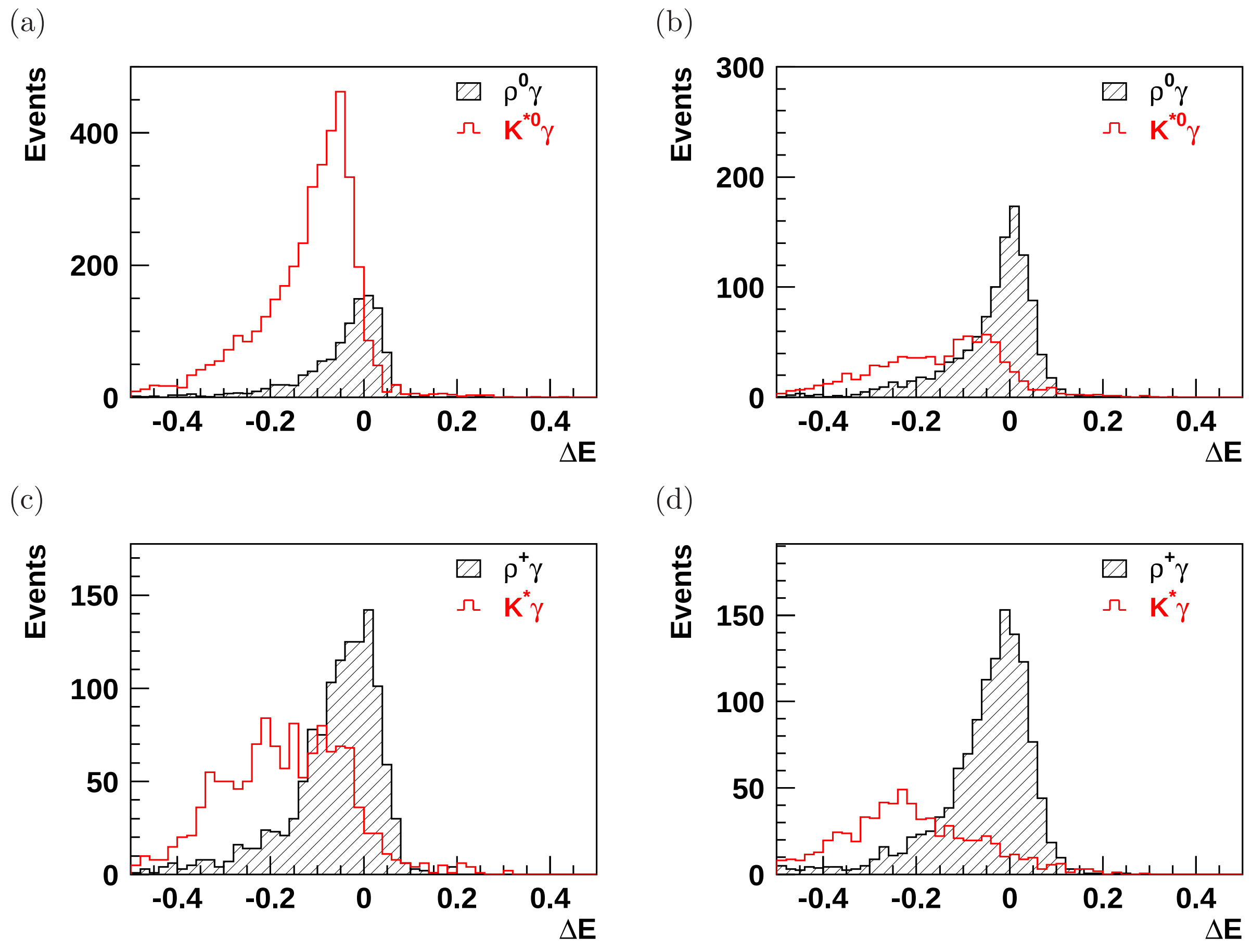

*Figure 8.41: Expected $\Delta E$ distribution at 7.5 ab$^{-1}$ for (a) $B^0 \to \rho^0\gamma$ with Belle PID configuration (B1+F1), (b) $B^0 \to \rho^0\gamma$ with TOP and ARICH (B2+F2), (c) $B^+ \to \rho^+\gamma$ with Belle PID configuration (B1+F1), (d) $B^+ \to \rho^+\gamma$ with TOP and ARICH (B2+F2).*

Tables 8.6 and 8.7 tabulate the FOM for four detector configurations. The $\Delta E$ distributions for two PID configurations—F1+B1 with the Belle devices and F2+B2 with the ARICH and TOP—are shown for $\rho\gamma$ signals together with $K^*\gamma$ backgrounds in Fig. 8.41. Comparing the histograms for $B^0 \to \rho^0\gamma$ in this figure, the suppression of the $K^*\gamma$ background is quite significant with the TOP and ARICH. The $K^*\gamma$ component below $-0.2$ GeV contains events that arise from misreconstruction rather than a simple misidentification between $K$ and $\pi$. In Fig. 8.41(b), where the misidentification is highly suppressed by the good PID devices, such misreconstructed events are more apparent. For the $B^+ \to \rho^+\gamma$ case in the same figure, we also observe a significant improvement in the analysis, although the effect is smaller if compared to the $\rho^0\gamma$ channel: misreconstructed events already represent a significant fraction of the background in the Belle configuration. Tables 8.8 and 8.9 show the "luminosity gain" of the B2+F2 configuration relative to the B1+F1 configuration, defined as the square of the ratio of the FOM; it is the luminosity ratio that would be necessary to obtain the same sensitivity with the Belle PID configuration. From Table 8.8, the TOP detector gives us a significant luminosity gain in the $B^0 \to \rho^0\gamma$ mode, and the ARICH detector further improves the physics sensitivity: up to $\sim$ 80% gain. In the $B^+ \to \rho^+\gamma$ mode, we obtain 20% higher sensitivity with the new PID system.

| | F1 | F2 |
|---|---|---|
| B1 | 1.00 | 1.12 |
| B2 | 1.60 | 1.83 |

*Table 8.8: Luminosity gains for $B^0 \to \rho^0\gamma$ for four PID configurations.*

| | F1 | F2 |
|---|---|---|
| B1 | 1.00 | 1.07 |
| B2 | 1.12 | 1.20 |

*Table 8.9: Luminosity gains for $B^+ \to \rho^+\gamma$ for four PID configurations.*

## 8.7 Schedule

Figure 8.42 shows the ARICH construction schedule. After the photon detector technology choice made in December 2009, we continue R&D on HAPD improvement and will finalize its specification for bidding. According to the producer, the HAPD production will take 2.5 years; this could be on a critical path in the ARICH overall schedule. In parallel, readout electronics production will be planned in 2012 after a further two years of R&D to arrive at the final version. The aerogel radiator is not a critical component for the timeline. After some further work on optical improvements, the producer for mass-production will be selected in the coming years. For the mechanical structure, the baseline design has to proceed in detail after the outer radius of the end-cap is decided. In 2010-2011, a 1/6-size mock-up will be produced to verify many technical issues such as radiator tile alignment, HAPD arrangement cabling space, and so on. After this, the mechanical container will be ordered.
Modules consisting of an HAPD and the front-end readout system will be produced in 2012-2013, and they will be installed into the container in the 3rd to 4th quarter of 2013. The attachment of the ARICH to the E-CsI container will be done at the beginning of 2014; the end-cap apparatus will be installed in the spring of 2014. The cosmic ray test in situ as well as ARICH commissioning is expected in the summer of 2014.

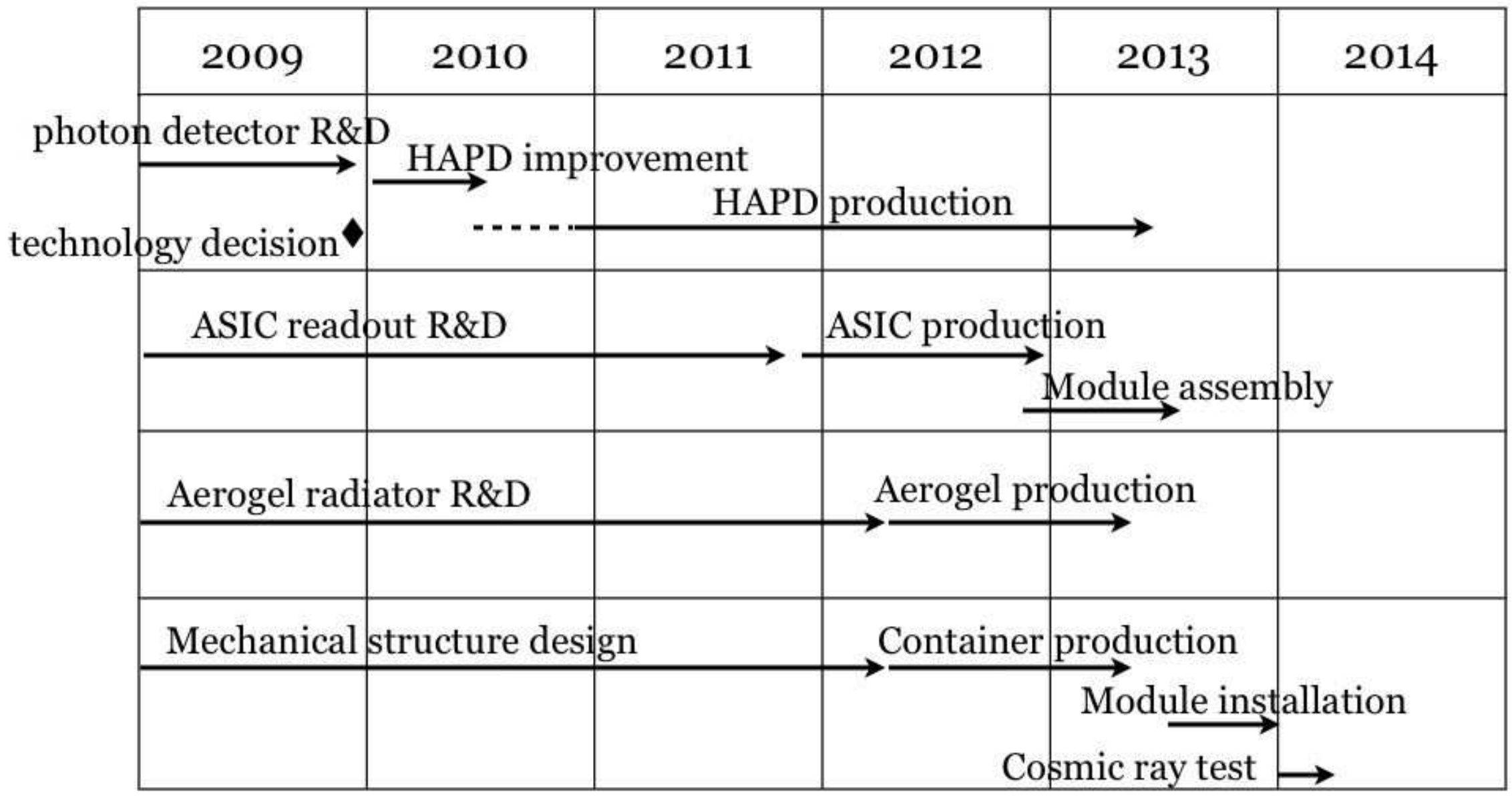

*Figure 8.42: ARICH construction schedule by calendar year.*

# Chapter 9

# Calorimeter (ECL)

## 9.1 Introduction

Since one third of $B$-decay products are $\pi^0$'s or other neutral particles that provide photons in a wide energy range from 20 MeV to 4 GeV, a high resolution electromagnetic calorimeter is a very important part of the Belle II detector. CsI(Tl) was chosen as the scintillation crystal material for the Belle calorimeter due to its high light output, relatively short radiation length, good mechanical properties and moderate price. The main tasks of the calorimeter are:

- detection of photons with high efficiency,
- precise determination of the photon energy and angular coordinates,
- electron identification,
- generation of the proper signal for trigger,
- on-line and off-line luminosity measurement.
- $K_L^0$ detection together with the KLM (Ch. 10).

## 9.2 The Belle electromagnetic calorimeter

The Belle electromagnetic calorimeter (ECL) consists of a 3 m long barrel section with an inner radius of 1.25 m and annular endcaps at $z = 1.96$ m (forward) and $z = -1.02$ m (backward) from the interaction point. The calorimeter covers the polar angle region of $12.4^\circ < \theta < 155.1^\circ$, except for two gaps $\sim 1^\circ$ wide between the barrel and endcaps.
The barrel part has a tower structure that projects to a region near the vicinity of the interaction point. It contains 6624 CsI(Tl) crystals of 29 distinct shapes. Each crystal is a truncated pyramid of an average size about $6 \times 6$ cm$^2$ in cross section and 30 cm ($16.1X_0$) in length. The endcaps consists of 2112 CsI crystals of 69 shapes. The total number of crystals is 8736, with a total mass of about 43 tons.
Each crystal is wrapped with a layer of 200-$\mu$m thick Gore-Tex porous teflon and covered by a laminated sheet of 25-$\mu$m thick aluminum and 25-$\mu$m thick mylar. For scintillation light readout, two $10 \times 20\,\mathrm{mm}^2$ Hamamatsu Photonics S2744-08 photodiodes are glued to the rear surface of the crystal via an intervening 1-mm thick acrylite plate. An LED attached to the plate can inject light pulses into the crystal volume to monitor the optical stability. A preamplifier is attached to each photodiode, which provides two independent output signals from each crystal. These two pulses are summed at an external shaper board. To monitor the stability of the electronics, the test pulses can be injected into the inputs of the preamplifiers.
The barrel crystals are installed in a multi-cell structure formed by 0.5 mm thick aluminum septum walls stretched between the inner and outer cylinders. Eight crystals, four rows in $\theta$ and two columns in $\varphi$, are inserted in each cell and fixed in position from behind support jigs. The overall support structure is gas tight and flushed with dry air to provide a low humidity (5%) environment for the crystals. The heat generated by the preamplifiers (about 3 kW in total) is extracted by a water cooling system. The endcap support structure is similar to the barrel one. The calorimeter is described in detail in reference [1].
The preamplifier attached to the crystal is followed with a shaper board placed in the crates around the detector, as well as a digitizing and trigger module in the electronics hut. The shaper board contains $CR-(RC)^4$ active filters with shaping time $\tau = 1\,\mu$s and MQT300A chips, which convert the input charge integrated over certain gate time to three time intervals that are measured by a multi-hit TDC (LeCroy 1877S). The corresponding ranges are: 0.06 MeV/bin, 0.5 MeV/bin and 4 MeV/bin. An auto-range selection option provides readout of only one range with optimal sensitivity. In addition to the main channel, the shaper board contains fast shaper amplifiers ($\tau = 200$ ns), which generate signals used for triggering and timing [2].
The average output signal of the crystals, as measured by using calibration with cosmic ray muons, is about 5000 photoelectrons per MeV, while the noise level was about 200 keV prior to irradiation.
The intrisic energy resolution of the calorimeter, as measured in a prototype [3], can be approximated as:

$$\frac{\sigma_E}{E} = \sqrt{\left(\frac{0.066\%}{E}\right)^2 + \left(\frac{0.81\%}{\sqrt[4]{E}}\right)^2 + (1.34\%)^2}, \tag{9.1}$$

where $E$ is in GeV and the first term represents the electronics noise contribution.

## 9.3 Performance with increasing luminosity

The Belle detector has been operating since 1999 and, throughout this period, the calorimeter has demonstrated high resolution and good performance [4]. The photon energy resolution measured for the $e^+e^- \to \gamma\gamma$ process, averaged over the entire calorimeter, is $\sigma_E/E$=1.7%. The photon energy range for this process is $3.5\,\text{GeV} < E_\gamma < 8\,\text{GeV}$, depending on the photon angle. The azimuthal angular resolution for these photons is 0.23°. The $\gamma\gamma$ invariant mass for $\pi^0$ and $\eta$ mesons is determined with a resolution of 4.8 MeV/c$^2$ and 12 MeV/c$^2$, respectively. These results are in good agreement with Monte Carlo simulation.
However, the Belle II experiment, operating with a SuperKEKB luminosity of up to $8 \times 10^{35}$ cm$^{-2}$s$^{-1}$, puts new, severe requirements on the calorimeter. An obvious concern is the anticipated degradation of the thallium-doped CsI crystal performance parameters due to the high radiation dose.
The absorbed dose collected by the crystals in the course of the Belle exeriment is shown in Fig. 9.1. By summer 2009, the measured integrated dose (i.e., for an integrated luminosity of 900 fb$^{-1}$) is about 100 rad for the barrel crystals and about 400 rad for the highest-dose endcap crystals. This has caused the light-output deterioration shown in Fig. 9.2. The light output loss is around 7% in the barrel and up to 13% in the endcap region closest to the beam pipe.
These results are in good agreement with previous measurements of the crystal radiation hardness [5, 6]. Since these studies show the loss of the light output to be less than or about 30% at 3.6 krad, an increase of the absorbed dose by one order of magnitude in Belle II will not pose a serious problem.
Another effect of the radiation background is the increase of the photodiode dark current due to neutron bombardment. At present, the dark current increase in the barrel is not substantial—less than 10 nA—while it is rather large in the endcap calorimeter—up to 200 nA. Thus, at higher luminosity, the elevated dark current may be a problem.
An important issue requiring careful treatment is the so-called pile up noise, caused by soft background photons with an average energy less than 1 MeV. Fluctuation in the number of these photons arriving during the integration time prior to readout contributes to the total noise level. The measured noise level, $\sigma$, is near the 1 MeV level in the endcaps at a luminosity of $L = 10^{34}\,\mathrm{cm^{-2}s^{-1}}$.
High-energy photon backgrounds produce random cluster candidates in the calorimeter. At present, each triggered event contains an average of 6 background clusters (3 in the barrel and 3 in the endcaps) with an energy exceeding 20 MeV. By simple extrapolation, assuming that the current ECL hardware and software is unchanged, we will have several tens of background clusters at ten times higher luminosity. This results in a substantial combinatorial background for the event reconstruction. Our approaches to address this problem are described in the following sections.

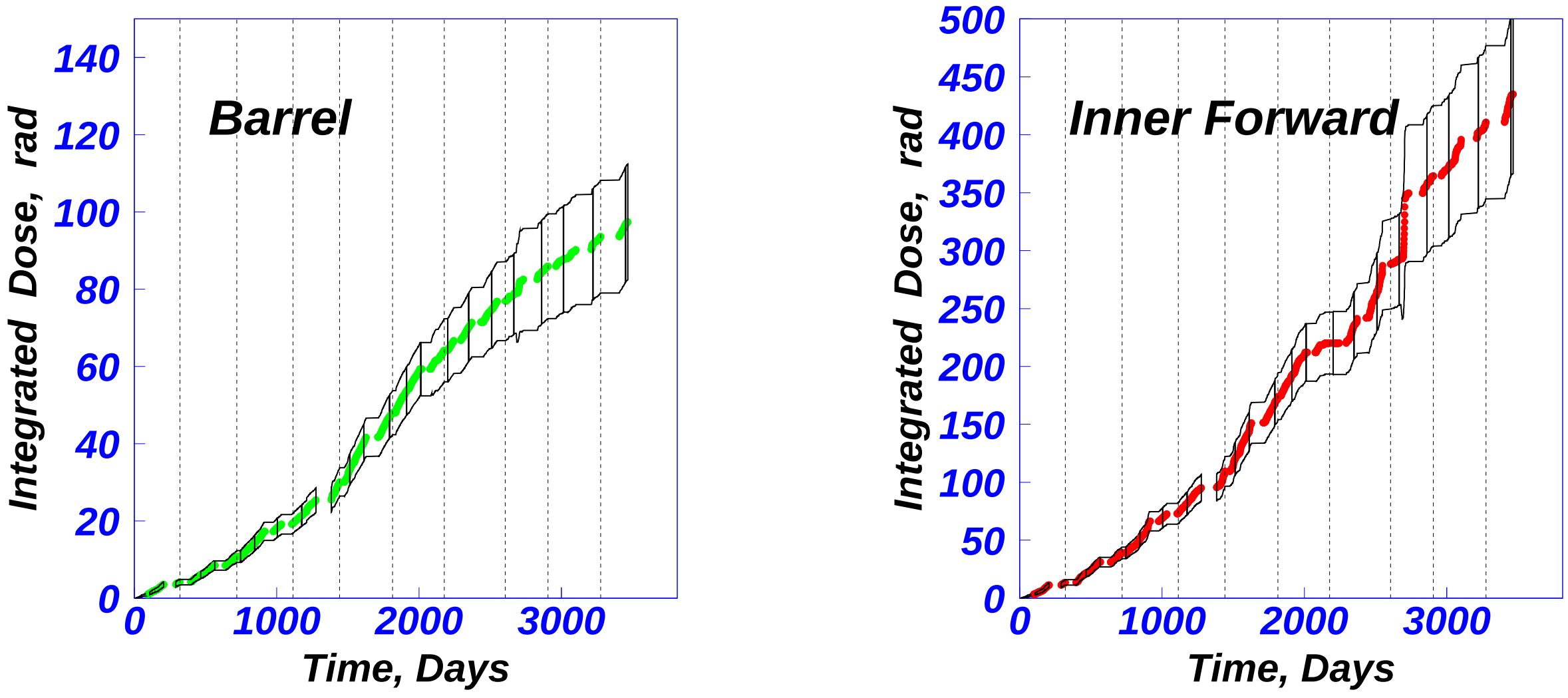

*Figure 9.1: Absorbed dose recieved by the CsI(Tl) crystals during the Belle experiment.*

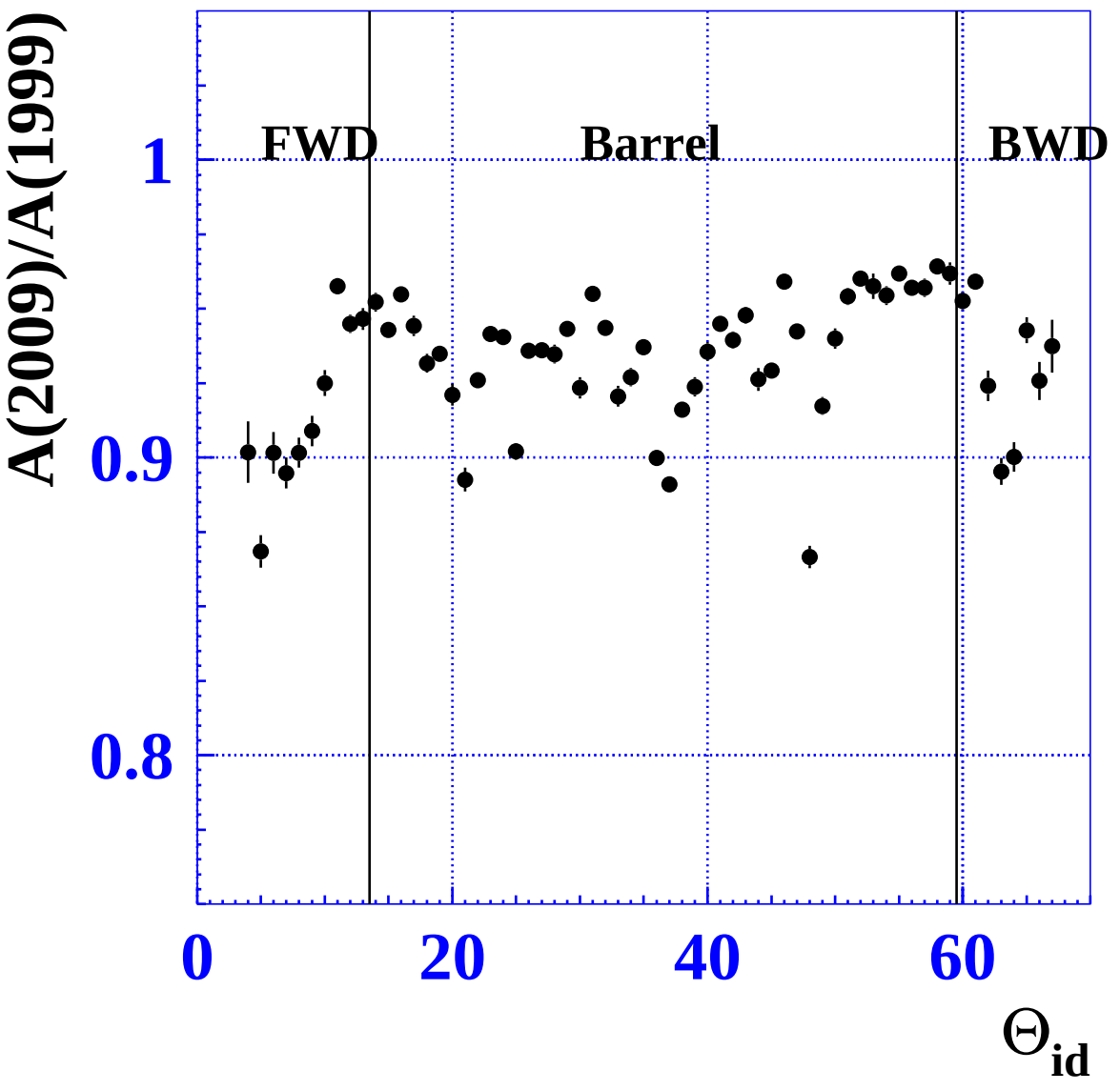

*Figure 9.2: Decrease in crystal light output after ten years of Belle operation.*

## 9.4 General electronics upgrade

Modification of the calorimeter electronics follows the strategy in the other Belle II subsystems. The main idea is to shorten the shaping time and to use pipelined readout with waveform processing. In the barrel region, the shaping time is changed to 0.5 $\mu$s and each crystal counter output signal is digitized continuously at 2 MSa/s. (Further reduction of the shaping time is not effective, since the CsI(Tl) scintillation decay time is 1 $\mu$s.) Upon receipt of a trigger, 16 sample points within the event window are fit to a signal function $F(t) = A_0 \times f(t - t_0)$, where the pulse height $A_0$ and the event time $t_0$ are free parameters. The shape of the signal response function, $f(t)$, is evaluated using separate measurements. The on-line fit determines $A_0$ and $t_0$ by minimizing

$$\chi^2 = \sum_{i,j}(A_i - F(t_i))S_{ij}^{-1}(A_j - F(t_j)), \tag{9.2}$$

where $A_i$ is the sampled amplitude at time $t_i$ and $S_{ij}$ is the predetermined covariance matrix.
To realize the scheme described above, the existing shaper-QT modules will be replaced with newly developed shaper-digitizer modules. Expected improvements from this modification were estimated by calculation, as well as the prototype tests.

### 9.4.1 Signal reconstruction

To estimate the effects of beam background for the current calorimeter and to test its performance with the improved readout, a MC simulation was performed, taking into account the signal shape and the background measured under the highest-luminosity conditions in Belle.
To simulate the pipelined readout, we use the obtained signal amplitudes ($A_i$) that are recorded in 16 sample points around the maximum of the signal and with a sample time interval $\Delta t = 0.5\,\mu$s. For each set of generated data $\{A_i\}$, the magnitude $A_0$ and time $t_0$ are determined by the fit procedure described above. Figure 9.3(left) shows the distribution of the reconstructed timing ($t_0$) with respect to the trigger timing ($t_{\mathrm{trig}}$) for both signal and fake clusters. In Belle II, the trigger timing is determined by one of several detector subsystems, including the ECL.

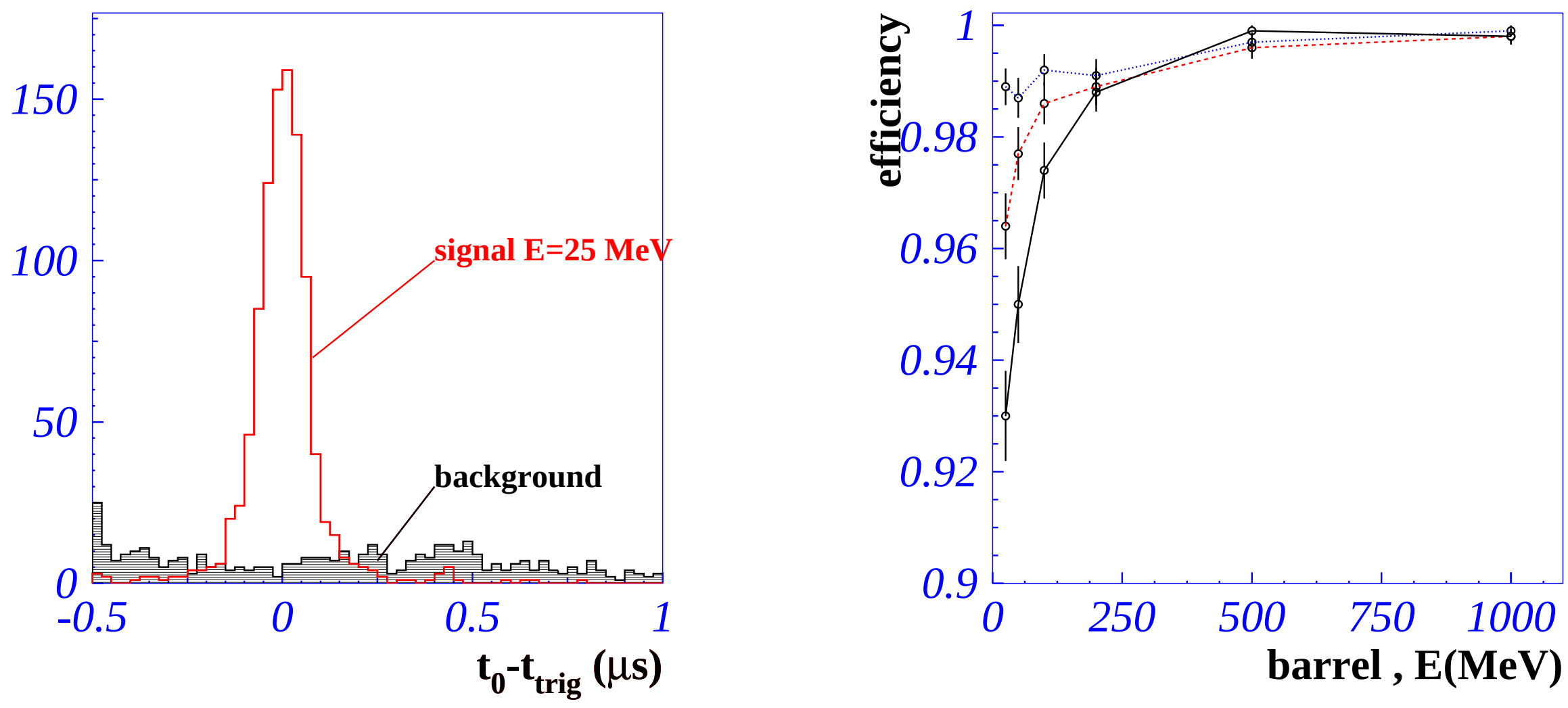

*Figure 9.3: Left: distribution of the event times determined by pulse shape fit for signals and fake clusters. Right: efficiency vs. photon energy (above 20 MeV) for true photons in the ECL barrel after the in-time cut on $|t_0 - t_{\rm trig}|$, where the dotted blue line corresponds to 5 additional background clusters, the dashed red line to 10 clusters, and the solid black line to 20 clusters.*

Applying an energy-dependent in-time cut of $|t_0 - t_{trig}| < 5000\,\text{ns}/(E/\text{MeV})$, we suppress the number of fake photons by a factor of 7 while keeping the efficiency for true photons at more than 93% for photons above 20 MeV. However, with the factor of 20 larger background expected in Belle II, the pile-up noise will be a factor of 3 higher. In this case, we will need to increase the photon energy threshold to 30 MeV to maintain the same level of combinatorial background.

### 9.4.2 Electronics for CsI(Tl) crystal with PIN-PD readout

A block diagram of the ECL electronics is presented in Fig. 9.4. At the beginning of the Belle II experiment, the existing CsI(Tl) crystals, PIN-Photo-diodes, preamplifiers and the cables to reach the shaper will be reused. However, the Shaper-QT modules used in Belle are replaced with new Shaper-DSP modules.

Each signal channel contains a shaper circuit and an 18-bit flash ADC (Analog Devices AD7641) that digitizes the signal at a 2-MHz clock frequency. The ADC data are processed by the digital filtering algorithm implemented in a XILINX FPGA. The data processing is initiated by arrival of a trigger signal and yields three parameters: amplitude, time, and data quality, which are calculated according to the algorithm described above.

This data from the Shaper-DSP modules are sent to a Collector module that combines the data from 8-12 Shaper-DSP modules and sends this via high-speed serial link utilizing the XILINX ROCKET-I/O resources to a COPPER FINESSE module, wich is the back end (common) interface to the data acquisition system (Ch. 13). Each collector module also contains a circuit for test pulse generation, to calibrate the response of each channel.

Each Shaper-DSP module channel also includes fast shaping ($\tau_d = 0.2\,\mu$s) and gain adjustment circuits for calorimeter-trigger purposes. A fast analog sum signal, which is a combination of 16 gain-corrected fast shaping channels, is delivered to the FAM (Flash-ADC trigger Module), where a trigger cell signal is generated when a threshold ($\sim 100$ MeV) is exceeded. These trigger cell signals are collected and processed by the TMM (Trigger and Monitoring Module), where ECL subtrigger decisions are made and sent to the Global Decision Logic.
A schematic of the shaper circuit is shown in Fig. 9.5. Signals from the two preamplifiers are summed at the stage U1 and the sum pulse is processed by the differential part at differentiating amplifier U2 ($\tau_d \approx 0.5\,\mu$s). Component U3 provides subtraction of the certain fraction of the integrated signal to suppress the tail caused by long CsI(Tl) scintillation decay components. The signal is formed by two second-order active integration stages, U4 and U5, with a shaping time of $\tau \approx 0.5\,\mu$s. Figure 9.6 shows the output signal with and without the long component suppression.
A simplified block diagram of the pulse feature extraction logic is shown in Fig. 9.7. The digital signal processing is performed with a XILINX XC3S1500-FG456 FPGA, which has enough resources to realize the required waveform analysis in real time. SDRAM chips are used to store precomputed coefficients for the digital signal processing.
The waveform analysis algorithm proceeds as follows. First, we tabulate the signal function $f(t)$ and its deriviative $f'(t)$ to obtain the tables of values $f_{ki} = f(t_{ki}) = f(k\delta t + i\Delta t)$, where $\Delta t$ is the digitizing interval, and $\delta t$ is the timing bin width of tabulation that is 100 times smaller than $\Delta t$, $i$ and $k$ are the corresponding bin indexes. Linearizing $A_0 f(t)$ in Eq. 9.2,

$$Af(t_{ki} - t_0) = A_0 f(t_{ki}) - t_0 A_0 f'(t_{ki}) = A_0 f_{ki} + B f'_{ki}\,,$$

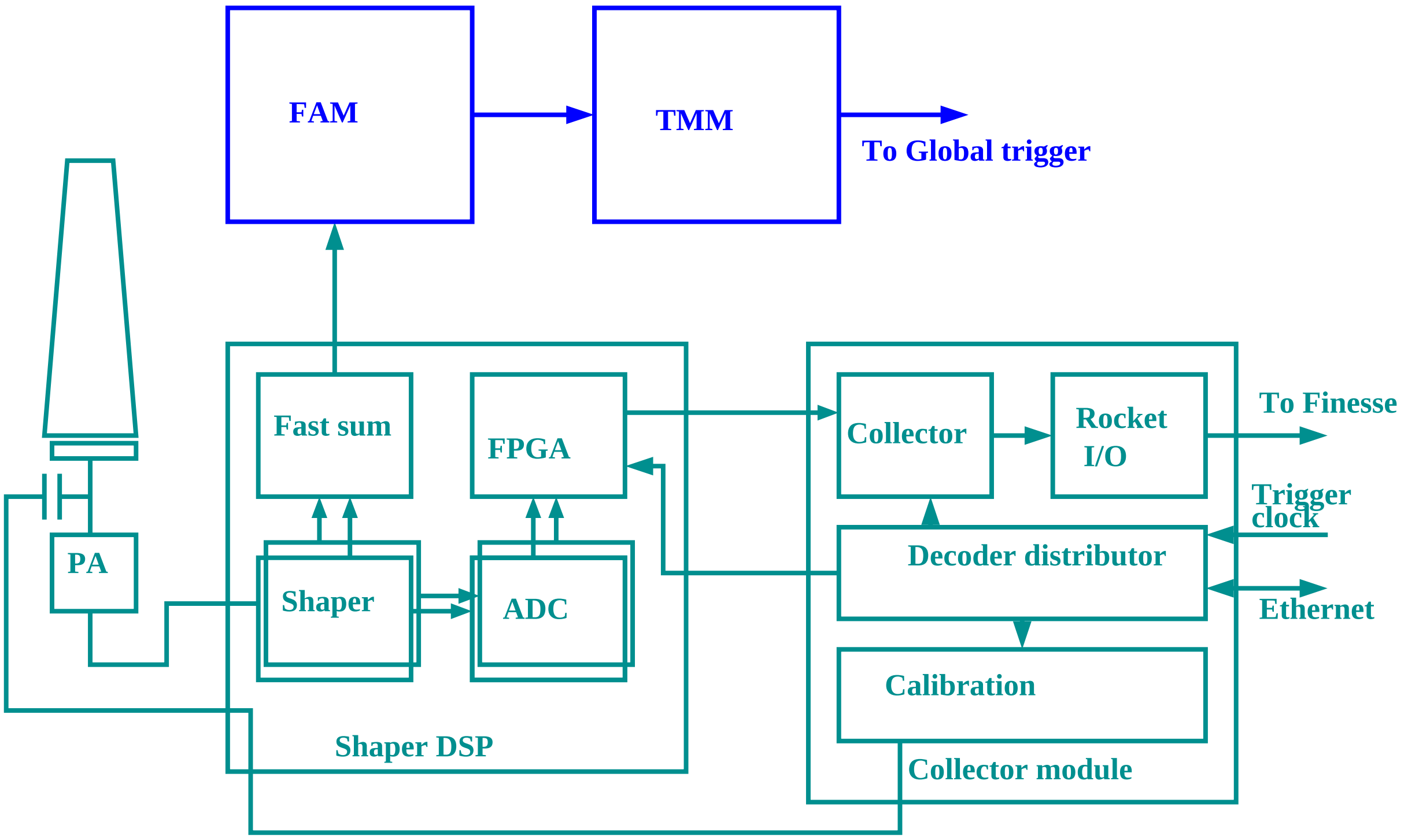

*Figure 9.4: Block diagram of the electronics used to read out the CsI(Tl) crystals in Belle II.*

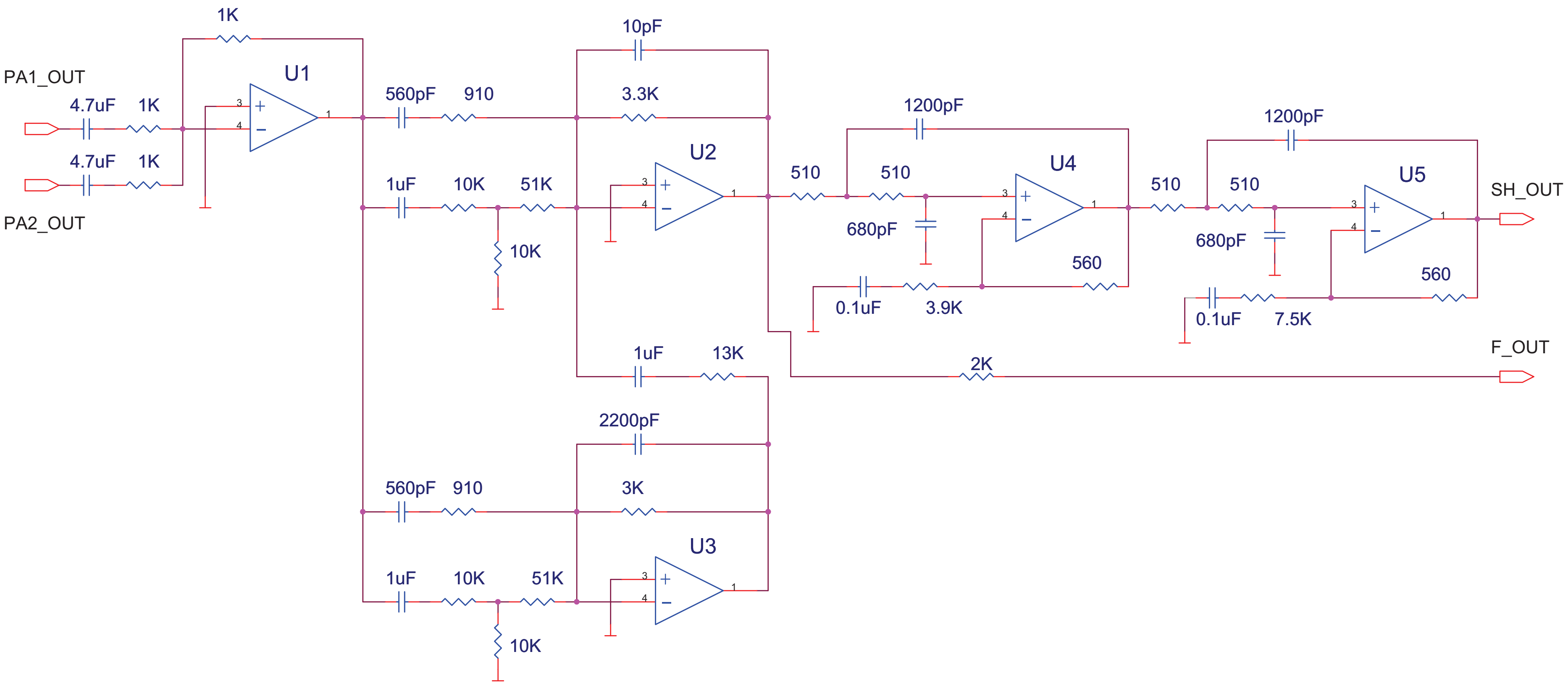

*Figure 9.5: Circuit schematic of the shaping amplifier chain.*

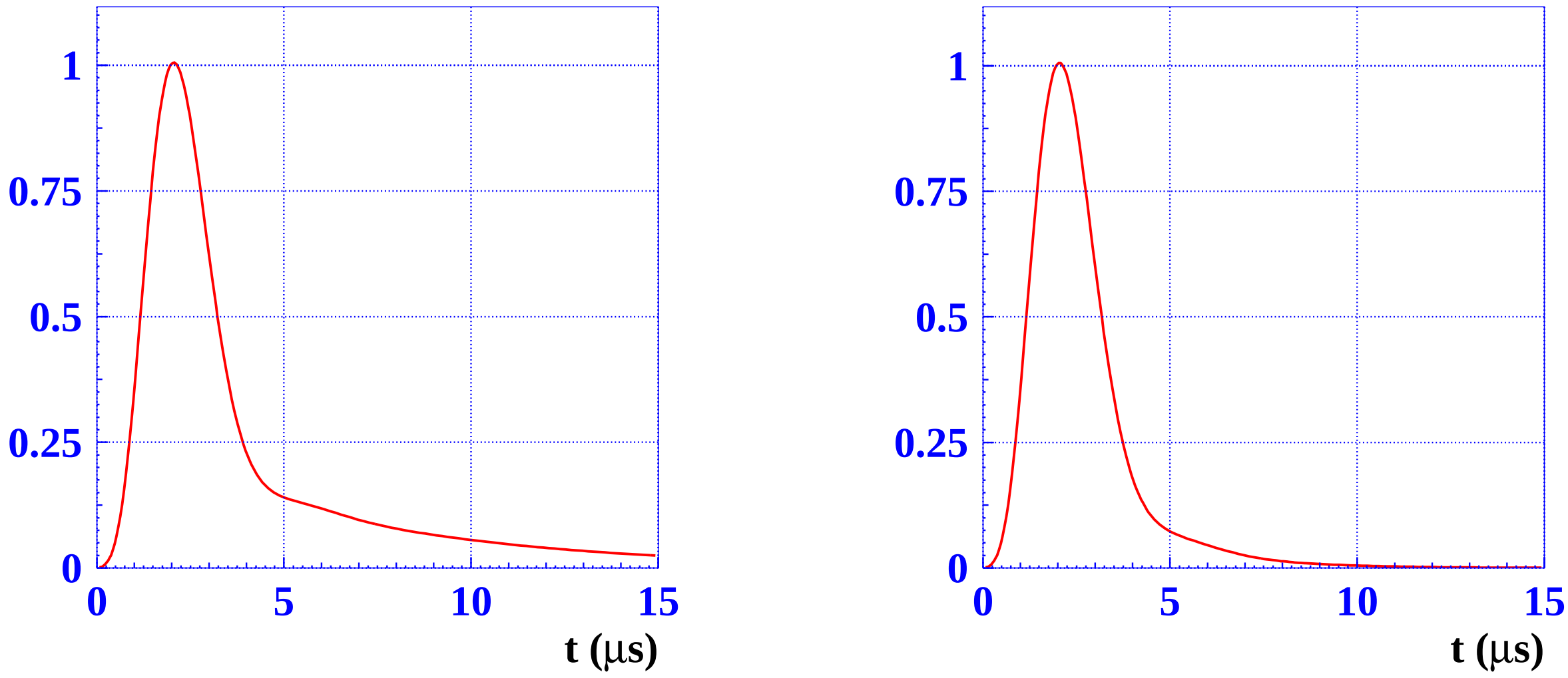

*Figure 9.6: Shape of the CsI(Tl) scintillation signal without (left) and with (right) tail suppression.*

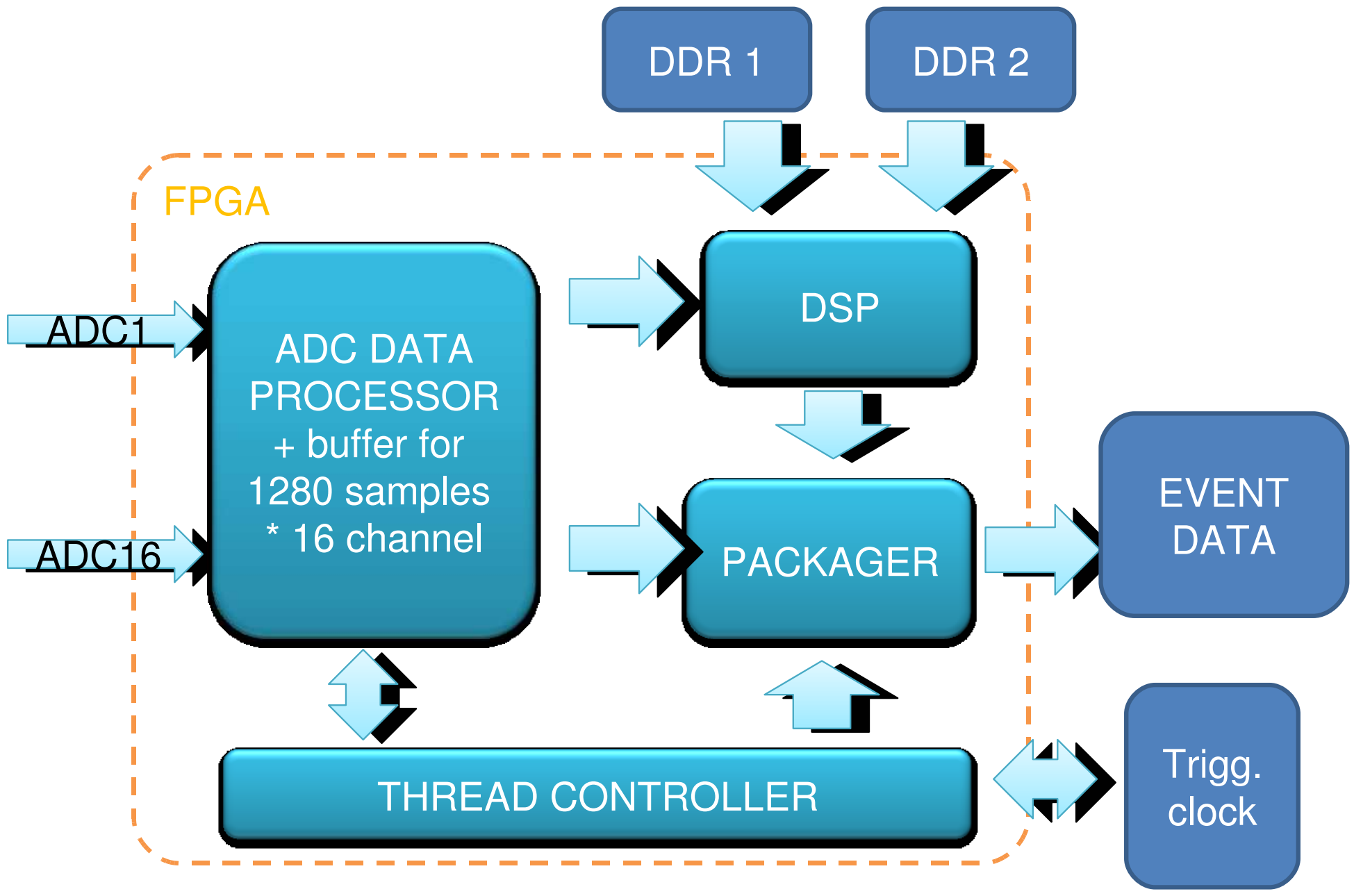

*Figure 9.7: Block diagram of digital signal processing logic implemented in XILINX FPGAs for the energy and time feature extraction.*

we can rewrite Eq. 9.2, incorporating the pedestal $p$ as well:

$$\chi^2 = \sum_i (A_i - A_0 f_{ki} - B f'_{ki} - p)(S^{-1})_{ij}(A_j - A_0 f_{kj} - B f'_{kj} - p)\,. \tag{9.3}$$

The minimization of this expression is equivalent to the linear-system solutions:

$$
\begin{aligned}
\sum_{i,j} f_{ki} S^{-1}_{ij} (S^{-1})_{ij} (A_j - A_0 f_{kj} - Bf'kj - p) = 0 \qquad & A_0 = \sum_i \alpha_i^k A_i \\
\sum_{i,j} f'_{ki} S^{-1}_{ij} (S^{-1})_{ij} (A_j - A_0 f_{kj} - Bf'kj - p) = 0 \quad & \Rightarrow B = \sum_i \beta_i^k A_i,\ t_0 = -B/A_0 \\
\sum_{i,j} S^{-1}_{ij} (S^{-1})_{ij} (A_j - A_0 f_{kj} - Bf'kj - p) = 0 \qquad & p = \sum_i \gamma_i^k A_i
\end{aligned}
\tag{9.4}
$$

where $\alpha_i^k$, $\beta_i^k$, $\gamma_i^k$ are expressed via tabulated values $f_{ki}$, $f'_{ki}$ and a covariance matrix $S^{-1}$. Three parameters are obtained from these solutions: the amplitude $A_0$, the peak time $t_0$, and the pedestal $p$. The extracted event time $t_0$ allows the calculation of a new $k$-index for the next iteration. Three iterations are enough for the convergence of the solution. For the first iteration, the trigger time is used as the initial pulse time. As one can see from Eq. 9.4, one division and one multiply-and-accumulate (MAC) operation are needed in each iteration. Fixed-point arithmetic and a precomputed array of coefficients ($\alpha_i^k$, $\beta_i^k, \gamma_i^k$) are sufficient to realize the real time implementation of this algorithm in the chosen FPGA.

The FPGA is capable of processing up to 8 events in parallel. ADC data are stored in an input cyclic buffer inside the FPGA. Its capacity is 8192 18-bit words (512 samples $\times$ 16 channels), enough to delay event processing by up to 290 $\mu$s. The input buffer must not be overwritten while the digital signal processor is busy with the previous events.

The DSP is highly parallelized, as shown in Fig. 9.8 (left), where the two independent cores are depicted. Each core has its own computing resources: MAC, division unit, and interface to one

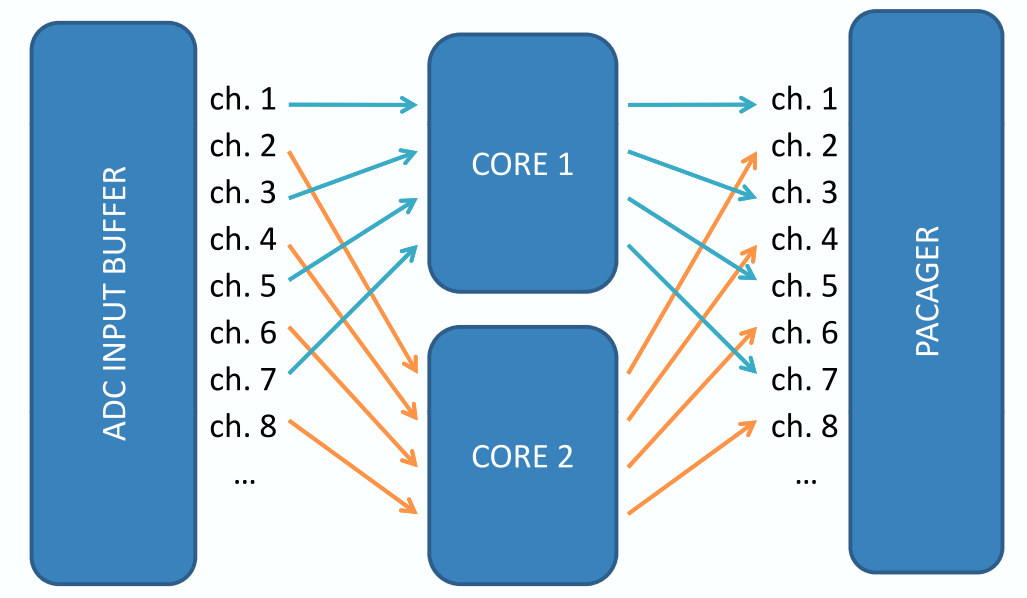

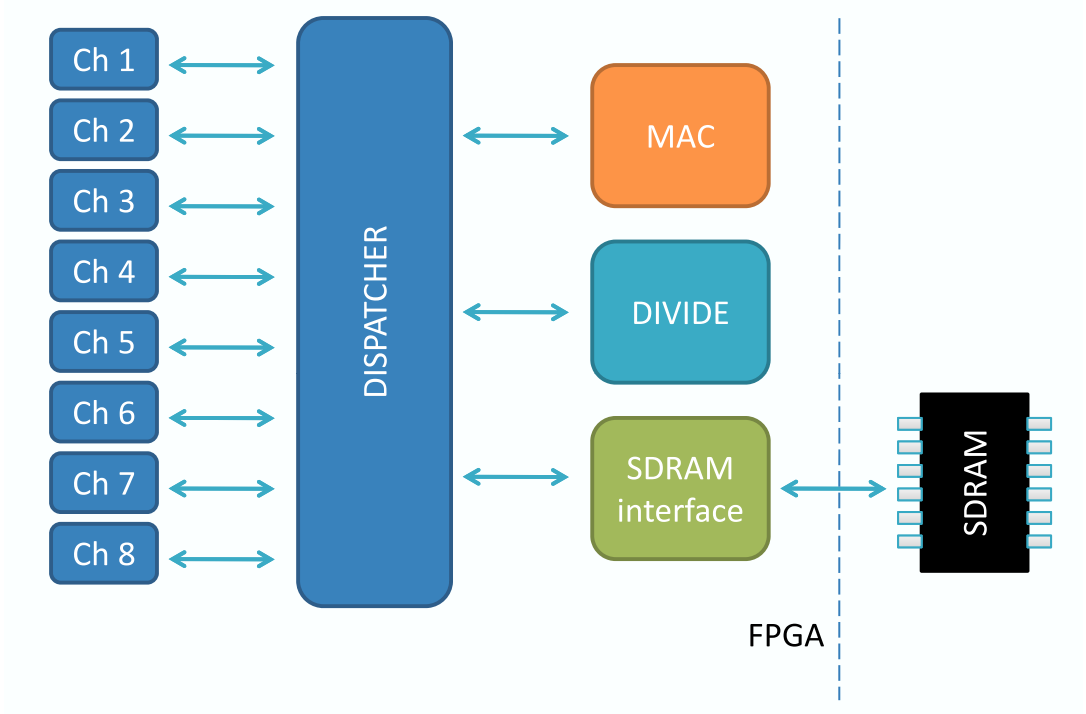

*Figure 9.8: Left: DSP cores. Right: DSP core details.*

SDRAM chip, and up to 8 channels are processed in parallel by each core, which utilizes its resources with priority given to earlier events. In this way, a MAC operation can be performed for one channel, while division and the fetching of coefficients are performed for the others.

There are two amplitude thresholds, THR_LOW and THR_HIGH (0 < THR_LOW < THR_HIGH). The former is used for zero-suppression. If the pulse amplitude is smaller than THR_HIGH (which is typically around 5 MeV), the parameter $t_0$ is fixed in the pulse shape fit at the trigger time value. This constraint improves the energy resolution for low amplitude pulses. The processed data are packed in the specified event data format and sent to a FIFO. For debug and calibration purposes, the ADC samples can be attached to reconstructed data for the chosen fraction of events.

The shaper-DSP modules and FAM modules are designed for a 9U VME module and the collector for a double-width 6U VME module. The modules are located in the same racks that are used for Belle's TKO crates (containing shaper-QT modules [1]). One VME crate contains 8-12 Shaper-DSP modules, one FAM and one collector module. In total, there are 52 VME crates.

#### 9.4.2.1 Performance

The preamplifier and shaper have nonlinearity less than 0.3% for output signals from preamplifiers up to 4 V, corresponding to a 10-GeV energy deposition.
If an energy deposition in the crystal exceeds 19 GeV—which can happen at the beam injection—the preamplifier becomes overloaded and does not respond to the input signals during a dead time of $t_{\text{dead}}[\mu\text{s}] \approx (E[\text{GeV}] - 19) \times 2.7$. Even for a 100-GeV energy deposition, this injection-overload dead time is only about $220\,\mu$s, which corresponds to about 2% at a 100-Hz injection rate.
For the most severe background conditions, we expect 30% occupancy. In this case, the FPGA algorithm can continue to process data for a trigger rate of up to 30 kHz.

### 9.4.3 Electronics Test

The performance of the Wave Form Analysis (WFA) circuitry and algorithm have been tested with the existing CsI(Tl) crystals [7]. Eight prototype modules of the new electronics were produced and installed into the ECL DAQ temporarily. The shaper and digitizer circuitry were mounted on a TKO-standard module to match the current detector electronics infrastructure. Digitized signals from the modules were sent to FINESSE cards on a COPPER [8] module. This module contained the FPGAs that implemented the signal signal processing algorithm of Sec. 9.4.2. The shape of the reference signal and noise correlation information were stored in an array uploaded to the FPGA memory. The results of the pulse fit to the amplitude, time and fit quality information were calculated and recorded as a 32 bit word (18 bits for amplitude, 12 bits for time and 2 bits for quality). To test the reliability of the FPGA algorithm, raw information was recorded together with fitted information for a portion of the events and the fitting algorithm was performed again offline. No online-offline difference was found in more than 50 million events.
During October 16-23, 2008, the ECL was operated in a configuration with 120 channels being readout by the new shaper-digitizer boards. About 965 pb$^{-1}$ of integrated luminosity were collected at the energy of the $\Upsilon(4S)$ resonance. Measured time distributions for low (5 MeV) and high (100 MeV) energy depositions in the crystals are presented in Fig. 9.9.
Incoherent and coherent noise is measured with and without beam. The incoherent noise without beam was 330–410 keV, depending on polar angle. With beam, the incoherent noise increases due to pile-up noise to about 500–600 keV, as shown in Fig. 9.10. The impact of the pile-up noise with the new electronics is reduced compared with the case of the Belle electronics. Closest to the beampipe, the pile-up noise suppression is about a factor of 1.5. The coherent noise within one module was about 70 keV.
The dependence of time resolution on a crystal's energy deposition is shown in Fig. 9.10 (right). It is about 100 ns for a 5 MeV energy deposition and only 3 ns at 1 GeV. The beam background has a uniform time distribution; applying energy-dependent cuts on the time can be used to suppress fake clusters. For example, a roughly 3 $\sigma$ cut can suppress the rate of fake clusters by a factor of 7, while maintaining an efficiency of about 97%. These values are in good agreement with estimates of the expected perfomance improvement.

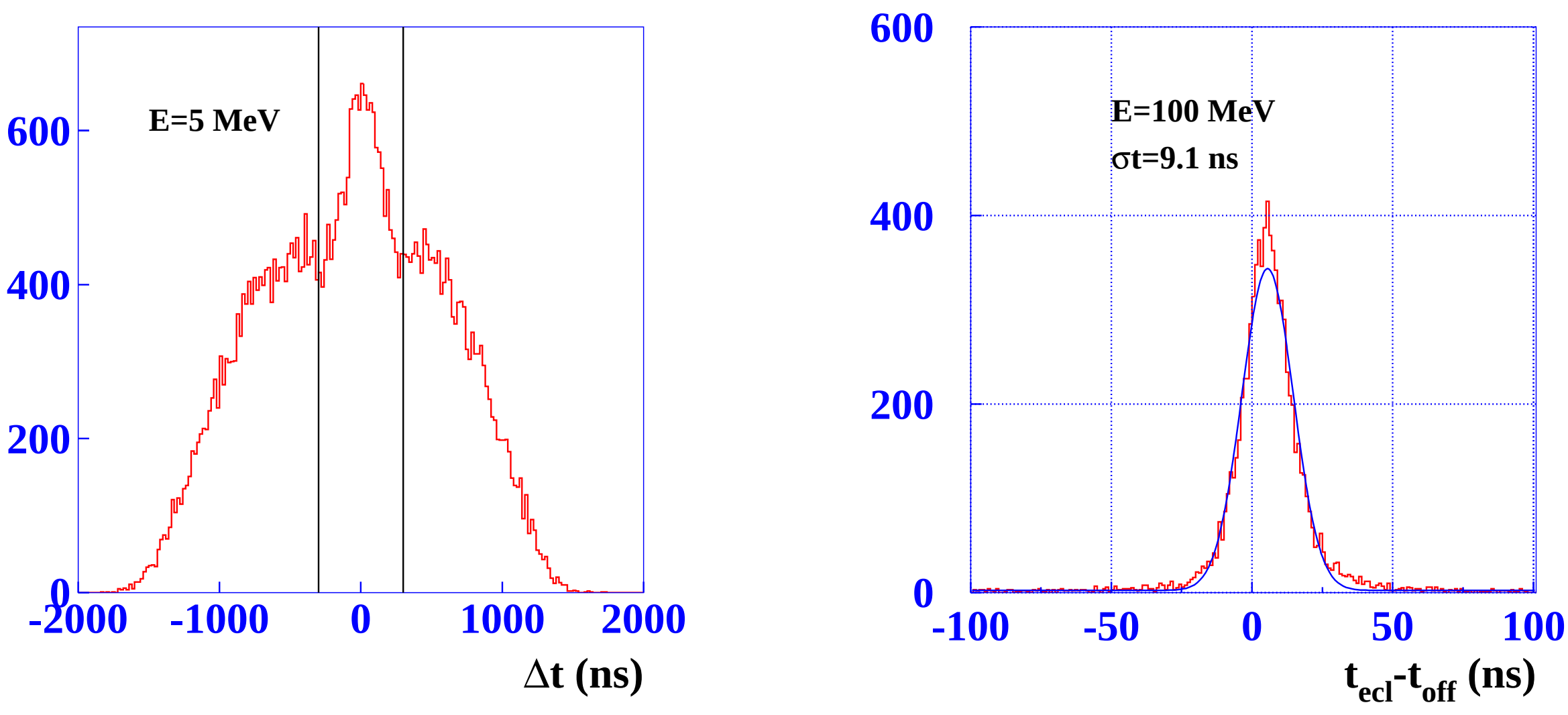

*Figure 9.9: Measured time distributions for 5 MeV (left) and 100 MeV (right) cluster energy depositions.*

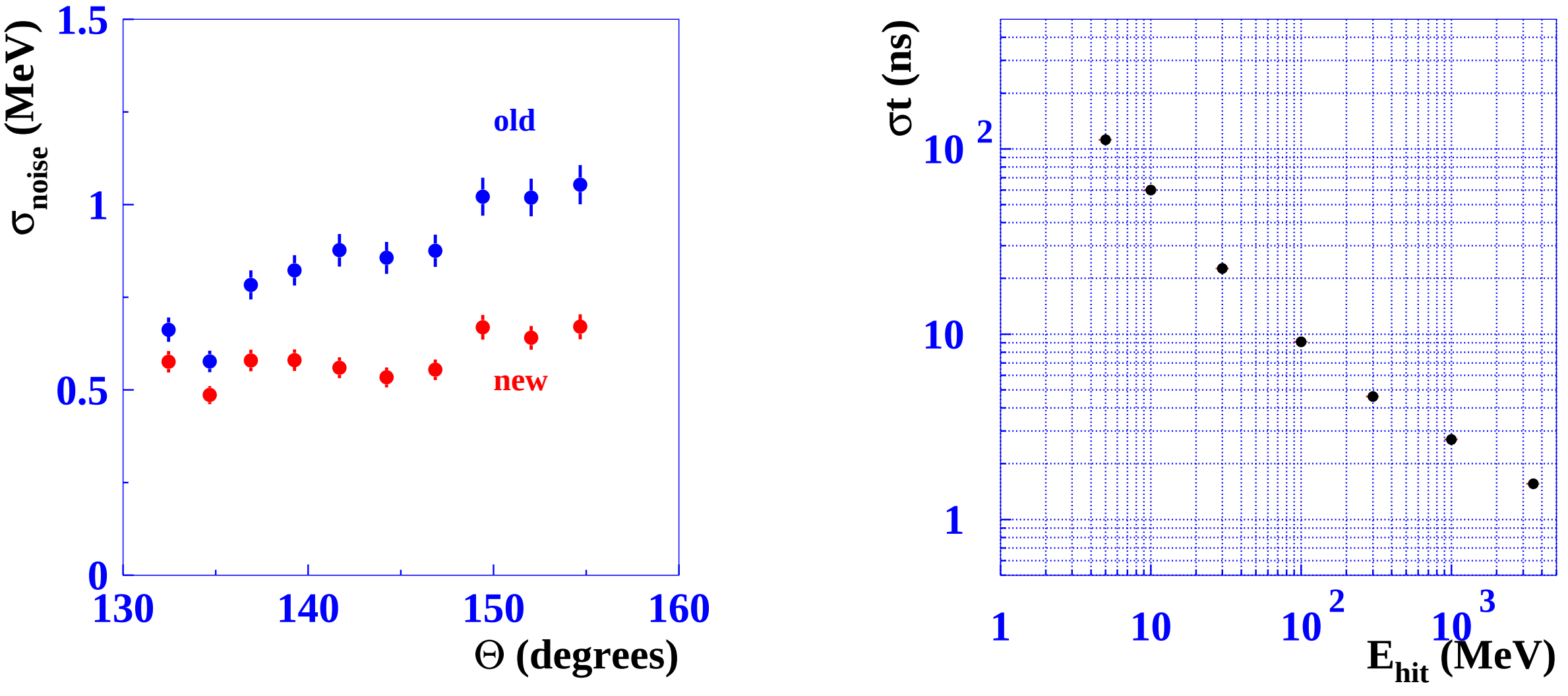

*Figure 9.10: Left: pile-up noise dependence on polar angle for new and old electronics. Right: measured time resolution as a function of single-crystal hit energy.*

### 9.4.4 Dead time during injection

SuperKEKB will operate with continuous injection (Ch. 2). For a brief interval after each injection pulse, the beam is excited and produces more background in the detector. Belle's DAQ copes with this by blocking triggers for about 4 ms after an injection pulse. However, at a 100-Hz injection in SuperKEKB, such a veto time would correspond to 40% dead time. To reduce this, the following veto scheme is proposed. The DAQ is blocked for $4\,\mathrm{ms} \pm 0.5\,\mu s$ *for the injected bunch only,* as it is the most copious source of background. For the other bunches, the DAQ is blocked for a much shorter time of about 150 $\mu s$.
We have studied the feasibility of this scheme in Belle using a special run without the injection veto. A signal from the backward end-cap calorimeter was used to study the energy deposition during injection.

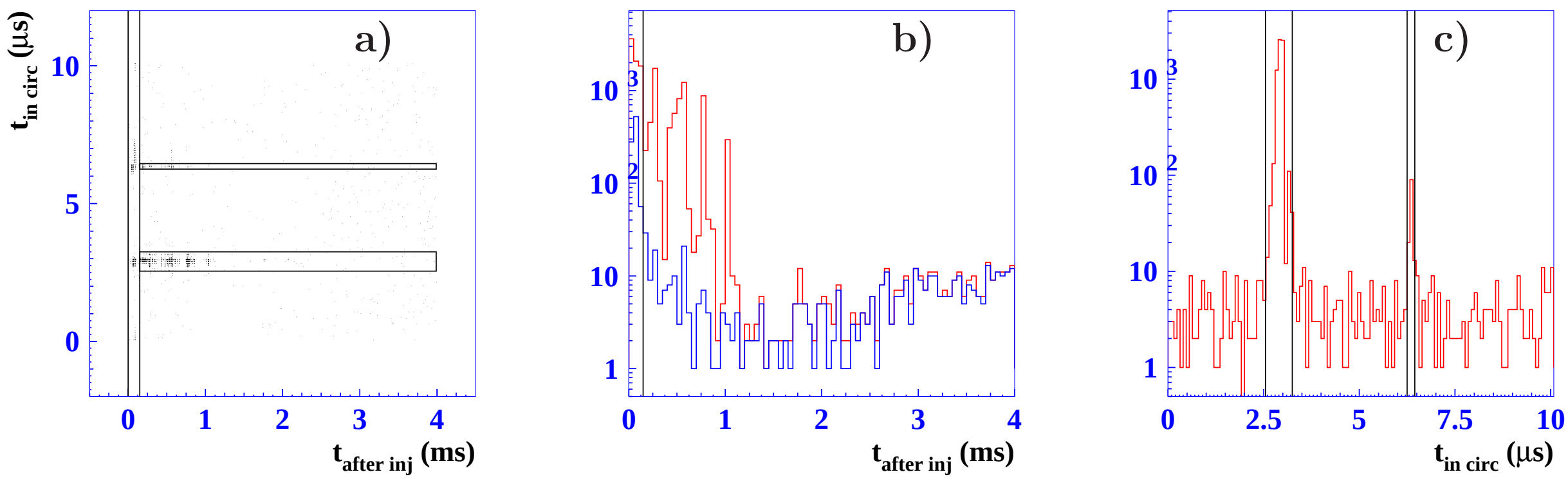

*Figure 9.11: Trigger time distributions for LER injection. a) Scatter plot of time within a revolution period vs. time after injection. b) Time-after-injection distribution (red: all events; blue: excluding the two horizontal bands in (a)). c) Time-in-revolution distribution for $t_{\mathrm{after\ inj}} <$ 150 $\mu$s.*

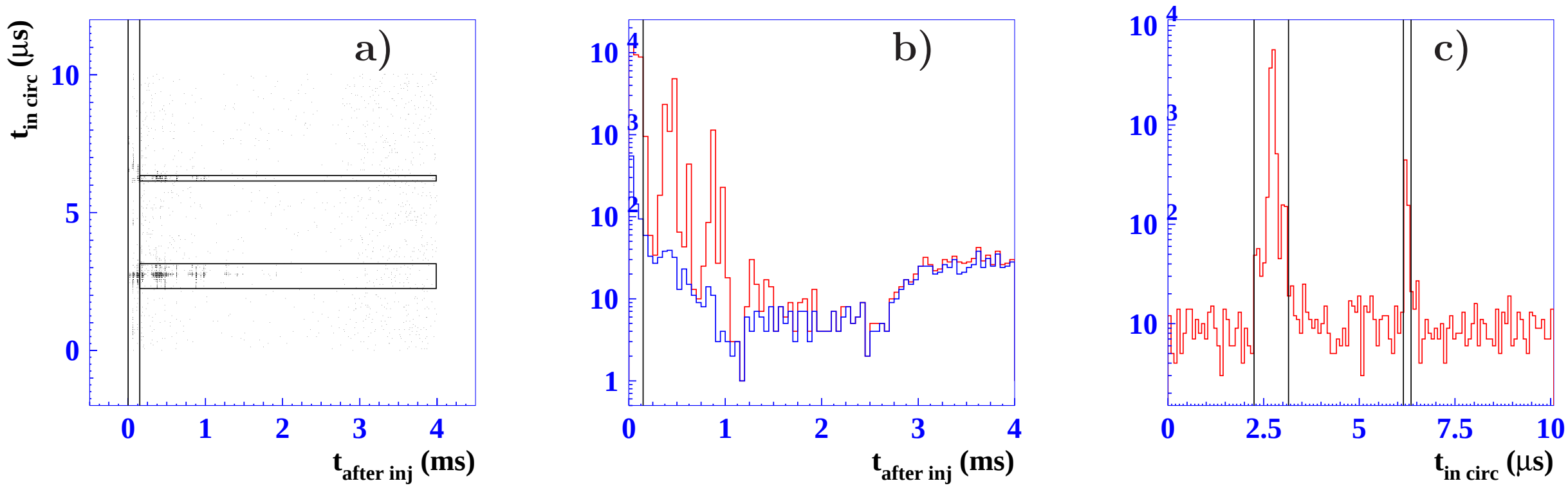

*Figure 9.12: Trigger time distributions for HER injection. a) Scatter plot of time within a revolution period vs. time after injection. b) Time-after-injection distribution (red: all events; blue: excluding the two horizontal bands in (a)). c) Time-in-revolution distribution for $t_{\mathrm{after\ inj}} <$ 150 $\mu$s.*

Figures 9.11 and 9.12 show the distributions of the trigger time within one revolution period ($t_{\mathrm{in\ circ}}$) and the trigger time after injection ($t_{\mathrm{after\ inj}}$). Most of the triggers come within 150 $\mu$s after injection. Later triggers are clustered around the time when the injected bunch passes through the interaction point. A less prominent cluster of late triggers appears at a $t_{\text{in circ}}$ time of 3.45 $\mu$s beyond the IP-crossing time. If we veto any post-injection trigger with $t_{\text{after inj}} <$ 150 $\mu$s or within the boundaries of these two clusters in Figs. 9.11(a) and 9.12(a), the dead time corresponds to an acceptable 5%.
The energy distributions for the time regions of Figs. 9.11(a) and 9.12(a) are shown in Figs. 9.13 and 9.14, respectively. More than 97% of the energy is deposited during first 150 $\mu s$. The energy deposited by the injected bunch is less than 2.7%, which would not cause preamplifier saturation. So if the injection-background conditions are similar in Belle and Belle II, we can accept this new veto scheme. As one improvement, we will develop a more sophisticated algorithm for the energy reconstruction in the possible presence of a tail from the injected bunch. The trigger channel should also be designed with efficient tail suppression after a fast signal to avoid energy shift during injection.

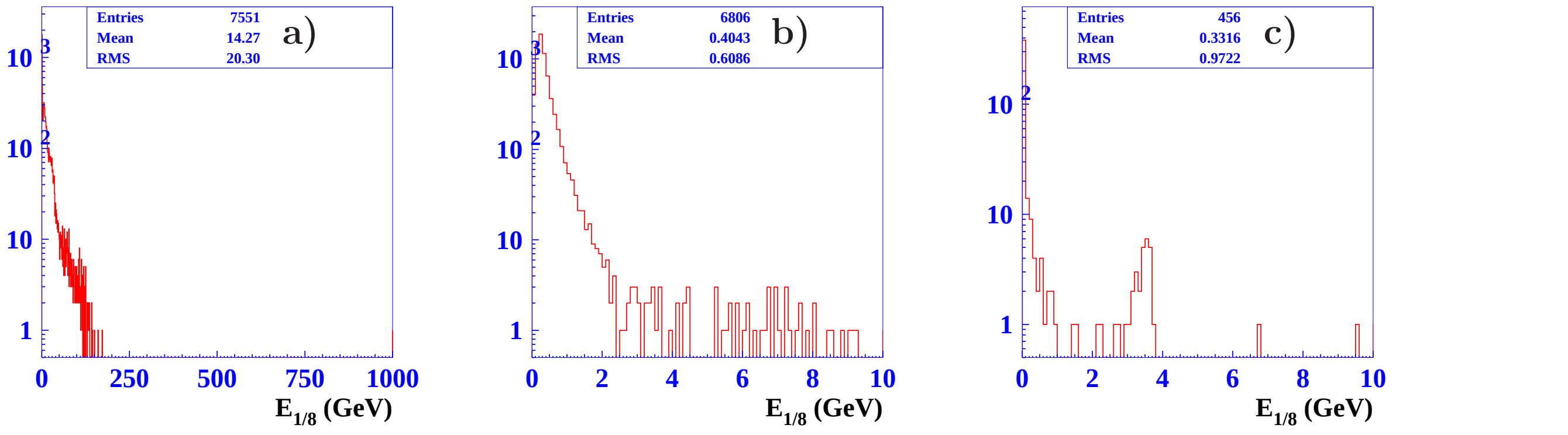

*Figure 9.13: Energy distribution in 1/8 backward end-cap for LER injection. a) for events in the vertical band of Fig. 9.11(a). (∼97% of lost energy); b) for events in either horizontal band of Fig. 9.11(a) extended leftward to $t_{\text{after inj}} > 0$ ($\sim$ 2.7% of lost energy); c) for events outside the vertical and horizontal bands of Fig. 9.11(a) (∼0.2% of lost energy).*

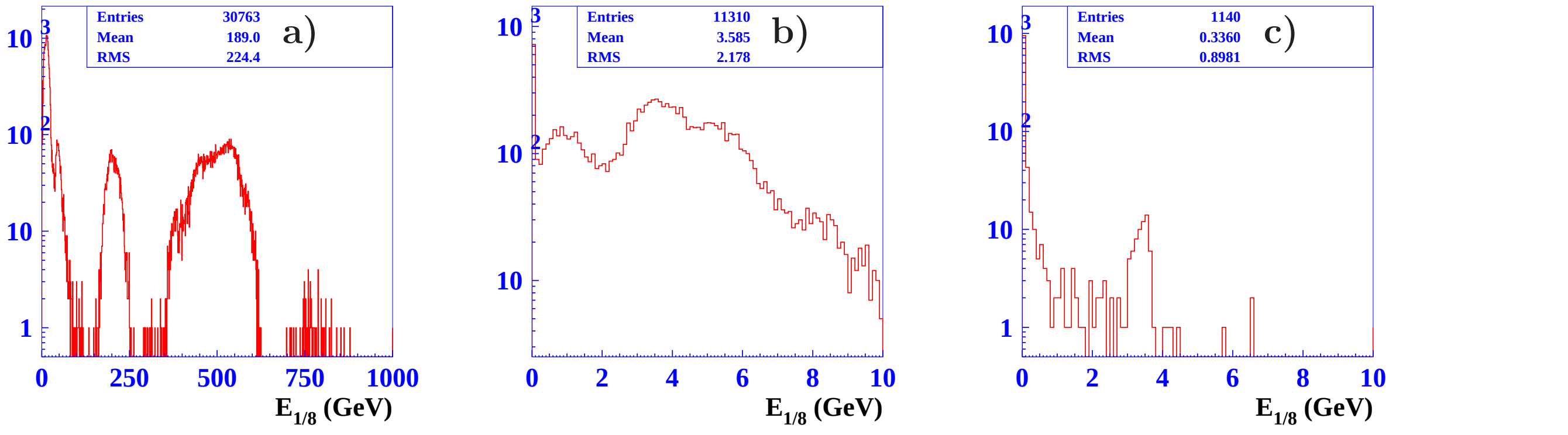

*Figure 9.14: Energy distribution in 1/8 backward end cap for HER injection. a) for events in the vertical band of Fig. 9.12(a). (∼99% of lost energy); b) for events in either horizontal band of Fig. 9.12(a) extended leftward to $t_{\text{after inj}} > 0$ ($\sim$ 0.7% of lost energy); c) for events outside the vertical and horizontal bands of Fig. 9.12(a) (∼0.01% of lost energy).*

### 9.4.5 Electronics Schedule

Design and production prototype evaluation of the shaper-DSP module will have to be finalized in 2010. During 2011-2013, the mass production and tests of the required 600 modules will be done.
The collector module will be designed and tested during 2010-2012, following the development of the readout DAQ modules. The production of 55 modules is expected in 2013. Module installation will be finished by the end of 2013. The DAQ tests will start in the beginning of 2014.
To cope with the PIN-PD leakage current increase, the bias voltage resistors, which are 100 MΩ combined, will be replaced by lower-resistance ones in the endcap region.

## 9.5 Options for the endcap upgrade

A drastic way to improve the performance of the ECL is the replacement of the slow CsI(Tl) crystals with faster ones, at least in the endcaps. Three options of such an endcap upgrade are presently under consideration. All options were examined by a special Review Committee assigned by the Executive Board of the Belle II collaboration. The following conclusions of this Committee were provided for each of these options.

1. Pure CsI crystals are used instead of Tl doped crystals [7]. Scintillation light readout is made with vacuum photopentodes. The existing endcap mechanical structure is kept. This option has been explored for the last five years and now it is well prepared for construction. Reducing the number of photosensors per crystal to one from the existing two for PIN-PD readout represents a weakness of this proposal, since there is zero redundancy. Furthermore, the long-term stability of the photopentode has yet to be confirmed. Pure CsI plus multiple APDs can provide redundancy, though further signal-to-noise ratio improvement is needed.

2. PWO2 (improved lead tungstate) plus Avalanche Photodiodes (APD): this is being actively studied by the PANDA experiment [9] and could be promising. Production costs and the need for cooling to achieve good performance is a concern and requires further study. These much denser crystals could provide better forward/backward $\pi^0$ detection at high momentum and would be an improvement in basic ECL performance.

3. $Bi_4Si_3O_{12}$ (BSO) scintillating crystal plus APD: a potentially interesting option, though rather immature in terms of detailed studies. Cooling is not required, though costs and readiness for large-scale production are unclear. Again, it would represent an improved gamma isolation and is worth considering, if a strong improved physics performance case can be made.

In the latter two cases, in addition to the need for a compelling physics case, further and detailed mechanical support studies and engineering are needed.
Based upon these conditions, the Review Committee formulated the following recommendations.

> *Considering this state of development, and the physics and economic situation, our recommendation is therefore the scenario to defer upgrade of the ECL endcaps until a solid physics reach case has been made and the degradation due to backgrounds mandates an upgrade. During this time innovation can continue. At such a later point in time any mature technology, with solid costing and delivery schedule, that* *justifies physics return on investment, should be earnestly evaluated. We note in summary that this timescale could be accelerated if the backgrounds observed end up being significantly higher than those considered in the studies presented.*

### 9.5.1 Pure CsI for the endcap

#### 9.5.1.1 Primary motivation

A mature option to use pure CsI crystals of the same shape and size as the presently-used CsI(Tl) counters is proposed for the endcap. An advantage of this crystal choice is a short scintillation decay time and moderate cost as compared to other scintillation crystals. Since the physical properties of pure CsI are the same as CsI(Tl), the present sizes of the calorimeter elements as well as the mechanical structure may be kept.
However, the light output of pure CsI crystals is approximately one tenth that of the doped crystals. Thus, to keep the energy noise equivalent value at the same level, one needs to use photodetectors with internal gain. A suitable solution is to use vacuum photopentodes (PP), i.e., photomultiplier tubes with three dynodes. Since the emission wavelength of pure CsI is about 300 nm, the photosensor should be UV sensitive. Such a 2-inch diameter device, with a low output capacitance of $C \approx 10\,\mathrm{pF}$, has been developed recently by Hamamatsu Photonics. The low output capacitance is attractive because the electric noise depends on that value. These newly developed PPs have a quantum efficiency of about 20–25% at 300 nm and an internal gain factor of 120–200 in zero magnetic field. We tested several pure CsI crystals of the same size as those presently used in the Belle endcap calorimeter.
The obtained preamplifier's equivalent noise charge (ENC) is about 980 electrons, while the PP output anode signal varies from 20000 to 30000 electrons per MeV at the PP anode. The ENC corresponds to an equivalent noise energy (ENE) of 40-60 keV. Taking into account the decrease of the PP gain due to the magnetic field, we can expect an ENE of about 120-200 keV in the Belle II 1.5 T magnetic field.
Shorter scintillation decay times and matched, shorter shaping times of the electronics suppresses the pile-up noise by a factor of about 5.5 in comparison to the present endcap calorimeter. Estimated pile-up noise behavior as a function of background for the current and upgraded electronics is shown in Fig. 9.15. The fake photon rate is suppressed by a factor of more than 100 due to improved timing provided by the new electronics. Thus, even in the case of 20 times larger background, the resulting fake rate is expected to be lower than in the current barrel calorimeter.
The advantages of this option follow.

- Low pile-up noise and good energy and spatial resolution can be obtained while maintaining efficient fake cluster suppression.
- The physical characteristics and the radiation length of the crystals are the same as for present CsI(Tl) so that the same granularity of the calorimeter can be kept and the same container can be used for the endcaps.
- Pure CsI crystals of acceptable quality can be produced by three manufacturers: Kharkov (Ukraine), Saint Gobain (France), and Beijing (China). The required number of the crystals can be produced within two years following mass production preparation.
- A large amount of R&D work on pure CsI counters with PP readout has already been performed.

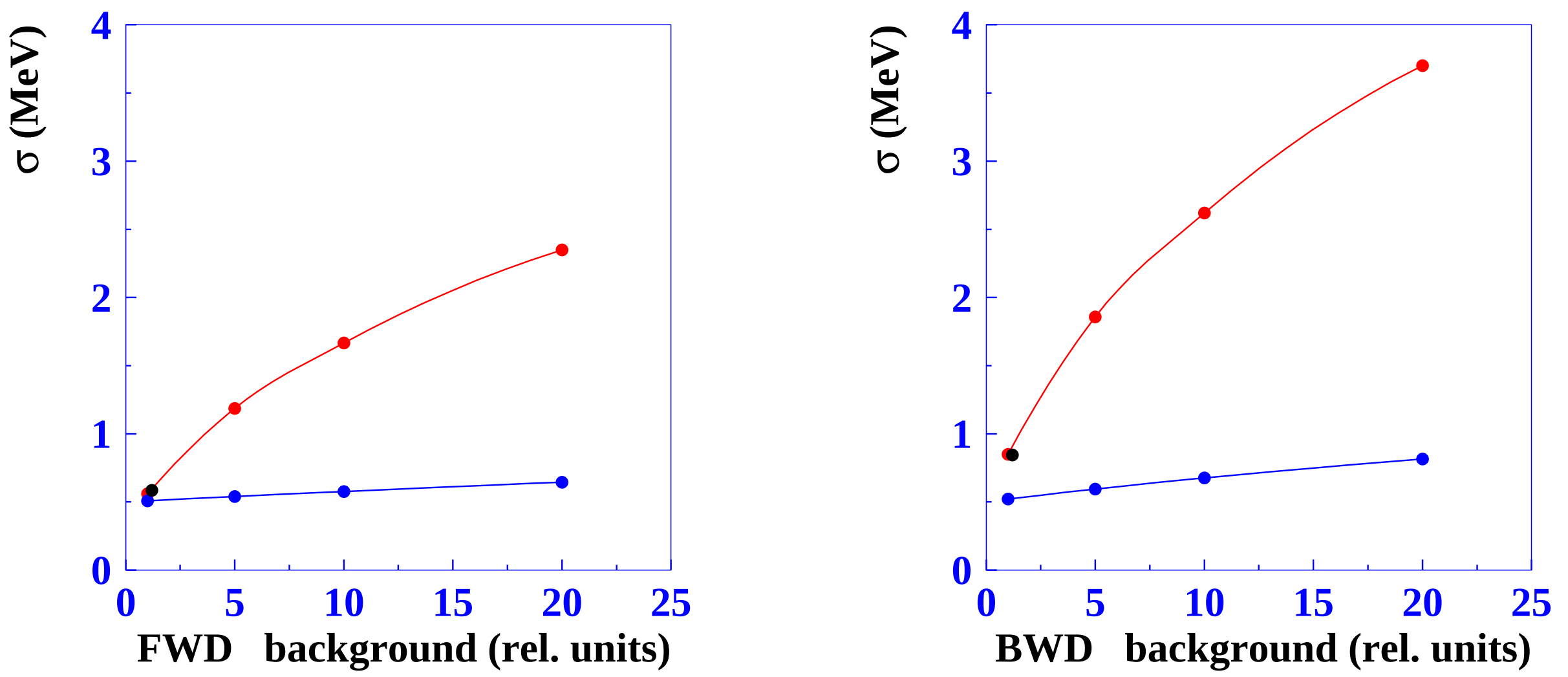

*Figure 9.15: Dependence of the pile-up noise per crystal for the current (red) and modified (blue) endcap calorimeters. The background at Belle (for a luminosity of* $10^{34}\,\mathrm{cm}^{-2}\mathrm{s}^{-1}$*) is taken for unity.*

There are disadvantages:

- Usage of a single photodetector per counter represents a single-point failure and therefore less robustness in comparison with the present scheme, employing two PIN photodiodes per counter.
- The PP gain factor depends on the magnetic field and calibration is needed more often. This gain factor can also vary slightly with background conditions.
- The PP depth of 58 mm requires rebuilding of the supporting structure inside container.

Each CsI crystal is wrapped by one layer of a white porous 200-$\mu$m Teflon sheet and placed in a 40-$\mu$m aluminized mylar envelope. The light of each crystal is read by a 2-inch photopenthode (PP) that is fixed to the large-end face of the crystal with optical grease.
The counters are installed in the existing endcap containers. Due to the increased thickness of the new PP sensor, with respect to the low-profile of the existing PIN photodiodes, the assembled counters are longer (360 mm) compared to the present counters (319 mm). Therefore, the support structure of the containers must be modified. The final design of the counter and location of the electronics (connection board and HV feed) are coupled to the supporting structure design and will be done together.
A first version of the preamplifier has been developed. The input transistor is the same as in the existing Belle calorimeter. The modification includes a change of the output cascade to the differential transmitter. The signal from the preamplifier is sent to the intermediate electronics via the existing twisted pair cable. The power consumption of the new preamplifier and voltage divider is about 190 mW/channel, similar to the present 140 mW/channel.
The downstream electronics will be built using the same basic architecture as for the barrel (Sec. 9.4). The shaping and digitization part is proposed to be implemented on VME boards. Each VME board contains 16 channels. Each channel includes a differential receiver, CR-(RC)$^4$ shaper with shaping time $\tau = 30$ ns, and two 14-bits fast flash ADCs. The use of two ADCs provides an effective 18-bit digitization for the full dynamical range. Digitization of the shaper output signals will be performed using a common 43-MHz reference clock. The digitized data are processing in FPGAs to obtain the amplitude, time and quality bits for each cluster candidate. The feature-extracted information from 8 (10) boards are aggregated by the collector module and sent to the COPPER with ROCKET-IO. In total, the endcap electronics includes 144 (80+64) VME boards located in 16 (8+8) crates.

#### 9.5.1.2 Study of the radiation hardness

We performed radiation hardness studies of pure CsI crystals from the Kharkov, Saint Gobain and SIC companies. Most of the crystals show no essential degradation up to 10 krad of gamma irradiation and reactor-neutron irradiation with the flux of up to $10^{12}\,\text{cm}^{-2}$. The results for $\gamma$ irradiation are presented in Fig. 9.16. Since one of the crystals showed significant degradation, an additional test of the radiation hardness should be performed.

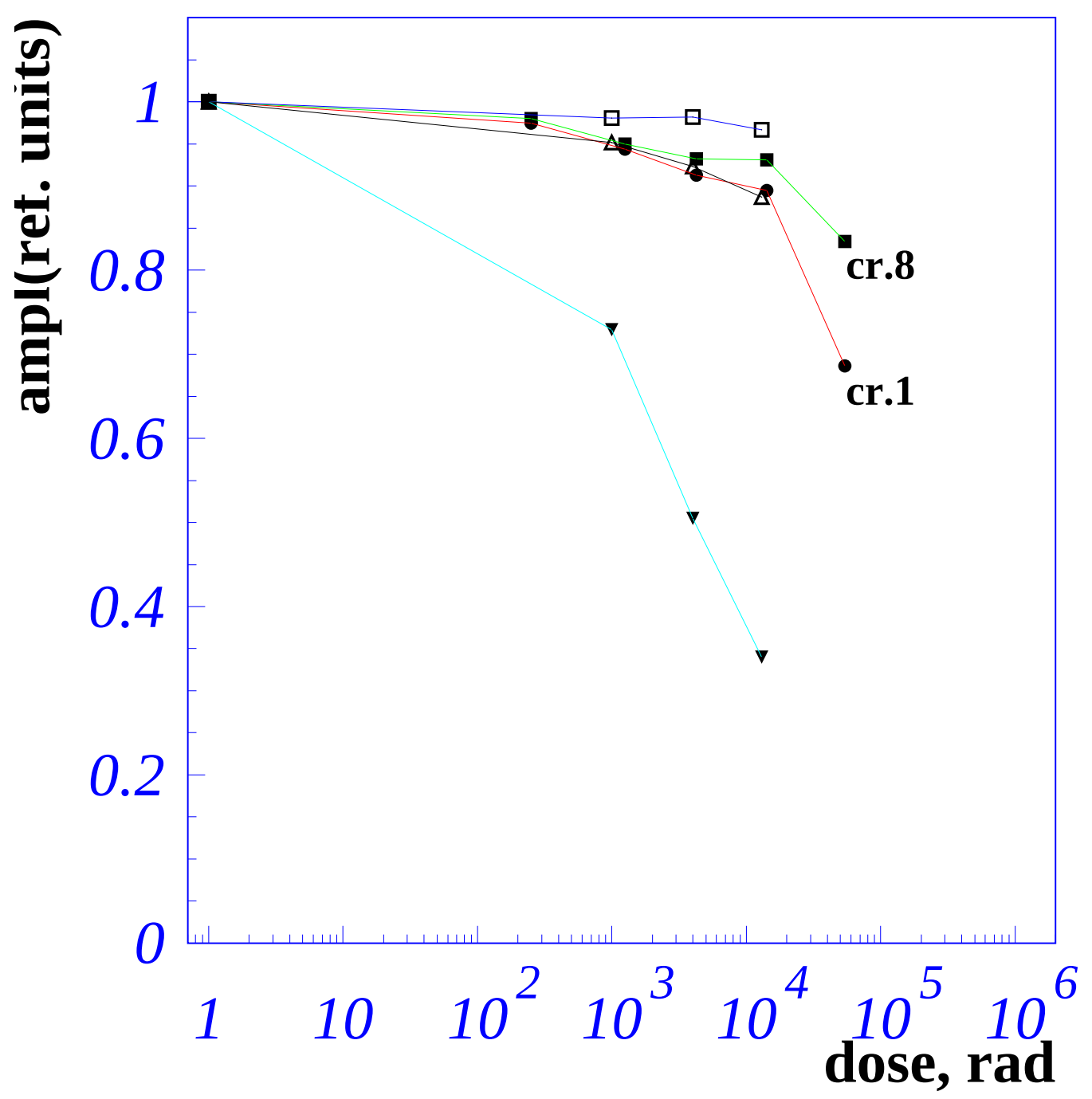

*Figure 9.16: Radiation hardness measurement results of Kharkov crystals [10].*

#### 9.5.1.3 Tests with prototype and other R&D on pure CsI

An accelerator measurement of 20 pure CsI crystals coupled to PPs and readout electronics providing waveform analysis was carried out at the BINP photon beam. The obtained energy resolution (Fig. 9.17) is consistent with both present calorimeter energy resolution and MC prediction. A time resolution better than 1 ns for deposited energies greater than 20 MeV has been obtained [11].

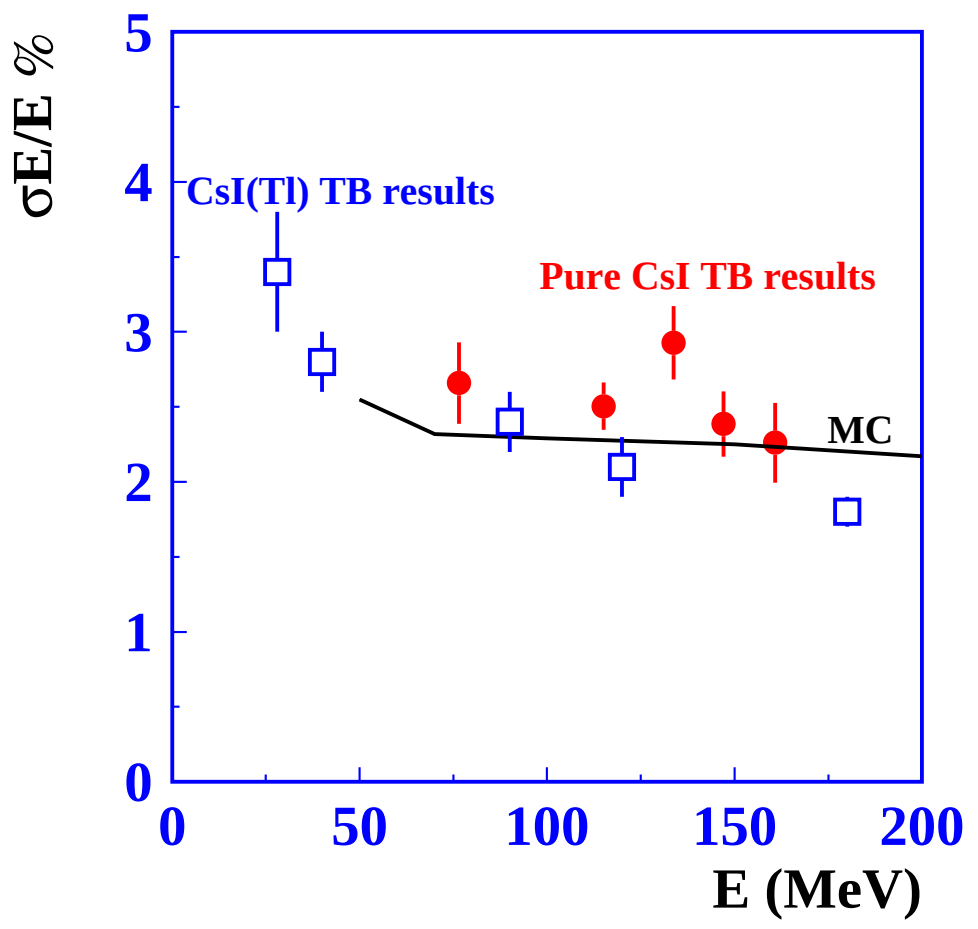

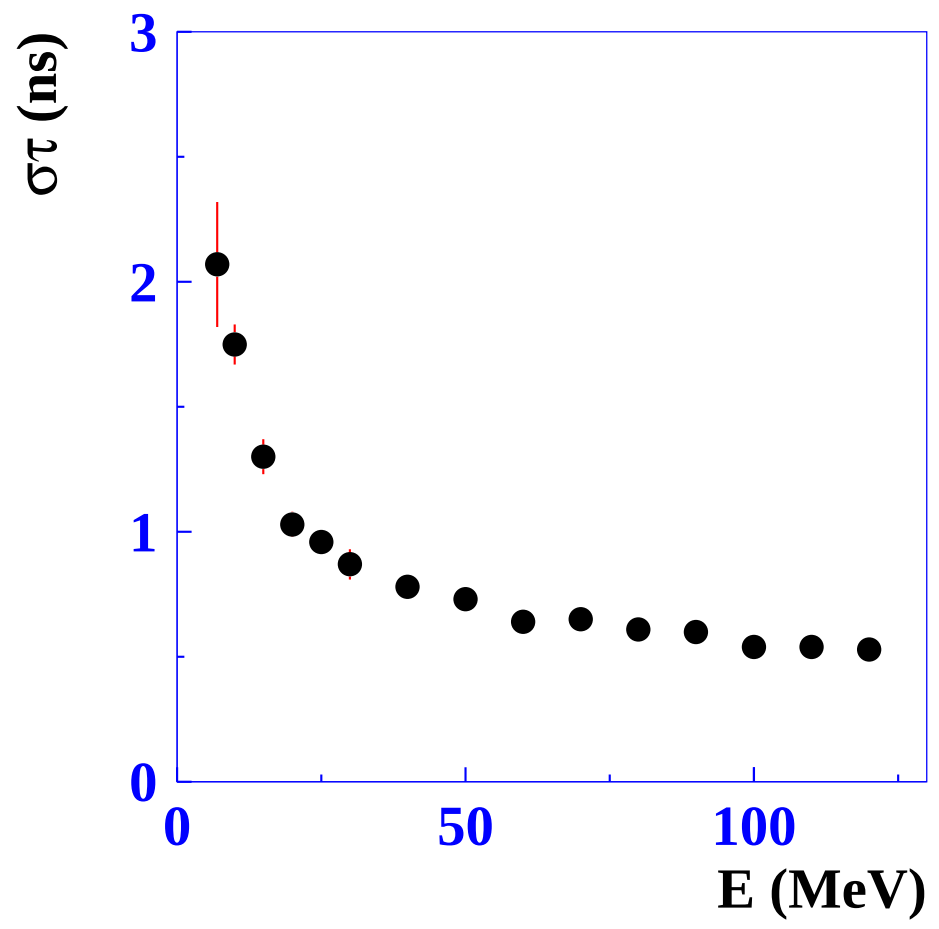

*Figure 9.17: Energy and time resolution for photons obtained in the test beam measurements at the BINP photon beam line.*

We have also carried out a direct comparison of the pile-up noise for the counter based on pure CsI and one of the CsI(Tl) counter from the Belle detector. The pile-up noise was emulated using gamma rays from a $^{60}$Co radioactive source. The background intensity varied by changing the distance between the cobalt source and the crystals. The pile-up noise was calculated from the width of the test pulse signal. Figure 9.18 shows the effective noise of the pure CsI counter as a function of that for the CsI(Tl) counter.

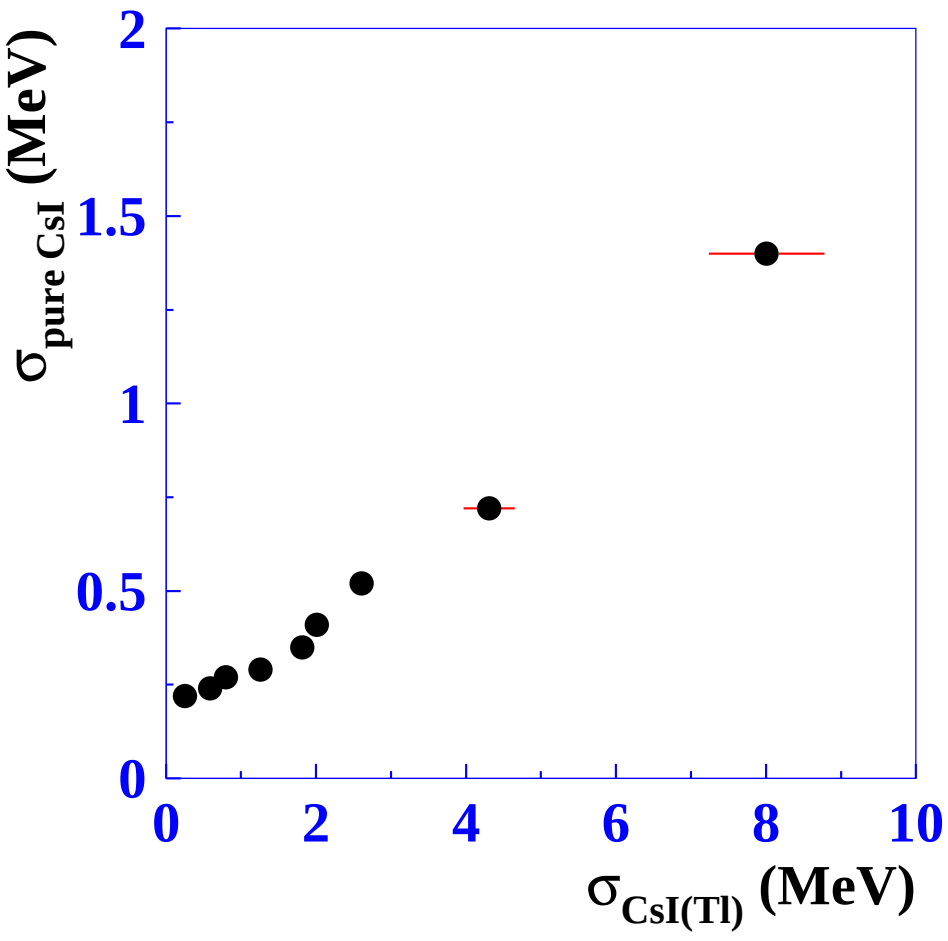

*Figure 9.18: The pile-up of pure CsI and new electronics counter and that for the standard Belle CsI(Tl) counter readout using the present MQT-TDC electronics.*

The gain factor reduction of photopentodes in a magnetic field have been measured. The results are shown in Fig. 9.19. When subjected to a magnetic field, the gain factor of the photopentode decreases but remains reasonably high.

An aging test of the PPs is being carried out. We have studied PP sensitivity depending on the integrated anode charge. The signal from a pure CsI crystal was used as a reference light source. For calibration we measure the maximum of the cosmic particle spectrum as a function of time. The result of this study is shown in Fig. 9.20. A blue LED was used to provide the illumination to a rate about 100 times larger than we expect at Belle II. This intensity corresponds to anode current of 35 $\mu$A without magnetic field and about 9 $\mu$A with magnetic field. We saw a sensitivity increase of 10% after 15 C of integrated anode charge (with 3 C corresponding to one year of Belle II operation with magnetic field). With further integrated charge, the increase appears to saturate. There is about a 2% increase of PP gain during intense illumination that should be studied more carefully. In a 1.5-T magnetic field, the same sensitivity increase of about 10% and absence of the degradation was observed.

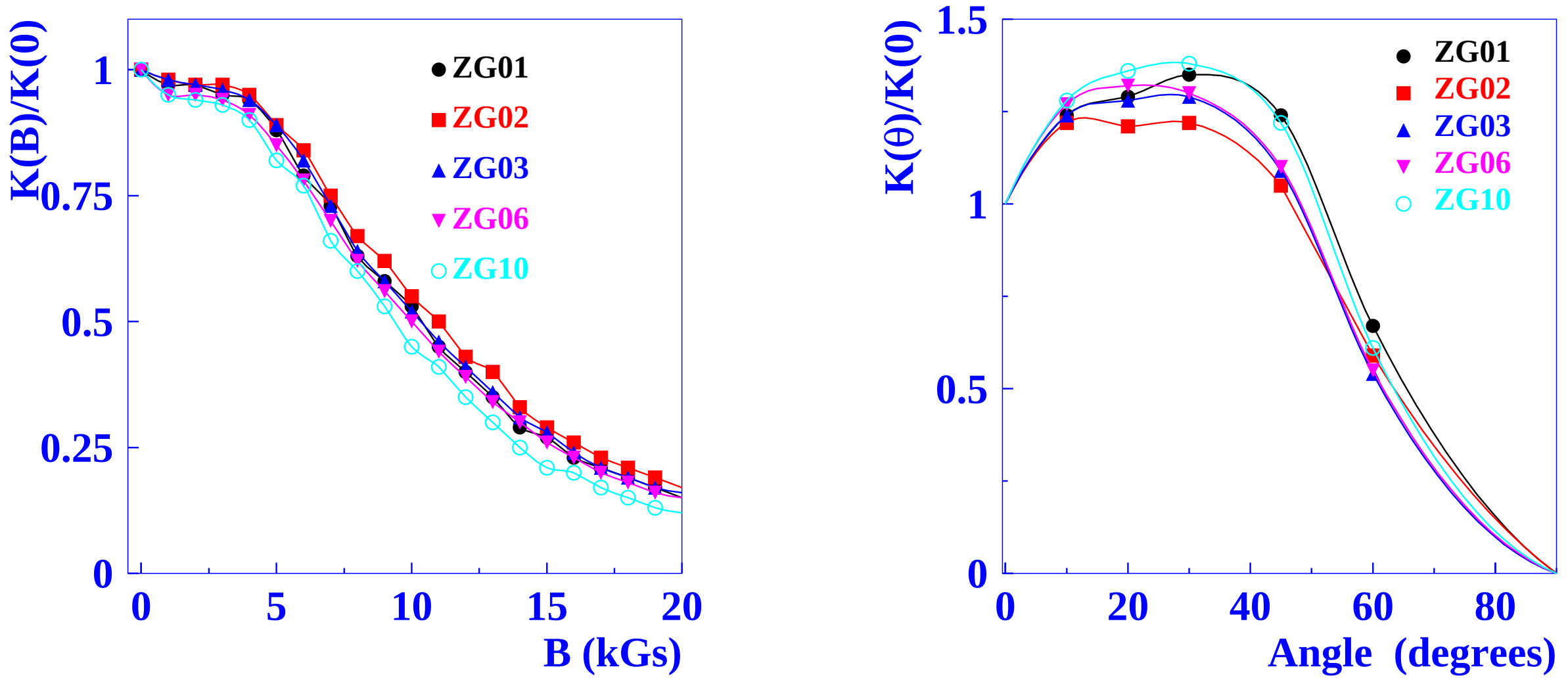

*Figure 9.19: Left: gain factor dependence on the magnetic field along the photopentode axis. Right: gain factor dependence on the angle between the 1.5-T magnetic field direction and the photopentode axis.*

#### 9.5.1.4 Simulation study

Use of faster-scintillator crystals in the endcaps allows us to suppress the background clusters more efficiently and reduce the pile-up noise smearing. To estimate the calorimeter performance for different options, we chose to study $B^- \to \tau^- \bar{\nu}$ decay since this mode is very sensitive to the missing energy accuracy.

For the background simulation, we used the Belle GEANT3 program package with the overlaid-background component ("addbg") scaled for different options and a modified pile-up noise contribution.

Simulations of the signal MC and generic MC were performed for the cases shown in Table 9.1. The tagged $B$ was reconstructed in the $\bar{D}^0\pi^+$ mode with $\bar{D}^0 \to K^+\pi^-$. The requirements $M_{\rm bc} > 5.27$ GeV/$c^2$, $|\Delta E| < 50$ MeV, and $|M_{D^0} - 1.865| < 15$ MeV/$c^2$ were applied for tagged $B$ selection. The $\tau^-$ lepton was reconstructed in the $\pi^0\pi^-\nu$ mode. To select signal events, the missing energy plot was analyzed. The missing energy includes a sum of the ECL cluster energy that excludes the clusters matched to the reconstructed tracks and the photons matched to $\pi^0$. Figure 9.21 shows the missing energy distribution for the signal and generic MC in case of a 20

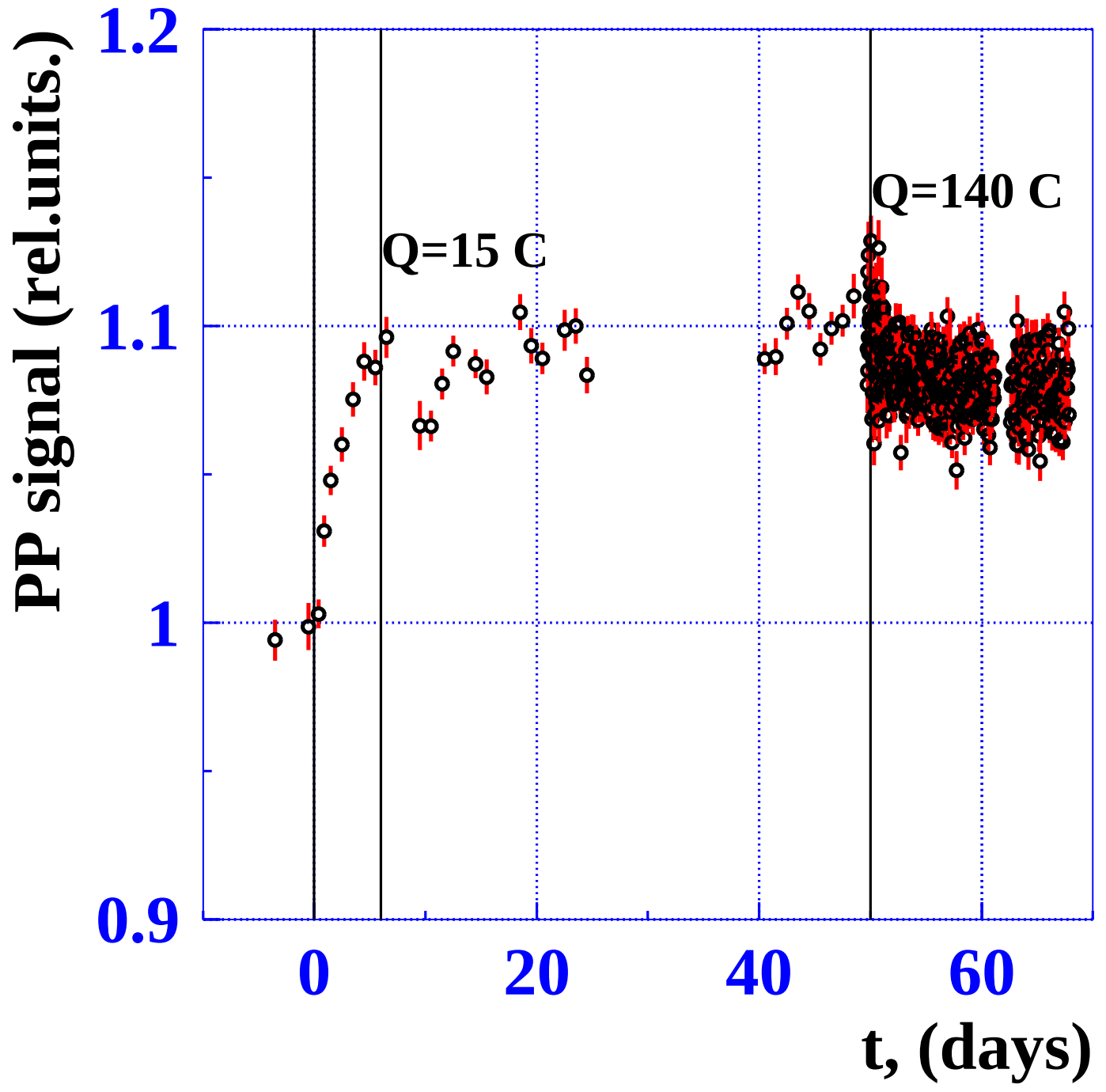

*Figure 9.20: The PP sensitivity dependence on time during the illumination of the light. Three days correspond to about 1 year Belle II operation.*

| | | barrel | | end cap | |
|---|---|---|---|---|---|
| | | addbg | $\sigma_{pile}$ | addbg | $\sigma_{pile}$ |
| 1 | Present calorimeter $\times 1$ | 1 | $\sigma_{pile}(\theta)$ | 1 | $\sigma_{pile}(\theta)$ |
| 2 | Present calorimeter $\times 10$ | 10 | $\sigma_{pile}(\theta) \times \sqrt{10}$ | 10 | $\sigma_{pile}(\theta) \times \sqrt{10}$ |
| 3 | New electronics $\times 10$ | 2 | $\sigma_{pile}(\theta) \times \sqrt{5}$ | 2 | $\sigma_{pile}(\theta) \times \sqrt{5}$ |
| 4 | Pure CsI $\times 10$ | 2 | $\sigma_{pile}(\theta) \times \sqrt{5}$ | 0 | $\sigma_{pile}(\theta)/\sqrt{3}$ |

*Table 9.1: Simulated overlaid-background and pileup noise options; $\sigma_{pile}(\theta)$ is the pileup noise measured at a luminosity of $2 \times 10^{34}$ $cm^{-2}s^{-1}$.*

MeV cluster energy threshold. As expected, the pure-CsI case has the narrowest signal peak (i.e., best resolution) for the missing energy.
To estimate the sensitivity of the method, we fit the generic MC distribution with the sum of background $B(E)$ and signal $S(E)$ functions: $F = N_S S(E) + N_B B(E)$. We then studied the dependence of the $N_S$ statistical error on the cluster energy threshold. An example of such fit and the dependence of the sensitivity on the threshold energy are shown in Fig. 9.22. The results of the study show that the replacement of the end-caps with pure CsI crystals will allow us to improve the sensitivity for $B \to \tau\nu$ decay by $\sim$25%, assuming ten times larger background than in Belle. This difference in sensitivity is magnified with larger background levels.

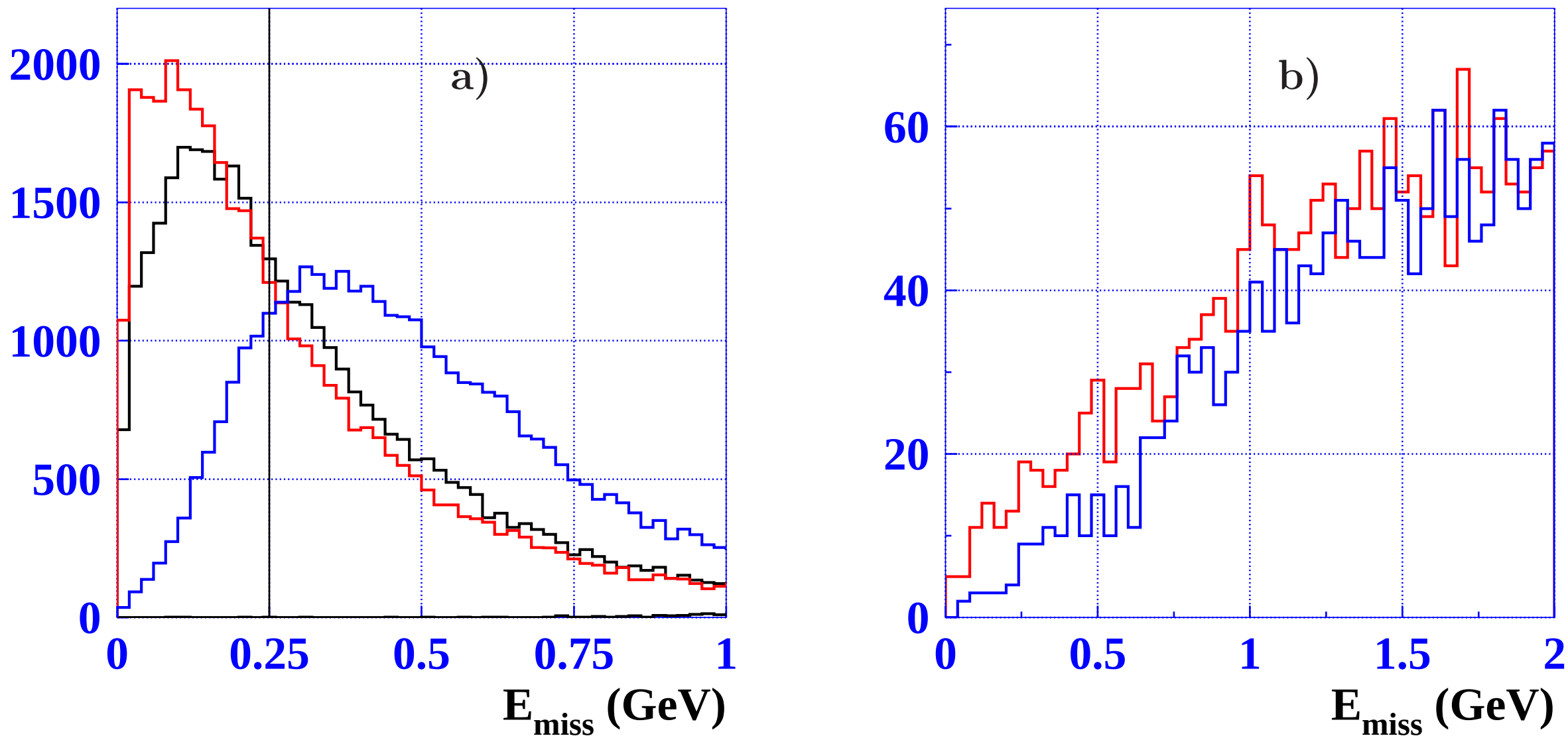

*Figure 9.21: Missing energy distributions: a) for the signal MC; b) for the generic MC. Black: case 1; blue: case 3; red: case 4.*

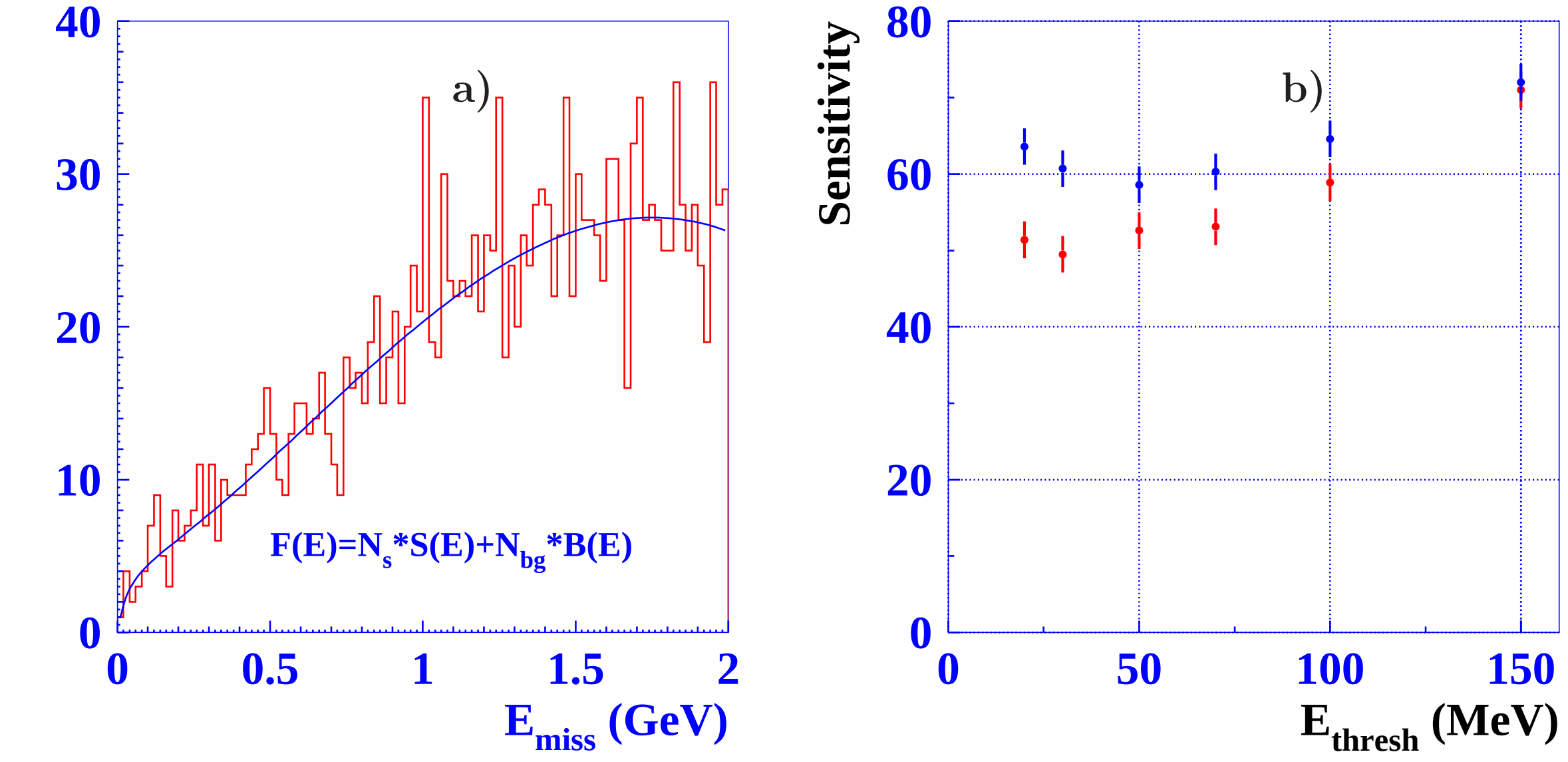

*Figure 9.22: a) Fit of the missing energy distribution for case 4 ($E_{thresh}$ = 20 $MeV$). b) Dependence of the sensitivity on the threshold energy.*

### 9.5.2 The APD readout option

#### 9.5.2.1 Motivation

Thanks to the recent semiconductor photon sensor development, large area (1 $cm^2$) APDs have become available. Such semiconductor photosensors are naturally insensitive to the magnetic field. The Hamamatsu S8664-series APD is widely used, for example, in the PWO calorimeters for CMS experiment and the PANDA experiment. In the catalog for this series, the Hamamatsu S-8664-1010 device has the largest sensitive area of $1\times1$ $cm^2$. Since it is a compact device, we can put a few pieces on the scintillator crystal readout surface to realize a redundancy for robustness in the practical operation of the calorimeter.

#### 9.5.2.2 Performance

The electromagnetic shower development depends only on the choice of the scintillating crystal. Therefore, the energy and position resolutions for high energy photons and neighboring-shower separation are independent of the photosensor selection. In the $\Upsilon$ energy region, the photon energy is typically a few hundred MeV. The corresponding energy resolution is predominantly affected by the equivalent noise energy of each channel. Attaching one S8664-1010 sensor on an actual-size pure CsI crystal, a cosmic ray test has been performed. The resultant equivalent noise energy is described in Sec. 9.5.2.2.2.

**9.5.2.2.1 Advantages** Since the Hamamatsu S-8664-1010 APD is a compact device, one scintillator crystal can be equipped by 2–4 APDs on its readout surface. This redundancy gives us the operational robustness in order to avoid an entirely dead channel due to APD or signal-line failure. Its compactness enables us to use the same (or quite similar) preamplifier casing and counter-fixing mechanical structure inside the endcap calorimeter container as in Belle. This minimizes the needed change of the mechanical structure and other infrastructure, especially when combined with pure CsI crystal.

**9.5.2.2.2 Concerns** The device capacitance of the APD, 270 pF, is larger than that for the photopentode (10 pF) or PIN-photodiode (80 pF), so that the noise due to the device capacitance is larger than for these alternatives. To estimate the equivalent noise energy of the APD option with pure CsI, we constructed a prototype counter by attaching one APD on a real-size crystal and performed a cosmic ray test. The prototype counter was placed inside a thermostatic box to maintain its temperature at 25°C and the applied high voltage was 455 V (15 V below its breakdown). The preamplifier was based on the existing one for Belle's PIN-photodiode. The preamplifier output signal was fed to the CAMAC module equipped with a $\tau$=30 ns shaper circuit and 43 MHz 12 bit FADC. The obtained pulse height spectrum is shown in Fig. 9.23. The cosmic particles transversely penetrating the crystal are triggered, so the energy deposit is approximately 30 MeV for passage through 5.5 cm thick CsI. The corresponding average pulse height is 120 FADC counts. The noise level is 8 FADC counts, based on the width of a Gaussian fitted to the pulse height spectrum for random triggers. Hence, the estimated equivalent noise energy is 2 MeV. This is significantly higher than the requirement of $\sim$0.6 MeV. The possible test to improve it is mentioned in Sec. 9.5.2.3. Note that the S8664 APD has a lower quantum efficiency (QE) of $\sim$35% at the wavelength of pure CsI (330 nm), compared to its QE of 80% at 500 nm.

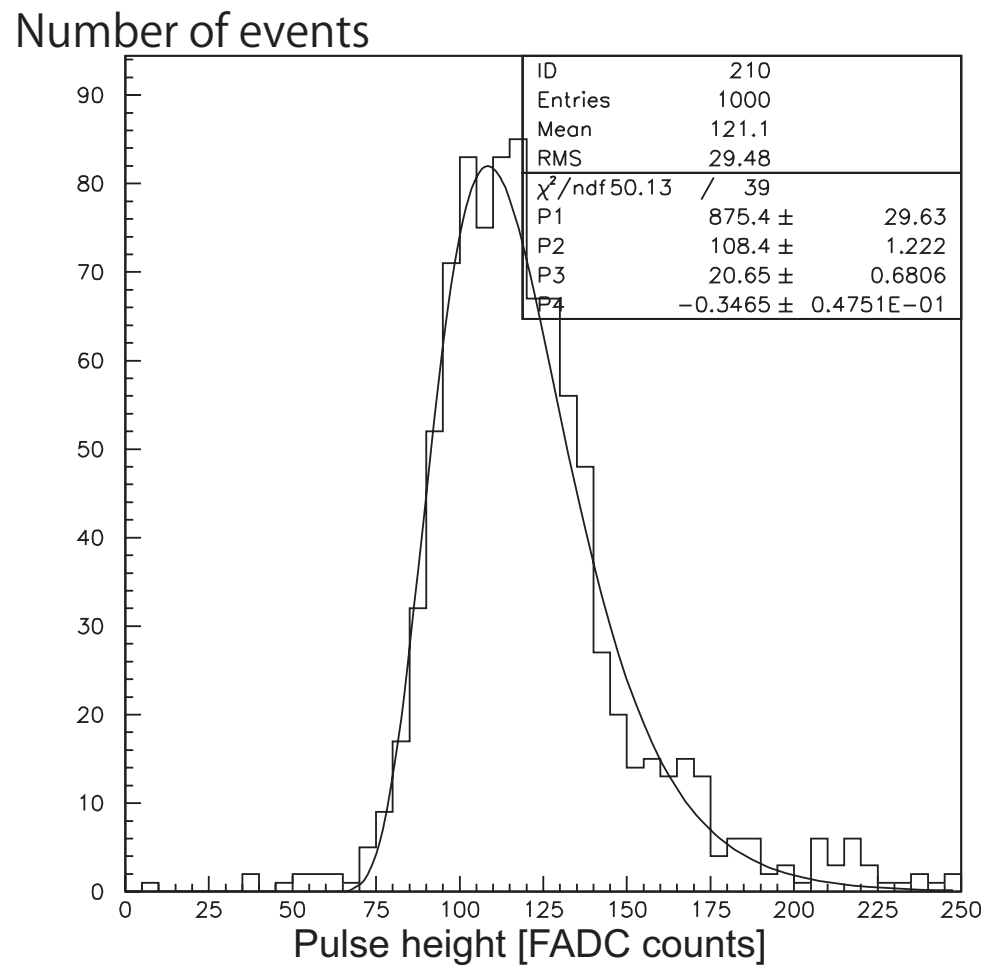

*Figure 9.23: Pulse height spectrum obtained by the cosmic run of the prototype conter of the APD readout option. The average pulse height is 120 FADC counts.*

#### 9.5.2.3 Remaining research and development items

In the first test of the APD readout with pure CsI crystal, the existing preamplifier for the CsI(T$\ell$) crystal was used by adding the driver circuit for differential signal transfer to the $\tau$=30 ns shaper input. In addition, another preamplifier developed for the KEDR experiment has been tested. The KEDR-type preamplifier, with a 5-pF feedback capacitor, is found to have a factor of 1.7 higher signal charge collection efficiency. This means that the existent preamplifier's open-loop-gain is not high enough at the $\tau$=30 ns signal shaping. Therefore, using a preamplifier with higher open-loop-gain for the corresponding frequency as well as smaller feedback capacitor to obtain a higher absolute gain, will result in a significantly larger APD signal. This will allow us to decrease the APD operating voltage, which will lower the leakage current—whose fluctuation contributes to the noise. (A lower operating voltage is also strongly preferred from the viewpoint of long term stability.)

When the applied voltage is kept constant, the gain of the APD strongly depends on the temperature. Nowadays, in many APD applications in high energy physics, nuclear physics, astrophysics and medical instruments, the temperature is sensed by an appropriate device and is used in a feedback circuit to manipulate the applied voltage to maintain constant APD gain. For example, the Matsusada HAPD-T series compact power supply unit has the voltage temperature correction functionality as well as the proper electrode pin to be connected to the temperature sensing diode. CAEN also provides the high voltage power supply modules equipping temperature correction functionality. Those might be designed, tested, produced and implemented commonly with the Particle identification detector system (PID) if the Hybrid Avalanche Photo-diode (HAPD) is selected as the photon detector for PID.

Finally, we would like to determine if the Hamamatsu S-8664-1010 APD can achieve our needed 18-bit-equivalent linearity.

#### 9.5.2.4 Cost

According to the estimate by Hamamatsu, mass production of 8000 S8664-1010 APDs can be done within two years. The price quotes from Hamamatsu are 28 kyen/piece for 4000 APDs and 24 kyen/piece for 8000 APDs. Since the PP price is 76 kyen/piece for 2000 pieces, attaching a few pieces of S8664-1010 APD results in similar or even a little smaller cost with respect to the baseline PP option.

### 9.5.3 BSO crystals with APD readout

#### 9.5.3.1 Motivation

$Bi_4Si_3O_{12}$ (BSO) scintillating crystal has been recently developed. It has comparable light output with pure CsI and a relatively short scintillation decay time of 100 ns. The 480-nm wavelength of the scintillation light is well-matched to many photon sensors. The radiation length ($X_0$) and Molière radius ($R_{\mathrm{M}}$) are 1.2 cm and 2.4 cm, respectively, both significantly shorter than for CsI. Thus, BSO provides better two-shower separation for high momentum $\pi^0 \to \gamma\gamma$ decays. Large-size crystal ingots, up to 70 mm diameter and 200 mm length, have been produced. In 2009, the Oxide Corp. [12] succeeded in the test production of four pieces of $2 \times 2 \times 20\,\mathrm{cm}^3$ blocks (Fig. 9.24).

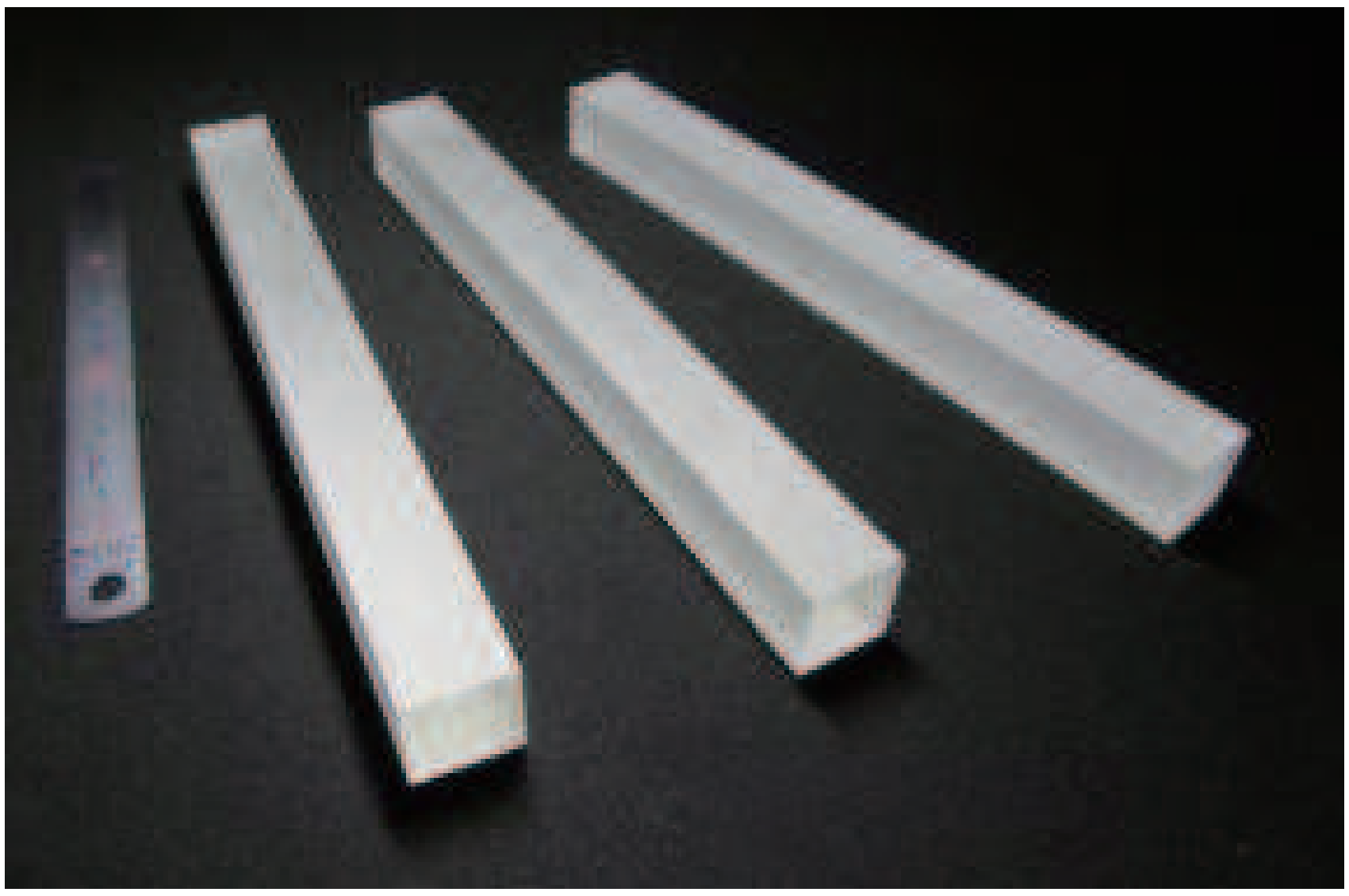

*Figure 9.24: The $Bi_4Si_3O_{12}$ (BSO) crystals with $2 \times 2 \times 20\,\mathrm{cm}^3$ dimensions produced by Oxide.*

#### 9.5.3.2 Expected performance

The BSO scintillator's Molière radius of 2.4 cm is about two thirds that of CsI. This contributes to improve high momentum $\pi^0$ reconstruction, for example from $B^0 \to \pi^0\pi^0$. To get a quantitative estimation, a simulation study is necessary. For lower energy photons, the equivalent noise energy of each channel is important. Since the amount of scintillation light is almost the same as for pure CsI, the photopentode (PP) can be used for the photosensor. The light output has been measured and compared with pure CsI by PMT readout using a $1\times1\times2\,\mathrm{cm}^3$ crystal sample, as shown in Fig. 9.25. Using two APDs on a single BSO crystal, we expect an equivalent noise energy of 0.7 MeV. This would be adequate to have reasonable energy resolution. To verify this expectation, we are performing a cosmic ray test by attaching one Hamamatsu S-8664-1010 APD on the $2.2\times2.2\times18$ cm$^3$ BSO scintillator borrowed from Prof. H. Shimizu of Tohoku University.

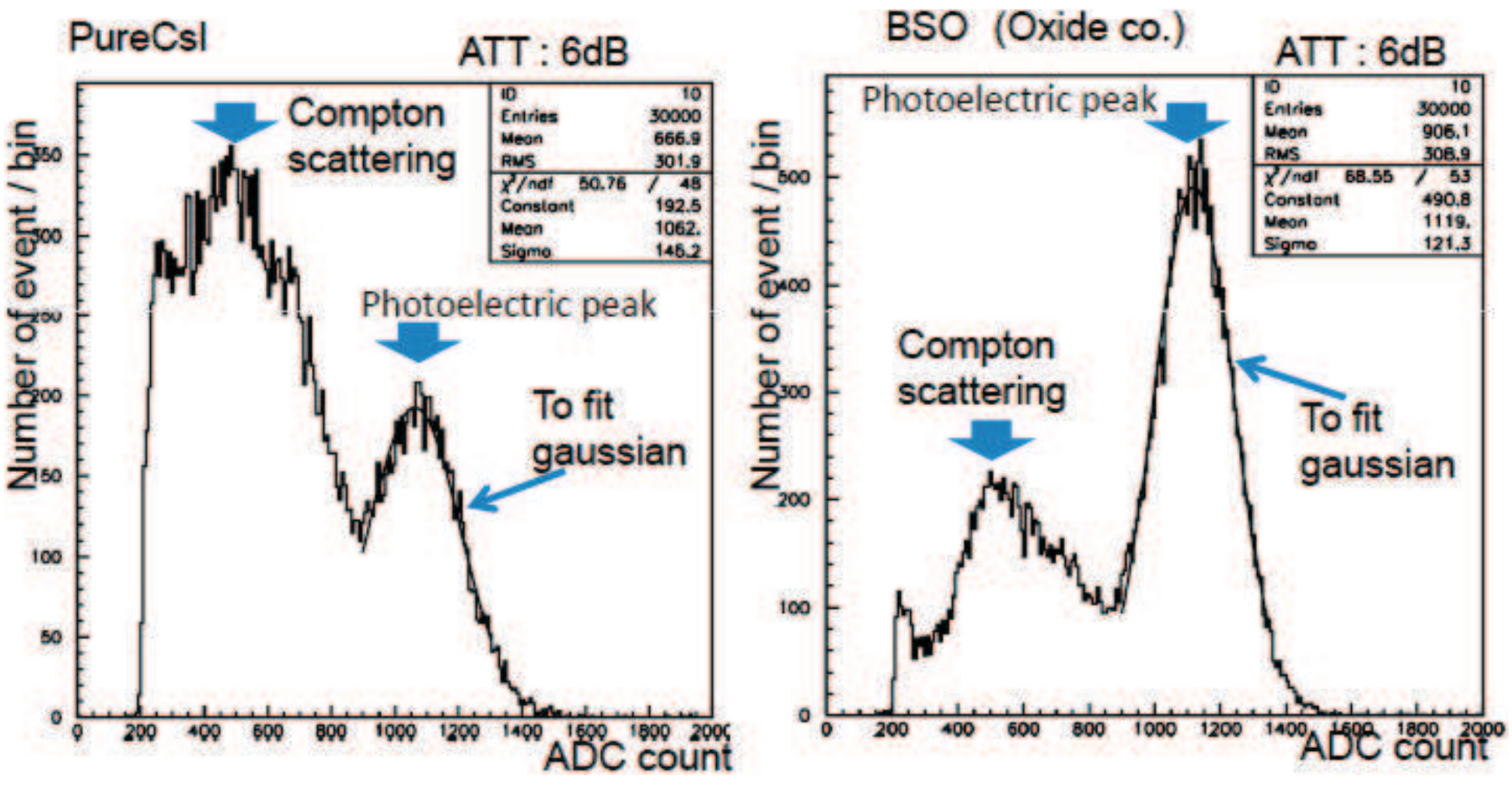

*Figure 9.25: Light output measurement for Pure CsI (left) and BSO (right) using 662 keV $\gamma$-ray emitted by a $^{137}$Cs source. The readout is done by Hamamatsu H3167 PMT and LeCroy 2249W charge sensitive ADC. The scintillator dimensions are $1 \times 1 \times 2\,\text{cm}^3$. The full-energy peak by photoelectric absorption shows that both scintillators provide similar light output.*

Radiation hardness of BSO seems to be adequate. After gamma irradiation of several Gy, no significant light output degradation was observed. Neutron radioactivation was found to be small: less than 0.1 $\mu$Sv/h even after $10^{11}$ n/cm$^2$ irradiation. The light output degradation is $\sim$20% for $10^{12}$ n/cm$^2$ dose.

#### 9.5.3.3 Remaining research and development items

As mentioned in Sec. 9.5.3.2, the simulation study to quantitatively check the improvement in high energy photon and $\pi^0$ detection is necessary. Since BSO crystal is harder than CsI, it is difficult to tap holes in the crystal. Therefore, the mechanical counter construction—including crystal wrapping and attaching APDs and preamplifiers—must be newly designed. Also, crystal containers and crystal-fixing structures must be newly designed with very careful finite element analysis calculation to ensure their strength. These structural issue would be the limiting factor to make this option ready in time for the inauguration of Belle II.

#### 9.5.3.4 Cost

Corresponding to smaller Molière radius and radiation length of BSO, a suitable crystal dimension would be $\sim 4.5 \times 4.5 \times 20\,\text{cm}^3$. To cover the existing endcap area, $\sim$3000 BSO crystals are needed to fill both endcaps. According to the test-production experience of $\sim$0.75 Myen per $2 \times 2 \times 20\,\text{cm}^3$ crystal, it is thought to be feasible to make one $\sim 4.5 \times 4.5 \times 20\,\text{cm}^3$ crystal for $\sim$0.35 Myen in mass production, taking into account the fact that the test production includes extra initial costs such as pots and rejects to find the optimum crystal growth. Photosensor and readout electronics costs scale with the number of crystals. In addition, because of the different crystal density, the endcap crystal containers need to be newly built. This is estimated to cost 100 Myen = 1 Oku-yen. The expected costs are summarized in Table 9.2.

| Item | Cost/unit | number | Oku-Yen |
|---|---|---|---|
| Crystal | 0.35 Myen | ∼3000 | 10 |
| APD | 56 kyen | ∼3000 | 1.7 |
| Preamp. | ∼10 kyen | ∼3000 | 0.3 |
| Electronics | | | 1.3 |
| Mechanical structure | | | 1.0 |
| Test bench | | | 0.1 |
| Assemble | | | 0.3 |
| Total | | | 14.7 |

*Table 9.2: Summary of expected cost for BSO option.*

### 9.5.4 PWO-II option for the endcap upgrade

#### 9.5.4.1 Motivations

R&D on a PWO-II option was initiated in response to issues with the present pure CsI option for the Belle II endcap calorimeter upgrade. Existing crystal technologies were comprehensively surveyed to determine which technology is the best option from purely scientific and technical viewpoints. The technical details were presented in a separate document and the results and concerns raised in the response to the internal committee's review are summarized here.

#### 9.5.4.2 Expected Performance

- Strengths

  We expect the energy resolution of 1 GeV photons to be in the range of 1–2% for $2 \times 2 \times 20\,\mathrm{cm}^3$ PWO-II crystals when using the waveform readout of the signal. Unfortunately, the proponents had no time to demonstrate this directly; it is based on the results achieved in recent studies by the PANDA collaboration. The radiation hardness of PWO-II crystals ($10^6$–$10^7$ rad) is excellent and more than adequate. Since the peaking time of PWO-II is also short (30 ns) and the transverse size of the shower in PWO is smaller than in CsI (the Molière radii being $R_M(\mathrm{PWO}) = 2$ cm and $R_M(\mathrm{CsI}) = 3.6$ cm, for a gain of $(R_M(\mathrm{CsI})/R_M(\mathrm{PWO}))^2 = 3.2$), the pile-up is also significantly smaller than for pure CsI. The smaller transverse area also helps in $\pi^0$ separation for pion momenta up to 4 GeV/$c$. The PWO-II crystal length can be 20 cm (rather than 30 cm for CsI) because of the difference in radiation length. Their finer segmentation should also improve the gamma and $\pi^0$ efficiency.

- Weaknesses of alternatives

  The main weakness of the PWO-II option is the lack of experience with the large scale operation of a cooling system operating at $-25°$C. A full study has been carried out by PANDA, but our Belle II proponents have no direct experience. Even if it is claimed that lower temperature operation has no technical show-stoppers, based on PANDA's experience, an alternative to this issue is to run the endcap detectors at some higher temperature, such as $0 \sim 10°$C, to avoid the complication of sub-zero cooling system. To compensate for the lower light yields due to this, we plan to study the case of the two-APD readout configuration.

Ongoing R&D suggests that the crystals from SICAS can have even higher (∼double) light yields though, at the moment, there is an issue of light uniformity and stability of the production of SICAS crystals that needs to be understood. If this can be resolved, there is a chance to operate these crystals even at room temperature.

#### 9.5.4.3 Increased Robustness/Reduced background sensitivity

The mechanical properties PWO crystals have been studied by CMS and PANDA, who showed no significant differences compared with other standard crystals. Therefore, the mechanical robustness of the crystal modules themselves is not a major issue. The only concern is the design of a new container to house the denser material. This design effort has not yet been started.

#### 9.5.4.4 Cost and proposed cost sharing

Note: this cost estimation is for the forward endcap option only. The extrapolation to the backward endcap is straightforward.

- Crystal

  For the forward endcap, we expect to have approximately 8,000 PWO-II crystals. Based upon a small production volume quotation 480 Euro/crystal, we estimate the amortized cost for the crystal production to be 3,840,000 Euro (or 516,000,000 Yen).

- Photosensors

  For the APD photosensors, based upon a small-volume quote of 56,000 Yen/sensor for the S8664-1010 from Hamamatsu, a total photosensor cost of 224,000,000 Yen is estimated.

- Electronics and Infrastructure

  For the readout electronics and infrastructure, including the new container, there is no cost estimate for the PWO-II option. Using quoted numbers from other options, we expect 150,000,000 Yen for electronics and 130,000,000 Yen for the mechanical structure.

- Total Cost

  Summing all these, the total cost is in the range of 10 Oku-Yen. The proponents will apply for a grant of up to 0.8 Oku-yen equivalent for this forward-endcap upgrade from Korean Agency for National Research Foundation. We hope that the main funding source will be the Japanese Belle II budget.

#### 9.5.4.5 Remaining R&D items

- Beam tests with full size crystals in realistic Belle II environment
- Two APDs/crystal option
- Radiation tolerance tests for crystals and photosensors
- Electronics and readout
- Light improvement with SICAS crystal production for room temperature operation

#### 9.5.4.6 Schedule

In the near term, we plan to perform

- a full beam test with an array of 5×5 PWO-II crystals, and
- SICAS crystal characterization.

These are urgent R&D steps planned for the first half of 2010. We would like to note that the usual crystal production rate at the Bogoroditsk Plant of Technochemical Products (BTCP) for the PANDA experiment was 800/month.

# Chapter 10

# $K^0_{\mathrm{L}}$ and $\mu$ detection (KLM)

## 10.1 Overview

The $K_L$ and muon detector (KLM), shown in side view in Fig. 10.1, consists of an alternating sandwich of 4.7-cm thick iron plates and active detector elements located outside the superconducting solenoid. The iron plates serve as the magnetic flux return for the solenoid. They also provide 3.9 interaction lengths or more of material, beyond the 0.8 interaction lengths of the calorimeter, in which $K_L$ mesons can shower hadronically.

The octagonal barrel covers the polar angle range from 45° to 125°, while the endcaps extend this coverage from 20° to 155°. There are 15 detector layers and 14 iron plates in the barrel and 14 detector layers and 14 iron plates in each endcap.

Muons and non-showering charged hadrons (that decay in flight or do not interact hadronically) with a momentum above $\sim 0.6\,\mathrm{GeV}/c$ traverse the KLM until they escape (if their momentum exceeds $\sim 1.5\,\mathrm{GeV}/c$, depending on the polar angle) or range out due to electromagnetic energy deposition. They travel along nearly straight lines through the KLM. $K_L$ mesons that interact in the ECL or the iron plates create a hadronic shower that can be detected in either the ECL alone, the KLM alone, or both.

The Belle KLM, based on glass-electrode resistive plate chambers (RPC) [1, 2], has demonstrated good performance during the entire data taking period of the Belle experiment. The principle of operation of RPCs is described in Sec. 10.2.

The long dead time of the RPCs during the recovery of the electric field after a discharge significantly reduces the detection efficiency under high background fluxes. At the present KEKB luminosity, the background occupancy in the barrel KLM is estimated to be consistent with the cosmic ray flux. Thus, even with the much higher beam background expected for SuperKEKB, the barrel RPCs can be operated successfully.

In the endcaps, the background is worse because of the limited shielding of neutrons and other particles that are generated externally along the beam lines. At the background rate in the Belle KLM endcaps of about 0.1–0.3 $\mathrm{Hz/cm}^2$, the RPC efficiency has already sagged to between 90% and 95% [3] and the two outermost layers (in each endcap) have been turned off since 2002. In SuperKEKB, the background rate in the endcaps is expected to be a factor of 20 to 40 higher, resulting in an RPC efficiency of below 50%. In addition, the proportion of background to true $K_L$ clusters seen by the endcap RPCs would rise in SuperKEKB. (Our background studies are described in Sec. 10.4.) Therefore, the endcap RPCs will be retired and replaced with scintillators in Belle II (Sec. 10.5).

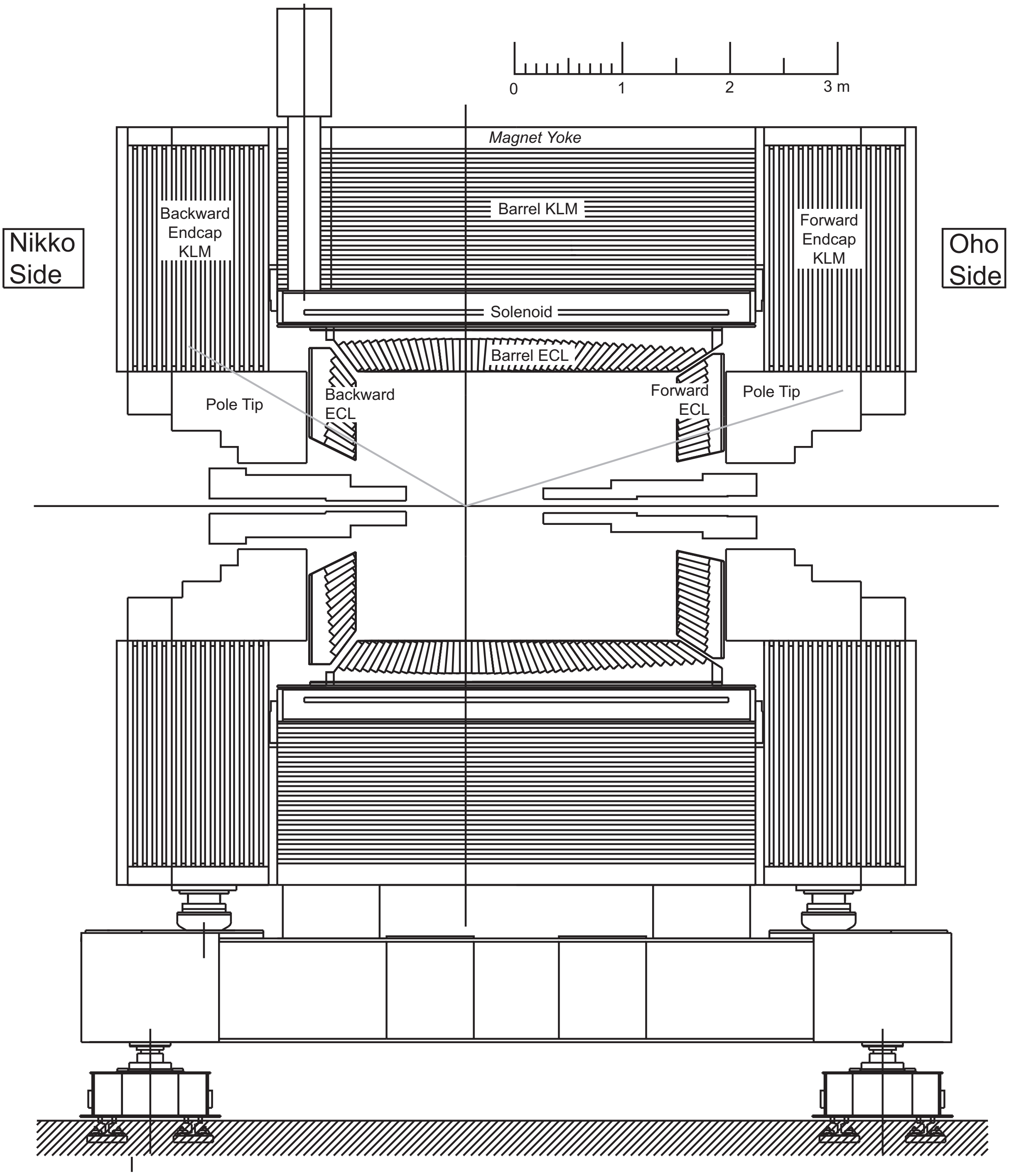

*Figure 10.1: Side view of the KLM, located outside the ECL and solenoid. The gray lines mark the nominal polar angular acceptance of Belle II.*

## 10.2 Resistive Plate Chambers

In the Belle KLM, charged particles are detected by glass electrode RPCs. The glass electrodes, although technically insulating, do conduct electricity but with a very high bulk resistivity of $\sim 5 \times 10^{12}\,\Omega \cdot \text{cm}$. The electrodes are two parallel sheets of float glass (73% silicon dioxide, 14% sodium oxide, 9% calcium oxide, and 4% uncategorized compounds for the 2.4-mm thick barrel glass; 70–74% silicon dioxide, 12–16% sodium oxide, 6–12% calcium oxide, 0–2% aluminum oxide, and 0–4% magnesium oxide for the 2.0-mm thick endcap glass) separated by 1.9-mm thick noryl spacers, epoxied in place. High voltage is distributed to the electrodes via a thin layer of carbon-doped paint (conducting carbon-loaded tape in the endcaps) on the outer surfaces of each electrode; this layer has a surface resistance of $10^{6-8}\,\Omega/\Box$ and is transparent to fast pulses. The energized electrodes maintain a quiescent uniform electric field of up to 4.3 kV/mm in the gas-filled gap. The current drawn from the high voltage supply corresponds to roughly $1\,\mu\text{A/m}^2$ of electrode area, and most of this flows through the noryl spacers. We use a gas mixture of 62% HFC-134a, 30% argon, and 8% butane-silver (the latter being a mixture of 70% $n$-butane and 30% isobutane).

A throughgoing charged particle ionizes the gas molecules along its path. The electric field then accelerates the electrons toward the anode and the ions toward the cathode. In the strong electric field, the electrons initiate more ionizations, leading to a streamer between the electrodes. The electrode charge in the $\sim 2\,\text{mm}^2$ spot nearest the streamer then flows across the gap, momentarily reducing the electric field there. The gas enhances the initial streamer formation and suppresses afterpulsing in the surrounding area from ultraviolet photons ejected from the electrode atoms. With a weaker electric field, streamer formation is inhibited and the electrodes do not discharge nor deaden. Nevertheless, a small signal from the primary ionization is seen in this "proportional mode."

The streamer (or proportional-mode pulse) is imaged on a plane of external pickup strips, each about 5 cm in width. These strips, separated from an outer conducting ground plane by dielectric foam, behave like a transmission line with a characteristic impedance of about $50\,\Omega$. Orthogonal pickup strip planes are mounted on either side of a pair of RPCs to form a superlayer. Figure 10.2 shows an exploded view of the superlayer geometry. Because the RPC materials are electrically transparent to fast signals, a pulse that forms in either RPC is imaged on both readout planes and then travels along the transmission line and twisted-pair cables to discriminators and serializers mounted on the periphery of the iron yoke. The redundancy in this design results in a superlayer detection efficiency of up to 99% at the low ambient rates seen in Belle (Fig. 10.3).

During the normal high-electric-field operation, the amplitude of a streamer's image pulse ($\sim 100\,\text{mV}$ into a $50\,\Omega$ termination) is large enough that it can be discriminated without amplification. (A proportional-mode pulse would need a preamplifier with a gain of 10 at the RPC and before the discriminator.) The pulse width is under 50 ns FWHM; the intrinsic time resolution is a few ns. The discriminator threshold is typically set to near 40 mV (70 mV) for the barrel (endcap) RPCs due to the presence (absence) of an impedance-matching resistor at the junction between the transmission line and the $100\,\Omega$ twisted-pair signal cable.

Each of the two RPCs in one module has independent high voltage and gas connections so that, in the event that a problem develops in one RPC, the other continues to function (but with a single-RPC muon detection efficiency of between 90% and 95%).

The layout of one barrel RPC and of one octant of a barrel superlayer is shown in Fig. 10.4. The corresponding geometries for the endcap are shown in Fig. 10.5.

The area of each barrel module ranges from $2.2 \times 1.5\,\text{m}^2$ to $2.2 \times 2.7\,\text{m}^2$, except near the solenoid's helium chimney (Fig. 10.1), where the modules are shorter by 0.63 m. Each full-size module weighs an average of 110 kg. The two RPCs in a module are placed so that the gas-gap spacers do not overlap geometrically. In each module, there are 48 pickup strips to measure $z$ (or, equivalently, $\theta$) and 36 (48) pickup strips in layers 0–6 (7–14) to measure $\phi$. Each strip is roughly 5 cm wide; this matches the scale of multiple scattering of a muon passing through the iron.

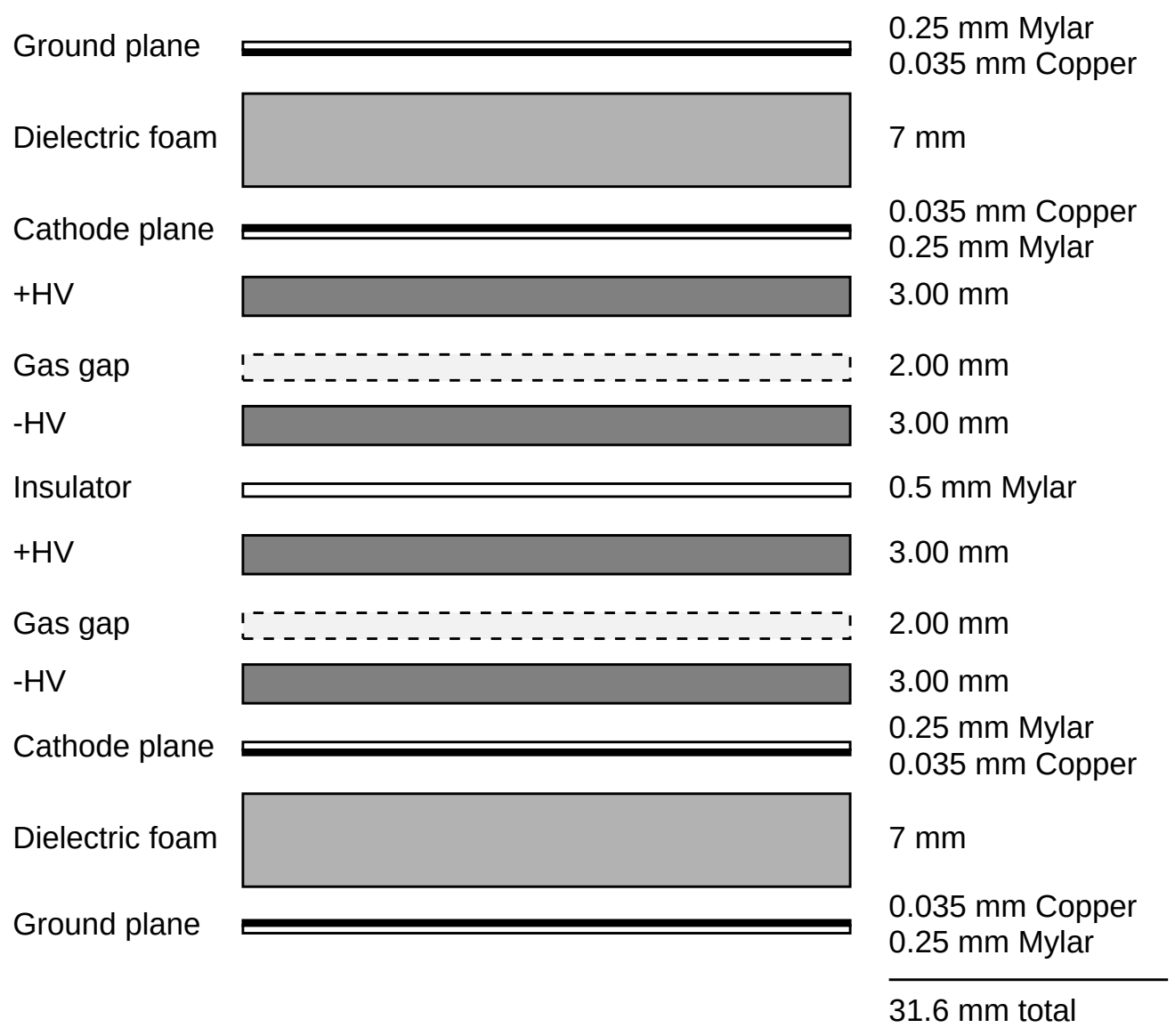

*Figure 10.2: Exploded cross-section of an RPC superlayer.*

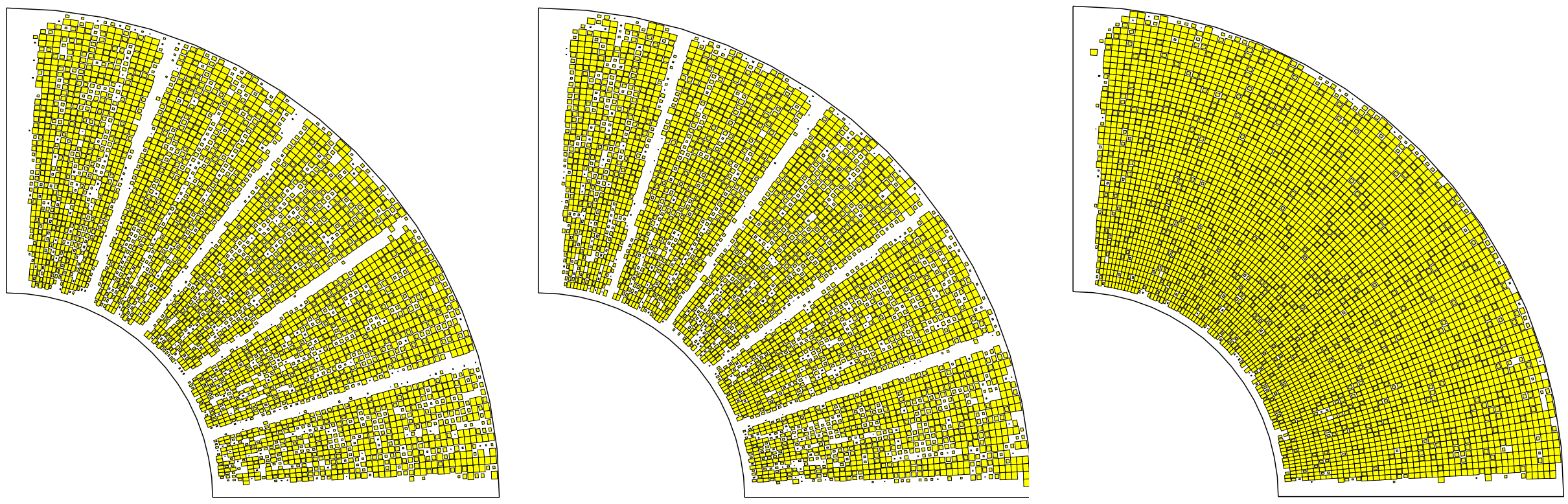

*Figure 10.3: Two-dimensional efficiency map in an endcap superlayer module, with efficiency proportional to the size of a box, for (a) only the five top-layer RPCs, (b) only the five bottom-layer RPCs, and (c) all ten RPCs.*

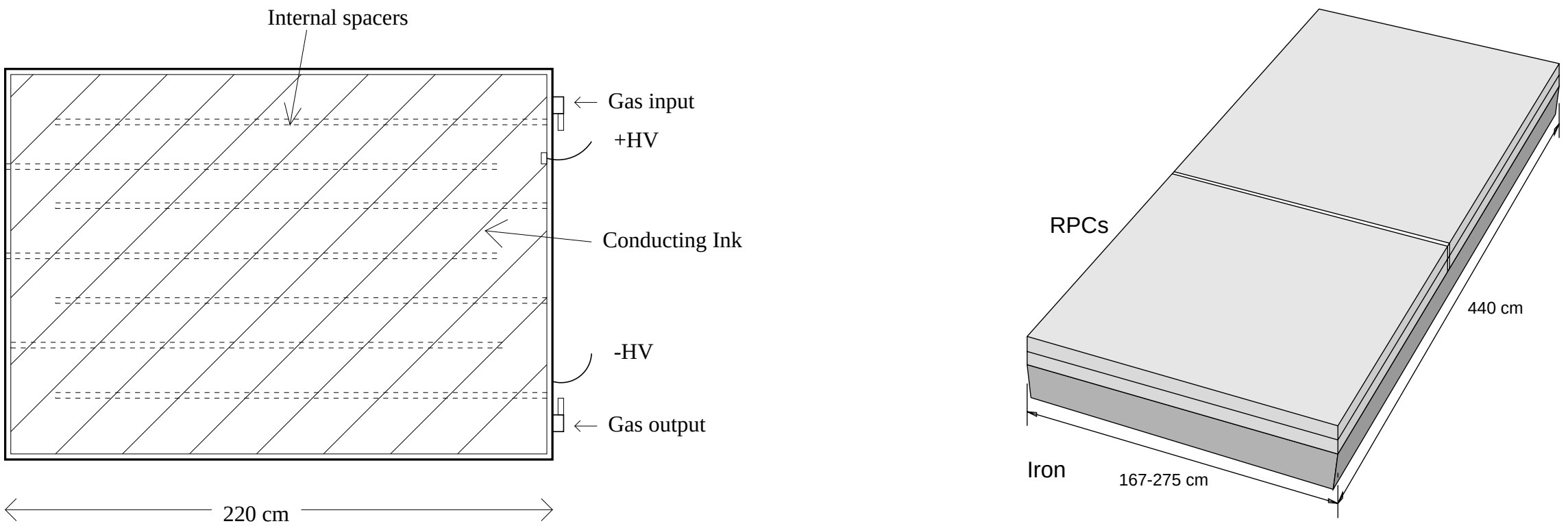

*Figure 10.4: Left: top view of a barrel RPC, showing the internal spacers and service ports. Right: isometric view of one octant of a barrel superlayer, with two side-by-side modules above an iron flux-return plate.*

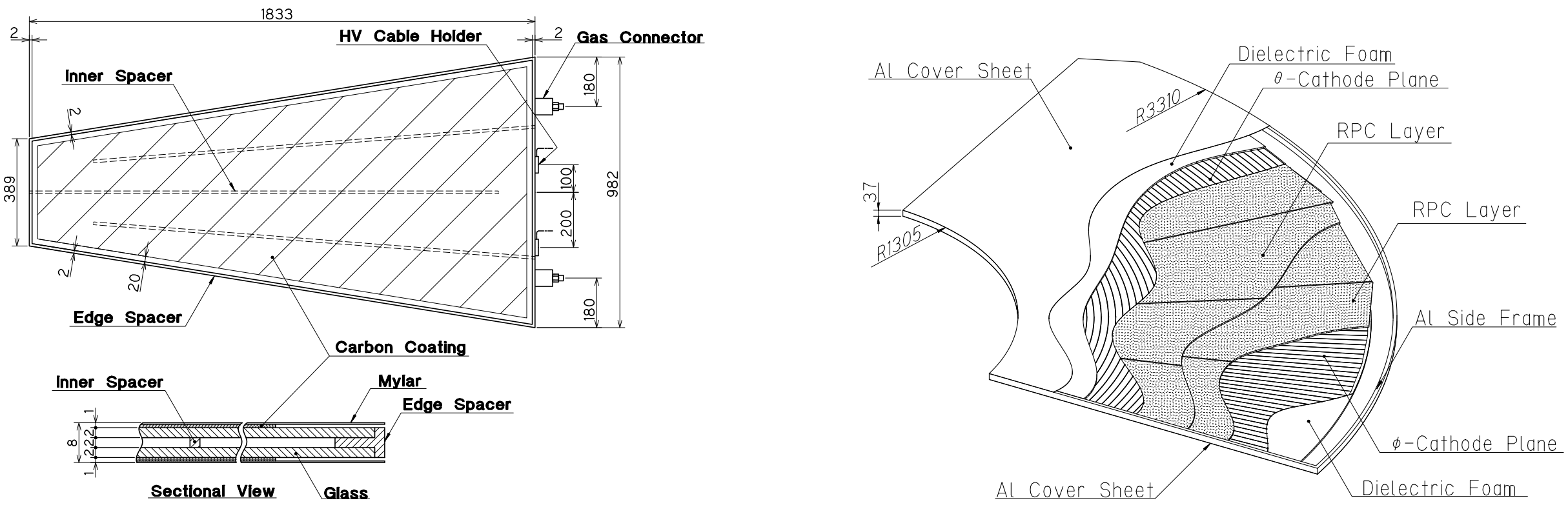

*Figure 10.5: Left: top and side view of an endcap RPC, showing the internal spacers and service ports. Right: cutaway isometric view of an endcap superlayer module.*

The area of a wedge-shaped endcap RPC is about 1.26 m$^2$, with a slightly smaller area for two out of ten RPCs placed in a module to avoid geometric overlap of the RPC borders in the two planes. In each module, there are 48 coaxial-arc cathode strips of width 3.6 cm to measure $\theta$ and 96 radial-trapezoid cathode strips of average width 3.3 cm to measure $\phi$. There are no termination resistors on these strips. Gas, high voltage and signal lines penetrate the module frame at one of its outer corners. There are dead regions of a few centimeters' width at the radial borders of each module (Fig. 10.3).

The spatial resolution of a superlayer is 1.1 cm (one standard deviation) when either one or two adjacent cathode strips fire, and rises to 1.7 cm for the infrequent case where three strips fire (due to UV-photon-induced afterpulsing of the neighboring electrode area).

## 10.3 Muon and $K_L$ identification

### 10.3.1 Muon identification

Muon identification [4] begins with the reconstruction of a charged track in the CDC. Each track is extrapolated outward beyond the outermost CDC hit as if it were a pion; the extrapolation uses the simulation geometry but only considers mean energy loss in determining the range of the track; fluctuations in energy loss as well as the average amount of multiple scattering are accumulated in the track parameters' evolving covariance matrix during this extrapolation. The track is considered to be within the KLM acceptance if it crosses at least one RPC layer; this requires a momentum of at least 0.6 GeV/$c$. (The pion hypothesis is used initially because this benefits the initial extrapolation into the particle-identification detector and calorimeter.)

If a KLM hit is found near the crossing of the extrapolated track with a detector layer, that hit is associated with the track. The outermost layer crossed by the extrapolated track defines the predicted range of the track; the actual range is determined by the outermost layer with an associated hit.

If the predicted and actual ranges disagree substantially, the track is classified as a hadron and not treated further ("pre-rejection"). Otherwise, the extrapolation is restarted at the entry point into the KLM, but now using a muon hypothesis. In this re-extrapolation, we use the associated hits with a Kalman filtering/fitting technique to steer the track as it is swum through the KLM. The difference between the predicted and measured ranges as well as the goodness of fit of the transverse deviations of the associated hits from the re-extrapolated track provide the two variables used in a likelihood ratio to test the hypothesis that the track resembles a muon rather than a charged hadron.

Typically, a physics analysis requires a value of 0.9 or greater for this normalized likelihood ratio to classify a track as a muon. The muon detection efficiency plateaus at 89% above 1 GeV/$c$, while the hadron fake rate is about 1.3% (but rises with lower momentum to as high as 3.8% at 0.7 GeV/$c$). The fake muons are pions that have neither decayed in flight (to a softer muon) nor suffered an inelastic hadronic interaction. (Kaons that might have this behavior are positively identified by the PID subsystem and are therefore much less likely to contribute to the hadron fake rate.)

### 10.3.2 $K_L$ identification

Hits in the KLM that are within a 5° opening angle of each other (measured from the interaction point), whether in the same layer or not, are grouped together into a cluster. After all clusters have been formed, the charged track veto is applied. Each track is extrapolated to its entrance into the KLM (as described earlier), then a straight line is drawn between this entrance point and the interaction point. If this line is within 15° of the line between the cluster centroid and the interaction point, the cluster is discarded. If the cluster is aligned with a reconstructed neutral ECL cluster to within 15°, then the ECL and KLM cluster are associated, and the ECL's cluster direction overrides that of the KLM. Finally, a cut on the cluster size is imposed: to be classified as a KLM-only $K_L$ candidate, the cluster must have hits in at least two distinct layers; to be classified as a KLM+ECL $K_L$ candidate, the cluster must have a hit in the ECL and at least one KLM hit.

Because of large fluctuations in the shower development of a $K_L$–nuclear collision, the number of hits within a cluster that is associated with a $K_L$ meson is a poor measure of the incident energy. Therefore, this cluster gives only the direction of the $K_L$ candidate. The resolution is 3° for KLM-only candidates or 1.5° for KLM+ECL candidates. The $K_L$ detection efficiency rises almost linearly with momentum from zero at 0 GeV/$c$ to a plateau of 80% at 3 GeV/$c$; it is nearly flat in polar angle within the detector's acceptance.

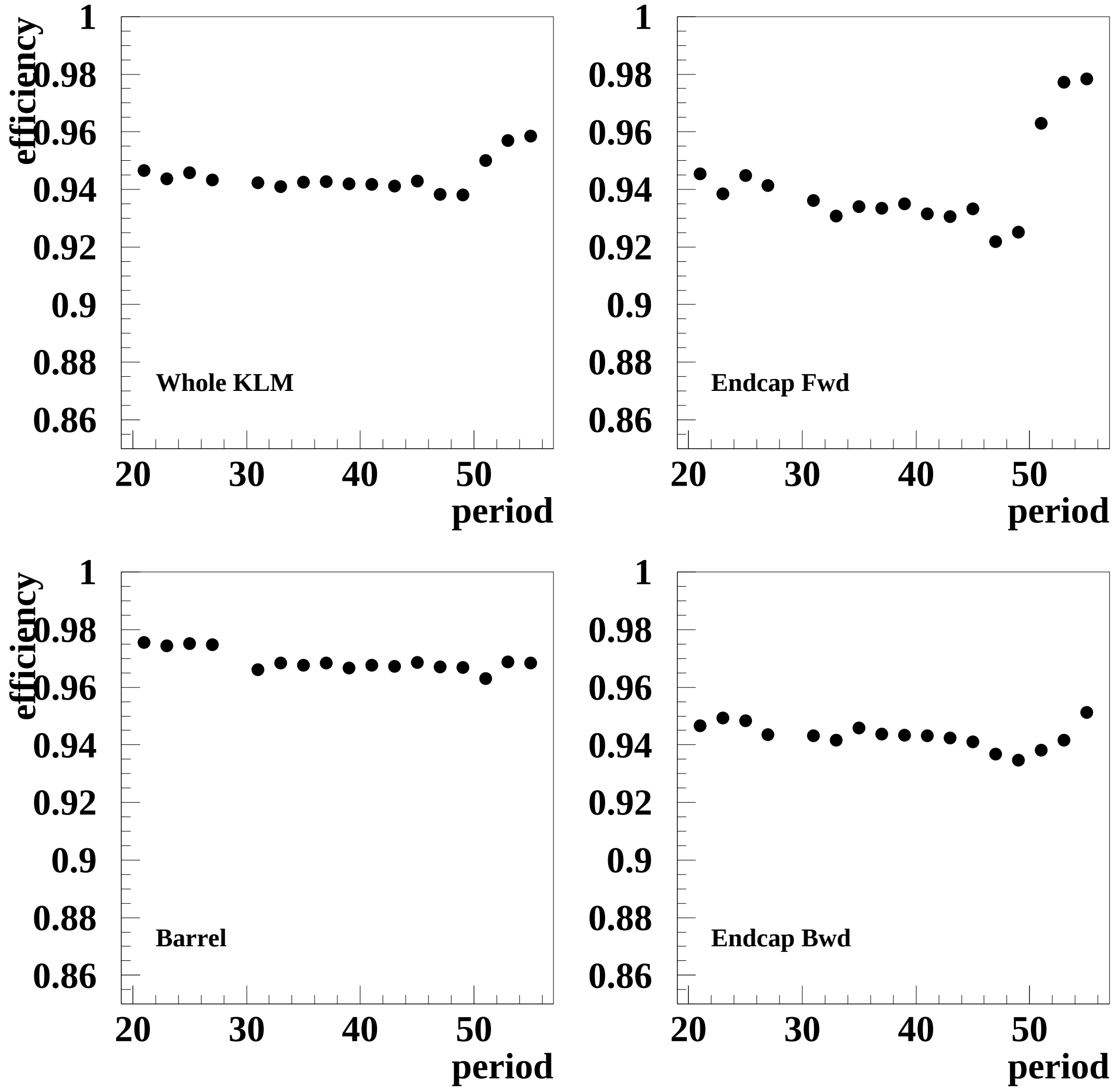

*Figure 10.6: Experimental-number dependence of the muon detection efficiency in Belle for the whole KLM, the forward endcap, the barrel, and the backward endcap.*

## 10.4 Backgrounds in the endcap RPCs

Since muon identification requires hit information from the KLM, the muon identification efficiency is a good monitor of KLM performance. The long-term history of the muon identification efficiency in Belle is shown in Fig.10.6. The efficiency for the barrel RPCs is stable, while that for the endcap RPCs shows a gradual degradation.

Due to the intrinsic dead time associated with the recovery of the RPC electric field near a discharge, the particle detection efficiency depends on the ambient hit rate per unit area.

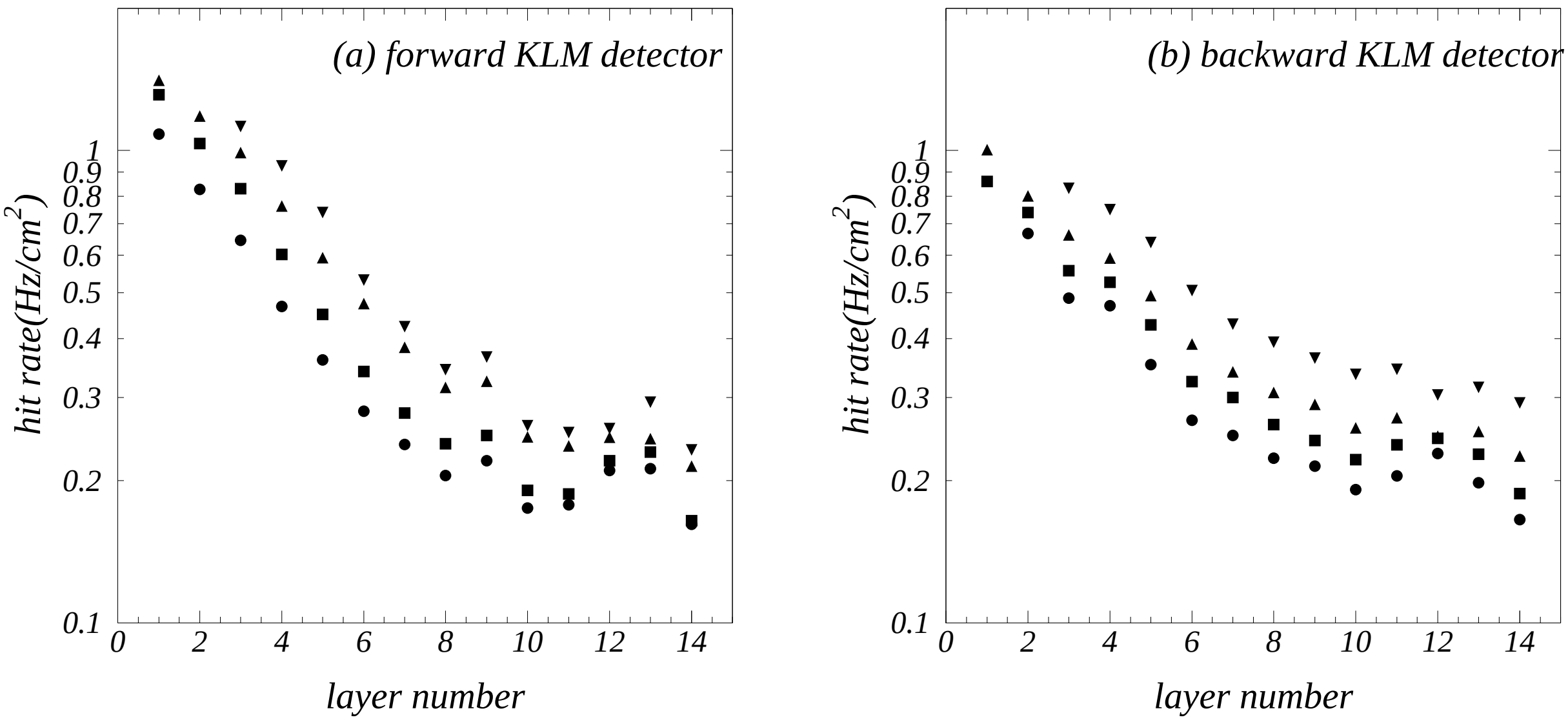

*Figure 10.7: Hit rates in each layer of the endcap RPC for four LER current ranges: 400–600 mA (●), 500–650 mA (■), 600–800 mA (▲), and 900–1300 mA (▼). Left and right plots are for the forward and backward endcap, respectively. In these plots, layer number increases from the outside inward, starting at #1 for the outermost layer.*

Figure 10.7 shows the distribution of this quantity for each endcap-RPC superlayer module during KEKB operation for four intervals of the LER beam current, using RPC hits in the area of $1.044 \times 10^5\,\mathrm{cm}^2$ that overlaps the interior of the concrete shielding structure surrounding the beam lines. The hit rate per unit area exceeds $1\,\mathrm{Hz/cm}^2$ in the outer-layer RPCs, resulting in a sag in the RPC efficiency to between 70 and 80%. Figure 10.8 shows the hit rate per unit area for endcap layer 3 (counting from the outside in) for the same LER currents as in Fig. 10.7. The correlation between the measured flux and the LER current demonstrates the association of this flux with beam-induced background.

To identify the source of the ambient hits, we performed a simulation study using Geant4 in which we generated either low-energy neutrons or bremsstrahlung photons originating upstream along the beamline. For the transportation of low energy neutrons in matter, we used data-driven models based on the data formats of ENDF/B-V1 [5]. Since the energy coverage of these models is from 0.025 keV (thermal) to 20 MeV, the parametrization-driven models based on the GHEISHA package were used for the more energetic neutrons. Figures 10.9(a) and 10.10(a) show the distributions of expected hit rates in each endcap RPC layer for selected values of the mean energy of the incident neutrons and photons, respectively. (The energy of incident particles is distributed exponentially.)

To compare the simulation and experimental results, the data in Fig. 10.7 (for real data) and Figs. 10.9(a) and 10.10(a) (for simulated particles) are fitted to the empirical function $y(\ell) = p_1 e^{-p_2 \ell} + p_3$, where $\ell$ is the layer number and $p_1$, $p_2$, and $p_3$ are free parameters. The exponential term in the function corresponds physically to the attenuation of the particle with increasing depth into the endcap iron flux return. From the fits to the real data, the exponential damping parameter $p_2$ falls between 0.29 and 0.40. Figures 10.9(b) and 10.10(b) show the fitted value of $p_2$ for each energy of the simulated neutrons and photons, respectively. Clearly, the exponential damping rates seen in the real data lie far below those extracted for photons of any energy. On the other hand, the real rates are quite consistent with the damping rates extracted for neutrons of energy between 10 and 100 keV.

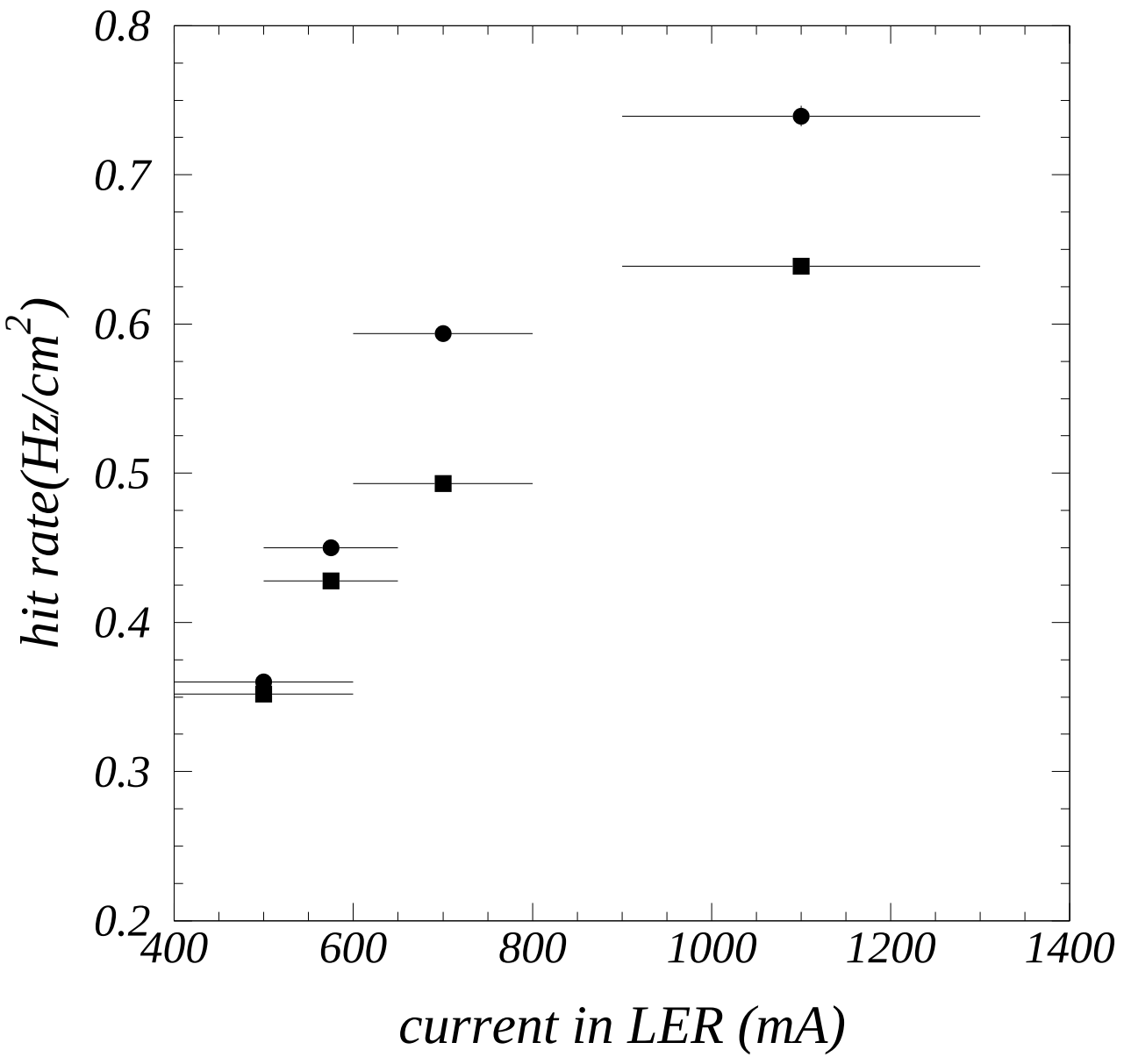

*Figure 10.8: Hit rate per unit area in forward (■) and backward (•) endcap RPC layer 3—counting inward—for the four LER current intervals of Fig. 10.7. The error bars indicate the range of the beam current interval.*

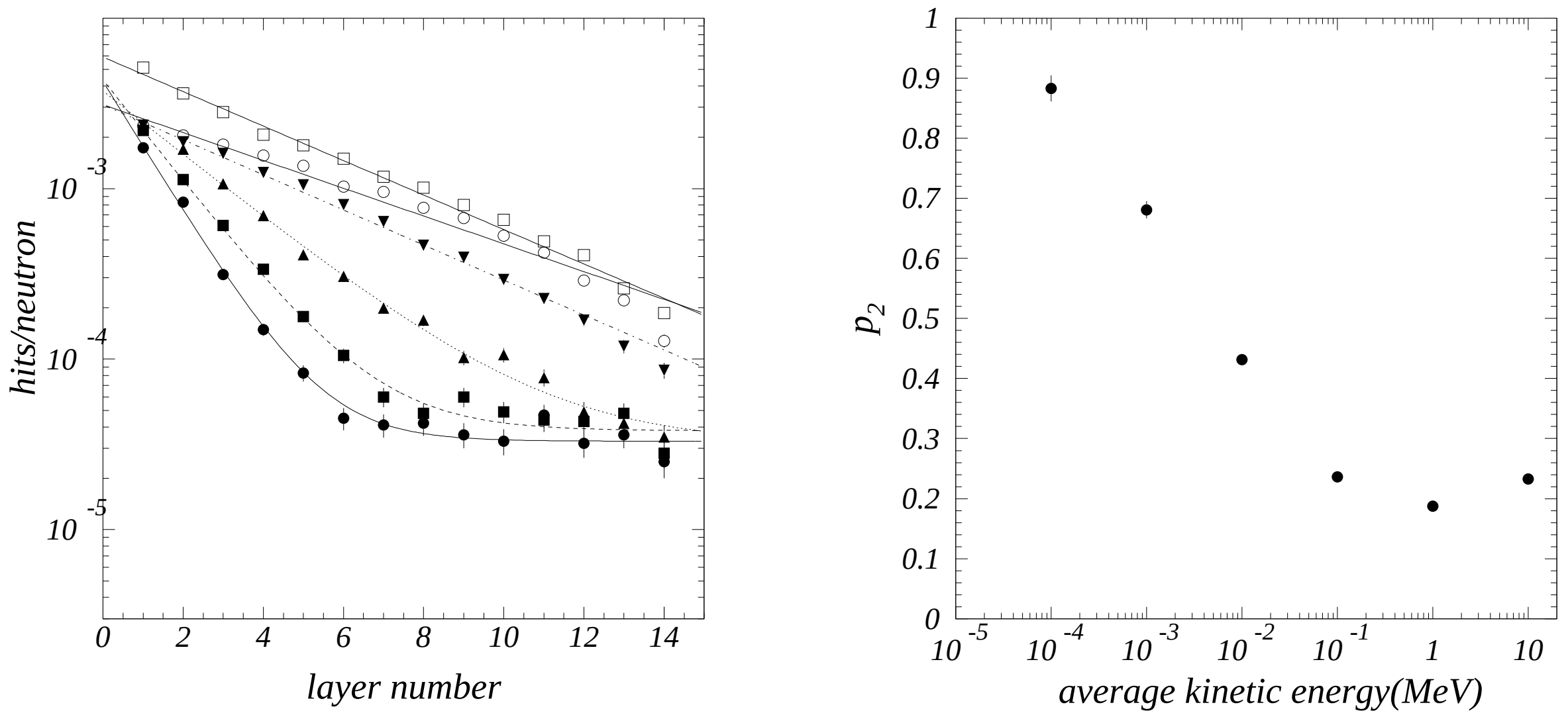

*Figure 10.9: Left: hits per generated neutron as a function of layer number—counting inward—for incident mean energy of 100 eV (•), 1 keV (■), 10 keV (▲), 100 keV (▼), 1 MeV (○), and 10 MeV (□). The curves are fits to the empirical function $p_1 e^{-p_2 \ell} + p_3$. Right: fitted exponential damping parameter $p_2$ as a function of simulated incident neutron energy.*

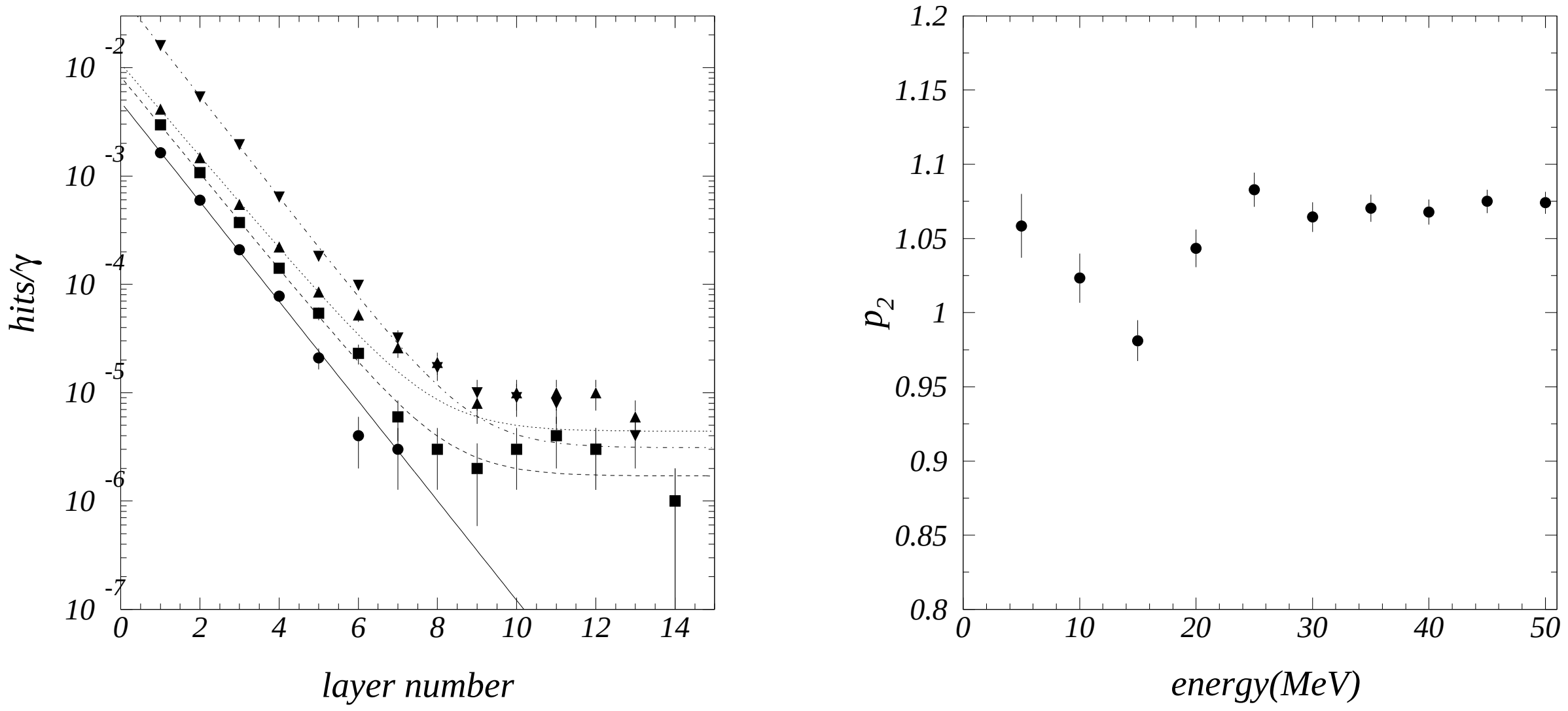

*Figure 10.10: Left: hits per generated photons as a function of layer number—counting inward—for incident mean energy of 5 MeV (●), 10 MeV (■), 15 MeV (▲), and 50 MeV (▼). The curves are fits to the empirical function $p_1 e^{-p_2 \ell} + p_3$. Right: fitted exponential damping parameter $p_2$ as a function of simulated incident photon energy.*

We conclude from the above studies that the ambient flux illuminating the endcap RPCs originates outside the Belle detector, is associated with the accelerator beams, and is dominated by neutrons with an energy of about 10 to 100 keV.

To test this conclusion, we placed neutron shielding in the form of polyethylene sheets on the outside surface of the Belle endyoke and within the concrete shield surrounding the beamline (Fig. 10.11). We observed a partial recovery in the endcap muon detection efficiency (Fig. 10.6) starting at experiment number 51 for the forward endcap and at number 55 for the backward endcap.

In spite of the efficiency recovery over most of the endcap RPC area after the installation of the polyethylene shield, we noticed that there was no improvement in the area nearest the beam line and surmised that there was an unshielded neutron source in the beamline near the Belle detector. To test this, an additional polyethylene shield (Fig. 10.12) was installed on the lower azimuth of the tapered part of the forward yoke. Later, the gap between the original planar shield and this tapered shield was filled with polyethylene bricks.

The ambient noise reduction associated with the installation of this additional shielding was measured by the ratio of after and before rates (in experiments 67 and 61, respectively, for the tapered shield, and in experiments 69 and 67, respectively, for the gap shield) in each of thirty areal sections of the endcap RPCs, with radial boundaries at $r$ = 135, 150, 165, 180, 195, and 210 cm and six azimuthal sections labelled "Covered," "Transition," and "Reference" (Fig. 10.13), according to their overlap with the new shielding. The Reference region was expected to display no effect from the new shielding, and so would monitor the influence of other environmental factors such as beam current or beamline vacuum condition. Each ratio was normalized by the number of events in the before and after experiments to obtain a value near 1.

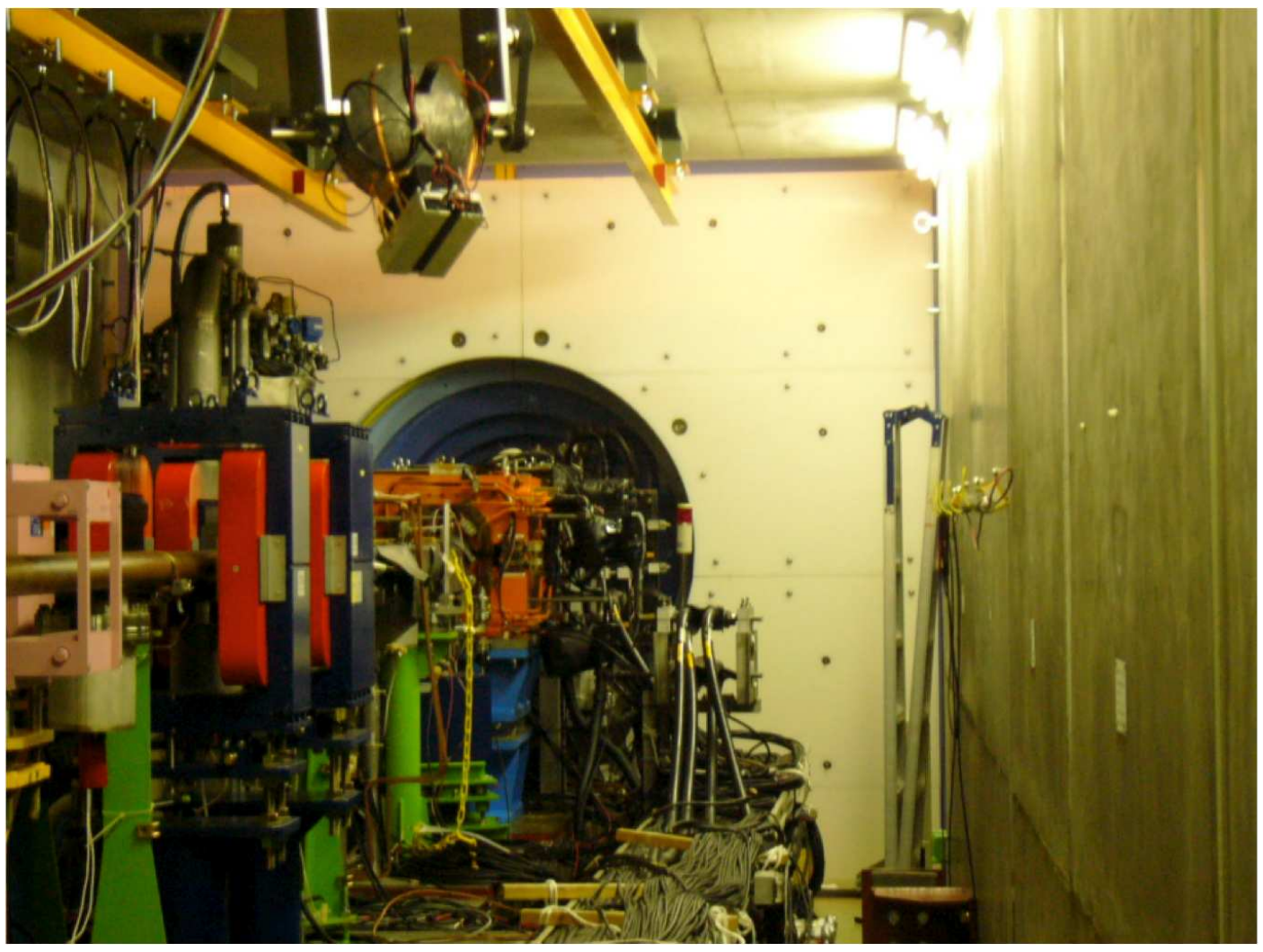

*Figure 10.11: Polyethylene shield.*

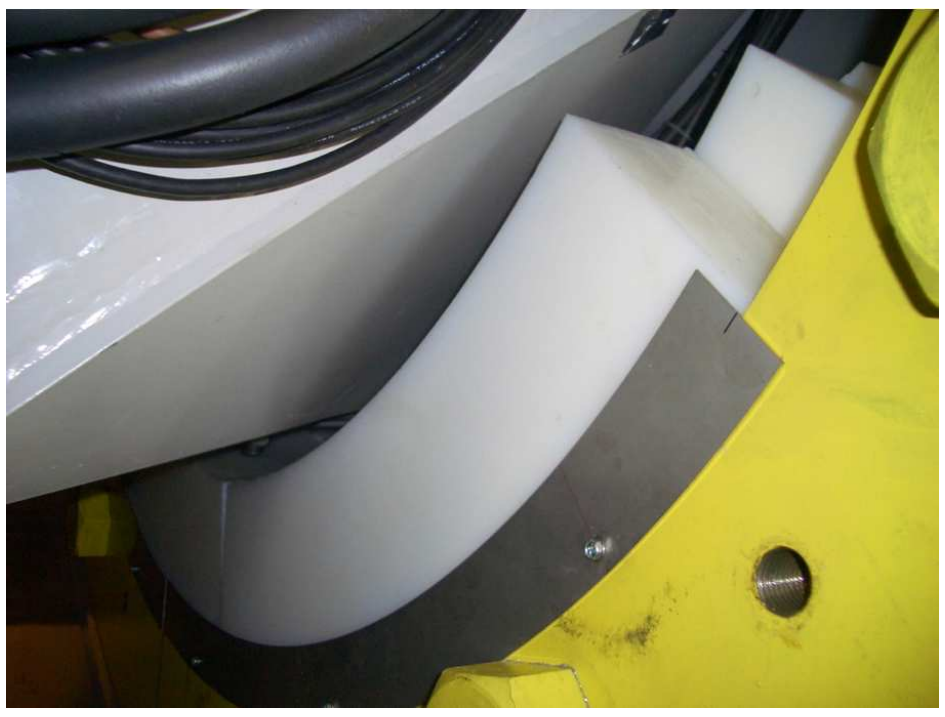

*Figure 10.12: Tapered shield.*

The quadrant boundaries of the endcap RPCs appear as dead regions near $x = 0$ and $y = 0$ Fig. 10.13 The tapered and gap shields are expected to reduce the ambient rate in the lower-azimuth quadrants (sector 2, with $x < 0$ and $y < 0$, and sector 3, with $x > 0$ and $y < 0$).

For the tapered shield alone, the after-to-before normalized ratios are shown in each of the thirty areal regions for each RPC layer in Fig. 10.14. The double ratios of Covered-to-Reference and Transition-to-Reference, shown in Fig. 10.15, indicate that the ambient-rate reduction is most prominent in the small-radius regions.

The patterns of minima as a function of radial region in Fig. 10.15 were speculated to be due to the shading by the tapered shield of particles incident from a pointlike source. We extracted the minimum for each radial region (by a fit to a double Gaussian as a function of layer number), then fitted these minima to a straight line as a function of radius. The minima and this fitted straight line is shown in Fig. 10.16; the intercept lies at the QC1E beamline magnet.

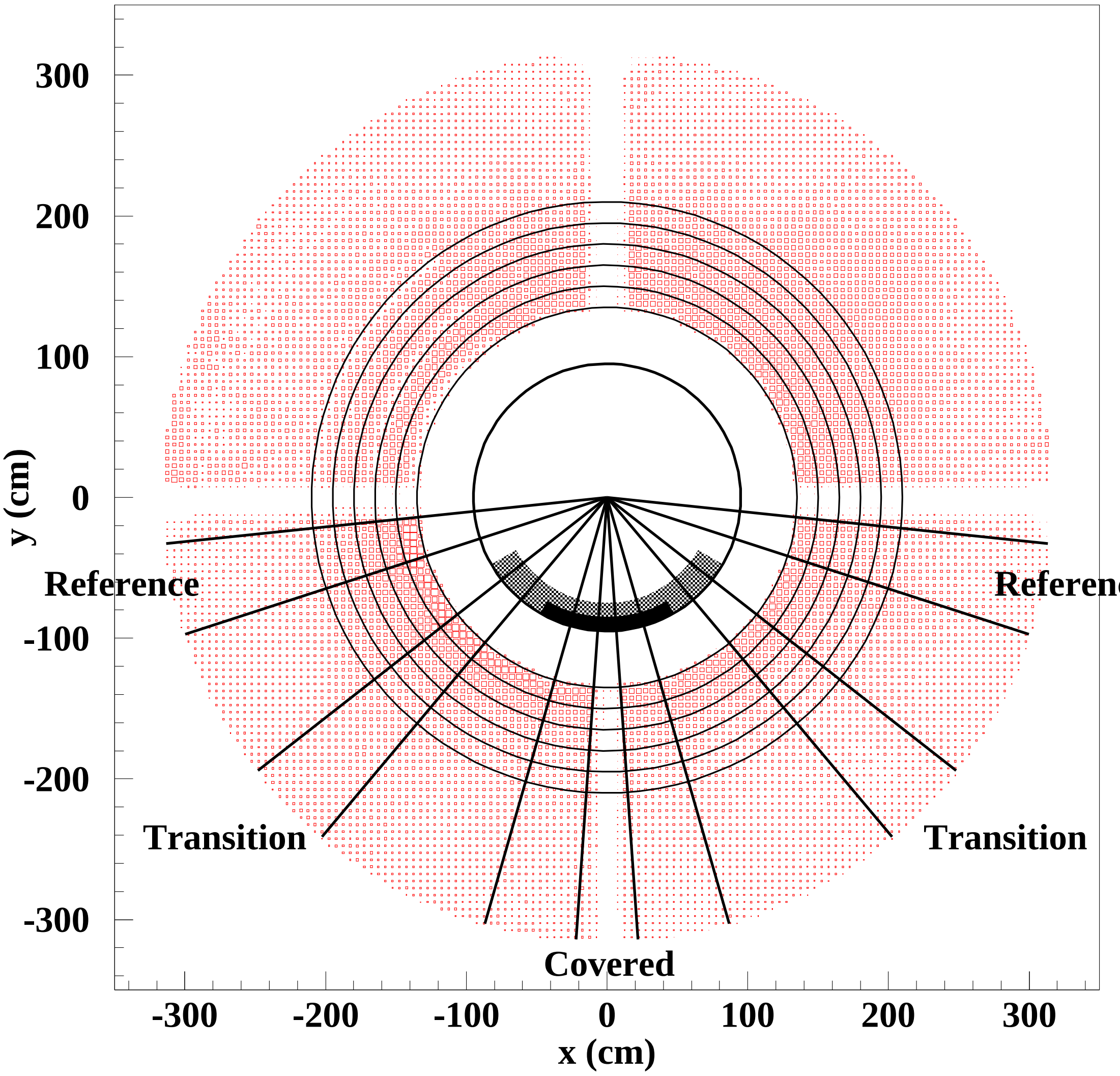

*Figure 10.13: Radial and azimuthal segmentation of the endcap RPCs to test the amelioration of the ambient hit rate by the tapered polyethylene shield (in gray) and, later, the interstitial polyethylene bricks (in black). The area of each small square is proportional to the ambient hit rate in its vicinity.*

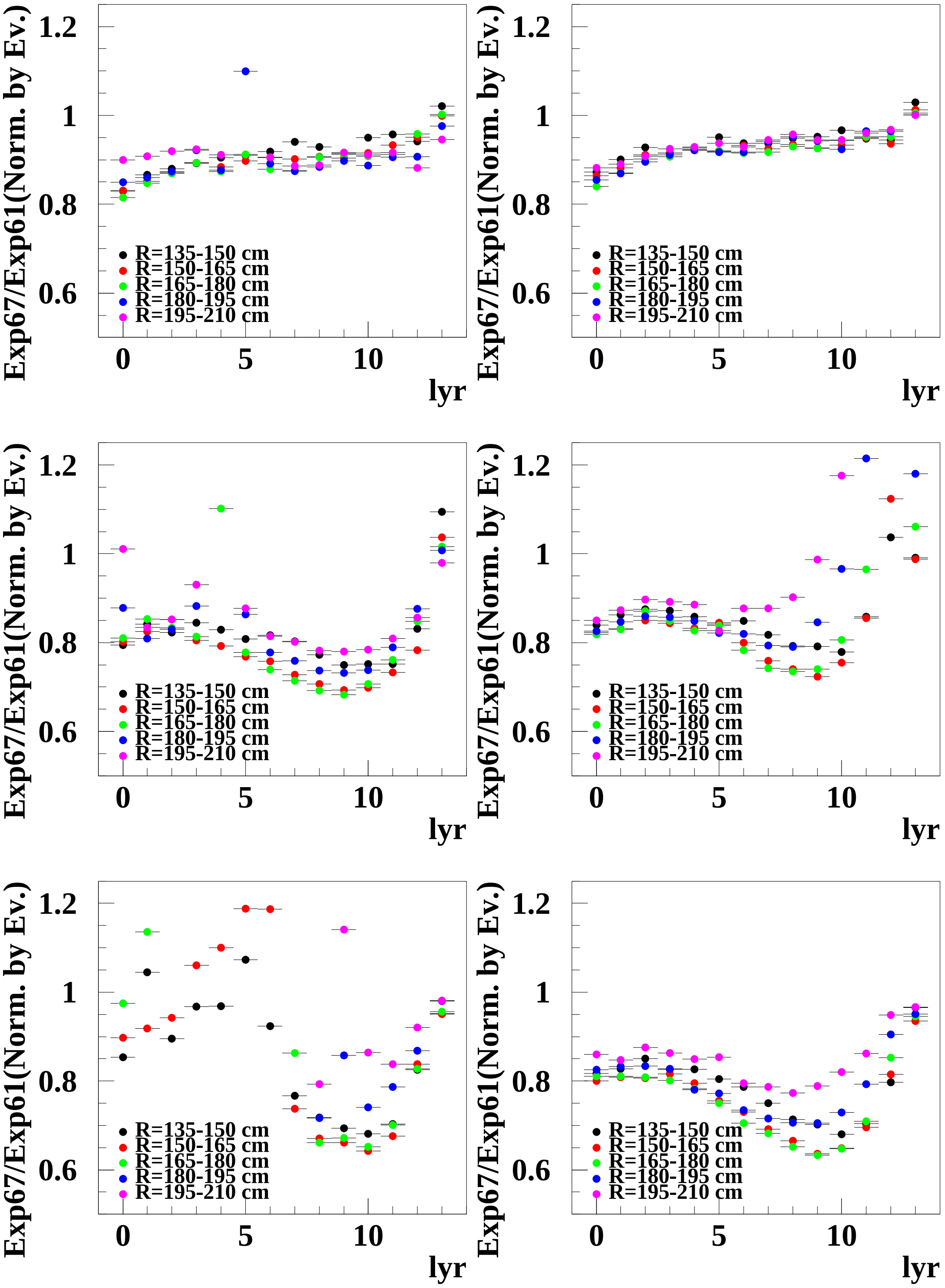

*Figure 10.14: After-to-before ambient-rate normalized ratios for the tapered shield as a function of endcap layer number. Left column: endcap sector 2; right: sector 3. Top row: Reference azimuthal wedge; middle: Transition; bottom: Covered.*

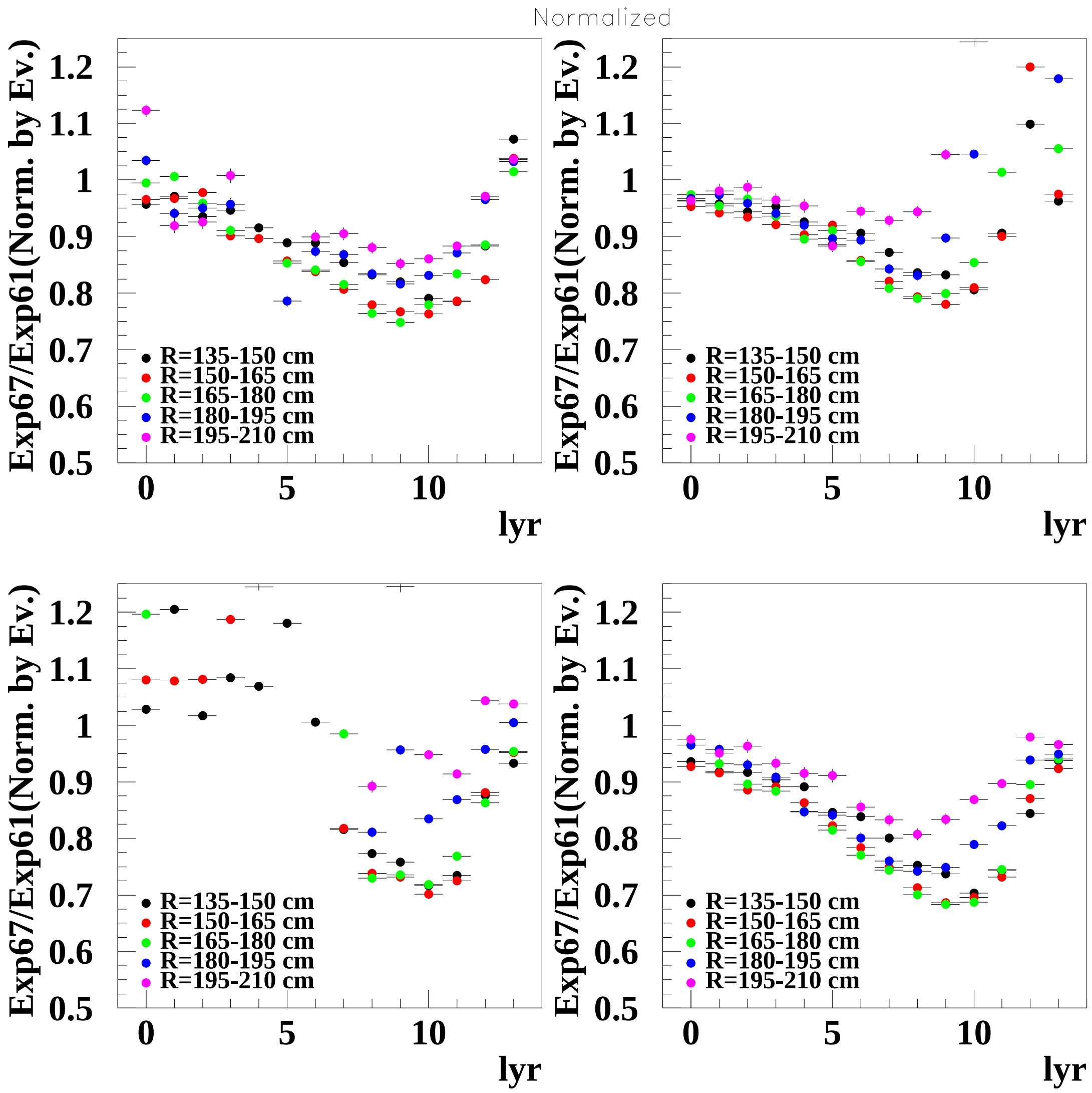

*Figure 10.15: After-to-before ambient-rate normalized double ratios for the tapered shield, as a function of endcap layer number. Left column: endcap sector 2; right: sector 3. Top row: Transition-to-Reference double ratio; bottom: Covered-to-Reference.*

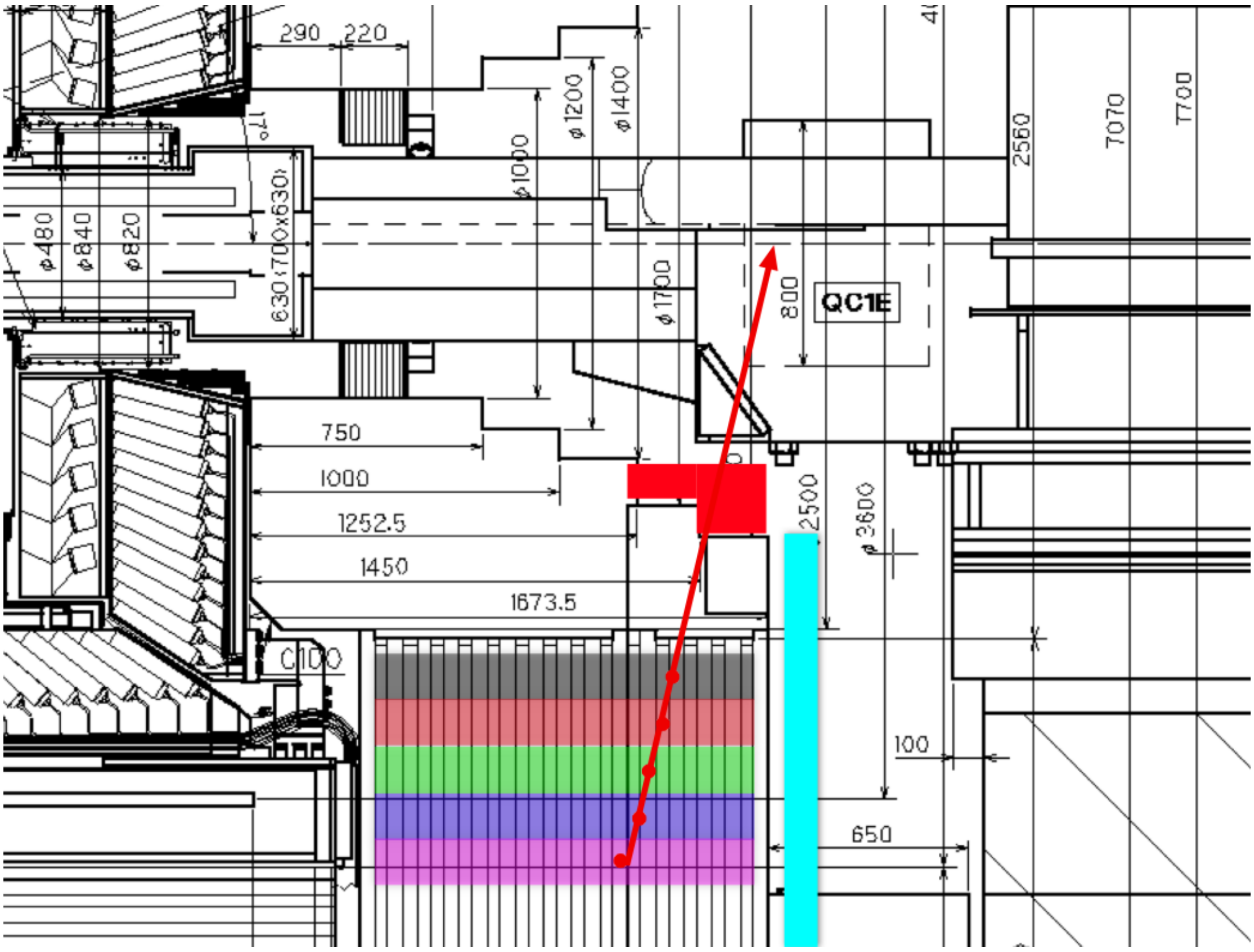

*Figure 10.16: Fitted minimum-rate points in each layer and their straight-line extrapolation to $r = 0$. The red shape is the tapered shield; the cyan shape is the planar shield.*

The same technique was applied after installation of the gap shield to measure its effect. The corresponding results are shown in Figs. 10.17 and 10.18. From the upper half of Fig. 10.18, we observe no change in the Transition region due to installation of the gap shield: this is not surprising, since the gap shield does not overlap with either Transition region. The lower half of this figure, however, shows a dramatic reduction in the ambient noise rate, particularly for the larger-radius regions. From the increased width (in layers) of the minimum as a function of radius, it appears that the gap acted like a pinhole camera for the incident particles before it was plugged with the polyethylene bricks.

One overriding conclusion from these studies is that low-$Z$ neutron shielding is quite effective in reducing the ambient rate in the endcaps: regardless of the detector technology, as much of this shielding as possible should be installed in Belle II.

### 10.4.1 Capability to operate at SuperKEKB luminosity

Though we have not identified all sources of background neutrons, it is likely that they are generated by secondary interactions following processes such as the backscattering of a radiative-Bhabha photon, beam-gas collisions, and the Touschek process (Ch. 2).

We extrapolate the measured hit rates in each region of the Belle KLM to Belle II by fitting these data to the sum of four terms, each linearly proportional to a beam-related quantity: the luminosity $\mathcal{L}$ (correlated with the backscattering of radiative-Bhabha photons), the product of vacuum pressure and LER current $V_c \times I_{\rm LER}$ (proportional to the rate of beam-gas collisions in the LER), the product of vacuum pressure and HER current $V_c \times I_{\rm HER}$ (proportional to the rate of beam-gas collisions in the HER), and the square of the LER current $I^2_{\rm LER}$ (proportional to the rate of background associated with Touschek scattering within each bunch of the small-bunch LER beam). The expected hit rates in SuperKEKB, with an instantaneous luminosity of $\mathcal{L} = 5 \times 10^{35}\,{\rm cm}^{-2}{\rm s}^{-1}$, are listed in Table 10.1
The expected efficiencies are obtained by converting these extrapolated hit rates according to the dependences shown in Fig. 10.19. Differences in the characteristics of the glass used in the barrel and endcap RPCs account for the distinct dependence of efficiency on ambient rate. The slope of the efficiency vs hit rate for the barrel (endcap) is about $-0.08$ ($-0.5$), with hit rate measured in Hz/cm$^2$.
The expected efficiencies are listed in Table 10.4.1. The innermost layers of the barrel exhibit a somewhat reduced efficiency, but are still operable in Belle II. On the other hand, all layers of the endcap are projected to be completely inefficient. We therefore consider a full replacement of the endcap KLM and the replacement of few innermost barrel layers with scintillator strip detector described in the next section. We note that the low-$Z$ material of scintillator layers provides additional neutron background shielding of the remaining RPC which is not taken into account in the above estimates for being conservative.

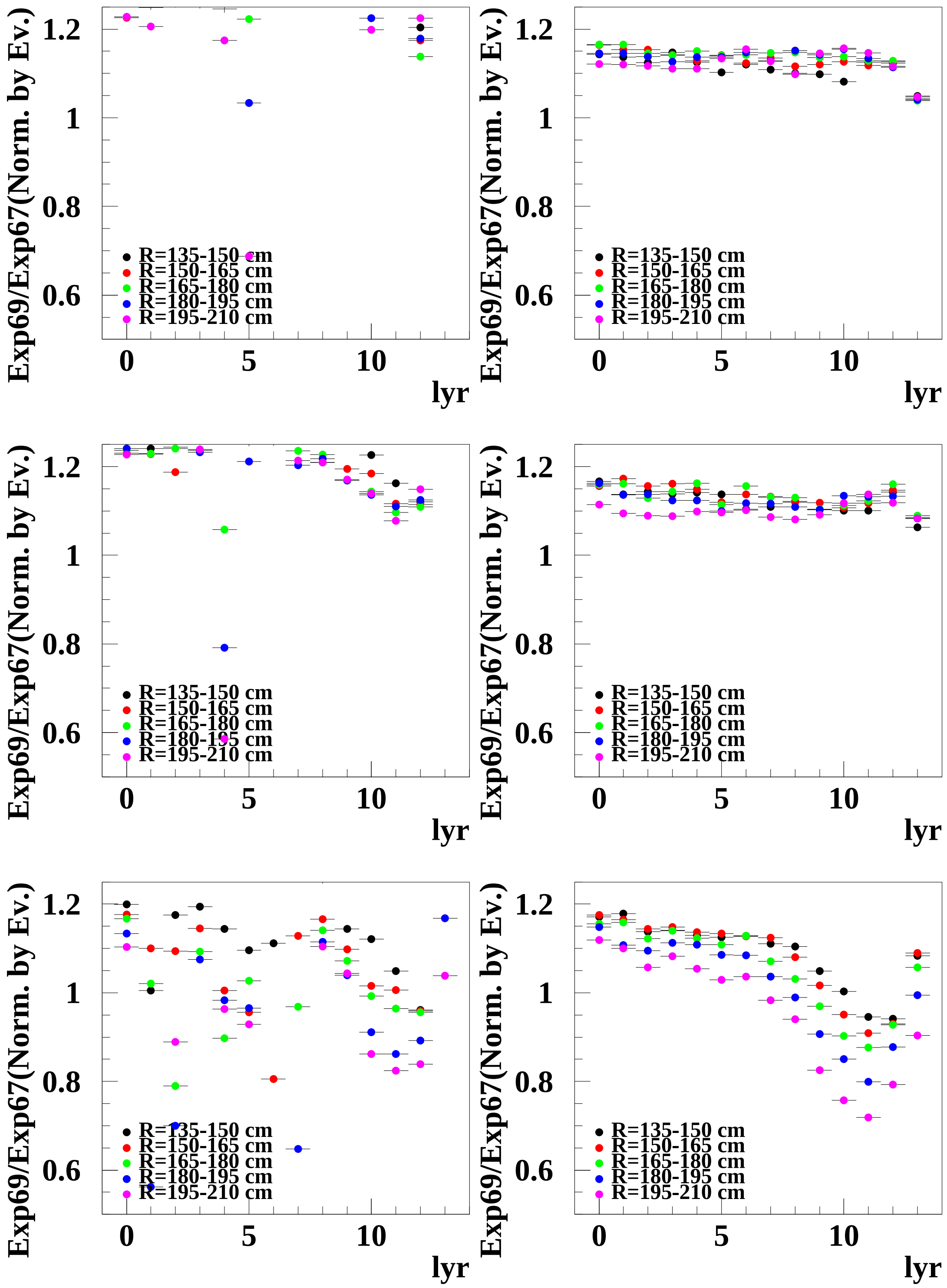

*Figure 10.17: After-to-before ambient-rate normalized ratios for the gap shield as a function of endcap layer number. Left column: endcap sector 2; right: sector 3. Top row: Reference azimuthal wedge; middle: Transition; bottom: Covered.*

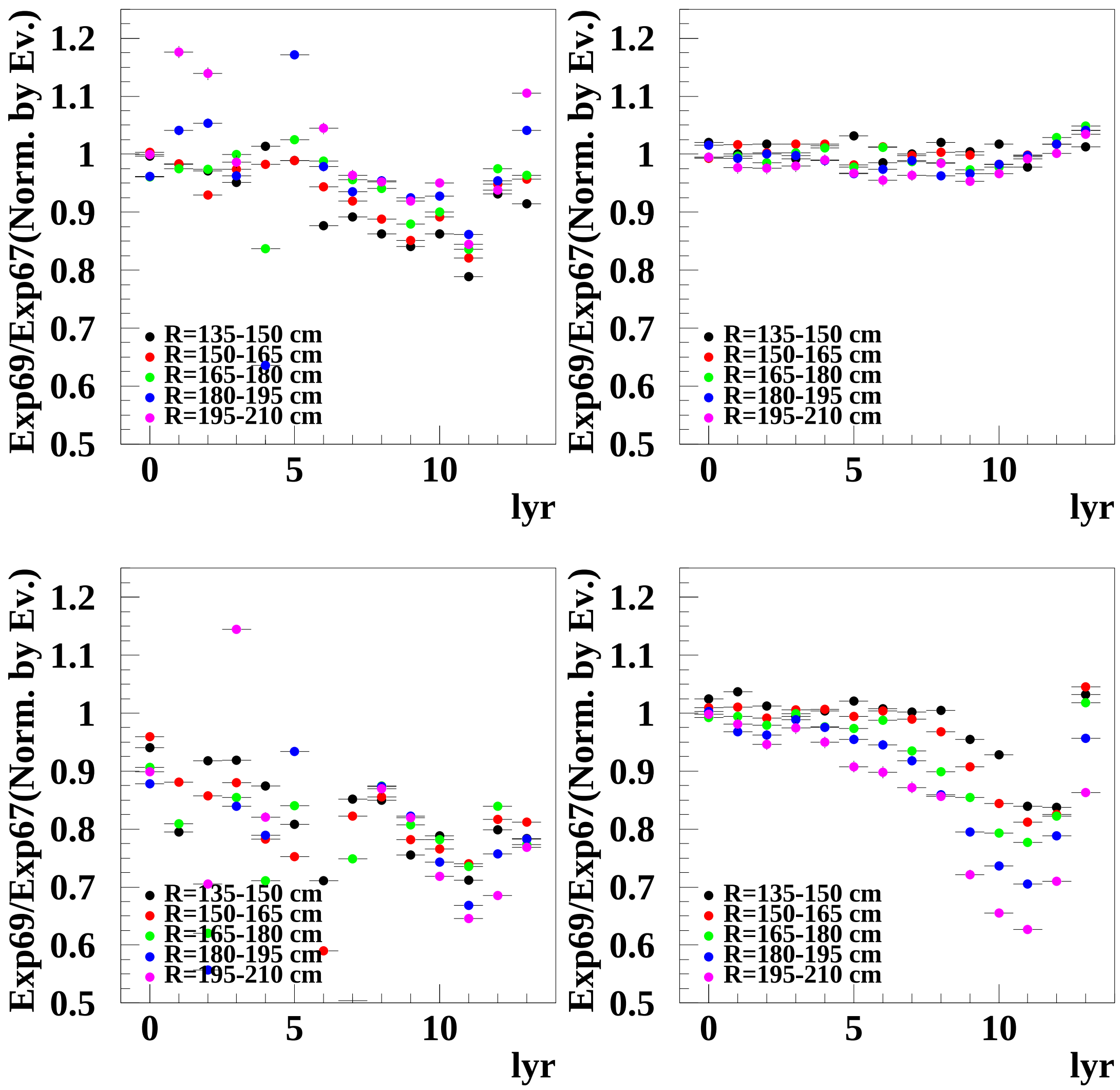

*Figure 10.18: After-to-before ambient-rate normalized double ratios for the gap shield, as a function of endcap layer number. Left column: endcap sector 2; right: sector 3. Top row: Transition-to-Reference double ratio; bottom: Covered-to-Reference.*

| Layer | Barrel | | Endcap forward | | Endcap backward | |
|---|---|---|---|---|---|---|
| | KEKB | SuperKEKB | KEKB | SuperKEKB | KEKB | SuperKEKB |
| 0 | 1.07 | 3.6 | 0.29 | 2.4 | 0.31 | 3.4 |
| 1 | 0.74 | 2.3 | 0.25 | 2.4 | 0.31 | 2.9 |
| 2 | 0.43 | 1.6 | 0.23 | 2.4 | 0.31 | 2.8 |
| 3 | 0.28 | 1.1 | 0.23 | 2.0 | 0.31 | 2.8 |
| 4 | 0.21 | 0.67 | 0.23 | 2.2 | 0.32 | 2.8 |
| 5 | 0.16 | 0.60 | 0.26 | 2.7 | 0.35 | 2.9 |
| 6 | 0.14 | 0.63 | 0.24 | 2.7 | 0.34 | 1.5 |
| 7 | 0.13 | 0.43 | 0.26 | 3.3 | 0.36 | 2.6 |
| 8 | 0.12 | 0.73 | 0.27 | 3.1 | 0.38 | 3.0 |
| 9 | 0.11 | 0.47 | 0.21 | 3.9 | 0.41 | 2.8 |
| 10 | 0.09 | 0.29 | 0.36 | 4.7 | 0.46 | 3.5 |
| 11 | 0.13 | 0.39 | 0.47 | 5.3 | 0.50 | 3.0 |
| 12 | 0.09 | 0.44 | 0.55 | 3.7 | 0.49 | XX |
| 13 | 0.12 | 0.42 | 0.56 | XX | 0.55 | XX |
| 14 | 0.09 | 0.48 | N/A | N/A | N/A | N/A |

*Table 10.1: Ambient rate (Hz/cm$^2$) measured in KEKB and extrapolated to SuperKEKB.*

## 10.5 Scintillator Endcaps

Because of the projected inefficiency of RPCs at high ambient rate, the Belle II endcaps are instrumented with scintillator strips [6]. Charged particles detection with such detectors, where the scintillation light is trapped in embedded wavelength-shifting (WLS) fibers and delivered to photomultipliers or other sensors, is a well-established technique [7, 8, 9].
In Belle II, the limited space and strong magnetic field do not allow us to use photomultipliers.

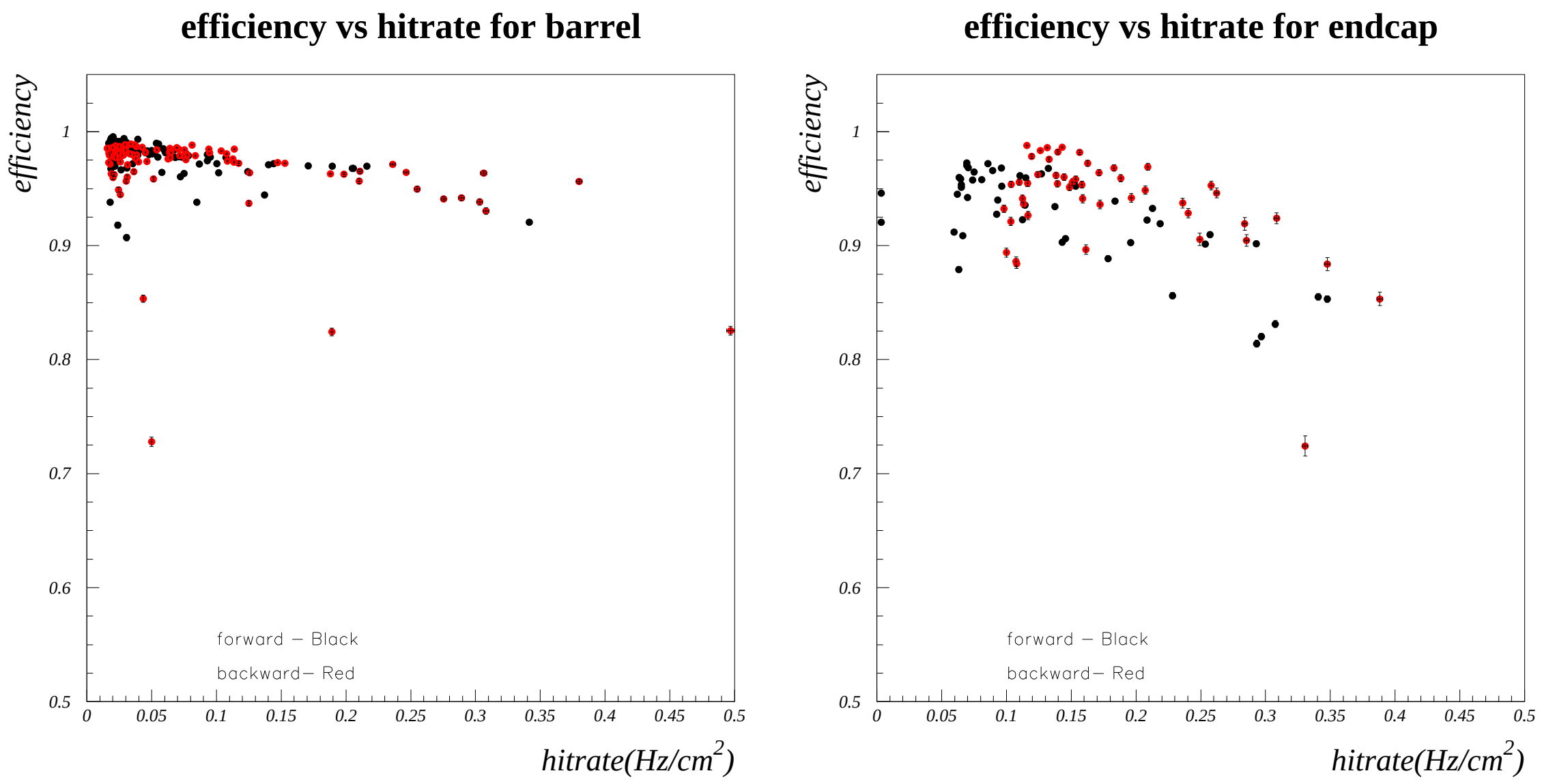

*Figure 10.19: Efficiency vs ambient rate for the barrel (left) and endcap (right) RPCs.*

| Layer | Barrel | | Endcap forward | | Endcap backward | |
|---|---|---|---|---|---|---|
| | KEKB | SuperKEKB | KEKB | SuperKEKB | KEKB | SuperKEKB |
| 0 | 0.91 | 0.70 | 0.91 | 0.0 | 0.90 | 0.0 |
| 1 | 0.94 | 0.81 | 0.93 | 0.0 | 0.90 | 0.0 |
| 2 | 0.96 | 0.87 | 0.94 | 0.0 | 0.90 | 0.0 |
| 3 | 0.98 | 0.91 | 0.94 | 0.0 | 0.90 | 0.0 |
| 4 | 0.98 | 0.94 | 0.94 | 0.0 | 0.89 | 0.0 |
| 5 | 0.99 | 0.95 | 0.92 | 0.0 | 0.88 | 0.0 |
| 6 | 0.99 | 0.95 | 0.93 | 0.0 | 0.89 | 0.0 |
| 7 | 0.99 | 0.96 | 0.92 | 0.0 | 0.87 | 0.0 |
| 8 | 0.99 | 0.94 | 0.92 | 0.0 | 0.86 | 0.0 |
| 9 | 0.99 | 0.96 | 0.90 | 0.0 | 0.85 | 0.0 |
| 10 | 0.99 | 0.98 | 0.87 | 0.0 | 0.82 | 0.0 |
| 11 | 0.99 | 0.97 | 0.82 | 0.0 | 0.80 | 0.0 |
| 12 | 0.99 | 0.96 | 0.78 | 0.0 | 0.81 | 0.0 |
| 13 | 0.99 | 0.97 | 0.77 | 0.0 | 0.76 | 0.0 |
| 14 | 0.99 | 0.96 | N/A | N/A | N/A | N/A |

*Table 10.2: RPC efficiency measured in KEKB and extrapolated to SuperKEKB.*

As a photodetector, we consider multipixel silicon photodiodes operating in the Geiger mode, which were developed in Russia [10, 11, 12, 13] in the 1990s. They are now produced by many companies that use different names for their products: SiPM, MRS APD, MPPC, MAPD, etc. Here, we use the generic term Silicon PhotoMultiplier (SiPM). These sensors are compact and operate in a strong magnetic field.
The first experience of large-scale SiPM use was obtained by ITEP in the production of the 8,000 scintillator tiles with WLS fiber and SiPMs for the CALICE hadron calorimeter prototype [14]. Now, the contemporaneous T2K experiment uses SiPM readout in many components of the near detector (FGDs and SMRD), with 65,000 channels in total [9]. This technology is mature and our group has a lot of direct experience with it. The scintillator-based endcaps meet the physics requirements of Belle II, and can cope with a background rate that is at least two orders of magnitude higher than in Belle and well above that expected in Belle II.

### 10.5.1 General layout

Scintillator-based superlayers are installed in all 14 existing gaps in the magnet yoke in both forward and backward endcaps. The whole system consists of 16,800 scintillator strips with WLS readout, arranged in two orthogonal planes to form a superlayer within each gap. The independent operation of two planes reduces the combinatorial background in comparison with the present RPC superlayer design, where one background hit produces signals in both readout planes.
The scintillator strips have a cross section of (7–10) $\times$ 40 mm and a length of up to $\sim$ 2.8 m. The strip width is a compromise between the desire to limit the total number of channels and need for reasonable spatial resolution for muon and $K_L^0$ reconstruction. This granularity is similar to the average cathode-strip granularity of the RPCs. Individual scintillators are covered with a diffuse reflective coating. The strip has a groove in the center to accommodate a WLS fiber. Scintillator light is caught by this fiber and transported to the SiPM. The WLS fiber is read out from one side. The far end of the WLS fiber is mirrored to double the light yield at the SiPM. The WLS fiber is glued to the scintillator to increase the efficiency of the light collection. The SiPM is coupled to the fiber end. It is fixed and aligned with the fiber using a plastic housing. One layer of scintillator strips, arranged in a quadrant that fits into the existing gap in the iron yoke, is shown in Fig. 10.20. The sector frame is identical to the existing one used for the RPC mounting. The dead zone around the inner arc is estimated to be $\sim$ 1% of the total sector area, due to the inscribing of the rectangular structure into the circle. The inner dead zone is approximately the same as in the present RPCs. Around the outer circle, the dead zone is 4%, because of location here of cables and the eschewal of very short scintillators. The acceptance loss at large radii is not critical for muon or $K_L^0$ reconstruction. In the central area of the sector, the small dead zones ($<$ 1%) are due to support structures. In addition, the insensitive area between strips due to the reflective cover is only 0.3%. In total, the geometrical acceptance of the proposed system is slightly better than that of the present RPCs.

### 10.5.2 Scintillator strips

The scintillator strips are made of polystyrene doped with PTP (1.5%) and POPOP (0.01%). The strips are produced by the extrusion technique. This technique provides the possibility to produce long strips. Different producers use slightly different ingredients and production techniques, but generally they achieve quite similar quality and request a similar price. Among the known producers are Fermilab and "Uniplast" (Vladimir, Russia), which produced the scintillator strips for the T2K experiment [9], and "Amkris-Plast" (Kharkov, Ukraine), which produced the strips for the Opera experiment [8]. We tested the characteristics of the scintillator strips from three producers. All strips had a width of 40 mm. The thickness of the Fermilab and Kharkov strips was 10 mm, while that of the Vladimir strips (surplus stock for T2K) was 7 mm. In all strips, the Kuraray WLS multi-cladding 1.2-mm diameter fibers were glued using SL-1 gel produced at SUREL, St. Petersburg. This fiber provides more light output than other fibers and has a large attenuation length. This fiber was used in many detectors with similar scintillator plastic and geometry [7, 8, 9]. The light emission spectrum of the Kuraray Y11 fiber matches well with the SiPM spectral efficiency.

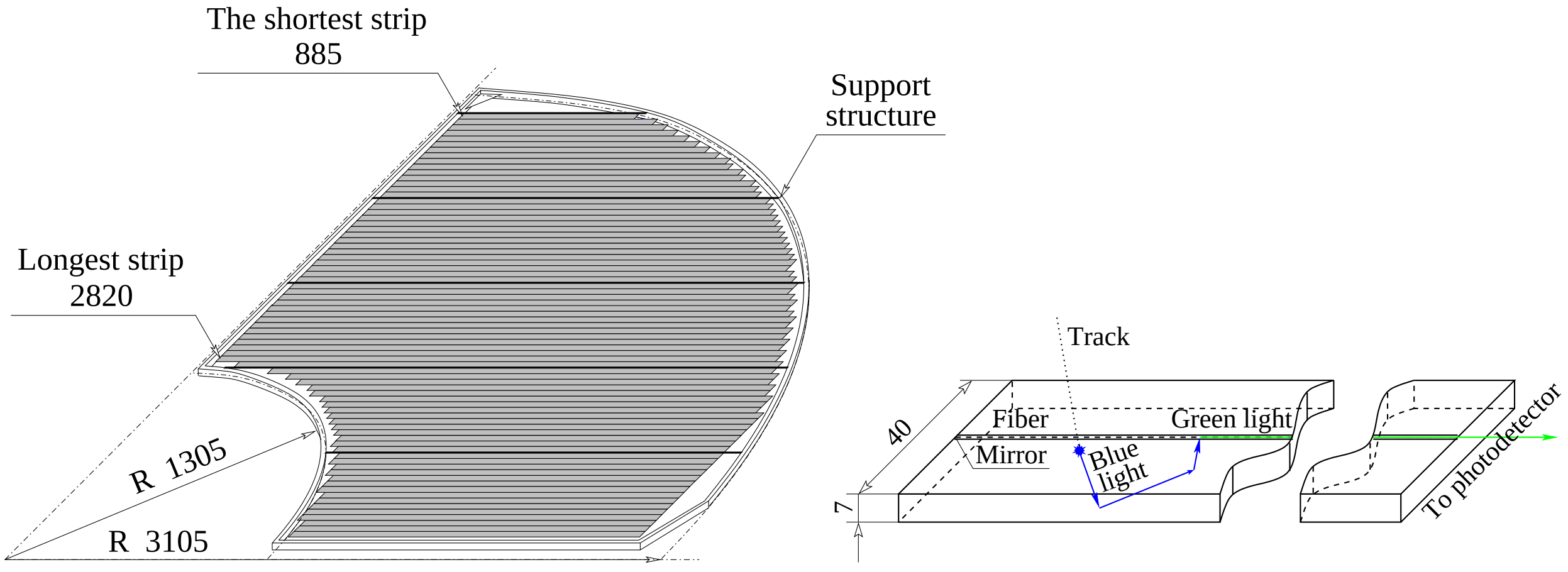

*Figure 10.20: a) One layer formed by scintillator strips. b) Scintillator light detection in the strip.*

The strips have been tested using a cosmic-ray test bench. The cosmic trigger is provided by a pair of short trigger strips ($L = 16$ cm) placed above and below the strip under consideration. The trigger strips were moved along the tested strip to measure the light yield depending on the distance to the photodetector. All strips were tested using the same SiPM to avoid a possible spread in the SiPM efficiency. To determine the average number of detected photons, we correct the measured number of fired pixels by the cross-talk (factor $1/(1 + \delta) \sim 0.85$) and conservatively use a truncated mean of the Landau distribution, discarding 10% of the lower part of the distribution and 30% of the higher part and then taking the average. In this way, the Landau tail is discarded. The averaged number is then corrected for the slanted incidence of the cosmic muons using a toy Monte Carlo simulation. In the case of the Vladimir scintillator, the reflective coating is very thin ($\sim 50\,\mu$m, compared with $\sim 150\,\mu$m for the FNAL and Kharkov strips). Their light yield is improved by about 15% by using a reflective layer of white paper (Tyvek).

The distribution of the average number of detected photons at the SiPM is shown in Fig. 10.21(a). For comparison, the light yield of the FGD of the T2K experiment, collected from the Fermilab strips ($10 \times 10\,\text{mm}^2$), is shown in Fig. 10.21(b). As can be seen from these distributions, we achieve a slightly better light yield in spite of our four times larger strip width. (The light yield is approximately proportional to $1/\sqrt{\text{width}}$.) This is due to the larger diameter of our fiber and the gluing of this fiber to the strip.

We measure the time resolution using the cosmic trigger for a strip equipped with two SiPMs, one at each end of the fiber. From the time difference dispersion, we derive the time resolution of $\sigma_t \approx 0.7$ ns. The measured velocity of the light propagation in the WLS Y11 Kuraray fiber

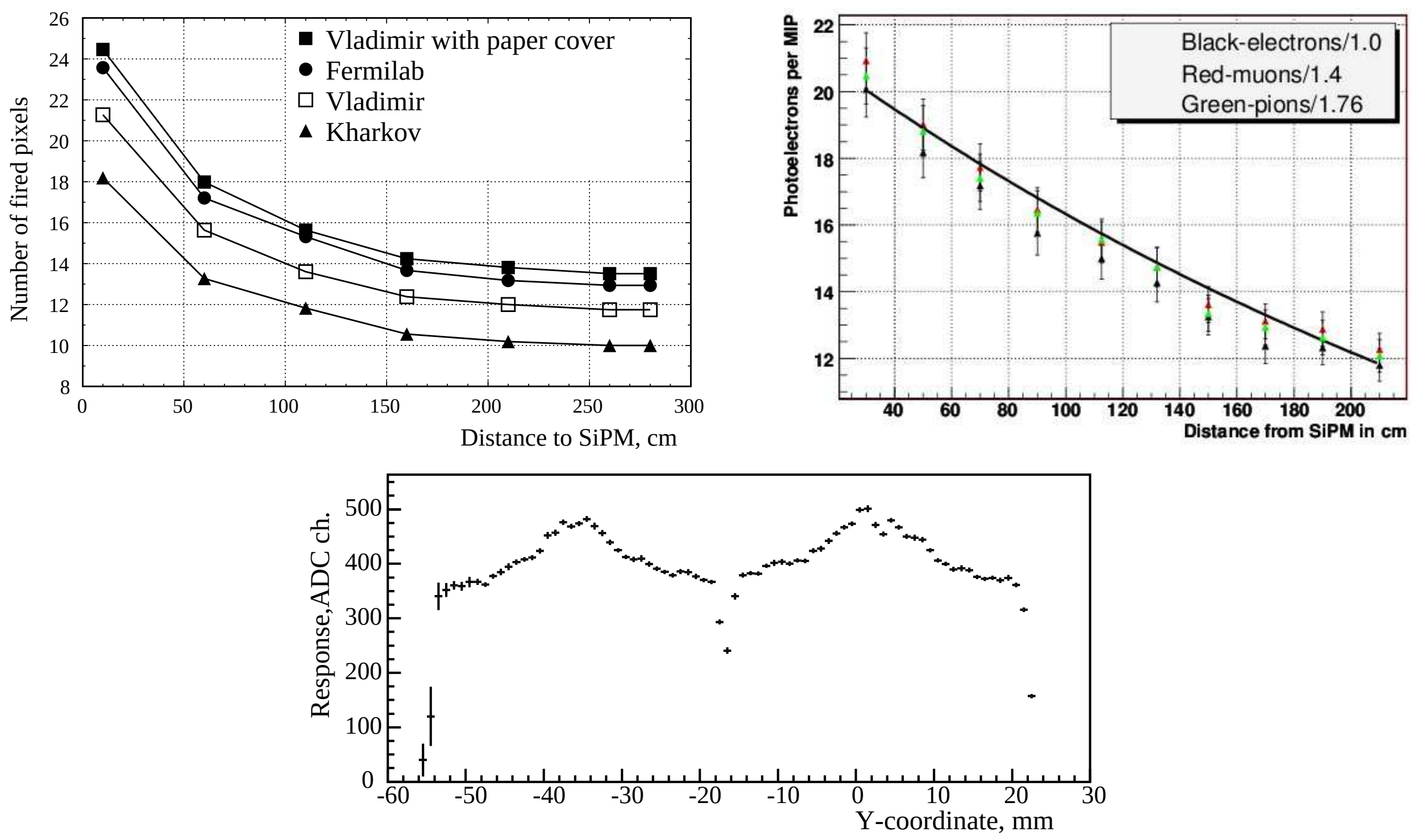

*Figure 10.21: The distribution of the average number of detected photons at SiPM as a function of the distance to the photodetector for a) our study b) FGD at T2K experiment. c) ADC response, which is proportional to the light yield, for two neighboring strips.*

with the diameter of 1.2 mm is $v = 17\,\mathrm{cm/ns}$.

The transverse uniformity of the strip response was measured at the ITEP proton beam. Figure 10.21(c) (upper plot) shows the ADC response, which is proportional to the light yield, for two strips connected to each other. The nonuniformity of the response is $\sim 15\%$. From these results, we conclude that the light yield is sufficient no matter where a particle hits the strip.

The long-term stability of the light collection efficiency of the strips with glued WLS fibers was tested using the test module of 96 strips (1 meter long) that was manufactured in 2006. The strips were equipped with CPTA's SiPMs. The test module was used for the measurement of the neutron background in the KEKB tunnel over a six-month span. In 2009, these strips were studied again. No degradation of the light yield, within the 10% accuracy of the measurements, was observed in the three years since their production.

### 10.5.3 SiPMs

A SiPM is a matrix of tiny photodiodes (pixels) connected to a common bus and working in the Geiger mode. Typically, there are $\sim 100 - 1000$ pixels in an area of $1 \times 1\,\mathrm{mm}^2$. Connection to the common bus is done usually by a $\sim 10^6\,\Omega$ resistor. This resistor limits the Geiger discharge developed in a pixel by photo or thermal electrons. In this mode, the SiPM response depends on the number of fired pixels and is proportional to the initial light until the number of fired pixels is much less than the total number of pixels.

As compared with conventional vacuum photomultipliers, SiPMs have low operational voltage of a few tens of Volts. This is higher than the breakdown threshold by several volts. The values of the overvoltage and the capacitance of a single pixel determine the photodetector gain (typically of order of $10^6$). The values of the quenching resistor connected in series to each pixel and the capacitance of a single pixel determine the dead time of a pixel, which is typically $\sim 100\,\mathrm{ns}$. The photon detection efficiency ($PDE$) depends on overvoltage and reaches 25–30% for modern devices. SiPMs with a smaller number of pixels have higher $PDE$ because of the smaller non-sensitive area between pixels. SiPMs are not affected by a magnetic field of up to 4 T [15]. Among the disadvantages of SiPMs are a high level of noise ($\sim 10^6\,\mathrm{Hz/mm^2}$ at the threshold of one p.e.), an optical inter-pixel crosstalk producing a tail of higher amplitudes at random spectrum, and a high sensitivity of the SiPM response to ambient temperature (the break-down voltage depending on temperature according to $\sim 60\,\mathrm{mV/K}$ for Hamamatsu and $\sim 20\,\mathrm{mV/K}$ for CPTA).

A comparison of SiPMs from different manufacturers is depicted in Fig. 10.22. The photon detection efficiency (for green light), gain, crosstalk, and noise rate are shown as a function of overvoltage for detectors produced by MEPhI/PULSAR, CPTA (Russia), Hamamatsu (Japan). SiPMs produced by CPTA and Hamamatsu meet our requirements of high MIP registration efficiency, low crosstalk, and low noise frequency. Both producers have the experience of producing thousands of SiPMs and can produce SiPMs matched with the 1.2-mm diameter fiber. (Both CPTA and Hamamatsu were considered as the main producers for T2K experiment.)

SiPMs demonstrate a very high long-term stability. The experience in the operation of the 7620-channel CALICE hadron calorimeter prototype during three years of beam tests at CERN and FNAL [14, 16] showed no significant problem. Only eight dead channels (0.1%) were observed, while the characteristics of remaining channels did not change. The long-term stability of the Hamamatsu SiPMs was tested in the T2K experiment, where $\sim 65,000$ channels were studied during more than one year of operation. They observed failure in only $\sim 20$ channels—less than 0.03%—mostly because of mechanical damage.

### 10.5.4 Radiation hardness

Scintillator counters with WLS fiber are used in many existing detectors, even at hadron machines where the radiation dose is a few orders of magnitude higher than those expected at SuperKEKB. In particular, the light collection at the electromagnetic calorimeter for the HERA-B experiment is based on the Vladimir scintillator tiles read out by Kuraray Y11 fibers [17]. During six years of operation, no deterioration of the electromagnetic calorimeter performance was observed. However, the proposed method of light detection by SiPM is a new method that required detailed radiation hardness tests. In particular, the radiation hardness of the SiPM is measured to be high in case of irradiation with electrons and photons, while neutrons and protons cause significant damage after a moderate integrated dose of $\sim 1\,\mathrm{krad}$ [14]. The flux of harmful particles through the SiPM results in defect formation in the thin area from which charge carriers are collected. This results in an increase of the SiPM noise rate and, consequently, the dark current. The SiPM dark current grows linearly with the particle flux as in other silicon detectors.

The damage effects from proton and neutrons are energy dependent and, without knowledge of the beam background spectra, it is difficult to predict the radiation damage. Therefore, to measure the possible damage to SiPMs due to the radiation level at SuperKEKB, we performed radiation tests directly in the KEKB tunnel. Eight Hamamatsu SiPMs and eight CPTA SiPMs were placed in the KEKB tunnel for six weeks during accelerator operation. The integrated neutron dose in the tunnel, measured with Luxel dosimeters (J-type badges), was $\sim 1.2\,\mathrm{Sv}$.

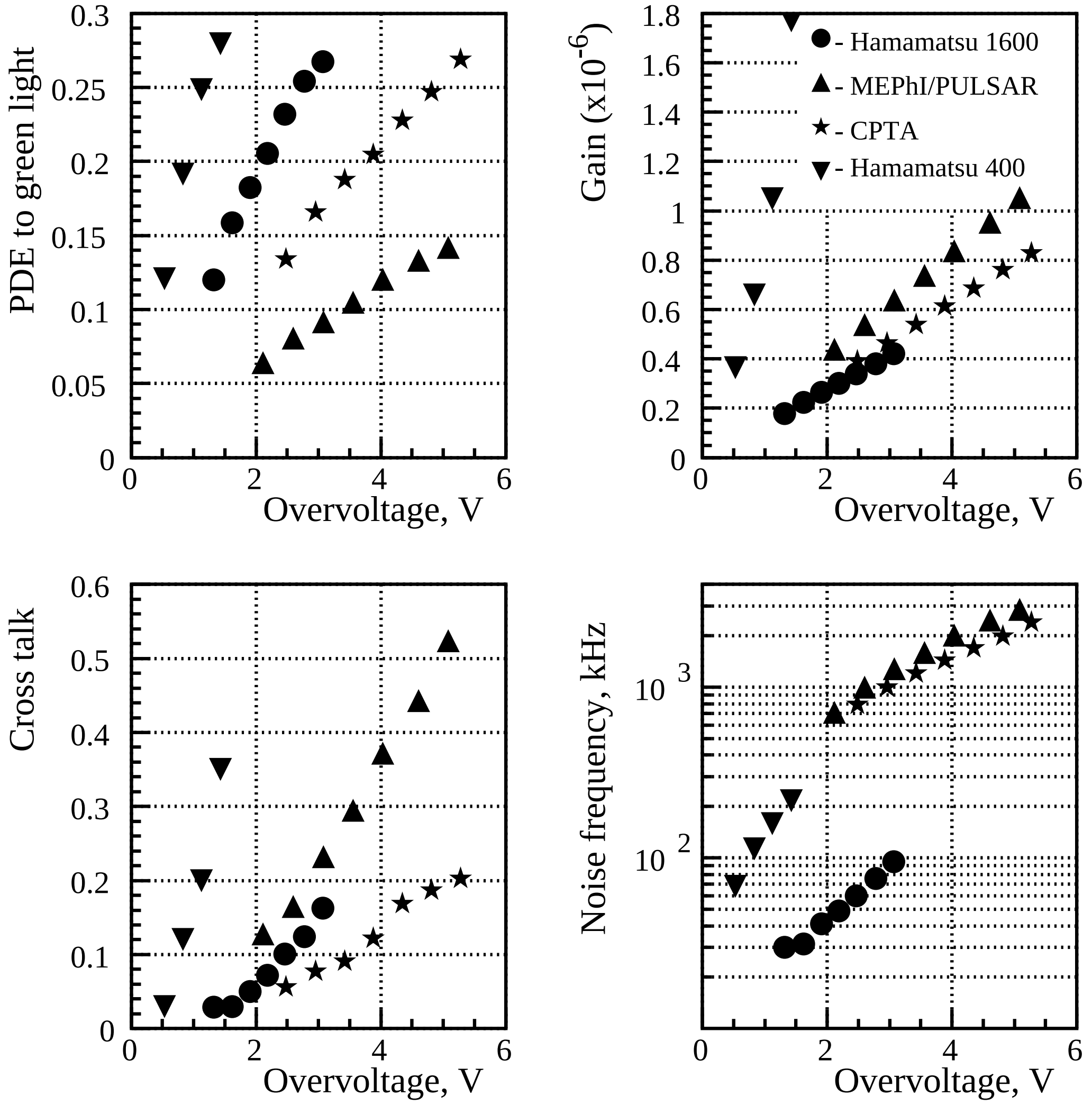

*Figure 10.22: Overvoltage dependence of a) efficiency, b) gain, c) crosstalk and d) noise frequency for SiPMs from different manufacturers: Hamamatsu 1600 pixels (∘), MEPhI/PULSAR (▲), CPTA (⋆), Hamamatsu 400 pixels (▼).*

After irradiation, the SiPM dark current increased for both manufacturers' devices by a factor of 3.6, with a small dispersion ($\sim$ 15%) in the eight specimens.
The neutron dose was also measured near the proposed installed position of the SiPM using Luxel dosimeters. The measured dose varies from 0.8 to 2.0 mSv/week at the luminosity of $\mathcal{L} \sim 1.7 \times 10^{34}\,\mathrm{cm}^{-2}\mathrm{s}^{-1}$. Assuming the neutron dose increases linearly with the luminosity, the expected neutron dose integrated over ten years of operation at $\mathcal{L} \sim 8 \times 10^{35}\,\mathrm{cm}^{-2}\mathrm{s}^{-1}$ does not exceed 40 Sv. The extrapolated factor of dark current increase after ten years of operation of SuperKEKB is therefore 150, resulting in an expected final dark current of $\sim 14\,\mu$A ($\sim 200\,\mu$A) for Hamamatsu's (CPTA's) SiPM. The difference between the Hamamatsu and CPTA SiPMs can be explained by the different thickness of the sensitive zone, to which both initial noise and the damage effect are proportional.
The study of SiPM properties after irradiation was done at ITEP. The SiPMs were subjected to fast irradiation in the 200-MeV secondary proton beam of ITEP synchrotron ($\sim 10^{10}\,p/\mathrm{cm}^2/\mathrm{hr}$). One month after the irradiation, the Hamamatsu SiPM dark current was $16\,\mu$A for two specimens subjected to an integrated proton flux of $2.0 \times 10^{11}\,p/\mathrm{cm}^2$ and $6.5\,\mu\mathrm{A}$ for two other specimens subjected to an integrated proton flux of $7.4 \times 10^{10}\,p/\mathrm{cm}^2$.

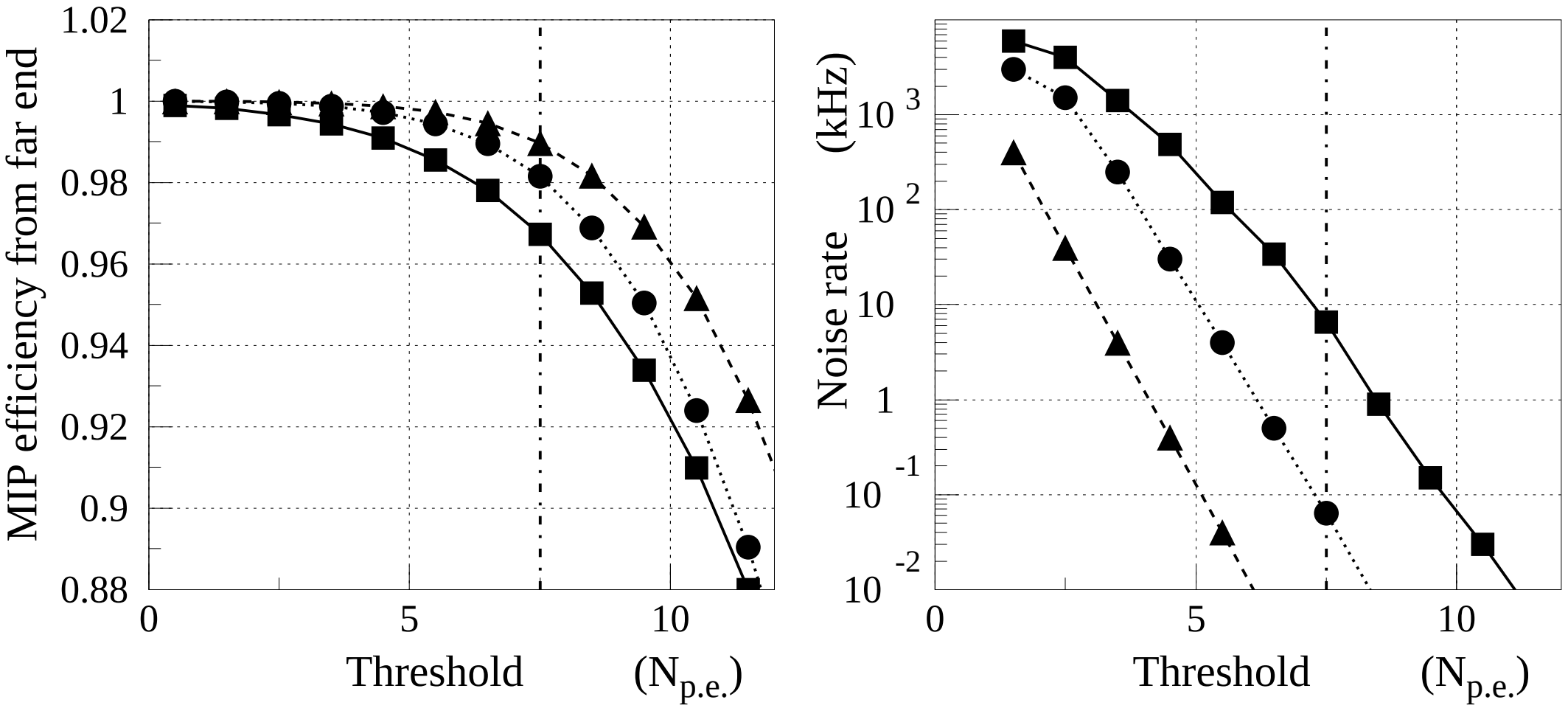

*Figure 10.23: Dependence on the threshold in number of photoelectrons of a) the detection efficiency for MIPs striking the strip's far end, and b) the SiPM internal noise rate. The triangles correspond to the virgin SiPM, the circles to the SiPM irradiated with a dose of* $7.4 \times 10^{10}\,p/\mathrm{cm}^2$ *(five years of operation at SuperKEKB), and the squares to the SiPM irradiated with a dose of* $2.0 \times 10^{11}\,p/\mathrm{cm}^2$ *dose (more than ten years of operation at SuperKEKB).*

We studied the light yield from MIPs passing through the strip at the far end ($\sim 3$ m from the photodetector) detected with virgin and irradiated SiPMs. The average number of photoelectrons from the MIP signal using irradiated and control SiPMs is unchanged, though, for the irradiated SiPMs, the MIP signal is slightly smeared because of noise. The dependence of the efficiency for the far end of the strip and the noise rate on the threshold (in number of fired pixels) is presented in Fig. 10.23. We conclude that the light detection efficiency is not significantly changed after irradiation, while the internal noise rate increases significantly. However, at the threshold of $\sim$ 7.5 fired pixels, the background rates from the increased SiPM noise remains smaller ($\lesssim$ 10 kHz) than the physical background rate due to neutron hits in the scintillator ($\lesssim$ 200 kHz). The MIP detection efficiency is similar to that of irradiated CPTA's SiPM, but the latter's noise rate is too high (100–500 kHz).

We therefore conclude that the Hamamatsu's SiPM radiation hardness is sufficient for successful KLM operation at the designed luminosity for at least ten years. The significantly increased dark current, however, should be considered in the design of electronics, HV supplies, slow control, and the calibration procedure.

### 10.5.5 Measurement of the background neutron rate

As RPC and scintillator have different responses to the beam background, the extrapolation of the measured background rates in the RPCs (Sec. 10.4) to the expected rate in the scintillator endcaps at SuperKEKB is difficult. Therefore, we have performed measurements of the neutron flux using a scintillator test module installed in the KEKB tunnel near the KLM endcap. The test module contained four layers, each surrounded by a shielding box for protection from external light and assembled from 24 strips. Each scintillator strip with a WLS fiber and SiPM readout had a size of $1000 \times 40 \times 10\,\mathrm{mm}^3$. During KEKB operation, the signals from 96 channels were collected by a random trigger using a CAMAC ADC. The gate was set to be 100 ns.
The shielding box could not protect the module from numerous charged tracks and showers. To discriminate the neutron signals from other sources, different hit patterns in the four layers were used. The significant part of the observed events (where at least one SiPM signal exceeds the threshold of 0.5 MIP) has multiple hits from charged tracks or showers. We require the veto in the two outer layer and require a single hit in on of the two inner layers. The calculated neutron rate at 0.5 MIP threshold at the luminosity of $\mathcal{L} \sim 1.4 \times 10^{34}\,\mathrm{cm}^{-2}\mathrm{s}^{-1}$ was $6\,\mathrm{Hz/cm}^2$. We use this number to estimate the occupancies in the readout electronics (maximum 160 kHz from the longest strip, 60 kHz averaged) and trigger, as well as to estimate the expected backgrounds for $K_L^0$ reconstruction.

### 10.5.6 Electronics

The SiPM has excellent time resolution and high output-rate capability. Signals with different number of photoelectrons are well separated, which gives the possibility of accurate calibration of the device. To use these advantages fully, the front-end electronics should comply with SiPM characteristics and provide fast processing of the SiPM signals with low noise and short dead time.
The core of the front-end electronics is the 16-channel TARGET ASIC chip (Ch. 13), a waveform sampling device ("oscilloscope on a chip"). Amplifiers, fine HV control, and an ammeter to monitor SiPM aging are also installed on the readout board. An FPGA is used for slow control, analysis of the digitalized signal, and implementation of the interface with COPPER boards.
Two possible layouts are considered, each with its advantages and drawbacks:

1. Each SiPM is equipped with a preamplifier, while the rest of the electronics is located outside of the detector. If a preamplifier fails (e.g., because of irradiation), it may be replaced only during long shutdowns, when a partial disassembly of the endcap is possible. Many cables are needed to connect every SiPM with the readout board in this case.

2. Electronics is completely installed outside of the detector. One twisted pair in a ribbon connects each SiPM, reducing number of cables to the minimum. In this layout, failed parts are replaced during regular maintenance. However, pick-up noise may pose a problem, so that readout board should be located as close to the detector as possible.

The second layout option was tested in the KEKB tunnel during accelerator operation. No significant pick-up noise was observed under the present conditions. A schematic sketch of this layout is shown in Fig. 10.24( a).
The amplified input signal, with a gain of $\sim 15$, is digitized with a sampling rate of $\sim 1\,\mathrm{GHz}$ by the TARGET. The FPGA scans this stream, searching for a typical SiPM signal pattern. Figure 10.24(b) shows the amplified SiPM one-photoelectron pulse, fitted by a simple function. The fit provides accurate time and amplitude information. Due to the steep leading edge of the pulse, the time resolution is very good ($\sigma \sim 1\,\mathrm{ns}$), which is useful for the rejection of combinatorial background. We are interested in two characteristic SiPM outputs: small-amplitude signals, which are due mainly to noise and are used for calibration, and large-amplitude signals that are associated with muons and $K_L^0$ mesons.

### 10.5.7 Slow control and monitoring

To ensure good data quality, the slow control and monitoring uses hardware measurements as well as fast data analysis. The SiPM response is sensitive to the ambient temperature, so ten temperature sensors are placed near the SiPMs in each quadrant. Temperatures are permanently recorded. The possibility of adjusting the SiPM bias voltages to compensate for drifts in temperature is under study.

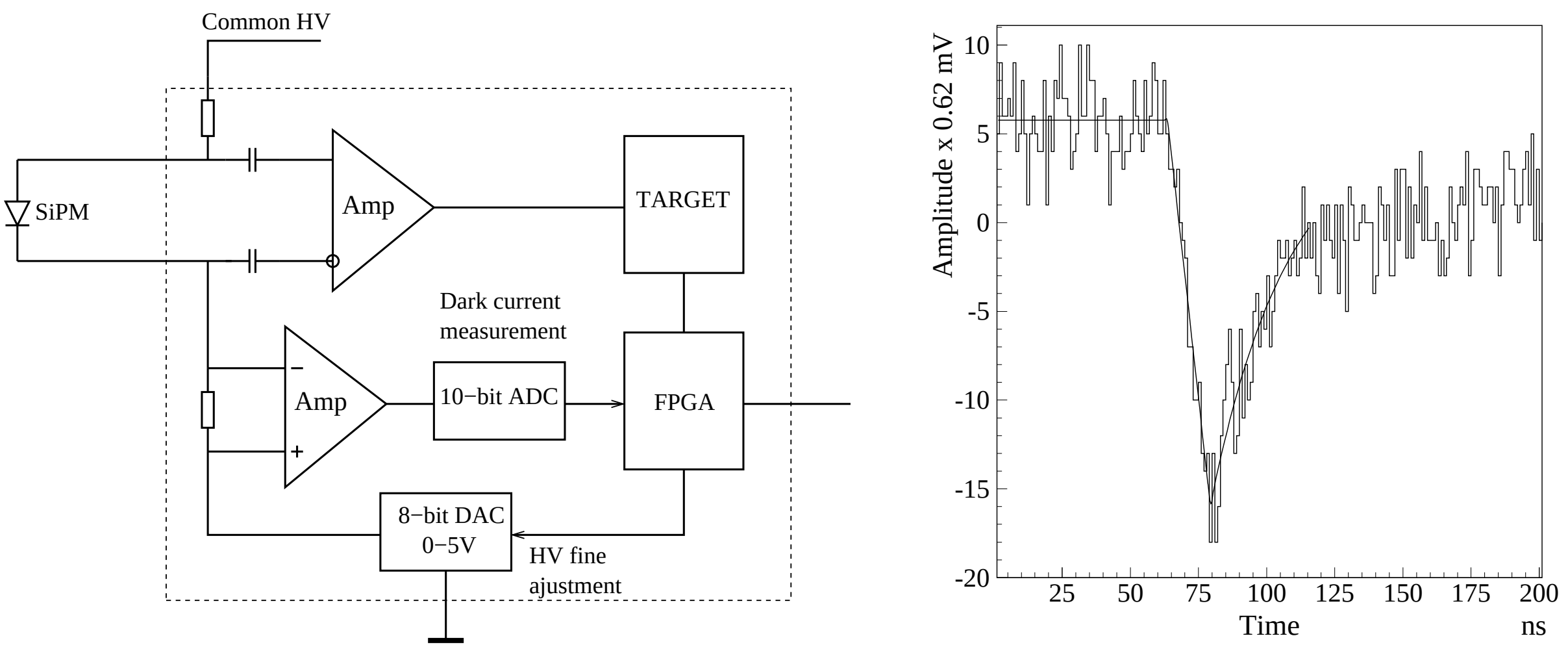

*Figure 10.24: a) Schematic sketch of the front-end electronics for the endcap KLM. b) Amplified SiPM pulse readout for one photoelectron by the TARGET ASIC chip.*

The SiPM current is recorded for each channel. During luminosity runs, the currents are sensitive to the background level and spatial distribution (after the channel-by-channel dark current subtraction). Dead (disconnected) channels manifest themselves as zero-current readings. The SiPM currents from calibration runs are used to monitor the long term stability and correlated with the accumulated radiation dose.

Ten thousand random triggers during each calibration run provide reliable pedestal measurements for each channel. The pedestal width is a good measure of the SiPM and electronics noise. With a virgin SiPM, it is possible to determine the position of the one-photoelectron peak and hence the gain. This determination will probably disappear after a few years of operation as the SiPM dark current increases with radiation damage. The muon detection efficiency for every channel can be determined on-line. At $\mathcal{L} = 8 \times 10^{35}$ cm$^{-2}$s$^{-1}$, one expects more than 1000 muons per strip per hour. KLM online histograms show the current and efficiency for every channel as well as the muon trigger rate and angular distribution.

### 10.5.8 Production

In this section, we describe the assembly procedure for the option with Vladimir's scintillator and Hamamatsu's SiPM, which seems the most attractive. However, the assembly procedure is identical for any other producers.

#### 10.5.8.1 Manufacturing of strips

The cross section of the strip is $7 \times 40\,\text{mm}^2$; the length varies from 50 to 280 cm depending on the strip position inside the layer (Fig. 10.20). At the Vladimir enterprise, scintillator plates are extruded with a thickness of 7 mm and a width of about 25 cm. The edges of the extruded plates are slightly rounded. The plates are sawed into 40-mm wide strips using special machinery developed at Vladimir. After sawing, the strips are subjected to chemical etching that produces a diffuse reflective covering of their surfaces. The thickness of the covering is about 50 $\mu$m. A 1.3-mm wide and 3-mm deep groove is milled in each strip to hold the WLS fiber.

#### 10.5.8.2 Assembly of strips

The WLS fiber is fixed at the strip ends with a two-component epoxy, Bicron BC600. In the middle part, the groove with the fiber is filled with an optical gel SL-1 produced at SUREL, St. Petersburg. The gel is cheaper and more elastic than BC600, which is useful to compensate for differences in the thermal expansion coefficients of the gel and the scintillator.
One WLS fiber end is connected to a photosensor while the other is mirrored using a silver-shine paint. The reflectivity of the paint is measured to be above 90%. The SiPM is attached to the WLS fiber using the connector shown in Fig. 10.25. The connector is made of plastic and is produced using casting in Vladimir. The fiber is glued into the hole in the male part of the connector and this male part is glued to the strip. We use BC600 for the gluing. The WLS fiber end sticking out of the optical connector is cut away using a milling machine. The SiPM is placed into the female part of the connector above a spring made of micro-porous material. When the connector is closed, the spring presses the SiPM face against the WLS fiber end. The female part of the connector has holes for the SiPM pins. A twisted pair connects the SiPM pins with the preamplifier.
We plan to perform tests of all strips using a cosmic-ray test stand.

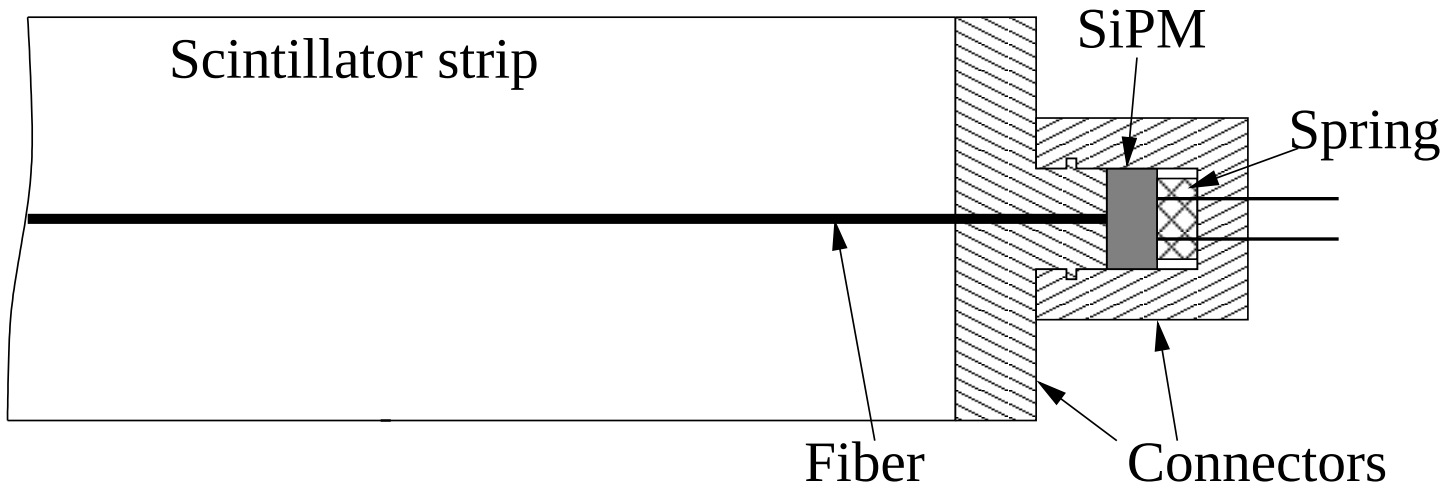

*Figure 10.25: Schematic cutaway view of the strip end.*

#### 10.5.8.3 Layer Segment

The $x$ and $y$ planes of the superlayer are each divided into five segments (Fig. 10.20). The segments of the $x$ and $y$ planes are identical. Each segment consists of 15 strips. The strips are glued to a 3-mm thick plastic plate using Plexus MA550 glue. Glue from the same producer was used for the FGD detector at the T2K experiment [18]. The MA550 glue includes $TiO_2$ dye that improves the light reflectivity of the strip. The other side of the segment, which contains the grooves with the WLS fibers, is covered by white paper to further improve the reflectivity.

#### 10.5.8.4 Support structure

The segments are supported by aluminum I-beams that run along the segment boundaries (Fig. 10.26). The I-beams supporting the $x$- and $y$-plane segments are welded together into a grid. The segments are inserted into the grid and fixed using wedges. The grid with segments is attached to the outer frame that is located along the perimeter of the layer. We plan to use the existing RPC frames from the Belle endcaps. The photosensors, cables, and amplifiers (optionally) are installed after the segments, I-beams, and outer frame are assembled.

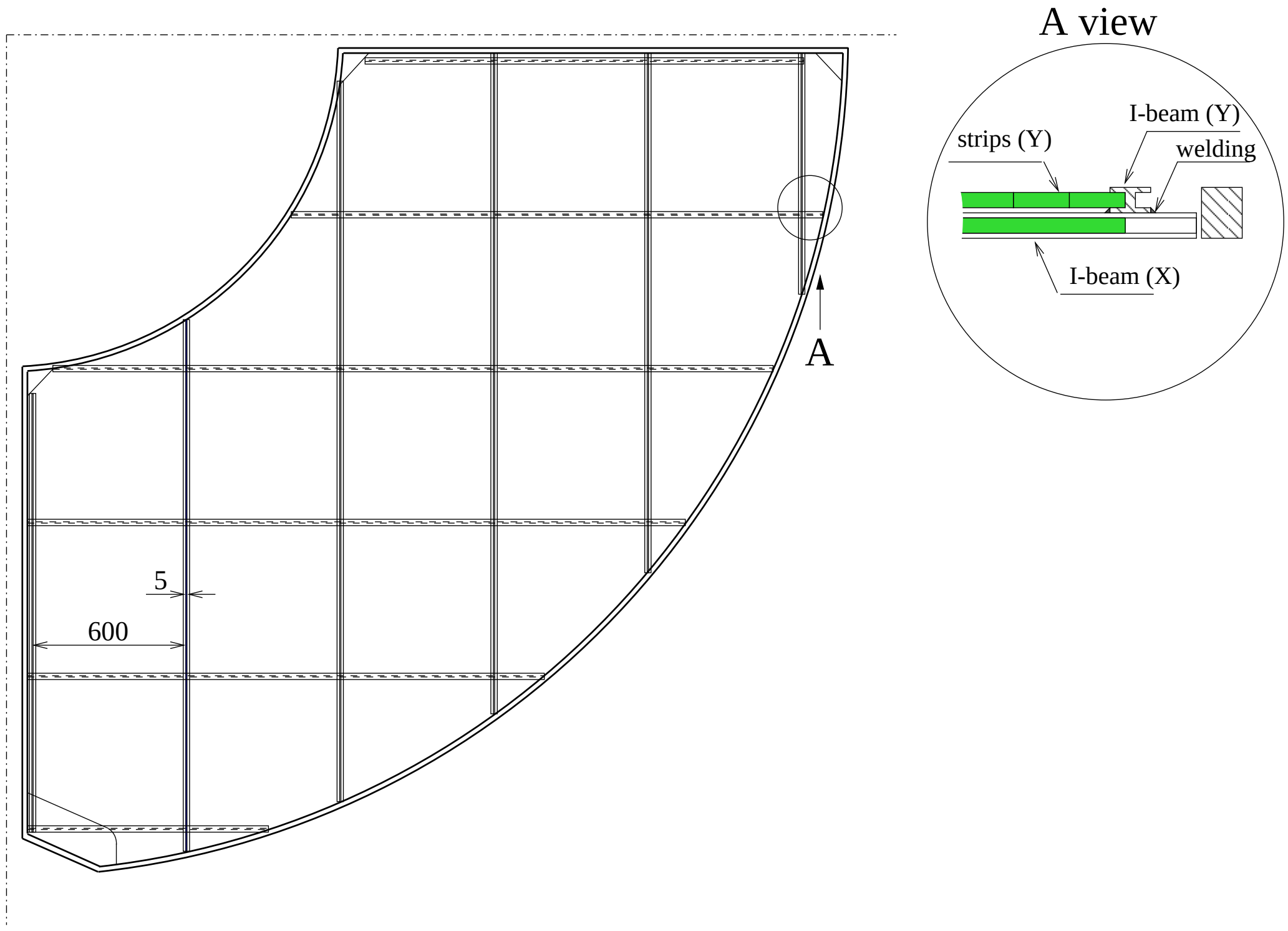

*Figure 10.26: Supporting structure for one layer: grid from I-beams and outer frame.*

Finally, the frame is covered by 1.5-mm thick aluminum plates to protect the detectors from the external light. Again, we plan to use existing RPC-module plates.

#### 10.5.8.5 Mounting at KEK

The segments are produced at ITEP, while the I-beam grids are produced at KEK. All SiPMs are tested at KEK before they are connected to the scintillator strips. After the assembly and cabling of one superlayer, it is tested using the cosmic-ray stand. The frames are installed into the gaps of the iron absorber with a crane using the same procedure as for the RPCs. We plan to use flat signal cables that are placed between the module and the cover plates.

#### 10.5.8.6 Schedule

The extrusion of all strips at Vladimir enterprise will take about three months. Further operations at Vladimir—cutting the scintillator plates into strips, matting of the strips, and milling the groove for the WLS fiber—will take about half a year. All these operations can be done in parallel with the strip assembly at ITEP. The goal is to assemble 75 strips per day, which will allow us to complete the mass production within one year. Four technicians at ITEP will glue the fibers into the grooves and two more technicians will glue the strips into the segments. The other operations, such as cutting the fiber, mirroring the fiber ends, *etc.*, will require one more technician. The shipment of the strips into KEK in four batches will allow an early start of the assembly work at KEK. Assembly of two quadrants per day seems feasible, for which three persons are required. In this case, the assembly of the complete system at KEK will require half a year.

### 10.5.9 Physics performance

To estimate the muon reconstruction efficiency, the charged-hadron fake rate, the $K_L^0$ meson reconstruction efficiency, and the fake-$K_L^0$ veto efficiency, we perform a Monte Carlo detector study. To estimate the backgrounds, we extrapolate the neutron background rate measurement with the scintillator test module as well as the Belle data collected with random triggers.

#### 10.5.9.1 Monte Carlo simulation

The endcap geometry is modelled within the Geant4 package. The energy deposition of signal hits in each strip is simulated by Geant4. The SiPM response to a given energy deposition is simulated using the empirical functions obtained from our test measurements. The arrival time of the signal to the SiPM is smeared according to the measured time resolution of the detector. We assume that the multiple hits at the same strip are correctly resolved by the readout electronics if their arrival time difference is greater than 10 ns; otherwise, they are merged into one hit. The hit information is stored if its amplitude exceeds the threshold of 7.5 p.e..

The background induced by neutrons is simulated by adding the random hits to the events. The energy spectrum and rate of the hits from background neutrons are estimated from the direct background measurements in the KEKB tunnel using scintillator KLM test module (Sect. 10.5.5). The background hits are distributed uniformly in time while their spatial distribution is tuned to match that of the Belle random-trigger data. We use the same procedure to digitize the background hits and the signal ones.

#### 10.5.9.2 Hits reconstruction

The raw-hit reconstruction procedure is similar to that used in the Belle reconstruction software. We exploit the better time resolution that can be achieved with new electronics and DAQ.

Neighboring hit strips in the same plane with time difference of 10 ns or less are grouped to form a 1D hit whose geometric coordinate is defined as the equal-weighted mean of the grouped strips. In principle, the SiPM amplitude could be used for the weighting factor to improve the position resolution. The crossing of 1D hits in the orthogonal planes of one superlayer form a 2D hit with $x$ and $y$ coordinates inherited from the 1D hits, a $z$ coordinate located between the two scintillator planes, and a time that averages the two 1D times after correction for the propagation delay in the WLS fiber.

#### 10.5.9.3 Muon identification

The granularity of the detector is not changed, while the expected efficiency of the scintillator-based detector is slightly higher the the RPC-based detector, so we expect that the muon identification performance should be similar or slightly better in our design. The detailed study requires modification of the reconstruction programs and will be done later.

#### 10.5.9.4 $K_L^0$ reconstruction

The $K_L^0$ clusters are reconstructed as a group of 2D hits within a cone of 5° half-angle relative to the IP that are compatible in time with having been produced by a single nuclear shower. As in the RPC-based algorithm, we require 2D hits in two or more superlayers. The time of the cluster is defined as the average of the times of its individual 2D hits, projected to the innermost superlayer with a hit. The efficiency of finding a $K_L^0$ cluster in the standalone KLM (where the ECL is ignored) is shown in Fig. 10.27 a) as a function of $K_L^0$ momentum.

We measure the $K_L^0$ angular resolution by comparing the angle between the true $K_L^0$ momentum and the direction of the reconstructed $K_L^0$ cluster [Fig. 10.27(b)]. The angular resolution is found to be $\sim 10\,$mrad in the wide-momentum region, which is approximately the same as in the present framework.
We could also take advantage of the good scintillator time resolution to determine the $K_L^0$ momentum using a time-of-flight determination, at least for medium-momentum kaons. The obtained momentum resolution as a function of the $K_L^0$ momentum is shown in Fig. 10.27(c). We observe that the time measurement provides useful information for kaons with momenta up to 1.5 GeV/$c$.

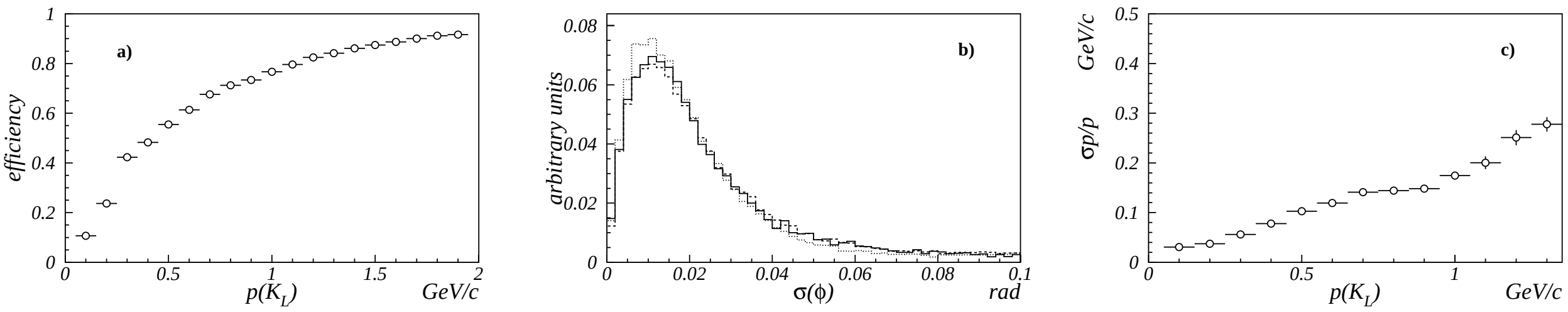

*Figure 10.27: $K_L^0$ reconstruction: a) efficiency; b) angular resolution; c) momentum resolution.*

The main source of $K_L^0$ fake candidates is the accidental coincidence of 1D hits induced by neutrons. After SiPM irradiation, some small contribution to fake candidates comes from the increased SiPM noise. We find $< 0.01$ fake $K_L^0$ candidates per event at the designed luminosity. Based on the simulation, we conclude that the proposed detector provides better performance for $K_L^0$ reconstruction than the RPC-based one in Belle.

### 10.5.10 Cost estimate

A cost estimate for the option with Vladimir strips and Hamamatsu SiPMs is presented in Table 10.3.

*Table 10.3: Cost estimate for the scintillator KLM with Hamamatsu SiPM readout.*

| | | |
|---|---|---|
| Strips | 9500kg | 285k$ |
| WLS fibers | 34km | 136k$ |
| Optical gel | 400kg | 28k$ |
| Epoxy Bicron BC600 | 50kg | 15k$ |
| Silver paint | 20kg | 2k$ |
| Glue | 400kg | 8k$ |
| Machinery for gluing | | 30k$ |
| SiPM | #17000 | 340k$ |
| SiPM housing | #17000 | 12k$ |
| Preamplifier | #17000 | 34k$ |
| Power supplies | | 70k$ |
| Cables | 80km | 50k$ |
| Readout board | #1100 | 110k$ |
| Labour (strips) | | 200k$ |
| Transportation | | 80k$ |
| Mechanical structure | | 40k$ |
| Labour (structure & mounting) | | 50k$ |
| Total | | 1490k$ |

# Chapter 11

# Detector Solenoid and Iron Structure

## 11.1 Iron Yoke

The iron structure of the Belle detector (Table 11.1) serves as the return path for the solenoid's magnetic flux and an absorber for the KLM. It also provides the overall support for all of the detector components. It consists of a fixed barrel part (Fig. 11.1) and movable end-cap parts (Fig. 11.2), both on a base stand. The barrel part consists of eight KLM blocks and 200-mm thick flux-return plates surrounding the outermost layers of of the KLM blocks. Neighboring KLM blocks are joined using fitting blocks. Each end-cap part can be retracted for access to the inner detectors.

The weight of the iron yoke is 608 and 524 (= 262 × 2) tonnes for the barrel yoke and end-cap yokes, respectively.

The precision of the construction was confirmed to be ±0.3 mm to ±0.5 mm for the machined parts and ±1 mm to ±2 mm for the assembly of the octagonal blocks. The final inner diameters of the octagon in which most of the sub detectors are installed were measured to be ±4.5 mm from the design value. The location of the structure with respect to the accelerator was confirmed in place within ±0.5 mm with the telescope survey's 0.2 mm precision.

The guideline of the mechanical design against earthquakes was established to follow the reported seismic wave in the Miyagi-ken oki earthquake ($M7.4$) on June 12, 1978. From the analysis, the equivalent maximal horizontal static force was defined to be $0.3G$, which should be applied all the mechanical tolerance evaluation.

## 11.2 Solenoid Magnet

A superconducting solenoid provides a magnetic field of 1.5 T in a cylindrical volume 3.4 m in diameter and 4.4 m in length [1]. The coil is surrounded by a multi-layer structure consisting of iron plates and calorimeters, which is integrated into a magnetic return circuit. The main coil parameters are summarized in Table 11.2. The overall structure of the cryostat and a schematic drawing of the coil cross section are shown in Fig. 11.3.

For the last ten years, the power supply that had been used in the TRISTAN-era TOPAZ experiment was used also for the Belle solenoid. In 2009, it was replaced with a new unit. Although the performance of the output power is same (5000 A/15 V), the ripple of the current is halved to 150 mV p-p and the stability is improved by an order of magnitude, from $1.8 \times 10^{-4}$ to $1.5 \times 10^{-5}$.

*Table 11.1: Main parameters of the iron structure.*

| Items | Parameters |
|---|---|
| Belle Iron Yoke | |
| Height | 9.57 m |
| Beam level | 5.72 m |
| Total Weight | 11740 $kN$ |
| Barrel yoke | |
| Shapes | octagonal |
| Material | S10C iron |
| Height | 7.7 m |
| Width | 7.7 m |
| Length | 4.4 m |
| Total Weight | 6240 kN |
| Number of iron plates | 15 |
| Thickness of iron plate | 47 mm |
| Thickness of gap | 44 mm |
| End Yoke | |
| Material | S10C iron |
| Height | 7.7 m |
| Width | 7.7 m |
| Length in beam direction | 1321 mm |
| Total Weight | 5254 $kN$ |
| Number of iron plates | 15 |
| Thickness of iron plate | 47 mm |
| Thickness of gap | 44 mm |

## 11.3 Cryogenic System

The fundamental flow diagram for the Belle II cryogenic system is shown in Fig. 11.4. Most of the hardware, such as the compressor, the refrigerator and the sub-cooler, will be reused, since the heat load of the solenoid is the same as in Belle. From the viewpoint of efficiency, the operating conditions should be improved as follows. In the normal mode, 120 W of heater power is supplied steadily into the liquid helium buffer tank in the sub-cooler to offset redundant cooling power. Since the $COP$ (Coefficient of Performance) is 0.01, if the heater power can be reduced by 90 W, the overall power would be reduced by 90 kW, corresponding to almost 1 MY/month of cost reducttion, given the cost of electricity. A power-saving operation has been developed for the J-PARC neutrino cryogenic system [2], and a similar improved cryogenic operation should be possible in Belle II.

#### 11.3.0.1 Control and Interlock

Except for the power supply and monitor of the magnet, all the control of the cryogenic system are established by the Distributed Control System by Hitachi Ltd. All the sequence control was specially programed for the Belle solenoid operation and operated thru the terminal in the Nikko cryogenics control room.
The normal operation sequence consists of the following 11 modes: initialization of valve condi-tion, start of compressor, gas circulation and charge, circulating purification, solenoid pre-cooling 1, start of turbine, solenoid pre-cooling 2, normal cooling operation, stop of turbine and liquid withdrawal, raising system temperature, and system shutdown. There are several recovery sequences defined for solenoid quench, abnormal stop of compressor, and abnormal status in turbine to protect the equipment from possible damage. Another sequence is defined for the period while raising the current in the solenoid magnet to control the flow of liquid helium at the current lead.

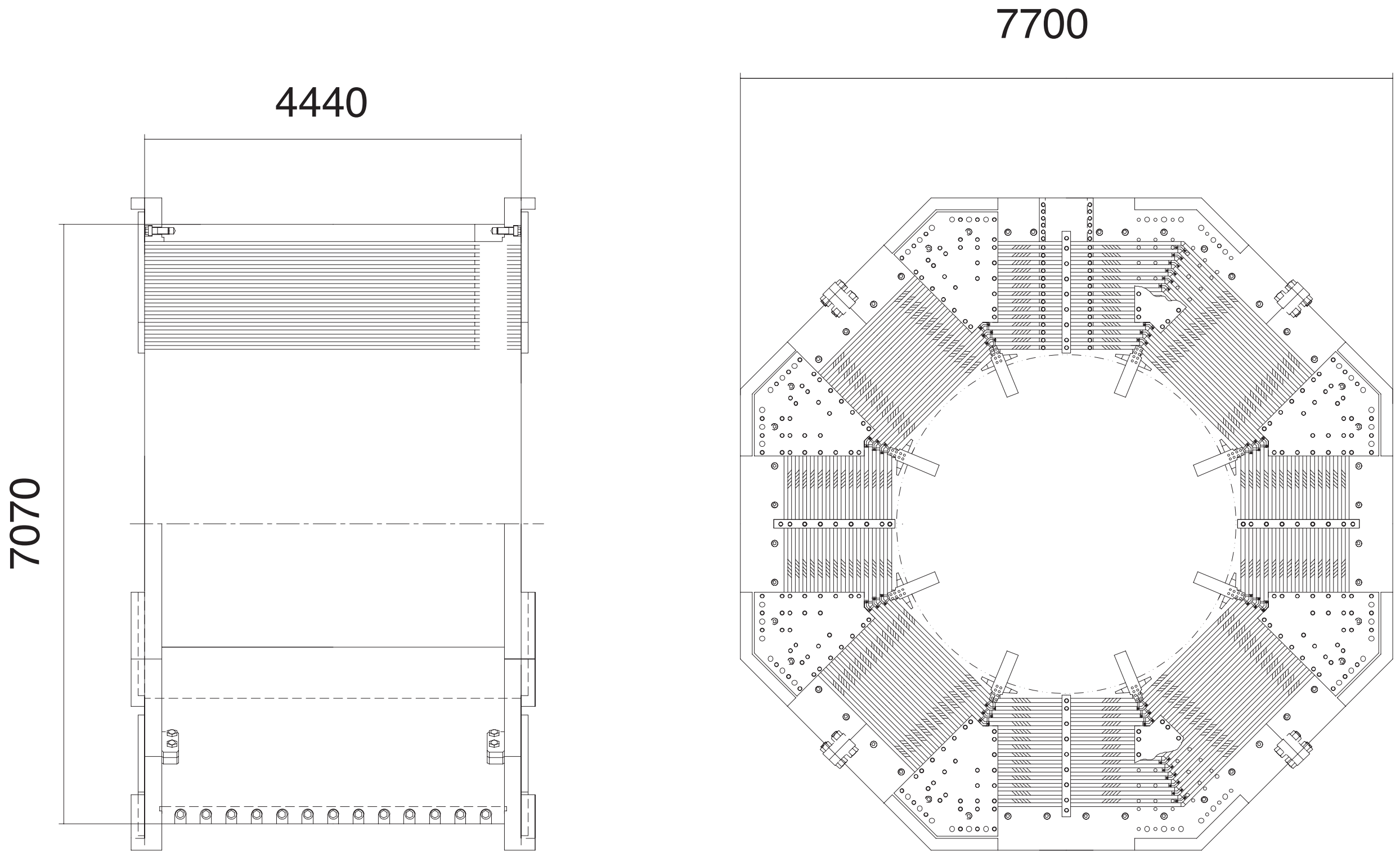

Figure 11.1: Barrel part of the iron yoke.

Since the system clock is 1 Hz—much slower than the response necessary for the quench protection—another interlock system works independently inside the power supply to shut down and dump the current flowing in the solenoid coil when a quench is detected.

#### 11.3.0.2 Old components to be maintained

As mentioned for the power supply, there are several existing components that will be replaced, overhauled, or maintained. They are summarized in Table 11.3

A field map of 100,000 points was made over a period of one month [3]. Figure 11.5 shows a contour plot of the field measured inside the tracking volume for the nominal magnet settings. The field strength is shown in Fig. 11.6 as a function of $z$ for various radii. The tracking performance was evaluated from the reconstructed $J/\psi$ mass peak in $B$ meson decays. The fitted value of the peak mass was 3.089 GeV/$c^2$, compared with the accepted value of 3.0969 GeV/$c^2$. Thus, the uncertainty in the absolute calibration of the present measurement is estimated to be approximately 0.25%.

Table 11.2: Main parameters of the solenoid coil.

| Items | Parameters |
|---|---|
| Cryostat | |
| Radius: outer/inner | 2.00 m/1.70 m |
| Central field | 1.5 T |
| Total weight | 23 $t$ |
| Effective cold mass | $\sim 6\ t$ |
| Length | 4.41 m |
| Coil | |
| Effective radius | 1.8 m |
| Length | 3.92 m |
| Conductor dimensions | $3 \times 33$ mm$^2$ |
| Superconductor | NbTi/Cu |
| Stabilizer | 99.99 % aluminum |
| Nominal current | 4400 A |
| Inductance | 3.6 H |
| Stored energy | 35 MJ |
| Typical charging time | 0.5 h |
| Liquid helium cryogenics | Forced flow two-phase |
| Cool down time | $\leq 6$ days |
| Quench recovery time | $\leq 1$ day |

Table 11.3: Major components of the cryogenic system.

| Items | Status |
|---|---|
| Power supply | replaced |
| Cryogenics control system | needs replacement |
| Turbine for refrigerator | needs overhaul |
| Control unit for Turbine | replacement purchased |
| Chiller for Turbine | replacement purchased |
| Compressor | needs overhaul |
| Chiller for compressor | needs replacement |
| Evacuation unit for refrigerator | replacement purchased |
| Evacuation unit for solenoid cryostat | needs overhaul |
| Liq N2 tubing of radiation shield | needs examination |
| Automated valves | need replacement |
| Transducers/sensors | need replacement/adjustment |
| Data loggers | need replacement |

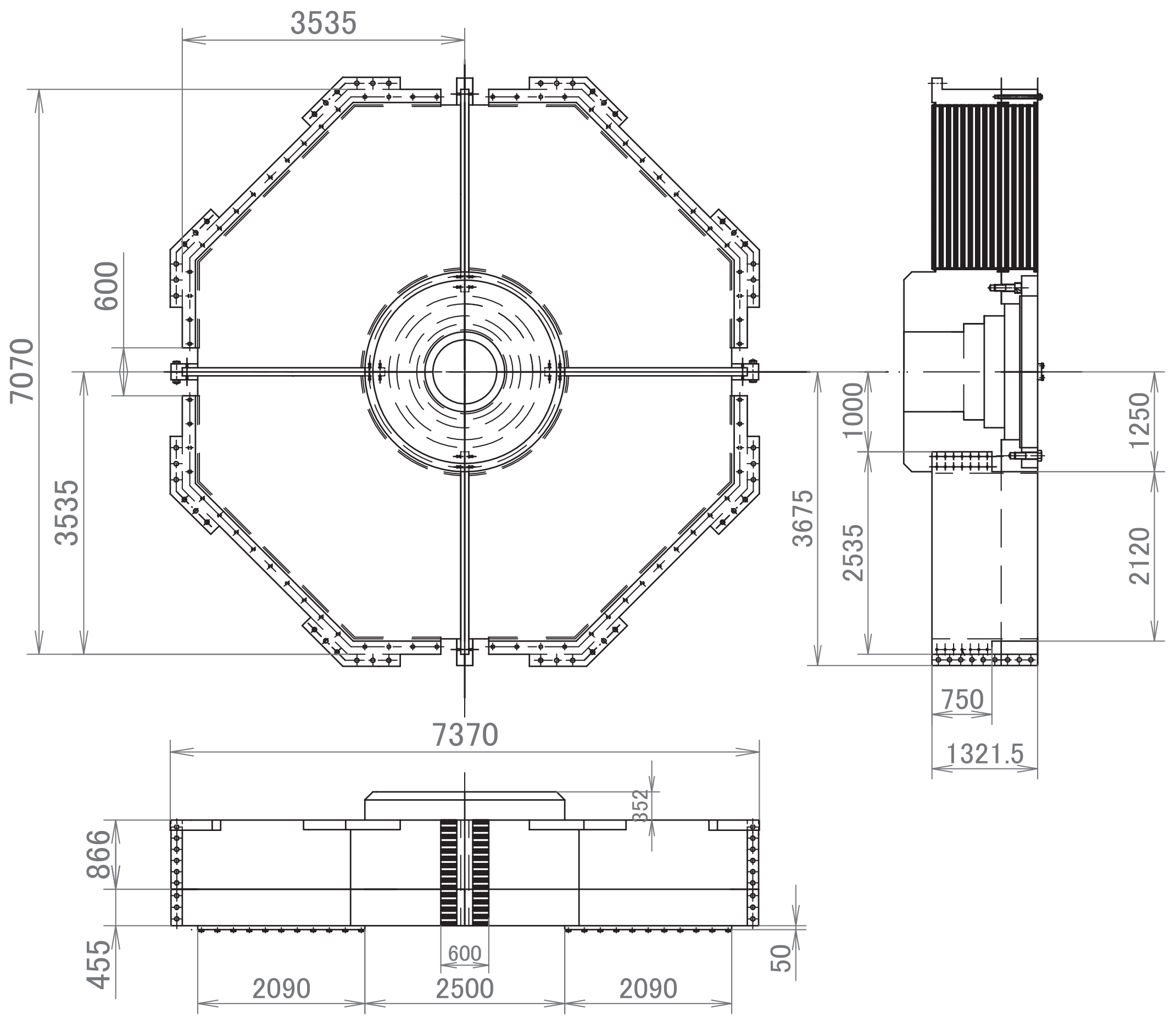

*Figure 11.2: The iron endyoke.*

## 11.4 Possible rotation of the Belle II structure

There is a possibility that SuperKEKB will require a dramatic change in the configuration of the IR with respect to the Belle solenoid axis. In the KEKB IR, the axis is aligned along the LER direction, which is tilted by 22 mrad from the KEKB tunnel axis, as shown in Fig. 11.7. In the new configuration, the entire 1500 ton Belle II structure might be aligned to the median line of the LER and HER directions. This will be accomplished by rotating the structure by several tens of milli-radians, as shown in Fig. 11.8. From the technical point of view, the rotation could be made at the floor level with special hydraulic jacks and rollers to be inserted under the platform on which the structure is mounted, as depicted in Fig. 11.9. The final decision will be made during the optimization of the accelerator design.

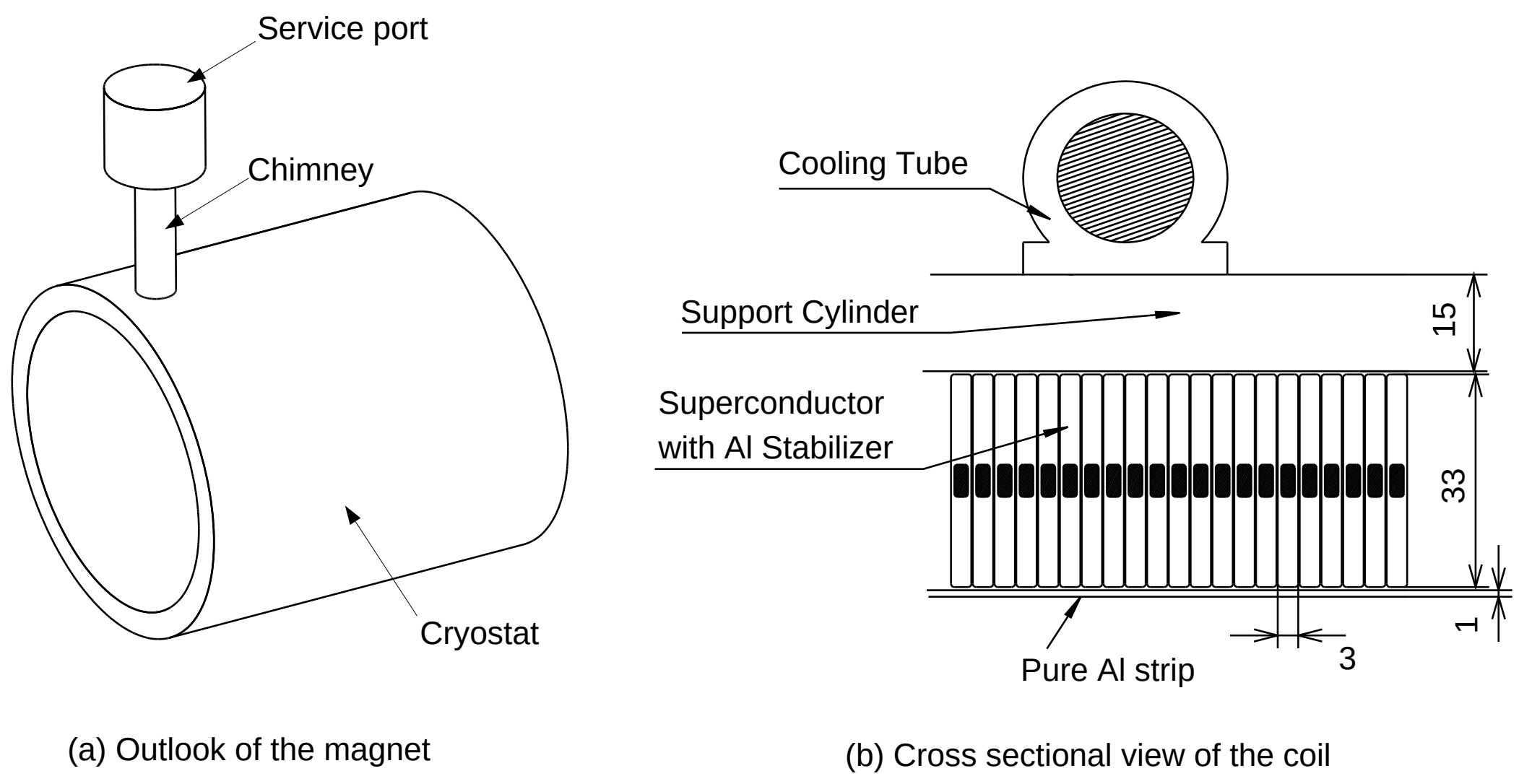

*Figure 11.3: Perspective view of the solenoid and cross sectional view of the coil.*

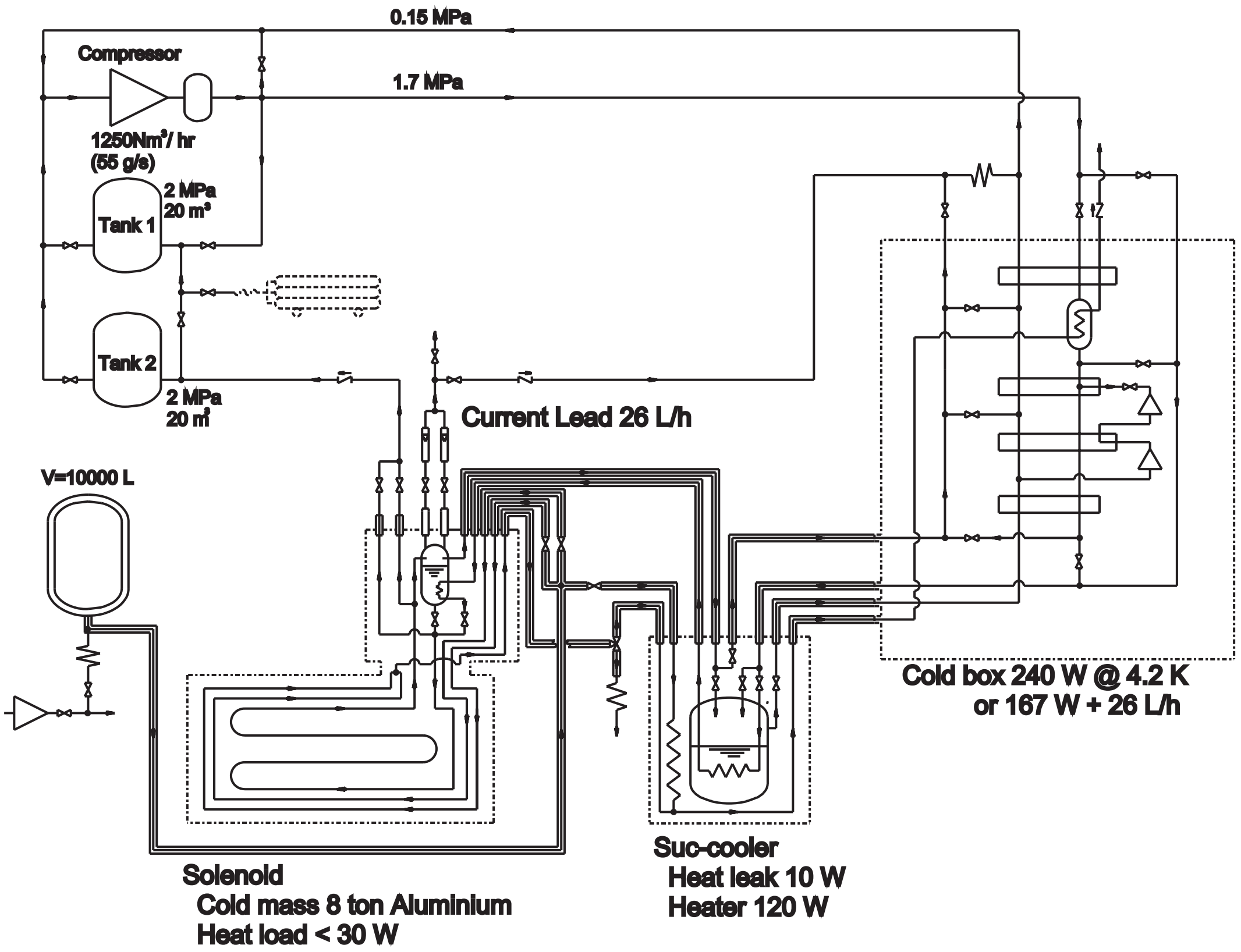

*Figure 11.4: Flow diagram of the cryogenic system.*

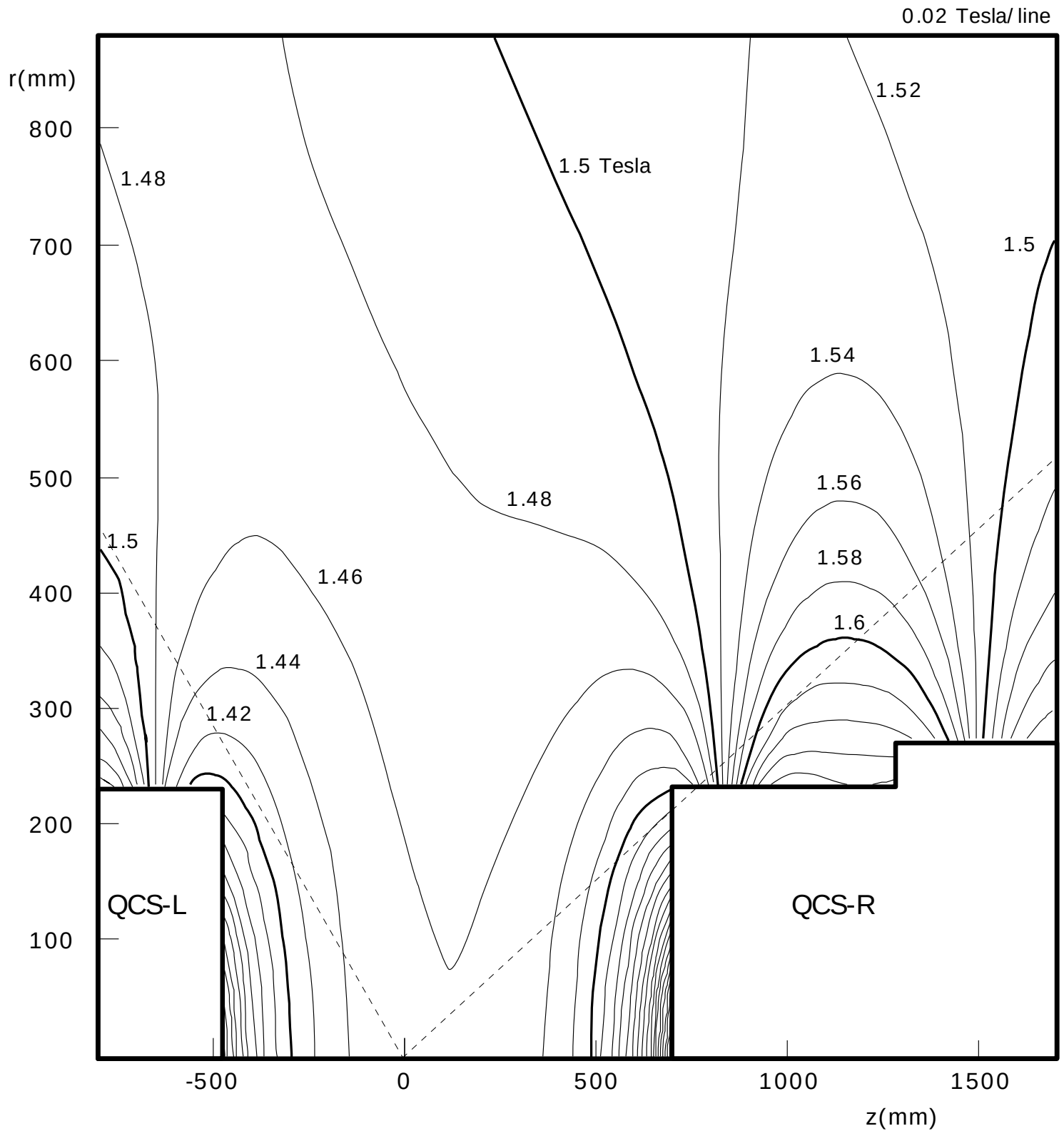

*Figure 11.5: Contour plot of the magnetic field measured in the Belle coordinate system with the origin at the interaction point.*

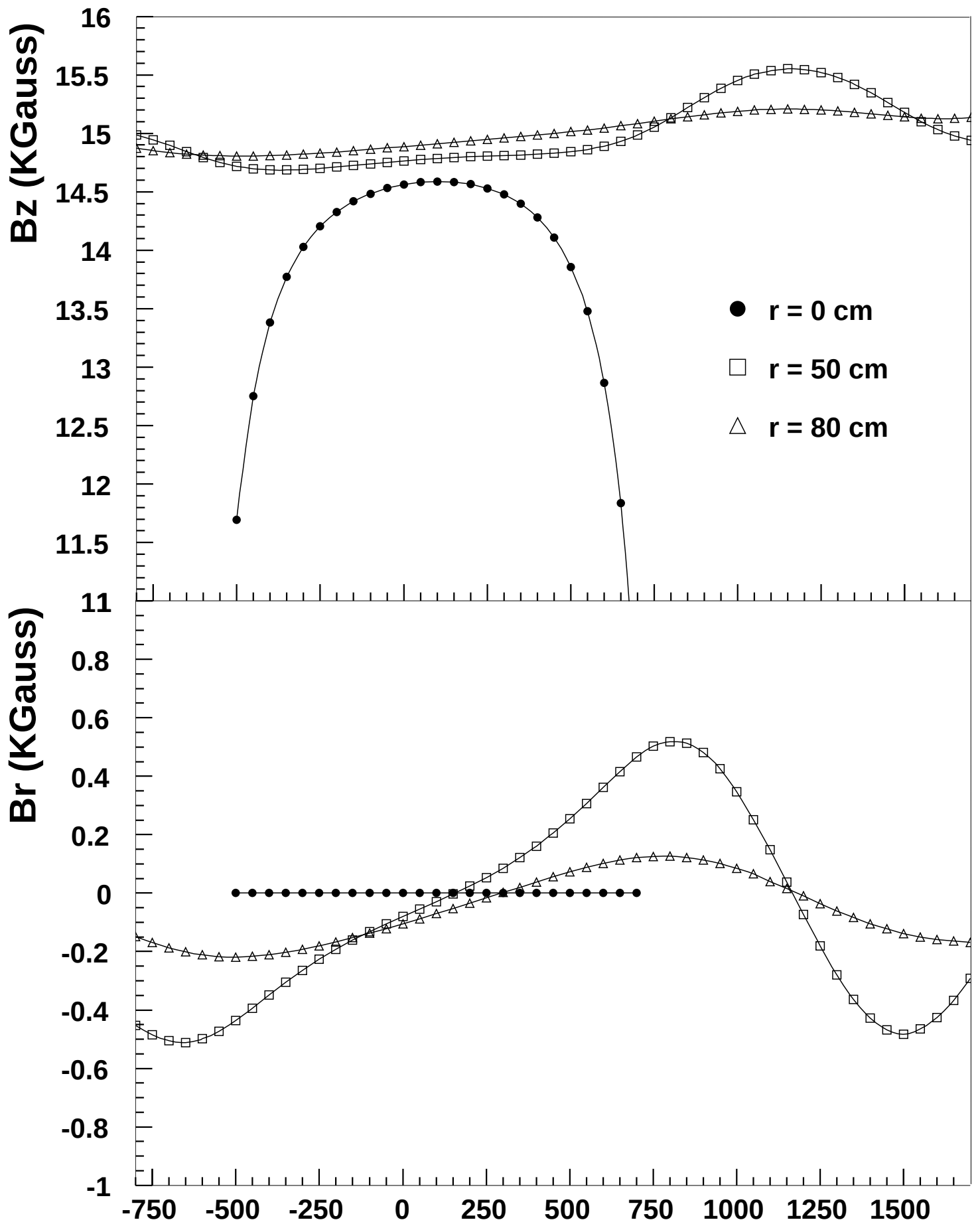

*Figure 11.6: Field strength (top: axial component; bottom: radial component) as a function of z for r = 0, 50, and 80 cm.*

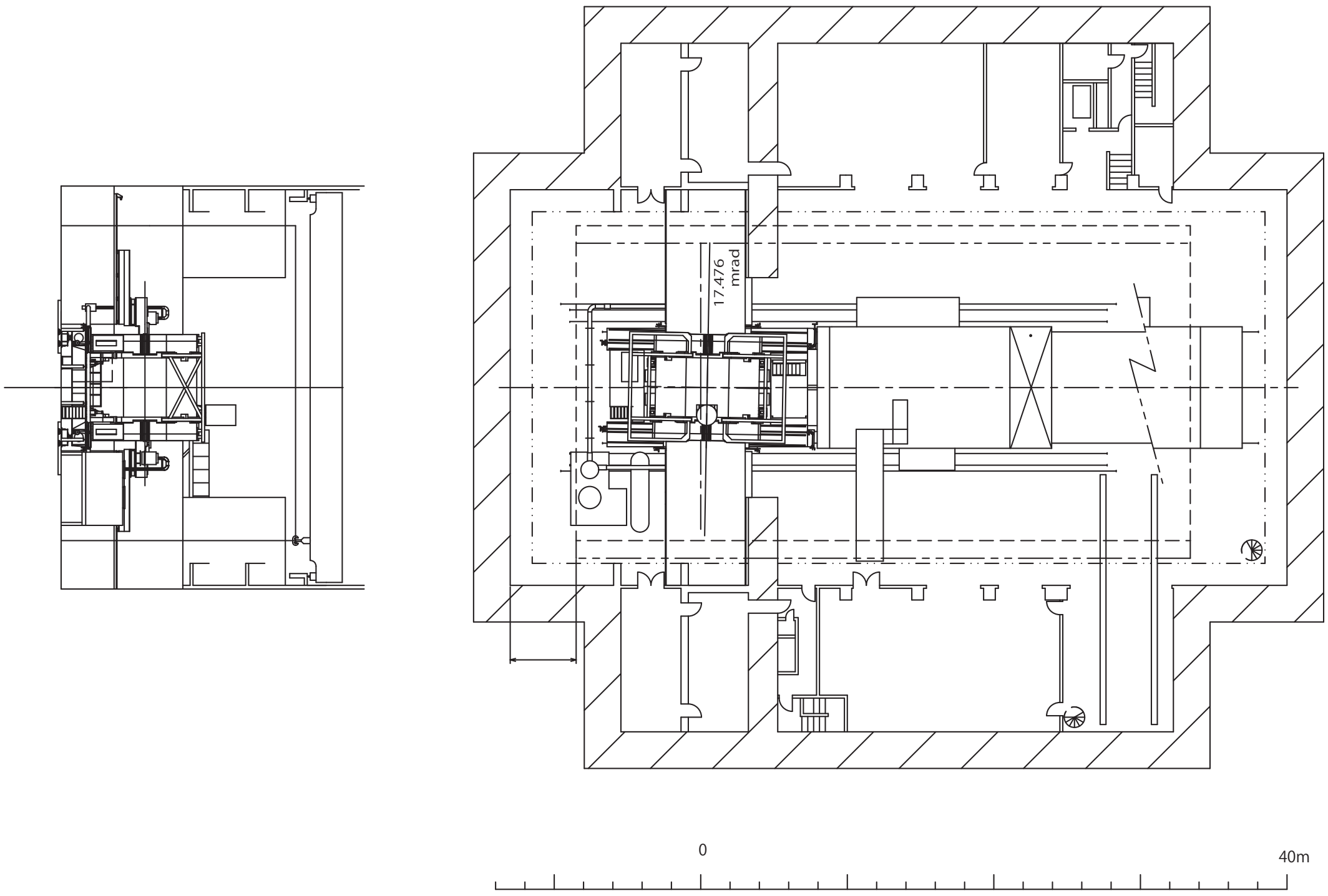

Figure 11.7: Plan view of the IR and Tsukuba experimental hall.

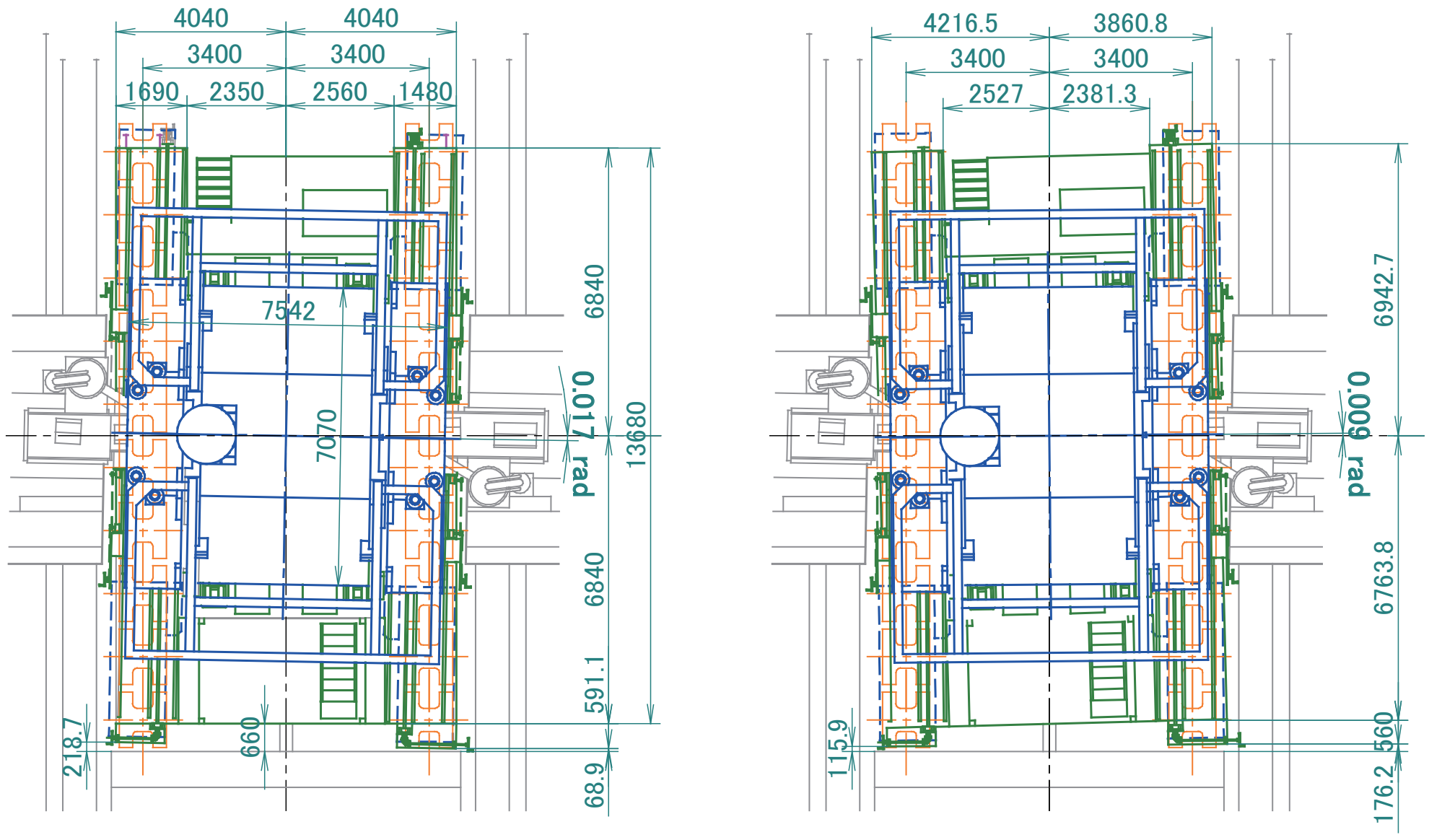

Figure 11.8: Comparison of the present and rotated detector placement.

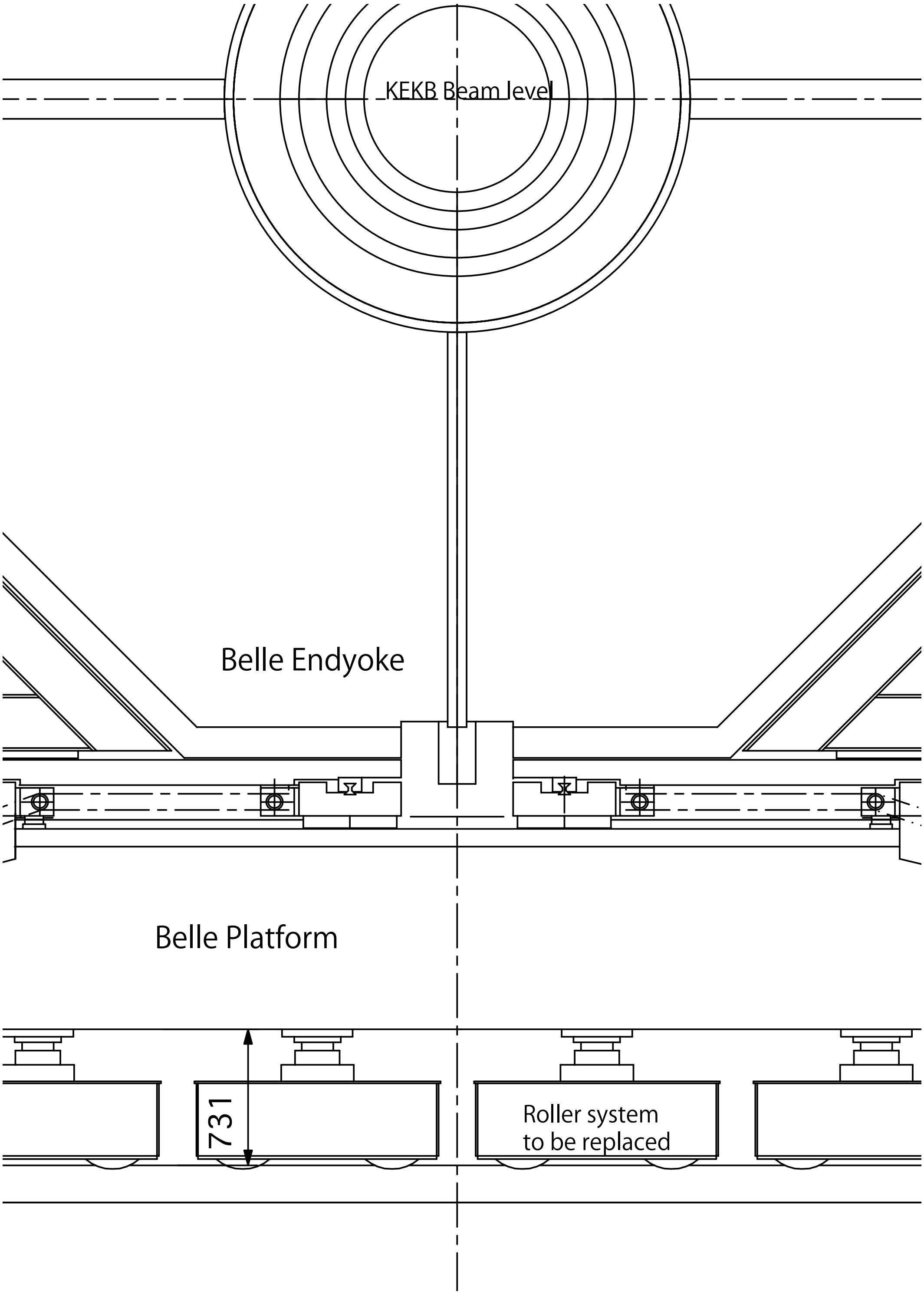

*Figure 11.9: Detector support structures*

# Chapter 12

# Trigger

The total cross sections and trigger rates at the goal luminosity of $8 \times 10^{35}\,\mathrm{cm}^{-2}\mathrm{s}^{-1}$ for several physical processes of interest are listed in Table 12.1. Samples of Bhabha and $\gamma\gamma$ events will be used to measure the luminosity and to calibrate the detector responses. Since the Bhabha and $\gamma\gamma$ cross sections are very large, these triggers are pre-scaled by a factor of 100 or more; this is straightforward due to their distinct signatures.

| Physics process | Cross section (nb) | Rate (Hz) |
|---|---|---|
| $\Upsilon(4S) \to B\bar{B}$ | 1.2 | 960 |
| Hadron production from continuum | 2.8 | 2200 |
| $\mu^+\mu^-$ | 0.8 | 640 |
| $\tau^+\tau^-$ | 0.8 | 640 |
| Bhabha ($\theta_{\mathrm{lab}} \geq 17^\circ$) | 44 | 350 $^{(a)}$ |
| $\gamma\gamma$ ($\theta_{\mathrm{lab}} \geq 17^\circ$) | 2.4 | 19 $^{(a)}$ |
| $2\gamma$ processes ($\theta_{\mathrm{lab}} \geq 17^\circ$, $p_t \geq 0.1\,\mathrm{GeV}/c$) | $\sim 80$ | $\sim 15000$ |
| Total | $\sim 130$ | $\sim 20000$ |

$^{(a)}$ rate is pre-scaled by a factor of 1/100

*Table 12.1: Total cross section and trigger rates with $L = 8 \times 10^{35}\,\mathrm{cm}^{-2}\mathrm{s}^{-1}$ from various physics processes at $\Upsilon$(4S).*

The requirements for the trigger system are

0. high efficiency for hadronic events from $\Upsilon(4S) \to B\bar{B}$ and from continuum;
1. a maximum average trigger rate of 30 kHz;
2. a fixed latency of about 5 $\mu$s;
3. a timing precision of less than 10 ns;
4. a minimum two-event separation of 200 ns; and
5. a trigger configuration that is flexible and robust.

To meet these requirements, we adopt the Belle triggering scheme [1] with new technologies. In the Belle triggering scheme, the trigger system consists of sub-trigger systems and one final-decision logic. A sub-trigger system summarizes trigger information on its sub-system, and sends it to the final-decision logic, which then makes combinations of sub-triggers and issues a trigger when its criteria are satisfied. This was quite successful in the Belle experiment to achieve high efficiency for hadronic events. In Belle II, we use this concept but replace all components and connections with new technologies. Each component has a Field Programmable Gate Array (FPGA) so that the trigger logic is configurable rather than hard-wired. All data flow along high speed serial links, not parallel (ribbon) cables, which enables us to funnel a huge amount of information—the equivalent of $\mathcal{O}(1000)$ channels—to one FPGA.

The schematic overview of the Belle II trigger system is shown in Fig. 12.1. The CDC sub-trigger provides the charged track information (momentum, position, charge, multiplicity, and so on). The ECL sub-trigger gives energy deposit information, energy cluster information, Bhabha identification, and cosmic-ray identification. The Barrel PID (BPID) sub-trigger gives precise timing and hit topology information. The Endcap PID (EPID) sub-trigger is expected to give precise timing information.[1] The KLM sub-trigger gives muon track information. The Global Decision Logic (GDL) receives all of this sub-trigger information and makes the final decision. A positive decision is sent to SEQ as a trigger signal. The total latency in the trigger system is about $5\,\mu$s.[2] To achieve high efficiency for hadronic events, we use two independent sub-triggers: CDC and ECL.

In Belle II, the background condition is expected to be worse than in Belle due to the higher instantaneous luminosity and the beams' smaller transverse dimensions. In addition, from Belle's experience, we expect much worse background condition at the beginning of the accelerator operation when the vacuum condition and other accelerator parameters are not yet optimal. To satisfy the trigger system's requirements in the face of such backgrounds, it is important to have several independent, effective trigger strategies to reduce the Level 1 trigger rate. In this way, we can tune the collection of triggers to minimize the deadtime of the data acquisition operation without sacrificing our efficiency for recording events of physics interest. These strategies are described in the following sections.

## 12.1 CDC 2D Trigger

The CDC sub-trigger (Fig. 12.2) finds and characterizes the charged tracks detected by the drift chamber. Because of the limited solid angle coverage of the CDC, this sub-trigger is not sensitive to charged tracks in the far forward or backward regions. In the present design, we measure $p_t$, $\lambda$, $d_z$, $\phi$, and the charge of tracks. The CDC sub-trigger should be sensitive only to the charged tracks coming from the vicinity of the interaction point ($|d_z| \lesssim 4\,\text{cm}$ and $|d_r| \lesssim 4\,\text{cm}$.).

### 12.1.1 CDC Trigger Front-end

Discriminated signals from each of 48 CDC wires are generated in the front-end board in each pulse of the 1 GHz clock. These precise wire-hit signals are down-sampled by a 62.5 MHz clock. A one-bit hit state of the 48 wires on this board is sent to the multiplexer (Fig. 12.3) by the high speed serial links, whose data transfer capacity is 12.5 ($3.125 \times 4$) Gbps. The actual data rate is 3 Gbps. The multiplexer can receives four sets of such serial links. Each line in Fig. 12.3 corresponds to a superlayer in the CDC. There are three to six multiplexers per superlayer, depending on the number of sense wires.

[1] This is under discussion.

[2] The precise value of the latency is not yet decided.

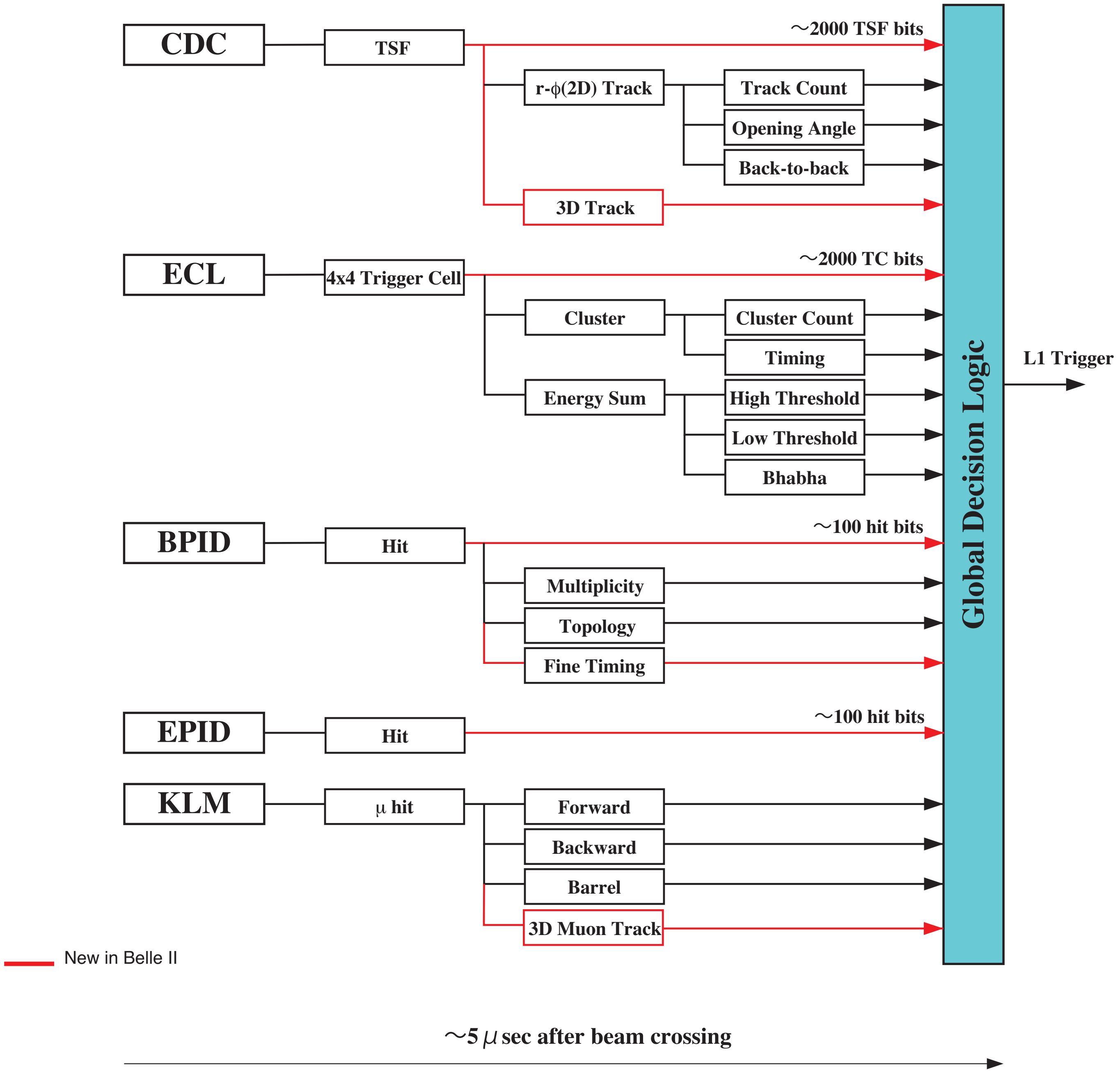

*Figure 12.1: Schematic overview of the trigger system. The output from the five sub-trigger systems are sent to the Global Decision Logic (GDL). The final trigger decision is made in the GDL. The lines in red are newly added information paths in the Belle II trigger system.*

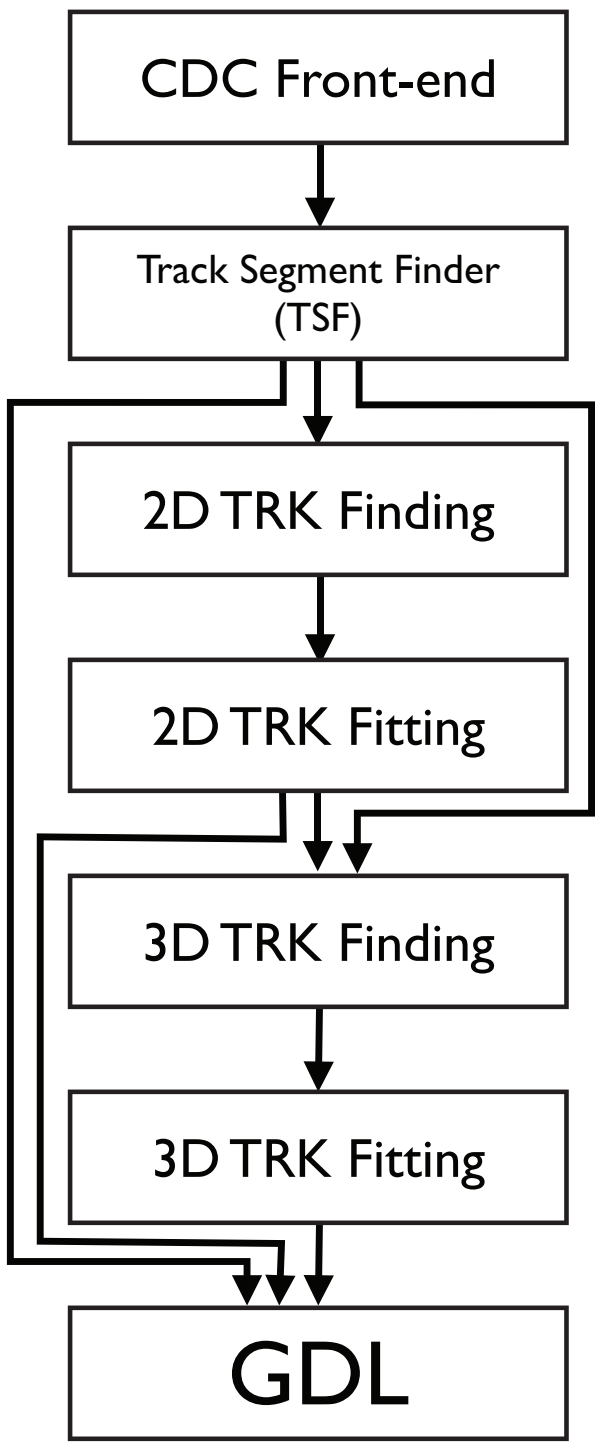

*Figure 12.2: Logical overview of the CDC sub-trigger system.*

### 12.1.2 Track Segment Finder

The merged wire hit information by the multiplexer is sent to the next stage, the Track Segment Finder (TSF). The TSF is realized using one Universal Trigger Board (UT3) per CDC superlayer, with nine such boards in total.
In each TSF, the geometric regions for Track Segments (TS) are defined (Fig. 12.4). The number of TS in a superlayer depends on the number of wires in a single wire layer. Each TS overlaps the neighboring TS except for the innermost cell of the innermost TS or for the center cell for other TS. There are 2336 TS in total. At each tick of the 62.5 MHz clock, the pattern of wire hits in each TS is examined. The hit patterns expected for a charged track are predefined and stored in a memory. By the memory look-up method, the presence of a TS hit is determined. The TS hit information (2336 bits total) is sent to the 2D Track Finding stage (Fig. 12.2) at each clock tick.

### 12.1.3 2D Track Finding

In the 2D track finding stage, we use a conformal transformation and Hough transformation [2] to search for charged tracks. The schematic image of the conformal transformation is shown in Fig. 12.5. A point $(x, y)$ in the plane transverse to the CDC's axial wires is transformed into $(X, Y)$ in the conformal plane using the following equations.

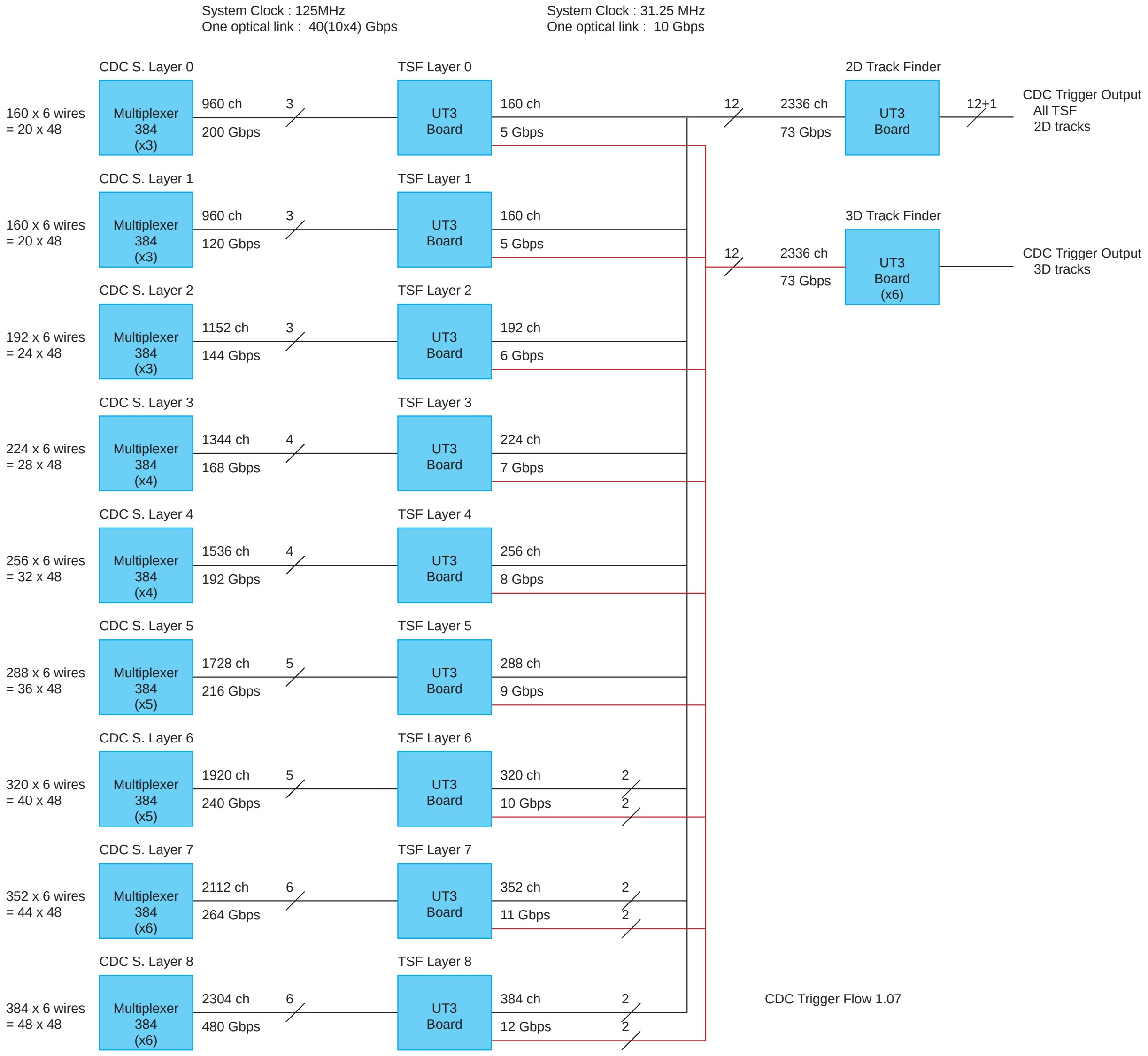

*Figure 12.3: Hardware configuration of the CDC sub-trigger system. The data flows from left to right. The CDC front-end is omitted here.*

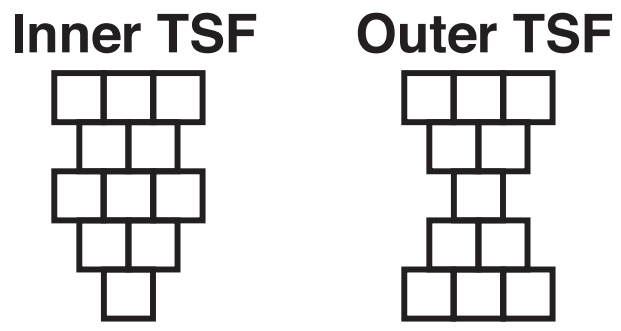

*Figure 12.4: The geometric shape of a track segment (TS). Each square corresponds to a wire cell in the CDC (the sense wire being at the center of the square); the IP is downward. Left: a TS for the innermost superlayer. Right: a TS for other superlayers.*

$$X = \frac{2x}{x^2 + y^2} \quad \text{and} \quad Y = \frac{2y}{x^2 + y^2}. \tag{12.1}$$

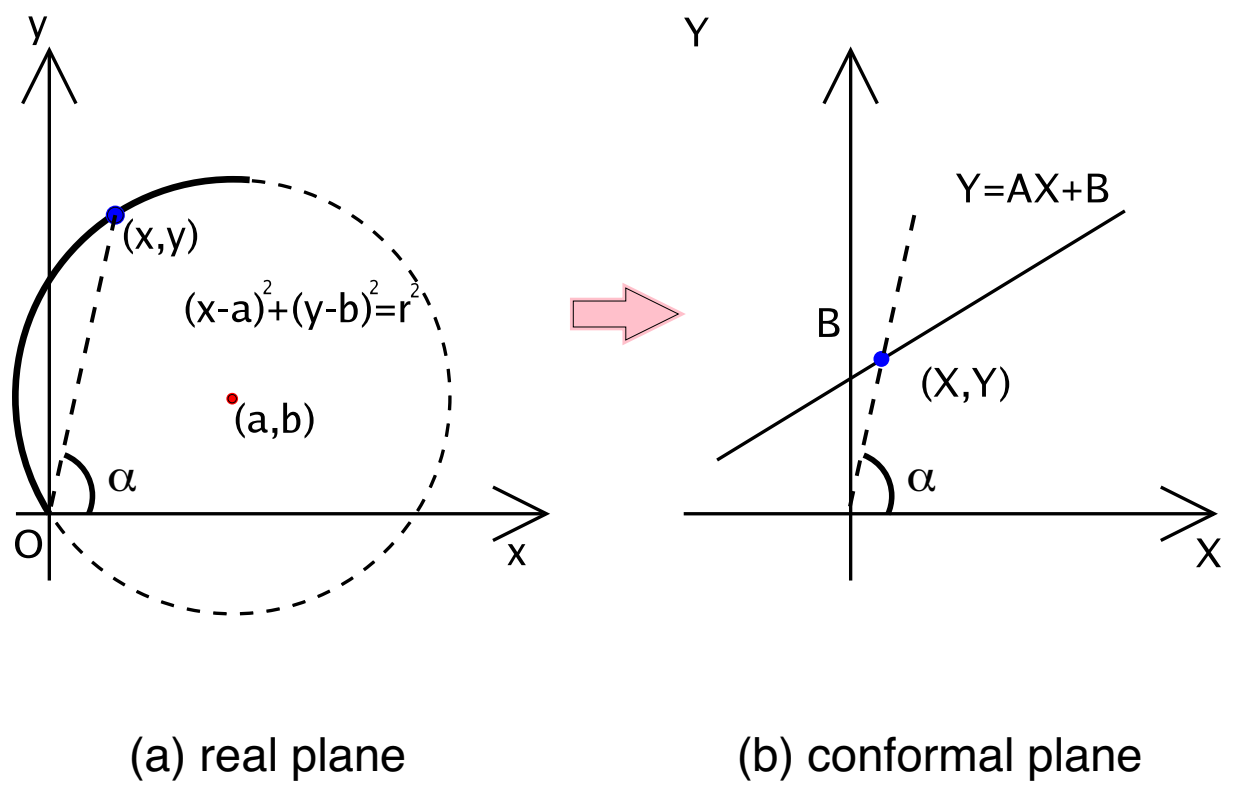

*Figure 12.5: Schematic view of the conformal transformation: (a) in CDC transverse plane, (b) in the conformal plane.*

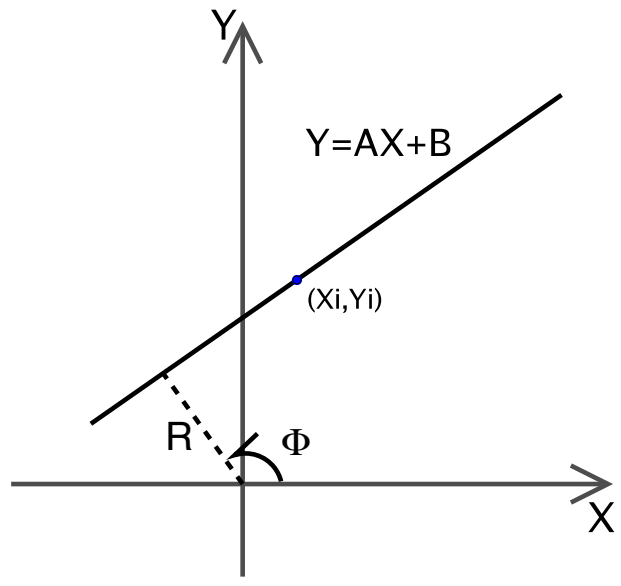

*Figure 12.6: The distance $R$ and angle $\Phi$ of the point of closest approach to the origin of a straight line in the conformal plane.*

In this transformation, the angle $\alpha$ of point $(x, y)$ on the arc (Fig. 12.5) relative to the CDC origin is preserved. An arc of a circle passing through the CDC origin is transformed into a straight line segment in the conformal plane. This line segment is parametrized by the quantities $R$ and $\Phi$ (Fig. 12.6); these are related to the arc's radius and center via

$$R = \sqrt{\frac{B^2}{A^2+1}} = \sqrt{\frac{1}{a^2+b^2}} = \frac{1}{r}, \quad (12.2)$$

$$\tan\Phi = -\frac{1}{A} = \frac{b}{a}. \quad (12.3)$$

The polar equation of the straight line though the point $(X, Y)$ in the conformal plane is

$$R = X\cos\Phi + Y\sin\Phi. \quad (12.4)$$

The plane with $(R, \Phi)$ on orthogonal axes is called the Hough plane. A single point in the conformal plane has an infinite number of straight lines through it and, therefore, an infinite number of curves through it in the Hough plane. However, a collection of points $(X_i, Y_i)$ that lie on a straight line in the conformal plane are mapped by Eq. 12.4 to curves that pass through a common point $(R, \Phi)$ in the Hough plane. We therefore obtain a solution of the straight line in the conformal plane as the intersection point of the curves in the Hough plane.

We performed $r - \phi$ fits to track hits in the Belle II CDC configuration using the `tsim-cdc` package. This package provides the drift-time integrated TS hits based on a Geant4 simulation of the track in the CDC. The `tsim-cdc` package assumes 100% hit efficiency. The `tsim-cdc` algorithm is coded in `c++` with normal float-pointing precision; the Hough-finding minimization is based on the linear regression method.

An event display of a simulated event with a $\mu^-\mu^-$ like-charge pair is shown in Fig. 12.7. The upper left figure shows the TS hits in the CDC's transverse plane, with two curved tracks passing through the interaction point $(0, 0)$. These TS hits are then transformed into two distinct straight lines in the conformal plane and, finally, into the family of curves in the Hough plane (one curve per TS hit). The two intersections of these Hough-plane curves, at $R \simeq 0.015$ and $\Phi \simeq 1.9,\ 5.1$, represent the solutions of the reconstructed tracks.

The distribution of reconstructed transverse momenta for single-muon events is shown in Fig. 12.8; each muon was generated with $p_t = 0.5\,\mathrm{GeV}/c$. There is no apparent bias observed in the fit to this distribution. We also fitted single-muon samples with other input $p_t$ values and found no obvious bias in the fits. However, as expected, the fitted $p_t$ resolution rises with $p_t$.

## 12.2 CDC 3D Trigger

We describe here a possible implementation of a three-dimensional (3D) CDC track trigger system, based on the measurement of the axial coordinate of a track at its closest approach to the interaction point (IP). This coordinate, labelled $z_0$, lies near the IP for a track originating from an $e^+e^-$ collision but is not localized for a track associated with any background process. Figure 12.9 shows the $z_0$ distribution from real data recorded with a random trigger by the Belle experiment. From various simulation studies and our accumulated understanding of the background near the IP, we are confident that the peak at $z_0 = 0\,\mathrm{cm}$ is associated with $e^+e^-$ collisions while the broad distribution at other values of $z_0$ corresponds to beam-gas and beam-wall backgrounds.

Note that Fig. 12.9 reflects the physics-to-background distribution after ten years of KEKB operation; the background condition was much worse in the first year. Therefore, during the early stages of Belle II, an efficient rejection of beam-gas and beam-wall backgrounds will prove quite valuable in maintaining high physics performance with minimal deadtime. This can be done if $z_0$ information for each track is available within the global Level 1 trigger latency of $\sim 5\,\mu$s. After accounting for the latencies associated with the TSF and GDL, only $\sim 2\,\mu$s is available to calculate $z_0$. This is beyond the capability of conventional CPUs, so our design uses FPGAs for $z_0$ reconstruction.

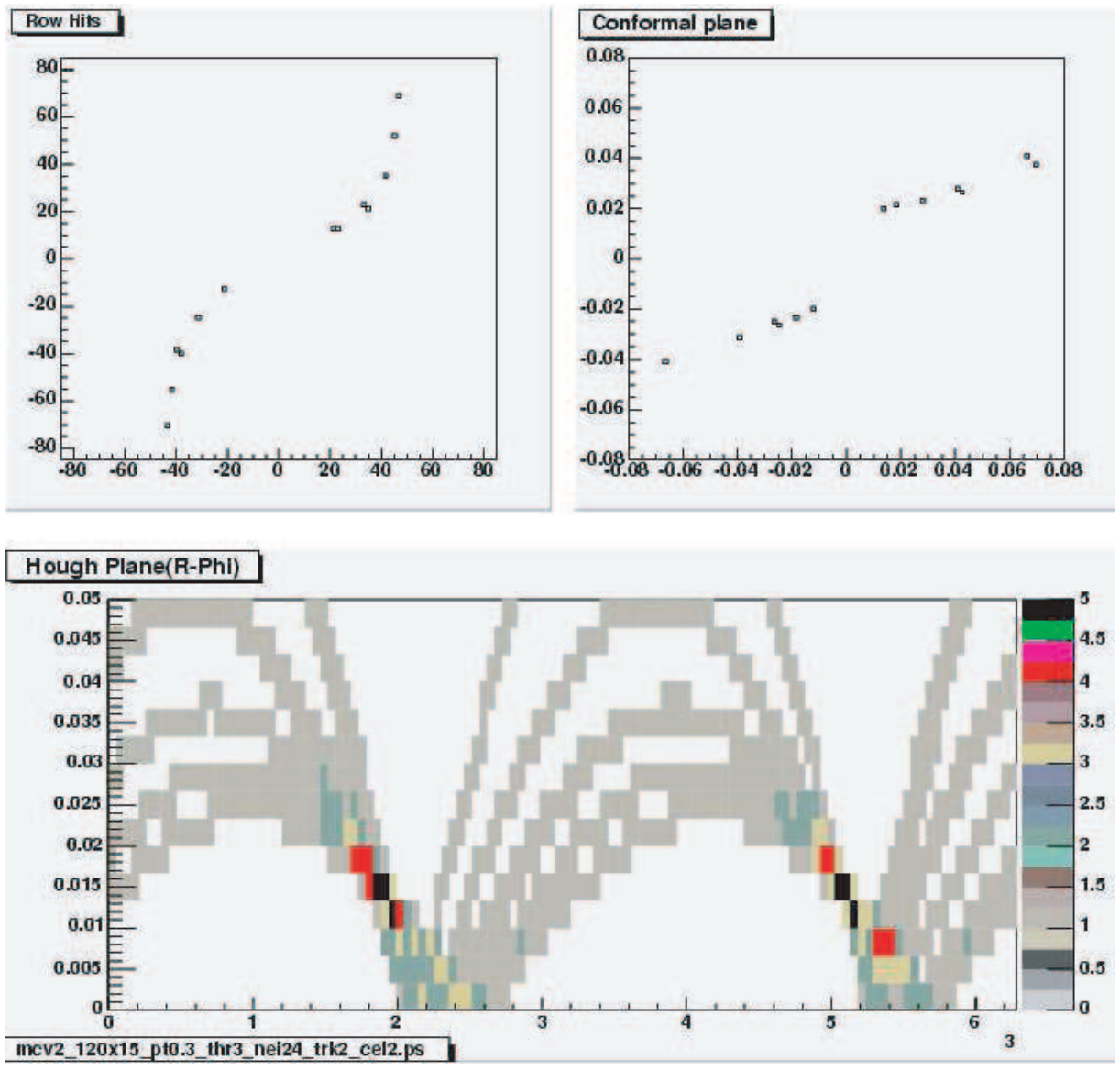

*Figure 12.7: An event display of the 2D track finding stage for a simulated $\mu^-\mu^-$ event. Upper left: TS axial hits (dots) in the $r$-$\phi$ plane in the CDC; the horizontal (vertical) axis is $x$ ($y$). Upper right: the corresponding TS axial hits (dots) in the conformal plane; the horizontal (vertical) axis is $X$ ($Y$). Bottom: the corresponding TS axial hits (pixellated curves) in the Hough plane; the horizontal (vertical) axis is $R$ ($\Phi$).*

This type of computation has been proven to work since 2004 in the BaBar experiment [3]. They use the TS information from their drift chamber's stereo wires to calculate $z_0$ for each track candidate. Their computation is done purely within an FPGA, running at 120 MHz for the core part of the algorithm.

We will implement 3D tracking in the Belle II Level one trigger with a concept based on BaBar's. There are substantial differences in implementation due to the fact that the information available in our TS is somewhat different; we may have to run the FPGA at a higher clock frequency to carry out the computation within the aforementioned latency.

### 12.2.1 Strategy

The 3D track reconstruction is a two-step process. The first part involves "finding" a track candidate; the second requires "fitting" the track to get the 3D information of the track under consideration. We will re-use the 2D track-finding algorithm described in Sec. 12.1. To improve the fitter performance, we may augment the 2D finder with a fast auxiliary algorithm that provides an initial value of the polar angle of the track candidate under consideration. The overall 3D algorithm structure, including the 2D finder and fitter, is illustrated in Fig. 12.10. The 3D fitter is itself a two-step process: an $r$-$\phi$ fit followed by an $r$-$z$ fit.

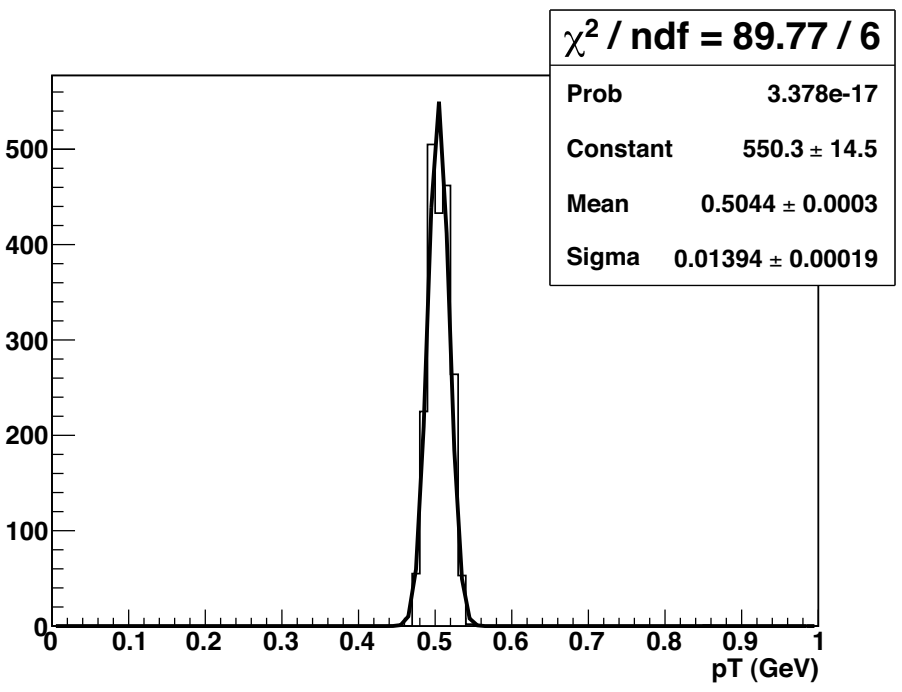

*Figure 12.8: Transverse momentum distribution from the 2D fit of tracks in single-muon events. Each muon was generated with $p_t = 0.5\,\mathrm{GeV}/c$.*

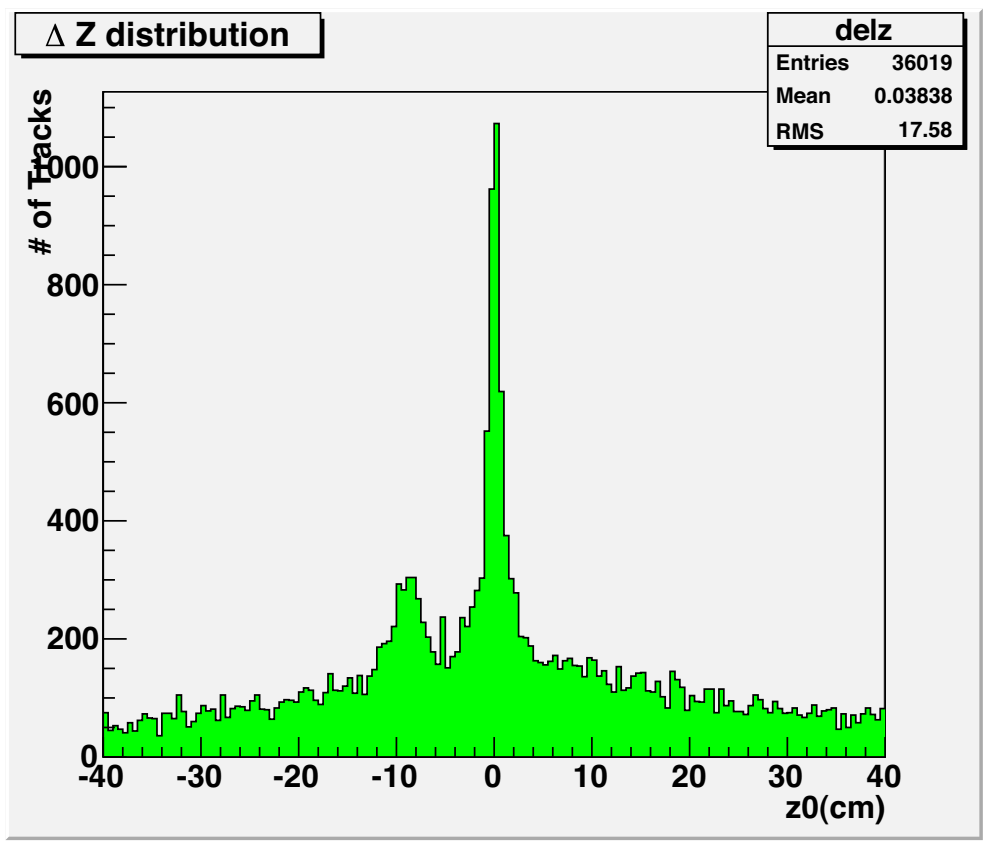

*Figure 12.9: The offline $z_0$ distribution for random-triggered data in Belle's experiments #57 and 59. The dominant peak at $z_0$=0 cm is from $e^+e^-$ collisions; the smaller peak at $z_0$=−10 cm is due to beam collisions with a beamline structure at that location.*

### 12.2.2 $r$-$\phi$ fit

In the first step of the 3D fit, the 2D $r$-$\phi$ track fitter provides a list of track candidates, each with its own list of TS hits. This duplicates the functionality of the standalone 2D track fitter described in Sec. 12.1. In this step, one can average two `U-V` stereo TS hits to form an additional pseudo-axial TS. According to our studies, using such pseudo-axial TS hits does not change the $r$-$\phi$ fit itself but does improve the $z_0$ resolution in the next step—the $r$-$z$ fit—by $\sim$ 5–10%. The results of the $r$-$\phi$ fit are passed to the $r$-$z$ fit.

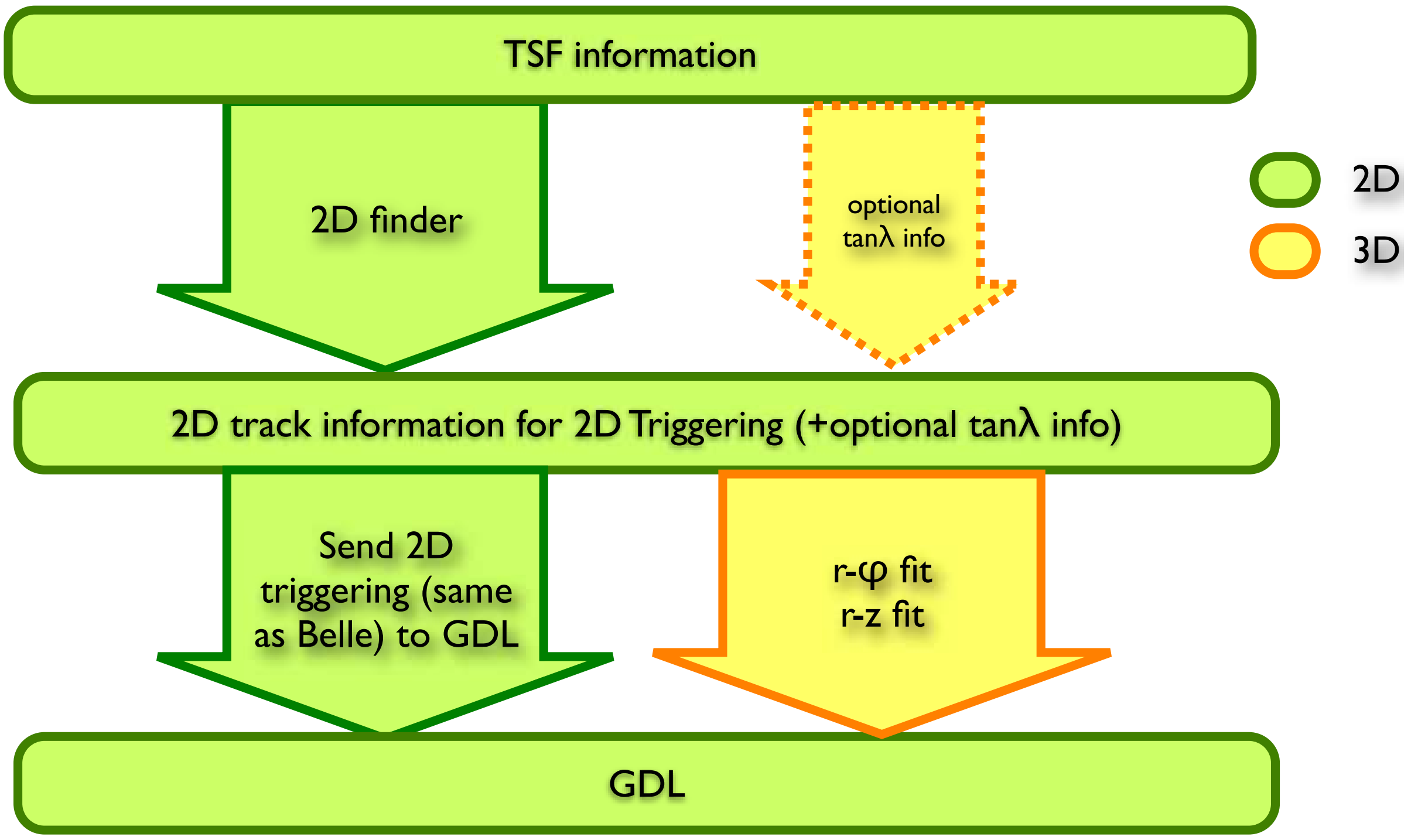

*Figure 12.10: The track-reconstruction algorithm structure. In the initial stage, the 2D finder may be supplemented with an algorithm (dashed-boundary arrow) that provides polar angle information to the fitter.*

### 12.2.3 $r$-$z$ fit

The second step of the 3D fit uses the stereo TS information. A charged track from the IP appears as a cycloid in the $r$-$z$ projection; this can be approximated as a straight line for all TS hits from an energetic track or for the TS hits near the IP from a soft track. For a given stereo TS with index $i$,

$$r_i \tan(\phi_{\text{fit}} - \phi_i) = z_i \tan\theta_{\text{stereo}} \tag{12.5}$$

is satisfied, where $r_i$ is the radius, $\phi_i$ is the azimuthal angle, and $z_i$ is the axial coordinate of the stereo TS; $\theta_{\text{stereo}}$ is the angle of stereo wire of the stereo TS with respect to axial wires. The azimuthal angle $\phi_{\text{fit}}$ is provided by the $r$-$\phi$ fit; $\phi_{\text{fit}} - \phi_i$ should be the residual from the stereo angle effect. We construct a goodness of fit

$$\chi^2 = \sum_{\text{stereo}} \frac{(z_i - z_0 - r_i \cot\theta)^2}{\sigma_i^2} \tag{12.6}$$

where the free parameters $z_0$ and $\theta$ represent the axial coordinate (nearest the IP) and the polar angle, respectively, of the track candidate. The "uncertainty" $\sigma_i$ is defined as

$$\sigma_i = \frac{r_i \times s}{\tan \theta_{\text{stereo}}} \tag{12.7}$$

where $s$ is an arbitrary scale factor. By minimizing this $\chi^2$, the $r$-$z$ fit is performed in the $r$-$z$ plane using the straight-line approximation to the cycloid. Note that the resolution of the fitted $z_0$ depends on the precision of $\phi_i$ measurement, according to Eq. 12.5.
The distribution of fitted $z_0$ values for a sample of single-muon events (generated at the origin with $p_t = 0.5\,\text{GeV}/c$ and a range of polar angles) is shown in Fig. 12.11. The Gaussian resolution of this distribution is slightly less than 10 cm and, from our studies using other generated event samples, is roughly independent of transverse momentum.

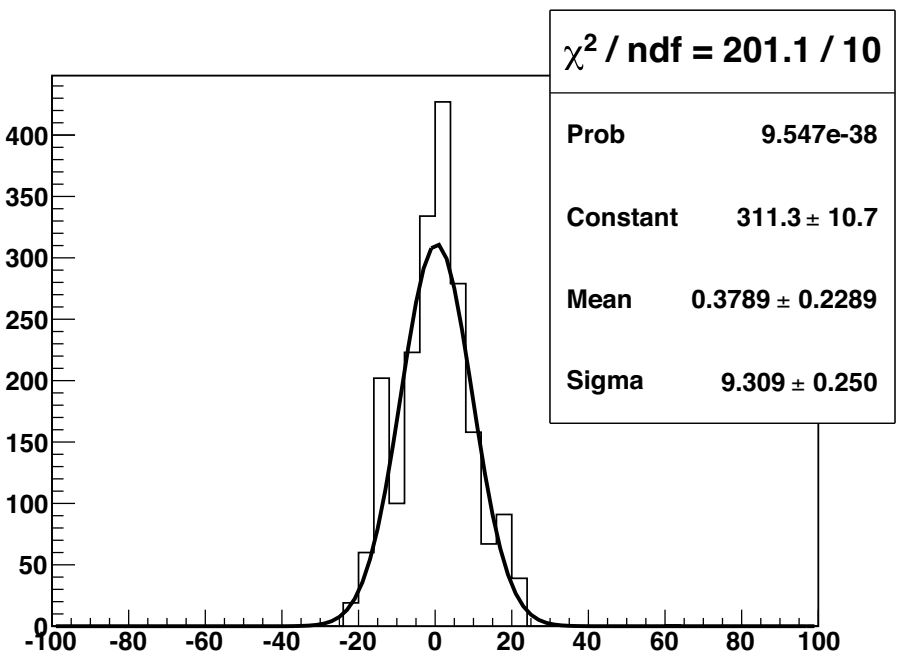

*Figure 12.11: The $z_0$ distribution (in cm) of the $r$-$z$ fit to the track in single-muon events.*

### 12.2.4 Possible improvements

Comparing with Fig. 12.9, the $z_0$ resolution of $\sim 10$ cm is about twice the desired value needed to discriminate IP-originating tracks from backgrounds. It is clear that improvements are needed for this 3D track trigger to be effective. Three areas of investigation are presented below.

#### 12.2.4.1 TS information

The Belle II TSF sends a single bit per TS, stating whether it has fired or not. On the other hand, the BaBar TSF sends the estimated $\phi$ location as well as its error, representing the possible range of $\phi$ for the TS [4]. To implement something similar, the Belle II TSF firmware must be altered. Before doing so, we will investigate the degree of improvement in the $z_0$ resolution using the `tsim-cdc` simulation.

#### 12.2.4.2 Geometry of stereo superlayers

The Belle II CDC superlayers are configured as `AUAVAUAVA` from the inner radius outward, with a total of 5 axial and 4 stereo superlayers. The BaBar configuration is `AUVAUVAUVA`, with a total of 4 axial and 6 stereo superlayers. BaBar's drift chamber has 50% more stereo layers, and `U-V` superlayers are adjacent. If the Belle II configuration were changed to `AUVAUVAUV`, to have 3 axial and 6 adjacent stereo layers, the `tsim-cdc` simulation predicts an improved $z_0$ resolution of 7.4 cm. Further investigation will clarify whether this improvement is associated with the increased number of stereo superlayers or to their adjacency, and to model other superlayer configurations.

#### 12.2.4.3 Usage of DSP chip hardware

So far, our track-fitting studies have been carried out with `c++` and floating-point calculations. We expect that, in implementing the present algorithm into FPGA, the $z_0$ resolution is degraded somewhat due to integer-only calculations and the usage of look-up tables in some steps. We may also have to sacrifice precision to fit within the Level 1 trigger's latency (including TSF and GDL overhead). We plan to host at least one DSP chip—to permit floating-point calculations—in the next prototype of the trigger board to avoid the degradation inherent in fixed-point calculations in the 3D fitter.

### 12.2.5 Summary

A very rough, qualitative sketch of the 3D fitter algorithm is presented here. We plan to implement the algorithm in FPGA by the end of the year 2013.

## 12.3 ECL Trigger

The calorimeter is a very important component to generate fast signals for a fully efficient trigger for both neutral- and charged-particle oriented physics events. Two trigger schemes are envisaged: a total energy trigger and an isolated-cluster counting trigger. These are complementary: the former is sensitive to physics events with high electromagnetic energy deposit while the latter is sensitive to multi-hadronic physics events that have low energy clusters and/or minimum ionizing particles. In addition, the system should identify Bhabha and $\gamma\gamma$ events, which are needed to measure the online luminosity; this information is only available from the calorimeter system.

### 12.3.1 Trigger scheme

We have ten years' experience in the operation of the similar Belle calorimeter trigger system [5]. The dominant issue to be considered in Belle II is the increased trigger rate due to the much higher luminosity and beam background. We plan to adopt the Flash-ADC (FADC) and FPGA scheme. The former digitizes a trigger analog signal called the Trigger Cell (TC), formed by $4 \times 4$ crystals, and the latter determines the amplitude and timing of the TC and performs the final ECL sub-trigger decision by a firmware algorithm that is easily modifiable. The merits, compared with the Belle trigger system, are (1) the simplicity of the electronics hardware and signal cabling and (2) the ease of upgrading in the case of the endcap calorimeter replacement. At present, three kinds of trigger modules are planned: Fast shaper in the main shaper board; FADC trigger module (FAM); and Trigger and Monitor module (TMM). The block diagram of ECL trigger readout electronics system is shown in Fig. 12.12.
The allocated trigger timing latency for the ECL sub-trigger is about 4 $\mu$s. Good timing resolution is needed in the endcap region that is not covered by the barrel PID trigger that provides the most precise trigger timing. The ECL timing latency is less than the requirement, as estimated in Table 12.2, and the timing resolution is expected to be same as in Belle: 20 ns at the $1\sigma$ level.

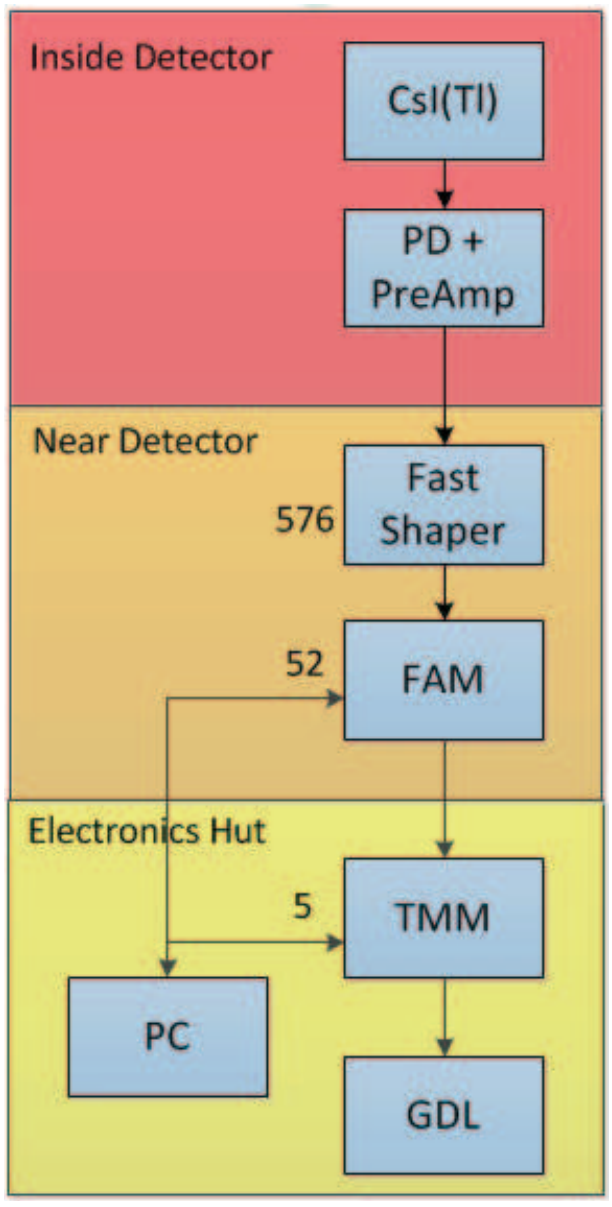

*Figure 12.12: Block diagram of ECL trigger readout electronics system*

To construct the total energy and isolated-cluster triggers, we will prepare 26 final trigger bits, described in Table 12.3. Other useful trigger bits will be added after trigger simulation (TSIM) study.

The beam-gas background is expected to be high at early stage of operation, while the other backgrounds that scale with beam current or luminosity will grow later. To cope with this evolution, several tuning knobs are incorporated: (1) TC threshold, (2) total energy threshold, (3) number of isolated clusters, (4) the possibility of excluding a part of endcap from the trigger, and (5) enabling/disabling the cosmic veto. Option (4) will require a detailed trigger simulation study. We plan to run without the cosmic veto because the machine luminosity is one order of magnitude higher than KEKB; thus, cosmic events do not harm the DAQ bandwidth, while enabling the cosmic veto might sacrifice some fraction of interesting low-multiplicity processes such as initial state radiation (ISR) events, two-photon collisions and tau pair production. On the other hand, since the cosmic veto can also suppress a part of beam background events, we may enable it during extreme beam-background conditions.

### 12.3.2 Trigger electronics

#### 12.3.2.1 Fast shaper

Sixteen fast-shaping signals from individual crystals, with a shaping time of 200 ns, are merged in the fast shaper circuit. This is implemented on the main shaper board to form an analog trigger sum called the trigger cell (TC). This reduction in the number of signals reduces the trigger timing latency and suppresses coherent electronics noise in the high beam background environment. In the fast shaper, each crystal's signal is corrected before the merge for the crystal-to-crystal variation in light yield (for equal-energy incident particles).
WWe have performed a test of the noise level, signal shape, and pulse-gain adjustment on a prototype of the main shaper, including the fast shaper circuit. Figure 12.13(b) shows the noise level of 1.58 channels, equivalent to 10 MeV, for the TC; this is well below the threshold of 100 MeV corresponding to the passage of a minimum-ionizing particle (MIP) through the length of a crystal. We have confirmed that the fast shaper is almost ready for mass production.

| Item | Latency (ns) |
|---|---|
| Peaking time of TC | 700 |
| ADC pipeline @ FAM | 100 |
| Peak finding process @ FAM | 300–400 |
| Programmable delay @ FAM | 300 |
| Gbit transfer( 200 bits) | 100 |
| Optical cable length | 200–300 |
| Trigger input alignment @ TMM | 100 |
| Stage-1 to stage-2 bit transfer @ TMM | 700 |
| Trigger decision @ TMM | 200–300 |
| Total latency | 2700–3000 |

*Table 12.2: Estimate of ECL trigger timing latency based on expected performance*

| Item | Number of bits |
|---|---|
| Trigger timing (Final, Fwd, Barrel, Bwd) | 4 |
| Total Energy (>0.5, 1.0, 3.0 GeV) | 3 |
| Isolated cluster | 4 |
| Bhabha-type | 11 |
| OR-ed Bhabha | 1 |
| Barrel Bhabha | 1 |
| Prescale Bhabha | 1 |
| Cosmic veto | 1 |
| TC hit pattern | 576 |
| Total | 26+576 |

*Table 12.3: Final ECL trigger output to GDL*

#### 12.3.2.2 FAM

The FAM module determines the pulse height and time at the peak of each TC signal. Each FAM has 40% pulse gain reduction attenuators, baseline adjustment and anti-aliasing filters in front of four 12-bit and 100-MHz sampling quad-FADCs (ADS6424). Channel-by-channel digitized pulse information from the FADC is recorded by digital signal processing logic, implemented in an FPGA (XC5VLX50T). The FPGA logic finds the pulse's peak value by examining the sliced pulse values within a 400-ns window that straddles the peak. The pulse-peak value and time of each TC are recorded in a data buffer memory inside the FPGA. When any pulse peak out of 12 TC signals is larger than a predetermined (downloaded) threshold, the FAM aligns the stored records for the 12 TC information to the earliest above-threshold signal, then sends this information (including the 12-bit peak value, the 1-bit discriminator value, and the aligned time value with a 300-ns programmable delay) for all 12 TCs to a TMM module in the Electronics Hut via an optical link. The trigger timing is determined by the arrival at the TMM. Figures 12.14 and 12.15 show the block diagram and photograph of FAM prototype board, respectively. Altogether, 52 FAM modules are needed to cover the 576 TC signals.

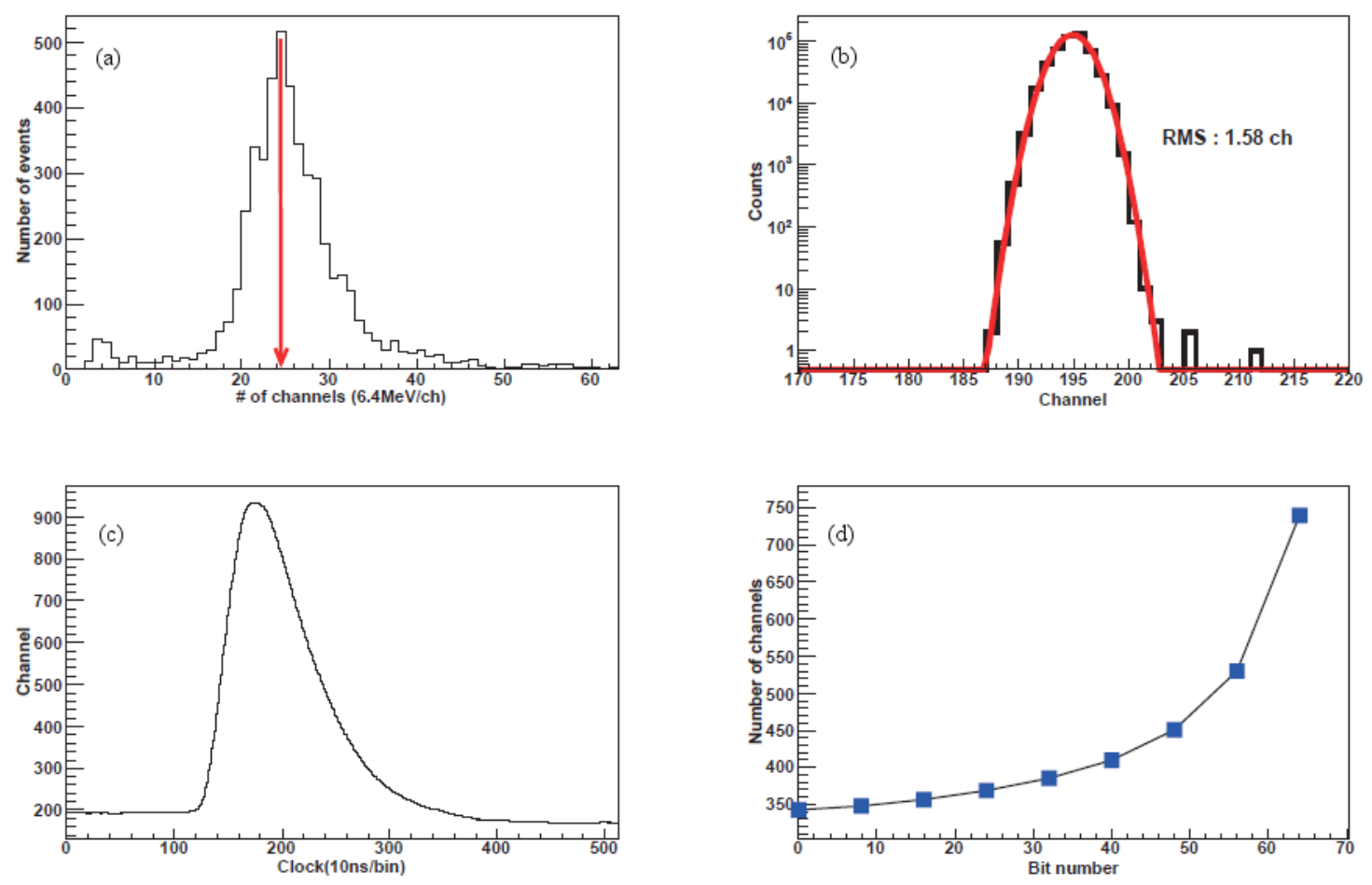

*Figure 12.13: Measurement of fast shaper characteristics at the cosmic stand. (a) Cosmic ray energy distribution, where the peak corresponds to a MIP passing through the 30-cm length of a crystal. (b) Pedestal distribution of TC. (c) TC analog signal shape by test pulse. (d) Individual-crystal gain adjustment.*

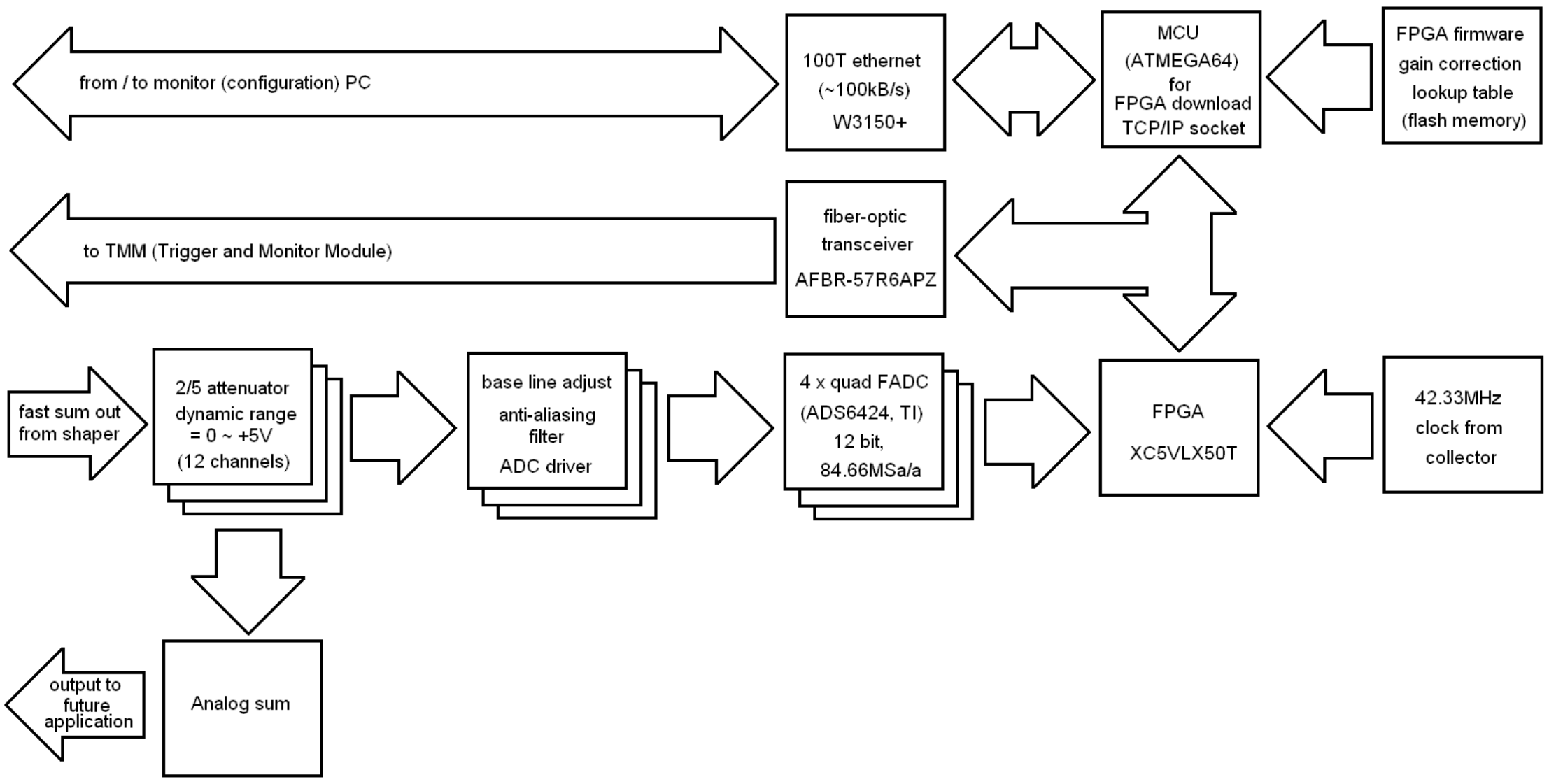

*Figure 12.14: Block diagram of 6U-VME type FAM prototype board*

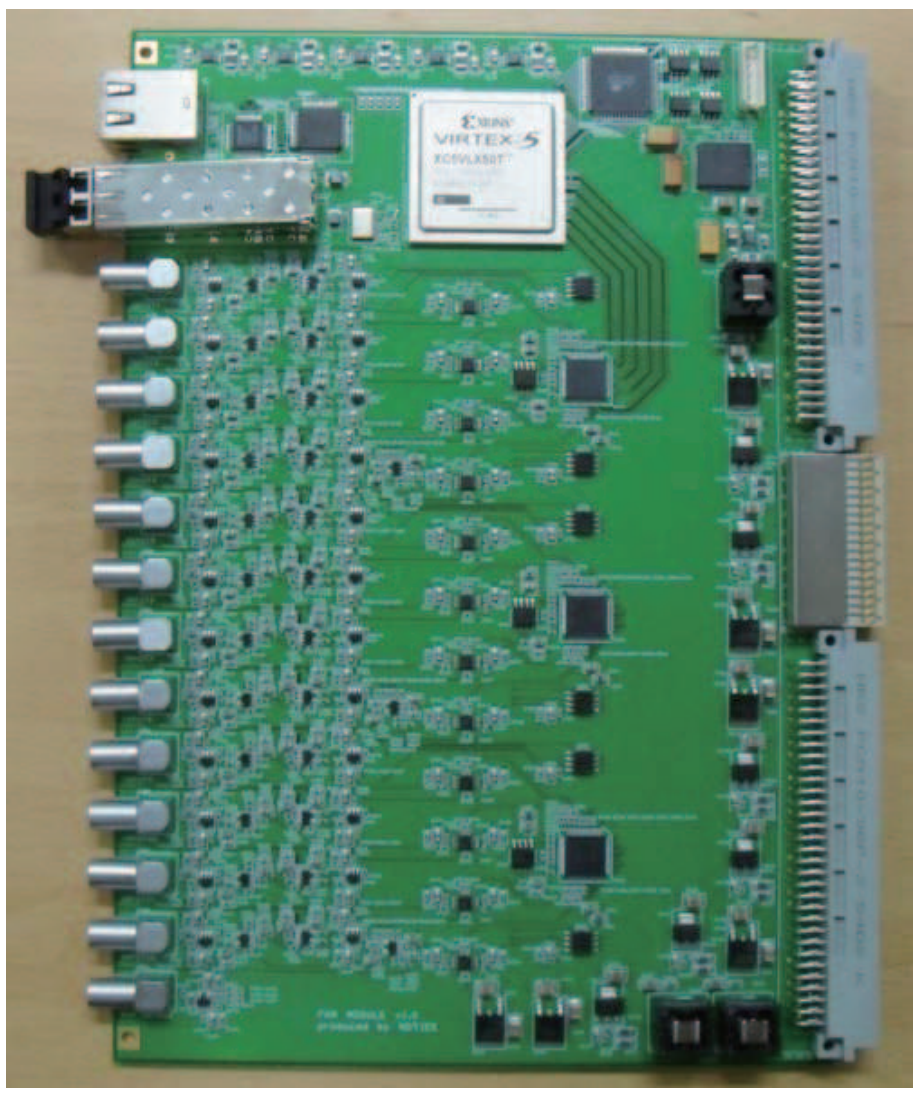

Figure 12.15: A photograph of 6U-VME type FAM prototype board

#### 12.3.2.3 TMM

The TMM module receives the digitized and time-aligned FAM signals and determines the final ECL trigger outputs. A two-stage TMM is needed for the 52 FAM inputs because of the limited front-panel space on each TMM board. Both stages use the same module but have distinct firmware algorithms on the FPGA. The first-stage module receives 13 FAM signals and generates two kinds of output. One output is sent to second-stage module and the other to the Global Decision Logic (GDL). The module has an FPGA (XC5VLX50T) and performs input signal alignment based on the external clock provided from the beam-collision clock system. The TC hit pattern output—the discriminated TC signals—are sent directly to the GDL to be used for matching with the CDC trigger hit pattern. The single second-stage TMM receives the TC information from all of the first-stage modules and calculates the final ECL trigger outputs in its FPGA, then sends this to the GDL. Based on the current design architecture, 4+1 TMM modules are needed. A pre-prototype of the TMM, shown in Fig. 12.16, has been produced.

### 12.3.3 Schedule

The fast shaper will be prepared according to main shaper production schedule. Until 2012,we will perform trigger integration tests with the entire readout electronics chain setup and final versions of FAM and TMM modules will be ready. Mass-produced modules will be installed by 2013. In parallel with hardware production, the FPGA firmware algorithms of these modules will be developed in the VHDL hardware description language. To optimize the trigger algorithms, we will continue our studies with the TSIM simulator package.

## 12.4 Barrel PID Trigger

The Time-Of-Propagation (TOP) counter that will be used for the barrel PID (Ch. 7) has an intrinsically good time resolution of under a nanosecond. However sub-100 ps timing is usually only obtained in offline reconstruction, and is more than is needed at the trigger level, where a couple-ns time resolution for the GDL trigger would help to reduce the data volume from out-of-time hits in the SVD. An algorithm for determining the event time resolution from tracks that hit the TOP detector is studied below, based upon a full GEANT4 simulation of the detector configuration described in Ch. 7. Single-photon hit times are time-encoded using available programmable logic devices, and the resource requirements are evaluated and shown to be acceptable.

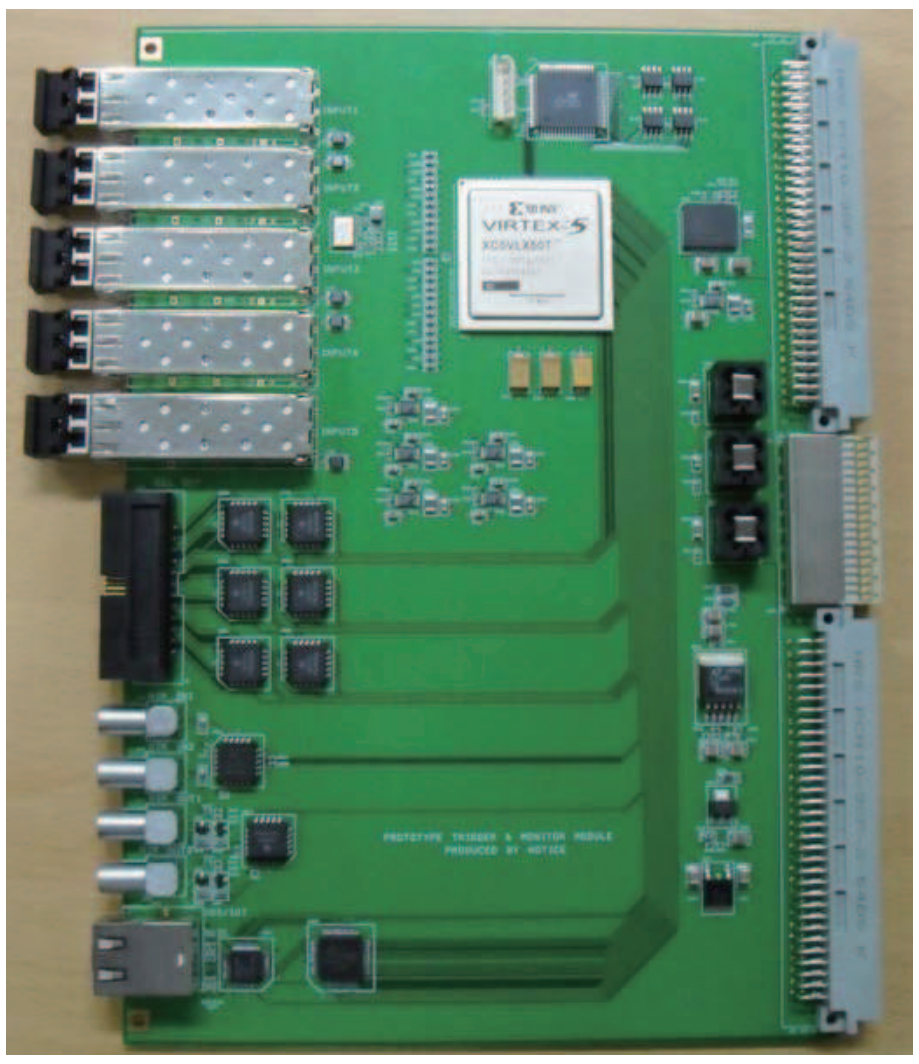

*Figure 12.16: A photograph of the 6U-VME TMM pre-prototype board. The final design will be implemented on a 9U-VME board to accommodate the input/output connectors.*

### 12.4.1 Trigger Configuration

The readout electronics and integrated trigger functionality are shown for the detector and front-end readout modules in Fig. 12.17. Giga-bit fiber links carry the position and time information of hit signals from the 16 detector staves to dedicated trigger processing boards in a flow diagram indicated in the figure.

### 12.4.2 Trigger Performance Estimates

A preliminary study based on Monte Carlo generated data was used to identify the most effective detection techniques that could be efficiently implemented in real-time hardware. As seen in Fig. 12.18, the detected pattern of photons in both space and time depend upon the track impact position and this information can be used to evaluate the track flight length and determine and event time.
Statistical estimates of the obtained timing errors are reported, where a comparison was made between a system that uses time and space information of detected pixels versus timing only to estimate the trigger timing error. Such a time estimate is computed as follows:

1. For all incoming data, we record the timing difference between the first pixel received and all the others, as well as their spatial position.

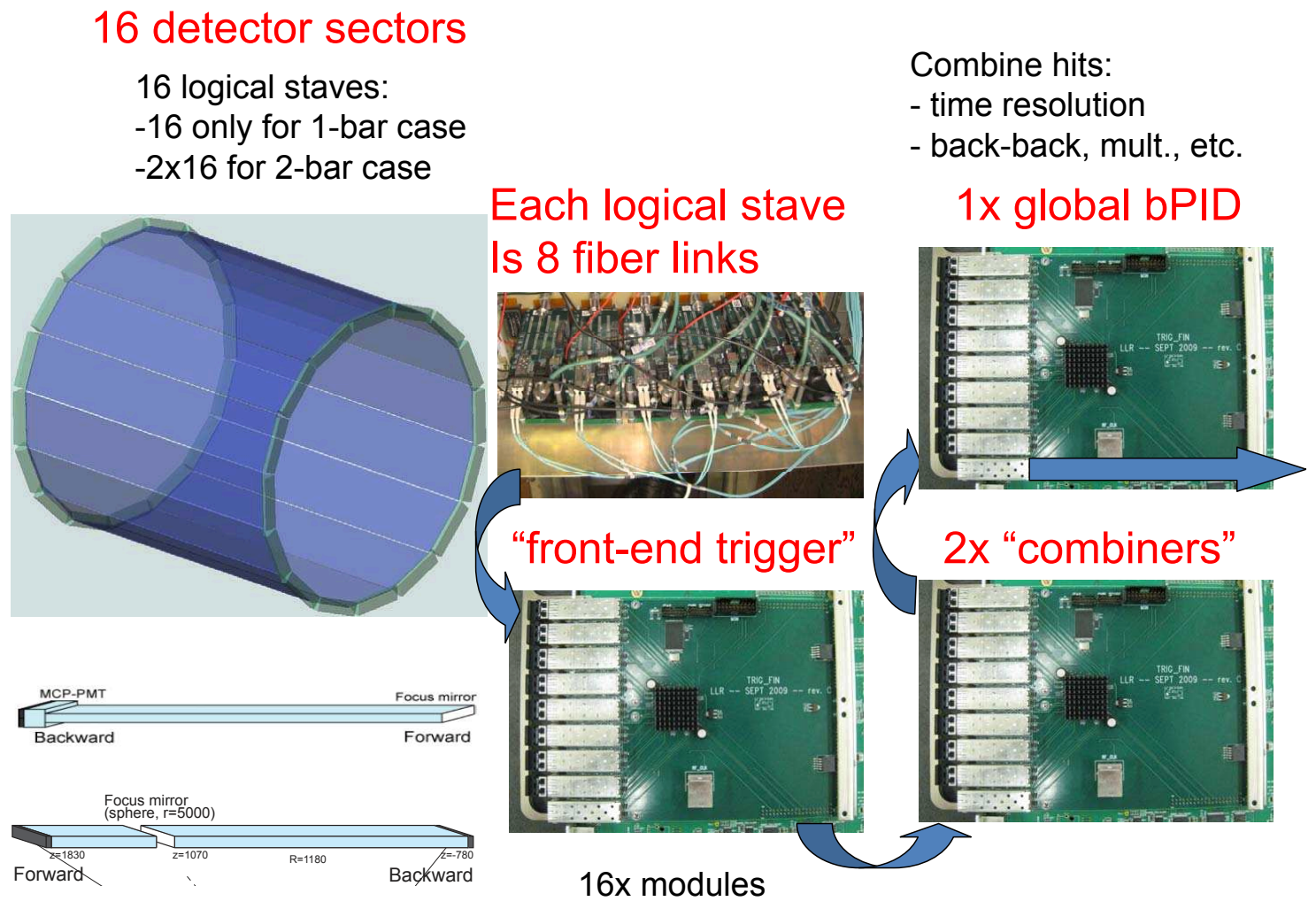

Figure 12.17: Overview of the barrel PID trigger system.

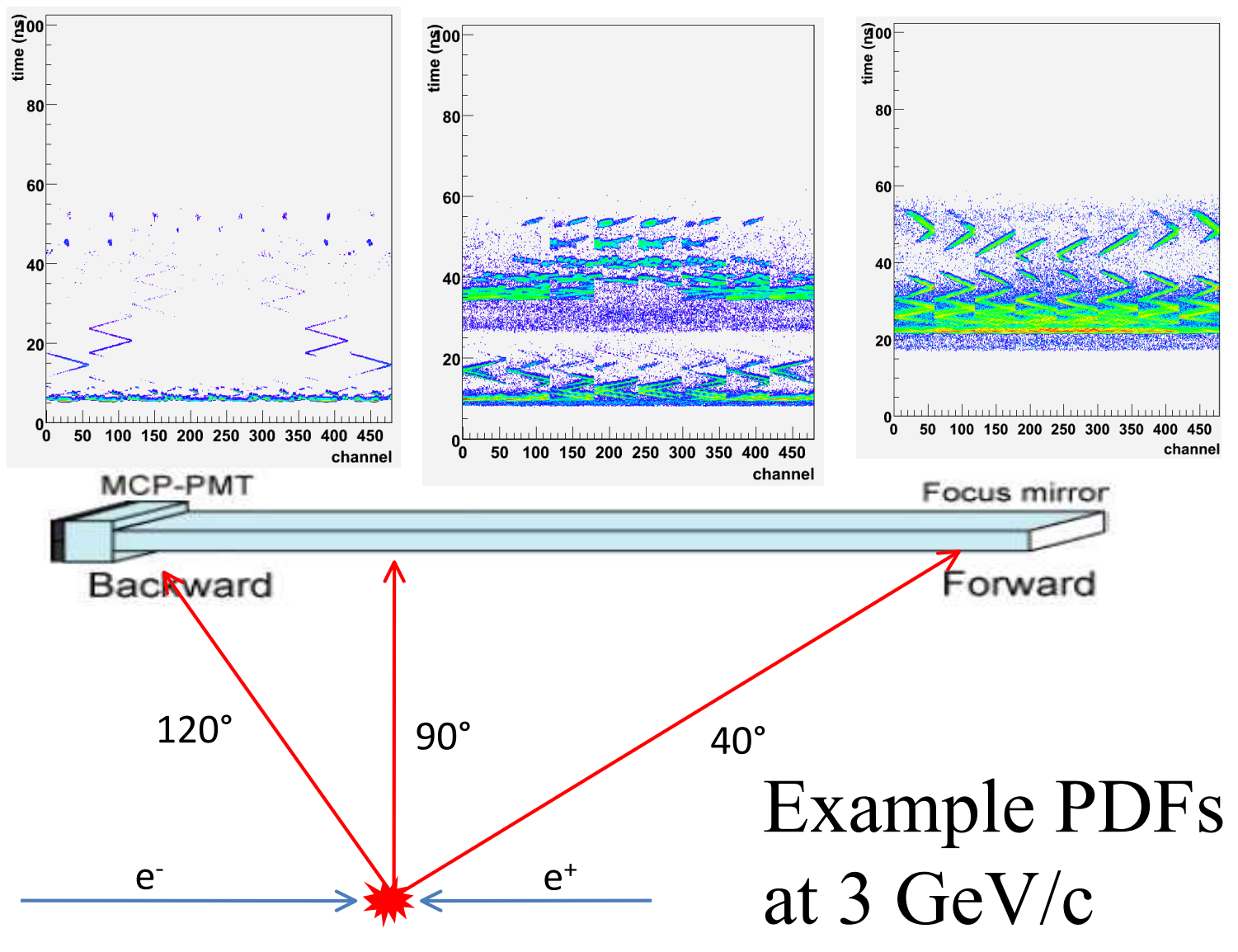

Figure 12.18: A graphical representation of the information available to differentiate the time (position) of hits on the barrel PID detector.

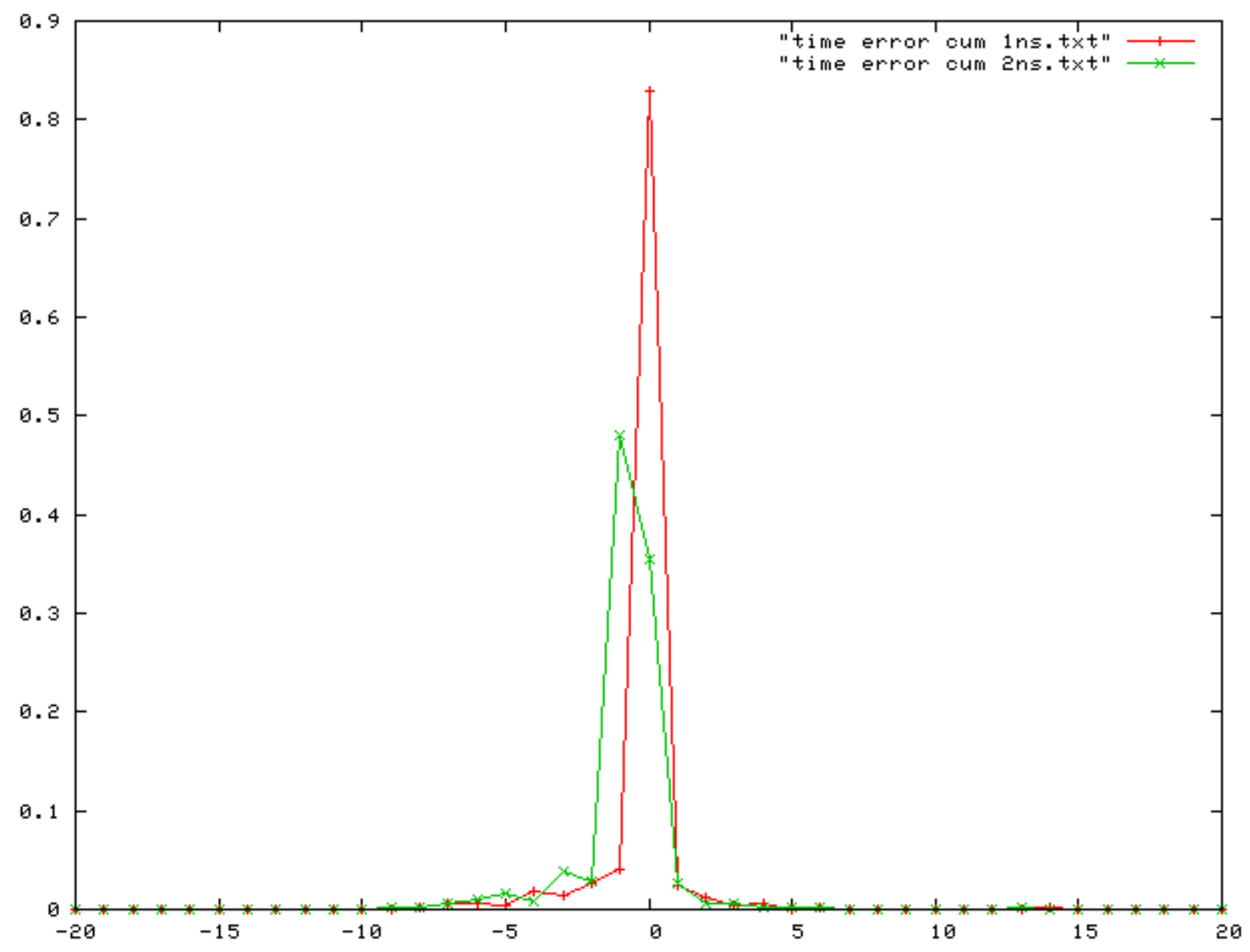

*Figure 12.19: Estimate of timing errors: use of time and space information.*

2. We compute the log-probability of the timing and spatial profile for 10 different positions of interaction in the bar. For the timing case, this in fact requires computing also—for each of the 10 positions—the optimal shift of the time PDF with respect to the time of the first sample; this is equivalent to computing $n$ different log-probabilities and picking the highest value.

3. For the case in which both space and time information is used, the two log-probabilities are added and the maximum value is used as an estimate for the position; otherwise, only the maximum of the 10 log-probability for timing is used.

4. The "shifting" time corresponding to the optimal position is used as an estimate for the time of interaction.

Experiments were run on 5000-track samples using Probability Distribution Functions computed on a distinct set of 5000-track events. For this initial study, all tracks are 2.5 GeV/$c$ kaons and incident on the center of the long axis of the 1-bar detector configuration. There should be minimal difference in the result with respect to pions or higher momentum tracks. Interactions at the bar extremities will introduce somewhat larger errors, though full reconstruction studies indicate that the effect should be small. The results follow.

- Figure 12.19: error distribution in case both time and space information is used in the estimate. The two curves correspond to an input time resolution of 2 ns and 1 ns. The average error is $\approx 0.8$ ns and $\approx 0.2$ ns, respectively (the estimate being slightly asymmetrical—a consequence of the asymmetries of the angles). What is more important, the standard deviation is $\approx 1.9$ ns and $\approx 1.7$ ns, respectively. Furthermore, the curves are not described by a single Gaussian due to the long tails, and $\approx 90\%$ of the errors are within $\pm 2$ ns and $\pm 1$ ns, respectively.

- Figure 12.20: in this case the previous 1-ns distribution is compared with the distribution obtained using the time information only. In this case, the asymmetry is slightly more pronounced ($\approx 0.25$ ns) and the standard deviation higher ($\approx 2.1$ ns), yet more than 85% of the cases have errors within 1 ns.

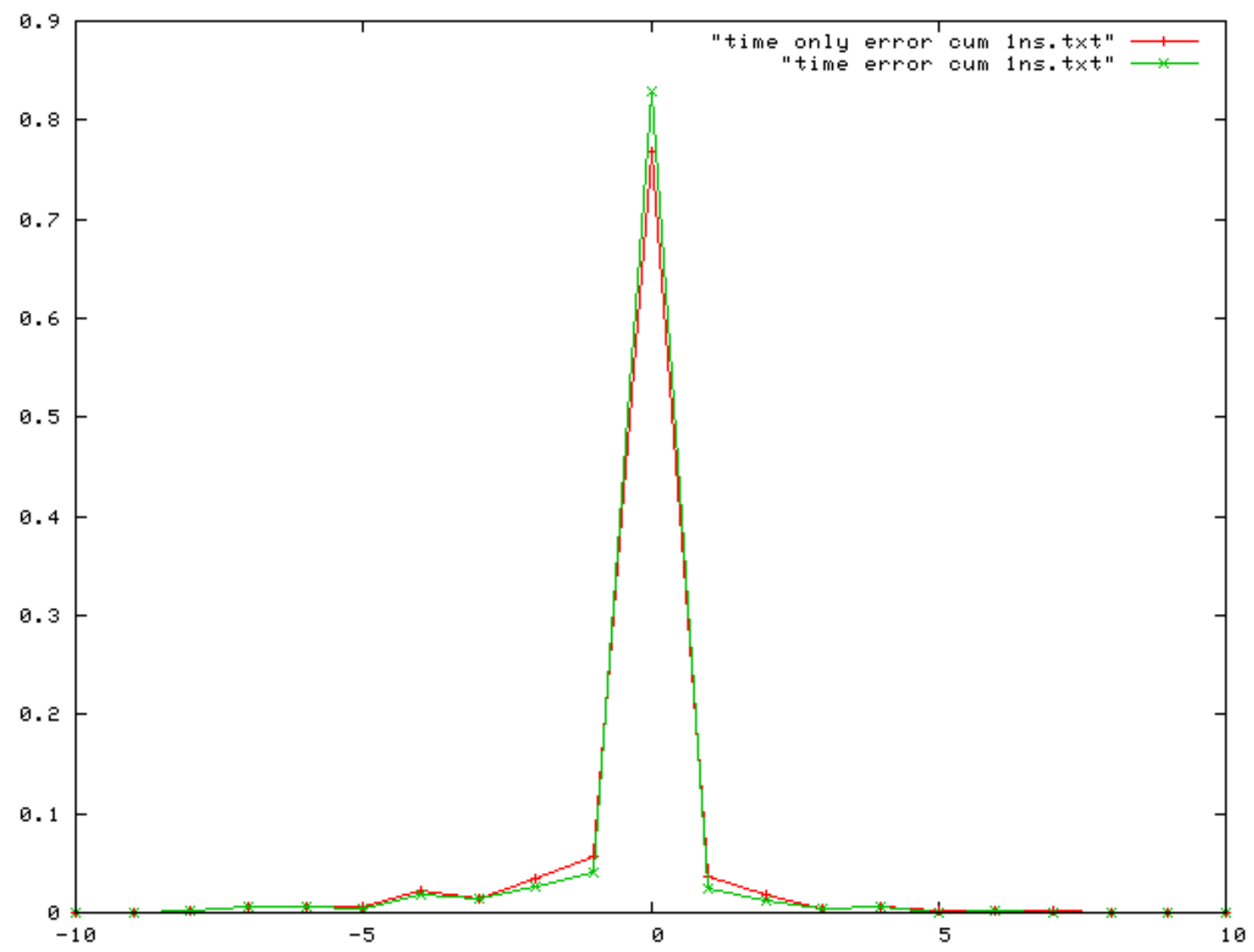

*Figure 12.20: Estimate of timing errors: use of time information only.*

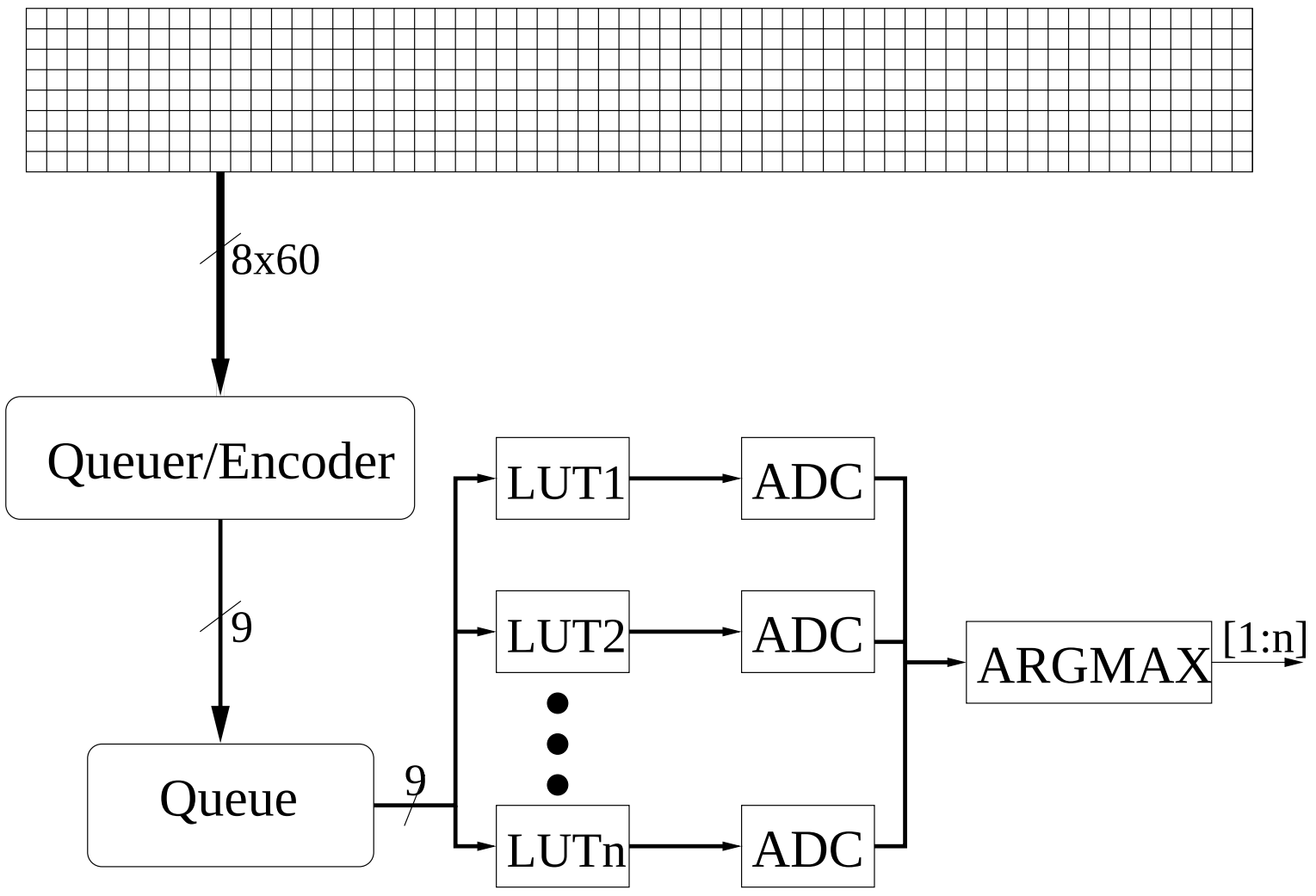

*Figure 12.21: Barrel PID Trigger: Conceptual Block Diagram.*

These studies support the conclusion that an accurate identification of the interaction time can be obtained based on the timing information only, thus reducing the trigger algorithm complexity.

### 12.4.3 Proposed Trigger Block Diagram

From the block diagram of Fig. 12.21, a succinct description of the behavior of the system is as follows:

1. The $60 \times 8$ detector (or a FF-image of it), provides at every clock cycle a sampled view of

the arriving photons (if any). This occurs at the detector front-end.

2. A queuer/encoder orders the hits and encodes them to 9 bits of resolution. This operation is implicitly performed by the communication protocol between the front-end board (that receives and packetizes the detector information) and the trigger board. As the communication is time asynchronous due to a random time lag introduced by the protocol, a synchronization is performed to indicate when all hits corresponding to a possible event are received. This mechanism uses both a timeout (tuned to the average features of the signals) and information about maximum effective delay between interaction events.

3. A queue models the communication between the two boards. All detector hits are enqueued by arrival order, and received in order by the trigger board. This guarantees that a signal with a timestamp older than the previously enqueued signal represents later events, and generates a hard decision point to conclude the triggering decision.

4. At the trigger board, a series of $n$ lookup tables (LUTs) select, in parallel, the correct $\log(p)$ corresponding to the position of the pixel for their own PDF.

5. In parallel, these are added and accumulated to existing values.

6. A maximum is computed among all the accumulated sums, and the index of the maximum returned.

The core of the detection consists of the calculation of the most likely time of interaction based on the probability distribution of different interaction positions on the quartz bar. As the absolute time of first interaction is not known, it is necessary to calculate a maximal value of the correlation between the received times and the PDFs of each of the 10 positions in which the interaction position on the bar is quantized. In practice, the first sample received constitutes the reference point for times, while all the successive samples are evaluated in terms of their relative difference with respect to the first, and compared with the PDFs that correspond to each position on the bar. The highest value of the highest correlation is used to identify the position, and the delay with respect to the first reception time. Some of the computation can be performed in parallel: in practice, at least all 10 correlations are computed at the same time. In the worst case scenario for throughput, when the time correlations are computed sequentially, the throughput equals the correlation length multiplied by the cycle time (correlations of around 32 need to be performed). The decision on the best estimate is therefore provided after a maximum of 32 cycles from the receipt of the last sample. The best throughput solution is analyzed below: in such a case, maximum throughput is possible as the entire correlation is computed in parallel, with the LUTs storing the PDFs being duplicated.
Further work is still required to optimize the performance and will be evaluated with extensive Monte carlo simulations to optimize the parameters for ideal detection.

### 12.4.4 Resource Estimation

A possible detailed implementation of the trigger algorithm is shown in Fig. 12.22.

#### 12.4.4.1 Block Diagram

In the diagram, the signals may be interpretated as follows:

- Sample_Timing_Info: time when a sample arrives.

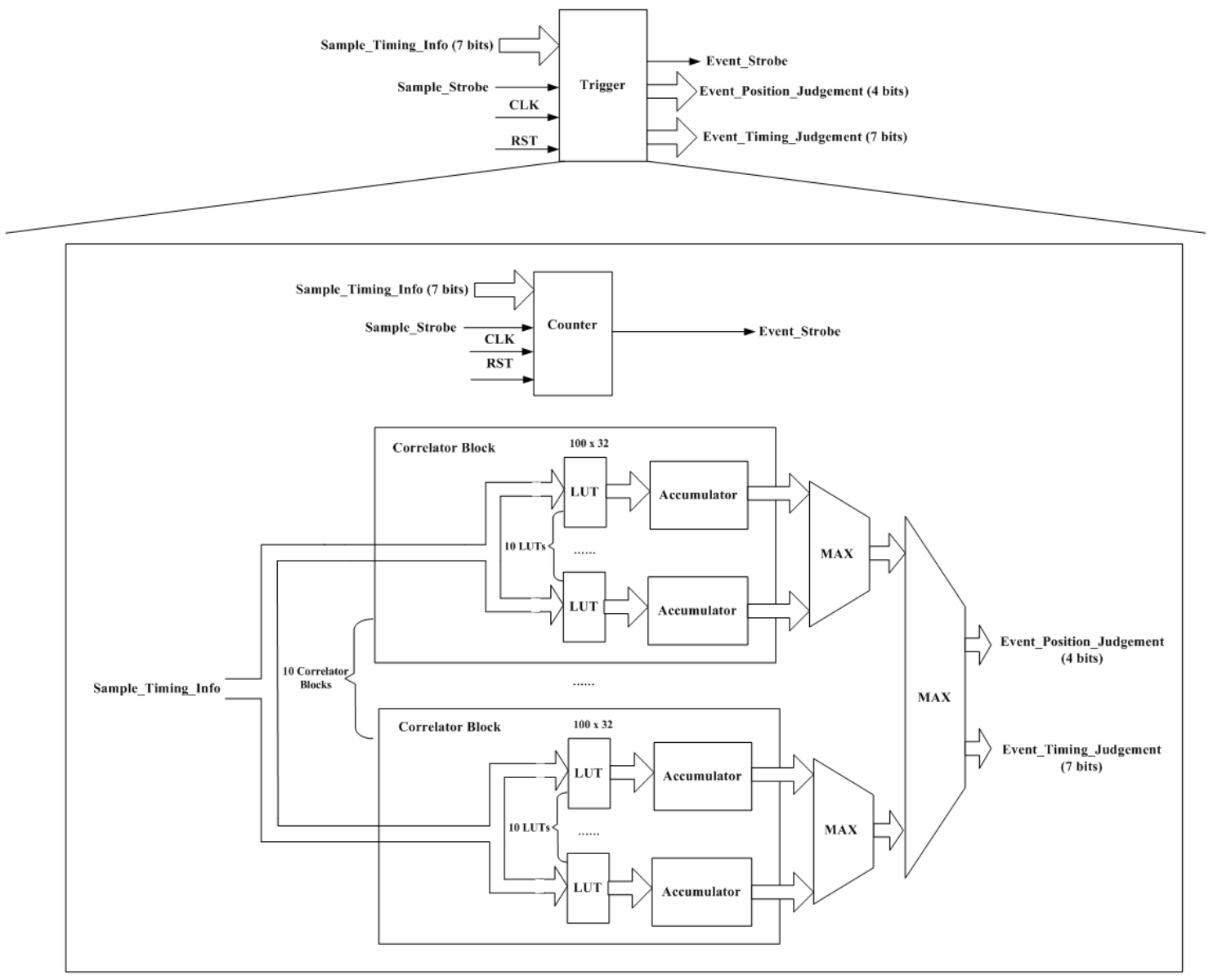

*Figure 12.22: Trigger: Detailed Block Diagram.*

- Sample_Strobe: Indicates whether Sample_Timing_Info is valid.
- Event_Strobe: Indicates detection of an event.
- Event_Position_Judgement: Position of the bar interaction.
- Event_Timing_Judgement: Time of interaction.

#### 12.4.4.2 Resource usage and timing performance

An estimate of the resources required to implement a completely parallel estimator is:

- Block Rams: $100 \times 32 \times 10 \times 10 = 320,000\,\text{bits}$
- Flip-Flops : $7 \times 10 \times 10 + 32 \times 10 \times 10 + 128 \times 32 + 128 \times 7 = 4992\,\text{bits}$
- Adders: 200 32-bit adders (comparators).

If Xilinx Virtex 4 series FPGAs are used, the design requires 2496 slices for the flip-flops and 3200 slices for the adders.
Even the smallest Virtex 4 is able to implement this design.
**Latency:** Since all the correlations are done simultaneously, the latency is only determined by the MAX operation which would need 7 clock cycles if implemented as a pipeline.
**Throughput:** The throughput is 1. This means the output speed is the same as the input speed of the samples.

## 12.5 KLM Trigger

In the Belle experiment, the KLM sub-trigger was mainly used for the logic to take $\mu$ pair events, which is necessary for the detector calibrations. Because the KLM sub-trigger was independent from the CDC sub-trigger, for example, it was very useful to measure the efficiencies of other elements of the trigger logic. For these reasons, the KLM sub-trigger in the Belle II experiment will be very useful. The trigger logic boards developed for other sub-trigger system, with their programmable FPGAs and high-speed serial links, can be adopted for the KLM sub-trigger. In fact, three-dimensional tracking of a muon may be possible in the KLM sub-trigger if enough information is read out. Such 3D tracking is expected to increase the trigger efficiency of very forward/backward low-multiplicity events such as $e^+e^- \to \mu^+\mu^-$ and $e^+e^- \to \tau^+\tau^-$ while being insensitive to backgrounds. The feasibility of the KLM sub-trigger is under discussion.

## 12.6 GDL

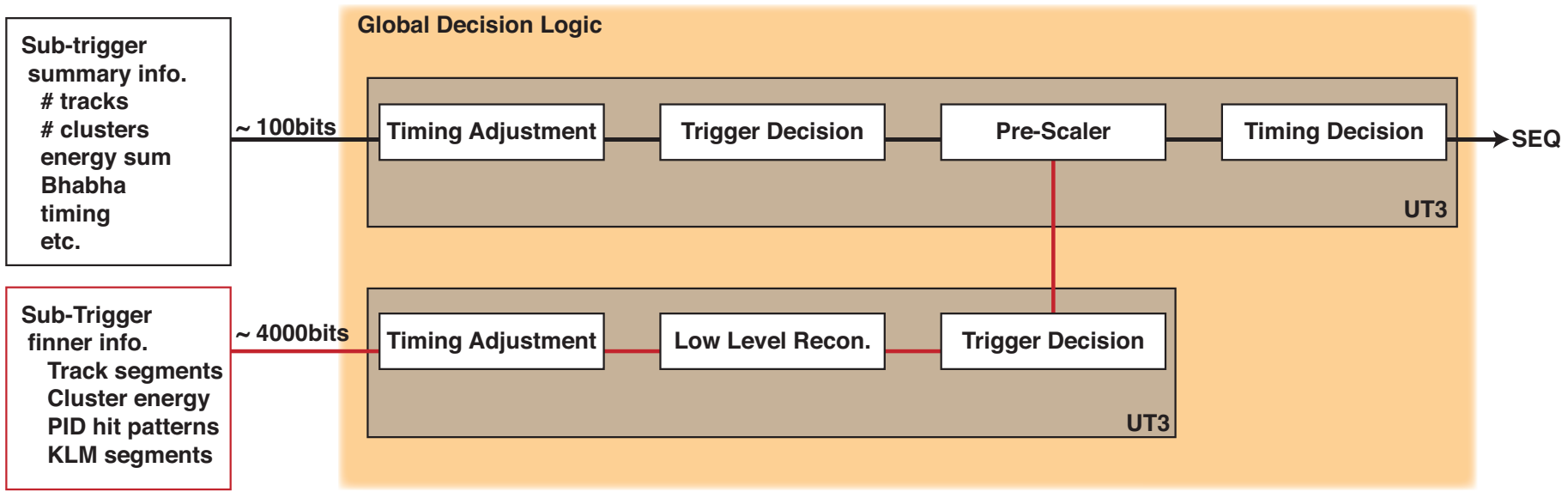

*Figure 12.23: Schematic view of GDL.*

The global decision logic (GDL) is the final arbiter of the Level 1 trigger decision. It receives all sub-trigger information, makes logical calculations and pre-scaling, and then issues the Level 1 trigger on appropriate timing. The GDL works as a pipeline so there must be no dead-time logically. We will use a 62.5 MHz (or higher) clock as the system clock. The schematic view of GDL is shown in Fig. 12.23. GDL receives two types of information:

0. summary information from sub-triggers ($\mathcal{O}(100)$ bits);
1. fine information from sub-triggers ($\mathcal{O}(4000)$ bits).

The summary information is used in the main logic that identifies events of physics interest and for calibration triggers. (These inputs are very similar to the sub-trigger summary information of the Belle experiment.) The fine sub-trigger information consists of, for example, the track segments (TS) in the CDC sub-trigger and the energy deposits (TC) in the ECL sub-trigger. This information is used for low-level reconstruction within the GDL: the association of a charged track with an ECL cluster, the identification of a neutral ECL cluster, etc. The detail of the low-level reconstruction is under simulation study now.
The GDL processes the sub-trigger information in four stages. In the first stage, the GDL applies a timing adjustment to compensate for the different latencies of each sub-trigger. In the second stage, the GDL makes the trigger decision. The following capabilities are under consideration.

0. Physics triggers
   - $\Upsilon(4S)$ and continuum: "three-track," "total energy," and "four-cluster"
   - $\tau$ pair: "two-track"

1. Calibration triggers
   - Bhabha : "bhabha"
   - $\gamma$-$\gamma$ : "gamma-gamma"
   - $\mu$ pair : "two-track"
   - Random trigger : "random"
2. Veto logic
   - beam injection : "inj"
   - two photon events : "two-photon"

The "three-track" ("two-track") logic is satisfied by the presence of more than two tracks (one track) in the CDC sub-trigger; the "total energy" logic is satisfied by an energy deposit of greater than 1 GeV in the ECL sub-trigger; the "four-cluster" logic is satisfied by the presence of more than three isolated clusters in the ECL sub-trigger. These comprise the main triggers for physics events. They are almost independent so that quite high efficiency for the physics events is expected. For $\tau$ pair events, we have only the "two-track" logic and the efficiency is limited because of the coverage of the CDC sub-trigger. (This coverage may be extended for muon tracks by the implementation of a 3D KLM sub-trigger.)
For the detector calibration purposes, the GDL must trigger cleanly on Bhabha events, $\gamma$-$\gamma$ events, and $\mu$ pair events; it must also provide a random trigger. The "bhabha" logic is formed using the ECL sub-trigger inputs; the "gamma-gamma" logic is formed by the "bhabha" logic without any matching tracks in the CDC sub-trigger. The "random" logic is a delayed "bhabha" trigger, to take events proportional to the luminosity. Events taken by the "random" trigger are used to record minimum-bias events that are overlapped with simulated physics events to mimic the beam-related background in the Monte Carlo simulation.
The "inj" signal is active during accelerator beam injection, when the beams are unstable and deliver high background to the Belle II detector. When active, "inj" vetoes all physics triggers. The "two-photon" logic vetoes two-photon events.
In the third stage, the GDL reduces the overall trigger rate by prescaling the calibration triggers. Bhabha events, for example, are necessary to measure the luminosity and to calibrate the ECL system. However, the cross section for this process is so high that the GDL would generate 100 or more Bhabha triggers for each hadronic-physics trigger in the absence of prescaling. To address this imbalance, the GDL issues only one "bhabha" trigger for every $N$ that it finds; $N$ is called the prescale factor. The GDL logic permits us to choose the prescale factor for each of the calibration triggers.
In the fourth and final stage, the GDL makes the trigger-timing decision using information from the BPID and ECL sub-triggers. We expect the best timing precision from the BPID sub-trigger (on the order of 1 ns) and somewhat worse resolution from the ECL sub-trigger (about 30 ns). The timing decision logic issues the Level 1 trigger after a fixed delay upon receipt of the BPID timing signal. In the absence of the BPID timing signal, this logic issues the Level 1 trigger after a fixed delay upon receipt of the ECL timing signal. If neither timing signal arrives, the GDL issues the Level 1 trigger after a fixed delay from the time that the trigger was generated.

# Chapter 13

# Data Acquisition System

## 13.1 Global Design

The goal of the data acquisition (DAQ) system is to read out detector signals upon the Level-1 (L1) trigger decision given by the trigger system. The system transfers the data from the front-end electronics through several steps of data processing, and finally to the storage system. The main components of the data flow are the unified data link called the Belle2Link, the common readout platform called COPPER, the event builder system, and the high level trigger (HLT) system.

Figure 13.1 shows the global design of the Belle II DAQ system. The detector front-end boards with digitizers are placed near or inside the detector structure and the digitized signals are transferred into COPPER systems through long optical fibers using the Belle2Link. A simple data reduction is performed on each front-end electronics board or on the receiver module of COPPER, while the data formatting and module-level event building is done on COPPER using the on-board CPU. The further event building and reduction are done on the readout PCs and the event builder, and finally processed by the HLT farms for the software event selection.

The Belle II DAQ system is designed with the following policy:

1. **Smooth transition from Belle**

   The existing Belle DAQ system already includes the key component, the COmmon Pipeline Platform for Electronics Readout (COPPER), which is designed to be used in the Belle II environment. The digitization system of Belle has been upgraded from a system based on FASTBUS TDC to a system based on COPPER modules. Experiences acquired during the transition will be indispensable at the construction stage. Keeping the COPPER system as the basis of the DAQ system certainly makes the transition to Belle II smoother.

2. **Unification**

   In the Belle DAQ system, the readout system for all the detector subsystems except the SVD is implemented using the charge-to-time (Q-to-T) conversion combined with a multi-hit FASTBUS TDC. This "unified" detector readout scheme has drastically reduced the development and maintenance cost in hardware, software and human resources. The DAQ design for Belle II inherits this concept. The use of COPPER is the baseline of this unification. In addition, the data link from the front-end electronics placed near the detector to COPPER, which we call the Belle2Link, is unified to use the same FPGA logic in the transmitter side, the same hardware, firmware and software for the receiver side with a common data transfer protocol. The transmitter side logic also includes the timing specific logic that enables the unified control over the entire system at the level of the frontend boards. The other unification is the software framework for the data flow, which runs on each COPPER, readout PC, event builder node, and high level trigger node with a common data transfer protocol. A new software framework, "roobasf" [1], is being developed for the common use in DAQ and offline. The "roobasf" is to be used at the all levels of DAQ nodes as well as offline.

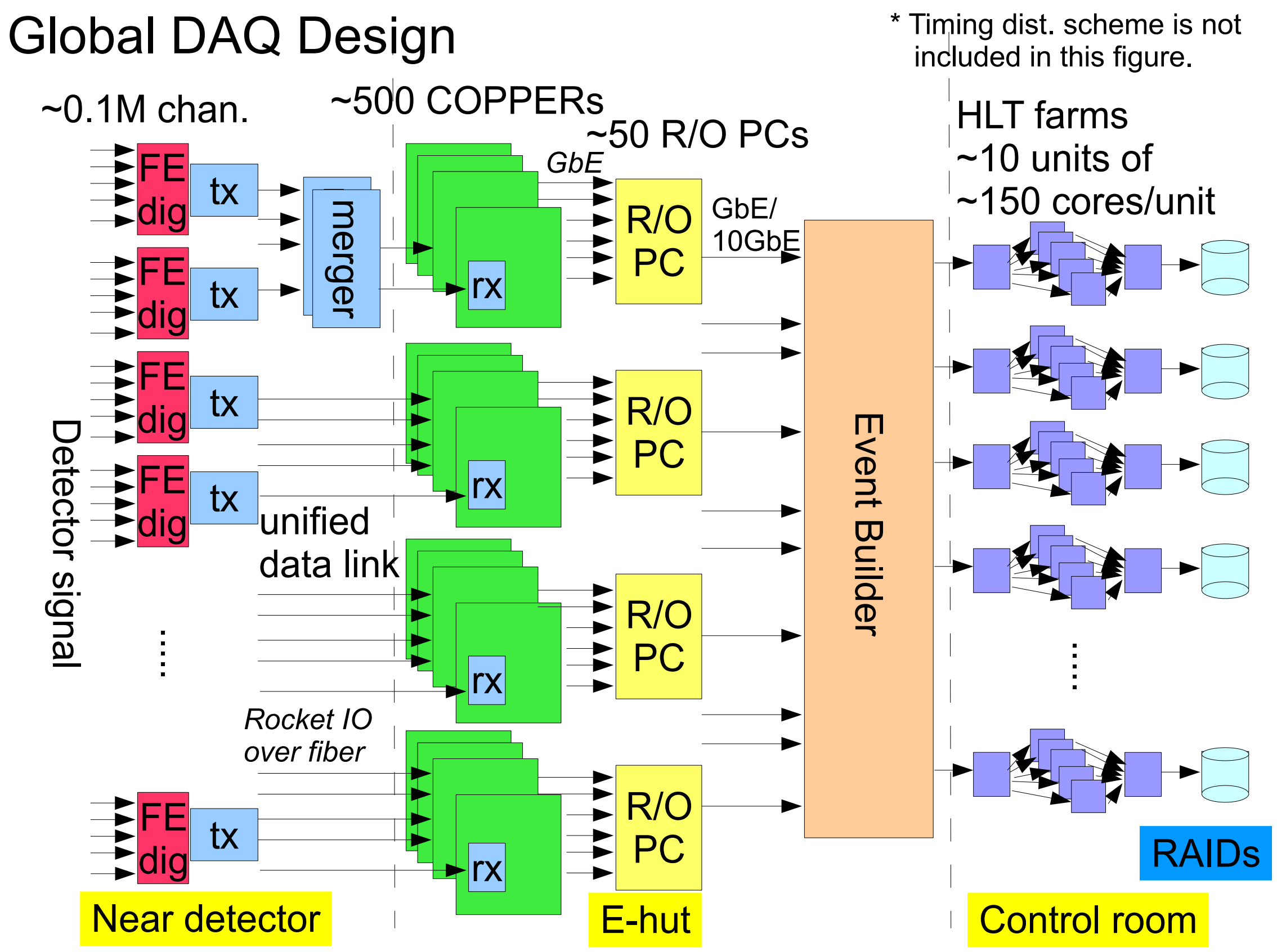

*Figure 13.1: Conceptual design of Belle II DAQ system.*

**3. Scalability**

The luminosity at the beginning of the Belle II experiment will be several times lower than the design, and it will take a few years to reach the design luminosity. On the other hand, the trigger rate may be at or near the design level from the beginning due to the initial vacuum condition. Therefore, the total bandwidth of Belle2Link may stay almost constant. Since it is hard to change later, the bandwidth of Belle2Link is designed to have a large margin over the full-luminosity expectation and is capable of handling an increase in the occupancy. However, even if the trigger rate is high at the beginning because of the high rate of junk background triggers, the required processing power for such events at the high level trigger is significantly less than those for multi-track $B$-decay events, and less CPU power is needed. It matches with the budget profile, in which the CPU power is added later when needed and when the CPU cost is drastically reduced. Therefore, the high level trigger and related down-stream components are designed to have a modular architecture so that the processing power is scalable by adding more units.

## 13.2 Requirements

Challenges in the Belle II DAQ system are the expected high trigger rate and the required processing time; either could lead to a high deadtime, corresponding to an effective waste of luminosity. There are constraints from the digitization hardware of state-of-the-art technologies for which some of the relevant parameters cannot be modified. Other requirements are related to the data size, which determines the necessary number of datalinks, size of the network, and processing power at the later stages.

### 13.2.1 Timing Requirements

We set our timing-related requirements mostly based on the constraints given by the SVD readout system. Here, we do not base our parameters on the PXD system, since it takes significantly longer to read out the signal and the signals from more than one trigger often overlap in the data stream. Also, there remain unsettled issues in the PXD data acquisition, including its maximum trigger rate capability, handling of overlapped triggers, and the techniques to reduce the data bandwidth.
The nominal average L1 trigger rate is expected to be 20 kHz at the instantaneous luminosity of $8 \times 10^{35}\,\mathrm{cm}^{-2}\mathrm{s}^{-1}$. We set the design average trigger rate of the data acquisition system to be 30 kHz, which has a 50% margin.
Little DAQ deadtime should occur while processing all of the triggers in the pipeline from the frontend to the backend subsystems. In reality, there are limitations due to buffer sizes and data-moving sequences, and the actual deadtime at the very frontend is determined by four parameters: the average trigger rate, minimum interval between two triggers, number of triggers that can be processed in the pipeline, and the latency of the readout system. The most severe constraints comes from the SVD readout system, which handles multiple triggers in the pipeline subject to the following constraints:

- Minimum interval between two triggers — 190 ns (6 clocks in 31 MHz)
- Number of triggers in pipeline — 5
- Latency of the readout system — 26.6 $\mu$s (840 clocks in 31 MHz).

With these conditions, the expected deadtime fraction is calculated to be 3.4% for the 30 kHz trigger rate that follows a Poisson distribution, or a much higher deadtime fraction for a higher trigger rate. This is the part of the reason for the choice of 30 kHz trigger rate for the design.
There is an additional and irreducible deadtime fraction in order to mask the background triggers due to the continuous beam injection. The trigger is masked for at least 150 $\mu$s after the injection over the entire ring and a longer window of a few ms after the injected beam bunch (Sec. 9.4.4). At the anticipated beam injection rate of 50 Hz, this adds a few more percent to the deadtime fraction.

### 13.2.2 Datalink and Data Size Requirements

The data size requirements are based on requests from subdetectors that are summarized in Table 13.1. The frontend datalink called Belle2Link bundles multiple channels to reduce the number of lines between the detector and backend DAQ system to a manageable level, and is operated at a much lower rate than the capable maximum bandwidth. The number of datalinks

*Table 13.1: Estimated average occupancy and data size and required number of subcomponents such as the number of Belle2Link for data transfer and number of COPPER modules. In addition to the listed subdetectors, trigger information is also planned to be read out in a similar way.*

| | #ch | occ. [%] | #link | /link [B/s] | #COPPER | ch size [B] | ev size [B] | total [B/s] | /COPPER [B/s] |
|---|---|---|---|---|---|---|---|---|---|
| PXD | 8M | 1 | 40 | 182M | — | 4 | 320k | 7.2G | — |
| SVD | 243456 | 1.9 | 80 | 6.9M | 80 | 4 | 18.5k | 555M | 6.9M |
| CDC | 15104 | 10 | 300 | 0.6M | 75 | 4 | 6k | 175M | 2.3M |
| BPID | 8192 | 2.5 | 128 | 7.5M | 8 | 16 | 4k | 120M | 15M |
| EPID | 77760 | 1.3 | 138 | 0.87M | 35 | 0.5 | 4k | 120M | 15M |
| ECL | 8736 | 33 | 52 | 7.7M | 13 | 4 | 12k | 360M | 30M |
| BKLM | 21696 | 1 | 86 | 9.7M | 6 | 8 | 2K | 60M | 10M |
| EKLM | 16800 | 2 | 66 | 19.5M | 5 | 4 | 1.4k | 42M | 8.4M |

is mostly determined by the hardware configuration, while the bandwidth in the later stage (except PXD, handled separately) is arranged according to the bandwidth.
The data size requirements are based on the limitations of the pipeline buffers as well as the requests from subdetectors. Here, we need three numbers: the average size, the maximum size during the nominal data taking, and the absolute maximum size. The average size determines the design of the overall structure and the size of the system, while the nominal maximum size determines the allocation of buffers. At this moment,

- Maximum data size per event per datalink — 16 kbyte

is requested for subdetectors. This is due to the buffer size limitation in Belle2Link.
The data size in Table 13.1 still has a large uncertainties which comes from the estimation of the detector occupancy. For example, the data size of PXD in the table is estimated by assuming an occupancy of 1%, however, it can easily be doubled depending on the background situation. In our DAQ design, therefore, we take a safety factor of 2 and assume the data size from PXD to be 1MB, while 100kB as a total size from other detectors.

## 13.3 Timing distribution and handshake

The main task of the timing distribution system is to distribute the reduced RF clock for the system-wide synchronization, the trigger signal, and several fast control signals to the entire data acquisition system including frontend readout boards and the COPPER readout system in a unified way. In the FPGA of readout boards, there is a unified logic core that receives and handles these signals. As some of the readout boards are located inside the detector, the longest path will be 30 to 40 m from the central timing source to the destination. By placing a middle station on or near the detector, the longest single-cable path for signal transmission is halved.
The clock signal is distributed as a 127 MHz clock, which is one-quarter of the RF clock. The high precision clock source is provided by the barrel PID system, which requires the smallest jitter clock. After distribution, the necessary clock speeds for readout are generated inside the FPGA. The clock speeds requested by subdetectors include 127, 101, 63, 42, 32, and 2 MHz (the RF clock divided by 4, 5, 8, 12, 16, and 256, respectively), and they are synchronized to the revolution signal of the accelerator.

The trigger signal has to be blocked when the readout systems cannot handle it due to their limitations as described in the requirement section, i.e., two trigger signals should not be issued with an interval less than 190 ns and no more than 5 triggers should be delivered to be processed in the pipeline at the same time. The first condition is always fulfilled due to the restriction in the L1 trigger system. For the second condition, the system has to know when the processing for the first trigger has completed and is ready to issue the sixth trigger (or stop blocking the trigger). The method proposed here is based on following three key conditions:

1. The end of processing of an event by every readout system is notified to the timing system, and this information is quickly collected by a central trigger distribution system.
2. The data is sent through the datalink to the COPPER with no flow control (i.e., no back pressure towards the front-end side).
3. The buffer-full status at the COPPER is notified to the timing system to block the trigger. This buffer can become full when the COPPER CPU is busy. Since it is computer driven from this point on, the latency of processing is unpredictable.

To distribute and collect necessary signals, we use a small-latency serial-link. This serial-link is not required to have the high bandwidth of the Belle2Link. We do not require a clock recovery feature, which is usually one of the latency sources for serial links, since we distribute a clock signal to all the nodes of the distribution tree. To maintain flexibility and compactness, serialization and deserialization are performed inside the FPGA. One limitation is that one serial data sequence has to be shorter than the 200-ns trigger interval. The actual trigger timing, which is synchronized to the 127 MHz clock at the L1 trigger system, is recovered from the phase information embedded in the serial data.

Currently, 32-bit word serialized to the line rate of 254 Mbps (length of 126 ns), or 24-bit word encoded with 8b10b and serialized to the line rate of 159 Mbps (length of 189 ns) is under consideration. Encoding such as 8b10b or another homemade one has to be made to guarantee a sufficient number of polarity flips for the AC coupling connection, the bit synchronization in the word, and the insertion of control words in the stream.

The clock and serial-link signal is distributed over an off-the-shelf CAT7 LAN cable, directly driven by the LVDS driver of the FPGA. Various types and lengths of LAN cable have been tested for signal integrity. We found that a 15-m flat LAN cable or a 30-m thick CAT7 cable can be used without inserting additional driver or receiver chips in front of the FPGA. The clock jitter increases up to about 50 ps, but this could be reduced to 20 ps using the built-in PLL circuit of the FPGA. The maximum serial link rate is based on a test that successfully established a serial data link at 254 Mbps over a 15-m cable using the serializer and deserializer function of the Xilinx Virtex5 FPGA. It is confirmed that an additional equalizer circuit is not necessary for our requirements.

In the single LAN cable, there will be a distributed clock signal, a distributed serial-link signal in which the trigger timing information is embedded, and a collecting serial-link signal. The distributed signals can be fanned out without encoding and decoding, and this will be the operation mode during a run. The same line can be configured to decode and encode at every step and, by doing so, one can use the line for the node-by-node control. The encoding patterns will be chosen so that the trigger timing signal and the response signal can be decoded within one clock cycle to reduce the latency.

The clock and trigger signal distribution is made by a FTSW (Frontend-Timing-SWitch) module shown in Fig. 13.2, which has 24 RJ-45 ports; one port is used as input and 20 ports as output.

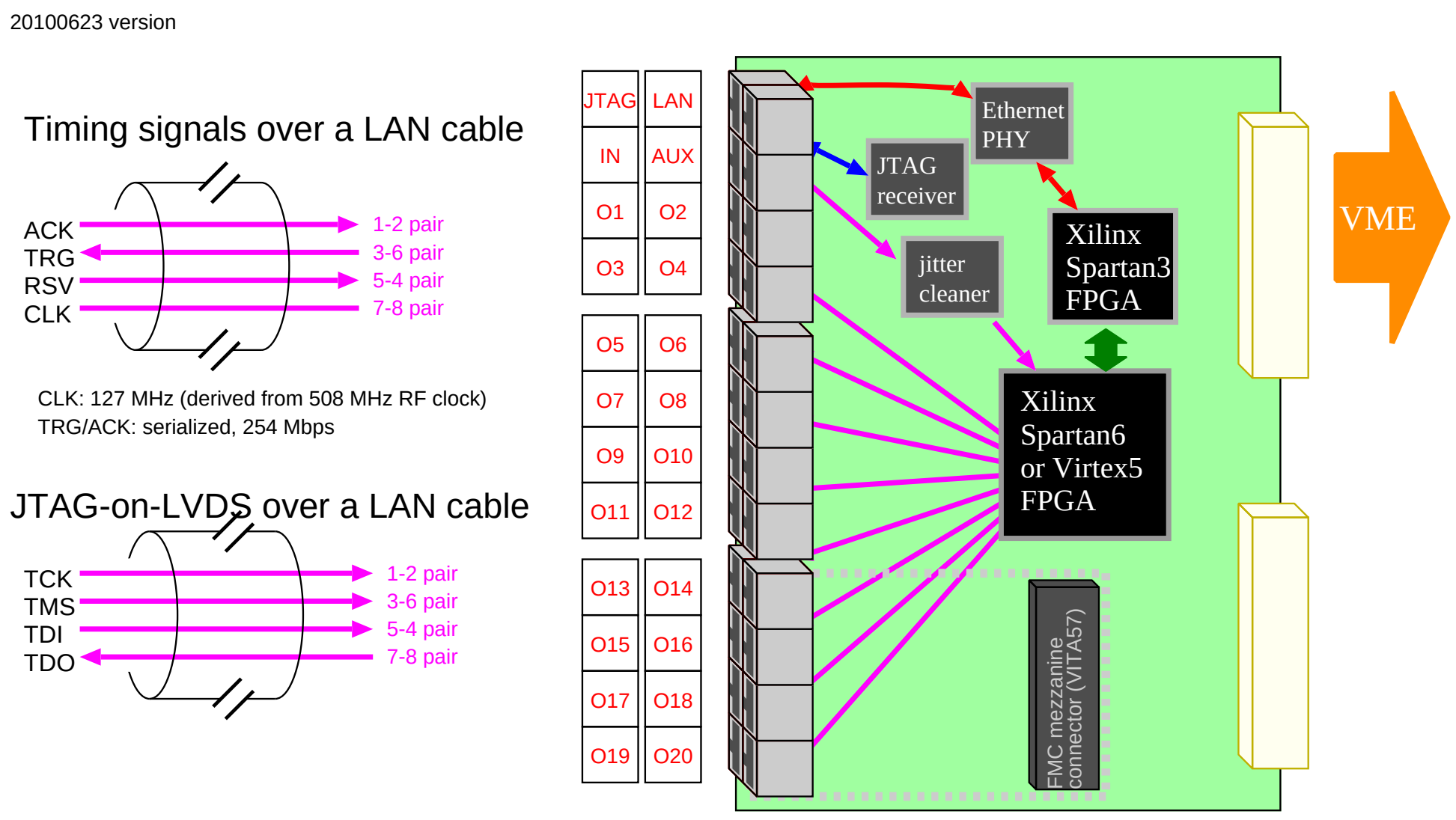

*Figure 13.2: LAN cable usage and the Frontend Timing Switch module.*

This module is a double-width 6U VME board, and is also designed for standalone operation without a VME crate. Cascading this module three times will cover all the readout boards.
It is being discussed whether one can drive the Belle2Link using the distributed reduced RF clock, for simplicity and to avoid extra noise source inside the detector. The quality of the recovered clock is at least within the specification of the Xilinx RocketIO.

## 13.4 Belle2Link

Belle II DAQ system uses the RocketIO GTP technology over optical fibers for the data transmission between front-end electronics (FEE) and the backend system. GTP is a cost-effective version of the RocketIO which can handle up to 3.125 Gbps line rate on LXT series of the Xilinx Virtex5, Virtex6 and Spartan6 FPGAs and can also intercommunicate with older versions of RocketIO in Virtex II pro and Virtex4 FPGAs. For simplicity of development of the front-end systems and reliability of the operation, an unified high speed link which we call the Belle2Link has been defined for use in all connections between each of the subsystems. The Belle2Link provides

1. unification of the hardware design,
2. unification of the firmware design,
3. electrical isolation,
4. high speed data transmission,
5. versatileness for different input data rates, and
6. home brew transmission protocol.

The Belle2Link comprises the high speed transmission lines based on the RocketIO technology over optical fibers and various interfaces to the FEE(IF_FE), to the trigger timing distribution system (IF_TTC) and to the FIFO buffer of COPPER (IF_CO) (Fig. 13.3). The Belle2Link accepts data in FIFO_FE via IF_FE from FEE for transmission and receives trigger and timing signals from timing distribution system via IF_TTC in the sender part (BLSNDR), and at the receiver part (BLRSVR) it saves the received data in the FIFO_BL and moves data after necessary processing to the FIFO_CO in IF_CO which is on the COPPER platform (Fig. 13.4). As the unified link for all sub-detector systems, Belle2Link provides a single type (or at most two different types) of the receiver FINESSE boards and all the unified firmware for sender part and receiver part.

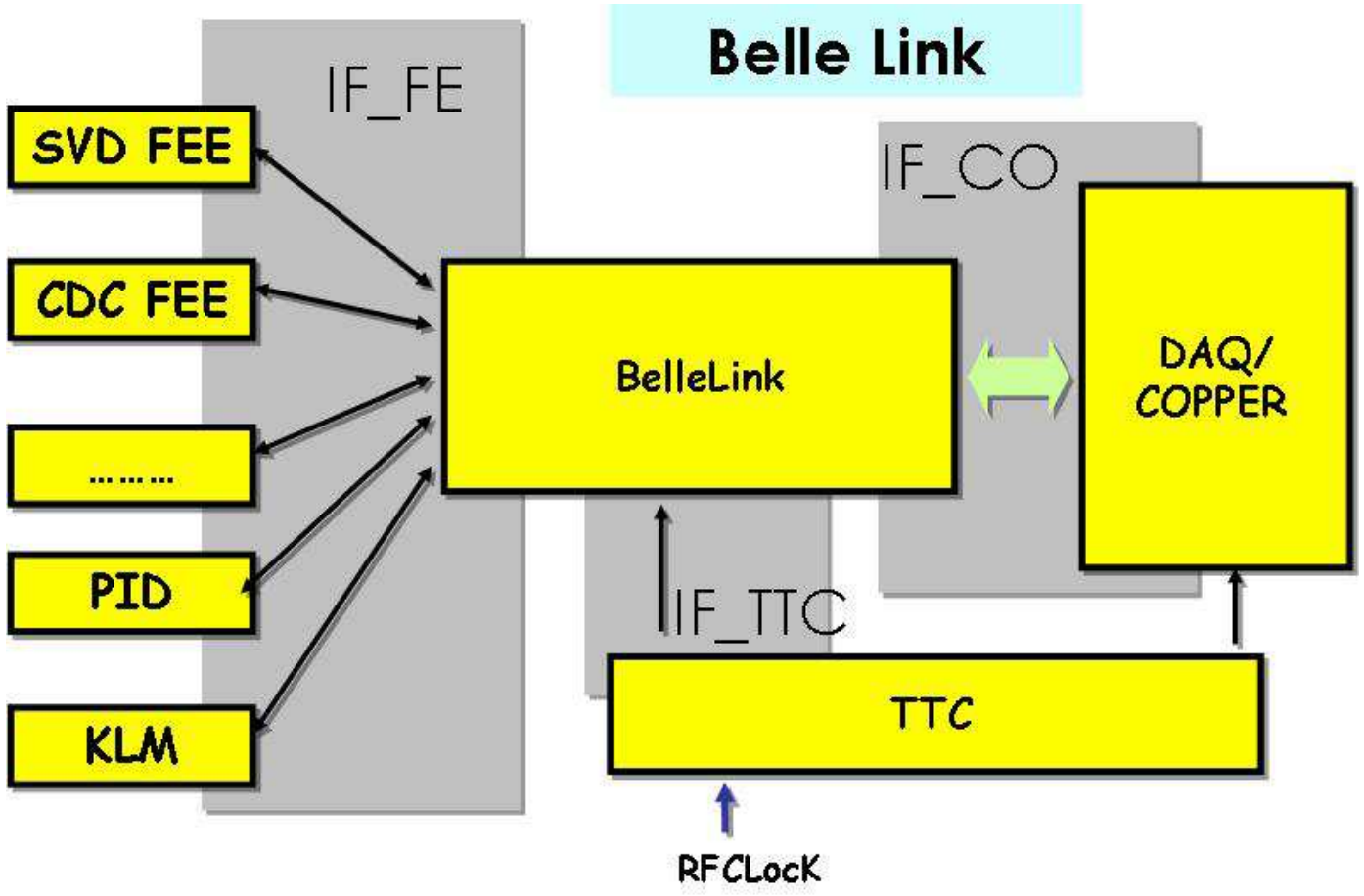

*Figure 13.3: Belle2Link connecting front-end electronics and back-end COPPER based DAQ system.*

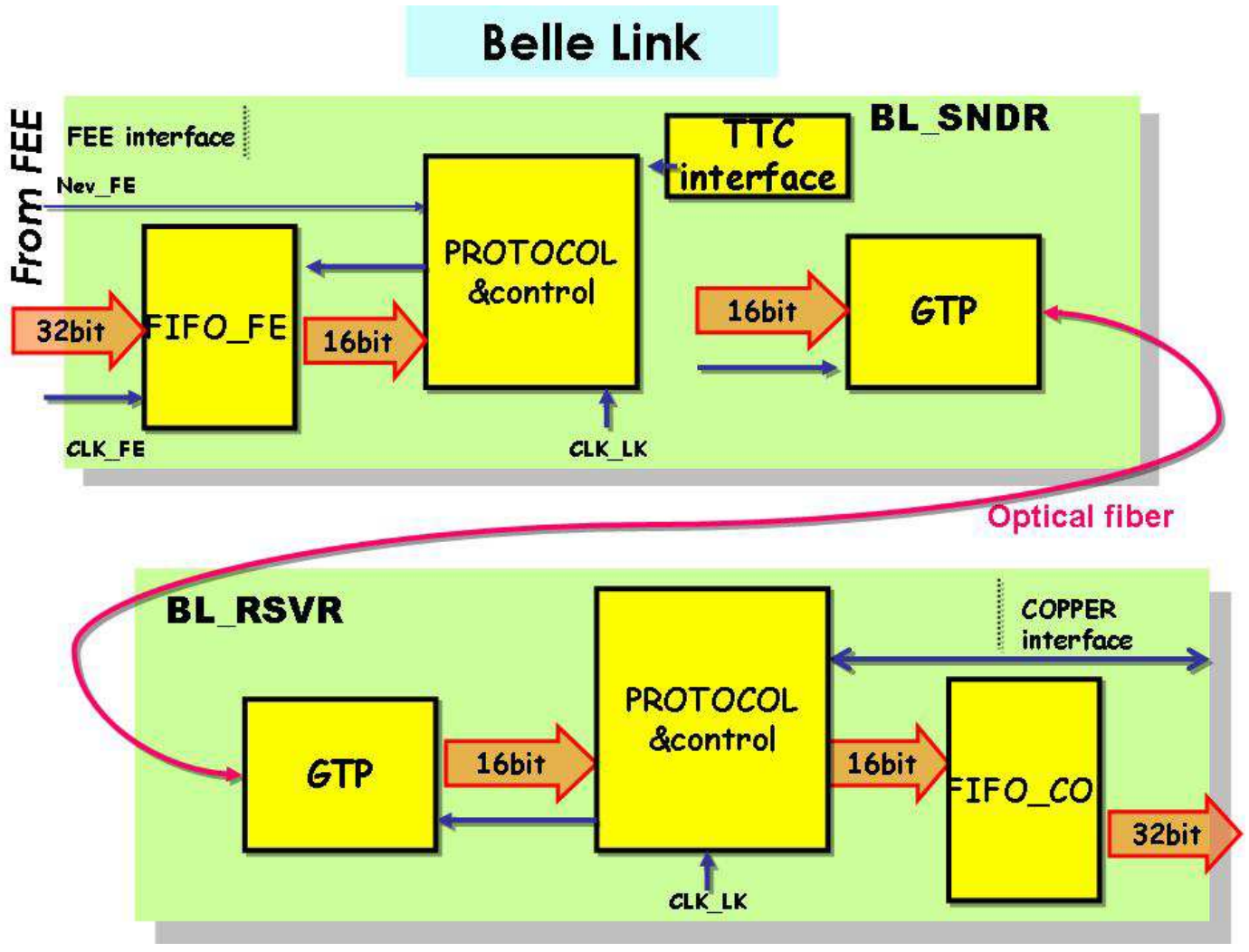

*Figure 13.4: Block diagram of the Belle2Link .*

### 13.4.1 Unification in hardware

The BLSNDR is implemented in a Virtex-5 FPGA in the FEE readout boards and shares its resources. The FEE readout of each sub detector systems provide the FPGA and the BLSNDR part of the unified Belle2Link firmware will be integrated with the readout firmware. When multiple readout channels have to be combined before sending out through the Belle2Link, necessary signal mergin has to be performed before the BLSNDR. A unique frequency will be chosen for the high precision clock CLK_LK for the RocketIO transmission for all Belle2Link lines. This CLK_LK is preliminarily chosen to be 200MHz for a line rate of 2Gbps and be produced by a high precision crystal for the prototype of the FEE board and FINESSE receiver board (BLRB). BLRD board is a FINESSE board which fits the general requirements of high speed transmission and also some data processing for system like B-PID and electrically and mechanically fit to the COPPER board.

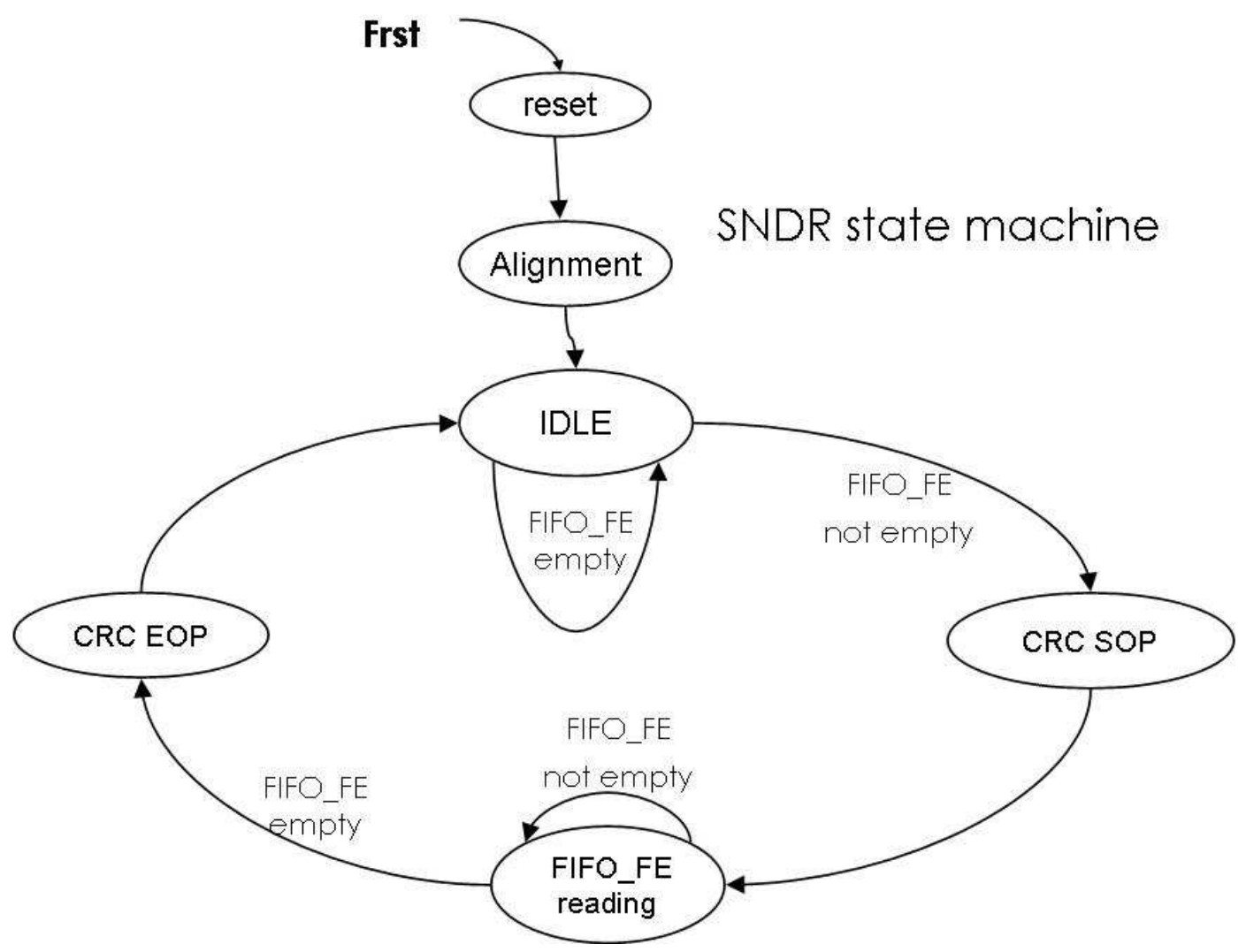

*Figure 13.5: The state machine of the sender part.*

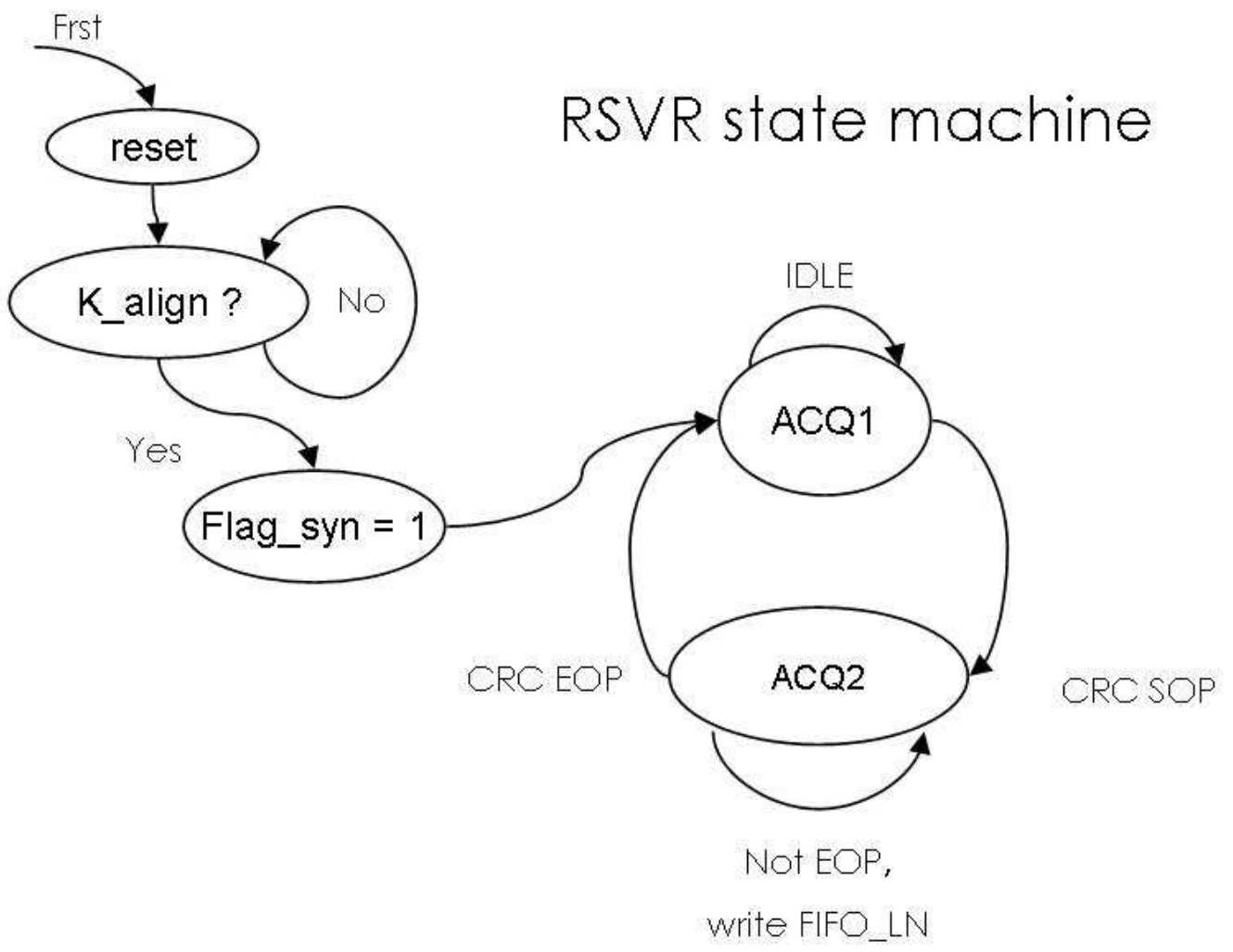

*Figure 13.6: The state machine of the receiver part.*

### 13.4.2 Unification in firmware and protocol

The circuitry of the Belle2Link will be implemented by firmware in the Xilinx FPGA. The development of the link will be based on Xilinx ISE 11 development software. The firmware for BLSNDR will be unified and provided to all system for integration to the system. The firmware for BLRSVR will also be unified and can be online configured for use in system like in B-PID. The working status can be checked by registers and displayed on some LED on the front panel. The interface with FEE is a FIFO (FIFO_FE). The FIFO_FE is a 32bit write and 16bit read FIFO and will be filled with data by FEE readout under control of clock CLK_FE and then be moved to GTP under sender clock CLK_LN after adding some control and error checking data by BLSNDR state machine (Fig. 13.5).
In the BLRSVR, another state machine checks for the GTP line (Fig. 13.6). The data on the line will be first received and stored in a FIFO_LN under control of clock CLK_LN. If it is error free this data will be moved to a FIFO_LN (N_EV will be incremented). If there is an error occurred in the transmission, an error flag is added in the header part of data frame for later use in the online data monitoring. In most of the systems the data in FIFO_LN will be direct moved to another fifo FIFO_WR, but for systems like B-PID the date must be first moved to a pipeline(Fig. 13.7), and part of the data in a time window will be fetch after receives a trigger and be processed accordingly, then be stored in FIFO_WR in the data form showed in table 13.9. Special error bits are allocated to indicate if a data transmission error occurred.
Another state machine (Fig. 13.8) checks the NEV_WR, if there is an event in the FIFO_WR, then send the data for this event to FIFO_CO under control of clock CLK_CO and decrease the NEV_WR if the COPPER is ready to read. This condition should not occur since the almost full status of FIFO_CO (FFUL signal) is asserted and received by the timing receiver (TT-RX) that is attached on the COPPER in order to block further triggers when there is still room to receive predefined number of all events in the trigger distribution and readout chain.
After FFUL is asserted, the number of event NEV_WR is counted and if it exceeds the predefined number a BUSY signal is generated to TT-RX. The BUSY signal is removed two CLK_CO clocks after FFUL is cleared. The error checking scheme will be kept for the online checking of the transmission between Belle2Link and COPPER.

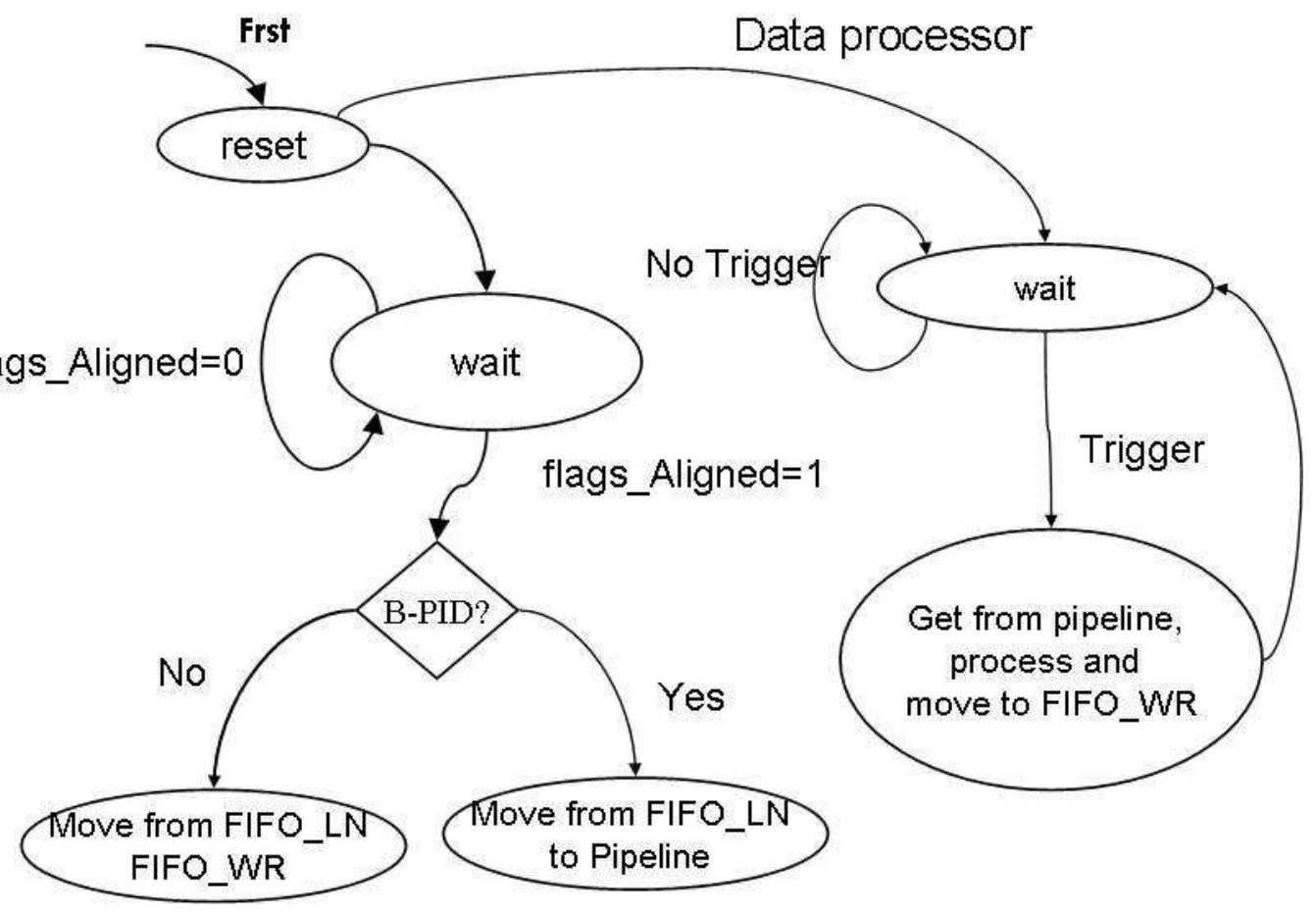

*Figure 13.7: The state machine of the data processor.*

### 13.4.3 Clocks

Three key clocks will be needed by Belle2Link, CLK_FE which is converted from central trigger and timing distribution, CLK_LK for the high speed transmission which is a high precision clock from a crystal on the board, and CLK_CO received from TT-RX via COPPER connector. The clocks for data transmission on the RocketIO will be composed by the DCS in Virtex-5 core. For the prototype system, we have used a CLK_LK of 200MHz and successfully operated at a line rate of 2Gbps. This clock can be changed to a 125 MHz (or other frequency up to 156.25 MHz) for a line rate of 2.5 Gbps (or up to 3.125 Gbps) which will be further studied.

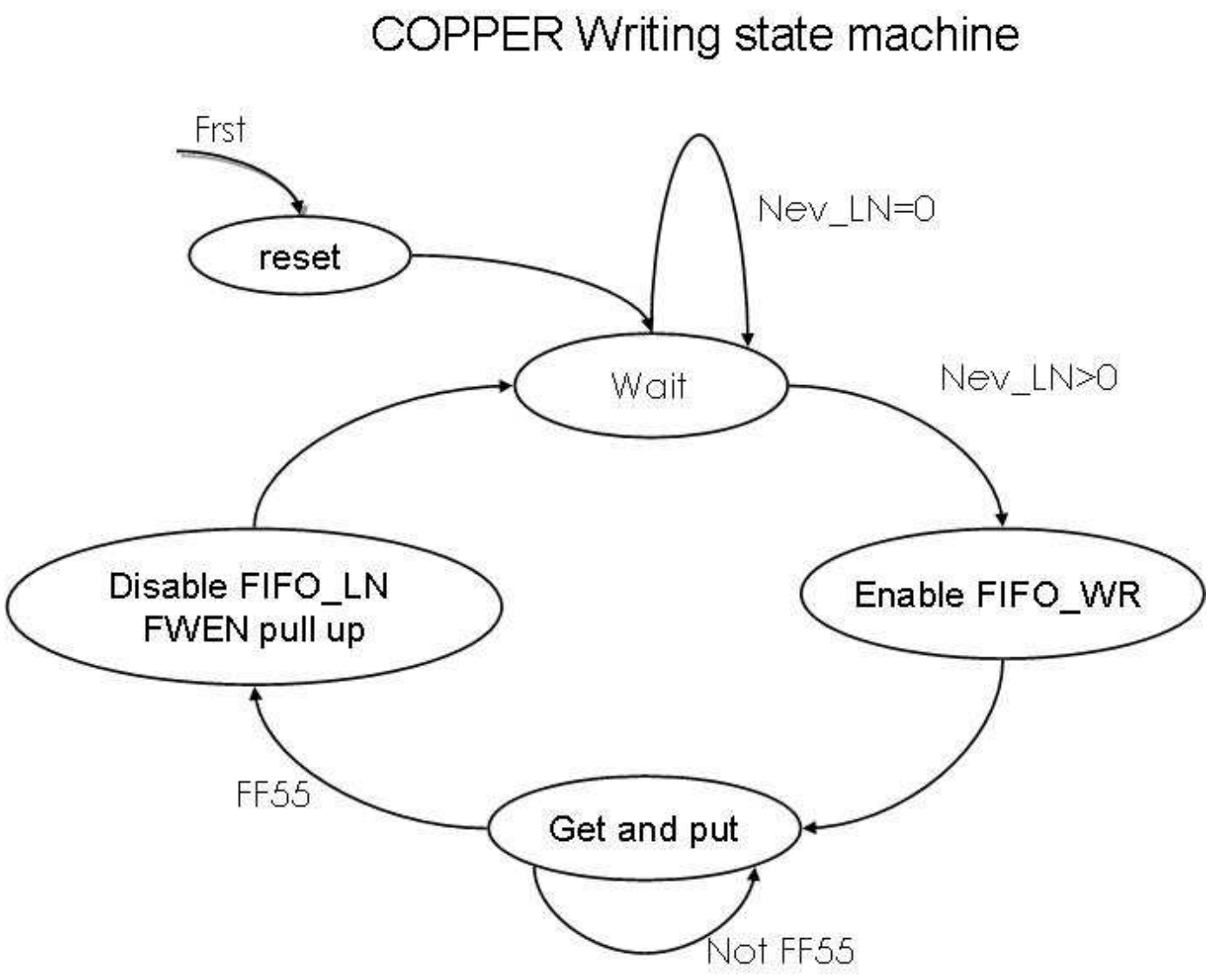

*Figure 13.8: The state machine of the data writing to COPPER.*

| | | |
|---|---|---|
| OxFFAA (16) | RESERV | ERR(2) |
| *FINESSE event count* (24) | | *TTRX-ev* (8) |
| *Data #0* (32) | | |
| *Data #1* (32) | | |
| ...... | | |
| *Data #n* (32) | | |
| OxFF55 (16) | Check-sum (16) | |

*Figure 13.9: Data format to be written to COPPER.*

### 13.4.4 Further studies

The protocol will include bidirectional communication for online slow control command and data transmission. This is needed since the Belle2Link is the only connection for example to set the parameters of the frontend readout board.

## 13.5 Detector front-end interface

The transmission side of the unified Belle2Link is implemented in an FPGA that is used mainly for subdetector-specific purposes. To share the same FPGA, we define the requirements for the board for the datalink purpose as given in Fig. 13.10. Implementing the datalink part into a separate chip or another daughter card may help to clearly define the interface, but it is unacceptable in terms of space for most, if not all, of the subdetector systems.

First, we define the FPGA type to be a Xilinx Virtex5 FPGA with a RocketIO GTP transceiver (LXT series). We are also investigating the possibility to use the less expensive Spartan6 FPGA.

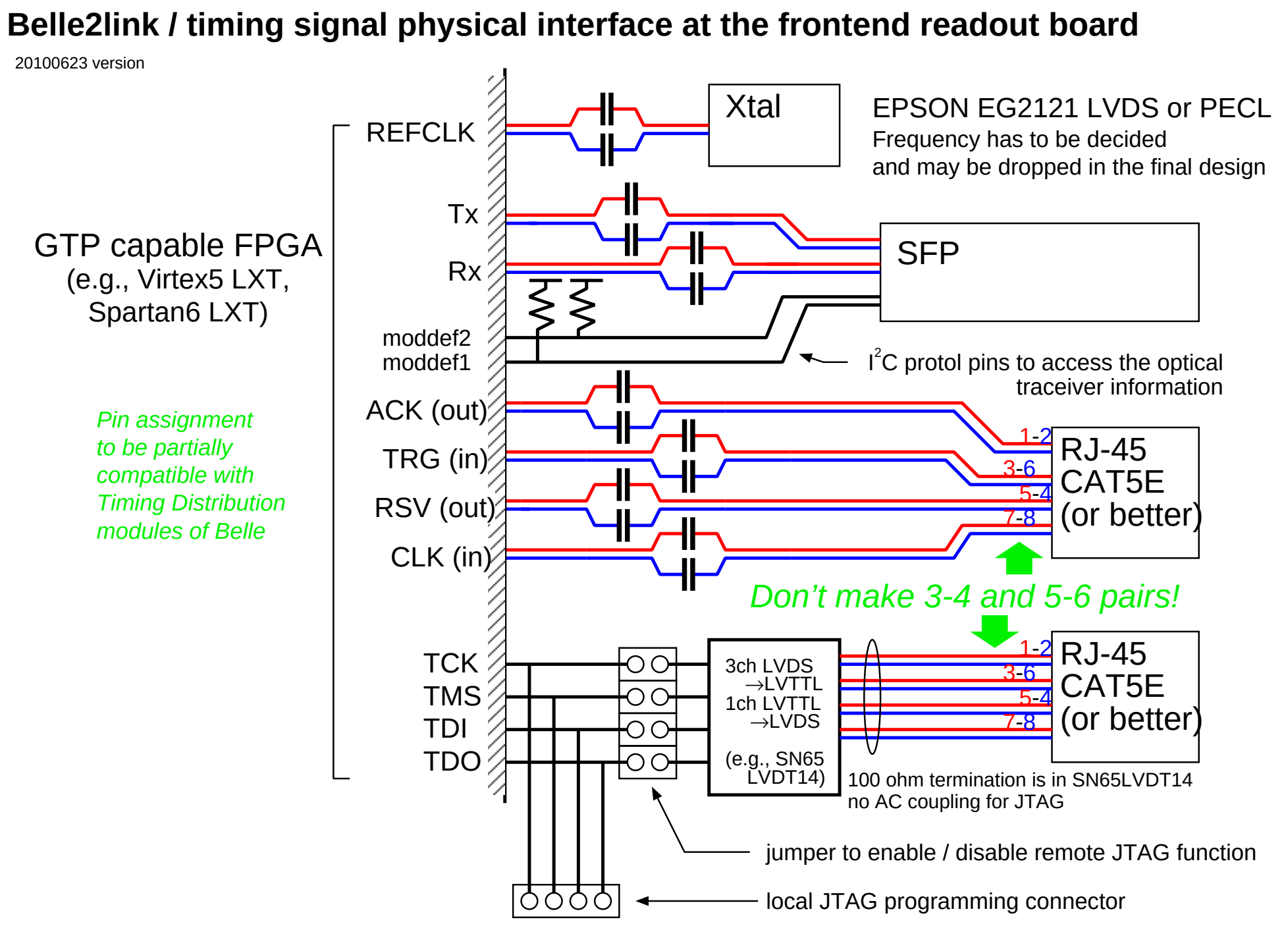

*Figure 13.10: Definition of the Belle2Link and timing signal interface at the frontend.*

Second, we define the necessary connectors. For the GTP datalink, a SFP connector and a high quality Xtal oscillator are placed on the board and appropriately connected to the FPGA. To receive the timing signal, we use an RJ-45 connector with the LAN cable pair definition (1–2, 3–6, 5–4 and 7–8 pins are paired).

In addition to these two connectors, we are investigating the possibility of having a unified JTAG extension connector and a set of necessary driver and receiver chips to permit remote reprogramming of the FPGA. The main reason is that we worry about the stability of a flash memory as the firmware storage device in the radiation-hard environment.

For the unified datalink purpose, two firmware cores, one for the datalink and the other for timing-signal handling, will be included in the FPGA. The specification of the set of the signals from and to these cores will be similar to the set of signals between FINESSE and COPPER, but they are yet to be defined. For the timing signal, the reduced RF clock(s) of the requested frequencies, trigger timing signal, trigger tag, and reset signal at the run start will be provided. For the data transfer, the data bus, the writing clock and other functions are still undefined. It has been requested that the many parameters be downloadable from COPPER through the datalink to the FPGA, e.g., to set the threshold or gain of the digitization circuits, and a set of small local data and address line for such registers as is available for FINESSE and COPPER would be useful.

## 13.6 COPPER Readout System

To reduce the total cost of development and maintenance of the readout system, we designed the system based on COPPER (COmmon Pipelined Platform for Electronics Readout) [2]. It is a general purpose pipelined readout platform developed at KEK and is designed so that the detector interface is mounted on the platform as daughter cards called FINESSE. The Belle2link receiver (BLRSVR) is equipped as a FINESSE card to receive the data from the front-end via optical cables. Up to four FINESSE cards can be mounted on a COPPER board.
A commercial CPU card (PrPMC) is also mounted on the platform and the data from FINESSE cards are fed through the PCI bus. The data formatting and reduction are performed by the CPU and the output are then sent to the event builder through the FastEthernet connection.

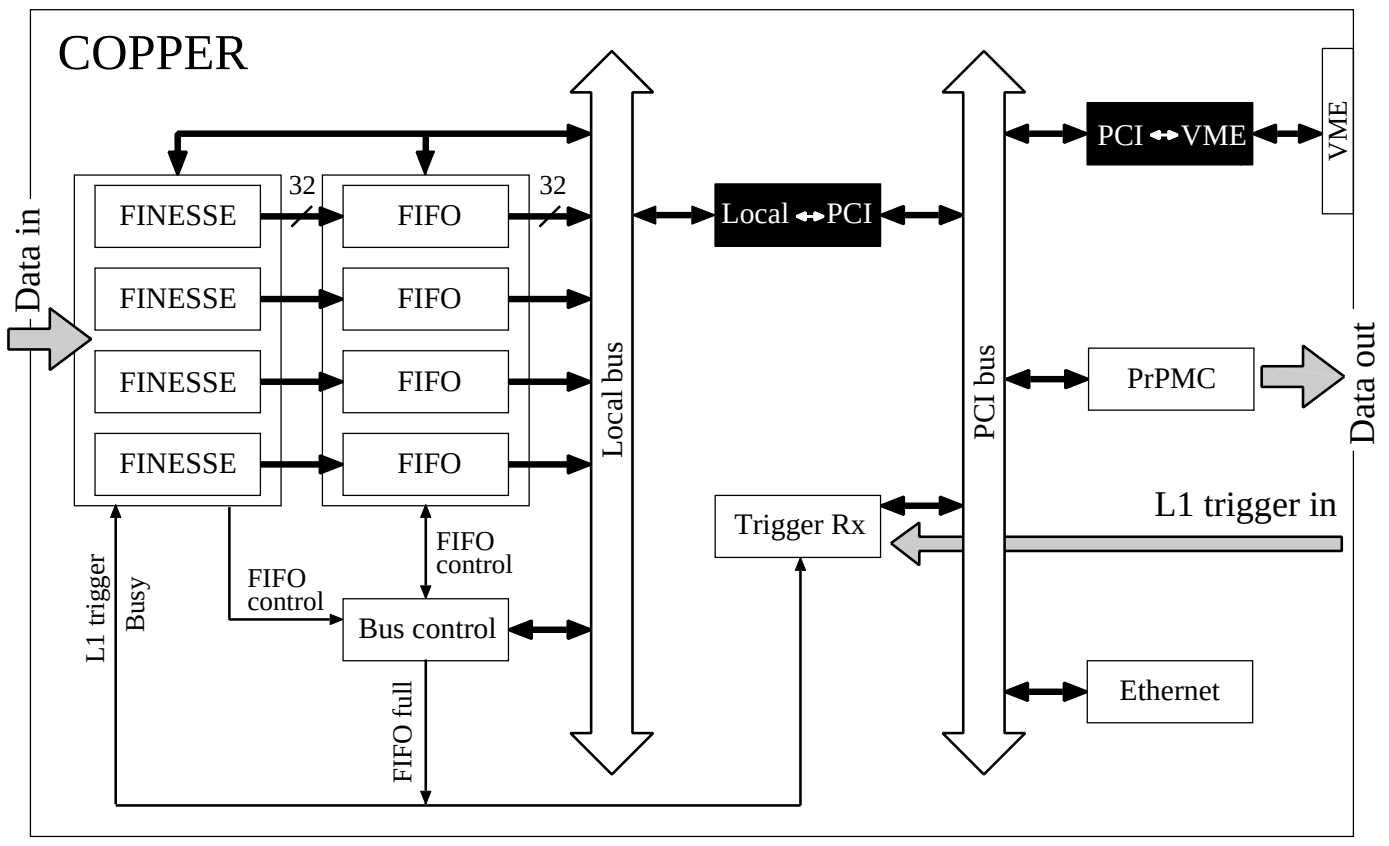

*Figure 13.11: Block diagram of the COPPER board, with four FINESSE cards indicated.*

### 13.6.1 Data flow in the COPPER readout system

Figure 13.11 shows a block diagram of the COPPER readout platform together with the FINESSE interface cards. A typical data stream from the detector to the event building PC, through the FINESSE cards and the COPPER board, is as follows.
The detector signals received by Belle2link receiver are fed into four 1024-kB FIFOs corresponding to each FINESSE card on the COPPER board in 32-bit data width. The data transfer to the FIFO is performed asynchronously with the L1 trigger timing.
The FIFOs are connected to the 32-bit local bus on the COPPER board. The event data size in each FIFO is counted by the local-bus controller. Once the data size exceeds a predefined threshold (typically a few hundred events), a PCI interrupt is issued by the local-bus controller to a CPU card through a local-to-PCI bus bridge to indicate the arrival of data in the FIFOs. The data stored in the FIFOs are then transferred to the memory on the CPU card by the chained direct-memory-access (DMA), initiated by a local-to-PCI bus bridge. During the DMA procedure, the data from four FIFOs are combined and formed as event records. The data formatting and data size reduction/feature extraction are performed on the CPU card using the unified software framework. Finally, the formatted data are sent to the readout PC through a 100 Mbps or 1000 Mbps ethernet link.

### 13.6.2 FINESSE

The FINESSE is a $76.0 \times 186.0\,\mathrm{mm}^2$ daughter card [Fig. 13.12(top)] that is mounted on a COPPER board. Up to four FINESSE cards are mounted in this way, with a 2-mm gap between neighboring FINESSE cards. The height of the FINESSE card is required to be less than 10 mm, as indicated in Fig. 13.12(bottom).
The FINESSE card and the COPPER board are connected with 116 pins on three connectors. Table 13.2 lists the connector pin assignments. The data drain from the FINESSE card to the

*Table 13.2: The list of the connector pin assignments between the FINESSE card and the COPPER board.*

| **pin#** | **row-A** | **row-B** | **row-C** |
|---|---|---|---|
| 1 | FF[00] | −3.3V | IRSTB |
| 2 | FF[01] | −3.3V | IENA |
| 3 | FF[02] | −3.3V | IO2 |
| 4 | FF[03] | −3.3V | TYP[0] |
| 5 | FF[04] | −3.3V | TYP[1] |
| 6 | FF[05] | −3.3V | TYP[2] |
| 7 | FF[06] | GND | TYP[3] |
| 8 | FF[07] | GND | TAG[0] |
| 9 | FF[08] | GND | TAG[1] |
| 10 | FF[09] | GND | TAG[2] |
| 11 | FF[10] | GND | TAG[3] |
| 12 | FF[11] | GND | TAG[4] |
| 13 | FF[12] | +3.3V | TAG[5] |
| 14 | FF[13] | +3.3V | TAG[6] |
| 15 | FF[14] | +3.3V | TAG[7] |
| 16 | FF[15] | +3.3V | LD[0] |
| 17 | FF[16] | +3.3V | LD[1] |
| 18 | FF[17] | +3.3V | LD[2] |
| 19 | FF[18] | GND | LD[3] |
| 20 | FF[19] | GND | LD[4] |
| 21 | FF[20] | GND | LD[5] |
| 22 | FF[21] | GND | LD[6] |
| 23 | FF[22] | GND | LD[7] |
| 24 | FF[23] | GND | LA[0] |
| 25 | FF[24] | +5V | LA[1] |
| 26 | FF[25] | +5V | LA[2] |
| 27 | FF[26] | GND | LA[3] |
| 28 | FF[27] | GND | LA[4] |
| 29 | FF[28] | −5V | LA[5] |
| 30 | FF[29] | −5V | LA[6] |
| 31 | FF[30] | GND | LWR |
| 32 | FF[31] | GND | CSB |
| 33 | GATE | +12V | TRG+ |
| 34 | FRST | GND | TRG− |
| 35 | FWEN | −12V | REV+ |
| 36 | FWCLK | GND | REV− |
| 37 | FFUL | – | RCK+ |
| 38 | NWFF | – | RCK− |
| 39 | ABRT | – | SCK+ |
| 40 | BUSY | – | SCK− |

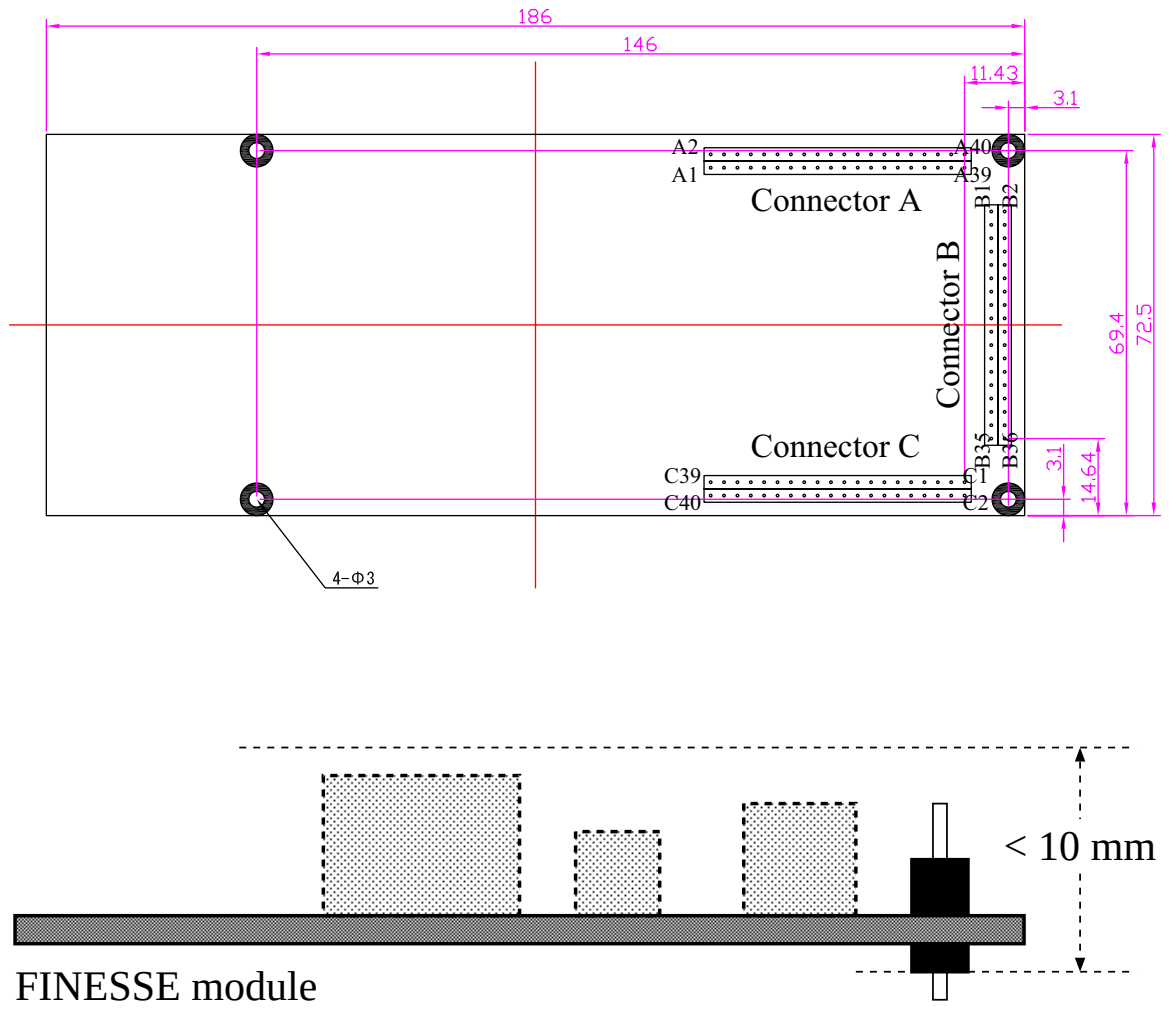

*Figure 13.12: The dimensions of the FINESSE card.*

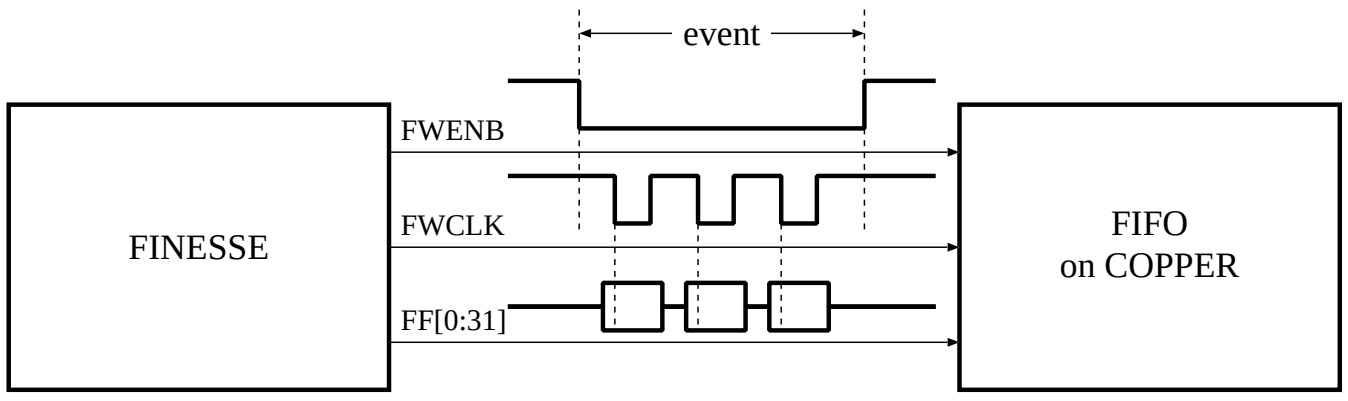

*Figure 13.13: The usage of the `FWENA` and `FWCLK` signals.*

FIFO is maintained by:

| | | |
|---|---|---|
| → | `FF[0:31]` | 32-bit-width data drain |
| → | `FWENA` | indication of a event start/end |
| → | `FWCLK` | write action to the FIFO |
| ← | `FFUL` and `NWFF` | FIFO full condition on the COPPER board |

where the → symbol indicates a signal from the FINESSE card to the FIFO and vice versa for the ← symbol. The usage of the `FWENA` and `FWCLK` signals is shown in Fig. 13.13. The `FWENA` signal indicates a single start/end to the local-bus controller on the COPPER board so that the event separation can be properly treated. At every activation of the `FWENA` signal, the local-bus controller assumes the event sequence starts. All `FF[0:31]` signals at `FWCLK` timings until the `FWENA` signal negated are considered to belong to a same event record.

The FINESSE is connected to the local bus of the COPPER, and is accessible from the CPU card to monitor its status and to configure its registers. The communication between the FINESSE card and COPPER is maintained by:

| | | |
|---|---|---|
| → | `LD[0:7]` | 8-bit-width address bus |
| → | `LA[0:6]` | 7-bit-width data bus |
| ← | `LWR` | data direction |
| ← | `CSB` | FINESSE card selector |

The register map in the FINESSE is listed in Table 13.3.

The FINESSE card also communicates with the timing system. The communication is maintained through the timing interface card (TT-RX) via following signals:

*Table 13.3: The register map of the FINESSE card on the local bus.*

| Address | Description | Mandatory |
|---|---|---|
| 0x00 - 0x77 | user defined | no |
| 0x78 | CSR (active `H`) | no |
| | bit 0: card initialization | yes |
| | bit 1: busy-out emulation | no |
| | bit 2: reserved | yes |
| | bit 3: `FWCLK` emulation | no |
| | bit 4: `FEWNB` emulation | |
| | bit 5: reserved | |
| | bit 6: abort signal emulation | |
| | bit 7: enable bit for bit-1 to bit-6 | |
| 0x79 | data emulation for `FF[0:7]` | no |
| 0x7a | firmware version number | no |
| 0x7b | serial number [7:0] | no |
| 0x7c | serial number [15:8] | no |
| 0x7d | card type [7:0] | yes |
| 0x7e | card type [15:8] | yes |
| 0x7f | reserved | yes |

$\leftarrow$ `TAG[0:7]` 8-bit event tag
$\leftarrow$ `TRG` level-1 trigger timing for backup use
$\rightarrow$ `BSYB` response to `TRG` for backup use
$\leftarrow$ `REV` revolution signal
$\leftarrow$ `SCK` 42.3 MHz system clock

The TT-RX labels event tag to every L1 trigger signal. The FINESSE card is required to include the event tag in the data record.
The electric power, at $\pm 3.3\,$V, $\pm 5\,$V, and $\pm 12\,$V, is provided to FINESSE through the row-B connector.

### 13.6.3 COPPER

The COPPER is a 9U-VME size board with dimension of $366.7 \times 400.0\,\mathrm{mm}^2$. A photograph of the COPPER board is shown in Fig. 13.14.
A COPPER board is equipped with four FINESSE slots, four 1024 kB FIFOs connected to each FINESSE slot (4-slots $\times$ 2 $\times$ IDT 72V36110L15PF), an A32D32 local bus operated at 33 MHz, a local-bus controller (ALTERA Cyclone III), an A32D32 PCI bus operated at 33 MHz, a local-PCI bus bridge (PLX 9054), three PrPMC slots (for the PrPMC, TT-RX card, and generic purpose), one network port directly connected to the PrPMC network port, and one 100BaseT (Intel 82559 for the COPPER-II board) or 1000BaseT (Intel 82541 for the COPPER-3 board) network port on the PCI bus.
The COPPER board has the capability to interface with the VME A32D32 bus as a slave module. We only use VME bus to reset the COPPER board using the `sysreset` signal; no detector signal transfer is intended over the VME bus. The VME bus is bridged with the PCI bus by a 4 kB dual port memory (ALTERA EP20K100QI240-2V).
A COPPER board has three connectors on its rear side; two for the VME bus, and one for electric power supply. The pin assignments of the power connector are listed in Table 13.4. The

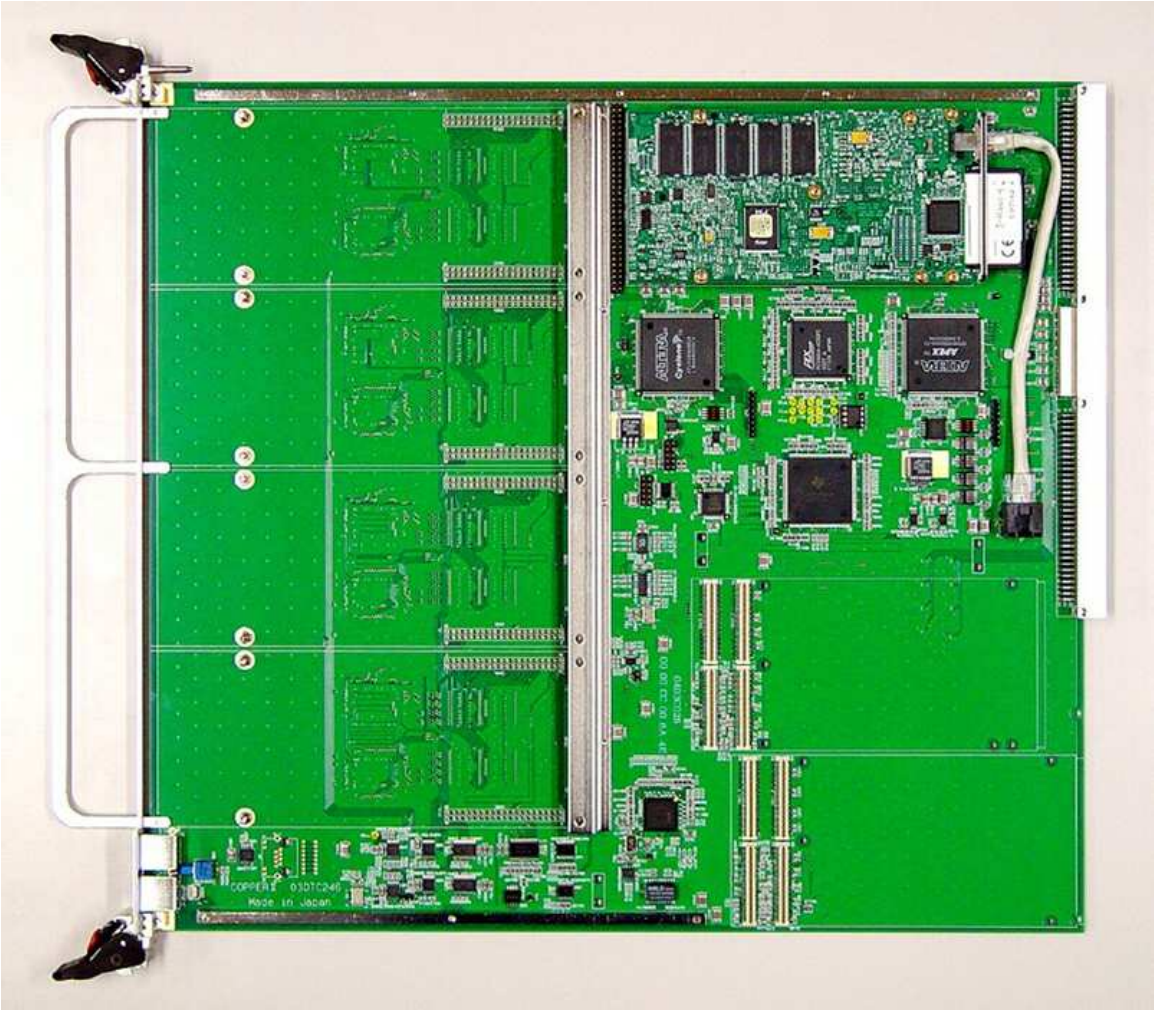

*Figure 13.14: A photograph of the COPPER board.*

*Table 13.4: The pin assignments of the COPPER power connector. Two all-GND rows (row-z on the left to the row-a and row-f on the right on the row-e) are omitted.*

| | **row-a** | **row-b** | **row-c** | **row-d** | **row-e** |
|---|---|---|---|---|---|
| 1 | GND | GND | GND | GND | GND |
| 2 | GND | GND | GND | GND | GND |
| 3 | GND | GND | GND | GND | GND |
| 4 | +3.3V | +3.3V | +3.3V | +3.3V | +3.3V |
| 5 | +3.3V | +3.3V | +3.3V | +3.3V | +3.3V |
| 6 | +3.3V | +3.3V | +3.3V | +3.3V | +3.3V |
| 7 | +3.3V | +3.3V | GND | GND | GND |
| 8 | GND | GND | GND | GND | GND |
| 9 | GND | GND | GND | GND | GND |
| 10 | GND | GND | GND | −3.3V | −3.3V |
| 11 | −3.3V | −3.3V | −3.3V | −3.3V | −3.3V |
| 12 | −3.3V | −3.3V | −3.3V | −3.3V | −3.3V |
| 13 | GND | GND | GND | GND | GND |
| 14 | −5V | −5V | −5V | −5V | −5V |
| 15 | GND | GND | GND | GND | GND |
| 16 | S[1]+ | S[1]− | GND | S[2]+ | S[2]− |
| 17 | S[3]+ | S[3]− | GND | S[4]+ | S[4]− |
| 18 | S[5]+ | S[5]− | GND | S[6]+ | S[6]− |
| 19 | S[7]+ | S[7]− | GND | S[8]+ | S[8]− |

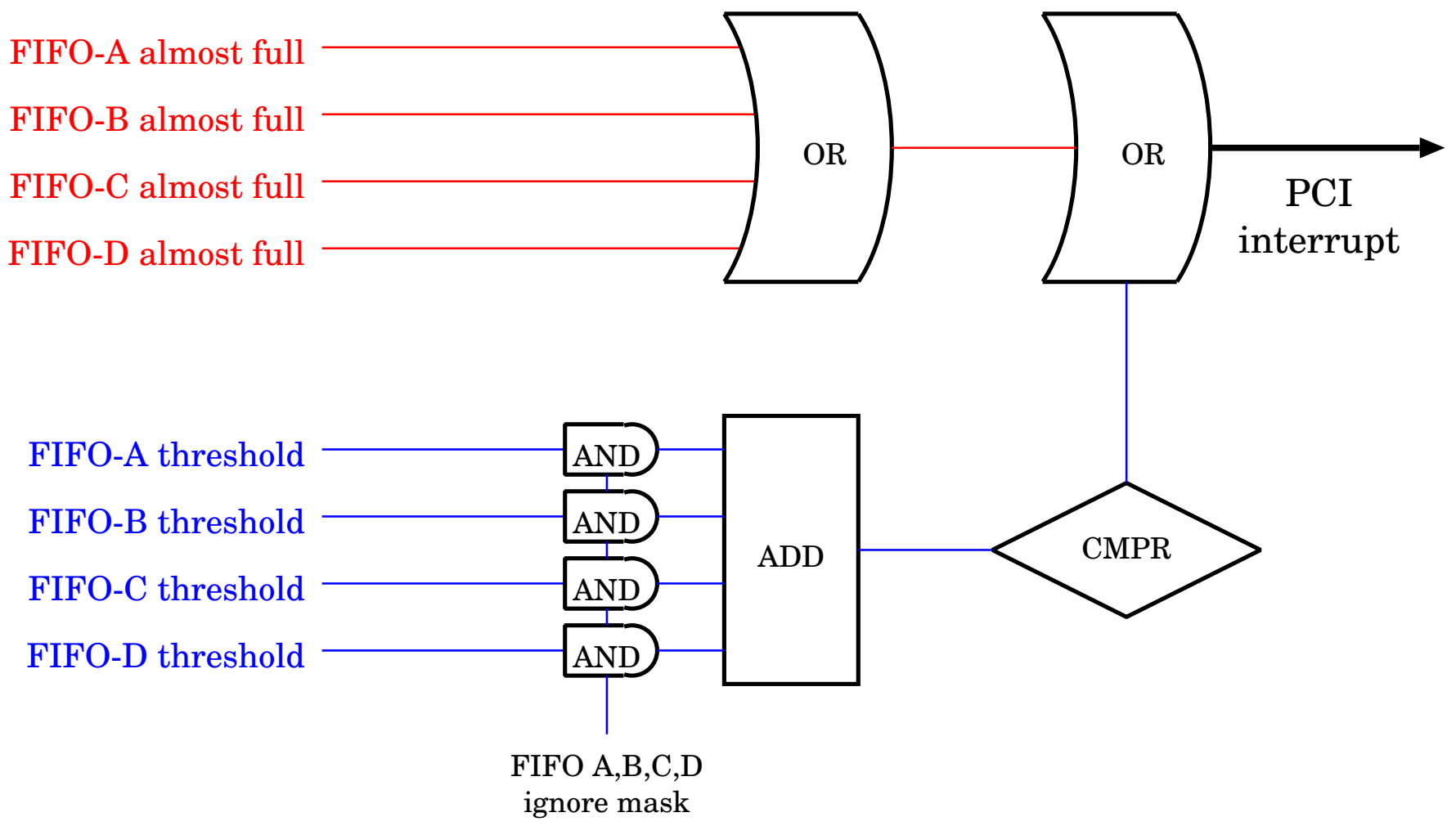

*Figure 13.15: The logic in the local-bus controller on the COPPER to issue the PCI interrupt to initiate the data transfer from the FIFOs to the PrPMC main memory.*

`S[1:8]`± signals are not used in Belle II.

An interrupt is generated at the arrival of data in the FIFO and its condition is configurable to match with the characteristics of data from each detector. The logic is shown in Fig. 13.15. Each FIFO issues the `almost-full` signal when the data size in the FIFO exceeds a predefined configurable watermark (the default being 1023 kB). Once one of four FIFOs gets almost full, the PCI interrupt for the data transfer is immediately issued. When the data size in the FIFO exceeds another watermark, typically configured smaller than the `almost-full` watermark (the default being 768 kB), the FIFO issues the `excess-threshold` signal. This signal can be buffered before the PCI interrupt until other FIFOs also issues their `excess-threshold` signal as well.

### 13.6.4 The VME crate housing the COPPER boards

The COPPER board is housed in a VME 9U crate with a special backplane equipped with two normal VME connectors and one special power connector per slot. The crate can house up to 17 COPPER boards.

The CPU card on each COPPER is supposed to be network-booted with a unique host name derived from the slot-ID and the crate-ID. Dip switches on the COPPER board are used to specify the slot and crate IDs, at which it resides.

### 13.6.5 CPU card

There are commercially available CPU cards following the industrial standard called PrPMC. COPPER is designed to accept a CPU card compatible with this standard. In the Belle DAQ system, we use the RadiSys EPC-6315, which is equipped with an Intel Pentium III processor (800 MHz) and 256 MB of main memory. However, the commercial life of this product is near its end, and the market trend of the PrPMC-compatible CPUs is hazy. We have thus decided to develop our own PrPMC to ensure a sufficiently long product lifetime.

Table 13.5 lists major features of the PrPMC under development. Figure 13.16 shows a prototype CPU board with an area of $170 \times 170\,\mathrm{mm}^2$; this must be shrunk to the PMC size. The CPU card is network-booted and operated by Linux.

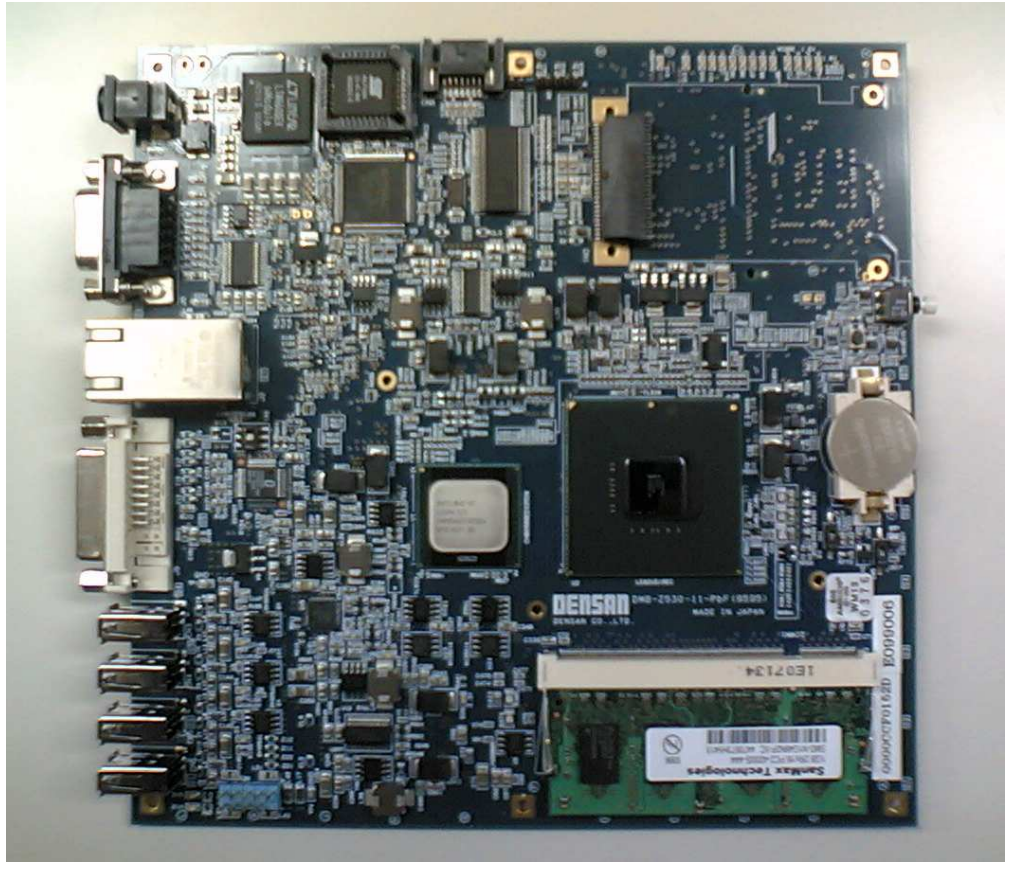

Figure 13.16: A prototype CPU board with $170 \times 170\,\mathrm{mm}^2$ area.

Table 13.5: The list of major features of the new PrPMC under development.

| | |
|---|---|
| CPU | Intel Atom 1.6 GHz processor (Z530P) |
| main memory | 4 × 128 MB DDR2 SDRAM |
| chip-set | Intel US15WP |
| on-board storage | 4 GB Solid State Disk |
| network I/F | Intel 82574L (1000BaseT) |
| user I/F | 1 × serial port<br>PS/2 keyboard and mouse<br>2 ch × USB<br>VGA |
| operating system | RHEL 3 or CentOS 3 or later<br>bootable from SSD, USB, and network |
| electric power | less than 15 W |

### 13.6.6 Data format of the COPPER readout system

Figure 13.17 shows the format of the raw data output from the FINESSE card. The `FFAA` and `FF55` are event header and event trailer, respectively. The `FINESSE event counter` is a 24-bit event counter incremented upon the L1 trigger signal from the TT-RX card. The `ttrx-ev` is an event tag provided by the `TAG[0:7]` signal. The `checksum` is an 16-bit xor checksum from the `FFAA` to the `FF55`.
Figure 13.18 shows the format of the data record sent to the CPU card, which is formatted by the local-bus controller. The `FFFFFAFA` and `FFFFF5F5` are event header and event trailer, respectively.

### 13.6.7 Readout PC

The data from COPPERs housed in a VME 9U crates are sent to a Readout PC through a 100Base-T or 1000Base-T network connection via a network switch. There data are already formatted by the software processing on the CPU on COPPER and directly fed into the event building framework described in the next section. The first level data monitoring is done on the Readout PCs using the unified software framework.

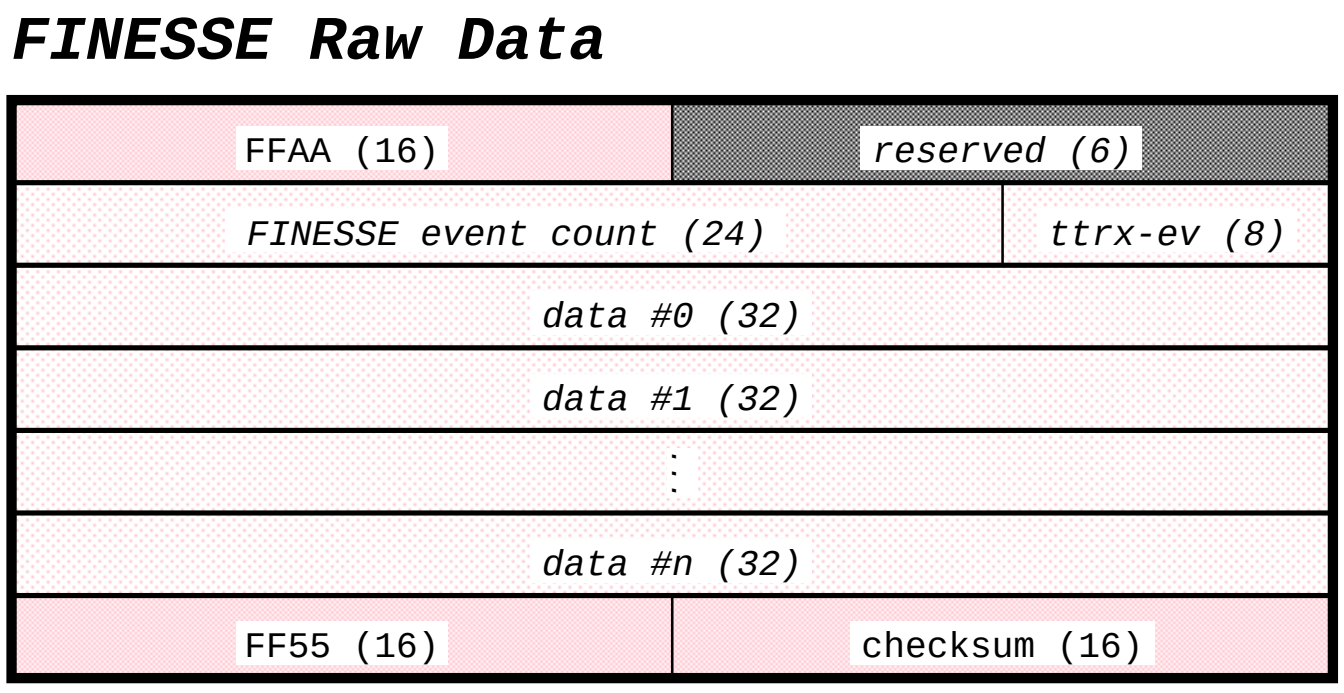

*Figure 13.17: The raw data format output from the FINESSE card.*

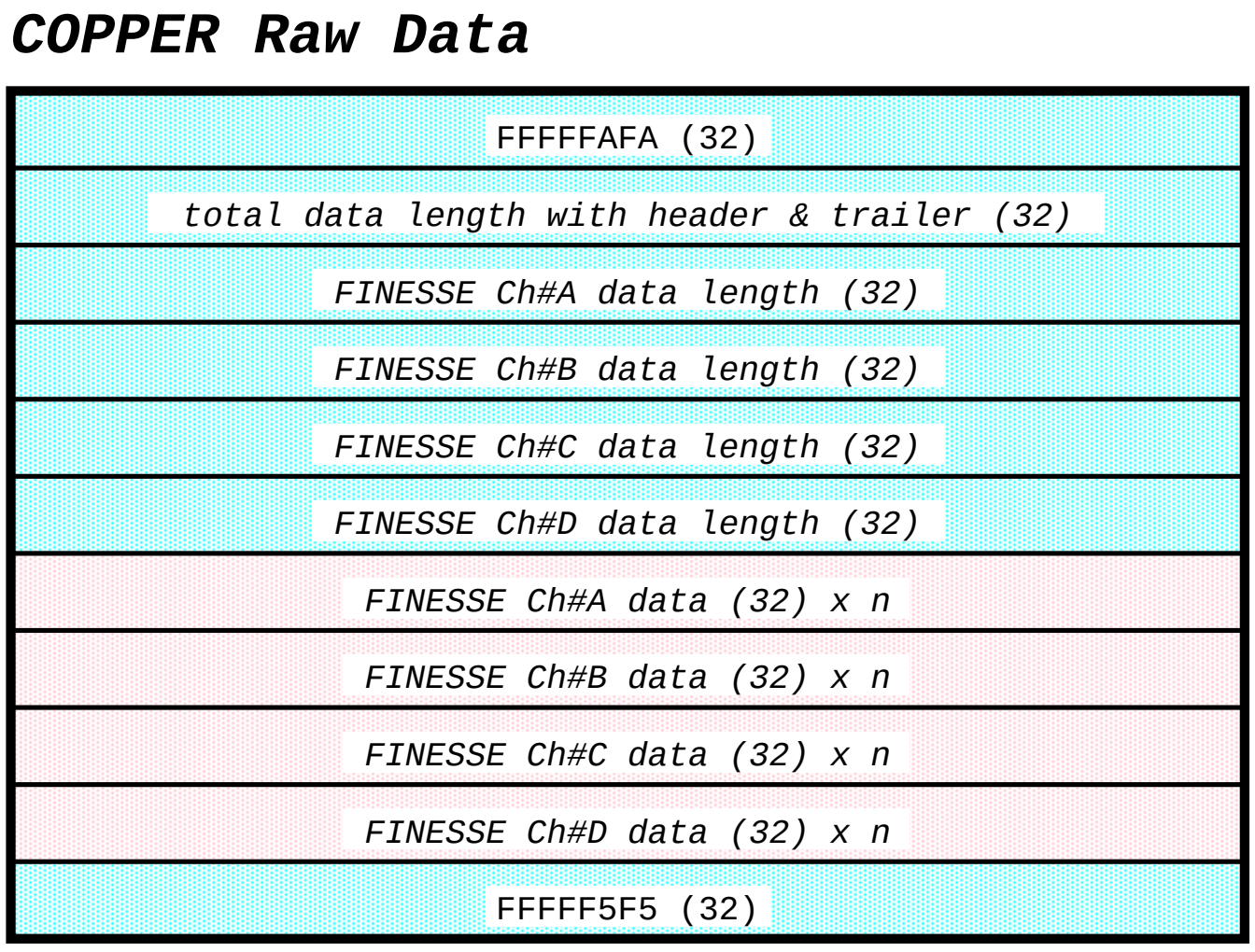

*Figure 13.18: The format of raw data fed into CPU card.*

We will have up to 50 Readout PCs in total to read out all Belle II detector components.

### 13.6.8 Data transfer performance

We study data transfer performance inside the COPPER board. Figure 13.19 shows a block diagram of the test setup for the performance study.
The FINESSE emulator card outputs predefined data sequence to the FIFO on the COPPER board at every arrival of the emulated L1 trigger signal from the TT-RX card. The data size per event can be adjusted to one of $2^n$ bytes ($n = 4$ or $8 \ldots 12$).
We used a CPU card called PSL09 for the test, which is equipped with an Intel PentiumM processor (1.1 GHz) and 256 MB of main memory. The performance of the CPU card is expected to be similar to that of the new CPU card that is being developed.
The data transfer performance is studied by measuring the total throughput inside the COPPER board by varying the generated data size and the trigger rate. Table 13.6 lists the results of the performance study. In case of the trigger at 50.0 kHz (marked with the $^\star$ in the table), the CPU idle time is measured to be 78.4%. The throughputs marked with the $^\dagger$ are close to the PCI-bus limit of 133 MB/s. We conclude that the COPPER and its CPU ensures enough performance

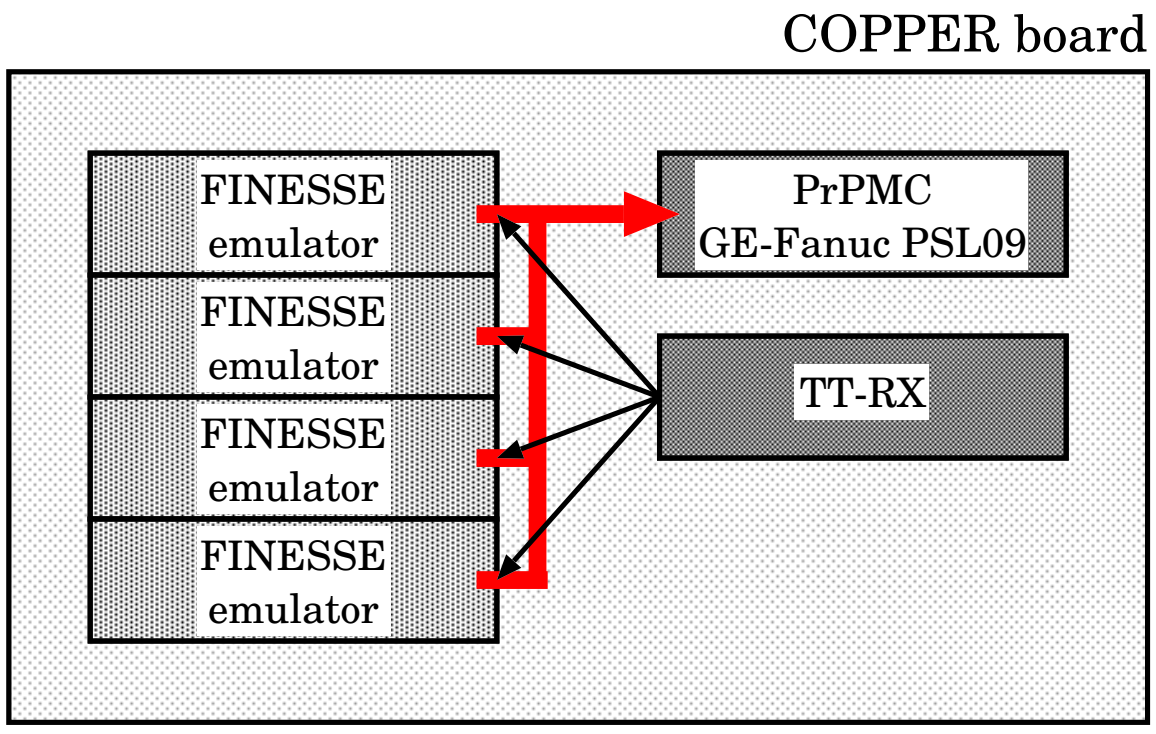

*Figure 13.19: Block diagram of the test setup for the performance study of the COPPER readout system.*

*Table 13.6: List of measured throughputs inside the COPPER board with varying the generated data size and the level-1 trigger rate.*

| accepted rate [kHz] | data size [kB/ev] | throughput [MB/s] |
|---|---|---|
| 50.0 ⋆ | 1.1 | 53 |
| 70.0 | 1.1 | 74 |
| 102.3 | 1.1 | 108 |
| 61.9 | 2.1 | 128 † |
| 31.6 | 4.2 | 128 † |
| 15.9 | 8.3 | 127 † |
| 8.0 | 16.5 | 129 † |

for the operation at the L1 trigger rate of 50 kHz.

## 13.7 Event Builder

The task of the event builder is to collect data pieces from all detectors, merge them into one block, and send it to the high level trigger for the data reduction. The number of event builder inputs in the Belle II DAQ system is estimated to be about 50, corresponding to the number of readout PCs, and the number of outputs is of an order of 10, corresponding to the number of HLT units.

The event builder in the present Belle DAQ system is based on the concept of "switchless event building," where the event is built step by step using a number of PCs connected point-to-point via Gigabit Ethernet[3]. Each PC receives event fragments from all upstream nodes, assembles them into one record, and then sends it to a downstream node. These steps are repeated in multiple layers of PCs. The PC in the final layer receives all event fragments from all detectors and builds one complete event.

In the Belle II DAQ system, the estimated data rate reaches 9 GB/s at most under the average trigger rate of 30 kHz. The total flow rate is almost 500 times higher than that of Belle. The approach based on the switchless event building is not cost-effective for such a high rate because it would require far too many PCs.

The other approach to build the event builder is to use a large network switch as already adopeted by many other experiments. Modern network switches are equipped with many gigabit ethernet ports that can connect a large number of PCs. When a trigger signal issued, data fragments from ∼50 readout PCs are transmitted simultaneously to the event builder when a trigger is issued. Since these data have to be sent to one receiver PC node to build the event, a heavy network congestion is expected in the receiver. It causes the TCP restransmission which deteriorates the data transfer performance. To avoid the TCP restransmission caused by the congestion, the network swich has to have enough large buffers for each network port.

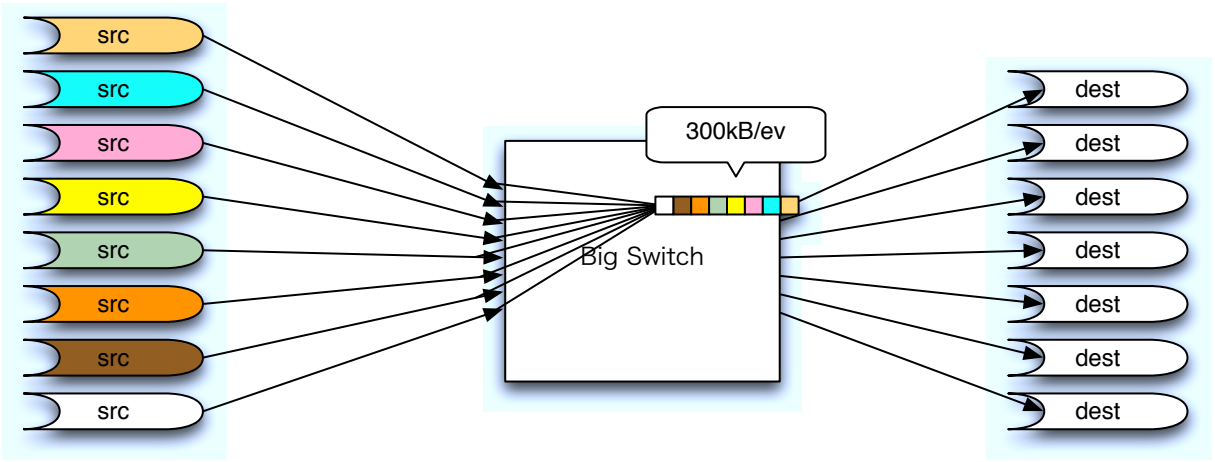

*Figure 13.20: Network switch for the event builder requires deep buffer at each output port.*

Recently large scale switches that are equipped with a deep buffer for each port become commercially available from several vendors, such as Force10, HP, and ARISTA. One has a 16MB buffer for each port, which can be found in ARISTA products.

We studied the network performace using an ARISTA 7120T switch and observed the occurance of the TCP restansmission. The testbench consists of four transmitter PCs and one receiver PC. They are connected together by GbE links through the ARISTA switch. The transmitter PCs send data packets to the receiver PC at a full speed. The result shows no TCP retransmission is observed and the data transmission rate of four transmitter PCs are correctly balanced. We are continuing the test by increaing the number of transmitters to 16 or more. If this scheme is confirmed to work in such a large configuration, we will adopt the large network switch option for our eventbuilder as shown in Fig. 13.20.

As a backup, we are also considering a layered event builder design combining PCs and small network switches as building blocks of the barrel shifter[4, 5]. A large buffer can easily be implemented on PCs to avoid the TCP retransmission in this design, however, its cost might be slightly higher than that of the network switch. Using a switch component with 16 inputs $\times$ 16 output, a large switch with 64 inputs $\times$ 64 outputs can be implemented as shown in Fig. 13.21. Network switches are used in the upstream layer while downstream layer consists of PCs, because the event data at the upstream layer is partially built and required buffer depth is shorter than that at downstream. In order to reduce the number of PCs, one PC is desired to house many GbE links, but it consumes more CPU power to transmit data.

We studied the CPU consumption of a PC connected to four transmitter PCs and four receiver PCs connected through GbE. It is measured to be 75% of 1 CPU core of the PC (Intel W5590), and we conclude that this option is feasible. The case with 10GbE link is also tested and found that it provides a faster transmission with a little more CPU consumption. Since a faster transmission from the switch to PC reduces the probability of the buffer full, we might use 10GbE link from a 16×16 switch component to PCs.

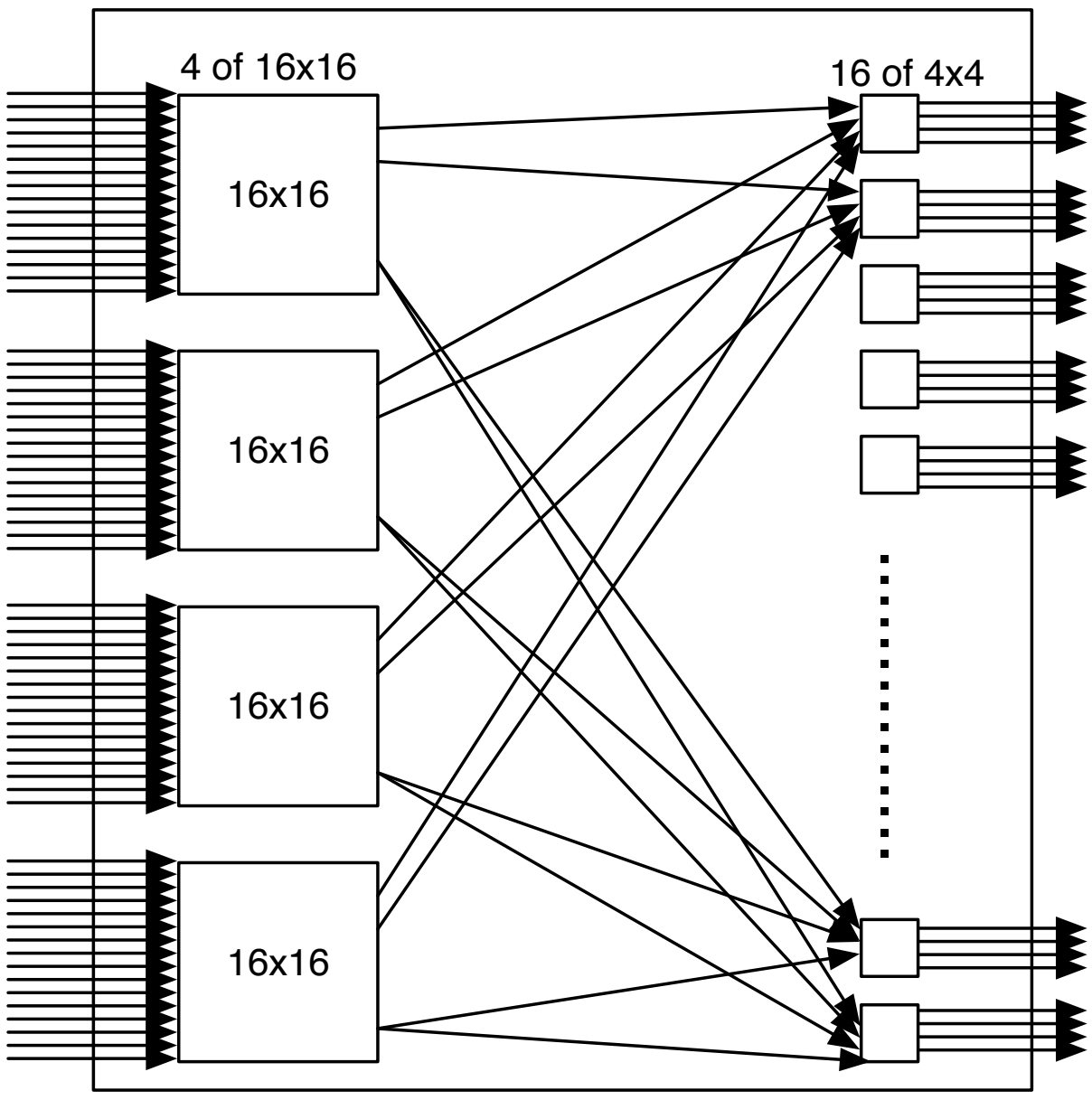

*Figure 13.21: 64×64 event builder by 16×16 and 4×4.*

## 13.8 Data Processing Framework

### 13.8.1 Data processing in Belle II DAQ

In the Belle II data stream, the following real-time data processing tasks are required at each stage:

1. Detector front-end and data-link:
   a) the data reduction by zero suppression, the feature extraction from wave-form, etc., and b) the data collection and formatting at the channel level.
2. COPPER:
   a) additional data compression by sophisticated software processing, b) data collection and formatting at the module level, and c) data monitoring.
3. Readout PC:
   a) data collection, formatting, and reduction at the subsystem level, and b) data monitoring.
4. High Level Trigger:
   a) event selection by full event reconstruction, and b) detailed data monitoring.

Except for the front-end processing that is performed on FPGAs, the processing platforms are all Linux-based. The total processing power of these platforms is huge.

- COPPER: Each COPPER module is equipped with a Linux-operated Intel Atom chip running at 1.6 GHz. The expected data rate per module is $10^6$ words/s. We will have an order of a few hundred COPPERs for the DAQ.
- Readout PC: This is possibly a PC server equipped with dual Intel Core i7 running at 3.3 GHz, which is equivalent to eight cores. A total of $\sim 50$ servers are connected to COPPERs. A typical data rate per core is estimated to be $\sim 2 \times 10^6$ words/s.

- High Level Trigger (HLT): The HLT consists of multiple units of PC clusters in which one cluster consists of $\sim 20$ PC servers with dual Core i7. With 10 units of PC clusters, the total number of cores is estimated to be $\sim 1600$, with a data rate of $\sim 0.5 \times 10^6$ words/s.

### 13.8.2 Unified framework for data processing

Since the data processing after the Belle2link is performed on the same Linux platform, it is desired to have a common software framework for this processing. We prefer that this framework be compatible with the Belle II offline software so that the software development can be done identically in the offline environment.

The DAQ software framework is required to have the software bus structure, where various software components developed independently are plugged into the framework and work together to process a common data stream. For the common data access by different software components, the data should be formatted and managed using the same scheme.

Such a framework already exists in the Belle DAQ system. It is composed of the B.A.S.F. framework [6] and the Panther [7] data management system. B.A.S.F. and Panther are developed as the standard offline analysis framework for Belle. To use them in Belle's DAQ data processing, the I/O interface of B.A.S.F. is modified so that it can send and receive event data through the UNIX socket to pass the data stream over the network, and the execution control of the framework is implemented externally through NSM [8]. This framework is commonly used on COPPERs, event builder PCs, and reconstruction farm nodes. The processing software for these platforms is developed solely in the offline environment and then ported to the DAQ elements without any modifications; this is found to be useful to minimize the development cost.

For the common framework in Belle II, compatibility with readily available modern offline software technology is key. The Belle II software should be fully object-oriented and based on ROOT IO, so the framework must be capable of reading and writing ROOT IO objects. A new framework named "roobasf" [1] is being developed for this purpose. This framework is designed to meet the requirements of real-time processing; the socket I/O interface can be easily implemented as one of the selectable I/O packages for roobasf. Therefore, the unified DAQ framework for Belle II can be realized by combining roobasf with the NSM interface. Figure 13.22 shows the design of the unified data processing framework for the Belle II DAQ.

### 13.8.3 Object-oriented data flow

The object-oriented approach to data handling in Belle II is essential to realize the unified and offline-compatible software environment. ROOT IO is the basis of our object-oriented data handling, and the raw data from the detector frontends are required to be formatted in ROOT objects at the very early stage.

Figure 13.23 shows the schematic view of the object-oriented data flow. The raw data from detectors are fed into COPPERs through the unified data link and formatted in ROOT objects. The unified framework with roobasf runs on each of DAQ components and the processing results are passed through UNIX socket connections.

To send an object's data through a socket, the object has to be "serialized" to a plain byte stream by the "streamer." ROOT has the capability of generating the streamer code automatically from the class definition. However, since the raw data structure can vary widely across subdetectors, the processing time for the streaming may become critical for complicated structures. A careful treatment is required in the design of the class structure of the raw data to optimize the streamer performance.

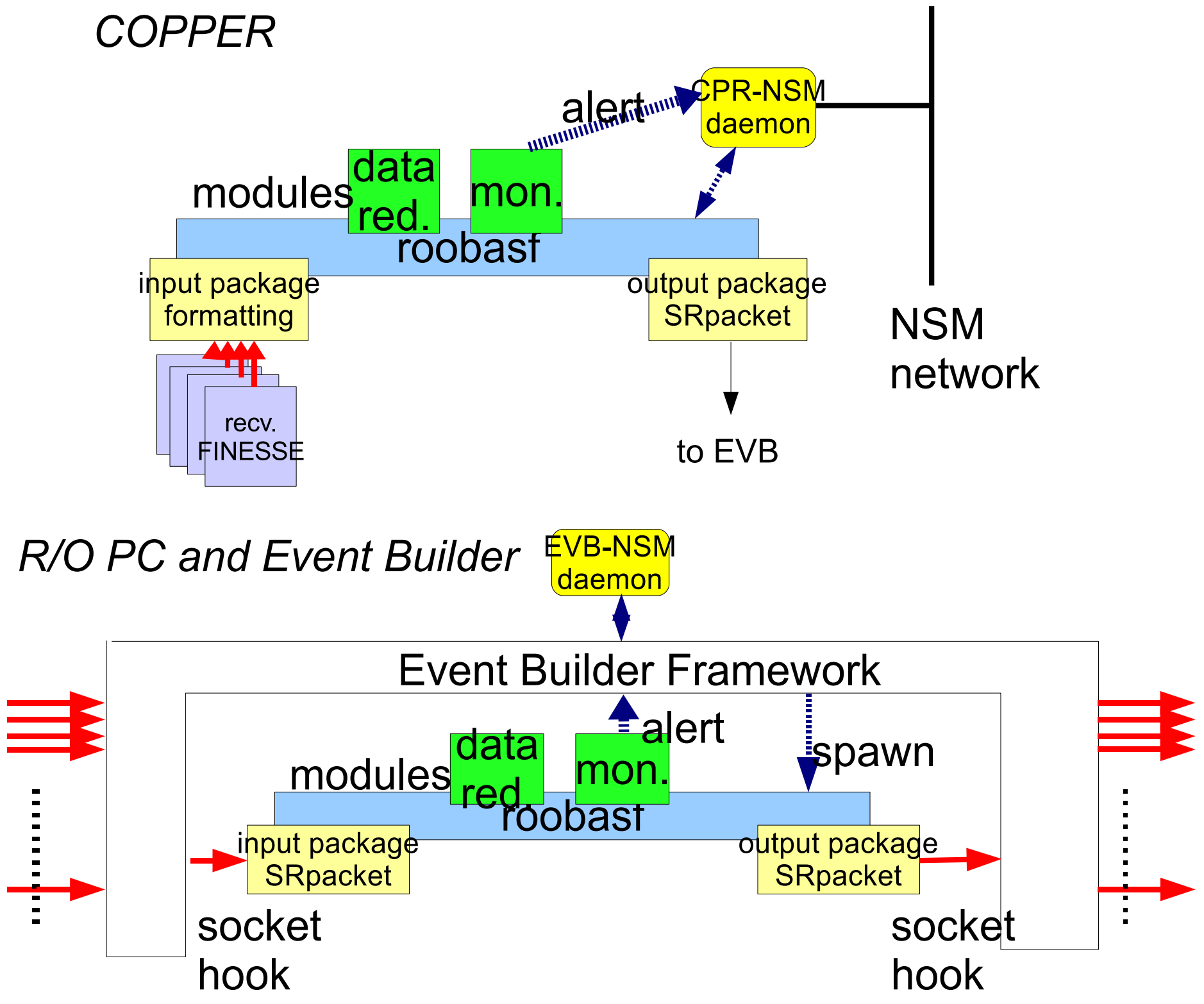

Figure 13.22: Unified processing framework

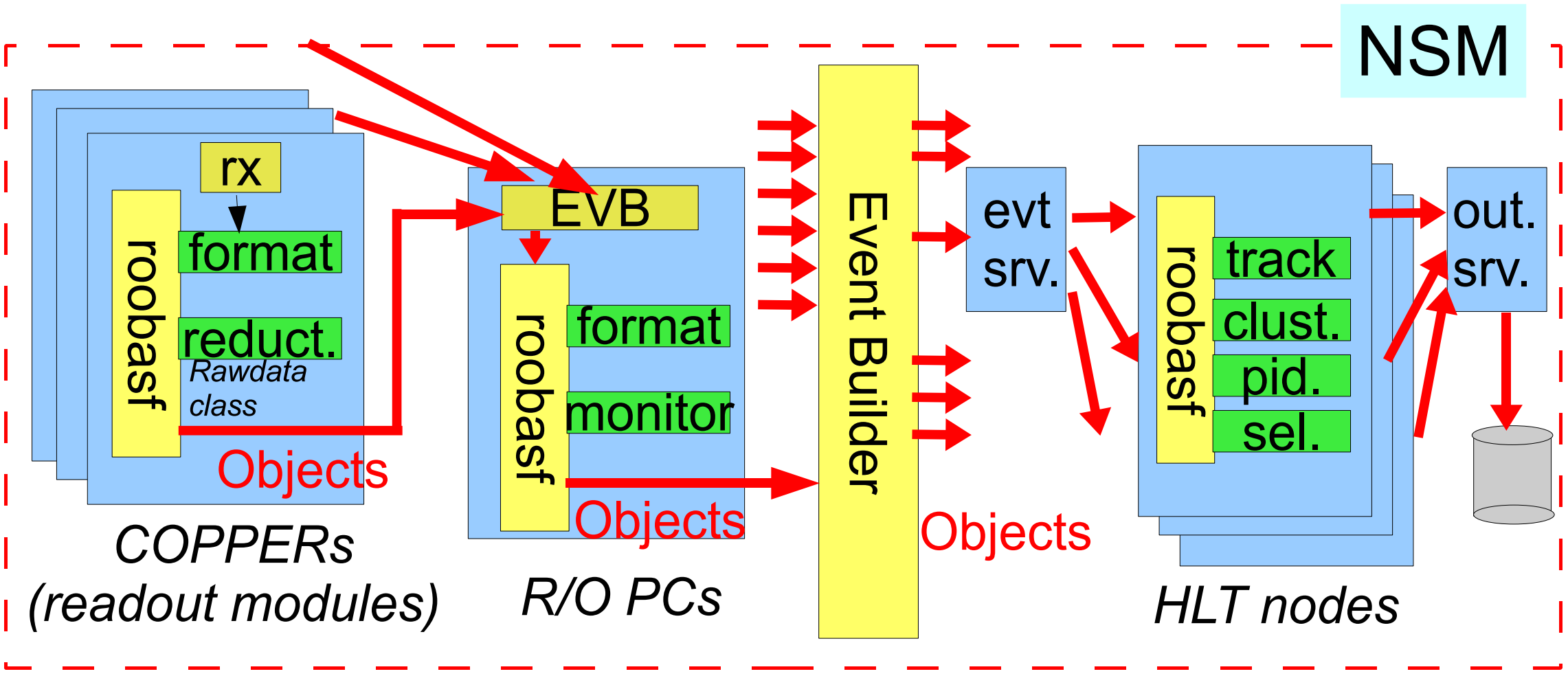

Figure 13.23: Object oriented data flow in DAQ

## 13.9 High Level Trigger

### 13.9.1 Requirements and Global Design

In the high level trigger (HLT), a full reconstruction using the event data from all detectors is performed and the software trigger for the event is made using the physics-level event selection software.

Based on the experience with the real-time reconstruction farm (RFARM) [9] in the Belle DAQ, two levels of software trigger are being considered for the HLT. The first is the so-called "Level 3" trigger. After a fast track reconstruction (using the CDC data only) and a fast ECL cluster reconstruction, this trigger makes a rough cut on the track multiplicity, the event vertex position, and the total energy deposition. Its expected reduction factor is estimated to be around 50%, according to the Belle's experience. The pre-selection by this trigger is useful to shorten the average processing time for an event.

The second is the physics-level event selection using the full event reconstruction results. Almost 100% of the physics analyses in Belle are performed using the skimmed data set with the hadronic event selection (for $B$ and $D$ physics analyses) or with the low-multiplicity selection (for $\tau$-lepton and two-photon physics). These event selection codes are used offline in the Belle data processing, and their reduction factors are 14.2% and 9.6%, respectively, under typical beam conditions for the events that have passed Level 3. We also need some pre-scaled samples of Bhabha-scattering and $\mu$-pair events for monitoring purposes; their pre-scaling rate can be as low as $\sim 1\%$. With all of these selections, an overall reduction factor of $\sim 25\%$ beyond Level 3 is expected. By combining these two reductions, the rate reduction factor by the HLT selection is estimated to be $\sim 12.5\%$. In the design of the DAQ system, we assume the reduction factor by HLT to be $1/5 = 20\%$ considering the safety margin.

The actual trigger software on HLT is supposed to be the same offline software used in the offline reconstruction to avoid introducing additional systematics different from that in offline processing. The tracking and clustering software are developed as modules which can be directly plugged into the HLT framework as well as offline framework. The physics analysis modules such as the hadronic event selection and the low multiplicity skim are used as the real trigger software. These software modules are well tested using the Monte Carlo simulations in offline and then ported to HLT without any modifications. Fig. 13.24 shows an expected data processing chain at HLT and offline.

The trigger processing of an event data should be done, ideally, on a single CPU core in the PC clusters of HLT. Its processing time (i.e., the latency) depends on the performance of the full reconstruction software. Our experience at Belle shows the average (peak) time for all the L1-triggered events is about 0.5 (10) s/event on a single Intel Xeon core running at 3.0 GHz. Therefore, about 15,000 such cores are required to process the Belle II events at a 30 kHz trigger rate. By using a PC server housing sixteen cores, the total number of such servers required for the HLT is about 1000, which can be accommodated in about 25 racks. We may consider the use of GPUs to increase the processing power and to reduce the processing latency.

### 13.9.2 Cluster Configuration

The accelerator luminosity is expected to increase gradually and the HLT processing power is to be implemented on a demand basis. The HLT is modularized into a set of small PC cluster units. Each unit is equipped with about 20 PC servers with $> 10$ cores per server running at a $>3$ GHz clock, and is expected to process events at a rate of $\sim 1$ kHz. We will start from a configuration with ∼ 10 units, and add add units whenever more processing power becomes necessary.
Figure 13.25 shows the structure of one PC cluster unit. L1-triggered events are fed into the event distributor node through a 10GbE connection. The 20 PC servers (processing node) are connected via network switch and the events are distributed to the nodes via a GbE connection. A full event selection chain is performed on roobasf in each node and the selected events are sent to the output server node via the GbE connection. The output server node is connected to a fast RAID system to record the raw data of selected events. In some cases, the processing results are also recorded together with the raw data. The events are recorded in the "streamed" format.

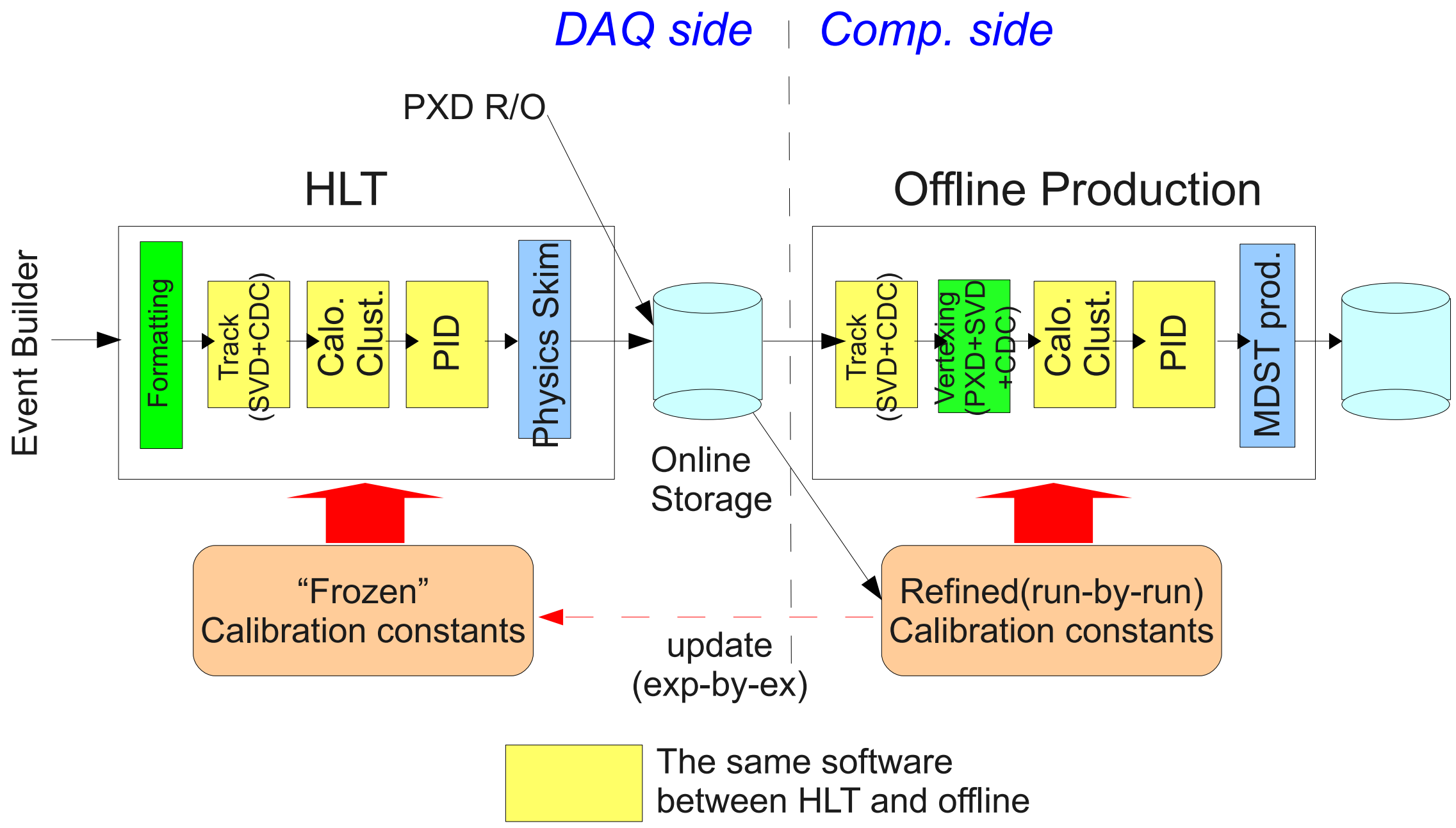

Figure 13.24: The data processing chain at HLT and offline.

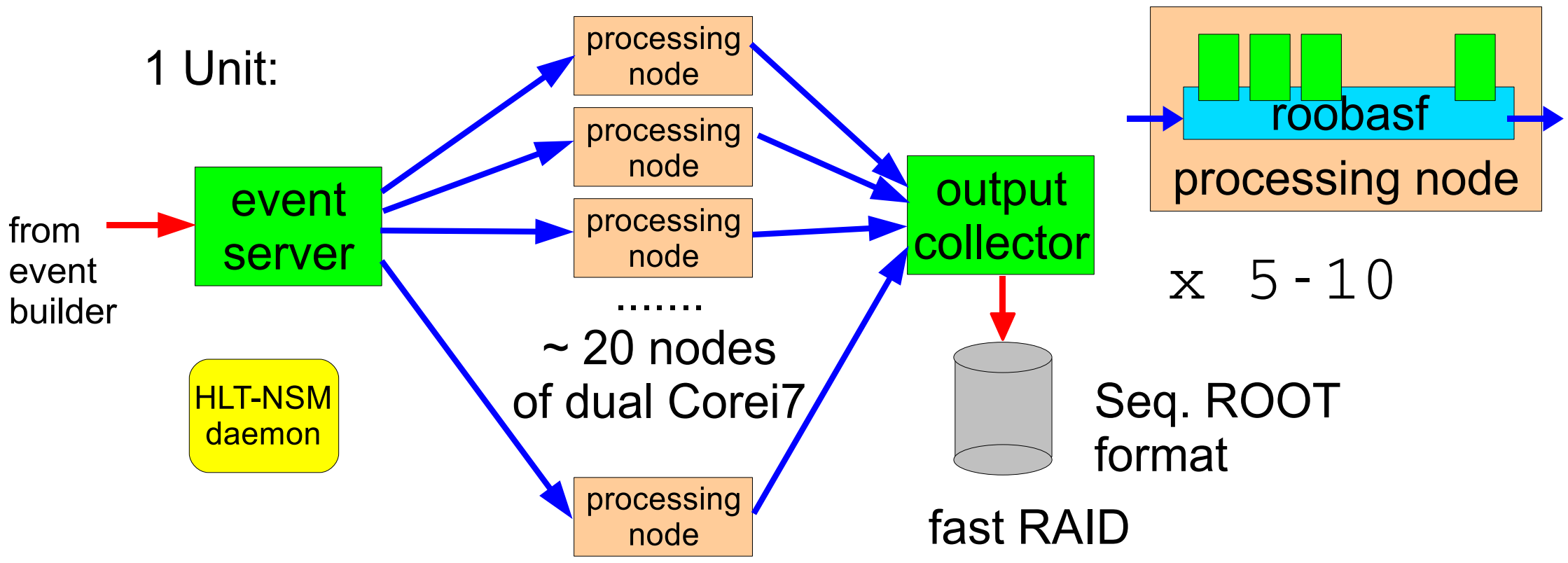

Figure 13.25: The High Level Trigger.

### 13.9.3 HLT software framework

To process the 30 kHz event data flow without delay, parallel processing is essential. Since the parallelization of the processing of a single event requires a special software treatment and is challenging to implement, we consider instead the trivial event-by-event parallel processing in HLT.
Two types of parallel processing mechanisms are implemented in the HLT software framework. The first is the mechanism to utilize multiple CPU cores in a PC server. The Linux operating system on the server treats the cores as SMP processors. A mechanism to distribute events to these processors and run the actual processing software on each of them in parallel is necessary to utilize the multiple cores. The other is the parallel processing to make use of the PC cluster with the 20 servers connected via a network. The event data from the event distributor connected to the event builder are distributed to different servers through the network and processed in parallel. The processing results are gathered again by collector nodes in a reverse way and recorded in the high speed RAID system. The software framework to control such a distributed parallel processing in the cluster is necessary.
This "type 1" parallel processing is already implemented in the core of roobasf and each HLT processing node utilizes the capability. The second type will be implemented by extending the type 1 parallel processing in roobasf. The "event server" and "output server" processes in the roobasf core can accept data from and feed data to the network socket; the events from the event distributor node are directly fed into the roobasf core on each node and vice versa to the output collector node. The current design of this "type 2" parallel processing in the HLT is shown in Fig. 13.26.
There are several candidates for the overall control scheme of the PC cluster for type 2 parallel processing. The baseline design is to recycle the scheme used for the Belle RFARM, which is based on NSM [8].

## 13.10 Slow control

The slow control system must

- control the start and stop sequence of the entire data acquisition system,
- collect the status and messages of the data acquisition subsystems,
- control the HV system, and
- collect various monitor information.

These functions are realized by a software package called the "Network Shared Memory" (NSM). This software handles point-to-point message passing using TCP and shared memory over a network segment using UDP broadcast.
The expected number of systems to be controlled is not expected to increase over that in Belle, so the existing system can be used without modification.

## 13.11 PXD integration

Since the expected event size from the PXD is very large ($\sim$ 1 MB/event), it is quite difficult to manage its readout using COPPERs. Therefore, its integration into our DAQ is mismatched with our global design of Fig. 13.1. In addition, it is also very hard to leave such a huge amount of data in offline storage as is because of the limitation in the storage and computing cost. It behooves us to consider methods for data reduction from the PXD by a factor of ten or more before this stream is combined with the data from other subsystems.

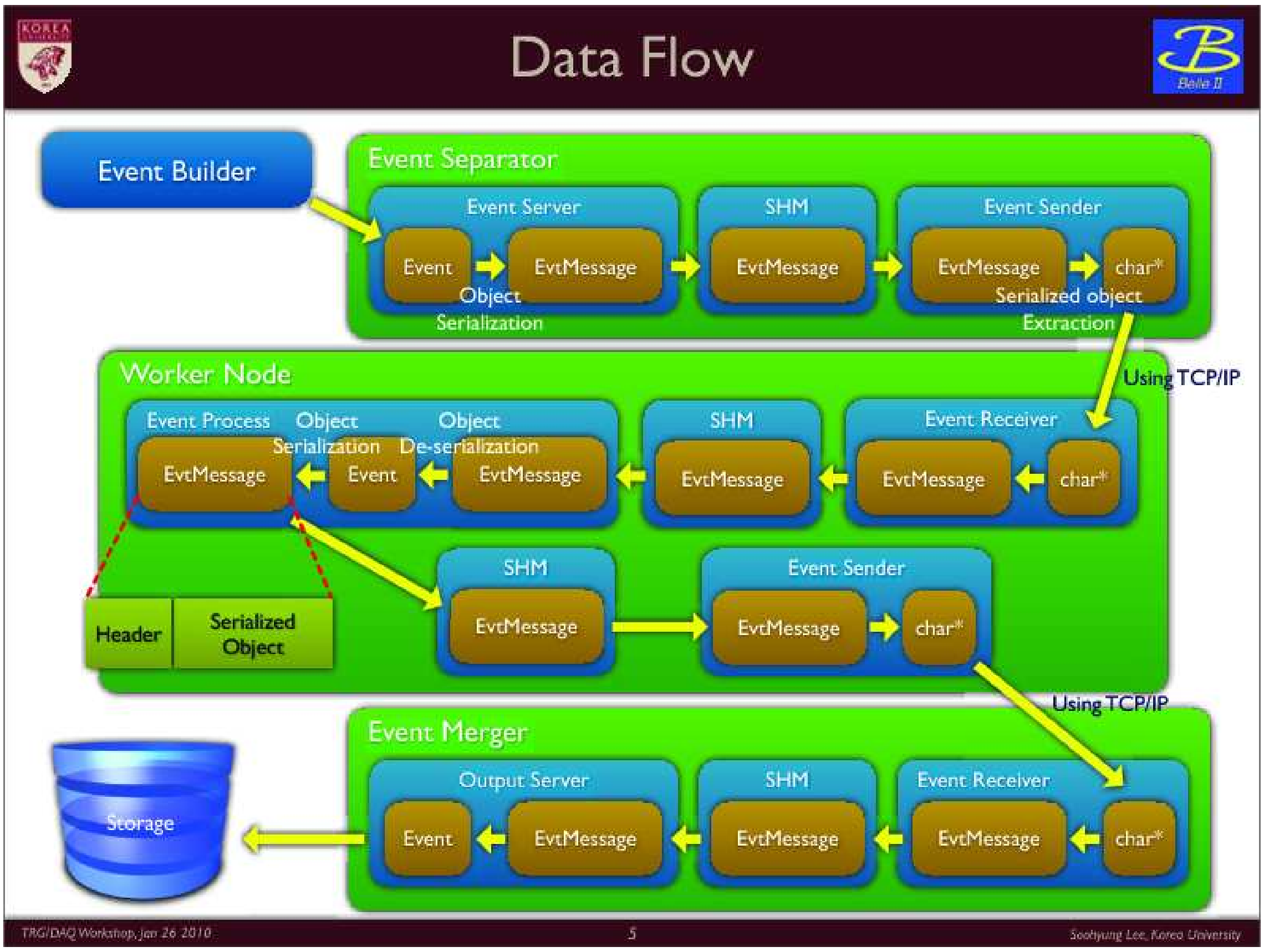

*Figure 13.26: Type 2 parallel processing in HLT*

There are two ways to reduce the PXD data flow rate. One is to reduce the size of each event. The raw data from the PXD contains many noise hits in addition to the hits associated with real tracks. If we send only the PXD hits that are near bona fide tracks to the event builder, the data size is dramatically reduced. The other way is to reduce the number of events sent by the PXD. If we send the PXD data only for triggers tighter than L1 trigger (after HLT selection, for example), the PXD data flow rate is reduced.

Figure 13.27 shows these two solutions for the PXD integration into the DAQ: the upper figure shows the option 1, while the lower shows option 2.

Here, we describe each integration option.

- Option 1

  A number of special modules in a ATCA crate[10], which were originally developed for use in the HADES and PANDA experiments, receive data from the PXD DHH over the RocketIO link together with the SVD and possibly CDC, where the SVD and CDC data are branched from the main data stream. In each module, five Virtex 4 FPGAs perform the track finding, track reconstruction, and association with PXD hits. The track finding and reconstruction is done using the SVD and CDC hit signals. The module send out only the PXD hits associated with tracks to the event builder over the GbE link.

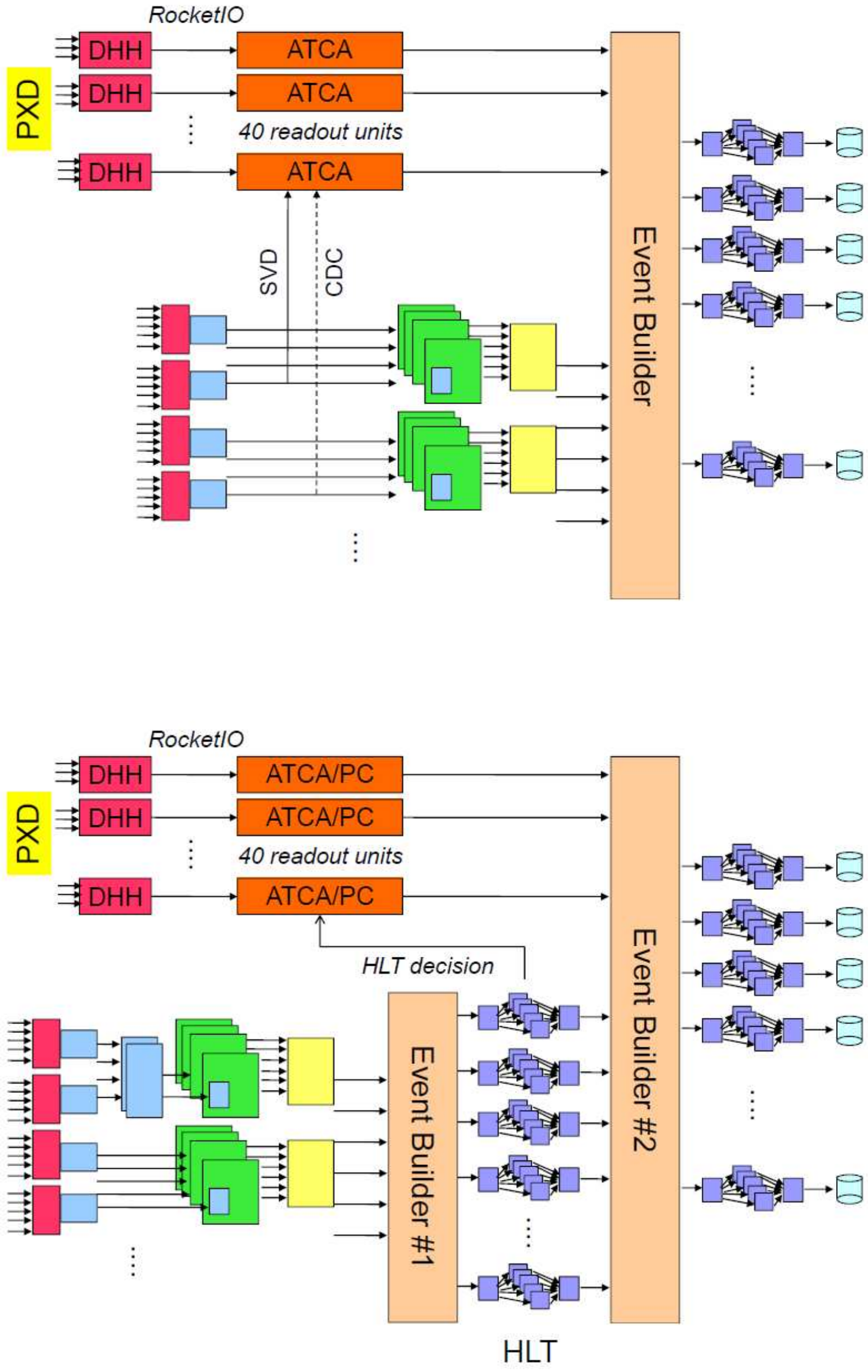

Figure 13.27: Two options for the integration of the PXD data stream into the DAQ system. Top: baseline option, in which the size of each PXD event is reduced. Bottom: substitute option, in which only selected PXD events are read out.

- Option 2 and 2'
  The PXD data are fed into either the same ATCA boards as in option 1 (called Option 2) or into PCs (called Option 2') from the PXD DHH over the RocketIO link, whereas the SVD and CDC data are not fed into them. Instead, the processing result of the HLT, containing the event tag and parameters of the tracks in the event, are sent to the ATCA boards (or PCs). Since the full event reconstruction is performed by the HLT using the offline software, the track parameters can be obtained therefrom without additional software. The track parameters are sent to all ATCA boards (or PCs) and the hit-track association is performed. The selected hits are sent to the second-level event builder through GbE connection so as to combine with the data from other detectors.

  In this option, because of the large latency—up to 5 seconds—of the HLT processing, a large memory to buffer the PXD data stream during this latency is required. The memory size is estimated to be at least 3 GB for each of the 40 DHH data streams.

  The design of the second-level event builder has different aspects from the one described in the previous section. Since the event sequence in the HLT output is disordered because of the parallel processing nature, and also the event number from the output of each HLT unit is not the same, a different approach to the event building is required. The design is now in progress.

## 13.12 Data reduction

There are two aspects to data reduction: the reduction of the event size and the reduction of the trigger rate. The data size reduction is expected to be done primarily by the FPGA processing in the front-end readout. The data size after this reduction is estimated to be 100 kB/event, as previously described. We can expect more reduction by the software processing on COPPERs and readout PC's; however, we currently do not consider the latter reduction in the design of the DAQ system.

The reduction of the PXD event size depends on the integration option. In the case of option 1 where the track reconstruction is done by the fast FPGA processing, we currently estimate the reduction factor to be 1/5 due to the limited performance of the hardware tracking. In the case of option 2 or 2', the factor is expected to become 1/10 or less thanks to the precise measurement of track parameters made possible by the HLT software processing. The event size from PXD after the reduction is, therefore, estimated to be 200 kB for option 1 and 100 kB for option 2.

The reduction of the trigger rate solely relies on the HLT in our DAQ design, as described in the previous section. The reduction factor is estimated to be 1/5 or less.

Table 13.7 shows the estimated data flow for several cases. The detail of the data reduction steps is also shown in Figs. 13.28 and 13.29 for the two PXD integration schemes.

| | Worst I | Worst II | Modest I | Modest II |
|---|---|---|---|---|
| Event size of PXD | 1 MB | 1 MB | 0.5 MB | 0.5 MB |
| Event size of other detectors | 100 kB | 100 kB | 50 kB | 50 kB |
| PXD reduction factor | 1/5 | 1/10 | 1/5 | 1/10 |
| Event size of PXD after reduction | 200 kB | 100 kB | 100 kB | 50 kB |
| Total event size | 300 kB | 200 kB | 150 kB | 100 kB |
| Level1 trigger rate | 30 kHz | 30 kHz | 20 kHz | 20 kHz |
| HLT reduction | 1/5 | 1/5 | 1/5 | 1/5 |
| Rate at storage | 6 kHz | 6 kHz | 4 kHz | 4 kHz |
| Bandwidth at storage | 1.8 GB/sec | 1.2 GB/sec | 600 MB/sec | 400 MB/sec |

*Table 13.7: Estimates of data flow for several cases. Worst I and II cases are obtained with the assumptions of the maximum event size with a safety factor 2 and the trigger rate of 30 kHz. Modest I and II are with the typical event size and trigger rate. I and II are the cases for the option 1 and 2(2') PXD integration schemes, respectively.*

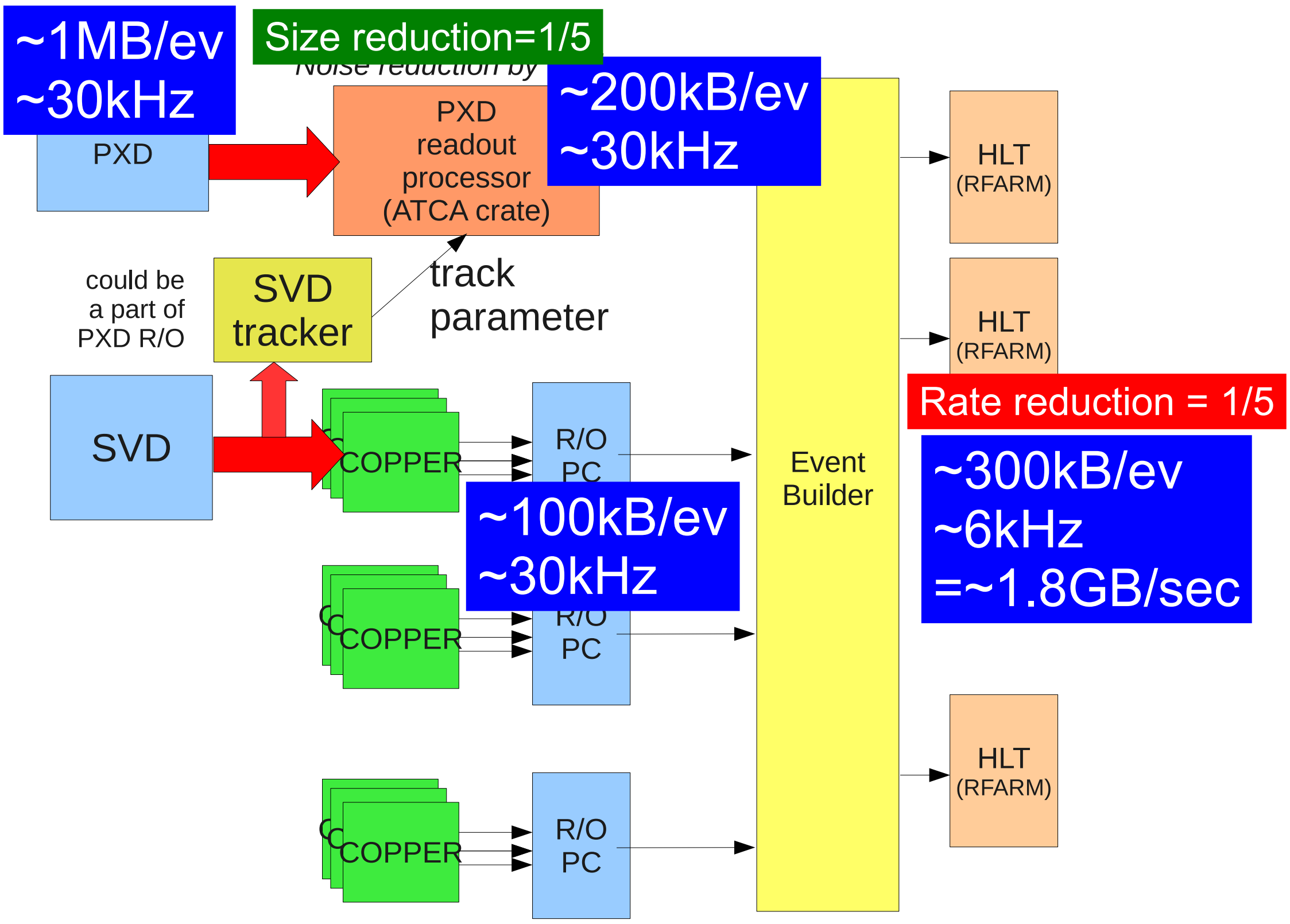

*Figure 13.28: Data reduction steps for option 1 PXD integration (Worst I).*

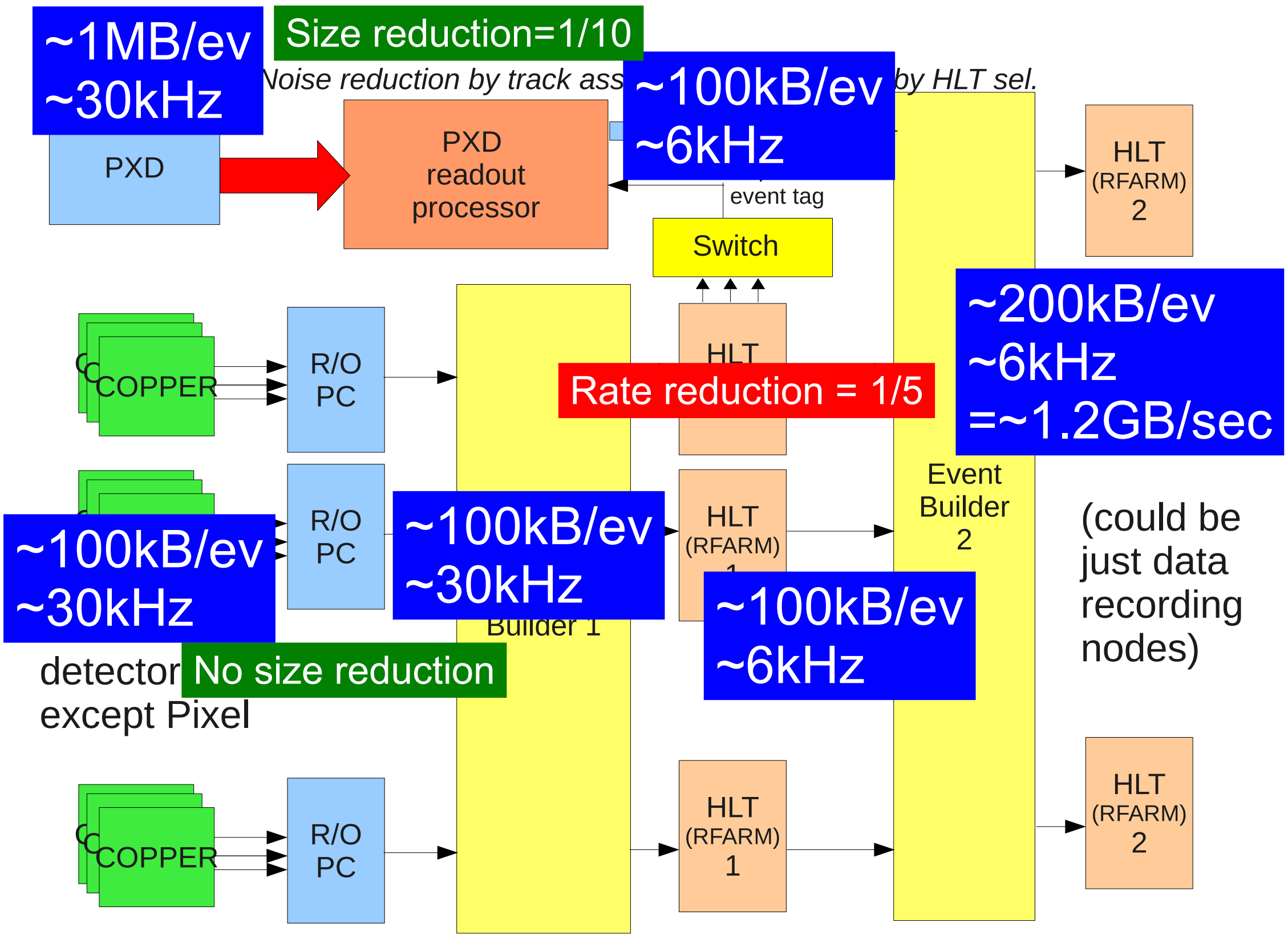

Figure 13.29: Data reduction steps for option 2 PXD integration (Worst II).

# Chapter 14

# Computing and Software

## 14.1 Overview

The Belle experiment [1], over its ten-year operation [2], was supported by a centralised computing facility at KEK, where almost all data processing, MC production and physics analysis was performed. The system was well designed for the amount of data and worked smoothly.
However, the Belle II computing system has to handle an amount of data eventually corresponding to 50 times the Belle level by the end of 2020. This means an amount of raw data of the order of $10^{10}$ events per year. It is required that this data be processed without any delay to the experiment data acquisition, in addition to the production of MC events corresponding to at least 3 times of real data. Moreover, computing power for physics analysis has to be provided. To ensure sufficient computing resources to carry out our physics program, we have to enable all Belle II member institutes to contribute. Therefore, a distributed computing model based on the grid will be adopted. KEK will host the main center that is responsible for raw data processing. Grid sites allow users to produce ntuples from skimmed datasets; they also handle MC production, possibly complemented by Cloud Computing facilities. Finally, users analyze ntuples on local resources. The computing model is discussed in detail in Sec. 14.2.
The data management system is essential to enable users to effectively and smoothly analyze datasets. Due to the limited resources and short time-line in the preparation of Belle II computing, distributed computing concepts from other successful experiments have been adopted. The key ideas reused are metadata and project structure from D0 and CDF experiments, keeping the system as simple as possible, and the use of grid (gLite) services as much as possible, as in the LHC experiments. The technical details of the data management system are described in Sec. 14.3.
C++ has been adopted as the programming language for the offline software, but the use of python is also allowed when it shows clear advantages. Physics analysis is supported by providing a framework that allows users to 'plug in' appropriate reconstruction modules for their specific needs. To realize this, a common software framework, where each module can handle the event data through an unified method such as ROOT I/O based object persistency, is desired. Other processes, such as DST processing, simulation, and data skimming, are done within this common framework. The simulation and reconstruction tools are developed mainly by Belle II computing group members, supplemented with the standard tools in the field of high energy physics. Offline software issues, including the geometry handling method, are covered in Sec. 14.4.
In Sec. 14.5, the required human resources to support and maintain the proposed distributed computing system and associated software are discussed. Additionally, the requirements for computing hardware based on the design luminosity of the SuperKEKB accelerator are esti-mated.
In summary, Belle II has decided to employ a world-wide distributed computing system and to use state-of-the-art tools and concepts in the offline software. To develop, establish and operate this system successfully, substantial resources are needed, both human and hardware, at the main computer facility at KEK and at grid sites at the local institutes of the Belle II collaborators.

## 14.2 Computing Model

This section details the Computing Model for the Belle II experiment. The purpose of the model is to provide all members of the collaboration easy access to experiment data and facilities on which to process it, in addition to ensuring the resources for processing and storage are distributed accordingly.

### 14.2.1 Basic Considerations and Outline

To reach our goal of publishing high-quality physics results in a timely manner, the Belle II computing system has to accomplish several tasks. The raw data taken by the detector and recorded by the data acquisition system has to be stored. After calibration constants dependent on the condition of the accelerator and detector have been determined, the raw data is processed and the derived higher-level information is stored in smaller files (mDST). Subsets composed of selected events are the input to physics analyses, which usually create files containing even higher-level, more condensed analysis-specific information in a user-defined format. While the production of these files, called ntuples, is typically done only a few times, the ntuples themselves are processed frequently, requiring the turn-around time for this step to be fast.
In addition to the real data, high-statistics samples of simulated data (Monte Carlo, or MC) are needed to derive physics quantities from the real data. Generic MC samples corresponding to several times the statistics of real data are created so that they reproduce the conditions in real data as faithfully as possible. Much smaller samples of simulated signal events are generated for individual analyses. The output of the MC simulation are mDST files, from which ntuples are produced that are analyzed in the same way as is done for real data.
To accomplish these tasks, the Belle II computing model uses a grid-based approach, with a dispersed set of facilities connected by a common software layer (the 'middleware'). This is a fundamental change compared to the computing model of Belle, where almost all computing resources for raw data processing, MC production and physics analysis were provided by KEK. With an anticipated data sample of about 50 times the size of the Belle data over ten years, the computing resources required to process and analyze it increase faster than the projected performance of CPUs and storage devices. Therefore, we cannot expected KEK to provide all computing resources for the whole Belle II collaboration. In addition to sharing the responsibility for the construction of the detector, we have to share the allocation of computing resources among the Belle II institutions.
However, not all tasks described above are equally suited for a distributed system. Because the raw data is produced at KEK—the location of the Belle II detector—this site is distict from other centers. It will play the role of a main center and be responsible for the processing and storage of raw data as well as the operation of central services. Thus, it has to provide a high availability of at least 98%. As MC production can be distributed easily, we will use grid sites for this task. We also envision analyzing the produced MC samples and replicated real data samples on the grid. The distribution of grid sites should reflect the geographical distribution of Belle II collaborators. The fractions of resources will be determined by a Memorandum of Understanding. Together with KEK, the remote grid sites form the bulk of the Belle II computing resources and their deployment and operation is coordinated centrally. For efficiency and scalability, experiment-specific services will most likely to be distributed globally to selected grid sites.

While ntuple-level analysis could be done on the grid as well, a fast turn-around time usually requires having the ntuples available on resources local to the user. These local resources are ideally grid-enabled, but we explicitly include non-grid resources, like private clusters at institutes, desktops or laptops, for which we will provide means to install the software needed for ntuple analysis and for access to the Belle II grid system.

The classification into main center, grid, and local resources describes the tasks for which the resources are used. Physical sites may contribute to more than one task. For example, KEK will provide resources for raw data processing, MC production and data analysis, and ntuple-level analysis. Figure 14.1 illustrates the relation between physical sites and their function.

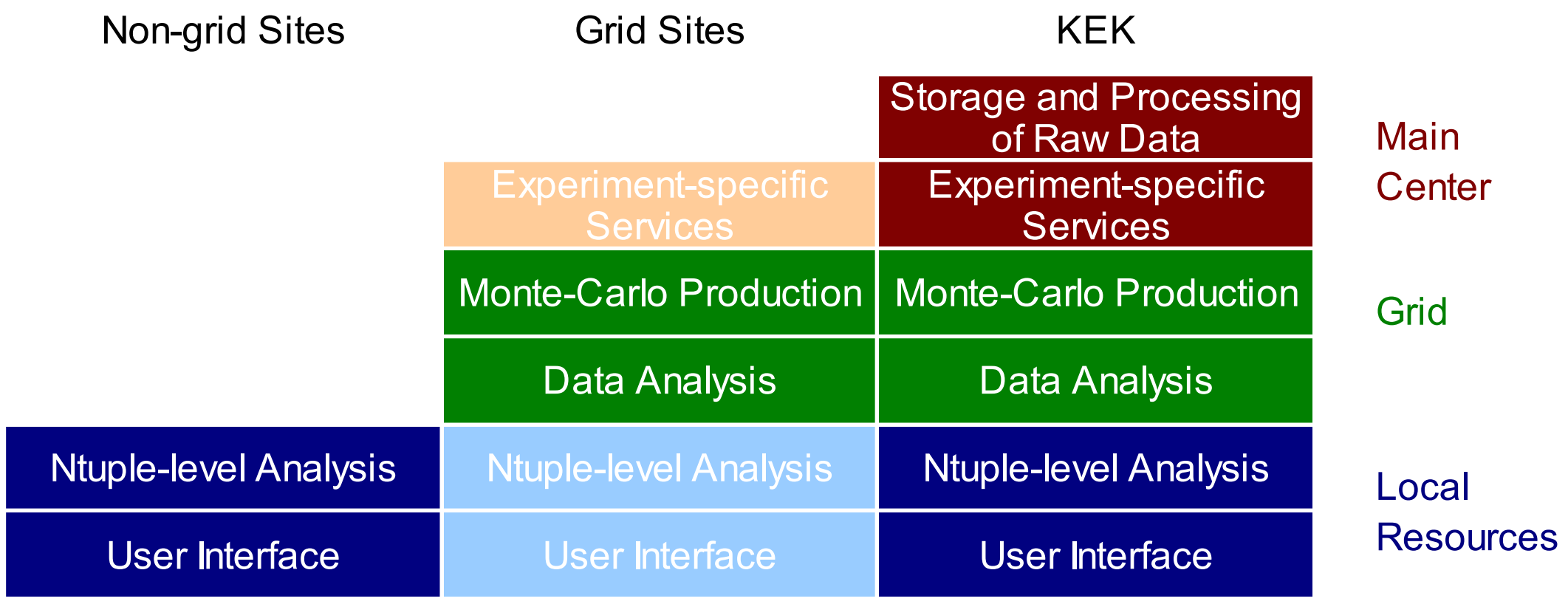

*Figure 14.1: Tasks of computing facilities.*

We will use grid technology, based on the EGEE middleware gLite [3], to manage the increased complexity due to a distributed computing model. This allows us to benefit from the infrastructure that was developed and is maintained for the LHC experiments. By the time the Belle II experiment will start taking data, the LHC experiments will have collected data for a few years. We will profit from their experience and will have well-established and mature solutions at hand.

While the operation of a distributed computing system requires more human resources than a centralized one, the application of standard grid technologies will help to minimize the additional effort. Moreover, many Belle II member institutes already have running grid sites and the addition of support for Belle II requires much less effort than the setup and operation of dedicted remote computing systems, a solution that was adopted for the distributed MC production at Belle.

### 14.2.2 Cloud Computing

An important fact we have to consider in the design of the computing model and the planning of resources is the variation of resource demands with time. Two effects on different scales are involved. First, there is a steady increase of required resources because of the growing data size.

This is taken into account in the planning on a yearly basis. Second, there are variations on a shorter time scale of months or weeks.
One reason for the short-term variability is that the raw data processing and the MC production can only start after a data taking period (which usually lasts several months) has finished and calibration constants have been determined. To have the data and MC ready for analysis in a timely manner, it is advantageous to save CPU resources during the data taking phase and spend them when the processing phase starts. However, CPU resources can not be easily saved because idle CPU time is lost and can not be regained. To some extent, this issue can be solved by fair-share mechanisms when several virtual organizations on the grid share CPU resources at the same site. However, it is unlikely that the resource demands of all virtual organizations balance each other at all times.
Another case of varying resource demands comes from the physics analysis. Experience has shown that the request for CPU resources usually increases before important physics conferences. In this particular case, the fair-share mechanism will probably not provide additional resources because the conference deadline typically applies to all high energy physics experiments.
Computing system are usually designed to match the expected peak demand. This has the risk of a resource shortage if the demand exceeds the expectation. This situation is illustrated in Fig. 14.2 a). Cloud computing can provide a solution by offering the possibility to purchase CPU resources for a limited amount of time. This mitigates the problem of a CPU resource shortage due to unexpectedly high demand, as financial resources can be transformed into CPU resources almost instantaneously. Figure 14.2 b) demonstrates the advantages of cloud computing.

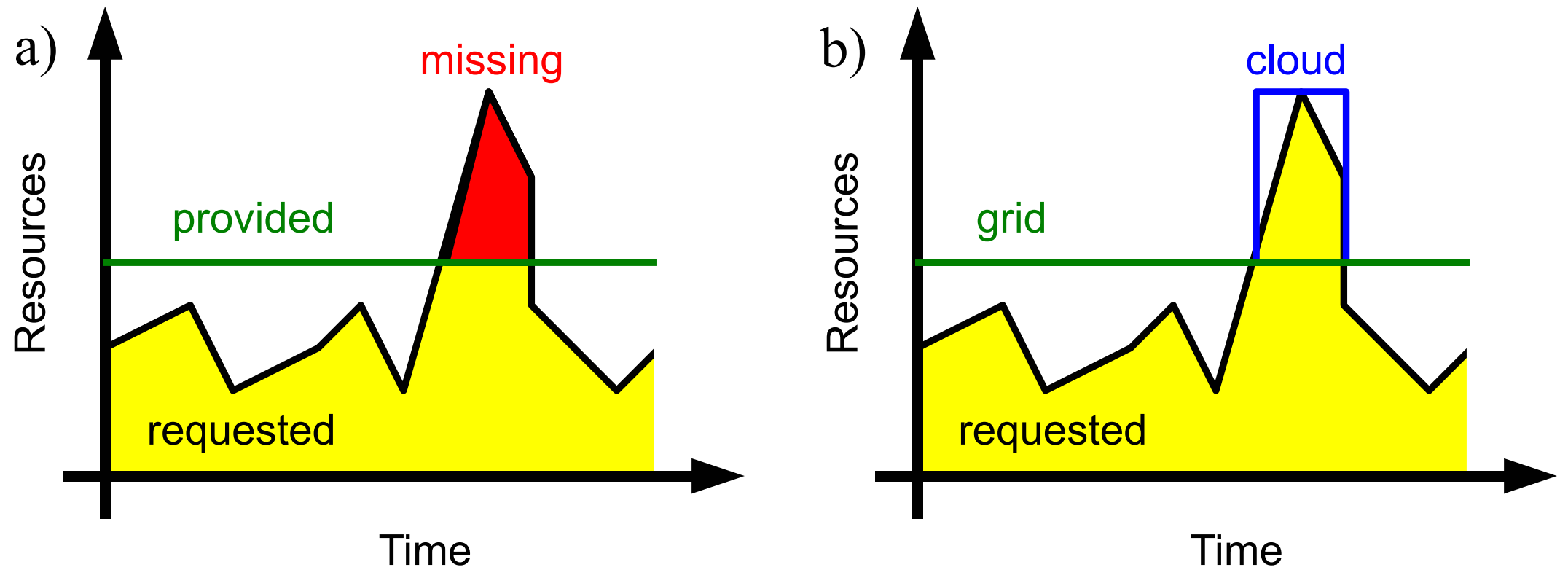

*Figure 14.2: Comparison of a classical system a grid sites (a) and a system utilizing cloud resources in addition (b) for the case of varying resource demands.*

In contrast to publicly funded grid sites, the cloud computing facilities are usually operated by commercial companies. This may change in the future, as a clear trend towards cloud computing technologies can be observed in the grid computing development. Although more and more cloud facilities funded by governments may emerge, the aspect of buying resources from a company that aims for financial profit deserves special attention. There is, in particular, the issue of a potential vendor lock-in, which we must avoid. For this reason, we do not plan to use cloud resources for permanent storage of real data or MC. In the worst case of a disappearing vendor, relying on cloud storage could result in the unacceptable risk of data loss.
The risk of a vendor lock-in is less severe for CPU resources, as no commitment to buy from the same vendor in the future is involved. However, we have to design our software in a way that the technical issues of a vendor change are solvable with reasonable effort. The technical aspects of making cloud computing resources available to Belle II are discussed in Sec. 14.3.6. An evaluation of the cost of cloud resources is presented there as well.
To summarize our strategy concerning cloud computing, we see that this technology has a high potential and we feel obliged to keep the possibility of using it for Belle II. While our baseline computing resources will be provided by grid sites, cloud computing will be an option for peak demands in MC production or physics analysis.

### 14.2.3 Raw Data Processing

As outlined in Sec. 14.5, the expected raw data size is of the same order of magnitude or even larger than the one of LHC experiments. The storage and processing of raw data therefore require considerable resources. Given the amount of data, the cost of storage is an important parameter. Currently and in the foreseeable future, the best ratio of cost/PB is provided by tape systems. Compared to disks, tapes also have the advantages of being storable in libraries and consuming no power when idle. Their drawbacks are their inherently sequential access and the high latency associated with the mechanical mounting of tapes in an available tape drive. Since raw data is stored only once and read only very few times for the (re)processing, and because this is done in a managed way, these limitations are acceptable. Thus, we use magnetic tape libraries as the storage technology for raw data.
The raw data has a well-defined source: the Belle II detector. Therefore, the raw data processing chain has to start at KEK. While HEP experiments traditionally perform all data processing steps at the site where the detector is located, LHC experiments decided to ship raw data to remote Tier1 sites for reprocessing.
Given the similar order of magnitude of total raw data size compared to LHC experiments, we explored a distributed raw data processing model as well. However, we decided against this approach for the following reasons.
The distribution of raw data shares the load of storage and processing among sites, but it does not reduce the overall resource requirements. On the contrary, the maintenance effort for the operation of a raw data processing center scales with the number of sites. Each site has to operate a tape system, which has been shown in the past to be a labor-intensive task. The requirements are exacerbated if the raw data has to be imported almost synchronously with its production at the detector. An operation with an efficiency close to 100%, 24 hours a day, 7 days a week, is needed in such a case. Such a quality of service has to be ensured for the tape system and for the network connection. Moreover, the network traffic caused by the distribution of raw data is huge. Therefore, the storage and processing of raw data is concentrated at KEK in our computing model.
Having no backup copy of the raw data increases the risk of raw data loss, for example in the case of a fire. This risk has been taken by most pre-LHC experiments, including Belle. Since the reconstructed mDST data will be copied to remote sites, we will still have all data for analysis available in case of a complete data loss at KEK. The possibility of reprocessing would be lost in this situation but, given the additional effort of a raw data backup, this risk seems acceptable. Of course, if the resources for a raw data backup would become available to us, we would revise our decision.
In addition to the tape system with appropriate disk buffers, the CPU resources to process the raw data are needed at KEK. The number of required CPU cycles depends on the time period in which we want to finish the reconstruction. This enters our computing model via the parameter $T_{Data}$, the time anticipated for the processing of the data taken in one year. In the interest of a timely analysis, this parameter should be as small as possible.
The possibility of reprocessing data in case of updated calibration constants or new reconstruction code has to be taken into account in the computing model as well. One option is to utilize the idle CPU resources between phases of processing new raw data. As the reprocessing will likely happen more frequently in the start-up phase of the Belle II experiment, we may store the raw data during this period not only on tape, but also on disk, to achieve a fast turn-around time.
The first step in the processing chain is the determination of calibration constants. Only a small fraction of the total data sample is needed to perform this task. The calibration data files (DST) will be written to disk in parallel with the storage of raw data on tape. After the calibration constants have been determined and written to the central database, the raw data is read from tape and the reconstruction software is run on it. The output of the reconstruction are mDST files that are stored on disk. Several samples of preselected events, called skims, are created and stored in mDST format on disk as well. The selection criteria are adjusted to different classes of analyses. The mDST files are the input for all physics analyses, described in Sec. 14.2.5.

### 14.2.4 Monte-Carlo Production

For most analyses, it is essential to have an accurate simulation of signal and background events. To provide this, generic, run-dependent MC datasets are produced. To limit the systematic uncertainty associated with the number of simulated events, a MC sample corresponding to $N_{streams}$ times the data size is created, where $N_{streams}$ is a parameter of the computing model and should be at least 3. At Belle, $N_{streams}$ is usually set to 10, a reasonable value to have sufficient MC statistics.
In contrast to the raw data processing, the production of MC data requires no input data, except for some constants and a few background files. This makes this task very well suited for a distributed environment. We plan to share the MC production work among the grid sites. The production will be planned and managed by a MC production coordinator in collaboration with the Distributed Computing and Data Management Group Convener. If appropriate, cloud computing resources may be used as well.
In addition to the integrated luminosity and $N_{streams}$, the needed CPU resources depend on the time period, $T_{MC}$, in which the MC corresponding to one year of data taking is produced. The parameter $T_{MC}$ should be of the same order as $T_{Data}$ to avoid delay in physics analyses due to missing MC samples.
The output of the simulation jobs are files in the same format as processed real data. The MC mDST files are stored on disk at the storage element close to the site where they are produced. Cloud computing jobs need a nearby storage element, too. Their output will not be stored on the cloud, but transferred to storage elements at grid sites. Like the data mDST files, the generic MC samples are the input for physics analyses.
Some analyses need further specific samples of high statistics simulated signal events. If these signal MCs are of interest for several people, they will be produced at one or more grid sites in a managed way, like generic MC. The output is made available to all collaborators on a nearby storage element. The production of very specific signal MC samples is treated as part of the physics analysis and is not coordinated centrally.

### 14.2.5 Physics Analysis

The aim of the Belle II experiment is to produce physics results. Thus we have to provide efficient, reliable, and user-friendly access to data and CPU resources so that people can focus on the physics analysis.
Since the processing of mDST data files is an essential part of all analyses, we will replicate mDST data skims to grid sites. The replication process is managed by the Distributed Computing and Data Management Group Convener and the Site Representatives, but not triggered automatically by the user. The decision of which dataset to be copied where depends on the size of the site and the physics topic(s) the users at that center are interested in. If it seems beneficial, generic and signal MC samples can be replicated in a managed way as well.
The usual method to analyze the data and MC samples is the creation of ntuples. For this task, the user submits grid jobs that run at the grid sites. A front end will be provided to facilitate the bulk submission of jobs to the centers that have the dataset of interest. We pursue two approaches: a simple push model and a more sophisticated pull model that allows for a better steering of assigned resources.
It is essential for an analysis that each input file of a dataset is processed exactly once. To achieve this, we consider two options: a static and a dynamic assignment of files to jobs. The details of both options, as well as of the job submission tool, are described in Sec. 14.3.
The output of the analysis job, the ntuple, is eventually stored on resources local to the user that initiated it. This could mean that local storage space has to be made accessible from grid jobs, typically via a grid storage service; alternatively, grid sites may provide limited temporary storage to buffer the job output until the user can copy it to their local storage space.
The analysis of the ntuples can then be done entirely on local resources. No grid infrastructure is needed at this stage.

### 14.2.6 Summary

The Belle II computing model assumes the following structure based on grid technologies:

- The main center is located at KEK and is responsible for the raw data storage and processing. In addition, it provides several central experiment-specific services, like the conditions database, and grid services, like the file catalog and the virtual organization management. The main center is expected to have 24x7 operations, designed for extreme reliability. Staff dedicated to maintaining grid services, supporting users, managing the virtual organization, liaising with other sites are required.

- Grid sites provide computing resources geographically close to the users. KEK will have the role of a grid site, in addition to the role of the main center. Grid sites provide resources for the managed production and storage of generic MC samples. They also host replicas of data skims and allow users to produce ntuples from data and MC samples.

- Ntuple-based analyses are usually performed on local resources (desktops and institute clusters).

Additional processing resources for MC production and analysis may be provided by cloud computing in case of peak demands. Figure 14.3 illustrates the Belle II computing model.

## 14.3 Distributed Computing and Data Management

### 14.3.1 Overview

Belle II will make extensive use of distributed computing solutions built upon gLite[3] middleware. By doing this, we leverage the many hundreds of man-years of development from the EGEE [4] and WLCG [5] projects and their predecessors.

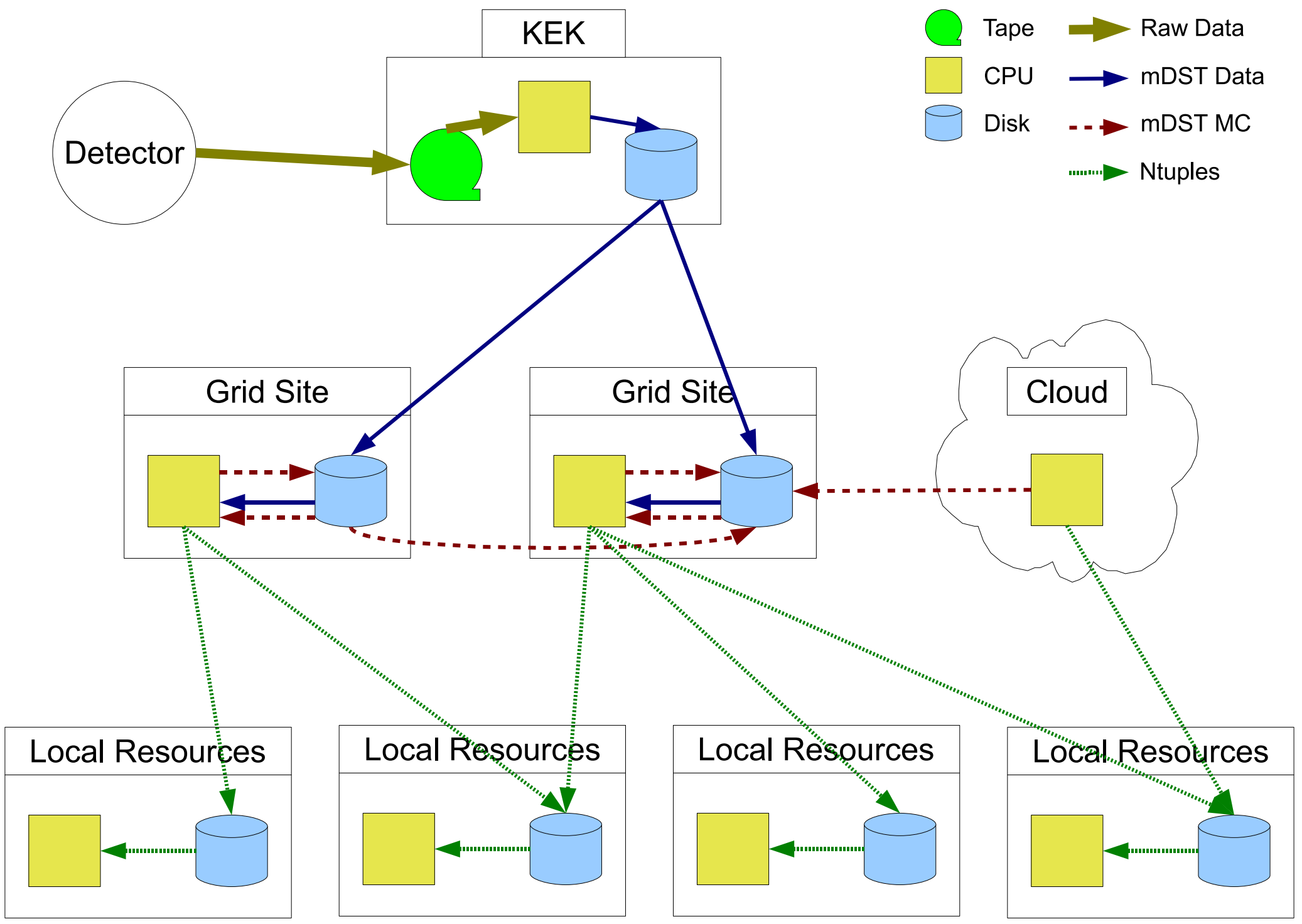

*Figure 14.3: Concept of the Belle II computing model.*

The gLite middleware enables both data and CPU resources to be located at a variety of sites around the world. It also enables the movement of data between storage resources and allows jobs to run at sites with CPU resources.
We plan to deploy most pieces of the gLite middleware used by the LHC experiments. In addition, we plan to develop a new piece of middleware, the "Project Server," to steer particular tasks amongst Belle II's distributed compute and data resources.
To simplify the model and ensure only data of interest is located at grid sites, we do not intend to automatically distribute data. Instead, administrators will migrate selected data to sites at the request of users.
Since the centers will provide substantial computational resources beyond that of KEK, we expect that data skims in high demand may be duplicated across the Belle II network.
Consequently, it is essential that all data, whether real or generated from MC, along with data-skims be registered in a central database, called the LCG File Catalogue (LFC). The LFC will record the location of the every physical location (PFN) against a logical reference (LFN/GUID). In addition, each logical data item (files, datasets) will have associated metadata recorded in an AMGA metadata catalog. The metadata catalog will greatly increase the ability of users to quickly find data of interest.

### 14.3.2 Analysis Model

An important part of physics analyses, and raw data processing, is the processing of a set of selected files with a given piece of (user-written) analysis code. We call this task a project. To obtain correct physics results, it is essential that each input file is processed exactly once. If files are missed or files are processed twice, the obtained result of the analysis is most likely wrong. Thus, we have to design a system that minimizes the risk of failing this requirement.
As the number of files belonging to a project is usually large, their processing is done in several jobs. From the system's point of view, all these jobs are independent and unrelated. However, from the user's point of view, they belong to one project. This means we have to provide a component on top of the grid job layer that allows the user to interact with it on the level of a project. This includes the job submission, the monitoring of the progress, and the recovery of failed jobs. The last part is of particular importance to fulfill the requirement of processing each file exactly once.
Our strategy for the development of the project component is to start with a simple system and then investigate further options that would increase the convenience for the user, but also the complexity of the system.
The first problem we have to solve is to identify the right files that should be used as input for a project. For this task, we set up up an AMGA metadata server, described in the next section. It allows the user to get a list of files, in form of GUIDs, that match certain physics-motivated criteria. For example, the user may wish to "run over all full reconstruction skims of data collected between June 2013 and January 2015."
The next issue is to determine where the files are located on the grid and, thus, to which sites the jobs that process these files should be submitted. The first part is solved by the LCG File Catalog (LFC). It matches GUIDs to physical locations of files in form of Storage URLs (SURL). To shield the user from the interaction with the AMGA server and the LFC and to manage the grid jobs on a project level, we will provide a project client. The client will be part of the Belle II offline software and installed on the local machine of the user. Authorization will be performed with proxy certificates. When the user wants to start a project, they call the client and provide the following information:

- A string with a metadata query to identify the input files.
- A tarball containing the analysis code.
- The name of the script inside the tarball that should be executed.
- The release version of the Belle II software that should be used.
- The desired number of jobs, or alternatively the desired number of input files per job.

The project client then talks to AMGA and the LFC to get the list of GUIDs and matching SURLs. Based on the SURL, it determines the sites where the data file are available. Then it submits the desired number of jobs to these grid sites. All jobs may be submitted to one site or they may be split across several sites. Optionally, the user can provide a list of preferred sites. The job submission system will be based on some lightweight service for job management operation at the Computing Element (CE) level. A candidate for this is the CREAM (Computing Resource Execution And Management) [6] service.
The input files are distributed equally to all jobs. The assignment of files to jobs is realized by creating a script for each job and setting an environment variable in this script to the assigned SURLs. The script that is run at the CE then executes the user code after setting up the right software version. An input module in the analysis framework will evaluate the environment variable and process the assigned files.
To store the output file(s) on an SE and register them in the LFC and AMGA, a tool will be provided within the Belle II software. It selects either the nearest SE or a predefined one that the user specifies at job submission. A fall-back solution, if the selected SE is unavailable, may be implemented.
In addition to submitting the grid jobs, the project client will store the list of GUIDs, selected SURLs, and corresponding job identifiers in a local file. This allows one to identify the project via the local project file.
Now the user can query the project client for the status of their jobs by providing the name of the project file. From the project file, the client reads in the job identifiers and determines their status via standard grid mechanisms. The status of successfully finished and failed jobs will be recorded in the project file to avoid further unnecessary queries.
Based on the status information in the project file, the client can also easily either resubmit all failed jobs or generate a new project containing only the unprocessed files. This recovery mechanism simplifies the bookkeeping effort for the user enormously.
Figure 14.4 summarizes the procedure of starting a project. Not shown in the diagram is the conditions database used to provide the calibration and alignment constants. This database will be distributed to each site.

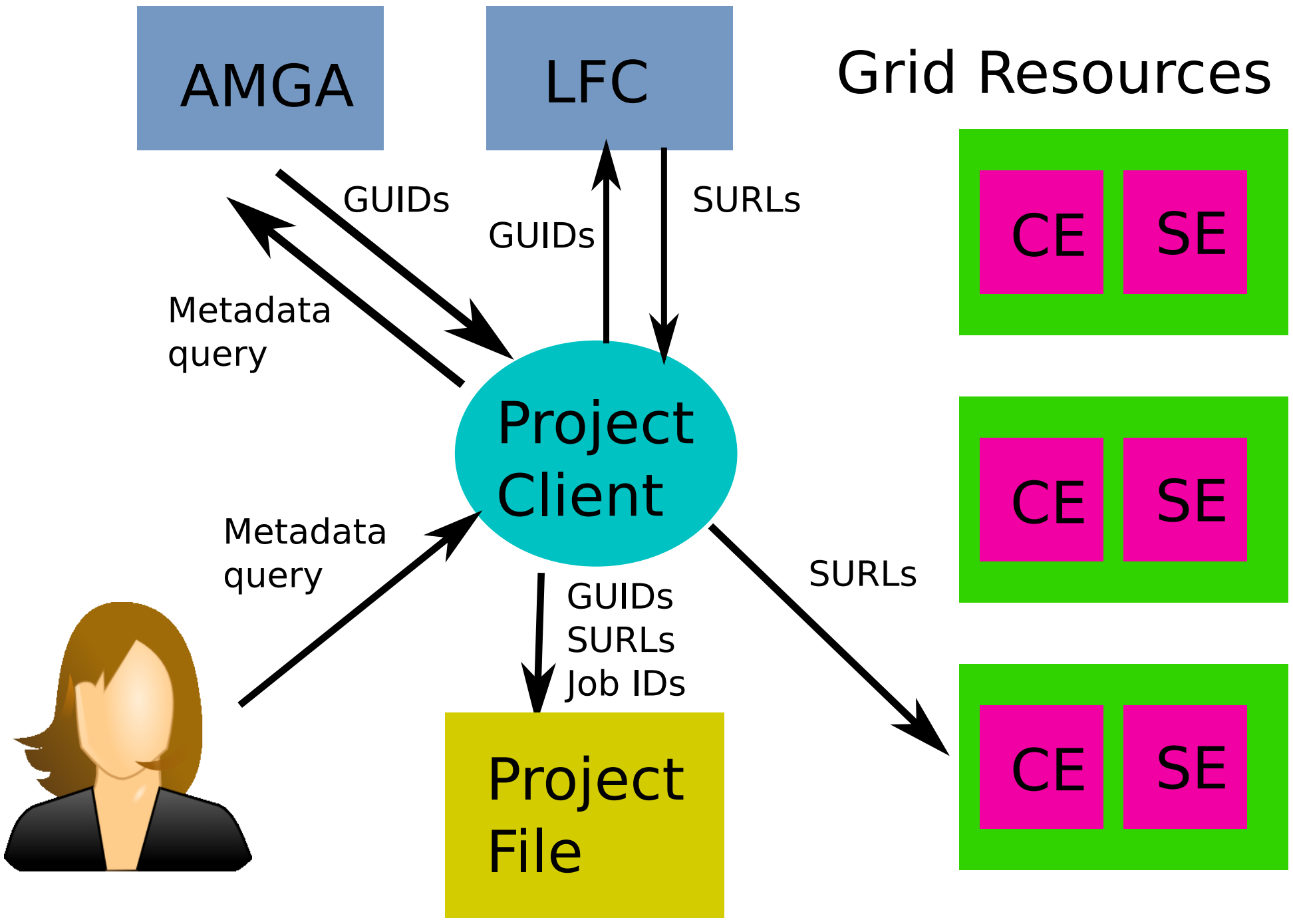

*Figure 14.4: Outline of user interaction with the project client to submit jobs to grid sites for the analysis of a dataset defined by a metadata query.*

As an extension of this baseline system, we investigate the idea of a dynamic assignment of input files to jobs, inspired by the success of this concept in the SAM [7] system employed by the CDF and D0 experiments. To realize this mechanism, the project bookkeeping has to be moved from a local file to a central service known as the project server.
Now, the project client just passes on the metadata query to the project server when the user starts a project. The project server contacts AMGA and the LFC as the client did in the simple model. Instead of a local file, a database is used to store the list of GUIDs. The list is identified by a unique number, the project ID. It takes over the role of the project file name. The project server returns the project ID to the client together with a list of sites on which the data is available.
The client prepares a single job script with the project ID in an environment variable. The desired number of jobs is submitted to the site (or sites) indicated by the project server with the same single job script. The procedure up to this point is illustrated in Fig. 14.5.

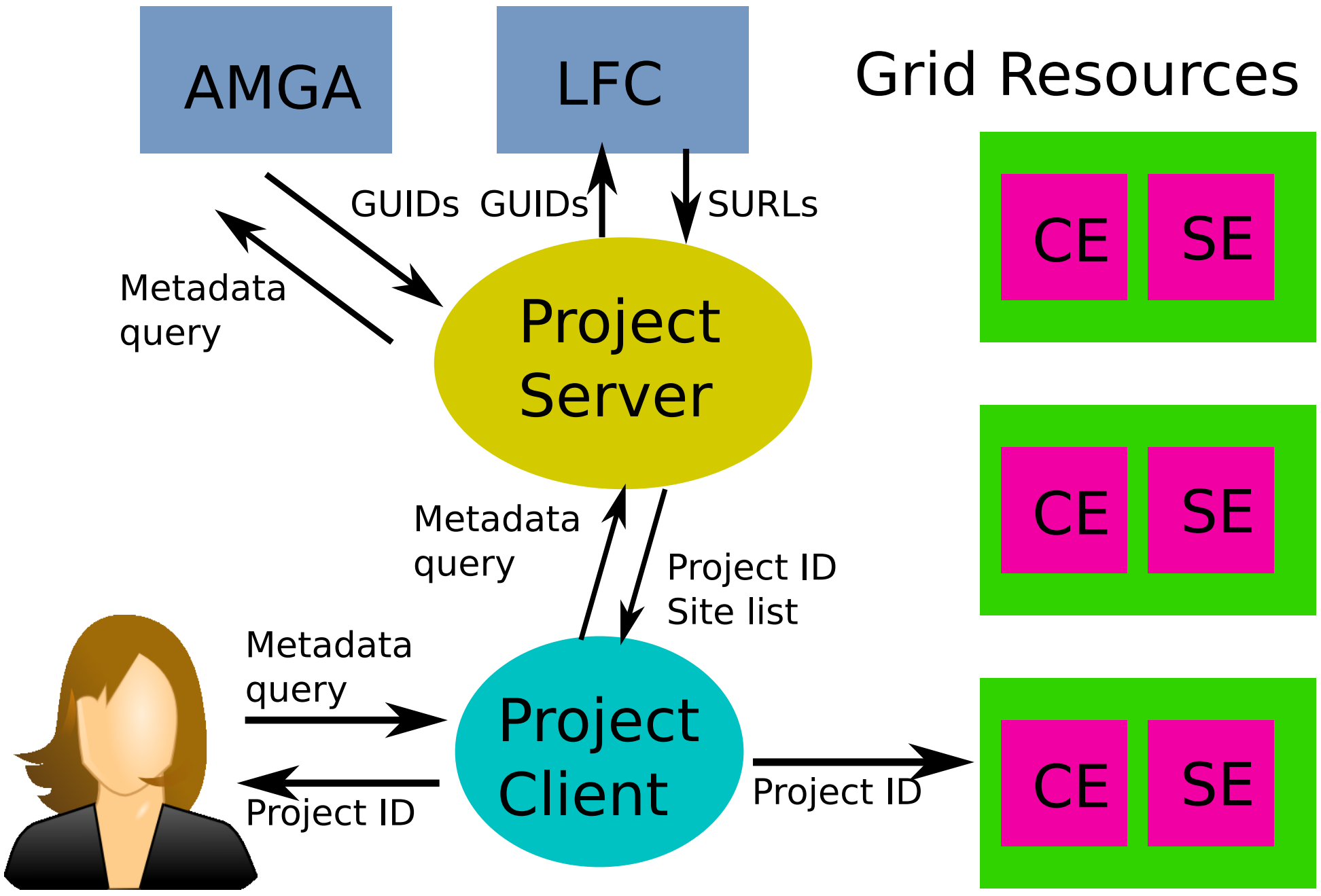

*Figure 14.5: Start of an analysis project in the case of a dynamic assignment of files to jobs.*

Upon execution at the CE, an input module of the analysis framework contacts the project server with the project ID obtained from the environment variable. From the project ID, the server knows which input files should be processed. It picks a file that is available at the site where the job runs and returns the SURL to it. This mechanism of dynamic assignment of a file to a running job is depicted in Fig. 14.6.
In the database, it is noted that the file was given to a job by assigning the job ID to the file. The analysis job processes the delivered file and contacts the project server again when it is done with the file. The successful (or unsuccessful) processing of the file is recorded in the database and another file name is returned if requested. Once the job has processed a maximal number of input files or there are no unprocessed files left, the analysis program stops, the output file is written to a SE, and then the job informs the project server about a successful completion of its task.
The project client returns the project ID to the user when the project is started. With this ID, the user can request the status of their project from the client or ask it to generate a recovery project. The client then talks to the project server to accomplish both tasks as shown in Fig. 14.6. From the user's point of view, the static and dynamic file assignment systems differ only by the way how projects are identified (local file vs. ID).
The extended approach is certainly more complex and introduces a further possible single point of failure, the project server. But it has several advantages:

- The job submission preparation is much easier. It is no longer necessary to have many individual jobs with a static assignment of input files to jobs. All jobs are identical. They just have to know the project ID, which is common to all jobs.

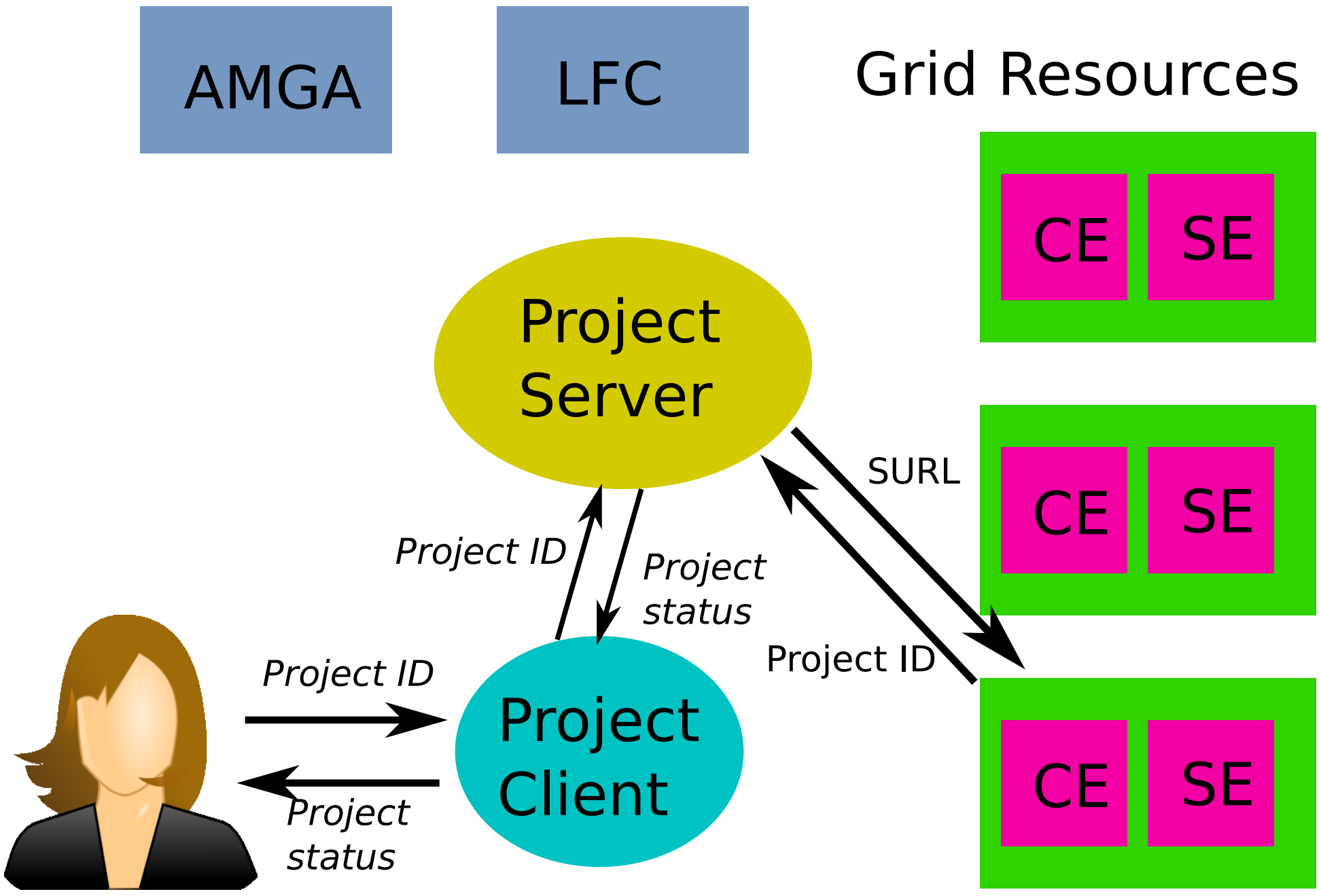

*Figure 14.6: Interaction between a grid job and the project server to dynamically assign an input file, and (in italic font) the handling of a user query to obtain the status of a project.*

- A more fine-grained monitoring is possible. The progress can not only be measured in terms of finished jobs, but also in terms of processed files.
- It is ensured that only properly running jobs get data assigned. If a site is misconfigured so that the analysis program cannot be executed, no files are assigned to jobs at this site. Thus no recovery of these jobs is needed. In case of a resubmission, in the simple model the recovery procedure would not work for misconfigured sites. In the system with dynamic file assignment, only properly working sites are used for processing.
- The overall time to finish a project is reduced. In the diverse grid environment, the time that jobs wait in a batch system queue and the performance of the execution host can vary significantly. In the simple model, the end of a project is reached when the slowest job has finished all its input files. In case of a dynamic file assignment, there is an automatic load balancing between jobs that start earlier or run on a faster host and jobs starting later or on a slower host. The former will request and process more files than the later ones. In this way, the time to finish a project is no longer determined by the slowest job.

Both systems, with static and dynamic file assignment, can be easily extended for the purpose of run-dependent MC production, as explained in Sec. 14.3.5. In this case, the file is just replaced by the input information for the MC job: experiment number, run number, and total number of events that should be generated (Fig. 14.9).

#### 14.3.2.1 Pilot Model

Keeping track of the availability and status information of every resource in a virtual organisation in a timely fashion is a challenge. Furthermore, using this data to schedule jobs optimally from a central location is an NP-complete problem [8].

Recently, many large grid users have moved toward a late-matching job submission model, or pilot model. In this approach, 'pilot' jobs containing a script to obtain real jobs from a central task queue once executing on a worker node are submitted to all grid sites. This moves the matching of resources to jobs to the worker nodes themselves, greatly reducing the load on central services.

There are several auxillary benefits to this approach, including allowing the assignment of priority to user jobs on a virtual-organisation wide basis, verification of environment suitability and confirmation of resource availability. The only disadvantage of the pilot model is that it requires additional components for the job management that are not provided by the generic grid middleware. However, many experiments have working solutions, and we are evaluating how these can be adapted to Belle II.

### 14.3.3 Metadata

An essential requirement for the Belle II computing system is the ability to select a set of files for processing according to physics criteria. For example, users may want to analyze the data of certain data taking periods, called experiments. Other frequently used selection criteria are the type of preselected sample, called a skim, and the version of code with which the data was reconstructed. Such information has to be known for each file to decide whether it should be included in the input data of the processing jobs using it. Information about the type of data contained in a file is called metadata.

For the Belle II experiment, we need a reliable and efficient metadata service that translates selection criteria into a set of files. There is already a metadata service at the Belle experiment. However, the existing metadata service has problems with performance, scalability, and robustness, which makes it inappropriate for Belle II. Moreover, the solution applied by Belle is not intended to be used in a distributed environment. Therefore, we design our metadata service based on AMGA (Arda Metadata catalog for Grid Application) [9] in order to solve these problems.

AMGA is the official metadata service for EGEE, which is designed to allow high performance access to metadata suitable for the demands of applications in high energy physics. AMGA is designed to ensure good performance and scalability, grid-based authentication, database connection pooling, a hierarchical table structure and a replication mechanism. We have designed an AMGA schema, with which the metadata service has been implemented for the Belle II experiment. In addition, we have implemented a user-friendly command line tool for metadata queries.

The metadata is extracted from the data files that are written to storage resources and used to fill the AMGA metadata catalog. Using the catalog, a user query is translated into a file list and a list of sites where the data is available. This information is used to submit jobs that process the requested data as shown in Figs. 14.4 and 14.5.

In addition to this file-level metadata, we are also investigating the idea of using AMGA for the selection of events based on event-level metadata. This may considerably speed up analyses that select only a small fraction of events in a large dataset. Unfortunately, the large number of events increases the size of the metadata catalog by at least one order of magnitude compared to the file-level only solution.

#### 14.3.3.1 Metadata Schema Definition

AMGA uses a relational database back-end to store the metadata. Several database products are supported, as shown in Fig. 14.7. AMGA provides a tree-like structure to arrange different kinds of metadata. The leaves of the structure contain the attributes of the metadata and are implemented by tables in the relational database.

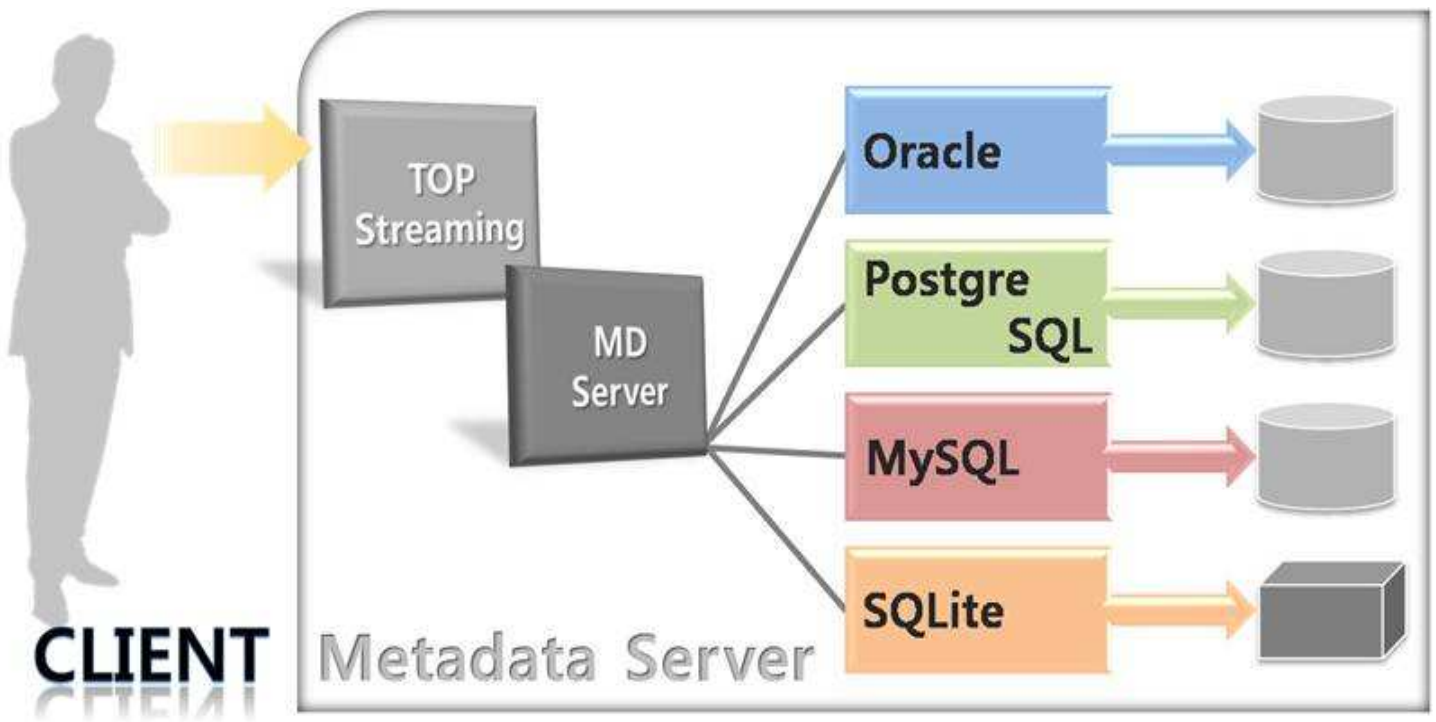

Figure 14.7: Possible AMGA back-ends.

For the directory structure of the file catalog (LFC) we envision the following scheme:

`/belle2/data/e`*[exp #]*`/`*[dataset name]*`/`
`/belle2/MC/generic/e`*[exp #]*`/`*[dataset name]*`/`
`/belle2/MC/signal/e`*[exp #]*`/`*[dataset name]*`/`
`/belle2/user/`*[username]*`/`*[dataset name]*`/`

where *[exp #]* is the number of the data-taking period (experiment), and *[dataset name]* is a name identifying the type of data. The extent to which users will be allowed to register their files in the metadata catalog is not yet decided.
The LFC structure is not well suited for AMGA because it would create too many database tables, which may degrade the search performance. For example, assuming that there are several tens of experiments and dataset types, then there could be more than ten thousand tables created. In PostgreSQL [10], one of the AMGA back-ends, each table is stored in a file and all the tables in a database are managed in a single directory. If there are ten thousand files in a directory, then accessing a file in the directory will consume a large amount of time, since most file systems search through filenames sequentially.
To solve the above problem, the depth of the directory structure is reduced by one layer. The implemented metadata structure is:

| | |
|---|---|
| `/belle2/data/e`*[exp #]*`/FC` | file-level metadata for real data |
| `/belle2/MC/generic/e`*[exp #]*`/FC` | file-level metadata for generic MC data |
| `/belle2/MC/signal/e`*[exp #]*`/FC` | file-level metadata for signal MC data |
| `/belle2/dataset` | metadata for datasets |
| `/belle2/skim` | metadata for skims |
| `/belle2/user_info` | metadata for users |
| `/belle2/site` | metadata for sites |
| `/belle2/software` | metadata for software versions |
| `/belle2/config/path` | path information for metadata |

The attributes of the file-level metadata are listed in Table 14.1. It includes a LFN attribute, which is redundant because the LFC service has a mapping between GUID and LFN. At this time, it is included to allow us to test the metadata service using the existing Belle data, which do not support the grid services.

*Table 14.1: File-level attributes.*

| Attribute | Type | description | example |
|---|---|---|---|
| guid | varchar(40) | grid unique ID (hex format) | |
| lfn | varchar(1024) | logical file name | aaaaa.root |
| status | varchar(16) | good/bad/and so on | good |
| events | int | total number of events | 3913 |
| datasetid | int | dataset ID | 5 |
| stream | int | stream number | 0 |
| runH | int | highest run number | 1490 |
| eventH | int | highest event number | 3913 |
| runL | int | lowest run number | 1490 |
| eventL | int | lowest event number | 1 |
| parentid | int(128) | IDs of parent files | 0 |
| softwareid | int | ID of software version | 1 |
| siteid | int | ID of site where it was created | 1 |
| userid | int | ID of a user who created it | 1 |
| log_guid | varchar(40) | GUID of log file (hex format) | |

The metadata attributes on datasets, skims, users, software versions, and sites can be found in Ref. [11].

#### 14.3.3.2 Estimation of Metadata Size

To judge the feasibility of this metadata scheme, in particular for the event-level option, it is important to estimate the total size of the data that has to be stored in a database. Table 14.2 summarizes the following rough calculation.

According to our estimate of required hardware resources presented in Sec. 14.5.2, we will have about 400 PB of real data (see Table 14.8) and about 50 PB of MC data (see Table 14.9) in 2020. In order to efficiently use our resources, we aim to store the data in large files of similar size. For our calculation, we assume an average file size of 4 GB. Taking these numbers, we expect to have about 100M data and 12.5M MC files for which we have to store metadata in the AMGA database. From the metadata attributes listed above, a size of 600 bytes per file entry in the database is determined.

The expected total number of physics events ($b\bar{b}$, continuum, $\tau^+\tau^-$, 2-photon) in 2020 is about $500\times10^9$, according to Tables 14.6 and 14.4. If we produce MC events corresponding to six times the amount of hadronic events, we obtain $1500 \times 10^9$ events in 2020. Thus, an efficient storage of event-level metadata is essential. This is achieved by using the variable-bit data format that is only supported by PostgreSQL. If Oracle or MySQL were used as a back-end, the size would increase by about a factor of six. With the variable-bit data format, a metadata size of 12 bytes per event was reached.
While the size of the file-level metadata seems manageable, the feasibility, benefits and drawbacks of the event-level metadata option require further investigation.

*Table 14.2: Estimation of metadata size.*

| | # of files | Size for file level | # of events | Size for event level |
|---|---|---|---|---|
| data | $100 \times 10^6$ | 56 GB | $0.5 \times 10^{12}$ | 5.5 TB |
| MC | $12.5 \times 10^6$ | 7 GB | $1.5 \times 10^{12}$ | 16 TB |

#### 14.3.3.3 Metadata Access

There are two main roles in the access of metadata: manager and user. The manager role is responsible for creating conforming metadata when a file is created, and is the only role with write access to the database. Once the metadata is created, it is not expected to be modified under normal operation. The metadata will be replicated to many grid sites, depending on the AMGA replication configuration. If the manager of metadata deletes an entry, it should be removed at the replica sites automatically. Normal users only have read-only access to the metadata catalog. If metadata for user files is allowed to be registered in the central file-level metadata catalog, the above scenarios will be modified.
Users can retrieve information from either a command line metadata access tool or via a web portal. A prototype of the command line tool has already been implemented based on AMGA client version 2.0. The prototype allows a user to query either the file-level or the event-level catalog at their choosing. The prototype applies strong security measures based on grid certificates. Its usage will be optimized by detailed studies of end-user access patterns.
Convenient tools should be provided for the metadata manager, allowing the management of many files with minimal effort, since there may be thousands of files created at the same time. The LFC service will interface with the metadata catalog, using the GUID attribute in the file-level metadata.

#### 14.3.3.4 Metadata Server

One of the notable features of AMGA is the ability to replicate the metadata catalog, or parts of it, to remote sites. The replication was successfully tested with a master database at KISTI and a slave at Melbourne, using metadata of the Belle experiment.
For the event-level metadata case, the size of the metadata becomes too large for a single database. A multi-server setup is considered to guarantee scalability and performance. It employs a redirection system, as illustrated in Fig. 14.8, to allow end-users to access the full data in their analysis.

### 14.3.4 Data Management System

We will employ the standard data management systems used by the WLCG, although we will not employ automatic data distribution to remote grid sites. As raw data is generated by the experiment, it is stored in the main center SE and registered with the LFC and AMGA. Remote grid sites may choose to transfer all or some of the derived data sets created from the main center processing of the raw data.
All subsequent data generations or replications to any SE will also be registered in the LFC and AMGA. We envision numerous derived data skims. These will be replicated around the distributed computing system to match the needs and interests of physicists. Popular data skims like the Full Reconstruction skim may be replicated on all sites.
It is clear that both the LFC and AMGA play a critical role in the correct and efficient operation of the computing model. Experience from the WLCG shows both have proved to be stable, reliable and scalable to our needs.

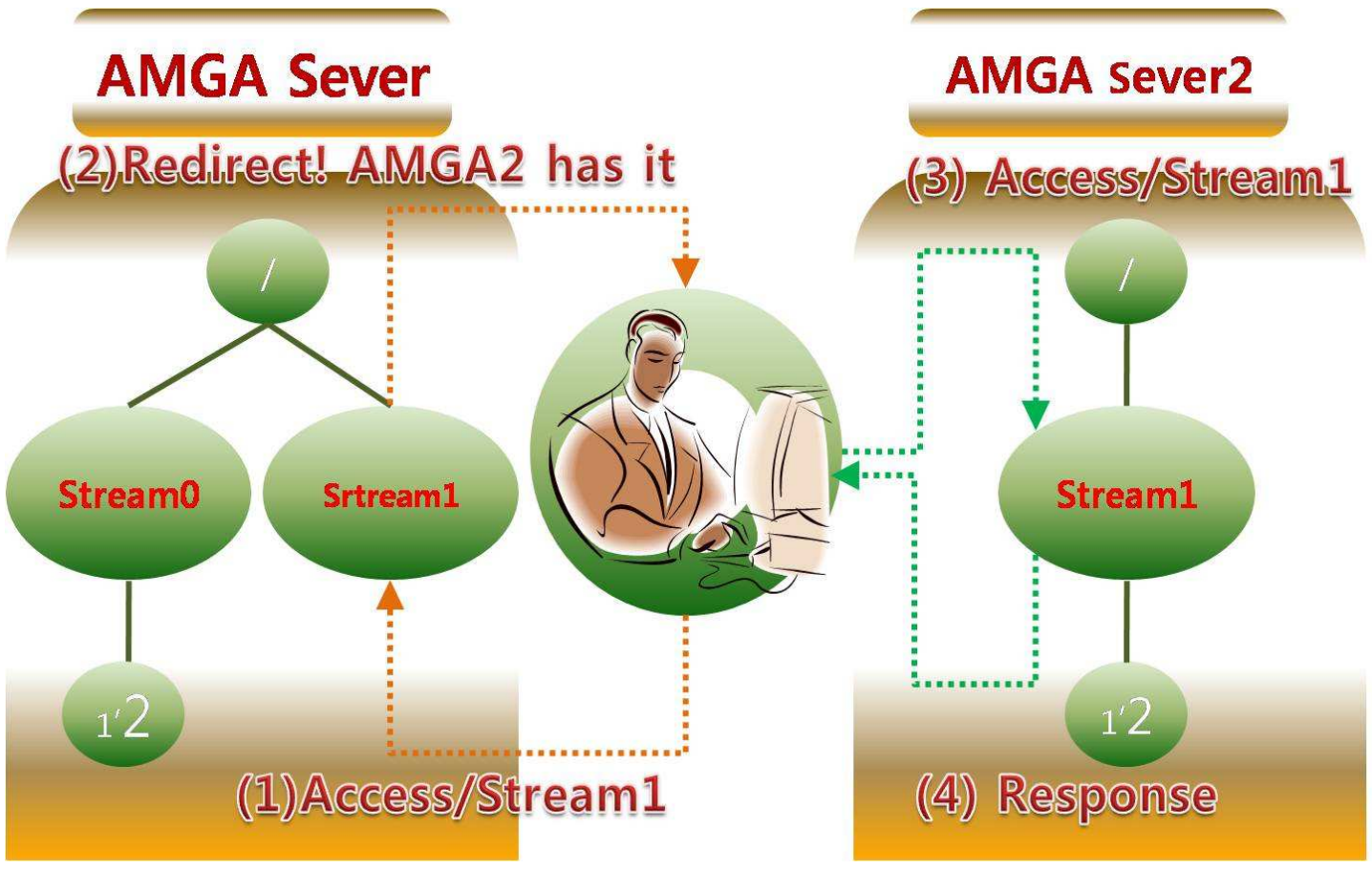

*Figure 14.8: The concept of the redirection system.*

### 14.3.5 Monte Carlo Production

MC simulation is the most CPU intensive and least data intensive aspect of the Belle II computing requirements. We wish to employ the least-cost but still effective solution. Consequently, we plan to employ both data-grid and virtualized cloud resources to generate our MC data. A number of governments are investigating making virtualized cloud resources available at no (or low) cost to the academic community. The idea would be use as many resources as we can find at low cost, with a fall-back to employ commercial cloud providers for time-critical needs.
We plan to employ the same project-based control flow for MC production on both grid or cloud provider. This situation is depicted in Fig. 14.9.

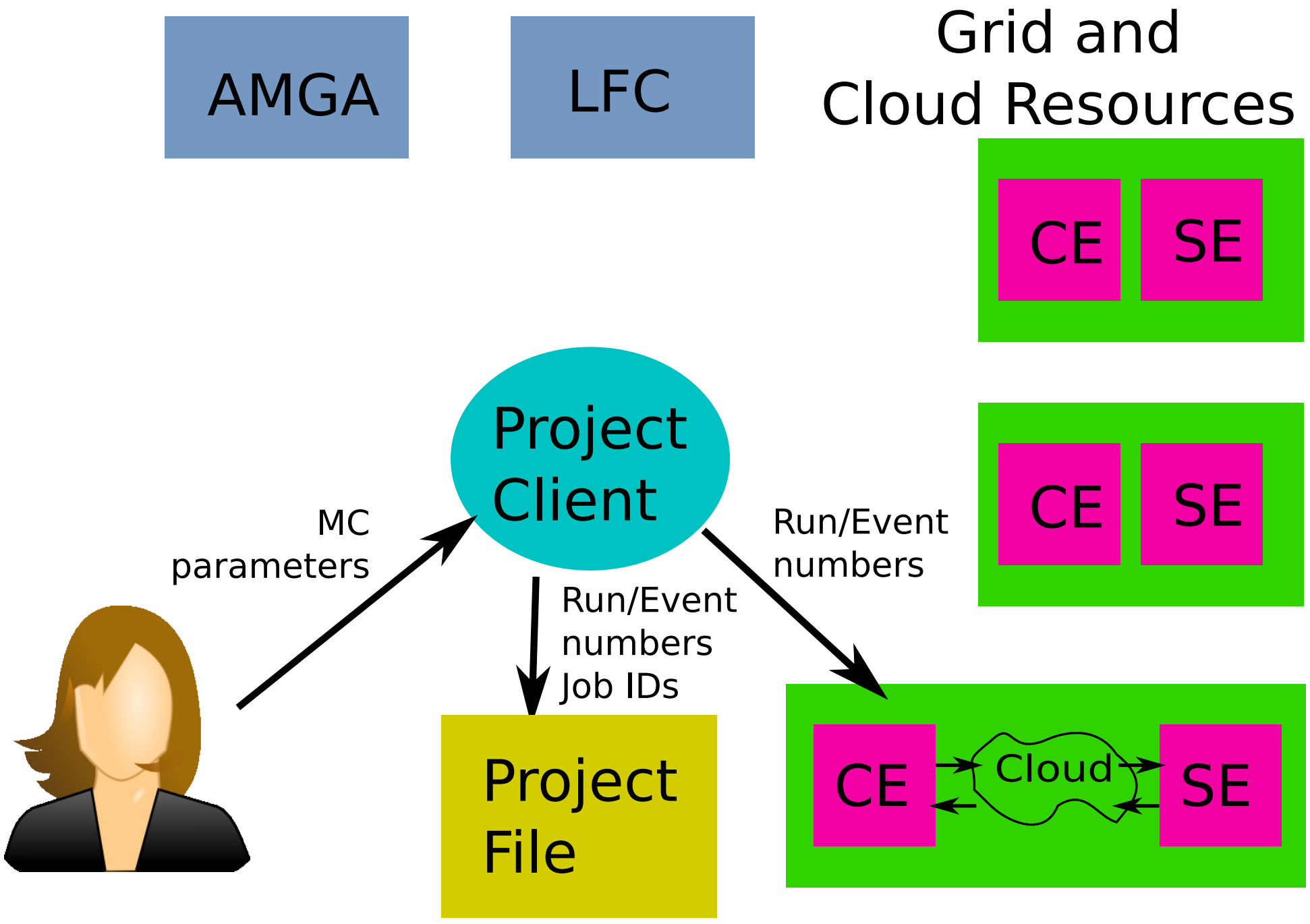

*Figure 14.9: MC production on both grid and cloud via the project system.*

The user requests MC production with a specific physics model, run number range, numbers of events, etc. The project client once again translates this request to a physical collection of resources and submits the requested number of jobs. In the static model, each job has fixed parameters for the simulation (experiment and run number, number of events per run) assigned. In the dynamic model, the job receives this information at run time from the project server.
An additional requirement for MC production is that appropriate "Random overlay data files" are provided. These are real data collected along with physics events but recorded with a random trigger and with no filtering software. These data were successfully employed to simulate real-beam background for the Belle experiment; we plan to use this technique for Belle II (see also Sec. 14.4.6.4).
The generated data is registered with AMGA and the LFC and stored in the SEs. It is thus made available to users who can easily locate and use the data as needed.

### 14.3.6 Cloud Computing and Virtualization

Cloud computing's recent rise to popularity is derived from the novel utilisation of a software technique called virtualisation. In virtualisation, host software running on physical hardware creates an abstracted, simulated computer environment—a virtual machine—for some guest software, usually an operating system. Due to the abstraction, from the perspective of the guest software, it has an entire computer platform to itself. Access to physical hardware is mediated by the host software, of which there are two types (shown in Fig. 14.10). Type 1 has the host software integrated into the host operating system, whereas Type 2 has the host software as an application running on the host operating system. The benefits of either approach is outside the scope of this document.

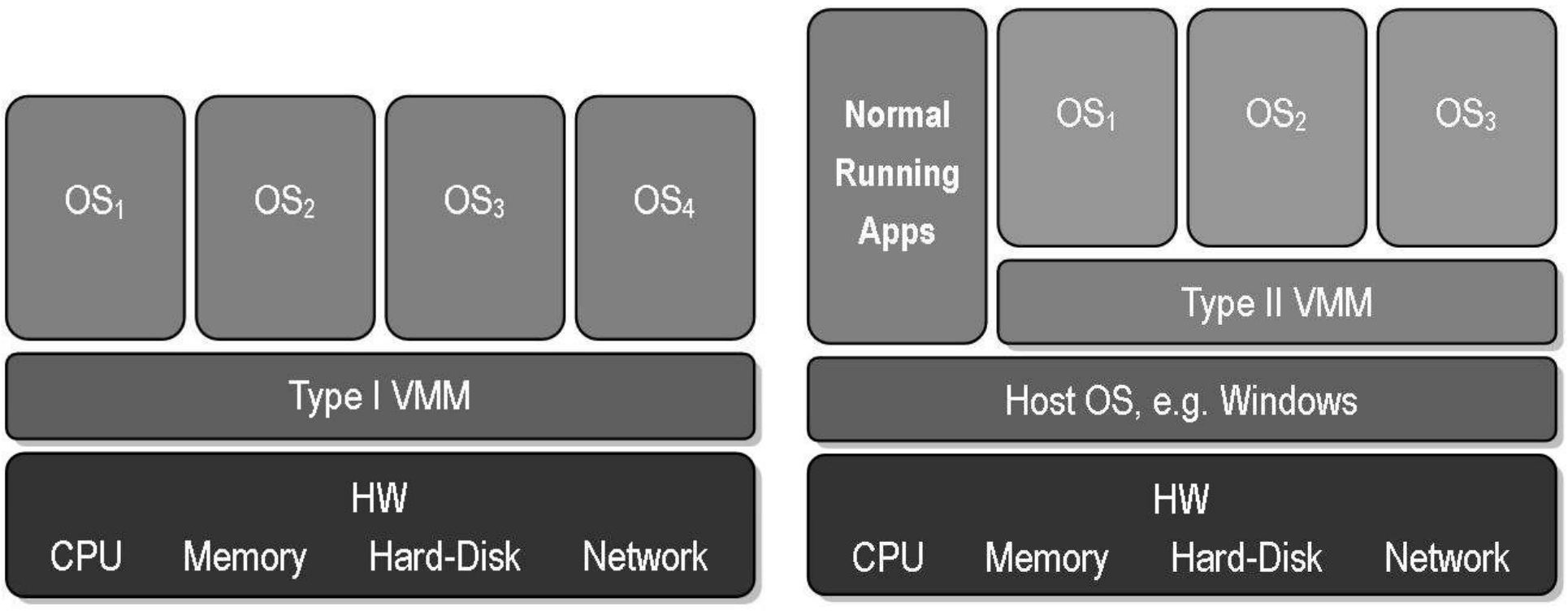

*Figure 14.10: Components in virtualisation.*

There are more than a dozen providers of cloud services in existence today. However, for the purposes of the Belle II experiment, where we require a specific operating system and custom libraries, many are less appropriate due to restrictions on access to the virtual machines created. Therefore, we utilise the set of providers known as Cloud Infrastructure-as-a-Service Providers. When you use a cloud service, you are in effect running software within a complex virtual machine—software that appears like an entire system, but is actually one of potentially many running on the same physical hardware. Various types of virtual machines, which vary in specification and cost—typically paid for by the hour—are available.
Though the hardware resources can be purchased easily from the cloud in this manner, additional software is required to control this process. Previously, we ran several successful tests using an open source product known as Lifeguard that we modified.

Subsequently, we have been working with the creators of DIRAC (Distributed Infrastructure with Remote Agent Control) to utilise their stable product as a solution for our cloud computing needs, in addition to integrating grid and local cluster resources.

#### 14.3.6.1 Prototype solution

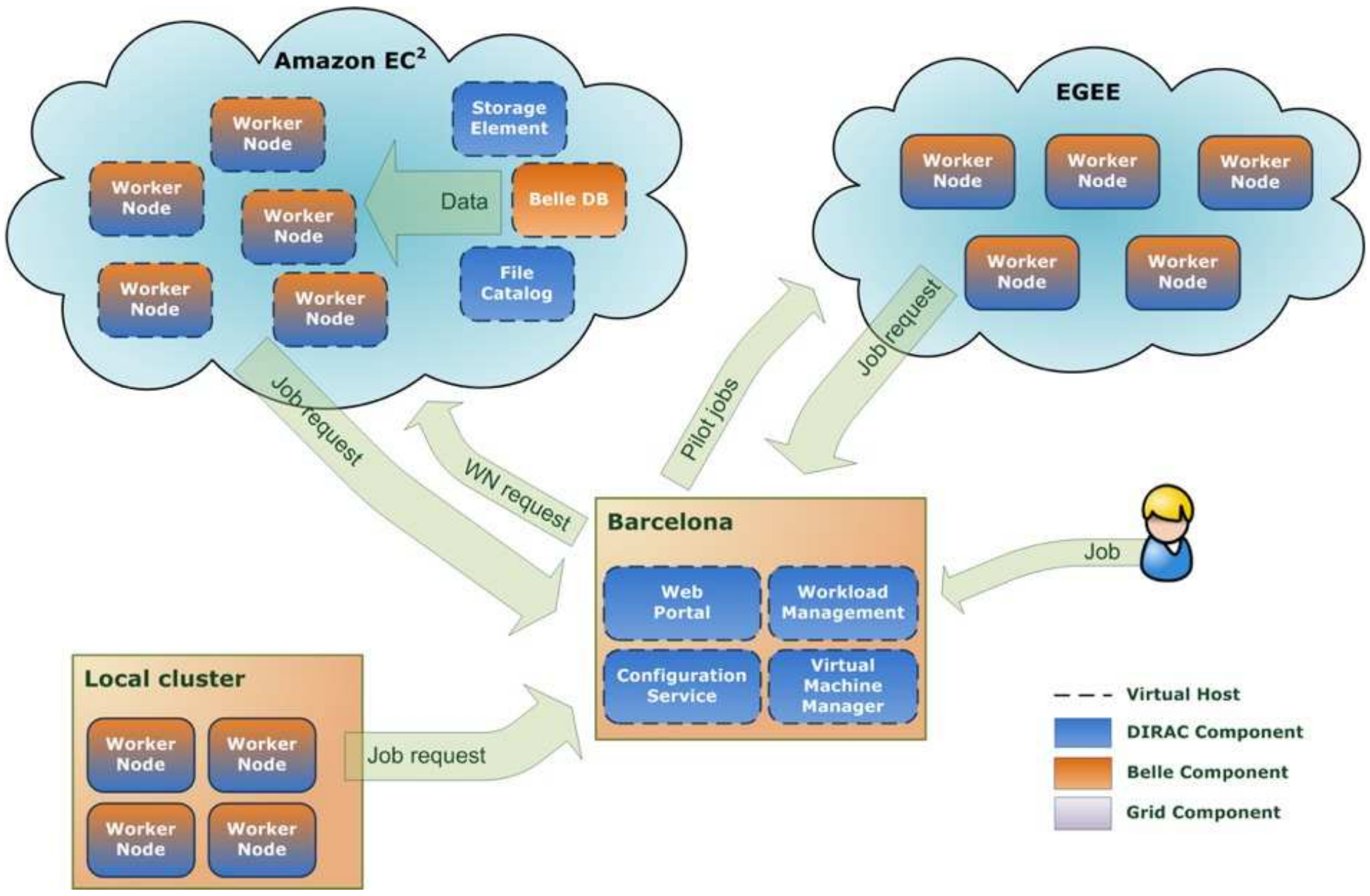

*Figure 14.11: Architecture of cloud-computing solution*

As cloud computing is a relatively new technology, we have felt the need to test our concepts in action prior to considering it as a solution. The results of our testing to date using the production of MC data on the Amazon Elastic Compute Cloud (Amazon EC2) are described below.

Figure 14.11 gives an architectural view of the most recent exercise, where a new DIRAC component was created, providing ability to control cloud resources.

#### 14.3.6.2 Control Flow

The most important part of control that is unique to the cloud is the management of the fleet of virtual machines running on the cloud. Since each virtual machine costs money when it is running, it is important to only run a new machine when necessary, terminate idle machines promptly and minimise overhead such as time wasted during start-up.

In our solution, this is controlled by a piece of software known as the "Virtual Machine Manager," which uses information from the central task queue in the Workload Management DIRAC component to make these decisions.

The Virtual Machine Manager constantly monitors the queue, and takes action by starting and stopping EC2 instances based on the number of jobs in the queue and several configuration options, including:

- the maximum number of instances to have running;

- time increments by which to step up the number of instances running;
- timeout values before shutting down idle or unresponsive instances

Once an instance is started, it will first download and setup the International Grid Trust Federation Certificate Authority configuration that is used to secure connections with grid resources. Following this, DIRAC is installed and the Virtual Machine Manager and Job Agents are started. Subsequently, the jobs will download the files they need from storage elements defined in the file catalog, simulate and generate Monte Carlo events, then upload the result files to grid storage elements. Status information about virtual machines and running jobs, including an excerpt of the job output, can be gathered from the Web Portal.

#### 14.3.6.3 Data Flow

As previously mentioned, the avoidance of vendor lock-in with regards to data storage is paramount. Therefore, we only use cloud facilities for temporary storage.
In our tests so far, we have used MC production. Recall from Sec. 14.3.5 that this task takes in data of different types and produces output files in the form of mDST. We need to have all of these inputs easily accessible to cloud nodes.
We have tested three methods of staging input to/from cloud virtual machines.

- Temporary use of Amazon Simple Storage Solution (S3)
- Running a dedicated virtual machine as a storage element
- Direct-copy to/from grid to cloud resources

Of these, only the first was deemed intractable: S3 was unable to sustain the transfer rates required, and the cost overhead and use of provider-specific APIs is undesirable.
The second option is achievable as each virtual machine has a local disk of varying size. The largest on Amazon is 1.7TB, which is enough to be a staging area for the relatively small Monte-Carlo input data. Transfer rates of 40 Mbps per connection were measured.
Directly using grid resources for data input and output reduces the overhead of cloud services, and the modifications required to allow computing jobs to run on the cloud. The weakest point is the typically lengthy network connection to the external resources. However, despite this, we were able to achieve over 400 Mbps inbound and outbound.

#### 14.3.6.4 Results of Testing

*Table 14.3: Cost of full production runs on EC2.*

| | | Cost in $ | | | | |
|---|---|---|---|---|---|---|
| Run | Number of Events | CPU | Storage | Transfer | Total | per $10^4$ events |
| 1 | 752,233 | 80.00 | 0.20 | 6.65 | 86.85 | 1.16 |
| 2 | 1,473,818 | 108.11 | 0.25 | 7.12 | 115.37 | 0.78 |
| 3 | 10,000,998 | 724.80 | 1.42 | 39.96 | 766.18 | 0.76 |
| 4 | 120,000,000 | 5198.60 | 0.0 | 415.68 | 5614.28 | 0.46 |
| 5 | 101,478,333 | 1566.76 | 0.0 | 349.52 | 1916.28 | 0.20 |

Table 14.3 summarizes the costs of the MC production on the cloud. It can bee seen that the total cost is dominated by the investment in CPU cycles and that it scales roughly with the number of produced events.
Since we first started testing, the prices at EC2 have dropped by 15%, and—perhaps more importantly—a new product ("spot instances") was released that allowed us to access significantly discounted prices. The use of spot-instance pricing, which prices virtual machine hours based on current market demand, allowed us to achieve savings of up to 50% (Run 5 in Table 14.3). Further tests are needed to verify that we can achieve these costs savings at the scale required for Belle II.

## 14.4 Offline Software

### 14.4.1 Introduction

After having written the data to tape or disk, the offline software takes the data and performs the final processing steps. Among them are the reconstruction of tracks, MC simulation, and physics analysis. The offline software is designed as a framework, providing a software environment to the user that allows access to all functions built into the framework. User code can be added to extend the functionality. A typical user-written module would be code that performs a physics analysis.
This section discusses the way the offline software source code is managed, provides an overview of the framework itself and takes a closer look on the event data model, the geometry description, the simulation and reconstruction. At the end, a possible solution for an event display is given.

### 14.4.2 Code Management

As in other HEP experiments, the Belle II offline software is developed by a group of dedicated people with different levels of experience who are distributed around the world. To create a reliable, user-friendly, and well-maintainable software environment, we take the following organizational and technical measures.
All Belle II code is maintained in a central Subversion repository [12] located at KEK. The code is structured in packages. A package is a container for source files that are later compiled and linked into a shared library with the same name as the package. Each package has an assigned librarian who is responsible for the code in the given package. By default, only the librarian has the ability to change the package, but they can grant the right to commit code to other developers.
All Belle II members have read access to the full repository. A repository browser (ViewVC [13]) for inspecting and comparing different versions of source files in a web browser is provided as well.
To help developers to identify integration issues between packages, nightly code builds will be done automatically. In addition, regular integration builds will be performed that serve as a basis for further code developments. For the integration build, the release coordinator will collect package versions from the package librarians. The package versions are identified by tags that are assigned by the librarian to a version of code they consider ready for a build.
This procedure of collecting tags from package librarians is also used to build releases. In contrast to the deadline-driven integration builds, the releases are feature-driven. Moreover, a series of quality tests have to be passed before code is released to users. For example, verification that the software compiles and runs on different architectures and operating systems is undertaken. We plan to support the following systems:

- SL5 64bit, because it is used on the grid.
- CentOS 64bit, because it is used at KEK.
- Ubuntu 32bit and 64bit, because it is used by many people on their desktops and laptops, and because it is a debian-based system, in contrast to the first two redhat-based ones.
- MacOS, mainly to ensure portability of our software as a measure of software quality assurance, but also to permit users to run the software on their laptops.

The supported compiler is gcc 4. We may try other compilers, like the intel compiler, to check the portability of our code.
Documentation is generated from the source code with the doxygen tool [14] for all releases and builds. Further documentation can be provided on a wiki or in form of manuals.
To keep code maintainable, we have agreed on several essential issues that affect how code is contributed. We agreed to use C++, the de-facto HEP standard, as the only programming language. In cases where a scripting language has clear advantages, python can be used. A consensus about the coding conventions has been established as well. The overall guideline is to keep the code as simple as possible. A formatting tool, astyle [15], is used to help developers unifying the layout of source code. Furthermore, we will do code reviews and implement unit tests to improve the software quality.
To facilitate the installation of the Belle II software on grid sites, institute clusters, and user's desktops and laptops, dependencies on external software should be minimized. Any addition of a third party library has to be well-motivated. The external packages will be provided via the same channel as our software. Both can be obtained from svn and compiled locally. In addition, we will provide tarballs and perhaps rpm/deb packages. For the management of software installations on grid sites, we plan to adapt the ATLAS tool [16].

### 14.4.3 Framework

The data analysis framework is the basis of the Belle II software. Various kinds of software components are be written as "modules" for the framework, and they are plugged into the framework on demand. The application programs used in Belle II, such as the full event reconstruction, the event skim, and the user analysis codes, are realized by combining a set of modules on the framework. The framework reads an event record from a file, then processes it by executing a chain of modules, and stores the resulting event record into a file. The histograms, ntuples and other data required to be accumulated in each module are also managed by the framework.
A new analysis framework called "roobasf" is being developed for use in Belle II. To benefit from the experience made in other frameworks, several software frameworks from other HEP experiments were evaluated. Therefore, the Belle II software framework combines the proven concepts of the previous Belle framework together with ideas taken from the GAUDI, ILC and ALICE software frameworks. This section describes the requirements for the framework in Belle II and the roobasf design that satisfies these requirements.

#### 14.4.3.1 Object I/O

The event structure (event model) of Belle II is described as "objects" in C++, and as a consequence, the framework is required to have the capability to read and write C++ objects in data files. Since the event model can evolve as the software developments goes, the I/O capability should not be restricted to a certain event model. To store C++ objects in a file, the objects have to be converted to a plain byte stream by "streamer" code, but it is desired to avoid writing the codes by hand, which introduces an extra burden in the software coding.
To satisfy these requirements for the object I/O, we adopted ROOT IO [17] as the object I/O model for our framework. The ROOT IO is a part of the ROOT package, which is a collection of object-oriented analysis tools. It has the object I/O capability, which can generates streamer codes automatically by reading the C++ class definition files using *rootcint* or *genreflex*. The ROOT IO object model is already the de-facto standard in many HEP experiments and is designed to keep up with the evolution of the event model ("schema evolution"). By using ROOT IO for object I/O in the "roobasf" framework, the requirements are satisfied and the development of the event model are well separated from the framework.

#### 14.4.3.2 Software bus

The framework accepts a set of modules plugged in it and controls the chain execution of them. A simple software bus architecture is suitable for managing the chain. The execution chain is required to be stopped and restarted at any point by dumping and restoring the intermediate objects in a data file (check-pointing). Figure 14.12 shows the software bus architecture for the framework.

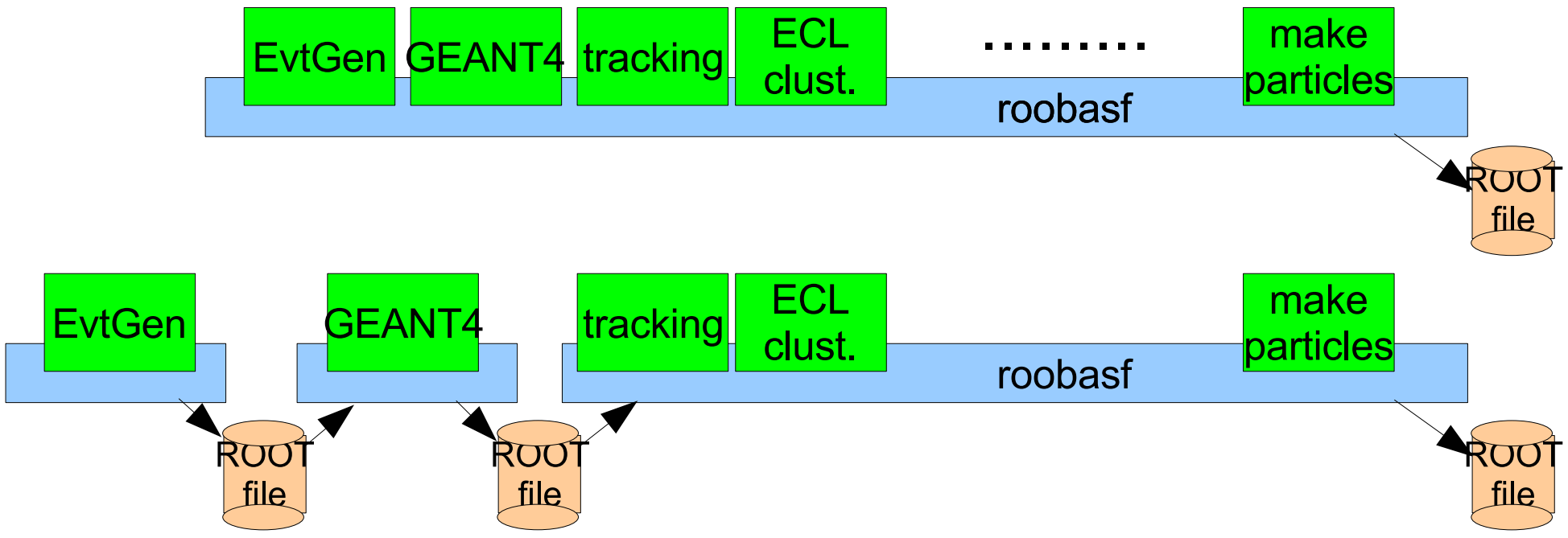

*Figure 14.12: Software bus concept.*

Roobasf recycles the design for the previous B.A.S.F. [18] framework used in the Belle experiment to implement the software bus. The modules are plugged into the framework via the dynamic link and therefore each module can be compiled and debugged independently of the framework. The well-defined event model and module functionality can provide the check-pointing capability easily with the ROOT IO.

#### 14.4.3.3 Belle Compatibility

There is a substantial accumulation of software experience for the $B$-physics experiment in the so-called Belle library. The maximal use of the Belle library in Belle II is desired to avoid re-inventing similar software. The compatibility with the Belle software also enables the smooth transition from Belle to Belle II for the software developers. Therefore, we require full compatibility with the Belle library in the new framework.
Roobasf realizes two levels of Belle compatibility. One is the module-level compatibility. The modules written for the B.A.S.F. framework can be plugged into roobasf without any modifica-tions. The other is the capability to read Belle's Panther data files by linking the legacy I/O package. In some cases, it is useful to compare the performance of newly developed software on roobasf with that of Belle. However, the ability to write the processing output in Panther format is deliberately omitted, to urge people to move to object I/O based data handling.

#### 14.4.3.4 DAQ compatibility

The software used in Belle II DAQ (Ch. 13) has a very close relationship with the offline software. In particular, the high level trigger (HLT) software (Sec. 13.9) is supposed to be identical to the offline event reconstruction code, and the same analysis framework is desired to be used in DAQ.
The I/O package of roobasf is implemented as a shared library with a set of functions to provide low-level I/O. The package is dynamically linkable/changeable. An I/O package dedicated to the DAQ use, which can send/receive data through the UNIX socket, is provided to manage the data flow in the DAQ system. The data between the I/O package and the event processing core are passed via ring buffers so as to absorb the difference in the event processing time.

#### 14.4.3.5 Parallel processing

The data processing is required to handle a huge amount of data and some kind of "parallel processing" is necessary to analyze these data in a reasonable period. The parallel processing capability is desired to be implemented inside the framework and concealed from users. It is applied at the event level.
Recent CPU chips house multiple cores and these cores are normally seen as multiple CPUs in an SMP server by an operating system like Linux. The parallel processing in the framework is desired to make use of the multi-core CPU. In addition, the parallel processing using a PC cluster, where a large number of processing nodes are connected via network, is also required. This is necessary in the HLT for the real-time full-event reconstruction, and also for the data reproduction with the updated software and/or calibration constants in off-line.
The event-by-event B.A.S.F. parallel processing models (shared-memory BASF and networked dBASF and RFARM) are the starting point of the parallel processing design in roobasf. Figure 14.13 and 14.14 show the data flow in roobasf for the parallel processing. The event records in a file are read by the "event server" and placed in a ring buffer on the shared memory. A number of event processes pick up event records in the buffer and process them in parallel. The resulting outputs are then stored in a separate ring buffer and collected by the output server to store in an output file. The event server and output server can be connected to an external node through a UNIX socket for the parallel processing with a PC cluster.
Roobasf has both single and parallel processing modes. When operated in the single process mode, all the processing is done within a single process; this is useful for the debugging of module codes before the mass-processing.

#### 14.4.3.6 Histogram management

The histograms, ntuples and ROOT trees are defined in the modules on the framework independently. They are typically stored in a single file for the further analysis.
Roobasf is designed to manage the histograms using ROOT calls. The histograms can be defined using the standard ROOT functions in each module independently. The histogram manager of roobasf takes care of the output histogram file in which the histograms are stored, and it manages the calling sequence of histogram definitions by each module. It also takes care of collection of histograms after the parallel processing is performed.

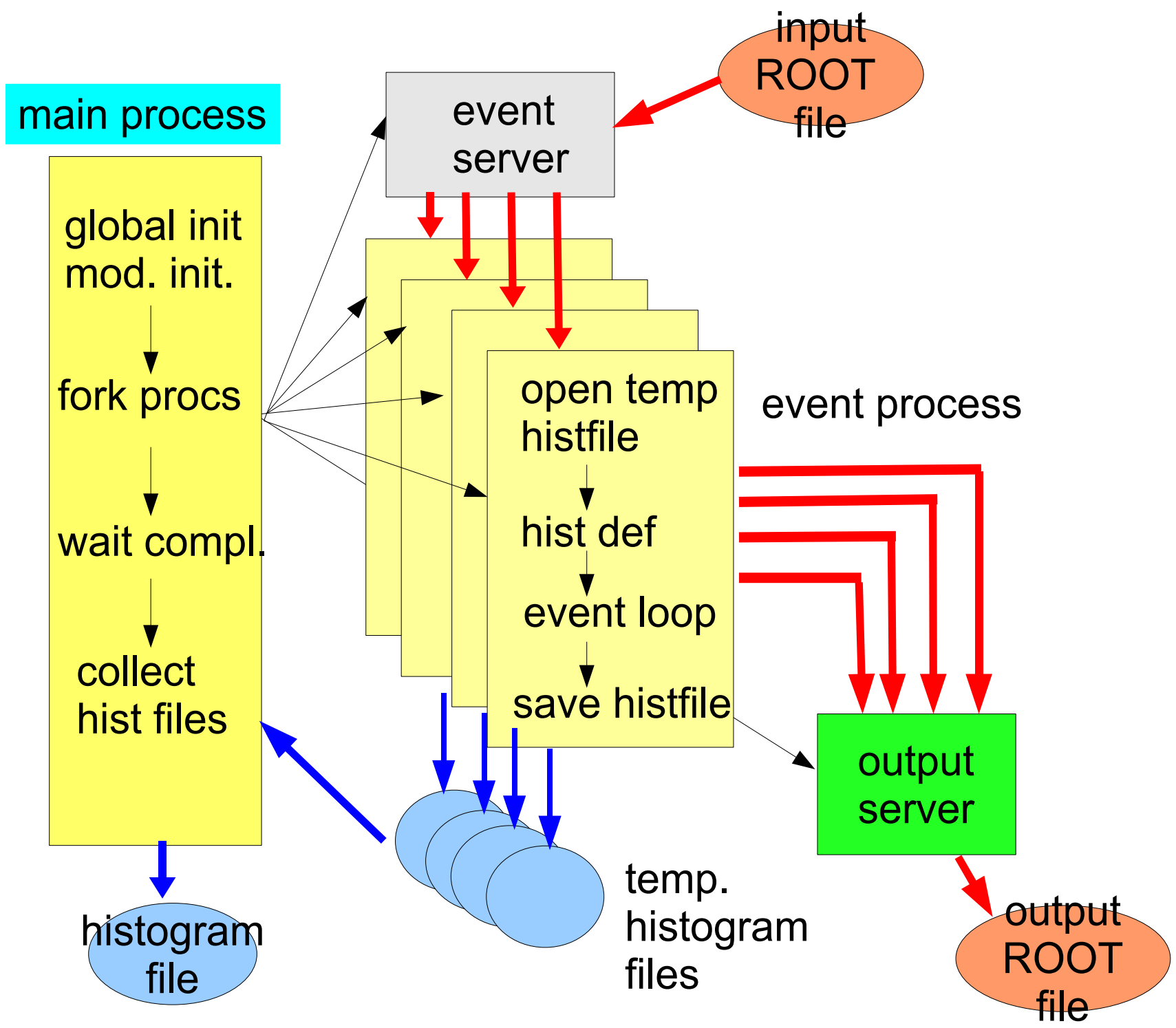

*Figure 14.13: Parallel processing in roobasf.*

#### 14.4.3.7 Scripting

Module execution in a framework can widely vary depending on the application, and has to be described in a robust and versatile scripting language. The B.A.S.F. framework used a custom command parser for this purpose that was closely tied to the framework; its cooperation with external programs was difficult. It is desired to use some "common" scripting language to drive the module execution on the framework cooperating with other analysis tool like ROOT.
We adopt python [19] as the scripting language of roobasf. The framework execution can be described by the python-roobasf interface. Python has become very popular in HEP, and other HEP software applications are now equipped with a Python interface, like PyROOT. By using Python as the scripting language for our framework, roobasf can easily co-operate with ROOT through PyROOT for the full data analysis. Figure 14.15 shows the example of the python script for roobasf.

### 14.4.4 Event Data Model

The event data model describes the in-memory representation and the storage of all data that comes from the detector. It is essential that this be done in a unified way, as it defines an interface for the different sub-detector groups and the tracking and software groups. In Belle II, this will be done in an object-oriented manner.

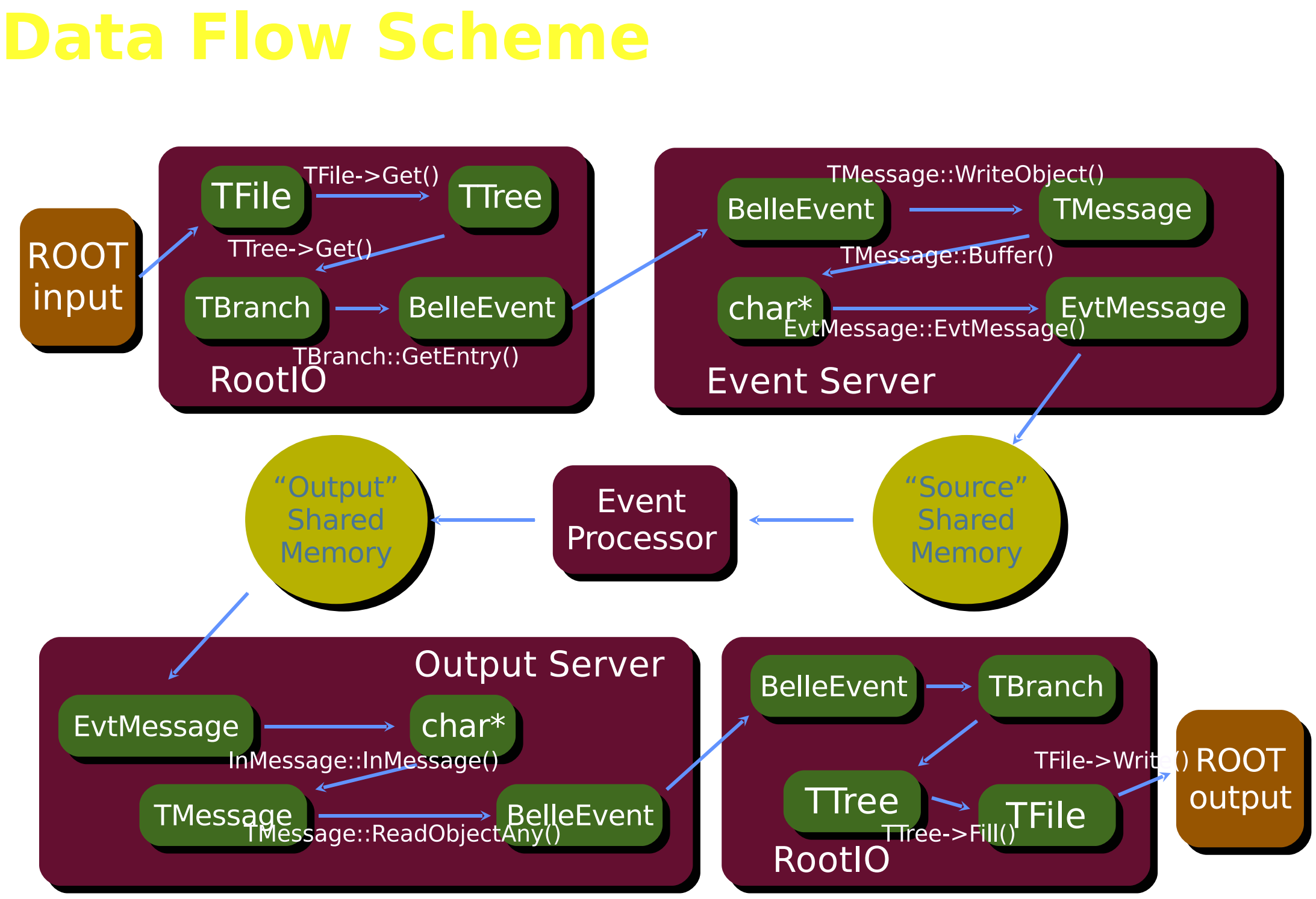

*Figure 14.14: Event data flow in roobasf.*

```
import PyRooBasf
f = Framework()
fix_mdst = Fix_mdst()
main = Path()
main.add(fix_mdst)
main.add_condition('<:0:exit')
f.add(main)
i = Input(exp=57,
          run=range(1,200),
          type='HadronB')
f.process(i)
```

*Figure 14.15: Example of a python steering script for the roobasf framework.*

For the reader's understanding, we define the terms persistency and serialization. The process of converting an object in memory to a string of bytes is called serialization. The serialized object can then be sent across the network or saved into a file. The process of saving an object into a database or a file using serialization is called persistency.
We describe the event data model in C++ instead of a meta-language, since this programming language is used across the entire framework. We want to fully exploit the object-oriented features of C++ and store data in lists, trees or maps. The data event model is designed to fully utilize the standard template library (STL). This avoids the overhead of reimplementing very well optimized code. As a starting point, we assessed the existing Belle Panther banks for data content and LCIO for object-oriented structure. To ease analysis, we are planning to use ROOT as a persistency provider. ROOT provides tools to write arbitrary objects to a file and to read in the persisted objects into a program, which can be used in the online software to create a uniform framework.
The core element of the framework is the event class. To make it easy to write a whole event, the class inherits from the ROOT TObject class, which allows standard ROOT methods (e.g., write) to be used for serialization. The serialized object can then be stored in a file or sent across the network. The event class holds collections of data classes that represent the sub-detector information. The collection class is based on the STL vectors, allowing fast access to the data classes. All data classes are required to inherit from a BelleObject class. This ensures that all data objects can be saved in a collection. The design of the data classes is driven by the needs of the sub-detector groups. For example, for the storage of the Monte Carlo Particle information, a decay tree seems natural: pointers between the Monte Carlo Particle objects are used to form a tree structure. To ensure that future changes in the persistency model do not affect the existing code base, the use of BelleII pointers is suggested.
In addition to data objects, the collections in the event class should be able to store relations. All three kinds of relations (one-to-one, one-to-many, many-to-many) can occur in the data model. Therefore, a Relation object containing a "from" and a "to" member variable is suggested. To create a one-to-many relation, one needs to create multiple Relation objects where all "from" member variables point to the same data object. Sometimes one cannot define unbiased relations. In this case, a weight can be applied to a relation. For example, consider a track that is misreconstructed using two hits from one Monte-Carlo particle and three hits from another particle. In this case, we could create two Relation objects: one with a weight of 40% and the other with a weight of 60%.
The ROOT persistency must save not only the member variables but also the pointers to other classes. As a default, ROOT follows the pointers to other objects and also saves the referenced objects. This leads to problems if you want to store two sets of objects in different files that have common elements. Our plan is to avoid such a scenario.
The persistency layer should, however, allow us to write out only those collections that are of interest. This allows us to define sets of collections that are saved; for example, analysis and alignment sets. It is also helpful to join streams of event data. Commonly, this is needed when noise and background is merged with MC simulation. Therefore, the event data model must be designed in a way to easily allow merging and discarding of collections in a given event.

### 14.4.5 Geometry and Conditions Database

The description of the detector geometry is an integral part and represents an essential ingredient for the different stages of the offline software. In the following section, the requirements imposed on a geometry handling system are discussed and the approach taken for the Belle II detector is presented in detail.

#### 14.4.5.1 Requirements for a geometry handling system

From the simulation of the passage of particles through the detector to the simulation of the response of the sub-detector hardware and finally the reconstruction (e.g., tracking) algorithms, nearly all tools of the offline software chain need a geometrical description of the Belle II detector. However, each tool requires a different view of the detector geometry. The full simulation of the detector (described in Sec. 14.4.6) takes a very detailed geometrical description of all detector parts as an input. Usually, the detailed geometrical description uses a collection of primitives (boxes, tubes, spheres, etc.) to compose a complicated detector structure. In addition, material properties like radiation length, specific energy loss are assigned to each primitive. The simulation of the sub-detector hardware response, called digitisation (Sec. 14.4.6), requires only a subset of the full geometrical description of the detector: hardware parameters and the transformation matrices from global space (usually defined by the experimental hall) to the local space of a sub-detector volume. An example is the digitisation of the PXD sub-detector: the track-finding and -fitting algorithms need a simplified view of the detector geometry. This is given as a collection of geometry primitives but, unlike the full detector simulation, does not require the same degree of detail. Usually, it is sufficient to describe the detector roughly by using basic shapes like planes, boxes, tubes, cylinders, cones, etc.

#### 14.4.5.2 Basic geometry handling scheme

To make sure that all tools of the framework have access to the same version of the detector description, it has to be stored centrally. From this central storage, the geometry for the full simulation and for the reconstruction is created and the parameters for the digitisation are taken. This leads to the basic architecture for the Belle II geometry handling system shown in Fig. 14.16.

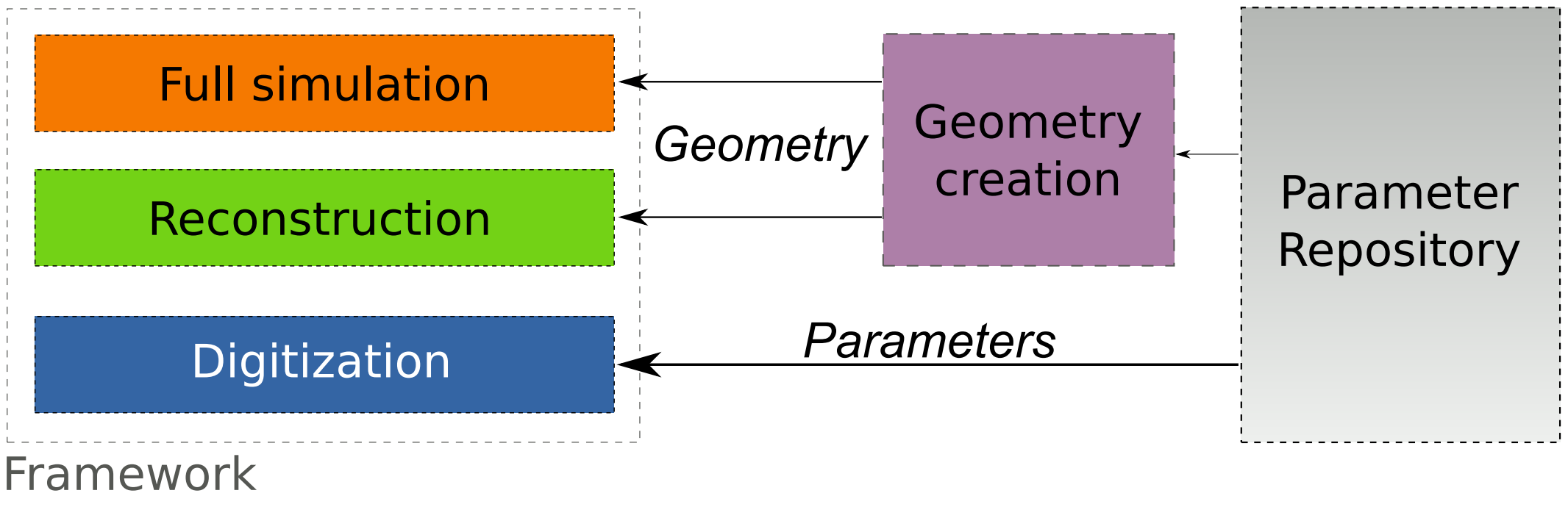

*Figure 14.16: Basic architecture of the Belle II geometry handling system*

The central repository stores all parameter values needed for fully describing the Belle II detector. The concrete geometry used for the simulation and the reconstruction algorithms is created using C++ source code from the stored parameters. The digitisation algorithms take the parameters directly from the repository.

Storing parameter values instead of a concrete geometry in the central repository allows for a simple and generic way to handle an evolving geometry. For example, the position of the measuring parts of the vertex detector have to be known with high precision. Various forces, like gravitation, may change the position of these parts over time. The slow drift of the vertex detector from its initial position can be measured and stored as time varying parameters in the central repository. These parameters are then used for correcting the initial values during the construction of the concrete detector geometry for the simulation and reconstruction. They are also taken into account for direct parameter retrieval and the construction of transformation matrices. Other examples for time varying parameters are calibration constants or slow control values.

#### 14.4.5.3 Implementation of parameter storage and access

The central repository is realized using XML documents for the basic detector parameters and a PostgreSQL database [10] for the condition parameters. XML documents have the advantage of being human readable and highly extensible. Furthermore, they are widely used in particle physics and in industry, leading to the availability of high-quality libraries, tools and software (both open-source and commercial) to write, read and manage XML documents. For storing large datasets, like condition parameters, a database is more suitable. Therefore, a PostgreSQL database is used in addition to XML documents. This leads to the following scheme: a time-independent parameter is stored directly in an XML document and a condition parameter is stored in an XML document as a link pointing to the location in the database where all available time spans for that parameter are saved. Figure 14.17 shows a schematic drawing of the parameter storage solution.

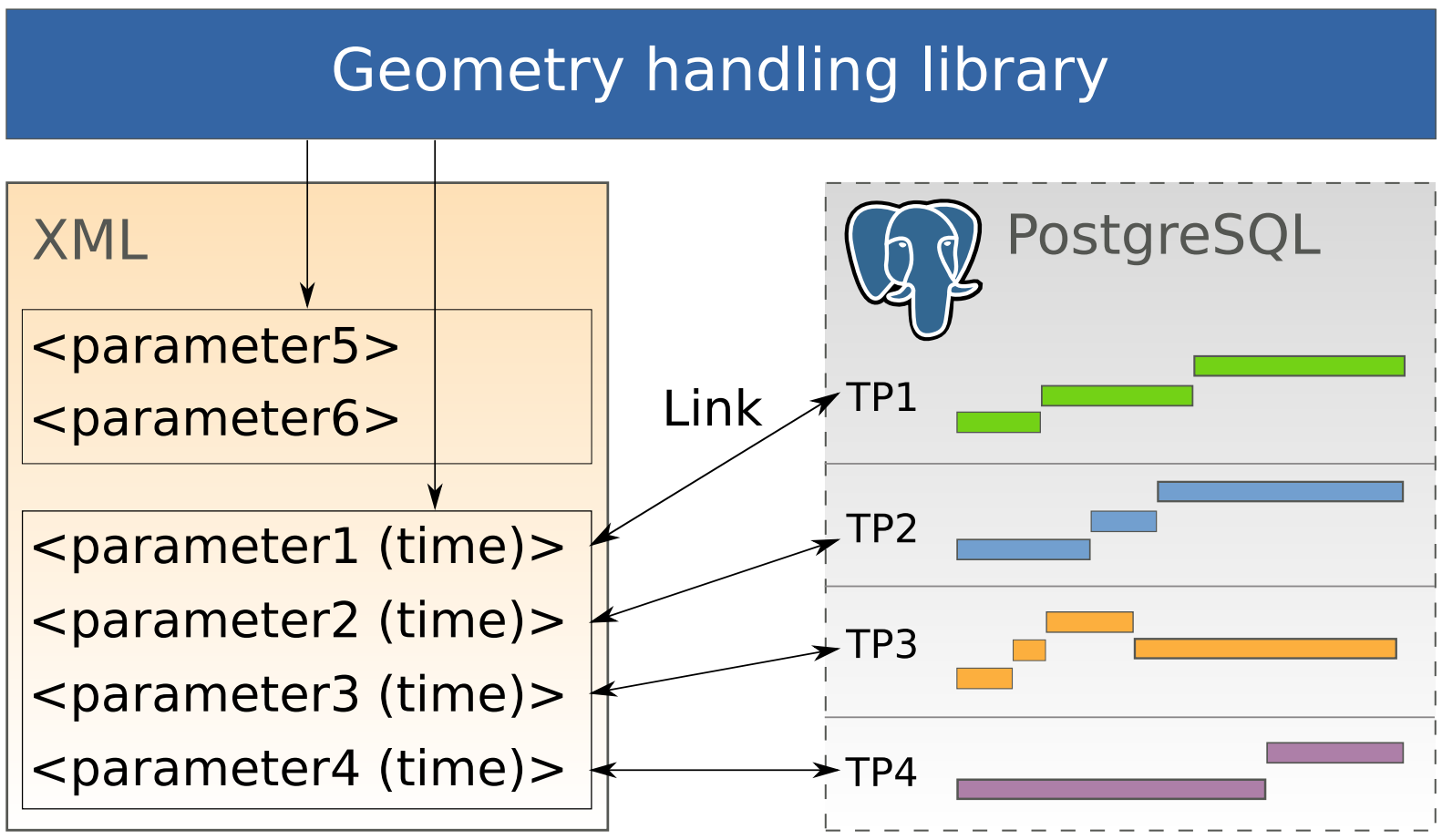

*Figure 14.17: The parameter storage solution.* $\langle parameter5 \rangle$ *and* $\langle parameter6 \rangle$ *are time independent parameters, which are returned directly.* $\langle parameter1 \rangle$ *to* $\langle parameter4 \rangle$ *are time dependent parameters. Their time spans are calculated from the given time and their values are fetched from the PostgreSQL database. The parameter entries TP1 to TP4 in the database correspond to the elements* $\langle parameter1 \rangle$ *to* $\langle parameter4 \rangle$ *in the XML document.*

The combination of XML-based parameter storage and database condition parameter storage is hidden from the user by a common interface. This interface allows the user to send queries for parameters to the geometry handling system without having to worry about the time dependence of certain parameters. Figures 14.18 and 14.19 show the process diagram for a typical time-dependent and -independent parameter query, respectively.
Using a defined query language, which will be described in detail later, the user sends a request for a parameter to the geometry handling system. If the requested parameter is not a condition parameter, its value is directly taken from the XML document and returned to the user. If it is a condition parameter, the system follows the link to the location in the database and returns the value of the time span evaluated using the given time value. This time value can either be set manually by the user or automatically. The latter is especially useful for MC production.
The parameters describing the Belle II detector are stored in a tree-based hierarchical structure. The root of the tree is the Belle II detector and the first level of branches describe the various sub-detectors. Each sub-detector consists of an arbitrary number of branches, where each branch groups parameters having a similar type or describing the same part of a sub-detector. The actual parameter values are then saved as leaves of a branch. The tree structure is realized in an XML document using elements as basic building blocks. Each element can either have an arbitrary number of child elements or store a parameter value. Every XML document consists of exactly one root element which serves as parent for all other elements. An arbitrary number of elements, being either child of the root element directly or of another element, can then be added to the XML document.
As explained in Sec. 14.4.5.1, the detector geometry description is based on the fact that the full detector is composed from different sub-detectors. Therefore, the parameters of each sub-detector are stored in a separate XML document and combined into a single document using the XInclude technology [20, 21]. This keeps the size of each sub-detector XML document reasonable small and allows the definition of different detector models, where each model is given by a specific set of sub-detectors.
To facilitate the process of locating and retrieving parameters from an XML document, query statements very similar to file paths are proposed. This approach is motivated by the hierarchical structure of the detector parameters. The language used to express a query statement in the geometry handling system, is known as the XML Path Language or shortly XPath. XPath is a standardized language developed by the W3C [22].

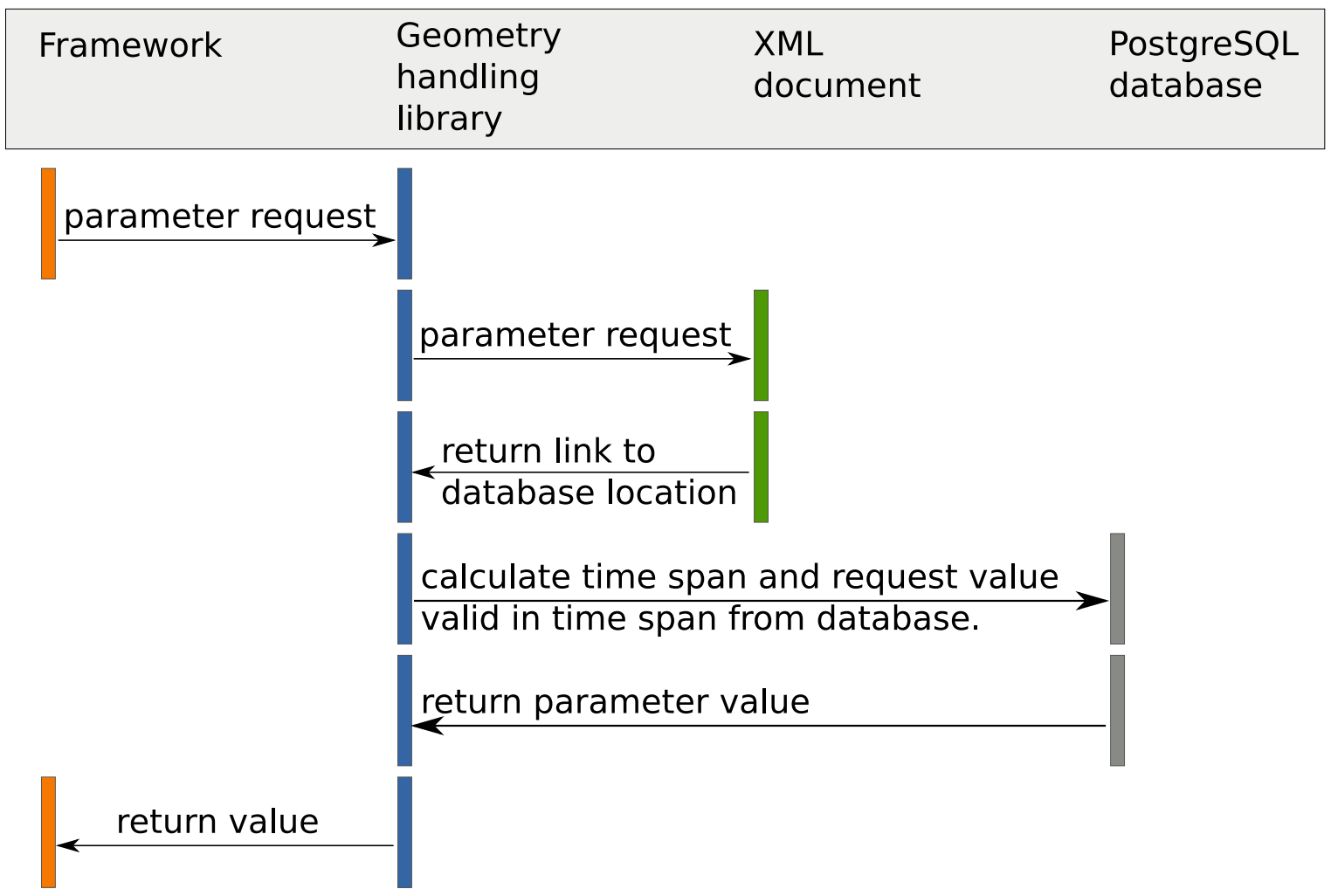

*Figure 14.18: Process diagram of a query for a time-dependent parameter.*

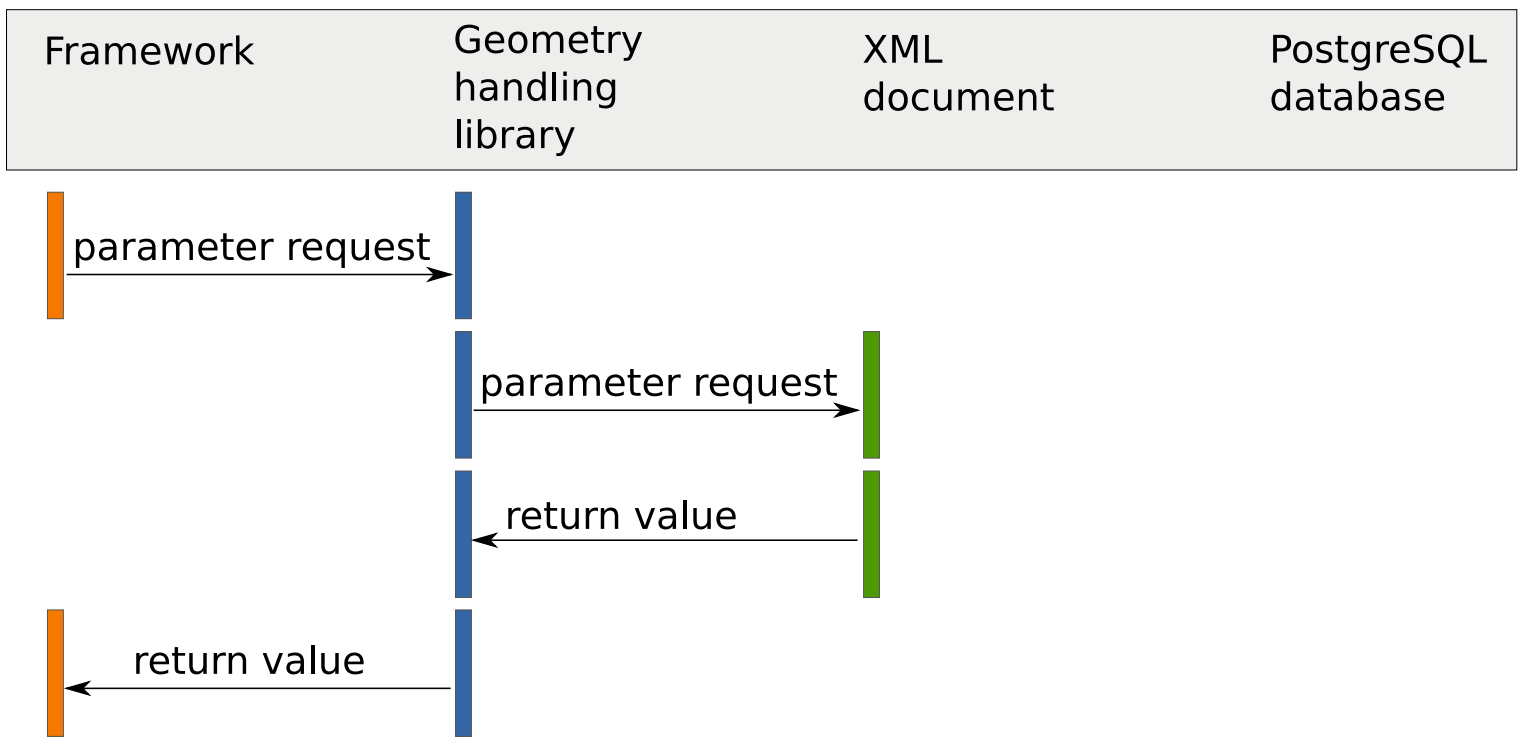

*Figure 14.19: Process diagram of a query for a time-independent parameter.*

#### 14.4.5.4 Creating geometry objects from parameters

Both the detector simulation and the tracking algorithms depend on a detector description based on actual geometry objects like planes, cubes etc. The geometry handling system, explained so far, only stores parameter values and optionally their development over time. The missing link between parameter values and geometry objects is filled by C++ code. Figure 14.20 shows the basic architecture of the geometry object building system.

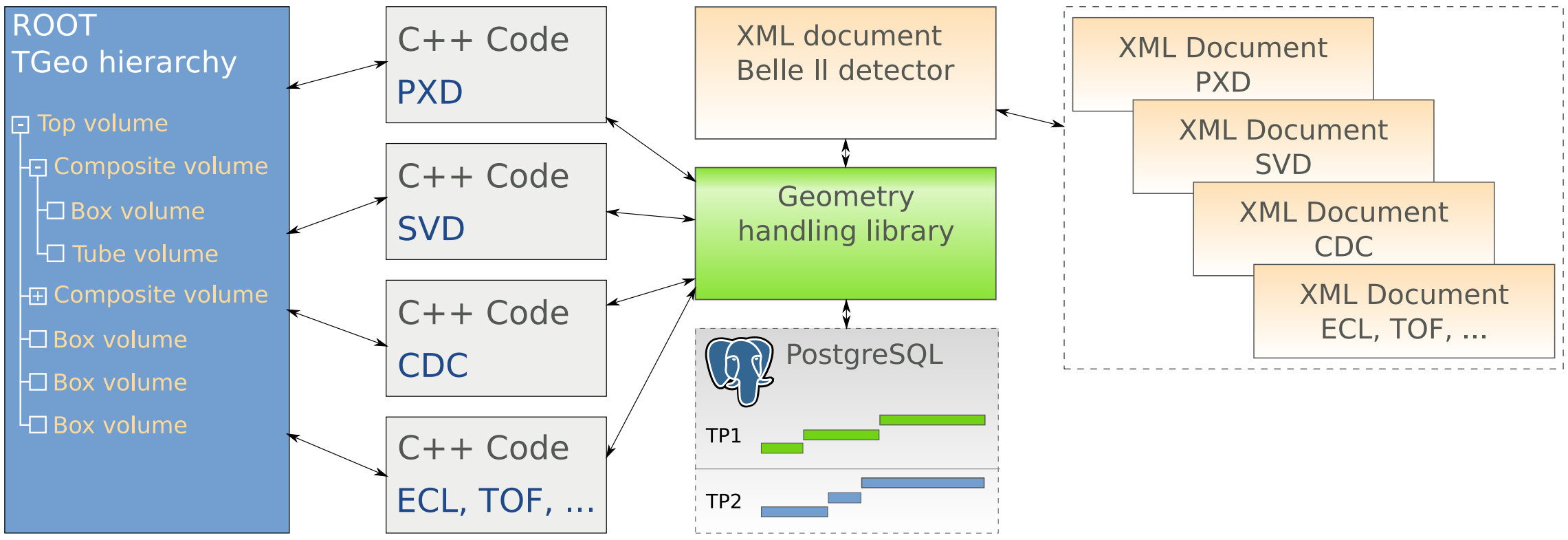

*Figure 14.20: The basic architecture of the geometry object building system.*

Each sub-detector has C++ code associated with it, which is aware of the available parameters for that specific sub-detector. The C++ code requests the parameter values that are needed in order to build the actual geometry of the sub-detector from the geometry handling library. The geometry handling library automatically discriminates between time-independent and -dependent parameters and returns the values from the XML document or from the PostgreSQL database, respectively. Then, using the parameter values, each sub-detector C++ code creates the actual geometry objects and writes the created geometry into an object hierarchy. The geometry objects and their relationships within the hierarchy are implemented using ROOT [17] TGeo objects, a common standard for geometry objects. The hierarchy is shared between all sub-detectors, leading to a single geometry object hierarchy for the whole Belle II detector. It can then either be kept in memory, and directly used by the simulation/reconstruction software, or saved to disk for later usage.

### 14.4.6 Simulation

In high energy physics experiments, the term "simulation" refers to three steps: event generation, the simulation of passage of particles through the detector, and digitisation. The first step involves the generation of events for the various physics studies, for example $B$ meson decays or beam background simulation. The second step performs the simulation of the interaction of individual particles passing through the detector, described by all its different materials and geometries, and records the particles' energy deposits in "sensitive" volumes. The third step simulates the response of individual sensitive-detector components, taking into account detailed physics processes of signal generation, electronics effects, and final hit creation. All three steps will work as loadable modules within the software framework explained in Sec. 14.4.3 and will exchange information through the event data structure described in Sec. 14.4.4 Here, each simulation step is described in detail.

#### 14.4.6.1 Event Generator Software

In event generation, the information (the production point, momentum, particle species, and so on) of particles produced for a given process by the event generator is stored in ROOT I/O and fed into the detector simulation.
One of the purposes of the Belle II experiment is to search for new physics through the precise measurements of $CP$ asymmetry with $B$ mesons. To simulate the physics of $B$ decays, EvtGen [23], which was initially developed by the CLEO collaboration and is maintained at SLAC, has been used in many experiments, including Belle. We will continue to use EvtGen, updating the decay channels for the Belle II experiment, unless a better event generator is found.
For continuum ($e^+e^- \to q\bar{q}$) events, EvtGen and Pythia are used.
Tau production and decay are modelled using the combination of KKMC [24] and tauola [25]. The former is a package for $\tau$-pair production in $e^+e^-$ collisions and the latter for the decay of $\tau$ produced in the generator.

#### 14.4.6.2 Detector Simulation Software

For detector simulation, a package based on the Geant4 toolkit [26] is being developed, written in C++ and designed to exploit advanced software techniques and object-oriented technology. The flow of the program (Fig. 14.21) is intended to be controlled via a steering file, where all simulation parameters can be defined by the user. The program itself is then run either in batch mode, where no interaction with the user is expected or in interactive mode, where the program flow is driven by Geant4 UI (User Interface) commands. If no steering file is provided or not all required parameters are given by the user, the default values accessible via a singleton class `G4 Control` are used instead. The main inputs of the simulation are the geometry description and the primary particle(s) information. The first is accessible from the central geometry repository via its interface (Sec. 14.4.5); the latter can be described either by single particle parameters—used by `G4 Primary Generator`—or in an event generator file that is read using a suitable `G4 Generator` interface. The output of the chain is recorded in an object-based format described by the event data model (Sec. 14.4.4) with a predefined persistency. In our case, the persistency is based on ROOT I/O libraries [17].
The basic idea of the Geant4 design philosophy can be described in the following way. The primary unit of an experimental run, represented by `G4 Run`, is an event (`G4 Event`) and consists of a set of primary particles produced in an interaction and a set of detector responses to these particles. Before each event is processed, it contains only primary vertices and primary particles defined in a steering file or provided by an external physics generator. After the event is processed, it is augmented with created hits, generated trajectories and created secondary particles. The hits are snapshots of a physical interaction in sensitive parts of the detector and are saved in hits collections.

#### 14.4.6.3 Digitization Software

Digitization software performs for each detector component (PXD, SVD, CDC, etc.) a detailed simulation of in-detector physics processes together with electronics effects, taking into account the desired detector geometry and magnetic field information accessible via the geometry interface. In this way, it transforms one or more individual Geant4 hits into so-called

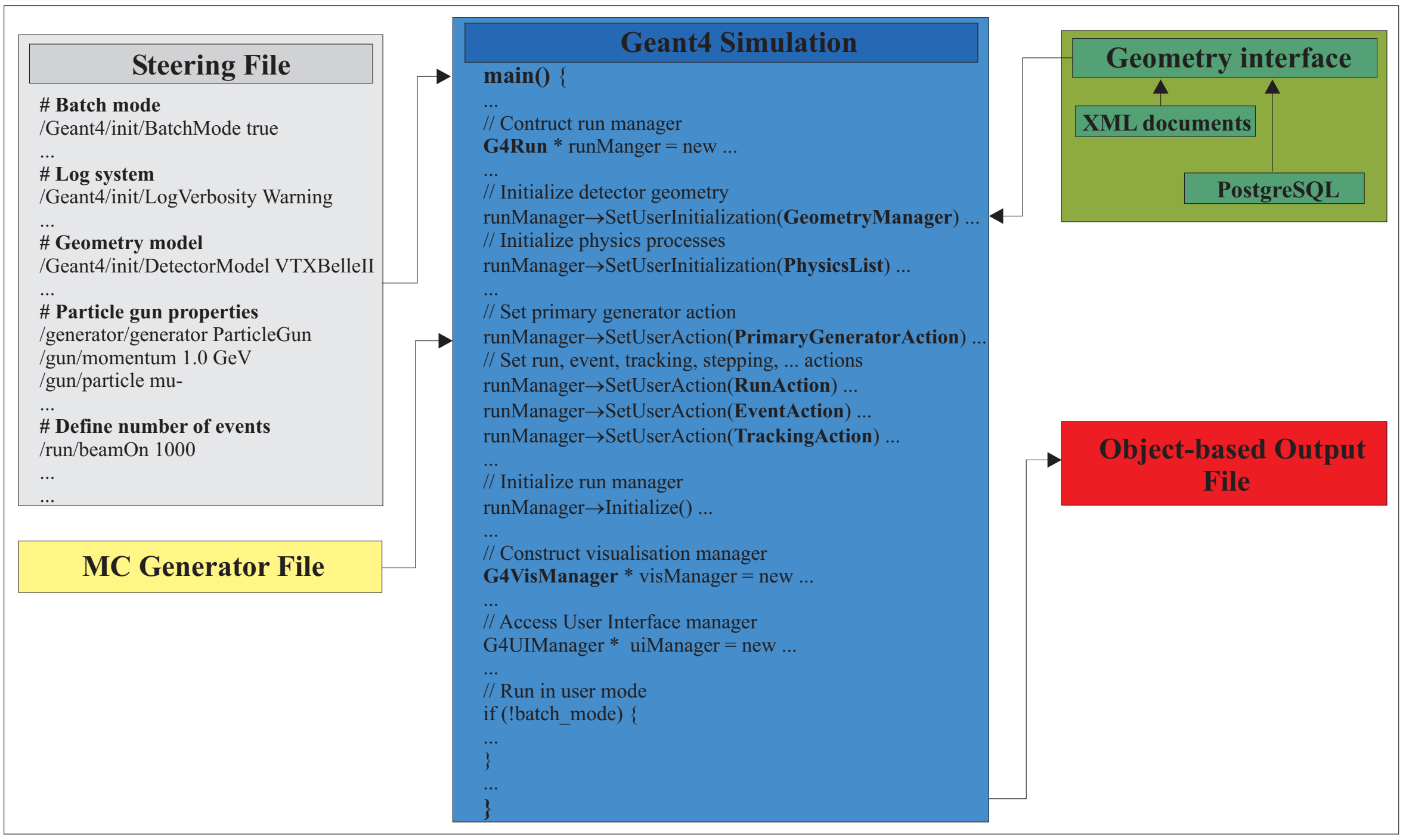

*Figure 14.21: The scheme of detector simulation software based on the Geant4 toolkit.*

digits—measured detector signals. The resulting simulated data has the same format as real raw data and can be processed further by the reconstruction modules of each detector component. For detector design studies, special reconstruction algorithms may be used that are faster than the default reconstruction code for real data or that can read the underlying "truth" information about the simulated event.

As the digitization procedure requires knowledge of how each sub-detector works and which physics processes are relevant for signal formation, the software is completely written by an expert from the sub-detector group in C++; a modular approach is used instead of a complete integration of the code within the Geant4 toolkit. An advantage of such an approach is that, during the detector design phase, one can easily reprocess all the data with different settings, for example voltages or noise characteristics, and avoid having to run the whole Geant4 simulation again, which might be very time- and CPU-consuming. Another advantage is that, for one sub-detector, several modules with different level of detail may exist and a user then selects just the desired one. For instance, MC data production with full detector simulation requires much less detail in digitization than detailed studies of a sub-detector prototype in a testbeam environment.

As an example, a brief description of the SVD digitization is given here (see Fig. 14.22). After a particle crosses a detector volume, electron-hole pairs are generated along its path (calculated from Geant4 hits) and naturally grouped into $e$–$h$ pair clusters based on the spatial precision required. These clusters are then drifted in the presence of an electric field to collecting electrodes-strips. The electrons drift to one side, the holes to the other. They are diffused by multiple collisions, modeled by Gaussian smearing, and shifted by a Lorentz angle that is calculated based on the information about the ambient magnetic field. Due to the fact that individual strips are connected to the circuit by DC/AC coupling and might mutually communicate by inter-strip capacitance, the collected charge is redistributed and a signal is also generated on adjacent strips. Finally, the noise on each strip is randomly generated and added.

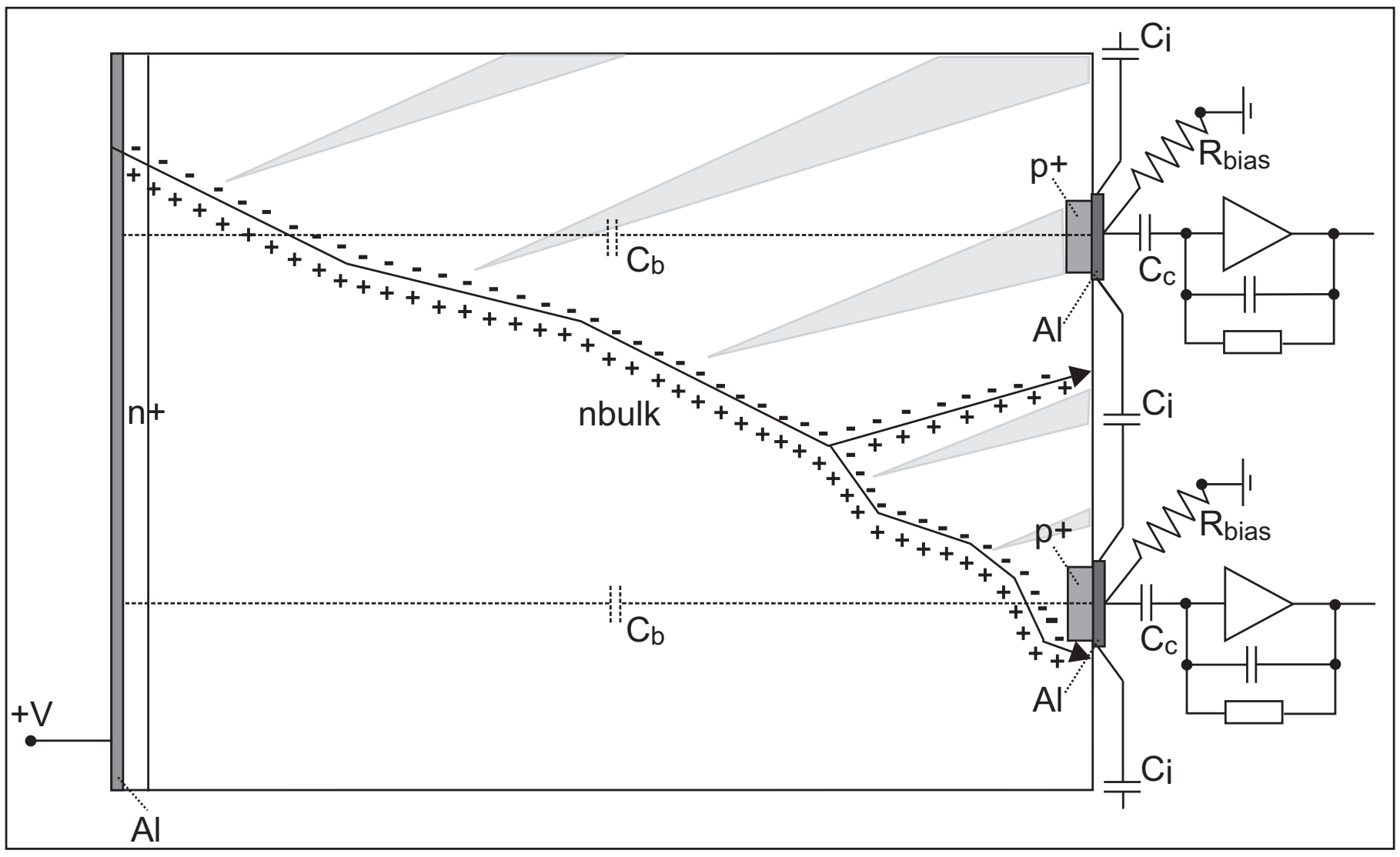

*Figure 14.22: Signal digitization in silicon microstrip detectors: 1D projection, with delta-electron generated.*

#### 14.4.6.4 Beam Background Effect

To simulate the beam background effect, two approaches are pursued. One is to generate the beam background particles based on the inputs from the accelerator design with consideration of the beam-gas interaction, synchrotron radiation, Touschek effect, and so forth. This is useful to evaluate the performance during the sub-detector design phase before real data is available. However, it is quite time-consuming to simulate all such particles, including the low energy photons, and quite difficult to reproduce the realistic condition of the acclerator run-by-run once the experiment starts. An alternative approach makes use of the real data that is taken with a trigger that is not synchronized with the collision timing. In Belle, such random-triggered events are embedded to the MC and works well to reproduce the run-dependent experimental conditions. However, this is not available before the experiment starts. Therefore, we need both approaches in Belle II. The former is treated as part of the event generators and the latter as part of the simulation tools. Both should work as an independent loadable module in the Belle II software framework.

### 14.4.7 Reconstruction

The reconstruction of events acquired in the track-sensitive parts of a detector is a chain of tasks, each of which contributes to detector abstraction and data reduction. Starting with raw measurements, these tasks are calibration, local pattern recognition in individual subdetectors (track search), global pattern recognition (track finding), track parameters fitting (based on an appropriate track model), vertex finding, and geometric and kinematic vertex fitting, ending with the estimation of the event parameters relevant for subsequent physics analyses.
Calibration and (in case of the CDC) local pattern recognition are tasks of subdetector-specific software, and therefore will not be discussed in this section. However, they will be assumed to be reliably available.
This section deals with the next steps of the reconstruction chain. Their implementations will be based on established state-of-the art algorithms and techniques, and will take advantage of available abstract interfaces to the framework (Sec. 14.4.3), to the data model (Sec. 14.4.4), and to the geometry & materials information (Sec. 14.4.5). Access to the "simulation truth" (Sec. 14.4.6) and interfaces to the event graphics (Sec. 14.4.8) are required for software testing and optimization studies.
The reconstruction effort is still in a very early phase. This description of our approaches to the tasks mentioned above is intended to show the broad range and sophistication of techniques available; it is not exhaustive. For a thorough review, see Ref. [27].

#### 14.4.7.1 Track Finding

The task of global pattern recognition, a.k.a. track finding, is to determine which hits in the track-sensitive parts of a detector belong to a common track. Information from a preceding local pattern recognition, yielding a track candidate with associated hits, is used if available.
There is a large variety of algorithms for track finding. Which one to choose depends, among other things, on the track density, the background noise level, the distance to the interaction point (IP), and the subdetector in question. For instance, the silicon detector hits have high spatial resolution but an appreciable amount of multiple scattering, whereas the CDC hits have little multiple scattering but lower spatial resolution. Some track finding algorithms rely on the circular shape of the track in the transverse projection, and those will work only in regions where the magnetic field is reasonably homogeneous.
Due to the smaller track density in the outer subdetectors, we propose to apply an "inwards/outwards" strategy:

- Track candidates found by local pattern recognition in the CDC are extrapolated inwards to the silicon detector PXD/SVD. Tracks that can be found by this technique have a transverse momentum high enough to cross most or all of the CDC.

- Each PXD/SVD hit is tested against each extrapolation. Those hits that can be associated with a track candidate are removed from the hit collection.

- The remaining PXD/SVD hits are the input to a local pattern recognition in the silicon detector. Various techniques can be used in this "silicon stand-alone track search." Among them are the conformal transformation [27], the Hough transform [28], the elastic net [29, 30], the combinatorial Kalman filter [31] and related methods of progressive track search.

- In the final step, track candidates found in this way are extrapolated from the silicon detector outwards into the CDC. If unused CDC hits are found that fit to the extrapolation, they are associated with this track candidate.

We end up with a set of global track candidates, each consisting of a crude estimate of its parameters together with an ordered list of associated PXD/SVD/CDC hits, and a set of unassociated hits. A more complex approach allows hits to be shared among two or more track candidates; in this case, the final hit association is relegated to the next step, track fitting.

#### 14.4.7.2 Track Fitting

The trajectory of a charged particle moving freely in a static magnetic field is determined by the Lorentz force law. It is uniquely described by five parameters, representing the initial conditions at (for example) the point of intersection with a chosen reference surface: two positions, two directions, and a quantity depending on the momentum.

The global track candidates that have been found in the previous step are subjected to single track fitting based on some track model, with the virtual measurements and their errors being derived from the hit coordinates. The purpose of this step is twofold: (i) find a "best estimate" for the track parameters, yielding a 5-tuple and a corresponding $5 \times 5$ covariance matrix; and (ii) test and possibly update the track hypothesis, i.e., the association of hits to this track, by identifying and removing or down-weighting "outliers" (true measurements with errors larger than shown, or false ones not belonging to the track at all) and, if present, resolving ambiguous associations of hits. These two tasks are achieved by an iterative procedure.

The choice of an appropriate track model depends on the magnetic field. In Belle II, the field is sufficiently close to being homogeneous in the central tracking detectors, thus implying a helix track model. Moreover, the detector layout is approximately rotational symmetric with respect to an axis parallel to the magnetic field. The track parameters suggested to be used internally in this step are given in the appendix of Ref. [11].

The free particle trajectory is disturbed by material effects in the detector. Multiple Coulomb scattering is a stochastic process that is well described by the Rossi-Greisen formula with Highland's correction term [32].[1] Since the material budget is concentrated in thin layers (PXD, SVD, CDC walls), those may be approximated by zero geometrical thickness; in this case, no lateral displacement takes place, and only the two direction parameters are affected in the course of track fitting.

The other material effect is energy loss. For minimal ionizing particles, this is dominated by ionization; if taken into account (in particular for low momentum tracks), it may be described deterministically by the Bethe-Bloch formula with Fermi plateau [27]. For electrons, energy loss is dominated by bremsstrahlung, a stochastic process that can be described by the Bethe-Heitler model [34]. As it is extremely non-Gaussian, this poses problems that can be solved by special modifications of the fitting algorithm, e.g., the Gaussian Sum Filter [35] or alternatives [36].

State-of-the-art in track fitting are techniques based on the Kalman filter [37], a linear estimator that is statistically equivalent to the least squares method, but is superior in coping with the requirements of a complex modular detector. It is applied on a linear approximation to the track model, the expansion point being a "reference track" sufficiently close to the true track; the crude estimate from track finding may be a good choice. If necessary, a Newton method can be applied, with the new reference track being the one fitted in the previous iteration.

The best estimate of the track parameters is achieved only at the end of the filter, with all measurements having contributed. To obtain best estimates all along the trajectory, the filter must be complemented by a smoother. This can be achieved with help of a second filter, running in the opposite direction. The smoothed $\chi^2$s may serve as test criteria for outlier measurements; in the realistic case of multiple outliers, however, the power of these tests is limited.

Linear estimators are inherently sensitive to the influence of outliers. Robustification can be achieved by using non-linear estimators, which in the case of track fitting are most easily implemented as iterative extensions of the Kalman filter: measurements with "big" normalized residuals are down-weighted in the next iteration. Such adaptive filters are largely resistant against wrong measurements; thus, they are able to defer the final association of hits from the

[1] except for extremely thin material with few scatterings; for details see [33].

track finding step to the track fitting step, where more complete information about the track is available for the test criteria. This final association need not be a “hard” one, as the final weights of the measurements can be anywhere between 0 and 1.
There exist several adaptive filter methods. In particular, we intend to explore the Deterministic Annealing Filter (DAF) [38] that has successfully been implemented and refined by the CMS experiment at LHC.
In addition, a detector-independent skeleton toolkit for track reconstruction exists (GENFIT [39]); its usefulness will be evaluated once a proven implementation is available. Another challenge is flexible track propagation in complex sequences of cylindrical and plane detectors, as in the forward region of PXD/SVD; a detector-independent implementation (LDT [40]) may serve as an example.

#### 14.4.7.3 Alignment

During the detector alignment, the detector geometry is corrected for small displacements of the sensitive elements in the detector. As a general rule, the precision of the alignment should be significantly better than the intrinsic resolution of the sensitive elements. To achieve this, various strategies can be used. Sensor positions can be measured in the lab or in situ by lasers. To obtain the ultimate precision, however, reconstructed tracks have to be used.
The alignment task can be performed in several stages. The first is the internal alignment of the individual subdetectors. This is followed by the global alignment, in which the position and orientation of the entire subdetectors with respect to each other is determined. This has the advantage that the number of alignment parameters that are estimated concurrently is reduced. A possible disadvantage is the problem of achieving a consistent global alignment while keeping the internal alignments fixed. It may therefore be necessary to envisage a final adjustment of all alignment parameters of all subdetectors.
The required frequency of realignment depends mainly on the mechanical stability of the various components. The mounting of the PXD detector on the beam pipe may make it necessary to do one or even several global realignments per day.
With a sufficiently large number of tracks, the statistical errors of the estimated alignment parameters can be made as small as required. The challenge is to control the systematic errors to the required level [41]. This is because, for any kind of tracks, there are several degrees of freedom that are not constrained, usually referred to as weak modes. A mixture of tracks that is as diverse as possible is required to constrain all or at least most of the weak modes.
The requirements of past and current collider experiments on alignment performance, both in terms of precision and in terms of the sheer number of parameters to be estimated, have instigated the development of a generic algorithm called Millepede [42, 43, 44]. It has been used and is being used by several large experiments. Clearly, Millepede is an excellent candidate for the alignment task of Belle II. It is worth noting, however, that the two largest LHC experiments, ATLAS and CMS, which have the most difficult alignment tasks, do not rely on a single alignment method, but have implemented several methods. This is extremely useful for debugging and cross-validation [45, 46, 47].
Alignment can be complemented by a precise estimation of the material budget, using reconstructed tracks. This is particularly important for low momentum tracks, which are strongly affected by energy loss and multiple Coulomb scattering. Estimation methods for this task are currently under development.

#### 14.4.7.4 Vertexing

Vertex reconstruction consists of (i) vertex finding and track bundling (a pattern recognition task); (ii) geometric vertex fitting (statistical estimation of the vertex position and the track parameters at the vertex); and (iii) re-fitting with the kinematic constraints of energy-momentum conservation. These tasks are, in practice, intertwined.

The virtual measurements of the vertex fit are the tracks fitted in the previous step, and extrapolated inwards into the beam-tube; thus, no material effects need to be taken into account any more. The track parameters and covariance matrix may be defined on a reference surface (e.g., the inside of the beam-tube), or as perigee parameters with respect to a fixed pivot point close to the beam interaction (see appendix of Ref. [11]). In the case of a primary vertex, the beam interaction profile, if known, may be included as another virtual measurement.

A vertex fitted from $n$ tracks consists of $(3+3n)$ parameters: 3 position coordinates, and 3 track parameters at the vertex for each track. For a subsequent kinematic fit, the full $(3+3n)\times(3+3n)$ covariance matrix is required; if this is not the case, only some $3 \times 3$ sub-matrices need to be computed.

Geometric vertex reconstruction starts with a Kalman filter [37, 48], with the linear expansion point being chosen from the beam interaction profile and/or crude track intersections. A smoother is required for obtaining the best estimates of all tracks added before the last one. The smoothed $\chi^2$s may be used as test criteria for outliers, i.e., tracks not belonging to this vertex; however, the power of these tests suffers for the same reasons as discussed in the case of track fitting above.

The remedy are robustified, non-linear, adaptive filters—in particular, the Deterministic Annealing Filter (DAF) [49], which down-weights the influence of outlier tracks on the fitted vertex. It introduces an annealing schedule that serves two purposes: it prevents the fit from falling into a local minimum, and allows progressively moving from “soft” to ultimate “hard” assignments. This leads in a natural way to a technique of vertex finding, called Adaptive Vertex Reconstruction (AVR [50]): if the DAF is applied to all tracks, with the proper setting of the annealing schedule, it converges to the vertex with the largest number of tracks. The tracks attached to this vertex are removed from the track collection, and the DAF is applied to the remaining tracks. This is iterated until no more vertices can be found. The AVR will be evaluated against other vertex finders, in particular, the topological vertex finder ZvTop [51].

The DAF can be further generalized by introducing a Multi-Vertex Filter (MVF) [52], simultaneously fitting $n$ tracks to $m$ competing vertices by “soft assignment” of each track to more than one vertex.

The inclusion of electron tracks fitted with the Gaussian Sum Filter (GSF) causes the vertex fit also to be modeled as a Gaussian sum. This may result in an exponential bloat unless compensated for by an appropriate “collapsing strategy” [53]. The GSF is fully compatible with the adaptive methods (DAF, MVF).

Kinematic fitting imposes the constraints of energy-momentum conservation on each vertex. The method used is that of Lagrangian multipliers [27, 54]. It can be achieved either as a separate re-fit step after pure geometric vertex fitting, or directly by a simultaneous geometry & kinematic fit. Problems arise in cases of insufficient information (unknown masses, unseen particles, etc). Complex cascade decay chains are difficult to handle in a general way. Kinematic fitting may more likely be subject to the requirements of a subsequent physics analysis.

At Belle and Belle II, vertex reconstruction is of utmost importance for physics studies like $CP$ violation in $B\bar{B}$ systems, which rely on precise measurements of $\Delta z$ between the two $B$ decay vertices. This can only be achieved if the high-resolution PXD and SVD detectors are complemented by sophisticated reconstruction, based on statistical methods fully exploiting the information available.
Vertex reconstruction, being the last step in the event reconstruction chain, may be boxed in a detector-independent toolkit. Such a toolkit exists (RAVE [55, 56]). Its algorithms are source-code compatible with those implemented for the CMS experiment at LHC. They include the Kalman filter, the DAF, the MVF, the GSF, and the AVR; an interface to ZvTop also exists. Kinematic vertex reconstruction is implemented but is limited to the Kalman filter; it allows for user-defined decay topologies. Implementing RAVE for Belle II will require only minimal effort for interfacing to our framework's track propagator and data model.
Geometric vertex finding and fitting in RAVE are state-of-the-art. Its kinematic vertex reconstruction, however, will require a critical comparison with that of the old Belle framework. Given the importance of kinematic fitting for the analysis of Belle II data, a symbiosis between RAVE and ExKFitter could be feasible to attain the same functionality and quality in kinematics.

### 14.4.8 Event Display

The data, flowing through the offline software, is usually visualized by the user via histograms or scatterplots or by direct printing of the values. But there are occasions where a more pleasing visualization is advantageous. In the following, examples for data visualization and the requirements they impose on a visualization solution are given. Finally, the approach taken for the Belle II offline software is presented.

#### 14.4.8.1 Examples of data visualization needs

The need for proper data visualization starts in the early phase of the offline software life cycle. During the development of algorithms, it is mandatory to consistently check that the algorithms behave correctly. Besides the usual histogram/direct printing approach, event-visualization techniques are extremely helpful in doing this. For example, the development of pattern recognition/tracking algorithms is greatly supported by a three-dimensional visualization of tracks, their associated hits and the tracking geometry. In particular, it is often enlightening to compare the simulated tracks with those found by the pattern recognition/tracking algorithms visually. Also, using visual tools to build the detector from its subdetectors can illuminate problems associated with the overlap of detector parts, uncovered gaps in the acceptance region, or cabling issues. Later, during the running phase of the detector, event visualization is helpful to support the surveillance and monitoring of the data acquisition and to provide a visual feedback for the control room.
The fields of application mentioned above require a simple-to-use yet powerful software to visualize event-level data of the Belle II experiment. Therefore, its main purpose is to display the following items in a three-dimensional view and in different two-dimensional projections:

- detector geometry
- tracks (simulated and reconstructed)
- vertices (simulated and reconstructed)
- reconstructed hits
- additional information like energy deposit, particle id ...

The event display should be easily adjustable for various types of input data, should provide tools to work and display this data in a physics-oriented way and allow the creation of high-quality pictures for talks and public relations purposes. Therefore, a modular system was chosen for the event display software. The next section provides a more in-depth view of the solution found for the Belle II experiment.

#### 14.4.8.2 The GenericEventViewer

The event display developed for the Belle II experiment carries the name **GenericEventViewer** (GeV). As the name suggests, the GeV was developed with a generic data- and task-independent concept in mind. Therefore, the GeV is not based on the common software framework. Rather, due to GeV's modular design, the software framework will be tightly integrated into the event display. This integration will allow the user to load a framework steering file into the Event Display and run it there. The user has then the possibility to change parameter values of framework modules in the steering and will experience a visual feedback immediately. The same holds for changes made to the source code of a framework library or module. The high modularity of the GenericEventViewer is achieved by using plugins. In general, plugins are user-written extensions to an already existing application and widely used in large software projects to keep source code complexity moderate. In the language of software design, an application that can be extended by plugins is called a plugin host. It provides a well-defined interface for accessing and modifying data, stored inside the application, and a mechanism to dynamically link plugins at runtime into the application. New features can be added to the application without having to recompile the application itself. In addition, plugins automatically benefit from updates made to the core functions of the host application. Depending on the operating system, plugins are stored in shared object files (*.so, Linux) or dynamic link libraries (*.dll, Windows). The GenericEventViewer contains a newly developed, operating system independent, plugin system capable of loading and executing user written plugins. There are different types of plugins:

- **Import plugins**
  Load data into the GenericEventViewer by converting the data to geometry objects.
- **Export plugins**
  Save the geometry data to files, which can then be used by external software.
- **Render plugins**
  Render plugins create pretty looking pictures from the geometry data. Usually external render engines are used.

Being the most important plugin type, the import plugin mechanism is now presented in more detail. Using the import plugin, data from various sources can be brought into the GenericEventViewer. Inside the import plugin, data is read from a source and converted to geometry objects. For example, hits in a subdetector are represented by three-dimensional points, tracks as helices and detector geometry as a collection of boxes, cylinders and spheres. The GeV provides all necessary geometry types and transformations (rotation, translation). All created geometry objects are stored in a list. From this list, different hierarchical representations, called views, are derived. This concept allows GeV to present the same objects in different kinds of hierarchies to the user. For example, one can think of a hierarchy showing the geometrical relationship of particles/tracks and another hierarchy concentrating on the parent/child relationship of particles.

Each geometry object can carry additional information, called MetaData. This allows GeV to attach information like momentum, vertex position and particle type to tracks or energy deposit and position to hits. Using this MetaData, the user of the event viewer can then interactively apply cuts on certain values (e.g., momentum) to limit the number of objects visualized.

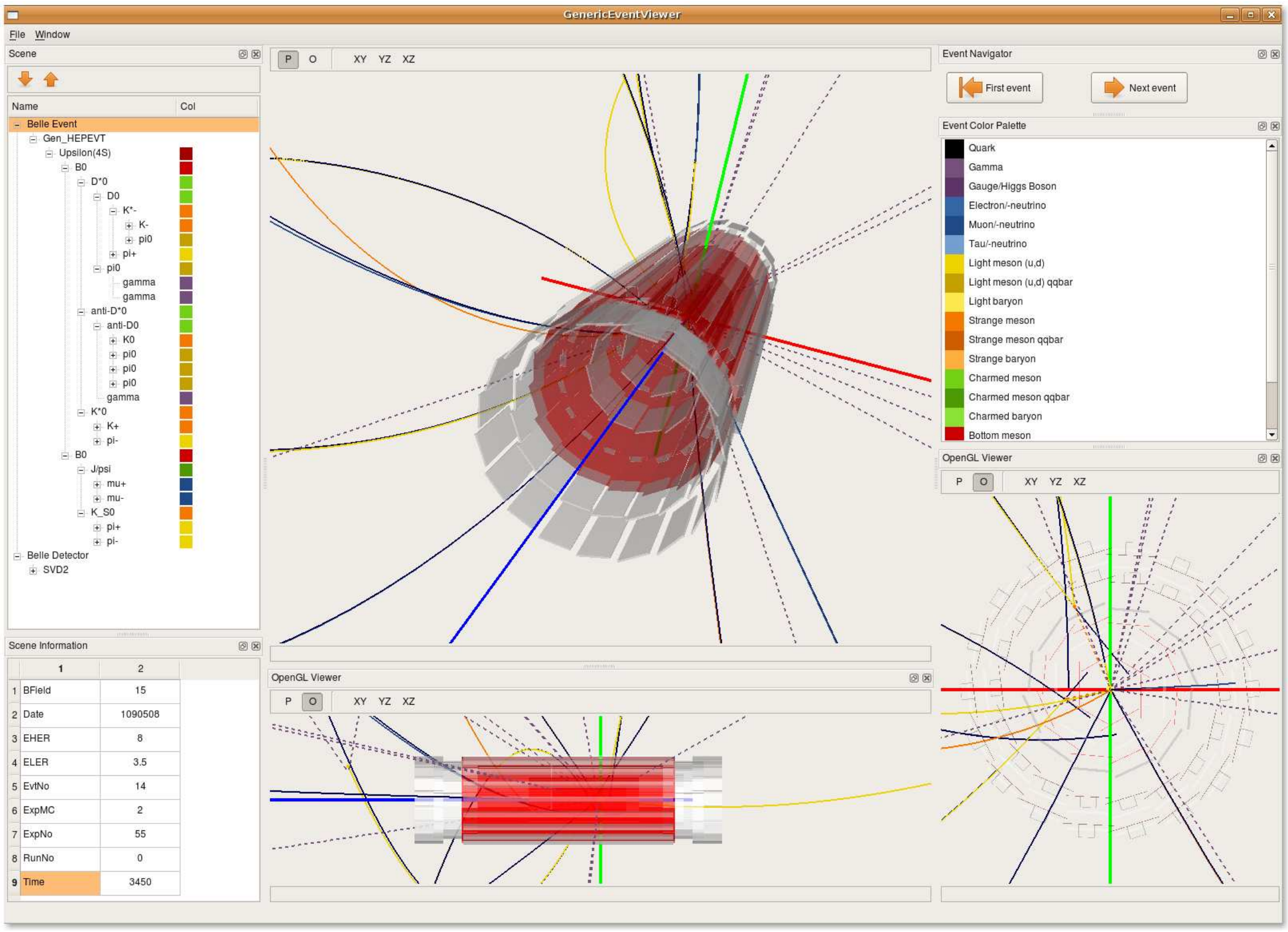

*Figure 14.23: Screenshot of the GenericEventViewer. Shown are the tracking subdetectors and the particle trajectories from a simulated $B\bar{B}$ event*

All objects are shown in a three-dimensional view (Fig. 14.23), which can be rotated, panned and zoomed in realtime. OpenGL is used for hardware accelerated and operating system independent drawing of the three dimensional shapes. It is possible to select objects to display their MetaData or perform per-object oriented tasks like hiding objects or applying specific actions to them. The process of selecting objects is carried out either by selecting the appropriate entry in a tree widget, which represents the internal object hierarchy, or by simply selecting the objects in the three-dimensional view.

Special emphasis is put on a particle physics oriented visualization solution. Therefore, the GenericEventViewer supports various three-dimensional drawing modes (projection, orthogonal, fish-eye), two-dimensional representations (e.g., $x$–$y$, $r$–$z$) and advanced event selection features based on the geometrical topology of an event.

## 14.5 Resource Requirements

To develop, maintain, and operate the components of the Belle II computing system described in the previous sections, considerable human resources are required. Because of the large expected data size, there is a high demand of computational, storage, and network resources as well. The anticipated requirements for both types of resources are described in detail in this section.

### 14.5.1 Human Resources

The tasks that have to be covered to develop and run the Belle II computing system can be roughly classified into the following categories:

- Code development
- Operation of central (Belle II specific) services
- Operation of grid sites and grid services
- Data processing and MC production
- Support and training
- Management

Individual roles assigned to the task categories mentioned above are listed and described in detail in the Belle II Computing TDR [11]. An overview of the roles and their relations is given in Fig. 14.24. There is no exact one-to-one matching between roles and persons. One person can have several roles, and one role can be shared by more than one person.
In addition to the categorization of tasks according to the type of work, the computing group is organized in working groups focusing on particular topics that can involve and provide a connection between code development and operational tasks. The working groups are

- **Core Software Group**: develops and maintains the analysis framework (Sec. 14.4.3), the event data model (Sec. 14.4.4), the geometry and conditions database (Sec. 14.4.5), and the event display (Sec. 14.4.8). It is also responsible for the operations of the central conditions database.
- **Distributed Computing and Data Management Group**: responsible for the planning, deployment and operation of computing, storage, and network resources. It also has to develop and operate a system to make these resources available to the Belle II collaboration (Sec. 14.3).
- **Simulation Group**: mainly consists of experts from detector groups who develop the simulation software for the Belle II detector components (Sec. 14.4.6).
- **Tracking Group**: based on the local detector component reconstruction provided by detector experts this group develops and maintains global tracking and alignment code (Sec. 14.4.7).

Some topics are not covered by groups, but by individuals. For example there are **Coordinators** who takes care of the code management, the calibration, the raw data processing, and the MC production.
Decisions on priorities of competing tasks will be taken by the **Computing Steering Committee**, composed of the Distributed Computing and Data Management Group Convener, the Data Processing, MC Production, Physics, and Computing Group Coordinators.

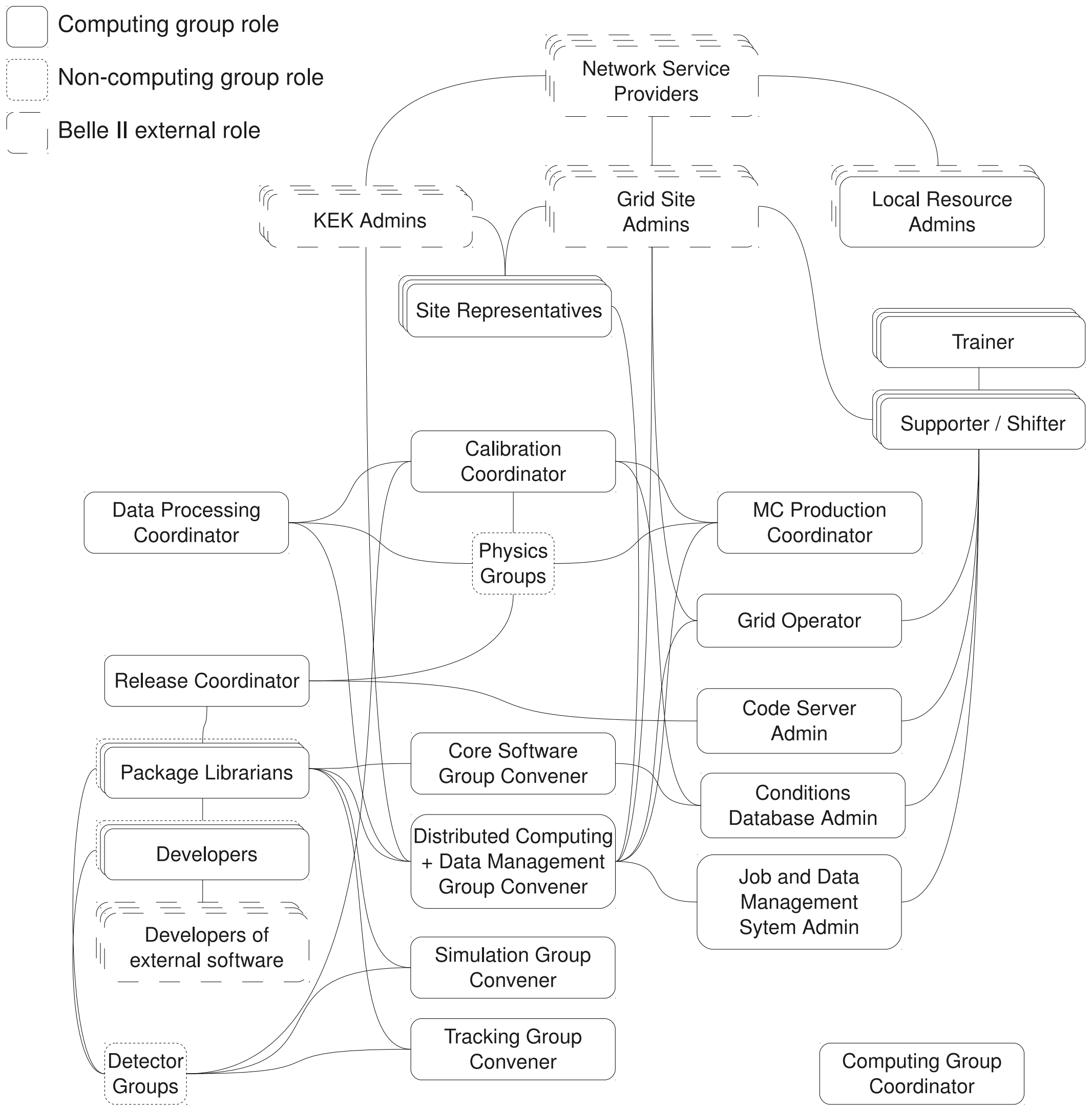

*Figure 14.24: Roles involved in the development, maintenance, and operation of the Belle II computing system. Envisaged management flows are indicated by lines. Not all relations are shown.*

### 14.5.2 Hardware Resources

As described in Sec. 14.2, a hierarchical structure is considered for the Belle II distributed computing system. KEK should be the main computer center, where all raw data is processed and stored. All produced DST and mDST files from the raw data are also archived at KEK. Here, the DST file is stored in RAW data format and used for the calculation of calibration constants for each sub-detector and sometimes for systematic error studies. On the other hand, mDST files have only reconstructed track/cluster/particle-ID information in an analysis-friendly format. Analysis users can make any analysis using this mDST file format.
The remote grid sites are responsible for the MC production and for the user analysis. Furthermore, the produced MC data in mDST format is stored at the site where it was created. In particular, KEK serves not only as the main center but also as a grid site for MC production and user analysis. In this section, we estimate the requirement for the hardware resources to establish the proposed computing model.

#### 14.5.2.1 Expected event rate and expected event data size

The expected event rate can be roughly calculated from the instantaneous luminosity and the event production cross section. The SuperKEKB accelerator will be operated at the energy of the $\Upsilon(4S)$ resonance (= 10.58 GeV/$c^2$) and the $b\bar{b}$ production cross section at this energy is roughly equal to 1 nb. Though the target instantaneous luminosity is $8 \times 10^{35}$/cm$^2$/s, the SuperKEKB accelerator will at the early stage of the experiment be operated with a lower luminosity, which will then gradually increase.
In this estimation, we assume the instantaneous luminosity at around 2016 to be $2 \times 10^{35}$/cm$^2$/s and the running-time of the accelerator to be roughly two thirds of a year. The rate for $b\bar{b}$ events is calculated to be $\sim$ 200 Hz and the number of $b\bar{b}$ events is to be $4 \times 10^9$ events per year. The current Belle experiment has accumulated an integrated luminosity of $\sim$ 950 fb$^{-1}$ in ten years of operation, and this corresponds to roughly $1 \times 10^9$ $b\bar{b}$ events. That is, four times the entire Belle data will be accumulated in one year even in the early stage of the experiment.
However, $b\bar{b}$ events are not the only events in the Belle II experiment. We need to take into account other $q\bar{q}$ events ($\sim$ 3 nb) as well as $\tau^+\tau^-$ pair production ($\sim$ 1 nb). For these physics events, we need to handle on the order of $10^{10}$ events per year.
For the moment, the amount of beam background due to lost beam particles, beam-gas interaction, synchrotron radiation, and intra-beam effects is still being estimated, and we do not quote a final value for our estimation. However, according to the Trigger group, after applying the High Level Trigger, the event rate to the storage is estimated to be 6 kHz at an instantaneous luminosity of $8 \times 10^{35}$/cm$^2$/s in the worst case. Of this, 4 kHz comes from physics events and correspond to the cross section 5nb. The remaining 2 kHz arises from the beam background. We follow this Trigger group's estimation and adopt it in our calculation (see Table 14.4).

*Table 14.4: Cross section for each physics event.*

| | cross section (nb) |
|---|---|
| $b\bar{b}$ | 1 |
| $q\bar{q}$ (continuum) | 3 |
| $\tau^+\tau^-$ | 1 |
| Subtotal (physics) | 5 |
| background (cross-section equivalent) | 2.5 |
| total | 7.5 |

The projected integrated luminosity per year is quoted from the luminosity prospect provided by SuperKEKB and summarized in Table 14.5.
Based on Tables 14.4 and 14.5, we can calculate the expected number of events for each fiscal year, summarized in Table 14.6.
The total event size is estimated by the Trigger group and ranges from 100kB to 300kB. The event size easily varies by the data size from PXD in the current estimation. We take the worst case, 300kB, for our resource estimation.

*Table 14.5: Expected annual integrated luminosity. (See also the luminosity prospect in Fig. 14.25.)*

| Fiscal year (Apr.1-Mar.31) | 2014 | 2015 | 2016 | 2017 | 2018 | 2019 |
|---|---|---|---|---|---|---|
| Projected $\mathcal{L}_{int}$/year [$\text{ab}^{-1}$] | 1 | 2 | 8 | 11 | 13 | 13 |
| Total $\mathcal{L}_{int}$ | 1 | 3 | 11 | 22 | 35 | 48 |

*Table 14.6: Expected number of events by the end of each fiscal year.*

| Fiscal year (Apr.1-Mar.31) | 2014 | 2015 | 2016 | 2017 | 2018 | 2019 |
|---|---|---|---|---|---|---|
| $b\bar{b}[\times 10^9]$ (events/year) | 1 | 2 | 8 | 11 | 13 | 13 |
| Hadronic $[\times 10^9]$ (events/year) | 4 | 8 | 32 | 44 | 52 | 52 |
| All $[\times 10^9]$ (events/year) | 7.5 | 15 | 60 | 83 | 98 | 98 |
| Integrated $[\times 10^9]$ (events) | 7.5 | 23 | 83 | 165 | 263 | 360 |

#### 14.5.2.2 Resource requirements for DST/MC production and user analysis

KEK, as a main computer facility, is responsible for the processing and storage of the raw data. We adopted the event rate and event data size estimated by the Trigger group in our calculation. But there are still many uncertainties in the event data size, effect of the beam background and optimization of the High Level Trigger, and so on. To take these into account, we apply a safety factor of 2 to the storage space for the raw data, which will be estimated more realistically in the near future.

The procedure of processing of the raw data and preparing the mDST files for physics analysis is referred to as a DST production, as explained in Sec. 14.2. Usually, the DST production follows a certain completed experimental running period, since the final detector-calibration constants are required for the raw data processing. We assign a time period of $T_{Data} = 5$ months for the processing stage in the resource estimation. The time estimate for the processing of a single event is based on the current Belle CPU resources and the obtained value is 8.33 HepSPEC $\cdot$ sec/event [57], which includes also the effect of the hyperthread technology. (A typical HepSPEC value for Intel Xeon 3.6GHz 2core is roughly 14.)

After the processing of the raw data, the physics skimming process is applied to produce the mDST files. The ratio of events produced after the physics scheme is assumed to be 1/3 and the size of the mDST is assumed to be 40 kB/event, which is again based on the Belle experiment. This event size could be larger if the amount of data from the PXD can not be reduced enough. For the calibration and sub-detector studies, like the determination of the alignment constants for the PXD and/or the SVD, the DST files are needed as well. However, it seems that 2% of the processed raw data saved as DST files is enough for this purpose.

We need MC events corresponding to at least 3 times the real data to reduce the systematic error associated with statistical fluctuations in the produced MC events. In the Belle experiment, 10 streams of the MC were prepared. Therefore, as a baseline we adopt $N_{streams} = 6$ streams of the MC. To reproduce the condition when the data was taken, the MC event is produced with the measured detector constants and with the embedding of real beam background that was taken during the experiment. Most of these are prepared after the experiment running period and then the MC production starts at the grid computing facilities. We set a period of $T_{MC} = 5$ months for the MC production. The size of the mDST for MC events is assumed to be 40 kB/event, which is the same for the real data. The mDST for MC events consists of the physics analysis oriented information, e.g., track and particle ID, as well as the MC generator level information. Roughly 10% of the mDST event is occupied by the generator information; in other words, the size of the mDST for MC becomes larger than that for data by 10%. However, this 10% increase is smaller than other uncertainties in the estimation of the storage space for raw data. Therefore, we use the same value for the mDST event size for both data and MC cases.

Table 14.7: Summary of the input parameters for the resource estimation.

| parameter | setting value |
|---|---|
| Raw data size [kB/event] | 300 |
| DST [kB/event] | 300 |
| mDST(data) [kB/event] | 40 |
| mDST(MC) [kB/event] | 40 |
| safety factor for the storage space of raw data | 2 |
| DST production period [months/year] | 5 |
| MC production period [months/year] | 5 |
| CPU resources for DST prod. [HepSPEC·sec/event] | 8.33 |
| CPU resources for MC prod. [HepSPEC·sec/event] | 16.44 |
| # of MC streams | 6 |
| DST event output fraction | 0.02 |
| a fraction of CPU power for user analysis wrt. DST production | 0.5 |
| a fraction of storage space for user analysis wrt. mDST | 0.5 |
| Conversion factor of the CPU power to power consumption [W/HepSPEC] | 3.83 |
| Conversion factor of the storage space to power consumption [kW/PB] | 24 |

In the current Belle experiment, the CPU is shared by the DST/MC production and user analysis. This situation will not change in Belle II: 50% of the CPU resources assigned to the DST production is assumed to be used for user analysis. For the storage, we just assume that a disk space corresponding to 50% of mDST is used by user analysis.

For the conversion factors to evaluate the power consumption due to the CPU power and the storage space, we adopted 3.83 W/HepSPEC based on the current typical 8-core machine and 24 kW/PB based on the existing RAID server, respectively.

By combining these inputs, summarized in Table 14.7, we obtained the requested power of the CPU and amount of the storage for the DST production (Table 14.8), MC production (Table 14.9) and user analysis (Table 14.10).

*Table 14.8: Expected amount of storage and CPU power for DST production.*

| Fiscal year (Apr.1-Mar.31) | 2014 | 2015 | 2016 | 2017 | 2018 | 2019 |
|---|---|---|---|---|---|---|
| **Raw data** | | | | | | |
| size per year [PB] | 4 | 8 | 32 | 46 | 54 | 54 |
| Total size [PB] | 4 | 12 | 45 | 90 | 143 | 196 |
| **mDST** (per one version) | | | | | | |
| size per year [PB] | 0.2 | 0.4 | 1.5 | 2.0 | 2.4 | 2.4 |
| Total size [PB] | 0.2 | 0.6 | 2.0 | 4.0 | 6.4 | 8.7 |
| **DST** (per one version) | | | | | | |
| size per year [PB] | 0.04 | 0.1 | 0.3 | 0.5 | 0.5 | 0.5 |
| Total size [PB] | 0.04 | 0.1 | 0.5 | 0.9 | 1.4 | 2.0 |
| CPU [kHepSPEC] | 5 | 10 | 38 | 52 | 62 | 62 |

*Table 14.9: Expected amount of storage and CPU power for MC production.*

| Fiscal year (Apr.1-Mar.31) | 2014 | 2015 | 2016 | 2017 | 2018 | 2019 |
|---|---|---|---|---|---|---|
| **MC data** | | | | | | |
| size per year [PB] | 0.9 | 1.8 | 7.0 | 9.6 | 11.4 | 11.4 |
| Total size [PB] | 0.9 | 2.6 | 9.6 | 19 | 31 | 42 |
| CPU [kHepSPEC] | 30 | 60 | 240 | 330 | 390 | 390 |

*Table 14.10: Expected amount of storage and CPU power for user analysis.*

| Fiscal year (Apr.1-Mar.31) | 2014 | 2015 | 2016 | 2017 | 2018 | 2019 |
|---|---|---|---|---|---|---|
| **user area** | | | | | | |
| size per year [PB] | 0.1 | 0.2 | 0.7 | 1.0 | 1.2 | 1.2 |
| Total size [PB] | 0.1 | 0.3 | 1.0 | 2.0 | 3.2 | 4.4 |
| CPU [kHepSPEC] | 2.4 | 7.1 | 26 | 52 | 83 | 114 |

#### 14.5.2.3 Resource requirements for KEK

Here, a more concrete estimation of the required resources for KEK not only as a main computer facility but also as a grid site is presented.

The role of the main computer facility is the processing of the raw data, the archiving of the produced mDST and DST data as well as the raw data, and the migration of the mDST data to remote grid sites. Because the raw data is not processed so frequently except in the early stage of the experiment, we assume that the raw data is stored on tape. On the other hand, mDST is heavily accessed by users. Therefore, it is useful to keep the mDST data on disk.
When the reconstruction software is upgraded, e.g., upon implementation of better tracking and clustering algorithms, or we find that there is a serious bug in the code, we need to reprocess the raw data and iterate the physics skimming process. In that case, we have to store the new reprocessed data, while keeping the old one. Therefore, we set the number of mDST versions at should be kept at 2.
The network bandwidth is estimated such that the mDST data can be transfered to four remote sites in parallel to its production.
KEK also plays a role as a grid site and has a responsibility to produce and to keep a certain fraction of MC events, which is proportional to the number of assigned people to the KEK grid site. We shall assume 25% for this fraction in this estimation. We assume that the MC production will take place after DST production, because we need information of beam energy, IP profile, and number of $B\bar{B}$ events as inputs for MC production to reproduce the experimental condition as much as possible. In that sense, we can effectively utilize the CPU power that is assigned to DST production. This overlap reduces the need for additional CPU resources. In this estimation, we estimated the CPU resources that are required for DST production and those for MC production, first. After comparing these two requirements, the larger resource value is adopted to be a required CPU resources for KEK.
Another task of the grid facility is to transfer copies of mDST datasets from KEK and keep them on disk. However, for KEK, because all mDST datasets are kept as a concequence of the DST production and no transfer is needed, we do not consider the network bandwidth in terms of MC production.
Table 14.11 shows the resource requirements for KEK.

*Table 14.11: Resource requirements for KEK.*

| Fiscal year (Apr.1-Mar.31) | 2014 | 2015 | 2016 | 2017 | 2018 | 2019 |
|---|---|---|---|---|---|---|
| Total tape size [PB] | 4 | 12 | 45 | 90 | 143 | 196 |
| Total disk size [PB] | 0.9 | 2.6 | 9.5 | 19 | 30 | 41 |
| CPU [kHepSPEC] for DST prod. | 5 | 10 | 38 | 52 | 62 | 62 |
| CPU [kHepSPEC] for MC prod. + User Analysis | 8 | 17 | 67 | 96 | 118 | 126 |
| Adopted CPU [kHepSPEC] | 8 | 17 | 67 | 96 | 118 | 126 |
| WAN (outward from KEK)[Gbit/s] | 0.5 | 0.9 | 3.7 | 5.1 | 6.0 | 6.0 |
| Power (CPU) [kW] | 31 | 64 | 255 | 366 | 453 | 483 |
| Power (Disk) [kW] | 10 | 29 | 107 | 214 | 340 | 466 |

#### 14.5.2.4 Resource requirements for a typical remote grid site

The role of the grid computer facility is the production and archiving of MC events. The amount of MC is proportional to the number of analysis users assigned to this facility relative to the total number of the Belle II members. Furthermore, the remote site serves also as an analysis facility for these users. In the exemplary calculation presented here, we assume the fraction of the Belle II members at the typical remote grid site is 15%.
Another important task of the regional computing facility is to keep the copy of the mDST datasets that are transferred from KEK. Thanks to this, we can reduce the load of the network at KEK when users access the mDST frequently during the physics analysis.
The current best estimation is summarized in Table 14.12.

*Table 14.12: Resource requirements for a typical remote grid site.*

| Fiscal year (Apr.1-Mar.31) | 2014 | 2015 | 2016 | 2017 | 2018 | 2019 |
|---|---|---|---|---|---|---|
| Total disk size [PB] | 0.5 | 1.4 | 5.0 | 10 | 16 | 22 |
| CPU [kHepSPEC] | 5 | 10 | 40 | 57 | 71 | 76 |
| WAN (inward from KEK)[Gbit/s] | 0.1 | 0.2 | 0.9 | 1.3 | 1.5 | 1.5 |
| Power (CPU) [kW] | 19 | 39 | 153 | 220 | 272 | 290 |
| Power (Disk) [kW] | 11 | 33 | 121 | 242 | 384 | 527 |

## 14.6 Schedule

To optimize the detector design and to check the sensitivity to the physics of interest, a full detector simulation that works within a common software framework is essential. We also need an event data model of each sub-detector to handle objects produced in the simulator. As written in Sec. 14.4, we plan to have our own software framework so-called roobasf, on which development has been progressing. The implementation of the GEANT4-based detector simulation in this new framework has started. At this moment, a few sub-detectors related to the tracking have decided the outline of their event data model and installed the basic geometry into the simulator. For other sub-detectors, we will finish installing the geometry as soon as possible with help from each sub-detector group.
Regarding the reconstruction tool, we have initiated an activity to develop a new track finding/fitting tool and we aim to release the 0-th version by 2013. In parallel, we will release a benchmark of pattern recognition based on the track finding tool developed for the current Belle. Until the new tracking tool is ready, the benchmark version can be used as a tracking tool for Belle II. For the particle ID tools, the TOP and A-RICH groups have already developed stand-alone reconstruction tools. They will now start to incorporate these tools within the new software framework.
We have finished the design of the distributed Belle II computing system and summarized it in this chapter. For now, we have started several tests for the basic components of the proposed design. The started tests include the prototype of the data handling system utilizing the AMGA metadata catalogue, the construction of the Belle II virtual organization on the grid, a test for the establishment of the job handling, and a challenge for the cloud computing. We would like to finish these tests by the middle of the Japanese fiscal year (JFY) 2011.
We also have to take into consideration when the new computing system will be installed at KEK. Because of the expiry of the existing contract, the Belle computing system will be replaced with a new system at the end of JFY 2011 (February or March in 2012). Though we are now discussing the contract itself, the specification needs to be prepared by the summer of 2010 (Fig. 14.25). In this specification, we have to consider not only the hardware resources required for the Belle II computing system by JFY 2015 or 2016 when the follow-on replacement will take place but also the transfer of the large amount of legacy data stored by the Belle experiment from the current system to the new one. In any case, the detailed schedule of the Belle II computing system will be adjusted to the contract, the bidding process, and the KEK-wide computing system.
Once the new computing system is installed, we would like to start the dress rehearsal of the Belle II computing system immediately. At that stage, we will have at least two year prior to the first collision. Furthermore, we would like to start the systematic MC mass production to test the computing system and also the reconstruction procedure.

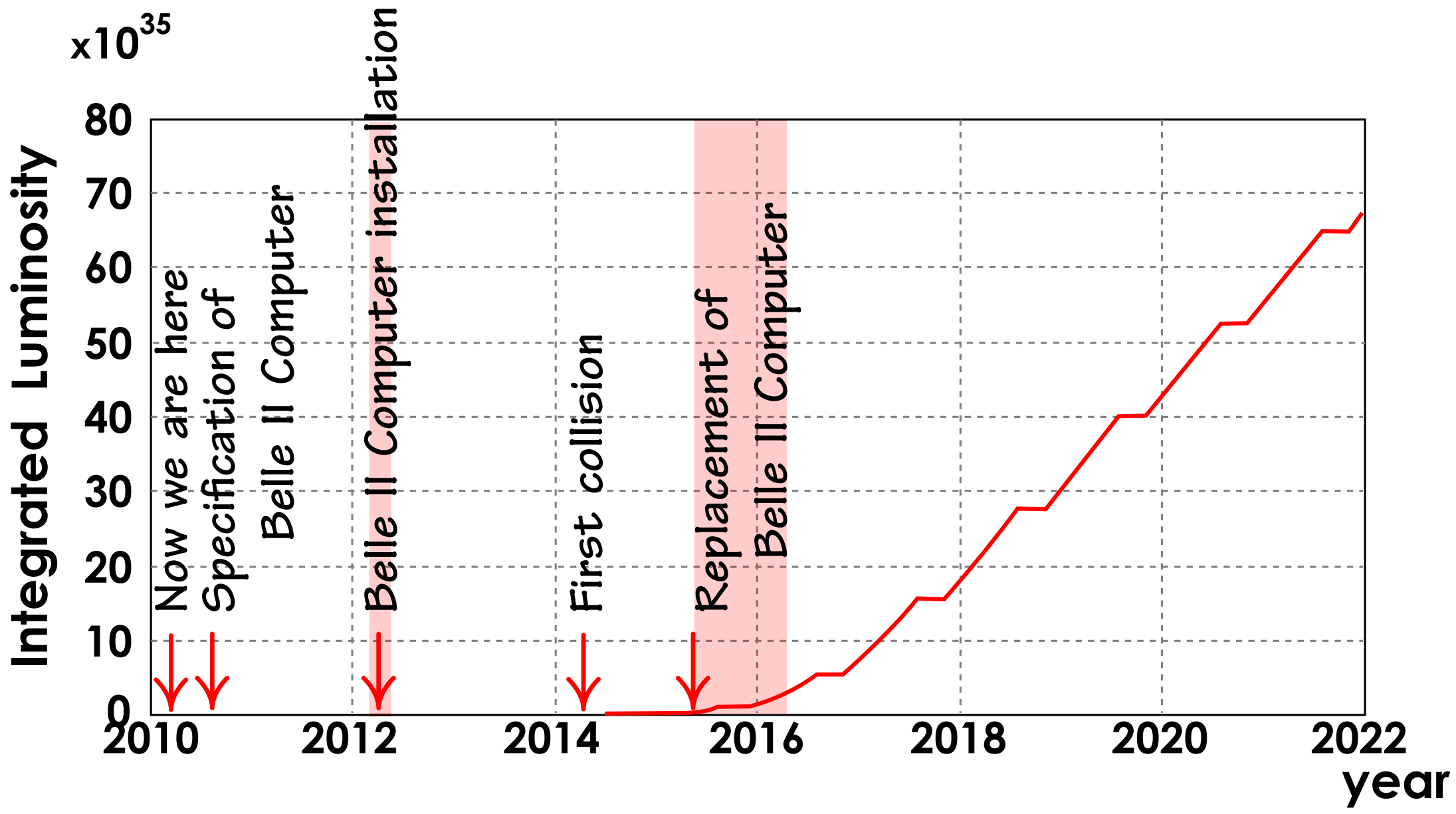

*Figure 14.25: Important milestones for Belle II computing with a luminosity prospect*

# Chapter 15

# Organization

## 15.1 Collaboration members and organization

To pursue the R&D and design of the Belle II detector, a new collaboration has been formed. It consists of 44 institutions as of January 2010. Table 15.1 lists all member institutions (as of January 2010) together with the number of collaborators at each institution; the number of students is given in parentheses. The collaboration now consists of nearly 300 people from 44 institutions in 13 countries.
Open collaboration meetings are held regularly three times per year. By summer 2009, all the administrative bodies of the Belle II Collaboration have been formed. As shown in Fig. 15.1, the two main bodies are the Institutional Board and the Executive Board. The project is led by the Spokesperson and Project Manager.

**The Institutional Board** gathers representatives of all collaborating institutes. This body defines and approves all major policy principles and documents (bylaws, etc.). It appoints a Nominating Committee, responsible for nominating individuals to stand for election as Spokesperson. The Board is also responsible for carrying out the election of the Spokesperson.

**The Executive Board** advises the Spokesperson, appoints the detector subsystem coordinators and may appoint special committees when additional expertise is required. All decisions of the Executive Boards must be ratified by the Institutional Board.

**The Spokesperson** is the scientific representative of the Collaboration and is responsible for all scientific, technical and organizational affairs of the Collaboration. The Spokesperson is elected in a general collaboration-wide election.

**The Project Manager** assists the Spokesperson in keeping the KEK management apprised of all Collaboration affairs and to informing the Collaboration of relevent KEK matters.
Individual aspects of the project are coordinated by the Physics Coordinator(s), Technical Coordinator and the Software and Computing coordinator(s), reporting to the Spokesperson.

## 15.2 Cost and schedule

The current estimates of the cost of detector components are in given in Table 15.2. The construction schedule is shown in Fig. 15.2.

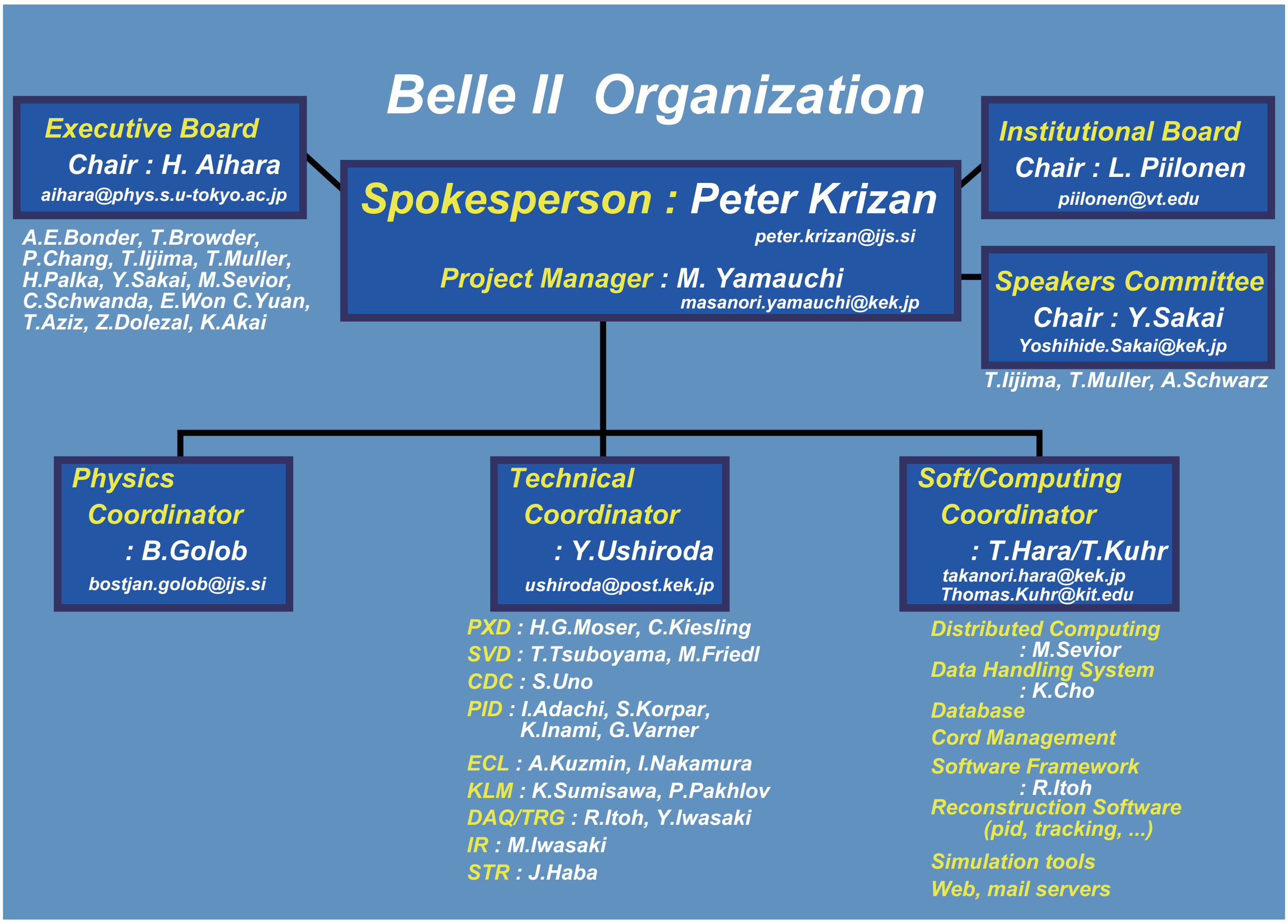

Figure 15.1: Belle II Organization

*Table 15.1: Institutions in the Belle II Collaboration (October 2010)*

| **Country** | **Persons** | **Institute** | **Persons** |
|---|---|---|---|
| Australia | 8 | University of Sydney | 4 |
| | | University of Melbourne | 4(2) |
| Austria | 11 | Austrian Academy of Sciences (HEPHY), Vienna | 11(3) |
| China | 17 | Institute of High Energy Physics, Chinese Acad. Sci. Beijing | 10 |
| | | University of Science and Technology of China | 7(5) |
| Czech Rep. | 6 | Charles University in Prague | 6(2) |
| Germany | 53 | University of Bonn | 9(4) |
| | | University of Göttingen | 2(1) |
| | | University of Heidelberg | 4(2) |
| | | Karlsruhe Institute of Technology | 15(4) |
| | | Ludwig Maximilians Univ. and Excell. Cl. Universe, Munich | 2(0) |
| | | Technical University and Excellence Cl. Universe, Munich | 3(1) |
| | | Max-Planck-Institut für Physik, Munich | 18(7) |
| India | 10 | Indian Institute of Technology, Guwahati | 2 |
| | | Indian Institute of Technology, Madras | 2 |
| | | Institute of Mathematical Sciences (Chennai) | 1 |
| | | Panjab University | 3(1) |
| | | Tata Institute of Fundamental Research | 2 |
| Korea | 34 | Gyeongsang National University | 1 |
| | | Hanyang University | 6(4) |
| | | Korea Institute of Science and Technology Information | 3 |
| | | Korea University | 6(2) |
| | | Kyungpook National University | 7(3) |
| | | Seoul National University | 6(2) |
| | | Yonsei University | 3 |
| Poland | 12 | Henryk Niewodniczanski Inst. of Nucl. Phys. PAN, Cracow | 12(1) |
| Russia | 31 | Budker Institute of Nuclear Physics, Novosibirsk | 16(3) |
| | | Institute for High Energy Physics, Protvino | 4(1) |
| | | Institute for Theoretical Experimental Physics, Moscow | 11(1) |
| Slovenia | 13 | Jozef Stefan Institute, Ljubljana | 6(1) |
| | | University of Ljubljana | 5(1) |
| | | University of Maribor | 1 |
| | | University of Nova Gorica | 1 |
| Taiwan | 21 | Fu Jen Catholic University | 4 |
| | | National Central University | 2 |
| | | National United Univ | 2 |
| | | National Taiwan University | 13(2) |
| U.S.A. | 28 | Pacific Northwest National Laboratory | 4 |
| | | University of Cincinnati | 6 |
| | | University of Hawaii | 11(3) |
| | | Virginia Polytechnic Institute and State Univ. | 5(2) |
| | | Wayne State University | 2(1) |
| Japan | 102 | Nagoya University | 13(4) |
| | | Nara University of Education | 1 |
| | | Nara Women's University | 9 |
| | | Niigata University | 4(2) |
| | | Osaka City University | 4 |
| | | Toho University | 2 |
| | | Tohoku University | 15(3) |
| | | Tokyo Metropolitan University | 4(2) |
| | | University of Tokyo | 4(2) |
| | | KEK | 46 |
| Total | 346 | 52 institutes from 13 countries | 346 |

*Table 15.2: Estimated cost of detector components (1 Oku yen=$10^8$yen $\approx$ 1 M$).*

| Component | Estimated cost (Oku yen) |
|---|---:|
| Beam pipe | 1.0 |
| PXD | 3.5 |
| SVD | 3.0 |
| CDC | 3.8 |
| B-PID | 7-8 |
| E-PID | 4-5 |
| ECL (no crystals) | 3.5 |
| KLM | 1.4 |
| TRG | 0.9 |
| DAQ | 4.3 |
| Structure | 4.5 |
| Total | 37.9 |

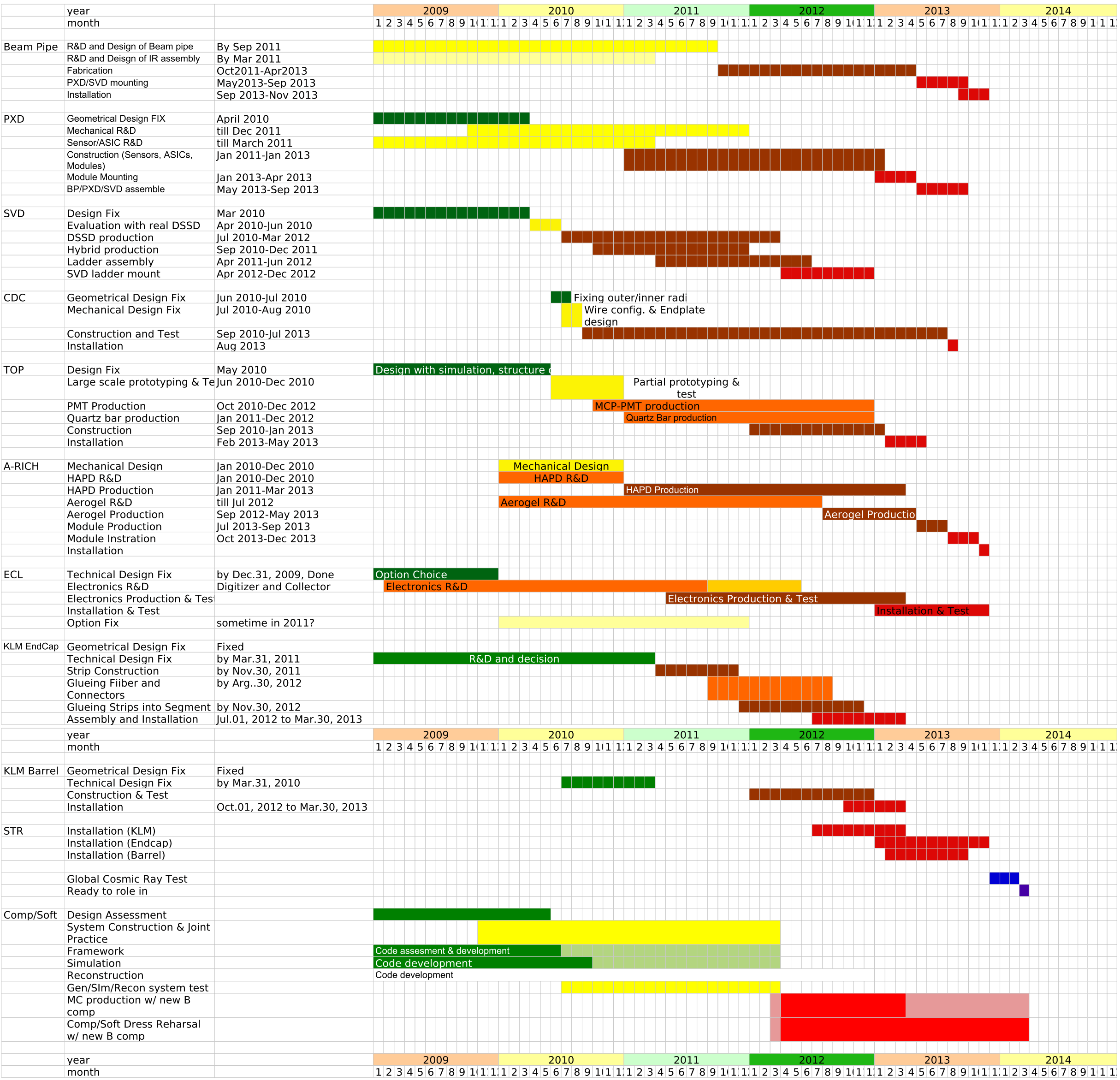

Figure 15.2: Belle II Construction schedule